SLAC-R-709
December, 2004

# The Discovery Potential of a Super B Factory

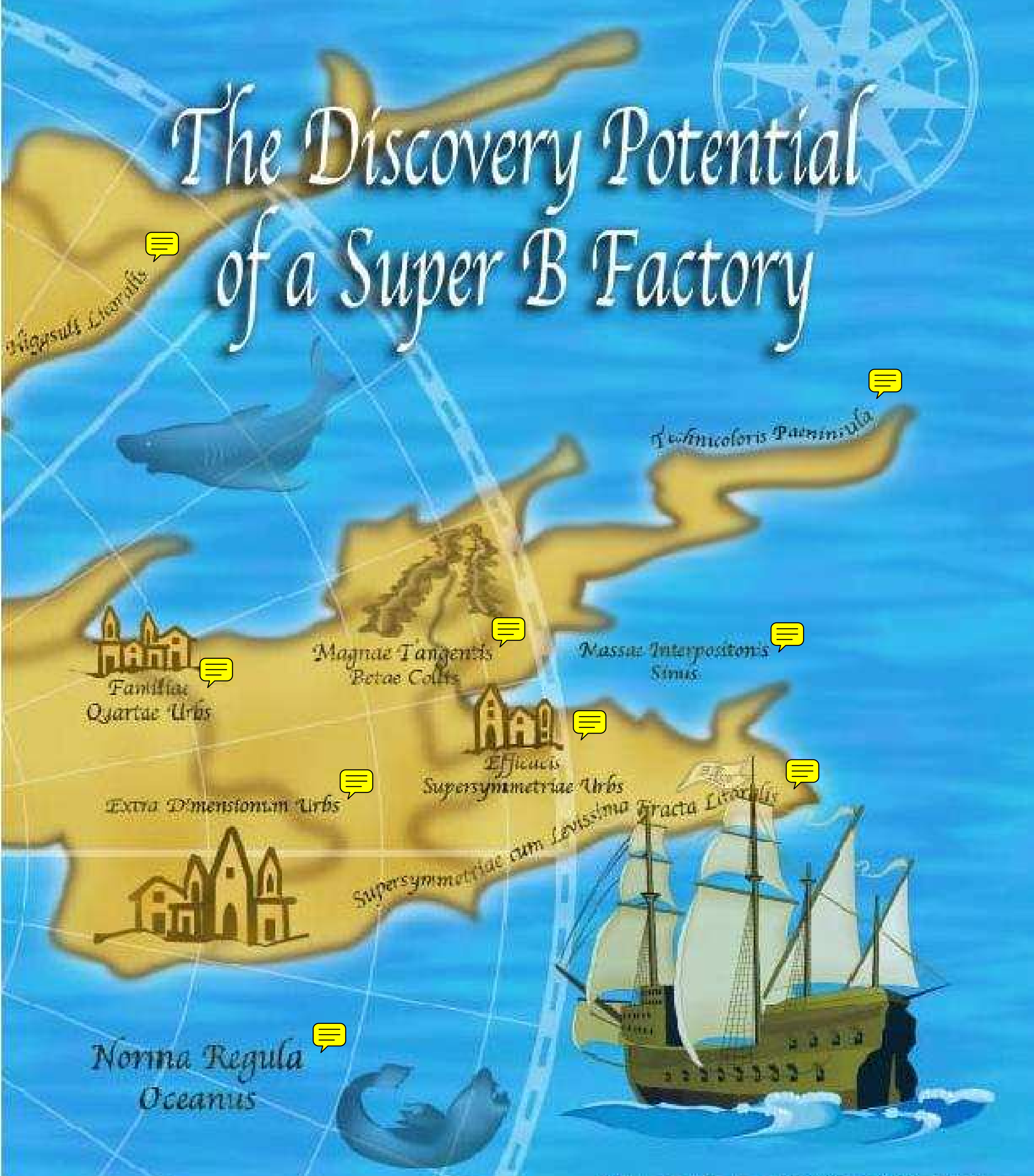

JoAnne L. Hewett and David G. Hitlin, Editors

# The Discovery Potential of a Super $B$ Factory

## Proceedings of the 2003 SLAC Workshops

**Editors:**

**JoAnne Hewett**
*SLAC*

**David G. Hitlin**
*California Institute of Technology*

**Conveners:**

**Yuval Grossman**
*Technion*

**JoAnne Hewett**
*SLAC*

**Gudrun Hiller**[a]
*University of Munich*

**Tobias Hurth**
*CERN*

**Urs Langenegger**[b]
*University of Heidelberg*

**Zoltan Ligeti**
*Lawrence Berkeley National Laboratory*

**Yasuhiro Okada**
*KEK*

**Aaron Roodman**
*SLAC*

**Anders Ryd**[c]
*California Institute of Technology*

**Abner Soffer**
*Colorado State University*

**Iain Stewart**
*Massachusetts Institute of Technology*

[a] Present address: CERN
[b] Present address: ETH, Zurich
[c] Present address: Cornell University

# Contents

# 1

# Introduction

## 1.1 Overview

These Workshop Proceedings present a comprehensive exploration of the potential of a Super $B$ Factory, an asymmetric $e^+e^-$ $B$ Factory capable of producing an integrated luminosity of 10 ab$^{-1}$/year, to explore flavor physics beyond the Standard Model. In the next decade, we expect that the Large Hadron Collider at CERN and, perhaps, the International Linear Collider, will open the door to new phenomena that will fundamentally change our understanding of elementary particle physics. A radical shift in what is considered an important problem, similar that which took place after the "November Revolution" of 1974, is likely to result. The study of heavy flavor physics is today, by consensus, an important problem, quite central to the HEP program. The question we attempt to answer herein is whether, in the advent new world post 2010, there is a role for the continued study of heavy flavor physics, *i.e.*, the study of the decays of heavy quarks and leptons. Will heavy flavor physics still be central to the core concerns of the field? Many believe that there is a clear affirmative answer to this question; hence the title of the Proceedings: *The Discovery Potential of a Super $B$ Factory.*

The asymmetric $e^+e^-$ $B$ Factories PEP-II and KEK-$B$, and their associated experiments *BABAR* and Belle, have been in operation since 1999. Both of these enterprises, building on the foundation of results from ARGUS, CLEO, and the LEP experiments, as well as from experiments at hadron accelerators, have been remarkably successful, both technically and scientifically.

The current $B$ Factories' design peak luminosities ($3 \times 10^{33}\,\mathrm{cm}^{-2}\mathrm{s}^{-1}$ for PEP-II and $1 \times 10^{34}\,\mathrm{cm}^{-2}\mathrm{s}^{-1}$ for KEK-$B$) were very ambitious, and it is fair to say, were regarded with some skepticism in certain quarters. PEP-II, however, reached design luminosity in a remarkably short time, and has now exceeded its design performance by a factor of three. KEK-$B$, with a more ambitious design objective, has also exceeded its design performance, and currently operates at even higher luminosity. These accelerators and experiments also operate at unprecedentedly high efficiency, with yearly integrated luminosity totals for a given peak instantaneous luminosity that are forty percent higher than was anticipated on the basis of previous experience (see below).

The scientific productivity of PEP-II/*BABAR* and KEK-$B$/Belle has been no less remarkable, with wide-ranging pioneering studies of $CP$ violation in the $B$ meson system that have, for the first time, demonstrated that the $CP$-violating phase of the three generation Standard Model is capable of explaining all $CP$-violating phenomena thus far observed in the $K$ and $B$ meson systems. This new triumph of the Standard Model is, however, bittersweet. It reemphasizes the power of the elegant and economical Standard Model *ansatz*, but it leaves the crucial question of the origin of the matter-antimatter asymmetry of the universe unanswered. This presents an opportunity for a fruitful dialog in the next decade between studies at the LHC and ILC and those at a Super $B$ Factory.

*BABAR* and Belle have each published more than one hundred papers in refereed journals, covering ground-breaking $CP$ violation measurements, studies of rare decay phenomena and high precision measurements in $B$ and $D$ meson and $\tau$ lepton decays. Their productivity continues unabated; the next few years will certainly bring a host of beautiful new results, and, perhaps, even a few surprises. There are already hints of results that disagree with the Standard Model in areas where one might expect measurable New Physics effects, although none of these are as yet of adequate statistical significance.

Current plans call for the $B$ Factory programs to run through most of this decade. With anticipated increases in peak luminosity performance, this will provide an increase in the size of total data samples, now each of order 250-300 fb$^{-1}$, to 700–1000 fb$^{-1}$. That is the limit of what can be achieved by incremental upgrades to the PEP-II and KEK-$B$ colliders, as the existing storage rings will have reached the maximum circulating currents they can sustain, and further improvements in lattice optics will not be feasible. The sensitivity required for a meaningful exploration of physics beyond the Standard Model requires much larger data samples; hence the target for the Super $B$ Factory of 10 ab$^{-1}$, *i.e.*, 10,000 fb$^{-1}$, per year.

After five years' experience, we have an excellent understanding of the actual physics performance of *BABAR* and Belle, and their ability to elucidate the full range of $CP$-violating effects in $B$ decays, searching for New Physics in rare decays, and making precision measurements of CKM parameters. We have also learned a great deal from PEP-II and KEK-$B$ operation at these high luminosities, including a quite detailed understanding of the (differing) backgrounds in the two experiments. This experience gives us a solid basis for contemplating significant upgrades to these colliders and experiments, which would open up exciting new scientific opportunities.

A Super $B$ Factory, an asymmetric $e^+e^-$ collider with a luminosity of the order of $7{\times}10^{35}$ cm$^{-2}$s$^{-1}$, would be a uniquely sensitive probe of the flavor couplings of New Physics beyond the Standard Model. A series of workshops at KEK and SLAC over the last few years have explored in detail the physics case for a Super $B$ Factory that could provide data samples nearly two orders of magnitude larger than those currently projected by *BABAR* or Belle, as well as issues of collider and detector design. These Proceedings summarize two workshops on physics issues, held at SLAC in May and October, 2003 [1]. Hence they focus on an exploration of the physics landscape. Technical questions are, of course, under active study, and have been the subject of other workshops [2], [3] in which both physics and technical matters have been explored in some detail [4] [5].

The potential of a Super $B$ Factory to explore the effects of New Physics in the flavor sector encompasses two somewhat different strategies:

- measuring branching fractions, $CP$-violating asymmetries, and other detailed kinematic distributions in very rare $B$, $D$, and $\tau$ decays in which there are clear potential signatures of New Physics, and
- pushing the most precise predictions of the Standard Model to their limits, by measuring the sides and angles of the unitarity triangle to the ultimate precision warranted by theoretical uncertainties, in hopes of unearthing a discrepancy with theory.

The primary objective of a Super $B$ Factory is to produce the very large data samples that will allow us to explore very rare $B$, $D$, and $\tau$ decays, at a sensitivity in which New Physics effects are likely to manifest themselves through higher order (loop) Feynman diagrams. A large variety of phenomena can be affected by New Physics. In some cases, the Standard Model predicts that $CP$ asymmetries in different $B$ decay modes are identical, whereas particular New Physics schemes predict that these asymmetries can differ by tens of percent. The pattern of departure from equality is characteristic of particular models. Certain very rare decays are predicted to be either absent or very small in the Standard Model, but can be enhanced by New Physics. In other cases, kinematic distributions can be substantially modified from those accurately predicted by the Standard Model. Thus access to the study of very rare decays may show the effects of New Physics through loop diagrams, and can be crucial in clarifying the nature of the New Physics in the flavor sector. In these Proceedings, we will, for definiteness, use SUSY and extra dimensions as examples of New Physics, although other proposed Beyond-the-Standard Model physics can also show up in heavy quark and heavy lepton decays.

Measurements of unitarity triangle-related quantities can be improved quite substantially before reaching the expected limiting precision of lattice QCD calculations. The precision of $\sin 2\beta$ measurements has now reached 5%; other measurements related to the unitarity triangle construction are more difficult and are, consequently, less precise. Most such measurements are very far from being statistics-limited, and are not yet approaching the limits of theory. A few measurements, such as the extraction of the absolute values of CKM matrix elements from semileptonic $B$ decays, will reach the practical limit of theoretical precision either before, or early in, the Super $B$ Factory era. Making these precision measurements is an important objective that may well yield clues to physics beyond the Standard Model. A primary objective of the workshops was therefore to probe the limits of Standard Model theoretical predictions, as well as the statistical and systematic constraints on experimental measurements.

## 1.2 Measurement Capabilites of a Super $B$ Factory

The physics opportunities available with data samples of 10 to 50 ab$^{-1}$ are the subject of these Proceedings. With our conventional "Snowmass Year" constant (1 year=$10^7$ seconds), it takes a $10^{36}$cm$^{-2}$s$^{-1}$ machine to generate 10 ab$^{-1}$/year. This has been the working assumption in Super $B$ Factory studies dating back to 2001 [4]. The Snowmass Year of $10^7$ seconds was adopted as a standard for $e^+e^-$ comparisons at the Snowmass meeting of 1988, based on then-current CESR/CLEO performance. This constant was meant to account for the difference in peak and average luminosity, the dead time of the experiment, time lost to accelerator and detector breakdowns, *etc.*. The PEP-II/*BABAR* and KEK-$B$/Belle complexes however, have quite substantially improved on previous performance. The dominant effect is the introduction of trickle injection, now used by both PEP-II and KEK-$B$, which allows continuous integration of data at peak luminosity. Trickle injection has the added benefit that, since the current in the collider is quite constant, the machine temperature is more stable, producing noticeable improvements in stability of operation. The experiments are also very efficient (*BABAR* is more than 97% efficient), so that a greater fraction of machine luminosity is recorded. Figure 1-1) shows a years worth of recent operation of PEP-II. Taking these improvements into account, a more appropriate Snowmass Year constant is $1.4 \times 10^7$ seconds/year. This means that it is possible to produce 10 ab$^{-1}$/year with an instantaneous luminosity of $7 \times 10^{35}$cm$^{-2}$s$^{-1}$, which is therefore the current design goal for SuperPEP-II. Hence, in the physics reach tables, we have tabulated the precision on measured quantities at the 3, 10, and 50 ab$^{-1}$ levels to explore a lower range of peak luminosity, a typical one year sample at the nominal upgrade level, and an asymptotic sample.

The tables also include 1-year sensitivities for the hadron experiments LHC$b$ and $B$TeV, where these estimates are available. It should be noted that the experimental sensitivities are based on a conventional 1988 Snowmass Year; there is no reason to adjust the Snowmass Year constant for the hadron experiments. We have not included estimates of the capabilities of CMS and ATLAS in the tables; these experiments have similar sensitivity to LHC$b$ and $B$TeV in some cases, less in others.

Several comments are in order. The hadron experiments can measure $\alpha$, $\beta$ and $\gamma$ in the standard unitarity triangle modes, generally with a one-year sensitivity somewhat less than that of a 3 ab$^{-1}$ sample. They cannot make measurements of the unitarity triangle sides, as these require absolute measurements of semileptonic or purely leptonic branching fractions, which are difficult in a hadronic collider environment. The hadron experiments can make measurements in the $B_s$ system, which, the $e^+e^-$ experiments running at the $\Upsilon(4S)$ cannot.

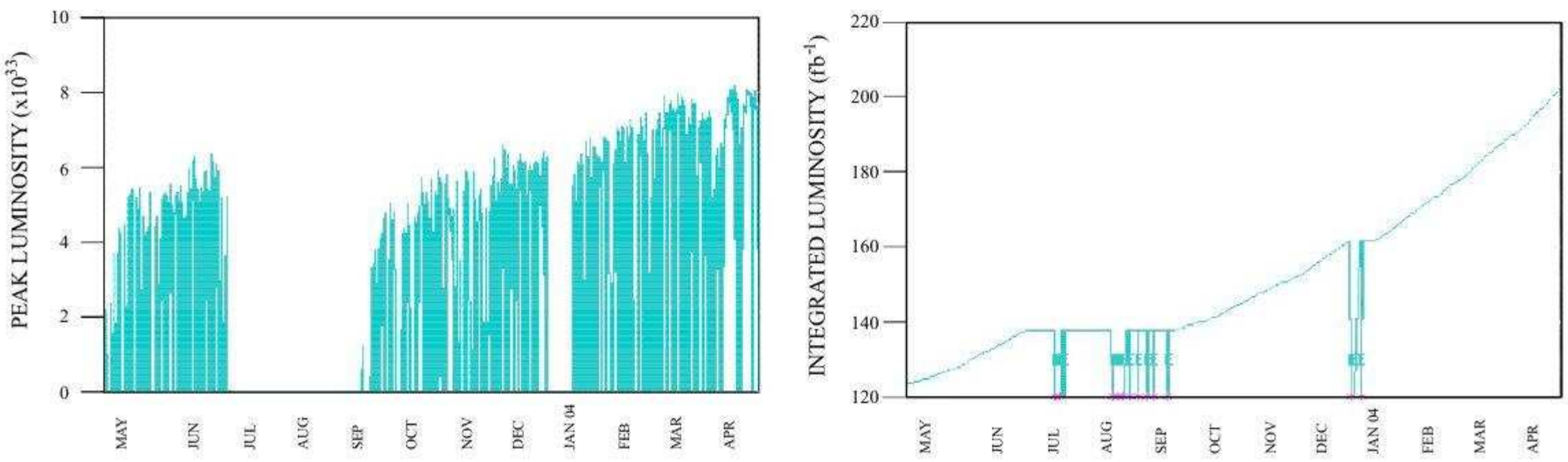

**Figure 1-1.** *Peak luminosity (left) and integrated luminosity (right) of PEP-II for the period 23 April, 2003 to 23 April, 2004. Comparison of the peak and integrated luminosities for this one-year period leads to the conclusion that the "Snowmass Year" constant for PEP-II and BABAR is* $1.4 \times 10^7$*, a forty percent improvement over the the classical Snowmass Year.*

### 1.2.1 New Physics

The major motivation for studying very large samples of $B$ or $\tau$ decays is the discovery potential for New Physics. In this brief discussion, we will use supersymmetry as a shorthand for all types of physics beyond the Standard Model. Other Standard Model extensions, such as extra dimensions or left-right symmetric models, can produce a similar range of effects, but the language of supersymmetry is convenient for developing an understanding of where New Physics effects are likely to be measurable.

Figure 1-2 is based on a paper by Ciuchini, *et al.* [6], a SPIRES TOPCITE50 selection, that estimates the size of loop contributions to $CP$ asymmetries in a model-independent mass-insertion calculation valid for any low energy SUSY extension of the Standard Model. The mass insertion can, in principle, connect any two generations, but there are already substantial constraints on these couplings. The effect of the second to third (23) generation coupling is potentially the largest by far, followed by the first to third (13) generation coupling. The left graph shows the minimum and maximum difference of the $CP$ asymmetries in $\pi^0 K^0_S$ and $J/\psi K^0_S$, as a function of the 13 insertion mass; the right graph shows the minimum and maximum difference of the $CP$ asymmetries in $\phi K^0_S$ and $J/\psi K^0_S$, as a function of the 23 insertion mass. This calculation tells us two important things. First, it sets the scale of measurement sensitivity for $CP$ asymmetries needed to reveal New Physics effects, and second, it shows that the most interesting effects are likely to appear in the 23 coupling, *i.e.*, in $b \to s$ transitions. It is the congruence of the latter point and the *BABAR* and Belle measurements of $CP$ asymetries in $\phi K^0_S$ and related decay modes that have attracted so much recent attention (34 theory papers in SPIRES since 2000).

This sensitivity to high-mass insertions depends on the precision of the measurement of the appropriate $CP$ asymmetry and on the precision of our knowledge of the expected asymmetry within the Standard Model. If we take $\phi K^0_S$ as an example, a 5% measurement of the difference of the $CP$ asymmetry from that in $J/\psi K^0_S$ indicates a 23 mass in the range of 800 GeV. The current conservative limit set by data on the effects of rescattering on the the $\phi K^0_S$ asymmetry is $\sim 30\%$, but this is expected to be reduced to $\sim 5\%$ with large data samples. Thus a precise measurement of the asymmetry provides a window on interesting SUSY mass scales. The $\pi^0 K^0_S$ asymmetry is likely to have similar Standard Model theory uncertainties. A more precise asymmetry measurement is required, however, as SUSY effects are smaller: a 5% measurement of the asymmetry difference would indicate a 13 mass scale in the range of 300 GeV, which is still quite relevant.

LHC$b$ and $B$TeV should also be sensitive to the effects of a 23 mass insertion in $B_s$ decays. However, sensitivity to such effects in $B_s$ decays, except in one instance, offers no particular advantage with respect to a Super $B$ Factory, and many disadvantages. The process with an advantage is the second-order transition involved in $B_s\overline{B_s}$ mixing, which directly involves the $b \to s$ coupling and, can therefore be affected by a 23 mass insertion. This question will likely be addressed by CDF and DO before the new generation of hadronic experiments comes online. LHC$b$ and $B$TeV should be in a good position to extend these measurements, and to make precision measurements of heavy-light lifetime differences, decay polarizations, $CP$ asymmetries. Beyond the mixing process, most other $B_s$ decays are primarily spectator decays; the presence of an $s$ quark in the parent is irrelevant. Transitions such as $b \to s\gamma$ and $b \to s\ell^+\ell^-$ are in general more easily studied in detail at an asymmetric $B$ Factory.

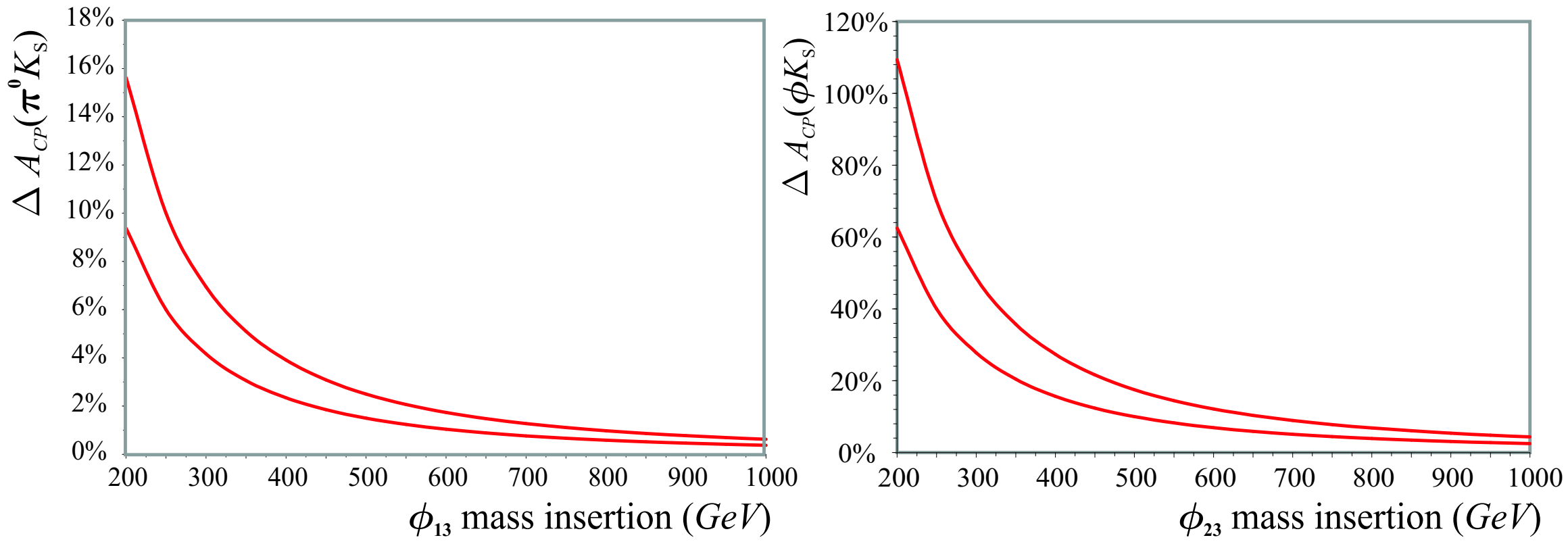

**Figure 1-2.** *Model-independent estimate by Ciuchini, et al. of the difference in $A_{CP}$ between (left) $B^0 \to J/\psi K^0_S$ and $B^0 \to \pi^0 K^0_S$, governed by a 13 mass insertion, and (right) $B^0 \to J/\psi K^0_S$ and $\phi K^0_S$, governed by a 23 mass insertion, as a function of the mass insertion scale. The upper and lower curves represent the largest and smallest effect expected.*

### $CP$ Violation in Rare Decays

The centerpiece of the search for New Physics is likely to be the study of $CP$-violating asymmetries in rare $B$ decays in which penguin amplitudes play a prominent role, such as $\phi K^0_S(K^0_L)$, $\eta' K^0_S$ or $\pi^0 K^0_S$. Table 1-1 shows the measurement precision for a variety of rare decays. With this precision, how much integrated luminosity is required to clearly demonstrate a $CP$ asymmetry different from that in $J/\psi K^0_S$ and how does the capability of an asymmetric $B$ Factory compare to that of an experiment at a hadron accelerator?

Consider a 20% difference in $A_{CP}(J/\psi K^0_S) - A_{CP}(\phi K^0_S)$, which corresponds to 23 insertion in the mass range $\sim 350 - 450$ GeV. If we are to establish this 20% difference at the $5\sigma$ level, *i.e.* that $A_{CP}(\phi K^0_S) = 0.60 \pm 0.03$, we need, at the current per event sensitivity, 30 ab$^{-1}$. In other words, we would have a first indication of an effect in a year or so of running, and would clearly establish the effect in about three years.

The radiative penguin decays $b \to s\gamma$ provide a particularly clean environment for searching for New Physics. Direct $CP$ violation in these decays is expected to be $\approx 0.5\%$ in the Standard Model, but could be an order of magnitude larger if there are New Physics contributions to the penguin loop. Recent inclusive and exclusive measurements are just beginning to constrain such contributions. These measurements are statistics limited, and will continue to be so until at least 10 ab$^{-1}$. With larger samples it would be interesting to measure the direct $CP$ asymmetry in $b \to d\gamma$ decays, where the Standard Model prediction is -12%. *BABAR* has also shown that it is feasible to measure time-dependent $CP$ violation in $B^0 \to K^{*0}(\to K^0_S\pi^0)\gamma$, where the sine term is related to the helicity of the photon. In the Standard Model the sine term is suppressed by $m_s/m_b$ compared to $\sin 2\beta$. This measurement, which is sensitive to New Physics couplings with the opposite helicity, will continue to be statistics limited up to 50 ab$^{-1}$. An alternative method of studying the photon polarization in $b \to s\gamma$ is the Dalitz plot distribution of the $K\pi\pi$ system in $B^0 \to K\pi\pi\gamma$, which also requires a large statistics sample.

**Table 1-1.** *Measurement precision for $CP$ asymmetries in rare decays sensitive to New Physics. The current BABAR central values are assumed when measurements exist.*

| **$CPV$ in Rare $B$ Decays** | | **$e^+e^-$ Precision** | | | **1 Yr Precision** | |
|---|---|---|---|---|---|---|
| **Measurement** | **Goal** | **3/ab** | **10/ab** | **50/ab** | **LHC$b$** | **$B$TeV** |
| $S(B^0 \to \phi K^0_S)$ | $\approx 5\%$ | 16% | 8.7% | 3.9% | 56% | 22% |
| $S(B^0 \to \eta' K^0_S)$ | $\approx 5\%$ | 5.7% | 3% | 1% | - | - |
| $S(B^0 \to K^0_S\pi^0)$ | | 8.2% | 5% | 4% | - | - |
| $S(B^0 \to K^0_S\pi^0\gamma)$ | SM: $\approx 2\%$ | 11% | 6% | 4% | - | - |
| $A_{CP}$ $(b \to s\gamma)$ | SM: $\approx 0.5\%$ | 1.0% | 0.5% | 0.5% | - | - |
| $A_{CP}$ $(B \to K^*\gamma)$ | SM: $\approx 0.5\%$ | 0.6% | 0.3% | 0.3% | - | - |
| $CPV$ in mixing ($\lvert q/p \rvert$) | | $< 0.6\%$ | | | - | - |

### Rare Decay Branching Fractions

Many rare $B$ decay modes can potentially give access to physics beyond the Standard Model via measurements other than of $CP$-violating asymmetries. Some examples of these modes are listed in Table 1-2. Typically, these decays do not occur at tree level and consequently the rates are strongly suppressed in the Standard Model. Substantial enhancements in the rates and/or variations in angular distributions of final state particles could result from the presence of new heavy particles in loop diagrams, resulting in clear evidence of New Physics. Moreover, because the pattern of observable effects in highly model-dependent, measurements of several rare decay modes can provide information regarding the source of the New Physics.

**Table 1-2.** *Measurement precision for rare decays sensitive to New Physics.*

| **Rare $B$ and $\tau$ Decays** | | $e^+e^-$ **Precision** | | | **1 Yr Precision** | |
|---|---|---|---|---|---|---|
| **Measurement** | **Goal** | **3/ab** | **10/ab** | **50/ab** | **LHC$b$** | **$B$TeV** |
| $\lvert V_{td}\rvert/\lvert V_{ts}\rvert \sim \sqrt{\frac{\mathcal{B}(b\to d\gamma)}{\mathcal{B}(b\to s\gamma)}}$ | | 19% | 12% | 5% | - | - |
| $\mathcal{B}(B \to D^*\tau\nu)$ | $\mathcal{B} = 8 \times 10^{-3}$ | 10% | 5.6% | 2.5% | - | - |
| $\mathcal{B}(B \to s\nu\overline{\nu})$ $(K^{-,0}, K^{*-,0})$ | 1 exclusive: $\sim 4 \times 10^{-6}$ | $\sim 1\sigma$ (per mode) | $> 2\sigma$ (per mode) | $> 4\sigma$ (per mode) | - | - |
| $\mathcal{B}(B_d \to \text{invisible})$ | | $< 2 \times 10^{-6}$ | $< 1 \times 10^{-6}$ | $< 4 \times 10^{-7}$ | - | - |
| $\mathcal{B}(B_d \to \mu\mu)$ | $\sim 8 \times 10^{-11}$ | $< 3 \times 10^{-8}$ | $< 1.6 \times 10^{-8}$ | $< 7 \times 10^{-9}$ | 1-2 ev | 1-2 ev |
| $\mathcal{B}(B_d \to \tau\tau)$ | $\sim 1 \times 10^{-8}$ | $< 10^{-3}$ | $\mathcal{O}(10^{-4})$ | | - | - |
| $\mathcal{B}(\tau \to \mu\gamma)$ | | | $< 10^{-8}$ | | - | - |

The ratio of the branching fractions of $b \to d\gamma$ and $b \to s\gamma$ decays is directly related to the ratio of CKM matrix elements $V_{td}/V_{ts}$. It is interesting to measure $V_{td}/V_{ts}$ in penguin processes as well as through $B_d/B_s$ mixing, since New Physics enters these amplitudes in different ways. The ratio of the exclusive decays $B \to \rho\gamma$ and $B \to K^*\gamma$ can be accurately measured, but the precision of the determination of $V_{td}/V_{ts}$ is limited by theoretical uncertainties of $\approx 12\%$ in the ratio of the form factors. A measurement of the ratio of the inclusive decays does not suffer from this uncertainty, but is experimentally rather challenging, and requires a large data sample.

Searches for $B \to s\nu\overline{\nu}$, either inclusively or exclusively, are extremely difficult due to the presence of the two final state neutrinos. The required sensitivity can, however, be obtained using the recoil method, in which a signal mode (in this case the exclusive $B \to K\nu\overline{\nu}$ and $K^*\nu\overline{\nu}$ modes) is sought in the recoil against a fully reconstructed hadronic $B$ decay. Assuming Standard Model branching fractions, extrapolation of current analyses suggest that we would expect a signal of 10 events in each of the four modes $(K^{0,-}, K^{0,*-})$, although with a substantial background, with 3 ab$^{-1}$ of data. A statistically significant signal would emerge in the combination of modes with approximately 10 ab$^{-1}$ even using a simple cut-and-count analysis.

The decays $B_d \to \ell\ell$ ($\ell = e, \mu, \tau$) are somewhat less promising, in the sense that it appears impossible to reach the predicted Standard Model branching fractions even with more than 50 ab$^{-1}$ of data. Moreover, $B_d \to \mu\mu$ is expected to be accessible at both LHC$b$ and $B$TeV, and these experiments will also be able to access $B_s \to \mu\mu$, which is expected to provide a more stringent test of New Physics. However, even 10 ab$^{-1}$ of data will improve the existing limits on these modes by an order of magnitude, and an $e^+e^-$ $B$ Factory does have the advantage of also being able to search for $B_d \to e^+e^-$ and the (extremely challenging) $B_d \to \tau^+\tau^-$ mode.

### $B \to s\ell^+\ell^-, K\ell^+\ell^-, K^*\ell^+\ell^-$ Decays

The exclusive decays $K^{(*)}\ell^+\ell^-$ and inclusive $s\ell^+\ell^-$ have been intensively studied theoretically, as they provide a potentially unique window on New Physics. For example, in the Standard Model, the forward/backward asymmetry $A_{FB}$ of the lepton pair has a zero at lepton pair mass $\hat{s}_0 = 0.14$ GeV. In extensions of the Standard Model, this zero may be approached from the opposite direction, or may be altogether absent. This region of lepton pair invariant mass represents only a small fraction of the allowed kinematic region of these rare decays, so a large data sample is required to make this measurement. The measurement of $A_{FB}$ can be done at hadronic experiments, but only in the exclusive modes involving muons. Theoretical predictions are typically more precise for inclusive processes, which can only be measured at a Super $B$ Factory. It is very important to compare $A_{FB}$ in muon and electron modes, as this asymmetry can be changed by the presence of a new neutral Higgs. Table 1-3 summarizes the achievable measurement precision.

**Table 1-3.** *Measurement precision for $B \to s\ell^+\ell^-,\ K\ell^+\ell^-,\ K^*\ell^+\ell^-$ decays.*

| $B \to s\ell^+\ell^-,\ K^{(*)}\ell^+\ell^-$ **Decays** | $e^+e^-$ **Precision** | | | **1 Yr Precision** | |
|---|---|---|---|---|---|
| **Measurement** | **3/ab** | **10/ab** | **50/ab** | LHC$b$ | $B$**TeV** |
| $\mathcal{B}(B \to K\mu^+\mu^-)/\mathcal{B}(B \to Ke^+e^-)$ | $\sim 8\%$ | $\sim 4\%$ | $\sim 2\%$ | - | - |
| $A_{CP}(B \to K^*\ell^+\ell^-)$ (all) | $\sim 6\%$ | $\sim 3\%$ | $\sim 1.5\%$ | $\sim 1.5\%$ | $\sim 2\%$ |
| (high mass) | $\sim 12\%$ | $\sim 6\%$ | $\sim 3\%$ | $\sim 3\%$ | $\sim 4\%$ |
| $A_{FB}(B \to K^*\ell^+\ell^-) : \hat{s}_0$ | $\sim 20\%$ | $\sim 9\%$ | $\sim 9\%$ | $\sim 12\%$ | |
| $A_{FB}(B \to s\ell^+\ell^-) : \hat{s}_0$ | $\sim 27\%$ | $\sim 15\%$ | $\sim 7\%$ | | |
| $: C_9,\ C_{10}$ | $36-55\%$ | $20-30\%$ | $9-13\%$ | | |

## 1.2.2 Unitarity Triangle Measurements

The major objective for *BABAR* and Belle was a precision measurement of $\sin 2\beta$ ($\sin 2\phi_1$), as a unique overconstrained test of the Standard Model, with the addition of $\sin 2\alpha$ ($\sin 2\phi_3$) and $\gamma$ ($\phi_2$) measurements as the program matured. We now have a measurement of $\sin 2\beta$ to $\sim 5\%$ precision, with further substantial improvements on the way, and we are making interesting determinations of $\sin 2\alpha$ and $\gamma$ as well.

**Table 1-4.** *Measurement precision for sides of the unitarity triangle. $|V_{cb}|$ is omitted, as it will be theory/systematics limited before we enter the ab$^{-1}$ regime.*

| **Unitarity Triangle - Sides** | | $e^+e^-$ **Precision** | | | **1 Yr Precision** | |
|---|---|---|---|---|---|---|
| **Measurement** | **Goal** | **3/ab** | **10/ab** | **50/ab** | **LHC$b$** | $B$**TeV** |
| $\lvert V_{ub}\rvert$ (inclusive) | syst=5-6% | 2% | 1.3% | | - | - |
| $\lvert V_{ub}\rvert$ (exclusive) $(\pi, \rho)$ | syst=3% | 5.5% | 3.2% | | - | - |
| $f_B$: $\mathcal{B}(B \to \mu\nu)$ | SM: $\mathcal{B} \sim 5 \times 10^{-7}$ | $3\sigma$ | $6\sigma$ | $> 10\sigma$ | - | - |
| $f_B$: $\mathcal{B}(B \to \tau\nu)$ | SM: $\mathcal{B} \sim 5 \times 10^{-5}$ | $3.3\sigma$ | $6\sigma$ | $> 10\sigma$ | - | - |
| $f_B$: $\mathcal{B}(B \to \ell\nu\gamma)$ | SM: $\mathcal{B} \sim 2 \times 10^{-6}$ | $> 2\sigma$ | $> 4\sigma$ | $> 9\sigma$ | - | - |
| $\lvert V_{td}\rvert/\lvert V_{ts}\rvert$ $(\rho\gamma/K^*\gamma)$ | Theory 12% | $\sim 3\%$ | $\sim 1\%$ | | - | - |

### Measuring the sides of the Unitarity Triangle

Table 1-4 summarizes the projected uncertainties on measurements of the sides of the unitarity triangle for various sample sizes at a Super $B$ Factory. With tens of ab$^{-1}$, new methods for CKM element determination, some with smaller systematic uncertainities, become feasible. The leptonic decays $B \to \ell\nu(\gamma)$ ($\ell = e, \mu, \tau$) give a theoretically clean determination of $|V_{ub}| f_B$ and, with the exception of $B \to e\nu$, have branching fractions which are well within the reach of a Super $B$ Factory. Due to the presence of multiple unobserved neutrinos in the final state, $B \to \tau\nu$ searches require full reconstruction of the accompanying $B$ using the so-called "recoil method", resulting in a substantially reduced selection efficiency compared with $B \to \mu\nu$ searches which do not use this method. Consequently, these two modes are expected to produce comparable sensitivity to $|V_{ub}| f_B$, in spite of the fact that their Standard Model branching fractions differ by two orders of magnitude. With a data sample of $\sim 10\,\mathrm{ab}^{-1}$, these two modes could each independently give determinations of $|V_{ub}| f_B$ to better than 10%. The radiative modes $B \to \ell\nu\gamma$ ($\ell = e, \mu$) are also potentially accessible using the recoil method, giving an additional determination of $|V_{ub}| f_B$, although with somewhat larger theoretical uncertainties.

This improvement in measurement precision is an excellent match to expected improvements in theoretical calculations. Lattice QCD calculations have made great strides in the past few years, and now with unquenched calculations and improved lattice actions appear to be on course for making usefully precise calculations. This synergy is a major motivation for the CLEO-$c$ program; further, both CLEO-$c$ measurements and improved QCD lattice calculations are crucial to achieving the ultimate precision in unitarity triangle determinations at a Super $B$ Factory.

The expected precision of lattice calculations, as presented at the Workshop, is shown in Table 1-5. Other projections of the rate of progress and the asymptotic limiting precision vary depending on the projector, but it is likely that theoretical inputs to unitarity triangle constraints will reach the several percent level on a time scale commensurate with a Super $B$ Factory reaching limiting experimental precision.

**Table 1-5.** *Expected improvement in the precision of calculation of lattice QCD parameters in the coming decade.*

| **Lattice QCD** | **Uncertainty (%)** | | | |
|---|---|---|---|---|
| **Quantity** | **Now** | **1-2 years** | **3-5 yrs** | **5-8 years** |
| $f_B$ | 15 | 9 | 4 | 3 |
| $f_B\sqrt{B_B}$ | 15-20 | 12 | 5 | 4 |
| $f_{B_s}/f_{B_d}$ | 6 | 3 | 2 | 1 |
| $\xi$ | 7 | 6 | 2 | 1.5 |
| $B \to \pi l\nu$ | 15 | 11 | 5 | 3 |
| $B \to D\ell\nu$ | 6 | 4 | 1.6 | 1.2 |

The experimental precision with $10\,\mathrm{ab}^{-1}$ samples is a good match to theory limits for the unitarity triangle. This program, which will make heavy use of recoil techniques unique to $e^+e^-$, is well motivated. By extending the precision of these measurements, and by employing new techniques, we can both refine and extend the overconstrained tests of the unitarity triangle pioneered by *BABAR* and Belle. There is potential here to discover New Physics, such as a fourth generation or an extra $Z^0$ boson, in the $B$ unitarity triangle. Additional, perhaps more likely, routes to isolating New Physics effects are described below.

### Measuring $\beta$

The precision of the measurement of $\sin 2\beta$ in $c\bar{c}s$ modes will continue to be statistics-limited until the $\sim 10\,\mathrm{ab}^{-1}$ regime, by which time $\beta$ will be known to a fraction of a degree. If reducing systematics further remains an interesting goal, then further improvement can be obtained by using lepton tags only. $\sin 2\beta$ is one of the theoretically cleanest measurements that can be made in flavor physics; it should be pursued to the sub-percent level. $\sin 2\beta$ is also the benchmark for measurements of $CP$ violation in the much rarer $s\bar{s}s$ modes that appear so promising for isolating New Physics effects.

**Table 1-6.** *Measurement precision for angles of the unitarity triangle.*

| **Unitarity Triangle - Angles** | $e^+e^-$ **Precision** | | | **1 Yr Precision** | |
|---|---|---|---|---|---|
| **Measurement** | **3/ab** | **10/ab** | **50/ab** | **LHC$b$** | **$B$TeV** |
| $\alpha\ (\pi\pi)\ (S_{\pi\pi},\ B \to \pi\pi\ \mathcal{B}'s + \text{isospin})$ | $6.7^\circ$ | $3.9^\circ$ | $2.1^\circ$ | - | - |
| $\alpha\ (\rho\pi)$ (isospin, Dalitz) (syst$\geq 3^\circ$) | $3^\circ$, $2.3^\circ$ | $1.6^\circ$, $1.3^\circ$ | $1.0^\circ$, $0.6^\circ$ | $2.5^\circ$-$5^\circ$ | $4^\circ$ |
| $\alpha\ (\rho\rho)$ (Penguin, isospin) (stat+syst) | $2.9^\circ$ | $1.5^\circ$ | $0.72^\circ$ | - | - |
| $\beta\ (J/\psi K^0_S)$ (all modes) | $0.6^\circ$ | $0.34^\circ$ | $0.18^\circ$ | $0.57^\circ$ | $0.49^\circ$ |
| $\gamma\ (B \to D^{(*)}K)$ (ADS+$D \to K_S\pi^+\pi^-$) | | $2-3^\circ$ | | $10^\circ$ | $< 13^\circ$ |
| $\gamma\ (B \to D^{(*)}K)$ (all methods) | | $1.2-2^\circ$ | | | |

**Measuring $\alpha$**

The measurement of the angle $\alpha$ is complicated by the presence of penguin amplitudes, which undermine the ability of a measurement of the $CP$ asymmetry in, for example, $B^0 \to \pi^+\pi^-$ to directly determine $\sin 2\alpha$. Several techniques have been proposed to isolate the effect of penguin amplitudes, thereby allowing the extraction of $\alpha$, as opposed to a penguin-contaminated $\alpha_{eff}$. This can be done in the $\pi\pi$, $\rho\pi$, and $\rho\rho$ final states. A common feature of all these techniques is that they require very large data samples in order to measure very small branching fractions (such as the separate branching fractions for $B^0 \to \pi^0\pi^0$ and $\overline{B}^0 \to \pi^0\pi^0$) and/or to resolve (typically, four-fold) ambiguities. Table 1-6 shows that these methods can yield an ultimate precision of a few degrees for $\alpha$; but at least 10 ab$^{-1}$ is generally needed to resolve ambiguities. It is worth noting that the hadron experiments have mostly studied measurement capabilities with the $\rho\pi$ mode; the promising $\rho\rho$ channel, with two $\pi^0$ mesons in the final state, may be less accessible.

**Measuring $\gamma$**

Table 1-6 lists two expected statistical errors on the measurement of $\gamma$ with 10 ab$^{-1}$. The more conservative error estimate of $2^\circ - 3^\circ$ is obtained employing only methods and decay modes that have already been observed and used in $\gamma$-related measurements. These are $B^\pm \to DK^\pm$ with $D^0 \to K^-\pi^+$, $D^0 \to CP$-eigenstates, and $D^0 \to K_S\pi^+\pi^-$.The sensitivities are estimated from the quoted experimental references. The range in the estimates is due to current uncertainties in the ratio between the interfering $b \to u$ and $b \to c$ amplitudes, taken to be between 0.1 and 0.2. The mode $D^0 \to K_S\pi^+\pi^-$ is especially important in that it reduces the 8-fold asymmetry to a 2-fold asymmetry.

The less conservative estimate of $1.2^\circ - 2^\circ$ is based on cautious assumptions about the sensitivities that could be obtained with modes that have yet to be fully explored experimentally. One category of such modes is additional multi-body $D$ decays to final states such as $\pi^+\pi^-\pi^0$, $K^+K^-\pi^0$, $K_SK^+K^-$, $K_SK^+\pi^-$, $K_S\pi^+\pi^-\pi^0$, $K^+\pi^-\pi^0$, and $K^+\pi^-\pi^+\pi^-$. The second category is $B^0 \to DK^{(*)0}$, with a time-dependentand time-independentanalysis. The third category is $B \to D^+K_S\pi^-$ decays [16], which also requires a time-dependent analysis.

It should be noted that there are additional modes and methods, not included in the $\gamma$ sensitivity estimates of Table 1-6, that can be used to measure $\gamma$. These methods presently suffer from difficulties in obtaining a clean extraction of $\gamma$, which will likely be resolved in the future. Examples include $\sin(2\beta+\gamma)$ in $B \to D^{(*)+}\pi^-$, $B \to D^{(*)+}\rho^-$, where *BABAR* has already published measurements of $CP$ asymmetries and constraints on $\gamma$. It is not clear, however, whether the ratio between the interfering $b \to u$ and $b \to c$ amplitudes can be measured with sufficient precision for these measurements to be competitive with the $B \to DK$ measurements at high luminosity. The estimates also exclude the possible contribution of $B^\pm \to DK^\pm\pi^0$, where experimental issues are yet to be resolved and the level of interference is not yet known.

### 1.2.3 Physics Performance Projections

Various benchmark physics measurements, discussed in Sections 1.2.1 and 1.2.2, have been used to illustrate the physics reach of a Super $B$ Factory on the basis of integrated samples. In some cases, comparisons with hadronic experiments are also possible. A set of assumptions has been made concerning the pace at which these projects reach their design luminosity goals, as summarized in Table 1-7. These assumptions then form the basis for time varying projections of effective tagged samples in a number of important channels (observed yield weighted by effective tagging efficiency). Finally, the effective tagged sample sizes, when combined with measured or simulated single event sensitivities can be used to project the errors on benchmark observables. In the case of the $e^+e^-$ collider options the samples are assumed to be continuations of the event samples obtained at PEP-II through the end of the already planned program. In all cases, PEP-II is assumed to cease operations at the point when installation of the upgraded collider must begin. Integrated luminosities in these periods are taken from the published PEP-II plan.

Figure 1-3 shows the time evolution of the error on the sine coefficient for time-dependent $CP$ violation in various $b \to s$ Penguin modes. In this case the error reaches below 0.04 in most case within two years, which is the regime that is relevant for definitive demonstration of potential New Physics in such modes. Figure 1-7 shows the error evolution for the electromagnetic Penguin mode $B^0 \to K^0_s\pi^0\gamma$.

Figure 1-4 shows the effective tagged sample accumulations and expected evolution of the error on the sine amplitude in time-dependent $CP$ asymmetries for $B^0 \to \pi^+\pi^-$. In this case both LHC$b$ and $B$TEV are capable of measurements, as well as a Super $B$ Factory. However, only a Super $B$ Factory is capable of doing the complete isospin analysis of the two-body modes in order to obtain the correction from the determination of $\alpha_{eff}$ in the $CP$ asymmetry from the charged mode to the unitarity angle $\alpha$. The evolution of this correction as a function of time is shown in Figure 1-5. The error on $\sin 2\alpha$ falls below 0.05 within two years of startup for the Super $B$ Factory.

Figure 1-6 shows a comparable measurement of $\sin 2\alpha$, which reaches below 0.04 within the same period.

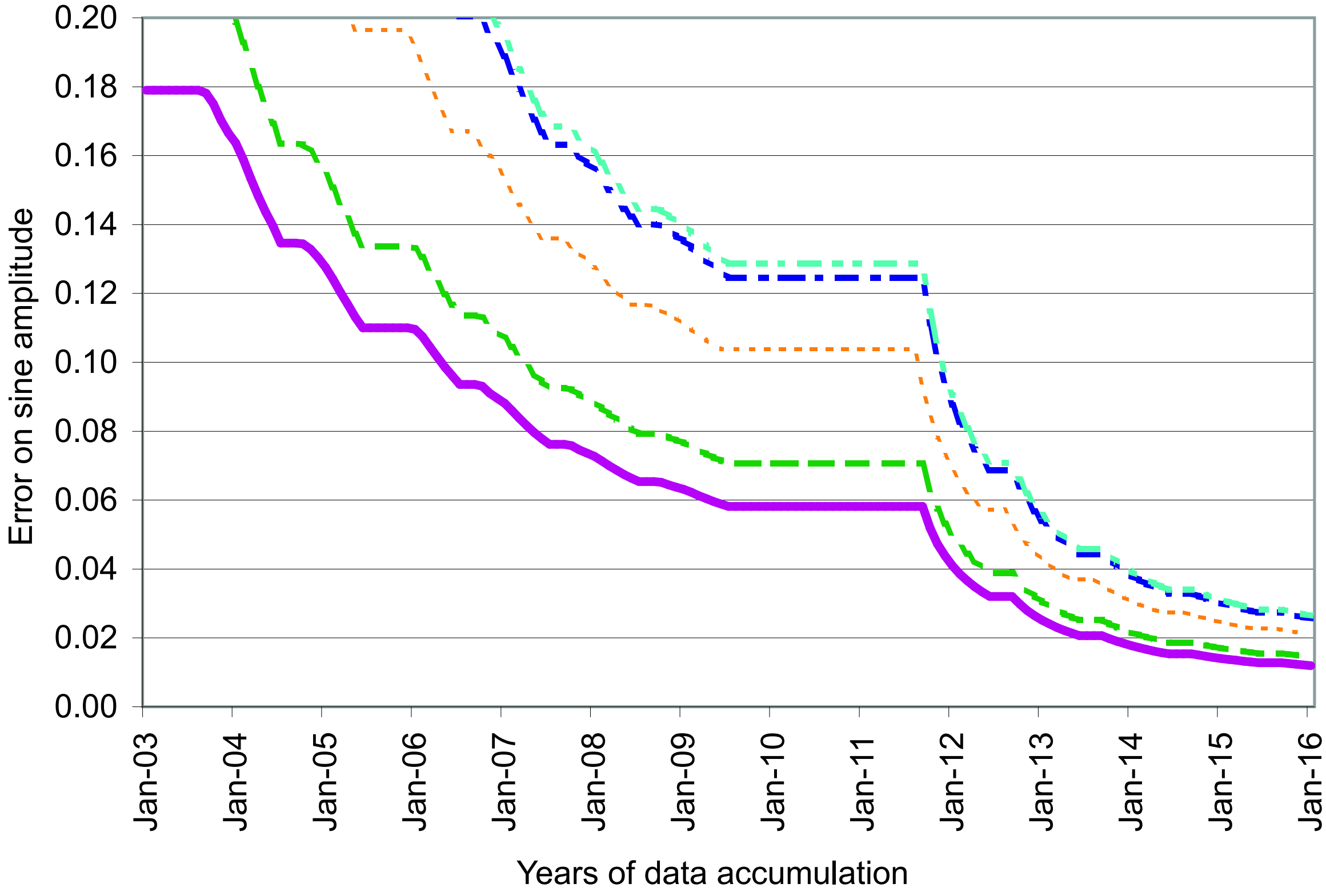

**Figure 1-3.** *Projected time development for the error on the sine coefficient in a fit to the time-dependent $CP$ asymmetry for $B$ decays in various $b \to s\bar{s}s$ Penguin channels: $B^0 \to f_0 K^0_S$ (——), $B^0 \to K^0_S \pi^0$ (— — —), $B^0 \to \varphi K^0_S$ (—), $B^0 \to \eta' K^0_S$ (——), and $B^0 \to K^+K^-K^0_S$ (——) at the Super $B$ Factory. Per event errors are based on current BABAR analysis.*

**Table 1-7.** *Startup efficiencies and initial peak luminosities assumed for a Super $B$ Factory at SLAC, and the hadron accelerator-based experiments LHC$b$ and $B$TeV in making projections of integrated data samples and measurement precision.*

| Facility | Start Date | Initial efficiency | Duration (years) |
|---|---|---|---|
| LHC$b$ | 1/2008 | 50% | 2 |
| $B$TeV | 1/2010 | 50% | 2 |
| Super $B$ Factory | 10/2011 | 50% | 1 |
| | 10/2012 | 100% | 1 |
| | 10/2013 | 140% | indefinite |

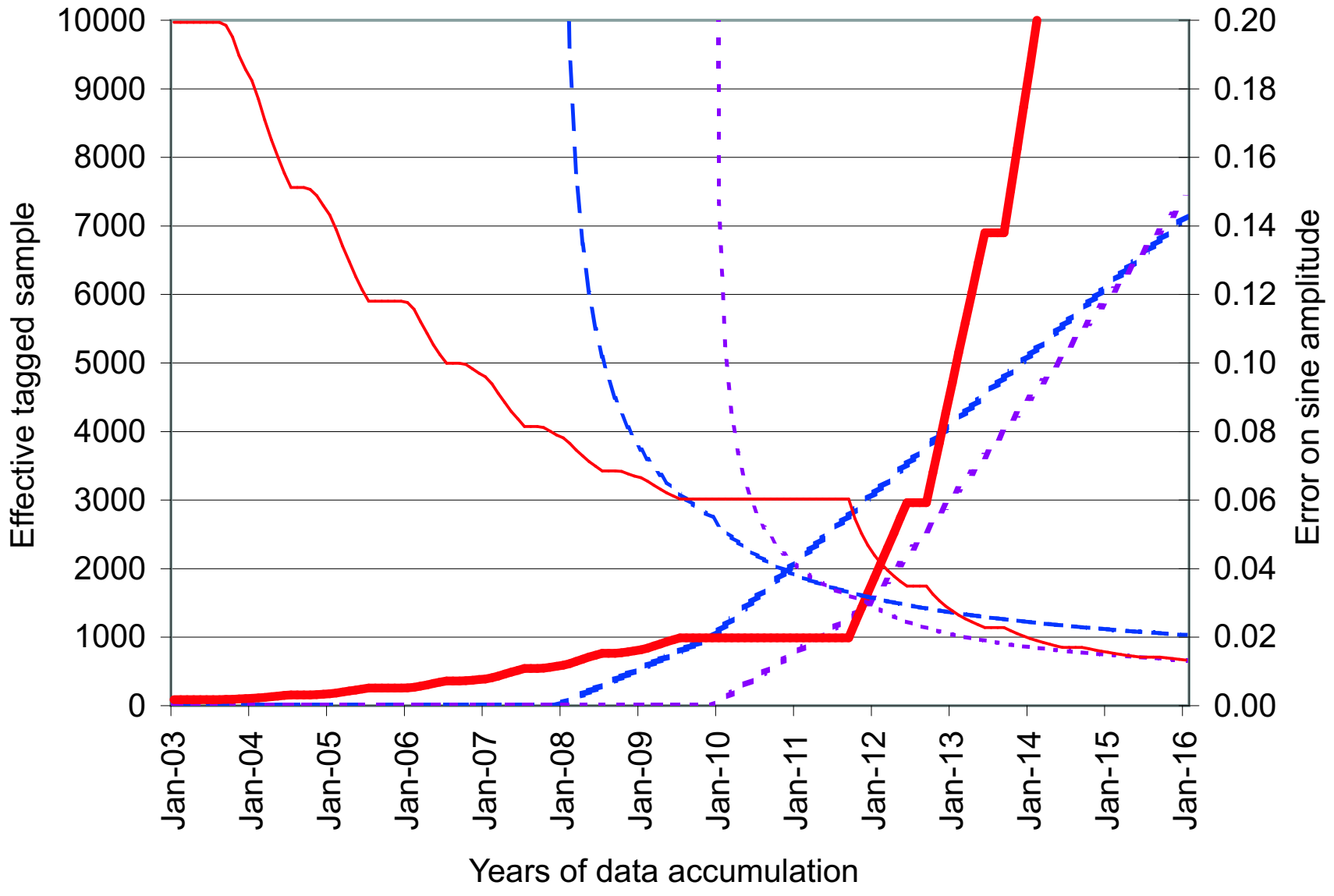

**Figure 1-4.** *Projected time development for the size of the effective tagged sample (dashed red line) and the error on the amplitude for the sine coefficient (solid red line) for a fit to the time-dependent $CP$ asymmetry in a $B^0 \to \pi^+\pi^-$ sample accumulated at a Super $B$ Factory . Tagging efficiencies and per event errors are based on current BABAR analysis. Also shown are effective tagged samples and the error on the amplitude for the sine coefficient for LHC$b$ (dashed blue curves) and $B$TeV (dotted purple curves), based on their simulations of tagging efficiencies and per event sensitivities.*

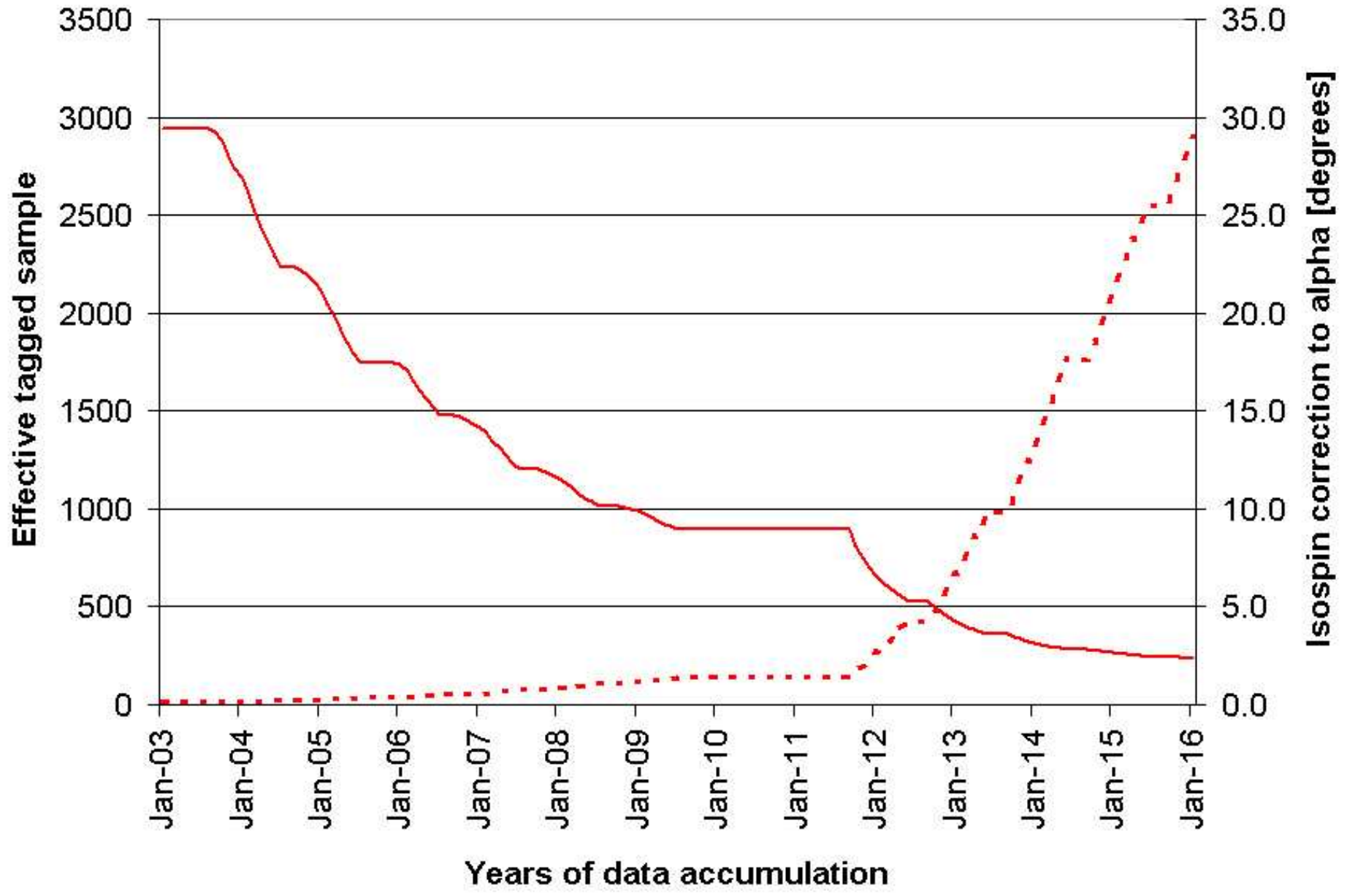

**Figure 1-5.** *Projected time development for the size of the effective tagged sample ( - - - ) and the correction to the effective value of alpha ( —— )as determined from a full isospin analysis of the two-body $B$ decay modes for a Super $B$ Factory. Tagging efficiencies and per event errors are based on current BABAR analysis.*

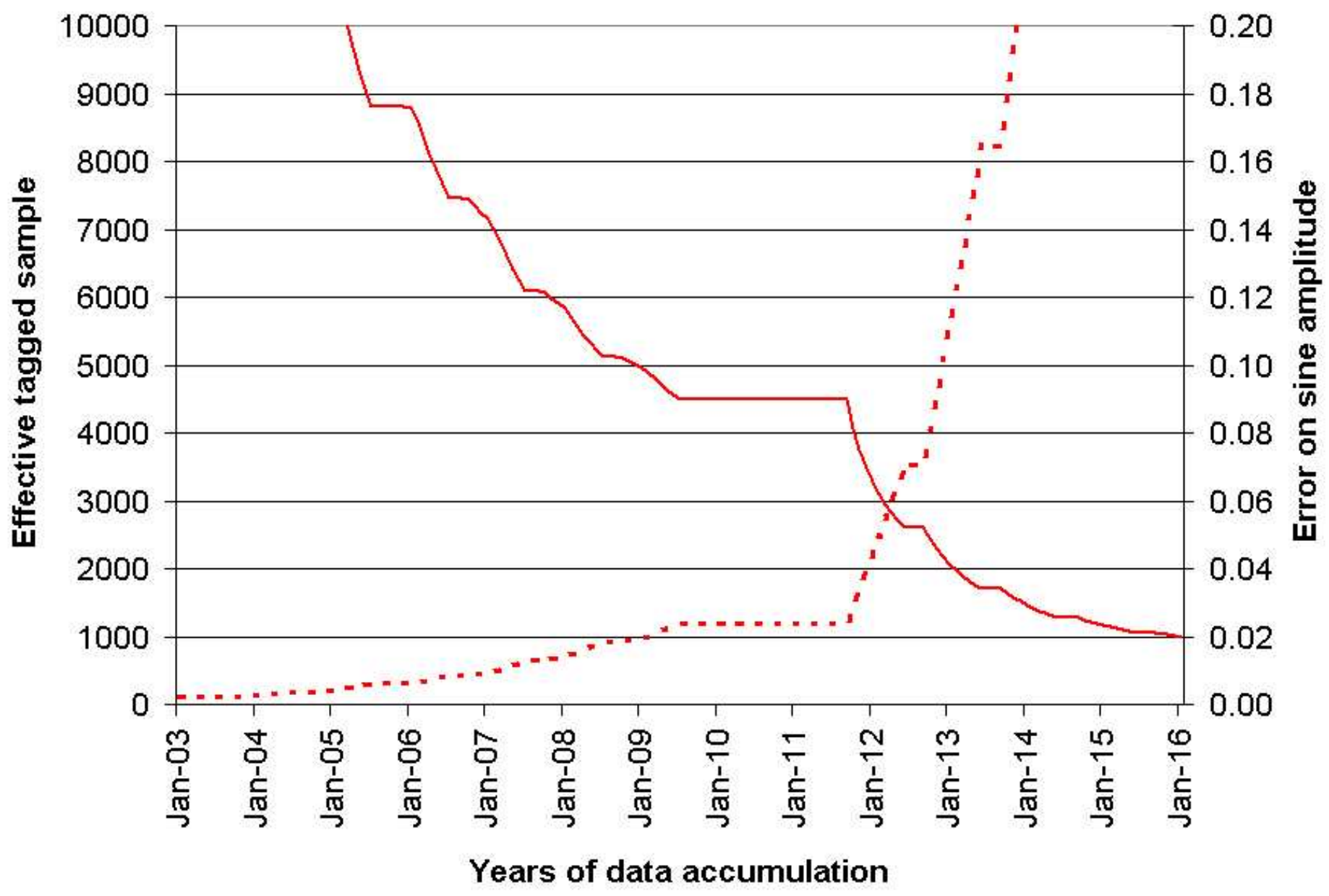

**Figure 1-6.** *Projected time development for the size of the effective tagged sample (- - - ) and the error on the amplitude for the sine coefficient (——)from a fit to the time-dependent $CP$ asymmetry in a $B^0 \rightarrow \rho^+\rho^-$ sample accumulated at Super $B$ Factory. Tagging efficiencies and per event errors are based on current BABAR analysis.*

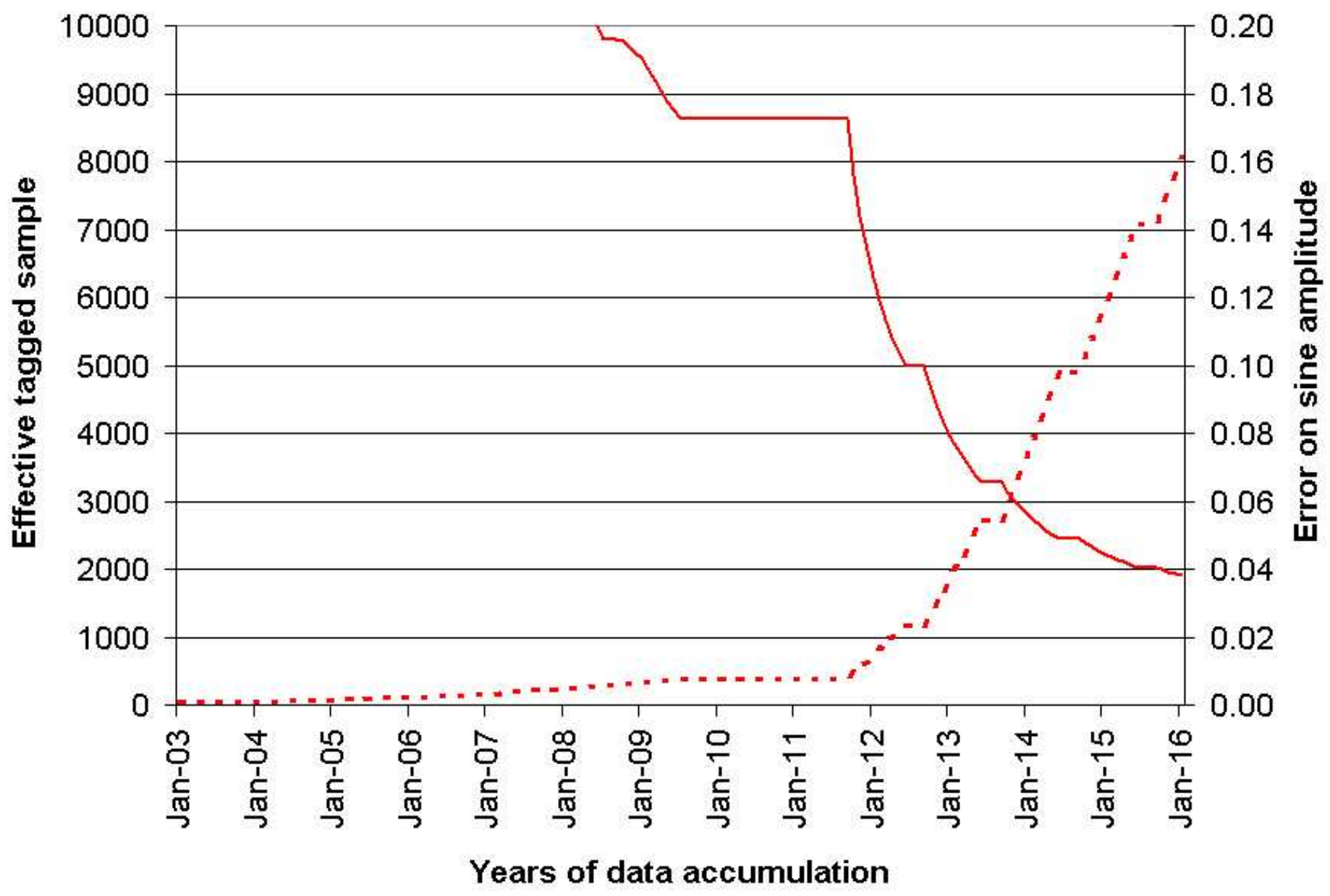

**Figure 1-7.** *Projected time development for the size of the effective tagged sample (- - - ) and the error on the amplitude for the sine coefficient (——)from a fit to the time-dependent $CP$ asymmetry in a $B^0 \rightarrow K^0_S\pi^0\gamma$ sample accumulated at a Super $B$ Factory. Tagging efficiencies and per event errors are based on current BABAR analysis.*

## 1.3 Conclusions

As the subject of flavor physics is at this point quite extensive, organizing the presentation of an intrinsically interrelated set of topics in $B$ and $D$ physics is a complex task. We have approached this task as follows.

This volume addresses four principal areas. Chapter 2 discusses rare decays of $B$ and $D$ mesons and $\tau$ leptons, with an emphasis on a detailed discussion of the precision to which Standard Model calculations of rare decays are known, and the resulting sensitivity of a variety of measurements in isolating and interpreting signals of New Physics. Chapter 3 deals with the many and varied approaches to determining the angles of the unitarity triangle by measuring the $CP$-violating phases of the CKM matrix, through the measurement of $CP$ asymmetries or the isolation of phases using constructions that often involve rather small $B$ meson branching fractions. Chapter 4 addresses the determination of the sides of the unitarity triangle by measuring semileptonic and purely leptonic $B$ meson decays. Measurements at the precision promised by improvements in lattice QCD calculations in the next decade provide a stringent set of overconstrained tests of the CKM *ansatz* and are a method of finding evidence for the existence of a fourth quark generation or of extra $Z^0$ bosons. Chapter 5 discusses model-independent analyses aimed at isolating New Physics, and presents a variety of examples using supersymmetric and extra dimension models.

Results continue to pour out of the asymmetric $B$ Factories. In the year since these Super $B$ Factory Workshops there has been substantial experimental and theoretical progress that is not covered in these Proceedings. There are even intriguing hints in the data from BABAR Belle, and CDF of discrepancies with the Standard Model. Existing $e^+e^-$ facilities are capable of doubling or tripling current data samples, but it will take a Super $B$ Factory to provide the sensitivity to explore in detail effects in flavor physics due to the New Physics we expect to encounter at the LHC. The LHC, the ILC and a Super $B$ Factory each will make a unique contribution to the exploration of physics beyond the Standard Model. Having pioneered the use of $CP$ violation in unique tests of the Standard Model, we are now poised to employ $CP$ violation as a unique diagnostic tool for the exploration of New Physics.

## 1.4 Acknowledgements

These Workshops were supported by the SLAC Research Division.

We wish to thank the authors of the set of excellent articles that comprise these Proceedings, as well as the attendees of the May and October, 2003 Workshops who made both meetings so stimulating. The Conveners of the four working groups did superb editorial work. Thanks are also due to Sharon Jensen and Suzy Vascotto for excellent editorial assistance, to Terry Anderson for the design of the Proceedings cover, to Latin scholars Gregory Dubois-Felsmann, Gerald Eigen, Manfred Eigen and Marcello Giorgi, and to David MacFarlane for the physics performance projections.

# 2

# Rare Decays

**Conveners:** G. Hiller, T. Hurth, A. Ryd

**Authors:** T. Abe, J. Albert, A. Ali, G. Burdman, M. Convery, S. Dasu, F. Di Lodovico, Th. Feldmann, G. Hiller, T. Hurth, G. Isidori, C. Jessop, P. Koppenburg, V. Koptchev, F. Krüger, J. Libby, E. Lunghi, Y. Parkhomenko, D. Pirjol, M. Purohit, S.H. Robertson, A. Rubin, A. Ryd, A. Soni, H. Staengle, G. Wilkinson, D.C. Williams, S. Willocq, D. Wyler

## 2.1 Overview

Rare decays play a crucial role in the discovery potential of a Super $B$ Factory. As they involve loop-suppressed flavor-changing neutral currents, they are highly-sensitive probes for new degrees of freedom beyond the Standard Model.

We present herein a comprehensive study of a large variety of measurements of rare decays that are likely to be of crucial interest at a time when hadronic $B$ experiments such as LHC$b$ or $B$TeV are already operating. Most of the measurements discussed in this chapter require a clean, well-understood experimental environment that only a Super $B$ Factory can provide.

The chapter is organized as follows. In the first three general sections, we review the present status of rare decay searches and measurements, summarize the theoretical tools employed, and discuss the prospects of reducing present theoretical uncertainties by the time a Super $B$ Factory is in operation. Each rare decay mode is then analyzed in detail from the experimental and theoretical point of view. Sensitivity to New Physics in various rare decays is also discussed in a general way; model-independent analyses of New Physics and specific model studies can be found in Chapter 5. Future prospects for measurements of purely hadronic 'rare' $B$ decays are discussed in Chapter 3. The extraction of CKM elements from rare decays in the context of the Standard Model is discussed in Chapter 5. This chapter concludes with a short summary on the impact of rare $B$ and $D$ meson decay studies on the search for physics beyond the Standard Model

## 2.2 Present Status of Rare $B$ Decays

≻– A. Ali –≺

### 2.2.1 Motivation

Rare $B$ decays such as $b \to s\gamma$, $b \to d\gamma$, $b \to s\ell^+\ell^-$, $b \to d\ell^+\ell^-$, $B_d^0 \to \ell^+\ell^-$, $B_s^0 \to \ell^+\ell^-$ are flavor-changing-neutral-current (FCNC) processes in which a $b$-quark transforms due to weak interactions either into an $s$-quark ($b \to s$ transition) or into a $d$-quark ($b \to d$ transition). They are characterized by the quantum number flow $|\Delta B| = 1$, $\Delta Q = 0$. The other examples of the FCNC processes in the $B$ meson sector are the particle-antiparticle mixings $B^0\overline{B}^0$ and $B_s^0\overline{B}_s^0$. For the mixings, the quantum number flow is characterized by $|\Delta B| = 2$, $\Delta Q = 0$. As in the Standard Model, all electrically neutral particles ($\gamma$, $Z^0$, $H^0$, and the gluons) have only diagonal couplings in the flavor space, FCNC transitions are forbidden at the tree level and are allowed only through induced (loop) effects. This is the essence of the Glashow-Iliopoulos-Maiani (GIM) mechanism [1], which governs all FCNC processes in the Standard Model.

A number of inferences can be drawn from this observation: FCNC transitions in $B$ decays probe the underlying fundamental theory at the quantum level and hence they are sensitive to masses much higher than that of the $b$ quark. In the Standard Model, this higher scale is characterized by the top quark mass $\sim$ 175 GeV; the virtual top quark contribution dominates the $|\Delta B| = 2$, $\Delta Q = 0$ transitions. In the case of rare $B$ decays, in general, light-quark contributions are also present, which have to be included for a satisfactory phenomenological description of the observed phenomena, but the role of the top-quark-induced amplitudes remains crucial. By virtue of this, FCNC transitions enable us to determine the CKM matrix elements in the third row of this matrix, namely $V_{td}$, $V_{ts}$ and $V_{tb}$. Of these, $|V_{tb}|$ has been measured by the CDF collaboration in the production and decay of the top quark, $p\overline{p} \to tX$, $t \to b\,W^+$, yielding $|V_{tb}| = 0.96^{+0.16}_{-0.23}$ [2]. However, FCNC processes are at present the only quantitative probes of the other two CKM matrix elements $V_{ts}$ and $V_{td}$ [3]. Current measurements yield $|V_{td}| = (8.5 \pm 1.0) \times 10^{-3}$, which results from the measured mass difference $\Delta M_{B_d} = (0.502 \pm 0.006)$ ps$^{-1}$ [4] and the lattice-QCD-based estimates of the pseudoscalar coupling $f_{B_d}\sqrt{B_{B_d}} = (210 \pm 24)$ MeV [5], whereas $V_{ts} = -(47 \pm 8) \times 10^{-3}$ [6], resulting from the next-to-leading order calculations of the branching ratio for the inclusive $B \to X_s\gamma$ decay and experiment, discussed below.

In beyond-the-Standard Model scenarios, FCNC processes are sensitive to new particles with masses up to $\mathcal{O}(1\text{TeV})$, such as the Higgses, charginos, stops and neutralinos in supersymmetric theories. The best-studied case to date is the decay $B \to X_s\gamma$, which has become the standard candle of flavor physics, and provides important constraints on the parameters of models beyond the Standard Model [7]. Close on the heels of the radiative decays are the FCNC semileptonic decays $B \to X_s\ell^+\ell^-$ (and their exclusive modes such as $B \to (K, K^*)\,\ell^+\ell^-$). The first goal of the experiments has already been achieved, in that all the inclusive and exclusive decays (for $\ell^+\ell^- = e^+e^-$, $\mu^+\mu^-$) have been measured by the *BABAR* and BELLE experiments at the current $B$ Factories [8]. Within the present experimental and theoretical precision, these measurements are in agreement with Standard Model estimates at NLO accuracy [9, 10]. Being the first measurements probing the electroweak penguin sector of the $B$ mesons, only the integrated decay rates in $B \to (X_s, K, K^*)\,\ell^+\ell^-$ are thus far well-established. Measurements of the invariant dilepton mass and the hadron mass $M_{X_s}$ are, however, sparse. They will greatly improve in precision at a Super $B$ Factory, where the full force of the increased luminosity will be brought to bear on the precise measurements of the Dalitz distributions in these decays. In particular, measurements of the dilepton invariant mass spectra and the forward-backward asymmetries for $B \to (K^*, X_s)\ell^+\ell^-$ [11] would determine the effective Wilson coefficients of the underlying effective theory [12, 10]. Given a fundamental theory, such as the Standard Model or a supersymmetric theory, the Wilson coefficients can be calculated quite precisely taking into account QCD renormalization effects. These can then be extracted from the data, taking into account the residual power and radiative corrections, thereby allowing Super $B$ Factory experiments to test the Standard Model precisely in the electroweak penguin sector and carry out a focused search of physics scenarios beyond the Standard Model. It should be emphasized that, as opposed to the electroweak precision tests, where physics beyond the Standard Model enters in most observables only as part of the loop corrections, and hence such effects are small, in many rare $B$ (and $K$) decays, the contribution of physics beyond the Standard Model can be comparable contribution to that in the Standard Model, and in some cases the former could be much larger. Hence, precision studies of the flavor-changing rare $B$ (and $K$) decays provide sensitivity to high scales, such that the signal (from New Physics) to background (from the Standard Model) ratios are much more favorable than is the case in the flavor-diagonal electroweak precision tests.

The purely leptonic decays, $B_d^0(B_s^0) \to \ell^+\ell^-$, apart from being a precision test of the Standard Model [13, 14], provide potentially sensitive probes of an extended scalar (Higgs) sector. The current experimental upper bounds [15, 16], while they are orders of magnitude away from the Standard Model branching ratios, however, do probe the large-$\tan\beta$ region in supersymmetric theories [17, 18, 19, 20, 21, 22, 23, 24].

The decays $B \to (X_s, X_d)\nu\overline{\nu}$ are arguably the cleanest probe of the short-distance contribution in rare $B$ decays [25], but, lacking sufficient kinematic constraints to construct the final state, they present a great challenge even at $B$ Factories. The current upper limit on $\mathcal{B}(B \to X_s\nu\overline{\nu}) < 6.4 \times 10^{-4}$ from the ALEPH collaboration [26], is more than an order of magnitude away from the estimates in the Standard Model [13, 14] $\mathcal{B}(B \to X_s\nu\overline{\nu}) = (4.0 \pm 1.0) \times 10^{-5}$. The branching ratios of the exclusive decays $B \to (K, K^*)\nu\overline{\nu}$ have larger theory uncertainty due to the form factors. Since the Standard Model branching ratios $\mathcal{B}(B \to K\nu\overline{\nu}) = 3.8^{+1.2}_{-0.6} \cdot 10^{-6}$ and $\mathcal{B}(B \to K^*\nu\overline{\nu}) = 1.3^{+0.4}_{-0.3} \cdot 10^{-5}$ [27] are not small (compared to, *e.g.*, $B \to K^{(*)}\ell^+\ell^-$ decays), with a reconstruction efficiency of $O(10^{-3})$, one can probe them with a sample of $10^9$ $B$'s. The current bound $\mathcal{B}(B^- \to K^-\nu\overline{\nu}) < 7.0 \cdot 10^{-5}$ [28] is an order of magnitude away from the Standard Model but already provides interesting constraints on scenarios beyond the Standard Model, *e.g.*, [29, 24].

Precision studies of the radiative and semileptonic $b \to d$ transitions will be undertaken at the current and Super $B$ Factoriesin the decays $B \to (X_d, \rho, \omega)\gamma$ and $B_d^0 \to (X_d, \pi, \rho, \omega)\ell^+\ell^-$, respectively. The threshold for these transitions has now been reached by the current $B$ Factory experiments. As this contribution is being written, the Belle collaboration has just announced the first measurement of the exclusive $b \to d$ radiative transition through the decays $B \to (\rho, \omega)\,\gamma$ [30]. The branching ratio $\mathcal{B}(B \to \rho/\omega\,\gamma) = (1.8^{+0.6}_{-0.5} \pm 0.1) \times 10^{-6}$ is in agreement with the Standard Model-based estimates [31, 32, 3]. Together with the measured branching ratios for $B \to K^*\gamma$ decays, the ratio $\mathcal{B}(B \to \rho/\omega\,\gamma)/\mathcal{B}(B \to K^*\gamma)$ constrains the CKM-Wolfenstein parameters $\overline{\rho}$ and $\overline{\eta}$ [33], and, as worked out in detail in [34], they must be included in overall CKM unitarity fits in the Standard Model. A first determination of the ratio $|V_{td}/V_{ts}|$ from radiative $B$ decays is being discussed here, yielding [35] $0.16 \leq |V_{td}/V_{ts}| \leq 0.29$ (at 68% C.L.), which is in agreement with the indirect estimates of the same in the Standard Model obtained using CKM unitarity fits $0.18 \leq |V_{td}/V_{ts}| \leq 0.24$. At a Super $B$ Factory, these decays will be measured with great precision, and the challenging measurements of the isospin-violating and *CP*-violating asymmetries in the $B \to \rho\gamma$ decays will also be undertaken. Both of these asymmetries provide a determination of the angle $\alpha$ [31, 32]. In that respect, the radiative decays $B \to (\rho, \omega)\gamma$ are complementary to the hadronic decays $B \to \pi\pi$, $B \to \rho\pi$ and $B \to \rho\rho$ being currently studied by the Belle [36] and *BABAR* [37, 38] collaborations, which will become very precise in the era of the Super $B$ Factory and in experiments at the hadron colliders, $B$TeV and LHC$b$. The inclusive decay $B \to X_d\gamma$ is theoretically cleaner than its exclusive counterparts discussed earlier but experimentally a good deal more challenging. The estimated branching ratio $\mathcal{B}(B \to X_d\gamma) \simeq 1.3 \times 10^{-5}$ in the Standard Model [39] is typically a factor 30 smaller than the rate for the dominant decay $B \to X_s\gamma$, and hence one requires very good control on the $s$ quark rejection to suppress this background. Using the $s$ quark mistag efficiency $\omega_s = 30\%$, it has been estimated that a 15% measurement of $|V_{td}/V_{ts}|$ from the measurement of the ratio $\mathcal{B}(B \to X_d\gamma)/\mathcal{B}(B \to X_s\gamma)$ would require a data sample of $\mathcal{O}(10)$ (ab)$^{-1}$ at a Super $B$ Factory.

The third motivation in the precision studies of rare $B$ decays is that they provide almost ideal situations to develop and test quantitative theoretical tools. Leptonic, semileptonic and radiative decays, being simpler as far as the strong interactions are concerned, are theoretically more tractable than their nonleptonic counterparts. We have in mind here processes such as $B \to K^*\gamma$, $B \to \rho\gamma$, $B \to (K, K^*)\,\ell^+\ell^-$ and $B \to \ell\nu_\ell\gamma$, that are less challenging theoretically than nonleptonic decays such as $B \to \pi\pi$ and $B \to \pi K$. Radiative and semileptonic decays undergo calculable perturbative QCD and power corrections (in $1/m_b$ and $1/m_c$) and teach us about the non-trivial aspects of the effective theory relevant for the heavy-to-light hadronic transitions. These include, among others, factorization, treatment of large logarithms that are usually present in processes with an intermediate scale, the light-cone distribution amplitudes for the $B$ meson as well as the light hadrons, and the shape functions which determine the energy-momentum profile of the final states. Radiative and semileptonic $B$ decays also provide case studies for the formulation of the soft-collinear effective theory to undertake precise theoretical calculations in $B$ decays.

In the following, we briefly discuss the theoretical and experimental status of some principal measurements in rare $B$ decays that have already been undertaken in current experiments, and briefly mention some of the benchmark measurements in this area at that can be made at the planned experimental facilities.

### 2.2.2 Inclusive radiative rare $B$ decays

We start by discussing the general structure of the amplitudes in radiative and semileptonic rare $B$ decays and their dependence on CKM matrix elements. The transitions $b \to s\gamma$ and $b \to s\,\ell^+\ell^-$ involve the CKM matrix elements from the second and third columns of this matrix, with the unitarity constraint taking the form $\sum_{u,c,t} \lambda_i = 0$, where $\lambda_i = V_{ib}V_{is}^*$. This equation yields a unitarity triangle which is highly squashed, as one of the sides of this triangle $\lambda_u = V_{ub}V_{us}^* \simeq A\lambda^4(\overline{\rho} - i\overline{\eta})$ is doubly Cabibbo-suppressed, compared to the other two sides $\lambda_c \simeq -\lambda_t = A\lambda^2 + ....$ Here, $A$, $\lambda$, $\overline{\rho}$ and $\overline{\eta}$ are the CKM-Wolfenstein parameters. Hence, the transitions $b \to s\gamma$ and $b \to s\,\ell^+\ell^-$ are not expected to yield useful information on the parameters $\overline{\rho}$ and $\overline{\eta}$, which define the apex of the unitarity triangle of current interest. The test of unitarity for the $b \to s$ transitions in rare $B$ decays lies in checking the relation $\lambda_t \simeq -\lambda_c$, which holds up to corrections of order $\lambda^2$. The impact of the $b \to d\gamma$ and $b \to d\,\ell^+\ell^-$ decays on the CKM phenomenology is, however, quite different. These transitions involve the CKM matrix elements in the first and third columns, with the unitarity constraints taking the form $\sum_{u,c,t} \xi_i = 0$, with $\xi_i = V_{ib}V_{id}^*$. Now, all three matrix elements are of order $\lambda^3$, with $\xi_u \simeq A\lambda^3(\overline{\rho} - i\overline{\eta})$, $\xi_c \simeq -A\lambda^3$, and $\xi_t \simeq A\lambda^3(1 - \overline{\rho} + i\overline{\eta})$. This equation leads to the same unitarity triangle as studied through the constraints $V_{ub}/V_{cb}$, $\Delta M_{B_d}$ (or $\Delta M_{B_d}/\Delta M_{B_s}$). Hence, the transitions $b \to d\gamma$ and $b \to d\,\ell^+\ell^-$ lead to complementary constraints on the CKM parameters $\overline{\rho}$ and $\overline{\eta}$, as illustrated in the following.

A theoretical framework for analyzing the $b \to s\gamma$ transition is set by the effective interaction Hamiltonian:

$$\mathcal{H}_{\rm eff} = -\frac{4G_F}{\sqrt{2}} V_{ts}^* V_{tb} \sum_{i=1}^{8} C_i(\mu) O_i \, . \tag{2.1}$$

The definition of the operators $O_i$ is given in Ref. [6]. Perturbative calculations (see Refs. [40, 41] and references therein) are used to find the Wilson coefficients in the $\overline{\rm MS}$ scheme, at the renormalization scale $\mu_b \sim m_b$

$$C_i(\mu_b) = C_i^{(0)}(\mu_b) + \frac{\alpha_s(\mu_b)}{4\pi} C_i^{(1)}(\mu_b) + \left(\frac{\alpha_s(\mu_b)}{4\pi}\right)^2 C_i^{(2)}(\mu_b) + \dots . \tag{2.2}$$

Here, $C_i^{(n)}(\mu_b)$ depend on $\alpha_s$ only via the ratio $\eta \equiv \alpha_s(\mu_0)/\alpha_s(\mu_b)$, where $\mu_0 \sim m_W$. In the leading order (LO) calculations, everything but $C_i^{(0)}(\mu_b)$ is neglected in Eq. (2.2). At the next-to-leading order (NLO), one takes $C_i^{(1)}(\mu_b)$ into account. The Wilson coefficients contain information on the short-distance QCD effects due to hard gluon exchanges. Such effects enhance the perturbative branching ratio $\mathcal{B}(b \to s\gamma)$ by roughly a factor of three [42]. This formalism applies to $b \to d\gamma$ as well. The corresponding operators $O_i$ are also given in Ref. [6]. The matching conditions $C_i(\mu_0)$ and the solutions of the RG equations, yielding $C_i(\mu_b)$, coincide with those needed for the $b \to s\gamma$ process.

The inclusive branching ratio $\mathcal{B}(\overline{B} \to X_s\gamma)$ was first measured by the CLEO collaboration in 1995 [43]. Since then, it has also been measured by the *BABAR* [44], CLEO [45], Belle [46] and ALEPH [47] collaborations. These measurements were averaged in 2003 [4] to yield

$$\mathcal{B}(\overline{B} \to X_s\gamma) = (3.48 \pm 0.36) \times 10^{-4} \, . \tag{2.3}$$

Recently, Belle [48] has reported an inclusive measurement of the photon energy spectrum in $B \to X_s\gamma$ in the photon energy interval 1.8 GeV $\leq E_\gamma^* \leq$ 2.8 GeV in the center-of-mass frame. The Belle $E_\gamma$-spectrum is similar to the one measured by CLEO [45], but the precision of the branching ratio $\mathcal{B}(B \to X_s\gamma)$ in the Belle measurement is less affected by the theoretical errors, as the fraction of events satisfying the Belle cut is estimated to be about 95%. Correcting for this, the inclusive branching ratio is [48]:

$$\mathcal{B}(\overline{B} \to X_s\gamma) = (3.59 \pm 0.32^{+0.30}_{-0.31}\ ^{+0.11}_{-0.07}) \times 10^{-4}\,, \tag{2.4}$$

where the errors are statistical, systematic, and theoretical. The measurements (2.3) and (2.4) are to be compared with the Standard Model calculations to NLO accuracy, obtained using the $\overline{\mathrm{MS}}$ scheme for the quark masses [49, 50], and the pole quark masses [51], respectively

$$\mathcal{B}(\overline{B} \to X_s\gamma)|_{\overline{\mathrm{MS}}} \;=\; (3.73 \pm 0.30) \times 10^{-4}, \tag{2.5}$$

$$\mathcal{B}(\overline{B} \to X_s\gamma)|_{\mathrm{pole-quark\ mass}} \;=\; (3.35 \pm 0.30) \times 10^{-4}. \tag{2.6}$$

The theoretical uncertainty in the branching ratio from the scheme dependence of the quark masses $\Delta\mathcal{B}(\overline{B} \to X_s\gamma) = 0.38 \times 10^{-4}$ is larger than the rest of the parametric uncertainty. The difference between the two theoretical branching ratios is actually a NNLL effect; removing this uncertainty requires a full NNLO calculation. Parts of these contributions incorporating the two-loop matching of the operators $O_1 - O_6$ [52], the fermionic NNLL corrections in $b \to sg$ [53] and three-loop matching of the dipole operators $O_7$ and $O_8$ for $b \to s\gamma$ and $b \to sg$ [54] are already available, but the crucial part resulting from the three-loop corrections to the matrix element of the four-quark operators $O_1$ and $O_2$ remains to be done. Of course, there are also other contributions that are still missing in this order. At a Super $B$ Factory, the experimental errors on $\mathcal{B}(\overline{B} \to X_s\gamma)$ will be reduced from the present $\pm 10\%$ to a few per cent. Hence, there is a strong motivation to reduce the theoretical errors on the Standard Model-based estimates as well. While this will take a while, another estimate in the NLO accuracy (with updated input parameters) is suggested by Hurth, Lunghi and Porod [55], by using the ratio $m_c/m_b = 0.23^{+0.08}_{-0.05}$, where the asymmetric errors cover the current dispersion in the value of this ratio in the two quark mass schemes being discussed, yielding:

$$\mathcal{B}(\overline{B} \to X_s\gamma) = (3.79^{+0.36}_{-0.53}) \times 10^{-4}\ . \tag{2.7}$$

Thus, whether Eq. (2.5), (2.6) or (2.7) is used for the NLO Standard Model-based estimate, within the experimental and theoretical errors, the Standard Model agrees well with the present measurements (2.3) and (2.4). This quantitative agreement allows very stringent constraints to be placed on the parameters of a theory beyond the Standard Model, such as supersymmetry [56, 57, 22].

Concerning the determination of the CKM factor $\lambda_t$ from the $b \to s\gamma$ decay, we note that when the theoretical result is reevaluated without use of the CKM unitarity in the dominant contributions (*i.e.,* everywhere except for three small ($< 2.5\%$) corrections), a comparison with the experiment leads to the following constraint on the CKM matrix elements [6]:

$$|\ 1.69\ \lambda_u\ +\ 1.60\ \lambda_c\ +\ 0.60\ \lambda_t\ |\ \ =\ (\ 0.94\ \pm\ 0.07\ )\ |V_{cb}|. \tag{2.8}$$

After using the numerical values of $\lambda_c \simeq |V_{cb}| = (41.0 \pm 2.1) \times 10^{-3}$ and $\lambda_u$ from the PDG [2], this equation yields [6]:

$$\lambda_t = V_{tb}V_{ts}^* \simeq -(47.0 \pm 8.0) \times 10^{-3}, \tag{2.9}$$

corresponding to a precision of about 17%. This is consistent with the unitarity relation $\lambda_c \simeq -\lambda_t$. Its accuracy will improve at a Super $B$ Factory, providing a determination of $\lambda_t$, hence of $V_{ts}$, to an accuracy of better than 10%, limited essentially by theoretical errors.

Contrary to $\mathcal{B}(\overline{B} \to X_s\gamma)$, a measurement of the branching ratio $\mathcal{B}(\overline{B} \to X_d\gamma)$, would provide us with useful constraints on the Wolfenstein parameters $\overline{\rho}$ and $\overline{\eta}$ [39]. To get the theoretical estimate of the isospin-averaged branching ratio $\langle\mathcal{B}(B \to X_d\gamma)\rangle$, one calculates the ratio of the branching ratios $\langle\mathcal{B}(B \to X_d\gamma)\rangle/\langle\mathcal{B}(B \to X_s\gamma)\rangle$. Then, using the central values of the CKM parameters $(A, \overline{\rho}, \overline{\eta}) = (0.82, 0.22, 0.35)$ and $\langle\mathcal{B}(B \to X_s\gamma)\rangle = 3.5 \times 10^{-4}$, implies $\langle\mathcal{B}(B \to X_d\gamma)\rangle \simeq 1.3 \times 10^{-5}$ in the Standard Model. Thus, with $\mathcal{O}(10^8)$ $B\overline{B}$ events collected so far at the $B$ Factories, $\mathcal{O}(10^3)$ $B \to X_d\gamma$ decays have already been produced. However, as discussed elsewhere in this Proceedings, extracting them from the background remains a non-trivial issue. Hence, no limits on the branching ratio for the inclusive decay $B \to X_d \gamma$ are currently available.

Apart from the total branching ratios, the inclusive decays $\overline{B} \to X_{s(d)}\gamma$ provide us with other observables that might be useful for the CKM phenomenology. First, the $\overline{B} \to X_s\gamma$ photon spectrum, in particular the moments of the photon energy, enable to extract the HQET parameters $\lambda_1$ (kinetic energy of the $b$ quark) and $\overline{\Lambda}$ (mass difference $m_B - m_b$) that are crucial for the determination of $V_{ub}$ and $V_{cb}$. Second, $CP$-asymmetries contain information on the CKM phase. These asymmetries can be either direct (*i.e.*, occur in the decay amplitudes) or induced by the $B^0\overline{B}^0$ mixing.

The mixing-induced $CP$ asymmetries in $\overline{B} \to X_{s(d)}\gamma$ are very small ($\mathcal{O}(m_{s(d)}/m_b)$) in the Standard Model, so long as the photon polarizations are summed over. It follows from the structure of the dominant operator $O_7$ in the Standard Model that the photons produced in the decays of $B$ and $\overline{B}$ have opposite *circular* polarizations. Thus, in the absence of New Physics, observation of the mixing-induced $CP$ violation would require selecting particular *linear* photon polarization with the help of matter-induced photon conversion into $e^+e^-$ pairs [58]. Theoretical prospects for measuring the photon polarization in $\overline{B} \to X_s\gamma$ are discussed by Pirjol in Section 2.10.

The Standard Model predictions for the direct $CP$ asymmetries are [39]

$$A_{CP}(\overline{B} \to X_s\gamma) \equiv \frac{\Gamma(\overline{B} \to X_s\gamma) - \Gamma(B \to X_{\overline{s}}\gamma)}{\Gamma(\overline{B} \to X_s\gamma) + \Gamma(B \to X_{\overline{s}}\gamma)} \simeq 0.27\,\lambda^2\overline{\eta} \sim 0.5\%, \tag{2.10}$$

$$A_{CP}(\overline{B} \to X_d\gamma) \equiv \frac{\Gamma(\overline{B} \to X_d\gamma) - \Gamma(B \to X_{\overline{d}}\gamma)}{\Gamma(\overline{B} \to X_d\gamma) + \Gamma(B \to X_{\overline{d}}\gamma)} \simeq \frac{-0.27\,\overline{\eta}}{(1-\overline{\rho})^2 + \overline{\eta}^2} \sim -13\%, \tag{2.11}$$

where $\overline{\rho} = 0.22$ and $\overline{\eta} = 0.35$ have been used in the numerical estimates. As stressed in Ref. [39], there is considerable scale uncertainty in the above predictions, which would require one-loop corrections to the existing theoretical results. The smallness of $A_{CP}(\overline{B} \to X_s\gamma)$ is caused by three suppression factors: $\lambda_u/\lambda_t$, $\alpha_s/\pi$ and $m_c^2/m_b^2$. Recent updates given in [55] are compatible with the earlier estimates [39]. The Standard Model predictions (2.10) and the ones given in [55] are consistent with the (currently most stringent) bound on this quantity from the Belle collaboration [59]:

$$-0.107 < A_{CP}(\overline{B} \to X_s\gamma) < +0.099 \qquad \text{at 90\% C.L.}, \tag{2.12}$$

and rule out any sizable direct $CP$ asymmetry in this decay mode. The search for a weak phase in the $B \to X_s\gamma$ transition will be set forth at a Super $B$ Factory with sensitivities of a few percent.

### 2.2.3 Exclusive radiative $B$ Meson decays

The effective Hamiltonian acting between the $B$ meson and a single-meson state (say, $K^*$ or $\rho$ in the transitions $B \to (K^*, \rho)\,\gamma$) can be expressed in terms of matrix elements of bilinear quark fields inducing heavy-light transitions. These matrix elements are dominated by strong interactions at small momentum transfer and cannot be calculated perturbatively. They have to be obtained from nonperturbative methods, such as the lattice-QCD and QCD sum rules. As the inclusive branching ratio $\mathcal{B}(B \to X_s\gamma)$ in the Standard Model is in striking agreement with data, the role of the branching ratio $\mathcal{B}(B \to K^*\gamma)$ is that it will teach us a lot about the QCD dynamics, such as the behavior of the perturbation series in $\alpha_s$ and $1/m_b$, quantitative tests of the factorization properties of the $B \to K^*\gamma$ hadronic matrix element, and the form factor governing the electromagnetic penguin transition, $T_1^{K^*}(0)$. Moreover, the Standard Model can be tested precisely through the isospin and $CP$ violations in the decay rates.

In the following, we focus on the exclusive decay $B \to K^*\gamma$. The discussion of the $B \to (\rho, \omega)\gamma$ modes is presented in Section 2.8. In Table 2.2.3 we present all the available experimental measurements on $B \to K^*\gamma$ decays from CLEO [63], Belle [61] and *BABAR* [62], with the current averages taken from [4]. These are to be compared with the theoretical calculations for the branching ratios calculated in the next-to-leading order [31, 32, 64] using the QCD-factorization framework [65]. An updated analysis based on [31] (neglecting a small isospin violation in the decay

**Table 2-1.** *The Standard Model-based predictions for branching ratios, isospin-violating ratio and $CP$ asymmetry for the decays $B \to K^*\gamma$ and comparison with the BABAR and Belle data.*

| Observable | Theory (SM) | Experiment |
|---|---|---|
| $\mathcal{B}(B^0 \to K^{*0}\gamma)$<br>$\mathcal{B}(B^\pm \to K^{*\pm}\gamma)$ | $(6.9 \pm 2.9) \times 10^{-5}$ [35] | $(3.97 \pm 0.21) \times 10^{-5}$ [4]<br>$(4.06 \pm 0.27) \times 10^{-5}$ [4] |
| $\Delta_{0+}(B \to K^*\gamma)$ | $(8^{+2}_{-3})\%$ [60] | $(3.4 \pm 4.4 \pm 2.6 \pm 2.5)\%$ [61]<br>$(5.1 \pm 4.4 \pm 2.3 \pm 2.4)\%$ [62] |
| $\mathcal{B}(\overline{B} \to X_s\gamma)$ | $(3.79^{+0.36}_{-0.53}) \times 10^{-4}$ [55], | $(3.48 \pm 0.36) \times 10^{-4}$ [4] |
| $A_{CP}(B \to K^*\gamma)$ | $< 0.5\%$ | $(-1.4 \pm 4.4 \pm 1.2)\%$ [61] |

widths) yields [35]:

$$\mathcal{B}(B \to K^*\gamma) = (6.9 \pm 0.9) \times 10^{-5} \left(\frac{\tau_B}{1.6 \text{ ps}}\right) \left(\frac{m_{b,\text{pole}}}{4.65 \text{ GeV}}\right)^2 \left(\frac{T_1^{K^*}(0, \overline{m}_b)}{0.38}\right)^2 = (6.9 \pm 2.9) \times 10^{-5}\,, \quad (2.13)$$

where the default value for the form factor $T_1^{K^*}(0, \overline{m}_b)$ is taken from the LC-QCD sum rules [66] and the pole mass $m_{b,\text{pole}} = (4.65 \pm 0.10)$ GeV is the one loop-corrected central value obtained from the $\overline{\text{MS}}$ $b$-quark mass $\overline{m}_b(m_b) = (4.26 \pm 0.15 \pm 0.15)$ GeV in the PDG review [2]. Using the ratio

$$R(K^*\gamma/X_s\gamma) \equiv \frac{\mathcal{B}(B \to K^*\gamma)}{\mathcal{B}(B \to X_s\gamma)} = 0.117 \pm 0.012\,, \quad (2.14)$$

the agreement between the QCD-factorization-based estimates and the data requires $T_1^{K^*}(0, \overline{m}_b) \simeq 0.27 \pm 0.02$. The allowed phenomenological values of $T_1^{K^*}(0, \overline{m}_b)$ are about $25\%$ below the current estimates of the same from the LC-QCD approach $T_1^{K^*}(0, \overline{m}_b) = 0.38 \pm 0.05$.

Attempts to bridge the factorization-based theory and experiment in $B \to K^*\gamma$ decays are underway. Along this direction, SU(3)-breaking effects in the $K$- and $K^*$-meson light-cone distribution amplitudes (LCDA's) have recently been re-estimated by Ball and Boglione [67]. This modifies the input value for the Gegenbauer coefficients in the $K^*$-LCDA and the contribution of the hard spectator diagrams in the decay amplitude for $B \to K^*\gamma$ is reduced, decreasing, in turn, the branching ratio by about 7% [35]. The effect of this correction on the form factor $T_1^{K^*}(0, \overline{m}_b)$, as well as of some other technical improvements [67], has not yet been worked out. Updated calculations of this form factor on the lattice are also under way [68], with preliminary results yielding $T_1^{K^*}(0, \overline{m}_b) \sim 0.27$, as suggested by the analysis in (2.14), and considerably smaller than the ones from the earlier lattice-constrained parametrizations by the UKQCD collaboration [69]. Finally, the Sudakov logarithms, due to the presence of an intermediate scale of $\mathcal{O}(\sqrt{\Lambda_{\text{QCD}} M_B})$ characterizing the virtuality of a nested gluon in the calculation of the matrix element in $B \to K^*\gamma$, have recently been resummed to all orders of the perturbation theory in the phenomenologically significant chromomagnetic operator $O_8$ [70]. The resummation effects decrease the matrix element $\langle K^*\gamma|O_8|B\rangle$ by about 4% and hence are not sufficient by themselves to bring down the Sudakov-improved theoretical branching ratio by the required amount. Understanding the experimental decay rates for $B \to K^*\gamma$ remains an open theoretical problem.

Other important observables are the $CP$ asymmetry $A_{CP}(B \to K^*\gamma)$ and the isospin-violating ratio

$$\Delta_{0+} \equiv \frac{\Gamma(B^0 \to K^{*0}\gamma) - \Gamma(B^+ \to \overline{K}^{*+}\gamma)}{\Gamma(B^0 \to K^{*0}\gamma) + \Gamma(B^+ \to \overline{K}^{*+}\gamma)}\,. \quad (2.15)$$

Experimental results and theoretical predictions are summarized in Table 2.2.3. The determinations of the isospin-violating ratios are consistent with the Standard Model-based estimate and rule out any significant isospin breaking in the respective decay widths, anticipated in some beyond-the-Standard Model scenarios. Likewise, the $CP$ asymmetry in $B \to K^*\gamma$ decays, which in the Standard Model is expected to be of the same order of magnitude as for the inclusive decay (2.10) and ([55]), is completely consistent with the present experimental bounds. At a Super $B$ Factory, this $CP$ asymmetry can be probed at the level of 2%, with $\mathcal{O}(1)$ (ab)$^{-1}$ data, which would still not probe the Standard Model expectation $< 0.5\%$, but would provide sensitivity to the presence of new weak phases, for example, in supersymmetric theories [71].

### 2.2.4 Semileptonic decays $b \to s\ell^+\ell^-$ and $B \to (K, K^*)\ell^+\ell^-$

First measurements of the inclusive semileptonic decays $B \to X_s\,\ell^+\ell^-$ and some exclusive decay modes such as $B \to (K, K^*)\,\ell^+\ell^-$ have already been made by the *BABAR* and Belle experiments at the $B$ Factories at SLAC and KEK. Below, we review the phenomenology of these decays and quantify the *rapport* between the experiments and the Standard Model.

The theoretical framework to study the semileptonic decays is the same as that of the radiative decays, namely the effective Hamiltonian approach, where the operator basis has to be extended to include the four-Fermi semileptonic operators. In the context of the Standard Model, there are two such operators, called $\mathcal{O}_9$ and $\mathcal{O}_{10}$:

$$O_9 = \frac{e^2}{g_s^2}\,(\bar{s}_L\gamma_\mu b_L)\sum_\ell \underbrace{(\bar{\ell}\gamma^\mu\ell)}_{\text{V}}, \qquad O_{10} = \frac{e^2}{g_s^2}\,(\bar{s}_L\gamma_\mu b_L)\sum_\ell \underbrace{(\bar{\ell}\gamma^\mu\gamma_5\ell)}_{\text{A}}, \tag{2.16}$$

with their associated Wilson coefficients $C_9(\mu)$ and $C_{10}(\mu)$. Here, $e$ and $g_s$ are the electromagnetic and strong coupling constants,respectively, and $L(R)$ stands for the left (right) chiral projection. In the semileptonic decays being considered here, there is a strong contribution whose presence is due to the long-distance resonant amplitude $B \to (X_s, K, K^*)\,(J/\psi, \psi^*, \ldots) \to (X_s, K, K^*)\,\ell^+\ell^-$. This contribution, which can be modeled in terms of the Breit-Wigner functions for the resonances [11] or calculated in terms of a dispersion relation [72], can be essentially removed by putting a cut on the invariant dilepton mass near the resonance mass $s = (p_{\ell^+} + p_{\ell^-})^2 = m_{J/\psi}^2, m_{\psi'}^2, \ldots$. We shall assume that this can be done quite efficiently in the ongoing and planned experiments. However, the non-resonant $c\bar{c}$ contribution, entering through the so-called charm penguins, remains. This is included in the calculations of the $b \to s\ell^+\ell^-$ matrix elements which contains, in addition, the short-distance part of the amplitude from the (virtual) top quark.

There are two quantities of principal experimental interest: (i) The dilepton invariant mass (DIM) spectrum, and (ii) the forward-backward (FB) charge asymmetry $\overline{A}_{\rm FB}(s)$ [11]. The current $B$ Factory experiments yield information only on the DIM-spectrum; the measurement of the forward-backward asymmetry will be undertaken, in all likelihood, at the Super $B$ Factory and in experiments at the hadron colliders, such as the LHC$b$ and $B$TeV. Both of these measurements are needed to test the Standard Model precisely in the electroweak sector and to determine the effective Wilson coefficients. In what follows, we first summarize the main theoretical developments in calculating the rates and distributions in the inclusive $B \to X_s\ell^+\ell^-$ decay, and then discuss the current measurements. The exclusive $B \to (K, K^*)\,\ell^+\ell^-$ decays are then reviewed and the Standard Model-based estimates compared with the current data.

The lowest order calculation of the DIM-spectrum in the $B \to X_s\ell^+\ell^-$ decay was performed in Ref. [73] in the context of the Standard Model and its minimal extension to the case with two Higgs doublets. In this order, the one-loop matrix element of the operator $O_9$ depends on the renormalization scheme. This scheme-dependence is removed by calculating the NLL corrections to the anomalous dimension matrix (hence the Wilson coefficient $C_9$). The $\mathcal{O}(\alpha_s)$ perturbative corrections to the matrix elements of the operator $O_9$ were calculated in Ref. [74]. Inclusion of the matching conditions at the NLL level reduced the scale-dependence in the top quark mass $(\mu_W)$ in the DIM-spectrum to about $\pm 16\%$ [75, 76].

The next step was to implement the leading power corrections in $\Lambda_{\rm QCD}/m_b$ to enable a transition from the partonic decay rates and distributions calculated for the $b \to s\ell^+\ell^-$ process to the corresponding rates and distributions in the hadronic $B \to X_s\ell^+\ell^-$ decay. This was done using the operator product expansion (OPE) and the heavy quark effective theory (HQET) [77]. The first $\mathcal{O}(\Lambda_{\rm QCD}^2/m_b^2)$ corrections to the decay rate and the DIM-spectrum were undertaken in Ref. [78]; these were subsequently rederived and corrected in Ref. [79]. The corresponding power corrections to the FB-asymmetry in the $B \to X_s\ell^+\ell^-$ decay were also first calculated in Ref. [79], and those in the hadron mass $(X_s)$ and hadron moments were derived in Refs. [80, 81]. The power corrections in the decay rate and the DIM-spectrum were extended to include the $\mathcal{O}(\Lambda_{\rm QCD}^3/m_b^3)$ corrections in Ref. [82], while the $\mathcal{O}(\Lambda_{\rm QCD}^2/m_c^2)$ power corrections to the DIM-spectrum and the FB-asymmetry due to the intermediate charm quark were calculated in Ref. [25] using the HQET approach. Of these, the corrections up to $\mathcal{O}(\Lambda_{\rm QCD}^2/m_b^2)$ and $\mathcal{O}(\Lambda_{\rm QCD}^2/m_c^2)$ have been implemented in the analysis of data on $B \to X_s\ell^+\ell^-$ discussed later.

In the recent past, several steps in the re-summation of the complete NNLL QCD logarithms have been undertaken. The counting is such that this corresponds to the calculation of the complete $\mathcal{O}(\alpha_s)$ corrections in this process. They are itemized below:

- Two loop $\mathcal{O}(\alpha_s^2)$ matching corrections to the Wilson coefficients $C_i(M_W)$ were obtained in Ref. [52]. They reduced the $\mu_W$-dependence discussed above but the decay rate remained uncertain by $\pm 13\%$ due to the lower scale $(\mu_b)$-dependence.
- Two loop $\mathcal{O}(\alpha_s^2)$ matrix element calculations, yielding $\langle O_{1,2}(m_b)\rangle$, were obtained in Ref. [83, 84]. With their inclusion, the lower scale $(\mu_b)$-dependence in the DIM-spectrum was reduced to $\pm 6\%$.
- Two loop $\mathcal{O}(\alpha_s^2)$ matrix element calculations yielding $\langle O_9(m_b)\rangle$, were obtained in Ref. [85]. Dominant $\mathcal{O}(\alpha\alpha_s)$ effects up to NNLL were also calculated in this paper.

The only missing piece is the two loop $\mathcal{O}(\alpha_s^2)$ calculations of the matrix elements $\langle O_{3-6}(m_b)\rangle$. However, their Wilson coefficients are too small to have any appreciable effect in the decay rates and distributions.

This work has been put to good use in calculating the DIM-spectrum in the NNLL accuracy for $\hat{s} = s/m_b^2 < 0.25$ [83, 84], which has been recently confirmed and extended to the full DIM-spectrum [86]. The FB-asymmetry in $B \to X_s\ell^+\ell^-$ to NNLL accuracy has also been recently completed [87, 88].

Taking into account the various input parametric uncertainties, the branching ratios for $B \to X_s\, e^+e^-$, $B \to X_s\,\mu^+\mu^-$ and $B \to X_s\,\ell^+\ell^-$, which is the average over $e^+e^-$ and $\mu^+\mu^-$, are given in Table 2-2. Note that the inclusive measurements from Belle and *BABAR*, as well as the Standard Model rates, include a cut on the dilepton invariant mass $M_{\ell^+\ell^-} > 0.2$ GeV. Within the current experimental and theoretical uncertainties, there is good agreement between the Standard Model-based estimates and data. At the Super $B$ Factory, the DIM-spectrum will be measured precisely, which will provide information on the possible contribution from physics beyond the Standard Model.

As with the DIM-spectrum, the NNLL contributions stabilize the scale $(= \mu_b)$-dependence of the forward-backward asymmetry; the small residual parametric dependence is dominated by $\delta(m_c/m_b)$ for $\hat{s} = 0$ [88]

$$A_{\rm FB}^{\rm NLL}(0) = -(2.51 \pm 0.28) \times 10^{-6}; \qquad A_{\rm FB}^{\rm NNLL}(0) = -(2.30 \pm 0.10) \times 10^{-6}. \tag{2.17}$$

Apart from the FB-asymmetry $A_{\rm FB}(\hat{s})$, the FB-asymmetry zero $A_{\rm FB}(\hat{s}_0) = 0$ is a precise test of the Standard Model, correlating $\widetilde{C}_7^{\rm eff}$ and $\widetilde{C}_9^{\rm eff}$. Inclusion of the NNLL corrections causes a significant shift in $\hat{s}_0^{\rm NLL}$ [88, 87] and the resulting theoretical error is around $5\%$. Detailed studies of the FB-asymmetry in the decay $B \to X_s\ell^+\ell^-$ will be undertaken at a Super $B$ Factory. This is a precision test of the Standard Model and may reveal possible New Physics.

There are, as yet, no measurements of the direct $CP$ asymmetries in the rate for $B \to X_{(s,d)}\ell^+\ell^-$ decays. Theoretical studies have been done (see Refs. [89] and [90]), the latter in the NNLL approximation. The Standard Model predicts direct $CP$ violation for $b \to s$ transitions to be tiny, due to the double Cabibbo suppression of the weak phase, hence there is room for New Physics effects. These asymmetries can be searched for with sufficient statistics, *i.e.*, at a Super $B$ Factory. The ratio of $B \to X_d\ell^+\ell^-$ and $B \to X_s\ell^+\ell^-$ rates can also be used to extract $|V_{td}/V_{ts}|$ [89].

**Table 2-2.** *The Standard Model-based predictions from Ref. [10] for the branching ratios of the decays $B \to (K, K^*, X_s)\,\ell^+\ell^-$ and comparison with the BABAR and Belle data, both in units of $10^{-6}$. Experimental averages are taken from [4]. For the inclusive modes, both data and theory require $m_{\ell^+\ell^-} > 0.2$ GeV.*

| Decay mode | Theory (SM) | Expt. (Belle & *BABAR*) |
|---|---|---|
| $B \to K\ell^+\ell^-$ | $0.35 \pm 0.12$ | $0.55 \pm 0.08$ |
| $B \to K^* e^+e^-$ | $1.58 \pm 0.52$ | $1.25 \pm 0.39$ |
| $B \to K^* \mu^+\mu^-$ | $1.2 \pm 0.4$ | $1.19 \pm 0.31$ |
| $B \to X_s \mu^+\mu^-$ | $4.15 \pm 0.70$ | $7.0 \pm 2.1$ |
| $B \to X_s e^+e^-$ | $4.2 \pm 0.70$ | $5.8 \pm 1.8$ |
| $B \to X_s \ell^+\ell^-$ | $4.18 \pm 0.70$ | $6.2 \pm 1.5$ |

The exclusive decays $B \to K\ell^+\ell^-$ and $B \to K^*\ell^+\ell^-$ have already been measured by *BABAR* and Belle. Their branching ratios are given in Table 2-2 together with the theoretical estimate in the Standard Model, calculated using the form factors from the LC-QCD sum rule approach [9]. Within current errors there is agreement between the Standard Model and experiments. This comparison will become very precise at a Super $B$ Factory. Future high luminosity measurements will also access the forward-backward asymmetry in $B \to K^*\ell^+\ell^-$ and search for its zero, very similar to the case of inclusive decays.

Form factors are the biggest source of theory error in the description of exclusive semileptonic decays. Effective field theory tools and SU(3) relations with $B \to (\pi, \rho)\ell\nu_\ell$ decays (once they are precisely measured) greatly improve the theoretical precision, at least in some kinematic range, *e.g.*, the low dilepton mass for LEET/SCET relations or in specific observables. For example, the position of the zero of the forward-backward asymmetry in $B \to K^*\ell^+\ell^-$ decays is insensitive to hadronic effects and its experimental study can distinguish between the Standard Model and physics beyond the Standard Model.

## 2.3 Theoretical tools

≻— T. Hurth and E. Lunghi —≺

The effective field theory approach serves as a theoretical framework for both inclusive and exclusive modes. The standard method of the operator product expansion (OPE) allows for a separation of the $B$ meson decay amplitude into two distinct parts, the long-distance contributions contained in the operator matrix elements and the short-distance physics described by the so-called Wilson coefficients. The latter do not depend on the particular choice of the external states. New physics can manifest itself only by changing the numerical values of these coefficients or introducing new operators. Within the OPE, all particles with mass larger than the factorization scale (in the Standard Model, these are the $W$ boson and the top quark) are integrated out, *i.e.,* removed from the theory as dynamical fields.

In the following, we discuss, as an example, the theoretical framework for $b \to s/d\,\gamma$ transitions. These theoretical tools are also used in all other rare decays, with specific modifications.

The effective Hamiltonian for radiative $b \to s/d\gamma$ transitions in the Standard Model can be written as

$$\mathcal{H}_{eff} = -\frac{4G_F}{\sqrt{2}} \left[ \lambda_q^t \sum_{i=1}^{8} C_i(\mu) O_i(\mu) + \lambda_q^u \sum_{i=1}^{2} C_i(\mu)(O_i(\mu) - O_i^u(\mu)) \right] \tag{2.18}$$

where $\mathcal{O}_i(\mu)$ are dimension-six operators at the scale $\mu \sim \mathcal{O}(m_b)$; $C_i(\mu)$ are the corresponding Wilson coefficients. Clearly, only in the sum of Wilson coefficients and operators, within the observable $\mathcal{H}$, does the scale dependence cancels out. $G_F$ denotes the Fermi coupling constant and the explicit CKM factors are $\lambda_q^t = V_{tb}V_{tq}^*$ and $\lambda_q^u = V_{ub}V_{uq}^*$. The unitarity relations $\lambda_q^c = -\lambda_q^t - \lambda_q^u$ were already used in (2.18).

The operators can be chosen as (we only write the most relevant ones):

$$O_2 = (\bar{s}_L \gamma_\mu c_L)(\bar{c}_L \gamma^\mu b_L)\,, \tag{2.19}$$

$$O_2^u = (\bar{s}_L \gamma_\mu u_L)(\bar{u}_L \gamma^\mu b_L)\,, \tag{2.20}$$

$$O_7 = e/g_s^2 m_b (\bar{s}_L \sigma^{\mu\nu} b_R) F_{\mu\nu}\,, \tag{2.21}$$

$$O_8 = 1/g_s m_b (\bar{s}_L \sigma^{\mu\nu} T^a b_R) G^a_{\mu\nu}\,, \tag{2.22}$$

where the subscripts $L$ and $R$ refer to left- and right- handed components of the fermion fields. In $b \to s$ transitions the contributions proportional to $\lambda_s^u$ are rather small, while in $b \to d$ decays $\lambda_d^u$ is of the same order as $\lambda_d^t$ and they play an important role in $CP$ and isospin asymmetries (for a complete list of operators see [41]).

While the Wilson coefficients $C_i(\mu)$ enter both inclusive and exclusive processes and can be calculated with perturbative methods, the calculational approaches to the matrix elements of the operators differ in both cases. Within inclusive modes, one can use quark-hadron duality in order to derive a well-defined heavy mass expansion (HME) of the decay rates in powers of $\Lambda_{\rm QCD}/m_b$. In exclusive processes, however, one cannot rely on quark-hadron duality and has to face the difficult task of estimating matrix elements between meson states, which leads to large theoretical uncertainties in spite of recent developments such as the method of the QCD-improved factorization and the soft collinear effective theory. The latter methods, in general, do not allow a quantification of the important $1/m_b$ corrections to the heavy quark limit.

In the inclusive modes, the hadronic matrix elements are dominated by the partonic contributions. Bound state effects of the final states are eliminated by averaging over a specific sum of hadronic states. Moreover, long-distance effects of the initial state are also accounted for, through the heavy mass expansion in which the inclusive decay rate of a heavy $B$ meson is calculated using an expansion in inverse powers of the $b$ quark mass. In particular, it turns out that the decay width of the $\overline{B} \to X_s\gamma$ is well-approximated by the partonic decay rate, which can be calculated in renormalization-group-improved perturbation theory:

$$\Gamma(\overline{B} \to X_s\gamma) = \Gamma(b \to X_s^{\rm parton}\gamma) + \Delta^{\rm nonpert.} \tag{2.23}$$

Non-perturbative effects, $\Delta^{nonpert.}$, are suppressed by inverse powers of $m_b$ and are well under control thanks to the Heavy Mass Expansion (HME); they can be further estimated through the application of the Heavy Quark Effective Theory (HQET). In fact, the optical theorem relates the inclusive decay rate of a hadron $H_b$ to the imaginary part of certain forward scattering amplitudes

$$\Gamma(H_b \to X) = \frac{1}{2m_{H_b}} \Im \langle H_b \mid \mathbf{T} \mid H_b \rangle \,, \tag{2.24}$$

where the transition operator $\mathbf{T}$ is given by $\mathbf{T} = i \int d^4x \, T[H_{\rm eff}(x) H_{\rm eff}(0)]$. It is then possible to construct an OPE of the operator $\mathbf{T}$, which is expressed as a series of *local* operators – suppressed by powers of the $b$ quark mass and written in terms of the $b$ quark field:

$$\mathbf{T} \stackrel{\rm OPE}{=} \frac{1}{m_b} \big( \mathcal{O}_0 + \frac{1}{m_b} \mathcal{O}_1 + \frac{1}{m_b^2} \mathcal{O}_2 + ... \big) \,. \tag{2.25}$$

This construction is based on the parton–hadron duality, using the facts that the sum is done over all exclusive final states and that the energy release in the decay is large with respect to the QCD scale, $\Lambda_{\rm QCD} \ll m_b$. With the help of the HQET, namely the new heavy-quark spin-flavor symmetries arising in the heavy quark limit $m_b \to \infty$, the hadronic matrix elements within the OPE, $\langle H_b \mid \mathcal{O}_i \mid H_b \rangle$, can be further simplified. The crucial observations within this well-defined procedure are the following: the free quark model turns out to be the first term in the expansion constructed in powers of $1/m_b$, and therefore the dominant contribution. This contribution can be calculated in perturbative QCD. Second, in the applications to inclusive rare $B$ decays one finds no correction of order $1/m_b$ to the free quark model approximation, and the corrections to the partonic decay rate start with $1/m_b^2$ only. The latter fact implies a rather small numerical impact of the nonperturbative corrections to the decay rate of inclusive modes.

The dominant perturbative contributions to the $\overline{B} \to X_{s,d}\gamma$ decay rate are based on the following three calculational steps: as the heavy fields are integrated out, the top and $W$-mass dependence is contained in the initial conditions of the Wilson coefficients $C_i(\mu)$, determined by a matching procedure between the full and the effective theory at the high scale $\mu \sim m_W$ (Step 1). By means of RG equations, the $C_i(\mu)$ are then evolved to the low scale $\mu \sim m_b$ (Step 2). Finally, the corrections to the matrix elements of the operators are evaluated at the low scale (Step 3). The dominant short-distance QCD corrections enhance the partonic decay rate by a factor of more than 2 and lead to large logarithms of the form $\log(m_b/m_W)$.

In the context of exclusive decays, we face the difficult task of estimating matrix elements between meson states. The naive approach to the computation of exclusive amplitudes consists in writing the amplitude $A \simeq C_i(\mu_b)\langle O_i(\mu_b)\rangle$ and parametrizing $\langle O_i(\mu_b)\rangle$ in terms of form factors.

A promising approach is the method of QCD-improved factorization that has recently been systemized for nonleptonic decays in the heavy quark limit [65, 91]. This method allows for a perturbative calculation of QCD corrections to naive factorization and is the basis for the up-to-date predictions for exclusive rare $B$ decays. However, within this approach, a general, quantitative method to estimate the important $\Lambda_{\rm QCD}/m_b$ corrections to the heavy quark limit is missing. More recently, a more general quantum field theoretical framework for the QCD-improved factorization was proposed - known as soft collinear effective theory (SCET) [92, 93, 94, 95, 96, 97].

Let us consider processes involving the decay of a heavy meson into fast moving light particles ($B \to \gamma e \nu$, $B \to (\rho, K^*)\gamma$, $B \to K\pi$, ...) and indicate with $Q \sim \mathcal{O}(m_b)$ their typical large energy. The idea is to isolate *all* the relevant degrees of freedom necessary to correctly describe the infrared structure of QCD below the scale $Q$ and associate independent fields to each of them. It is possible to identify two distinct *perturbative* modes, called hard ($p^2 \sim Q^2$) and semi-hard ($p^2 \sim \Lambda_{\rm QCD} Q$). These modes are produced, for instance, in interactions of energetic light particles with the heavy quark and the $B$ meson spectator, respectively. These two modes do not appear in the initial and final states and, therefore, must be integrated out. We do not wish to entertain here a comprehensive discussion of the technicalities involved in this step. It will suffice to say that the resulting theory (also called SCET$_{II}$ in the literature) contains only *nonperturbative* degrees of freedom with virtualities $\mathcal{O}(\Lambda_{\rm QCD}^2)$ and that hard and semi-hard modes are reflected in the coefficient functions in front of the operators of that (SCET$_{II}$) theory. We note that these coefficients depend, in general, on energies of order $Q$ and $\Lambda_{\text{QCD}}$. Moreover, the hierarchy $\Lambda_{\text{QCD}} \ll Q$ allows for an expansion in the small parameter $\lambda = \Lambda_{\text{QCD}}/Q$.

Given a process, one has to construct the most general set of ($\text{SCET}_{II}$) operators at a given order in $\lambda$, and show that all the possible gluon exchanges can be reabsorbed, at all orders in perturbation theory, into form factors and meson light-cone wave functions. The resulting amplitude is a convolution of these nonperturbative universal objects with the coefficient functions encoding the contribution of hard and semi-hard modes. Questions regarding the convergence of these convolution integrals lead to a deeper understanding of factorization in SCET. From the analyses presented in Refs. [98, 99, 100] it clearly emerges that the presence of an end-point singularity in the matrix element of a given $\text{SCET}_{II}$ operator signals a breakdown of factorization (technically it was seen that it is not possible to choose an infrared regulator that preserves factorization).

The few form factors that describe the transition $B \to M$ (where $M$ denotes a pseudo-scalar or vector meson) can be written as [101]:

$$F_i^{B\to M} = C_i\, \xi^{B\to M} + \phi_B \otimes T_i \otimes \phi_M + \mathcal{O}\left(\frac{\Lambda}{m_b}\right) \tag{2.26}$$

where $\xi^{B\to M}$ is the so-called non-factorizable (or soft) contribution to the form factors (actually there is one soft form factor for the decay into pseudoscalar meson and two for the decay into vector mesons); $\phi_{B,M}$ are the $B$ and $M$ meson light-cone wave functions; $C_i$ are Wilson coefficients that depend on hard scales; and $T_i$ are perturbative hard scattering kernels generated by integrating out hard and semi-hard modes. In Ref. [102] the factorization formula Eq. (2.26) has been proved at all orders in perturbation theory and at leading order in $\Lambda_{\text{QCD}}/m_b$, using SCET techniques. The strength of Eq. (2.26) is that it allows us to express several independent QCD form factors in terms of only one soft form factor (two in the case of vector mesons) and moments of the light-cone wave functions of the light pseudo-scalar (vector) and $B$ mesons.

Let us now briefly discuss the form of factorization for the decays $B \to V\gamma$ (with $V = K^*,\ \rho$) as a specific example. At leading order, only the operator $O_7$ contributes and its matrix element between meson states is given by an expression similar to (2.26). The choice of using either the full QCD form factor $T^{B\to V}$ or the soft one $\xi_\perp$ is clearly a matter of taste (note that nonperturbative methods, such as lattice-QCD and light-cone QCD sum rules, only give information on the full QCD form factors, and not on the soft contributions alone). The advantage of the QCD-improved factorization approach is evident in the computation of the next-to-leading order (in $\alpha_s$) corrections. In fact, one can show that the matrix elements of the operators $O_2$ and $O_8$, which are expected to contribute at this order, are given by the matrix element of $\mathcal{O}_7$ times a computable hard scattering kernel. Moreover, spectator interactions can be computed and are given by convolutions involving the light-cone wave functions of the $B$ and $V$ mesons. It must be mentioned that light-cone wave functions of pseudo-scalar and vector mesons have been deeply studied using light-cone QCD sum rules methods [103, 104, 105, 106]. On the other hand, not much is known about the $B$ meson light-cone distribution amplitude, whose first negative moment enters the factorized amplitude at NLO. Since this moment enters the factorized expression for the $B \to \gamma$ form factor as well, it might be possible to extract its value from measurements of decays like $B \to \gamma e\nu$, if it can be shown that power corrections are under control [107].

Finally, let us stress that a breakdown of factorization is expected at order $\Lambda_{\text{QCD}}/m_b$ [91, 60, 108]. In Ref. [60], in particular, the authors have shown that in the analysis of $B \to K^*\gamma$ decays at subleading order an infrared divergence is encountered in the matrix element of $\mathcal{O}_8$. Nevertheless, some very specific power corrections might still be computable. Indeed, this is the case for the annihilation and weak exchange amplitudes in $B \to \rho\gamma$ at the one-loop level.

## 2.4 Theoretical uncertainties and possible improvements

≻– Th. Feldmann –≺

Rare $B$ decays are a perfect tool for studying the small non-diagonal entries in the CKM matrix, and to find indirect effects from beyond-the-Standard Model contributions to $b$ decay amplitudes. Present experiments at $B$ Factories have already measured many rare decay modes, from which Standard Model parameters can be extracted and New Physics scenarios can be constrained.

Super $B$ Factoriesare intended to improve the sensitivity to small branching ratios and/or small deviations from Standard Model predictions by a significant increase in luminosity. This naturally raises the question of the extent to which theoretical uncertainties for the observables of interest are under control, such that the gain in luminosity directly translates into improved bounds on electroweak and New Physics parameters. This becomes even more relevant in view of the competition with present and future $B$ physics experiments at hadron colliders and of the possible direct detection of New Physics particles at the LHC.

In the following, we briefly summarize the basic theoretical limitations, and the recent progress that has been achieved on theoretical uncertainties.

### 2.4.1 Perturbative and non-perturbative QCD effects

The main limitations for accurate theoretical predictions of $B$ decay observables come from our incomplete quantitative understanding of strong interaction effects. In practice, the different faces of QCD (perturbative regime at short distances, hadronic effects at large distances) lead to two sources of theoretical uncertainties:

- Truncation of perturbative expansion at some order in the strong coupling constant (including the perturbative summation of large logarithms).
- Dependence on nonperturbative hadronic parameters.

Improvement on the first point is mainly a technical issue related to the practical calculation of partonic processes at high orders in perturbation theory. An important aspect is the systematic separation ("factorization") of short-distance and long-distance QCD dynamics, which can be achieved by operator product expansion (OPE) or effective field theory methods, exploiting the fact that the $b$ quark mass is large with respect to the QCD scale (heavy quark mass expansion). Recently, theoretical progress has been achieved for $B$ decays into light energetic hadrons. The diagrammatic approach to QCD factorization introduced in [65, 91, 109], has been formalized in terms of the so-called "soft-collinear effective theory" (SCET, see *e.g.*, [92, 93, 110, 96, 99, 111]). This provides a well-defined scheme in which to calculate heavy-to-light decay amplitudes in the heavy-quark-mass limit. The numerical relevance of $1/m_b$ power corrections should be considered as a possible quantitative limitation of that approach, at present.

The theoretical description of nonperturbative QCD effects is a more critical point. A main challenge is to provide a reliable estimate of systematic uncertainties for the different theoretical methods at hand: Numerical simulations of QCD correlation functions on space-time lattices have the advantage of calculating hadronic observables "from first principles". In practice, however, with current computers and algorithms, several extrapolations and approximations must be controlled. The most severe approximation is perhaps the neglect of dynamical fermions ("quenched approximation"). Another problem is the simulation of realistic light and heavy quark masses on finite-size lattices and the implementation of chiral symmetry. Recent progress, at least for a certain class of observables, has been reported in [112], where a particular approximation to implement dynamical fermions is proposed.

QCD sum rules are based on the assumption of parton-hadron duality, and provide another useful nonperturbative method to determine hadronic parameters. Error estimates in this approach follow from an empirical analysis of the stability of the predictions with respect to variations of the sum rule parameters (continuum threshold, Borel mass). The discussion of QCD factorization in the framework of SCET has initiated new investigations of exclusive matrix elements in QCD sum rules [113, 114, 115, 116], which may help to reduce the systematic uncertainties for some of the input parameters relevant to rare $B$ decays.

Phenomenological models have the advantage of being based on physical intuition. They often provide a reasonable order-of-magnitude estimate. However, the procedure to assign theoretical uncertainties by considering sufficiently many different models, is not very systematic.

Since, so far, theoretically reliable and numerically accurate estimates for (at least most of) the relevant hadronic input parameters are not available, precision tests with rare $B$ decays require observables that are to a good approximation insensitive to nonperturbative effects.

### 2.4.2 Inclusive decays

From the theoretical point of view, the simplest examples are inclusive decay rates. To first approximation, they can be expressed in terms of partonic rates, including large radiative corrections to the Born-level cross section, which are calculable in perturbative QCD.

Non-perturbative hadronic effects are suppressed by at least two powers of the involved heavy quark masses ($m_b$ or $m_c$). The present strategy is to determine the hadronic parameters from experimental data. Predictive power is obtained by using the OPE and the heavy-quark mass expansion to express the hadronic corrections in terms of HQET (heavy quark effective theory) parameters. In this way nonperturbative corrections in different inclusive decay channels can be related. In practice, experimental analyses involve phase-space cuts, which requires additional (model-dependent) nonperturbative effects to be taken into account.

The typical size of present theoretical uncertainties for the inclusive decays $\overline{B} \to X_s\gamma$ and $\overline{B} \to X_s\ell^+\ell^-$ (for small lepton-pair invariant mass) is about 10%. (To put this number into perspective, we remind the reader that, because of the additional $1/16\pi^2$ suppression factor in loop diagrams, a 10% effect in rare $B$ decays should be compared with $10^{-3}$ accuracy in tree-level electroweak processes.) Since a good part of that uncertainty is of a perturbative nature, it may still be improved by calculating the next order in perturbation theory (at present: next-to-leading logarithmic accuracy for $\overline{B} \to X_s\gamma$, and next-to-next-to leading logarithmic accuracy for $\overline{B} \to X_s\ell^+\ell^-$). This is technically involved, but feasible. In the case of $\overline{B} \to X_s\gamma$ the calculation of higher-order perturbative effects should also resolve the sizable scheme-dependence with respect to the treatment of the charm-quark mass. For more details see [41] and references therein.

### 2.4.3 Exclusive decays—Type I

From the experimental point of view, exclusive decays are simpler to measure than inclusive decays. However, exclusive branching ratios in general depend on hadronic input parameters already to leading approximation. A well-known example is the decay rate for $B \to K^*\gamma$, which involves (among others) a $B \to K^*$ transition form factor that induces theoretical uncertainties of several tens of percent (see [64, 32, 31], and also Section 2.8). We will refer to these types of observables (which also include the rates for semileptonic transitions $B \to \pi$, $B \to \rho$, $B \to \gamma$ *etc.*) as Type I. They are not directly useful for precision tests of flavor parameters. However, the hadronic quantities measured in these decays often provide important input to other decay modes. We should emphasize that some of the hadronic effects in exclusive decays are not "naively" factorizable (*i.e.*, not included in the definition of hadronic decay constants or two-particle transition form factors).

One example for a Type I observable is the first inverse moment $\lambda_B^{-1}$ of the $B$ meson distribution amplitude that could be extracted (with some uncertainty) from the decay $B \to \gamma l\nu$ [117, 118, 107, 119]. This moment, in turn, enters the theoretical predictions for many exclusive heavy-to-light decays in the QCD-factorization approach.

Another example is the $B \to \pi$ form factor, which represents one important nonperturbative ingredient in the calculation of nonleptonic $B \to \pi\pi$ and $B \to \pi K$ decays. In this case, the measurement of the differential decay rate for $B \to \pi l \nu$ should not be considered as a determination of $|V_{ub}|$ (for a given estimate of the $B \to \pi$ form factor ), but rather as a measurement of the $B \to \pi$ form factor (for a given value of $|V_{ub}|$ from other sources).

By comparing experimental and theoretical results for “type-I” observables, one can try to improve the theoretical methods that deal with nonperturbative QCD effects. Eventually, this should lead to an improvement of theoretical uncertainties for other observables as well.

### 2.4.4 Exclusive decays—Type II

Observables with reduced sensitivity to hadronic uncertainties can be obtained from appropriate ratios of decay rates where—to leading approximation—the dependence on hadronic input parameters drops out. The set of possible “type-II” observables can be enlarged by including approximate symmetries of QCD, like isospin, flavor SU(3) or heavy quark symmetries. The theoretical uncertainty for Type II observables, induced by radiative corrections, $1/m_b$ corrections, flavor symmetry corrections *etc.*, is typically of order 15-20%.

Particularly robust predictions can be obtained by considering Type II observables within the QCD-factorization approach. For instance, a well-known phenomenological strategy to extract CKM angles from $B \to \pi\pi$ and $B \to \pi K$ decays is to use flavor symmetries to determine the unknown hadronic effects entering the ratio of penguin and tree amplitudes directly from experimental data. QCD factorization can then be used to quantify the corrections from flavor violation. The related theoretical uncertainties can be estimated in a reliable way, because the neglected effects are suppressed by two small parameters, $(m_s - m_{u,\mathrm{d}})$ and $1/m_b$ (see Section 5 in [109]).

Another prominent example, which is particularly interesting for future $B$ physics experiments, is the forward-backward asymmetry zero in $B \to K^* \ell^+ \ell^-$ (see [120, 9, 64] and Section 2.16). Here the leading dependence on hadronic form factors drops out, thanks to new symmetries [121] in the large-energy limit for the outgoing $K^*$ meson. Another important quantity is the ratio of branching ratios for $B \to K^* \gamma$ and $B \to \rho\gamma$ (see [31, 34, 122] and Section 2.8). It has also been proposed to relate the rare decays $B \to \gamma\gamma$ ($B \to \gamma \ell^+ \ell^-$) and $B \to \gamma l \nu$ (see [118, 123, 124]), or $B \to K \nu \overline{\nu}$ and $B \to K \ell^+ \ell^-$ (see Section 2.20).

The time-dependent *CP* asymmetries in $B^0 \to \phi K^0_S$ and $B^0 \to \eta' K^0_S$ also belong to Type II, since the hadronic uncertainties can be constrained from experimental data in other decay channels using SU(3)-flavor symmetry [125]. Furthermore, in the QCD factorization approach [126] one finds no dynamical mechanism to enhance the CKM-suppressed amplitudes that could be responsible for a *CP* asymmetry in $B^0 \to \phi K^0_S$ and $B^0 \to \eta' K^0_S$ different from that in $B^0 \to J/\psi K^0_S$. In any case, the present discrepancy between the central experimentally measured values for $B^0 \to \phi K^0_S$ and $B^0 \to \eta' K^0_S$ (ignoring the large experimental uncertainties) and the well-understood $B^0 \to J/\psi K^0_S$ decay cannot be explained by QCD effects alone.

In contrast, the observed large branching ratio for $B^0 \to \eta' K^0_S$ (a Type I observable) seems to be in line with theoretical expectations, once the rather large (and uncertain) perturbative and nonperturbative QCD corrections are taken into account [127].

### 2.4.5 Exclusive decays—Type III

Another interesting option is to consider observables that, in the Standard Model, are suppressed by small or tiny coefficients. As a consequence, New Physics contributions to such Type III observables may compete with sizable hadronic uncertainties.

A classic example is the decay $B_s \to \mu^+ \mu^-$. It has an additional suppression factor $m^2_\mu / m^2_b$, which leads to a very small branching ratio compared with other radiative $B$ decays, of the order of $10^{-9}$ in the Standard Model. On the other hand, in New Physics models with enhanced scalar and pseudoscalar $b \to s \ell^+ \ell^-$ operators, it can receive large contributions. While this decay mode is not accessible at $e^+e^-$ $B$ factories, a competitive observable, namely the deviation of the ratio of branching ratios for $B \to K \mu^+ \mu^-$ and $B \to K e^+ e^-$ from unity, can be studied at Super $B$ Factories(see [128, 129] and Section 2.16.3).

**Table 2-3.** *Classification of some important observables in rare exclusive $B$-decays according to their theoretical uncertainties. The third column denotes the main hadronic effect for Type I observables, some of the main sources of theoretical uncertainties for Type II observables, and the additional suppression factors for Type III observables in the Standard Model, respectively.*

| | Decay mode | Observable | Remarks |
|---|---|---|---|
| Type I | $B \to \pi(\rho)\ell\nu$ | diff. branching ratio | transition form factor |
| | $B \to K^*(\rho)\gamma$ | branching ratio | form factor, non-factorizable effects |
| | $B \to K^*(\rho, K)\ell^+\ell^-$ | diff. branching ratio | form factors, non-factorizable effects |
| | $B \to \gamma\ell\nu$ | diff. branching ratio | $\lambda_B^{-1}$ moment, non-factorizable effects |
| | $B^0 \to \eta' K_S^0$ | branching ratio | form factor, non-factorizable effects |
| Type II | $B \to \pi\pi$ *etc.* | branching ratio/$A_{CP}$ | $\mu$ dep., $1/m_b$ corr., $\lambda_B$, $m_c$, ... |
| | $B \to K^*\ell^+\ell^-$ | $A_{\rm FB}$ | $\mu$ dependence, $1/m_b$ corr., $\lambda_B$, ... |
| | $B \to K^*(\rho)\gamma$ | $\mathcal{B}[B \to K^*\gamma]/\mathcal{B}[B \to \rho\gamma]$ | $F^{B\to K^*}/F^{B\to\rho}$, ... |
| | $B^0 \to \eta'(\phi)K_S^0$ | time-dependent $A_{CP}$ | SU(3)$_{\rm F}$ violation, $1/m_b$ corr., ... |
| Type III | $B_s \to \mu^+\mu^-$ | branching ratio | suppressed by $m_\mu^2/m_b^2$ |
| | $B \to K\ell^+\ell^-$ | $1 - \frac{\mathcal{B}[B\to K\mu^+\mu^-]}{\mathcal{B}[B\to Ke^+e^-]}$ | suppressed by $m_\mu^2/m_b^2$ |
| | $B \to K^*\gamma$ | direct $CP$ asymmetry | suppressed by $\lambda_u/\lambda_c$ |

Other interesting Type III observables are the direct $CP$ asymmetry in $B \to K^*\gamma$ which is CKM-suppressed in the Standard Model (see [32]), or exotic channels like $B \to$ invisible (see Section 2.23).

The isospin asymmetry between charged and neutral modes in $B \to K^*\gamma$ and $B \to K^*\ell^+\ell^-$ decays may be considered as Type III. However, to compete with the rather large hadronic uncertainties, one needs New Physics effects with a significant enhancement of penguin coefficients, which should also lead to sizable modifications in nonleptonic decays (see [60, 108] and Section 2.16.2).

## 2.4.6 Remarks

We illustrate the above discussion in Table 2-3. For some observables the classification as Type I, II, or III may not be clear-cut, but rather may depend on one's personal interpretation of the reliability of theoretical approaches, as well as on the kind of New Physics model one is aiming at. In any case, the different strong interaction effects and the uncertainties that they induce have to be taken seriously, if we want to extract precise information about the flavor sector in and beyond the Standard Model from rare $B$ decays. Achieving reasonable theoretical uncertainties, in particular in exclusive decay modes, requires the combined effort of theory and experiment.

## 2.5 Prospects for Inclusive $b \to (s, d)\gamma$ Measurements

>– C. Jessop and J. Libby –<

The prospects for inclusive radiative decay measurements at a Super $B$ Factory are discussed in this section. Three topics are covered: the measurement of the inclusive $b \to s\gamma$ branching fraction, $\mathcal{B}(b \to s\gamma)$, the measurement of the inclusive $b \to d\gamma$ branching fraction, $\mathcal{B}(b \to d\gamma)$, and measurements of direct $CP$ violation. Each section will give a brief review of current measurements, followed by a discussion of how these can be extended, and possibly augmented, in the Super $B$ Factory regime. The measurement of the photon energy spectrum is discussed in Section 4.4.2. of this report.

### 2.5.1 $\mathcal{B}(b \to s\gamma)$

The desire to measure the inclusive decay rate $b \to s\gamma$ arises from the greater accuracy of theoretical predictions compared to exclusive channels. However, the experimental difficulties of inclusive measurements lead to significant systematic uncertainties, that must be controlled. To date there have been several measurements of $\mathcal{B}(b \to s\gamma)$ [43]-[47]; Table 2-4 summarizes measurements made at the $\Upsilon(4S)$ resonance.

**Table 2-4.** *Measurements of $\mathcal{B}(b \to s\gamma)$ made at the $\Upsilon(4S)$ resonance. The first uncertainty is statistical, the second is the experimental systematic and the third is the theoretical systematic. The difference between the 'lepton tag' and 'sum–of–exclusive' BABAR measurements is explained in the text.*

| Experiment | $\mathcal{B}(b \to s\gamma)$ $[\times 10^{-4}]$ |
|---|---|
| BELLE [46] | $3.36 \pm 0.53 \pm 0.42 \pm 0.52$ |
| CLEO [45] | $3.21 \pm 0.43 \pm 0.27^{+0.18}_{-0.10}$ |
| *BABAR* (lepton tag) [130] | $3.88 \pm 0.36 \pm 0.37^{+0.43}_{-0.28}$ |
| *BABAR* (sum–of–exclusive) [44] | $4.3 \pm 0.5 \pm 0.8 \pm 1.3$ |

The principal challenge is selecting the small signal in the presence of large backgrounds from continuum $q\overline{q}$ production and inclusive $B\overline{B}$ production. Figure 2-1(a) shows the signal compared to the backgrounds as a function of the center-of-mass energy of the photon, $E^*_\gamma$. The signal lies beneath a $q\overline{q}$ background which is approximately three orders of magnitude larger. Furthermore, there is a large background from $B\overline{B}$ decays at photon energies below $2.2\ \mathrm{GeV}$. The source of most of the background are asymmetric $\pi^0$ decays. There are also backgrounds from $\eta$, $\omega$, $\eta'$ and $J/\psi$ decays, from hadronic interactions, primarily $\overline{n}$, in the calorimeter, and from electrons produced in semileptonic $B$ decays in which the track is not reconstructed, or is not matched to the electromagnetic cluster. All analyses use photon cluster cuts, $\pi^0$ and $\eta$ vetoes, and shape variables to reduce the backgrounds. The methods for further reduction of background vary among the analyses:

- exclusive reconstruction of the $X_s$ system in different modes containing a $K^\pm$ or a $K^0_S$ with one to three pions [46, 44];
- 'pseudo-reconstruction', which calculates the probability of a detected photon combined with a $K^\pm$ or $K^0_S$ and 1 to 4 pions being consistent with the $B$ meson mass [45]; and
- lepton tagging of the non–signal $B$–decays [45, 130].

In the sum-of-exclusive mode analysis, the remaining background is subtracted using a fit to sidebands; the value of $\mathcal{B}(b \to s\gamma)$ is then calculated by a weighted sum of the results for each mode. The 'pseudo-reconstruction' and lepton tagged methods remove the remaining continuum background using the results of the measurement performed on off-resonance data scaled by $(\mathcal{L}/s)_{\rm on}/(\mathcal{L}/s)_{\rm off}$, where $\mathcal{L}$ is the integrated luminosity and $\sqrt{s}$ is the center-of-mass energy; the off-resonance subtraction is the dominant source of statistical uncertainty. The simulation is used to remove the remaining $B\overline{B}$ background for the 'pseudo-reconstruction' and lepton tagged methods.

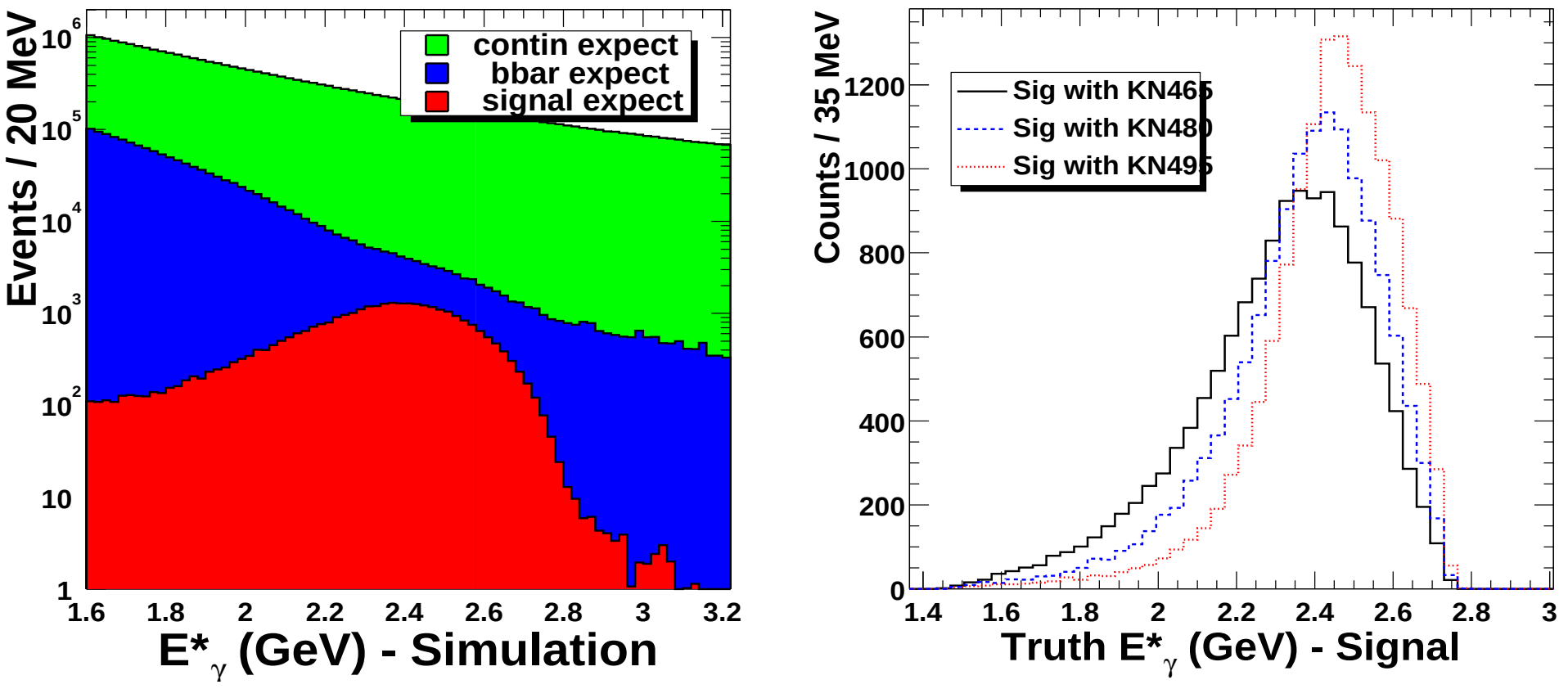

(a) (b)

**Figure 2-1.** *(a) The $E^*_\gamma$ distribution of signal, continuum (contin) and $B\overline{B}$ events (bbar) when only a high energy photon is required. (b) The expected $E^*_\gamma$ distribution of signal events for $m_b = 4.65$ (solid) , 4.80 (dashed) and 4.95 $\mathrm{GeV}/c^2$ (dotted).*

The theoretical uncertainty common to all these measurements arises from the extrapolation of the measured value B($b \to s\gamma$) to below the experimental cut placed on $E^*_\gamma$. The analyses have a differing value of the $E^*_\gamma$ cut between 2.0 to $2.1\ \mathrm{GeV}$. The fraction of the spectrum below the experimental cutoff is sensitive to the *ansatz* used to parameterize the spectrum shape in the signal model. The shape of the spectrum can be parameterized in terms of QCD quantities such as the $b$ quark mass, $m_b$, and the Fermi momentum of the confined $b$ quark. Figure 2-1(b) shows the expected spectrum, normalized to the same branching fraction, for three different values of $m_b$; the spectra were generated using the Kagan and Neubert *ansatz* [131]. The theoretical dependence is significantly reduced by lowering the value of the $E^*_\gamma$ cutoff as far as possible. However, reducing the $E^*_\gamma$ cut in the 'pseudo-reconstruction' and lepton tagged analyses is difficult due to the significant increase in the subtracted $B\overline{B}$ background, which would inflate what is already the largest source of experimental systematic error. In the sum-of-exclusive modes analyses the low $E^*_\gamma$ states are the highest hadronic mass states, which also have a larger average multiplicity. The increased multiplicity leads to a large combinatorial background which, when coupled with the decreasing cross section of the signal, makes the final states difficult to reconstruct above the background.

The sum-of-exclusive measurement has other large systematic uncertainties related to the fraction of modes which are reconstructed, fragmentation and the signal extraction fit. All of these will be improved in the future but they will be limiting factors in the measurement of $\mathcal{B}(b \to s\gamma)$ using this method. A new version of the *BABAR* lepton tag analysis is currently being finalized. Validation and correction of the $B\overline{B}$ background in simulation is the largest undertaking in the analysis; reducing the uncertainty in the $B\overline{B}$ subtraction leads to a reduction of the $E^*_\gamma$ cut–off, which yields a smaller overall systematic uncertainty. Therefore, to improve inclusive measurements of $\mathcal{B}(b \to s\gamma)$ using tagging at a Super $B$ Factory will require a very detailed understanding of the inclusive production of $\pi^0$, $\eta$, $\omega$, $\eta'$, and $\overline{n}$ in the data and their modeling in the $B\overline{B}$ simulation. The tracking and track-cluster matching inefficiencies must also be accurately measured to estimate the background from semi-electronic $B$ decays. An additional concern at a Super $B$ Factory will be higher rates of beam-related backgrounds with the increased instantaneous luminosity; it is likely, however, that the continuum subtraction should adequately account for these. It will be important to have a significant amount of off–resonance running at a Super $B$ Factory to ensure that the statistical error does not become dominant; at least the current value of $\mathcal{L}_{\rm on}/\mathcal{L}_{\rm off}$ of 8.5 at *BABAR* will have to be maintained.

One new method that may be very productive in the Super $B$ Factory era has been studied by *BABAR*. The $B_{\rm RECO}$ sample described in Section 4.2.1 is used to select a pure sample of $B\overline{B}$ events, from which events with a high energy photon are selected. The photon combined with the remaining reconstructed particles in the event is then used to calculate $m_{ES}$; the fitted $m_{ES}$ distribution of all candidates is used to extract the signal yield. The drawback of this method is the small efficiency for reconstruction of the $B_{\rm RECO}$ sample, which is 0.4% in the current analysis. From this enriched sample of $B\overline{B}$ events, around 40% of the $B \to X_s\gamma$ decays are reconstructed, after further $B\overline{B}$ combinatorial background suppression criteria. Preliminary results show that a statistical uncertainty of 6.5% would be expected for a data sample corresponding to an integrated luminosity of $1\ {\rm ab}^{-1}$. The systematic uncertainties are still under investigation, particularly related to the modeling of the $B\overline{B}$ background. Also to be explored is the use of kinematic constraints from the two fully reconstructed $B$-decays and the initial $e^+e^-$ state to enhance the resolution of the photon energy in the rest frame of the $B$ meson. The main advantage of this method is that it may result in a small enough background to lower the $E_\gamma^{B-{\rm rest}}$ threshold well below $2.0\ {\rm GeV}$.

### 2.5.2 $\mathcal{B}(b \to d\gamma)$ and $|V_{td}|/|V_{ts}|$

A feasibility study of measuring $\mathcal{B}(b \to d\gamma)$ inclusively has been done for this report. (The prospects for exclusive measurements of $B \to \rho\gamma$ and $B \to \omega\gamma$ at a Super $B$ Factory are discussed in Section 2.7.) In fact, the quantity measured is the ratio of B($b \to d\gamma$)/B($b \to s\gamma$) which is equal to $|V_{td}|^2/|V_{ts}|^2$ to within a theoretical correction, which is predicted to be of the order of 10% with an uncertainty of 5% for $E_\gamma^{B-{\rm rest}} > 1.6\ {\rm GeV}$ [55]. An experimental strategy has been considered where a selection similar to the *BABAR* lepton tagged measurement is used [44], which does not distinguish between $b \to s\gamma$ and $b \to d\gamma$. After this a strangeness tag is run which will use the kaons in the final state to tag events as $b \to s\gamma$, the absence of kaons correlated with the $X_s$ system would classify the event to be $b \to d\gamma$. The details of the strangeness tag algorithm have not been considered; therefore, different values of the mistag rate of the algorithm, $\omega_s$, are considered to see whether a statistically significant result is achievable with the large data sets available at a Super $B$ Factory.

The ratio of the measured number of $b \to d\gamma$ events to $b \to s\gamma$ events at a given $E_\gamma^*$ cut, assuming that the energy spectrum is the same, is equal to $|V_{td}|^2/|V_{ts}|^2$. (In this study the theoretical correction has been ignored because it has not been computed at the experimental value of the $E_\gamma^*$ cut.) In terms of experimental quantities, $|V_{td}|^2/|V_{ts}|^2$ can be expressed as:

$$\frac{|V_{td}|^2}{|V_{ts}|^2} = \frac{(1-\omega_s)(N_d - N_d^{\rm bkg}) - \omega_s(N_s - N_s^{\rm bkg})}{(1-\omega_d)(N_s - N_s^{\rm bkg}) - \omega_d(N_d - N_d^{\rm bkg})}\ ,$$

where $N_{d(s)}$ is the number of selected events without (with)a strangeness tag, $N_{d(s)}^{\rm bkg}$ is the number of background events without (with) a strangeness tag, and $\omega_d$ is the mistag rate of $b \to d\gamma$ events as $b \to s\gamma$.

The first step in estimating the sensitivity is optimizing the $E_\gamma^*$ cut. Given that the measurement is a ratio, there is no systematic uncertainty due to extrapolation to lower values of $E_\gamma^*$, as in the measurement of the absolute value of the branching fraction; the best statistical sensitivity to was found with $2.3 < E_\gamma^* < 2.7\ {\rm GeV}$. The other inputs to the initial estimates of the sensitivity are: $\omega_s = 0.33$, the most optimistic case with only $K_L^0$ and $K_S^0 \to \pi^0\pi^0$ missing; $\omega_d = 0.05$ due to $s\overline{s}$ popping and association of a kaon from the other $B$ decay; an uncertainty on the $B\overline{B}$ background of 5%; an on–to–off resonance luminosity ratio of 8.5, the current *BABAR* value; 50% of background is strangeness tagged; and $\frac{|V_{td}|^2}{|V_{ts}|^2} = 0.04$. Figure 2-2(a) shows the expected sensitivity as a function of integrated luminosity for three values of $\omega_s$; even with a 50% mistag rate a 20% error on $\frac{|V_{td}|}{|V_{ts}|}$ could be achieved with $10\ {\rm ab}^{-1}$. The other inputs were varied, leading to uncertainties between 10% and 20% for a $10\ {\rm ab}^{-1}$ data set. The asymptotic limit in the uncertainty at large luminosities is dominated the knowledge of the $B\overline{B}$ background, which is illustrated in Fig. 2-2(b).

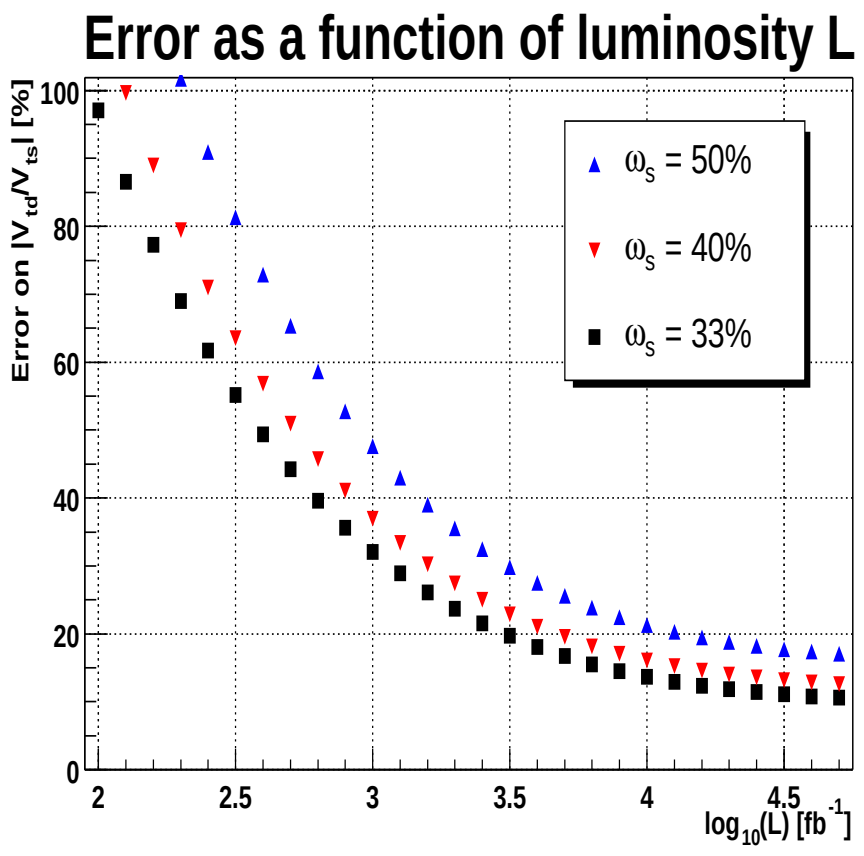

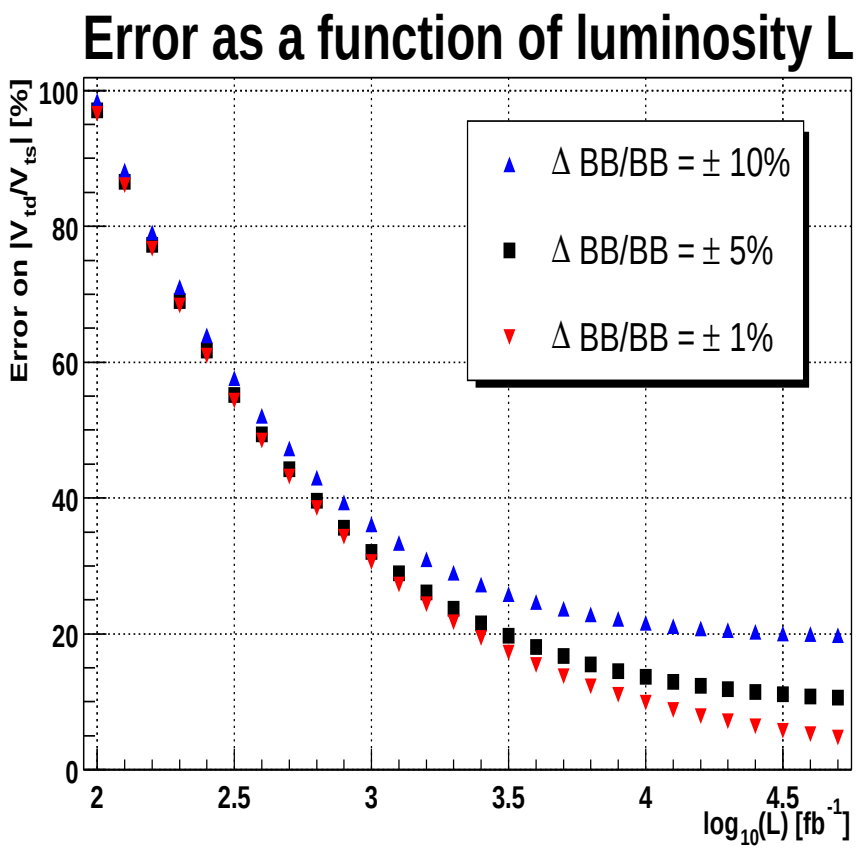

**Figure 2-2.** *The expected uncertainty on $|V_{td}|/|V_{ts}|$ as a function of integrated luminosity.* $10\ \mathrm{ab}^{-1}$ *is equal to 4 on the* $\log_{10}$ *scale. The uncertainty is shown for different values of (a)* $\omega_s$ *and (b) fractional uncertainty on the* $B\overline{B}$ *background* $(\Delta B\overline{B}/B\overline{B})$.

The conclusion of this study is that an interesting measurement of $\frac{|V_{td}|}{|V_{ts}|}$ may be possible with the very large data sets available at a Super $B$ Factory. Further study of the strangeness tag implementation, as well as the possible use of $K^0_L$, is required to better ground the conclusions. Also, a theoretical estimate of the correction required for the experimental cut on $E^*_\gamma$ is needed.

### 2.5.3 Inclusive $A_{CP}$

The direct $CP$ asymmetry parameter $A_{CP}$ can also be measured inclusively. There are two different $A_{CP}$ parameters of interest: that for $b \to s\gamma$ and that for the combination of $b \to s\gamma$ and $b \to d\gamma$. The two parameters are complementary probes of New Physics, as described in Section 2.6.3 and in [55]. The measurements of $A_{CP}$ to date are:

$$A_{CP} = (-0.079 \pm 0.108(\mathrm{stat.}) \pm 0.022(\mathrm{syst.})),$$

by CLEO [132], which used ‘pseudoreconstruction’ and lepton tag methods, and

$$A_{CP} = (-0.004 \pm 0.051(\mathrm{stat.}) \pm 0.038(\mathrm{syst.})),$$

by Belle [59], which used a sum-of-exclusive final states. The sum-of-exclusive modes and ‘pseudoreconstruction’ methods use the flavor of the kaons in the final state to self tag the flavor of the decaying $b$ quark, whereas the lepton tag method measures the flavor of the non–signal $B$ decay from the lepton’s charge. The sum-of-exclusive and ‘pseudoreconstruction’ methods have very little contamination from $b \to d\gamma$, because kaons are required in the final state. The lepton tagged measurement does not place any flavor requirements on the signal hadronic system, so it measures $A_{\{CP}$ in the sum of $b \to s\gamma$ and $b \to d\gamma$.[1] Therefore, the CLEO measurement is not a pure measurement of $A_{\{CP}$ in $b \to s\gamma$ because lepton tags were also used, however the contamination is small, since the statistical precision is dominated by the ‘pseudoreconstructed’ events. Measurements of $A_{\{CP}$ using kaons have a very small mistag rate, $\omega$, of around 0.5%. The lepton-tagged measurement of $A_{\{CP}$ is significantly diluted by $B^0\overline{B}^0$ mixing, which leads to a contribution to $\omega$ of $\chi_d/2 = 0.091$ [2], where $\chi_d$ is the time-averaged mixing probability. Furthermore, there is a contribution from cascade decays in which the tag lepton is from the decay of a charmed particle, which has the wrong sign to identify the decaying $b$ quark. The forthcoming *BABAR* lepton tag analysis estimates $\omega$ to be around 13%.

[1] In the lepton tagged branching fraction measurement a correction of 4.2% is made to the measured value of $\mathcal{B}(b \to s\gamma)$ to account for the $b \to d\gamma$ component.

The main systematic uncertainties arise from any bias due to detector charge asymmetries in the kaon or lepton tagging. These can be measured from control samples which are statistically limited; therefore at a Super $B$ Factory the systematic error should continue to decrease from the current value of around 1%. Other systematic uncertainties, such as a small amount of direct $CP$ violation in the background, may become dominant if they can not be evaluated using a suitable control sample.

Table 2-5 gives extrapolations of the expected precision on $A_{CP}$ at a Super $B$ Factory from current BABAR analyses. The sum-of-exclusive modes and lepton tag methods measure the $b \to s\gamma$ and the combined $b \to s\gamma$ and $b \to d\gamma$ $A_{CP}$, respectively. The $\mathcal{O}(1\%)$ uncertainty with a $10\,\mathrm{ab}^{-1}$ data set would provide an excellent test of New Physics models with and without minimal flavor violation.

**Table 2-5.** *The expected statistical and systematic uncertainties on $A_{CP}$ with different integrated luminosities, $\mathcal{L}$. The uncertainties on $A_{CP}$ for $b \to s\gamma$ alone and for the combination of $b \to s\gamma$ and $b \to d\gamma$ are given.*

| | $\Delta A_{CP}(b \to s\gamma)$ | | $\Delta A_{CP}$ $(b \to s\gamma$+ $b \to d\gamma)$ | |
|---|---|---|---|---|
| $\mathcal{L}$ $[\mathrm{ab}^{-1}]$ | Statistical | Systematic | Statistical | Systematic |
| 0.1 | 0.050 | 0.015 | 0.10 | 0.010 |
| 1 | 0.016 | 0.005 | 0.03 | 0.003 |
| 10 | 0.005 | 0.002 | 0.01 | 0.001 |

## 2.6 Theoretical Prospects for the Inclusive Modes $b \longrightarrow (s,d)\,\gamma$

>– T. Hurth –<

### 2.6.1 The inclusive mode $b \longrightarrow s\gamma$

The rare decay $\overline{B} \to X_s\gamma$ is dominated by perturbative contributions and is, therefore, a theoretically clean decay mode (see section 2.3). The theoretical prediction for the $B \to X_s\gamma$ decay rate is usually normalized by the semileptonic decay rate in order to get rid of uncertainties related to the CKM matrix elements and the fifth power of the $b$ quark mass. Moreover, an explicit lower cut on the photon energy in the bremsstrahlung correction has to be made. At the partonic level, one starts with

$$R_{\rm quark}(\delta) = \frac{\Gamma[b \to s\gamma] + \Gamma[b \to s\gamma gluon]_\delta}{\Gamma[b \to X_c e \overline{\nu}_e]}, \tag{2.27}$$

where the subscript $\delta$ means that only photons with energy $E_\gamma > (1-\delta)E_\gamma^{\rm max} = (1-\delta)\frac{m_b}{2}$ are counted. The ratio $R_{\rm quark}$ is divergent in the limit $\delta \to 1$, owing to the soft photon divergence in the subprocess $b \to s\gamma gluon$. In this limit only the sum of $\Gamma[b \to s\gamma]$, $\Gamma[b \to s\, gluon]$ and $\Gamma[b \to s\gamma\, gluon]$ is a reasonable physical quantity, in which all divergences cancel. It is suggestive to give up the concept of a 'total' decay rate of $b \to s\gamma$ and to compare theory and experiment using the same energy cut.

The QCD corrections due to hard-gluon exchange are by far the dominant corrections to the electroweak one-loop contribution. These perturbative corrections have been calculated to NLL precision. All present predictions are based on the original NLL calculations presented in [133, 134, 51] and on independent checks of these calculations [135, 136, 137]. The impact of these NLL corrections are significant, leading to a shift of the central value of about $20\%$ and a reduction of the scale dependence from about $25\%$ to about $5\%$. In the meanwhile also subleading two-loop electroweak corrections were calculated and found to be less than $2\%$ [138]. The nonperturbative corrections mentioned above also play only a subdominant role: the $1/m_b^2$ corrections correspond to the OPE for $\mathrm{T}(\mathcal{O}_7^\dagger\mathcal{O}_7)$ and can be estimated to have an impact well below $10\%$. HQET estimates the effect to be of order $+1\%$ [78]. There are additional nonperturbative effects if one also takes into account the operator $\mathcal{O}_2$. Nonperturbative corrections due to $T(\mathcal{O}_7^\dagger\mathcal{O}_2)$ can also be analysed in a model-independent way and scale with $1/m_c^2$. Due to small coefficients in the expansion also their impact is very small, around $+3\%$ [25]. A systematic analysis of terms like $\Gamma^{(O_2,O_2)}_{\overline{B}\to X_s\gamma}$ at first order in $\alpha_s(m_b)$ is still missing. Rigorous techniques such as OPEs do not seem to be applicable in this case. However, these contributions have to be under control if one reaches the experimental accuracy possible with a Super $B$ Factory.

This large calculational enterprise leads to the present theoretical predictions. A recent phenomenolgical analysis [55] gives, for $E_\gamma > 1.6\,\mathrm{GeV}$:

$$\mathcal{B}[\overline{B} \to X_s\gamma] = \left(3.61\ \ {}^{+0.24}_{-0.40}\Big|_{\frac{m_c}{m_b}} \pm 0.02_{\rm CKM} \pm 0.24_{\rm param.} \pm 0.14_{\rm scale}\right) \times 10^{-4}\,, \tag{2.28}$$

for $E_\gamma > m_b/20$

$$\mathcal{B}[\overline{B} \to X_s\gamma] = \left(3.79\ \ {}^{+0.26}_{-0.44}\Big|_{\frac{m_c}{m_b}} \pm 0.02_{\rm CKM} \pm 0.25_{\rm param.} \pm 0.15_{\rm scale}\right) \times 10^{-4}\,. \tag{2.29}$$

The dominant error is due to the $m_c/m_b$ dependence. It is induced by the large renormalization scheme ambiguity of the charm mass. There are at least two issues that need further studies:

Since the charm quark in the matrix element $\langle O_1 \rangle$ are dominantly off-shell, it is argued in [49] that the running charm mass should be chosen instead of the pole mass. The latter choice was used in all previous analyses [134, 51, 131, 139]:

$$m_c^{\rm pole}/m_b^{\rm pole} \qquad \Rightarrow \qquad m_c^{\overline{\rm MS}}(\mu)/m_b^{\rm pole}, \ \ \mu \in [m_c, m_b]. \tag{2.30}$$

Numerically, the shift from $m_c^{\rm pole}/m_b^{\rm pole} = 0.29 \pm 0.02$ to $m_c^{\overline{MS}}(\mu)/m_b^{\rm pole} = 0.22 \pm 0.04$ is rather important and leads to a $+11\%$ shift of the central value of the $\overline{B} \to X_s\gamma$ branching ratio. Since the matrix element starts at NLL order and, thus, the renormalization scheme for $m_c$ is an NNLL issue, one should regard the choice of the $\overline{\mathrm{MS}}$ scheme as an educated guess of the NNLL corrections. Nevertheless, the new choice is guided by the experience gained from many higher order calculations in perturbation theory. Moreover, the $\overline{\mathrm{MS}}$ mass of the charm quark is also a short-distance quantity which does not suffer from nonperturbative ambiguities, in contrast to its pole mass. Therefore the central value resulting within this scheme is definitely favored. However, one has to argue for a theoretical uncertainty in $m_c^{\overline{\mathrm{MS}}}(\mu)/m_b^{\mathrm{pole}}$, which also includes the value of $m_c^{\mathrm{pole}}$. This is done in the above theoretical predictions by using a large asymmetric error in $m_c/m_b$ that fully covers any value of $m_c/m_b$ compatible with any of these two determinations:

$$\frac{m_c}{m_b} = 0.23^{+0.08}_{-0.05}\,. \tag{2.31}$$

The dominant uncertainty due to the renormalization scheme dependence is a perturbative error that could be significantly reduced by a NNLL QCD calculation. Such a calculation would also further reduce the scale uncertainty given in the theoretical predictions above. Needless to say, the parametric error can also be further reduced by independent experiments. Thus, a theoretical error around $5\%$ seems possible. At that stage a further study of the nonperturbative corrections seems to be appropriate in order to make sure that they are under control at this level of accuracy.

The uncertainty regarding the fraction of the $\overline{B} \to X_s\gamma$ events above the chosen lower photon energy cut-off $E_\gamma$ quoted in the experimental measurement, also often cited as model dependence, should be regarded as a purely *theoretical* uncertainty: in contrast to the 'total' branching ratio of $\overline{B} \to X_s\gamma$, the photon energy spectrum cannot be calculated directly using the heavy mass expansion, because the OPE breaks down in the high-energy part of the spectrum, where $E_\gamma \approx m_b/2$. However, a partial resummation of an infinite number of leading-twist corrections into a nonperturbative universal shape function is possible. At present this function cannot be calculated, but there is at least some information on the moments of the shape function, which are related to the forward matrix elements of local operators. An important observation is that the shape of the photon spectrum is practically insensitive to physics beyond the Standard Model (see Fig. fig:Toymodel. This implies that we do not have to assume the correctness of the Standard Model in the experimental analysis. A precise measurement of the photon spectrum would allow a determination of the parameters of the shape function. Moreover, the universality of the shape function, valid to lowest order in $\Lambda_{\mathrm{QCD}}/m_b$, allows us to compare information from the endpoint region of the $\overline{B} \to X_s\gamma$ photon spectrum and of the $\overline{B} \to X_u\ell\nu$ charged-lepton spectrum up to higher $1/m_b$ corrections. Thus, one of the main aims in the future should therefore be a precise measurement of the photon spectrum. It is clear, that a lower experimental cut in the photon spectrum within the measurement of $\overline{B} \to X_s\gamma$ decreases the sensitivity to the parameters of the shape function and that the ideal energy cut would be $1.6$ GeV. In this case, however, a better understanding of the $B\overline{B}$ background is necessary. In the last Belle measurement the photon cut was already pushed to $1.8$ GeV [48].

The important role of the $\overline{B} \to X_s\gamma$ decay in the search for New Physics cannot be overemphasized (for a recent review, see [41]), as it already leads to stringent bounds on various supersymmetric extensions of the Standard Model (see for example [56, 57, 140, 141, 142, 143]). Also, in the long run, after New Physics has been discovered via the direct search, this inclusive decay mode will play an even more important role in analyzing in greater detail the new underlying dynamics.

### 2.6.2 The inclusive mode $b \to d\gamma$

Most of the theoretical improvements on the perturbative contributions and the power corrections in $1/m_b^2$ and $1/m_c^2$, carried out in the context of the decay $\overline{B} \to X_s\gamma$, can straightforwardly be adapted to the decay $\overline{B} \to X_d\gamma$; thus, the NLL-improved decay rate for $\overline{B} \to X_d\gamma$ decay has greatly reduced the theoretical uncertainty [39]. But as $\lambda_d^u = V_{ub}V_{ud}^*$ for $b \to d\gamma$ is not small with respect to $\lambda_d^t = V_{tb}V_{td}^*$ and $\lambda_d^c = V_{cb}V_{cd}^*$, one also has to take into account the operators proportional to $\lambda_d^u$ and, moreover, the long-distance contributions from the intermediate $u$ quark in the penguin loops might be important. However, there are three *soft* arguments that indicate a small impact of these nonperturbative contributions: first, one can derive a model-independent suppression factor $\Lambda_{\mathrm{QCD}}/m_b$ within these long-distance contributions [25]. Then, model calculations, based on vector meson dominance, also suggest this conclusion [144, 145]. Furthermore, estimates of the long-distance contributions in exclusive decays $\overline{B} \to \rho\gamma$ and $\overline{B} \to \omega\gamma$ in the light-cone sum rule approach do not exceed 15% [146]. Finally, it must be stressed that there is no spurious enhancement of the form $\log(m_u/\mu_b)$ in the perturbative contribution, as was shown in [147, 148]. All these observations exclude very large long-distance intermediate $u$ quark contributions in the decay $\overline{B} \to X_d\gamma$. Nevertheless, the theoretical status of the decay $\overline{B} \to X_d\gamma$ is not as clean as that of $\overline{B} \to X_s\gamma$.

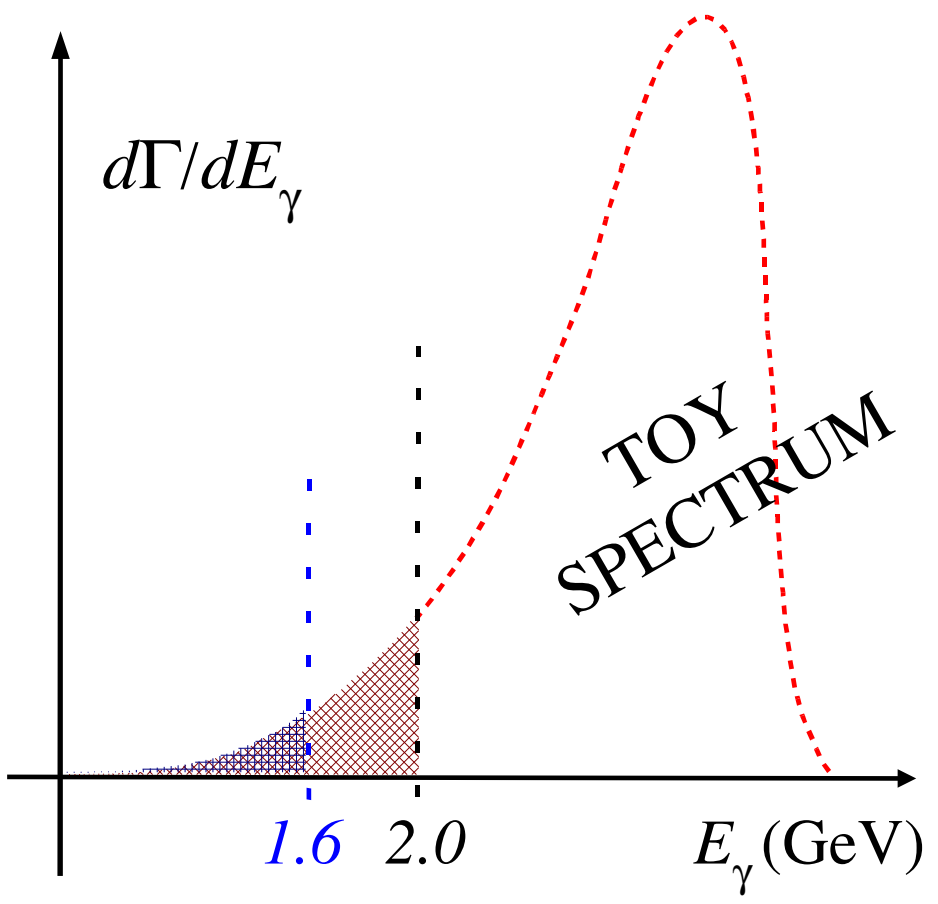

**Figure 2-3.** *Schematic photon spectrum of $B \to X_s\gamma$.*

While $b \to s$ transitions such as $\overline{B} \to X_s\gamma$ do not gave an impact on CKM phenomenology, due to of the flatness of the corresponding unitarity triangle, $b \to d$ transitions give important complementary information on the unitarity triangle, which is also tested by the measurements of $V_{ub}/V_{cb}$, $\Delta M_{B_d}$, and $\Delta M_{B_d}/\Delta M_{B_s}$. Thus, a future measurement of the $\overline{B} \to X_d\gamma$ decay rate will help to significantly reduce the current allowed region of the CKM-Wolfenstein parameters $\rho$ and $\eta$.

The branching ratio of $\overline{B} \to X_d\gamma$ might also be of interest in a New Physics context, because, while it is CKM-suppressed by a factor $|V_{td}|^2/|V_{ts}|^2$ in the Standard Model, this may not be the case in extended models. We also emphasize that in the ratio

$$R(d\gamma/s\gamma) \equiv \frac{\mathcal{B}(\overline{B} \to X_d\gamma)}{\mathcal{B}(\overline{B} \to X_s\gamma)}, \tag{2.32}$$

a substantial portion of the theoretical uncertainties cancel out. It is therefore of particular interest for CKM phenomenology and for the search for New Physics.

As discussed above, the measurement of the $\overline{B} \to X_d\gamma$ is rather difficult, but within the reach of the Super $B$ Factories. A recent update of the theoretical prediction was presented in [55]. For $E_\gamma > m_b/20$ one gets:

$$\mathcal{B}[\overline{B} \to X_d\gamma] = \left(1.46\ {}^{+0.15}_{-0.23}\Big|_{\frac{m_c}{m_b}} \pm 0.16_{\rm CKM} \pm 0.10_{\rm param.} \pm 0.06_{\rm scale}\right) \times 10^{-5}\,, \tag{2.33}$$

$$\frac{\mathcal{B}[\overline{B} \to X_d\gamma]}{\mathcal{B}[\overline{B} \to X_s\gamma]} = \left(3.86\ {}^{+0.11}_{-0.18}\Big|_{\frac{m_c}{m_b}} \pm 0.43_{\rm CKM} \pm 0.09_{\rm param.} \pm 0.15_{\rm scale}\right) \times 10^{-2}\,. \tag{2.34}$$

Note that the errors on the ratio $R_{ds} = \mathcal{B}[\overline{B} \to X_d\gamma]/\mathcal{B}[\overline{B} \to X_s\gamma]$ are dominated by CKM uncertainties. But it should be emphasized, that on top of the mentioned sources of error, the $B \to X_d\gamma$ mode is affected by the presence of nonperturbative $u$ quark loops whose effect is expected to be at most around $10\%$, according to the arguments presented here.

### 2.6.3 Direct *CP* violation

The direct *normalized* $CP$ asymmetry of the inclusive decay modes represent another interesting observable [149]-[161]:

$$A_{CP} = \frac{\Gamma(\overline{B} \to X_{s/d}\gamma) - \Gamma(B \to X_{\overline{s}/\overline{d}}\gamma)}{\Gamma(\overline{B} \to X_{s/d}\gamma) + \Gamma(B \to X_{\overline{s}/\overline{d}}\gamma)} \,. \tag{2.35}$$

The Standard Model predictions are essentially independent of the photon energy cut-off, and read (for $E_\gamma = 1.6\,\mathrm{GeV}$) [55]:

$$A_{CP}(\overline{B} \to X_s\gamma) \;=\; \left(0.44 \;\; {}^{+0.15}_{-0.10}\big|_{\frac{m_c}{m_b}} \;\pm 0.03_{\mathrm{CKM}} \;{}^{+0.19}_{-0.09}\big|_{\mathrm{scale}}\right)\% \,, \tag{2.36}$$

$$A_{CP}(\overline{B} \to X_d\gamma) \;=\; \left(-10.2 \;\; {}^{+2.4}_{-3.7}\big|_{\frac{m_c}{m_b}} \;\pm 1.0_{\mathrm{CKM}} \;{}^{+2.1}_{-4.4}\big|_{\mathrm{scale}}\right)\% \,. \tag{2.37}$$

The two dominant errors are the perturbative scale ambiguity and the renormalization scheme dependence of the charm mass, which are both of a perturbative nature, and can be reduced by a NNLL calculation, which is also desirable for the prediction of the branching ratio, as discussed above. The additional parametric uncertainties are subdominant.

The two $CP$ asymmetries are connected by the relative factor $\lambda^2\,((1-\rho)^2+\eta^2)$. Moreover, the small Standard Model prediction for the $CP$ asymmetry in the decay $\overline{B} \to X_s\gamma$ is a result of three suppression factors: an $\alpha_s$ factor needed in order to have a strong phase, a CKM suppression of order $\lambda^2$ and a GIM suppression of order $(m_c/m_b)^2$, reflecting the fact that in the limit $m_c = m_u$ any $CP$ asymmetry in the Standard Model would vanish.

The application of quark–hadron duality is, in general, problematic within a semi-inclusive measurement of $CP$-violating effects, if only 50% or 70% of the total exclusive modes are detected. In fact, the strong rescattering phases responsible for the presence of $CP$ violation can be different for each exclusive channel. It is impossible to reliably quantify the resulting systematic uncertainty without a detailed study of the individual modes and of their direct $CP$ asymmetries. Therefore, a fully inclusive measurement of the so-called *untagged* direct $CP$ asymmetry, the sum of the unnormalized $CP$ asymmetries in the $b \to s$ and the $b \to d$ sector, is favored. Moreover, this quantity allows for a very stringent Standard Model test and is very sensitive to new $CP$ phases beyond the Standard Model. Such a measurement is possible because the experimental efficiencies within the inclusive $b \to s$ and $b \to d$ modes are expected to be equal.

The unnormalized $CP$ asymmetry for the sum of the partonic processes $b \to (s+d)\gamma$ vanishes in the limit of $m_d = m_s = 0$, as was first observed in Ref. [162]. This is still valid for the weaker condition $m_d = m_s$, which corresponds to the so-called $U$ spin limit. However, if the down and the strange quark were degenerate, the Standard Model would be completely $CP$-conserving, because any $CP$ violation in the Standard Model is proportional to the quark mass differences, especially to $(m_d - m_s)$. Thus, the $U$ spin limit at the quark level does not make much sense with respect to $CP$ asymmetries. However, this symmetry should be used only with respect to the influence of the strong interactions on the hadronic matrix elements (in particular on the strong phases), while the down and strange quark masses are different. The unitarity of the CKM matrix implies

$$J = \mathrm{Im}(\lambda_u^{(s)}\lambda_c^{(s)*}) = -\mathrm{Im}(\lambda_u^{(d)}\lambda_c^{(d)*}) \,, \tag{2.38}$$

where $\lambda_q^{(q')} = V_{qb}V^*_{qq'}$. As a consequence, the following relation for the rate asymmetries is found, in the $U$ spin limit of the hadronic matrix elements and for real Wilson coefficients:

$$\Delta\Gamma(\overline{B} \to X_s\gamma) + \Delta\Gamma(\overline{B} \to X_d\gamma) = \Delta\Gamma_s + \Delta\Gamma_d = 0 \,, \tag{2.39}$$

where $\Delta\Gamma_q = \Delta\Gamma(\overline{B} \to X_q\gamma) = \Gamma(\overline{B} \to X_q\gamma) - \Gamma(B \to X_{\overline{q}}\gamma)$.

$U$ spin-breaking effects can be estimated within the heavy mass expansion, even beyond the partonic level [163, 164]:

$$\Delta\Gamma(\overline{B} \to X_s\gamma) + \Delta\Gamma(\overline{B} \to X_d\gamma) = b_{\rm inc}\,\Delta_{\rm inc}, \tag{2.40}$$

where the right-hand side is written as a product of a 'relative $U$ spin breaking' $b_{\rm inc}$ and a 'typical size' $\Delta_{\rm inc}$ of the $CP$-violating rate difference. In this framework one relies on parton-hadron duality and one can compute the breaking of $U$ spin by keeping a non-vanishing strange quark mass. A rough estimate of $b_{\rm inc}$ gives a value of the order of $|b_{\rm inc}| \sim m_s^2/m_b^2 \sim 5\times 10^{-4}$, while $|\Delta_{\rm inc}|$ is the average of the moduli of the two $CP$ rate asymmetries. Thus, one arrives at the following estimate within the partonic contribution [163]:

$$|\Delta\mathcal{B}(B \to X_s\gamma) + \Delta\mathcal{B}(B \to X_d\gamma)| \sim 1\times 10^{-9}\,. \tag{2.41}$$

Going beyond the leading partonic contribution within the heavy mass expansion, one has to check if the large suppression factor from the $U$ spin breaking, $b_{inc}$, is still effective in addition to the natural suppression factors already present in the higher order terms of the heavy mass expansion [164]. In the leading $1/m_b^2$ corrections, the $U$ spin breaking effects also induce an additional overall factor $m_s^2/m_b^2$. In the nonperturbative corrections from the charm quark loop, which scale with $1/m_c^2$, one finds again the same overall suppression factor, because the effective operators involved do not contain any information on the strange mass. Also the corresponding long-distance contributions from up quark loops, which scale with $\Lambda_{\rm QCD}/m_b$, follow the same pattern [164]. Thus, in the inclusive mode, the right-hand side in (2.41) can be computed in a model-independent way, with the help of the heavy mass expansion, and the $U$ spin breaking effects can be estimated to be practically zero [2]. Therefore, the prediction (2.41) provides a very clean Standard Model test, whether generic new $CP$ phases are active or not. Any significant deviation from the estimate (2.41) would be a direct hint to non-CKM contributions to $CP$ violation. This implies that any measurement of a non-zero untagged $CP$ asymmetry is a direct signal for New Physics beyond the Standard Model. For example, a Super $B$ Factory with an integrated luminosity of $10\,{\rm ab}^{-1}$ will allow this Standard Model prediction to be tested with an experimental accuracy of around $1\%$.

As was analysed in [55], the *untagged* direct $CP$ asymmetry also allows for a clear discrimination between scenarios beyond the Standard Model with minimal or general flavor violation: MFV models are characterized by the requirement of expressing all flavor-changing interactions in terms of powers of the Yukawa matrices. If one assumes the CKM phase to be the only $CP$ phase present at the grand unification scale, one finds that the untagged $CP$ asymmetry receives only very small contributions, at most $0.5\%$. Clearly, this class of models cannot be distinguished from the Standard Model with the help of this observable. If one allows for general $CP$ phases at the grand unification scale and takes the EDM bounds into account, only asymmetries below the $2\%$ level survive. One finds a strict proportionality between the untagged ($\overline{B} \to X_{s+d}\gamma$) and tagged ($\overline{B} \to X_s\gamma$) $CP$ asymmetries. The task of distinguishing these two MFV scenarios is beyond the possibilities of the existing $B$ Factories, but will be within the reach of future Super $B$ Factories. In the model-independent approach with generic new flavor violation [55], the untagged $CP$ asymmetry can be as large as $\pm 10\%$, once the recent experimental data on the $CP$ asymmetries are taken into account [165]. One also finds that in this general scenario the tagged and untagged asymmetries are again strictly proportional to each other. Moreover, assuming New Physics in the $d$ sector only, one finds untagged $CP$ asymmetries not larger than $2\%$: this implies that the untagged $CP$ asymmetry is not really sensitive to New Physics effects in the $d$ sector [55]. With the expected experimental accuracy of the Super $B$ Factory, a clear distinction between a minimal and a more general flavor model will be possible through a measurement of the untagged $CP$ asymmetry.

[2] The analogous Standard Model test within exclusive modes is rather limited, because $U$ spin-breaking effects cannot be calculated in a model-independent way. Estimates [32, 164] lead to the conclusion that the $U$ spin breaking effects are possibly as large as the rate differences themselves.

## 2.7 Experimental prospects for $B \to (K^*, \rho, \omega)\,\gamma$

≻– M. Convery –≺

The exclusive radiative penguin decay modes $B \to K^*\gamma$ and $B \to \rho\gamma$, which we will take to include $B^0 \to \rho^0\gamma$, $B^+ \to \rho^+\gamma$ and $B^0 \to \omega\gamma$, offer unique experimental challenges and opportunities at a Super $B$ Factory. One of the three $B \to \rho\gamma$ modes will be the first decay of the type $b \to d\gamma$ to be discovered,[3] and precision measurements will yield information on the CKM element $V_{td}$. The measurement of $A_{CP}$ and $\Delta_{0-}$ in $B \to K^*\gamma$ provides opportunities to search for New Physics in the $b \to s\gamma$ transition. In this section, we describe the current *BABAR* analyses of these modes, and discuss possible improvements and extrapolations at a Super $B$ Factory.

### 2.7.1 $B \to \rho\gamma$ analysis

Measurement of $\mathcal{B}(B \to \rho\gamma)$ represents a significant analysis challenge. $B \to \rho\gamma$ suffers from large continuum backgrounds as well as background from $B \to K^*\gamma$, which has a branching fraction 50 to 100 times larger. Continuum background may be rejected with event shape and similar variables. Optimization of these variables the is key to the sensitivity of $B \to \rho\gamma$ analyses. $B \to K^*\gamma$ is separable only with $\Delta E$ and hadronic PID. The $\Delta E$ separation is typically less than $2\sigma$, which places a premium on good particle identification. The current *BABAR* analysis [166], which focuses on these two aspects, will be described here.

**Particle ID for $B \to \rho\gamma$:** The $K^*$ and $\rho$ daughters from $B \to K^*\gamma$ and $B \to \rho\gamma$ have typical momenta $1 < p_{\rm lab} < 3{\rm GeV}/c$. In this region $\pi/K$ separation comes only from the DIRC, where the separation is good, and any misidentification comes from non-Gaussian effects. It is therefore not advantageous to do fits to Cherenkov angle PDFs, as is done in the charmless two-body analyses. Rather, we optimize selection criteria for pion selection. This problem is somewhat different from the usual one of kaon selection. In fact, we find that a significant improvement in kaon misidentification can be obtained by requiring that the number of photons observed in the DIRC be consistent with the number expected for a pion. This is in addition to the usual requirement that the measured Cherenkov angle be closer to the one expected for pion than kaon. Figure 2-4 shows the performance achieved by the pion selector. Since the kaon misidentification rate is typically 1%, the $B \to K^*\gamma$ background is reduced to levels about equal to the expected $B \to \rho\gamma$ level. In combination with the $\Delta E$ difference, this renders $B \to K^*\gamma$ background nearly negligible. Significant degradation in the particle identification capabilities would likely make it necessary to reject the $K^*$ background using an $m_{K\pi}$ cut, which reduces signal efficiency considerably.

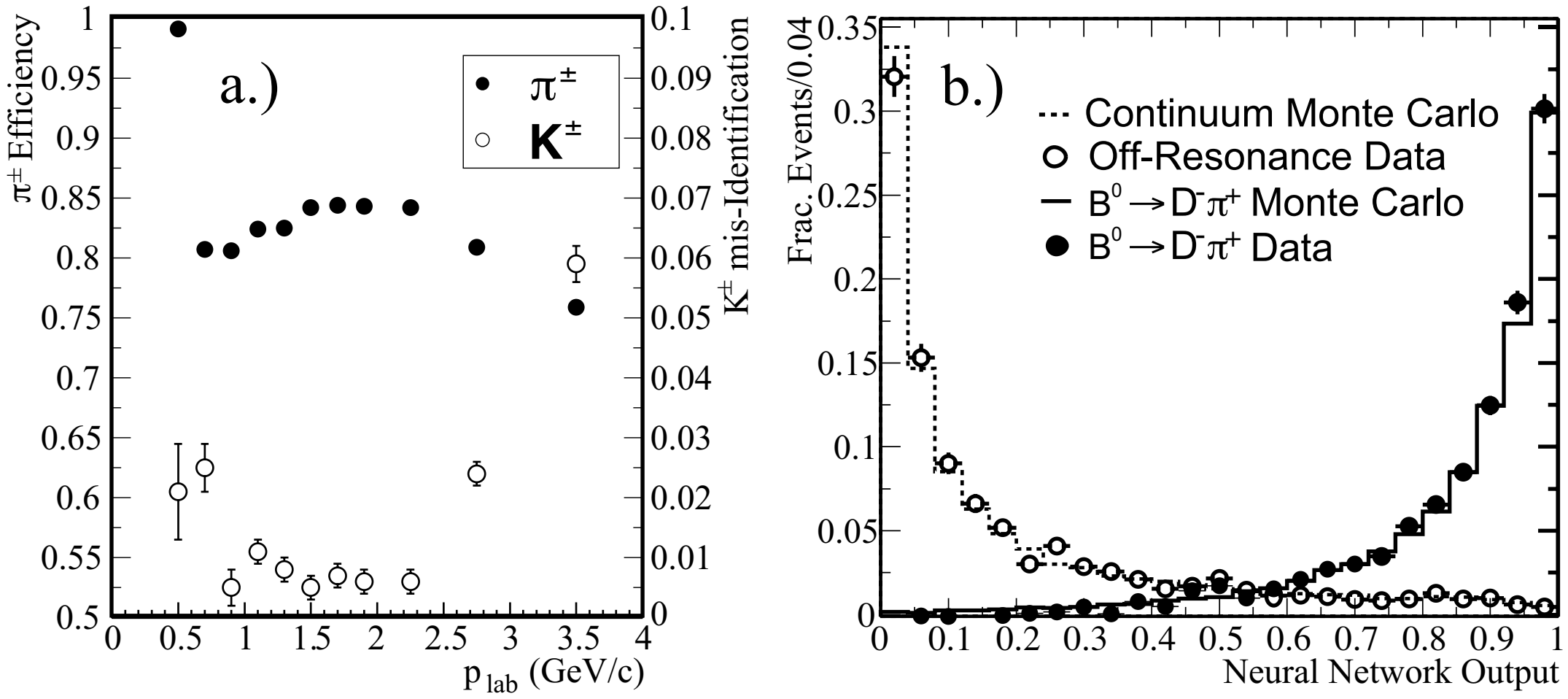

**Figure 2-4.** *Performance of the pion selector (left) and neural net (right) in the $B \to \rho\gamma$ analysis.*

[3]At Moriond '04, Belle claimed 3.5 $\sigma$ evidence for observation of the combination of the three $B \to \rho\gamma$ modes [30].

**Continuum background rejection:** Continuum background rejection is achieved in the current $B \to \rho\gamma$ analysis by combining a number of variables together in a neural net. In addition to the familiar variables of $\cos\theta_T^*$, $\cos\theta_H$, $\cos\theta_B$ and the "CLEO energy cones", we also include $R_2'$, $\Delta z$, and flavor tag information. Figure 2-4 shows the performance of this neural net for signal and continuum background.

**Current *BABAR* analysis results:** The current *BABAR* analysis, which is based on 78 fb$^{-1}$ obtains the results shown in Table 2-6. No evidence for these decays was found, and limits were set.

**Extrapolations to higher luminosity:** We assume that the statistical error on the branching fraction measurement will improve as $\mathcal{L}^{-1/2}$, and that the systematic error is composed of one part that similarly improves and one part that remains constant at 5%. Figure 2-5 shows this extrapolation. We see that with the current analysis, it will require almost 700 fb$^{-1}$ to see a 3-$\sigma$ Standard Model signal in $B^0 \to \rho^0\gamma$. The situation improves if we are able to improve the continuum background rejection by a factor of two, while maintaining the same signal efficiency. In this scenario, a 3-$\sigma$ signal could be observed with approximately 300 fb$^{-1}$. One finds that the measurement becomes systematically dominated at about 2 ab$^{-1}$. One also finds that the measurement of $V_{td}/V_{ts}$ becomes dominated by theoretical uncertainty at a similar point. Combining the three modes together reduces the required luminosity by roughly a factor of two.

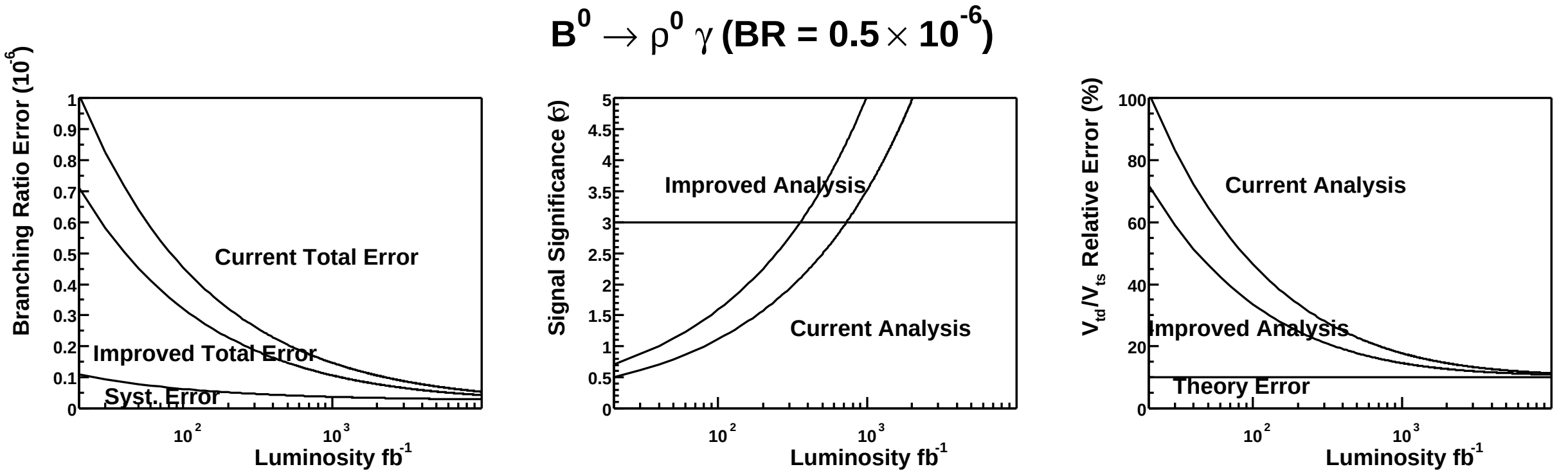

**Figure 2-5.** *Extrapolation to higher luminosity of the measurement of $\mathcal{B}(B^0 \to \rho^0\gamma)$*

Predictions for $A_{CP}$ in $B \to \rho\gamma$ are large. Reference [31] predicts $A_{CP}^{\pm} = 0.10^{+0.03}_{-0.02}$. Systematic errors on this quantity are rather small and an interesting measurement will become possible with about 10 ab$^{-1}$ of data.

The isospin-violation parameter, in contrast, is expected to be smaller. Reference [31] predicts $\Delta(\rho\gamma) = 0.04^{+0.14}_{-0.07}$. Something like 100 ab$^{-1}$ would be required for a significant measurement of this quantity.

**Table 2-6.** *Results of the current BABAR analysis of $B \to \rho\gamma$. The last line shows the limit for $B^0 \to \rho^0\gamma$ and $B^+ \to \rho^+\gamma$ combined under the assumption that $\Gamma(B \to \rho\gamma) = \Gamma(B^+ \to \rho^+\gamma) = 2 \times \Gamma(B^0 \to \rho^0\gamma)$,*

| Mode | Yield (Events) | Bias (Events) | Upper Lim. (Events) | $\epsilon$ (%) | $\mathcal{B}$ $(10^{-6})$ | $\mathcal{B}$ 90% UL $(10^{-6})$ |
|---|---|---|---|---|---|---|
| $B^0 \to \rho^0\gamma$ | $4.8^{+5.7}_{-4.7}$ | [−0.5,0.8] | 12.4 | $12.3 \pm 1.0$ | $0.4^{+0.6}_{-0.5}$ | 1.2 |
| $B^+ \to \rho^+\gamma$ | $6.2^{+7.2}_{-6.2}$ | [−0.1,2.0] | 15.4 | $9.2 \pm 1.2$ | $0.7^{+0.9}_{-0.8}$ | 2.1 |
| $B^0 \to \omega\gamma$ | $0.1^{+2.7}_{-2.0}$ | [−0.3,0.5] | 3.6 | $4.6 \pm 0.6$ | $0.0^{+0.7}_{-0.5}$ | 1.0 |
| $B \to \rho\gamma$ | | | | | | 1.9 |

### 2.7.2 $B \to K^*\gamma$ analysis

In contrast to the $\mathcal{B}(B \to \rho\gamma)$ analysis, where the key point is reducing continuum background, in $B \to K^*\gamma$ it is mostly concerned with reducing systematic error. Even so, the measurements of branching fractions are nearly systematics dominated in the current *BABAR* analysis, based on $81\,\text{fb}^{-1}$ [167]. The systematic errors can be roughly divided into those coming from signal efficiency and from $B$ background that mimics signal. The former is measured with control samples in data and the latter is controlled by using $\Delta E$ in the fit.

Fortunately, the systematics partially cancel in the more theoretically-interesting ratios $A_{CP}$ and $\Delta_{0-}$.

**Current *BABAR* analysis:** The results for the current *BABAR* analysis are shown in Table 2-7. Taking into account the correlation of systematics between modes, we obtain: $\Delta_{0-} = 0.051 \pm 0.044(\text{stat.}) \pm 0.023(\text{sys.}) \pm 0.024(R^{+/0})$, where the last error takes into account the experimental uncertainty in the ratio: $R^{+/0} \equiv \Gamma(\Upsilon(4S) \to B^+B^-)/\Gamma(\Upsilon(4S) \to B^0\overline{B}^0)$.

**Table 2-7.** *Results of the current BABAR analysis of $B \to K^*\gamma$.*

| Mode | $\mathcal{B} \times 10^{-5}$ | Combined $\mathcal{B} \times 10^{-5}$ | $A_{CP}$ | Combined $A_{CP}$ |
|---|---|---|---|---|
| $K^+\pi^-$ | 3.92±0.20±0.23 | } 3.92±0.20±0.24 | -0.069±0.046±0.011 | } -0.013±0.036±0.010 |
| $K^0_S\pi^0$ | 4.02±0.99±0.51 | | | |
| $K^+\pi^0$ | 4.90±0.45±0.46 | } 3.87±0.28±0.26 | 0.084±0.075±0.007 | |
| $K^0_S\pi^+$ | 3.52±0.35±0.22 | | 0.061±0.092±0.007 | |

**Extrapolations to higher luminosity:** The measurements of branching fractions are essentially systematics limited with current data sets. Due to cancelation of systematics, however, $\Delta_{0-}$ is still statistics-limited. It is hoped that systematics can be further improved by a factor of two, to about the 1% level. We presume that the systematic error is composed of one part that improves as $\mathcal{L}^{-1/2}$ and one part that remains fixed at 1%. This would then allow a significant measurement of $\Delta_{0-}$ with something less than $1\,\text{ab}^{-1}$. Note that improvements would also be necessary in the measurement of $R^{+0}$. Figure 2-6 shows the extrapolation of the error on this quantity to higher luminosity.

Measuring $A_{CP}$ in $B \to K^*\gamma$ is rather straightforward; the only significant systematics come from detector matter-antimatter asymmetries. These are currently understood at the 1% level. The limiting systematic is the charge-asymmetry of the hadronic interaction of kaons with the detector material. To get much below 1% systematic, this asymmetry would probably have to be measured in *BABAR* data with a kaon control sample. No viable technique for doing this measurement has yet been found, so for this extrapolation, we presume the systematic error remains 1%. Figure 2-6 shows the extrapolation of the error on this quantity.

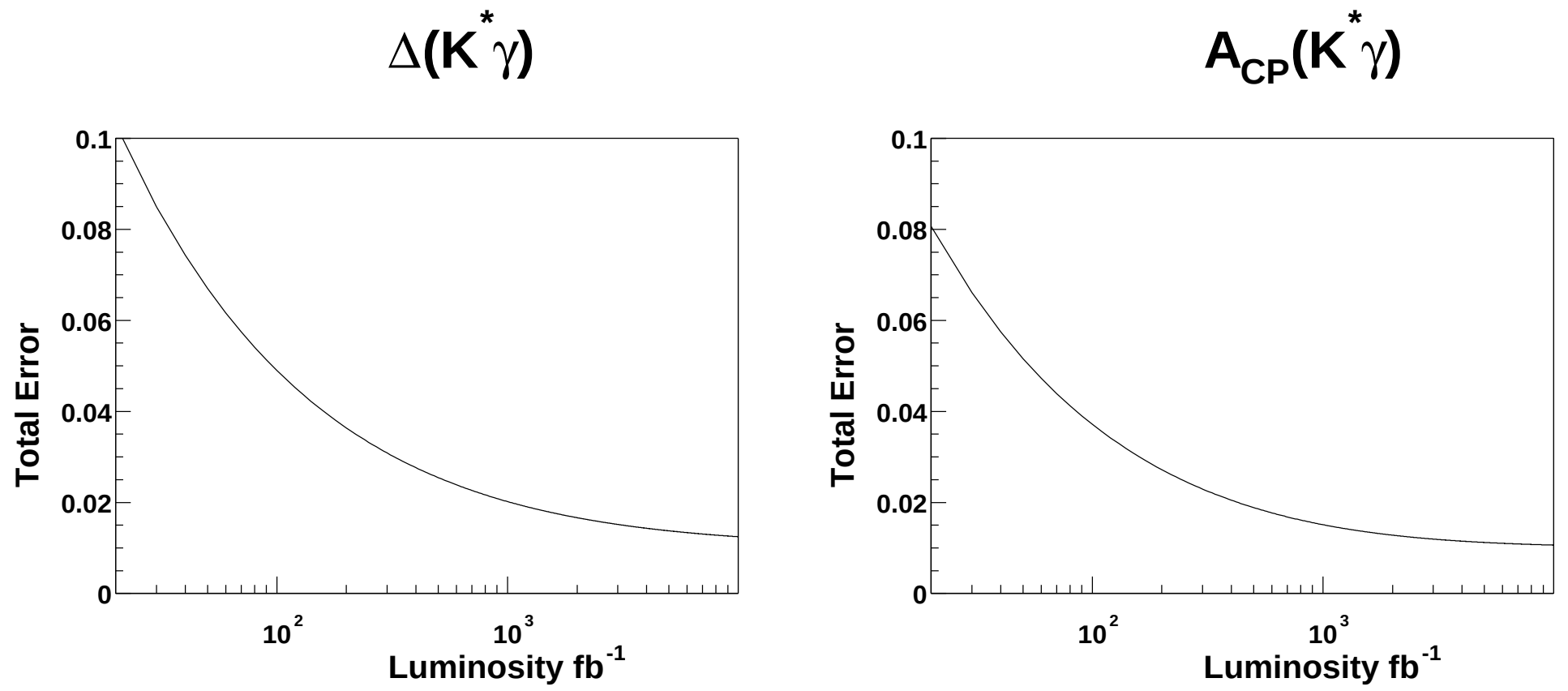

**Figure 2-6.** *Extrapolation to higher luminosity of $\Delta_{0-}(B \to K^*\gamma)$ (left) and $A_{CP}(B \to K^*\gamma)$ (right)*

## 2.8 Theoretical Prospects for $B \to (K^*, \rho)\,\gamma$

≻— A. Ali, E. Lunghi and A. Y. Parkhomenko —≺

### 2.8.1 Phenomenology of $B \to (\rho, \omega)\,\gamma$ decays

We now discuss the $B \to \rho\gamma$ and $B \to \omega\gamma$ decays whose measurements have just been announced by the Belle collaboration [30]. The potential impact of the decays $\mathrm{B} \to \rho\gamma$ and $\mathrm{B} \to \omega\gamma$ on the CKM phenomenology was first worked out in Refs. [168, 169] using the leading order estimates for the penguin amplitudes. Since then, annihilation contributions have been estimated in a number of papers [170, 146, 171], and the next-to-leading order corrections to the decay amplitudes have also been calculated [31, 32]. Deviations from the Standard Model estimates in the branching ratios, isospin-violating asymmetry $\Delta^{\pm 0}$ and $CP$-violating asymmetries $A_{CP}(\rho^{\pm}\gamma)$ and $A_{CP}(\rho^0\gamma)$ have also been worked out in a number of theoretical scenarios [172, 34, 173]. These CKM-suppressed radiative penguin decays were searched for by the CLEO collaboration [63], and the searches have been set forth at the $B$ Factory experiments Belle [174] and *BABAR* [166].

Recently, the Belle collaboration have presented evidence for the observation of the decays $B^+ \to \rho^+\gamma$, $B_d^0 \to \rho^0\gamma$ and $B_d^0 \to \omega\gamma$ (and their charge conjugates) [30]. Their observation, based on an integrated luminosity of 140 fb$^{-1}$, lacks statistical significance in the individual channels, but combining the data in the three decay modes and with their charged conjugate modes, yields a signal at $3.5\sigma$ CL [30]:

$$\overline{\mathcal{B}}_{\rm exp}[B \to (\rho,\omega)\,\gamma] = (1.8^{+0.6}_{-0.5} \pm 0.1) \times 10^{-6}. \tag{2.42}$$

This result updates the previous upper bounds [174] by the Belle collaboration, while the upper bound from the *BABAR* collaboration (at 90% C.L.) [166]:

$$\overline{\mathcal{B}}_{\rm exp}[B \to (\rho,\omega)\,\gamma] < 1.9 \times 10^{-6}, \tag{2.43}$$

remains to be updated.

The measurements from Belle and the upper limit from *BABAR* on the $B \to (\rho,\omega)\gamma$ decays given in (2.42) and (2.43), respectively, can be combined with their respective measurements of the $B \to K^*\gamma$ decay rates to yield the following ratios:

$$R_{\rm exp}[(\rho,\omega)\gamma/K^*\gamma] < 0.047, \qquad (BABAR) \tag{2.44}$$

$$R_{\rm exp}[(\rho,\omega)\gamma/K^*\gamma] = 0.042 \pm 0.013, \qquad (\text{Belle}) \tag{2.45}$$

where $R_{\rm exp}[(\rho,\omega)\gamma/K^*\gamma] = \overline{\mathcal{B}}_{\rm exp}[B \to (\rho,\omega)\,\gamma]/\overline{\mathcal{B}}_{\rm exp}(B \to K^*\gamma)$.

The branching ratios for $B \to \rho\gamma$ have been calculated in the Standard Model at next-to-leading order [31, 32] in the QCD factorization framework [65]. As the absolute values of the form factors in $B \to K^*\gamma$, $B \to \rho\gamma$ and $B \to \omega\gamma$ decays are quite uncertain, it is useful to calculate, instead, the ratios:

$$R_{\rm th}(\rho\gamma/K^*\gamma) = \frac{\mathcal{B}_{\rm th}(B \to \rho\gamma)}{\mathcal{B}_{\rm th}(B \to K^*\gamma)} = S_\rho \left|\frac{V_{td}}{V_{ts}}\right|^2 \frac{(M_B^2 - m_\rho^2)^3}{(M_B^2 - m_{K^*}^2)^3}\,\zeta^2\,\left[1 + \Delta R(\rho/K^*)\right], \tag{2.46}$$

$$R_{\rm th}(\omega\gamma/K^*\gamma) = \frac{\overline{\mathcal{B}}_{\rm th}(B_d^0 \to \omega\gamma)}{\overline{\mathcal{B}}_{\rm th}(B_d^0 \to K^{*0}\gamma)} = \frac{1}{2} \left|\frac{V_{td}}{V_{ts}}\right|^2 \frac{(M_B^2 - m_\omega^2)^3}{(M_B^2 - m_{K^*}^2)^3}\,\zeta^2\,\left[1 + \Delta R(\omega/K^*)\right], \tag{2.47}$$

where $m_\rho$ and $m_\omega$ are the masses of the $\rho$ and $\omega$ mesons, $\zeta$ is the ratio of the transition form factors, $\zeta = \overline{T}_1^{\rho}(0)/\overline{T}_1^{K^*}(0)$, which we have assumed to be the same for the $\rho^0$ and $\omega$ mesons, and $S_\rho = 1$ and $1/2$ for the $\rho^{\pm}$ and $\rho^0$ meson, respectively. To get the theoretical branching ratios for the decays $B \to \rho\gamma$ and $B_d^0 \to \omega\gamma$, the ratios (2.46) and (2.47) should be multiplied with the corresponding experimental branching ratio of the $B \to K^*\gamma$ decay. Explicit expressions for the NLO corrections $\Delta R^{\pm,0}$ and a detailed description of the input parameters can be found in Refs. [31, 35].

**Table 2-8.** *Theoretical estimates [35] for branching ratios, $CP$ asymmetries and isospin-violating ratio for exclusive $b \to d\gamma$ decays.*

| | $B^\pm \to \rho^\pm\gamma$ | $B_d^0 \to \rho^0\gamma$ | $B_d^0 \to \omega\gamma$ | $B \to (\rho,\omega)\gamma$ |
|---|---|---|---|---|
| $\Delta R$ | $0.116 \pm 0.099$ | $0.093 \pm 0.073$ | $0.092 \pm 0.073$ | |
| $R_{\rm th}$ | $0.0334 \pm 0.0103$ | $0.0164 \pm 0.0049$ | $0.0163 \pm 0.0049$ | $0.033 \pm 0.010$ |
| $\mathcal{B}_{\rm th}$ | $(1.35 \pm 0.42) \times 10^{-6}$ | $(0.66 \pm 0.20) \times 10^{-6}$ | $(0.65 \pm 0.20) \times 10^{-6}$ | $(1.38 \pm 0.42) \times 10^{-6}$ |
| $A_{CP}^{\rm dir}$ | $(-11.6 \pm 3.3)\%$ | $(-9.4^{+4.2}_{-3.8})\%$ | $(-8.8^{+4.4}_{-3.9})\%$ | |
| $\Delta(\rho,\gamma)$ | | | | $(1.1 \pm 3.9) \times 10^{-3}$ |

The theoretical uncertainty in the evaluation of the $R_{\rm th}(\rho\gamma/K^*\gamma)$ and $R_{\rm th}(\omega\gamma/K^*\gamma)$ ratios is dominated by the imprecise knowledge of $\zeta = \overline{T}_1^{\rho}(0)/\overline{T}_1^{K^*}(0)$ characterizing the SU(3) breaking effects in the QCD transition form factors. In the SU(3) symmetry limit, $\overline{T}_1^{\rho}(0) = \overline{T}_1^{K^*}(0)$, yielding $\zeta = 1$. The SU(3) breaking effects in these form factors have been evaluated within several approaches, including the LCSR and Lattice QCD. In the earlier calculations of the ratios [31, 34], the following ranges were used: $\zeta = 0.76 \pm 0.06$ [31] and $\zeta = 0.76 \pm 0.10$ [34], based on the LCSR approach [169, 170, 175, 176, 177] which indicate substantial SU(3) breaking in the $B \to K^*$ form factors. There also exists an improved Lattice QCD estimate of this quantity, $\zeta = 0.9 \pm 0.1$ [68]. In the present analysis, we use $\zeta = 0.85 \pm 0.10$.

Within the Standard Model, measurements of the isospin-breaking and $CP$ asymmetries in the decay rates will provide a precise determination of the angle $\alpha$. They are of interest for searches beyond-the-Standard Model in the $b \to d$ radiative transitions. Of these, the isospin-breaking ratios in the decays $B \to \rho\gamma$ are defined as

$$\Delta(\rho\gamma) \equiv \frac{1}{2}\left(\Delta^{+0} + \Delta^{-0}\right), \qquad \Delta^{\pm 0} = \frac{\Gamma(\mathrm{B}^\pm \to \rho^\pm\gamma)}{2\Gamma(\mathrm{B}^0(\overline{\mathrm{B}}^0) \to \rho^0\gamma)} - 1\,. \tag{2.48}$$

They have been calculated in the NLO accuracy including the annihilation contributions [31, 32, 34]. Likewise, the $CP$ asymmetry defined as

$$A_{CP}^{\pm}(\rho\gamma) = \frac{\mathcal{B}(B^- \to \rho^-\gamma) - \mathcal{B}(B^+ \to \rho^+\gamma)}{\mathcal{B}(B^- \to \rho^-\gamma) + \mathcal{B}(B^+ \to \rho^+\gamma)} \quad , \quad A_{CP}^{0}(\rho\gamma) = \frac{\mathcal{B}(\overline{B}_d^0 \to \rho^0\gamma) - \mathcal{B}(B_d^0 \to \rho^0\gamma)}{\mathcal{B}(\overline{B}_d^0 \to \rho^0\gamma) + \mathcal{B}(B_d^0 \to \rho^0\gamma)} \tag{2.49}$$

has also been calculated in the NLO order [31, 32, 34].

We summarize in Table 2-8 the Standard Model-based estimates for all the observables introduced above (See Ref. [35] for the values of the theoretical parameters and the definition of the averaged $B \to (\rho,\omega)\gamma$ mode).

## 2.8.2 Impact of $\overline{R}_{\rm exp}[(\rho,\omega)/K^*]$ on the CKM unitarity triangle

In this section we present the impact of the $B \to (\rho,\omega)\,\gamma$ branching ratio on the CKM parameters $\overline{\rho}$ and $\overline{\eta}$. For this purpose, it is convenient to rewrite the ratio $\overline{R}_{\rm th}[(\rho,\omega)\gamma/K^*\gamma]$ in the form in which the dependence on the CKM-Wolfenstein parameters $\overline{\rho}$ and $\overline{\eta}$ is made explicit:

$$\begin{aligned}\overline{R}_{\rm th}[(\rho,\omega)\gamma/K^*\gamma] \;=\; & \frac{\lambda^2\zeta^2}{4}\,\frac{(M_B^2 - m_\rho^2)^3}{(M_B^2 - m_{K^*}^2)^3}\left[2\,G(\overline{\rho},\overline{\eta},\varepsilon_A^{(\pm)}) + G(\overline{\rho},\overline{\eta},\varepsilon_A^{(0)})\right] \\ & + \frac{\lambda^2\zeta^2}{4}\,\frac{(M_B^2 - m_\omega^2)^3}{(M_B^2 - m_{K^*}^2)^3}\,G(\overline{\rho},\overline{\eta},\varepsilon_A^{(\omega)}),\end{aligned} \tag{2.50}$$

where the function $G(\overline{\rho}, \overline{\eta}, \varepsilon_A)$ encodes both the LO and NLO contributions:

$$G(\overline{\rho}, \overline{\eta}, \varepsilon) = [1 - (1-\varepsilon)\,\overline{\rho}]^2 + (1-\varepsilon)^2 \overline{\eta}^2 + 2\,\mathrm{Re}\left[G_0 - \overline{\rho}\,G_1(\varepsilon) + (\overline{\rho}^2 + \overline{\eta}^2)\,G_2(\varepsilon)\right], \tag{2.51}$$

and the numerical values of the functions $G_i$ ($i = 0, 1, 2$) and of the parameters $\varepsilon_A^{(i)}$ can be found in Ref. [35].

To undertake the fits of the CKM parameters, we adopt a Bayesian analysis method. Systematic and statistical errors are combined in quadrature and the resulting $\chi^2$-function is then minimized over the following parameters: $\overline{\rho}$, $\overline{\eta}$, $A$, $\hat{B}_K$, $\eta_1$, $\eta_2$, $\eta_3$, $m_c(m_c)$, $m_t(m_t)$, $\eta_B$, $f_{B_d}\sqrt{B_{B_d}}$, $\xi$. Further details can be found in Ref. [35, 178].

**Table 2-9.** *The 68%* CL *ranges for the CKM-Wolfenstein parameters, CP-violating phases, $\Delta M_{B_s}$ from the CKM-unitarity fits.*

| | | | | | |
|---|---|---|---|---|---|
| $\overline{\rho}$ | $[0.10\,,\,0.24]$ | $\sin(2\alpha)$ | $[-0.44\,,\,+0.30]$ | $\alpha$ | $[81\,,\,103]^\circ$ |
| $\overline{\eta}$ | $[0.32\,,\,0.40]$ | $\sin 2\beta$ | $[0.69\,,\,0.78]$ | $\beta$ | $[21.9\,,\,25.5]^\circ$ |
| $A$ | $[0.79\,,\,0.86]$ | $\sin 2\gamma$ | $[0.50\,,\,0.96]$ | $\gamma$ | $[54\,,\,75]^\circ$ |
| $\Delta M_{B_s}$ | $[16.6\,,\,20.3]$ ps$^{-1}$ | | | | |

We present the output of the fits in Table 2-9, where we show the 68% CL ranges for the CKM parameters, $A$, $\overline{\rho}$ and $\overline{\eta}$, the angles of the unitarity triangle, $\alpha$, $\beta$ and $\gamma$, as well as $\sin 2\phi_i$ (with $\phi_i = \alpha, \beta, \gamma$) and $\Delta m_{B_s}$. The 95% CL allowed region in the $\overline{\rho}-\overline{\eta}$ plane is shown in Fig. 2-7 (shaded area). Here we also show the 95% CL range of the ratio $\overline{R}_{\rm exp}[(\rho,\omega)\,\gamma/K^*\gamma] = \overline{\mathcal{B}}_{\rm exp}[B \to (\rho,\omega)\,\gamma]/\overline{\mathcal{B}}_{\rm exp}(B \to K^*\gamma)$. We find that the current measurement of $\overline{R}_{\rm exp}[(\rho,\omega)\,\gamma/K^*\gamma]$ is in comfortable agreement with the fits of the CKM unitarity triangle resulting from the measurements of the five quantities ($R_b$, $\epsilon_K$, $\Delta m_{B_d}$, $\Delta m_{B_s}$, and $a_{J/\psi\,K^0_S}$). The resulting contour in the $\overline{\rho}-\overline{\eta}$ plane practically coincides with the shaded region, and hence is not shown. We conclude that due to the large experimental error on $\overline{R}_{\rm exp}[(\rho,\omega)\,\gamma/K^*\gamma]$, but also due to the significant theoretical errors, the impact of the measurement of $B \to (\rho,\omega)\gamma$ decays on the profile of the CKM unitarity triangle is currently small. That this is expected to change in the future is illustrated by reducing the current experimental error on $\overline{R}_{\rm exp}[(\rho,\omega)\,\gamma/K^*\gamma]$ by a factor of three, which is a realistic hope for the precision on this quantity from the $B$ Factory experiments within several years. The resulting (95% CL) contours are shown as dashed-dotted curves, which result in reducing the currently allowed $\overline{\rho}-\overline{\eta}$ parameter space. This impact will be enhanced if the theoretical errors, dominated by $\Delta\zeta/\zeta$, are also brought under control.

E.L. and A.Y.P. are partially supported by the Swiss National Funds.

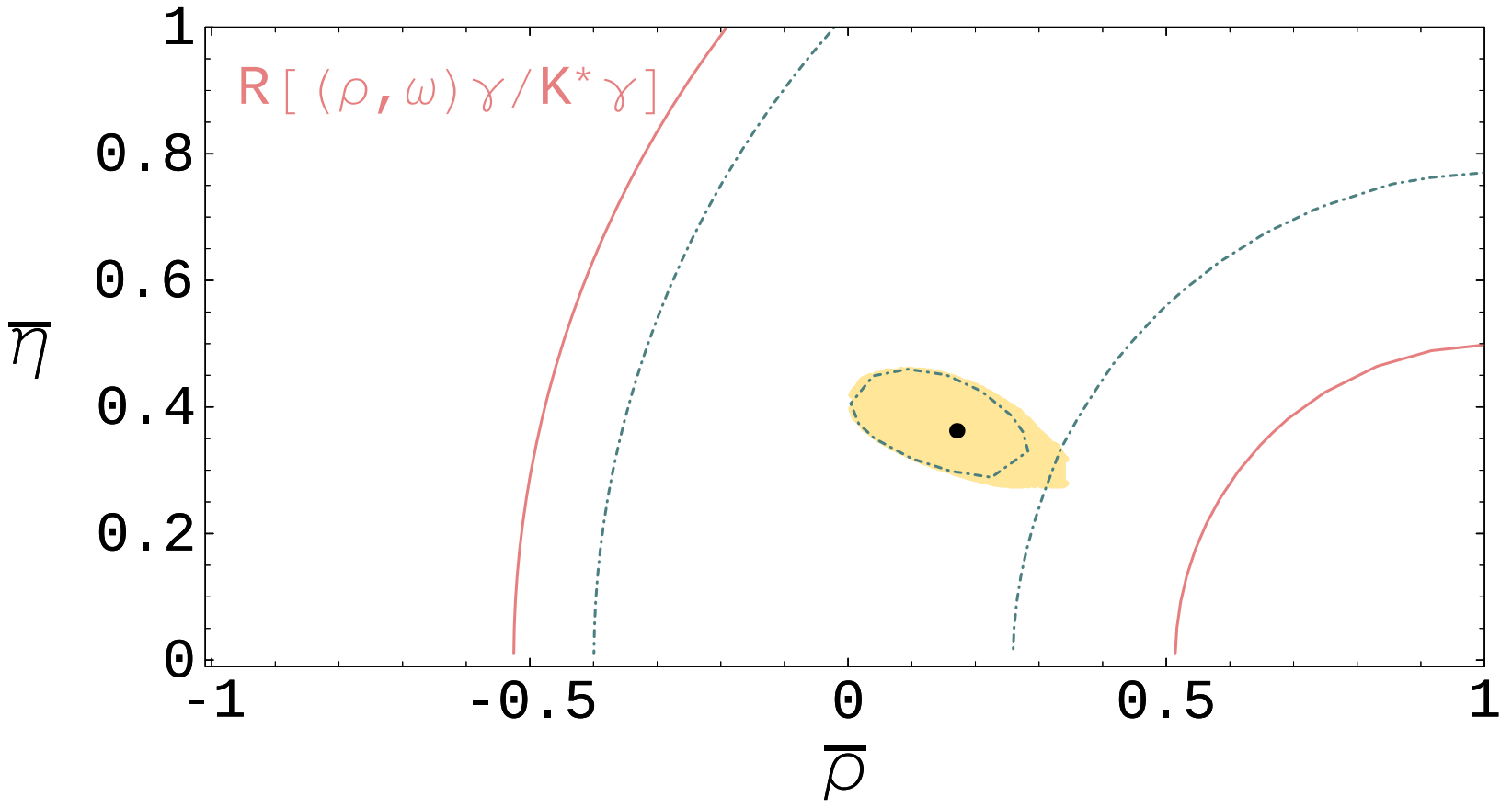

**Figure 2-7.** *Allowed $\overline{\rho} - \overline{\eta}$ regions following from the six measurements ($R_b$, $\epsilon_K$, $\Delta M_{B_d}$, $\Delta M_{B_s}$, $a_{\psi K^0_S}$, and $\overline{R}_{\rm exp}[(\rho,\omega)\,\gamma/K^*\gamma]$), corresponding to 95% C.L., with the dot showing the best-fit values. The shaded region shows the current profile. The two outer (solid) curves give the 95% C.L. constraints in the $\overline{\rho}$ - $\overline{\eta}$ plane from the current measurement of $\overline{R}_{\rm exp}[(\rho,\omega)\,\gamma/K^*\gamma]$. The inner (dashed-dotted) curves are the 95% C.L. constraints from an assumed measurement of $\overline{R}_{\rm exp}[(\rho,\omega)\,\gamma/K^*\gamma]$ with the current central value but the experimental errors reduced by a factor 3. The contour shows the potential impact of this assumed measurement in the $\overline{\rho}$ - $\overline{\eta}$ plane.*

## 2.9 The Time-Dependent $A_{CP}$ in $B^0 \to K^{*0}\gamma,\ (K^{*0} \to K^0_S\pi^0)$

≻– J. Libby –≺

*BABAR* is finalizing a measurement of the time-dependent $CP$-violating asymmetry parameters of the decay $B^0 \to K^{*0}\gamma,\ (K^{*0} \to K^0_S\pi^0)$ [4]. At the $\Upsilon(4S)$, the distribution of $\Delta t$, the proper time difference between the decay time of the $B$ meson to the $K^{*0}\gamma$ final state and the decay of the other $B$ meson to a self–tagging final state, is given by:

$$\mathcal{P}(\Delta t) = \frac{e^{-|\Delta t|/\tau}}{4\tau} \times \left[1 \pm \left(S_{K^{*0}\gamma}\sin(\Delta m_d \Delta t) - C_{K^{*0}\gamma}\cos(\Delta m_d \Delta t)\right)\right],$$

where the upper (lower) sign corresponds to the tagging $B$ meson decaying as a $B^0$ ($\overline{B}^0$), $\tau$ is the $B^0$ lifetime averaged over the two mass eigenstates, $\Delta m_d$ is the mixing frequency, $C_{K^{*0}\gamma}$ is the magnitude of direct $CP$ violation and $S_{K^{*0}\gamma}$ is the magnitude of mixing-induced $CP$ violation. In the Standard Model, the mode is nearly self-tagging since the parity-violating weak interaction leads, most of the time, to the photon from $B^0$ decay having opposite helicity to that from $\overline{B}^0$ decay. Thus, the Standard Model expectations for $C_{K^{*0}\gamma}$ and $S_{K^{*0}\gamma}$ are zero and $2m_s/m_b \sin 2\beta \sim 0.03$, respectively, where $m_s$ is the mass of the $s$ quark and $m_b$ is the mass of the $b$ quark. The small value of $S_{K^{*0}\gamma}$ accounts for the rate of helicity flip [180, 181]. New Physics might enhance the rate of wrong-helicity decays, leading to an increase in the value of $S_{K^{*0}\gamma}$.

The branching ratio for $B^0 \to K^{*0}\gamma$ is $4.3 \times 10^{-5}$ [2] and the branching fraction of $K^{*0} \to K^0_S\pi^0$ is 1/9, yielding an overall branching fraction of around $5 \times 10^{-6}$. The background arises mainly from combinatorics in continuum events, and from other $B \to X_s\gamma$ decays in which one particle in the final state is missing, leading to the $X_s$ system being reconstructed as a $K^{*0}$. The signal selection requires a high energy photon and a $K^{*0}$ candidate, composed of a $K^0_S$ and a $\pi^0$, which combine to lie within broad range of $m_{ES}$ and $\Delta E$ about the nominal values for a $B$ meson decay. The continuum and $B\overline{B}$ backgrounds are suppressed by cuts on the $K^{*0}$ helicity and the thrust angle. Also, a Fisher discriminant, $\mathcal{F}$, which combines event shape variables, is used to separate background in the final likelihood fit to extract the $CP$ parameters. The fit uses probability density functions, PDFs, of the parameterized distributions of signal and both background types in $\Delta E$, $m_{ES}$, $K^{*0}$ mass and $\mathcal{F}$. In addition, it uses flavor tag information from the other $B$ decay [182]. The other PDF for the fit is $\mathcal{P}(\Delta t)$, which requires a measurement of $\Delta t$, and is convolved with a $\Delta t$ resolution function. Given that the signal $B$ decay contains only neutral particles in the final state a novel method of vertex reconstruction was used to measure the decay time. The small transverse displacement of the $B$ meson in the laboratory frame is exploited by constraining it to decay at the interaction point in the transverse plane. Therefore, the intersection of the flight direction of the $K^0_S$ with the beam axis defines the decay vertex position. The additional uncertainty introduced by ignoring the transverse flight direction is included by inflating the error on the transverse position of the interaction point. This method of vertex reconstruction has already been used to measure time dependent $CP$ violation in $B^0 \to K^0_S\pi^0$ [183] and has been validated on control samples of $B^0 \to J/\psi K^0_S$. The $B$ decay vertex resolution depends strongly on the radius at which the $K^0_S$ decays, which dictates the amount of information from the silicon tracker, SVT, used to reconstructed the $K^0_S$ daughters. The events are classified according to the amount of SVT information used; only events in classes with good $\Delta t$ resolution are included in the fit. Furthermore, the classes used have differing parameteriz ations of the $\Delta t$ resolution in the fit.

The analysis has been performed on $113\ \text{fb}^{-1}$ of data and the statistical uncertainty on $S_{K^{*0}\gamma}$ ($C_{K^{*0}\gamma}$) is 0.63 (0.32). Therefore, using the same method at a Super $B$ Factory, the statistical uncertainty on $S_{K^{*0}\gamma}$ ($C_{K^{*0}\gamma}$) would be 0.21 (0.11) with $1\,\text{ab}^{-1}$ of data, and 0.07 (0.04) with $10\,\text{ab}^{-1}$ of data. The systematic uncertainty is currently estimated to be 0.14 on both $S_{K^{*0}\gamma}$ and $C_{K^{*0}\gamma}$; this is dominated by the uncertainties on the yield and $CP$ asymmetry of the $B\overline{B}$ background, which are evaluated very conservatively. As the statistical error approaches the level of the systematic error, the uncertainties from the $B\overline{B}$ background will be better constrained by measurements made on control samples.

[4] This analysis is now available [179]; the result is $S_{K^*\gamma} = 0.25 \pm 0.63 \pm 0.14$ and $C_{K^*\gamma} = -0.57 \pm 0.32 \pm 0.09$.

## 2.10 Prospects for measuring photon polarization in $b \to s\gamma$

>–D. Pirjol –<

Rare radiative $b \to s\gamma$ decays have been extensively investigated both as a probe of the flavor structure of the Standard Model and for their sensitivity to any New Physics beyond the Standard Model (for a recent review, see *e.g.*, [41]). In addition to the rather well-predicted inclusive branching ratio, there is a unique feature of this process within the Standard Model that has drawn only moderate theoretical attention, and has not yet been tested. Namely, the emitted photons are left-handed in radiative $B^-$ and $\overline{B}^0$ decays and are right-handed in $B^+$ and $B^0$ decays.

This prediction holds in the Standard Model to within a few percent, up to corrections of order $m_s/m_b$, for exclusive and inclusive decays. On the other hand, in certain extensions of the Standard Model, an appreciable right-handed component can be induced in $b \to s\gamma$ decays. This is the case in the MSSM with unconstrained flavor structure, where the gluino-squark loops can produce a right-handed photon [184]. Another possibility is the left-right symmetric model with gauge group $\mathrm{SU}(2)_L \times \mathrm{SU}(2)_R \times \mathrm{U}(1)$, where the same effect is introduced by $W_R - W_L$ mixing [185]. A measurement of the photon helicity in $b \to s\gamma$ will help to constrain these models and set bounds on the properties of New Physics particles.

Several methods have been suggested to measure the photon helicity in $b \to s\gamma$ processes. In the first method [180], the photon helicity is probed through mixing-induced $CP$ asymmetries. The sensitivity to the polarization is introduced through interference between $B^0$ and $\overline{B}^0$ decays into a common state of definite photon polarization. However, measuring asymmetries at a level of a few percent, as expected in the Standard Model, requires about $10^9$ $B$ mesons, which would only make it feasible at a Super $B$ Factory, see Section 2.9 for an experimental study. In a second scheme one studies angular distributions in $B \to \gamma(\to e^+e^-)K^*(\to K\pi)$, where the photon can be virtual [186, 187, 188] or real, converting in the beam pipe to an $e^+e^-$ pair [181]. This is discussed in detail in Section 2.17. The efficiency of this method is comparable to that of the previous method. A somewhat different method, proposed in [189], makes use of angular correlations in both exclusive and inclusive $\Lambda_b \to X_s\gamma$ decays.

An especially promising method [190, 191, 192] for measuring the photon polarization makes use of angular correlations in the strong decay of a $K_{\mathrm{res}}$ resonance produced in $B \to K_{\mathrm{res}}\gamma$. The dominant $B \to K^*\gamma$ mode cannot be used for this purpose, since the $K^*$ polarization information is not observable in its two-body strong decay $K^* \to K\pi$; it is impossible to form a $T$-odd quantity from just two vectors $\vec{q}$ (photon momentum in the $K_{\mathrm{res}}$ frame) and $\hat{n}$ (the direction parameterizing the final state $|K(\hat{n})\pi(-\hat{n})\rangle$.

A nonvanishing asymmetry is possible, however, in three-body strong decays: $K_{\mathrm{res}} \to K\pi\pi$, where $K_{\mathrm{res}}$ represents the lowest excitations of the $K$ meson, with quantum numbers $J^P = 1^-, 1^+, 2^+$, some of which have been seen in rare radiative decays. The Belle, CLEO and *BABAR* Collaborations observed the decay $B \to K_2^*(1430)\gamma$ with branching ratios shown in Table 2.10. Similar branching ratios are expected from theoretical estimates for decays into $K_1(1400)$ and $K_1(1270)$ [193].

**Table 2-10.** *Measurements of the branching ratio for $B \to K_2^*(1430)\gamma$ (in units of $10^{-5}$).*

| Decay | *BABAR* [194] | Belle [195] | CLEO [196] |
|---|---|---|---|
| $\mathcal{B}(B^0 \to K_2^{*0}(1430)\gamma)$ | $1.22 \pm 0.25 \pm 0.11$ | $1.3 \pm 0.5 \pm 0.1$ | $1.66^{+0.59}_{-0.53} \pm 0.13$ |
| $\mathcal{B}(B^- \to K_2^{*-}(1430)\gamma)$ | $1.44 \pm 0.40 \pm 0.13$ | | |

These states decay strongly to three-body final $K\pi\pi$ states[5]. Neglecting a small nonresonant contribution, these decays are dominated by interference of a few channels (see Fig. 2-8)

$$K^+_{\rm res} \to \left\{ \begin{array}{c} K^{*+}\pi^0 \\ K^{*0}\pi^+ \\ \rho^+ K^0 \end{array} \right\} \to K^0\pi^+\pi^0\,, \qquad K^0_{\rm res} \to \left\{ \begin{array}{c} K^{*+}\pi^- \\ K^{*0}\pi^0 \\ \rho^- K^+ \end{array} \right\} \to K^+\pi^-\pi^0\,. \tag{2.52}$$

We focus only on $K\pi\pi$ modes containing one neutral pion, which receive contributions from two distinct $K^*\pi$ intermediate states. These two contributions are related by isospin symmetry and contribute with a calculable relative strong phase which can be parameterized in terms of Breit-Wigner forms. The contribution of the $K\rho$ state has to be added as well, thereby introducing an uncertainty.

This uncertainty is minimal for decays proceeding through the $J^P = 1^+$ $K_1(1400)$ resonance. This state decays predominantly to $K^*\pi$ in a mixture of $S$ and $D$ waves, with a branching ratio of 95% [2]. To a good approximation one can neglect the $D$ wave component, allowing a parameter-free computation of the asymmetry. The smaller $D$-wave component and the $K\rho$ contribution can also be included using the data on partial wave amplitudes and phases measured by the ACCMOR Collaboration [197].

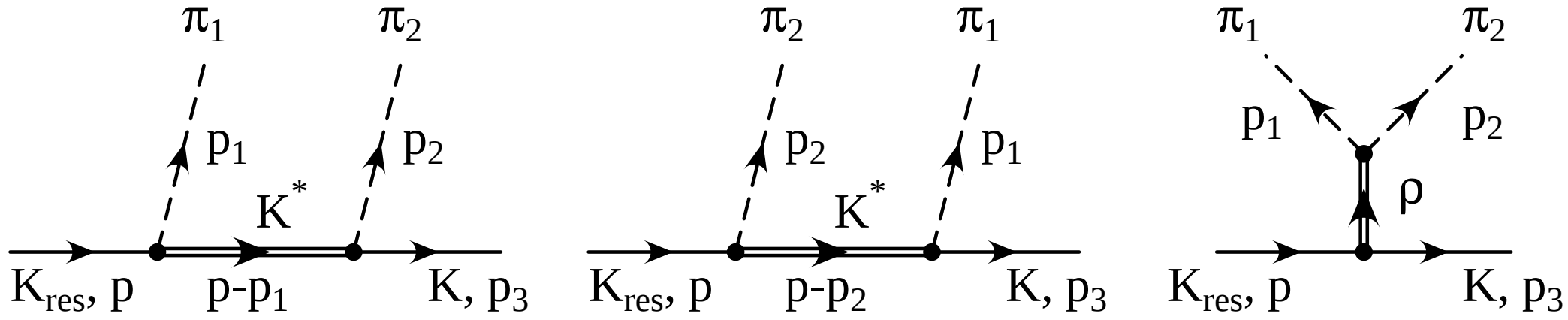

**Figure 2-8.** *Pole graphs contributing to the strong decay $K_{\rm res} \to K\pi\pi$, proceeding through resonant $K^*\pi$ and $K\rho$ intermediate states.*

A realistic computation of the $B \to K\pi\pi\gamma$ decay distribution in the region of $K\pi\pi$ invariant mass $M_{K\pi\pi} = 1.2 - 1.5$ GeV should take into account the interference of contributions from the few $K$ resonances with masses in this range. These include $K_1(1400)$ ($J^P = 1^+$), $K^*(1410)$ ($J^P = 1^-$) and $K_2^*(1430)$ ($J^P = 2^+$). Reference [192] contains detailed results for the Dalitz plot and angular distributions in $B \to K\pi\pi\gamma$ decays, including interference effects from multiple $K$ resonances.

We quote here only the result for the distribution in $s = (p_K + p_{\pi_1} + p_{\pi_2})^2$ and $\tilde\theta$, which is sensitive to the photon polarization (Eq. (44) in [192]). For this purpose it is convenient to work in the rest frame of the resonance $K_{\rm res}$. The angle $\tilde\theta$ is taken to be between the opposite of the photon momentum $-\vec q$ and the normal to the $K\pi\pi$ decay plane defined as $\vec p_{\rm slow} \times \vec p_{\rm fast}$, where $\vec p_{\rm slow}$ and $\vec p_{\rm fast}$ are the momenta of the slower and faster pions. With these definitions one has [192]

$$\begin{aligned} \frac{{\rm d}^2\Gamma}{{\rm d}s{\rm d}\cos\tilde\theta} &= \frac14 |c_1|^2 |B_{K_1}(s)|^2 \left\{1 + \cos^2\tilde\theta + 4P_\gamma R_1 \cos\tilde\theta\right\} \\ &+ \frac14 |c_2|^2 |B_{K_2^*}(s)|^2 \left\{\cos^2\tilde\theta + \cos^2 2\tilde\theta + 12 P_\gamma R_2 \cos\tilde\theta \cos 2\tilde\theta\right\} + |c_3|^2 |B_{K_1^*}(s)|^2 \sin^2\tilde\theta \\ &+ {\rm Im}\left[c_{12} B_{K_1}(s) B^*_{K_2^*}(s)\right] \frac12 (3\cos^2\tilde\theta - 1) + P_\gamma {\rm Re}\left[c'_{12} B_{K_1}(s) B^*_{K_2^*}(s)\right] \cos^3\tilde\theta\,, \end{aligned} \tag{2.53}$$

where the first three terms are produced by decays through $K_{\rm res}$ resonances with $J^P = 1^+, 2^+$ and $1^-$, and the last terms come from $1^+ - 2^+$ interference, respectively. We denoted here the Breit-Wigner form $B(s) = 1/(s - M^2 - i\Gamma M)$ corresponding to a $K$ resonance with parameters $(M, \Gamma)$.

[5] Note that the $K_2^*(1430)$ was seen only in two-body channels in Refs. [194, 195, 196]. The only three-body channel analysis was done in [195], which measured $\mathcal{B}(B \to K^+\pi^+\pi^-\gamma) = (2.4 \pm 0.5^{+0.4}_{-0.2}) \times 10^{-5}$.

The hadronic parameters $R_{1,2}$ appearing in Eq. (2.53) can be computed with relatively small model dependence as explained above, which gives [191, 192] $R_1 = 0.22 \pm 0.03$, $R_2 = 0.01 - 0.05$. Using these values, measurements of the angular distribution (2.53) can be used to extract the photon polarization parameter $P_\gamma$ [192].

Selecting only $K\pi\pi$ events with invariant mass around the $J^P = 1^+$ resonance $K_1(1400)$, the first term in the angular distribution (2.53) can be expected to dominate. This predicts an up-down asymmetry of the photon momentum direction relative to the normal to the $K\pi\pi$ plane $A_{\rm up-down} = \frac{3}{2} R_1 P_\gamma$. The significant value of this asymmetry makes this channel particularly attractive.

Assuming an exclusive branching ratio $\mathcal{B}(B \to K_1(1400)\gamma) = 0.7 \times 10^{-5}$ and taking the final state in (2.52) to be detected through the $K^+\pi^-\pi^0$ and $K^0_S\pi^+\pi^0$ modes, implies that about $2 \times 10^7$ $B\overline{B}$ pairs are required to measure 80 $K\pi\pi\gamma$ events which should be sufficient for a $3\sigma$ confirmation of a left-handed photon in $b \to s\gamma$ decay. Such a measurement should be feasible at the existing $B$ Factories in the near future, and will become a precision measurement of the photon polarization at a Super $B$ Factory.

## 2.11 Experimental prospects for $B \to K\gamma\gamma$

≻ S. Dasu, F. Di Lodovico, and A. Rubin ≺

The rare flavor-changing neutral current decay $b \to s\gamma\gamma$ is one of the processes that is sensitive to New Physics contributions and has been theoretically studied in detail, see Section 2.12. This quark level transition was previously studied in the exclusive decay $B_s \to \gamma\gamma$ at LEP and an experimental limit $\mathcal{B}(B_s \to \gamma\gamma) < 1.5 \times 10^{-4}$ was set by L3 collaboration [198] at 90% confidence level. At a $B$ Factory, this transition can be studied in $B \to K^{(*)}\gamma\gamma$ modes. The recent prediction for the exclusive $B \to K\gamma\gamma$ branching fraction is at least $2.7 \times 10^{-7}$ in the Standard Model, with a cut on $\sqrt{s_{\gamma\gamma}} > 3$ GeV [212].

Although in the Standard Model, this transition has a branching ratio of about $10^{-6} - 10^{-7}$, it can be appreciably different in two higgs doublet models [199, 209, 200]. Although $b \to s\gamma\gamma$ is suppressed by $\alpha_{\rm em}$ with regard to $b \to s\gamma$, it continues to be of interest because any New Physics contribution to this decay may manifest itself differently in experimental observables, due to the presence of additional diagrams. In addition to the usual observables, the rate and $A_{CP}$, forward-backward asymmetry, $A_{FB,s\gamma_{\rm low}}$, of the $s$ quark and the softer of the two photons, $\gamma_{\rm low}$ can play a role in the search for New Physics. Note that this process is of the same order in the electro-weak couplings as $b \to s\ell^+\ell^-$ although it is more difficult to study experimentally.

Following the techniques established in the *BABAR* analysis of $B \to K^*\gamma$, a Monte Carlo feasibility study of measuring the exclusive decay $B^{\pm} \to K^{\pm}\gamma\gamma$ was performed. Similar to $B \to K^*\gamma$ the continuum backgrounds for this process are due to ISR photons or photons from $\pi^0$ and $\eta$ decays. Requireming of two high energy photons in the event, both with energies, $1.0 \leq E_{\gamma_{\rm low}} \leq 3.5$ GeV and $1.5 \leq E_{\gamma_{\rm gigh}} \leq 3.5$ GeV, suppresses the background considerably. It was important to suppress the secondary photons from $\pi^0$ and $\eta$ decays by cutting strictly on the invariant mass of pairs of photons detected in the event. The continuum background, particularly that due to the initial state radiation photon in combination with a misidentified $\pi^0$ or $\eta$, is rather large. A neural network that used event shape variables helped reduce these backgrounds to a manageable level. Only four out of $7.34 \times 10^8$ generated continuum events survive these cuts in the signal region $5.273 \leq m_{ES} \leq 5.284$ GeV and $-0.08 \leq \Delta E \leq 0.09$ GeV. In addition to the continuum background there is a large $B\overline{B}$ background. Random combinations of $K^{\pm}$ and two photons in the decays of generic $B\overline{B}$ can mimic the signal. Often one or more of the photons are from $\pi^0$ or $\eta$ decays. Therefore, tight cuts requiring that the selected photons do not form a $\pi^0$ or $\eta$ when combined with any other photon in the event reduces this background. We found that 36 out of $3.36 \times 10^8$ $B^+B^-$ events and one out of $3.48 \times 10^8$ $B^0\overline{B}^0$ events survive these strict cuts within the signal region, $5.273 \leq m_{ES} \leq 5.284$ GeV and $-0.08 \leq \Delta E \leq 0.09$ GeV. We have looked at the generator information and found that all the $B^+B^-$ events surviving the cuts contain one $B$ that decayed into $X_s\gamma$. In almost all of these events, the kaon and one photon came from this $X_s\gamma$ decay, whereas the second photon was faked by a $\pi^0$ or $\eta^0$ decay from the other $B$. This signal peaks in the beam energy constrained variable $m_{ES}$. Although, the $\Delta E$ distribution for these events is somewhat different from the signal events which are peaked at zero, it is quite difficult to extract the signal cleanly. We have used a simple phase space model to generate 6666 $B^{\pm} \to K^{\pm}\gamma\gamma$ events. Of these, 389 events survive the cuts chosen to suppress the continuum and $B\overline{B}$ backgrounds, yielding only 5.9% signal efficiency. This study indicates that with a 200 fb$^{-1}$ *BABAR* data sample that will be available in 2004, a branching fraction limit of $\mathcal{B}(B_d \to K\gamma\gamma) < 7 \times 10^{-7}$ at 90% confidence level can be set.

We are devising ways to reduce the $B\overline{B}$ background further by making additional requirements on the second $B$ from which a fake photon is selected. The ultimate step in such a process would be to fully reconstruct the second $B$, thereby incurring a large efficiency reduction, $\epsilon \approx 10^-3$. Although this technique is clean, we will not be able to measure processes with $\approx 10^{-7}$ branching fraction using this technique. Therefore, we are investigating partial reconstruction of the second $B$, which may be devised with efficiencies $\approx 10^{-2}$. With ultimate data set of 10 ab$^{-1}$ one should be able to measure the process $B \to K\gamma\gamma$.

## 2.12 Double Radiative $b \to (s,d)\,\gamma\gamma$ Decays - Theory Aspects

$\succ$ G. Hiller $\prec$

The double radiative decays $B_{s,d} \to \gamma\gamma$ and $B \to (K, K^*, X_s)\gamma\gamma$ have so far received less attention than $b \to s\gamma$ induced decays. The reasons are the more complicated hadronic physics involved, the $\alpha_{em}$ suppression of the rate, and the obvious correlation with the single photon mode. All these points can at least partially be neutralized and turned into advantages:

- With Standard Model branching ratios of $\mathcal{O}(10^{-8} - 10^{-6})$ and possible New Physics enhancements of up to an order of magnitude the diphoton modes are accessible at a Super $B$ Factory.
- New physics in the penguin operators $\mathcal{O}_i \propto \bar{s}b\bar{f}f$ with fermions $f$ can alter the diphoton decay rates substantially whereas this is only a 2-loop effect in $b \to s\gamma$ decays.
- More observables beyond the rate can be studied in $B \to (K^{(*)}, X_s)\gamma\gamma$ decays.
- The $B \to \gamma\gamma$ decays can teach us about hadronic input to other modes.

Rare $b \to (s,d)\gamma\gamma$ decays have been studied in the Standard Model [201, 202, 203], with an effective Hamiltonian theory at leading log [204, 210, 205] and within QCD factorization [124]. The decays have been analyzed in several New Physics models, such as the 2HDM [199, 209, 200], the MSSM [206] and the $R$-parity-violating (RPV) MSSM [207]. Properties of the diphoton modes are summarized in Table 2-11. Note that in the MSSM the $B_s \to \gamma\gamma$ branching ratio can be enhanced over the Standard Model value by at most $\sim 1.3$ [206], which is similar to the 2HDM. This moderate impact of New Physics on $b \to s\gamma\gamma$ decays results from the experimental constraint on the $B \to X_s\gamma$ branching ratio, and is generic to models that predominantly modify the dipole operators $O_{7\gamma(8g)}$. On the other hand, the RPV-MSSM induces sizable contributions to $b \to s\gamma\gamma$ decays from sneutrino exchange in the 1PI contribution [207]. Since the respective 4-Fermi operators appear in the $b \to s\gamma$ rate only at higher order, that is, at two loops, they are unconstrained by single photon decays. In the following we briefly summarize the highlights of the individual decay modes.

**Table 2-11.** *Standard Model branching fractions and current upper bounds at 90 %* CL *for double radiative rare $b$ decays (see the original works for cuts applied for $\mathcal{B}(B \to (K^{(*)}, X_s)\gamma\gamma)$ estimates). Also given is the maximum enhancement of the branching ratios in the 2HDM and the RPV MSSM with respect to the Standard Model. †We removed the $\eta_c$-contribution.*

| Modes | Standard Model | Exp. bounds | 2HDM | RPV MSSM |
|---|---|---|---|---|
| $B_d \to \gamma\gamma$ | $3.1^{+6.4}_{-1.6} \times 10^{-8}$ [124] | $1.7 \times 10^{-6}$ [208] | – | – |
| $B_s \to \gamma\gamma$ | $1.2^{+2.5}_{-0.6} \times 10^{-6}$ [124] | $1.48 \times 10^{-4}$ [198] | 2 [209] | 16 [207] |
| $B \to X_s\gamma\gamma$ | $\sim (3.7 - 5.1) \times 10^{-7}$ [210, 200] | – | 2-3 [200] | 5 [207] |
| $B \to K\gamma\gamma$ | $\sim (0.5 - 5.6) \times 10^{-7}$ [211, 212] | – | – | – |
| $B \to K^*\gamma\gamma$ | few $\times 10^{-7}$ [213]† | – | – | – |

$\boldsymbol{B \to \gamma\gamma}$ **decays:** The Standard Model branching ratios for $B_{s,d} \to \gamma\gamma$ decays have a large uncertainty, which stems from the hadronic $B$ meson parameter $\lambda_B \sim \mathcal{O}(\Lambda)$. All other sources of theory error, *e.g.,* the $\mu$-scale, the $B$ meson decay constant $f_B$ and CKM elements are subdominant, of order $\pm 50\%$, and not included in Table 2-11, see [124] for details. Here, $\lambda_B$ plays the role of the spectator mass in previous calculations. In the framework of QCD factorization, $\lambda_B$ is a universal parameter that enters also other $B$ decays. The enhanced sensitivity of $B \to \gamma\gamma$ modes might be a way for its experimental determination. Long-distance effects via $B_s \to \phi\gamma \to \gamma\gamma$ [204] and $B_s \to \phi J/\psi \to \phi\gamma \to \gamma\gamma$ [205] are suppressed, because the intermediate vector bosons are sufficiently off-shell. They are power corrections in QCD factorization. Further, $CP$ asymmetries can be studied, *i.e.*,

$$r^q_{CP} \equiv \frac{|A_q|^2 - |\overline{A}_q|^2}{|A_q|^2 + |\overline{A}_q|^2} \tag{2.54}$$

where $A_q = A(\overline{B}_q \to \gamma\gamma), \overline{A}_q = A(B_q \to \gamma\gamma)$. In the Standard Model $r^d_{CP} \simeq -5\%$ with the dominant $\mathcal{O}(1)$ uncertainty arising from the scale dependence, followed by that from $\lambda_B$ [6].

**Inclusive $b \to s\gamma\gamma$ decays:** The inclusive three-body decay allows to study spectra such as distributions in the diphoton invariant mass $m_{\gamma\gamma}$ or in the angle between the photons. Further, a forward-backward asymmetry similar to the one in $b \to s\ell^+\ell^-$ decays, can be constructed as

$$A_{FB} = \frac{\Gamma(\cos\theta_{s\gamma} \geq 0) - \Gamma(\cos\theta_{s\gamma} < 0)}{\Gamma(\cos\theta_{s\gamma} \geq 0) + \Gamma(\cos\theta_{s\gamma} < 0)} \tag{2.55}$$

where $\theta_{s\gamma}$ denotes the angle between the $s$ quark and the softer photon [199].

The $b \to s\gamma\gamma$ amplitude has IR divergences for vanishing photon energies, which cancel with the virtual electromagnetic corrections to $b \to s\gamma$ [210]. Since we are interested in $b \to s\gamma\gamma$ with hard photons (rather than in $b \to s\gamma$ plus bremsstrahlung corrections), a cut on the photon energies *e.g.*, $E_\gamma > 100$ MeV is used for the estimates given in Table 2-11, or the minimum energy required for the experiment to detect photons. There is sensitivity to low energy physics, *i.e.*, the strange quark mass from the 1PR diagrams [210, 200]. Long-distance effects via $B \to X_s\eta_c \to X_s\gamma\gamma$ can be removed by cuts in $m_{\gamma\gamma}$ [210].

**$B \to (K, K^*)\gamma\gamma$ decays:** Very few calculations of $B \to K\gamma\gamma$ decays are available. They invoke phenomenological modeling of the cascade decays $B \to K^*\gamma \to K\gamma\gamma$ [211, 212] and $B \to \eta_x K \to K\gamma\gamma$, where $\eta_x = \eta, \eta'$ and $\eta_c$ [212] for the 1PR contributions. The irreducible contributions are obtained assuming factorization. The Standard Model branching ratios are estimated as $\mathcal{B}(B \to K\gamma\gamma) \simeq (0.5 - 0.7) \times 10^{-7}$ [211] with $|m_{K\gamma} - m_{K^*}| > 300$ MeV, $E_\gamma > 100$ MeV and $\mathcal{B}(B \to K\gamma\gamma) \simeq (2.7 - 5.6) \times 10^{-7}$ with $m_{\gamma\gamma} \gtrsim m_{\eta_c} + 2\Gamma_{\eta_c}$ [212]. The decay $B \to K^*\gamma\gamma$ is treated similarly in [213]. Note that the sum rule $\sum_{H=K,K^*} \mathcal{B}(B \to H\gamma\gamma) < \mathcal{B}(B \to X_s\gamma\gamma)$ puts constraints on the theory.

Recently, $B \to K^{(*)}\gamma\gamma$ decays have been investigated model-independently with an expansion in scales of the order $m_b$ [214]. In the region of phase space where the operator product expansion is valid, the Standard Model branching ratios induced by short-distance physics turn out to be quite small, order $10^{-9}$.

[6]Note that $r_{CP}$ is defined differently from the $CP$ asymmetries in Ref.[124], which require a determination of the photon polarization. We thank Gerhard Buchalla for producing the numerical value of $r_{CP}$ for us. It corresponds to the central values given in Table 2 of Ref. [124].

## 2.13 Experimental Aspects of the Inclusive Mode $b \to s\ell^+\ell^-$

>– T. Abe, V. Koptchev, H. Staengle, and S. Willocq –<

The electroweak penguin $b \to s\ell^+\ell^-$ decay is a flavor-changing neutral current process and is very sensitive to physics beyond the Standard Model [10, 41]. Therefore, the study of $b \to s\ell^+\ell^-$ decays is particularly interesting at a Super $B$ Factory. Several observables have been studied for these decays: branching fraction, dilepton mass and hadronic mass spectra, and forward-backward asymmetry. These probe physics beyond the Standard Model. The large event samples anticipated at a Super $B$ Factory provide excellent statistical accuracy but it is important to consider potentially limiting theoretical uncertainties. The forward-backward asymmetry proves to be an excellent tool to search for New Physics, since theoretical uncertainties are small, and large deviations from the Standard Model are expected in some of its extensions.

In this section, we discuss the measurement of the $b \to s\ell^+\ell^-$ branching fraction and forward-backward asymmetry with Super $B$ Factory luminosity, based on status of the current analysis at *BABAR* and Belle. First, the analysis method is described. Then, we consider branching fraction measurements at the Super $B$ Factory. Finally, we discuss the measurement of the forward-backward asymmetry.

### 2.13.1 Analysis method

Both *BABAR* [215] and Belle [216] use a "sum over exclusive modes" technique, which is a semi-inclusive approach, because a fully inclusive approach suffers from large backgrounds and has yet to be developed for such a measurement. We reconstruct the hadronic system as one $K^{\pm}$ or $K^0_S \to \pi^+\pi^-$ decay, and up to three pions with at most one $\pi^0$. This allows about 60% of the inclusive rate to be measured. The technique provides powerful kinematical constraints to suppress backgrounds, while it introduces some dependence on the hadronization model and on the knowledge of the particle content of the inclusive final state. The studies presented here were performed using the *BABAR* analysis, with no cut on the dilepton mass. We use fully-simulated Monte Carlo events assuming 90% muon identification efficiency and scale the branching fraction and forward-backward asymmetry results for luminosities of $500\,\mathrm{fb}^{-1}$, $1000\,\mathrm{fb}^{-1}$, $10\,\mathrm{ab}^{-1}$, and $50\,\mathrm{ab}^{-1}$. Total luminosities of $500\,\mathrm{fb}^{-1}$ to $1000\,\mathrm{fb}^{-1}$ are expected to be collected by *BABAR* and Belle; $10\,\mathrm{ab}^{-1}$ and $50\,\mathrm{ab}^{-1}$ are for the Super $B$ Factory after one and five years of operation at design luminosity, respectively.

### 2.13.2 Branching fraction measurement

The control of systematic errors is a key issue for the branching fraction measurement at higher luminosity. The systematic uncertainties can be classified in three categories: signal yield, detector model, and signal model. For the current *BABAR* and Belle analyses, they amount to 11%, 11%, and 13%, respectively. The uncertainty in the signal yield should scale as $1/\sqrt{N}$. For detector modeling, the same $1/\sqrt{N}$ rule is assumed. However, we may not assume that signal model systematics will scale in the same way. The signal model systematic error originates from the uncertainty in the fraction of exclusive decays ($K\ell\ell$ and $K^*\ell\ell$), hadronization and Fermi motion. Currently, the uncertainty in the fraction of exclusive decays is the dominant source of systematic error but future measurements will certainly improve and reduce the size of this uncertainty. To reduce hadronization uncertainties, one could use inclusive $B \to J/\psi X$ data to calibrate the signal model, eventually achieving uncertainties of $\sim 1-2$%. As for the Fermi motion, improved measurements of the photon spectrum in $b \to s\gamma$ decays could reduce the error down to the 1% level. Relative uncertainties are summarized in Table 2-12. In this table, both statistical and systematic errors are shown. The precision in the branching fraction integrated over all dilepton masses is expected to reach interesting levels of sensitivity by the end of *BABAR* and Belle, comparable to the theoretical uncertainty of 17%.

Besides the control of systematic errors, we expect a $\simeq 2\%$ statistical error at a Super $B$ Factory, a value much lower than the current theoretical error. Part of the 17% theoretical uncertainty is due to long distance contributions from $c\bar{c}$ states. Branching fractions in restricted dilepton mass regions are predicted with higher levels of precision. For example, the study in Ref. [86] indicates a theoretical error of about 12%. The branching fraction uncertainties in the restricted region are also summarized in Table 2-12. It is clear that the interest in the measurement of the partial branching fraction increases with increasing luminosity.

**Table 2-12.** *Summary of relative uncertainties of ($b \to s\ell^+\ell^-$) branching fraction measurements at various luminosities. The lower bound on systematic errors assumes pure, and perhaps unrealistic, $1/\sqrt{N}$ scaling.*

| Signal yield | Integrated luminosity | | | |
|---|---|---|---|---|
| $X_s e^+e^- + X_s\mu^+\mu^-$ | 500 fb$^{-1}$ | 1000 fb$^{-1}$ | 10 ab$^{-1}$ | 50 ab$^{-1}$ |
| All $\hat{s}$, (exc. $J/\psi$ veto) | $\sigma_{\rm stat} = 10\%$ $7\% < \sigma_{\rm syst} < 14\%$ | $\sigma_{\rm stat} = 7\%$ $5\% < \sigma_{\rm syst} < 14\%$ | $\sigma_{\rm stat} = 2.1\%$ $1.5\% < \sigma_{\rm syst} < 6\%(?)$ | $\sigma_{\rm stat} = 1.0\%$ $0.7\% < \sigma_{\rm syst} < 6\%(?)$ |
| $0.05 < \hat{s} < 0.25$ | $\sigma_{\rm stat} = 16\%$ | $\sigma_{\rm stat} = 11\%$ | $\sigma_{\rm stat} = 3.4\%$ | $\sigma_{\rm stat} = 1.5\%$ |
| $0.65 < \hat{s}$ | $\sigma_{\rm stat} = 22\%$ | $\sigma_{\rm stat} = 15\%$ | $\sigma_{\rm stat} = 5.0\%$ | $\sigma_{\rm stat} = 2.3\%$ |

### 2.13.3 Forward-backward asymmetry

The forward-backward asymmetry is defined as $A_{\rm FB} \equiv (N_F - N_B)/(N_F - N_B)$, where $N_F$ ($N_B$) is the number of decays with the positive lepton along (opposite) the $b$ quark direction in the dilepton rest frame. For the forward-backward asymmetry measurement, hadronic final states containing only $K^\pm$ or $K^0_S$ are removed, because for these modes the asymmetry is expected to be zero in the Standard Model. The Standard Model predicts the asymmetry to be negative at low dilepton mass and to become positive at high dilepton mass. We are particularly interested in the measurement of the zero point of the asymmetry, since the prediction is robust [87]. After checking that the momentum reconstruction does not affect the asymmetry, we estimate the zero point value with a luminosity of 10 ab$^{-1}$. Figure 2-9 shows the forward-backward asymmetry for pure signal (the subtraction of backgrounds results in an increase of the statistical errors by a factor of approximately two). We obtain $\hat{s}_0(\equiv m^2_{\ell\ell}/m_b) = 0.141 \pm 0.020$ and $\hat{s} = 0.14 \pm 0.04$ for pure signal and background-subtracted signal, respectively, where $m_{\ell\ell}$ is dilepton invariant mass and $m_b(= 4.8 \text{ GeV})$ is the $b$ quark mass. The error is statistical only. Here we should mention that the background asymmetry is not zero and needs further study.

Next we study the error in the asymmetry as a function of luminosity. We measure the asymmetry above and below $\hat{s}_0$. Table 2-13 summarizes the results. A decisive measurement of $A_{\rm FB}$ clearly needs a Super $B$ Factory.

### 2.13.4 Summary

Inclusive $b \to s\ell^+\ell^-$decays offer new sensitivity to extensions of the Standard Model. Measurements of the branching fraction and dilepton mass spectrum should reach interesting sensitivities by the end of *BABAR* and Belle (1000 fb$^{-1}$). The degree of improvement at a Super $B$ Factory depends on the control of systematic uncertainties for the measurement of the branching fraction in the whole dilepton mass range and for restricted "perturbative" ranges. The lepton forward-backward asymmetry $A_{\rm FB}$ is particularly powerful and a Super $B$ Factory is needed to reach interesting sensitivity.

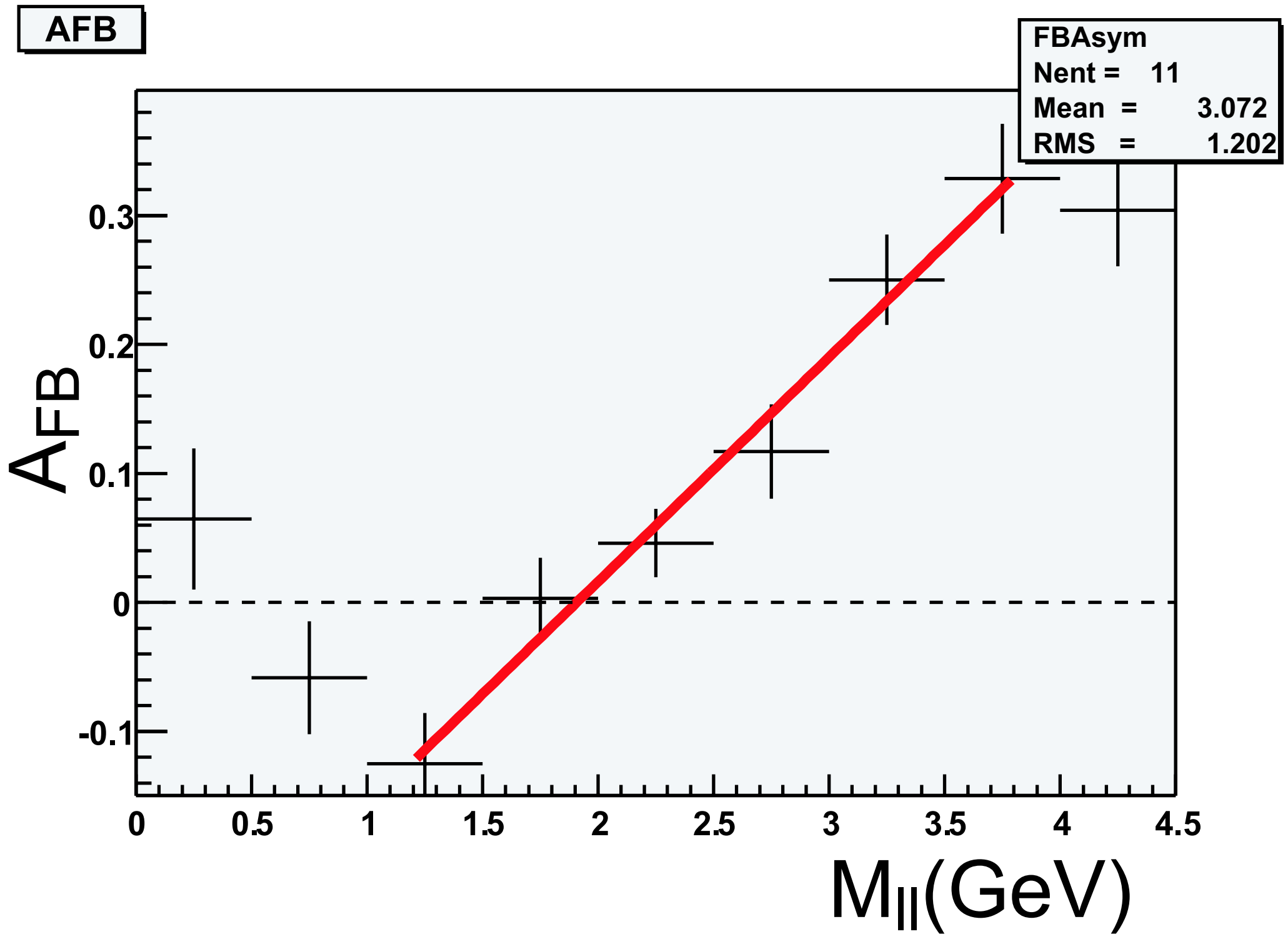

**Figure 2-9.** *The forward-backward asymmetry as a function of dilepton mass for pure signal case with the luminosity of 10 ab$^{-1}$.*

**Table 2-13.** *Anticipated measurements of $A_{\rm FB}$ for pure signal (upper row) and after background subtraction (lower row) for $\hat{s} < \hat{s}_0$ and $\hat{s} > \hat{s}_0$, with $\hat{s}_0 = 0.162 \pm 0.008$ (NNLL) [87].*

| $A_{\rm FB}$ | Integrated luminosity | | | |
|---|---|---|---|---|
| $X_s e^+e^- + X_s\mu^+\mu^-$ | 500 fb$^{-1}$ | 1000 fb$^{-1}$ | 10 ab$^{-1}$ | 50 ab$^{-1}$ |
| $\hat{s} < \hat{s}_0$ | $-0.02 \pm 0.11$ | $-0.02 \pm 0.08$ | $-0.017 \pm 0.024$ | $-0.017 \pm 0.011$ |
| | $-0.02 \pm 0.17$ | $-0.02 \pm 0.12$ | $-0.017 \pm 0.039$ | $-0.017 \pm 0.017$ |
| $\hat{s}_0 < \hat{s}$ | $0.17 \pm 0.09$ | $0.17 \pm 0.07$ | $0.173 \pm 0.021$ | $0.173 \pm 0.009$ |
| | $0.17 \pm 0.22$ | $0.17 \pm 0.16$ | $0.173 \pm 0.050$ | $0.173 \pm 0.022$ |

## 2.14 Theoretical Prospects for Inclusive Modes: $b \longrightarrow (s, d)\, \ell^+\ell^-$

### 2.14.1 Dilepton mass spectrum and forward-backward asymmetry

≻– T. Hurth –≺

Precise measurements of the dilepton spectrum and of the forward–backward asymmetry in the inclusive decay process $\overline{B} \to X_s\ell^+\ell^-$ allow for important tests for New Physics and for discrimination between different New Physics scenarios (for a recent review, see [41]). In comparison to the $\overline{B} \to X_s\gamma$ decay, the inclusive $\overline{B} \to X_s\ell^+\ell^-$ decay presents a complementary, albeit, more complex, test of the Standard Model.

As with all inclusive modes, the inclusive rare decay $\overline{B} \to X_s\ell^+\ell^-$ is very attractive, because, in contrast to most of the exclusive channels, it is a theoretically clean observable dominated by the partonic contributions. Non-perturbative effects in these transitions are small and can be systematically accounted for, through an expansion in inverse powers of the heavy $b$ quark mass. In the specific case of $\overline{B} \to X_s\ell^+\ell^-$, the latter statement is applicable only if the $c\bar{c}$ resonances that show up as large peaks in the dilepton invariant mass spectrum (see Fig. 2-10) are removed by appropriate kinematic cuts. In the *perturbative windows*, namely in the region below resonances and in the one above, theoretical predictions for the invariant mass spectrum are dominated by the purely perturbative contributions, and a theoretical precision comparable with the one reached in the inclusive decay $\overline{B} \to X_s\gamma$ is possible. In the high $q^2$ ($\equiv M^2_{\ell^+\ell^-}$) region, one should encounter the breakdown of the heavy mass expansion at the endpoint. Integrated quantities are still defined, but one finds sizable $\Lambda^2_{\rm QCD}/m_b^2$ nonperturbative corrections within this region.

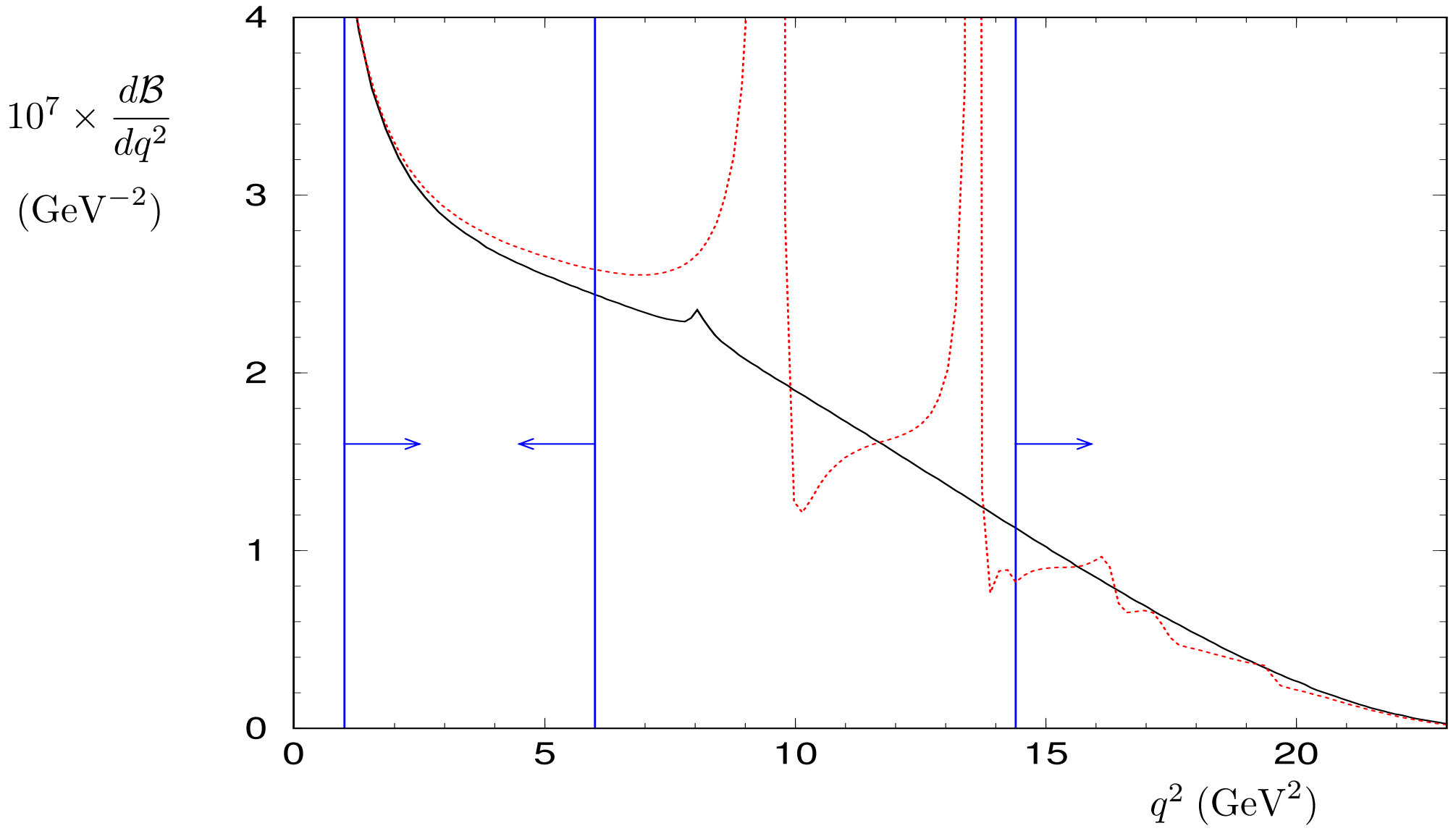

**Figure 2-10.** *NNLL predictions of $d\mathcal{B}(\overline{B} \to X_s\ell^+\ell^-)/dq^2$: partonic result with full $m_c$ dependence for $\mu$=5 GeV with (dotted line)and without (full line) factorizable $c\bar{c}$ corrections.*

Regarding the choice of precise cuts in the dilepton mass spectrum, it is important that one directly compares theory and experiment using the same energy cuts and avoids any kind of extrapolation.

Perturbative QCD corrections lead to a sizable modification of the pure short-distance electroweak contribution, generating large logarithms of the form $\alpha_s^n(m_b) \times \log^m(m_b/M_{\rm heavy})$, where $M_{\rm heavy} = O(M_W)$ and $m \leq n$ (with $n = 0, 1, 2, ...$), which have to be resummed. These effects are induced by hard–gluon exchange between the quark lines of the one-loop electroweak diagrams. A computation of the NNLL terms is needed if one aims at a numerical accuracy below $10\%$, similar to the one achieved by the NLL calculation of $\overline{B}\to X_s\gamma$. Thanks to the joint effort of several groups [52, 84, 87, 88, 217, 85, 86] the NNLL calculations have now been finalized.

The decay $\overline{B}\to X_s\ell^+\ell^-$ is particularly attractive because of kinematic observables such as the invariant dilepton mass spectrum and the forward–backward (FB) asymmetry. They are usually normalized by the semileptonic decay rate in order to reduce the uncertainties due to bottom quark mass and CKM angles and are defined as follows ($\hat{s}=q^2/m_b$)

$$R^{\ell^+\ell^-}_{\rm quark}(\hat{s})=\frac{d}{d\hat{s}}\Gamma(b\to X_s\ell^+\ell^-)/\Gamma(b\to X_c e\overline{\nu}), \tag{2.56}$$

$$A_{\rm FB}(\hat{s})=\frac{1}{\Gamma(b\to X_c e\overline{\nu})}\times\int_{-1}^{1}d\cos\theta_\ell\,\frac{d^2\Gamma(b\to X_s\ell^+\ell^-)}{d\hat{s}\;d\cos\theta_\ell}\mathrm{sgn}(\cos\theta_\ell), \tag{2.57}$$

where $\theta_\ell$ is the angle between $\ell^+$ and $B$ momenta in the dilepton centre-of-mass frame. These observables in the NNLL accuracy can be expressed as

$$\begin{aligned}R(\hat{s})=\frac{\alpha_{\rm em}^2}{4\pi^2}\left|\frac{V_{tb}^*V_{ts}}{V_{cb}}\right|^2\frac{(1-\hat{s})^2}{f(z)\kappa(z)}\bigg\{&4\left(1+\frac{2}{\hat{s}}\right)|C_7^{\rm eff}(\hat{s})|^2\left(1+\frac{\alpha_s}{\pi}\tau_{77}(\hat{s})\right)\\&+(1+2\hat{s})\left[|C_9^{\rm eff}(\hat{s})|^2+|C_{10}^{\rm eff}(\hat{s})|^2\right]\left(1+\frac{\alpha_s}{\pi}\tau_{99}(\hat{s})\right)\\&+12\,\Re\left[C_7^{\rm eff}(\hat{s})C_9^{\rm eff}(\hat{s})^*\right]\left(1+\frac{\alpha_s}{\pi}\tau_{79}(\hat{s})\right)+\frac{\alpha_s}{\pi}\delta_R(\hat{s})\bigg\}\ ,\end{aligned} \tag{2.58}$$

$$\begin{aligned}A_{\rm FB}(\hat{s})=-\frac{3\alpha_{\rm em}^2}{4\pi^2}\left|\frac{V_{tb}^*V_{ts}}{V_{cb}}\right|^2\frac{(1-\hat{s})^2}{f(z)\kappa(z)}\bigg\{&s\,\Re\left[C_{10}^{\rm eff}(\hat{s})^*C_9^{\rm eff}(\hat{s})\right]\left(1+\frac{\alpha_s}{\pi}\tau_{910}(\hat{s})\right)\\&+2\,\Re\left[C_{10}^{\rm eff}(\hat{s})^*C_7^{\rm eff}(\hat{s})\right]\left(1+\frac{\alpha_s}{\pi}\tau_{710}(\hat{s})\right)+\frac{\alpha_s}{\pi}\delta_{\rm FB}(\hat{s})\bigg\}\ ,\end{aligned} \tag{2.59}$$

where the definitions of the various functions can be found, for example, in [86]. The effective Wilson coefficients[7] $C_i^{\rm eff}$ have the advantage of encoding all dominant matrix-element corrections, leading to an explicit $\hat{s}$ dependence for all of them.

Before discussing the numerical predictions for the integrated branching ratios, it is worthwhile to emphasize that regions of low- and high-dilepton mass have complementary virtues and disadvantages. These can be summarized as follows ($q^2=M^2_{\ell^+\ell^-}$):

*Virtues of the low $q^2$ region:* reliable $q^2$ spectrum; small $1/m_b$ corrections; sensitivity to the interference of $C_7$ and $C_9$; high rate.

*Disadvantages of the low $q^2$ region:* difficult to perform a fully inclusive measurement (severe cuts on the dilepton energy and/or the hadronic invariant mass); long-distance effects due to processes of the type $\overline{B}\to J/\psi\,X_s\to X_s+X'\ell^+\ell^-$ not fully under control; non-negligible scale and $m_c$ dependence.

*Virtues of the high $q^2$ region:* negligible scale and $m_c$ dependence due to the strong sensitivity to the Wilson coefficient $|C_{10}|^2$; easier to perform a fully inclusive measurement (small hadronic invariant mass); negligible long-distance effects of the type $\overline{B}\to J/\psi\,X_s\to X_s+X'\ell^+\ell^-$.

*Disadvantages of the high $q^2$ region:* $q^2$ spectrum not reliable; sizable $1/m_b$ corrections; low rate.

Given this situation, future experiments should try to measure the branching ratios in both regions and report separately the two results. These two measurements are indeed affected by different systematic uncertainties (of a theoretical nature) but they provide different short-distance information.

[7] We note that slightly different definitions of effective Wilson coefficients are used in the literature.

In order to obtain theoretical predictions that can be confronted with experiments, it is necessary to convert the $\hat{s} = q^2/m_b^2$ range into a range for the measurable dilepton invariant mass $q^2$. Concerning the low $q^2$ region, the reference interval $q^2 \in [1,6]\ \mathrm{GeV}^2$ is most suitable. The lower bound on $q^2$ is imposed in order to cut a region where there is no new information with respect to $\overline{B} \to X_s\gamma$ and where we cannot trivially combine electron and muon modes. Then the NNLL QCD prediction for the low $q^2$ region is given by [86]:

$$\begin{aligned} R_{\rm cut}^{\rm low} &= \int_{1\ \mathrm{GeV}^2}^{6\ \mathrm{GeV}^2} dq^2 \frac{d\Gamma(\overline{B} \to X_s \ell^+\ell^-)}{\Gamma(\overline{B} \to X_c e\nu)} = 1.48 \times 10^{-5} \times \\ &\times \left[ 1 \pm 8\%\big|_{\Gamma_{\rm sl}} \pm 6.5\%\big|_{\mu} \pm 2\%\big|_{m_c} \pm 3\%\big|_{m_b(\rm cuts)} + (4.5 \pm 2)\%\big|_{1/m_b^2} - (1.5 \pm 3)\%\big|_{c\bar{c}} \right] \\ &= (1.52 \pm 0.18) \times 10^{-5}\ . \end{aligned} \quad (2.60)$$

The impact of the NNLL QCD contributions is significant. The large matching scale $\mu_W$ uncertainty of $16\%$ of the NLL result was removed; the low-scale uncertainty $\mu_b$ of $13\%$ was cut in half; and also the central value of the integrated low dilepton spectrum was significantly changed by more than $10\%$ because of the NNLL corrections. The uncertainty is now dominated by the parametric errors which can be improved by additional independent measurements.

Concerning the high-dilepton mass region, a suitable reference cut is $q^2 > 14.4\,\mathrm{GeV}^2$, which leads to the following NNLL prediction [86]:

$$\begin{aligned} R_{\rm cut}^{\rm high} &= \int_{q^2 > 14.4\ \mathrm{GeV}^2} dq^2 \frac{d\Gamma(\overline{B} \to X_s \ell^+\ell^-)}{\Gamma(\overline{B} \to X_c e\nu)} = 4.09 \times 10^{-6} \times \\ &\times \left[ 1 \pm 8\%\big|_{\Gamma_{\rm sl}} \pm 3\%\big|_{\mu} + 0.15 \left( \frac{m_b - 4.9\ \mathrm{GeV}}{0.1\ \mathrm{GeV}} \right) - (8 \pm 8)\%\big|_{1/m_b^{(2,3)}} \pm 3\%\big|_{c\bar{c}} \right] \\ &= (3.76 \pm 0.72) \times 10^{-6}\ . \end{aligned} \quad (2.61)$$

Here the explicitly indicated $m_b$ dependence induces the largest uncertainty. At present this is about $15\%$. It reflects the *linear* $\Lambda_{\rm QCD}/m_b$ correction induced by the necessary cut in the $q^2$ spectrum. However, significant improvements can be expected in the near future in view of more precise data on other inclusive semileptonic distributions. The impact of the $1/m_b^2$ and $1/m_b^3$ corrections is surprisingly small, $\pm 8\%$, in view of the breakdown of the $1/m_b$ expansion in the kinematical endpoint (a detailed analysis of the $1/m_b$ corrections can be found in section 5 of [86]). The impact of the NNLL corrections for the high $q^2$ region is a $13\%$ reduction of the central value and a significant reduction of the perturbative scale dependence (from $\pm 13\%$ to $\pm 3\%$). There are two non-negligible sources of uncertainties, which are not explicitly included in Eqs. (2.60) and (2.61): the error due to $m_t$ (and the high-energy QCD matching scale) and the error due to higher-order electroweak and electromagnetic effects. The first type of uncertainty has been discussed in detail in [52], and it amounts to $\approx 6\%$. The impact of the dominant electroweak matching corrections was recently analyzed in [85] and is also found to be at the level of a few per cent.

Using the present world average $\Gamma(\overline{B} \to X_c e\nu) = (10.74 \pm 0.24)\%$, one finally obtains [86]:

$$\mathcal{B}(\overline{B} \to X_s \ell^+\ell^-;\ q^2 \in [1,6]\ \mathrm{GeV}^2) = (1.63 \pm 0.20) \times 10^{-6}\ , \quad (2.62)$$

$$\mathcal{B}(\overline{B} \to X_s \ell^+\ell^-;\ q^2 > 14.4\ \mathrm{GeV}^2) = (4.04 \pm 0.78) \times 10^{-7}\ . \quad (2.63)$$

It is clear that the theoretical errors in both predictions could be systematically improved in the future, owing to the present dominance of the parametric uncertainties.

In Fig. 2-11 we plot the (adimensional) normalized differential asymmetry, defined by

$$\overline{A}_{\rm FB}(q^2) = \frac{1}{d\mathcal{B}(\overline{B} \to X_s \ell^+\ell^-)/dq^2} \int_{-1}^{1} d\cos\theta_\ell\ \frac{d^2\mathcal{B}(\overline{B} \to X_s \ell^+\ell^-)}{dq^2\ d\cos\theta_\ell} \mathrm{sgn}(\cos\theta_\ell)\ . \quad (2.64)$$

Most of the comments concerning the errors and the complementarity of low and high $q^2$ windows discussed above also hold for the forward-backward asymmetry. In the low $q^2$ region the most interesting observable is not the integral of the asymmetry, which is very small because of the change of sign, but the position of the zero. As analysed by several authors (see Refs. [87, 88]), this is one of the most precise predictions (and one of the most interesting Standard Model tests) in rare $B$ decays. Denoting by $q_0^2$ the position of the zero, and showing explicitly only the uncertainties and nonperturbative effects larger than 0.5%, we find at the NNLL order

$$q_0^2 = 0.161 \times m_b^2 \times \left[ 1 + 0.9\%\big|_{1/m_b^2} \pm 5\%\big|_{\rm NNNLL} \right] = (3.90 \pm 0.25)\ {\rm GeV}^2\ . \tag{2.65}$$

As pointed out in Ref. [87], the $\mu$ dependence is, in this case, accidentally small and does not provide a conservative estimate of higher-order QCD corrections. The $5\%$ error in (2.65) has been estimated by comparing the result within the ordinary LL counting and within the modified perturbative ordering proposed in Ref. [84]. The phenomenological impact of the NNLL contributions on the forward-backward asymmetry is also significant [87, 88]. The position of the zero of the forward-backward asymmetry, defined by $A_{\rm FB}(\hat{s}_0) = 0$, is particularly interesting to determine relative sign and magnitude of the Wilson coefficients $C_7$ and $C_9$ and it is therefore extremely sensitive to possible New Physics effects. An illustration of the shift of the central value and the reduced uncertainty between NNL and NNLL expressions of $A_{\rm FB}(s)$, in the low-$\hat{s}$ region, is presented in Fig. 2-12. The complete effect of NNLL contributions to the forward-backward asymmetry adds up to a $16\%$ shift compared with the NLL result, with a residual error reduced to the 5% level. Thus, the zero of the forward-backward asymmetry in the inclusive mode turns out to be one of the most sensitive tests for New Physics beyond the Standard Model.

In the high $q^2$ window the forward-backward asymmetry does not change sign, therefore its integral represents an interesting observable. In order to minimize nonperturbative and normalization uncertainties, it is more convenient to consider a normalized integrated asymmetry. Applying the same $q^2$ cut as in (2.61), we define

$$(\overline{A}_{\rm FB})^{\rm high}_{\rm cuts} = \left[ \int_{q^2 > 14.4\ {\rm GeV}^2} dq^2\, \frac{d\mathcal{B}_{\rm FB}(q^2)}{dq^2} \right]^{-1} \int_{q^2 > 14.4\ {\rm GeV}^2} dq^2\, \frac{dA_{\rm FB}(q^2)}{dq^2}\ . \tag{2.66}$$

All parametric and perturbative uncertainties are very small in this observable at the NNLL order level. On the other hand, despite a partial cancellation, this ratio is still affected by substantial $\Lambda^2_{\rm QCD}/m_b^2$ and $\Lambda^3_{\rm QCD}/m_b^3$ corrections (which represent by far the dominant source of uncertainty). Separating the contributions of the various subleading (in $1/m_b$) operators, one finds [86]

$$(\overline{A}_{\rm FB})^{\rm high}_{\rm cuts} \;=\; 0.42 \times [1 - (0.17 \pm 0.11)_{\lambda_1} - (0.42 \pm 0.07)_{\lambda_2} - (0.08 \pm 0.08)_{\rho_1}] \;=\; 0{,}14 \pm 0.06\,. \tag{2.67}$$

The recently calculated new (NNLL) contributions have significantly improved the sensitivity of the inclusive $\overline{B} \to X_s \ell^+\ell^-$ decay in testing extensions of the Standard Model in the sector of flavor dynamics. However, with the present experimental knowledge the decay $\overline{B} \to X_s\gamma$ still leads to the most restrictive constraints as was found in [10]. Especially, the MFV scenarios are already highly constrained and only small deviations to the Standard Model rates and distributions are possible; therefore no useful additional bounds from the semileptonic modes beyond that already known from the $\overline{B} \to X_s\gamma$ can be deduced for the MFV models at the moment. Within the model-independent analysis, the impact of the NNLL contributions on the allowed ranges for the Wilson coefficients was already found to be significant. In this analysis, however, only the integrated branching ratios were used to derive constraints. It is clear that one needs measurements of the kinematic distributions of the $\overline{B} \to X_s\ell^+\ell^-$, the dilepton mass spectrum and the FB asymmetry in order to determine the exact values and signs of the Wilson coefficients. In Fig. 2-13, the impact of these future measurements is illustrated. It shows the shape of the FB asymmetry for the Standard Model and three additional sample points, which are all still allowed by the present measurements of the branching ratios; thus, even rather rough measurements of the FB asymmetry will either rule out large parts of the parameter space of extended models or show clear evidence for New Physics beyond the Standard Model. A high-statistics experiment can contribute significantly to this effort and take full advantage of the high sensitivity of the $b \to \ell^+\ell^-$ observables to possible new degrees of freedom at higher scales.

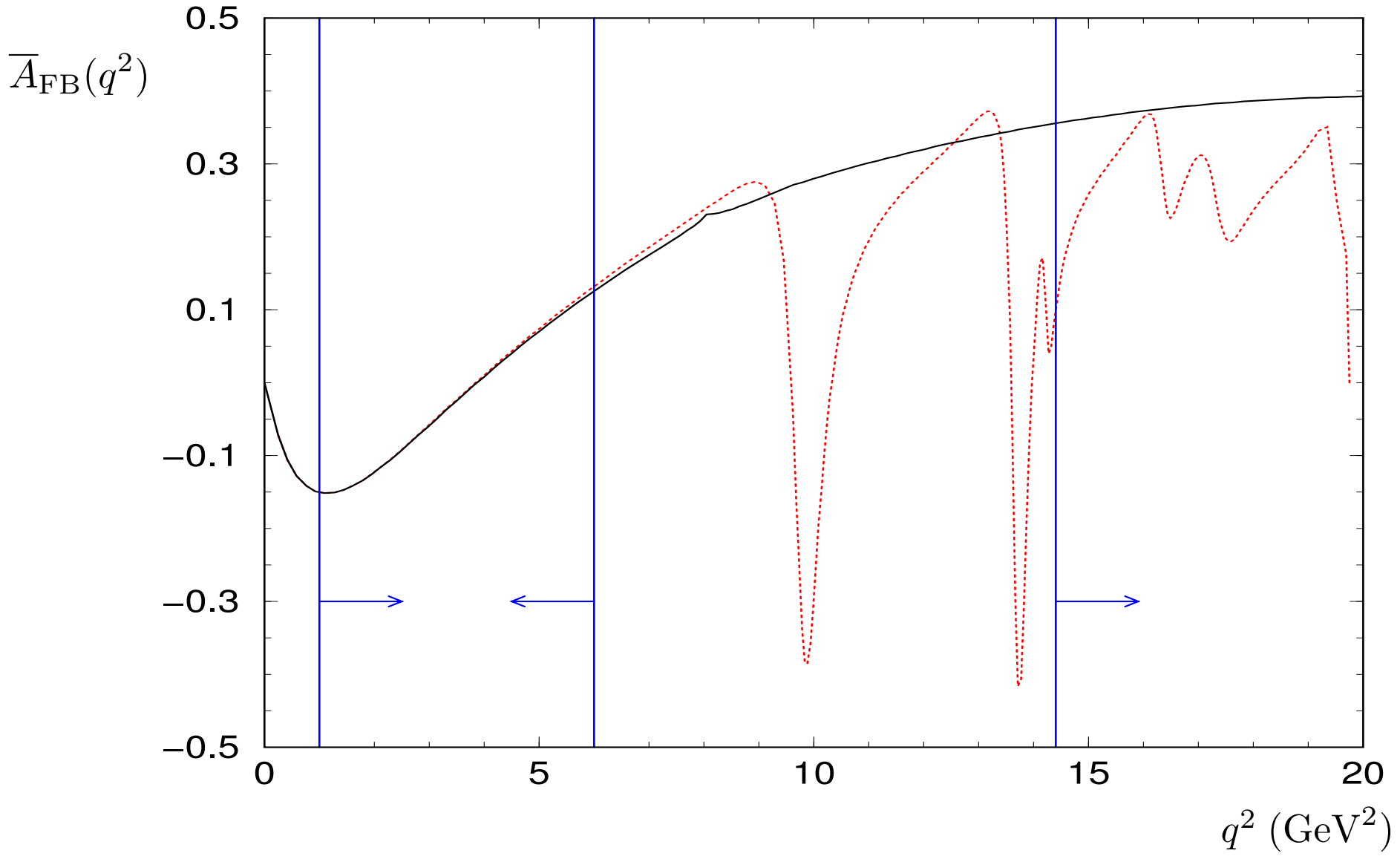

**Figure 2-11.** *NNLL perturbative contributions to the normalized FB asymmetry; partonic result with full $m_c$ dependence for $\mu$=5 GeV with (dotted line)and without (full line) factorizable $c\bar{c}$ corrections.*

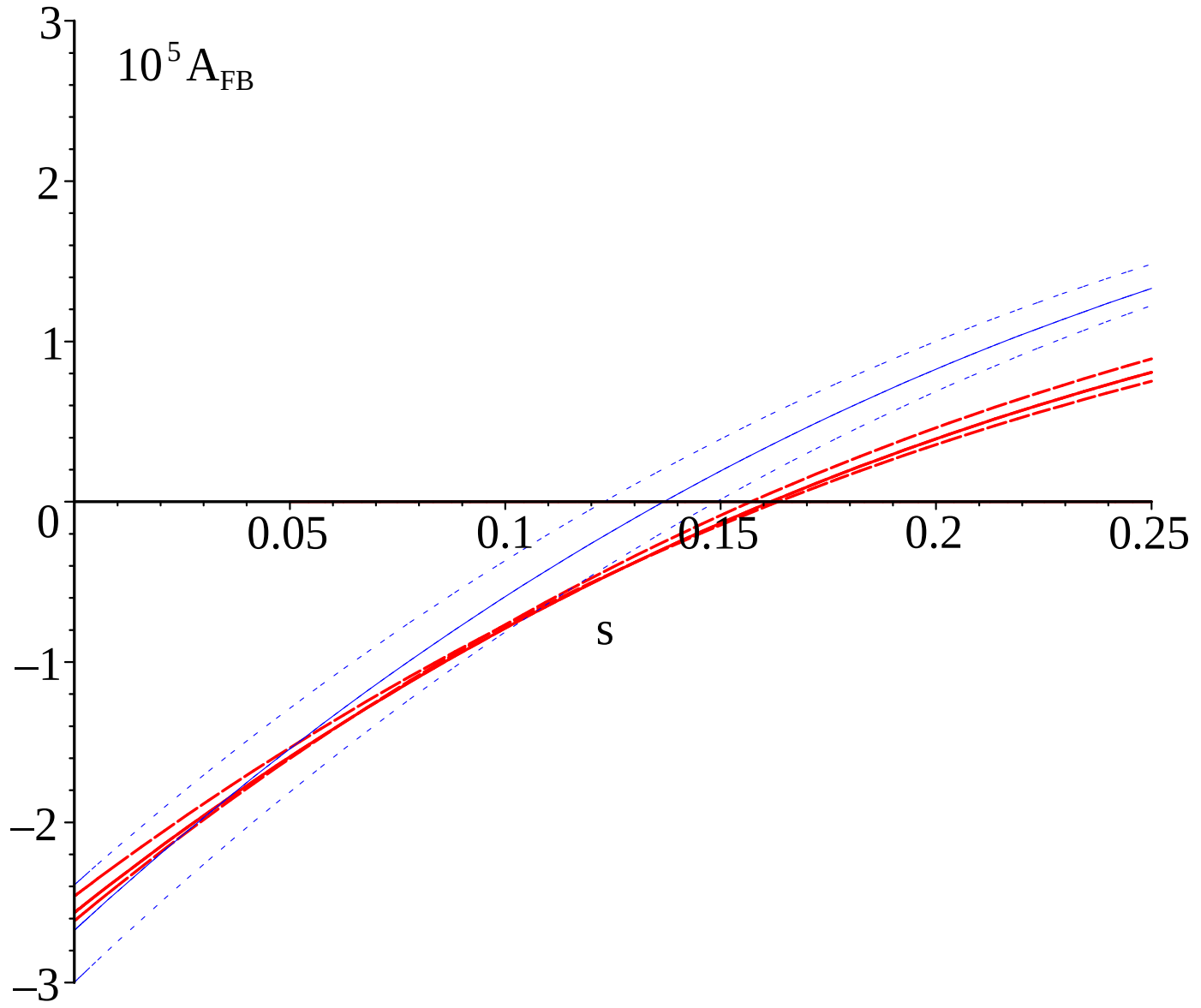

**Figure 2-12.** *Comparison between NNLL and NLL results for $A_{\rm FB}(s)$ in the low $s$ region. The three thick lines are the NNLL predictions for $\mu = 5$ GeV (full), and $\mu = 2.5$ and 10 GeV (dashed); the dotted curves are the corresponding NLL results. All curves for $m_c/m_b = 0.29$.*

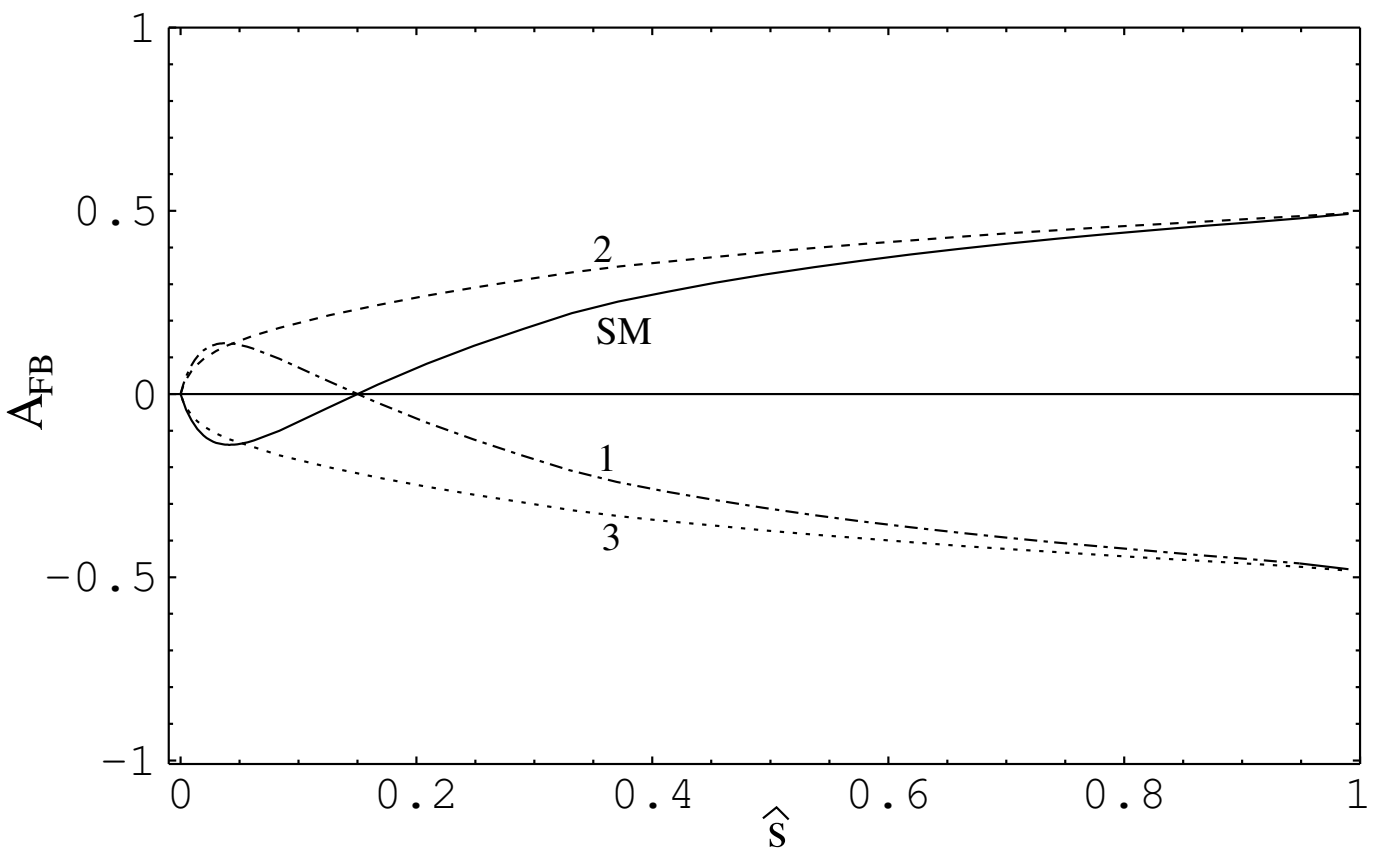

**Figure 2-13.** *Four different shapes of the normalized forward-backward asymmetry $\overline{A}_{\rm FB}$ for the decay $B \to X_s \ell^+\ell^-$. The four curves correspond to four sample points of the Wilson coefficients that are compatible with the present measurements of the integrated branching ratios.*

## 2.14.2 Inclusive modes: $b \to s\, e^+e^-$ versus $b \to s\, \mu^+\mu^-$

≻— G. Hiller —≺

The Standard Model branching ratios of $b \to s e^+e^-$ and $b \to s\mu^+\mu^-$ decays differ predominantly by phase space effects induced by the masses of the leptons in the final state. If the same kinematical cut on the dilepton mass is applied to both electron and muon modes, the remaining splitting between them is very small. In particular, the ratio $(H = X_s, K, K^*)$

$$R_H \equiv \int_{4m_\mu^2}^{q^2_{max}} dq^2 \frac{d\Gamma}{dq^2}(B \to H\mu^+\mu^-) / \int_{4m_\mu^2}^{q^2_{max}} dq^2 \frac{d\Gamma}{dq^2}(B \to He^+e^-)\ , \qquad q = p_{\ell^+} + p_{\ell^-} \tag{2.68}$$

with the same lower integration boundary for both numerator and denominator equals 1, up to corrections of order $m_\mu^2/m_b^2$. The finite lepton mass corrections reduce the ratio $R_{X_s}$ of the inclusive decays at the percent level for both full $q^2$ and below-the-$J/\psi$ ($q^2_{max} = 6\,{\rm GeV}^2$) integration regions [129]

$$R_{X_s}^{SM} = 0.99 \pm 0.01\ , \qquad R^{SM}_{X_s\, low\, q^2} = 0.98 \pm 0.01\ . \tag{2.69}$$

While the deviation from unity in $R_{X_s}$ is very small in the Standard Model, it is a correction comparable in size to some theoretical uncertainties in the branching ratios, see *e.g.,* Eq. (2.60). Hence, finite lepton mass effects should be taken into account in the $B \to X_s\mu^+\mu^-$ distributions.

In addition to the branching ratios, the observables $R_{X_s}$ are useful in testing the Standard Model and searching for New Physics that would distinguish between lepton generations. Such couplings can be induced, for example, by interactions involving neutral Higgs bosons. These contributions to $b \to s\ell^+\ell^-$ decays are tiny in the Standard Model - even for $\tau$'s - but can be substantial in the MSSM at large $\tan\beta$. Assuming that the new couplings are proportional to the respective fermion mass, the lepton flavor-dependent New Physics effects are suppressed in the electron modes, which therefore serve as a normalization [129, 218]. The ratios $R_{X_s}$ can be enhanced up to 1.08, 1.07 in the full and low dilepton mass region, respectively, while being consistent with current data on $b \to s\gamma, b \to s\ell^+\ell^-, B_s \to \mu^+\mu^-$ and $R_K$ [129]. In particular, order one effects in $R_{X_s}$ are already excluded by data on $B_s \to \mu^+\mu^-$ [129, 219] (also see Section 2.16.3 on the exclusive rare decays). While an enhancement of the $B \to X_s\mu^+\mu^-$ branching ratio of $\mathcal{O}(10\%)$ is within the current theoretical uncertainty of the Standard Model prediction, a corresponding effect in $R_{X_s}$ can be clearly distinguished from the Standard Model.

## 2.15 Experimental Prospects for $B \to K\ell^+\ell^-$ and $B \to K^*\ell^+\ell^-$

≻— A. Ryd —≺

The decays $B \to K\ell^+\ell^-$ and $B \to K^*\ell^+\ell^-$ proceed via a flavor-changing $b \to s$ transition, which, in the Standard Model, takes place only in through higher-order loop processes. Contributions from New Physics, *e.g.*, supersymmetry, enter at the same order, and can significantly affect these decays. Besides modifying the rate, these New Physics contributions can also affect kinematic distributions. In particular, in the case of the $B \to K^*\ell^+\ell^-$ decay, the lepton forward-backward asymmetry, $A_{\rm FB}$, is of great interest. The forward-backward asymmetry is expected in the Standard Model to change sign as a function of $q^2$ for $q_0^2 \approx 3.8$ GeV. The position of the zero of the forward-backward asymmetry can be predicted with rather small uncertainties as it is not strongly dependent on unknown hadronic form factors.

Both *BABAR* [220] and Belle [221] have measured the branching fractions for $B \to K\ell^+\ell^-$ and and $B \to K^*\ell^+\ell^-$. For the study presented here we use the average [222] of these measurements:

$$\mathcal{B}(B \to K\ell^+\ell^-) = (0.55^{+0.09}_{-0.08}) \times 10^{-6},$$
$$\mathcal{B}(B \to K^*\ell^+\ell^-) = (1.06^{+0.22}_{-0.20}) \times 10^{-6}.$$

This study will use the current full *BABAR* Geant4 MC to extrapolate to the higher luminosities at the Super $B$ Factory . For the event selection we use the the same selection as used in Ref. [220]. In this study we will only consider the final state $B \to K^{*0}(\to K^+\pi^-)\ell^+\ell^-$. Some additional statistics can be gained by using additional modes, however, the $S/B$ is worse in these modes. For the purpose of the study of the forward-backward asymmetry, the lepton selection is of great importance. For electrons we use laboratory momenta down to 0.5 GeVand for muons down to 1.0 GeV. These selection criteria directly map on to our ability to measure the forward-backward asymmetry at low $q^2$ as illustrated in Fig. 2-14. A measurement of the forward-backward asymmetry at low $q^2$ is challenging in the muon channel, as the acceptance cuts out those events that provide the most information on the asymmetry. It might be possible to lower the lepton momentum cut-off a little to optimize for the measurement of the forward-backward asymmetry. However, here we will employ the well-established selection criteria used in the branching fraction measurement. The efficiency of the *BABAR* muon system is rather poor, and for the purpose of extrapolating to the Super $B$ Factory we assume that the efficiency for muons is the same as that for electrons. This is consistent with what Belle observes.

Table 2-14 summarizes the expected yields based on the measured branching fractions and the statistical error. In Fig. 2-15 a plot is shown of the measured forward-backward asymmetry as a function of $q^2$ for a sample of 50 ab$^{-1}$. In the electron channel the zero of $A_{\rm FB}$ is clear.

In conclusion, we find that to establish the zero in $A_{\rm FB}$ we will need an integrated luminosity 50 ab$^{-1}$. The electron channel provides almost all the power, as the acceptance for the muons makes the determination of $A_{\rm FB}$ difficult.

**Table 2-14.** *Prediction for the number of reconstructed $B \to K^*\ell^+\ell^-$ candidates for a set of different luminosities.*

| Sample size (ab$^{-1}$) | Number of reconstructed events $B \to K^*e^+e^-$ | $B \to K^*\mu^+\mu^-$ |
|---|---|---|
| 0.1 | 14 | 7 |
| 0.5 | 70 | 35 |
| 1.0 | 140 | 70 |
| 10.0 | 1,400 | 700 |
| 50.0 | 7,000 | 3,500 |

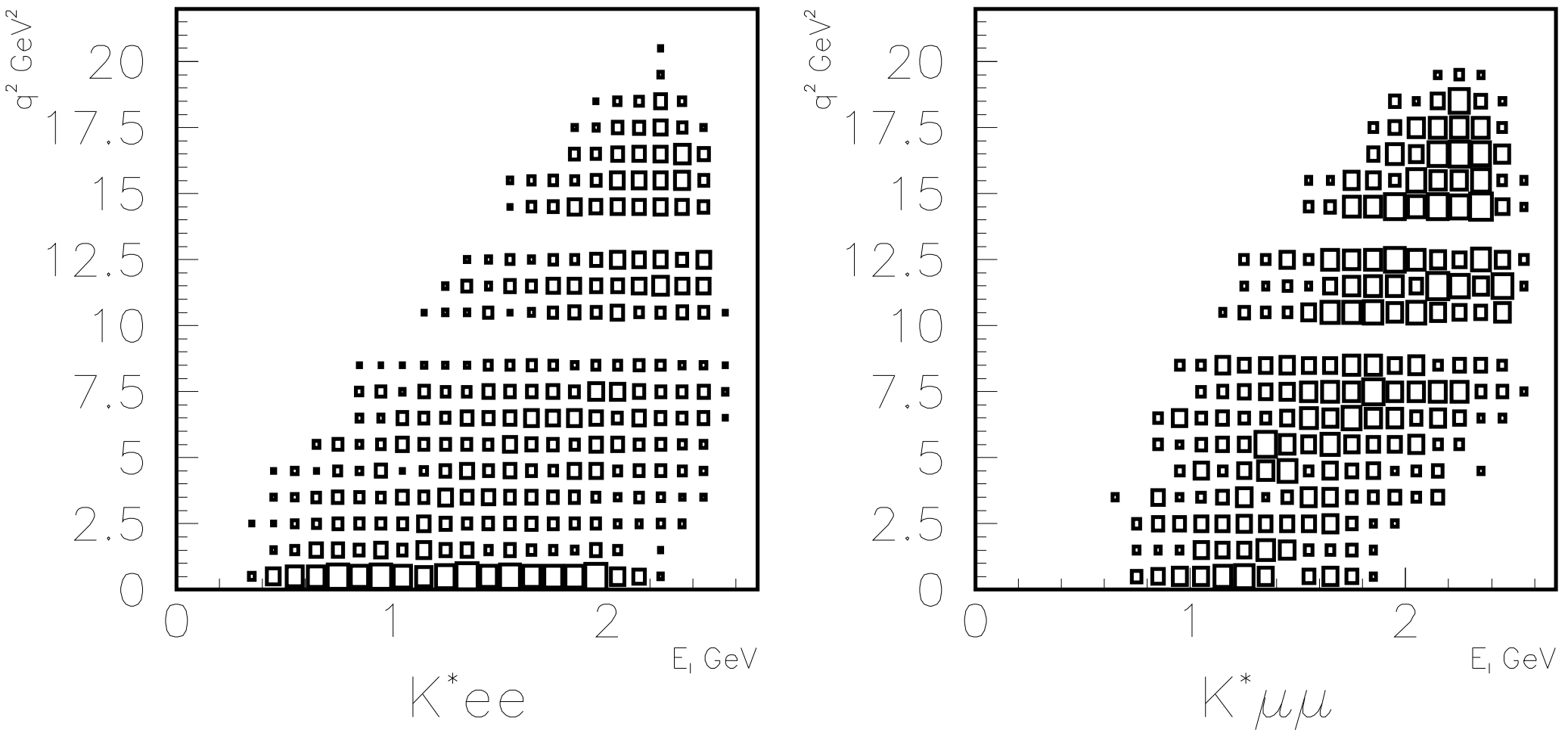

**Figure 2-14.** *The Dalitz plot, $q^2$ vs. $E_\ell$, for reconstructed events in the electron and muon channels for $B \rightarrow K^*\ell^+\ell^-$. For low $q^2$ it is evident that the lepton energy cut-off, 1 GeV for muons and 0.5 GeV for electrons in the lab frame, significantly reduces the phase space for muons, and removes those events having the greatest sensitivity to the forward-backward asymmetry for low $q^2$.*

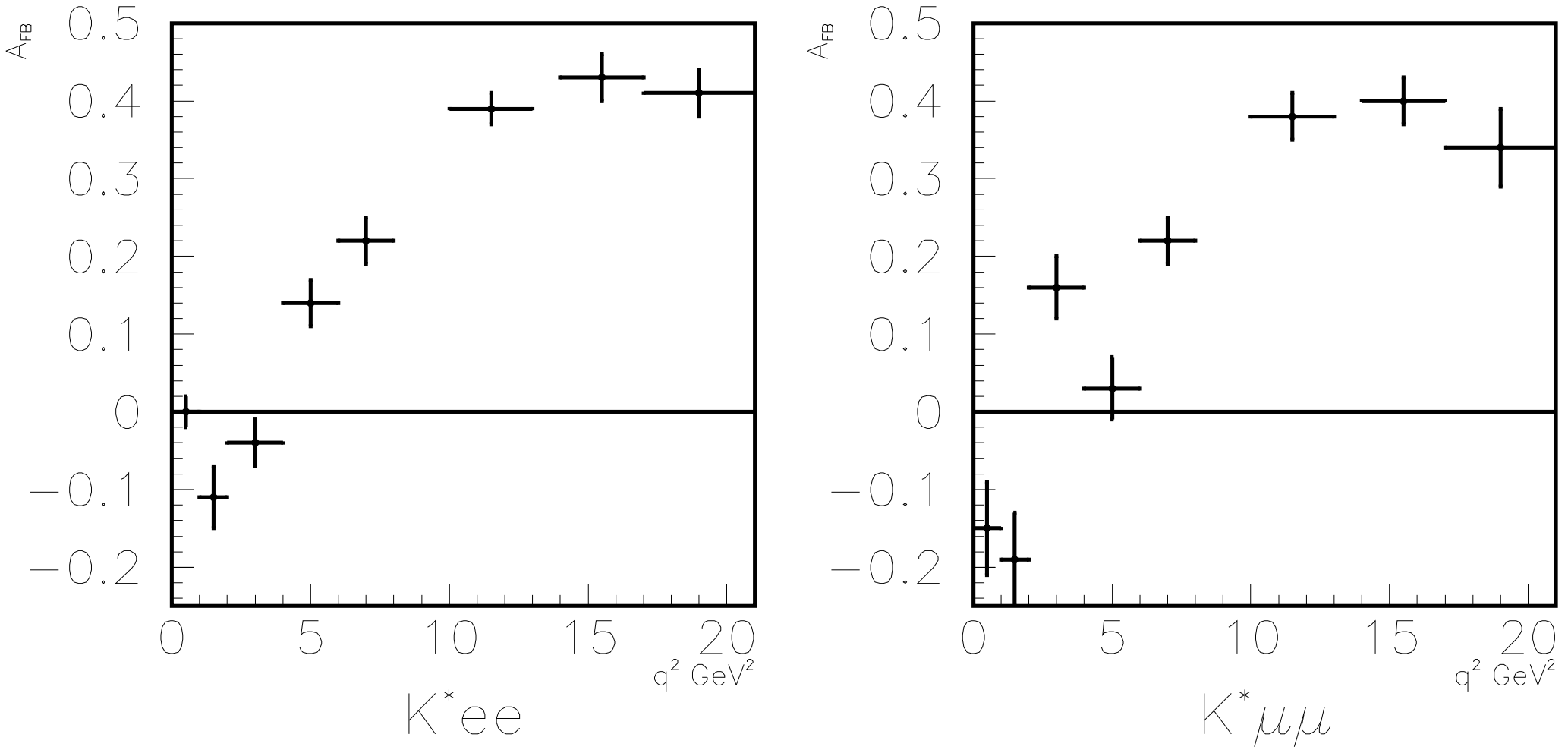

**Figure 2-15.** *The predicted forward-backward asymmetry in $B \rightarrow K^*e^+e^-$ and $B \rightarrow K^*\mu^+\mu^-$ in 8 different $q^2$ bins. The errors corresponds to a prediction for a 50 ab$^{-1}$ sample. (Note that the actual points in the muon mode has larger fluctuations as the signal Monte Carlo sample used corresponds on to about 25 ab$^{-1}$.)*

## 2.16 Theoretical prospects for $B \to (K, K^*)\, \ell^+\ell^-$

This section is devoted to exclusive $B \to (K, K^*)\ell^+\ell^-$ decays, where $\ell = e, \mu$. A brief overview has already been given in Section 2.2. We start in Section 2.16.1 with a discussion of the requisite hadronic matrix elements (form factors) and the symmetry relations that arise between them when the energy of the outgoing light meson is large. This framework is applied to $B \to (K, K^*)\, \ell^+\ell^-$ decays. Then a New Physics study with focus on asymmetries (isospin, forward-backward) Section 2.16.2 is followed by a Section 2.16.3 on the splitting between decays into an electron versus a muon pair. The angular analysis in $B \to (K^* \to K\pi)\ell^+\ell^-$ decays is discussed in Section 2.17, which is followed by Section 2.18 on the forward-backward asymmetry in $B \to K\ell^+\ell^-$.

### 2.16.1 Form factors, large energy relations

>– A. Ali –<

The semileptonic decays $B \to K\, \ell^+\ell^-$ and $B \to K^*\ell^+\ell^-$ are described by the following Lorenz decomposition of the matrix elements of the bilinear quark currents:

$$\langle K|\,(p_K)|\bar{s}\gamma^\mu b|B(p_B)\rangle \;=\; f_+(q^2)\left[p_B^\mu + p_K^\mu - \frac{M_B^2 - m_K^2}{q^2}\, q^\mu\right] + f_0(q^2)\, \frac{M_B^2 - m_K^2}{q^2}\, q^\mu, \tag{2.70}$$

$$\langle K|\,(p_K)|\bar{s}\sigma^{\mu\nu} q_\nu b|B(p_B)\rangle \;=\; \frac{iq^2 f_T(q^2)}{M_B + m_K}\left[p_B^\mu + p_K^\mu - \frac{M_B^2 - m_K^2}{q^2}\, q^\mu\right], \tag{2.71}$$

$$\langle K|^*\,(p_{K^*},\varepsilon)|\bar{s}\gamma_\mu b|B(p_B)\rangle \;=\; \frac{-2iV(q^2)}{M_B + m_{K^*}}\, \varepsilon_{\mu\nu\rho\sigma}\, p_B^\nu p_{K^*}^\rho \varepsilon^\sigma, \tag{2.72}$$

$$\begin{aligned}\langle K|^*\,(p_{K^*},\varepsilon)|\bar{s}\gamma^\mu\gamma_5 b|B(p_B)\rangle \;=&\; 2m_{K^*}\, A_0(q^2)\, \frac{(\varepsilon^* q)}{q^2}\, q^\mu + (M_B + m_{K^*})\, A_1(q^2)\left[\varepsilon^{*\mu} - \frac{(\varepsilon^* q)}{q^2}\, q^\mu\right] \\ &-\; A_2(q^2)\, \frac{(\varepsilon^* q)}{M_B + m_{K^*}}\left[p_B^\mu + p_{K^*}^\mu - \frac{M_B^2 - m_{K^*}^2}{q^2}\, q^\mu\right],\end{aligned} \tag{2.73}$$

$$\langle K|\,(p_{K^*},\varepsilon)|\bar{s}\sigma_{\mu\nu} q^\nu b|B(p_B)\rangle \;=\; 2T_1(q^2)\, \varepsilon_{\mu\nu\rho\sigma}\, p_B^\nu p_{K^*}^\rho \varepsilon^{*\sigma}, \tag{2.74}$$

$$\begin{aligned}\langle K|\,(p_{K^*},\varepsilon)|\bar{s}\sigma^{\mu\nu} q_\nu\gamma_5 b|B(p_B)\rangle \;=&\; -i\, T_2(q^2)\left[(M_B^2 - m_{K^*}^2)\, \varepsilon^{*\mu} - (\varepsilon^* q)(p_B^\mu + p_{K^*}^\mu)\right] \\ &-i\, T_3(q^2)\, (\varepsilon^* q)\left[q^\mu - \frac{q^2}{M_B^2 - m_{K^*}^2}\, (p_B^\mu + p_{K^*}^\mu)\right],\end{aligned} \tag{2.75}$$

where $q^\mu = p_B^\mu - p_K^\mu$ ( $q^\mu = p_B^\mu - p_{K^*}^\mu$) for the decay $B \to K\ell^+\ell^-$ ($B \to K^*\ell^+\ell^-$). Hence, these decays involve ten nonperturbative $q^2$-dependent functions (form factors), which introduce model-dependence in the decay rates and distributions. (In the limit of vanishing lepton mass $f_0$ and $A_0$ do not contribute to $B \to K\ell^+\ell^-$ and $B \to K^*\ell^+\ell^-$ decays.) However, restricting the dilepton mass $s = q^2$ so that the energy of the $K$ or $K^*$ in the decays $B \to (K, K^*)\ell^+\ell^-$, given as

$$E_{K,K^*} = \frac{M_B}{2}\left[1 - \frac{s}{M_B^2} + \frac{m_{K,K^*}^2}{M_B^2}\right], \tag{2.76}$$

remains large, the form factors introduced above obey the following relations [121]:

$$f_+(q^2) = \frac{M_B}{2E_K}\, f_0(q^2) = \frac{M_B}{M_B + m_K}\, f_T(q^2) = \xi_K(E_K), \tag{2.77}$$

$$\frac{M_B}{M_B + m_{K^*}}\, V(q^2) = \frac{M_B + m_{K^*}}{E_{K^*}}\, A_1(q^2) = T_1(q^2) = \frac{M_B}{2E_{K^*}}\, T_2(q^2) = \xi_\perp^{(K^*)}(E_{K^*}), \tag{2.78}$$

$$\frac{m_{K^*}}{E_{K^*}}\, A_0(q^2) = \frac{M_B + m_{K^*}}{E_{K^*}}\, A_1(q^2) - \frac{M_B - m_{K^*}}{M_B}\, A_2(q^2) = \frac{M_B}{2E_{K^*}}\, T_2(q^2) - T_3(q^2) = \xi_\parallel^{(K^*)}(E_{K^*}), \tag{2.79}$$

which are valid for the soft contribution to the form factors at large recoil, neglecting corrections of order $1/E_{K^{(*)}}$ and $\alpha_s$ [8] Thus, in the symmetry limit, only three FFs: $\xi_K$, $\xi_\perp^{(K^*)}$ and $\xi^{(K^*)}$ remain. Of these, the normalization of $\xi_\perp^{(K^*)}(q^2 = 0)$ is provided by the data on $B \to K^*\gamma$.

The relations given above are broken by the power and perturbative QCD effects. The leading symmetry-breaking corrections have been calculated, which bring in some unknown hadronic parameters but the approach remains quite predictive [64, 101]. In particular, it has been useful in developing a systematic approach in the decays $B \to (K^*, \rho)\gamma$ and in calculating the dilepton invariant mass (DIM) distribution and the FB-asymmetry in $B \to K^*\ell^+\ell^-$ over a limited kinematic range [64, 101, 225]. We discuss below some of the main results for the phenomenology in $B \to K^*\ell^+\ell^-$. In leading order, the forward-backward asymmetry for this decay can be expressed as

$$\frac{dA_{FB}(B \to K^*\ell^+\ell^-)}{d\hat{s}} \sim C_{10}[\mathrm{Re}(C_9^{eff})VA_1 + \frac{\hat{m}_b}{\hat{s}}C_7^{eff}(VT_2(1-\hat{m}_V) + A_1T_1(1+\hat{m}_V))] \,. \tag{2.80}$$

With the effective coefficients calculated, the forward-backward asymmetry has a characteristic zero in the Standard Model, which we denote by $\hat{s}_0$. The value of $\hat{s}_0$ is determined by the solution of the following equation:

$$\mathrm{Re}(C_9^{eff}(\hat{s}_0)) = -\frac{\hat{m}_b}{\hat{s}_0}C_7^{eff}(\frac{T_2(\hat{s}_0)}{A_1(\hat{s}_0)}(1-\hat{m}_V) + \frac{T_1(\hat{s}_0)}{V(\hat{s}_0)}(1+\hat{m}_V)) \,. \tag{2.81}$$

Model-dependent studies carried out in the context of form factor models had indicated that the uncertainties in the position of the zero are small [120]. The large energy framework provides a symmetry argument why the uncertainty in $\hat{s}_0$ is small. In the symmetry limit, using the relations given in (2.78) and (2.79), we have

$$\begin{aligned} \frac{T_2}{A_1} &= \frac{1+\hat{m}_V}{1+\hat{m}_V^2-\hat{s}}(1-\frac{\hat{s}}{1-\hat{m}_V^2}) \,, \\ \frac{T_1}{V} &= \frac{1}{1+\hat{m}_V} \,. \end{aligned} \tag{2.82}$$

Using them, the r.h.s. in (2.81) becomes independent of any form factors. Hence, in the symmetry limit, there is no hadronic uncertainty in $\hat{s}_0$, which is now determined by the solution of the following equation [9]

$$C_9^{eff}(\hat{s}_0) = -\frac{2m_bM_B}{s_0}C_7^{eff}. \tag{2.83}$$

Including the $O(\alpha_s)$ symmetry-breaking corrections leads to a shift[9] in $\hat{s}_0$ [101]

$$C_9^{eff}(\hat{s}_0) = -\frac{2m_bM_B}{s_0}C_7^{eff}\left(1 + \frac{\alpha_sC_F}{4\pi}[\ln\frac{m_b^2}{\mu^2} - L] + \frac{\alpha_sC_F}{4\pi}\frac{\Delta F_\perp}{\xi_\perp(s_0)}\right) \,, \tag{2.84}$$

where

$$L = -\frac{2E_{K^*}}{m_B - 2E_{K^*}}\ln\frac{2E_{K^*}}{m_B} \,, \tag{2.85}$$

and $\Delta F_\perp$ is a nonperturbative quantity. The term $\propto \Delta F_\perp/\xi_\perp(s_0)$ brings back the dependence on nonperturbative quantities, albeit weighted by the factor $\alpha_sC_F/(4\pi)$, and there is also a residual scale-dependence of $s_0$ on the scale $\mu$. So, the zero of the forward-backward asymmetry is not very precisely localized, *cf.* the conservative estimate [101] $s_0 = (4.2 \pm 0.6)$ GeV$^2$. The effect of the $\mathcal{O}(\alpha_s)$ corrections to the forward-backward asymmetry in $B^+ \to K^{*+}\ell^+\ell^-$ is shown in Fig. 2-17 in Section 2.16.2.

[8] The relation between $A_1$ and $V$, and likewise $T_1$ and $T_2$, holds to all orders in $\alpha_s$ at leading order in $1/E_{K^*}$ [223], which follows when the helicity of the light meson is inherited by the one of the outgoing quark [224].

[9] The main numerical difference between $s_0$ from Eq. (2.83) obtained in earlier analyses, *e.g.*, [9] and Eq. (2.84) stems from the use of $C_{7,9}^{eff}$ at higher order in the latter .

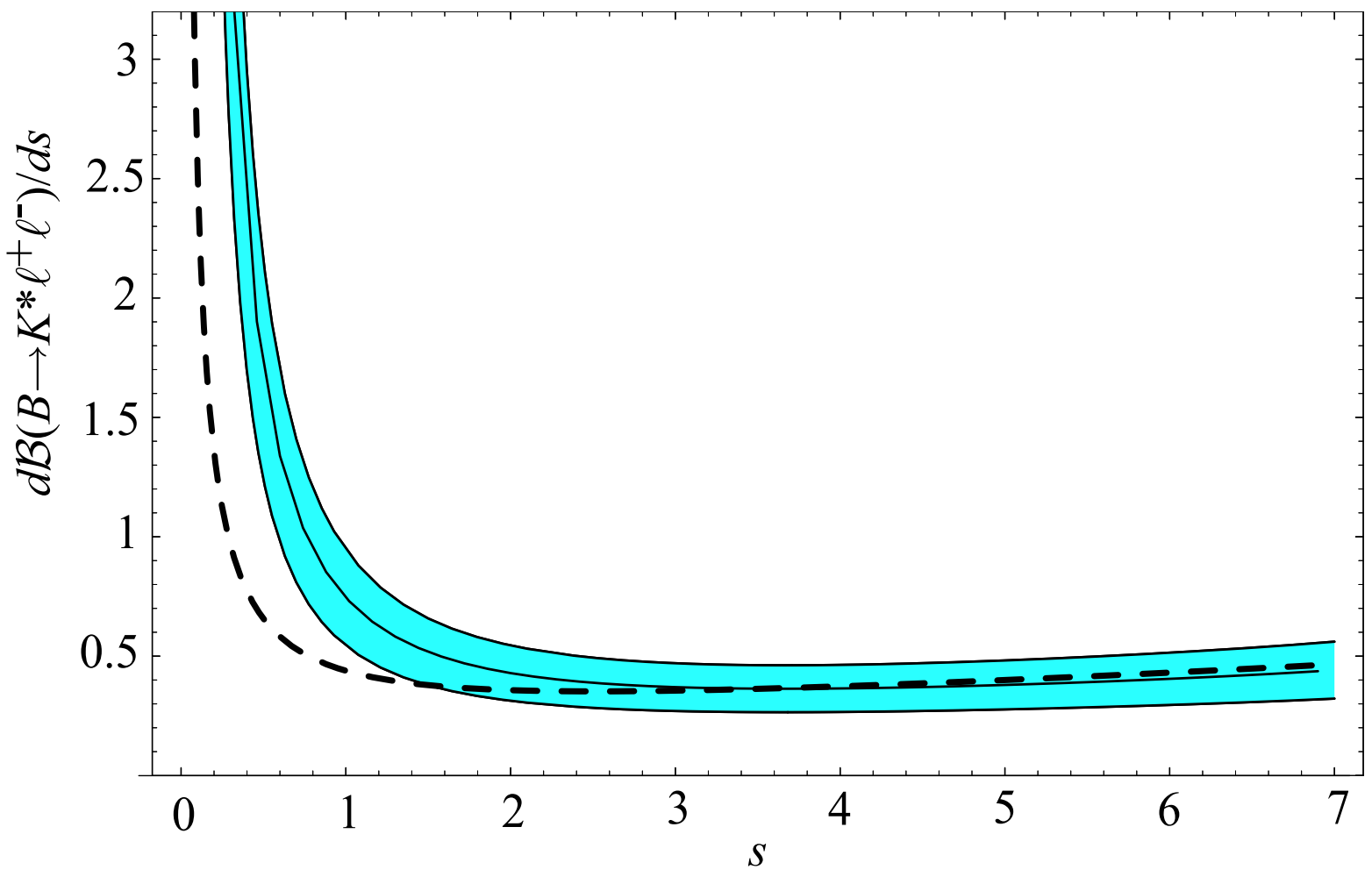

**Figure 2-16.** *Dilepton invariant mass distribution for $B \to K^* \ell^+ \ell^-$ in the Standard Model at NLO (shaded region), reflecting parametric uncertainties, and at leading order (dashed curve). (Figure from Ref. [225].)*

The DIM-distribution $d\mathcal{B}(B \to K^* \ell^+ \ell^-)/ds$ is shown in Fig. 2-16 for the low-$s$ region $s \leq 7$ GeV$^2$, where the energy of the $K^*$ is large enough ($E_{K^*} \sim O(m_B/2)$) for the theoretical calculations to remain valid. Comparison of the LO result (dashed curve) with the NLO result (shaded band) shows that except for $s < 1$ GeV$^2$, where the contribution from $C_7^{\rm eff}$ dominates, the DIM-distribution is rather stable against perturbative corrections. Theoretical uncertainties are dominated by nonperturbative quantities, in particular from the form factors and the results shown make use of the LC-QCD sum rule results from Ref [9]. It should be noted that HQET and SU(3) symmetry relate the decays $B \to (\pi, \rho)\, \ell \nu_\ell$ and $B \to (K, K^*) \ell^+ \ell^-$. Data on the transitions $B \to (\pi, \rho) \ell \nu_\ell$ are already available; they will become quite precise in forthcoming measurements from the current $B$ Factories. This data can be used together with estimates of SU(3)breaking to determine the remaining form factors in $B \to (K, K^*) \ell^+ \ell^-$. Thus, with precise measurements of the decays $B \to (\pi, \rho)\, \ell \nu_\ell$ and $V_{ub}$, one can make almost model-independent predictions for the $B \to (K, K^*)\, \ell^+ \ell^-$ decay rates and spectra. This analysis can be further refined by doing a helicity decomposition of the decays $B \to K^* \ell^+ \ell^-$ and $B \to \rho \ell \nu_\ell$ [225]. The angular distribution in the decay $B \to K^* (\to K\pi) \ell^+ \ell^-$ is discussed in Section 2.17.

The DIM-distribution for $B \to K \ell^+ \ell^-$ can be expressed as (dropping terms proportional to the lepton mass)

$$\frac{d\Gamma}{d\hat{s}} \sim |V_{ts}^* V_{tb}|^2 \left[ \left| C_9^{\rm eff} f_+(\hat{s}) + \frac{2\hat{m}_b}{1+\hat{m}_K} C_7^{\rm eff} f_T(\hat{s}) \right|^2 + |C_{10} f_+(\hat{s})|^2 \right] . \tag{2.86}$$

In the Standard Model, $|C_7^{\rm eff}| \ll |C_9^{\rm eff}|, |C_{10}|$, and also there is no kinematical enhancement for low values of $\hat{s}$ (as opposed to $B \to K^* \ell^+ \ell^-$). So, to a good approximation ($O(10\%)$), the dependence of the DIM distribution in $B \to K \ell^+ \ell^-$ simplifies, and one has

$$\frac{d\Gamma}{d\hat{s}} \sim |f_+(\hat{s})|^2 , \tag{2.87}$$

and, as noted above, this form factor $f_+(\hat{s})$ can be determined from the $B \to \pi \ell \nu_\ell$ decay and SU(3)-breaking. Thus, by using data, the LEET/SCET-technology and calculating the SU(3)-breaking effects in the form factors, rather precise phenomenological profiles of the exclusive decays $B \to (K, K^*) \ell^+ \ell^-$ can be obtained.

### 2.16.2 Sensitivity to New Physics in $B \to (K, K^*)\ell\ell$

≻— E. Lunghi —≺

The effective Hamiltonian governing the exclusive decays $B \to (K, K^*)\ell\ell$ is

$$\mathcal{H}_{\rm eff} = -4\frac{G_F}{\sqrt{2}} V_{tb} V_{ts}^* \Big\{ \sum_{i=1}^{10} C_i(\mu) O_i(\mu) + \sum_{i=S,P} C_i(\mu) O_i(\mu) \Big\} , \tag{2.88}$$

where $O_i(\mu)$ are dimension six operators at the scale $\mu_b \sim \mathcal{O}(m_b)$ and $C_i(\mu_b)$ are the corresponding Wilson coefficients. Within the Standard Model, only the coefficients $C_{1-10}$ receive sizable contributions. The scalar and pseudoscalar operators $O_{S,P}$ acquire non-vanishing coefficients in many extensions of the Standard Model and analyses of their effects are presented in Section 2.18 and 2.21. We refer to Ref. [226] for the definition of the operators and a discussion of the Wilson coefficients.

The matrix elements of the operators $O_i(\mu)$ between the hadronic states $B$ and $K^{(*)}$ must be parameterized in terms of form factors. Our present lack of control on hadronic uncertainties, that are of order $\mathcal{O}(30\%)$, affects seriously the possibility of using these decays as a probe for New Physics. According to the analysis presented in Ref. [10], the constraints induced by the current measurements of the branching ratios for these decays are weaker than the bounds coming from the corresponding inclusive modes (*i.e.*, $B \to X_s\ell\ell$).

In order to compete with the inclusive channels, it is necessary to consider observables in which the large form factor uncertainties partially drop out. In the following we focus on the forward–backward and the ($CP$-averaged) isospin asymmetries [64, 108]:

$$\frac{{\rm d}A_{\rm FB}}{{\rm d}q^2} = \frac{1}{{\rm d}\Gamma/{\rm d}q^2} \left( \int_0^1 {\rm d}(\cos\theta) \frac{{\rm d}^2\Gamma[B \to K^*\ell\ell]}{{\rm d}q^2 {\rm d}\cos\theta} - \int_{-1}^0 {\rm d}(\cos\theta) \frac{{\rm d}^2\Gamma[B \to K^*\ell\ell]}{{\rm d}q^2 {\rm d}\cos\theta} \right) \tag{2.89}$$

$$\frac{{\rm d}A_{\rm I}}{{\rm d}q^2} = \frac{{\rm d}\Gamma[B^0 \to K^{*0}\ell\ell]/{\rm d}q^2 - {\rm d}\Gamma[B^\pm \to K^{*\pm}\ell\ell]/{\rm d}q^2}{{\rm d}\Gamma[B^0 \to K^{*0}\ell\ell]/{\rm d}q^2 + {\rm d}\Gamma[B^\pm \to K^{*\pm}\ell\ell]/{\rm d}q^2} . \tag{2.90}$$

We focus here and in the remainder of the section on the $B \to K^*\ell\ell$ mode. This is because the Standard Model operator basis, that is $\mathcal{O}_{1-10}$, does not induce any forward-backward asymmetry in the decay into a pseudoscalar kaon; New physics contributions resulting in non-vanishing $C_S$ and $C_P$ are responsible for a non-vanishing forward-backward asymmetry in $B \to K\ell\ell$ decays. This possibility is entertained in Section 2.18.

The results summarized in this report have been obtained in the QCD factorization approach [91, 109] at NLO. Note, in particular, that the isospin asymmetry vanishes at tree level and is induced by computable non-factorizable $O(\alpha_s)$ contributions to the $B \to K^*\ell\ell$ amplitude. Details of the calculation are given in Ref. [108]. In the following, we concentrate on the region of the dilepton mass below the $c\bar{c}$ resonances ($q^2 < 4m_c^2 \sim 7\,{\rm GeV}$); in fact, the QCD factorization approach holds only if the energy of the final state kaon is of order $\mathcal{O}(m_b)$. We refer to Ref. [227] for a discussion of the high $q^2$ region.

The most interesting feature of the forward-backward asymmetry $dA_{\rm FB}/dq^2$ is the presence of a $q^2$ value at which the asymmetry vanishes. The precise location of this zero is not affected by the large form factors uncertainties; At leading order it is completely independent from hadronic quantities and is a simple function of the Wilson coefficients $C_7$ and $C_9$ [9]. A conservative estimate of the location of the asymmetry zero is [64]

$$q_0^2 = (4.2 \pm 0.6)\ {\rm GeV}^2 . \tag{2.91}$$

The left plot of Fig. 2-17 (taken from Ref. [64]) shows $dA_{\rm FB}/dq^2$ in the Standard Model. The yellow band reflects all theory uncertainties; note that NLO corrections move the asymmetry zero to higher $q^2$ values.

The isospin asymmetry has been measured at $q^2 = 0$ and found to be large: $A_I(B \to K^*\gamma) = 0.11 \pm 0.07$ [228, 229, 230]. For higher values of $q^2$ the asymmetry decreases in the Standard Model, see the right plot in Fig. 2-17, and eventually becomes negligibly small for $q^2 > 2\mathrm{GeV}$. Note that the $q^2 > 2\mathrm{GeV}$ region is also characterized by a particularly small theoretical uncertainty. Thus in this region, $A_I$ is sensitive to the presence of New Physics.

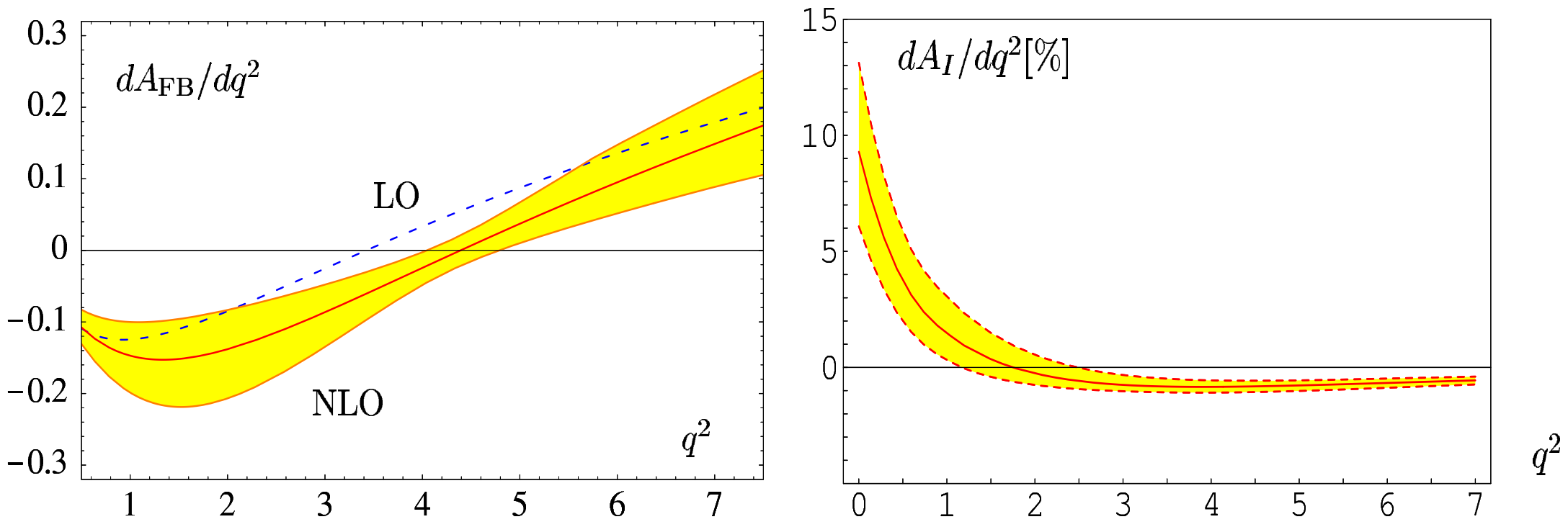

**Figure 2-17.** *Standard Model predictions for the forward-backward and isospin asymmetries for the decays $B \to K^*\ell\ell$. The bands reflect the theory uncertainties.*

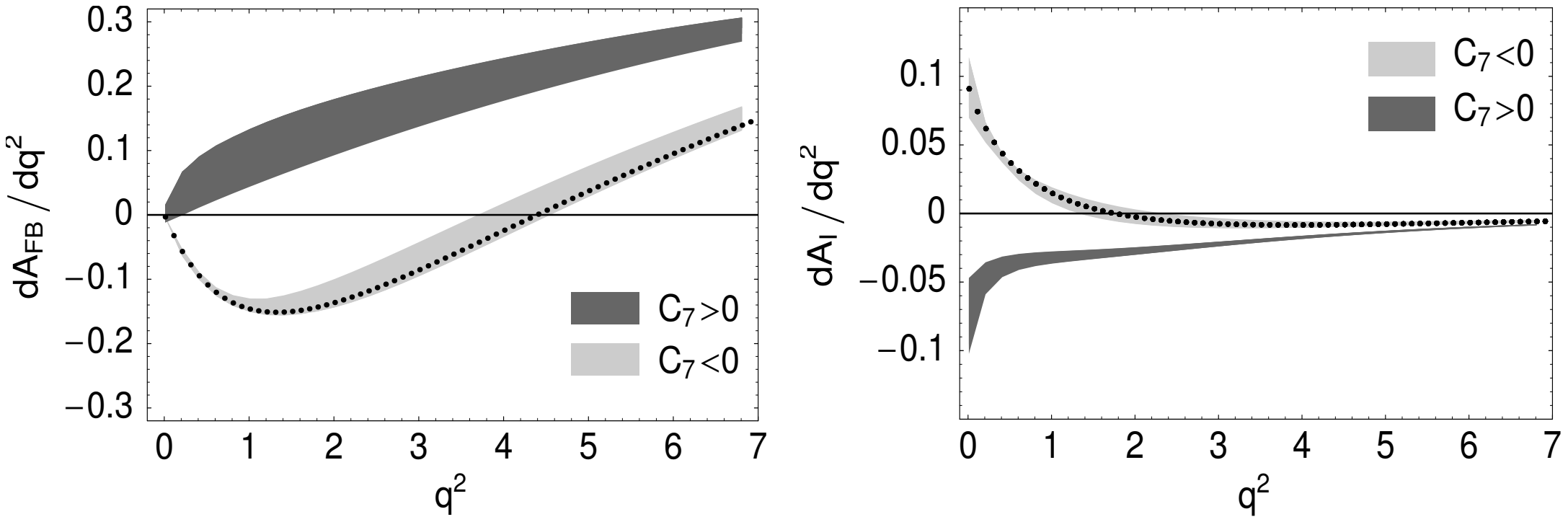

**Figure 2-18.** *Forward-backward and isospin asymmetries in the Minimal Flavor Violating MSSM for different signs of $C_7$ ($C_7^{SM} < 0$).*

As an example of a concrete extension of the Standard Model on these two observables, we present results for the Minimal Flavor Violating (MFV) MSSM. The reason for considering this subset of the full MSSM parameter space is that any effect observed in this restricted framework will also perforce appear in any more general framework. The model is constructed at the electroweak scale and all flavor change is induced by the CKM matrix. The results of the analysis are summarized in Fig. 2-18, taken from Ref. [108].

For both observables, the most striking effect is obtained for those points in the MFV parameter space for which the sign of the Wilson coefficient $C_7$ is opposite to that in the Standard Model; for the forward-backward asymmetry this implies, in particular, the absence of the zero and a change of sign in the very low $q^2$ region. The MFV result for the isospin asymmetry depends also on the sign of $C_7$; the allowed bands are much thinner than in the forward-backward asymmetry case, because $A_I$ is dominated by QCD penguins $\mathcal{O}_{3-6}$. The latter are modified only slightly in the MSSM with minimal flavor violation. A measurement of a large isospin asymmetry for moderate values of $q^2$ would point to physics beyond the Standard Model and to the MFV MSSM.

This work is partially supported by the Swiss National Fonds.

### 2.16.3 Electron versus muon modes as a probe of New Physics

≻— G. Hiller —≺

The ratios of branching ratios $R_{K^{(*)}}$ of $B \to K^{(*)}e^+e^-$ vs. $B \to K^{(*)}\mu^+\mu^-$ decays *with the same cuts* on the dilepton mass defined in Eq. (2.68) are sensitive probes of the flavor sector, see the discussion in Section 2.14.2 on the inclusive modes. The Standard Model predictions for the exclusive decays are very clean [129]

$$R_K^{SM} = 1 \pm 0.0001\,, \qquad R_{K^*}^{SM} = 0.99 \pm 0.002\,, \tag{2.92}$$

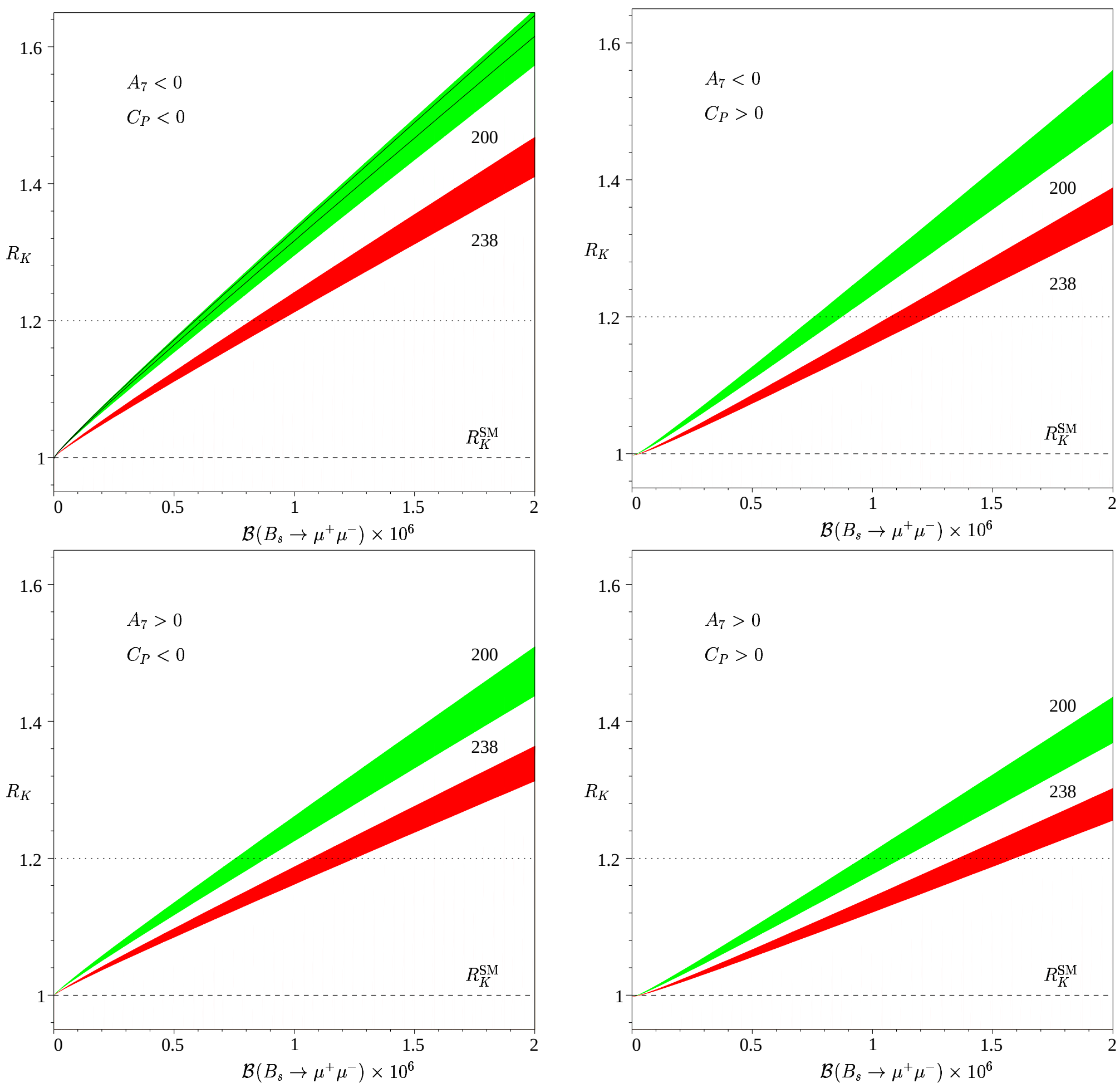

**Figure 2-19.** *Correlation between $R_K$ and the $B_s \to \mu^+\mu^-$ branching ratio for different signs of couplings $A_7$ and $C_P$, two values of $f_{B_s}$ in MeV and $A_{9,10} = A_{9,10}^{SM}$, see [129] for details. The shaded areas have been obtained by varying the $B \to K$ form factors according to [9] and $A_7$ within experimentally allowed ranges. In the upper left plot, the form factor uncertainty is illustrated for fixed $A_7 = A_7^{SM}$ and $f_{B_s} = 200$ MeV by solid lines. The dotted lines correspond to the $90\%$ C.L. upper limit on $R_K$. Dashed lines denote the Standard Model prediction for $R_K$.*

where the spectra have been integrated over the full dilepton mass region, *i.e.*, $q^2_{max} = (m_B - m_{K^{(*)}})^2$ and the form factors are varied according to [9]. As with inclusive decays, the ratios are equal to one, up to small kinematical corrections of order $m_\mu^2/m_b^2$.

The ratios $R_{K^{(*)}}$ can be affected by New Physics that couples differently to electrons and muons. Current $B$ Factory data yield $0.4 \leq R_K \leq 1.2$ @ 90 % C.L [129, 220, 221]. Correlation with $R_K$ and other rare $B$ decay data allow for an enhanced value of $R_{K^*}$ of up to 1.12 [129]. The impact of the new lepton-specific couplings can can be larger for $B \to K$ than for $B \to K^*$ and $B \to X_s$ decays. The reason is that, besides different hadronic matrix elements in the latter decays, the photon pole, which is absent from the decay into a pseudoscalar meson, dominates the rate for very low dilepton mass.

The bound $R_K \leq 1.2$ yield constraints on new couplings induced by scalar and pseudoscalar interactions comparable to the ones obtained from the current upper bound on the $B_s \to \mu^+\mu^-$ branching ratio $\mathcal{B}(B_s \to \mu^+\mu^-) < 5.8 \cdot 10^{-7}$ @ 90 % C.L. [16]. The correlation between these observables is shown in Fig. 2-19 for a generic model beyond the Standard Model, which would also apply to the MSSM with minimal flavor violation, see [129] for details. The dependence on the $B \to K$ form factors is weak. If one allows also for contributions from right-handed currents, the correlation breaks down and both observables become complementary: While $B_s \to \mu^+\mu^-$ constrains the difference of couplings with opposite helicity, $R_K$ constrains their sum. Hence, combined data can be used to exclude the possibility of large cancellations or, in other words: New Physics might hide in one mode or the other, but not in both [129].

## 2.17 Angular distribution in $B \to K^*(\to K\pi)\ell^+\ell^-$

>– F. Krüger –<

### 2.17.1 Theoretical framework

The four-body decay $\overline{B} \to K^{*0}(\to K^-\pi^+)\ell^+\ell^-$ into dileptons invites attention as a testing ground for the Standard Model and possible extensions [231, 186, 187, 188, 232, 181, 225]. The corresponding matrix element can be obtained from the effective Hamiltonian for the $b \to s\ell^+\ell^-$ transition.[10] Assuming an extended operator basis that contains the Standard Model operators [76, 75, 233] together with their helicity-flipped counterparts [234], the matrix element can be written as

$$M = \frac{G_F\alpha}{\sqrt{2}\pi} V_{tb}V_{ts}^*[F_V^\mu(\bar{l}\gamma_\mu l) + F_A^\mu(\bar{l}\gamma_\mu\gamma_5 l)], \tag{2.93}$$

with

$$F_V^\mu = C_9^{\rm eff}(\bar{s}\gamma^\mu P_L b) + C_9^{\rm eff\prime}(\bar{s}\gamma^\mu P_R b) - \frac{2m_b}{q^2}\bar{s}i\sigma^{\mu\nu}q_\nu(C_7^{\rm eff}P_R + C_7^{\rm eff\prime}P_L)b, \tag{2.94}$$

$$F_A^\mu = C_{10}(\bar{s}\gamma^\mu P_L b) + C_{10}'(\bar{s}\gamma^\mu P_R b). \tag{2.95}$$

Here $q = p_{l^+} + p_{l^-}$, $P_{L,R} = (1 \mp \gamma_5)/2$ and $C_i^{(\prime)}$ are the Wilson coefficients with $C_i^{\prime\,\rm SM} = 0$.

The hadronic part of the matrix element, which describes the transition $B \to K\pi$, can be parameterized in terms of the $B \to K^*$ form factors by using a narrow-width approximation [187]. The relevant form factors are defined as [235, 9]

$$\begin{aligned}\langle K^*(p_{K^*})|\,\bar{s}\gamma_\mu P_{L,R} b\,|\overline{B}(p)\rangle &= i\epsilon_{\mu\nu\alpha\beta}\epsilon^{\nu*}p^\alpha q^\beta \frac{V(s)}{M_B + M_{K^*}} \mp \frac{1}{2}\Big\{\epsilon_\mu^*(M_B + M_{K^*})A_1(s) - (\epsilon^*\cdot q)(2p-q)_\mu \\ &\times \frac{A_2(s)}{M_B + M_{K^*}} - \frac{1}{s}(\epsilon^*\cdot q)[(M_B + M_{K^*})A_1(s) - (M_B - M_{K^*})A_2(s) - 2M_{K^*}A_0(s)]q_\mu\Big\},\end{aligned} \tag{2.96}$$

$$\begin{aligned}\langle K^*(p_{K^*})|\,\bar{s}i\sigma_{\mu\nu}q^\nu P_{R,L} b\,|\overline{B}(p)\rangle &= -i\epsilon_{\mu\nu\alpha\beta}\epsilon^{\nu*}p^\alpha q^\beta T_1(s) \pm \frac{1}{2}\Big\{[\epsilon_\mu^*(M_B^2 - M_{K^*}^2) \\ &- (\epsilon^*\cdot q)(2p-q)_\mu]T_2(s) + (\epsilon^*\cdot q)\left[q_\mu - \frac{s}{M_B^2 - M_{K^*}^2}(2p-q)_\mu\right]T_3(s)\Big\},\end{aligned} \tag{2.97}$$

where $q = p - p_{K^*}$, $s \equiv q^2$ and $\epsilon^\mu$ is the $K^*$ polarization vector.

In the limit in which the initial hadron is heavy and the final light meson has large energy, relations between these form factors emerge [121, 236, 237, 101, 64]. This happens in the low dilepton invariant mass region $s \ll m_b^2$. As a consequence, the seven *a priori* independent form factors in (2.96) and (2.97) reduce to two universal form factors $\zeta_\perp(E_{K^*})$ and $\zeta_\parallel(E_{K^*})$ at leading order, see also Section 2.16.1 for details. The impact of corrections to the form factor relations of order $1/M_B$, $1/E_{K^*}$ and $\alpha_s$ [101, 64] for $\overline{B} \to K\pi\ell^+\ell^-$ decays are discussed in [238].

### 2.17.2 Transversity amplitudes

Neglecting the lepton mass there are three $K^*$-spin amplitudes $A_{\parallel,\perp,0}$. In the presence of right-handed currents, they are, for $m_s = 0$ [238]:

$$A_{\perp L,R} = N\sqrt{2}\lambda^{1/2}\Big\{[(C_9^{\rm eff} + C_9^{\rm eff\prime}) \mp (C_{10} + C_{10}')]\frac{V(s)}{M_B + M_K^*} + \frac{2m_b}{s}(C_7^{\rm eff} + C_7^{\rm eff\prime})T_1(s)\Big\}, \tag{2.98}$$

[10] We use $\overline{B} \equiv b\bar{d}$ and neglect the mass of the strange quark.

$$A_{\parallel L,R} = -N\sqrt{2}(M_B^2 - M_{K^*}^2)\left\{[(C_9^{\rm eff} - C_9^{\rm eff\prime}) \mp (C_{10} - C_{10}')]\frac{A_1(s)}{M_B - M_{K^*}} + \frac{2m_b}{s}(C_7^{\rm eff} - C_7^{\rm eff\prime})T_2(s)\right\}, \quad (2.99)$$

$$A_{0L,R} = -\frac{N}{2M_{K^*}\sqrt{s}}\left\{[(C_9^{\rm eff} - C_9^{\rm eff\prime}) \mp (C_{10} - C_{10}')]\left[(M_B^2 - M_{K^*}^2 - s)(M_B + M_{K^*})A_1(s) - \lambda\frac{A_2(s)}{M_B + M_{K^*}}\right]\right.$$
$$\left. + 2m_b(C_7^{\rm eff} - C_7^{\rm eff\prime})\left[(M_B^2 + 3M_{K^*}^2 - s)T_2(s) - \frac{\lambda}{M_B^2 - M_{K^*}^2}T_3(s)\right]\right\}, \quad (2.100)$$

where $\lambda = M_B^4 + M_{K^*}^4 + s^2 - 2(M_B^2 M_{K^*}^2 + M_{K^*}^2 s + M_B^2 s)$ and

$$N = \left[\frac{G_F^2\alpha^2}{3\times 2^{10}\pi^5 M_B^3}|V_{tb}V_{ts}^*|^2 s\lambda^{1/2}\right]^{1/2}. \quad (2.101)$$

The transversity amplitudes are related to the helicity amplitudes via $A_{\parallel,\perp} = (H_{+1} \pm H_{-1})/\sqrt{2}$, $A_0 = H_0$, see *e.g.*, [186, 188, 232].

With $\hat{s} = s/M_B^2$ and $\hat{m}_i = m_i/M_B$ and exploiting the *leading order* form factor relations valid at low $s$ [121, 236, 237, 101, 64] the above amplitudes become much simpler [238], *i.e.*,

$$A_{\perp L,R} = \sqrt{2}NM_B(1-\hat{s})\left\{[(C_9^{\rm eff} + C_9^{\rm eff\prime}) \mp (C_{10} + C_{10}')] + \frac{2\hat{m}_b}{\hat{s}}(C_7^{\rm eff} + C_7^{\rm eff\prime})\right\}\zeta_\perp(E_{K^*}), \quad (2.102)$$

$$A_{\parallel L,R} = -\sqrt{2}NM_B(1-\hat{s})\left\{[(C_9^{\rm eff} - C_9^{\rm eff\prime}) \mp (C_{10} - C_{10}')] + \frac{2\hat{m}_b}{\hat{s}}(C_7^{\rm eff} - C_7^{\rm eff\prime})\right\}\zeta_\perp(E_{K^*}), \quad (2.103)$$

$$A_{0L,R} = -\frac{NM_B}{\sqrt{\hat{s}}}(1-\hat{s})\left\{[(C_9^{\rm eff} - C_9^{\rm eff\prime}) \mp (C_{10} - C_{10}')] + 2\hat{m}_b(C_7^{\rm eff} - C_7^{\rm eff\prime})\right\}\zeta_\parallel(E_{K^*}). \quad (2.104)$$

In Eqs. (2.102)–(2.104) we have dropped terms of $\mathcal{O}(M_{K^*}^2/M_B^2)$. Within the Standard Model, we recover the naive quark-model prediction of $A_\perp = -A_\parallel$ (see, *e.g.*, [239]) in the $M_B \to \infty$ and $E_{K^*} \to \infty$ limit. In fact, the $s$ quark is produced in helicity $-\frac{1}{2}$ by weak $V-A$ interactions in the limit $m_s \to 0$, which is not affected by strong interactions in the massless case. Thus, the strange quark combines with a light quark to form a $K^*$ with helicity either $-1$ or $0$ but not $+1$. Consequently, at the quark level the Standard Model predicts $H_{+1} = 0$, and hence $A_\perp = -A_\parallel$, which is revealed as $|H_{-1}| \gg |H_{+1}|$ (or $A_\perp \approx -A_\parallel$) at the hadron level [224].

### 2.17.3 Differential decay rate in the transversity basis

If the spins of the particles in the final state are not measured and assuming the $K^*$ to be on its mass shell, the decay rate of $\overline{B} \to K^-\pi^+\ell^+\ell^-$ decays can be written as [187]

$$d^4\Gamma = \frac{9}{32\pi}I(s,\theta_l,\theta_{K^*},\phi)ds\, d\cos\theta_l\, d\cos\theta_{K^*}\, d\phi, \quad (2.105)$$

where

$$I = I_1 + I_2\cos 2\theta_l + I_3\sin^2\theta_l\cos 2\phi + I_4\sin 2\theta_l\cos\phi + I_5\sin\theta_l\cos\phi + I_6\cos\theta_l$$
$$+ I_7\sin\theta_l\sin\phi + I_8\sin 2\theta_l\sin\phi + I_9\sin^2\theta_l\sin 2\phi, \quad (2.106)$$

and with the physical regions of the phase space

$$0 \le s \le (M_B - M_{K^*})^2, \quad -1 \le \cos\theta_l \le 1, \quad -1 \le \cos\theta_{K^*} \le 1, \quad 0 \le \phi \le 2\pi. \quad (2.107)$$

The three angles $\theta_l, \theta_{K^*}, \phi$, which describe the decay $\overline{B} \to K^-\pi^+\ell^+\ell^-$, are illustrated in Figure 2-20.

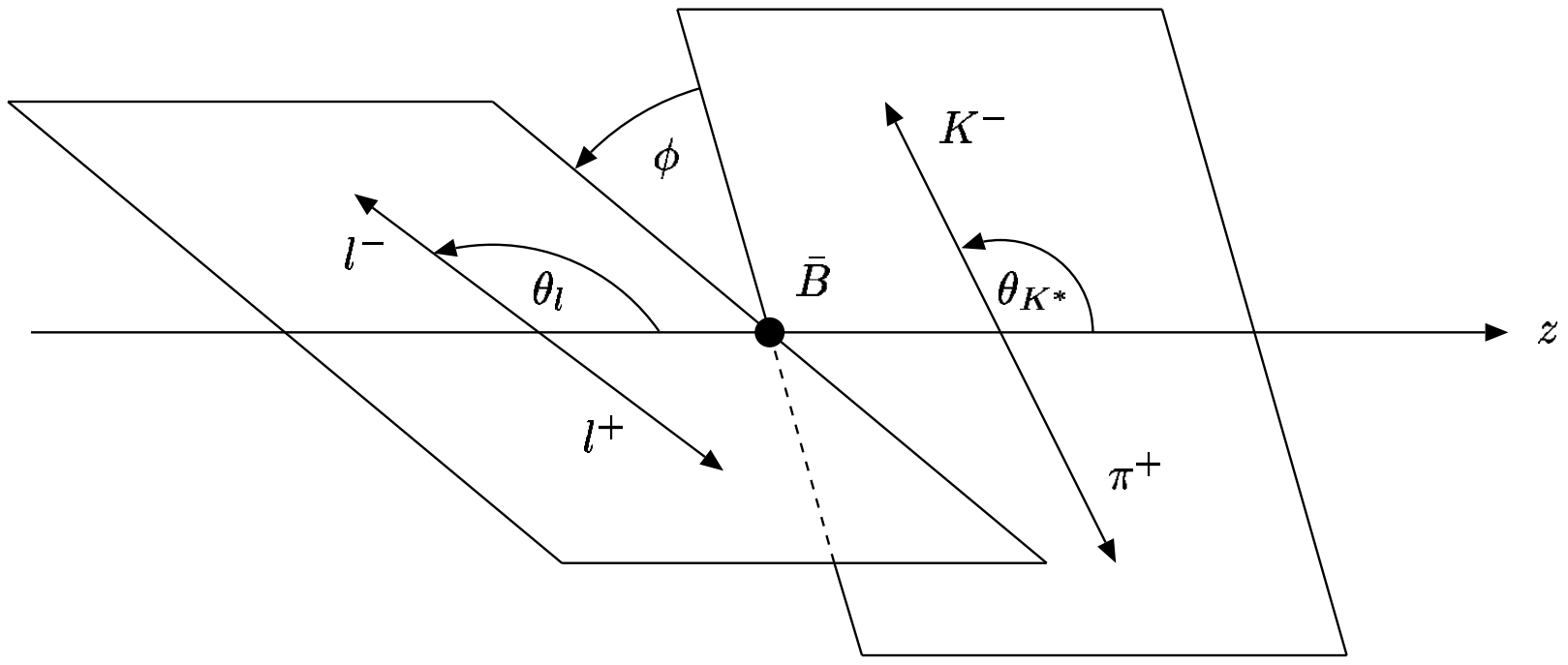

**Figure 2-20.** *Definition of the kinematic variables.*

The functions $I_{1-9}$ in terms of the transversity amplitudes are given by

$$\begin{aligned}
I_1 &= \left[\frac{3}{4}(|A_{\perp L}|^2 + |A_{\| L}|^2)\sin^2\theta_{K^*} + |A_{0L}|^2\cos^2\theta_{K^*}\right] + (L \to R),\\
I_2 &= \left[\frac{1}{4}(|A_{\perp L}|^2 + |A_{\| L}|^2)\sin^2\theta_{K^*} - |A_{0L}|^2\cos^2\theta_{K^*}\right] + (L \to R),\\
I_3 &= \frac{1}{2}(|A_{\perp L}|^2 - |A_{\| L}|^2)\sin^2\theta_{K^*} + (L \to R),\\
I_4 &= \frac{1}{\sqrt{2}}\mathrm{Re}(A_{0L}A^*_{\| L})\sin 2\theta_{K^*} + (L \to R),\\
I_5 &= \sqrt{2}\mathrm{Re}(A_{0L}A^*_{\perp L})\sin 2\theta_{K^*} - (L \to R),\\
I_6 &= 2\mathrm{Re}(A_{\| L}A^*_{\perp L})\sin^2\theta_{K^*} - (L \to R),\\
I_7 &= \sqrt{2}\mathrm{Im}(A_{0L}A^*_{\| L})\sin 2\theta_{K^*} - (L \to R),\\
I_8 &= \frac{1}{\sqrt{2}}\mathrm{Im}(A_{0L}A^*_{\perp L})\sin 2\theta_{K^*} + (L \to R),\\
I_9 &= \mathrm{Im}(A^*_{\| L}A_{\perp L})\sin^2\theta_{K^*} + (L \to R),
\end{aligned} \tag{2.108}$$

where the subscripts $L$ and $R$ denote a left-handed and right-handed $\ell^-$ in the final state.

## 2.17.4 Searching for New Physics via $K^*$ polarization

From the differential decay distribution in Eq. ((2.105)) one can construct various observables that allow tests of the Standard Model and its possible extensions. Here we consider the asymmetries

$$A_T^{(1)}(s) = \frac{-2\mathrm{Re}(A_\| A_\perp^*)}{|A_\perp|^2 + |A_\||^2}, \quad A_T^{(2)}(s) = \frac{|A_\perp|^2 - |A_\||^2}{|A_\perp|^2 + |A_\||^2}, \tag{2.109}$$

where $A_iA_j^* \equiv A_{iL}A^*_{jL} + A_{iR}A^*_{jR}$ $(i,j = \|,\perp,0)$. (The former observable was first discussed in Ref. [186] in terms of the helicity amplitudes.) From (2.102)–(2.104), it is clear that the form factors drop out in the asymmetries at *leading order* in $1/M_B$, $1/E_{K^*}$ and $\alpha_s$. In this approximation are $A_T^{(1)}(s) = 1$ and $A_T^{(2)}(s) = 0$ at small dilepton mass in the absence of right-handed currents.

In our numerical analysis we focus on muons in the final state and take $|C_7^{\mathrm{eff}\prime}|^2 + |C_7^{\mathrm{eff}}|^2 \leq 1.2|C_7^{\mathrm{eff,SM}}|^2$. This is consistent with the experimental result $\mathcal{B}(\overline{B} \to X_s\gamma) = (3.34 \pm 0.38) \times 10^{-4}$ [240] at 2 $\sigma$. Examples of New Physics scenarios that give sizable contributions to $C_7^{\mathrm{eff}\prime}$ include the left-right model [241] and a $\mathrm{SO}(10)$ SUSY GUT model

with large mixing between $\tilde{s}_R$ and $\tilde{b}_R$ [242]. Since the low dimuon mass region is dominated by the contributions to $O_7^{(\prime)}$, we neglect those of the helicity-flipped operators $O'_{9,10}$.

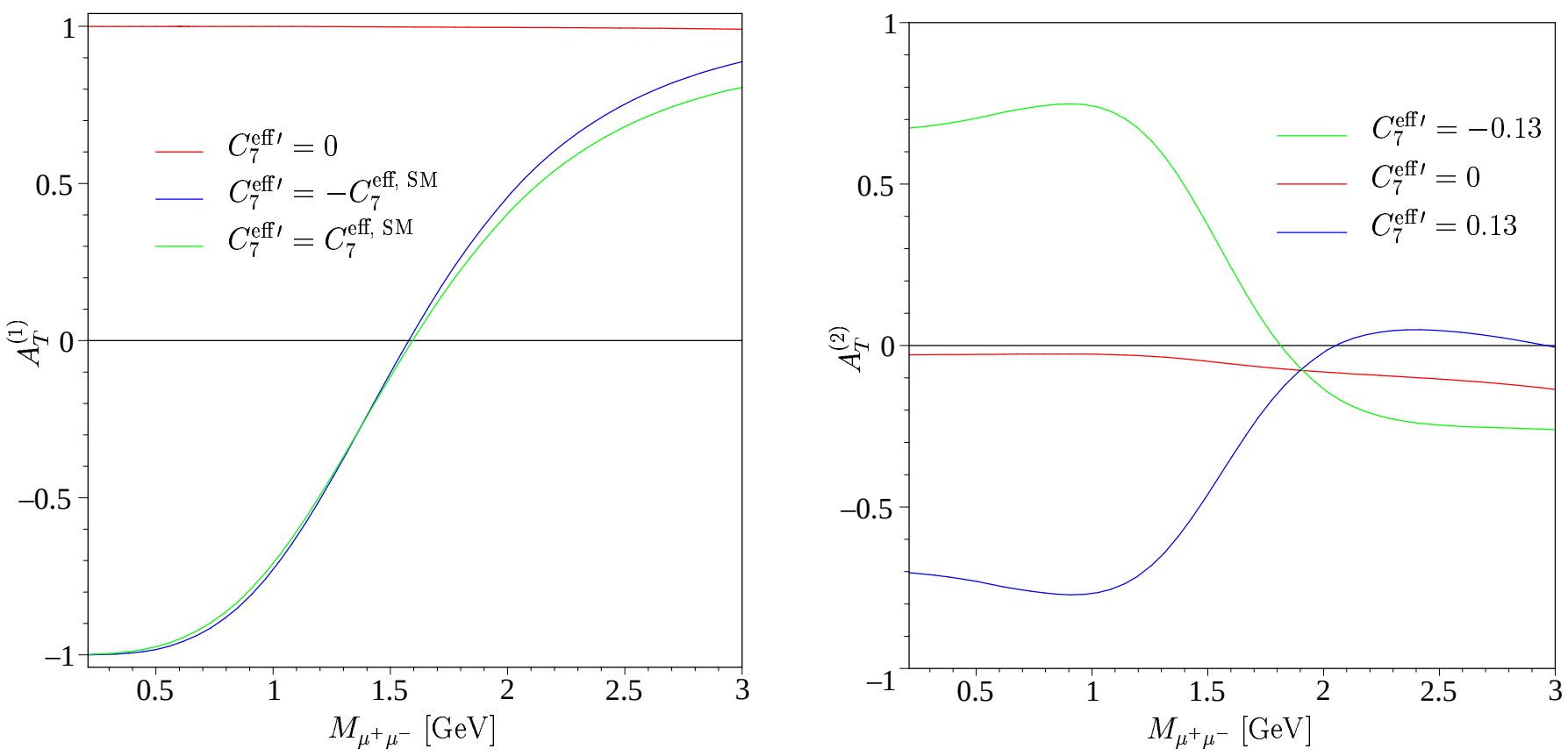

**Figure 2-21.** *The asymmetries $A_T^1$ (left plot) and $A_T^2$ (right plot) as a function of the dimuon mass $M_{\mu^+\mu^-}$ assuming $C_9^{\mathrm{eff}}$ and $C_{10}$ to be Standard Model-like. The lines with $C_7^{\mathrm{eff}\prime}=0$ correspond to the Standard Model, all other are New Physics scenarios. In the right plot we set $C_7^{\mathrm{eff}}=C_7^{\mathrm{eff,SM}}$.*

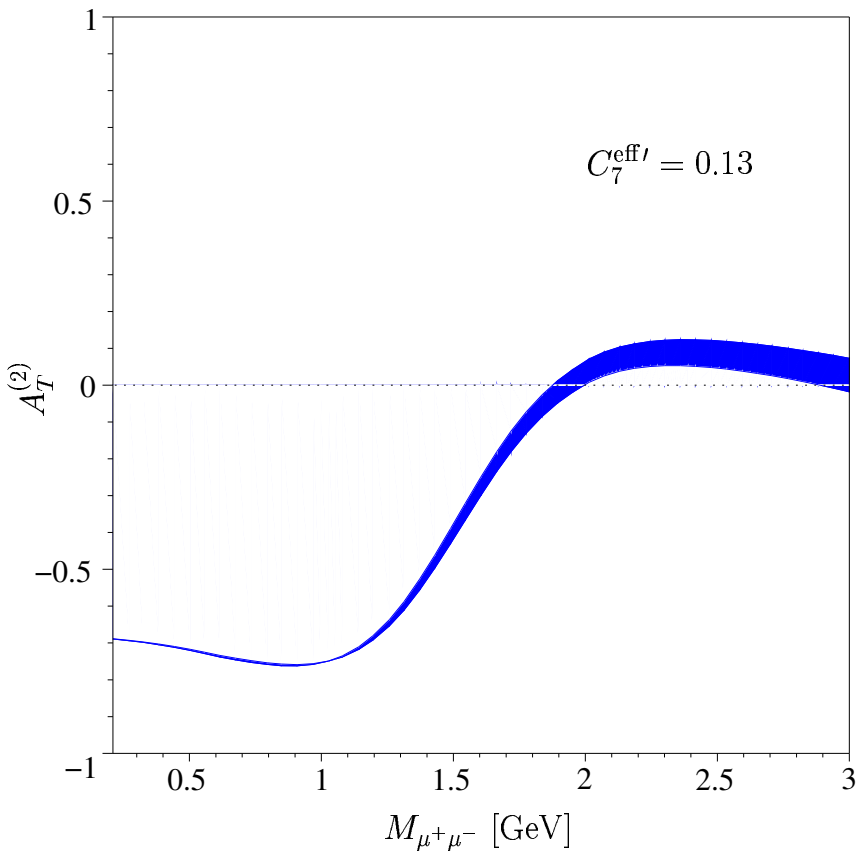

**Figure 2-22.** *Theoretical uncertainty of the asymmetry $A_T^{(2)}$ for low dimuon mass for $C_7^{\mathrm{eff}\prime}=0.13$. The shaded area has been obtained by varying the form factors according to [9].*

Figure 2-21 shows the asymmetries $A_T^{(1,2)}$ as a function of the dimuon invariant mass $\sqrt{s}=M_{\mu^+\mu^-}$ for the Standard Model (the curve with $C_7^{\mathrm{eff}\prime}=0$) and for different New Physics scenarios.

We have used the expressions in (2.98)–(2.100) together with the leading order form factor relations of [121, 236, 237], *i.e.*, kept $M_{K^*}$ finite in the kinematical factors in the transversity amplitudes. Large New Physics effects due to the helicity-flipped operator $O'_7$ can show up while being consistent with the inclusive $b\to s\gamma$ measurement. The asymmetries can have a zero in the presence of New Physics while the Standard Model predicts $A_T^{(1)}\approx 1$ and $A_T^{(2)}\approx 0$ for low dimuon mass. To get a quantitative idea of the theoretical uncertainties of the asymmetries, we show in Figure 2-22 $A_T^{(2)}$ obtained by using the minimum and maximum form factor sets of [9] for a beyond the Standard Model model with $C_7^{\mathrm{eff}\prime}=0.13$.

Both asymmetries $A_T^{(1,2)}$ are very sensitive to the non-Standard Model operator $O_7'$ in the dilepton mass region below the $J/\psi$ mass. Since these asymmetries can be predicted with small theoretical uncertainties, they provide complementary information on the structure of the underlying effective Hamiltonian describing the $b \to s\ell^+\ell^-$ transition.

### 2.17.5 Remarks on $CP$ violation in $B \to K\pi\ell^+\ell^-$

In [231, 187] it was shown that one can construct eight $CP$-violating observables by combining the differential decay rates of $d\Gamma(\overline{B} \to K^-\pi^+\ell^+\ell^-)$ and $d\overline{\Gamma}(B \to K^+\pi^-\ell^+\ell^-)$. These $CP$ asymmetries are either odd under $CP$ and even under naive $T$ transformations or they are $CP$-odd and $T$-odd observables. While the former is proportional to $\sim \sin\delta\sin\phi$, the latter involves the combination $\sim \sin\delta\cos\phi + \cos\delta\sin\phi$, where $\delta$ and $\phi$ are the strong and weak phases, respectively [243]. Furthermore, some of these $CP$ asymmetries can be determined even for an untagged mixture of $B$ and $\overline{B}$ [231, 187, 244]. An example of such a Dalitz-plot asymmetry is the sum of the forward-backward asymmetries of the lepton in $B$ and $\overline{B}$ decays [245, 246, 27].

Within the Standard Model, the $CP$ asymmetries in $\overline{B} \to K^-\pi^+l^+\ell^-$ were found to be unobservably small [187], so that any significant $CP$-violating effect would signal a non-standard source of $CP$ violation, see Table 2-15. Note that the $CP$ asymmetry that involves the function $I_7$ [see (2.108)] is zero in the Standard Model, as it is proportional to $\mathrm{Im}(C_{10}C_7^{\mathrm{eff}*})$.

**Table 2-15.** *Standard Model values of the $CP$-violating asymmetries $A_k$ in units of $10^{-4}$ (see [187] for details).*

| $A_{CP}$ | $A_3$ | $A_4$ | $A_5$ | $A_6$ | $A_7$ | $A_8$ | $A_9$ |
|---|---|---|---|---|---|---|---|
| 2.5 | $-0.6$ | $-1.9$ | 4.9 | $-4.3$ | 0 | 0.6 | $-0.04$ |

Since the strong phase in $\overline{B} \to K^-\pi^+\ell^+\ell^-$ decays is small below the $J/\psi$ mass, the $CP$-odd and $T$-even asymmetries are tiny, even in the presence of $O(1)$ non-standard $CP$-violating phases [247]. In the high dilepton mass region, on the other hand, these $CP$ asymmetries can be as large as $\sim 10\%$ in the presence of new sources of $CP$ violation. To conclude, the various $CP$ asymmetries in $\overline{B} \to K^-\pi^+\ell^+\ell^-$ decays provide a useful tool to search for New Physics.

The work of F.K. was supported by the Deutsche Forschungsgemeinschaft (DFG) under contract Bu.706/1-2.

## 2.18 Forward-backward asymmetry in $B \to K\ell^+\ell^-$

≻ F. Krüger ≺

To study the forward-backward asymmetry in a model-independent way, we write the most general matrix element for $\overline{B} \to K\ell^+\ell^-$ decays as [248]

$$M = \frac{G_F\alpha}{\sqrt{2}\pi} V_{tb}V_{ts}^*[F_S\bar{l}l + F_P\bar{l}\gamma_5 l + F_V p^\mu \bar{l}\gamma_\mu l + F_A p^\mu \bar{l}\gamma_\mu\gamma_5 l], \tag{2.110}$$

where $p^\mu$ is the four-momentum of the initial $B$ meson. The $F_i$'s are given by $(m_s = 0)$

$$F_S = \frac{1}{2}(M_B^2 - M_K^2)f_0(s)\frac{C_S}{m_b}, \tag{2.111}$$

$$F_P = -m_l\widetilde{C}_{10}^{\rm eff}\left\{f_+(s) - \frac{M_B^2 - M_K^2}{s}[f_0(s) - f_+(s)]\right\} + \frac{1}{2}(M_B^2 - M_K^2)f_0(s)\frac{C_P}{m_b}, \tag{2.112}$$

$$F_A = \widetilde{C}_{10}^{\rm eff}f_+(s), \quad F_V = \left[\widetilde{C}_9^{\rm eff}f_+(s) + 2\widetilde{C}_7^{\rm eff}m_b\frac{f_T(s)}{M_B + M_K}\right], \tag{2.113}$$

where $s = (p_{\ell^+} + p_{\ell^-})^2$. The $s$-dependent form factors and the Wilson coefficients $\widetilde{C}_i$ can be found in [9] and [52, 83, 84, 249, 87, 137], respectively, and for the definition of the corresponding operators, see [129].

After summing over lepton spins, the differential decay rate reads as

$$\frac{1}{\Gamma_0}\frac{d\Gamma(\overline{B} \to K\ell^+\ell^-)}{ds\, d\cos\theta} = \lambda^{1/2}\beta_l\Bigg\{s(\beta_l^2|F_S|^2 + |F_P|^2) + \frac{1}{4}\lambda(1 - \beta_l^2\cos^2\theta)(|F_A|^2 + |F_V|^2) + 4m_l^2M_B^2|F_A|^2$$
$$+ 2m_l[\lambda^{1/2}\beta_l{\rm Re}(F_SF_V^*)\cos\theta + (M_B^2 - M_K^2 + s){\rm Re}(F_PF_A^*)]\Bigg\}, \quad \Gamma_0 = \frac{G_F^2\alpha^2}{2^9\pi^5M_B^3}|V_{tb}V_{ts}^*|^2, \tag{2.114}$$

where $\beta_l = \sqrt{1 - 4m_l^2/s}$ and $\lambda = M_B^4 + M_K^4 + s^2 - 2(M_B^2M_K^2 + M_K^2s + M_B^2s)$; $\theta$ is the angle between the three-momentum vectors $\mathbf{p}_{l^-}$ and $\mathbf{p}_K$ in the dilepton center-of-mass system.

The term linear in $\cos\theta$ in (2.114) produces a FB asymmetry

$$A_{\rm FB}(s) = \frac{\displaystyle\int_0^1 d\cos\theta\frac{d\Gamma}{ds\, d\cos\theta} - \int_{-1}^0 d\cos\theta\frac{d\Gamma}{ds\, d\cos\theta}}{\displaystyle\int_0^1 d\cos\theta\frac{d\Gamma}{ds\, d\cos\theta} + \int_{-1}^0 d\cos\theta\frac{d\Gamma}{ds\, d\cos\theta}}, \tag{2.115}$$

which is given as

$$A_{\rm FB}(s) = \frac{2m_l\lambda\beta_l^2{\rm Re}(F_SF_V^*)\Gamma_0}{d\Gamma/ds}. \tag{2.116}$$

Within the Standard Model, the scalar function $F_S$ is suppressed by $C_S^{\rm SM} \propto m_lM_B/M_W^2$, so that $A_{\rm FB}^{\rm SM} \simeq 0$. Hence, the forward-backward asymmetry in $B \to K\ell^+\ell^-$ decays is a probe of possible new interactions outside the Standard Model.

To assess New Physics contributions to $A_{\rm FB}$ we consider the average FB asymmetry $\langle A_{\rm FB}\rangle$, which is obtained from (2.116) by integrating numerator and denominator separately over the dilepton mass. Here we focus on $\ell = \mu$ and assume the short-distance coefficients to be real. Then, adopting the notation of [129] and taking the minimum allowed form factors from [9], we find

$$\langle A_{\rm FB}\rangle = C_S(0.425 + 0.981A_9 + 1.827A_7)\left[\frac{10^{-9}}{\mathcal{B}(\overline{B}\to K\mu^+\mu^-)}\right], \tag{2.117}$$

where $A_i \equiv A_i^{\rm SM} + A_i^{\rm NP}$, with $A_7^{\rm SM}(2.5\ \text{GeV}) = -0.330$ and $A_9^{\rm SM}(2.5\ \text{GeV}) = 4.069$. In (2.117) we have used the Wilson coefficients in the NNLO approximation [52, 83, 84, 249, 87, 137] except for $C_S$, where we have taken the lowest order expression (see [129] for details).

Since the size of the forward-backward asymmetry in $\overline{B} \to K\ell^+\ell^-$ decays depends sensitively on $C_S$, it is useful to relate $\langle A_{\rm FB}\rangle$ to the branching ratio $\mathcal{B}(\overline{B}_s \to \mu^+\mu^-) \propto f_{B_s}^2(|C_S|^2 + |C_P|^2 + \cdots)$, where $f_{B_s}$ is the $B_s$ decay constant. The purely leptonic decay mode is discussed in detail in Section 2.21. An upper limit on its branching ratio which is currently $\mathcal{B}(\overline{B}_s \to \mu^+\mu^-) < 5.8\times 10^{-7}$ at 90% C.L.[16] implies an upper bound on $|C_{S,P}|$ [248, 129, 219]. In our numerical analysis we restrict ourselves to the case $C_S = -C_P$, which is realized, for instance, in the MSSM with large $\tan\beta$ [248, 250, 17, 251, 23, 219]. For simplicity, the remaining short-distance coefficients are assumed to be Standard Model-like. Figure 2-23 shows the forward-backward asymmetry as a function of the $\overline{B}_s \to \mu^+\mu^-$ branching fraction for both signs of $C_P$ (left plot). The shaded areas have been obtained by varying $f_{B_s}$ between 200 MeV and

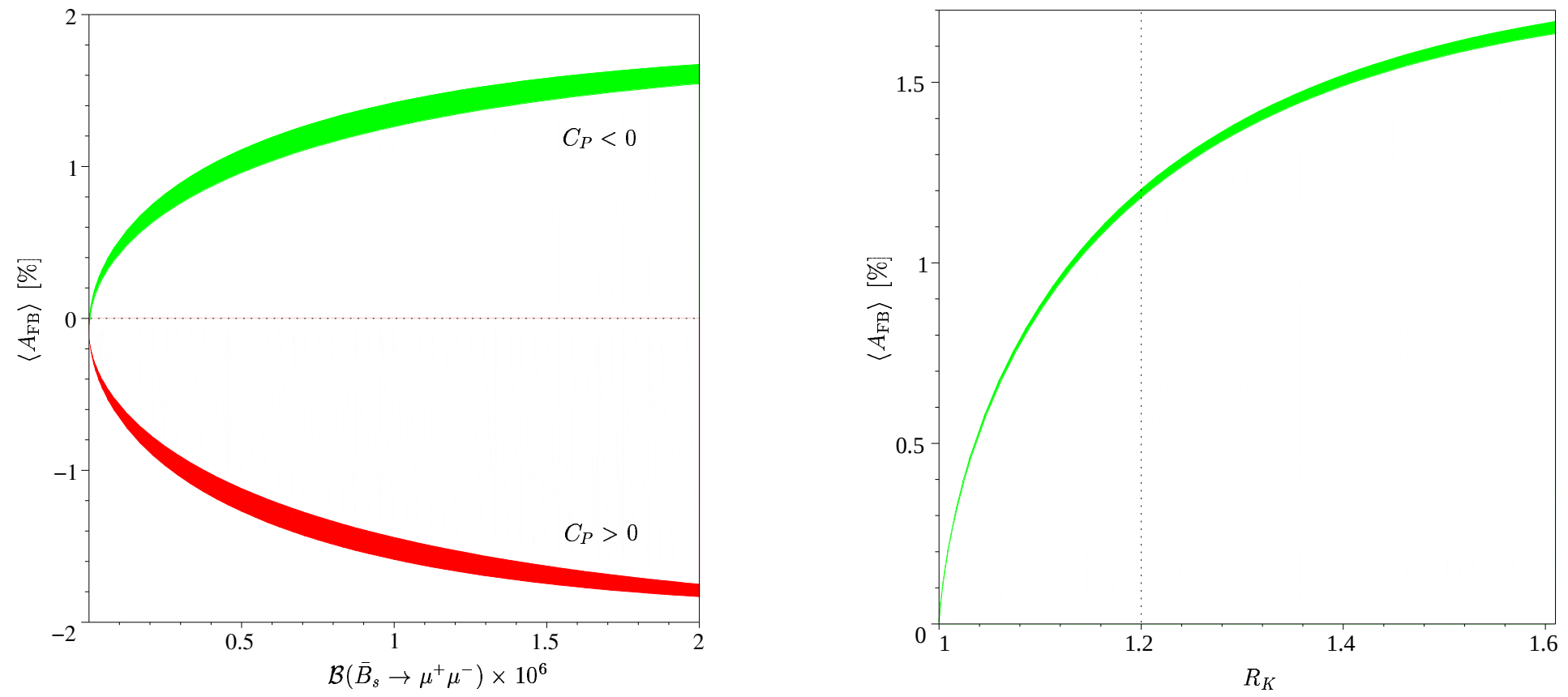

**Figure 2-23.** *The dependence of the average FB asymmetry in $\overline{B} \to K\mu^+\mu^-$ on the $\overline{B}_s \to \mu^+\mu^-$ branching ratio (left plot) and on $R_K \equiv \mathcal{B}(\overline{B} \to K\mu^+\mu^-)/\mathcal{B}(\overline{B} \to Ke^+e^-)$ (right plot) assuming $C_S = -C_P$. The dotted line in the right plot corresponds to the 90% C.L. experimental upper limit on $R_K$.*

238 MeV [252] and the $B \to K$ form factors according to [9]. As can be seen, the theoretical uncertainties of $< A_{\rm FB} >$ are rather small. Note that the two branches in the left plot of Figure 2-23 are slightly different due to the interference term $\mathrm{Re}(\widetilde{C}_{10}^{\rm eff} C_P^*)$ in the expression for the $B \to K\ell^+\ell^-$ branching fraction (see *e.g.*, [129]). While the asymmetry is rather sensitive to the sign of $C_P$, its absolute value is less than 2%, even if we vary $A_7$ in the ranges allowed by $b \to s\gamma$ [129, 10].[11] Taking into account the experimental limit for $\mathcal{B}(\overline{B}_s \to \mu^+\mu^-)$, the asymmetry is even smaller. Similarly to the constraint from $\overline{B}_s \to \mu^+\mu^-$ decays current data on $\overline{B} \to K\ell^+\ell^-$ decays [216, 215, 220, 221] exclude larger values of the forward-backward asymmetry. This can be seen from the right plot of Figure 2-23, where we have shown the forward-backward asymmetry versus the ratio $R_K \equiv \mathcal{B}(\overline{B} \to K\mu^+\mu^-)/\mathcal{B}(\overline{B} \to Ke^+e^-)$ with the current experimental upper bound $R_K \le 1.2$ (90% C.L.) [129], see also Section 2.16.3. The shaded area has been obtained by varying the form factors according to [9] for $C_P < 0$.

We conclude that the forward-backward asymmetry in $B \to K\mu^+\mu^-$ decays induced by scalar interactions, although small, can be a useful laboratory for studying possible extensions of the Standard Model once the required experimental sensitivity is gained. Finally, an appreciable forward-backward asymmetry can show up in $\overline{B} \to K\tau^+\tau^-$ decays, as a result of the overall factor $m_\tau$ in (2.116) and the lower rate of the $\tau^+\tau^-$ channel, together with $C_{S,P} \propto m_\tau$. Within the framework of the constrained MSSM, the average forward-backward asymmetry can be of $\mathcal{O}(10\%)$ [253].

[11] If we also allow for deviations of $A_{9,10}$ from their Standard Model values, the magnitude of $< A_{\rm FB} >$ is $\pm 4\%$ at most [248].

## 2.19 Experimental Prospects for $b \to s\nu\overline{\nu}$ and $B \to K\nu\overline{\nu}$

>– S. H. Robertson –<

Searches for rare decays in which the final state contains multiple neutrinos or other unobservable particles pose a significant experimental challenge. These decay modes lack significant kinematic contraints which can be used to suppress background processes and which can be used to uniquely identify the signal decay mode. The flavor-changing neutral current processes $b \to s\nu\overline{\nu}$ and $b \to d\nu\overline{\nu}$ are of considerable theoretical interest, see Section 2.20, because the Standard Model rates can be computed with small theoretical uncertainties, and because non-Standard Model contributions can lead to significant enhancements in the predicted rates. Currently, an experimental limit of $\mathcal{B}(b \to s\nu\overline{\nu}) < 6.4 \times 10^{-4}$ exists for the inclusive $b \to s\nu\overline{\nu}$ rate from ALEPH [26], and limits on the exclusive $B^- \to K^-\nu\overline{\nu}$ process are available from CLEO [254] and *BABAR* [28]. There are no published limits on either the inclusive or exclusive $b \to d\nu\overline{\nu}$ decay rates.

It is extremely difficult to perform an inclusive search for $b \to s\nu\overline{\nu}$ or $b \to d\nu\overline{\nu}$ in the experimental environment of a $B$ Factory. However, searches for specific exclusive decay modes, particularly in the case of $b \to s\nu\overline{\nu}$, may prove tractable. In an $\Upsilon(4S)$ environment, the experimental problem lies in distinguishing the observable decay daughter particles of the desired signal mode from other tracks and clusters in the event, and in deducing the presence of the two unobserved neutrinos in the final states.

### 2.19.1 $B^- \to K^-\nu\overline{\nu}$

The method adopted by *BABAR* for the $B^- \to K^-\nu\overline{\nu}$ search has been to attempt to exclusively reconstruct either of the two decaying $B^\pm$ mesons produced in the $\Upsilon(4S)$ decay into one of many known hadronic ($B^- \to D^{(*)0}X^-_{\rm had}$) or semileptonic ($B^- \to D^{(*)0}\ell^-\overline{\nu}$) final states, and then to search for evidence of a $B^- \to K^-\nu\overline{\nu}$ decay in the remaining particles in the event after the decay daughters of the reconstructed "tag $B$" have been removed. Details of the tag $B$ reconstruction process are described elsewhere in this Proceedings (see Section 4.2.1). If the tag $B$ reconstruction is successful, it is then expected that all tracks and clusters that were not identified as decay daughters of the tag $B$ are associated with decay daughters of the other $B$ meson in the event. Once the tag $B$ has been reconstructed, there are relatively few kinematic constraints which can be exploited in order to identify $B^- \to K^-\nu\overline{\nu}$ candidates. The signal selection is therefore essentially topological: after "removing" tracks and clusters associated with the tag $B$ reconstruction from the event, signal candidate events are required to have exactly one remaining track with charge opposite to that of the tag $B$ and satisfying kaon PID criteria. The number of remaining tracks in events with a reconstructed hadronic $B$ decay is plotted in Fig. 2-24.

Similarily, signal events are expected to contain no additional (observable) neutral particles, hence no additional energy deposition in the EMC is expected. In practice, signal events are typically found to contain one or more low energy neutral clusters, usually attributable to beam related backgrounds, detector "noise", or bremsstrahlung and/or hadronic split-offs from either the signal kaon or from daughter particles associated with the tag $B$. For the semileptonic tag $B$ sample, additional photons may also result from higher-mass open charm states (*e.g.*, $D^{*0}$) which have been incorrectly reconstructed as $B^- \to D^0\ell^-\overline{\nu}$ events. For the hadronic tag $B$ sample and the current *BABAR* detector performance, requiring that the total energy of all additional EMC clusters be less than $250$ MeV results in a loss of approximately $25\%$ of signal efficiency. Additional background rejection can be obtained by cutting on the CM frame momentum of the $K^\pm$ candidate track, which peaks towards high momenta for signal events but which has a relatively soft spectrum for background events (see Fig. 2-25). This, however, introduces a small uncertainty into the signal efficiency, due to the $B$ decay form factor model-dependence, and also potentially complicates the interpretation of limits on non Standard Model physics. The current *BABAR* analysis requires $p_K > 1.25$ GeV$/c$ in the CM frame.

The $B^- \to K^-\nu\overline{\nu}$ signal selection efficiency after applying all tag $B$ selection cuts is approximately $28\%$. Due to the large $B^- \to D^{(*)0}\ell^-\overline{\nu}$ branching fractions, the semileptonic tag sample has a factor 2 to 3 higher yield than the hadronic tag sample, however it also yields higher backgrounds and a lower efficiency for the $B^- \to K^-\nu\overline{\nu}$ signal mode once the tag $B$ reconstruction efficiency is factored out. The semileptonic tag analysis has achieved an overall efficiency of approximately $11 \times 10^{-4}$, while the hadronic tag analysis has achieved approximately $5 \times 10^{-4}$. If the existing analysis is scaled to higher luminosity, in a sample of $1\,\text{ab}^{-1}$ of data we would therefore expect $\sim 7$ signal events, assuming a Standard Model branching fraction of $\mathcal{B}(B^- \to K^-\nu\overline{\nu}) = 4 \times 10^{-6}$, with an expected background of $\mathcal{O}(100)$ events. However, significantly lower backgrounds can be obtained by increasing the $p^*_K$ (and other) selection cuts at the cost of some signal efficiency.

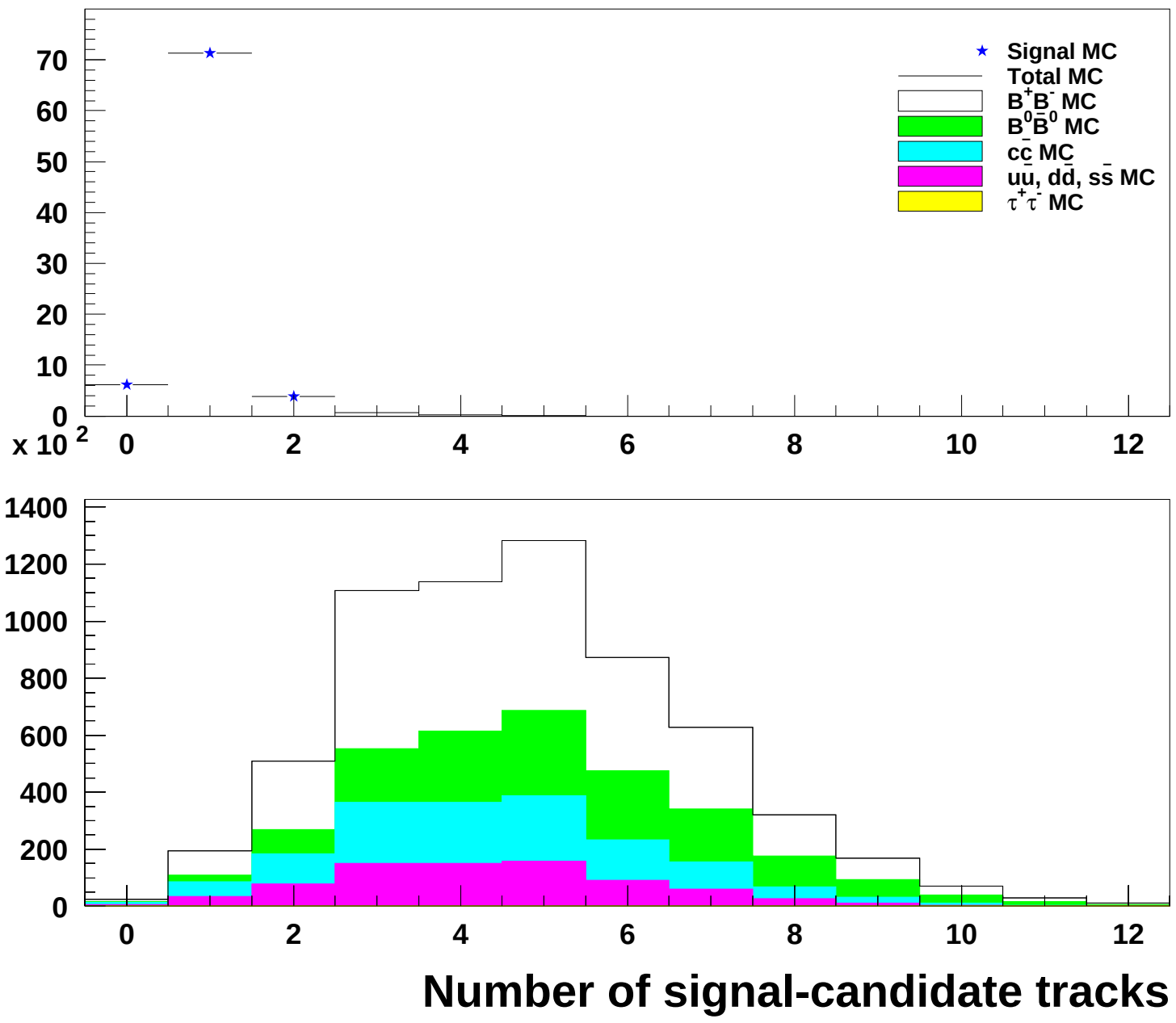

**Figure 2-24.** *The number of reconstructed charged tracks, after removing tracks associated with the reconstructed tag $B$, is plotted for $B^- \to K^-\nu\overline{\nu}$ signal MC (top) and for generic $B\overline{B}$ and continuum background MC (bottom). Both plots are normalized to $100$ fb$^{-1}$ and the signal MC assumes a branching fraction of $\mathcal{B}(B^- \to K^-\nu\overline{\nu}) = 4 \times 10^{-6}$.*

## 2.19.2 Exclusive $b \to d\nu\overline{\nu}$ and higher mass $b \to s\nu\overline{\nu}$ exclusive final states

In the case of $B^- \to \pi^-\nu\overline{\nu}$, the selection is identical, apart from the PID criteria applied to the signal candidate track. Removing the kaon PID requirement results in an increase in backgrounds of approximately an order-of-magnitude relative to the $B^- \to K^-\nu\overline{\nu}$ analysis, but with a signal efficiency that is similar to the $B^- \to K^-\nu\overline{\nu}$ analysis.

Current *BABAR* analyses have sought only to produce limits on the charged $B$ decay modes $B^- \to K^-\nu\overline{\nu}$ and $B^- \to \pi^-\nu\overline{\nu}$. Initial studies of the reconstruction efficiency and background levels associated with $B^- \to K^{*-}\nu\overline{\nu}$ (with $K^{*-} \to K^-\pi^0$), and, to a lesser extent, with $B^- \to \rho^-\nu\overline{\nu}$, have been performed. In both instances, permitting an additional $\pi^0$ in the event results in an intrinsically higher level of background and a reduced signal efficiency due to the $\pi^0 \to \gamma\gamma$ reconstruction efficiency ($\sim 2/3$). In the case of $B^- \to K^{*-}\nu\overline{\nu}$, a further efficiency reduction results from the $K^{*-} \to K^-\pi^0$ branching fraction, however, the background is also substantially reduced by the requirement that the $K^-\pi^0$ combination be consistent with a $K^*(892)$ invariant mass (see Fig. 2-26).

A similar procedure can be used for the neutral $B$ decay modes such as $B^0 \to K^0_S\nu\overline{\nu}$ and $B^0 \to \pi^0\nu\overline{\nu}$ although it is expected that a somewhat lower overall efficiency will be obtained, due to the lower branching fractions of the reconstructed tag $B$ modes and due to the lower intrinsic reconstruction efficiency for the neutral decay daughters of the signal decay.

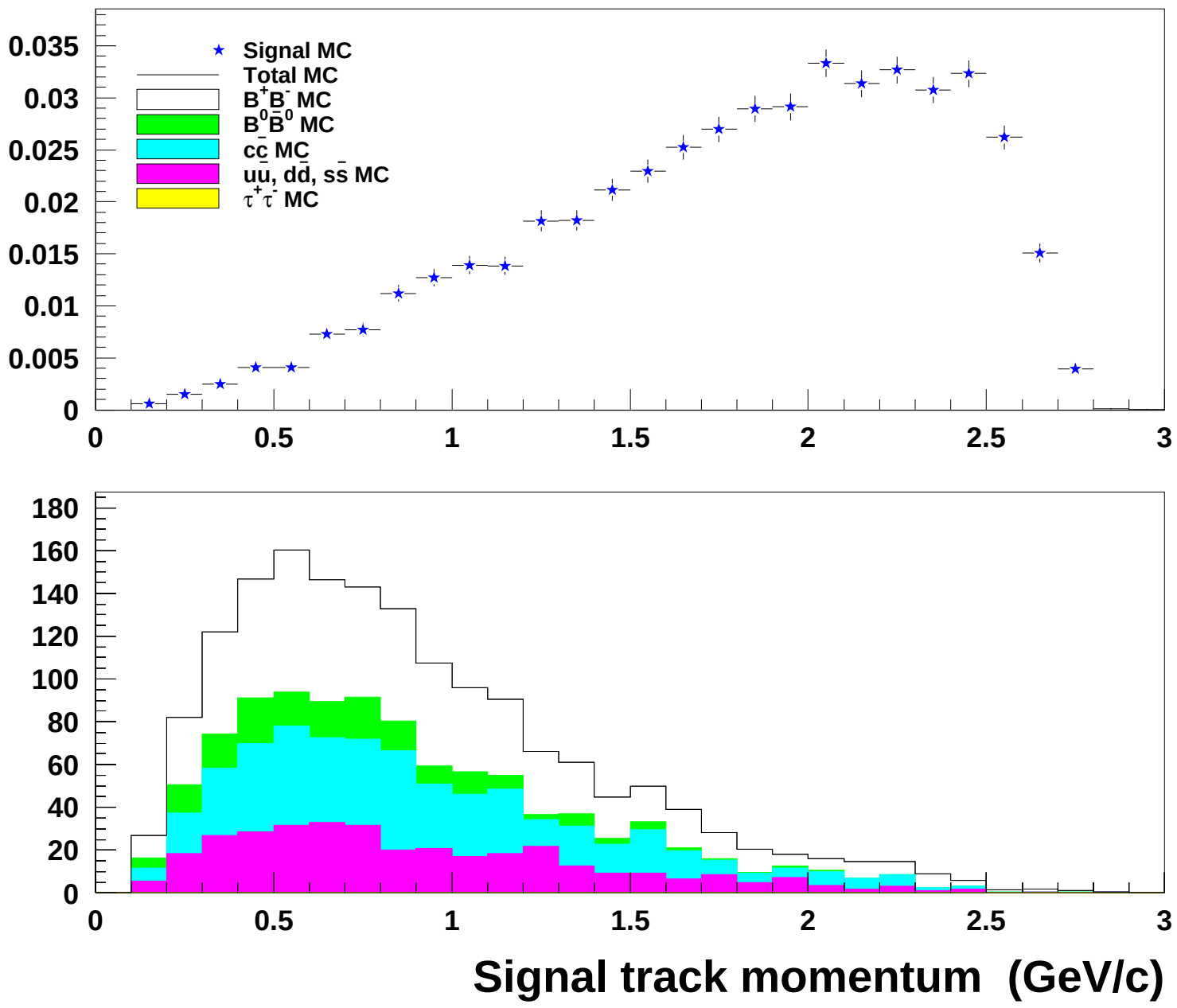

**Figure 2-25.** *The signal kaon momentum in the center of mass frame is plotted for $B^- \to K^-\nu\overline{\nu}$ signal MC (top) and for generic $B\overline{B}$ and continuum background MC (bottom). Both plots are normalized to $100$ fb$^{-1}$ and the signal MC assumes a branching fraction of $\mathcal{B}(B^- \to K^-\nu\overline{\nu}) = 4 \times 10^{-6}$.*

Although it is not possible to perform a fully inclusive search for $B \to X_s\nu\overline{\nu}$ using this method, it is conceivable that specific additional exclusive decay modes could be added in the future. It is not clear, however, at this point whether this would result in any significant improvement in experimental sensitivity.

## 2.19.3 Experimental considerations

Backgrounds can arise from three sources:

- Misreconstruction of the tag $B$ due to combinatorics or mismeasured tracks or clusters, leading to backgrounds from both charged and neutral $B$ decays, and from continuum sources. This background source depends both on detector performance and the specific tag $B$ mode that is being reconstructed. If sufficient data statistics are available, this background can be significantly reduced by reconstructing only tag $B$ modes which are known to have high purity, such as $B^- \to D^0\pi^-$ in which the $D^0$ is reconstructed in only the $D^0 \to K^-\pi^+$ mode, and/or by applying tight requirements on event shape variables and $B$ reconstruction quantities ($B$ vertex, $D$ mass, $m_{ES}$, *etc.*).

- Events in which the tag $B$ has been correctly reconstructed, but for which the additional tracks and clusters in the event resemble the low-multiplicity and missing energy signature of the signal mode due to particles that have passed outside of the detector acceptance or have otherwise failed to be identified in the detector, or due to the presence of additional spurious tracks and clusters resulting from detector noise, beam backgrounds, hadronic shower reconstruction, bremsstrahlung, *etc.*, which degrade the resolution of the missing energy and multiplicity selection variables. These background sources are dictated purely by the performance of the detector and event reconstruction software.

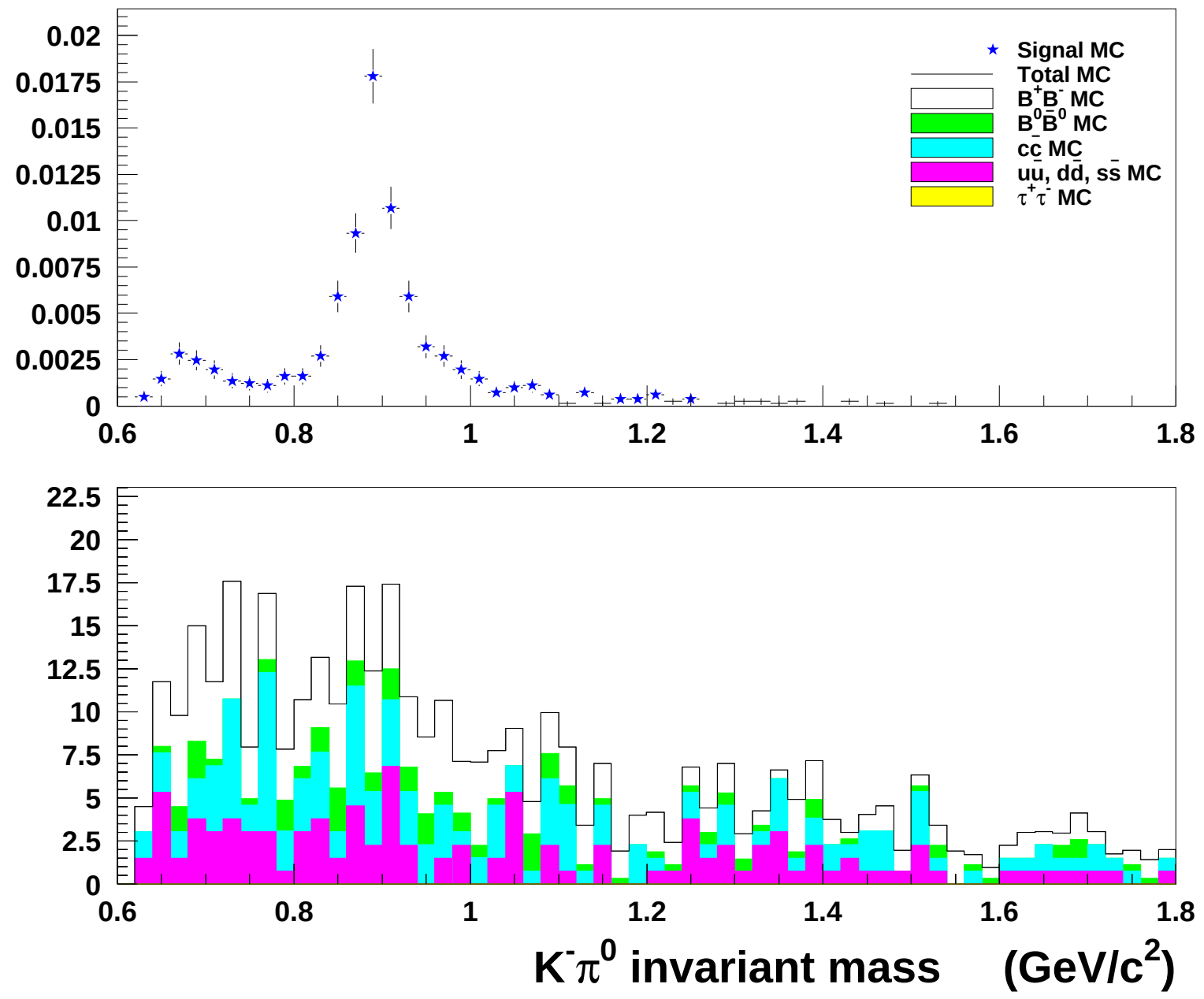

**Figure 2-26.** *The $K^-\pi^0$ invariant mass is plotted for $B^- \to K^{*-}\nu\overline{\nu}$ signal MC (top) and for background MC (bottom) after requiring that there is exactly one charged track identified as a kaon and exactly one $\pi^0$ candidate remaining after hadronic tag $B$ reconstruction. The lower plot is normalized to $100$ fb$^{-1}$, while the signal MC normalization is arbitrary.*

- "Irreducible" physics backgrounds from $B$ decays in which the tag $B$ has been correctly reconstructed, and for which the accompanying $B$ decay is topologically and kinematically indistinguishable from a signal decay. For $B^- \to K^{(*)-}\nu\overline{\nu}$ searches these backgrounds are currently negligible, and are not expected to contribute at a rate above the Standard Model prediction for exclusive $b \to s\nu\overline{\nu}$ modes. For $b \to d\nu\overline{\nu}$ searches, it is expected that irreducible physics backgrounds will arise from the Standard Model $B^- \to \tau^-\overline{\nu}$ process, however, it is likely that by the time this becomes an issue, either $B^- \to \tau^-\overline{\nu}$ will be well-measured, or both searches will be limited by detector performance issues.

In practice, real background events are usually the result events that have intrinsically low particle multiplicity and significant missing energy due to the presence of neutrinos and/or undetected $K^0_L$, which then lose additional particles that either pass outside of the detector acceptance, or somehow otherwise fail to be reconstructed. These events are usually topologically similar to signal events, but are indistinguishable from signal events only because of the imperfect performance of the *BABAR* detector. The ability to perform searches for modes such as $b \to s\nu\overline{\nu}$ and $b \to d\nu\overline{\nu}$ requires:

- the ability to exclusively reconstruct large samples of tag $B$ mesons with reasonable purity in hadronic and semileptonic final states;
- relatively good hermiticity of the detector, in order to minimize backgrounds resulting from events with particles that pass outside of the fiducial acceptance faking the missing energy signature of the signal mode;
- a relatively low rate of spurious charged tracks and/or calorimeter clusters which result from sources other than the direct decay of the signal candidate event. Sources of such spurious particles include reconstruction artifacts (*e.g.,* ghost tracks, looping tracks, unmatched SVT and DCH tracks, misassociation of EMC clusters with tracks), detector noise artifacts (*e.g.*, EMC noise clusters), "physics" artifacts (*e.g.*, bremsstrahlung clusters and photon conversions, hadronic split-off clusters) and particles not associated with the physics event (*e.g.*, beam-background-related tracks and clusters, cosmics).

- the ability to veto events containing $K_L^0$ mesons using the IFR. Events with one or more undetected $K_L^0$ mesons are currently a dominant source of background in these analyses.

### 2.19.4 $B^0 \to \tau^+\tau^-$

There are currently no experimental limits for $B^0 \to \tau^+\tau^-$, in spite of the fact that its sensitivity to third-generation couplings makes it the most theoretically interesting of the $B_d^0 \to \ell^+\ell^-$ modes [12]. This absence of experimental measurements is due to the lack of kinematic constraints that can be used to uniquely identify decays of $\tau$ leptons in a $B$ Factory environment. In contrast, the lepton flavor-violating modes $B^0 \to \tau^+e^-$ and $B^0 \to \tau^+\mu^-$ are comparatively straightforward experimentally, due to the presence of a mono-energetic high-momentum final state lepton.

Two experimental issues must be addressed for an analysis in a $B$ Factory environment. The first is to distinguish the (charged and neutral) signal $\tau$ decay daughters from other particles produced in the event. This can be accomplished by exploiting a tag-$B$ reconstruction method similar to that which is used in the $b \to s/d\nu\overline{\nu}$ searches described previously. Provided that the reconstruction process is sufficiently clean, the dominant backgrounds will be from $\Upsilon(4S) \to B^0\overline{B}^0$ decays in which one $B$ has been correctly reconstructed and the second $B$ decays in a manner which resembles a $B^0 \to \tau^+\tau^-$ "signature", characterized by low multiplicity and significant missing energy. Other sources of background are assumed here to be neglegible. $\tau^\pm$ decays to $\ell^\pm\nu\overline{\nu}$, $\pi^\pm\nu$, $\pi^\pm\pi^0\nu$ (via $\rho^\pm\nu$) and $3\pi^\pm\nu$ (via $a_1^\pm\nu$), totaling $\sim 80\%$ of the total $\tau$ branching fraction, are potentially usable for this analysis, however only the "one-prong" ($e, \mu, \pi, \rho$) have been considered to date, leading to a topological efficiency for $B^0 \to \tau^+\tau^-$ of $\sim 51\%$.

Since all final states contain a minimum of two (and as many as four) neutrinos, there are very few kinematic constraints. The event selection therefore proceeds by requiring exactly two charged tracks remaining after the tag $B$ reconstruction then requiring that there be little or no additional calorimeter activity, other than possibly a reconstructed $\pi^0$ attributable to a $\rho$ decay. PID criteria are used to categorize events by $\tau^+\tau^-$ decay mode. Backgrounds can result from events in which two or more charged particles pass outside the detector acceptance, or in which neutral particles fail to be reconstructed. In addition, physics backgrounds can arise from cascade $b \to c \to s$ transitions, which can result in topologies consisting of, for example, a $K_L^0$, two charged leptons and two neutrinos. If the $K_L^0$ is undetected, then this process will mimic a $B^0 \to \tau^+\tau^-$ decay. "Irreducible" physics backgrounds are observed in processes with branching ratios at the level of $\mathcal{O}(10^{-4})$. Some additional background rejection can be obtained by exploiting correlations in the angular and energy distributions of particles observed in $b \to c \to s$ using multvariate analysis techniques, however none of the possible $B^0 \to \tau^+\tau^-$ final states has proven to yield significantly lower backgrounds. Consequently, obtaining a branching fraction limit for $B^0 \to \tau^+\tau^-$ which is substantially below the level of $\mathcal{O}(10^{-4})$ will be extremely challenging. As is the case for $b \to s\nu\overline{\nu}$ searches, excellent detector hermeticity and the ability to veto $K_L^0$ mesons are essential for this analysis.

[12] There is an indirect bound from LEP data on $B \to \tau\nu$ decays $\mathcal{B}(B_d^0 \to \tau^+\tau^-) < 1.5\%$ [255].

## 2.20 Theoretical Prospects for $b \to s\nu\overline{\nu}$ and $B \to K(\pi)\nu\overline{\nu}$ Decays

>– T. Hurth and D. Wyler –<

The decays $\overline{B} \to X_{s(d)}\nu\overline{\nu}$ (and the corresponding exclusive decays) are particulary clean rare decays. Thus, these flavor-changing neutral-current amplitudes are extremely sensitive to possible new degrees of freedom, even if these appear well above the electroweak scale.

Because the neutrinos do not interact strongly or electromagnetically, only the short-range weak interactions 'connect' the hadrons and leptons, and they can be totally accounted for by standard perturbation theory. The notoriously difficult strong interactions only affect the hadronic side; as we will see, this makes them completely controllable and it appears possible to make predictions at the few per cent level. Because the rates are quite small and detection at hadronic facilities is virtually impossible, these decays are of prime interest to a Super $B$ Factory.

The decay of the $B$ meson into strange particles and two neutrinos proceeds through an (effective) flavor-changing neutral interaction. It is induced by $Z$ penguins and box diagrams with two $W$-bosons. These diagrams (in contrast to the photonic penguin diagrams, effective in the decay into two charged leptons) are characterized by a 'hard' GIM-suppression proportional to $(m_q^2/m_W^2)$.

We may consider (on the hadronic side) exclusive or inclusive decays. The inclusive ones are believed to be accounted for by quark-hadron duality and therefore to be essentially free of nonperturbative effects. On the other hand, exclusive decays involve complicated final-state corrections. In the case $B \to K(\pi)\nu\overline{\nu}$ only the form factors of the hadronic current enters. We argue that it can be found from relations to other decays and therefore also the exclusive decays can be predicted with high precision.

Let us start with the inclusive decay mode $\overline{B} \to X_{s,d}\nu\overline{\nu}$. The effective Hamiltonian reads

$$\mathcal{H}_{\rm eff}(\overline{B} \to X_s\nu\overline{\nu}) = \frac{G_F}{\sqrt{2}}\frac{\alpha}{2\pi\sin^2\theta_W}\, V_{tb}V_{ts}^*\, X(m_t^2/m_W^2)\,(\overline{s}\gamma_\mu P_L b)(\overline{\nu}\gamma^\mu P_L\nu) + h.c. \tag{2.118}$$

For the decay $\overline{B} \to X_d\nu\overline{\nu}$ obvious changes have to be made. The short-distance Wilson coefficient $X$ is governed by the hard (quadratic) GIM mechanism, which leads to $X(m_c^2/m_W^2)/X(m_t^2/m_W^2) \approx \mathcal{O}(10^{-3})$. Moreover, the corresponding CKM factors in the top and the charm contribution are both of order $\lambda^2$. As a consequence, the dependence of the amplitude $\overline{B} \to X_s\nu\overline{\nu}$ on the up and charm quark masses is negligible[13].

The NLL QCD contributions to the partonic decay rate were presented in [13, 14]. The perturbative error, namely the one due to the renormalization scale, was reduced from $O(10\%)$ at the LL level to $O(1\%)$ at the NLL level. Moreover, the nonperturbative contributions scaling with $1/m_b^2$ are under control and small [256, 9, 257]. Because of the absence of the photon-penguin contribution, the nonperturbative contributions scaling with $1/m_c^2$ can be estimated to be at the level of $10^{-3}$ at most [25].

After normalizing to the semileptonic branching ratio and summing over the three neutrino flavors, the branching ratio of the decay $\overline{B} \to X_s\nu\overline{\nu}$ is given by [226]:

$$\mathcal{B}(\overline{B} \to X_s\nu\overline{\nu}) = \mathcal{B}_{\rm exp}(\overline{B} \to X_c\ell\nu)\,\frac{3\alpha^2}{4\,\pi^2\sin^4\theta_W}\frac{|V_{ts}|^2}{|V_{cb}|^2}\frac{X^2(m_t^2/m_W^2)\,\overline{\eta}}{f(m_c^2/m_b^2)\,\kappa(m_c^2/m_b^2)}. \tag{2.119}$$

Using the measured semileptonic branching ratio and the phase-space factor of the semileptonic decay $f$, the corresponding QCD correction $\kappa$, the QCD correction of the matrix element of the decay $\overline{B} \to X_s\nu\overline{\nu}$, namely $\overline{\eta} = \kappa(0)$, and scanning the input parameters, one ends up with the theoretical prediction [226]:

[13] We note that the notion that 'the amplitude is dominated by the top-quark contribution' is slightly imprecise because of the presence of GIM-canceling mass-independent contributions.

$$\mathcal{B}(\overline{B} \to X_s\nu\overline{\nu}) = (3.4 \pm 0.7) \times 10^{-5}. \tag{2.120}$$

The replacement of $V_{ts}$ by $V_{td}$ in (2.119) leads to the case of the decay $\overline{B} \to X_d\nu\overline{\nu}$. Obviously all uncertainties cancel out in the ratio of the two branching ratios of $\overline{B} \to X_d\nu\overline{\nu}$ and $\overline{B} \to X_s\nu\overline{\nu}$. Thus, it allows for a clean direct determination of the ratio of the two corresponding CKM matrix elements.

The inclusive decay $\overline{B} \to X_s\nu\overline{\nu}$ is the theoretically cleanest rare $B$ decay, but also the most difficult experimentally. However, as discussed above, the lack of an excess of events with large missing energy in a sample of $0.5 \times 10^6$ $b\overline{b}$ pairs at LEP already allowed ALEPH to establish an upper bound on the branching ratio [258, 259], which is still an order of magnitude above the Standard Model prediction, but nonetheless leads to constraints on New Physics models [259]. For this purpose, the QCD corrections to the decays $\overline{B} \to X_{s,d}\nu\overline{\nu}$ in supersymmetric theories (MSSM) have recently been presented [260].

Let us move to the exclusive channel $B \to \pi\nu\nu$, with an estimated branching ratio of $10^{-7}$. Similar to the methods used in the decay $K \to \pi\nu\nu$ (see [13, 14]), we can relate it by isospin to the charged-current decay $B \to \pi\ell\nu$. A precise measurement of the form factor in that decay then yields the branching ratio for $B \to \pi\nu\nu$. We obtain, after summing over the three neutrino flavors:

$$\mathcal{B}(B^+ \to \pi^+\nu\nu) = R_{\rm iso}\frac{3\alpha^2}{2\pi^2 sin^4\theta_W}\mathcal{B}(B^+ \to \pi^0\ell\nu))\frac{|V_{tb}^*V_{ts}|^2}{|V_{ub}|^2}\, X^2(m_t^2/m_W^2)\,, \tag{2.121}$$

where $X$ is again the short-distance Wilson coefficient and the factor $R_{\rm iso}$ accounts for isospin violations. A similar expression follows for the neutral meson decay $B^0 \to \pi^0\nu\overline{\nu}$. The factor $R_{\rm iso}$ was discussed in [261] for the decays of the kaon. There are at least three sources of isospin violation: Mass effects (which are very small in the present case), neutral form factor suppression of about $4\%$ through $\pi - \eta$ mixing, and a suppression of around $2\%$ due to the absence of a leading log correction. Barring further corrections and short of a detailed analysis, we conclude that $R_{\rm iso} \simeq 0.94$. Using the (measured) $\mathcal{B}(B^+ \to \pi^0\ell\nu)$ kinematical distributions, one can also determine the $s$ dependence, but in view of the limited numbers of events expected, this is a rather academic point.

The decay of real interest is $B \to K^{(*)}\nu\nu$, where $K^{(*)}$ can be either $K$. For definiteness, we discuss only $K$, but $K^*$ may be treated analogously. Starting from the Hamiltonian (2.118), the invariant mass spectrum of the decay $B \to K\nu\overline{\nu}$ can be written as follows (see for example [262, 263]):

$$\frac{d\Gamma(B \to K\nu\overline{\nu})}{ds} = \frac{G_F^2\alpha^2 m_B^5}{2^8\pi^5 sin^4\theta_W}\,|V_{ts}^*V_{tb}|^2\,\lambda^{3/2}(r_K, s)\,f_+^2(s)\,X^2 \tag{2.122}$$

$$\text{where}\quad \lambda \equiv [1 + r_K^2 + s^2 - 2s - 2r_K(1+s)] \qquad s = q^2/m_B^2. \quad r_K = m_K^2/m_B^2 \tag{2.123}$$

The form factor $f_+$ (the definiton follows) can now be calculated by QCD sum-rule techniques with an estimated $30\%$ accuracy (see [262, 9]). For example, based on results presented in [9], Ref. [27] predicts $B \to K\nu\overline{\nu}$

$$\mathcal{B}(B \to K\nu\overline{\nu}) = (3.8^{+1.2}_{-0.6}) \times 10^{-6}. \tag{2.124}$$

It is of course desirable to obtain the form factor more accurately. Unlike the case of the $\pi$ discussed above, it cannot be related to a charged-current decay by isospin, and we need to obtain the form factors for $B$ to $K^{(*)}$ by other arguments.

The first possibility we discuss is to relate the decay $B \to K^{(*)}\nu\nu$ to the rare decay $B \to K^{(*)}\ell\ell$. It has already been seen at *BABAR* [220] and Belle [264] with the following branching ratios

$$\mathcal{B}(B \to K^*\ell^+\ell^-) = \begin{cases} (0.88^{+0.33}_{-0.29} \pm 0.10) \times 10^{-6} & (\textit{BABAR}) \\ (11.5^{+2.6}_{-2.4} \pm 0.8 \pm 0.2) \times 10^{-7} & (\text{Belle}) \end{cases}$$

and

$$\mathcal{B}(B \to K\ell^+\ell^-) = \begin{cases} (0.65^{+0.14}_{-0.13} \pm 0.04) \times 10^{-6} & (\textit{BABAR}) \\ (4.8^{+1.0}_{-0.9} \pm 0.3 \pm 0.1) \times 10^{-7} & (\text{Belle}) . \end{cases}$$

and is expected to have been measured with high precision by the time a Super $B$ Factory exists. In particular, the dependence on the invariant mass-squared of the two leptons will be determined. It gives experimental information on the nonperturbative form factors that we want to use in the decay with the two neutrinos.

The effective Hamiltonian is a product of the Wilson coefficients characteristic of the model and the operators whose matrix elements (form factors) we need. In $B \to K$ transitions three independent form factors enter, which are defined by the following matrix elements[14] ($q = p_B - p_K$):

$$\langle K(p_K)|\bar{s}\gamma_\mu b|B(p_B)\rangle = f_+(q^2)\,(p_B + p_K)_\mu + f_-(q^2)\,(p_B - p_K)_\mu \tag{2.125}$$

$$\langle K(p_K)|\bar{s}\sigma_{\mu\nu} b|B(p_B)\rangle = i\,s(q^2)\,[(p_B + p_K)_\mu(p_B - p_K)_\nu - (p_B - p_K)_\mu(p_B + p_K)_\nu] \tag{2.126}$$

There is also the scalar form factor $l(q^2)$, which is connected to the vector form factors via the equation of motion of the quark fields:

$$\langle K(p_K)|\bar{s}b|B(p_B)\rangle = l(q^2) \equiv \frac{1}{m_b - m_s}[f_+(q^2)\,(m_B^2 - m_K^2) + f_-(q^2)\,q^2] \tag{2.127}$$

In the decay $B \to K\ell\ell$, all three form factors contribute; the tensor form factor enters through photon exchange between the quarks and the (charged) leptons. In addition photon exchange also gives rise to 'long-distance' contributions from four-quark operators, which include a nonperturbative part. In contrast, the decay $B \to K\nu\overline{\nu}$ involves the vector form factor $f_+$ only. In order to use the data from $B \to K\ell\ell$ to determine the form factors of $B \to K\nu\overline{\nu}$ one must 'subtract' the extra effects.

If we neglect the masses of the leptons *i.e.,* terms proportional to $m_\ell^2/m_B^2$, the contribution of the form factor $f_-$, which is proportional to $q^\mu$, does not contribute to the invariant mass spectrum $d\Gamma/dq^2$ of the decay $B \to K\ell^+\ell^-$. The latter can then be written in this approximation in terms of the two form factors $f_+$ and $s$ only (see for example [129]):

$$\frac{d\Gamma(B \to K\ell^+\ell^-)}{ds} = \frac{G_F^2 \alpha^2 m_B^5}{2^{10}\pi^5}\,|V_{tb}V_{ts}^*|^2\, 2/3\,\lambda^{3/2}\,(|A'|^2 + |C'|^2), \tag{2.128}$$

$$\text{with } A' = \widetilde{C}_9^{\text{eff}}\, f_+(q^2) + 2m_b\,\widetilde{C}_7^{\text{eff}}\, s(q^2) \quad\text{and}\quad C' = \widetilde{C}_{10}^{\text{eff}}\, f_+(q^2). \tag{2.129}$$

The effective Wilson coefficients, $\tilde{C}_i^{\text{eff}}(i = 7, 9, 10)$, in the decay $B \to K\ell\ell$ are specific to the Standard Model and are basically known. They mainly represent short-distance physics; however, $\widetilde{C}_i^{\text{eff}}$ also receive contributions of the matrix elements of the operators including their nonperturbative parts, such as the effect of the $c\bar{c}$ resonances:

$$\widetilde{C}_i^{\text{eff}}(\hat{s}) = \widetilde{C}_i + \widetilde{C}_{\text{i,matrix}}^{\text{pert}}(\hat{s}) + \widetilde{C}_{\text{i,matrix}}^{\text{nonpert}}(\hat{s}) \tag{2.130}$$

The explicit formulae can be found for example in [129] as Eqs. (4.1)–(4.3) with the bremsstrahlung functions ($\omega_{7,9,79}$) set to zero. The first two (perturbative) contributions to the effective Wilson coefficients were calculated to NNLL QCD precision in [52, 84, 86]. These perturbative contributions are also relevant to the inclusive mode $\overline{B} \to X_s\ell^+\ell^-$. For the third, nonperturbative, contribution, phenomenological parametrizations were proposed in the literature. In [11], the long-distance contributions due to the $c\bar{c}$ intermediate states were included via a Breit-Wigner ansatz, while in [72] these corrections are estimated by means of experimental data on $\sigma(e^+e^- \to c\bar{c}\,hadrons)$ using a dispersion relation.

[14] A different parametrization of the three form factors are also often used and presented in the subsection on exclusive semileptonic rare decays. The relations between the two parametrizations are given by $f_T(q^2) = (M_B + M_K)s(q^2)$ and $f_0(q^2) = f_-(q^2)[q^2/(M_B^2 - M_K^2)] + f_+(q^2)$.

Our suggestion is to make use of relations that follow directly from the heavy quark symmetry of QCD, thereby obtaining more reliable predictions for $f_+$, and use more fundamental methods to calculate the long-distance effects. Both are expected to be most accurate near maximal $q^2 = (M_B - M_K)^2$ (zero recoil). The heavy quark methods apply best in this region, which is far away from any resonance and the photonic pole which make the long-distance effects so untractable[15].

The well-known Isgur-Wise relation [265, 266],

$$2M_B s(q^2) = f_+(q^2) - f_-(q^2) + \mathcal{O}(m_b^{-1/2}), \tag{2.131}$$

connects the tensor and the vector form factors and is most useful at low recoil. Moreover, the scaling of the form factors is given by

$$f_+ + f_- \sim m_b^{-1/2}\,, \quad f_+ - f_- \sim m_b^{1/2}\,, \quad l \sim m_b^{1/2}\,. \tag{2.132}$$

It follows from these relations that the two contributions to the decay rate of $B \to K\ell\ell$, which correspond to the vector and the tensor form factors $f_+$ and $s$, enter at the same order of $m_b$.[16] Thus, a direct relation between $s$ and $f_+$ is needed. When inserted into Eq. (2.129), the differential rate depends on $f_+$ only, which can then be determined and used to predict the distribution of $B \to K\nu\overline{\nu}$ with high accuracy.

A useful result is this respect was given by Grinstein and Pirjol [267]. They present a relation between $f_-$ and $s$ including the subleading $1/m_b$ corrections. Using chiral perturbation combined with heavy hadrons [268, 269, 270], the subleading form factors could be estimated. This relation, in principle, allows extraction of $s$ with an estimated accuracy of 10%, once $f_-$ is known. In the present case we need a corresponding relation between $s$ and $f_+$. While writing the present report, a new paper by Grinstein and Pirjol appeared [271] where alternative methods for deriving the form factor relations and a clever treatment of the long-distance contributions are presented.

Using the approach of [271], we are able to derive improved form factor relations for the case $B \to K$. Let us sketch the derivation. For more details and numerical estimates we refer the reader to [272]. We start from the QCD operator identity

$$i\partial^\nu(\overline{s}i\sigma_{\mu\nu}b) = (m_b + m_s)\overline{s}\gamma_\mu b - 2\overline{q}i\overleftarrow{D}_\mu b + i\partial_\mu(\overline{s}b)\,. \tag{2.133}$$

From this, we find the following relations between the $B \to K$ form factors:

$$q^2\, s(q^2) \;=\; (m_b + m_s)\, f_+(q^2) \;+\; 2\, d_+ \tag{2.134}$$

$$(M_B^2 - M_k^2)\, s(q^2) \;=\; -(m_b + m_s)\, f_-(q^2) \;-\, 2\, d_- \;+\; l(q^2)\,. \tag{2.135}$$

While the form factor $l$ was defined previously in Eq. (2.127), two new ones, $d_+$ and $d_-$, enter; they are defined by the matrix elements of the operator $si\overleftarrow{D}_\mu b$:

$$\langle K(p_K)|si\overleftarrow{D}_\mu b|B(p_B)\rangle = d_+(q^2)(p_B + p_K)_\mu + d_-(q^2)(p_B - p_K)_\mu\,. \tag{2.136}$$

These two form factor relations are exact. In leading order in $1/m_b$, they combine to reproduce the Isgur-Wise relation (2.131) (see below). In the next step, the new QCD operator $\overline{s}i\overleftarrow{D}_\mu b$ is matched on the heavy quark effective theory operators in order to isolate the leading terms in $m_b$. The corresponding relation is [271]

$$\overline{s}i\overleftarrow{D}_\mu b = D_0^{(v)}(\mu)m_b\overline{s}\gamma_\mu h_v + D_1^{(v)}(\mu)m_b\overline{s}v_\mu h_v + \overline{s}i\overleftarrow{D}_\mu h_v + \cdots \tag{2.137}$$

The Wilson coefficients $D_i^{(v)}(\mu)$ begin at $O(\alpha_s)$. Taking the matrix element of this relation between the $B$ and the $K$ mesons, and using analogous matching relations between the currents,

$$\overline{s}\gamma_\mu b \;=\; C_0^{(v)}(\mu)\overline{s}\gamma_\mu h_v + C_1^{(v)}(\mu)\overline{s}v_\mu h_v + \frac{1}{2m_b}\overline{s}\gamma_\mu i\not{D} h_v + \cdots\,, \tag{2.138}$$

$$\overline{s}b \;=\; E_0^{(v)}(\mu)\overline{s}h_v + \cdots\,, \tag{2.139}$$

[15] We thank Y. Grossman and G. Isidori for a collaboration on this point.

[16] We note that for $q^2 \approx q^2_{\rm max} = (M_B - M_K)^2$ the contribution of the tensor form factor is small, due to the small Wilson coefficient $\widetilde{C}_7^{\rm eff}$ in this kinematic region. Thus, for low recoil, the form factor $f_+$ can in principle be determined from the measured $q^2$-distribution of $B \to K\ell^+\ell^-$ with restricted accuracy. Nevertheless, the uncertainties due to the neglected tensor contribution and due to the unknown $\Lambda_{\rm QCD}/M_B$ corrections in the Isgur-Wise relation might be relatively large.

one arrives at the desired relation

$$q^2\, s(q^2) \;=\; m_b(1 + 2D_0^{(v)})\, f_+(q^2)\,(1\; + O(1/m_b)\,)\,. \tag{2.140}$$

Using (2.137), (2.138), and (2.139), the $1/m_b$ terms can be explicitly expressed through additional matrix elements of local operators such as $\overline{s} i \overleftarrow{D}_\mu h_v$. Thus, this formula separates the leading (in $m_b$) and next-to-leading terms, which are suppressed by $1/m_b$. It should be possible to calculate them with existing hadronic methods, such as lattice QCD to a $20\%$ accuracy, which together with the $\Lambda_{\rm QCD}/m_B$ suppression would result in about $4\%$ precision for the form factors near the endpoint. Similiarly, the proposed strategy may also be used to determine the form factors of the decay $B \to K^* \nu\overline{\nu}$. The derivation of improved Isgur-Wise relations is analogous [271].

Another method to predict the neutrino modes is to relate the decay $B \to K^{(*)}\nu\overline{\nu}$ to $B \to \pi(\rho)\ell\nu$ by SU(3). However, the SU(3) breaking effects are large, for example, using chiral perturbation theory; they were estimated to be as large as $40\%$ [256]. Better is the 'double' ratio method, which is based on heavy quark symmetry [273]. The idea is to compare the $B$ and $D$ decay form factors for $K$ and $\rho$. The result is

$$f(B \to K) \simeq f(B \to \rho) |\frac{f(D \to K)}{f(D \to K)}|^2 (\frac{(m_B - m_K)}{(m_B - m_\rho)})^2\,. \tag{2.141}$$

The corrections are proportional to $(m_s/\Lambda) \times (\Lambda/m_c) = m_s/m_c$ and are generically around $10\%$. A more detailed discussion of the two methods for accurate predictions of the golden modes $\overline{B} \to K^{(*)}\nu\overline{\nu}$ will be presented in [272].

## 2.21 Purely Leptonic Decays of Neutral $B$ Mesons

>– F. Krüger –<

### 2.21.1 Theoretical framework

The part of the effective Hamiltonian describing the $b \to q\ell^+ l^-$ ($q = s, d$) transition relevant for $\overline{B}{}^0_q \to \ell^+ l^-$ ($l = e, \mu, \tau$) decays reads as, *e.g.*, [233, 274]

$$\mathcal{H}_{\rm eff} = -\frac{4G_F}{\sqrt{2}} V_{tb}V_{tq}^* \Big\{ C_{10}O_{10} + C_S O_S + C_P O_P + C'_{10}O'_{10} + C'_S O'_S + C'_P O'_P \Big\}, \tag{2.142}$$

where

$$O_{10}^{(\prime)} = \frac{e^2}{g_s^2}(\bar{q}\gamma_\mu P_{L(R)} b)(\bar{\ell}\gamma^\mu\gamma_5\ell), \quad O_S^{(\prime)} = \frac{e^2}{16\pi^2}(\bar{q}P_{R(L)}b)(\bar{\ell}\ell), \quad O_P^{(\prime)} = \frac{e^2}{16\pi^2}(\bar{q}P_{R(L)}b)(\bar{\ell}\gamma_5\ell), \tag{2.143}$$

and $P_{L,R} = (1 \mp \gamma_5)/2$. The hadronic matrix elements of the operators $O_i$ are characterized by the decay constant of the pseudoscalar meson [275, 255, 276]

$$\langle 0|\bar{q}\gamma_\mu\gamma_5 b|\overline{B}_q(p)\rangle = ip_\mu f_{B_q}. \tag{2.144}$$

The numerical value of $f_{B_q}$ can be determined *e.g.*, from lattice QCD computations [252]

$$f_{B_s} = (217 \pm 12 \pm 11)\ \mathrm{MeV}, \quad f_{B_s}/f_{B_d} = 1.21 \pm 0.05 \pm 0.01. \tag{2.145}$$

Similar results are obtained from QCD sum rules [277]. Contracting both sides in (2.144) with $p^\mu$ and employing the equation of motion for $q$ and $b$ quarks results in

$$\langle 0|\bar{q}\gamma_5 b|\overline{B}_q(p)\rangle = -if_{B_q}\frac{M_{B_q}^2}{m_b + m_q}. \tag{2.146}$$

The matrix element in (2.144) vanishes when contracted with the leptonic vector current $\bar{\ell}\gamma_\mu l$ as it is proportional to $p^\mu = p^\mu_{\ell^+} + p^\mu_{\ell^-}$, which is the only vector that can be constructed. Furthermore, the tensor-type matrix element $\langle 0|\bar{q}\sigma_{\mu\nu}b|\overline{B}_q(p)\rangle$ must vanish, since it is not possible to construct a combination made up of $p^\mu$ that is antisymmetric with respect to the index interchange $\mu \leftrightarrow \nu$. Therefore, operators such as $(\bar{q}\gamma_\mu P_{L(R)}b)(\bar{\ell}\gamma^\mu\ell)$ and $(\bar{q}\sigma_{\mu\nu}P_{L(R)}b)(\bar{\ell}\sigma^{\mu\nu}P_{L(R)}\ell)$ do not contribute to the decay $\overline{B}{}^0_q \to \ell^+\ell^-$.

Summing over the lepton spins, the branching ratio has the form

$$\mathcal{B}(\overline{B}{}^0_q \to \ell^+\ell^-) = \frac{G_F^2\alpha_{\rm em}^2 M_{B_q}^5 \tau_{B_q} f_{B_q}^2}{64\pi^3}|V_{tb}V_{tq}^*|^2\beta_q\left\{\beta_q^2\left|\frac{C_S - C'_S}{m_b + m_q}\right|^2 + \left|\frac{C_P - C'_P}{m_b + m_q} + \frac{2m_\ell}{M_{B_q}^2}(A_{10} - A'_{10})\right|^2\right\}, \tag{2.147}$$

where $\tau_{B_q}$ is the $B_q$-lifetime and $\beta_q = (1 - 4m_l^2/M_{B_q})^{1/2}$. Further, $C_{S,P}^{(\prime)} \equiv C_{S,P}^{(\prime)}(\mu)$, $m_{b,q} \equiv m_{b,q}(\mu)$ and $A_{10}^{(\prime)} = 4\pi/\alpha_s(\mu)C_{10}^{(\prime)}(\mu)$, where $\mu$ is the renormalization scale.

### 2.21.2 Standard Model predictions for $B \to \ell^+\ell^-$

Within the Standard Model, the neutral Higgs boson contributions to $C_{S,P}^{(\prime)}$ are suppressed by $m_\ell m_{b,(q)}/M_W^2$, and hence are tiny. Since the neutral Higgs does not contribute to $A_{10}^{(\prime)}$, the dominant contributions to the decay $\overline{B}{}^0_q \to \ell^+\ell^-$ arise from $Z^0$-penguin diagrams and box diagrams involving $W^\pm$-bosons [278, 13, 14]. Using the NLO expression for $A_{10}$ from [278, 13, 14], we obtain $A_{10}^{\rm SM} = -4.213$ and $A'^{\rm SM}_{10} = 0$.

From (2.147) it follows that the Standard Model branching ratios scale like $\sim m_\ell^2$ due to helicity suppression. Consequently, the branching ratios for $\ell = e$ and $\mu$ are small. Furthermore, they suffer from theoretical uncertainties of 30%–50% [274, 248, 279, 140, 280], mainly due to the uncertainty on the $B$ meson decay constant [cf. (2.145)]. However, these uncertainties on the Standard Model branching ratios can be considerably reduced by exploiting the relation between the $B_q^0$–$\overline{B}_q^0$ mass difference and $\mathcal{B}(\overline{B}_q^0 \to \mu^+\mu^-)$ [281]. Taking $\Delta M_{B_d}$ from [2] and assuming $\Delta M_{B_s} = (18.0 \pm 0.5)\ \text{ps}^{-1}$, the Standard Model predictions are [281]

$$\mathcal{B}(\overline{B}_s^0 \to \mu^+\mu^-) = (3.4 \pm 0.5) \times 10^{-9}, \quad \mathcal{B}(\overline{B}_d^0 \to \mu^+\mu^-) = (1.0 \pm 0.1) \times 10^{-10}. \tag{2.148}$$

The corresponding branching ratios of the $e^+e^-$ modes can be obtained from (2.148) by scaling with $(m_e^2/m_\mu^2)$. The current experimental upper bound on $B_q$-decays from CDF is [16]

$$\mathcal{B}(\overline{B}_s^0 \to \mu^+\mu^-) < 5.8 \times 10^{-7} \quad (90\%\ \text{C.L.}) \tag{2.149}$$

$$\mathcal{B}(\overline{B}_d^0 \to \mu^+\mu^-) < 1.5 \times 10^{-7} \quad (90\%\ \text{C.L.}) \tag{2.150}$$

Belle sets the following 90% C.L. upper limits for the $B_d$-decays [15]

$$\mathcal{B}(\overline{B}_d^0 \to e^+e^-) < 1.9 \times 10^{-7}, \quad \mathcal{B}(\overline{B}_d^0 \to \mu^+\mu^-) < 1.6 \times 10^{-7}. \tag{2.151}$$

As far as the tau channel is concerned, detection difficulties may be offset by larger branching ratios $\mathcal{B}(\overline{B}_q^0 \to \tau^+\tau^-)/\mathcal{B}(\overline{B}_q^0 \to \mu^+\mu^-) \sim (m_\tau^2/m_\mu^2) = \text{few}\times 10^2$. Following [281], we obtain the Standard Model branching fractions

$$\mathcal{B}(\overline{B}_s^0 \to \tau^+\tau^-) = (7.2 \pm 1.1) \times 10^{-7}, \quad \mathcal{B}(\overline{B}_d^0 \to \tau^+\tau^-) = (2.1 \pm 0.3) \times 10^{-8}. \tag{2.152}$$

The current experimental information on the $\tau$ modes is rather poor. Indirect bounds can be inferred from LEP data on $B \to \tau\nu$ decays [255]

$$\mathcal{B}(\overline{B}_s^0 \to \tau^+\tau^-) < 5.0\%, \quad \mathcal{B}(\overline{B}_d^0 \to \tau^+\tau^-) < 1.5\%. \tag{2.153}$$

Another interesting observable is the ratio $R_{\ell\ell} \equiv \mathcal{B}(\overline{B}_d^0 \to \ell^+\ell^-)/\mathcal{B}(\overline{B}_s^0 \to \ell^+\ell^-)$. It has the advantage that the relative rates of $B_d$ and $B_s$-decays have a smaller theoretical uncertainty since $f_{B_d}/f_{B_s}$ can be determined more precisely than $f_{B_s}$ alone [cf. (2.145)]. A determination of $R_{ll}$ can provide information on $|V_{td}/V_{ts}|$ and probe the flavor structure of the Standard Model and beyond [279, 280, 282]. For example, in the Standard Model as well as in models where the Yukawa couplings are the only source of flavor violation, $R_{\ell\ell}$ is approximately

$$\frac{\tau_{B_d}}{\tau_{B_d}} \frac{M_{B_d}}{M_{B_s}} \frac{f_{B_d}^2}{f_{B_s}^2} \frac{|V_{td}|^2}{|V_{ts}|^2} \sim \mathcal{O}(10^{-2}). \tag{2.154}$$

Since $|V_{td}|^2/|V_{ts}|^2 = \lambda^2 R_t^2$, a measurement of the ratio of leptonic $B_d$ to $B_s$ decays will allow for a determination of the side $R_t$ of the unitarity triangle [280] and for a test of the minimal-flavor-violation (MFV) hypothesis.

### 2.21.3 Predictions for $B \to \ell^+\ell^-$ beyond the Standard Model

Before addressing $CP$-violating effects in $\overline{B}_q^0 \to \ell^+\ell^-$ decays we briefly discuss the implications of New Physics contributions to the scalar and pseudoscalar coefficients in (2.147). They can receive substantial contributions *e.g.,* in models with an extended Higgs sector, such as the two-Higgs-doublet model (2HDM) and SUSY [275, 255, 276, 248, 279, 140, 283, 128, 284, 17, 19, 260, 23, 22, 285, 286, 287]. (For recent reviews, see [288, 289].) In this class of New Physics models the scalar and pseudoscalar coefficients vanish when $m_\ell \to 0$, so that $C_{S,P}^{(\prime)} \propto m_\ell$. Yet, large values of $\tan\beta$, the ratio of the two vacuum expectation values of the neutral Higgs fields, may compensate for the suppression by the mass of the light leptons $e$ or $\mu$. Assuming the scalar and pseudoscalar contributions to be dominant in (2.147), we can set an indirect upper limit on $\mathcal{B}(\overline{B}_d^0 \to \tau^+\tau^-)$. Given the upper bound on $\mathcal{B}(\overline{B}_d^0 \to \mu^+\mu^-)$ in (2.151), we obtain

$$\mathcal{B}(\overline{B}_d^0 \to \tau^+\tau^-) \leq 3.4 \times 10^{-5} \left[\frac{\mathcal{B}(\overline{B}_d^0 \to \mu^+\mu^-)}{1.6 \times 10^{-7}}\right]. \tag{2.155}$$

which is very similar to the bound derived in [23]. We stress that this constraint applies only to those models in which $C_{S,P}^{(\prime)} \propto m_l$. (This is not the case *e.g.,* in generic SUSY models with broken $R$ parity [276].)

In the type-II 2HDM holds $C_S \sim \tan^2\beta$ [248, 128]. Given a charged Higgs boson mass of $260\ \mathrm{GeV}$, the branching ratio of $\overline{B}{}_s^0 \to \mu^+\mu^-$ amounts to $(1.4-4.8)\times 10^{-9}$ for $40 \leq \tan\beta \leq 60$ [248], which is comparable to the Standard Model prediction in (2.148). We therefore conclude that within the type-II 2HDM there are only moderate New Physics effects in $\overline{B}{}_q^0 \to \ell^+\ell^-$ decays.

On the other hand, in the high $\tan\beta$ region of the MSSM the leading contribution to the (pseudo)scalar coefficients is $\sim \tan^3\beta$, with $C_S \simeq -C_P$ [248, 279, 284, 17, 19, 260, 23, 22, 285, 286, 287]. As a result, the $\overline{B}{}_q^0 \to \ell^+\ell^-$ branching ratios can be enhanced by orders of magnitude with respect to the Standard Model expectations. Note that large branching ratios can occur even without any new flavor structure beyond the one in the CKM matrix. An interesting feature of $\overline{B}{}_q^0 \to \ell^+\ell^-$ decays are possible correlations between their branching ratios and $\Delta M_{B_q}$ [22, 285] and $R_K = \mathcal{B}(B \to K\mu^+\mu^-)/\mathcal{B}(B \to Ke^+e^-)$ [129], the latter being discussed in Section 2.16.3. In the context of the MSSM with MFV, the experimental lower bound on the $B_s - \overline{B}_s$ mass difference yields the upper limits $\mathcal{B}(\overline{B}{}_s^0 \to \mu^+\mu^-) < 1.2\times 10^{-6}$ and $\mathcal{B}(\overline{B}{}_d^0 \to \mu^+\mu^-) < 3.0\times 10^{-8}$ [22, 285]. Thus, an observation of a larger branching ratio would indicate the existence of non-minimal flavor violation [290], see also [286]. The MFV MSSM correlation between $\Delta M_{B_s}$ and $\mathcal{B}(\overline{B}{}_q^0 \to \mu^+\mu^-)$ also breaks down with an additional singlet Higgs, *i.e.,* in the next-to-minimal-supersymmetric Standard Model (NMSSM) [24].

### 2.21.4 *CP* violation

Since there is no strong phase in the purely leptonic decays, which is mandatory for a non-zero rate $CP$ asymmetry besides a $CP$-violating phase, direct $CP$ violation cannot occur in these modes. Thus, $CP$ violation can arise only through interference between mixing and decay.

We define the time-integrated $CP$ asymmetries as [287, 291, 292]

$$A_{CP}^{(B_q^0\to\ell_i^+\ell_i^-)} = \frac{\int_0^\infty dt\Gamma(B_q^0(t)\to\ell_i^+\ell_i^-) - \int_0^\infty dt\Gamma(\overline{B}{}_q^0(t)\to\ell_j^+\ell_j^-)}{\int_0^\infty dt\Gamma(B_q^0(t)\to\ell_i^+\ell_i^-) + \int_0^\infty dt\Gamma(\overline{B}{}_q^0(t)\to\ell_j^+\ell_j^-)}, \tag{2.156}$$

where $i, j$ ($i \neq j$) denote left-handed ($L$) and right-handed ($R$) leptons in the final state. Assuming the $B_q^0$–$\overline{B}{}_q^0$ mixing parameter $q/p$ to be a pure phase,[17] and neglecting the light quark masses as well as the primed Wilson coefficients in (2.142), one finds [291, 292]

$$A_{CP}^{(B_q^0\to\ell_L^+\ell_L^-)} = -\frac{2x_q\mathrm{Im}\lambda_q}{(2+x_q^2)+x_q^2|\lambda_q|^2}\ , \quad A_{CP}^{(B_q^0\to\ell_R^+\ell_R^-)} = -\frac{2x_q\mathrm{Im}\lambda_q}{(2+x_q^2)|\lambda_q|^2+x_q^2}, \tag{2.157}$$

where $x_q = \Delta M_{B_q}/\Gamma_{B_q}$ and

$$\lambda_q = \frac{M_{12}^{q*}}{|M_{12}^q|}\left(\frac{V_{tb}V_{tq}^*}{V_{tb}^*V_{tq}}\right)\frac{\beta_q C_S + C_P + 2m_l A_{10}/M_{B_q}}{\beta_q C_S^* - C_P^* - 2m_\ell A_{10}/M_{B_q}}. \tag{2.158}$$

Here, $M_{12}^q$ is the off-diagonal element in the neutral $B$ meson mass matrix. From (2.157) it follows that the maximum $CP$ asymmetry is $A_{CP}^{\max} = 1/(2+x_q^2)^{1/2}$. The dependence on the $B$ meson decay constant drops out in the $CP$ asymmetries, which are therefore, theoretically very clean. Taking $x_d = 0.76$ and a nominal value of $x_s = 19$ [2], we find that the maximum $CP$ asymmetry is small for $B_s$ decay ($\approx 5\%$) but considerably larger for $B_d$ decay ($\approx 62\%$).

Within the Standard Model, $CP$ violation in $\overline{B}{}_q^0 \to \ell^+\ell^-$ decays is experimentally remote since $A_{\mathrm{CP}} \sim \mathcal{O}(10^{-3})$ [291, 292]. However, New Physics with non-standard $CP$ phases can give appreciable $CP$ asymmetries, in particular for the tau channel. For example, within the $CP$-violating MSSM where $C_S = -C_P$ at large $\tan\beta$, the asymmetries for the dimuon final state amount to $|A_{CP}^{(B_d^0\to\mu_L^+\mu_L^-)}| \approx 0.7\%$ and $|A_{CP}^{(B_d^0\to\mu_R^+\mu_R^-)}| \approx 3\%$, taking into account New Physics contributions to $B^0\overline{B}^0$ mixing [287]. These small values are mainly due to a cancellation in $\beta_d C_S + C_P$ in (2.158), since $\beta_d(m_\mu) \approx 1$. In the $\tau^+\tau^-$ mode the $CP$-violating effects are larger since $\beta_d(m_\tau) \approx 0.7$ [291, 292]. Using the same input parameters as before, the $CP$ asymmetries $|A_{CP}^{(B_d^0\to\tau_L^+\tau_L^-)}|$ and $|A_{CP}^{(B_d^0\to\tau_R^+\tau_R^-)}|$ can reach about $9\%$ and $36\%$, respectively [287, 292]. Going beyond the MSSM with $C_S \neq -C_P$ may lead to large $CP$ asymmetries in the muon channel as well [287]. For example, a model where the relation $C_S = -C_P$ does not hold is the NMSSM [24].

[17] For a definition of $q$ and $p$, see [293].

The observation of an unexpectedly large $CP$ asymmetry in the purely leptonic decay modes would be a signal of New Physics and a pointer to the existence of $CP$-violating sources outside the CKM matrix. Of particular interest is the analysis of the $CP$ asymmetries with $\tau$'s in the final state.

# 2.22 Theoretical Prospects for $B \to (X_c, D, D^*)\tau\overline{\nu}_\tau$

≻– A. Soni –≺

## 2.22.1 Introduction

, $b \to c\tau\overline{\nu}_\tau$ mediated processes can provide useful constraints on models with an extended Higgs sector such as the two Higgs doublet model (2HDM) II or the MSSM. The leading order diagrams are shown in Fig. 2-27. The charged Higgs ($H$)-exchange diagram is driven by $\beta_H \equiv \frac{\tan\beta}{m_H}$, where $m_H$ denotes the Higgs mass and $\tan\beta$ is the ratio of the vevs of the Higgses giving mass to the up and down-sector. As is well known $B \to X_s\gamma$ decay is also very sensitive to charged Higgs exchange; however, in that case if the Higgs sector is part of a supersymmetric theory then it can be argued that the new contributions of the charged Higgs may cancel against those from other SUSY-particles. Such a cancellation cannot be invoked for the $B \to (X_c, D, D^*)\tau\overline{\nu}_\tau$ decays case at least in a $R$ parity-conserving SUSY scenario at tree level.

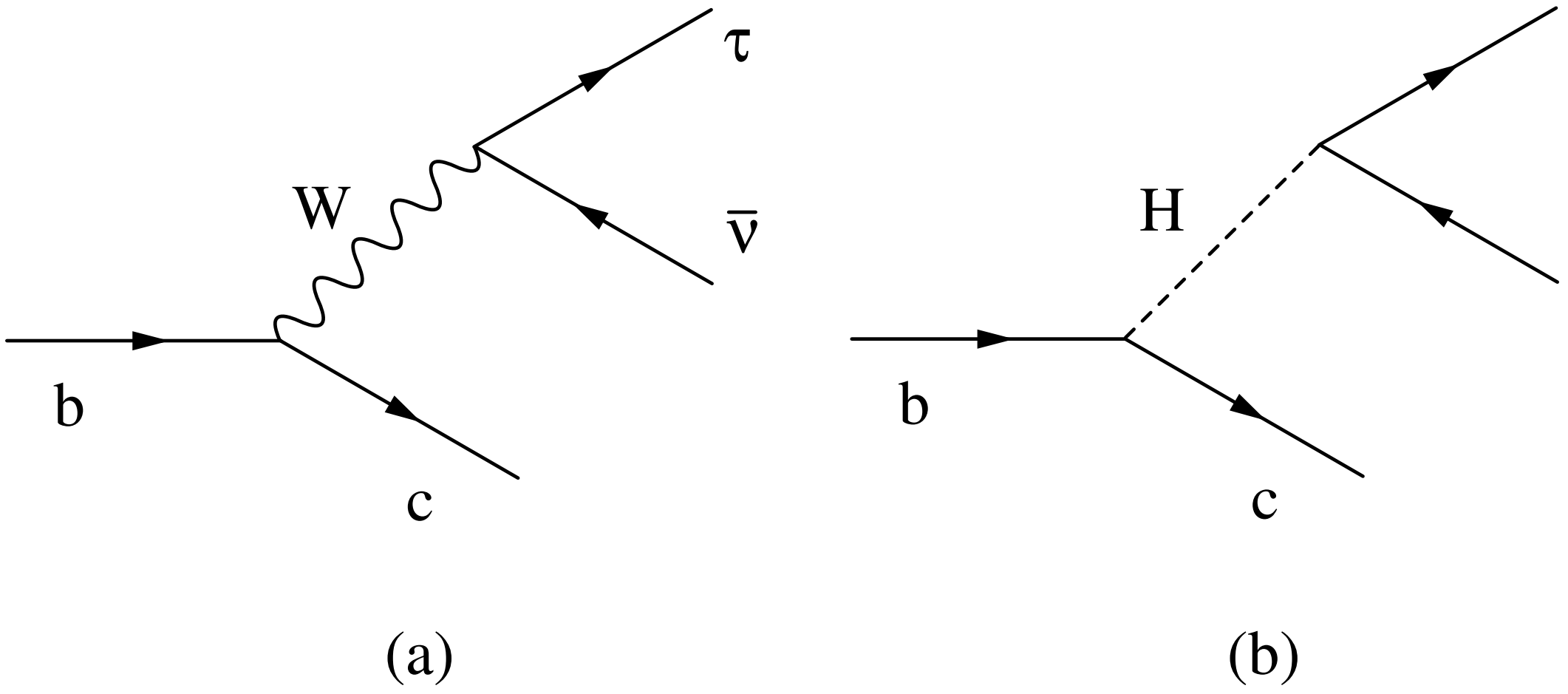

**Figure 2-27.** *The leading order Feynman diagrams for $b \to c\tau\overline{\nu}_\tau$ processes.*

Theoretical studies of the semileptonic decay rate into $D^{(*)}$ and inclusive $X_c$ final states have been performed in Refs. [294, 295, 296, 297, 298, 299]. It is useful to normalize the branching ratio and spectra to the Standard Model ones, *e.g.*,

$$R_M = \frac{\mathcal{B}(B \to M\tau\overline{\nu}_\tau)}{\mathcal{B}(B \to M\ell\overline{\nu}_\ell)} \quad \text{where } \ell = e, \mu \quad \text{and } M = X_c, D, D^* \tag{2.159}$$

since in the denominator the Higgs contribution is suppressed by a small lepton mass. Furthermore some theoretical uncertainties, *e.g.*, due to form factors, $V_{cb}$ and higher order corrections cancel at least partially in the ratio.

Decays into the pseudoscalar $D$ are most sensitive to scalar boson (Higgs) exchange [295, 298]. Also, the $q^2$ distribution, where $q = p_B - p_M$ and $p_B(p_M)$ denote the $B$ meson (final hadronic) momenta, is also significantly more sensitive than the total rate [298, 294]. A better reach in $\beta_H$ than the width has also the integrated longitudinal $\tau$ polarization in exclusive [295] and inclusive [296] decays. The decay $B \to D\tau\overline{\nu}_\tau$ is discussed in detail in the next Section 2.22.2.

QCD corrections bring in some model-dependence in the interpretation of $b \to c\tau\overline{\nu}_\tau$ measurements in terms of $\beta_H$. The $\mathcal{O}(\alpha_s)$-corrections have been calculated for the rate of inclusive $B \to X_c\overline{\tau}\nu_\tau$ decays in the 2HDM II and found to be moderate [297]. At LEP the $b \to c\overline{\tau}\nu_\tau$ branching ratio has been measured and the constraint $\beta_H < 0.53\,\text{GeV}^{-1}$ at 95 % C.L. [300] has been obtained. On the other hand, SUSY QCD corrections turn out to be substantial for large $\tan\beta$ and can weaken the bound significantly for some regions of the parameter space [299]. In the following

Sections the reach in $\beta_H$ is discussed neglecting $\mathcal{O}(\alpha_s)$-corrections since the corresponding analyses are not available. The bounds obtained are hence not valid in a general MSSM.

The transverse polarization of the $\tau$ lepton in semileptonic decays provides a unique and very sensitive probe of New Physics $CP$-odd phases present in the charged Higgs couplings [301, 302]. This is further discussed in Section 2.22.3.

### 2.22.2 Constraints from $B \longrightarrow D\tau\overline{\nu}_\tau$ decays

In addition to the form factors $F_0$ and $F_1$ that describe the semileptonic decays $B \to D\tau\overline{\nu}_\tau$ via $W$ exchanges, *i.e.*, the matrix element ($t = q^2/m_B^2$)

$$< D(p_D)|\overline{c}\gamma_\mu b|B(p_B) >= F_1(t)\left[(p_B + p_D)_\mu - \frac{m_B^2 - m_D^2}{q^2} q_\mu\right] + F_0(t)\frac{m_B^2 - m_D^2}{q^2} q_\mu\,, \tag{2.160}$$

one also needs a scalar form factor $F_s \propto < D|\overline{c}b|B >$ for the Higgs contributions. $F_s$ is related to $F_0$ by the equations of motion. A study of the semileptonic decays into electrons and muons should provide a very accurate determination of $F_1$. Heavy quark symmetry relates the form factors to the single Isgur-Wise [303] function. Although $1/m_Q$ corrections on the individual form factors are appreciable, their ratios receive only small corrections.

The dilepton invariant mass spectrum of $B \to D\tau\overline{\nu}_\tau$ decays can be written as [298]

$$\frac{d\Gamma}{dt} = \frac{G_F^2|V_{cb}|^2 m_B^5}{128\pi^3}\left[(1 + \delta_H(t))^2 \cdot \rho_0(t) + \rho_1(t)\right]\,, \tag{2.161}$$

where $\rho_{0(1)}$ are spin 0(1) contributions which involve $F_{0(1)}$, respectively. In the limit of vanishing lepton mass, only $\rho_1$ remains finite. The New Physics Higgs contribution $\delta_H$ is driven by $F_s$, see [298] for details.

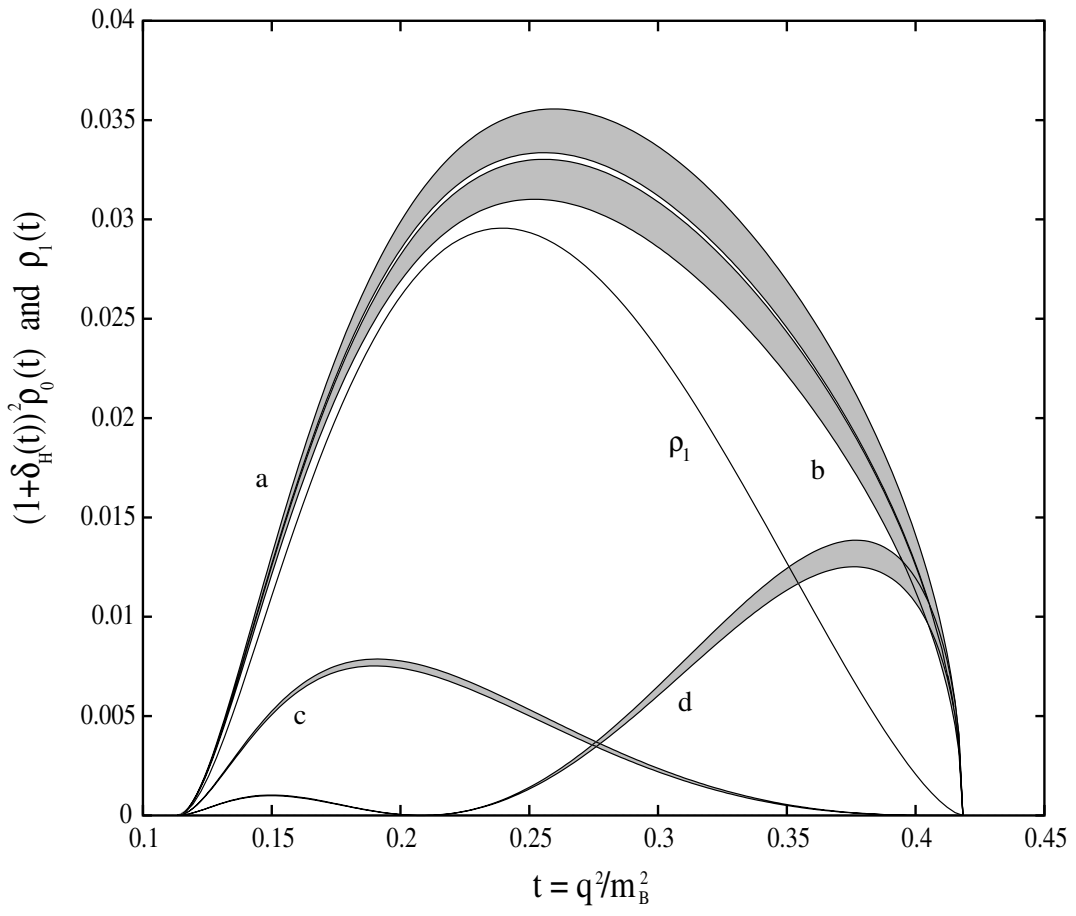

**Figure 2-28.** *The spin 0 contributions to the dilepton mass spectrum in $B \to D\tau\nu_\tau$ decays for $\tan\beta/m_H = 0, 0.06, 0.25, 0.35$GeV$^{-1}$ corresponding to lines a(=SM),b,c,d. The solid curve corresponds to the spin 1 contribution, $\rho_1$. Shaded regions denote form factor uncertainties. Figure taken from [298].*

Figure 2-28 shows the spin 0 and spin 1 contributions to the differential spectrum, where the spin 0 contributions for $\beta_H = 0, 0.06, 0.25, 0.35\,\mathrm{GeV}^{-1}$ are labeled as a, b, c and d respectively. The shaded regions indicate the residual theory uncertainties in the ratio $F_0(t)/F_1(t)$. For smaller values of $\beta_H$ the spectrum starts to look like the Standard Model (curve a), which corresponds to $m_H$ becoming extremely heavy. With sufficient data one may place a bound on $\beta_H < 0.06\,\mathrm{GeV}^{-1}$ from this method; for large $\tan\beta$ as favored due to the large value of $m_t/m_b$, say 30, this would translate into $m_H > 500\,\mathrm{GeV}$.

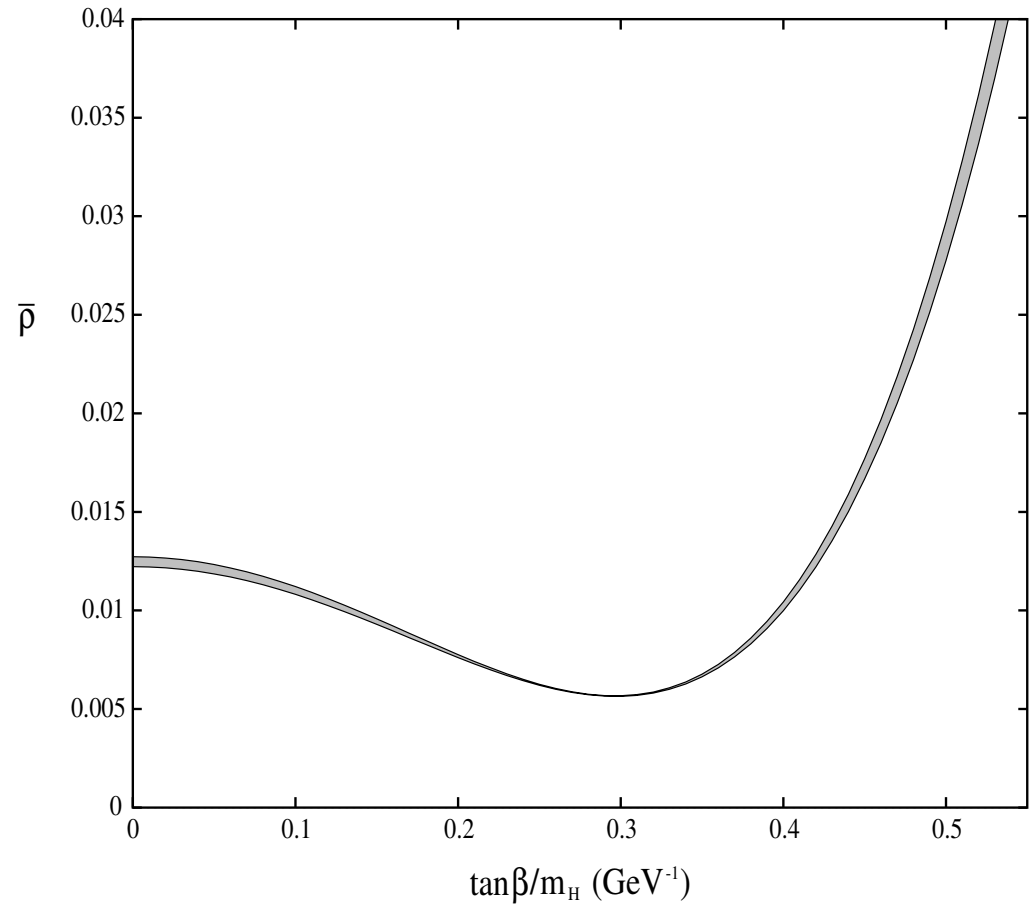

**Figure 2-29.** *The integrated width for $B \to D\tau\nu_\tau$ decays (normalized to $G_F^2|V_{cb}|^2 m_B^5/(128\pi^3)$ ) as a function of $\tan\beta/m_H$ [298].*

Given $\mathcal{B}(B \to D\tau\nu) \approx 0.5\%$, and assuming a 1% detection efficiency, would mean that with $10^9$ $B\overline{B}$-pairs one should see $10^5$ such events, so a detailed study of the differential spectrum should be feasible.

For comparison, Fig. 2-29 shows the $\beta_H$ reach for the $B \to D\tau\nu_\tau$ integrated width. Here $\beta_H = 0$ (origin) corresponds to the Standard Model value. For larger values of $\beta_H$ the Higgs contribution interferes destructively with the Standard Model, so that the integrated width reduces to half of the Standard Model value at around $\beta_H \approx 0.3\,\text{GeV}^{-1}$. For $\beta_H > 0.45\,\text{GeV}^{-1}$ the Higgs contribution starts to dominate the width. This is the level at which one can put bounds on $\beta_H$ from measurements of $R_D$, very similar to the current bound from $R_X$ data discussed in the Introduction. To show the contrast between this and the differential spectrum, note that the total width reduces by less than about 10% for $\beta_H = 0.06\,\text{GeV}^{-1}$. Experimental prospects for $B \to D, D^*\tau\overline{\nu}$ have been discussed in [304].

With sensitivities down to $\tan\beta/m_{H^\pm} > 0.06\,\text{GeV}^{-1}$ the spectra in $B \to D\tau\nu$ decays are competitive with the LHC reach for $H^\pm$ masses below $\sim 250$ GeV and moderate values of $\tan\beta$, see Fig. 15b in [305]. Note that in this region SUSY QCD corrections are not enhanced.

## 2.22.3 Transverse polarization of the $\tau$ and $CP$ violation

The transverse polarization of the $\tau$ in semileptonic decays

$$p_\tau^t \equiv \frac{\vec{S}_\tau \cdot (\vec{p}_\tau \times \vec{p}_M)}{|\vec{p}_\tau \times \vec{p}_M|} \tag{2.162}$$

where $\vec{p}_\tau(\vec{p}_M)$ denote the three-momentum of the $\tau$ (hadron) and $\vec{S}_\tau$ the $\tau$ spin is an extremely sensitive probe of a non-standard $CP$-odd phase from charged Higgs exchange [301, 302, 306]. Since in the Standard Model $p_\tau^t$ vanishes it serves as clean test of the CKM-paradigm of $CP$ violation. The transverse polarization is $T_N$-odd and can occur from tree level graphs [307]. In contrast, the partial rate asymmetry $A_{CP}(\Gamma)$ or the $\tau$-lepton energy asymmetry $< E_\tau >$, say between the $\tau^+$ and $\tau^-$ in $B^+$ vs. $B^-$ decays, are $T_N$-even and require $CP$-even phase(s).

In $B \to X_c\tau\nu_\tau$ decays the $W - H$ interference term contributing to $< E_\tau >$ and $A_{CP}(\Gamma)$ is proportional to $Tr[\gamma_\mu L(\not p_\tau + m_\tau)(L, R)\ \not p_\nu] \propto m_\tau/m_B$. This and the loop factor $\mathcal{O}(\pi/\alpha_s)$ tends to make $< p_\tau^t >$ larger compared to the other two asymmetries by $\sim \mathcal{O}(30)$, see [307] for details. On the other hand, experimental detection of $p_\tau^t$ via decay correlations in $\tau \to \pi\nu, \mu\nu\nu, \rho\nu$ *etc.* is much more difficult than measurements of the energy or rate asymmetry. With $\mathcal{B}(B \to X_c\tau\nu_\tau) = (2.48 \pm 0.26)\%$ [2], and assuming an effective efficiency for $p_\tau^t$ of 0.1%, then the detection of $< p_\tau^t > \approx 1\%$ with 3 $\sigma$ significance requires about $2 \times 10^9 B\overline{B}$ pairs.

Note that fake asymmetries due to final state interactions can arise if only $\tau^-$ *or* $\tau^+$ is studied; to verify that it is a true $CP$-violating effect one may need to study both particle and anti-particle decays. A non-vanishing ($CP$-odd ) $p^t_\tau$ switches sign from $\tau^-$ to $\tau^+$ final state leptons. Clearly rate and/or energy asymmetries should also be studied, especially if detection efficiencies for those are much higher.

This research was supported in part by USDOE Contract No. DE-AC02-98CH10886.

## 2.23 $B^0 \to$ invisible (+ gamma)

≻– J. Albert –≺

In addition to searching for new sources of $CP$ violation, which up to now have been fairly consistent with Standard Model predictions, a future Super $B$ Factorymust ensure that other manifestations of New Physics in $B$ decays cannot elude notice. Less than 50% of the total width of the $B$ is explained by known branching fractions; few constraints exist on decays beyond what is expected from the Standard Model.

There are presently no significant constraints on invisible decays of any particles that contain heavy flavor [18]. The Standard Model predicts infinitesimally small branching fractions for these decays. However, without causing any inconsistency with all current published experimental results, such decay rates could in principle be up to the order of 5%[2].

In the Standard Model, the lowest order decay processes for $B^0 \to$ invisible (+ $\gamma$) are second order weak decays (Fig. 2-30):

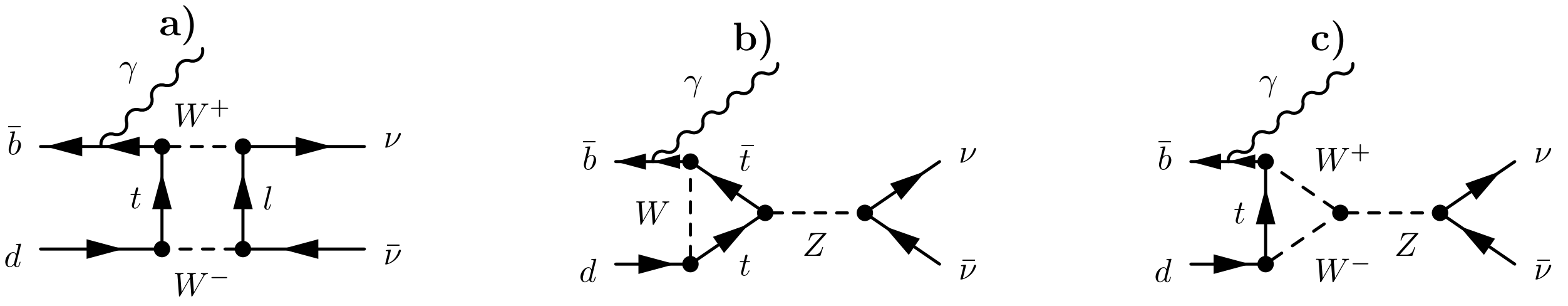

**Figure 2-30.** *The lowest-order Standard Model Feynman graphs for $B^0 \to$ invisible (+ gamma) decay: a) box diagram, b) $q\bar{q}$ weak annihilation diagram, and c) $W^+W^-$ weak annihilation diagram.*

Each of these diagrams is highly suppressed within the Standard Model. For the $\nu\bar{\nu}\gamma$ channel, the expected Standard Model branching fraction is at the $10^{-8} - 10^{-9}$ level [309, 310]. The $\nu\bar{\nu}$ channel has an additional helicity suppression and thus for all intents and purposes should never occur at all. The Standard Model branching fractions for these decays are well-predicted. For the $B^0 \to$ invisible channel, there should be no visible Standard Model contamination; an experimental observation would necessarily imply the existence of New Physics.

Significant rates for invisible $B^0$ decays can occur in several physical models, ranging from phenomenological models motivated by inconsistencies in neutrino experimental data with the Standard Model, to theoretical models motivated by attempts to resolve fundamental open questions, such as the hierarchy problem. An example of the former is described in Ref. [311]. This attempt to explain NuTeV's observation of an anomalous excess of dimuon events provides a model for the production of long-lived heavy neutral particles consistent with the NuTeV data [312]. They propose a supersymmetric model with a neutralino LSP that avoids tight LEP constraints on neutralino production by coupling to decays of $B$ mesons. Their model predicts invisible $B$ decays with a branching fraction in the $10^{-7}$ to $10^{-5}$ range, which is just below visibility with the current *BABAR* data sample. The SUSY production mechanism for invisible $B^0$ decays is shown above in Fig. 2-31. Figure 2-32 shows the MSSM phase-space corresponding to this model, which is completely consistent with LEP limits on neutralino production. Figure 2-33 shows the impact on the $B^0 \to$ invisible branching fraction compared with the expected number of dimuon events seen at NuTeV. In addition, models using large extra dimensions to solve the hierarchy problem can also produce significant, although small, rates for invisible $B$ decays. Examples of such models, and their predictions, may be found in Refs. [313, 314, 315].

### 2.23.1 Analysis overview

The current analysis at *BABAR* takes advantage of the fact that, at the $\Upsilon(4S)$, when one reconstructs a $B$ decay, one can be certain that there was a second $B$ on the other side. The essence of the analysis is to reconstruct a $B$ decay and, in the rest of the event, look for consistency with "nothing" or single gamma hypotheses. Similar to most rare decay analysis, this analysis is limited by statistics and thus it is critical to get as high an efficiency as possible. The efficiency is entirely dependent on the choice of tag algorithm, as the signal-side selection efficiency (for "nothing" or just a $\gamma$) is nearly independent of the tag that is used for the opposite $B$. We decided on the tag strategy used by the semileptonic-tag $B \to K\nu\overline{\nu}$ analysis [316] and by the semileptonic-tag $B \to \tau\nu$ analysis [317], due to its very high efficiency and its well-understood properties. The semileptonic tag approach relies on identifying a $D^{(*)\pm}l\nu$ candidate in one of three $D^0$ modes ($D^0 \to K\pi$, $K\pi\pi\pi$, or $K\pi\pi^0$) and one $D^\pm$ mode ($D^\pm \to K\pi\pi$). Since the branching fractions for these modes are very high, and the background rejection is due to both the lepton and the fully reconstructed $D^{(*)}$, this is an efficient and clean tag. As a further check to ensure Monte Carlo reproduces data for the recoil spectrum of the tag, we look at the additional channel "$B^\pm \to$ invisible" in data and Monte Carlo and check to make sure that the resulting branching fraction for this forbidden non-charge-conserving decay is consistent with zero.

[18] In the finalization of this write-up the *BABAR* collaboration published their search results; they obtained 90 % CL upper bounds on the branching ratios for $B^0 \to$ invisible as $22 \times 10^{-5}$ and $4.7 \times 10^{-5}$ for $B^0 \to \nu\bar{\nu}\gamma$ [308].

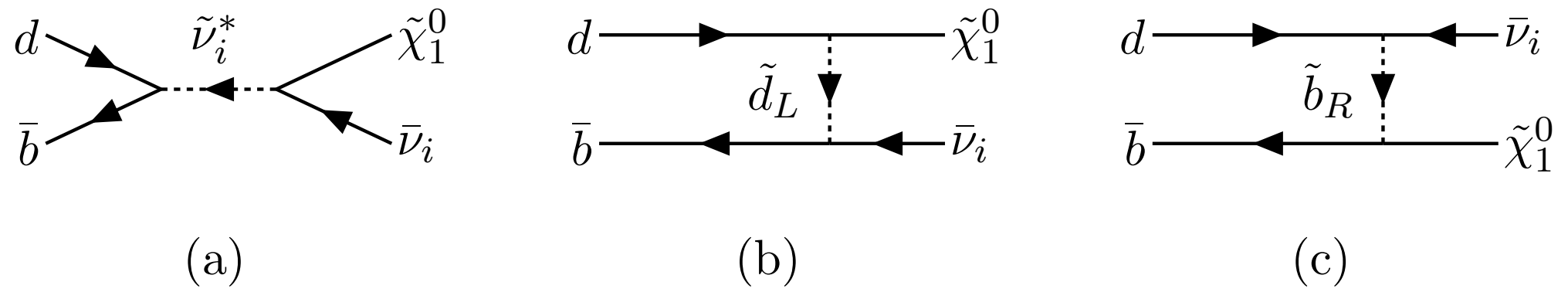

**Figure 2-31.** *From Ref. [311]: light neutralino production in $B$-meson decays: (a-c) $B_d^0 \longrightarrow \overline{\nu}_i \tilde{\chi}_1^0$.*

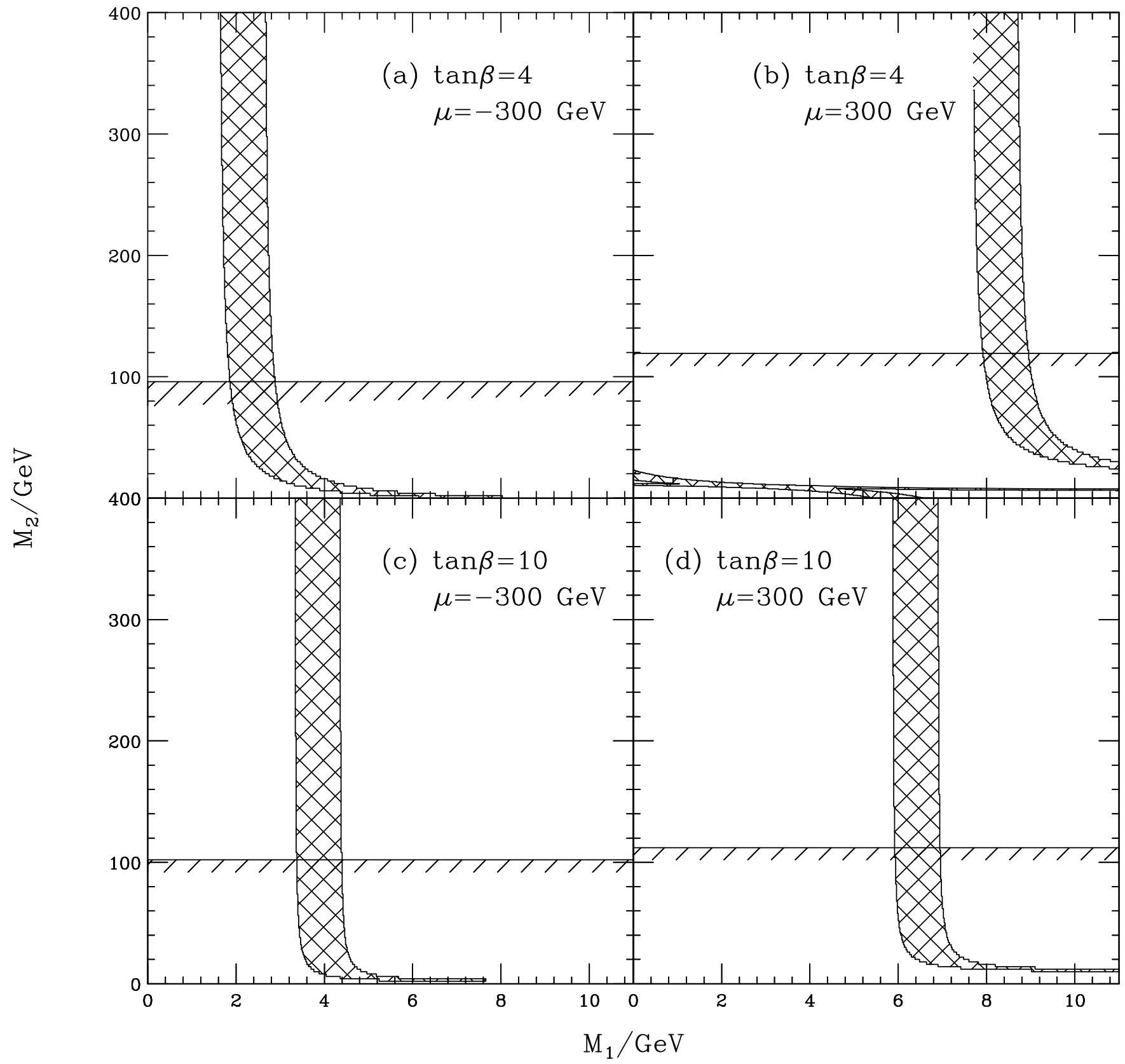

**Figure 2-32.** *From Ref. [311]: solutions in ($M_1$, $M_2$, $\mu$, $\tan\beta$) giving $4.5\,\mathrm{GeV} \leq \mathrm{M}_{\tilde{\chi}_1^0} \leq 5.5\,\mathrm{GeV}$ in the cross-hatched region. Points below the horizontal hatched line are excluded by the requirement that $M_{\tilde{\chi}_1^+} > 100\,\mathrm{GeV}$.*

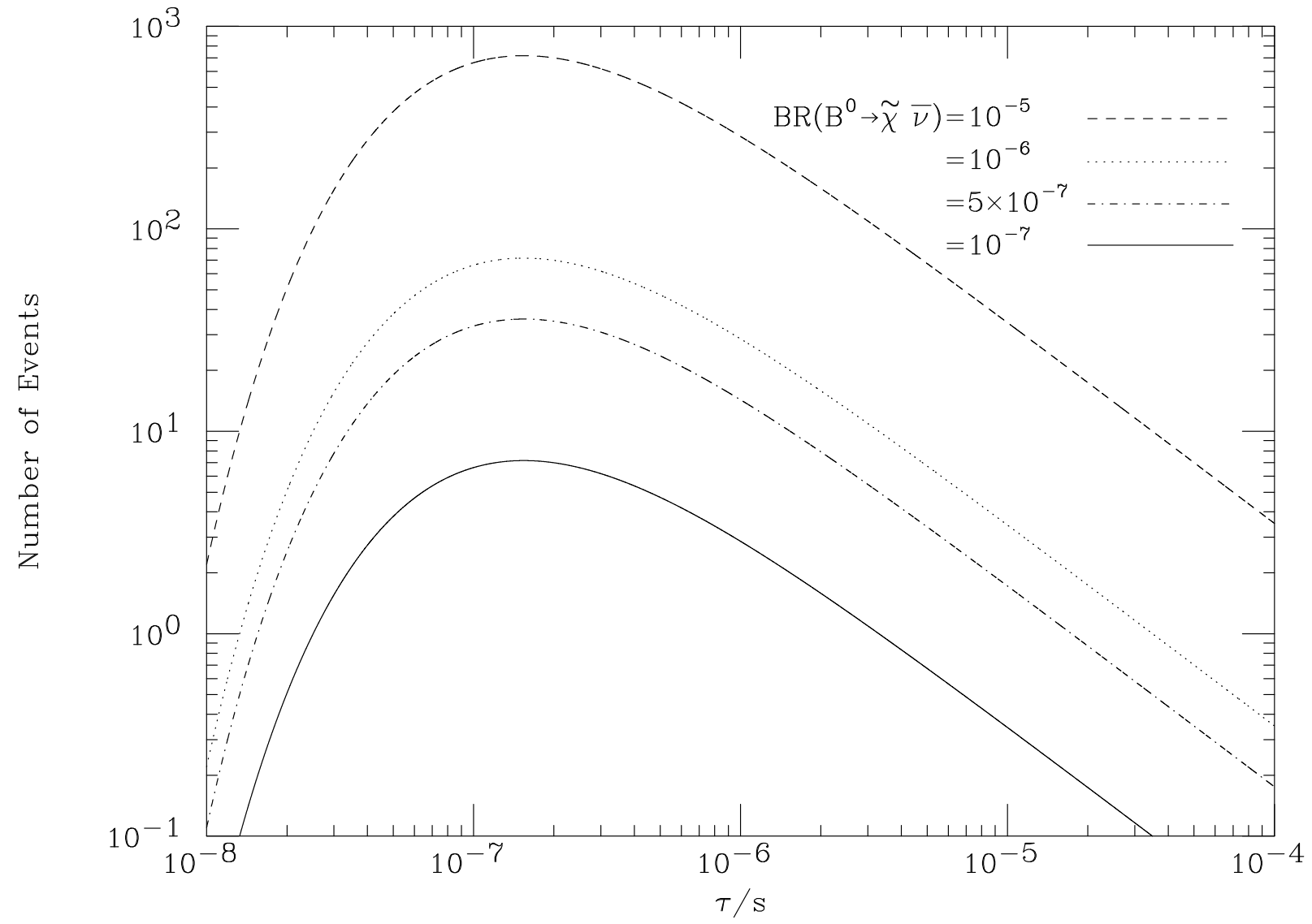

**Figure 2-33.** *From Ref. [311]: number of events in the NuTeV detector for neutralino production in $B$ meson decays as a function of the neutralino lifetime.*

After selecting events with a clean $D^{(*)\pm}l\nu$ tag, we choose events where the number of remaining charged tracks in the event is zero, and make a variety of cuts on the number of remaining photons, $\pi^0$'s, and $K^0_L$'s [318]. The total signal efficiency for each of the modes is each approximately $1 \times 10^{-3}$; the tag selection efficiency is the dominant limitation, being approximately $2 \times 10^{-3}$ [318]. Figure 2-34 shows distributions of the remaining energy in the electromagnetic calorimeter (EMC) in the event, after all tracks and neutral clusters associated with the tag $B$ have been removed. As seen on the right-side plots, peaking distributions are expected from signal. By either making a fixed cut in the remaining energy variable and subtracting the background expected from Monte Carlo in the data signal region, or by doing a full likelihood fit in data to a combination of the distributions from signal and background, one can determine the amount of signal in data. Both this "cut-based" and "likelihood-based" analysis strategies are pursued – the former as a check on the slightly tighter constraints provided by the latter.

We expect limits of approximately $8 \times 10^{-5}$ for each mode with the 82 fb$^{-1}$ sample we are using currently. Table 2-16 shows the expected limits for Super $B$ Factory integrated luminosities of 1, 10, and 50 ab$^{-1}$.

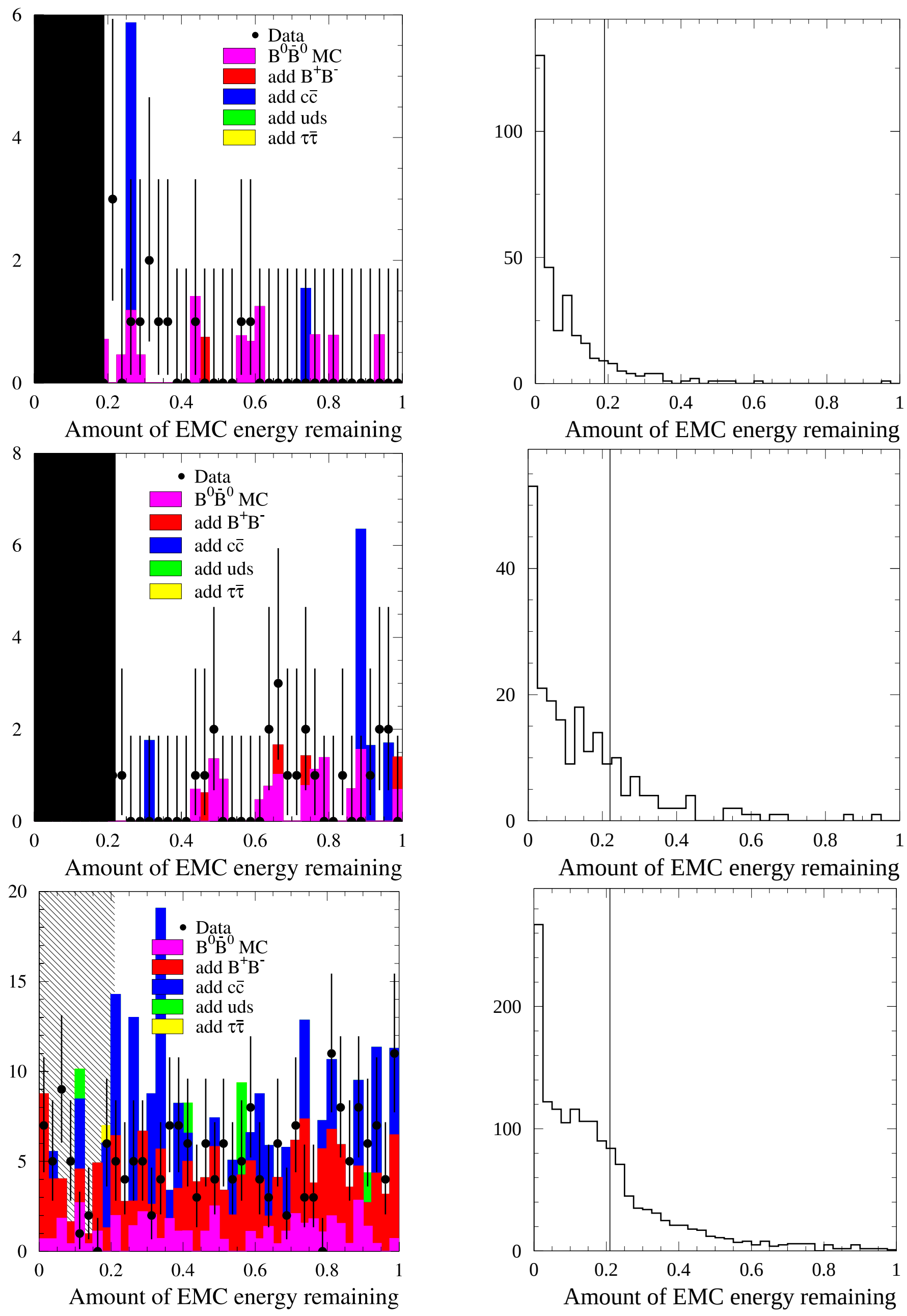

**Figure 2-34.** *Blinded signal plots comparing data and background Monte Carlo, and plots showing the peaking distribution in signal as expected from signal Monte Carlo, in the remaining energy variable. (Upper left) Comparison of data and background MC for the $B^0 \to$ invisible channel. (Upper right) Distribution of remaining energy expected from pure signal $B^0 \to$ invisible events (corresponding to a branching fraction of $3.85 \times 10^{-3}$). (Middle left) Comparison of data and background MC for the $B^0 \to$ invisible + gamma channel. (Middle right) Distribution of remaining energy expected from pure signal $B^0 \to$ invisible + gamma events (corresponding to a branching fraction of $4.10 \times 10^{-3}$). (Lower left) Comparison of data and background MC for the "$B^\pm \to$ invisible" calibration check. As expected in this validation channel, no significant signal is observed. (Lower right) Distribution of remaining energy expected from signal "$B^\pm \to$ invisible" events (corresponding to an effective branching fraction of $1.05 \times 10^{-2}$).*

**Table 2-16.** *Expected limits for $B^0 \to$ invisible and $B^0 \to$ invisible + $\gamma$ at a Super $B$ Factory.*

| $\mathcal{L}$ | 1 ab$^{-1}$ | 10 ab$^{-1}$ | 50 ab$^{-1}$ |
|---|---|---|---|
| Expected limit | $3 \times 10^{-6}$ | $1 \times 10^{-6}$ | $4 \times 10^{-7}$ |

## 2.24 Rare $B$ Decays at LHC$b$ and Other Hadron Experiments

>– G. Wilkinson and P. Koppenburg –<

We briefly review the potential of experiments at hadron machines in the field of rare $B$ decays. To enable comparison with experiments at the $\Upsilon(4S)$, emphasis is given to $B_d$ channels, although some results in $B_s$ modes are given. The discussion is centered on LHC$b$, with most of the results quoted from the recent re-optimization studies [319] of that experiment. Where appropriate, the complementary features of other experiments at hadron machines are indicated.

### 2.24.1 Introduction

LHC$b$ is an experiment which has been designed to fully exploit the very high cross-section for $b\bar{b}$ production ($\sigma_{b\bar{b}} \approx 500\,\mu$b) in 14 TeV $pp$ collisions at the LHC. The experiment is scheduled to begin operation at the start of LHC running and expects to continue data taking for several years at a constant local luminosity of $2 \times 10^{32}\ \mathrm{cm^{-2}s^{-1}}$. Unless stated otherwise, all event yields given below assume $10^7$ s operation in these conditions. The essential characteristics of the detector, and its potential in measuring $CP$-violating phases, are described elsewhere in this report [320]. Here those features of the experiment relevant for rare $B$ decays are emphasized. These are as follows:

- **Trigger**
  At the lowest level of triggering LHC$b$ looks for signatures of *single particles* (leptons, hadrons or photons) with high transverse momentum (thresholds of 1–5 GeV). The next trigger level relies on a vertex trigger. This strategy ensures good efficiency for a very wide range of $B$ decays, ranging from 38 % for $B_d \to K^*\gamma$ to 74 % for $B_d \to \mu\mu K^*$. More details may be found in [321].

- **Precise vertexing**
  The forward geometry of LHC$b$ together with the silicon strip Vertex Locator (VELO) allow secondary vertices to be reconstructed with excellent precision (typically $\sim 200\mu\mathrm{m}$ in the longitudinal direction). This provides a very powerful means of background suppression in any channel with charged tracks at the decay vertex.

- **Particle Identification**
  The RICH system of LHC$b$ provides reliable $\pi - K$ discrimination up to $p \sim 100\,\mathrm{GeV}$. This is extremely useful in the selection of many rare decays, for instance $B_d \to \mu\mu X$.

With these capabilities LHC$b$ has demonstrated a sensitivity to exclusive decays down to branching ratios of $10^{-9}$. Good performance is possible in radiative decays provided that a charged track vertex is present in the event.

$B$TeV is a similar experiment proposed for $\sqrt{s} = 2\,\mathrm{TeV}$ $p\bar{p}$ collisions at the Tevatron collider. The somewhat lower production cross-section at these energies will be largely compensated by deploying a pixel vertex trigger at the earliest level of triggering. A $\mathrm{PbWO_4}$ electromagnetic calorimeter is intended to enhance the performance in radiative decays.

ATLAS and CMS have the capabilities to conduct a wide-ranging $B$ physics program in the early, ‘low luminosity’ period of LHC operation. Lepton triggers provide a good sensitivity to decays such as $B \to \mu\mu X$. It may be possible to continue to search for extremely rare decays with very distinctive signatures, such as $B_{s,d} \to \mu\mu$ throughout the period of higher luminosity running.

## 2.24.2 $B \to \mu\mu X$

Events with two muons are "golden" modes for experiments at hadron colliders. Thanks to the large boost due to the 14 TeV collision energy at the LHC, both muons have a large energy. This allows the straightforward triggering and selection of such events. Moreover their energy is only marginally affected by the value of the dilepton mass ($m_{\mu\mu}$) or the decay direction defining the forward-backward asymmetry. Therefore the detection and selection efficiency is not correlated with these important observables.

LHC$b$ also exploits the high boost of the $B$-meson, and its very precise VELO, to isolate the $B$ decay vertex. A stringent constraint on the quality of the $B$ decay vertex fit allows to considerably reduce backgrounds from cascade $b \to \mu\nu c(\to \mu\nu s)$ decays or from events where both $b$ hadrons decay semileptonically. The good vertexing also allows a precise reconstruction of the $B$ mass, which further helps background suppression and enables a good purity to be obtained.

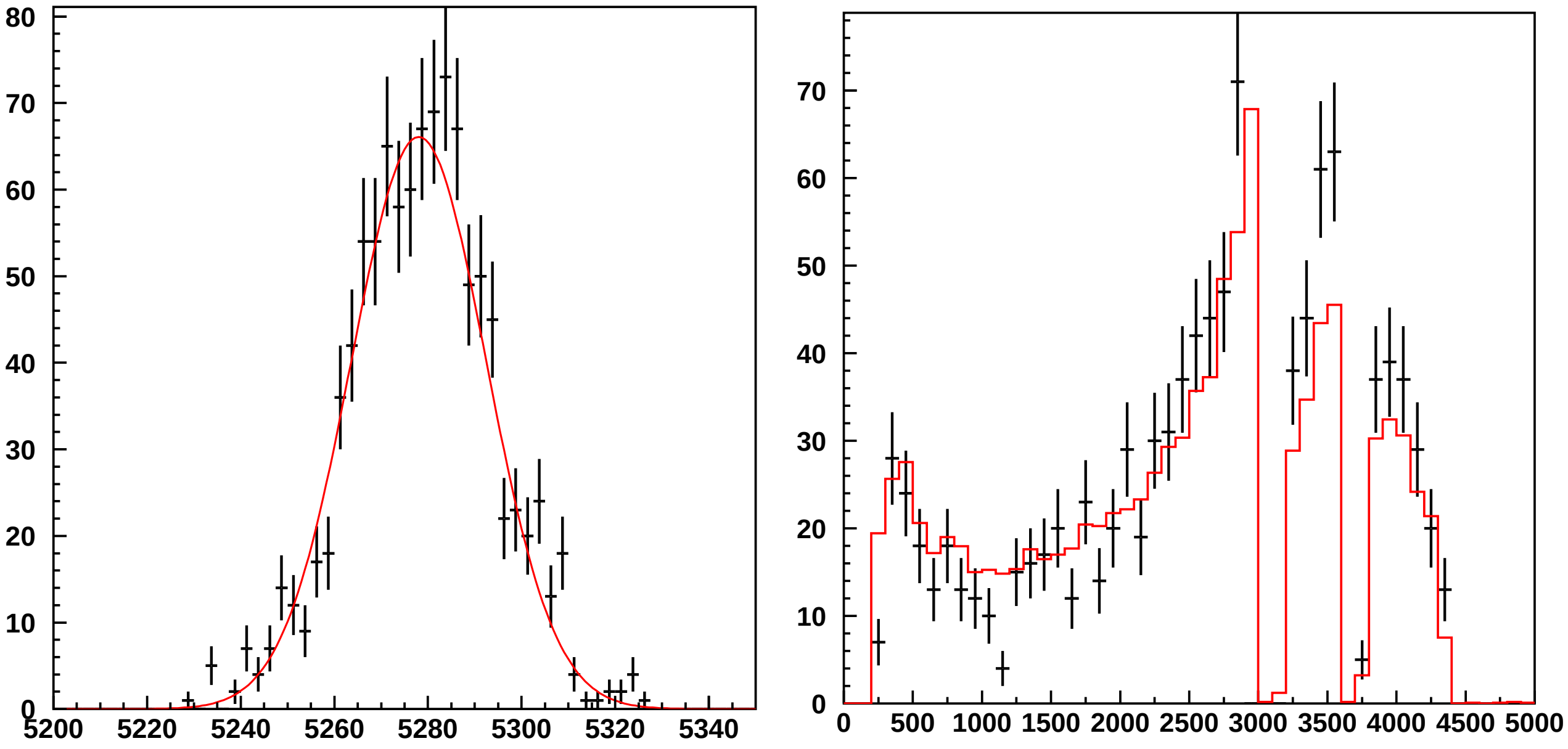

**Figure 2-35.** *LHCb simulation results for $B \to \mu\mu K^*$: $m_B$ (left), $m_{\mu\mu}$ (right). The units are* $\mathrm{MeV/c^2}$.

$B \to \mu\mu K^*$**:** A full GEANT simulation performed recently by LHC$b$ [322] estimates 10 MeV and 15 MeV mass resolutions for the $\mu\mu$ and the $B$ respectively (Fig. 2-35), and leads to the expectation of 4400 $B \to \mu\mu K^*(K^+\pi^-)$ (+c.c.) events per year. With the available Monte Carlo statistics, the background-to-signal ratio is estimated to be smaller than 2. These yields will allow a detailed study of the forward-backward asymmetry spectrum in the first year of data taking.

The zero-intercept of the forward-backward asymmetry ($\hat{s}_0$ defined in Fig. 2-36) can be determined with a precision of 0.01 allowing the determination of the ratio of Wilson coefficients $C_9^{\rm eff}/C_7^{\rm eff}$ with a 6% accuracy after two years.

Other experiments expect similar annual yields: 4000 events per year at CMS and 700 at ATLAS at $\mathcal{L} = 10^{33}\mathrm{cm}^{-2}\mathrm{s}^{-1}$, and 2500 at $B$TeV at $\mathcal{L} = 2\times10^{32}\,\mathrm{cm}^{-2}\mathrm{s}^{-1}$. All these experiments will also be able to measure the forward-backward asymmetry during their first years of operation.

**Semi-Inclusive and other modes:** LHC$b$ has also studied the semi-inclusive reconstruction of $B \to \mu\mu X_s$ and $B \to \mu\mu X_d$ decays. Channels have been considered with up to one charged or neutral ($K_S^0$) kaon and up to three charged pions. It is expected that 12000 $B \to \mu\mu X_s$ and 300 $B \to \mu\mu X_d$ events will be reconstructed each year.

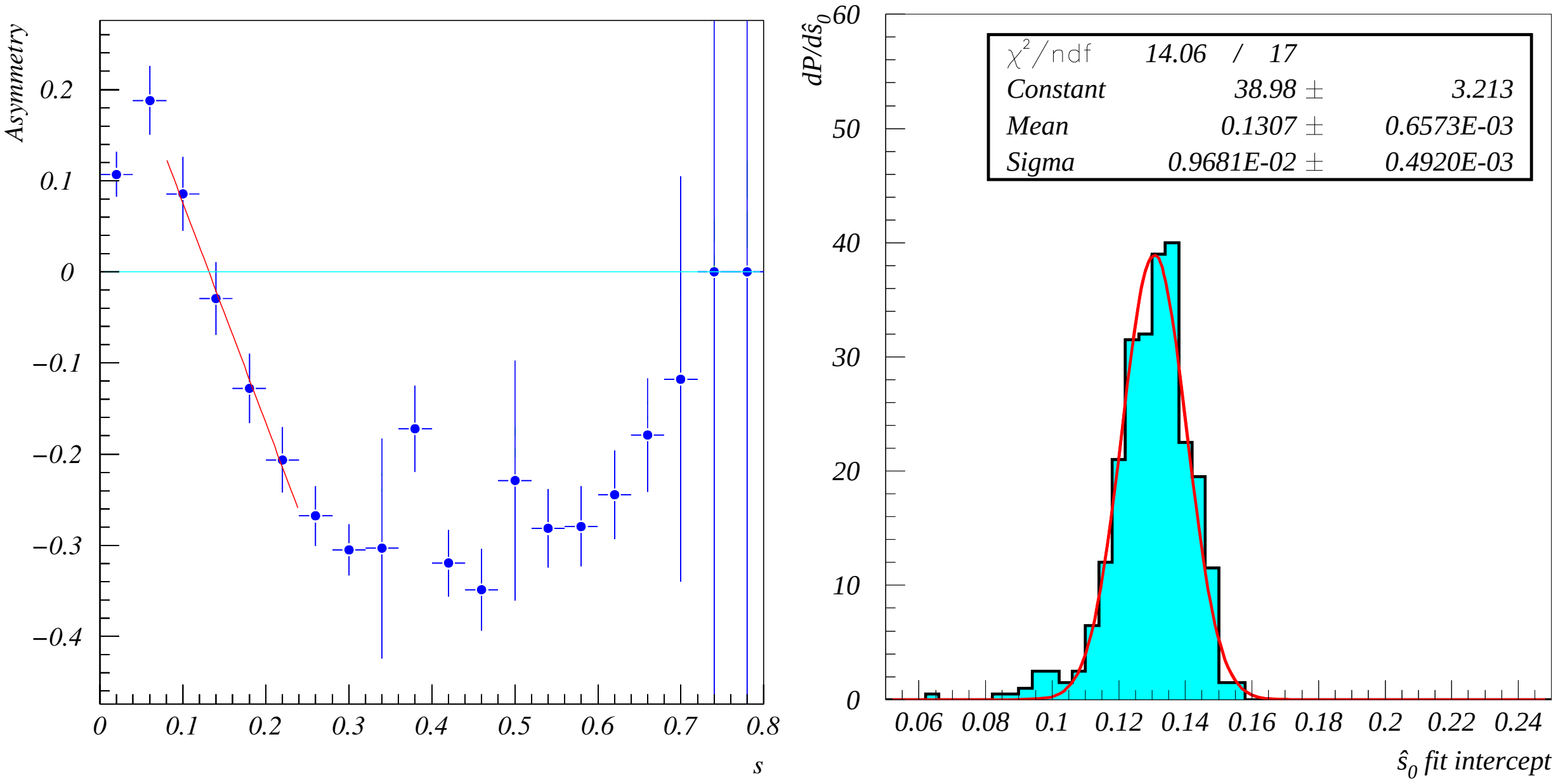

**Figure 2-36.** *Forward-backward asymmetry in $B_d \to \mu\mu K^*$ after two years of LHCb. Left: Typical reconstructed $A_{\rm FB}$ versus $\hat{s} = \left(\frac{m_{\mu\mu}}{m_B}\right)^2$ with linear fit to determine the zero-intercept $\hat{s}_0$. Right: $\hat{s}_0$ distribution for several 2-year pseudo-experiments.*

These numbers include about 4000 $B \to \mu\mu K$ and 100 $B_d^0 \to \mu\mu\rho$ events. Here the particle identification provided by the LHC$b$ RICH detector is of crucial importance in disentangling the two modes.

The reconstruction of the $B \to \mu\mu X_d$ modes allows the extraction of the CKM matrix elements ratio $|V_{td}|/|V_{ts}|$ with a 5% uncertainty (statistical error only) by the end of the LHC era. This precision is comparable in magnitude with what is expected from $B_s$ oscillation measurements.

$B$TeV has a similar strategy to reconstruct the inclusive $B \to \mu\mu X_s$ mode and expects a $20\sigma$ significant signal after one year [323] ATLAS and CMS also expect to observe the $B_d^0 \to \mu\mu\rho$ decay during their first years of operation [324].

The $B \to eeK^*$ decay has not yet been studied at hadron machines. The main problem is the energy-loss through bremsstrahlung, which causes the $B \to J/\psi K^*$ decay to be a major source of background. Yet it is plausible that this decay can be used at LHC$b$ and $B$TeV in selected dilepton mass ranges, above or far below the $\psi$ resonances.

## 2.24.3 Radiative decays

In the field of radiative decays, the reach of experiments at hadron colliders is limited to exclusive decays such as $B \to K^*\gamma$, $B_s \to \phi\gamma$ or $B \to \rho\gamma$. Thanks to the first-level electromagnetic trigger selecting photons with high transverse energy, the LHC$b$ experiment can achieve reasonable selection efficiencies for such channels. Here also the vertex detector plays an important rôle in selecting $K^* \to K\pi$ vertices well detached from the primary interaction vertex.

LHC$b$ expects to see $35\,000$ $B \to K^*\gamma$ decays per year with a background-to-signal ratio smaller than $0.7$. ATLAS expects about $10\,000$ events ($\mathcal{L} = 10^{33}\,{\rm cm}^{-1}{\rm s}^{-2}$). Because of its ${\rm PbWO}_4$ calorimeter $B$TeV may well obtain even higher yields. All these experiments should be able to measure the $B \to \rho\gamma$ mode as well and thus place some constraints on $|V_{td}|/|V_{ts}|$ through these decays.

### 2.24.4 $B \to \mu\mu$

The hadron collider experiments all plan to search for the leptonic decay $B_s \to \mu\mu$. CMS for instance has designed a trigger for such events, allowing the reconstruction of the $B_s$ mass with a $70\ \mathrm{MeV}$ precision already at trigger level. Using the full tracker, this resolution improves to $45\,\mathrm{MeV}$, allowing the selection of 26 events per year at the full LHC luminosity of $\mathcal{L} = 10^{34}\,\mathrm{cm}^{-2}\mathrm{s}^{-1}$. ATLAS expects 92 events in the same conditions but with a higher background and LHC$b$ 16 events at its nominal luminosity of $2 \times 10^{32}\ \mathrm{cm}^{-2}\mathrm{s}^{-1}$ [324, 325].

Whether the $B_d \to \mu\mu$ decay can be seen at the LHC is not yet certain. CMS studies suggest that this is feasible, provided it can be clearly distinguished from the close-lying $B_s \to \mu\mu$ peak.

### 2.24.5 Conclusions

The very high $b\bar{b}$ production cross-section at the hadron machines leads to an impressive performance in the search for and study of rare $B$ decays. Experiments at these facilities are particularly suited to the full reconstruction of exclusive modes having a charged track vertex. At the LHC many interesting physics topics can be studied in detail with one year's data sample, for example the forward-backward asymmetry of $B \to \mu\mu K^{*}$.

## 2.25 Rare decays in MFV models

>– G. Isidori –<

### 2.25.1 The basis of MFV operators

The Minimal Flavor Violation (MFV) hypothesis links the breaking of $CP$ and flavor symmetries in extensions of the Standard Model to the known structure of Standard Model Yukawa couplings [326, 327, 140, 328]. As shown in Ref. [140], this hypothesis can be formulated in terms of a renormalization-group-invariant symmetry argument, which holds independently of any specific assumption about the dynamics of the New Physics framework. The two main hypotheses are the following:

- The ordinary Standard Model fields (including at least one Higgs doublet) are the only light degrees of freedom of the theory.
- The three Yukawa couplings ($Y_D$, $Y_U$, and $Y_E$) are the only source of breaking the large flavor-symmetry group of the Standard Model fields: $\mathrm{U}(3)_{Q_L} \otimes \mathrm{U}(3)_{U_R} \otimes \mathrm{U}(3)_{D_R} \otimes \mathrm{U}(3)_{L_L} \otimes \mathrm{U}(3)_{E_R}$.

Combining these two hypotheses, or building effective gauge- and flavor-invariant operators in terms of $Y$ and Standard Model fields, we can construct the most general basis of new operators (with dimension $\geq 6$) compatible with the MFV criterion (see Ref. [140] and the New Physics chapter of this book).

As long as we are interested only in rare FCNC decays, this general formulation—assuming only one light Higgs doublet (or small $\tan\beta$)—is equivalent to the approach of Ref. [327]: all the non-standard effects can be encoded in the initial conditions of the ordinary Standard Model effective FCNC Hamiltonian

$$\mathcal{H}_{\rm eff}^{\Delta F=1} = \frac{G_{\rm F}\alpha}{2\sqrt{2}\pi\sin\theta_W} V^*_{3i}V_{3j}\sum_n C_n \mathcal{Q}_n \; + \; \text{h.c.} \tag{2.163}$$

basis In other words, all the $C_i(M_W^2)$ that are non-vanishing within the Standard Model should be considered as independent free parameters of the model. Note that the framework is still very predictive, since the same set of flavor-independent coefficients should describe FCNC amplitudes in $b \to d$, $b \to s$ and $s \to d$ transitions.

### 2.25.2 Bounds from rare decays

Rare FCNC decays are the best probe of the MFV scenario for two main reasons: i) in such processes the New Physics effect is naturally of the same order as the Standard Model contribution; ii) we have a direct access to the magnitude of the amplitude and not only to its phase (by construction, within the MFV framework, the weak phase of the amplitude is not sensitive to non-standard effects).

A detailed discussion of the phenomenological consequences of the MFV hypothesis on several rare decays can be found in Refs. [140, 328]. On general grounds, the initial conditions of the Wilson coefficients receive corrections of the type

$$\frac{\delta C_i(M_W^2)}{C_i^{\rm SM}(M_W^2)} = \mathcal{O}\left(\frac{\Lambda_0^2}{\Lambda^2}\right) , \qquad \Lambda_0 = \frac{\lambda_t \sin\theta_W M_W}{\alpha} \approx 2.4 \text{ TeV} , \tag{2.164}$$

where $\Lambda$ denotes the effective scale of the new degrees of freedom and $\Lambda_0$ is the typical scale associated to the Standard Model electroweak contribution. For this reason, an experimental determination of the $C_i(M_W^2)$ with a precision $p$, allow to set bounds of $\mathcal{O}(\Lambda_0/\sqrt{p})$ on the effective scale of New Physics. In observables for which the theoretical error is around or below $10\%$, precision experiments on rare decays could aim at probing effective scales of New Physics up to $\sim 10$ TeV. Such bounds would compete with the limits on flavor-conserving operators derived from electroweak precision tests. Thus, at this level of precision, there is a realistic chance of detecting deviations from the Standard Model.

It is worth recalling that all the $C_i(M_W^2)$ could be determined by experimental data on one type of $d_i \to d_j$ amplitudes only. Thus, in the presence of deviations from the Standard Model, the consistency of the MFV hypothesis could be tested experimentally by comparing different types of FCNC transitions (namely $b \to d$, $b \to s$ and $s \to d$).

Thus far, the only FCNC observable in which a $10\%$ error has been reached, both on the theoretical and the experimental sides, is the inclusive $B \to X_s\gamma$ rate. This precise information allows us to derive a significant constraint on the effective operator

$$\mathcal{O}_{F1} = eH^\dagger \left( \overline{D}_R Y_D Y_U^\dagger Y_U \sigma_{\mu\nu} Q_L \right) F_{\mu\nu} \, . \tag{2.165}$$

Defining its overall coefficient to be $1/\Lambda^2$, the present 99% CL bound is $\Lambda > 6.4$ (5.0) TeV in the case of constructive (destructive) interference with the Standard Model amplitude [140]. The bound could grow up to $\sim 10$ TeV with a 5% measurement of the rate, and, at the same time, a theoretical calculation at the NNLL level.

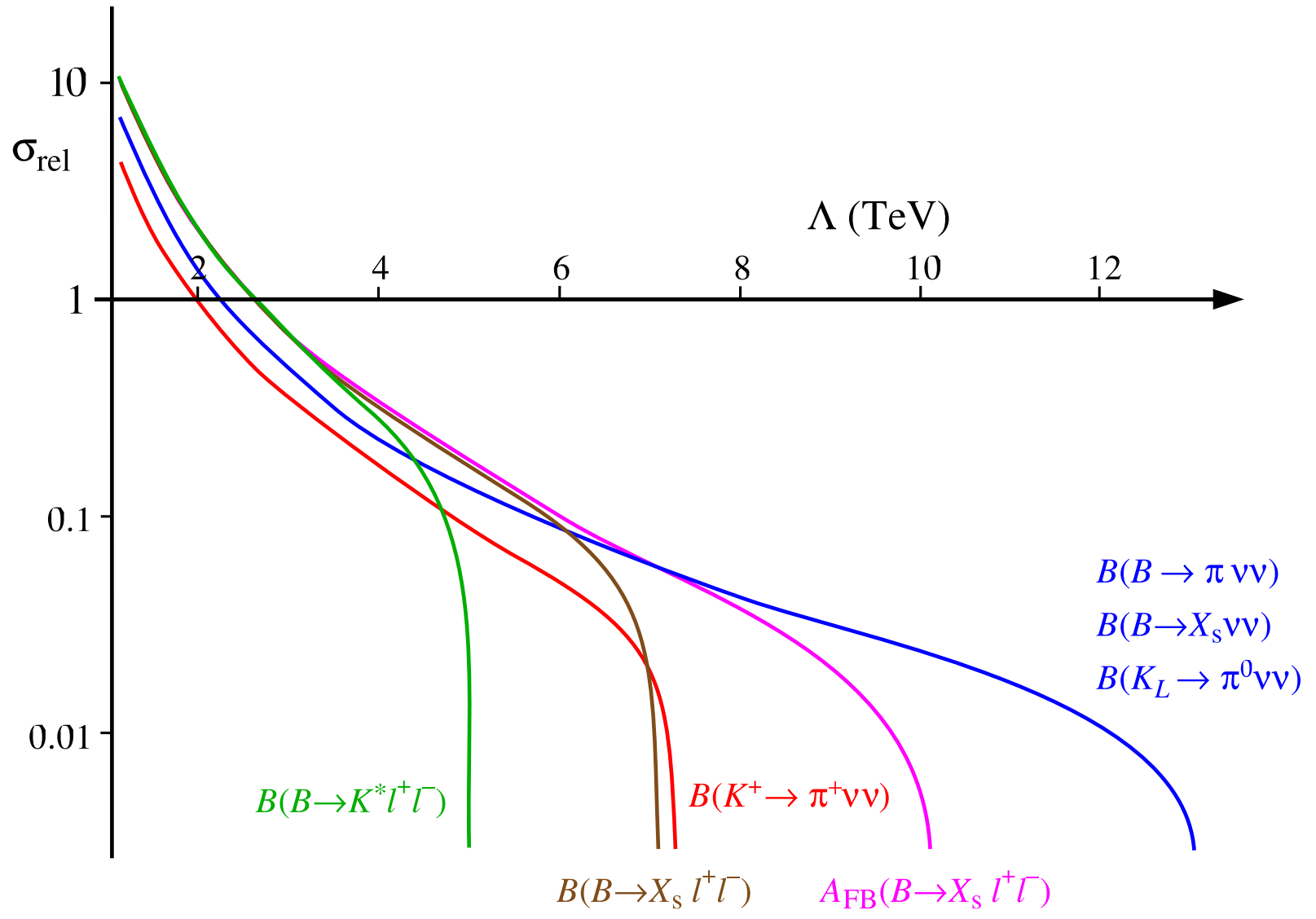

**Figure 2-37.** *Comparison of the effectiveness of different rare decay observables in setting future bounds on the scale of the representative operator $(\overline{Q}_L Y_U^\dagger Y_U \gamma_\mu Q_L)(\overline{L}_L \gamma_\mu L_L)$ within MFV models. The vertical axis indicates the relative precision of an hypothetic measurement of the observable, with central value equal to the Standard Model expectation. All the curves are obtained assuming a 1% precision on the corresponding overall CKM factor.*

The present bounds from other FCNC observables are much weaker, essentially because of larger experimental errors. In Fig. 2-37 we compare the potential sensitivity of future measurements. By means of the experimentally difficult but theoretically clean observables, such as the lepton forward-backward asymmetry in $B \to X_s\ell^+\ell^-$ or the $B \to X_s\nu\overline{\nu}$ rate, it would be possible to reach very high scales. On the contrary, in the most accessible observables the theoretical error provides a serious limitation. It must be stressed that the bounds on different operators cannot be trivially compared: the coefficients are expected to be of comparable magnitude; however, some differences are naturally foreseen. Then, for instance, the $B \to X_s\gamma$ bound on the operator (Eq. (2.165)) does not exclude the possibility of New Physics effects in $B \to X_s\ell^+\ell^-$, corresponding to an effective scale below 5 GeV for the operator $(\overline{Q}_L Y_U^\dagger Y_U \gamma_\mu Q_L)(\overline{L}_L \gamma_\mu L_L)$. Therefore, a systematic study of all the available observables is very important.

### 2.25.3 The large $\tan\beta$ scenario

By construction, the minimal basis of FCNC operators illustrated above (which coincides with the Standard Model basis), is valid for all the MFV models where there is only one light Higgs doublet. In models with more Higgs doublets, there is more freedom: the breaking of the SU(3) flavor subgroups, necessarily induced by the Yukawa couplings, can be separated from the breaking of (some of) the U(1) groups [140].

For instance, in two-Higgs doublet (2HD) models of type-II (such as the Higgs sector of the MSSM), the Yukawa interaction is invariant under an extra U(1) symmetry. The latter is necessary to forbid tree level FCNCs, which would arise if $H_U$ and $H_D$ can couple to both up and down type quarks. In this framework the smallness of the $b$ quark and $\tau$ lepton masses is naturally attributed to the smallness of $\langle H_D \rangle / \langle H_U \rangle = 1/\tan\beta$ and not to the hierarchy of the corresponding Yukawa couplings. As a result, within this framework $Y_D$ represents a new non-negligible source of flavor-symmetry breaking. This fact leads to a series of interesting consequences for all the helicity-suppressed rare decays that have been discussed in detail in the recent literature (see *e.g.,* Refs. [140, 250, 17, 20, 329, 23, 22, 288, 129] and references therein).

The main new feature is the enlargement of the basis of relevant FCNC operators with the inclusion of scalar operators, such as $\overline{b}_R s_L \overline{\mu}_R \mu_L$. Scalar FCNC amplitudes are present within the Standard Model, but they are negligible due to the smallness of down-type Yukawa couplings. This condition is no longer valid in 2HD models with large $\tan\beta$. Within this framework, scalar operators can induce spectacular effects (such as a two orders of magnitude enhancement of the rate) in $B \to \ell^+\ell^-$ decays. Interestingly, these enhancements are possible even if the FCNC amplitude is still proportional to the standard CKM factor $V_{3i}^* V_{3j}$ (because of the MFV hypothesis). Moreover, this interesting scenario does not pose any serious fine-tuning problem with the existing data, since the effect of scalar operators is still quite small in non-helicity-suppressed observables.

In principle, a sizable enhancement of the $B \to \ell^+\ell^-$ rates could be detected also at future hadronic machines. However, Super $B$ Factorieswould still play a very important role in clearly identifying this scenario with i) precision measurements of the non-helicity-suppressed decays; ii) experimental searches of the $B \to (X)\tau^+\tau^-$ modes [23]. In the non-helicity-suppressed modes one does not expect spectacular effects; however, the new scalar operators should induce $\mathcal{O}(10\%)$ breaking of lepton universality in FCNC processes of the type $B \to (K, K^*)\ell^+\ell^-$ [129]. Moreover, the MFV hypothesis implies a strict correlation between $b \to s$ and $b \to d$ amplitudes, which can be studied only at $B$ factories.

## 2.26 Experimental Prospects for Rare Charm Decays

>– M. Purohit and D.C. Williams –<

A Super $B$ Factory will produce copious amounts of charm particles. For example, given that the charm cross section near the $\Upsilon(4S)$ is approximately 1.3 nb, a total of 13 billion $c\overline{c}$ pairs would be expected for $10\ \mathrm{ab}^{-1}$ of luminosity. Combined with the low backgrounds characteristic of an $e^+e^-$ environment, such a Super $B$ Factory is an ideal place to search for rare charm decays.

### 2.26.1 Di-lepton searches

The CLEO Collaboration has published a set of limits [330] on $D^0$ di-lepton decays based on 3.85 $\mathrm{fb}^{-1}$ of data. The CLEO analysis can be scaled by luminosity to estimate the limits that can be obtained with higher luminosity. Most likely these estimates are conservative, since the event selection requirements used by CLEO were optimized to best match the size of their data set.

Branching ratio limits can be estimated from three numbers: the size of the charm meson sample, the event selection efficiency, and the size of the background. The event selection efficiency in the CLEO analysis ranged from 14% (for $D^0 \to e^+e^-$) to 1% (for $D^0 \to X\mu^+\mu^-$). Approximate background levels can be estimated from the mass plots in their paper. The 90% confidence level results are shown in Table 2-17, using a Bayesian calculation that incorporates Poisson statistical fluctuations in the size of the background. These estimates are compared against current limits [330, 331, 332, 333] and theoretical predictions [334].

**Table 2-17.** *Estimated 90% confidence limits on the branching fraction for various rare and forbidden charm meson di-lepton decays and Standard Model predictions.*

| | Experiment Limit ($\times 10^{-6}$) | | | Standard Model Prediction [334] |
|---|---|---|---|---|
| Decay Mode | Current | 100 $\mathrm{fb}^{-1}$ | 10 $\mathrm{ab}^{-1}$ | (Long Distance) |
| $D^+ \to \pi^+e^+e^-$ | 52 [331] | 5.2 | 0.47 | $2 \times 10^{-6}$ |
| $D^+ \to \pi^+\mu^+\mu^-$ | 9 [2] | 8.6 | 0.80 | $1.9 \times 10^{-6}$ |
| $D^+ \to \rho^+\mu^+\mu^-$ | 560 [2] | 58.8 | 5.56 | $4.5 \times 10^{-6}$ |
| $D^0 \to \pi^0e^+e^-$ | 45 [330] | 0.4 | $< 0.01$ | $8 \times 10^{-7}$ |
| $D^0 \to \rho^0e^+e^-$ | 100 [330] | 1.3 | 0.12 | $1.8 \times 10^{-6}$ |
| $D^0 \to \rho^0\mu^+\mu^-$ | 22 [332] | 7.8 | 0.70 | $1.8 \times 10^{-6}$ |
| $D^0 \to e^+e^-$ | 6 [331] | 1.0 | $< 0.1$ | $1 \times 10^{-13}$ |
| $D^0 \to \mu^+\mu^-$ | 4 [333] | 1.0 | $< 0.1$ | $1 \times 10^{-13}$ |
| $D^0 \to e^+\mu^-$ | 8 [331] | 8.0 | $< 0.1$ | 0 |
| $D^+ \to \pi^+e^+\mu^-$ | 34 [331] | 11.9 | 1.10 | 0 |
| $D^0 \to \rho^0e^+\mu^-$ | 49 [330] | 2.9 | 0.26 | 0 |

### 2.26.2 Radiative decays

The Standard Model branching fraction for various radiative charm meson decays has been estimated [335] to range from $10^{-4}$ to $10^{-6}$. A strong contribution from nonperturbative processes (vector dominance) introduces large uncertainties in these calculations and so a measurement of just the branching ratio is unlikely to uncover New Physics. The $\gamma$ spectrum from these decays, however, is potentially interesting, especially if the $CP$ asymmetry of this spectrum is measured. The large contribution from vector poles could be an advantage if the interference of New Physics produces $CP$ asymmetries in the pole shapes.

The CLEO Collaboration has placed limits on four photon radiative decay modes [336]. The sensitivity of this analysis can be extrapolated to higher luminosities using a method similar to that used for Table 2-17, assuming no signal. The results are shown in Table 2-18. Given the predicted Standard Model branching ratios, however, some of these decay modes should be observable by even the existing $B$ Factories. One example is the decay $D^0 \to \phi\gamma$, already reported by the Belle Collaboration in a preliminary analysis [337].

**Table 2-18.** *Estimated 90% confidence limits on the branching fraction for various charm meson $\gamma$ radiative decays and Standard Model predictions. Also shown is a preliminary branching fraction measurement from the Belle Collaboration.*

| | Sensitivity ($\times 10^{-5}$) | | | Standard Model | Belle |
|---|---|---|---|---|---|
| Decay | CLEO II | Estimated | | Prediction | Measurement |
| Mode | Limits [336] | 100 $\mathrm{fb}^{-1}$ | 10 $\mathrm{ab}^{-1}$ | ($10^{-5}$) [335] | ($10^{-5}$) [337] |
| $\phi\gamma$ | 19 | 0.1 | 0.01 | 0.1–3.4 | $2.6^{+0.7}_{-0.6}$ |
| $\omega\gamma$ | 24 | 0.6 | 0.06 | $\sim 0.2$ | — |
| $K^*\gamma$ | 76 | 0.5 | 0.05 | $\sim 0.01$ | — |
| $\rho\gamma$ | 24 | 0.2 | 0.02 | 0.1–0.5 | — |

## 2.26.3 The competition

Because the search for rare decays benefits from high statistics, hadron collider experiments are potential rivals of a Super $B$ Factory. Hadron colliders tend to produce more background and require more sophisticated triggers, both of which adversely affect rare decay searches. An example of a recent hadron collider result is from CDF, which has placed a 90% confidence limit of $2.5 \times 10^{-6}$ on the branching ratio for the decay $D^0 \to \mu^+\mu^-$ based on 65 $\mathrm{pb}^{-1}$ of data [338]. Luminosity projections [339] suggest an increase by a factor of 30 in statistics at the Tevatron by 2008; not sufficient to remain competitive with a Super $B$ Factory.

More serious competition can be expected from LHC experiments such as LHC$b$ ATLAS, and CMS. It is not clear at this time how much effort those collaborations will invest in charm physics.

## 2.27 Experimental Aspects of $D^0\overline{D}^0$ Mixing

⋟ M. Purohit and D.C. Williams ⋞

The search for $D^0 - \overline{D}^0$ mixing promises to be a fertile ground in the search for New Physics, since the Standard Model predicts that $r_{\rm mix} \sim 10^{-7}$ while several models of New Physics predict a higher rate [340].

### 2.27.1 Existing results

After early results from experiments including BCDMS, E615, ARGUS, E691, and Mark III, more recent results on $D^0 - \overline{D}^0$ mixing have come from E791, FOCUS, CLEO, and *BABAR*. We will cite results from E791 and FOCUS as examples of hadronic experiments and then attempt to link these to CDF mass plots to make predictions for the future. For future $e^+e^-$ experiments we can extrapolate from *BABAR* results.

Using semileptonic decays, the E791 Collaboration found [341] that $r_{\rm mix} < 0.5$%. In the $e$ and $\mu$ channels E791 had $\sim 1250$ right sign (RS) events and $\sim 500 - 600$ wrong-sign (WS) events. Using hadronic decays to $K\pi$ and $K\pi\pi\pi$, E791 obtained [342] various limits depending on assumptions about $CP$ violation. Assuming no $CP$ violation, E791 found $r_{\rm mix} < 0.4$%, assuming $CP$ violation in the interference term only $r_{\rm mix} < 1.1$% and allowing for $CP$ violation the results were $r_{\rm mix} < 1.1$%, 1.9% depending on the direction of the mixing. In the first direct comparison of the $D^0$ lifetime in the $KK$ and $K\pi$ decay modes, E791 obtained $y = (0.9 \pm 2.9 \pm 1.8)$%, where $y$ is defined as $\Delta\Gamma/2\Gamma$. (Similarly, $x$ is defined as $\Delta m/\Gamma$ and a strong phase difference between RS and WS decays rotates these to $x'$ and $y'$.) FOCUS has measured [343] a non-zero value $y = (3.4 \pm 1.4 \pm 0.7)$%, and should have results from their hadronic decay modes soon. Their semi-muonic result is $r_{\rm mix} < 0.131$% [344].

Using 57 fb$^{-1}$, *BABAR* has studied [345] $D^0$ decays in the $K^\pm\pi^\mp$ modes to obtain detailed limits in the $x'^2$-$y'$ plane. There are 120,000 RS events and 430 WS events. At the risk of oversimplification, we can say that the limit on $r_{\rm mix}$ is 1.3$\times10^{-3}$ assuming no $CP$ violation, and 1.6$\times10^{-3}$ allowing for $CP$ violation in the fit. Similarly, *BABAR* has measured both $y$ and $\Delta y$ [346], where $\Delta y$ is approximately the asymmetry in the $KK$ or $\pi\pi$ lifetimes from $D^0$ vs. from $\overline{D}^0$. The results based on 91 fb$^{-1}$ are $y = (0.8 \pm 0.4 \pm 0.5)$% and $\Delta y = (-0.8 \pm 0.6 \pm 0.2)$%. (As of this writing, Belle has published results based only on a smaller sample.)

### 2.27.2 The future

In the near future, we can expect both *BABAR* and Belle to publish updated results based on larger samples. CLEO-$c$ should also produce results in a few years. One can expect *BABAR*'s results to scale as 1/$\sqrt{N}$ and CLEO-$c$'s sensitivity to mixing is comparable to *BABAR* with its full data sample [347]. Therefore, with 50 ab$^{-1}$ we might expect a sensitivity to $r_{\rm mix}$ below $\sim 0.5 \times 10^{-4}$. Similarly, a 50 ab$^{-1}$ experiment should be sensitive to $y$ and $\Delta y$ at the $2.0 \times 10^{-4}$ level. Below we make projections for the limit on $r_{\rm mix}$ from hadronic experiments.

Large hadronic experiments are either in progress (CDF and D0) or gearing up to get data (LHC$b$ and $B$TeV). Fermilab expects to deliver 4 to 8 fb$^{-1}$ in Run II. LHC$b$ and $B$TeV are longer-term experiments that should have results comparable to each other. CDF has shown [348] a very preliminary mass peak with 5.8 pb$^{-1}$ in which they see 5515 $\pm$ 85 $D^{*+}$ events. With the full Run II sample they should get around 20 million RS events. However, the CDF signal to background ratio (S/B) would be lower than *BABAR*'s by a factor of $\sim$ 11, for two reasons: the width of the mass peak is about four times larger (800 keV vs. 200 keV) and the background levels are much higher. Indeed, only $\sim$ 20% of the CDF background seems to be predominantly $D$ mesons combined with random pions, and the rest seems to be largely due to combinatorics of random tracks [349].

Because of strong correlations between the terms linear and quadratic in time in the decay time distribution, one cannot easily estimate the limit on $r_{\rm mix}$ from the size of the data sets. However, judging from *BABAR*'s measured limit as compared to the limit to be expected from a simple $\sqrt{B}/S$ estimate, one might expect that with full Run II data CDF will get to a limit which is a factor of three to ten better than *BABAR*'s current result. Thus, they should be competitive with the full *BABAR* sample or a little better. $B$TeV and LHC$b$ hope to obtain 50 million $D^{*+}$'s per year [350]. Their background levels are even harder to estimate than CDF's. If we guess similar S/B ratios as CDF, then we might expect samples ten times those of CDF. Better particle ID might make their backgrounds somewhat lower than those of CDF. In any case, those hadronic experiments should then achieve the same sensitivity to $r_{\rm mix}$ as a 10 ab$^{-1}$ $e^+e^-$ experiment, but perhaps not competitive with a 50 ab$^{-1}$ $e^+e^-$ experiment. Note that these are extrapolations over many orders of magnitude from the present 5.8 pb$^{-1}$ CDF $D^{*+}$ mass peak and hence should be taken with a large grain of salt. It is not clear that the mass peak from hadronic experiments can be made much narrower; the lower S/B due to combinatoric background, however, could, perhaps, be somewhat reduced.

## 2.28 Theoretical Prospects for Rare Charm Decays

≻— G. Burdman —≺

The study of flavor changing neutral currents (FCNC) has been focused on processes involving $K$ and $B$ mesons such as $K^0$–$\overline{K}^0$ and $B^0\overline{B}^0$ mixing, and on rare decays involving transitions such as $s \to d\ell^+\ell^-$, $s \to d\nu\overline{\nu}$, $b \to s\gamma$, and $b \to s\ell^+\ell^-$. The analogous FCNC processes in the charm sector have received considerably less scrutiny. This is perhaps because, on general grounds, the Standard Model ( Standard Model ) expectations are very small for both $DD^0\overline{D}^0$ mixing and rare charm decays. For instance, no large non-decoupling effects arise from a heavy fermion in the leading one-loop contributions. This is in sharp contrast with $K$ and $B$ FCNC processes, which are affected by the presence of the virtual heavy top quark. In the Standard Model, $D$-meson FCNC transitions involve the rather light down-quark sector, which implies an efficient Glashow-Iliopoulos-Maiani (GIM) cancellation. If it turns out that the charm-quark mass is not heavy enough compared to a typical scale of hadronic effects, long-distance effects are likely to dominate. They will obscure the more interesting short-distance contributions that are the true test of the Standard Model. Large long-distance effects are expected in both $D^0\overline{D}^0$ mixing and FCNC charm decays. In the case of mixing, although the long-distance effects dominate over the Standard Model short-distance contributions, there could still be a significant window between these and the current experimental limits. The predictions of numerous extensions of the Standard Model lie in this window. In the case of rare charm decays, for some modes a window exists in which theoretical predictions are sufficiently under control to allow tests of the short-distance structure of the FCNC transition. This happens for $c \to u\ell^+\ell^-$ modes, and therefore we mainly concentrate on their potential. Radiative charm decays, such as those mediated by $c \to u\gamma$, are largely dominated by long-distance physics. Their experimental accessibility presents an opportunity to study purely nonperturbative effects. In the following we review the Standard Model predictions for the leptonic, semileptonic, and radiative decays in Section 2.28.1, and in Section 2.28.2 we study the potential for New Physics signals in $c \to u\ell^+\ell^-$ decays.

### 2.28.1 The Standard Model predictions

The short-distance contributions to the $c \to u$ transitions are induced at one loop in the Standard Model. It is convenient to use an effective description with the $W$ boson and the $b$ quark being integrated out as their respective thresholds are reached in the renormalization group evolution [351]. The effective Hamiltonian is given by [352, 353, 354]

$$
\begin{aligned}
\mathcal{H}_{\rm eff} &= -\frac{4G_F}{\sqrt{2}}\left[\sum_{q=d,s,b} C_1^{(q)}(\mu)O_1^{(q)}(\mu) + C_2^{(q)}(\mu)O_2^{(q)}(\mu)\right.\\
&\qquad \left. + \sum_{i=3}^{8} C_i(\mu)O_i(\mu)\right], \text{ for } m_b < \mu < M_W \\
\mathcal{H}_{\rm eff} &= -\frac{4G_F}{\sqrt{2}}\left[\sum_{q=d,s} C_1^{(q)}(\mu)O_1^{(q)}(\mu) + C_2^{(q)}(\mu)O_2^{(q)}(\mu)\right.\\
&\qquad \left. + \sum_{i=3}^{8} C_i'(\mu)O_i'(\mu)\right], \text{ for } \mu < m_b\,, \qquad (2.166)
\end{aligned}
$$

with $\{O_i\}$ being the complete operator basis, $\{C_i\}$ the corresponding Wilson coefficients, and $\mu$ the renormalization scale; the primed quantities are those for which the $b$ quark has been eliminated. In Eq. (2.166), the Wilson coefficients contain the dependence on the CKM matrix elements $V_{qq'}$. The CKM structure of these transitions differs drastically from that of the analogous $B$ meson processes. The operators $O_1$ and $O_2$ are explicitly split into their CKM components,

$$
O_1^{(q)} = (\overline{u}_L^\alpha\gamma_\mu q_L^\beta)(\overline{q}_L^\beta\gamma^\mu c_L^\alpha)\,, \qquad O_2^{(q)} = (\overline{u}_L^\alpha\gamma_\mu q_L^\alpha)(\overline{q}_L^\beta\gamma^\mu c_L^\beta)\,, \qquad (2.167)
$$

where $q = d, s, b$, and $\alpha$, $\beta$ are contracted color indices. The rest of the operator basis is defined in the standard way. The matching conditions at $\mu = M_W$ for the Wilson coefficients of the operators $O_{1-6}$ are given as

$$C_1^q(M_W) = 0, \qquad C_{3-6}(M_W) = 0, \qquad C_2^q(M_W) = -\lambda_q, \tag{2.168}$$

with $\lambda_q = V_{cq}^* V_{uq}$. The corresponding conditions for the coefficients of the operators $O_{7-10}$ read as follows

$$\begin{aligned}
C_7(M_W) &= -\frac{1}{2}\left\{\lambda_s F_2(x_s) + \lambda_b F_2(x_b)\right\}, \\
C_8(M_W) &= -\frac{1}{2}\left\{\lambda_s D(x_s) + \lambda_b D(x_b)\right\}, \\
C_9^{(\prime)}(M_W) &= \sum_{i=s,(b)} \lambda_i \left[-\left(F_1(x_i) + 2\overline{C}(x_i)\right) + \frac{\overline{C}(x_i)}{2s_w^2}\right], \\
C_{10}^{(\prime)}(M_W) &= -\sum_{i=s,(b)} \lambda_i \frac{\overline{C}(x_i)}{2s_w^2}.
\end{aligned} \tag{2.169}$$

In Eq. (2.169), we used $x_i = m_i^2/M_W^2$; the functions $F_1(x)$, $F_2(x)$, and $\overline{C}(x)$ are those derived in Ref. [355], and the function $D(x)$ is defined in Ref. [352].

At leading order, operators in addition to $O_7$, $O_9$, and $O_{10}$ contribute to the rate of $c \to u\ell^+\ell^-$. Even in the absence of the strong interactions, the insertion of $O_2^{(q)}$ at one-loop gives a contribution from lowest order mixing onto $O_9$ [233]. When the strong interactions are included, further mixing of the four-quark operators with $O_{7-10}$ occurs. The effect of these QCD corrections in the renormalization group running from $M_W$ down to $\mu = m_c$ is particularly important in $C_7^{\rm eff}(m_c)$, the coefficient determining the $c \to u\gamma$ amplitude. As was shown in Ref. [352], the QCD-induced mixing with $O_2^{(q)}$ dominates $C_7^{\rm eff}(m_c)$. The fact that the main contribution to the $c \to u\gamma$ amplitude comes from the insertion of four-quark operators inducing light-quark loops signals the presence of large long-distance effects. This was confirmed [352, 353] when these nonperturbative contributions were estimated and found to dominate the rate. Therefore, we must take into account effects of the strong interactions in $C_7^{\rm eff}(m_c)$. On the other hand, the renormalization group running does not affect $O_{10}$, *i.e.*, $C_{10}(m_c) = C_{10}(M_W)$. Thus, in order to estimate the $c \to u\ell^+\ell^-$ amplitude, it is a good approximation to consider the QCD effects only where they are dominant, namely in $C_7^{\rm eff}(m_c)$, whereas we expect these to be less dramatic in $C_9^{\rm eff}(m_c)$.

The one loop insertion of $O_2^{(q)}$ induces an effective coefficient for $O_9$

$$C_9^{(\prime)\rm eff} = C_9(M_W) + \sum_{i=d,s,(b)} \lambda_i \left[-\frac{2}{9}\ln\frac{m_i^2}{M_W^2} + \frac{8}{9}\frac{z_i^2}{\hat{s}} - \frac{1}{9}\left(2 + \frac{4z_i^2}{\hat{s}}\right)\sqrt{\left|1 - \frac{4z_i^2}{\hat{s}}\right|}\,\mathcal{T}(z_i)\right], \tag{2.170}$$

where we have defined

$$\mathcal{T}(z) = \begin{cases} 2\arctan\left[\frac{1}{\sqrt{\frac{4z^2}{\hat{s}} - 1}}\right] & \text{for } \hat{s} < 4z^2 \\ \ln\left|\frac{1+\sqrt{1-\frac{4z^2}{\hat{s}}}}{1-\sqrt{1-\frac{4z^2}{\hat{s}}}}\right| - i\pi & \text{for } \hat{s} > 4z^2, \end{cases} \tag{2.171}$$

and $\hat{s} \equiv s/m_c^2$, $z_i \equiv m_i/m_c$. The logarithmic dependence on the internal quark mass $m_i$ in the second term of Eq. (2.170) cancels against a similar term in the Inami-Lim function $F_1(x_i)$ entering in $C_9(M_W)$, leaving no spurious divergences in the $m_i \to 0$ limit.[19]

[19] Fajfer et al. [356] do not take the mixing of $O_9$ with $O_2$ into account. This results in a prediction for the short-distance components that is mainly given by these logarithms.

**The $c \to u\ell^+\ell^-$ decay rates:** To estimate the differential decay rate, we use the two loop QCD-corrected value of $C_7^{\rm eff}(m_c)$ [353] and compute $C_9^{\rm eff}(m_c)$ from Eq. (2.170) and $C_{10}(m_c) = C_{10}(M_W)$ from Eq. (2.169). We obtain the inclusive branching ratios for $m_c = 1.5$ GeV, $m_s = 0.15$ GeV, $m_b = 4.8$ GeV and $m_d = 0$ as

$$\mathcal{B}(D^+ \to X_u^+ e^+e^-)^{\rm (sd)} \simeq 2\times 10^{-8}\,, \qquad \mathcal{B}(D^0 \to X_u^0 e^+e^-)^{\rm (sd)} \simeq 8\times 10^{-9}\,. \tag{2.172}$$

The dominant contributions to the rates in Eq. (2.172) come from the leading order mixing of $O_9$ with the four-quark operators $O_2^{(q)}$, that is the second term in Eq. (2.170). When considering the contributions of various New Physics scenarios, one should remember that their magnitudes must be compared to the mixing of these operators. Shifts in the matching conditions for the Wilson coefficients $C_7$, $C_9$ and $C_{10}$, even when large, may not be enough to give an observable deviation.

**The $c \to u\gamma$ rate:** The short-distance $c \to u\gamma$ contribution to radiative charm decays was first studied in detail in Ref. [352], where it was found that the effects of the leading logarithms on $C_7^{\rm eff}(m_c)$ enhanced the branching ratio by several orders of magnitude. Even with such enhancement, the rates are very small. However, it was noted in Ref. [353] that the leading logarithmic approximation was still affected by a fair amount of GIM suppression because the quark mass dependence on the resummed expressions was still mild. Going to two loops in the matrix elements of the operators in Eq. (2.166), specifically in $O_2^{(q)}$, leads to a more substantial mass dependence that in turn breaks GIM more efficiently. These two-loop contributions dominate the short-distance radiative amplitude giving [353]

$$\mathcal{B}^{\rm (sd)}(D^0 \to X\gamma) \simeq 2.5\times 10^{-8}. \tag{2.173}$$

Although this represents a very large enhancement even with respect to the leading logarithmic approximation (about five orders of magnitude!), it is still small, especially when compared with the estimated size of long-distance contributions (see below).

**Exclusive semileptonic modes:** The exclusive modes corresponding to $c \to u$ transitions are known to be dominated by long-distance dynamics. This is true for both the radiative and the semileptonic decays. For the $D \to H\gamma$ exclusive modes (*e.g.,* $H = \rho$), long-distance physics dominates all observables. However, in $D \to H\ell^+\ell^-$ decays (*e.g.,* $H = \pi, \rho$), it is possible to escape the largest long-distance contributions by looking at regions of phase space away from resonances. We now discuss in some detail the computation of $D \to \pi\ell^+\ell^-$ and $D \to \rho\ell^+\ell^-$ as presented in Ref. [354]. For completeness, the exclusive radiative and neutrino modes are also discussed below.

As a crude first estimate of the contributions of long-distance physics, we can consider the resonance process $D \to HV \to H\ell^+\ell^-$, where $V = \phi, \rho, \omega$. We isolate contributions from this particular mechanism by integrating $d\Gamma/dq^2$ over each resonance peak associated with an exchanged vector or pseudoscalar meson. The branching ratios thus obtained are in the $\mathcal{O}(10^{-6})$ range [357].

This result suggests that the long-distance contributions overwhelm the short-distance physics and any New Physics that might be present. However, this is not always the case. A more thorough treatment requires looking at all the kinematically available regions in $D \to H\ell^+\ell^-$, not just the resonance region. The effect of these states can be thought of as a shift in the short-distance coefficient $C_9^{\rm eff}$ in Eq. (2.170), since $V \to \ell^+\ell^-$ selects a vector coupling to the leptons. This follows from Ref. [358], which incorporates in a similar manner the resonant contributions to $b \to q\ell^+\ell^-$ decays via a dispersion relation for $\ell^+\ell^- \to$ hadrons. The new contribution can be written as [358]

$$C_9^{\rm eff} \to C_9^{\rm eff} + \frac{3\pi}{\alpha^2}\sum_i \kappa_i \frac{m_{V_i}\Gamma_{V_i\to\ell^+\ell^-}}{m_{V_i}^2 - s - i m_{V_i}\Gamma_{V_i}}, \tag{2.174}$$

where the sum is over the various relevant resonances, $m_{V_i}$ and $\Gamma_{V_i}$ are the resonance mass and width, and the factor $\kappa_i \sim \mathcal{O}(1)$ is a free parameter adjusted to fit the nonleptonic decays $D \to HV_i$ with on-shell $V_i$. We obtain $\kappa_\phi \simeq 3.6$, $\kappa_\rho \simeq 0.7$, and $\kappa_\omega \simeq 3.1$. The latter result comes from assuming $B(D^+ \to \pi^+\omega) = 10^{-3}$, since a direct measurement is not available yet.

$D^+ \to \pi^+ e^+ e^-$**:** The main long-distance contributions come from the $\phi$, $\rho$, and $\omega$ resonances. The $\eta$ and $\eta'$ effects are negligible. The dilepton mass distribution for this decay takes the form [354]

$$\frac{d\Gamma}{ds} = \frac{G_F^2 \alpha^2}{192\pi^5} |\vec{p}_\pi|^3 \, |f_+(s)|^2 \left( \left| \frac{2m_c}{m_D} C_7^{\rm eff} + C_9^{\rm eff} \right|^2 + |C_{10}|^2 \right), \qquad (2.175)$$

where $s = m_{ee}^2$ is the square of the dilepton mass. Here we have used the heavy-quark spin-symmetry relations that relate the matrix elements of $O_7$ to the "semileptonic" matrix elements of $O_9$ and $O_{10}$ [265, 266]. An additional form factor is formally still present, but its contribution to the decay rate is suppressed by $(m_\ell/m_D)^2$ and is neglected here. Precise measurements of $D \to \pi\ell\nu$ will give us $f_+(q^2)$. In the meantime, we make use of the prediction of chiral perturbation theory for heavy hadrons (ChPTHH) [269, 268, 359], which at low recoil gives

$$f_+(s) = \frac{f_D}{f_\pi} \frac{g_{D^*D\pi}}{(1 - s/M_{D^*}^2)}. \qquad (2.176)$$

Here we use the recent CLEO measurement [360] $g_{D^*D\pi} = 0.59 \pm 0.1 \pm 0.07$, and we take $f_D = 200$ MeV. In

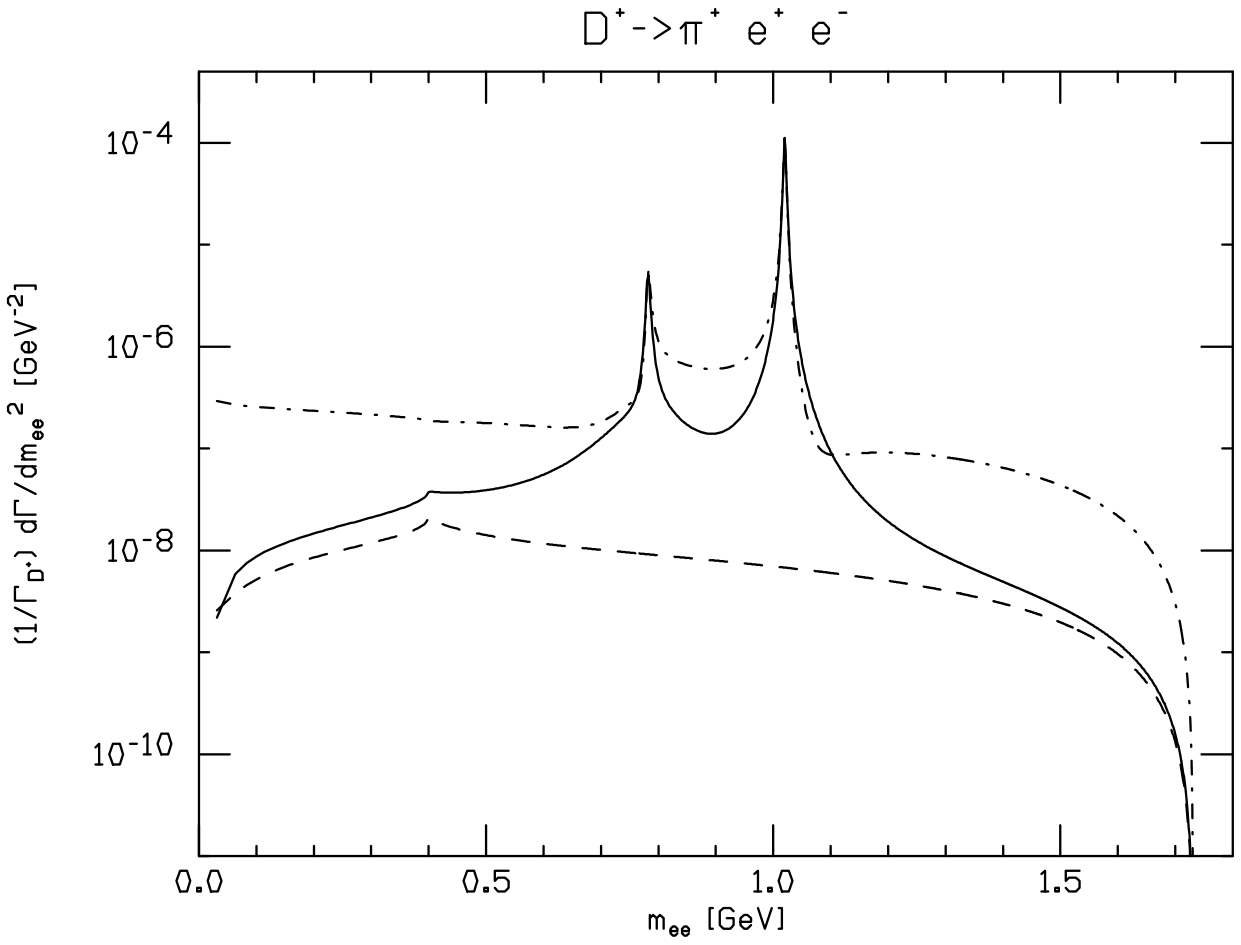

**Figure 2-38.** *The dilepton mass distribution for $D^+ \to \pi^+ e^+ e^-$ decays. The solid line shows the sum of the short and long-distance Standard Model contributions. The dashed line represents the short-distance contribution only. The dotted-dashed line includes the contribution of $R$-parity-violating terms in SUSY (see Section 2.28.2).*

Fig. 2-38, we present the dilepton mass distribution in $D^+ \to \pi^+ e^+ e^-$ decays. The two narrow peaks are the $\phi$ and the $\omega$, which sits on top of the broader $\rho$. The total rate results in $\mathcal{B}(D^+ \to \pi^+ e^+ e^-) \simeq 2 \times 10^{-6}$. Although most of this branching ratio arises from the intermediate $\pi^+\phi$ state, we see from Fig. 2-38 that New Physics effects as low as $10^{-7}$ can be observed as long as such sensitivity is achieved in the regions away from the $\omega$ and $\phi$ resonances, both at low and high dilepton mass.

$D^+ \to \rho^+ e^+ e^-$**:** As in the discussion of $D^+ \to \pi^+ e^+ e^-$ decays, we follow closely Ref. [354]. Because fewer data are currently available on the $D \to VV'$ modes, we take the values of the $\kappa_i$ in Eq. (2.174) from the fits to the $D^+ \to \pi^+ V$ case studied above. Again, once precise measurements of the $D \to \rho\ell\nu$ form factors are available, heavy quark spin symmetry relations can be used to turn these into $D \to \rho\ell^+\ell^-$ form factors. Lacking these at the moment, we use the extracted values from the $D \to K^*\ell\nu$ data [361, 362] and assume SU(3) symmetry [363]. The total integrated branching ratios are $\mathcal{B}(D^0 \to \rho^0 e^+ e^-) = 1.8 \times 10^{-6}$ and $\mathcal{B}(D^+ \to \rho^+ e^+ e^-) = 4.5 \times 10^{-6}$. Most of these rates comes from the resonance contributions. However, there is also a region—in this case confined to low values of $m_{ee}$ owing to the kinematics (see Fig. 2-39—where sensitive measurements could test short-distance physics. .

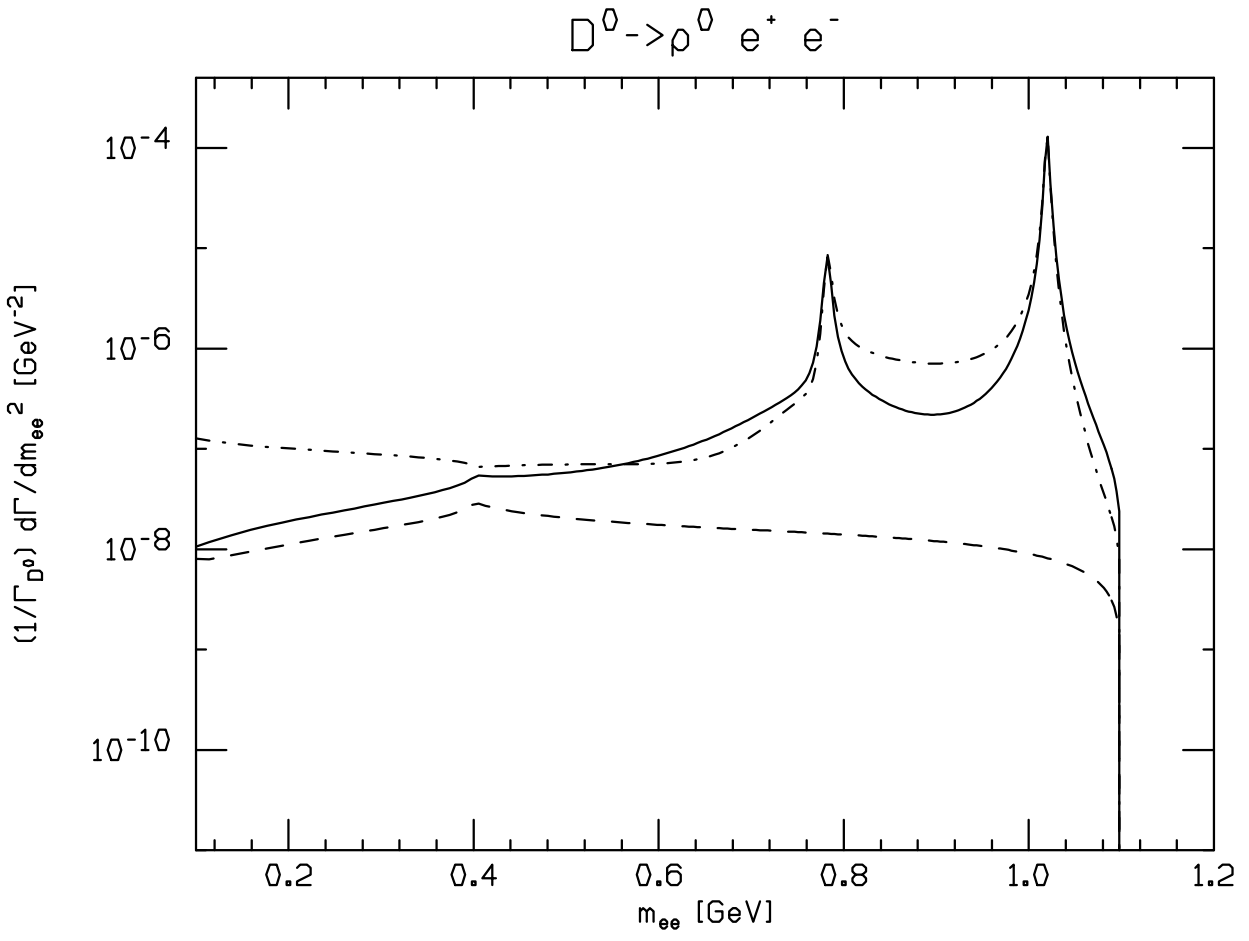

**Figure 2-39.** *The same as in Fig. 2-38, but for $D^0 \to \rho^0 e^+ e^-$ decays.*

The $\rho$ modes also contain angular information in the forward-backward asymmetry of the lepton pair. Because this asymmetry results from the interference between the vector and the axial-vector couplings of the leptons, it is negligible in the Standard Model , since vector couplings due to vector mesons overwhelm axial-vector couplings. This is true even away from the resonance region, since the coefficients $C_9^{\mathrm{eff}\,(\prime)}$ and $C_7^{\mathrm{eff}}$ get large enhancements due to mixing with $O_2$ and QCD corrections, whereas $C_{10}$—the axial-vector coupling—is not affected by any of these, which results in a very small interference.

For both the $\pi$ and $\rho$ modes, the sensitivity to New Physics effects is reserved for large $\mathcal{O}(1)$ enhancements because the long-distance contributions are still important even away from the resonances. In addition, some modes are driven almost exclusively by long-distance physics. Examples are $D^0 \to \overline{K}^{0(*)}\ell^+\ell^-$ and the radiative $D^0 \to \overline{K}^{0*}\gamma$ decays, dominated by $W$ exchange diagrams, as well as $D^\pm \to \overline{K}^{\pm(*)}\ell^+\ell^-$ and $D^\pm \to \overline{K}^{\pm*}\gamma$ decays, which contain both $W$ annihilation and exchange. The measurements of these modes, although not directly constraining New Physics, will help us understand long-distance physics. This may prove crucial to test the short-distance physics in the $\pi$ and $\rho$ modes. A complete list of predictions can be found in Ref. [364].

**Exclusive radiative decays** Exclusive decays mediated by the $c \to u\gamma$ transition are expected to be plagued by large hadronic uncertainties. As mentioned earlier, the large mixing of the $O_7$ operator with the four-quark operators, especially $O_2$, and the propagation of light quarks in the loops indicate the presence of potentially large nonperturbative effects. These are not calculable from first principles nor in a controlled approximation (other than lattice gauge theory). Moreover, even if lattice computations of these effects become available, they typically overwhelm the Standard Model short-distance contributions. Thus, modes such as $D \to \rho\gamma$ are not expected to be a probe of the short-distance structure of the Standard Model to the extent $B \to K^*\gamma$ can be if the transition form factor is known precisely.

On the other hand, one can try to estimate the size of the long-distance contributions and therefore the branching fractions of these modes. This is interesting in its own right; experimental observation of these modes will give us guidance in our otherwise limited understanding of these nonperturbative effects.

Several attempts have been made at estimating the long-distance contributions [352, 146, 365, 366]. An example is the decay $D^0 \to \rho^0\gamma$. We can identify two types of long-distance contributions: pole-mediated and vector-meson dominance (VMD) transitions. Pole contributions can be thought of as driven by "annihilation" diagrams with $\mathcal{B}(D^0 \to \rho^0\gamma)_{\mathrm{pole}} \leq$ few $\times\, 10^{-7}$ [352]. One can also use QCD sum rules to compute the annihilation contributions yielding $\mathcal{B}(D^0 \to \rho^0\gamma) \simeq$ few $\times\, 10^{-6}$ [146]. On the other hand, VMD contributions come from nonleptonic intermediate states. In our example, this corresponds to $D^0 \to \rho^0 V \to \rho^0\gamma$, where the neutral vector boson $V$ turns into an on-shell photon. Various methods have been used to compute the nonleptonic and $V \to \rho$ amplitudes [352, 365, 366]. A common assumption to estimate the VMD amplitude is factorization [367]. However, the contribution of the factorized nonleptonic amplitude vanishes when the photon is on-shell [368, 369]. This is a consequence of gauge invariance and is related to the fact that the mixing of four-quark operators with the photon penguin operator $O_7$ vanishes unless nonfactorizable gluons are exchanged. Thus, nonfactorizable contributions to the nonleptonic amplitude constitute the leading effect in the VMD amplitude. It is therefore possible that the VMD contributions to weak radiative decays of charm mesons are overestimated. At the same time, it is possible that the charm quark is not heavy enough for the nonfactorizable effects to be suppressed. The suppression is formally $O(\Lambda/m_c)$, with $\Lambda$ a typical scale of strong interactions. We conclude that uncertainties in these modes are very large. The Belle collaboration recently measured $\mathcal{B}(D^0 \to \phi\gamma) = (2.60^{+0.70+0.15}_{-0.61-0.17}) \times 10^{-5}$ [370], consistent with the upper end of the predictions in Ref. [352], which were obtained by making use of VMD plus the data on the relevant nonleptonic decay in addition to the pole contributions. If this trend is confirmed in $D^0 \to \phi\gamma$ decays, as well as other modes, it points in the direction of large nonfactorizable contributions. Experimental bounds are closing in on some of these predictions and will undoubtedly shed light on the size of these long-distance effects.

**Other Rare Charm Decay Modes:** In the previous subsections we focussed on $c \to u\gamma$ and semileptonic $c \to u\ell^+\ell^-$ decay modes. Here we briefly summarize some features of further rare charm decays.

**$D^0 \to \gamma\gamma$:** The Standard Model short-distance contributions can be obtained from the two-loop $c \to u\gamma$ amplitude. This results in $\mathcal{B}^{\rm sd}(D^0 \to \gamma\gamma) \simeq 3 \times 10^{-11}$ [354]. There are several types of long-range effects. Fajfer et al. [371] estimate these effects using ChPTHH to one-loop. This gives $\mathcal{B}^{\rm ld}(D^0 \to \gamma\gamma) \simeq (1 \pm 0.5) \times 10^{-8}$. Ref. [354] considers various long-distance effects and obtain similar results. The main contributions are found to come from VMD and the $K^+K^-$ unitarity contribution.

**$D^0 \to \ell^+\ell^-$:** The short-distance contributions to this mode are also extremely suppressed, not only by helicity but also by the quark masses in the loop. Unlike in $c \to u\gamma$ decays, the mixing with $O_2$ does not help. In Ref. [354], the branching ratio from the short-distance contribution only is estimated as $\mathcal{B}^{\rm sd}(D^0 \mu^+\mu^-) \lesssim 10^{-18}$. The most important source of long-distance effects is the two-photon unitary contribution, which gives

$$\mathcal{B}^{\rm ld}(D^0 \to \mu^+\mu^-) \simeq 3 \times 10^{-5}\, \mathcal{B}(D^0 \to \gamma\gamma). \tag{2.177}$$

**$D \to X\nu\overline{\nu}$:** Short- and long-distance contributions to $c \to u\nu\overline{\nu}$ processes in the Standard Model are extremely small, typically resulting in $\mathcal{B}(c \to u\nu\overline{\nu}) \lesssim 10^{-15}$ [354].

### 2.28.2 Rare charm decays and New Physics

Charm-changing neutral-current processes such as $D^0\overline{D}^0$ mixing and rare charm decays complement the constraints on extensions of the Standard Model obtained from processes initiated by down quarks, such as Kaon and $B$ meson transitions. Although bounds on $\Delta m_D$ are quite constraining in a variety of models, New Physics may still show itself in rare charm decays. We mainly review the potential for signals from supersymmetric theories (with and without $R$ parity conservation) and from new strong dynamics at the TeV scale. We briefly comment on the sensitivity to other models of New Physics, such as theories with extra dimensions and extended gauge and matter sectors.

**The minimal supersymmetric Standard Model:** The MSSM adds to the Standard Model description of loop-mediated processes contributions due to gluino-squark, chargino/neutralino-squark and charged Higgs-quark exchange. This last contribution carries the same CKM structure as the Standard Model loop and is proportional to the internal and external quark masses; thus, its effects in rare charm transitions are small and we neglect it here. The gluino-squark contribution proceeds via flavor-diagonal vertices proportional to the strong coupling constant and in principle dominates the CKM-suppressed, weak-scale strength chargino/neutralino-squark contributions. We therefore consider only the case of gluino-squark exchange here as an estimate of the potential size of SUSY effects in rare charm decays.

A typical squark-gluino contribution is depicted in Fig. 2-40.

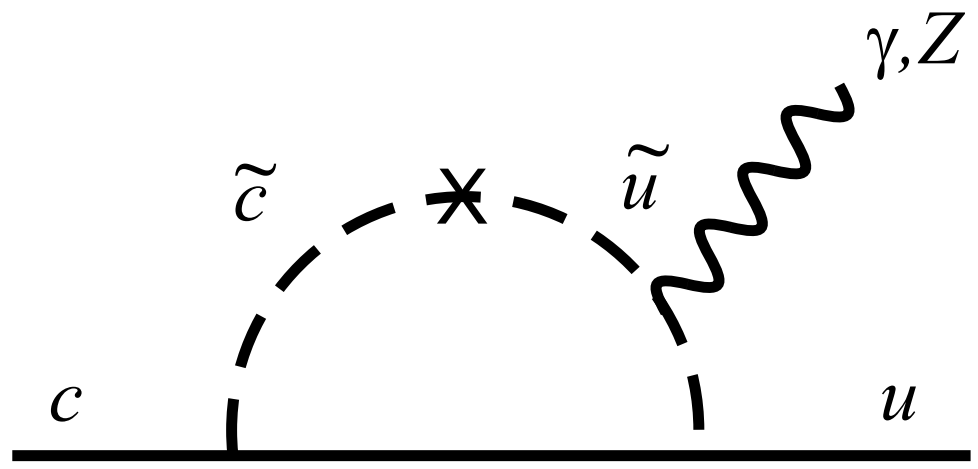

**Figure 2-40.** *A typical contribution to $c \to u$ FCNC transitions in the MSSM . The cross denotes one mass insertion $(\delta^{12})_{\lambda\lambda'}$, with $\lambda, \lambda' = L, R$.*

The corresponding effects in the $c \to u$ transitions were studied for $D \to X_u\gamma$ [372] and $D \to X_u\ell^+\ell^-$ decays [354]. Within the context of the mass insertion approximation [373], the effects are included in the Wilson coefficients corresponding to the decay $D \to X_u\ell^+\ell^-$ via

$$C_i = C_i^{\rm SM} + C_i^{\tilde g}\,, \tag{2.178}$$

for $i = 7, 9, 10$. Allowing for only one mass insertion, the gluino-squark diagrams [374, 372, 354] do not contribute to $C_{10}^{\tilde g}$, but only to $C_7^{\tilde g}$ and $C_9^{\tilde g}$. If we allow for two mass insertions, there are contributions to $C_{10}^{\tilde g}$ as well as additional contributions to $C_9^{\tilde g}$. In addition, the operator basis can be extended by the "wrong chirality" operators, obtained by switching the quark chiralities in $O_7$, $O_9$, and $O_{10}$.

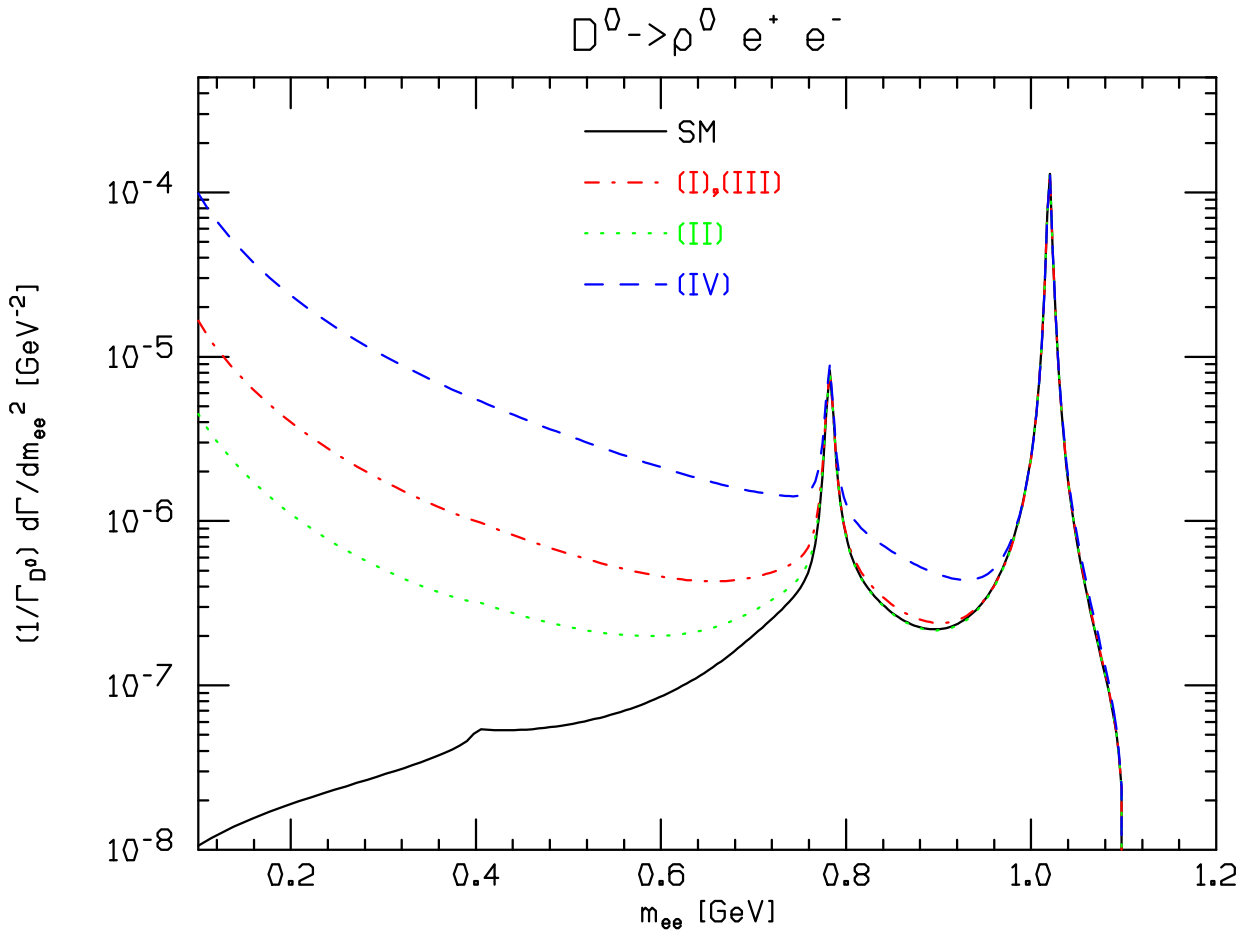

**Figure 2-41.** *The dilepton mass distribution for $D^0 \to \rho^0 e^+e^-$ decays in the MSSM with nonuniversal soft breaking effects. The solid line is the Standard Model. (I) $M_{\tilde g} = M_{\tilde q} = 250$ GeV; (II) $M_{\tilde g} = 2\,M_{\tilde q} = 500$ GeV; (III) $M_{\tilde g} = M_{\tilde q} = 1000$ GeV; (IV) $M_{\tilde g} = (1/2)\,M_{\tilde q} = 250$ GeV. Figure taken from [354].*

As noted in Refs. [372] and [374], in both $C_7^{\tilde g}$ and its chirality-flipped counterpart the term with mixed squark chirality labels introduces an enhancement factor $M_{\tilde g}/m_c$. In the Standard Model, the chirality flip that appears in $O_7$ occurs by a flip of one external quark line, resulting in a factor of $m_c$ included in the operator's definition.[20] However, in the gluino-squark diagram, the insertion of $(\delta_{12}^u)_{RL}$ forces the chirality flip to take place in the gluino line, thus introducing a factor of $M_{\tilde g}$ instead of $m_c$. This yields a significant enhancement in the short-distance contributions to the process $D \to X_u\gamma$ [372], which is unfortunately obscured by the large long-range effects. This is not the case in $c \to u\ell^+\ell^-$ processes.

[20] The $m_u$ term, proportional to the $(1-\gamma_5)$ in the operator, is neglected.

In order to estimate the effects in $c \to u\ell^+\ell^-$ transitions from the gluino contributions, we use the bounds given in Ref. [364]. In Fig. 2-41 we show the dilepton mass distribution for $D^0 \to \rho^0 e^+e^-$ decays in the Standard Model and in the MSSM for different quark and gluino masses. As can be seen the New Physics effect in the $\rho$ modes is quite pronounced and lies almost entirely in the low $m_{ee}$ region. Most of it comes from the helicity flip in the form of a $1/q^2$ enhancement. Since gauge invariance forces a cancellation of the $1/q^2$ factor in the pseudoscalar modes such as $D^0 \to \pi^0 e^+e^-$, the vector modes are more sensitive to these beyond the Standard Model effects. We conclude that rare charm decays are indeed sensitive to a generic extension of the Standard Model such as the MSSM.

**Supersymmetry with $R$ parity violation:** Imposing $R$ parity conservation in the MSSM prohibits baryon- and lepton-number–violating terms in the superpotential. However, other symmetries can be invoked to avoid rapid proton decay, such as baryon parity or lepton parity (see, *e.g.*, [375]), and hence allow for $R$-parity violation. The tree level exchange of down squarks results in the effective interaction [354]

$$\delta\mathcal{H}_{\rm eff} = -\frac{\tilde\lambda'_{i2k}\tilde\lambda'_{i1k}}{2m^2_{\tilde d^k_R}}\,(\overline{u}_L\gamma_\mu c_L)(\bar\ell_L\gamma^\mu\ell_L). \tag{2.179}$$

where the $R$ parity-violating couplings $\tilde\lambda'_{ijk}$ are defined in Ref. [354]. This corresponds to contributions at the high energy scale to the Wilson coefficients $C_9$ and $C_{10}$ given by

$$\delta C_9 = -\delta C_{10} = \frac{\sin^2\theta_W}{2\alpha^2}\left(\frac{M_W}{m_{\tilde d^k_R}}\right)^2\tilde\lambda'_{i2k}\tilde\lambda'_{i1k}. \tag{2.180}$$

Making use of the most recent bounds on $R$ parity violating coefficients (see Ref. [364]), we obtain predictions for the possible effects in rare charm processes. The effects for $\ell = e$ are rather small, as it can be seen from Fig. 2-39 for $D^0 \to \rho^0 e^+e^-$ decays. On the other hand, for $\ell = \mu$, we obtain

$$\delta C_9^\mu = -\delta C_{10}^\mu \leq 17.4\left(\frac{\tilde\lambda'_{22k}}{0.21}\right)\left(\frac{\tilde\lambda'_{21k}}{0.06}\right). \tag{2.181}$$

Since the bounds on $\tilde\lambda'_{2jk}$ are loose they lead to very large effects in the $\ell = \mu$ modes. In fact, the allowed values from other observables saturate the current experimental limits $\mathcal{B}^{exp}(D^+ \to \pi^+\mu^+\mu^-) < 1.5\times10^{-5}$ [331] and $\mathcal{B}^{exp}(D^0 \to \rho^0\mu^+\mu^-) < 2.2\times10^{-5}$ [376] resulting in [354]

$$\tilde\lambda'_{22k}\,\tilde\lambda'_{21k} < 0.004. \tag{2.182}$$

These large effects are observable away from the resonances.

In addition, the angular information in $D \to \rho\mu^+\mu^-$ decays can be used to confirm the New Physics origin of the large deviations in the rate. The forward-backward asymmetry for leptons is nearly zero in the Standard Model (see Section 2.28.1). New Physics contributions with $C_{10} \simeq C_9^{\rm eff}$ hence generate a sizable asymmetry. This is actually the case in $R$-parity-violating SUSY where the asymmetry in $D \to \rho\mu^+\mu^-$ decays can be large in the allowed parameter space [354].

The coefficients given in Eq. (2.180) also lead to a contribution to the two-body decay $D^0 \to \mu^+\mu^-$. The $R$-parity–violating contribution to the branching ratio then reads as

$$\mathcal{B}^{\not R_p}(D^0 \to \mu^+\mu^-) = \tau_{D^0}\,f_D^2\,m_\mu^2\,m_D\sqrt{1-\frac{4m_\mu^2}{m_D^2}}\;\frac{\left(\tilde\lambda'_{22k}\tilde\lambda'_{21k}\right)^2}{64\pi\,m^4_{\tilde d_k}}. \tag{2.183}$$

Applying the bound in Eq. (2.182) gives the constraint[21]

$$\mathcal{B}^{\not R_p}(D^0 \to \mu^+\mu^-) < 3.5 \times 10^{-7}\, \left(\frac{\tilde{\lambda}'_{22k}\tilde{\lambda}'_{21k}}{0.004}\right)^2 . \tag{2.184}$$

Thus, $R$ parity violation could give an effect that can be probed in these modes.

Finally, we consider the products of $R$ parity-violating couplings that lead to lepton flavor violation. For instance, the products $\tilde{\lambda}'_{11k}\tilde{\lambda}'_{22k}$ and $\tilde{\lambda}'_{21k}\tilde{\lambda}'_{12k}$ will give rise to $D^+ \to \pi^+\mu^+e^-$ and $D^0 \to \mu^+e^-$ decays. Updated constraints are given in Ref. [364].

**Strong dynamics I, technicolor models:** In standard technicolor theories, both fermions and technifermions transform under the new gauge interaction of extended technicolor (ETC) [377]. This leads to the presence of four-quark operators coming from the diagonal ETC generators and characterized by a mass scale $M$ bounded by $D^0\overline{D}^0$ mixing to be greater than $\sim 100$ TeV. However, additional operators are generated at low energies that are not suppressed by $M$. The condensation of technifermions leading to electroweak symmetry breaking leads to fermion mass terms of the form

$$m_q \simeq \frac{g^2_{\rm ETC}}{M^2_{\rm ETC}}\langle \overline{T}T\rangle_{\rm TC}. \tag{2.185}$$

Operators arising from the technifermion interactions have been shown [378] to give rise to FCNC involving the $Z$-boson,

$$\xi^2\frac{m_c}{8\pi v}\frac{e}{\sin 2\theta_W}\mathcal{U}_L^{cu}Z^\mu\,(\overline{u}_L\gamma_\mu c_L) \quad \text{and} \quad \xi^2\frac{m_t}{8\pi v}\frac{e}{\sin 2\theta_W}\mathcal{U}_L^{tu}\mathcal{U}_L^{tc*}Z^\mu\,(\overline{u}_L\gamma_\mu c_L), \tag{2.186}$$

where $\mathcal{U}_L$ is the unitary matrix rotating left-handed up-type quark fields into their mass basis and $\xi$ is a model-dependent quantity of $\mathcal{O}(1)$. The induced flavor conserving $Z$-coupling was first studied in Ref. [378] and FCNC effects in $B$ decays have been examined in Refs. [379] and [380].

The vertices in Eq. (2.186) induce contributions to $c \to u\ell^+\ell^-$ processes. These appear mostly as a shift in the Wilson coefficient $C_{10}(M_W)$,

$$\delta C_{10} \simeq \mathcal{U}^L_{cu}\,\frac{m_c}{2v}\,\frac{\sin^2\theta_W}{\alpha} \simeq 0.02, \tag{2.187}$$

where we assume $\mathcal{U}_L^{cu} \simeq \lambda \simeq 0.22$ (*i.e.*, one power of the Cabibbo angle) and $m_c = 1.4$ GeV. Although this represents a very large enhancement with respect to the Standard Model value of $C_{10}(M_W)$, it does not translate into a large deviation in the respective branching ratios. As we have seen, these are dominated by the mixing of the operator $O_2$ with $O_9$, leading to a very large value of $C_9^{\rm eff}$. The contribution in Eq. (2.187) represents only a few percent effect in the branching ratio with respect to the Standard Model. However, the effect is quite large in the region of low dilepton mass.

Furthermore, the interaction in Eq. (2.186) can also mediate $D^0 \to \mu^+\mu^-$ decays. The corresponding amplitude is given as

$$A_{D^0\to\mu^+\mu^-} \simeq \mathcal{U}^L_{cu}\,\frac{m_c}{2\pi v}\,\frac{G_F}{\sqrt{2}}\,\sin^2\theta_W f_D\, m_\mu\,. \tag{2.188}$$

This results in the branching ratio $\mathcal{B}^{\rm ETC}(D^0 \to \mu^+\mu^-) \simeq 0.6 \times 10^{-10}$, which, although still small, is not only several orders of magnitude larger than the Standard Model short-distance contribution but also more than two orders of magnitude larger than the long-distance estimates.

[21] In Ref. [354], this expression (Eq.(86)) was incorrectly given. Also, the branching ratio stated there did not reflect the bound from Eq. (122), but the less restrictive bounds to the individual couplings.

Finally, the FCNC vertices of the $Z$-boson in Eq. (2.186) also give large contributions to $c \to u\nu\overline{\nu}$ transitions. The enhancement is considerable and results in

$$\mathcal{B}^{\rm ETC}(D^+ \to X_u \nu\overline{\nu}) \simeq \xi^4 \left(\frac{\mathcal{U}_L^{cu}}{0.2}\right)^2 2 \times 10^{-9}. \tag{2.189}$$

**Strong dynamics II, topcolor:** In top-condensation models [381], the constituents of the Higgs are the third generation left-handed quarks as well as $t_R$. Hill [382] proposed that a new gauge interaction strongly coupled to the third generation quarks is responsible for top condensation. The topcolor interactions break at the TeV scale as $\mathrm{SU(3)}_1 \times \mathrm{SU(3)}_2 \to \mathrm{SU(3)}_{\rm color}$, leaving, besides the massless gluons, a set of color-octet gauge bosons (the top-gluons) leading to the Nambu–Jona-Lasinio effective interactions that result in top condensation. This leads to electroweak symmetry breaking as well as to a large "constituent" top mass.

Tilting the vacuum in the top directions to avoid a large $b$-quark mass is typically accomplished through additional Abelian interactions that leave a $Z'$ strongly coupled to third-generation fermions. In some models, the tilting is done by simply arranging that $b_R$ not couple to the topcolor interactions. The top-gluon interactions (as well as the $Z'$s if present) are nonuniversal, leading to FCNC at tree level. These arise after quarks are rotated to their mass eigenbasis by the rotations

$$U^i_{L,R} \to \mathcal{U}^{ij}_{L,R}\, U^j_{L,R}, \qquad D^i_{L,R} \to \mathcal{D}^{ij}_{L,R}\, D^j_{L,R}, \tag{2.190}$$

where $\mathcal{U}_{L,R}$ and $\mathcal{D}_{L,R}$ are unitary. The CKM matrix is then $V_{\rm CKM} = \mathcal{U}_L^\dagger\, \mathcal{D}_L$. Constraints on topcolor models are reviewed in [383]. The bounds from the down-quark sector impose severe constraints on the entries of $\mathcal{D}_{L,R}$ mainly coming from the exchange of bound states that couple strongly to the $b$-quark. There are several contributions to $\Delta m_D$. After these are taken into account, the potential effects in charm rare decays are rather moderate. After the transformations of quark fields in Eq. (2.190) have been performed, the exchange of top-gluons generates four-fermion couplings

$$\frac{4\pi\alpha_s \cot\theta^2}{M_G^2}\, \mathcal{U}^{tc*}\mathcal{U}^{tu}\, (\overline{u}\gamma_\mu T^a t)(\overline{t}\gamma^\mu T^a c) \tag{2.191}$$

where $\mathcal{U}^{ij} = \mathcal{U}^{ij}_L + \mathcal{U}^{ij}_R$ and $M$ is the mass of the exchanged color-octet gauge boson. The one-loop insertion of this operator results in contributions to the operators $\mathcal{O}_9$ and $\mathcal{O}_{10}$ in $c \to u\ell^+\ell^-$ as well as in the purely leptonic decays. These could lead to large deviations from the short-distance contribution of the Standard Model

$$\delta C_{10} \simeq 2\,\delta C_9 \simeq 0.01 \times \left(\frac{\mathcal{U}^{tc*}\mathcal{U}^{tu}}{\sin^5\theta_c}\right)\left(\frac{1\ \mathrm{TeV}}{M_G}\right)^2, \tag{2.192}$$

but the effects are rather modest in the branching ratios unless the quark rotation matrices are larger than expected. This would not be unnatural, for instance, for $\mathcal{U}_R$, since the rotations of right-handed quarks are not related to any known observable in the Standard Model .

**Other New Physics scenarios:** Extensions of the Standard Model, leading to effects in rare charm decays also tend to result in large contributions to $D^0\overline{D}^0$ mixing. In Ref. [354] a long list of these scenarios has been evaluated in detail. Generically, the New Physics effects are either negligible or amount to $\mathcal{O}(1)$ enhancement over the Standard Model short-distance contributions.

For instance, compact extra dimensions may lead not only to massive scalar and fermionic states but also to nonuniversal couplings of the Standard Model fermions to Kaluza-Klein (KK) excitations of gauge bosons that may induce flavor-violating loop effects. In general, the largest effects in rare charm decays are associated with massive neutral gauge bosons such as KK excitation of a $Z'$. They generate a FCNC current in the up-quark sector and then decay into either charged leptons or neutrinos. With masses starting around the TeV scale, these states could lead to $\mathcal{O}(1)$ enhancements in $c \to u\ell^+\ell^-$ modes, when compared to the Standard Model short-distance predictions. Thus, in the charged lepton modes this induces observable effects in the low $m_{\ell\ell}$ window. The enhancement in the $c \to u\nu\overline{\nu}$ modes could be several orders of magnitude above the Standard Model predictions, although they may be very difficult to observe.

Many other New Physics scenarios with additional matter and/or gauge fields lead to contributions to flavor physics and in particular, to rare charm decays. Most contributions that are potentially large are constrained by $D^0\overline{D}^0$ mixing. This is the case, *e.g.*, for models with extra down-type quarks and gauge bosons.

## 2.29 Summary

The study of rare decays provides great opportunities to test the flavor sector of the Standard Model and search for phenomena beyond. So far, the benchmark mode $b \to s\gamma$ has been measured, and is in quantitative agreement with the Standard Model at order ten percent. The decays $b \to s\ell^+\ell^-$ and $B \to (K, K^*)\ell^+\ell^-$, have been discovered at the existing $B$ Factories with branching ratios in the ballpark of the Standard Model values. They constitute a further crucial test once more data become available. Other FCNC $B$ decays have just been seen recently ($B \to \rho\gamma$) or only bounds exist like in the case of $B \to K\nu\overline{\nu}$, while some processes such as $B \to K\tau^+\tau^-$ are essentially unconstrained. Several purely hadronic rare $B$ decays have also been observed. Their present data show some intriguing anomalies at the $2-3\sigma$ level. The interpretation is controversial due to hadronic uncertainties and low statistics. It will be interesting to see whether this trend persists in the future with larger data samples, see Chapter 3 for a further discussion of hadronic $B$ decays.

To obtain a more precise and more complete map of the flavor landscape, a multitude of processes with branching ratios typically of order $10^{-7} - 10^{-4}$ must be measured,. In addition, it is vital to go beyond the pure study of rates, to measure complex kinematic distributions, some of which require flavor-tagging. In particular, $CP$, forward-backward, isospin, and polarization asymmetries are very sensitive to New Physics effects and have good control over theoretical uncertainties. Having a large set of complementary and overconstraining observables will allow us to detect subtle patterns and to distinguish between the Standard Model and the many different candidate extensions, see *e.g.*, [384]. Hence, the luminosity of a Super $B$ Factory must be substantially higher than that at current facilities. For instance, the time dependent $CP$ asymmetries in $B^0 \to (K^{0*} \to K^0_S\pi^0)\gamma$ decays are sensitive to right-handed currents in the $b \to s\gamma$ transition. This type of New Physics can hide in the branching ratio which constrains only the sum of the couplings with opposite helicities squared. To probe the helicity-flipped coupling down to the Standard Model value induced at $m_s/m_b$, one needs 10 $ab^{-1}$, which is more than 10 times larger than the anticipated accumulated luminosity of *BABAR* and Belle, but can be accumulated in one year at a Super $B$ Factory.

Precision measurements will also be undertaken at the hadron colliders Tevatron and the LHC. They will contribute to the physics of the $B_s$ mesons, in particular, mixing and the rare decay $B_s \to \mu^+\mu^-$. These experiments provide strong competition for the $B$ Factories in some particular $B$ decay modes, such as $b \to s\ell^+\ell^-$ and $B \to (K, K^*)\mu^+\mu^-$. Specific measurements, however, require the clean, well-understood, $e^+e^-$ experimental environment. One example is the time-dependent $CP$ study of $B^0 \to K^{0*}\gamma$ decay discussed above; other important examples are the fully inclusive measurements, such as $b \to s\gamma$ and $b \to s\ell^+\ell^-$ decays, and their corresponding kinematic distributions. For the same reasons, it is also important that the upgraded detectors have experimental sensitivity for decays into (semi)-invisibles such as neutrinos and $\tau$'s.

In this chapter, we have investigated the physics reach in studying rare $B$ and charm decays at a future Super $B$ Factory running at the $\Upsilon(4S)$ resonance. The theoretical and experimental prospects for measuring a variety of important decay modes have been analyzed, with emphasis on the experimental requirements. We did not discuss in detail methods to extract the coefficients of dipole and dileptonic operators from data on rare $b$ decays ("model-independent analysis"). This important topic is covered in Chapter 5. As has been stressed in [129], this program needs improved constraints on the $B \to X_s g$ branching ratio, which could be obtained along the lines of the former CLEO measurement [385].

We will briefly summarize here some highlights of the material discussed in this chapter on rare decays; details can be found in the respective sections. Restricting our view only to rare decays, this list demonstrates that the physics case for a Super $B$ Factory is based on a large variety of key observables:

- **Rare radiative decays:**
  - The measurements of the inclusive $b \to s\gamma$ branching ratio and the corresponding $CP$ asymmetry are important ingredients in model-independent analysis, and will serve as precision tests of the Standard Model.
  - The untagged direct $CP$ asymmetry in $b \to (s+d)\gamma$ represents a very clean test for new sources of $CP$ violation beyond the CKM phase.
  - $CP$ and isospin asymmetries in exclusive $B \to K^*\gamma$ and $B \to (\rho,\omega)\gamma$ decays provide additional complementary measurements.
  - Measurements of $B \to K^{**}\gamma$ and $\sin 2\beta(K_S^0\pi^0\gamma)$ allow us to study the chirality of the dipole operators.
  - In spite of sizable theoretical uncertainties, in some scenarios the double radiative decay $b \to q\gamma\gamma$ allows New Physics searches complementary to $b \to q\gamma$ decays.
- **Rare semileptonic decays:**
  - In the rare inclusive mode $b \to s, d\ell^+\ell^-$, the measurements of the kinematic distributions such as decay spectra, forward-backward asymmetries and $CP$ asymmetries provide important precision tests of the Standard Model. In particular, the existence and position of the zero of the forward-backward asymmetry represents one of the most sensitive observables in a New Physics search. The experimental information on $b \to s, d\ell^+\ell^-$, combined with the radiative $b \to s\gamma, g$ decays, allows a model-independent extraction of the Wilson coefficients $C_{7\gamma,8g,9,10}$.
  - The comparison of the electron and the muon modes in $b \to s\ell^+\ell^-$ transitions is sensitive to non-Standard Model Higgs exchange. This is complementary to the $\mathcal{B}(B_s \to \mu^+\mu^-)$ mode in constraining the pseudo)scalar coefficients $C_{S,P}$.
  - The angular analysis in $B \to (K^* \to K\pi)\ell^+\ell^-$ probes right-handed couplings without $CP$ violation. The zero of the forward-backward asymmetry in $B \to K^*\ell^+\ell^-$ is also a relatively clean observable.
  - The ratio of the branching ratios $B_d \to \mu^+\mu^-$ over $B_s \to \mu^+\mu^-$ is a clean probe of non-minimal flavor symmetry breaking.
  - Branching ratios of rare charm decays $D \to (\pi,\rho)e^+e^-$ in the Standard Model are dominated by non-perturbative effects. The study of the dilepton spectrum outside the resonance region, however, provides sensitivity to physics beyond the Standard Model.
- **Rare (semi)invisible decays including $\nu$'s and $\tau$'s:**
  - The inclusive mode $b \to s\nu\overline{\nu}$ is theoretically very clean, and is sensitive to new degrees of freedom, even if they are far above the electroweak scale. However, the measurement of the inclusive mode is very difficult.
  - The corresponding exclusive mode $B \to K^{(*)}\nu\overline{\nu}$ is less clean, but is still very interesting, as it can be used to constrain missing energy signatures, such as light $CP$-odd scalars or dark matter candidates.
  - The modes $B \to (K, K^*, X_s)\tau^+\tau^-$, $B \to \tau^+\tau^-$ are unique probes of non-Standard Model Higgs to $\tau$ couplings.
  - The $B \to (D, D^*, X_c)\tau\nu$ decays are sensitive to $\tan\beta/m_{H^\pm}$. The transverse $\tau$ polarization in $B \to X_c\tau\nu$ is a clean probe of $CP$ violation beyond the Standard Model .

A Super $B$ Factory would play an important and unique role in the study of the flavor sector in New Physics beyond the Standard Model. If there are new flavor structures beyond the CKM-Yukawa-pattern of the Standard Model to be discovered, a Super $B$ Factory will be indispensable for a detailed exploration, in particular through the clean measurements of rare decays. This is also true in New Physics scenarios in which the breaking of $CP$ and flavor symmetries is directly linked to the known structure of the Standard Model Yukawa couplings ('minimal-flavor-violation'). Here the indirect exploration of higher scales at a Super $B$ Factory using rare decay measurements can compete in precision with the direct search via flavor-conserving observables. Moreover, a Super $B$ Factory is an ideal tool to explore possible solutions of the well-known Standard Model flavor problem, which must be addressed in any viable New Physics scenario.

# 3

# Angles of the Unitarity Triangle

**Conveners:** Y. Grossman, A. Soffer, A. Roodman

**Authors:** J. Albert, D. Atwood, M. Ciuchini, A. Datta, D. Dujmic, U. Egede, A. Falk, S. Gopalakrishna, K. Graziani, A. Gritsan, M. Gronau, Y. Grossman, M. Hazumi, W. Hulsbergen, D. London, K. Matchev, M. Neubert, S. Oh, A.A. Petrov, H. Quinn, A. Roodman, J. Rosner, V.G. Shelkov, N. Sinha, R. Sinha, A. Soffer, A. Soni, J.D. Wells, J. Zupan

## 3.1 Introduction

$\succ$– M. Neubert –$\prec$

### 3.1.1 Introductory remarks–hopes and certainties

The physics potential of an $e^+e^-$ Super $B$ Factory must be evaluated on the basis of a vision of the high-energy physics arena in the next decade. By that time, the *BABAR* and Belle experiments will presumably have been completed, and each will have collected data samples in excess of 500 fb$^{-1}$, and hadronic $B$ factories will have logged several years of data taking. There are excellent prospects that many parameters of the unitarity triangle will have been determined with great precision and in multiple ways. Likewise, many tests of the flavor sector and searches for New Physics will have been performed using a variety of rare $B$ decays. A Super $B$ Factory operating at an $e^+e^-$ collider with luminosity of order $\mathcal{L} \approx 10^{36}\,\mathrm{cm}^{-2}\,\mathrm{s}^{-1}$ would be the logical continuation of the $B$ Factory program. If it is built, it will provide superb measurements of Standard Model parameters and perform a broad set of tests for New Physics. Such a facility could exhaust the potential of many measurements in the quark flavor sector, which could not be done otherwise.

However, it cannot be ignored that a Super $B$ Factory would come online in the LHC era. By the time it could start operation, the LHC will most likely (hopefully ...) have discovered new particles, such as one or more Higgs bosons, SUSY partners of the Standard Model particles, Kaluza–Klein partners of the Standard Model particles, new fermions and gauge bosons of a dynamical electroweak symmetry-breaking sector, or whatever else will be revealed at the TeV scale. The crucial question is, therefore, whether a Super $B$ Factory has anything to contribute to the physics goals of our community in this era. More specifically, can it complement in a meaningful way the measurements that will be performed at the energy frontier? And while energy-frontier physics will most likely attract most attention in the next decade or two, can a Super $B$ Factory do fundamental measurements that could not be done elsewhere (including earlier $B$ Factories)? Would it be indispensable to our community's goal to comprehensively explore the physics at and beyond the TeV scale?

Fortunately, there exist indeed some big, open questions in flavor physics, to which we would love to find some answers. Let me mention three of them:

**What is the dynamics of flavor?** The gauge forces in the Standard Model do not distinguish between fermions belonging to different generations. All charged leptons have the same electrical charge. All quarks carry the same color charge. In almost all respects the fermions belonging to different generations are equal—but not quite, since their masses are different. Today, we understand very little about the underlying dynamics responsible for the phenomenon of generations. Why do generations exist? Why are there three of them? Why are the hierarchies of the fermion masses and mixing angles what they are? Why are these hierarchies different for quarks and leptons? We have good reasons to expect that the answers to these questions, if they can be found in the foreseeable future, will open the doors to some great discoveries (new symmetries, forces, dimensions, . . . ).

**What is the origin of baryogenesis?** The existential question of the origin of the matter–antimatter asymmetry provides a link between particle physics and the evolution of the Universe. The Standard Model satisfies the prerequisites for baryogenesis as spelled out in the Sakharov criteria: baryon-number violating processes are unsuppressed at high temperature; $CP$-violating interactions are present due to complex couplings in the quark (and presumably, the lepton) sector; non-equilibrium processes can occur during phase transitions driven by the expansion of the Universe. However, quantitatively the observed matter abundance cannot be explained by the Standard Model (by many orders of magnitude). Additional contributions, either due to new $CP$-violating phases or new mechanisms of $CP$ violation, are required.

**Are there connections between flavor physics and TeV-scale physics?** What can flavor physics tell us about the origin of electroweak symmetry breaking, and, if the world is supersymmetric at some high energy scale, what can flavor physics teach us about the mechanism of SUSY breaking? Whereas progress on the first two "flavor questions" is not guaranteed (though it would be most significant), we can hardly lose on this question! Virtually any extension of the Standard Model that can solve the gauge hierarchy problem (*i.e.,* the fact that the electroweak scale is so much lower than the GUT scale) naturally contains a plethora of new flavor parameters. Some prominent examples are:

- SUSY: hundreds of flavor- and/or $CP$-violating couplings, even in the MSSM and its next-to-minimal variants
- extra dimensions: flavor parameters of Kaluza–Klein states
- Technicolor: flavor couplings of Techni-fermions
- multi-Higgs models: $CP$-violating Higgs couplings
- Little Higgs models: flavor couplings of new gauge bosons ($W'$, $Z'$) and fermions ($t'$)

If New Physics exists at or below the TeV scale, its effects should show up (at some level of precision) in flavor physics. Flavor- and/or $CP$-violating interactions can only be studied using precision measurements at highest luminosity. Such studies would profit from the fact that the relevant mass scales will (hopefully) be known from the LHC.

To drive this last point home, let me recall some lessons from the past. Top quarks have been discovered through direct production at the Tevatron. In that way, their mass, spin, and color charge have been determined. Accurate predictions for the mass were available before, based on electroweak precision measurements at the $Z$ pole, but also based on studies of $B$ mesons. The rates for $B^0\overline{B}^0$ mixing, as well as for rare flavor-changing neutral current (FCNC) processes such as $B \to X_s\gamma$, are very sensitive to the value of the top-quark mass. More importantly, everything else we know about the top quark, such as its generation-changing couplings $|V_{ts}| \approx 0.040$ and $|V_{td}| \approx 0.008$, as well as its $CP$-violating interactions (arg$(V_{td}) \approx -24^\circ$ with the standard choice of phase conventions), has come from studies of kaon and $B$ physics. Next, recall the example of neutrino oscillations. The existence of neutrinos has been known for a long time, but it was the discovery of their flavor-changing interactions (neutrino oscillations) that has revolutionized our thinking about the lepton sector. We have learned that the hierarchy of the leptonic mixing matrix is very different from that in the quark sector, and we have discovered that leptogenesis and $CP$ violation in the lepton sector may provide an alternative mechanism for baryogenesis.

In summary, exploring the flavor aspects of the New Physics, whatever it may be, is not an exercise in filling the Particle Data Book. Rather, it is of crucial relevance to answer fundamental, deep, questions about Nature. Some questions for which we have a realistic chance of finding an answer with the help of a Super $B$ Factory are:

- Do non-standard $CP$ phases exist? If so, this may provide new clues about baryogenesis.
- Is the electroweak symmetry-breaking sector flavor blind (minimal flavor violation)?
- Is the SUSY-breaking sector flavor blind?
- Do right-handed currents exist? This may provide clues about new gauge interactions and symmetries (left–right symmetry) at very high energy.

I will argue below that the interpretation of New Physics signals at a Super $B$ Factory can be tricky. But since it is our hope to answer some very profound questions, we must try as hard as we can.

The Super $B$ Factory workshops conducted in 2003 at SLAC and KEK showed that a very strong physics case can be made for such a machine. During these workshops it has become evident (to me) that a strength of a Super $B$ Factory is precisely that its success will not depend on a single measurement—sometimes called a "killer application". Several first-rate discoveries are possible, and even likely. It is the breadth of possibilities and the reach of a Super $B$ Factory that make a compelling physics case. As with electroweak precision measurements, we can be sure that New Physics effects must show up at some level of precision in flavor physics. The question remains, at which level? In the "worst-case scenario", in which we do not see any large signals of New Physics in $B$ meson studies, a Super $B$ Factory would play a role similar to that played by LEP for the understanding of electroweak symmetry breaking; it would impose severe constraints on model building in the post-LHC era.

### 3.1.2 CKM measurements—sides and angles

At a Super $B$ Factory , the goal with regard to CKM measurements is simply stated: achieve what is theoretically possible! Many smart theoretical schemes have been invented during the past two decades for making "clean" measurements of CKM parameters. We can safely assume steady theoretical advances in our field (the past track record is impressive). This will lead to ever more clever amplitude methods, progress in heavy-quark expansions and effective field theories, and perhaps breakthroughs in lattice QCD. Unfortunately, all too often these theoretical proposals are limited by experimental realities. With a Super $B$ Factory, it would finally become possible to realize the full potential of these methods. One of the great assets of such a facility, which is particularly valuable in the context of precision CKM physics, is the availability of huge samples of super-clean events, for which the decay of the "other $B$ meson" produced in $e^+e^- \to b\bar{b}$ at the $\Upsilon(4S)$ is tagged and fully reconstructed. Full reconstruction costs a factor 1000 or so in efficiency, which demands Super $B$ Factory luminosities. Once statistics is no longer of concern, the reduction in systematic error is a great benefit.

#### The sides $|V_{ub}|$ and $|V_{td}|$

A precision measurement of $|V_{ub}|$ with a theoretical error of 5% or less will require continued progress in theory. Determinations from exclusive semileptonic $B$ decays need accurate predictions for $B \to$ light form factors from lattice QCD or effective field theory. Determinations from inclusive $B$ decays need optimized cuts and dedicated studies of power corrections in the heavy-quark expansion. Recent advances using soft-collinear effective theory appear promising, but there is still much work left to be done. A Super $B$ Factory can provide vast, clean data samples of fully reconstructed decays, which would be an essential step toward eliminating the background from semileptonic decays with charm hadrons in the final state. It can also yield high-precision data on the $q^2$ dependence of form factors, and on the $B \to X_s\gamma$ photon spectrum down to $E_\gamma \sim 1.8$ GeV or lower. This would provide important constraints on theory parameters (*e.g.*, shape functions).

Another road toward measuring $|V_{ub}|$ is to study the leptonic decays $B \to \mu\nu$ or $B \to \tau\nu$, which would be accessible at a Super $B$ Factory. The rates for these processes are proportional to $f_B^2\,|V_{ub}|^2$. A lattice prediction for the $B$ meson decay constant can then be used to obtained $|V_{ub}|$. Alternatively, one can combine a measurement of the leptonic rate with that for the $B$–$\overline{B}$ mixing frequency to obtain the ratio $B_B^{-1/2}\,|V_{ub}/V_{td}|$, where the hadronic $B_B$ parameter would again have to be provided by lattice QCD. Such a determination would impose an interesting constraint on the parameters of the unitarity triangle.

A precision measurement of $|V_{td}|$ itself would require continued progress in lattice QCD. Rare radiative decays (or rare kaon decays) could also help to further improve our knowledge of this parameter.

**The angles $\beta = \phi_1$ and $\gamma = \phi_3$**

A Super $B$ Factory would allow us to exploit the full theory potential of various methods for model-independent extractions of $CP$ phases. We could finally do the measurements whose analyses require the least amount of theory input. In the Standard Model, it's really all about $\gamma$ (the unique $CP$ phase in $B$ decays), in various combinations with $\beta$ (the $CP$ phase in $B^0\overline{B}^0$ mixing). The importance of pursuing $\gamma$ measurements using different strategies (conventionally called measurements of $\alpha$ and $\gamma$) is that "$\gamma$ measurements" measure $\gamma$ in pure tree processes, whereas "$\alpha$ measurements" probe $\gamma$ in processes where penguins are present. Comparing the results obtained using these different methods probes for New Physics in penguin transitions, which are prominent examples of loop-induced FCNC processes in the Standard Model. The precision that can be reached on $\beta$ and $\gamma$ using various techniques accessible at a Super $B$ Factory is most impressive. A lot of marvelous physics can be done once such measurements will be at hand.

### 3.1.3 Searching for New Physics—never stop exploring

**Probing New Physics with CKM measurements**

The path is clear. If different determinations of unitarity-triangle parameters would turn out to be inconsistent, then this would signal the presence of some New Physics. For instance, it is interesting to confront the "standard analysis" of the unitarity triangle, which is primarily sensitive to New Physics in $B^0\overline{B}^0$ and $K^0\overline{K}^0$ mixing, with mixing-independent constructions using charmless hadronic decays such as $B \to \pi K$, $B \to \pi\pi$, $B \to \pi\rho$, and others. These studies, while not independent of theory, have already established $CP$ violation in the bottom sector of the CKM matrix (the fact that Im$(V_{ub}) \neq 0$ with the standard choice of phase conventions), while still leaving ample room for possible New Physics effects in $b \to s$ FCNC processes. (Some authors have argued that there are already some tantalizing hints of New Physics in $b \to s$ transitions sensitive to "electroweak penguin"-type interactions.)

It is also interesting to confront different determinations of $\beta$ with each other, such as the measurement of $\sin 2\beta$ from processes based on $b \to s\bar{c}c$ vs. $b \to s\bar{s}s$ or $b \to s\bar{q}q$ (with $q = u, d$) quark-level transitions. One of the burning issues today is whether there is something real to the "$\phi K_S^0$ anomaly" seen by Belle, but not confirmed by *BABAR*. With more precise data, many other decay modes can be added to obtain interesting information and perform non-trivial tests of the Standard Model.

Yet, let me stress that many more tests for New Physics can be done outside the realm of CKM measurements. Several of those involve rare hadronic $B$ decays. Others make use of inclusive decay processes. The general strategy is to look for niches where the "Standard Model background" is small or absent. One cannot overemphasize the importance of such "null (or close-to-null) measurements", as they provide direct windows to physics beyond the Standard Model. In comparison, the search for New Physics in CKM measurements always suffers from a large Standard Model background.

**Probing New Physics in exclusive decays**

Rare (charmless) hadronic $B$ decays are usually characterized by the presence of several competing decay mechanisms, often classified in terms of flavor topologies (trees, penguins, electroweak penguins, annihilation graphs, exchange graphs). These refer to the flow of flavor lines in a graph but do *not* indicate the possibility of multiple gluon exchanges. Therefore, reality is far more complicated. Until a few years ago, such nonleptonic decay processes were believed to be intractable theoretically. This has changed recently, thanks to the advent of QCD factorization theorems, perturbative QCD methods, and soft-collinear effective theory, which complement previous approaches based on flavor symmetries. Together, these approaches build the foundation of a systematic heavy-quark expansion for exclusive $B$ decays, much like heavy-quark effective theory provided the basis for such an expansion in the (much simpler) case of exclusive $B \to D^{(*)} l\nu$ decays. (The dispute between QCD factorization and pQCD practitioners is also beginning to be resolved, since the issue of Sudakov logarithms in heavy-to-light transition amplitudes is now under good theoretical control.)

With ever-improving theoretical control over exclusive $B$ decay processes, several possibilities for tests for New Physics become accessible. A partial list includes the measurement of $\sin 2\beta$ from the time-dependent $CP$ asymmetry in $B \to \phi K^0_S$ decays, probing electroweak penguins in rate measurements using $B \to \pi K^0_S$ decays, and searching for New Physics by measuring $CP$ asymmetries in $B \to K^*\gamma$ decays and the forward-backward asymmetry in $B \to K\ell^+\ell^-$ decays. While there will always be an element of theory uncertainty left in these analyses, in the cases above these uncertainties can be controlled with rather good precision, so that large deviations from Standard Model predictions would have to be interpreted as signs of New Physics. (Indeed, some intriguing "hints of anomalies" are seen in present data.)

#### Probing New Physics in inclusive decays

This is the more traditional approach, which profits from the availability of reliable theoretical calculations. Several methods have been discussed over the years, including precision measurements of the $B \to X_s\gamma$ branching ratio and $CP$ asymmetry, the $B \to X_s\, l^+\ell^-$ rate and forward-backward asymmetry, the inclusive $B \to X_s\nu\overline{\nu}$ decay rate, and some of the above with $X_s$ replaced with $X_d$. The mode $B \to X_s\nu\overline{\nu}$ is tough; it would definitely be Super $B$ Factory territory.

### 3.1.4 Interpreting New Physics–Measuring non-standard flavor parameters

The primary goal of a Super $B$ Factory would be to measure New Physics parameters in the flavor sector. In general, non-standard contributions to flavor-changing processes can be parametrized in terms of the magnitudes and $CP$-violating phases of the Wilson coefficients in a low-energy effective weak Hamiltonian. The main obstacle is that, in general, there can be many such coefficients! Ideally, we would like to probe and measure these couplings in a selective, surgical way, thereby measuring the fundamental coupling parameters of new particles. Equally important is to study the *patterns* of the New Physics, which may reveal important clues about flavor dynamics at very high (beyond-LHC) energy scales.

#### CKM measurements

A clean interpretation of New Physics signals in CKM measurements is difficult (if at all possible) due to the large Standard Model background. An important message is this: In the presence of New Physics, methods that are "clean" (*i.e.*, that do not rely on theory input) in the Standard Model in general become sensitive to hadronic uncertainties. This point is sometimes overlooked. Consider, as an example, the Gronau–London method for measuring $\gamma$ (or $\alpha$) from $B \to \pi\pi$ decays. In the Standard Model, one needs five measurements in order to extract the four unknown hadronic parameters $|P/T|$, $|C/T|$, $\delta_{P/T}$, $\delta_{C/T}$ along with $\gamma$. With New Physics present, there are six additional amplitude parameters and not enough observables to fix them. But things are, in fact, worse than that, for the six new parameters are linear combinations of New Physics parameters and a large number of hadronic parameters—the amplitudes and strong phases of the many $B \to \pi\pi$ matrix elements of the operators in the effective weak Hamiltonian. (It is a misconception to think that there is only one strong phase each for the $\pi\pi$ final states with isospin $I = 0$ or 2.)

The problem is, simply put, that CKM physics is hard. Consider how difficult it has been (and still is) to determine the four parameters of the CKM matrix, *for which there is no background*, since the CKM matrix is the only source of flavor violation in the Standard Model. With New Physics present, the Standard Model is a source of irreducible background for measurements in the flavor sector. In most cases, the subtraction of this background introduces large hadronic uncertainties.

#### Non-CKM measurements

In some cases, the Standard Model background can be strongly reduced or even eliminated, so that one can directly probe certain types of New Physics operators. Examples are decay observables sensitive to electroweak penguins, such as rate and $CP$ asymmetry measurements in $B \to \phi K_S^0$ and $B \to \pi K_S^0$ decays. The idea is to look for certain patterns of "isospin violation", which in the Standard Model are highly suppressed, because they only arise at second order in electroweak interactions ("electroweak penguins"). This fact offers a window for seeing New Physics effects with little Standard Model background. In many models, New Physics can fake the signature of electroweak penguin operators *without* an additional electroweak coupling involved ("trojan penguins"). This provides sensitivity to sometimes very large energy scales (up to several TeV). In other cases, such as $B \to VV$ modes or $B \to K^*\gamma$ decay, one can probe specific operators with non-standard chirality, thereby eliminating the Standard Model background altogether.

Searches for New Physics in inclusive decays are often simpler to interpret, as they are afflicted by smaller theoretical uncertainties in the relation between observables and Wilson coefficient functions. Still, in general it can be difficult to disentangle the contributions from (potentially many) new Wilson coefficients, as only a limited number of observables can be measured experimentally.

#### The importance of patterns of New Physics

Let me close this discussion on an optimistic note. Even if it is hard to cleanly disentangle the contributions from different New Physics operators, CKM measurements will play an important role in helping to distinguish between different *classes* of New Physics effects, such as New Physics in mixing vs. New Physics in decay amplitudes, or New Physics in $b \to s$ vs. $b \to d$ FCNC transitions. CKM measurements might indicate the existence of new $CP$-violating interactions or new flavor-changing interactions not present in the Standard Model. Also, they will help to differentiate between models with and without minimal flavor violation.

Studies of exclusive hadronic decays can help to distinguish between the "flavor-blind" transitions $b \to sg$ and $b \to s(\bar{q}q)_{\rm singlet}$ and "flavor-specific" $b \to s(\bar{q}q)_{\rm non-singlet}$ decays. We will also be in a position to check for the existence of right-handed currents and, more generally, probe for operators with non-standard chirality.

### 3.1.5 Conclusion

Precisely because we don't know what to expect and what to look for, it is the breadth of the physics program at a Super $B$ Factory that will guarantee success. The discovery of new particles at the LHC would help to interpret the possible findings of non-standard signals and guide further studies. Even finding no effects in some channels would provide important clues. Based on these consideration, it is my conviction that the physics case for a Super $B$ Factory is compelling. Such a facility would be an obvious choice to pursue if any of the "anomalies" seen in the present $B$ Factory data would ultimately turn out to be real effects of New Physics.

#### Disclaimer

I have presented some personal reflections on the physics potential and the physics case that can be made for a Super $B$ Factory. My thinking about such a facility has evolved over a period of several years, starting with a workshop in June 2000 in Glen Arbor, Lake Michigan that I helped organize. During this process, I have profited from numerous discussions with colleagues. I have also been influenced significantly by the splendid performance of the SLAC and KEK $B$ Factories and of the *BABAR* and Belle experiments. Many things that were nearly unthinkable even a few years ago now appear within reach. (It is characteristic that the title of our 2000 Workshop referred to a $10^{34}$ machine. In other words, the luminosity target has gone up by a factor 10 every two years!)

I have kept these introductory remarks brief. Much of the supporting material can be found elsewhere in this Proceedings.

## 3.2 $\gamma$ from $B \to DK$, $B \to D\pi$, and Variants

### 3.2.1 Sensitivity of the $B \to DK$ and $B \to D\pi$ decay modes

$\rightarrowtail$ A. Soffer $\leftarrowtail$

One of the main goals of $CP$ violation and $B$ physics in the LHC era will be to measure New Physics parameters that may not be accessible at the LHC. In this section we study how well this could be done with measurements of the CKM phase $\gamma$. The basic idea is to measure $\gamma$ using "Standard Model-only" methods that have negligible New Physics contributions, and compare the result to measurements that are sensitive to New Physics. We summarize the main methods for conducting Standard Model-only measurements of $\gamma$ and discuss some of their features, then make a rough estimate of the sensitivity of the measurement of $\gamma$ with the luminosity of a Super $B$ Factory. We present a brief comparison of these results with measurements that are sensitive to New Physics, giving an indication of the discovery reach of this machine. Finally, we explore some of the possibilities of New Physics contributions to the expectedly Standard Model-only $\gamma$ measurements.

#### Standard Model–only measurements of $\gamma$

The basic technique for measuring $\sin^2\gamma$ in a theoretically clean way is the method of Gronau and Wyler (GW) [1]. The idea is to measure 1) the magnitude $A$ of the $b \to c\overline{u}s$ amplitude $A(B^+ \to \overline{D}^0 K^+)$, by tagging the flavor of the $\overline{D}^0$ using its decay into a state such as $K^+\pi^-$; 2) the magnitude $a \sim A\, r_B$ of the $B \to X_u \ell \overline{\nu}_\ell$ amplitude $A(B^+ \to D^0 K^+)$, identified via the decay $D^0 \to K^-\pi^+$; and 3) the magnitude of the interference amplitude $A \pm ae^{i(\delta_B+\gamma)}$, which takes place when the charmed meson is observed decaying into a $CP$-even $(+)$ or $CP$-odd $(-)$ final state. Here $\delta_B$ is a strong phase difference between the interfering attitudes. These branching fractions and the branching fractions of the $CP$-conjugate $B^-$ modes provide enough information to extract the phase $\gamma$ up to an eight-fold ambiguity. The ambiguity stems from the symmetry of the observable under the three operations

$$\begin{aligned} S_{\rm ex} &\equiv \gamma \to \delta_B\ ,\ \ \delta_B \to \gamma \\ S_{\rm sign} &\equiv \gamma \to -\gamma\ ,\ \ \delta_B \to -\delta_B \\ S_\pi &\equiv \gamma \to \gamma + \pi\ ,\ \ \delta_B \to \delta_B + \pi. \end{aligned} \tag{3.1}$$

The exchange ambiguity $S_{\rm ex}$ can be resolved by combining measurements done with different modes, if each mode has a different value of $\delta_B$. As will be shown below, some variations of this method are able to resolve both $S_{\rm ex}$ and the sign ambiguity $S_{\rm sign}$. However, the $S_\pi$ ambiguity, which is associated with a symmetry of the amplitudes rather than their squares, may not be resolved without making some assumptions regarding the allowed values of $\delta_B$, or by using other measurements (such as $\sin 2\beta$) and taking them to be dominated by the Standard Model.

The GW method has a fair number of variations using different decay modes and techniques, most of which are described below. Each of the variations has various advantages and disadvantages. *A priori*, none of the methods is expected to be significantly more sensitive than the others. Therefore, the measurement of $\gamma$ must rely on at least several methods and decay modes to provide meaningful sensitivity and ambiguity resolution.

The first variation addresses that fact that the amplitude $A(B^+ \to (K^-\pi^+)_D K^+)$ is not equal to the sought-after amplitude $A(B^+ \to D^0 K^+)$, but has the additional contribution $A(B^+ \to \overline{D}^0 K^+)$, where the $\overline{D}^0$ undergoes a doubly-Cabibbo-suppressed decay to $K^-\pi^+$. As pointed out by Atwood, Dunietz, and Soni (ADS) [2], the interference between these two amplitudes is sufficient for measuring $\gamma$, provided one uses at least two $D$ decay final states with different values of the strong phase $\delta_D = {\rm Arg}(\overline{D}^0 \to K^-\pi^+) - {\rm Arg}(D^0 \to K^-\pi^+)$. Alternatively, $\cos\delta_D$ may be accurately measured at a charm factory [3], improving the measurement of $\gamma$ by reducing the number of parameters that need to be determined from the small $B^+ \to (K^-\pi^+)_D K^+$ sample.

We note that contrary to a common misconception, it is still very useful to use the original GW method for measuring $\gamma$ from the interference that takes place when the $D$ decays into $CP$-eigenstate final states, as long as both amplitudes contributing to the $B^+ \to (K^-\pi^+)_D K^+$ decay are taken into account. In fact, including the $CP$-eigenstate decays in the analysis may provide a significant increase of sensitivity with respect to the ADS method alone [4]. The statistical sensitivity of the measurement scales roughly as the smallest amplitude in the problem, which is of order $\min(\lambda^2, r_B)$ in the ADS method and of order $r_B\lambda$ in the GW method. Given the expectation $r_B \sim 0.1$, the sensitivities of both methods are of the same order.

The main difficulty in obtaining a statistically accurate measurement of $\gamma$ using $B \to DK$ is the small magnitude of the $B \to X_u \ell \overline{\nu}_\ell$ amplitude, which is both $|V_{ub}/V_{cb}|$- and color-suppressed. There have been several attempts to overcome this limitation. It should be pointed out that no single attempt results in a significantly more sensitive measurement on its own. However, this quest has resulted in additional modes and methods that add to the overall sensitivity of the $\gamma$ measurement. Several of these methods are described below.

Neglecting annihilation diagrams, several authors [5] have used isospin symmetry to relate the small $A(B^+ \to D^0 K^+)$ amplitude to $b \to c\overline{u}s$ color suppressed amplitudes, which are larger and therefore should be easier to measure. This is expected to slightly improve the measurement with respect to the GW method alone.

Another modification of the GW method is to measure $\gamma$ using singly-Cabibbo-suppressed $D$ decays to final states such as $K^{*\pm}K^\mp$ or $\rho^\pm\pi^\mp$ [6]. These modes are similar to the $CP$-eigenstate final states of the GW method, in that the ratio between the decay amplitudes of $D^0$ and $\overline{D}^0$ into these final states is $r_D \sim 1$. However, since $r_D \neq 1$, each final state provides two measurements, enabling one to do away with having to measure the $\mathcal{O}(a^2)$ branching fraction $\mathcal{B}(B^+ \to (K^-\pi^+)_D K^+)$. The measurement of $\gamma$ is then dependent only on terms of order $Aa$, rather than of order $a^2$, as in the GW or ADS methods. In addition, the variation of $\delta_D$ across the $D$ decay Dalitz plot breaks the $S_{\rm ex}$ and $S_{\rm sign}$ symmetries, leaving the measurement with only a two-fold ambiguity. As in the GW method, the smallest amplitude in this method is of order $r_B\lambda$, and so is expected to have similar sensitivity. A special case is when one uses only $CP$-odd and $CP$-even eigenstate decays of the $D$, making $r_D = 1$ and $\delta_D = \pi$ [7]. However, while no $\mathcal{O}(a^2)$ branching factions must be measured in this case, the measurement is still dependent on terms of order $a^2$, unlike the case of the non-$CP$-eigenstate decays.

A similar approach may be carried out with Cabibbo allowed $D$ decays, such as $D \to K^0_S\pi^+\pi^-$ [8]. The great advantage of high statistics is balanced by the fact that significant interference between $D^0$ and $\overline{D}^0$ decays occurs in only a small fraction of the Dalitz plot, finally resulting in a sensitivity that is most likely comparable to those of the other methods. The measurement of $\gamma$ with multi-body $D$ decays can be done in a model independent way, without making assumptions about resonances or other structure in the $D$ decay. Dividing the Dalitz plot into as few as four bins provides enough measurements to extract all the unknowns.

A different kind of modification of the GW method is to use multi-body $B$ decay modes, such as $B^+ \to DK^+\pi^0$ [9]. In this case, the $B \to X_u \ell \overline{\nu}_\ell$ amplitude has a color-allowed contribution, increasing $r_B$ from around $0.1 - 0.2$ to $0.4 - 0.8$. In addition, the variation of $\delta_B$ across the Dalitz plot resolves the $S_{\rm ex}$ and $S_{\rm sign}$ ambiguities. As in the case of the multi-body $D$ decays, Dalitz plot suppression balances this advantage to yield a sensitivity similar to those of the other methods. Monte Carlo studies suggest that a statistical error of order $2^\circ$ is possible with 10 ab$^{-1}$, given some assumptions.

Another way to measure $\gamma$ makes use of untagged neutral $B$ decays $B^0 \to DK^0_S$ [10]. Although the flavor of the decaying $B$ meson is not tagged, one can obtain enough observables to measure all the unknowns, including $\gamma$, by studying $D$ decays to three different modes, or two $B$ modes and two $D$ modes. Although both interfering amplitudes are color-suppressed, the sensitivity of the measurement of $\gamma$ is In general dominated by the magnitude of the smaller of the interfering amplitudes, which is similar in both charged and neutral $B$ decays. Tagging the flavor of the $B$ provides additional information that improves the measurement of $\gamma$, as well as additional combinations of CKM phases. However, the effective tagging efficiency at $B$ Factories is only about 30%. Therefore, using untagged $B^0$ decays along with tagged $B^0$ decays and $B^-$ decays is expected to significantly enhance the total sensitivity.

The second theoretically clean method to measure $\gamma$ makes use of interference between a $b \to u\overline{c}d$ amplitude and $B^0\overline{B}^0$ mixing followed by a $b \to c\overline{u}d$ amplitudes in the decay $B \to D^{(*)\pm}h^\mp$, where $h$ stands for a light hadron, such as $\pi$, $\rho$, or $a_1$ [11]. A time-dependent analysis of the decay provides a measurement of $\sin(2\beta+\gamma)$. The ratio between the interfering amplitudes is only about $r_B \sim 0.02$. Nonetheless, the high statistics one can obtain in these decays has enabled *BABAR* and Belle to make the first attempt to measure this weak phase with $B \to D^{(*)\pm}\pi^{\mp}$ [12]. Giving an idea of the sensitivity at a Super $B$ Factory, the total error *BABAR* obtained in $B \to D^{(*)\pm}\pi^{\mp}$ was $\sigma_{2r_B \sin(2\beta+\gamma)} = 0.023$. This error is almost entirely proportional to $1/\sqrt{N}$.

Although $r_B$ can be measured from the data for these modes, doing so requires sensitivity to the difference between $1-r_B^2$ and $1+r_B^2$. As this is impossible with current statistics, one has to assume SU(3) symmetry and take $r_B$ from the branching fraction of the $B \to X_u \ell \overline{\nu}_\ell$ decay $B^+ \to D_s^- \pi^+$, incurring a large theoretical error. This problem may be overcome with the vector-vector modes $B \to D^{*\pm}\rho^{\mp}$ and $B \to D^{*\pm}a_1^{\mp}$ [13]. In this case, interference between the different helicity amplitudes provides several $\mathcal{O}(r_B)$ terms, which are distinguishable by their different angular distributions, thus enabling the measurement of $r_B$ with much greater statistical significance. This comes at the price of a much more complicated time- and angle-dependent analysis.

One potentially serious problem with measuring $\sin(2\beta+\gamma)$ in $B \to D^{(*)\pm}\pi^{\mp}$ is the ambiguity between $\sin(2\beta+\gamma)$ and $\cos\delta_B$ [14]. The data suggest that both of these quantities may be around 1, in which case the measurement adds very little to our knowledge of the unitarity triangle even with integrated luminosities of several ab$^{-1}$. This problem should be solved for the vector-vector modes, where measurements by CLEO [15] indicate significant strong phase differences between the different helicity amplitudes in $B \to D^{*\pm}\rho^{\mp}$.

The problem should also go away in $B \to D^{(**)\pm}h^{\mp}$. Here, interference between different $D^{**}$ resonances (and between the resonances and continuum $D\pi\pi$ production) plays the role of interference between different helicity amplitudes in the vector-vector modes, enabling a much more accurate measurement of $r_B$ than with $B \to D^{(*)\pm}\pi^{\mp}$. In addition, the interfering resonances cause large variation of the strong phase as a function of the $D\pi\pi$ invariant mass, breaking the $\sin(2\beta+\gamma) \leftrightarrow \cos\delta$ ambiguity.

Similar to the idea of using $B^{\pm} \to D^0 K^{\pm}\pi^0$ to obtain a color-allowed $B \to X_u \ell \overline{\nu}_\ell$ amplitude, one can measure $\sin(2\beta+\gamma)$ in $B^0 \to D^{\pm}K_S^0\pi^{\mp}$ with an $r_B$ of about $0.4$ [16]. This method has similar advantages and disadvantages as the other multi-body modes. Finally, we mention the measurement of $\sin(2\beta+\gamma)$ in $B^0 \to DK^0$ [17]. In this case, all the measured modes are color suppressed and the analysis is time-dependent. However, one expects $r_B \sim 0.4$, consistent with current measurements [18] that suggest $r_B = 0.6 \pm 0.2$.

**$\gamma$ with Super $B$ Factory luminosity**

While there are still uncertainties regarding the values of $r_B$ and other relevant parameters, we have enough information to make a rough estimate of the Super $B$ Factory sensitivity to $\gamma$. Most of this information comes from toy Monte Carlo studies conducted when a new method is proposed or as part of an ongoing data analysis. In some cases, most notably $\sin(2\beta+\gamma)$, we have actual measurements upon which to base reliable estimates that include most of the experimental considerations. Unfortunately, due to the $\sin(2\beta+\gamma) \leftrightarrow \cos\delta_B$ ambiguity, it is difficult to extrapolate from the $\sin(2\beta+\gamma)$ error to the $\gamma$ error at this time.

Keeping these caveats in mind while adding up the estimates reported in the various papers, we conclude that an integrated luminosity of 10 ab$^{-1}$ is highly likely to yield a measurement of $\gamma$ with a statistical error of order $1-2^{\circ}$. Given the uncertainties involved in these estimates and the possibility that the flurry of new ideas we have seen recently continues in the next few years, an even smaller error is not out of the question. It remains to be seen whether the systematic error can be reduced to that level. This issue will become much better understood in the next year or two, as more analyses mature.

Discrete ambiguities can make the value of $\sigma_\gamma$ irrelevant [4]. However, with $10\,\text{ab}^{-1}$ and the ambiguity resolution capability of some of the methods surveyed here, it is clear that the $S_{\rm ex}$ and $S_{\rm sign}$ ambiguities will be completely resolved. This leaves the two-fold ambiguity of the $S_\pi$ symmetry, which is already forbidden by $\sin 2\beta$ and $\epsilon_K$. Therefore, $\gamma$ will be measured with essentially no discrete ambiguities in the Super $B$ Factory era.

**Comparison with measurements sensitive to New Physics**

Having cleanly measured $\gamma$ within the Standard Model, what New Physics-sensitive measurements can we compare this to in order to gain insight into the nature of the New Physics?

Measurements of $\gamma$ that involve interference between tree and penguin diagrams are sensitive to New Physics through the penguin loops. However, even with 10 ab$^{-1}$, the sensitivity of these measurements will be much larger than the $1-2^\circ$ expected from the measurements of Section 3.2.1. Therefore, comparison with penguin mode measurements of $\gamma$ is probably not the most useful way to study New Physics.

Another place where New Physics can contribute is in the box diagrams of $B^0\overline{B}^0$ and $B_s^0\overline{B}_s^0$ mixing. Mixing rates are related to CKM parameters through

$$\left|\frac{V_{td}}{V_{ts}}\right|^2 = \xi^2 \frac{\Delta m_d}{\Delta m_s}\frac{m_{B_s}}{m_{B_d}}. \tag{3.2}$$

The parameter $\xi$ will soon be calculable in lattice QCD to $\simeq 1-2\%$ [19], $\Delta m_d$ will be measured by the current generation of $B$ factories to about 1%, and $\Delta m_s$ should be measurable at hadronic machines to less than 1%. The $B_s^0$ ($B_d^0$) mass is already known to $0.05\%$ ($0.01\%$). From the measurements of these parameters, plus the unitarity relation $|V_{ts}| = |V_{cb}| + \mathcal{O}(\lambda^4)$, one can extract $|V_{td}|$, which is related to $\gamma$ through the unitarity triangle. This relation is a simple geometrical consequence of the fact that $\gamma \sim 90^\circ$. We see that it is reasonable to expect the relative errors of $|V_{td}|$ and $\gamma$ to be comparable in the Super $B$ Factory era. This makes the comparison of these parameters a good way to detect or study New Physics, as long as the New Physics contribution to the mixing amplitudes or to the breaking of the relation $|V_{ts}| \approx |V_{cb}|$ is no less than about 1% of the Standard Model.

**New Physics contamination in the measurements of $\gamma$**

While we said that the measurements of $\gamma$ are Standard Model-only, there is the possibility of some New Physics contribution to these measurements.

First, the formalisms of the $B \to D^0K$ measurements generally neglect the possibility of $D^0\overline{D}^0$ mixing and *CP* violation in the $D^0$ decay. Unaccounted for, these effects may bias the ADS measurement of $\gamma$ by $\mathcal{O}(1^\circ)$ and possibly even $\mathcal{O}(10^\circ)$ [20]. However, it is straightforward to take the effect into account in the equations, using measurements of or limits on $D^0\overline{D}^0$ mixing and $D^0$ decay *CP* violation as input. In addition, the effect on the non-ADS methods is smaller by about a factor of $r_B$.

Second, a charged Higgs contribution to the tree diagrams would appear just like a Standard Model charged current interaction. In this case, its existence will presumably be detected elsewhere. It would probably have different effects on mixing and $\gamma$ measurements, so measuring $\gamma$ would still be useful for studying New Physics.

**Summary**

To conclude, there are many methods and modes with which one can measure $\gamma$ in the Standard Model. With high confidence, we expect that conducting most of these measurements with 10 ab$^{-1}$ and combining the results will determine $\gamma$ to about $1-2^\circ$. More precise determination is possible and perhaps even likely, but more experimental experience is required before this can be stated with confidence. The $\gamma$ measurement will have essentially no discrete ambiguities. The measurement of $|V_{td}|$ will have a similar relative error, and so comparing the unitarity triangle constraints obtained with these two independently-measured parameters will yield sensitivity to New Physics at the level of $1-2\%$ of the Standard Model.

### 3.2.2 $B \to DK$ using Dalitz plot analysis

$\succ$– J. Zupan –$\prec$

There are many variants of the original Gronau-Wyler proposal [1] to extract $\gamma$ from the $B \to DK$ decays. Usually, several different decay modes of $D$ mesons are used, among them also the quasi-two-body $D$ decays with one or both of the particles in the final state a strongly decaying resonance (*e.g.*, $D^0 \to K^{*+}\pi^-$ [2, 21]). Since these are really many-body decays (for instance in the example mentioned, $K^{*+}$ decays strongly to $K^0\pi^+$ or $K^+\pi^0$ so that one in reality has a three body final state), one can pose the following questions:

- Can one use the complete phase space of such many-body $D$ decays for $\gamma$ extraction?
- Is it possible to avoid fits to Breit-Wigner forms in doing the Dalitz plot analysis?

As we show in the following, the answers to both of the questions are positive. Let us first discuss the first question on the list. To do so let us restrict ourselves to the following cascade decay [1]

$$B^- \to DK^- \to (K^0_S\pi^-\pi^+)_D K^-, \tag{3.3}$$

while the extension to the other multi-body final states can be found in [8], [22]. To pin down the notation let us define for the amplitudes

$$A(B^- \to D^0 K^-) \equiv A_B, \tag{3.4}$$
$$A(B^- \to \overline{D}^0 K^-) \equiv A_B r_B e^{i(\delta_B-\gamma)}. \tag{3.5}$$

Here $\delta_B$ is the difference of strong phases and $A_B$ is taken to be positive. The same definitions apply to the amplitudes for the $CP$ conjugate cascade $B^+ \to DK^+ \to (K^0_S\pi^+\pi^-)_D K^+$, except that the weak phase flips the sign: $\gamma \to -\gamma$ in (3.5).

For the $D$ meson decay we further define

$$\begin{aligned} A_D(s_{12}, s_{13}) \equiv A_{12,13}\, e^{i\delta_{12,13}} &\equiv A(D^0 \to K^0_S(p_1)\pi^-(p_2)\pi^+(p_3)) \\ &= A(\overline{D}^0 \to K^0_S(p_1)\pi^+(p_2)\pi^-(p_3)), \end{aligned} \tag{3.6}$$

where $s_{ij} = (p_i + p_j)^2$, and $p_1, p_2, p_3$ are the momenta of the $K^0_S, \pi^-, \pi^+$ respectively. Again $A_{12,13} \geq 0$, so that $\delta_{12,13}$ can vary between 0 and $2\pi$. In the last equality the $CP$ symmetry of the strong interaction together with the fact that the final state is a spin zero state has been used. With the above definitions, the amplitude for the cascade decay is

$$A(B^- \to (K^0_S\pi^-\pi^+)_D K^-) = A_B \mathcal{P}_D \big(A_D(s_{12}, s_{13}) + r_B e^{i(\delta_B-\gamma)} A_D(s_{13}, s_{12})\big), \tag{3.7}$$

where $\mathcal{P}_D$ is the $D$ meson propagator. Next, we write down the expression for the reduced partial decay width

$$\begin{aligned} d\hat{\Gamma}(B^- \to (K^0_S\pi^-\pi^+)_D K^-) = \Big(&A^2_{12,13} + r_B^2\, A^2_{13,12} \\ &+ 2 r_B\, \Re\Big[A_D(s_{12}, s_{13})\, A^*_D(s_{13}, s_{12})\, e^{-i(\delta_B-\gamma)}\Big]\Big) dp, \end{aligned} \tag{3.8}$$

where $dp$ denotes the phase space variables, over which one needs to integrate to make contact with experiment. The dependence on $\gamma$ enters in the interference term in (3.8), so that $\gamma$ can be easily extracted, if one knows (measures) the variation of both the moduli and the phases of the $D^0$ meson decay amplitudes $A_D(s_{12}, s_{13})$ along the Dalitz plot.

This can be accomplished by introducing mild model-dependent assumptions by performing a fit of the decay amplitude to a sum of Breit-Wigner functions and a constant term to the tagged $D$ data. Following the notations of Ref. [23] we write

$$\begin{aligned} A_D(s_{12}, s_{13}) &= A(D^0 \to K^0_S(p_1)\pi^-(p_2)\pi^+(p_3)) = \\ &= a_0 e^{i\delta_0} + \sum_r a_r e^{i\delta_r} \mathcal{A}_r(s_{12}, s_{13}), \end{aligned} \tag{3.9}$$

[1] In the following discussion we neglect $DzDzb$ mixing, which is a good approximation in the context of the Standard Model.

where the first term is the non-resonant term, while the rest are the resonant contributions with $r$ denoting a specific resonance. The functions $\mathcal{A}_r$ are products of Breit-Wigner functions and appropriate Legendre polynomials that account for the fact that $D$ meson is a spin $0$ particle. Explicit expressions can be found in Ref. [23].

One of the strong phases $\delta_i$ in the ansatz (3.9) can be put to zero, while others are fit to the tagged $D$ decay data together with the amplitudes $a_i$. The obtained functional form of $A_D(s_{12}, s_{13})$ can then be fed to Eq. (3.8), which is then fit to the Dalitz plot of the $B^\pm \to (K_S^0\pi^-\pi^+)_D K^\pm$ decay with $r_B$, $\delta_B$ and $\gamma$ left as free parameters. Thus only three variables, $r_B, \delta_B$, and $\gamma$, need to be obtained from the $B$ system. Note that this was the method used in [24] to obtain the first constraints on the $\gamma$ angle from the $B \to (K_S^0\pi^+\pi^-)_D K$ decay cascade.

The theoretical uncertainty now boils down to the question how well the $D$ decay amplitude is described by the fit to the Breit-Wigner forms. The related error can of course be reduced with increasing the sample of tagged $D$ decays, when more and more resonances can be introduced in the fit as well as if more sophisticated ansätze for the $s$ dependence of the Breit-Wigner forms are taken. Luckily, however, this question can be avoided altogether by performing a completely model-independent analysis.

In the following we will use the notation of [8], however, an equivalent formalism has been independently developed by Atwood and Soni in [22]. Starting from Eq. (3.8) we partition the Dalitz plot into $n$ bins and define

$$c_i \equiv \int_i dp\, A_{12,13}\, A_{13,12}\, \cos(\delta_{12,13} - \delta_{13,12}), \tag{3.10}$$

$$s_i \equiv \int_i dp\, A_{12,13}\, A_{13,12}\, \sin(\delta_{12,13} - \delta_{13,12}), \tag{3.11}$$

$$T_i \equiv \int_i dp\, A_{12,13}^2, \tag{3.12}$$

where the integrals are done over the phase space of the $i$-th bin. The variables $c_i$ and $s_i$ contain differences of strong phases and are therefore unknowns in the analysis. The variables $T_i$, on the other hand, can be measured from the flavor-tagged $D$ decays, and are assumed to be known inputs into the analysis.

Due to the symmetry of the interference term, it is convenient to use pairs of bins that are placed symmetrically about the $12 \leftrightarrow 13$ line, as shown in Fig. 3-1. Consider an even, $n = 2k$, number of bins. The $k$ bins lying below the symmetry axis are denoted by index $i$, while the remaining bins are indexed with $\bar{i}$. The $\bar{i}$-th bin is obtained by mirroring the $i$-th bin over the axis of symmetry. The variables $c_i, s_i$ of the $i$-th bin are related to the variables of the $\bar{i}$-th bin by

$$c_{\bar{i}} = c_i, \qquad s_{\bar{i}} = -s_i, \tag{3.13}$$

while there is no relation between $T_i$ and $T_{\bar{i}}$.

Together with the information available from the $B^+$ decay, we arrive at a set of $4k$ equations

$$\hat{\Gamma}_i^- \equiv \int_i d\hat{\Gamma}(B^- \to (K_S^0\pi^-\pi^+)_D K^-) = T_i + r_B^2 T_{\bar{i}} + 2r_B[\cos(\delta_B - \gamma)c_i + \sin(\delta_B - \gamma)s_i], \tag{3.14}$$

$$\hat{\Gamma}_{\bar{i}}^- \equiv \int_{\bar{i}} d\hat{\Gamma}(B^- \to (K_S^0\pi^-\pi^+)_D K^-) = T_{\bar{i}} + r_B^2 T_i + 2r_B[\cos(\delta_B - \gamma)c_i - \sin(\delta_B - \gamma)s_i], \tag{3.15}$$

$$\hat{\Gamma}_i^+ \equiv \int_i d\hat{\Gamma}(B^+ \to (K_S^0\pi^-\pi^+)_D K^+) = T_{\bar{i}} + r_B^2 T_i + 2r_B[\cos(\delta_B + \gamma)c_i - \sin(\delta_B + \gamma)s_i], \tag{3.16}$$

$$\hat{\Gamma}_{\bar{i}}^+ \equiv \int_{\bar{i}} d\hat{\Gamma}(B^+ \to (K_S^0\pi^-\pi^+)_D K^+) = T_i + r_B^2 T_{\bar{i}} + 2r_B[\cos(\delta_B + \gamma)c_i + \sin(\delta_B + \gamma)s_i]. \tag{3.17}$$

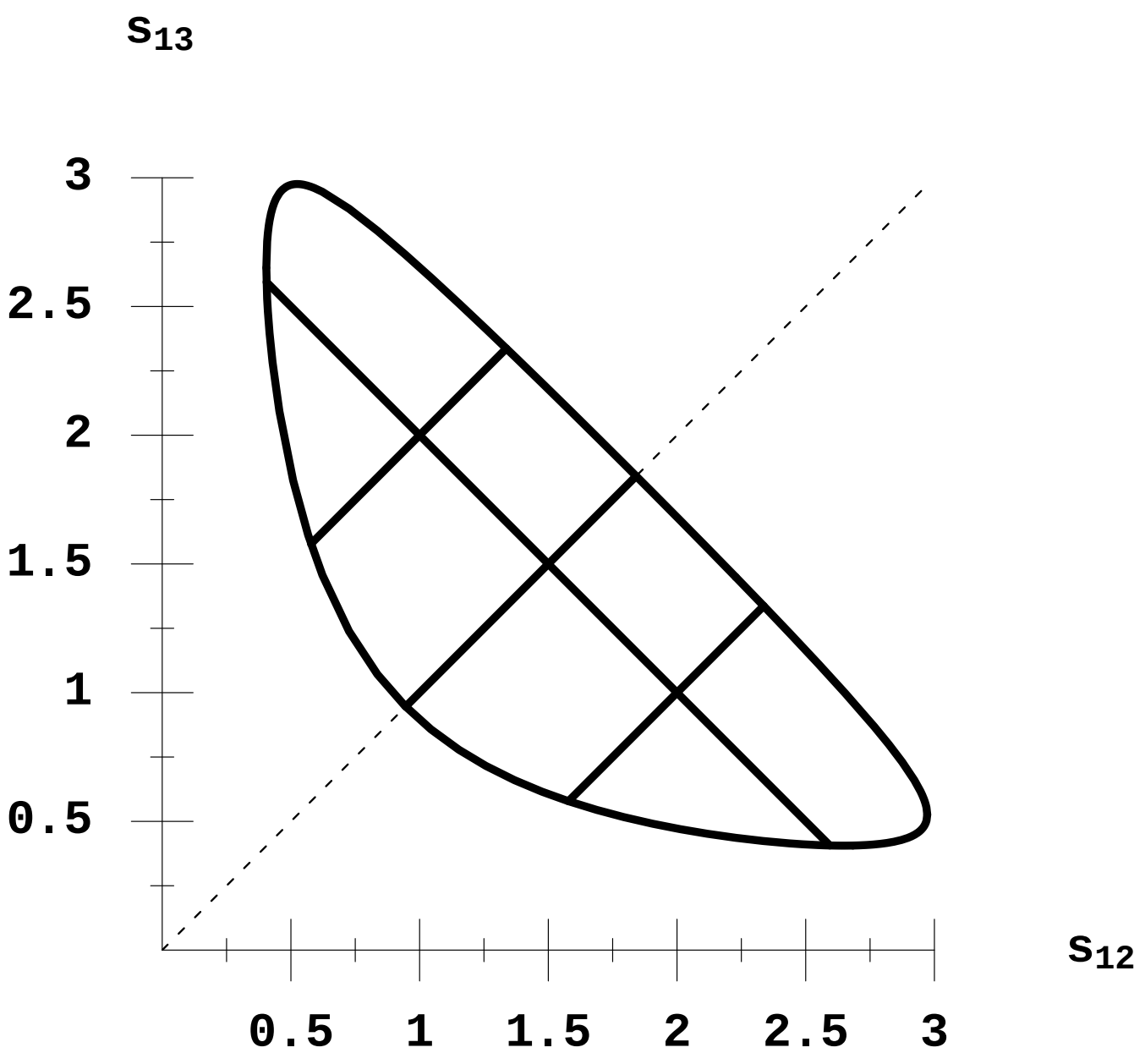

**Figure 3-1.** *The partitions of Dalitz plot as discussed in text. The symmetry axis is the dashed line. On the axes we have* $s_{12} = m^2_{K^0_S\pi^-}$ *and* $s_{13} = m^2_{K^0_S\pi^+}$ *in* $\mathrm{GeV}^2$.

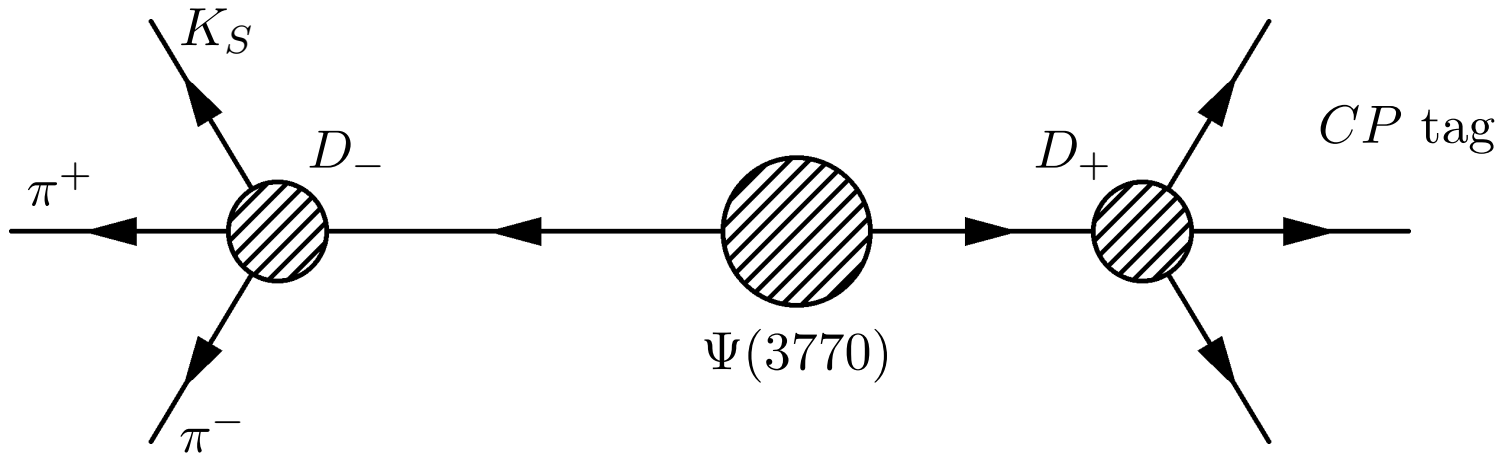

**Figure 3-2.** *The interference between the decays of* $D$ *mesons originating from* $\psi(3770)$ *allow for a measurement of* $c_i$ *and* $s_i$ *at charm factories. Shown is a decay allowing for determination of* $c_i$.

These equations are related to each other through $12 \leftrightarrow 13$ and/or $\gamma \leftrightarrow -\gamma$ exchanges. All in all, there are $2k+3$ unknowns in (3.14)-(3.17),

$$c_i,\ s_i,\ r_B,\ \delta_B,\ \gamma, \tag{3.18}$$

so that the $4k$ relations (3.14)-(3.17) are solvable for $k \geq 2$. In other words, a partition of the $D$ meson Dalitz plot to four or more bins allows for the determination of $\gamma$ without hadronic uncertainties.

So far, we have used the $B$ decay sample to obtain all the unknowns, including $c_i$ and $s_i$, which are actually parameters of the charm system. We now show that the $c_i$ and $s_i$ can be independently measured at a charm factory [3, 20, 25]. This is done by running the machine at the $\psi(3770)$ resonance, which decays into a $D\overline{D}$ pair. Let one of these decay into $K^0_S\pi^+\pi^-$ and the other into some general state $g$ (see Fig. 3-2. The partial decay width corresponding to the $i-$th bin of the $K^0_S\pi^+\pi^-$ Dalitz plot and the $j-$th bin of the $g$ final state's phase space is

$$\Gamma_{i,j} \propto T_i T^g_{\bar{j}} + T_{\bar{i}} T^g_j - 2(c_i c^g_j + s_i s^g_j), \tag{3.19}$$

where $T^g_j$, $c^g_j$, $s^g_j$ are defined as in (3.10)-(3.12). In particular, if one chooses $g = K^0_S\pi^+\pi^-$ and $j = i$ (or $j = \bar{i}$) one has in the last term $c_i^2 + s_i^2$. If, on the other hand, $g$ is a *CP* even (odd) eigenstate, $s^g_j = 0$, $T^g_j = T^g_{\bar{j}} = \pm c^g_j$ and in equation (3.19) the last term is linear in $c_i$. In this way one can measure $c_i$ as well as $s_i$ (the latter only up to a sign).

Some further remarks are in order:

- The observables $\hat{\Gamma}_i^{\pm}$ defined in (3.14)-(3.17) can be used to experimentally look for direct *CP* violation. Explicitly,
$$A_{CP}^{i,\bar{i}} \equiv \hat{\Gamma}^-_{i,\bar{i}} - \hat{\Gamma}^+_{\bar{i},i} = 4 r_B \sin\gamma \left[ c_i \sin\delta_B \mp s_i \cos\delta_B \right], \tag{3.20}$$
Nonzero $A_{CP}$ requires non-vanishing strong and weak phases. Due to the resonances the strong phases are expected to be large, therefore $A_{CP}$ is expected to be sizable as well.

- The model-independent method described above involves a four-fold ambiguity in the extracted value of $\gamma$. The set of equations (3.14)-(3.17) is invariant under each of the discrete transformations [2]
$$P_\pi \equiv \{\delta_B \to \delta_B + \pi, \gamma \to \gamma + \pi\}, \tag{3.21}$$
$$P'_\pi \equiv \{c_i \to -c_i, s_i \to -s_i, \gamma \to \gamma + \pi\} \tag{3.22}$$
$$P_- \equiv \{\delta_B \to -\delta_B, \gamma \to -\gamma, s_i \to -s_i\}. \tag{3.23}$$
The discrete transformation $P_\pi$ is a symmetry of the amplitude (3.7) and is thus an irreducible uncertainty of the method. The ambiguity due to $P_-$ can be resolved if the sign of $s_i$ is determined by fitting a part of the Dalitz plot to Breit-Wigner functions. Then the usual 8-fold ambiguity of the Gronau-Wyler method reduces to a two fold ambiguity.

- The presented formalism can be extended to the multibody $B$ decays, $B^- \to DX_s^- \to (K_S^0\pi^-\pi^+)_D X_s^-$, as well as to multibody $D$ decays with more than three particles in the final state [8].

- Unfortunately this formalism cannot be applied to a general multibody system. For the method to work in the $B \to DK$ case two ingredients were essential: (i) there is a separation of the $B$ and $D$ decay observables, so that $T_i$ in (3.14)-(3.17) can be measured separately from the tagged $D$ decays, and (ii) there are only two interfering amplitudes (*e.g.*, there are no penguin contributions). An example of the analysis where the outlined model independent method fails, is extraction of $\alpha$ from $B \to 3\pi$. Here isospin analysis is needed, so that there are many unknowns, *i.e.*, the integrals over the interference terms between all different penguin and tree invariant amplitudes with different isospin labels, $\int_i \mathrm{Tree_a Tree_b^*}$, $\int_\mathrm{i} \mathrm{Tree_a Peng_b^*}$. Therefore, there are just too few observables to fit all of them.

In conclusion, we have shown that the angle $\gamma$ can be determined without any model dependence from the cascade decays $B^\pm \to DK^\pm$, with $D$ decaying into a multibody final state. The theoretical uncertainties in this method are very small. In the formalism presented above the $DzDzb$ mixing has been neglected. If this mixing is *CP*-conserving, its effect is taken into account automatically (just replace $D^0$ and $\overline{D}^0$ in (3.6) with $D^0(t=0)$ and $\overline{D}^0(t=0)$, respectivelly, while everything else remains unchanged). The largest theoretical error is therefore due to possible *CP* violation in the $D$ decay, which, however, is highly CKM suppressed by $\lambda^5 \sim 5 \cdot 10^{-4}$.

### 3.2.3 $B \to DK$ with a $B_{CP}$ tag

>– A. Falk –<

Pair production of $B$ mesons at the $\Upsilon(4S)$ allows for the possibility of studying *CP*-tagged as well as flavor-tagged $B$ decays. A *CP* tagged decay is one in which the $B$ meson on the other side decays to a *CP* eigenstate such as $J/\psi K_S^0$. If approximately $10^6$ decays $B^0 \to J/\psi K_S^0$ could be reconstructed in a data set of ten to twenty ab$^{-1}$, as a naive extrapolation from *BABAR* might suggest, then this would yield correspondingly $10^6$ *CP* tagged $B^0$ decays on the other side.

The usefulness of a sample of *CP*-tagged decays is illustrated most easily by considering the $B_s$ meson, where *CP* violation may be neglected both in $B_s$ mixing and in tagging decays such as $B_s \to D_s^+ D_s^-$ [28]. The $B_s$ decays to

[2]Note that $P'_\pi$ was erroneously left out from Refs. [8],[26], however, this does not change the discussion about the ambiguity in $\gamma$ extraction.

$DK$ final states by the quark level processes $b \to c\overline{u}s$ and $b \to u\overline{c}s$, whose interference gives information about the CKM angle $\gamma$. With a particular strong phase convention, we may define amplitudes for the flavor tagged decays,

$$\begin{aligned} A_1 &= A(B_s \to D_s^- K^+) = a_1\,, & A_2 &= A(\overline{B}_s \to D_s^- K^+) = a_2 e^{-i\gamma} e^{i\delta}\,, \\ \overline{A}_1 &= A(\overline{B}_s \to D_s^+ K^-) = a_1\,, & \overline{A}_2 &= A(B_s \to D_s^+ K^-) = a_2 e^{i\gamma} e^{i\delta}\,, \end{aligned} \tag{3.24}$$

where $a_1$ and $a_2$ are taken to be real and $\delta$ is a strong phase. The $CP$ tagged decays are also defined,

$$A_{CP} = A(B_s^{CP} \to D_s^- K^+) \qquad \overline{A}_{CP} = A(B_s^{CP} \to D_s^+ K^-)\,. \tag{3.25}$$

Choosing a convention for the $CP$ transformation such that $B_s^{CP} \propto B_s + \overline{B}_s$, we then have $\sqrt{2}A_{CP} = A_1 + A_2$ and $\sqrt{2A}_{CP} = \overline{A}_1 + \overline{A}_2$. These triangle relations are illustrated in Fig. 3-3, from which it is clearly straightforward to extract $\gamma$.

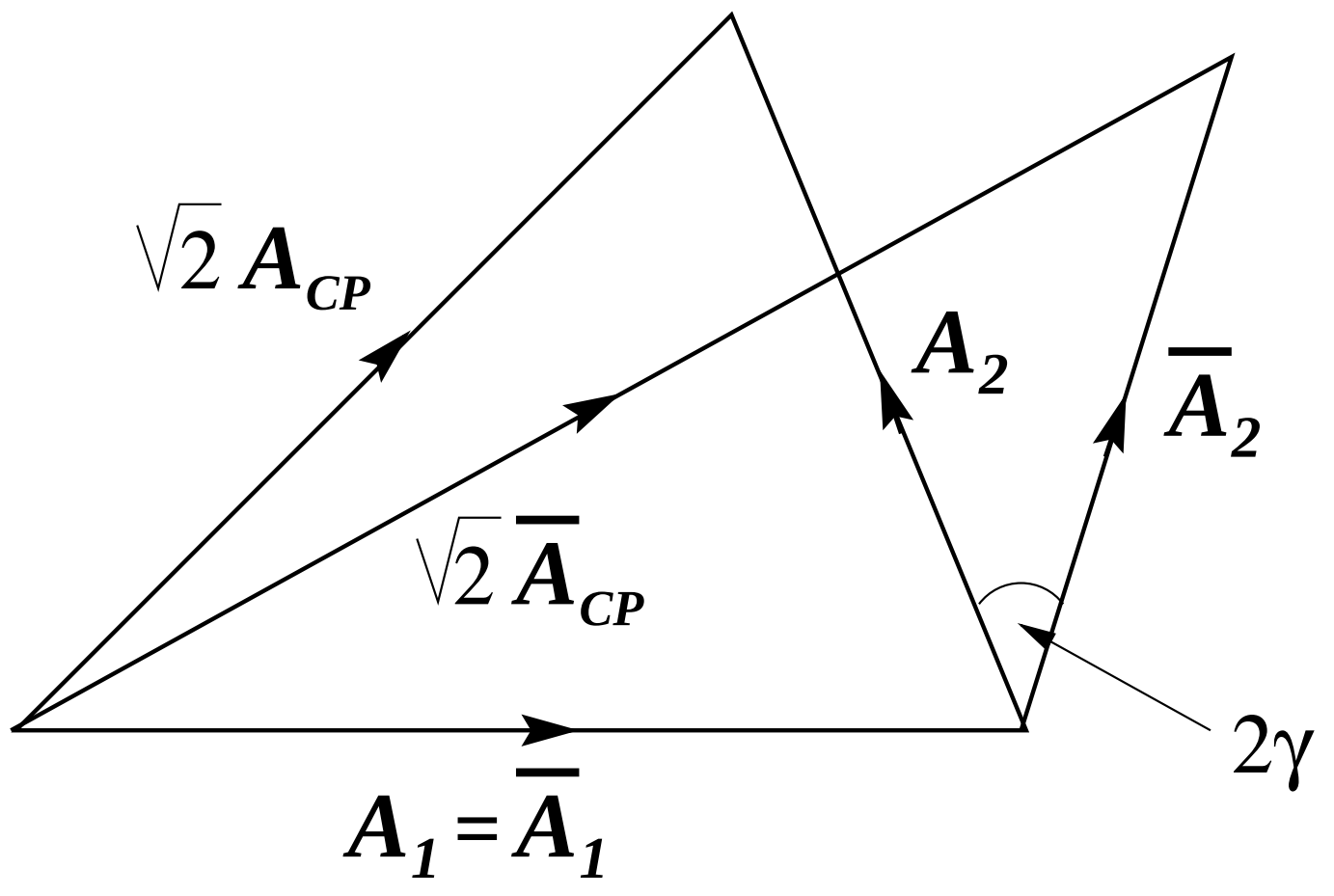

**Figure 3-3.** *Triangle relations for extracting $\gamma$ from $CP$ tagged $B_s$ decays*

The analytic result is given by

$$2\gamma = \arccos \alpha - \arccos \overline{\alpha}\,, \tag{3.26}$$

where $\alpha$ and $\overline{\alpha}$ are defined by

$$\alpha = \frac{2|A_{CP}|^2 - |A_1|^2 - |A_2|^2}{2|A_1||A_2|} \qquad \overline{\alpha} = \frac{2|\overline{A}_{CP}|^2 - |\overline{A}_1|^2 - |\overline{A}_2|^2}{2|\overline{A}_1||\overline{A}_2|}\,. \tag{3.27}$$

The squares of the amplitudes may be replaced by the relevant branching fractions in the ratios. There is an eightfold ambiguity in $\gamma$ over the range $0 \leq \gamma < 2\pi$.

For $CP$-tagged $B^0$ decays, the situation is complicated by $CP$ violation in the interference between $B^0$ mixing and the tagging decay $B^0 \to J/\psi K_S^0$. However, the effect can be incorporated into the analysis once $\sin 2\beta$ and $\Delta m_{B_d}$ are known. Define the time averaged quantities

$$C_{CP} = \frac{\Gamma}{2}\int_{-\infty}^{\infty} dt\,|A_{CP}|^2\,, \qquad \overline{C}_{CP} = \frac{\Gamma}{2}\int_{-\infty}^{\infty} dt\,|\overline{A}_{CP}|^2\,, \tag{3.28}$$

which are proportional to the time averaged $CP$-tagged branching ratios. Then the presence of $CP$ violation can be absorbed entirely into a dilution factor

$$R = \sqrt{(1 - 2\chi_d \sin^2 2\beta)^2 + (2\chi_d \sin 2\beta \cos 2\beta)^2}\,, \tag{3.29}$$

where $\chi_d = (\Delta m_{B_d})^2/2[(\Delta m_{B_d})^2 + \Gamma^2]$. Once again we have $2\gamma = \arccos\alpha - \arccos\overline{\alpha}$, where in this case

$$\alpha = \frac{2|C_{CP}|^2 - |A_1|^2 - |A_2|^2}{2R|A_1||A_2|} \qquad \overline{\alpha} = \frac{2|\overline{C}_{CP}|^2 - |\overline{A}_1|^2 - |\overline{A}_2|^2}{2R|\overline{A}_1||\overline{A}_2|} \,. \tag{3.30}$$

To extract $\gamma$ from $CP$-tagged $B^0$ decays, one would want to study the processes $B^0 \to (D^0, \overline{D}^0)K^0_S$, so that the two quark transitions would be of the same order in $\lambda$ and the amplitude triangles would not be squashed.

The accuracy in $\gamma$ that this method would yield depends not only on the accuracy with which the relevant branching ratios are measured, but on the actual values of $\gamma$, $\delta$ and $a_2/a_1$. The dominant experimental errors will be on the $CP$-tagged branching fractions, compared to which the flavor-tagged branching fractions can be assumed to be known precisely by the time this analysis would be performed. Similarly, the experimental errors on $\sin 2\beta$ and $\chi_d$ will be relatively unimportant.

Assuming, simply for the sake of argument, that the branching ratios for $B^0 \to DK^0_S$ and $B^0 \to D^*K^0_S$ are approximately $10^{-4}$, and that each final state could be reconstructed with 15% efficiency, an overall sample of $2 \times 10^6$ $CP$-tagged decays would yield roughly 10 $CP$-tagged events in each channel. Hence it is reasonable to expect that the $CP$-tagged branching fractions could be measured with a statistical accuracy $\Delta$ on the order of 30%. A measurement of this accuracy is unlikely to produce a competitive determination of $\gamma$. For the sake of completeness, in what follows the case of $\Delta = 10\%$ will also be considered, to explore what would be possible with an even larger data set.

Since the amplitudes $A_1$ and $A_2$ describe decays to the same final state, the strong phase difference $\delta$ cannot be generated by final state rescattering. Hence it is reasonable to hope that $\delta$ is no larger than $10^\circ$, and probably considerably smaller [28]. This is fortunate, because while the analysis does not require $\delta$ as an input, the accuracy $\Delta\gamma$ with which $\gamma$ is extracted degrades considerably for large $\delta$. (The actual dependence, in the $B^0$ case, is on $\overline{\delta} = \delta - \cot^{-1}[(1 - 2\chi_d \sin^2 2\beta)/2\chi_d \sin 2\beta \cos 2\beta] \approx \delta - 12^\circ$.) The analysis is also sensitive to the value of $a_2/a_1$, although $\Delta\gamma$ does not vary dramatically over the range $1/3 < a_2/a_1 < 3$. Nevertheless. it becomes much more difficult to extract $\gamma$ if $a_2/a_1$ is not of order one, in which case the amplitude triangles are squashed.

As an illustration, assume that $\overline{\delta} = 10^\circ$ and $a_2/a_1 = 1$. Then if $\gamma = 110^\circ$ and $\Delta = 10\%$, and the discrete ambiguity is resolved by other measurements, $\gamma$ can be extracted from this analysis with an uncertainty $\Delta\gamma = 3.7^\circ$. If instead the accuracy on the $CP$-tagged branching fraction is $\Delta = 30\%$, then $\Delta\gamma = 11^\circ$. If $\gamma = 70^\circ$ then the situation is worse: for $\Delta = 10\%$, $\Delta\gamma = 7^\circ$, and for $\Delta = 30\%$, $\Delta\gamma = 21^\circ$. If more than one final $CP$-tagged state can be used, the measurements are independent and can be combined. Although there are too many variables to say precisely how well one might do, it is clear that over a wide range of reasonable parameters, this method could yield a competitive, and theoretically clean, measurement of $\gamma$. However, this might well require the entire Super $B$ Factory data set.

If sufficient data were collected at the $\Upsilon(5S)$ , it would also be possible to extract $\gamma$ from $CP$-tagged $B_s$ decays. The $\Upsilon(5S)$ decays not only to the pair $B_s\,\overline{B}_s$, but to final states with one or two $B^*_s$, which then decays to $B_s$ by the emission of a magnetic photon. Although the photon is too soft to observe directly, its presence can be inferred from the boost of the $B_s$. Since angular momentum conservation forces the decay $\Upsilon(5S) \to B^*_s\overline{B}_s$ to occur in a $p$ wave, and the photon carries odd $CP$, the $CP$ values of the final $B_s$ and $\overline{B}_s$ pair arising from $B^*_s\overline{B}_s$ (or from its conjugate) are correlated (rather than anticorrelated, as with direct $\Upsilon(5S) \to B_s\overline{B}_s$). However, the final state with two vector mesons cannot be used for $CP$ tagging, because in that case the angular momentum of the pair can be 0, 1 or 2.

In the Standard Model, $CP$ is expected to be approximately conserved in the interference between $B_s$ mixing and $CP$ tagging decays such as $B_s \to D^+_s D^-_s$. (Of course, this assumption will be tested, and if it proves not to hold then an analysis analogous to the one for $B^0$ will be required.) Let us assume that $10\,\mathrm{ab}^{-1}$ is collected at the $\Upsilon(5S)$ . Then in the simplest, direct $\Upsilon(5S) \to B_s\overline{B}_s$ case, assuming the production cross section $\sigma(B_s\overline{B}_s)/\sigma(\Upsilon(4S))) \approx 10^{-2}$, the branching fractions $\mathcal{B}(B^{CP}_s \to D^+_s D^-_s) \approx 10^{-2}$ and $\mathcal{B}(B^{CP}_s \to D_s K \approx 2 \times 10^{-4}$, and a total reconstruction efficiency of 5%, one would collect approximately 50 $CP$-tagged decays in a single channel. This would yield a statistical error on the $CP$-tagged branching fraction of approximately $15\%$. The flavor-tagged rates would be measured simultaneously or, more likely, at LHC$b$ or $B$TeV.

Of course, at this point the production cross section is not well-known beyond model-dependent estimates [29, 30], the branching fractions have not been measured yet, this reconstruction efficiency may well be too optimistic, and there is no clear sense for how much running, if any, one might expect at the $\Upsilon(5S)$ . On the other hand, the number of $CP$-tagged events may be enhanced considerably by using the $B_s^*\overline{B}_s$ final state and by adding additional tagging and decay modes. In the end, it may well be possible to extract $\gamma$ with an accuracy of a few decays from $CP$-tagged $B_s$ decays as well as from $B^0$ decays.

### 3.2.4 Combined Strategies for $\gamma$ from $B \to KD^0$

>– D. Atwood and A. Soni –<

We report here on our studies on extraction of $\gamma$ using direct $CP$ violation in $B \to KD$ processes [2, 21, 22, 31]. In principle, these methods are theoretically very clean. The irreducible theory error originating from higher-order weak interactions is $\mathcal{O}(10^{-3})$ [32], *i.e.*, in all likelihood even smaller than the theory error in deducing the angle $\beta$ using time-dependent $CP$ asymmetry in $B \to J/\psi K_S^0$. However, $\gamma$ determination from $B \to KD$ is much harder than $\beta$ from $B \to J/\psi K_S^0$.

This study strongly suggests that the demands on integrated luminosity can be significantly alleviated if a combination of strategies is used. One interesting handle that we examined here which looks rather promising is to include $D^{*0}$ from $B \to K^{(*)} D^{0(*)}$. The formalism for the use of $D^0$ decays is identical after $D^{*0} \to \pi^0(\gamma) + D^0$.

Similarly including $K^*$ (via *e.g.*, $B^- \to K^{*-} D^0$) along with $B^- \to K^- D^0$ is helpful. Also it of course helps a great deal to use both $CP$ eigenstates [33] along with $CP$ non-eigenstates of $D^0$, whether they be doubly-Cabibbo-suppressed [2, 21] or singly-Cabibbo-suppressed [34].

#### Combined Strategies for $\gamma$

We will now consider various strategies to determine $\gamma$ using the number of events given in our core data sample of Table 3-1. In order to estimate luminosity requirements we include detection efficiencies and an overall factor ($R_{cut}$) for a hard acceptance cut:

$$R_{\pi^\pm} = 0.95 \qquad R_{K^\pm} = 0.8 \qquad R_{\pi^0} = 0.5 \qquad R_\gamma = 0.5 \qquad R_{\eta\to 2\gamma} = 0.5 \qquad R_{cut} = \frac{1}{6} - \frac{1}{3}$$

Thus we estimate that our core data sample of Table 3-1, where we have included a cut efficiency in the range $R_{cut} = [\frac{1}{6} \text{ to } \frac{1}{3}]$, will require about $N_\Upsilon = (3-6) \times 10^8$, corresponding to 300-600 fb$^{-1}$.

First, let us consider in isolation the case of $B^- \to K^-[D^0 \to K^+\pi^-]$. This rate, together with its charge conjugate gives us two distinct observables which are determined in terms of four unknown parameters: $\zeta_{KD}$, $\zeta_{K^+\pi^-}$, $b(KD)$ and $\gamma$. The two strong phases enter as the sum $\zeta_{tot} = \zeta_{KD} + \zeta_{K^+\pi^-}$, so in effect there are only three parameters $\{\zeta_{tot}, b(KD), \gamma\}$. We still cannot expect to extract $\gamma$ but, as discussed in [21], this data gives a bound on $\sin^2\gamma$.

To illustrate this, in Fig. 3-4 the thin solid line shows the minimum value of $\chi^2$ as a function of $\gamma$, given particular values of the strong phases and assumptions regarding the $CP$ eigenstate ($CPES$) modes used. For each value of $\gamma$ we minimize with respect to the other parameters $\{\zeta_{tot}, b(KD)\}$. One can see that given enough statistics, a bound on $\gamma$ may be obtained in the first quadrant. Clearly, the luminosity used to make this calculation is not sufficient to provide a useful bound. The $3\sigma$ bound (*i.e.*, where $\chi^2 \approx 9$) is only slightly above 0.

We can also consider the bound on $\gamma$ obtained from the use of the entire data set of $CPES-$ via the decay $D^0 \to CPES-$ [33] in isolation. There are more events of this type but the power of this data to bound $\gamma$ is not much greater since $A_{CP}$ is smaller (in general we expect the analyzing power of a particular mode to be $\sim A_{CP}^2$). The minimum $\chi^2$ in this case is shown with the dotted curve. Notice that taken in isolation the $CPES-$ data set seems to do worse than even the single $D^0 \to K^+\pi^-$ ($CPNES$) mode.

**Table 3-1.** *Initial "core data sample" of $300 - 600$ fb$^{-1}$ used in this simulation, where the number of events is assumed to be distributed among the given mode and its charge conjugate. The corresponding number of events for the three other initial $B^-$ decays: $D^{*0}K^-$, $D^0K^{*-}$ and $D^{*0}K^{*-}$ are assumed to be the same. (While this assumption is too optimistic for current detector technology, a 4-fold increase in statistics will easily be within the reach of a Super $B$ Factory.*

| Initial $B$ decay | Subsequent $D^0$ decay | Number of events |
|---|---|---|
| $B^- \to K^-D^0$ | $K^+\pi^-$ | 25 |
| $B^- \to K^-D^0$ | $K^{*+}\pi^-$ | 14 |
| $B^- \to K^-D^0$ | $K^+\pi^- + n\pi$ | 106 |
| $B^- \to K^-D^0$ | $CPES-$ | 827 |

Of course, both of these two data sets depend on the common parameter $b$ and if we have both sets of data together we obtain the results shown with thick solid curve which is an improvement on each of the data sets taken in isolation; in fact this thick solid curve gives a 3 $\sigma$ bound of $\gamma > 16^\circ$. As discussed in [2] since there the number of equations and observables is the same, there are ambiguous solutions which leads to the $\chi^2$ value being small over an extended range.

To improve the situation, we can also use data from all four decays of the form $B^- \to K^{(*)-}D^{(*)0}$. Note that each of these modes will have a different unknown value of $b$ and $\zeta$. In addition, the decay mode $K^{*-}D^{*0}$ has three polarization amplitudes which we will take into account by introducing a coherence factor $R$ into the fit since we are assuming that we are only observing the sum and we do not consider the additional information that could be determined from the angular distributions of the decays of the vectors as discussed in [35]. If we consider the single decay $D^0 \to K^+\pi^-$ we obtain the results shown by the dashed line which in this case gives a $3-\sigma$ bound on $\gamma$ of $\gamma > 23^\circ$. The dot dash curve shows the result where we have both the $D^0 \to K^+\pi^-$ and $D^0 \to CPES-$ data. In this case we obtain a $3\sigma$ determination of $\gamma$ (within the first quadrant) to be $60^{+15.5}_{-19.5}{}^\circ$. Using the additional data improved the situation both by providing more statistics and because the different data sets have different spurious solutions leaving only the correct solution in common.

For this dash-dot curve, it is instructive to examine the number of observables versus the number of unknown free parameters. First of all, for $D^0 \to K^+\pi^-$ there is the strong phase. For each of the four parent $B^-$ decays there is a strong phase. In the case of $B^- \to K^{*-}D^{*0}$ there is, in addition, a parameter $R$. Again, for each of the four parent decays there is the unknown branching ratio $b = \mathcal{B}(B^- \to K^{(*)-}\overline{D}^{*0})$ and finally the angle $\gamma$ giving a total of 11 parameters. On the other hand, for each combination of $B$ and $D$ decays there are two observables, $d$ and $\overline{d}$, giving a total of 16, so there is an overdetermination by 5 degrees of freedom.

As another example, consider the case where only two the four combinations of $B^- \to K^{(*)-}D^{(*)0}$ are observed with $D$ decay to $K^+\pi^-$ and $CPES-$, then the system is still overdetermined. In Fig. 3-4 the long dashed line takes into account only the two $B^- \to K^-D^0$ and $B^- \to K^-D^{*0}$ and so has 6 unknown parameters determined by 8 observables. Clearly having some overdetermination is helpful in obtaining a good determination of $\gamma$.

It is important to contrast thick solid curve with the long dashed one, in Fig. 3-4. Recall both of them have $D^0 \to K^+\pi^-$, $CPES-$. However, in case of the thick solid curve the $D^0$ originate only from $B^- \to K^-D^0$ whereas the long dashed curve is also getting the $D^0$ coming from $D^{*0} \to D^0 + \pi^0(\gamma)$. As a result whereas in the thick solid case there are 4 observables and 4 unknowns for the long-dashed case its 8 obervables for 6 unknowns. That ends up making a significant difference, as is evident from the figure; perhaps a lot more than one may naively expect just by doubling the number of $D^0$ or a factor of two in luminosity.

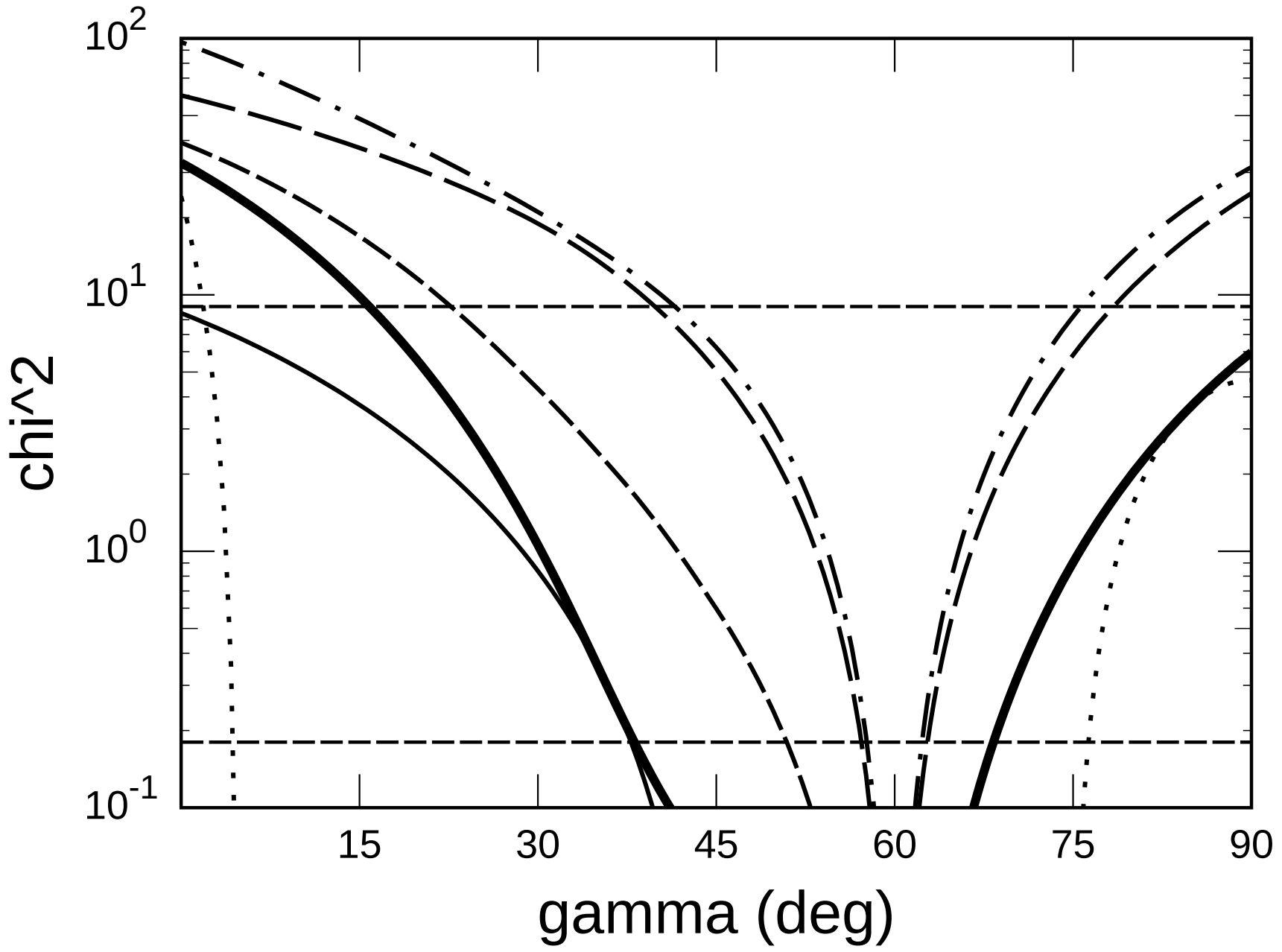

**Figure 3-4.** *The minimum value of $\chi^2$ is shown as a function of $\gamma$ for various combinations of data in the sample calculation. The thin solid line shows the result using just $B^- \to K^-[D^0 \to K^+\pi^-]$ data. The dotted line shows the result using just the $B^- \to K^-[D^0 \to CPES-]$ data. The thick solid curve shows the result taking both $B^- \to K^-[D^0 \to K^+\pi^-]$ and $B^- \to K^-[D^0 \to CPES-]$ data together. The dashed line shows the result using $B^- \to K^{(*)-}[D^{(*)0} \to K^+\pi^-]$. In the dash dotted curve, all four of the initial $B^-$ decays where the $D$ decays to the same two final states are considered. Thus the dashed dotted curve results from taking together data of the form $B^- \to K^{(*)-}[D^{(*)0} \to K^+\pi^-]$ and $B^- \to K^{(*)-}[D^{(*)0} \to CPES-]$. The long dashed curve only includes data from two parent $B^-$ decays, i.e., $B^- \to K^-D^0$ as well as $B^- \to K^-[D^{*0} \to D^0 + \pi^0(\gamma)]$ with either of the two $D^0$ decaying to $K^+\pi^-$ as well as $CPES-$.*

It is also instructive to compare the dash-dot curve, which clearly has substantially more data, with the long-dash one. Notice that quality of determination of $\gamma$ by the two data sets is about the same. This suggests that once the number of observables is sufficiently large as to overdetermine the parameters, further gains by including additional information lead only to modest gains.

### Summary

This study strongly suggests that the demands on integrated luminosity can be significantly alleviated if a combination of strategies is used. One interesting handle that we examined here which looks rather promising is to include $D^{*0}$ from $B \to K^{(*)}D^{*0}$. The formalism for the use of $D^0$ decays is identical to $B \to K^*D^0$ after $D^{*0} \to \pi^0(\gamma) + D^0$.

Similarly, including $K^*$ (via *e.g.*, $B^- \to K^{*-}D^0$) along with $B^- \to K^-D^0$ is helpful. Also it of course helps a great deal to use both $CP$ eigenstates [33] along with $CP$ non-eigenstates of $D^0$, whether they be doubly-Cabibbo-suppressed [2, 21] or singly-Cabibbo-suppressed [34].

Using Fig. 3-4 as a guide, we can anticipate possible determination of $\gamma$ at a Super $B$ Factory with a $3\sigma$ error of a few ($\approx$ 2) degrees.

While $B$ Factories with about $10^9$ $B$ pairs are likely to be able to make appreciable progress in determination of $\gamma$, a Super $B$ Factory with an integrated luminosity yielding $> 10^{10}$ $B$ pairs will be needed to extract $\gamma$ with an accuracy roughly commensurate with the intrinsic theory error that these methods allow. This in itself should constitute an important goal of $B$ physics in general, and a Super $B$ Factory in particular.

This research was supported by Contract Nos. DE-FG02-94ER40817 and DE-AC02-98CH10886.

## 3.3 $B \to K\pi$ and $K\pi\pi$

### 3.3.1 Theory

⋗–M. Gronau and J. Rosner–⋖

Current information on $\gamma = \mathrm{Arg}(V^*_{ub})$ from other CKM constraints is still in need of improvement, with $39^\circ < \gamma < 80^\circ$ at 95% CL [36]. Direct probes of $\gamma$ can tighten these bounds, possibly indicating New Physics effects in the event that an inconsistency with this range is observed. In order to study $\gamma$ directly in charmless two-body $B$ decays, which involve a $b$ to $u$ transition, one must generally separate strong and weak phases from one another. We describe several cases of $B \to K\pi$ decays in which progress has been made, and what improvements lie ahead. Some additional details are noted in earlier reviews [37, 38, 39] and in Refs. [40] and [41].

A great deal of information can be obtained from $B \to K\pi$ decay rates averaged over $CP$, supplemented with measurements of direct $CP$ asymmetries. In this manner, one probes tree-penguin interference in various processes. The data which are used in these analyses are summarized in Table 3-2 [42]. The $B^+$ to $B^0$ lifetime ratio is taken to be $\tau_+/\tau_0 = 1.078 \pm 0.013$, based on $\tau_+ = 1.653 \pm 0.014$ ps and $\tau_0 = 1.534 \pm 0.013$ ps [43]. Table 3-2 also contains contributions to the four $B \to K\pi$ decay processes of penguin ($P'$), electroweak penguin ($P'_{\rm EW}$), tree ($T'$) and color-suppressed tree ($C'$) amplitudes. These contributions are hierarchical and can be classified using flavor symmetries [44, 45, 46, 47]. Smaller contributions, from color-suppressed electroweak penguin amplitudes, annihilation and exchange amplitudes, are not shown in Table 3-2. All four $B \to K\pi$ decays are dominated by penguin amplitudes, which are related to each other by isospin. Tree amplitudes $T' + C'$ and electroweak penguin amplitudes $P'_{\rm EW}$ are subdominant and can be related to each other by flavor SU(3) [48]. SU(3) breaking in tree amplitudes is introduced assuming factorization.

**Table 3-2.** *Branching ratios and $CP$ asymmetries for $B \to K\pi$ decays [42].*

| Decay mode | Amplitude | $\mathcal{B}$ (units of $10^{-6}$) | $A_{CP}$ |
|---|---|---|---|
| $B^+ \to K^0\pi^+$ | $P'$ | $21.78 \pm 1.40$ | $0.016 \pm 0.057$ |
| $B^+ \to K^+\pi^0$ | $-(P' + P'_{\rm EW} + T' + C')/\sqrt{2}$ | $12.53 \pm 1.04$ | $0.00 \pm 0.12$ |
| $B^0 \to K^+\pi^-$ | $-(P' + T')$ | $18.16 \pm 0.79$ | $-0.095 \pm 0.029$ |
| $B^0 \to K^0\pi^0$ | $(P' - P'_{\rm EW} - C')/\sqrt{2}$ | $11.68 \pm 1.42$ | $0.03 \pm 0.37$ |

Several comparisons between pairs of processes can be made:

- $B^0 \to K^+\pi^-$ $(P' + T')$ vs. $B^+ \to K^0\pi^+$ $(P')$ [40, 49, 50, 51];
- $B^+ \to K^+\pi^0$ $(P' + P'_{\rm EW} + T' + C')$ vs. $B^+ \to K^0\pi^+$ $(P')$ [40, 48, 52, 53];
- $B^0 \to K^0\pi^0$ vs. other modes [40, 54, 55, 56, 57, 58].

We give the example of $B^0 \to K^+\pi^-$ in detail. The tree amplitude for this process is $T' \sim V_{us}V^*_{ub}$, with weak phase $\gamma$, while the penguin amplitude is $P' \sim V_{ts}V^*_{tb}$ with weak phase $\pi$. We denote the penguin-tree relative strong phase by $\delta$ and define $r \equiv |T'/P'|$. Then we may write

$$A(B^0 \to K^+\pi^-) = |P'|[1 - re^{i(\gamma+\delta)}] \,, \tag{3.31}$$

$$A(\overline{B}^0 \to K^-\pi^+) = |P'|[1 - re^{i(-\gamma+\delta)}] \,, \tag{3.32}$$

$$A(B^+ \to K^0\pi^+) = A(B^- \to \overline{K}^0\pi^-) = -|P'| \,. \tag{3.33}$$

In the last two amplitudes we neglect small annihilation contributions with weak phase $\gamma$, assuming that rescattering effects are not largely enhanced. A test for this assumption is the absence of a $CP$ asymmetry in $B^+ \to K^0\pi^+$, and a $U$ spin relation between this process and $B^+ \to \overline{K}^0 K^+$ [59], in which a corresponding amplitude with weak phase $\gamma$ is expected to be much larger. One also neglects small color-suppressed electroweak contributions, for which experimental tests were proposed in [60].

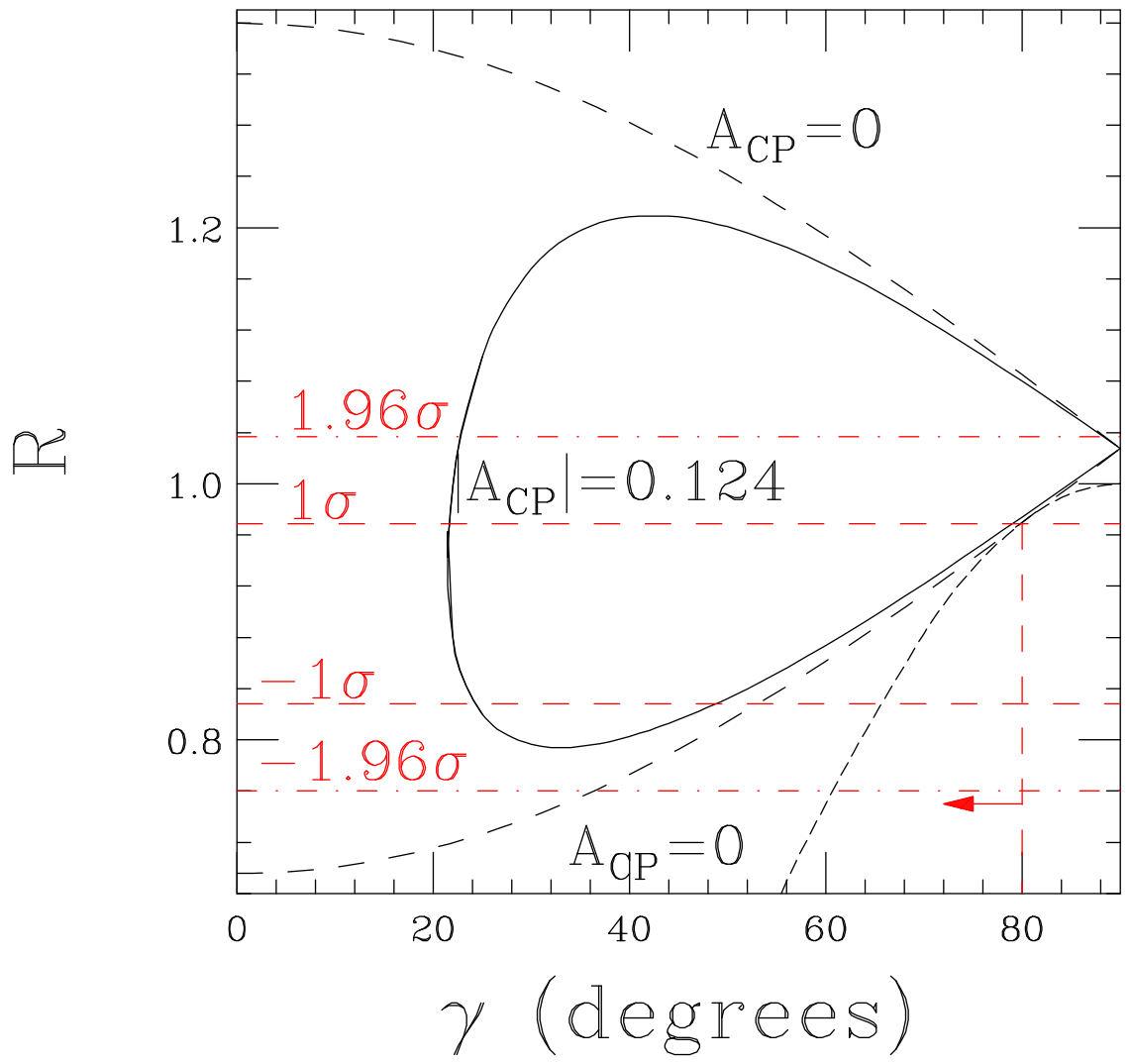

**Figure 3-5.** *Behavior of $R$ for $r = 0.166$ and $A_{CP} = 0$ (dashed curves) or $|A_{CP}| = 0.124$ (solid curve) as a function of the weak phase $\gamma$. Horizontal dashed lines denote $\pm 1\sigma$ experimental limits on $R$, while dot-dashed lines denote $95\%$ c.l. ($\pm 1.96\sigma$) limits. The short-dashed curve denotes the Fleischer-Mannel bound $\sin^2\gamma \leq R$. The upper branches of the curves correspond to the case $\cos\gamma\cos\delta < 0$, while the lower branches correspond to $\cos\gamma\cos\delta > 0$.*

One now forms the ratio

$$R \equiv \frac{\Gamma(B^0 \to K^+\pi^-) + \Gamma(\overline{B}^0 \to K^-\pi^+)}{2\Gamma(B^+ \to K^0\pi^+)} \\ = 1 - 2r\cos\gamma\cos\delta + r^2 \ . \qquad (3.34)$$

Fleischer and Mannel [49] pointed out that $R \geq \sin^2\gamma$ for any $r, \delta$ so if $1 > R$ one can get a useful bound. Moreover, if one uses

$$RA_{CP}(K^+\pi^-) = -2r\sin\gamma\sin\delta \qquad (3.35)$$

as well and eliminates $\delta$ one can get a more powerful constraint, illustrated in Fig. 3-5.

We have used $R = 0.898 \pm 0.071$ and $A_{CP} = -0.095 \pm 0.029$ based on recent averages [42] of CLEO, *BABAR*, and Belle data, and $r = |T'/P'| = 0.142^{+0.024}_{-0.012}$. In order to estimate the tree amplitude and the ratio of amplitudes $r$, we have used factorization in $B^0 \to \pi^-\ell^+\nu_\ell$ at low $q^2$ [61] and $\left|\frac{T'}{T}\right| = \frac{f_K}{f_\pi}\left|\frac{V_{us}}{V_{ud}}\right| \simeq (1.22)(0.23) = 0.28$. One could also use processes in which $T$ dominates, such as $B^0 \to \pi^+\pi^-$ or $B^+ \to \pi^+\pi^0$, but these are contaminated by contributions from $P$ and $C$, respectively. The $1\sigma$ allowed region lies between the curves $A_{CP} = 0$ and $|A_{CP}| = 0.124$. The most conservative upper bound on $\gamma$ arises for the smallest value of $|A_{CP}|$ and the largest value of $r$, while the most conservative lower bound would correspond to the largest $|A_{CP}|$ and the smallest $r$. Currently no such lower bound is obtained at a $1\sigma$ level. At this level one has $R < 1$, leading to an upper bound $\gamma < 80°$.

We note that for the current average value of $R$ the $1\sigma$ upper bound, $\gamma < 80°$, happens to coincide with that of Ref. [49]. This bound does not depend much on the value of $r$, for which we assumed factorization of $T$ in order to introduce SU(3) breaking. The upper bound on $\gamma$ varies only slightly, $\gamma < 78° - 80°$, for a wide range of values $r = 0.1 - 0.3$. On the other hand, a potential lower bound on $\gamma$ depends more sensitively on the value of $r$, and would result if small values of this parameter could be excluded. For instance, Fig. 3-5 shows that a value $r = 0.166$ implies $\gamma > 49^\circ$ at $1\sigma$. Thus, it is crucial to improve our knowledge of $r$.

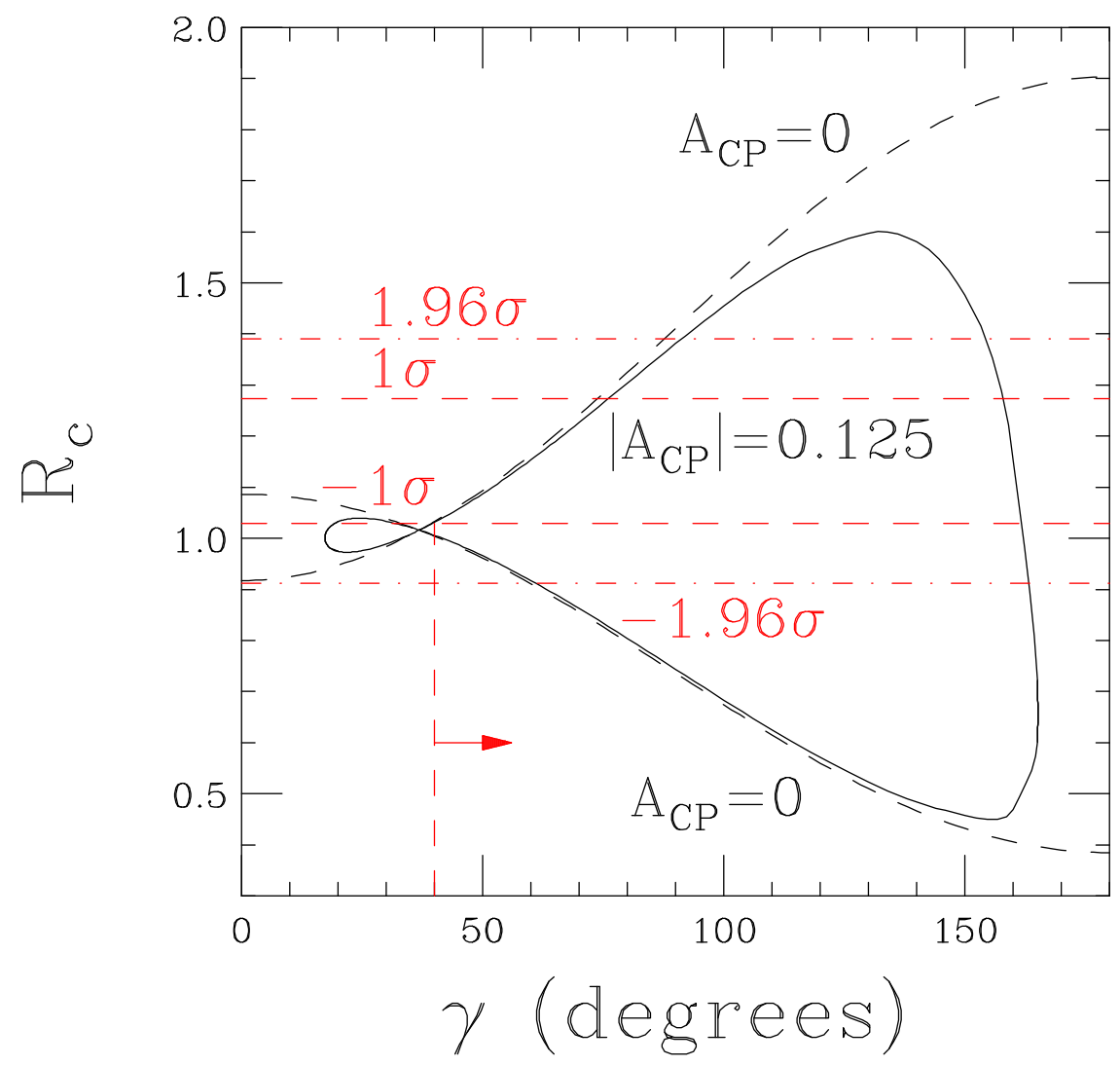

**Figure 3-6.** *Behavior of $R_c$ for $r_c = 0.21$ ($1\sigma$ upper limit) and $A_{CP}(K^+\pi^0) = 0$ (dashed curves) or $|A_{CP}(K^+\pi^0)| = 0.125$ (solid curve) as a function of the weak phase $\gamma$. Horizontal dashed lines denote $\pm 1\sigma$ experimental limits on $R_c$, while dotdashed lines denote 95% c.l. ($\pm 1.96\sigma$) limits. We have taken $\delta_{EW} = 0.80$ (its $1\sigma$ upper limit), which leads to the most conservative bound on $\gamma$. Upper branches of curves correspond to $\cos\delta_c(\cos\gamma - \delta_{EW}) < 0$, while lower branches correspond to $\cos\delta_c(\cos\gamma - \delta_{EW}) > 0$. Here $\delta_c$ is a strong phase.*

The process $B^+ \to K^+\pi^0$ also provides constraints on $\gamma$. The deviation of the ratio

$$R_c \equiv \frac{\Gamma(B^+ \to K^+\pi^0) + \Gamma(B^- \to K^-\pi^0)}{\Gamma(B^+ \to K^0\pi^+)} = 1.15 \pm 0.12 \tag{3.36}$$

from 1, when combined with $A_{CP}(K^+\pi^0) = 0.00 \pm 0.12$, $r_c = |(T' + C')/P'| = 0.195 \pm 0.016$ and an estimate of the electroweak penguin amplitude $\delta_{EW} \equiv |P'_{EW}|/|T' + C'| = 0.65 \pm 0.15$, leads to a $1\sigma$ lower bound $\gamma > 40^\circ$. Details of the method may be found in Refs. [37, 38, 40, 48, 52, 53]; the present bound represents an update of previously quoted values. The most conservative lower bound on $\gamma$ arises for smallest $A_{CP}$, largest $r_c$, and largest $|P'_{EW}|$, and is shown in Fig. 3-6. These values of $r_c$ and $|P'_{EW}|$ would also imply an upper bound, $\gamma < 77^\circ$, which demonstrates the importance of improving our knowledge of these two hadronic parameters.

Another ratio

$$R_n \equiv \frac{\Gamma(B^0 \to K^+\pi^-) + \Gamma(\overline{B}^0 \to K^-\pi^+)}{2\left[\Gamma(B^0 \to K^0\pi^0) + \Gamma(\overline{B}^0 \to \overline{K}^0\pi^0)\right]} = 0.78 \pm 0.10 \tag{3.37}$$

involves the decay $B^0 \to K^0\pi^0$. This ratio should be equal to $R_c$ since to leading order in $T'/P'$, $C'/P'$, and $P'_{EW}/P'$ one has

$$\left|\frac{P' + T'}{P' - P'_{\rm EW} - C'}\right|^2 \approx \left|\frac{P' + P'_{\rm EW} + T' + C'}{P'}\right|^2 \;, \tag{3.38}$$

but the two ratios differ by $2.4\sigma$. Possibilities for explaining this apparent discrepancy (see, *e.g.*, Refs. [40, 62]) include (1) New Physics, *e.g.*, , in the EWP amplitude, and (2) an underestimate of the $\pi^0$ detection efficiency in all experiments, leading to an overestimate of any branching ratio involving a $\pi^0$. The latter possibility can be taken into account by considering the ratio $(R_n R_c)^{1/2} = 0.96 \pm 0.08$, in which the $\pi^0$ efficiency cancels. As shown in Fig. 3-7, this ratio leads only to the conservative bound $\gamma \leq 88^\circ$. A future discrepancy between $R_c$ and $R_n$ at a statistically significant level implying New Physics effects would clearly raise questions about the validity of constraints on $\gamma$ obtained from these quantities.

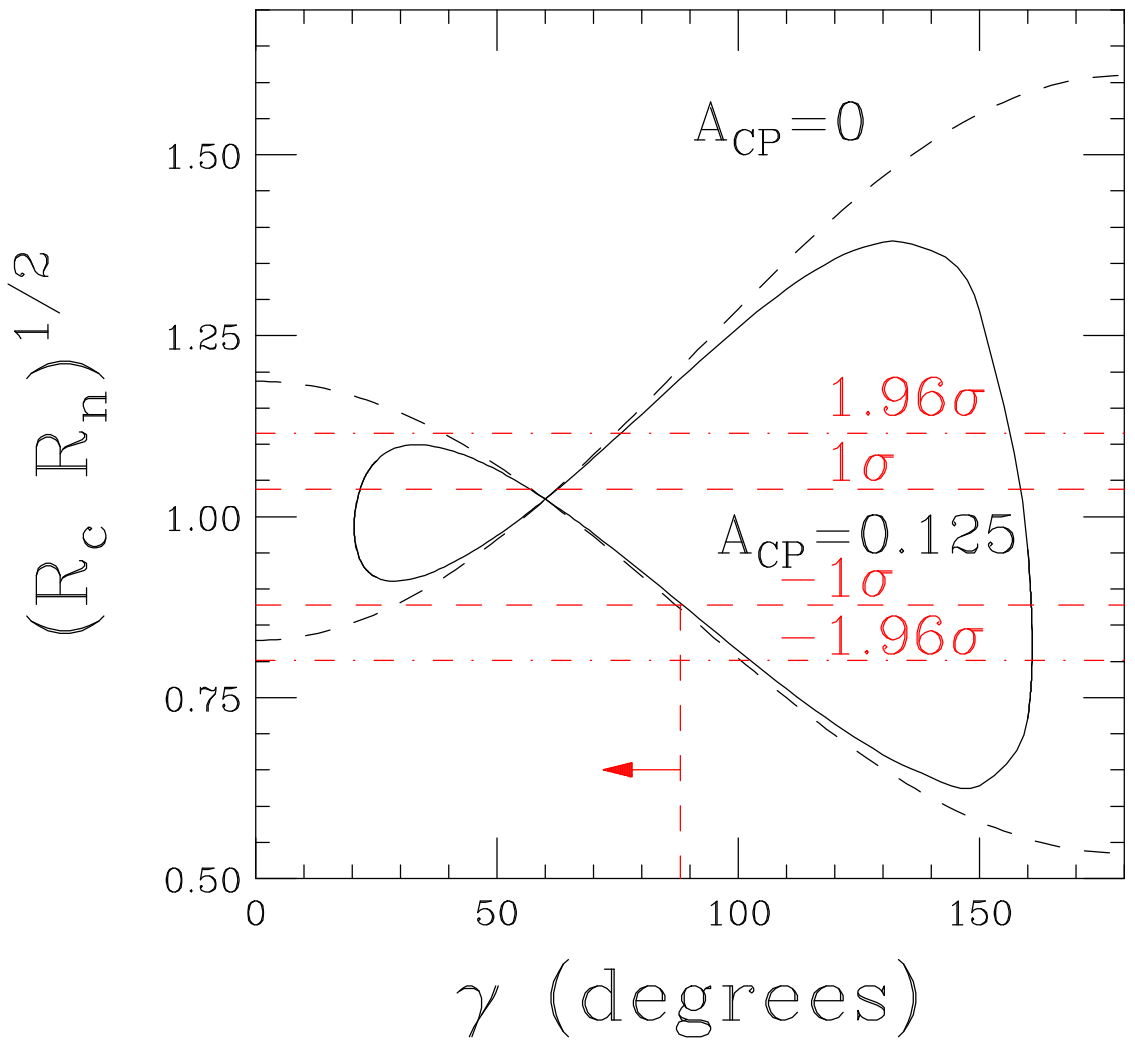

**Figure 3-7.** *Behavior of* $(R_c R_n)^{1/2}$ *for* $r_c = 0.18$ *(*$1\sigma$ *lower limit) and* $A_{CP}(K^+\pi^0) = 0$ *(dashed curves) or* $|A_{CP}(K^+\pi^0)| = 0.125$ *(solid curve) as a function of the weak phase* $\gamma$*. Horizontal dashed lines denote* $\pm 1\sigma$ *experimental limits on* $(R_c R_n)^{1/2}$*, while dotdashed lines denote 95% c.l. (*$\pm 1.96\sigma$*) limits. Upper branches of curves correspond to* $\cos\delta_c(\cos\gamma - \delta_{EW}) < 0$*, while lower branches correspond to* $\cos\delta_c(\cos\gamma - \delta_{EW}) > 0$*. Here we have taken* $\delta_{EW} = 0.50$ *(its* $1\sigma$ *lower limit), which leads to the most conservative bound on* $\gamma$*.*

Recently a time-dependent asymmetry measurement in $B^0(t) \to K^0_S \pi^0$ was reported [63]

$$S_{\pi K} = 0.48^{+0.38}_{-0.47} \pm 0.11 \ , \quad C_{\pi K} = 0.40^{+0.27}_{-0.28} \pm 0.10 \ , \tag{3.39}$$

where $S_{\pi K}$ and $-C_{\pi K}$ are coefficients of $\sin\Delta mt$ and $\cos\Delta mt$ terms in the asymmetry. In the limit of a pure penguin amplitude, $A(B^0 \to K^0\pi^0) = (P' - P'_{\rm EW})/\sqrt{2}$, one expects $S_{\pi K} = \sin 2\beta, C_{\pi K} = 0$. The color-suppressed amplitude, $C'$, contributing to this process involves a weak phase $\gamma$. Its effect was studied recently [41] by relating these two amplitudes within flavor SU(3) symmetry to corresponding amplitudes in $B^0 \to \pi^0\pi^0$. Correlated deviations from $S_{\pi K} = \sin 2\beta, C_{\pi K} = 0$, at a level of $0.1 - 0.2$ in the two asymmetries, were calculated and were shown to be sensitive to values of $\gamma$ in the currently allowed range. Observing such deviations and probing the value of $\gamma$ requires reducing errors in the two asymmetries by about an order of magnitude.

To summarize, promising bounds on $\gamma$ stemming from various $B \to K\pi$ decays have been discussed. So far all are statistics-limited. At $1\sigma$ we have found

- $R$ ($K^+\pi^-$ vs. $K^0\pi^+$) gives $\gamma \leq 80°$;
- $R_c$ ($K^+\pi^0$ vs. $K^0\pi^+$) gives $\gamma \geq 40°$;
- $R_n$ ($K^+\pi^-$ vs. $K^0\pi^0$) should equal $R_c$;
  $(R_c R_n)^{1/2}$ gives $\gamma \leq 88°$.

The future of most such $\gamma$ determinations remains for now in experimentalists' hands, as one can see from the figures. We have noted (see, *e.g.*, [50]) that measurements of rate ratios in $B \to K\pi$ can ultimately pinpoint $\gamma$ to within about $10°$. The required accuracies in $R$, $R_c$, and $R_n$ to achieve this goal can be estimated from the figures. For example, knowing $(R_c R_n)^{1/2}$ to within 0.05 would pin down $\gamma$ to within $10°$ if this ratio lies in the most sensitive range of Fig. 3-7. A significant discrepancy between the values of $R_c$ and $R_n$ would be evidence for New Physics.

It is difficult to extrapolate the usefulness of $R$, $R_c$, and $R_n$ measurements to very high luminosities without knowing ultimate limitations associated with systematic errors. The averages in Table 3-2 are based on individual measurements in which the statistical errors exceed the systematic ones by at most a factor of about 2 (in the case of $B^0 \to K^0\pi^0$) [42]. For $B^+ \to K^+\pi^0$ the statistical and systematic errors are nearly equal. Thus, the clearest path to improvements in these measurements is associated with the next factor of roughly 4 increase in the total data sample. Thereafter, reductions in systematic errors must accompany increased statistics in order for these methods to yield improved accuracies in $\gamma$.

In our study we used the most pessimistic values of the parameters $r$, $r_c$ and $\delta_{EW}$ leading to the weakest bounds on $\gamma$. The theoretical uncertainties in these parameters can be further reduced, and the assumption of negligible rescattering can be tested. This progress will rely on improving branching ratio measurements for $B \to K\pi$, $B \to \pi\pi$ and $B^0 \to \pi^-\ell^+\nu_\ell$, on an observation of penguin-dominated $B \to K\overline{K}$ decays, and on various tests of factorization which imply relations between $CP$-violating rate differences [64, 65].

A complementary approach to the flavor-SU(3) method is the QCD factorization formalism of Refs. [56, 57, 58]. It predicts small strong phases (as found in our analysis) and deals directly with flavor-SU(3) breaking; however, it involves some unknown form factors and meson wave functions and appears to underestimate the magnitude of $B \to VP$ penguin amplitudes. Combining the two approaches seems to be the right way to proceed.

### 3.3.2 $\gamma$ from $B \to K_S^0\pi\pi$

>– N. Sinha and R. Sinha –<

Time-dependent measurements of asymmetries of decay modes of $B^0$ into $CP$ eigenstates [66, 67, 68] allow weak phases to be extracted without any theoretical uncertainty from modes whose amplitudes have a single weak phase. Using the golden mode $B^0 \to J/\psi K_S^0$, the method has been successfully used to measure $\sin2\beta$. The decay mode $B^0 \to \pi^+\pi^-$ can be similarly used to extract $\sin2\alpha$. However, the presence of tree and penguin contributions in the amplitude complicates this measurement. Nevertheless, an isospin analysis still allows a possible measurement of $\sin2\alpha$ [69]. It is widely believed that $\gamma$ cannot be measured using similar time dependent techniques. As an alternative, several other methods have been developed [17, 1, 70, 2, 71, 72] to measure this weak phase. While $\gamma$ can be measured cleanly using some of these techniques at a later date, techniques [73] assuming flavor SU(3), are expected to provide the first estimates of angle $\gamma$.

In a recent paper [74] we proposed a method that uses the time-dependent asymmetry in the three body $K\pi\pi$ decay mode of the $B^0$. The $K\pi\pi$ modes with even isospin $\pi\pi$ states obey triangular isospin relations which allow us to obtain $\gamma$. The two body $K\pi$ modes also obey certain isospin relations; the various decay mode amplitudes form sides of a quadrangle. The isospin $3/2$ amplitude, which is free from gluonic penguin contributions, is not an observable but, in fact, the diagonal of the quadrangle. However, construction of triangles based on isospin analysis similar to that in Ref. [69] is again possible for the $K\pi\pi$ modes. If the direct $CP$ asymmetry for the charged $B$ decay mode is observed to be vanishingly small, then the tree and the electroweak penguin pieces of the weak Hamiltonian responsible for $\Delta I = 1$ transition have the same strong phase. This extra ingredient along with isospin analysis allows us to extract $\gamma$. Our technique is then free from approximations such as SU(3) symmetry, neglect of annihilation or re-scattering contributions. Further, our method is sensitive to the relative weak phase between the tree and penguin contribution, and as such will probe New Physics. Recently, several three-body non-charmed decay modes of the $B$ meson have been observed. In particular the branching ratios of the modes $B^0 \to K^0\pi^+\pi^-$ and $B^0 \to K^+\pi^-\pi^0$ have been measured [75, 76] to be around $5 \times 10^{-5}$. In fact, even with limited statistics, a Dalitz plot analysis has been performed and quasi two body final states have been identified.

The importance of these three-body decay modes was first pointed out by Lipkin, Nir, Quinn and Snyder [77]. Their analysis, however, did not incorporate the large electroweak penguin effects known to be present in these decays [78]. These decays are described by six independent isospin amplitudes $A(I_t, I_{\pi\pi}, I_f)$, where $I_t$ stands for the transition isospin, and describes the transformation of the weak Hamiltonian under isospin and can take only the values 0 and 1 in the Standard Model; $I_{\pi\pi}$ is the isospin of the pion pair and takes the value 0, 1, and 2 and $I_f$ is the final isospin and can take the values $1/2$ and $3/2$. Even values of $I_{\pi\pi}$ has the pair of pions in a symmetric state, and thus have even angular momenta. Similarly states with $I_{\pi\pi}$ odd must be odd under the exchange of two pions. A separation between $I_{\pi\pi} = \mathrm{even}$ and $I_{\pi\pi} = \mathrm{odd}$ should be possible through a study of the Dalitz plot.

The $I_{\pi\pi} = 0$ and 2 channels are described by the three amplitudes $A(0,0,\frac{1}{2})$, $A(1,0,\frac{1}{2})$, and $A(1,2,\frac{3}{2})$. It is straightforward to derive [77]:

$$\begin{aligned} A(B^{+(0)} \to K^{0(+)}(\pi^{+(-)}\pi^0)_{\mathrm{e}}) &= \pm X \\ A(B^{+(0)} \to K^{+(0)}(\pi^{+}\pi^{-})_{\mathrm{e}}) &= \mp\frac{1}{3}X \mp Y + Z \\ A(B^{+(0)} \to K^{+(0)}(\pi^{0}\pi^{0})_{\mathrm{e}}) &= \mp\frac{2}{3}X \pm Y - Z \,, \end{aligned} \tag{3.40}$$

where, $X = \sqrt{\frac{2}{5}}A(1,2,\frac{3}{2})$, $Y = \frac{1}{3}A(1,0,\frac{1}{2})$, and $Z = \sqrt{\frac{1}{3}}A(0,0,\frac{1}{2})$. The subscript "e(o)" represents the even(odd) isospin of the $\pi\pi$ system. It is easy to see that Eq. (3.40) implies the following two isospin triangles relations:

$$A(B^+ \to K^0(\pi^+\pi^0)_{\mathrm{e}}) = A(B^0 \to K^0(\pi^+\pi^-)_{\mathrm{e}}) + A(B^0 \to K^0(\pi^0\pi^0)_{\mathrm{e}}) \,, \tag{3.41}$$

$$A(B^0 \to K^+(\pi^-\pi^0)_{\mathrm{e}}) = A(B^+ \to K^+(\pi^+\pi^-)_{\mathrm{e}}) + A(B^+ \to K^+(\pi^0\pi^0)_{\mathrm{e}}) \,, \tag{3.42}$$

and also implies the relation,

$$A(B^+ \to K^0(\pi^+\pi^0)_{\mathrm{e}}) = -A(B^0 \to K^+(\pi^-\pi^0)_{\mathrm{e}}). \tag{3.43}$$

Decays corresponding to conjugate processes will obey similar relations. Comparison of the isospin-triangle represented by Eq. (3.41) and its conjugate allows the extraction of $\gamma$.

The decay $B(p_B) \to K(k)\pi(p_1)\pi(p_2)$, (where $p_B$, $k$, $p_1$ and $p_2$ are the four momentum of the $B$, $K$, $\pi_1$ and $\pi_2$ respectively) may be described in terms of the usual Mandelstam variables $s = (p_1 + p_2)^2$, $t = (k + p_1)^2$ and $u = (k + p_2)^2$. States with $I_{\pi\pi} = \mathrm{even}$ must be symmetric under the exchange $t \leftrightarrow u$. In what follows, we shall be concerned with differential decay rates $d^2\Gamma/(dtdu)$. These can be extracted from the Dalitz plot of the three body decays. A detailed angular analysis will permit extraction of even isospin $\pi\pi$ events. Note that $B \to K^0_S\pi^0\pi^0$ mode being symmetric in pions, always has pions in isospin even state.

For simplicity, we define the amplitudes $A^{+-}$, $A^{00}$ and $A^{+0}$, corresponding to the modes $B^0 \to K^0_S(\pi^+\pi^-)_{\mathrm{e}}$, $B^0 \to K^0_S(\pi^0\pi^0)_{\mathrm{e}}$, and $B^+ \to K^0_S(\pi^+\pi^0)_{\mathrm{e}}$, respectively. All observables, amplitudes and strong phases are to be understood to depend on $t$ and $u$; we will not denote this dependence explicitly. Using the unitarity of the CKM matrix, we separate these amplitudes into contributions containing the $V_{ub}$ and $V_{cb}$ elements respectively:

$$\begin{aligned} A^{+-} &= a^{+-}e^{i\delta_a^{+-}}e^{i\gamma} + b^{+-}e^{i\delta_b^{+-}} \\ A^{00} &= a^{00}e^{i\delta_a^{00}}e^{i\gamma} + b^{00}e^{i\delta_b^{00}} \\ A^{+0} &= a^{+0}e^{i\delta_a^{+0}}e^{i\gamma} + b^{+0}e^{i\delta_b^{+0}} \,. \end{aligned} \tag{3.44}$$

Note that the magnitudes $a^{+-}$, $b^{+-}$, $a^{00}$, $b^{00}$, $a^{+0}$ and $b^{+0}$ actually contain contributions from all possible diagrams (tree, color-suppressed, annihilation, $W$ exchange, penguin, penguin-annihilation and electroweak-penguin) and include the magnitudes of the CKM elements. Their explicit composition is irrelevant for this analysis, except for the fact that the isospin 3/2 amplitude $A^{+0}$ cannot get contributions from gluonic penguins. The amplitudes $\overline{A}^{+-}$, $\overline{A}^{00}$, $\overline{A}^{+0}$, corresponding to the conjugate process $\overline{B} \to \overline{K}\pi\pi$, can be written similarly, with the weak phase $\gamma$ replaced by $-\gamma$.

Figure 3-8 depicts the two triangles formed by the amplitudes $A^{+-}$, $A^{00}$ and $A^{+0}$ and the corresponding conjugate amplitudes in isospin space, along with the relative orientations. $\zeta(\overline{\zeta})$ are defined as the angle between $A^{+-}(\overline{A}^{+-})$ and $A^{+0}(\overline{A}^{+0})$ and the angle $2\tilde{\gamma}$ is the angle between $A^{+0}$ and $\overline{A}^{+0}$. The relative phase between $A^{+-}$ and $\overline{A}^{+-}$ (*i.e.*

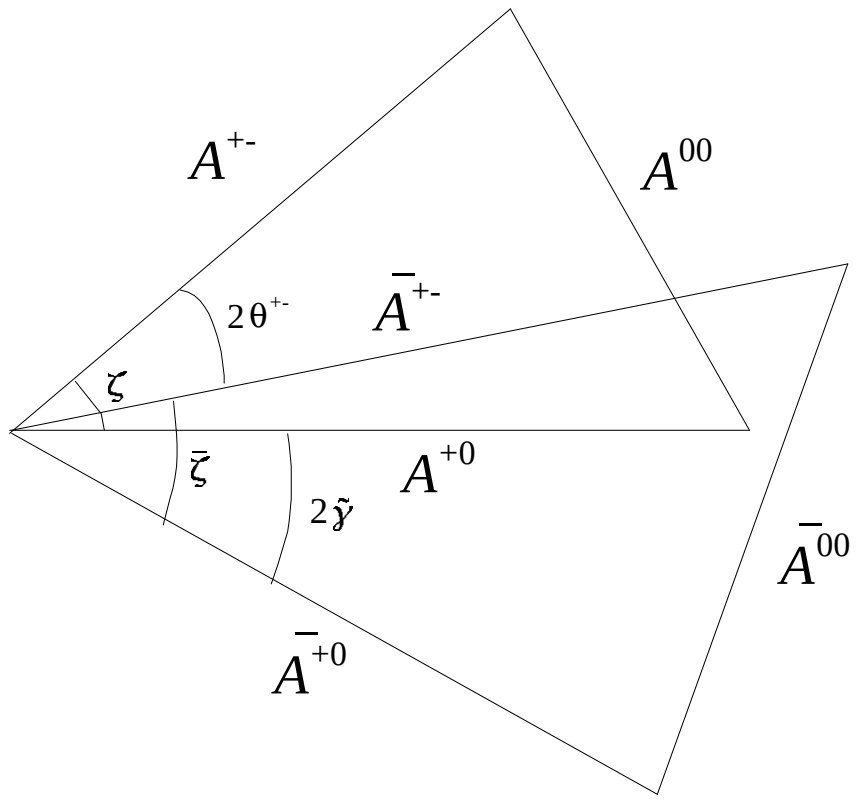

**Figure 3-8.** *The isospin triangles formed by the $B \to K\pi\pi$ amplitudes, as represented in Eq. (3.41) and that for the corresponding conjugate processes. Only one orientation of the conjugate triangle is depicted, this triangle could have been flipped around the base $\overline{A}^{+0}$.*

$\arg((A^{+-})^*\overline{A}^{+-})$) is defined as $2\theta^{+-}$. The coefficient of the $\sin(\Delta mt)$ piece in the time-dependent $CP$ asymmetry for the mode $B^0(t) \to K_S(\pi^+\pi^-)_{\rm e}$ will yield $\sin(2\theta^{+-} - 2\beta)$. Note that this measurement involves time-dependent asymmetry in the partial decay rate $d^2\Gamma^{+-}/{\rm d}t{\rm d}{\rm u}$ at a fixed $t$ and $u$.

With the knowledge of $\beta$, the angle $2\theta^{+-}$ may be regarded as an observable. In addition, measurement of six partial decay rates, $d^2\Gamma^{+0}/dtdu$, $d^2\Gamma^{+-}/dtdu$ and $d^2\Gamma^{00}/dtdu$ as well as their conjugates at the same $t$ and $u$ as used for $\theta^{+-}$ determination, now allows us to construct the two triangles in Fig. 3-8 with two fold ambiguity. We see, from Fig. 3-8, that the angle $2\tilde{\gamma}$ is related to $2\theta^{+-}$ as, $\zeta \pm \overline{\zeta} + 2\tilde{\gamma} = 2\theta^{+-}$. The 'plus–minus' sign ambiguity in the above reflects the possibility of same–side or opposite–side orientation of the triangles. Once $2\tilde{\gamma}$ is known, it is possible to determine $\gamma$. An additional requirement is that the amplitude $A^{+0} \equiv A(K^0(\pi^+\pi^0)_{\rm e})$ has a one single strong phase, $\delta = \delta_a^{+0} = \delta_b^{+0}$. This phase $\delta$ may be set equal to zero by convention. An experimentally verifiable consequence of this hypothesis would be the vanishing of direct $CP$-violating asymmetry for this charged $B$ mode.

Using the amplitudes $A^{+-}$, $\overline{A}^{+-}$, $A^{00}$ and $\overline{A}^{00}$ one can construct a maximum of seven independent observables (The amplitudes $A^{+0}$, $A^{-0}$ are not independent, as they can be obtained using isospin relations). The two triangles can be completely defined in terms of seven observables, the three sides of each of the triangles and a relative angle between the two triangles. The amplitudes under consideration involve the following eleven variables: $a^{+-}$, $b^{+-}$, $a^{00}$, $b^{00}$, $a^{+0}$, $b^{+0}$, $\delta_a^{+-}$, $\delta_b^{+-}$, $\delta_a^{00}$, $\delta_b^{00}$, and $\gamma$. These variables are connected by two isospin relations (see Eq. (3.41) and the corresponding relation for the conjugate process), which results in four constraints, reducing the number of independent variables to seven. Hence, all variables including $\gamma$, can be determined purely in terms of observables.

In order to determine $\gamma$, we express all the amplitudes and strong phases, in terms of observables and $\gamma$. The variables, $a^{ij}$ and $b^{ij}$ may be solved as a function of $\gamma$ and other observables as follows:

$$|a^{+-}|^2 = \frac{B^{+-}}{2\sin^2\gamma}\Big(1 - y^{+-}\cos(2\theta^{+-})\Big)\,, \tag{3.45}$$

$$|b^{+-}|^2 = \frac{B^{+-}}{2\sin^2\gamma}\Big(1 - y^{+-}\cos(2\theta^{+-} - 2\gamma)\Big)\,. \tag{3.46}$$

Similar solutions may be obtained for $a^{+0}$ ($a^{00}$) and $b^{+0}$ ($b^{00}$) with $B^{+-}$ replaced by $B^{+0}$ ($B^{00}$), $y^{+-}$ replaced by $y^{+0} = 1$ ($y^{00}$) and $2\theta^{+-}$ replaced by $2\tilde{\gamma}$ ($2\theta^{00}$) respectively. The branching ratio, $B^{+-} = \dfrac{(|\overline{A}^{+-}|^2 + |A^{+-}|^2)}{2}$; $y^{+-}$ is related to the direct asymmetry $a_{\rm dir}^{+-} = \dfrac{|\overline{A}^{+-}|^2 - |A^{+-}|^2}{|\overline{A}^{+-}|^2 + |A^{+-}|^2}$, through $y^{+-} = \sqrt{1 - (a_{\rm dir}^{+-})^2}$. Relations for $B^{00}$, $B^{+0}$ and

$y^{00}$ are similar. The angle $2\theta^{00}$ between $A^{00}$ and $\overline{A}^{00}$, need not be measured but can be determined from geometry of the two triangles and is given by,

$$\cos(2\theta^{00} - 2\tilde{\gamma}) = \frac{B^{00} - B^{+-} + |A^{+-}||\overline{A}^{+-}|\cos(2\theta^{+-} - 2\tilde{\gamma})}{|A^{00}||\overline{A}^{00}|}.$$

We define $\delta^{+-} = \delta_b^{+-} - \delta_a^{+-}$ and $\delta^{00} = \delta_b^{00} - \delta_a^{00}$, with $\delta^{+-}$ expressed in terms of $\gamma$ and observables as:

$$\tan\delta^{+-} = \frac{a_{\rm dir}^{+-}\tan\gamma}{1 - y^{+-}[\cos 2\theta^{+-} - \sin 2\theta^{+-}\tan\gamma]}\ , \tag{3.47}$$

with an analogous expression for $\tan\delta^{00}$. Our task now is to express the strong phases $\delta_a^{+-}$ and $\delta_a^{00}$ in terms of $\gamma$ and observables, just as we have done for the other variables. One finally intends to solve for $\gamma$, only in terms of observables.

The isospin triangle relation given by Eq. (3.41) and the similar relation for the conjugate process may be expressed as:

$$(a^{+-}e^{i\delta_a^{+-}} + a^{00}e^{i\delta_a^{00}})e^{\pm i\gamma} + (b^{+-}e^{i\delta_b^{+-}} + b^{00}e^{i\delta_b^{00}}) = (a^{+0}e^{\pm i\gamma} + b^{+0})\ . \tag{3.48}$$

The 'four' equations contained in Eq. (3.48) may be used to used to solve for $\cos\delta_a^{+-}$ and $\cos\delta_a^{00}$:

$$\cos\delta_a^{+-} = \frac{|a^{+0}|^2 + |a^{+-}|^2 - |a^{00}|^2}{2|a^{+0}||a^{+-}|}\ , \qquad \cos\delta_a^{00} = \frac{|a^{+0}|^2 + |a^{00}|^2 - |a^{+-}|^2}{2|a^{+0}||a^{00}|}\ , \tag{3.49}$$

as well as, obtain the relation,

$$|b^{+-}|^2 + |b^{00}|^2 + 2b^{+-}b^{00}\cos(\delta_b^{+-} - \delta_b^{00}) = |b^{+0}|^2\ . \tag{3.50}$$

Now $\delta_b^{+-} = \delta^{+-} + \delta_a^{+-}$ and $\delta_b^{00} = \delta^{00} + \delta_a^{00}$. Hence, Eq. (3.50) is expressed completely in terms of observables and $\gamma$. $\gamma$ can thus be determined cleanly, in terms of observables.

The CKM phase $\gamma$ can be determined simultaneously for several regions of the Dalitz plot. The ambiguities in the solution of $\gamma$ may thereby be removed. Having measured $\gamma$, $a^{+0}$ and $b^{+0}$ can be determined using equations similar to Eq. (3.46). We can thus determine the size of electroweak penguin contributions.

Current experimental data [75, 76] indicate that a statistically significant contribution in the $K_S^0\pi^+\pi^-$ mode, is from the $K^{*+}\pi^-$. It can be easily seen by a simple isospin analysis that $K^{*+}\pi^-$ final state *cannot* result in $K^0(\pi^+\pi^-)_{\rm o}$, but must contribute to $K^0(\pi^+\pi^-)_{\rm e}$ final state. If one takes the preliminary data of Ref. [76] seriously, then based on an integrated luminosity of $43.1\,{\rm fb}^{-1}$, there are $19.1^{+6.8}_{-5.9}$ $K^{*\pm}\pi^\mp$ events in a total of $60.3 \pm 11.0$ $K^0\pi^+\pi^-$ events. Certainly at the Super $B$ Factory, with an integrated luminosity of $10\,{\rm ab}^{-1}$, there will be enough $K(\pi^+\pi^-)_{\rm e}$ to allow a time-dependent measurement. Additional $K(\pi^+\pi^-)_{\rm e}$ events will occur at other regions of Dalitz plot. While $B^+ \to K_S^0\pi^+\pi^0$ has not yet been observed, the mode $B^0 \to K^+\pi^-\pi^0$ has been seen. The two amplitudes are related by Eq. (3.43). Again, if the $K^{*0}\pi^0$ contribution to the $K^+\pi^-\pi^0$ is significant, it must result in $K^+(\pi^-\pi^0)_{\rm e}$. Data from both $B^+ \to K_S^0\pi^+\pi^0$ and $B^0 \to K^+\pi^-\pi^0$ modes could be combined to improve statistics.

To conclude, the weak phase $\gamma$ can be measured using a time dependent asymmetry measurement in the three body decay, $B \to K\pi\pi$. A detailed study of the Dalitz plot can be used to extract the $\pi\pi$ even isospin states. These states obey certain isospin relations which along with the hypothesis of a common strong phase for the electroweak penguin and tree amplitudes in $B^+ \to K^0\pi^+\pi^-$, allow us to not only obtain $\gamma$, but also determine the size of the electroweak penguin contribution. The hypothesis made can be verified by a measurement of direct asymmetry for the charged $B$ mode. Unless the direct asymmetry is found to be sizable, this method allows extraction of $\gamma$ without any theoretical assumptions like SU(3) or neglect of any contributions to the decay amplitudes. By studying different regions of the Dalitz plot, it may be possible to reduce the ambiguity in the value of $\gamma$.

### 3.3.3 Measurement of the time-dependent $CP$ asymmetry in $B^0 \to K^0_S\pi^0$

>–W. Hulsbergen–<

**Introduction** In the Wolfenstein parameterization the leading penguin contributing to $B^0 \to K^0_S\pi^0$ decays is real and proportional to $P \sim V_{tb}V^*_{ts} \sim \lambda^2$, while the leading tree diagram is CKM suppressed ($T \sim V_{ub}V^*_{us} \sim \lambda^4 e^{-i\gamma}$, see Fig. 3-9). In the absence of the tree contribution, the Standard Model predicts $S_{K^0_S\pi^0} = \sin 2\beta$ and $C_{K^0_S\pi^0} = 0$. A recent estimate based on SU(3) flavor symmetry bounds the deviation with $S_{J/\psi K^0_S}$ to $\Delta S \in [-0.17, 0.18]$ [41]. This justifies a search for non-Standard Model contributions to the phase of the penguin diagram.

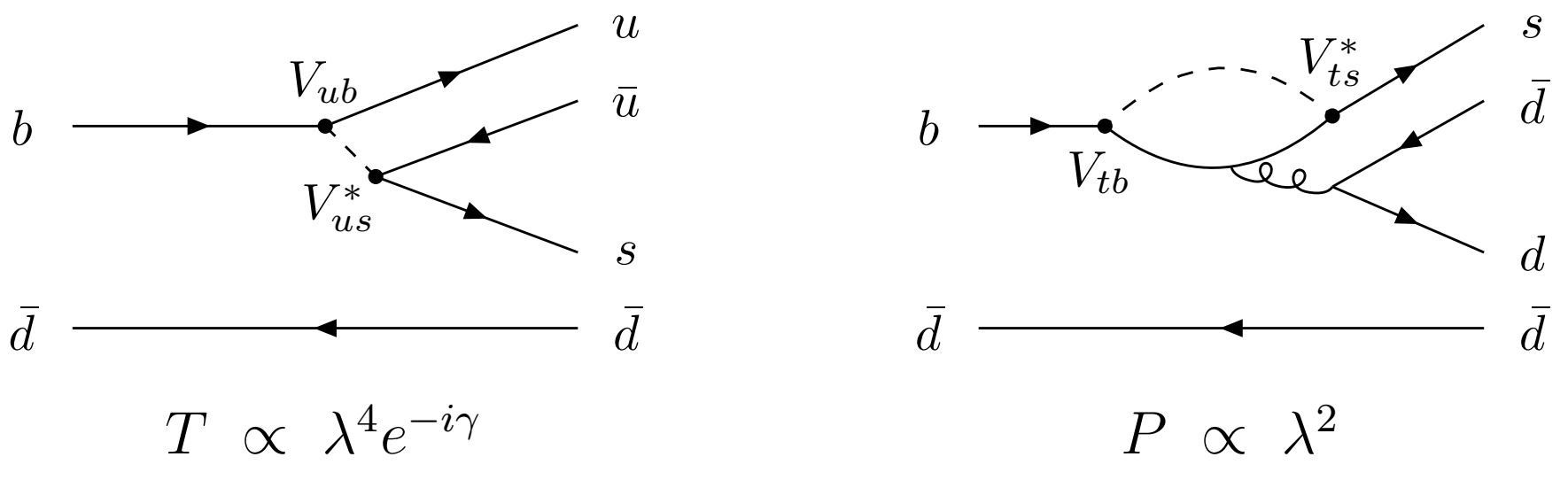

**Figure 3-9.** *Leading order tree (left) and penguin (right) diagrams for $B^0 \to K^0_S\pi^0$.*

The *BABAR* collaboration recently reported a first measurement of the time-dependent $CP$ asymmetry of $B^0 \to K^0_S\pi^0$ [79] on a data sample corresponding to an integrated luminosity of 113 fb$^{-1}$ at the $\Upsilon(4S)$ resonance. The $CP$ parameters extracted with $122 \pm 16$ signal events were

$$S_{K^0_S\pi^0} = 0.48^{+0.38}_{-0.47} \pm 0.06 \quad ,$$
$$C_{K^0_S\pi^0} = 0.40^{+0.27}_{-0.28} \pm 0.09 \quad .$$

where the first error is statistical and the second systematic. Here we summarize some of the experimental details that are relevant for an improved measurement of these parameters at a future Super $B$ Factory.

**$\Delta t$ reconstruction** Though $C_{K^0_S\pi^0}$ may be measured using $B$-flavor tagging alone, the extraction of $S_{K^0_S\pi^0}$ requires knowledge of the $B^0\overline{B^0}$ lifetime difference. The long lifetime of the $K^0_S$ and the lack of a second trajectory prohibit the reconstruction of the $B^0 \to K^0_S\pi^0$ decay vertex using the techniques employed in other time-dependent analyses such as $B^0 \to J/\psi K^0_S$. Instead, we must exploit the fact that the transverse decay length of the signal $B^0$ is small, such that its decay vertex can be obtained by intersecting the $K^0_S$ trajectory with the known interaction region (IR).

The viability of this reconstruction method is the consequence of the small size of the IR in $x$ ($\approx 200\ \mu$m) and $y$ ($\approx 4\ \mu$m) and the precise calibration of the IR position. To account for the transverse motion of the $B^0$ meson, the size of the IR in the transverse plane is increased with the rms of the $B^0$ transverse decay length distribution ($\approx 30\ \mu$m). The procedures for the reconstruction of the tag vertex and the extraction of $\Delta t$ are equal to those applied for the mainstream analyses [80].

Figure 3-10(a) shows the estimated uncertainty in the $z$ position of the $B$ vertex as a function of the transverse decay length of the $K^0_S \to \pi^+\pi^-$. This uncertainty is strongly correlated with the number of vertex detector (SVT) layers that the $K^0_S$ daughters traverse. For a meaningful accuracy on $\Delta t$ the $z$ uncertainty must be well below 1 mm, which implies that only events with a $K^0_S$ decay inside SVT layer 4 can be used in the time-dependent asymmetry measurement. The remaining $\sim 35\,\%$ of the events are only used for the measurement of $C$.

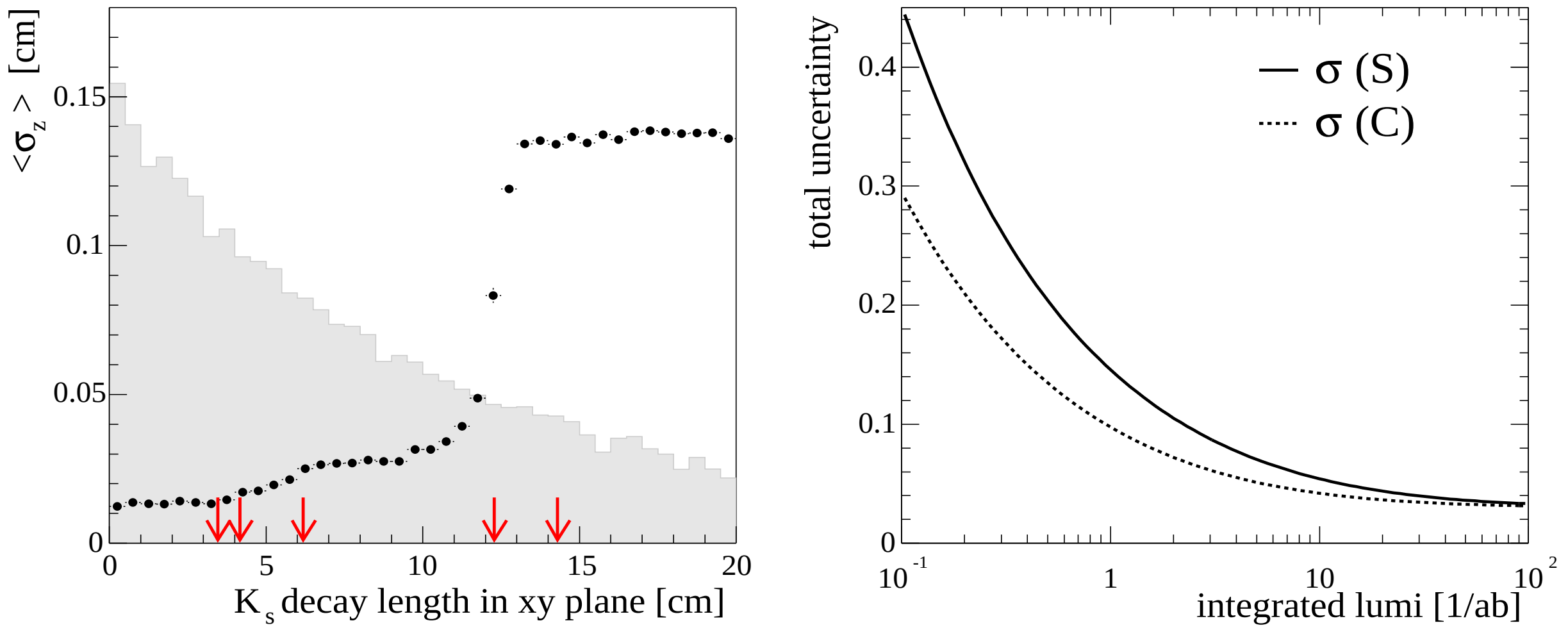

**Figure 3-10.** *(a) Average estimated uncertainty in $z_{B^0}$ (dots) as a function of the $K^0_S$ decay length. The arrows indicate the position of the five SVT layers. The superimposed histogram shows the decay length distribution (with arbitrary scale). (b) Total uncertainties in $S$ and $C$ as a function of the integrated luminosity, assuming a systematic uncertainty of* $0.03$ *and a constant signal-to-background ratio..*

**Extraction of the $CP$ parameters** The parameters $S$ and $C$ are extracted from the data with a maximum likelihood fit to $\Delta t$, $\sigma(\Delta t)$, tagging information and $B$ selection observables [81]. The $\Delta t$ resolution function is taken from the $B$ flavor sample, as for the $B^0 \to J/\psi K^0_S$ modes. This choice is motivated by the fact that Monte Carlo $\Delta t$ pull distribution for the $B^0 \to J/\psi K^0_S$ mode and the $B^0 \to K^0_S\pi^0$ mode are very similar, even though the latter includes the effects of the dedicated vertex reconstruction.

**Systematic uncertainties** Table 3-3 shows a breakdown of the systematic uncertainty on $S$ and $C$. The systematic uncertainty due to SVT alignment is estimated by reconstructing Monte Carlo data with different misalignment scenarios that reflect possible remaining distortions of the SVT. The assigned uncertainty is larger than that reported for the $B^0 \to J/\psi K^0_S$ mode (0.010 [82]), partially because the reconstruction of $B^0 \to K^0_S\pi^0$ is more sensitive to the alignment of the outer SVT layers.

A systematic uncertainty for the reconstruction method and a possible differences in resolution function is derived from a study of $B^0 \to J/\psi K^0_S$ decays. By excluding the $J/\psi$ decay products from the $B^0$ vertex, this decay can be reconstructed with the same method used for the signal mode. This allows for a direct comparison of the obtained values of $S$ and $C$ to those obtained with the nominal reconstruction. The data do not contain sufficient $B^0 \to J/\psi K^0_S$ events to result in a meaningful systematic, but are consistent with the systematic uncertainty derived from the Monte Carlo. Results from a recent study suggest that this systematic is related to a small bias in the $\Delta t$ scale, which can be reduced by taking the transverse motion of the $B^0$ into account.

The systematic uncertainty labeled 'PDF' comprises the uncertainties in the parameterization of the likelihood function. For our measurement this uncertainty is dominated by an observed tagging asymmetry in background events. This asymmetry constitutes a $2\sigma$ deviation from $0$ for events with a lepton tag. For the current analysis we have interpreted this deviation as a systematic uncertainty, although the effect could be real.

**Expected uncertainties on $10\,\mathrm{ab}^{-1}$** It is not self-evident that the systematic uncertainties discussed above are applicable to future measurements. The understanding of the vertexing method and the resolution function can be enhanced by using the large $B^0 \to J/\psi K^0_S$ and inclusive $K^0_S$ samples available at $10\,\mathrm{ab}^{-1}$. Systematic uncertainties due to the parameterization of the likelihood function will roughly scale with $1/\sqrt{N}$. Those uncertainties related to alignment should necessarily improve if meaningful results for other—high statistics—modes must be obtained. A recent analysis for $B^0 \to J/\psi\, K^0_S$ estimates the asymptotic systematic uncertainty in $\sin 2\beta$ at $0.021$ [83]. Therefore, a total systematic uncertainty of $0.03$ in both $S_{K^0_S\pi^0}$ and $C_{K^0_S\pi^0}$ seems not unrealistic.

**Table 3-3.** *Breakdown of the systematic uncertainty in $S$ and $C$.*

| | $\sigma(C)$ | $\sigma(S)$ |
|---|---|---|
| SVT alignment | 0.009 | 0.028 |
| vertexing method | 0.004 | 0.040 |
| PDF | 0.093 | 0.027 |
| total | 0.094 | 0.056 |

Figure 3-10(b) shows the expected total uncertainty in $S$ and $C$ as a function of the integrated luminosity, assuming a systematic uncertainty of $0.03$ and constant signal to background ratio. At $10\,\mathrm{ab}^{-1}$ the statistical uncertainties in $S$ and $C$ are $0.045$ and $0.029$, respectively, that is, comparable to the present uncertainties on the $S$ and $c$ parameters in the $J/\psi\, K^0_S$ mode.

# 3.4 $\alpha$ from $B \longrightarrow \pi\pi$ and Variants

## 3.4.1 Theoretical uncertainties in determining $\alpha$

$\succ$– H. Quinn –$\prec$

There are three sets of channels of interest for determining $\alpha$ to high accuracy. These are $\pi\pi$, $\rho\rho$ (the dominant channel) with angular analysis to isolate pure longitudinal contributions , and $\rho\pi$. In each case the question before us as we consider the value of a Super $B$ Factory is how accurately can we determine the theoretical uncertainties in extracting $\alpha$, as this will give a measure of the value of a high statistics measurement. As a rule of thumb, increased statistics is valuable only if the error is dominated by statistical error. Once theoretical uncertainties dominate the error, then statistical improvement cannot help. Of course, in looking to the future we must also take into account possible improvements of the theoretical calculations. All these analyses include channels with at least one neutral pion, which makes them challenging for the hadronic $B$ experiments, and so important targets for a $e^+e^-$ Super $B$ Factory.

In the first two cases, the measured asymmetries (of $\pi^+\pi^-$, $\rho^+\rho^-$ or $\rho^0\rho^0$) are proportional to

$$|\overline{A}_{ij}/A_{ij}|\ \sin(2\alpha+\delta_{ij}) \tag{3.51}$$

where $\delta_{ij}$ is the phase difference between $\overline{A}_{ij}$ and $A_{ij}$ after factorizing out the weak phase difference of the tree amplitude contributions, which is $-2\gamma$ in the standard Wolfenstein convention for the CKM elements. Here $i$ and $j$ represent the charges of the final particles. (In the $\pi^0\pi^0$ case one cannot hope to determine the time-dependent $CP$ asymmetry.)

If one ignores any contribution from electroweak penguin amplitudes and any isospin breaking effects, then one can use the measurements of the set of isospin-related channels [69] to measure the quantity $\delta_{ij}$ for each of these three channels and thus extract the value of $\alpha$ from each of the asymmetry measurements with no penguin contribution uncertainties. The residual theoretical uncertainties have been discussed very nicely by Ciuchini for the $\pi\pi$ case (see Section 3.4.2); the arguments in the $\rho\rho$ case are very similar. I will add here a few comments of my own on this situation. A recent comment points out a slight complication of an I=1 contribution in the $\rho\rho$ case, but argues that it can be constrained by experiment [84].

The electroweak penguin contributions proportional to the dominant operators $O_7$ and $O_8$ can be constrained using the fact that these operators are proportional to the operators $O_1$ and $O_2$ that mediate the tree diagrams [52]. Thus if one assumes that the remaining electroweak penguin operators, which have significantly smaller coefficients, are not inordinately enhanced, one can bound the impact of the electroweak penguins to produce an absolute shift of less than $\pm 0.02$ in $\sin 2\alpha$. In the $\rho\rho$ case, the impact of electroweak penguin effects in the charged $B$ decays gives the same correction to $\delta_{ij}$ for the two channels ($+-$ and $00$), so one cannot use the two asymmetry measurements to remove the uncertainty due to this correction.

The remaining question on electroweak penguins is whether one can better justify the assumption that $O_9$ and $O_{10}$ are not enhanced. Lattice QCD calculations could be helpful here. The continuation from the Euclidean region where the calculation is done to the physical Minkowski space introduces large uncertainties into the absolute value of this quantity. I am quite sure one cannot use the lattice to calculate the relative phases of these terms, but since all we want is a rough constraint on their magnitudes relative to the $O_7$ and $O_8$ terms, perhaps this can be achieved. I leave it to the lattice QCD experts to discuss this question further.

A larger uncertainty arises from isospin violation, predominantly that which manifests itself as either $\pi^0\eta$ or $\rho^0\omega$ mixing. This and other isospin-breaking effects have been been studied for the two pion case by Gardner [85]. Her analysis could be applied with little change to the $\rho\rho$ case; this work needs to be done. Her conclusion for the $\pi\pi$ case is discouraging: she finds large residual uncertainties, up to about 0.1 in sin($2\alpha$). This analysis needs to be updated, as measurements now available further constrain some of the quantities that were computed by model-dependent methods and used to make these estimates.

The remaining method to extract $\alpha$ is the analysis of the three pion Dalitz plot in terms of the $\rho\pi$ resonance contributions. It is by now very clear that the estimates in my paper with Snyder on the utility of this channel were overly optimistic. Just how good a result one can eventually get from this analysis is not clear. The theoretical uncertainty again lies in the use of isospin to constrain the number of independent amplitudes, and secondly in the assumption that the strong phase variation in these amplitudes as one moves around the rho bands is properly characterized by describing the $\rho$ resonance with a relativistic Breit-Wigner parametrization. Encouraging evidence that this is a reasonable expectation is given by the recent and very clever *BABAR* analysis of the phase shift change as one traces out the $K^*$ band in $B \rightarrow J/\psi K^*$. The match between *BABAR* and LASS data on this is quite remarkable. With a very large data sample one could make similar studies for the $\pi\pi$ phase shift in portions of the rho band that are not subject to significant interference effects. The issue of how isospin violations and electroweak penguin contributions affect this analysis has not been studied in detail, probably because we are far from the situation in which these are the dominant uncertainties, and indeed I think it is unlikely we will ever reach that point.

A second type of uncertainty is the question of how well one can parameterize other contributions to the three pion Dalitz plot, both those from non-resonant $B \rightarrow 3\pi$ decays and those from other resonant decays. The latter, and any interferences between them and the $\rho$ bands contain additional information, but at the price of adding parameters to the already-many-parameter fits. These are not, strictly speaking theoretical uncertainties. It is a matter of looking at what is needed to fit the data, once one has enough of it. The amount of data one needs to do a good job of this analysis is clearly large, and it grows as more parameters are needed to obtain a good fit. I think it is premature to try to determine the eventual accuracy of this measurement, as many of the questions can only be resolved as data accumulates in the next five to ten years. I am not optimistic that this analysis will eventually give the most accurate value for $\sin(2\alpha)$. However it probably will be able to provide sufficient information on the sign of $\cos(2\alpha)$ to reduce the ambiguity in $\alpha$ to a two-fold, rather than a four-fold degeneracy. For this reason this channel must continue to be pursued.

### 3.4.2 Isospin Breaking

$\succ$– M. Ciuchini –$\prec$

Popular methods for extracting the CKM angle $\alpha$ from the time-dependent $CP$ asymmetries in $b \rightarrow u$ transitions, such as $B \rightarrow \pi\pi$, $B \rightarrow \rho\pi$, $B \rightarrow \rho\rho$, rely on isospin (*i.e.*, flavor SU(2)) symmetry. This symmetry, corresponding to the exchange of up and down quarks, is known to be an approximate symmetry of the strong interactions, with violations as small as few percent in most cases. Nevertheless, also in view of the increasing accuracy of the present and future experiments, one can wonder whether isospin violation could hinder a precise determination of $\alpha$.

Let us discuss the Gronau-London isospin analysis in $B \rightarrow \pi\pi$ as the prototype of methods using flavor SU(2) symmetry [69]. Similar considerations apply to SU(2)-based bounds [86]–[88], such as the Grossman-Quinn bound, as well as to analyses using different final states.

Using only the isospin properties of the initial and final states with no assumption whatsoever on the interaction, one can write the following amplitude decomposition:

$$\begin{aligned} A(B_d \rightarrow \pi^+\pi^-) &= A_2^{3/2} - A_2^{5/2} - A_0^{1/2}\,, \\ A(B_d \rightarrow \pi^0\pi^0) &= \sqrt{2}(A_2^{3/2} - A_2^{5/2}) + \frac{1}{\sqrt{2}}A_0^{1/2}\,, \\ A(B^+ \rightarrow \pi^+\pi^0) &= \frac{3}{\sqrt{2}}A_2^{3/2} + \sqrt{2}A_2^{5/2}\,, \end{aligned} \tag{3.52}$$

where $A_I^{\Delta I}$ denotes the amplitude of a $\Delta I$ transition into a $\pi\pi$ final state with isospin $I$ (we remind the reader that the 2-pion state with I=1 is forbidden by Bose-Einstein symmetry).

Basically, the GL method exploits all the isospin-related $\pi\pi$ amplitudes to determine the unknown strong phase relating $\alpha$ to the coefficient of the sine term in the time-dependent $CP$ asymmetry. This strategy works (up to problems related to discrete ambiguities) under three assumptions:

1. The hadronic matrix elements are invariant under flavor SU(2) transformations;
2. There are no $\Delta I = 5/2$ transitions;
3. "Penguin" operators (*i.e.*, operators carrying the "penguin" weak phase, arg($V_{tb}^* V_{td}$) in our case) in the $\Delta B = 1$ weak effective Hamiltonian give rise to $\Delta I = 1/2$ transitions only.

Flavor symmetry actually enters only the first assumption, while the others concern the flavor structure of the weak interactions. However the GL method needs all of them. In fact, using the first two assumptions, one can rewrite the previous equations in the more usual form

$$\begin{aligned} A(B_d \to \pi^+\pi^-) &= A_2^{3/2} - A_0^{1/2}\,, \\ A(B_d \to \pi^0\pi^0) &= \sqrt{2}A_2^{3/2} + \frac{1}{\sqrt{2}}A_0^{1/2}\,, \\ A(B^+ \to \pi^+\pi^0) &= \frac{3}{\sqrt{2}}A_2^{3/2}\,. \end{aligned} \tag{3.53}$$

A consequence of this decomposition are the triangular relations

$$\begin{aligned} A(B^+ \to \pi^+\pi^0) - A(B_d \to \pi^0\pi^0) - \frac{1}{\sqrt{2}}A(B_d \to \pi^+\pi^-) = 0\,, \\ A(B^- \to \pi^-\pi^0) - A(\overline{B}_d \to \pi^0\pi^0) - \frac{1}{\sqrt{2}}A(\overline{B}_d \to \pi^+\pi^-) = 0\,, \end{aligned} \tag{3.54}$$

used by the GL method.

Finally, the third assumption is crucial to relate these two triangles. In fact, if the $\Delta I = 3/2$ amplitude has no contribution proportional to the "penguin" weak phase, the *CP*-conjugate amplitude can be easily rescaled using the "tree" phase only, so that

$$A(B^+ \to \pi^+\pi^0) = \tilde{A}(B^- \to \pi^-\pi^0) \equiv e^{2i\gamma} A(B^- \to \pi^-\pi^0)\,. \tag{3.55}$$

The above relation means that the two triangles in Eq. (3.54) have a side in common, once the second one is rescaled as above. In turn, this allows for the GL geometrical construction which gives the strong phase difference between $\tilde{A}(\overline{B}_d \to \pi^+\pi^-)$ and $A(B_d \to \pi^+\pi^-)$ needed to extract $\alpha$ from the time-dependent *CP* asymmetry [69].

The GL isospin analysis is considered very sound from a theoretical point of view, sharing the general confidence of theorists in the SU(2) flavor symmetry. However all the assumptions required by the GL analysis are violated to some extent. In fact:

1. Even considering strong interactions only, the SU(2) flavor symmetry is broken by the up-down mass difference. Furthermore, electromagnetic effects generate additional breaking terms;
2. There are no $\Delta I = 5/2$ operators in the weak effective Hamiltonian. However isospin breaking in the hadronic matrix elements give rise to an effective $\Delta I = 5/2$ amplitude;
3. While QCD penguin operators only mediate $\Delta I = 1/2$ transitions, the effective Hamiltonian contains also electroweak penguin (EWP) operators which are both $\Delta I = 1/2$ and $3/2$ and carry the "penguin" weak phase.

Therefore the real issue is how large these breaking effects are or, in other words, at which level of accuracy we should start worrying about them.

Let us discuss the effect of EWP operators first. They break the isospin symmetry as much as the other operators of the weak effective Hamiltonian. This breaking, being of electroweak nature, has nothing to do with the properties under flavor symmetries of the hadronic matrix elements. However the presence of EWP operators violates the assumption that the "penguin" weak phase appears only in $\Delta I = 1/2$ transitions. This implies that a simple rescaling is not enough to equate the two amplitudes in Eq. (3.55) and the GL construction is invalidated. In practice, the effect of the additional contributions coming from the EWP operators change the GL triangles into quadrilaterals. The size and orientation of the additional side depend on the relative modulus and phase of the EWP contribution with respect to the "tree" term entering $A_2^{3/2}$.

An interesting observation concerns $(V-A)\otimes(V-A)$ EWP operators (we remind the reader that there are four EWP operators in the $\Delta F = 1$ effective Hamiltonian usually denoted as $O_7$–$O_{10}$. The first two have a $(V-A)\otimes(V+A)$ Dirac structure, while the others are $(V-A)\otimes(V-A)$). Studying the flavor properties of the effective Hamiltionian, it was noted that these two operators are related to current-current operators $O_{1,2}$ by short-distance factors (namely the appropriate ratio of Wilson coefficients) [48, 89]. This means that no new hadronic parameter enters their matrix elements. They give simply a small and calculable correction to the "tree" amplitude. This remarkable result is actually not surprising. It is indeed known that $O_9$ and $O_{10}$ are not independent operators: they can be written in terms of QCD penguin and current-current operators [90]. Nevertheless, they are usually retained in the effective Hamiltonian to avoid that current-current operators could give both "tree" and "penguin" contributions. Therefore, if the $(V-A)\otimes(V+A)$ EWP operators are neglected, the GL isospin analysis can be easily recovered. Indeed, this assumption has been advocated in the literature, arguing that the Wilson coefficients $C_7$ and $C_8$ of the $(V-A)\otimes(V+A)$ EWP operators are numerically much smaller than their $(V-A)\otimes(V-A)$ counterparts at the weak scale [89]. However, arguments based on the numerical value of Wilson coefficients may be tricky. On the one hand, the renormalization-group running to lower scales increases the size of $C_{7,8}$ relative to $C_{9,10}$. On the other, matrix elements of $(V-A)\otimes(V+A)$ operators can be much larger than those of $(V-A)\otimes(V-A)$ operators. For example, the dominating EWP contribution to $\varepsilon'/\varepsilon$ in kaon decays comes from $O_8$ rather than $O_9$ or $O_{10}$. Of course, $B$ decays involve a rather different physics. Still, the matrix elements of $(V-A)\otimes(V+A)$ operators could be enhanced enough to compensate the short-distance suppression so that we likely cannot get rid of EWP operators so easily.

We are back to the problem of estimating the contribution of two hadronic amplitudes with different weak phases, both contributing to $A_2^{3/2}$. This cannot be done in a model-independent way. The advantage with respect to the original problem is that the EWP contribution is expected to be much smaller than the "tree" contribution. Therefore, larger theoretical uncertainties associated with model estimates can be more easily tolerated. However, the difficulty of quantifying the systematic uncertainties attached to a specific model remains. Moreover, in practice, all the available estimates are done within different realizations of factorization, going from the old naïve factorization to the recent QCD factorization, producing very similar results. The uncertainty introduced by the EWP operators in the extraction of $\alpha$ with the isospin analysis ranges from negligible to $5°$ [78, 47, 91, 56].

Further uncertainties are introduced by genuine isospin-breaking corrections originating from light quark mass and electromagnetic effects in the hadronic amplitudes. Both are expected to be reasonably small, being suppressed either by $(m_u\text{-}m_d)/\Lambda_{QCD}$ or $\alpha_e$. While the typical effect on the amplitudes is at the level of few percent, in some cases it can be substantially larger. In particular, the $u$-$d$ mass difference induces the $\pi^0$-$\eta$-$\eta'$ mixing which is estimated, although with very large uncertainties, to change the $A_0$ and $A_2$ amplitudes up to $10\%$ or even more [92, 93]. Genuine isospin-breaking corrections generate additional amplitudes $A_2^{5/2}$, breaking the triangular relations of Eq. (3.54), adding additional sides to the GL geometrical construction. The effect of the $\pi^0$-$\eta$-$\eta'$ mixing on the extraction of $\alpha$ has been studied in Ref. [94] using factorization. Within large uncertainties, the effect of these isospin-breaking terms has been estimated to induce an error on $\alpha$ of $\sim 5°$.

In summary, isospin-breaking effects in the extraction of $\alpha$ based on isospin relations can reasonably be neglected as long as the error is larger than $\sim 10°$. Indeed, for the GL analysis, the actual effect is likely between nil and $10°$, keeping in mind that the theoretical estimates are based on models and subject to uncertainties difficult to quantify. A high-precision determination of $\alpha$ based on SU(2) would require a theoretical control of the isospin-breaking terms missing at present and probably not attainable in the near future. Indeed the problem one has to face is nearly as difficult as the calculation of the full hadronic amplitudes.

### 3.4.3 Measurement of $\sin 2\alpha$ and $B^0 \to \pi^0\pi^0$

>– A. Roodman –<

The $CP$ violating asymmetry in the decay mode $B^0 \to \pi^+\pi^-$ depends both on the CKM angle $\alpha$ and on the interference between contributions from tree and penguin diagrams. The coefficient of the $\sin(\Delta m_d \Delta t)$ term in the asymmetry may be expressed as

$$S = \sqrt{1 - C^2}\sin\left\{2\alpha + Arg(\overline{A}/A)\right\}$$

where $C$ is the coefficient of the $\cos(\Delta m_d \Delta t)$ term, and $A$ and $\overline{A}$ are the decay amplitudes for $B^0$ and $\overline{B}^0$. The penguin pollution angle $\delta$ is then given by $\delta = Arg(\overline{A}/A)/2$.

This *penguin pollution* of the asymmetry can be determined experimentally by measuring the branching fractions for $B^0$ and $\overline{B}^0$ from all three $B \to \pi\pi$ decays. The decays are related by an isospin relation

$$\frac{1}{\sqrt{2}}A^{+-} = A^{+0} + A^{00}$$

between the amplitudes for $B^0$ decay and a similar relation for $\overline{B}^0$ decay [69]. In the limit of isospin symmetry, or ignoring electro-weak penguins, the $B^\pm \to \pi^\pm\pi^0$ amplitudes are equal, since there are no penguin amplitudes for this decay. The constraint on the penguin pollution angle can be understood using the triangle construction shown in Fig. 3-11, where the argument of the amplitude ratio $\overline{A}^{+-}/A^{+-}$ is given by the angle between the $A^{+-}$ legs of the two triangles. There is a four-fold ambiguity for the penguin pollution angle, $\delta$, corresponding to the two relative orientations of the two triangles, and to a positive or negative sign for $\delta$. Lastly, the presence of electro-weak penguin amplitudes, as an isospin breaking contribution, break the simple triangle relations. However, electro-weak penguins do seem to be rather small in $B \to \pi\pi$ decays.

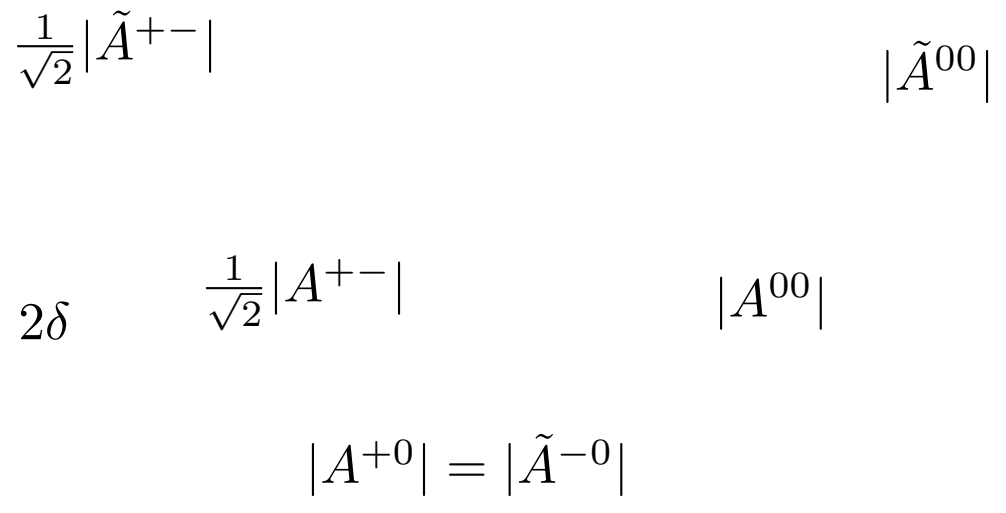

**Figure 3-11.** *Isospin Triangles for $B \to \pi\pi$. The amplitudes for the $\overline{B}^0$ triangle are rotated by $e^{2i\gamma}$ so that the bases of the two triangles overlap. The current world-averaged values are used.*

The current status of branching fractions and asymmetries in the $B \to \pi\pi$ modes are summarized in Table 3-4, with measurements from *BABAR* and Belle listed, along with averages from the Heavy Flavor Averaging Group [95]. Also shown are luminosity scaling expressions for each of the measurements. The most challenging measurement is the direct $CP$-violating asymmetry in the $B^0 \to \pi^0\pi^0$ decay, which has a small branching ratio and a large continuum background. Without a vertex, only a time-integrated measurement of the $C$ coefficient is possible, adding to the difficulty of this measurement. Prior estimates of the error scaling for $C_{\pi^0\pi^0}$, made for Super $B$ Factory studies, were $\sigma_{C_{\pi^0\pi^0}} \sim 10/\sqrt{\int \mathcal{L}}$, with $\mathcal{L}$ in units of $\text{fb}^{-1}$, assuming that only leptonic tags were used and that the background was a factor of two greater than is currently the case. Using all tagging sources and the current background levels should give an estimated 30% improvement in the error. Initial measurements of $C_{\pi^0\pi^0}$ have now been made by both *BABAR* and Belle [96, 97]. Their errors scale roughly as $\sigma(C_{\pi^0\pi^0}) \sim 8.3/\sqrt{\int \mathcal{L}}$, in good agreement with expectations.

The current limit on the penguin pollution angle $|\delta|$ is shown in Fig. 3-12, using the world averages. The Confidence Level as a function of $\delta$ is shown, using the CKMFitter package [36]. Effectively we scan over all values of $|\delta|$ and calculate a $\chi^2$ for a fit to the five amplitudes ($A^{+-}$, $\overline{A}^{+-}$, $A^{00}$, $\overline{A}^{00}$ and $A^{+0}$), given the five measurements, the fixed input value for $\delta$, and the two isospin relations as constraints. This $\chi^2$ is converted to a confidence level in the standard way. We can compare this result with the upper limit on $|\delta|$ found using the expression [86]

$$\sin^2\delta \leq \frac{\Gamma^{00}}{\Gamma^{+0}}.$$

With the current world averages, the 90% upper limit from the Grossman-Quinn bound is $38.5^o$, essentially the same upper limit found from the full isospin analysis, as shown in in Fig. 3-12. Once experimental errors are included, the more restrictive expression from Ref. [88] improves the upper limit by $1^o$. Lastly, extrapolating to $1\,\mathrm{ab}^{-1}$ or $10\,\mathrm{ab}^{-1}$, with the current central values, gives preferred regions for $|\delta|$ as shown in Fig. 3-12. In this treatment, the effects of isospin-violating electroweak penguin diagrams have been omitted. Both theoretical prejudice and current phenomenological fits point to small electroweak penguins, with an effect on the penguin pollution angle of a couple of degrees.

The current central values have a value of $C_{\pi^0\pi^0}$ close to the boundary of the physical region, such that the area of the $B^0$ triangle is quite small. In fact, given all other measurements the value of $C_{\pi^0\pi^0}$ is bounded by demanding that both triangles close, with limits given by

$$C^{\max}_{\pi^0\pi^0} = \frac{-\frac{1}{2}\Gamma^{+-}(1-C_{+-}) - \Gamma^{+0} + \Gamma^{00} + \sqrt{2\Gamma^{+0}\Gamma^{+-}(1-C_{+-})}}{\Gamma^{00}}$$

and

$$C^{\min}_{\pi^0\pi^0} = \frac{\frac{1}{2}\Gamma^{+-}(1+C_{+-}) + \Gamma^{+0} - \Gamma^{00} - \sqrt{2\Gamma^{+0}\Gamma^{+-}(1+C_{+-})}}{\Gamma^{00}}.$$

The current central values for these limits are $[-0.34 : 0.83]$, but with errors included no limit inside the physical region is obtained. With one triangle just barely closed, the two solutions for $|\delta|$ are close together. In the opposite limit, when $C_{\pi^0\pi^0} \approx -C_{\pi^+\pi^-}$, one solution is close to zero and coalesces with the mirror solution with $\delta < 0$. The value and error on the penguin pollution angle $|\delta|$ is shown in Fig. 3-13 for the current world averages of the other measurements, and with errors extrapolated to the level expected with $10\,\mathrm{ab}^{-1}$, as a function of the asymmetry in $B^0 \to \pi^0\pi^0$. We see that in much of the parameter space very well-separated solutions may be found.

**Table 3-4.** *Summary of current measurements for $B \to \pi\pi$, and luminosity scaling relations for the measured uncertainties. The world averages are from the HFAG [95], expect for $C_{\pi^+\pi^-}$ where I have included a scaling factor of two to cover the difference between the measurements. The branching ratios are in units of $10^{-6}$, and the error coefficient is to be used as $\sigma(\mathcal{B}, C) = \mathrm{Coeff}/\sqrt{\int \mathcal{L}}$ . These error coefficients are taken from the quality of the world averages, or the single dominant measurement.*

| Measurement | *BABAR* | Belle | World Average | Error Coeff. |
|---|---|---|---|---|
| $\mathcal{B}(B^0 \to \pi^+\pi^-)$ | $4.7 \pm 0.6 \pm 0.2$ | $4.4 \pm 0.6 \pm 0.3$ | $4.6 \pm 0.4$ | 5.4 |
| $\mathcal{B}(B^\pm \to \pi^\pm\pi^0)$ | $5.8 \pm 0.6 \pm 0.4$ | $5.0 \pm 1.2 \pm 0.5$ | $5.5 \pm 0.6$ | 10. |
| $\mathcal{B}(B^0 \to \pi^0\pi^0)$ | $1.17 \pm 0.32 \pm 0.10$ | $2.32\,^{+0.44}_{-0.48}\,^{+0.22}_{-0.18}$ | $1.51 \pm 0.28$ | 6.0 |
| $C_{\pi^+\pi^-}$ | $-0.09 \pm 0.15 \pm 0.04$ | $-0.58 \pm 0.15 \pm 0.07$ | $-0.32 \pm 0.23$ | 2.1 |
| $C_{\pi^0\pi^0}$ | $-0.12 \pm 0.56 \pm 0.06$ | $-0.43 \pm 0.51 \pm 0.16$ | $-0.28 \pm 0.39$ | 8.3 |

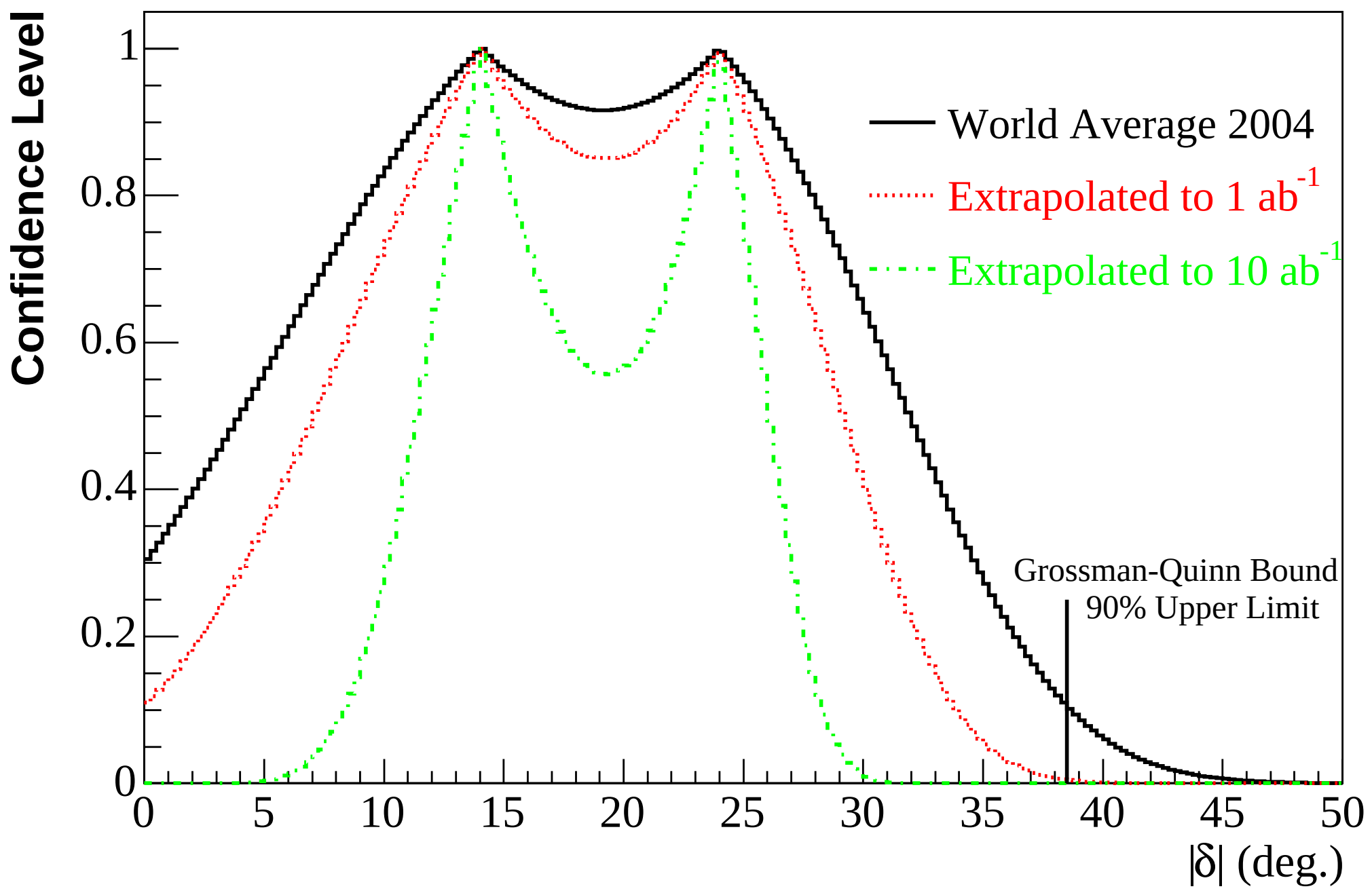

**Figure 3-12.** *Preferred region for the penguin pollution angle $|\delta|$ for the current world averages and errors. Also shown are the preferred regions with the errors on all measurements extrapolated to 1 ab$^{-1}$ and 10 ab$^{-1}$.*

While the exact error on $|\delta|$ will depend on the value of $C_{\pi^0\pi^0}$, a scaling expression for the typical error can be found by exploring the possible parameter space. We find that the expression $\sigma_\delta \sim 360^o/\sqrt{\int \mathcal{L}(\,\mathrm{fb}^{-1})}$ is a reasonable scaling expression. The error on the penguin pollution angle will dominate the error on $\alpha$ from the $B \to \pi\pi$ system, in the limit of high statistics.

With the large data samples possible at a Super $B$ Factory , it will be possible to accurately measure the CKM angle $\alpha$ with the $B \to \pi\pi$ decay, using the isospin construction to measure the penguin pollution. A similar, and much less penguin-polluted measurement, will be made with the decay $B^0 \to \rho^+\rho^-$, as well as another measurement with $B \to \rho\pi$. As in the comparison between $CP$ violation in $B^0 \to J/\psi K^0_S$ and $B^0 \to \phi K^0_S$, the comparison between the three different ways to measure $\alpha$, one with a significant penguin component, will provide an excellent test of the completeness of the CKM picture.

### 3.4.4 The prospects of measuring the CKM angle $\alpha$ with *BABAR*

>– V. G. Shelkov –<

#### Introduction

One of the most important goals for the current as well as next generations of $B$ Factories is to put a new set of constraints on the values of CKM angle $\alpha$. In this study we follow a quasi-two-body approach [98], and restrict the analysis to the two regions of the $\pi^\mp\pi^0 h^\pm$ Dalitz plot ($h = \pi$ or $K$) that are dominated by either $\rho^+ h^-$ or $\rho^- h^+$. More general approaches, like the full Dalitz plot analysis [99], have been proposed in the literature, and can be implemented once significantly larger data samples are available. We present here a simultaneous measurement of branching fractions and $CP$-violating asymmetries in the decays $B^0 \to \rho^\pm\pi^\mp$ and $B^0 \to \rho^- K^+$ (and their charge conjugates). The $\rho^{\pm}\pi^{\mp}$ mode provides a probe of both mixing-induced and direct $CP$ violation [100], whereas the self-tagging $\rho^- K^+$ can only exhibit direct $CP$ violation. The *BABAR* and Belle experiments have performed searches for $CP$-violating asymmetries in $B$ decays to $\pi^+\pi^-$ [101, 102], where the mixing-induced $CP$ asymmetry is related to the angle $\alpha \equiv \arg\left[-V_{td}V_{tb}^*/V_{ud}V_{ub}^*\right]$ of the Unitarity Triangle as it is for $\rho^{\pm}\pi^{\mp}$. However, unlike $\pi^+\pi^-$, $\rho^{\pm}\pi^{\mp}$ is not a $CP$ eigenstate, and four flavor-charge configurations $(B^0(\overline{B}^0) \to \rho^{\pm}\pi^{\mp})$ must be considered. Although this leads to a more complicated analysis, it benefits from a branching fraction that is nearly five times larger [103, 104]. Some examples of improvements in the precision of the experimental constraints on CKM angle $\alpha$ expected to materialize at Super $B$ Factoriesare then discussed.

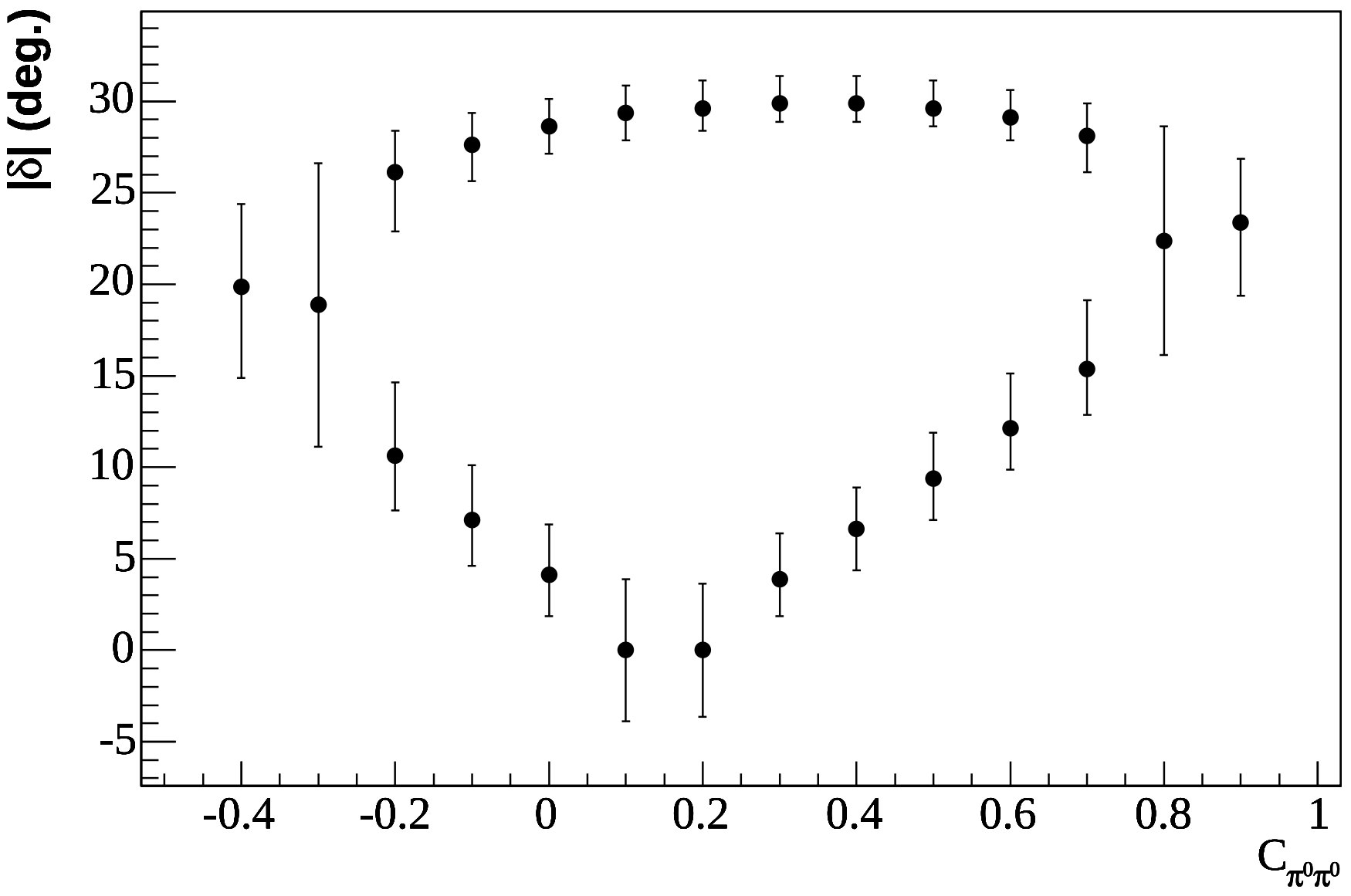

**Figure 3-13.** *Central value and one sigma error for the penguin pollution angle $|\delta|$ as a function of $C_{\pi^0\pi^0}$. For all other measurements the current world averages have been used with errors extrapolated to the level expected with $10$ ab$^{-1}$. When two of the solutions for $\delta$ coalesce, including the solutions with $\delta < 0$, only one solution with an error encompassing both of them is shown.*

**Theoretical framework**

In the Standard Model, $CP$-violating effects arise from a single complex phase in the three-generation Cabibbo-Kobayashi-Maskawa quark-mixing matrix [105]. One of the central, unresolved questions is whether this mechanism is sufficient to explain the pattern of $CP$ violation observed in nature.

With $\Delta t \equiv t_{\rho h} - t_{\rm tag}$ defined as the proper time interval between the decay of the reconstructed $B^0_{\rho h}$ and that of the other meson $B^0_{\rm tag}$, the time-dependent decay rates are given by

$$f^{\rho^{\pm}h^{\mp}}_{Q_{\rm tag}}(\Delta t) = (1 \pm A^{\rho h}_{CP})\frac{e^{-|\Delta t|/\tau}}{4\tau} \times \Big[1 + Q_{\rm tag}(S_{\rho h} \pm \Delta S_{\rho h})\sin(\Delta m_d \Delta t) - Q_{\rm tag}(C_{\rho h} \pm \Delta C_{\rho h})\cos(\Delta m_d \Delta t)\Big] , \quad (3.56)$$

where $Q_{\rm tag} = 1(-1)$ when the tagging meson $B^0_{\rm tag}$ is a $B^0(\overline{B}^0)$, $\tau$ is the mean $B^0$ lifetime, and $\Delta m_d$ is the $B^0\overline{B}^0$ oscillation frequency. The time- and flavor-integrated charge asymmetries $A^{\rho\pi}_{CP}$ and $A^{\rho K}_{CP}$ measure direct *CP* violation. For the $\rho\pi$ mode, the quantities $S_{\rho\pi}$ and $C_{\rho\pi}$ parameterize mixing-induced *CP* violation related to the angle $\alpha$, and flavor-dependent direct *CP* violation, respectively. The parameters $\Delta C_{\rho\pi}$ and $\Delta S_{\rho\pi}$ are insensitive to *CP* violation. $\Delta C_{\rho\pi}$ describes the asymmetry between the rates $\Gamma(B^0 \to \rho^+\pi^-)+\Gamma(\overline{B}^0 \to \rho^-\pi^+)$ and $\Gamma(B^0 \to \rho^-\pi^+)+\Gamma(\overline{B}^0 \to \rho^+\pi^-)$, while $\Delta S_{\rho\pi}$ is related to the strong phase difference between the amplitudes contributing to $B^0 \to \rho\pi$ decays. More precisely, one finds the relations $S_{\rho\pi} \pm \Delta S_{\rho\pi} = \sqrt{1-(C_{\rho\pi} \pm \Delta C_{\rho\pi})^2}\,\sin(2\alpha^{\pm}_{\rm eff} \pm \delta)$, where $2\alpha^{\pm}_{\rm eff} = \arg[(q/p)(\overline{A}^{\pm}_{\rho\pi}/A^{\mp}_{\rho\pi})]$, $\delta = \arg[A^{-}_{\rho\pi}/A^{+}_{\rho\pi}]$, $\arg[q/p]$ is the $B^0\overline{B}^0$ mixing phase, and $A^{+}_{\rho\pi}(\overline{A}^{+}_{\rho\pi})$ and $A^{-}_{\rho\pi}(\overline{A}^{-}_{\rho\pi})$ are the transition amplitudes of the processes $B^0(\overline{B}^0) \to \rho^+\pi^-$ and $B^0(\overline{B}^0) \to \rho^-\pi^+$, respectively. The angles $\alpha^{\pm}_{\rm eff}$ are equal to $\alpha$ in the absence of contributions from penguin amplitudes. For the self-tagging $\rho K$ mode, the values of the four time-dependent parameters are $C_{\rho K} = 0$, $\Delta C_{\rho K} = -1$, $S_{\rho K} = 0$, and $\Delta S_{\rho K} = 0$.

### Event selection

The data used in this analysis were accumulated with the *BABAR* detector [106], at the PEP-II asymmetric-energy $e^+e^-$ storage ring at SLAC. The sample consists of $(88.9 \pm 1.0) \times 10^6$ $B\overline{B}$ pairs collected at the $\Upsilon(4S)$ resonance ("on-resonance"), and an integrated luminosity of 9.6 fb$^{-1}$ collected about 40 MeVbelow the $\Upsilon(4S)$ ("off-resonance"). In Ref. [106] we describe the silicon vertex tracker and drift chamber used for track and vertex reconstruction, the Cherenkov detector (DIRC), the electromagnetic calorimeter (EMC), and their use in particle identification (PID).

We reconstruct $B^0_{\rho h}$ candidates from combinations of two tracks and a $\pi^0$ candidate. We require that the PID of both tracks be inconsistent with the electron hypothesis, and the PID of the track used to form the $\rho$ be inconsistent with the kaon hypothesis. The $\pi^0$ candidate mass must satisfy $0.11 < m(\gamma\gamma) < 0.16$ GeV/$c^2$, where each photon is required to have an energy greater than $50\,\mathrm{MeV}$ in the laboratory frame and to exhibit a lateral profile of energy deposition in the EMC consistent with an electromagnetic shower. The mass of the $\rho$ candidate must satisfy $0.4 < m(\pi^{\pm}\pi^0) < 1.3$ GeV/$c^2$. To avoid the interference region, the $B$ candidate is rejected if both the $\pi^+\pi^0$ and $\pi^-\pi^0$ pairs satisfy this requirement. Taking advantage of the helicity structure of $B \to \rho h$ decays ($h$ is denoted *bachelor track* hereafter), we require $|\cos\theta_\pi| > 0.25$, where $\theta_\pi$ is the angle between the $\pi^0$ momentum and the negative $B$ momentum in the $\rho$ rest frame. The bachelor track from the $\rho h$ decay must have a $e^+e^-$ center-of-mass (CM) momentum above $2.4\mathrm{GeV}/c$.

For $86\%$ of the $B^0 \to \rho h$ decays that pass the event selection, the pion from the $\rho$ has momentum below this value, and thus the charge of the $\rho$ is determined unambiguously. For the remaining events, the charge of the $\rho$ is taken to be that of the $\pi^{\pm}\pi^0$ combination with mass closer to the $\rho$ mass. With this procedure, $5\%$ of the selected simulated signal events are assigned an incorrect charge.

To reject background from two-body $B$ decays, the invariant masses of the $\pi^{\pm}h^{\mp}$ and $h^{\pm}\pi^0$ combinations must each be less than $5.14$ GeV/$c^2$. Two kinematic variables allow the discrimination of signal-$B$ decays from fake-$B$ decays due to random combinations of tracks and $\pi^0$ candidates. One variable is the difference, $\Delta E$, between the CM energy of the $B$ candidate and $\sqrt{s}/2$, where $\sqrt{s}$ is the total CM energy. The other variable is the beam-energy-substituted mass $m_{ES} \equiv \sqrt{(s/2+\mathbf{p}_i\cdot\mathbf{p}_B)^2/E_i^2 - \mathbf{p}_B^2}$, where the $B$ momentum $\mathbf{p}_B$ and the four-momentum of the initial state ($E_i$, $\mathbf{p}_i$) are defined in the laboratory frame. The $\Delta E$ distribution for $\rho\pi$ ($\rho K$) signal peaks around 0 ($-45$) MeV since the pion mass is always assigned to the bachelor track. We require $5.23 < m_{ES} < 5.29$ GeV/$c^2$ and $-0.12 < \Delta E < 0.15$ GeV, where the asymmetric $\Delta E$ window suppresses higher-multiplicity $B$ background, which leads to mostly negative $\Delta E$ values. Discrimination between $\rho\pi$ and $\rho K$ events is provided by the Cherenkov angle $\theta_C$ and, to a lesser extent, by $\Delta E$.

Continuum $e^+e^- \to q\overline{q}$ ($q = u, d, s, c$) events are the dominant background. To enhance discrimination between signal and continuum, we use a neural network (NN) to combine four discriminating variables: the reconstructed $\rho$ mass, $\cos\theta_\pi$, and the two event-shape variables that are used in the Fisher discriminant of Ref. [101]. The NN is trained in the signal region with off-resonance data and simulated signal events. The final sample of signal candidates is selected with a cut on the NN output that retains $\sim 65\%$ ($5\%$) of the signal (continuum).

Approximately $23\%$ $(20\%)$ of simulated $\rho\pi$ $(\rho K)$ events have more than one $\rho h$ candidate passing the selection criteria. In these cases, we choose the candidate with the reconstructed $\pi^0$ mass closest to the nominal $\pi^0$ mass. A total of 20,497 events pass all selection criteria. The signal efficiency determined from Monte Carlo (MC) simulation is $20.7\%$ $(18.5\%)$ for $\rho\pi$ $(\rho K)$ events; $31\%$ $(30\%)$ of the selected events are misreconstructed, mostly due to combinatorial-$\pi^0$ background.

We use Monte Carlo-simulated events to study the cross-feed from other $B$ decays. The charmless modes are grouped into eleven classes with similar kinematic and topological properties. Two additional classes account for the neutral and charged $b \to c$ decays. For each of the background classes, a component is introduced into the likelihood, with a fixed number of events. In the selected $\rho\pi$ $(\rho K)$ samples we expect $6 \pm 1$ $(20 \pm 2)$ charmless two-body background events, $93 \pm 23$ $(87 \pm 22)$ charmless three-body background events, $118 \pm 65$ $(36 \pm 18)$ charmless four-body background events, and $266 \pm 43$ $(54 \pm 11)$ $b \to c$ events. Backgrounds from two-, three-, and four-body decays to $\rho\pi$ are dominated by $B^+ \to \pi^+\pi^0$, $B^+ \to \rho^0\pi^+$, and longitudinally polarized $B^0 \to \rho^+\rho^-$ decays. The $\rho K$ sample receives its dominant two-body background from $B^+ \to K^+\pi^0$, and its dominant three- and four-body background from $B \to K^*\pi$ and higher kaon resonances, estimated from inclusive $B \to K\pi\pi$ measurements.

The time difference $\Delta t$ is obtained from the measured distance between the $z$ positions (along the beam direction) of the $B^0_{\rho h}$ and $B^0_{\rm tag}$ decay vertices, and the boost $\beta\gamma = 0.56$ of the $e^+e^-$ system [80, 81, 101]. To determine the flavor of the $B^0_{\rm tag}$ we use the tagging algorithm of Ref. [80, 81]. This produces four mutually exclusive tagging categories. We also retain untagged events in a fifth category to improve the efficiency of the signal selection and the sensitivity to charge asymmetries. Correlations between the $B$ flavor tag and the charge of the reconstructed $\rho h$ candidate are observed in various $B$-background channels and evaluated with MC simulation. We use an unbinned extended maximum likelihood fit to extract the $\rho\pi$ and $\rho K$ event yields, the $CP$ parameters and the other parameters defined in Eq. (3.56). The likelihood for the $N_k$ candidates $i$ tagged in category $k$ is

$$\mathcal{L}_k = e^{-N'_k} \prod_{i=1}^{N_k} \sum_{h}^{\pi,K} \left\{ N^{\rho h} \epsilon_k \mathcal{P}^{\rho h}_{i,k} + N^{q\bar{q},h}_k \mathcal{P}^{q\bar{q},h}_{i,k} + \sum_{j=1}^{N_B} \mathcal{L}^{B,h}_{ij,k} \right\} \tag{3.57}$$

where $N'_k$ is the sum of the signal and continuum yields (to be determined by the fit) and the fixed $B$-background yields, $N^{\rho h}$ is the number of signal events of type $\rho h$ in the entire sample, $\epsilon_k$ is the fraction of signal events tagged in category $k$, and $N^{q\bar{q},h}_k$ is the number of continuum background events with bachelor track of type $h$ that are tagged in category $k$. The total likelihood $\mathcal{L}$ is the product of likelihoods for each tagging category.

The probability density functions (PDFs) $\mathcal{P}^{\rho h}_k$, $\mathcal{P}^{q\bar{q},h}_k$ and the likelihood terms $\mathcal{L}^{B,h}_{j,k}$ are the product of the PDFs of five discriminating variables. The signal PDF is thus given by $\mathcal{P}^{\rho h}_k = \mathcal{P}^{\rho h}(m_{ES}) \cdot \mathcal{P}^{\rho h}(\Delta E) \cdot \mathcal{P}^{\rho h}(\mathrm{NN}) \cdot \mathcal{P}^{\rho h}(\theta_C) \cdot \mathcal{P}^{\rho h}_k(\Delta t)$, where $\mathcal{P}^{\rho h}_k(\Delta t)$ contains the measured physics quantities defined in Eq. (3.56), diluted by the effects of mistagging and the $\Delta t$ resolution. The PDF of the continuum contribution with bachelor track $h$ is denoted $\mathcal{P}^{q\bar{q},h}_k$. The likelihood term $\mathcal{L}^{B,h}_{j,k}$ corresponds to the $B$-background contribution $j$ of the $N_B$ categories.

The signal PDFs are decomposed into three parts with distinct distributions: signal events that are correctly reconstructed, misreconstructed signal events with right-sign $\rho$ charge, and misreconstructed signal events with wrong-sign $\rho$ charge. Their individual fractions are estimated by MC simulation. The $m_{ES}$, $\Delta E$, and NN output PDFs for signal and $B$ background are taken from the simulation except for the means of the signal Gaussian PDFs for $m_{ES}$ and $\Delta E$, which are free to vary in the fit. The continuum PDFs are described by six free parameters. The $\theta_C$ PDF is modeled as in Ref. [101]. The $\Delta t$-resolution function for signal and $B$-background events is a sum of three Gaussian distributions, with parameters determined from a fit to fully reconstructed $B^0$ decays [80, 81]. The continuum $\Delta t$ distribution is parameterized as the sum of three Gaussian distributions with common mean, two relative fractions, and three distinct widths that scale the $\Delta t$ event-by-event error, yielding six free parameters. For continuum, two charge asymmetries and the ten parameters $N^{q\bar{q},h}_k$ are free. A total of 34 parameters, including signal yields and the parameters from Eq. (3.56), are varied in the fit.

**Table 3-5.** *Summary of the systematic uncertainties.*

| Error source | $N^{\rho K}$ | $N^{\rho\pi}$ | $A_{CP}^{\rho K}$ | $A_{CP}^{\rho\pi}$ | $C_{\rho\pi}$ | $\Delta C_{\rho\pi}$ | $S_{\rho\pi}$ | $\Delta S_{\rho\pi}$ |
|---|---|---|---|---|---|---|---|---|
| | (events) | | (in units of $10^{-2}$) | | | | | |
| $\Delta m_d$ and $\tau$ | 0.1 | 0.1 | 0.0 | 0.0 | 0.4 | 0.4 | 0.2 | 0.1 |
| $\Delta t$ PDF | 1.2 | 1.9 | 0.4 | 0.2 | 1.4 | 0.8 | 1.5 | 1.2 |
| Signal model | 4.0 | 13.1 | 1.2 | 0.8 | 0.7 | 0.8 | 1.4 | 1.0 |
| Particle ID | 0.6 | 0.7 | 0.5 | 0.2 | 0.1 | 0.1 | 0.1 | 0.1 |
| Fit procedure | 8.0 | 15.7 | 0.4 | 0.2 | 0.4 | 0.4 | 0.4 | 0.3 |
| DCS decays | 0.0 | 0.3 | 0.0 | 0.1 | 2.2 | 2.2 | 0.8 | 0.7 |
| $B$ background | 16.0 | 14.2 | 7.9 | 2.8 | 3.0 | 3.5 | 2.1 | 1.8 |
| Total | 18.4 | 25.0 | 8.0 | 2.9 | 4.1 | 4.3 | 3.1 | 2.5 |

The contributions to the systematic error on the signal parameters are summarized in Table 3-5. The uncertainties associated with $\Delta m_d$ and $\tau$ are estimated by varying these parameters within the uncertainty on the world average [107]. The uncertainties due to the signal model are obtained from a control sample of fully reconstructed $B^0 \to D^-\rho^+$ decays. We perform fits on large MC samples with the measured proportions of $\rho\pi/\rho K$ signal, and continuum and $B$ backgrounds. Biases observed in these tests are due to imperfections in the PDF model; *e.g.,* unaccounted correlations between the discriminating variables of the signal and $B$ background PDFs. The biases are added in quadrature and assigned as a systematic uncertainty of the fit procedure. The systematic errors due to interference between the doubly-Cabibbo-suppressed (DCS) $\bar{b} \to \bar{u}c\bar{d}$ amplitude with the Cabibbo-favored $b \to c\bar{u}d$ amplitude for tag-side $B$ decays have been estimated from simulation by varying freely all relevant strong phases [108].

The main source of systematic uncertainty is the $B$-background model. The expected event yields from the background modes are varied according to the uncertainties in the measured or estimated branching fractions. Systematic errors due to possible nonresonant $B^0 \to \pi^+\pi^-\pi^0$ decays are derived from experimental limits [103]. Repeating the fit without using the $\rho$-candidate mass and helicity angle gives results that are compatible with those reported here. Since $B$-background modes may exhibit $CP$ violation, the corresponding parameters are varied within their physical ranges.

The maximum likelihood fit results in the event yields $N^{\rho\pi} = 428^{+34}_{-33}$ and $N^{\rho K} = 120^{+21}_{-20}$, where the errors are statistical. Correcting the yields by a small fit bias determined using the MC simulation ($3\%$ for $\rho\pi$ and $0\%$ for $\rho K$), we find for the branching fractions

$$\mathcal{B}(B^0 \to \rho^{\pm}\pi^{\mp}) = (22.6 \pm 1.8 \pm 2.2) \times 10^{-6}\,,$$
$$\mathcal{B}(B^0 \to \rho^- K^+) = (7.3\,^{+1.3}_{-1.2} \pm 1.3) \times 10^{-6}\,,$$

where the first errors are statistical and the second systematic. The systematic errors include an uncertainty of $7.7\%$ for efficiency corrections, dominated by the uncertainty in the $\pi^0$ reconstruction efficiency. Figure 3-14 shows distributions of $m_{ES}$ and $\Delta E$, enhanced in signal content by cuts on the signal-to-continuum likelihood ratios of the other discriminating variables. For the $CP$-violating parameters, we obtain

$$A_{CP}^{\rho\pi} = -0.18 \pm 0.08 \pm 0.03\,,\ A_{CP}^{\rho K} = 0.28 \pm 0.17 \pm 0.08\,,$$
$$C_{\rho\pi} = \quad 0.36 \pm 0.18 \pm 0.04\,,\quad S_{\rho\pi} = 0.19 \pm 0.24 \pm 0.03\,.$$

For the other parameters in the description of the $B^0(\overline{B}^0) \to \rho\pi$ decay-time dependence, we find

$$\Delta C_{\rho\pi} = 0.28^{+0.18}_{-0.19} \pm 0.04\,,\ \Delta S_{\rho\pi} = 0.15 \pm 0.25 \pm 0.03\,.$$

We find the linear correlation coefficients $c_{C,\Delta C} = 0.18$ and $c_{S,\Delta S} = 0.23$, while all other correlations are smaller. As a validation of our treatment of the time dependence we allow $\tau$ and $\Delta m_d$ to vary in the fit. We find $\tau =$ $(1.64 \pm 0.13)$ ps and $\Delta m_d = (0.52 \pm 0.12)\,\mathrm{ps}^{-1}$; the remaining free parameters are consistent with the nominal fit. The raw time-dependent asymmetry $A_{B^0/\overline{B}^0} = (N_{B^0} - N_{\overline{B}^0})/(N_{B^0} + N_{\overline{B}^0})$ in the tagging categories dominated by kaons and leptons is represented in Fig. 3-15.

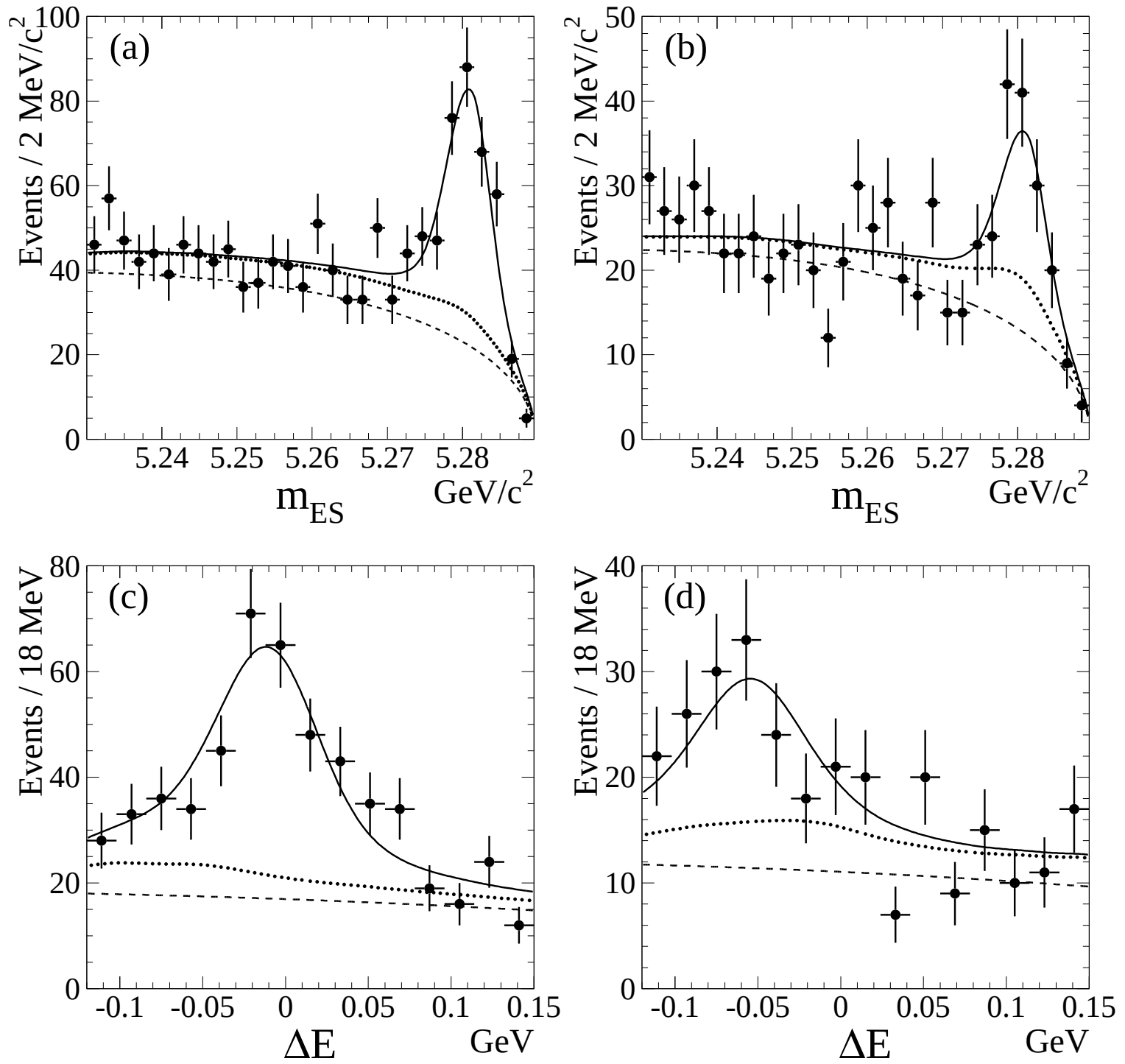

**Figure 3-14.** *Distributions of $m_{ES}$ and $\Delta E$ for samples enhanced in $\rho\pi$ signal (a,c) and $\rho K$ signal (b,d). The solid curve represents a projection of the maximum likelihood fit result. The dashed curve represents the contribution from continuum events, and the dotted line indicates the combined contributions from continuum events and $B$-related backgrounds.*

In summary, we have presented measurements of branching fractions and $CP$-violating asymmetries in $B^0 \to \rho^\pm\pi^\mp$ and $\rho^- K^+$ decays. We do not observe direct or mixing-induced $CP$ violation in the time-dependent asymmetry of $B^0 \to \rho^\pm\pi^\mp$ decays and there is no evidence for direct $CP$ violation in $B^0 \to \rho^- K^+$.

### Prospects for Super $B$ Factories

The precision of measured $CP$ parameters described in this note is statistics-dominated and thus will greatly benefit from much larger data samples of Super $B$ Factories At the same time, it is important to note that there is a number of issues which can not be resolved within quasi-two-body framework even a Super $B$ Factories One problem has to do with the translation of experimentally measured $A$,$C$,$\delta C$,$S$,$\delta S$ into constraints on CKM angle $\alpha$. Even after assuming that electroweak and annihilation diagrams are negligible, the remaining strong penguin pollution as well as unknown phases between contributing amplitudes tend to reduce the size of the exclusion region for $\alpha$. Qualitatively this can be called a problem of "multiple solutions". Another issue comes from the fact that the quasi-two-body approach doesn't take into account the effects of $\rho - vs - \rho$ interferences. We found a $-5\%$ average linear dependence of the fit bias on the generated parameters values, which is approximately the same for all $CP$ violation parameters. To get some idea on how much improvement is expected with more data, we quote some preliminary results from projections done by *CKMfitter* group [36]. In one specific example, we use measured $CP$ violation parameters, and assume zero penguin contribution (see Fig. 3-16). The unknown phase between two remaining tree amplitudes generates an eightfold ambiguity. In the second example, we drop "zero penguin" condition but employ information from $B^\pm \to \rho\pi$ decays, assume that $\mathcal{B}(B^0 \to \rho^0\pi^0)$ stays below experimental sensitivity, and use SU(2) flavor symmetries[3] (see Fig. 3-17). In general, it was found that unless the branching fraction $\mathcal{B}(B^0 \to \rho^0\pi^0)$ is small enough to be beyond the experimental sensitivity, very large statistics is needed to significantly constrain $\alpha$ from data alone, even with the help of the isospin analysis. It is effectively beyond the reach of present $B$ factories.

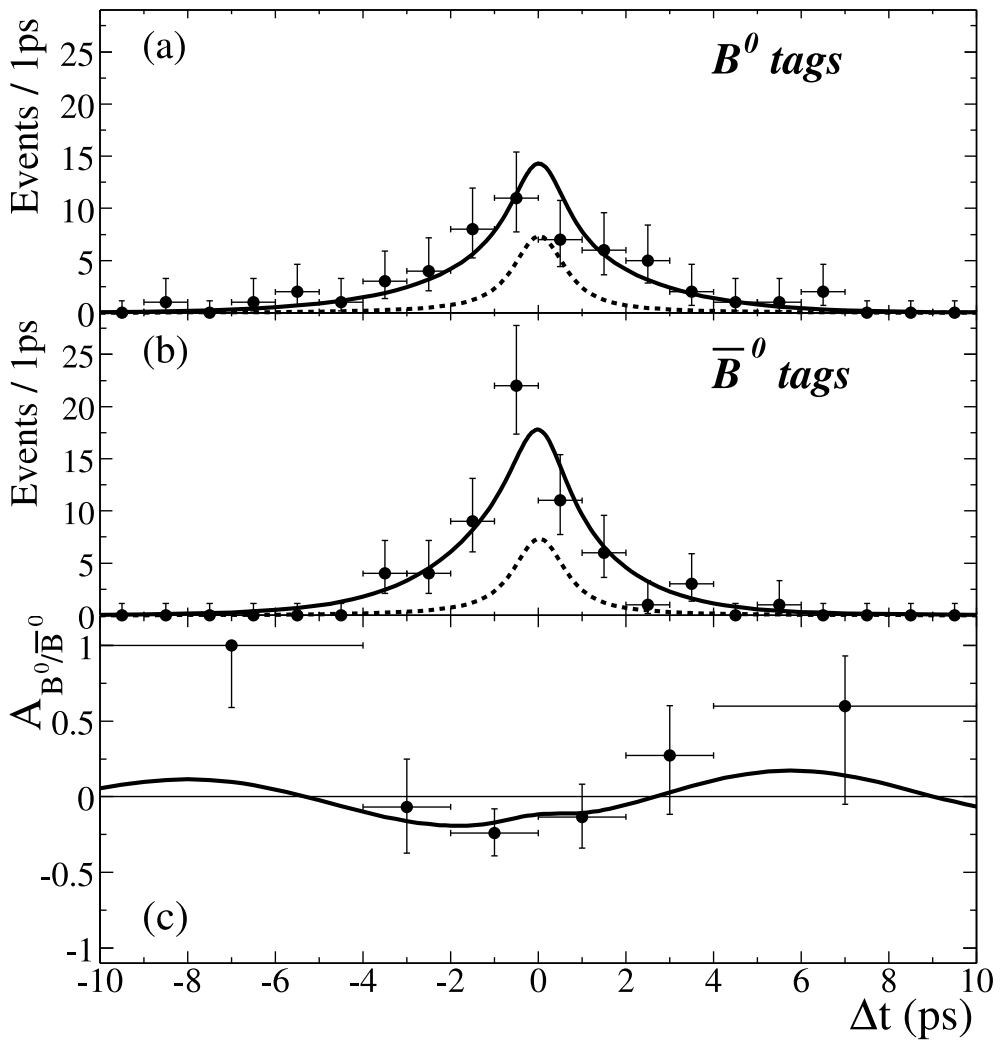

**Figure 3-15.** *Time distributions for events selected to enhance the $\rho\pi$ signal tagged as (a) $B^0_{\rm tag}$ and (b) $\overline{B}^0_{\rm tag}$, and (c) time-dependent asymmetry between $B^0_{\rm tag}$ and $\overline{B}^0_{\rm tag}$. The solid curve is a likelihood projection of the fit result. The dashed line is the total $B$- and continuum-background contribution.*

Some of the issues mentioned above could be better addressed by using full a Dalitz-plot analysis, which is beyond the scope of this paper.

[3] The details on SU(2) flavor decomposition of the neutral and charged $B \to \rho\pi$ amplitudes is given in Ref. [98].

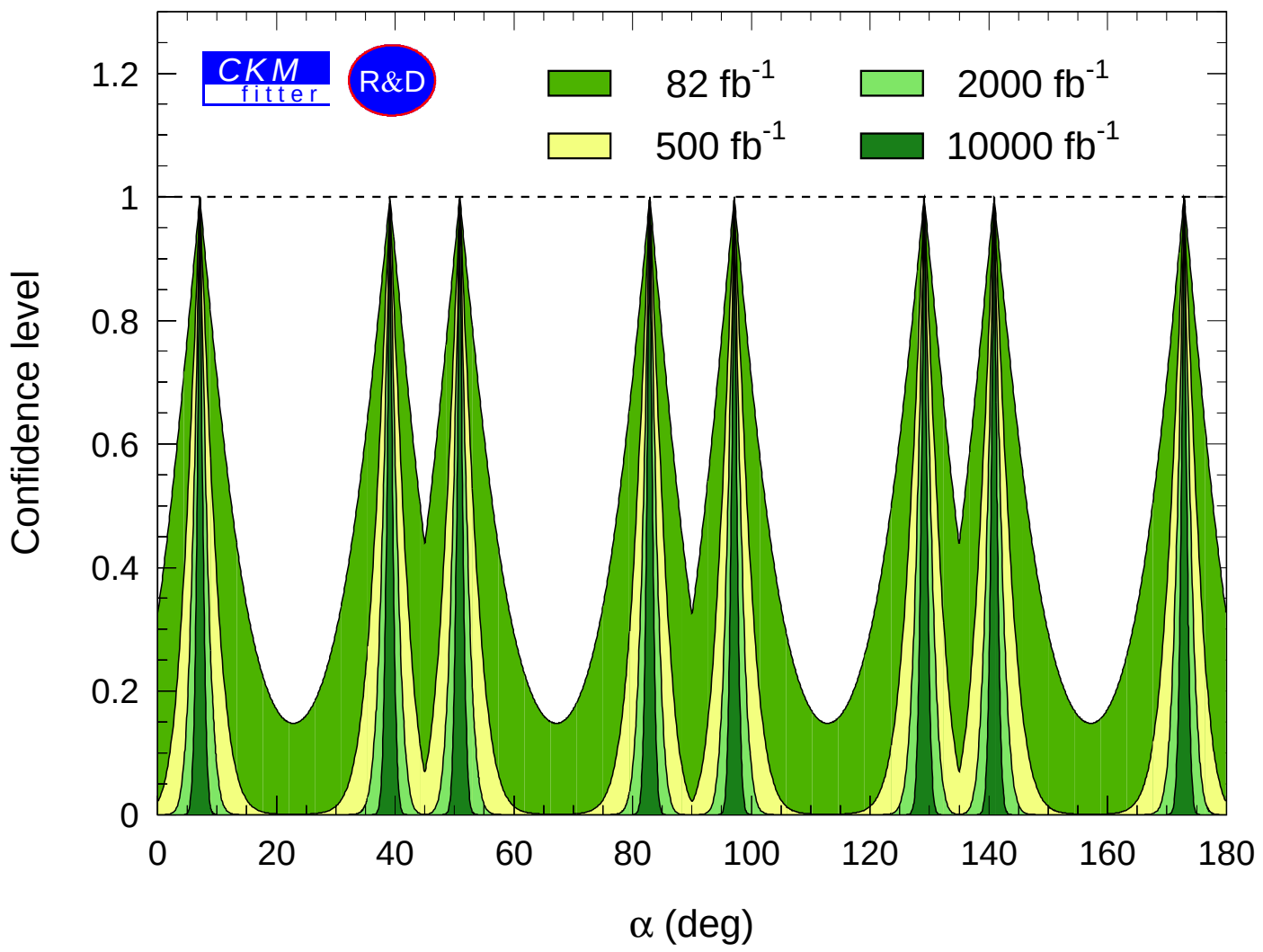

**Figure 3-16.** Prospective plots for the confidence level of $\alpha$, setting the penguin amplitudes to zero.

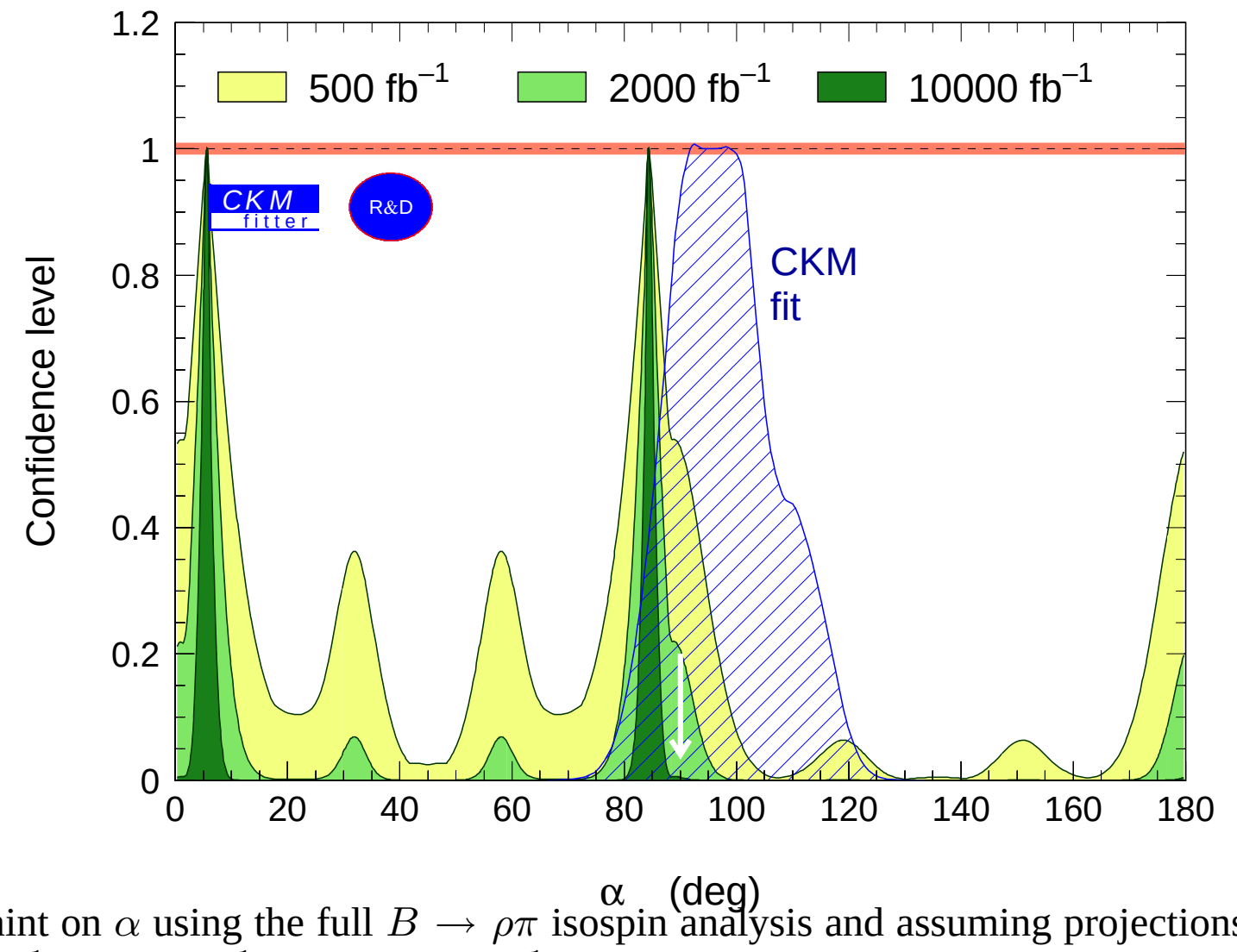

**Figure 3-17.** Constraint on $\alpha$ using the full $B \to \rho\pi$ isospin analysis and assuming projections into future integrated luminosities of 500 $\mathrm{fb}^{-1}$, 2000 $\mathrm{fb}^{-1}$ and 10000 $\mathrm{fb}^{-1}$. It is assumed in this scenario that the branching fractions of $B^0 \to \rho^0\pi^0$ is below the experimental sensitivity. The hatched area shows the constraint obtained from the CKM fit using the standard constraints (see [36]). The arrow indicates the true value of $\alpha$ used for simulations.

### 3.4.5 Experimental issues in $B \to VV$ decays

≻–A. Gritsan–≺

Charmless $B$ meson decays provide an opportunity to measure the angles of the unitary triangles constructed from the elements of the CKM quark-mixing matrix and to search for phenomena outside the Standard Model, including charged Higgs bosons and supersymmetric particles. The decays to two vector particles are of special interest because their angular distributions reflect both strong and weak interaction dynamics.

The decays $B \to \phi K^*$ and $\rho K^*$ are expected to proceed through $b \to s$ penguin transitions, with a smaller tree contribution to $B \to \rho K^*$ decays. Angular correlation measurements and asymmetries are particularly sensitive to phenomena outside the Standard Model potentially present in the penguin loops. These measurements provide wider set of observables for New Physics searches than the $B \to PP$ or $PV$ decays. The decays $B \to \rho\rho$ are expected to proceed through the tree-level $b \to u$ transition and through CKM-suppressed $b \to d$ penguin transitions. These are particularly promising modes for the CKM angle $\alpha$ studies and have the advantage of a larger decay rate and smaller uncertainty in penguin contributions compared to $B \to \pi\pi$.

The first evidence for the decays of $B$ mesons to pairs of charmless vector mesons was provided by the CLEO [109] and *BABAR* [110] experiments with the observation of $B \to \phi K^*$ decays. The CLEO experiment also set upper limits on the $B$ decay rates for several other vector-vector final states [111]. Both the *BABAR* [112, 113] and the Belle [114, 116] experiments recently improved the measurements of $B \to \phi K^*$ decays and reported observation of the $B \to \rho\rho$ and $\rho K^*$ decays, including the first results on polarization, and charge and "triple-product" asymmetries in the charmless vector-vector $B$ meson decays. We summarize the recent branching fraction measurements in Table 3-6. The results by *BABAR* and Belle are each based on the data sample of approximately 90 million $B\overline{B}$ pairs produced at $\Upsilon$(4S) resonance.

**Table 3-6.** *Summary of the recent branching fraction measurements (in units of $10^{-6}$) of the charmless vector-vector $B$ meson decays by the BABAR [112, 113], Belle [114, 116], and CLEO [109, 111] experiments.*

| Mode | *BABAR* | Belle | CLEO | PDG2002 [107] |
|---|---|---|---|---|
| $\phi K^{*+}$ | $12.7^{+2.2}_{-2.0} \pm 1.1$ | $6.7^{+2.1\,+0.7}_{-1.9\,-0.8}$ | $< 22.5$ | $10^{+5}_{-4}$ |
| $\phi K^{*0}$ | $11.2 \pm 1.3 \pm 0.8$ | $10.0^{+1.6\,+0.7}_{-1.5\,-0.8}$ | $11.5^{+4.5\,+1.8}_{-3.7\,-1.7}$ | $9.5^{+2.4}_{-2.0}$ |
| $\rho^0 K^{*+}$ | $10.6^{+3.0}_{-2.6} \pm 2.4$ | – | $< 74$ | $< 74$ |
| $\rho^0 K^{*0}$ | – | – | $< 34$ | $< 34$ |
| $\rho^+ K^{*-}$ | – | – | – | – |
| $\rho^+ K^{*0}$ | – | – | – | – |
| $\rho^+\rho^-$ | $25^{+7+5}_{-6-6}$ | – | – | $< 2200$ |
| $\rho^0\rho^+$ | $22.5^{+5.7}_{-5.4} \pm 5.8$ | $31.7 \pm 7.1^{+3.8}_{-6.7}$ | – | $< 1000$ |
| $\rho^0\rho^0$ | $< 2.1$ | – | $< 18$ | $< 18$ |

The experimental analysis of charmless vector-vector $B$ decays involves full reconstruction of the charged and neutral decay products including the intermediate states $\phi \to K^+K^-$, $K^{*0} \to K^+\pi^-$ and $K^0\pi^0$, $K^{*+} \to K^+\pi^0$ and $K^0\pi^+$, $\rho^0 \to \pi^+\pi^-$, $\rho^+ \to \pi^+\pi^0$, with $\pi^0 \to \gamma\gamma$ and $K^0 \to K^0_S \to \pi^+\pi^-$. The large number of channels with $\pi^0$ and the large fraction of misreconstructed events make these modes especially promising for the study in the clean $e^+e^-$ environment. The analysis with a maximum-likelihood fit technique allows extraction of the signal yields, asymmetries, and angular polarizations simultaneously. As evident from Table 3-6, many of these modes suffer from low statistics at present $B$ Factories, but promise precision measurements with the increase in statistics by two orders of magnitude expected at a Super $B$ Factory.

The asymmetries constructed from the number of $B$ decays with each flavor and with the triple product values are sensitive to $CP$ violation or to final state interactions. The triple product is defined as $(\mathbf{q_1} - \mathbf{q_2}) \cdot \mathbf{p_1} \times \mathbf{p_2}$, where $\mathbf{q_1}$ and $\mathbf{q_2}$ are the momenta of the two vector particles in the $B$ frame and $\mathbf{p_1}$ and $\mathbf{p_2}$ represent their polarization vectors. The triple product asymmetries provide complimentary measurement to direct $CP$ asymmetries and have the advantage of being maximal when strong phase difference is zero.

The most sensitive technique to extract the triple product asymmetries is the analysis of the full angular distributions which accounts for shapes of the observables. Figure 3-18 shows the expected sensitivity in the future measurement of asymmetry in $\mathrm{Im}(A_\perp A_0^*)/\Sigma|A_m|^2$ for $B \to \phi K^*$ decays, where $A_m$ are three contributing amplitudes. These measurements will reach precision level of ~1% with a few ab$^{-1}$ and will provide important constraints on physics beyond the Standard Model [117].

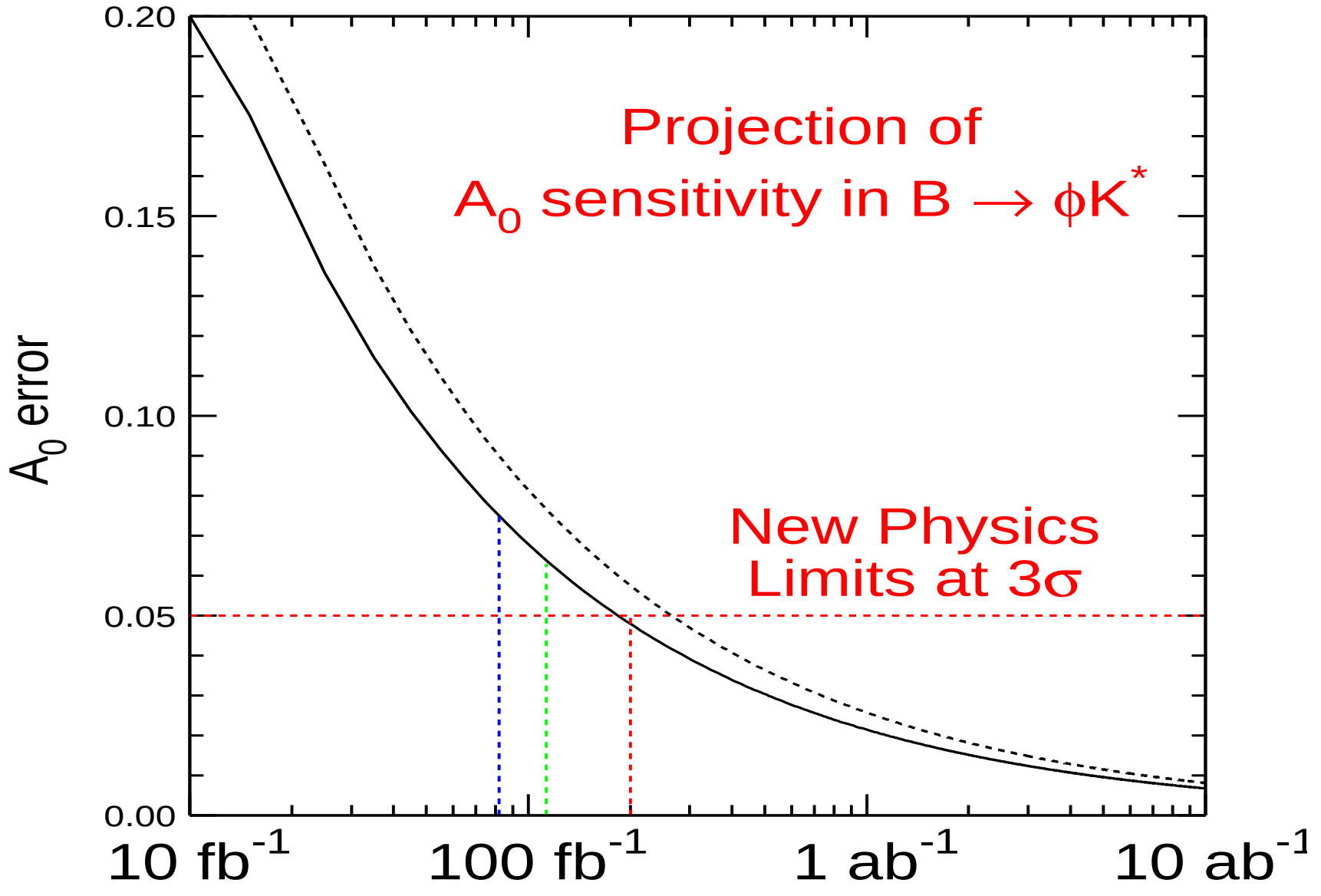

**Figure 3-18.** *Projection of expected sensitivity in the "triple product" asymmetry $A_0$ measurement of $\mathrm{Im}(A_\perp A_0^*)/\Sigma|A_m|^2$ in the $B \to \phi K^*$ decays. The dashed line shows expectation for $B^0 \to \phi K^{*0}$ channel alone based on observed ~100 events with 82 fb$^{-1}$. The solid line includes expectation from charged and neutral $B$ decay modes combined.*

The expected precision from the current $B$ Factory data sample on both $f_L = |A_0|^2/\Sigma|A_m|^2$ and $f_\perp = |A_\perp|^2/\Sigma|A_m|^2$ in $B \to \phi K^*$ decays is about 7%. This is also a very interesting measurement given the puzzle of relatively small value of $f_L \sim 0.6$ and potentially large fraction of $CP$-odd amplitude $f_\perp$. Both of these measurements will reach ~1% precision with the early data from a Super $B$ Factory. The $CP$-eigenstate mode $B^0 \to \phi K^{*0} \to \phi(K_S^0\pi^0)$ has reconstruction efficiency much lower than the dominant self-tagging mode $B^0 \to \phi K^{*0} \to \phi(K^+\pi^-)$, with about 6 reconstructed $CP$-eigenstate events compared to 100 self-tagging events. However, with a few ab$^{-1}$ a sample of $\sim 10^3$ events will allow measurements of time-dependent $CP$-violation effects with both $CP$-odd and $CP$-even components, which will provide deeper insight into the current puzzle of low $\sin 2\beta$ value measured with the $B^0 \to \phi K_S^0$ decays.

The decays $B \to \rho\rho$ have several advantages over $B \to \pi\pi$ in the CKM angle $\alpha$ measurements. The rates of the $B^0 \to \rho^+\rho^-$ and $B^+ \to \rho^0\rho^+$ decays appear to be larger than the corresponding rates of $B \to \pi\pi$ decays [107]. At the same time, the measurements of the $B \to \rho K^*$ and $\phi K^*$ and branching fractions do not show significant enhancement with respect to $B \to \pi K$ decays [107], both of which are expected to be dominated by $b \to s$ penguin diagrams. We can use flavor SU(3) to relate $b \to s$ and $b \to d$ penguins; the measured branching fractions indicate that the relative penguin contributions in the $B \to \rho\rho$ decays are smaller than in the $B \to \pi\pi$ case.

Another important advantage of the $B \to \rho\rho$ modes is that the $B^0 \to \rho^0\rho^0$ decay has four charged tracks in the final states, allowing vertexing and clean reconstruction as apposed to $B^0 \to \pi^0\pi^0$. The current experimental limit provided by *BABAR* [112, 113]:

$$\frac{\mathcal{B}(B^0 \to \rho^0\rho^0) \times f_L(B^0 \to \rho^0\rho^0)}{\mathcal{B}(B^+ \to \rho^0\rho^+) \times f_L(B^+ \to \rho^0\rho^+)} < 0.10 \quad (3.58)$$

corresponds to a $19°$ uncertainty (at 90% C.L.) on $\alpha$ due to penguin contributions in the time-dependent measurements with longitudinally-polarized $B^0 \to \rho^+\rho^-$ decays, assuming isospin relations analogous to those discussed in the context of $B \to \pi\pi$ and neglecting the nonresonant and $I = 1$ isospin contributions [118]. This limit is likely to improve if the true branching fraction of $B^0 \to \rho^0\rho^0$ is much lower than the current limit. However, if the branching fraction of $B^0 \to \rho^0\rho^0$ is around $10^{-6}$, a sample of a few thousand events will be reconstructed with the data from about one ab$^{-1}$. This will allow precision measurements of time-dependence with both $B^0 \to \rho^+\rho^-$ and $B^0 \to \rho^0\rho^0$ decays and this will resolve the penguin contribution uncertainty.

Among the experimental challenges in $B \to \rho\rho$ decays are potential nonresonant and $I = 1$ isospin contributions. The current experimental precision on nonresonant contribution is $\sim$10%, while $I = 1$ isospin contribution could be also tested experimentally, as proposed in [118]. The dominance of the $CP$-even longitudinal polarization makes the angular analysis in this decay relatively easy and a fit for only fraction of longitudinal polarization might be sufficient. While the $B^0 \to \rho^0\rho^0$ decay is relatively clean, the $B^0 \to \rho^+\rho^-$ decay with two neutral pions in the final state is a bit more challenging, but feasible, in the clean $e^+e^-$ environment, as demonstrated by *BABAR* [113].

In summary, charmless $B \to VV$ decays provide a much wider set of measurements than do $B \to PP$ or $PV$ decays. While the broader observable distributions make these analyses more challenging at the current $B$ Factories, they are the perfect match for a Super $B$ Factory. Among the main advantages of these modes are angular correlation measurements to search for and study phenomena outside the Standard Model, and potentially the most precise measurement of the CKM angle $\alpha$, using $B \to \rho\rho$ decays.

### 3.4.6 Experimental issues in $B \to \rho\rho$ decays

$\succ$– K. Graziani –$\prec$

The decay modes $B^0 \to \rho\rho$ provide another possibility for the measurement of $\alpha$. Though the different $CP$ components of the final state can be separated through an angular analysis, the first measurements performed by *BABAR* and Belle confirm the theoretical expectation of a dominating longitudinal (*i.e.*, $CP$-even) polarization:

| | | |
|---|---|---|
| $P(B^+ \to \rho^+\rho^0) =$ | (94.8 ±10.6±2.1) % (Belle,[114]) | $(97\,^{+3}_{-7} \pm 4)$ % (*BABAR*,[112]) |
| $P(B^0 \to \rho^+\rho^-) =$ | | $(98\,^{+2}_{-8} \pm 3)$ % (*BABAR*,[113]) |

For a given polarization, the Gronau–London isospin analysis of the modes $B^0 \to \rho^+\rho^-, \rho^0\rho^0$ and $B^+ \to \rho^+\rho^0$ is analogous to the $B \to \pi\pi$ case. The recent first observations of these modes show that the penguin pollution is smaller than in the $\pi\pi$ case, as already anticipated by the theory[115]:

| | | | |
|---|---|---|---|
| $\mathcal{B}(B^+ \to \rho^+\rho^0) \times 10^6 =$ | $31.7 \pm 7.1^{+3.8}_{-6.7}$ (Belle, [114]) | $22.5^{+5.7}_{-5.4} \pm 5.8$ | (*BABAR*,[112]) |
| $\mathcal{B}(B^0 \to \rho^+\rho^-) \times 10^6 =$ | | $25^{+7+5}_{-6-6}$ | (*BABAR*,[113]) |
| $\mathcal{B}(B^0 \to \rho^0\rho^0) \times 10^6 =$ | | <2.1 (90 % CL) | (*BABAR*,[112]) |

A Grossman–Quinn limit on $|\delta\alpha_{+-} = \alpha - \alpha^{eff}_{+-}|$ of about 17° can already be obtained from these data.

The $\rho^+\rho^-$ mode was first observed by *BABAR* from a data set of 81.9 fb$^{-1}$ (89M $B\overline{B}$). The observed sample of about 90 events allows to start measuring the time-dependent asymmetry, that is likely to soon provide the best measurement of $\alpha$ on the market.

The main uncertainties on the present analysis are introduced by:

- the yield and $CP$ asymmetry of the charmless $B$ background modes; this is expected to improve dramatically in the next years once the most relevant modes ($B^0 \to \rho\pi, \rho^+\rho^0, a_1\pi, a_1\rho, K^*\rho$) will be measured;
- the resolution function, the tagging efficiency and mistagging probability (for the $CP$ asymmetry); this is also expected to improve with the statistics;
- the fraction of misrecontructed signal events (about 40 % according to MC);
- the estimation of the efficiency (for the branching ratio);
- the vertex detector alignment (for the $CP$ asymmetry).

The three latter errors are hard to improve. The effect of beam background is presently negligible. At a Super $B$ Factory the huge rate of beam photons represents a potential danger. However, events with photons of energy $\lesssim$ 100 MeV are already suppressed by the cuts against the $q\bar{q}$ continuum background, so that we expect a limited loss of efficiency from this source. In Table 3-7 we extrapolate to an integrated luminosity of 10 ab$^{-1}$ the present statistical and "irreducible" systematic errors on the branching ratios, the polarization, and on the cosine and sine parameters of the $CP$ asymmetry for the longitudinal polarization component.

**Table 3-7.** *Extrapolated statistical and systematic errors on measurements of $B \to \rho\rho$ decays.*

| | $\mathcal{B}$ $(10^{-6})$ | Pol | $C_{long}$ | $S_{long}$ |
|---|---|---|---|---|
| Expected stat. error for $\mathcal{B} = 3 \times 10^{-5}$ | 0.4 | 0.003 | 0.03 | 0.04 |
| Syst. errors hard to improve | 4.5 | <0.01 | ∼0.01 | ∼0.02 |
| Naive estimate of total error | 5 | $\lesssim$0.01 | $\lesssim$0.04 | $\lesssim$0.05 |

The quoted total uncertainty on $S_{long}$ corresponds to an error smaller than 2° on $\alpha^{eff}$ (assuming the Standard Model-preferred value for $\alpha$).

The possibility of an interference between $\rho\rho$ and other resonant and non-resonant 4-pion final states constitutes an additional systematic uncertainty, that could disfavor this mode with respect to $B \to \pi\pi$. Various ways to evaluate its effect experimentally are presently under investigation, though with the present available statistics it is not possible to predict if the effect will be limiting at the Super $B$ Factory.

For the $\rho^+\rho^0$ mode, we expect more than $10^4$ events at 10 ab$^{-1}$; the branching ratio will be measured with negligible statistical error.

Finally, the $\rho^0\rho^0$ mode, having four charged tracks in the final state, is experimentally much easier than $\pi^0\pi^0$, and further, allows for a time dependent analysis. With loose selection cuts, we estimate an efficiency for this mode of about 35 %, namely fice times that for $\rho^+\rho^-$. The number of expected events is

$$N_{\rho^0\rho^0} \simeq 400 \times \frac{\mathcal{B}}{1\times 10^{-6}} \times \frac{\mathcal{L}}{1\times \mathrm{ab}^{-1}}$$

In the absence of a signal, a Grossmann-Quinn limit of $|\delta\alpha_{+-}| \lesssim 2°$ could be obtained with ∼2 ab$^{-1}$ of data. In the more likely scenario of a branching ratio of the order $10^{-6}$, a sizable sample is expected for 10 ab$^{-1}$, and the $CP$ asymmetry could be measured with an accuracy similar to the $\rho^+\rho^-$ case. This allows us to overconstrain the isospin triangle (Fig. 3-19) by measuring $\delta\alpha_{00} = \alpha - \alpha_{00}^{eff}$, resolving the ambiguities and setting a limit on the contribution of electroweak penguins.

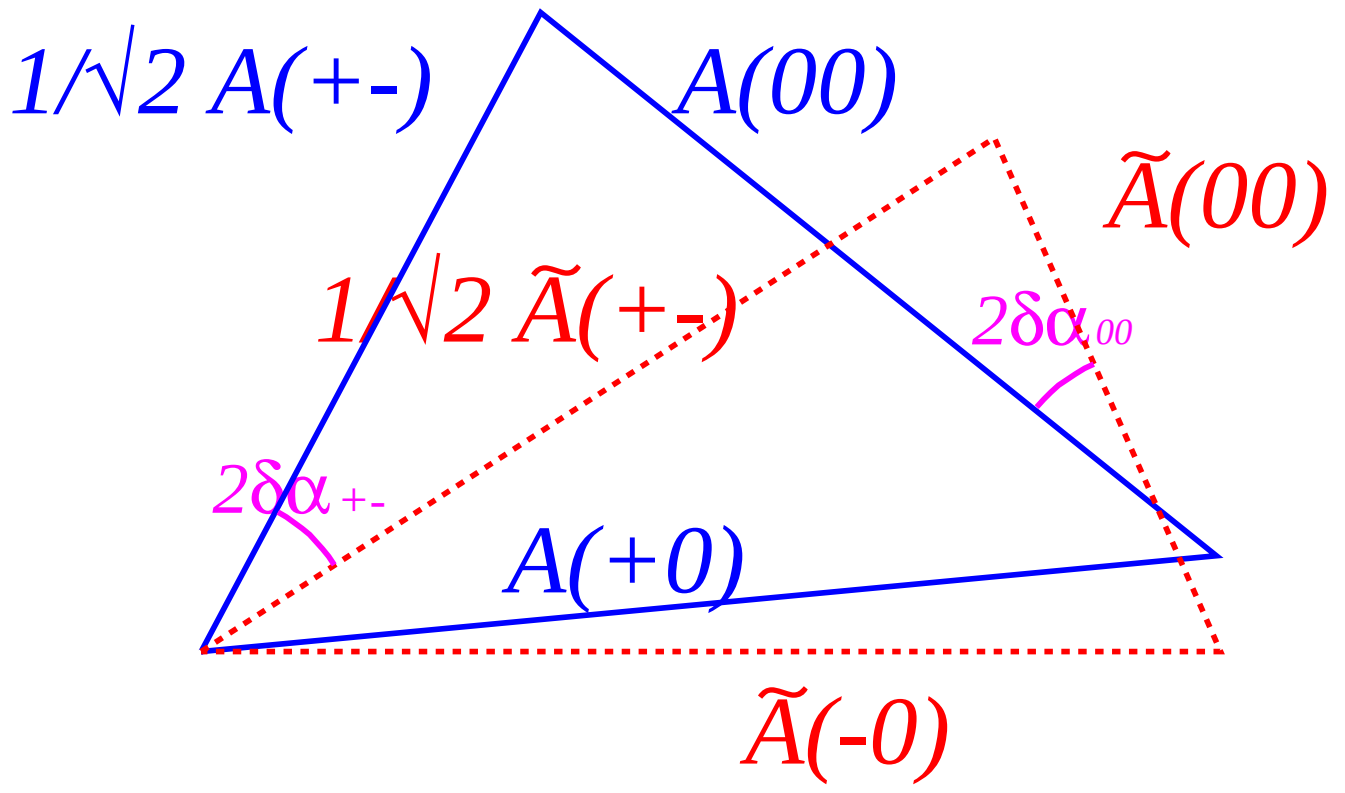

**Figure 3-19.** *Determination of the isospin triangle for the $\rho\rho$ modes. The length of all sides,* $\sin(2(\alpha + \delta\alpha_{+-}))$ *and* $\sin(2(\alpha + \delta\alpha_{00}))$ *can be measured at a Super $B$ Factory.*

In conclusion, we expect the measurement of $\alpha$ through the $\rho\rho$ mode to be very competitive at the Super $B$ Factory , provided that the effect of interference with other final states is under control. In this case, the isospin analysis will likely be limited by the SU(2) breaking effects discussed in the previous sections.

# 3.5 Measuring $\sin 2\beta$ With Penguins

## 3.5.1 Theory

>–Y. Grossman–<

**Introduction**

The time-dependent $CP$ asymmetries depend on two parameters, $S_f$ and $C_f$ ($f$ denotes here a final $CP$ eigenstate):

$$A_f(t) \equiv \frac{\Gamma(\overline{B}^0_{\rm phys}(t) \to f) - \Gamma(B^0_{\rm phys}(t) \to f)}{\Gamma(\overline{B}^0_{\rm phys}(t) \to f) + \Gamma(B^0_{\rm phys}(t) \to f)} = -C_f \cos(\Delta m_B\, t) + S_f \sin(\Delta m_B\, t)\,. \tag{3.59}$$

$CP$ violation in decay induces $C_f$, while $CP$ violation in the interference of decays with and without mixing induces $S_f$. (The contribution from $CP$ violation in mixing is at or below the percent level and can be safely neglected with the present experimental accuracy.) If the decay is dominated by a single weak phase, $C_f \approx 0$ and the value of $S_f$ can be cleanly interpreted in terms of $CP$-violating parameters of the Lagrangian. This is the case for decays which are dominated by the tree $b \to c\bar{c}s$ transition or by the gluonic penguin $b \to s\bar{s}s$ transition. If one neglects the subdominant amplitudes with a different weak phase, the $CP$ asymmetries in these two classes of decays are given by $S_f = -\eta_f \sin 2\beta$, where $\eta_f = +1(-1)$ for final $CP$-even (-odd) states and $\beta$ is one of the angles of the unitarity triangle. In particular, in this approximation, the $CP$ asymmetries in the both the tree $b \to c\bar{c}s$ transition and the gluonic penguin $b \to s\bar{s}s$ transition, are equal to each other, for example, $S_{\psi K} = S_{\phi K}$. A strong violation of such a relation would indicate New Physics in the decay amplitude [119]. Our aim here is to estimate or bound the Standard Model subleading effects lead to violation of the above statement.

**The problem with penguin decays**

The Standard Model amplitude for $b \to sq\overline{q}$ ($q = u, d, s$) penguin dominant decay modes can be written as follows:

$$A_f \equiv A(B^0 \to f) = V^*_{cb}V_{cs}\, a^c_f + V^*_{ub}V_{us}\, a^u_f\,. \tag{3.60}$$

The second term is CKM-suppressed compared to the first, since

$$\mathcal{I}m\left(\frac{V^*_{ub}V_{us}}{V^*_{cb}V_{cs}}\right) = \left|\frac{V^*_{ub}V_{us}}{V^*_{cb}V_{cs}}\right| \sin\gamma = \mathcal{O}(\lambda^2)\,, \tag{3.61}$$

where $\lambda = 0.22$ is the Wolfenstein parameter. It is convenient to define

$$\xi_f \equiv \frac{V^*_{ub}V_{us}\, a^u_f}{V^*_{cb}V_{cs}\, a^c_f}\,, \tag{3.62}$$

such that we expect $|\xi_f| \ll 1$. Then we rewrite the amplitude of Eq. (3.60) as

$$A_f = V^*_{cb}V_{cs}\, a^c_f\,(1 + \xi_f)\,. \tag{3.63}$$

A finite $\xi_f$ results in a deviation from the leading order result

$$-\eta_f S_f = \frac{\sin 2\beta - 2|\xi_f|\cos\delta_f \sin(2\beta+\gamma) - |\xi_f|^2 \sin(2\alpha)}{R}\,, \tag{3.64}$$

$$C_f = -\frac{2|\xi_f|\sin\delta_f \sin\gamma}{R}\,, \tag{3.65}$$

$$R \equiv 1 - 2|\xi_f|\cos\delta_f \cos\gamma + |\xi_f|^2\,, \tag{3.66}$$

where $\delta_f = \arg(a_f^u/a_f^c)$. It is useful to also present results valid to first order in $|\xi_f|$

$$-\eta_f S_f - \sin 2\beta \;=\; 2\cos 2\beta\, \sin\gamma\, \cos\delta_f\; |\xi_f|\,, \tag{3.67}$$

$$C_f \;=\; -2\sin\gamma\, \sin\delta_f\; |\xi_f|\,. \tag{3.68}$$

Our aim is to estimate or bound $\xi_f$.

One approach is to try to calculate $\xi_f$. For example, this was done using QCD factorization in [57]. They found

$$|\xi_{\eta' K_S^0}| \approx 0.06 - 0.09, \qquad \arg(\xi_{\eta' K_S^0}) \ll 1, \tag{3.69}$$

where the spread in the result is due to model dependence.

**SU(3) analysis**

In the following we discuss another way to estimate $\xi$ using SU(3) (or equivalently $U$ spin). [45, 120, 121, 122, 123, 124, 41]. The basic idea is to relate $b \to s$ to $b \to d$ penguin amplitudes. In the latter, the tree amplitude is enhanced, and thus there is larger associated sensitivity. Then, using SU(3), the tree amplitude in the $b \to d$ decay is related to that in $b \to s$ decay.

The crucial question, when thinking of the deviation of $-\eta_f S_f$ from $\sin 2\beta$, is the size of $a_f^u/a_f^c$. While $a_f^c$ is dominated by the contribution of $b \to s\bar{q}q$ gluonic penguin diagrams, $a_f^u$ gets contributions from both penguin diagrams and $b \to u\bar{u}s$ tree diagrams. For the penguin contributions, it is clear that $|a_f^u/a_f^c| \sim 1$. (The $a_f^c$ term comes from the charm penguin minus the top penguin, while the up penguin minus the top penguin contributes to $a_f^u$.) Thus, our main concern is the possibility that the tree contributions might yield $|a_f^u/a_f^c|$ significantly larger than one.

For final states with zero strangeness, $f'$, we write the amplitudes as

$$A_{f'} \equiv A(B^0 \to f') = V_{cb}^* V_{cd}\, b_{f'}^c + V_{ub}^* V_{ud}\, b_{f'}^u\,. \tag{3.70}$$

Neither term is CKM suppressed compared to the other. We use SU(3) flavor symmetry to relate the $a_f^{u,c}$ amplitudes to sums of $b_{f'}^{u,c}$.

Let us first provide a simple explanation of the method. Then we assume that the decays to final strange states, $f$, are dominated by the $a_f^c$ terms and that those to final states with zero strangeness, $f'$, are dominated by the $b_{f'}^u$ terms. Thus we can estimate $|a_f^c|$ and $|b_{f'}^u|$ from the measured branching ratios or upper bounds. The SU(3) relations then give upper bounds on certain sums of the $b_{f'}^c$ and $a_f^u$ amplitudes from the extracted values of $a_f^c$ and $b_{f'}^u$, respectively, leading to a bound on $|a_f^u/a_f^c|$, and consequently, on $|\xi_f|$.

Actually, the assumptions made in the previous paragraph can be avoided entirely [124, 41]. The SU(3) relations actually provide an upper bound on $|V_{cb}^* V_{cd}\, a_f^c + V_{ub}^* V_{ud}\, a_f^u|$, in terms of the measured branching ratios of some zero strangeness final states (or limits on them). Therefore, without any approximations, we can bound

$$\widehat{\xi}_f \equiv \left| \frac{V_{us}}{V_{ud}} \times \frac{V_{cb}^* V_{cd}\, a_f^c + V_{ub}^* V_{ud}\, a_f^u}{V_{cb}^* V_{cs}\, a_f^c + V_{ub}^* V_{us}\, a_f^u} \right| = \left| \frac{\xi_f + (V_{us}V_{cd})/(V_{ud}V_{cs})}{1+\xi_f} \right|. \tag{3.71}$$

If the bound on $\widehat{\xi}_f$ is less than unity, it gives a bound on $|\xi_f|$.

The SU(3) decomposition of $a_f^u$ and $b_{f'}^u$ is identical with that of $a_f^c$ and $b_{f'}^c$ although the values of the reduced matrix elements are independent for the $u$- and the $c$-terms. The contributions to $a_f^c$ and $b_{f'}^c$ come from penguin diagrams or the tree $b \to c\bar{c}q$ transition plus some form of rescattering (such as $D$-exchange) to replace the $c\bar{c}$ with lighter quark flavors. Aside from small electroweak penguin contributions, there is only an SU(3) triplet term in the Hamiltonian for these amplitudes. Neglecting electroweak penguins would result in additional SU(3) relations between the $a_f^c$ and $b_{f'}^c$ terms. We do not make such an approximation in our analysis, but it might be useful for other purposes.

In general we can write

$$a_f^u = \sum_{f'} x_{f'}\, b_{f'}^u, \tag{3.72}$$

where $x_{f'}$ are Clebsch-Gordan coefficients, which are calculated using group theory properties of SU(3). Then, using the relevant measured rates, we get

$$\widehat{\xi}_f \leq \lambda \sum_{f'} |x_{f'}| \sqrt{\frac{\mathcal{B}(f')}{\mathcal{B}(f)}}\,. \tag{3.73}$$

These bounds are exact in the SU(3) limit.

The SU(3) relations, together with the measurements or upper bounds on the rates for the non-strange channels, plus the measured rate for the channel of interest yield an upper bound on $\hat{\xi}_f$. In a few cases, where the SU(3) relation is such that $a_f^u$ is related to only one dominant $b_{f'}^u$ amplitude, we can not only bound $\hat{\xi}_f$, but actually estimate it.

**SU(3) relations**

The SU(3) relation has been worked out in detail for several modes [124, 41]. Using the tables in [124], relations for many other modes can be found. This will be important once the asymmetries in such modes are measured.

First, we present the bound for $\pi^0 K_S^0$. The SU(3) relation reads

$$a(\pi^0 K^0) = b(\pi^0\pi^0) + b(K^+K^-)/\sqrt{2}. \tag{3.74}$$

The available experimental data is

$$\begin{aligned} \mathcal{B}(B^0 \to \pi^0 K^0) &= (11.92 \pm 1.44) \times 10^{-6}, \\ \mathcal{B}(B^0 \to \pi^0 \pi^0) &= (1.89 \pm 0.46) \times 10^{-6}, \\ \mathcal{B}(B^0 \to K^+ K^-) &< 0.6 \times 10^{-6}\,, \end{aligned} \tag{3.75}$$

leading to

$$\hat{\xi}_{\pi K} < 0.13, \quad |S_{\pi K} - \sin 2\beta| < 0.19, \quad |C_{\pi K}| < 0.26\,. \tag{3.76}$$

We expect $\mathcal{B}(B^0 \to K^+K^-)$ to be much smaller than the present bound. If this is indeed the case we will be able to neglect it and we will get not only a bound on $\hat{\xi}_{\pi K}$, but an actual estimate. Note also that the bounds in Eq. 3.76 are correlated. This can be seen in Fig. 3-20, where the allowed values for $S_{\pi K}$ and $C_{\pi K}$ are plotted neglecting SU(3) breaking effects.

The bound for other modes we already studied are more complicated. Here we present only one additional bound:

$$\begin{aligned} \widehat{\xi}_{\eta' K_S^0} < \left|\frac{V_{us}}{V_{ud}}\right| \Bigg[ & 0.59\sqrt{\frac{\mathcal{B}(\eta'\pi^0)}{\mathcal{B}(\eta' K^0)}} + 0.33\sqrt{\frac{\mathcal{B}(\eta\pi^0)}{\mathcal{B}(\eta' K^0)}} + 0.14\sqrt{\frac{\mathcal{B}(\pi^0\pi^0)}{\mathcal{B}(\eta' K^0)}} \\ & + 0.53\sqrt{\frac{\mathcal{B}(\eta'\eta')}{\mathcal{B}(\eta' K^0)}} + 0.38\sqrt{\frac{\mathcal{B}(\eta\eta)}{\mathcal{B}(\eta' K^0)}} + 0.96\sqrt{\frac{\mathcal{B}(\eta\eta')}{\mathcal{B}(\eta' K^0)}}\,\Bigg]\,. \end{aligned} \tag{3.77}$$

At present, the experimental upper bounds on the relevant branching ratios gives

$$|\xi_{\eta' K_S^0}| < 0.36\,. \tag{3.78}$$

This number is much larger that what one expects to get once all the relevant rates are measured; then, this bound can improve to $\mathcal{O}(0.1)$. We do not present here other SU(3) relations that can be found in Ref. [124].

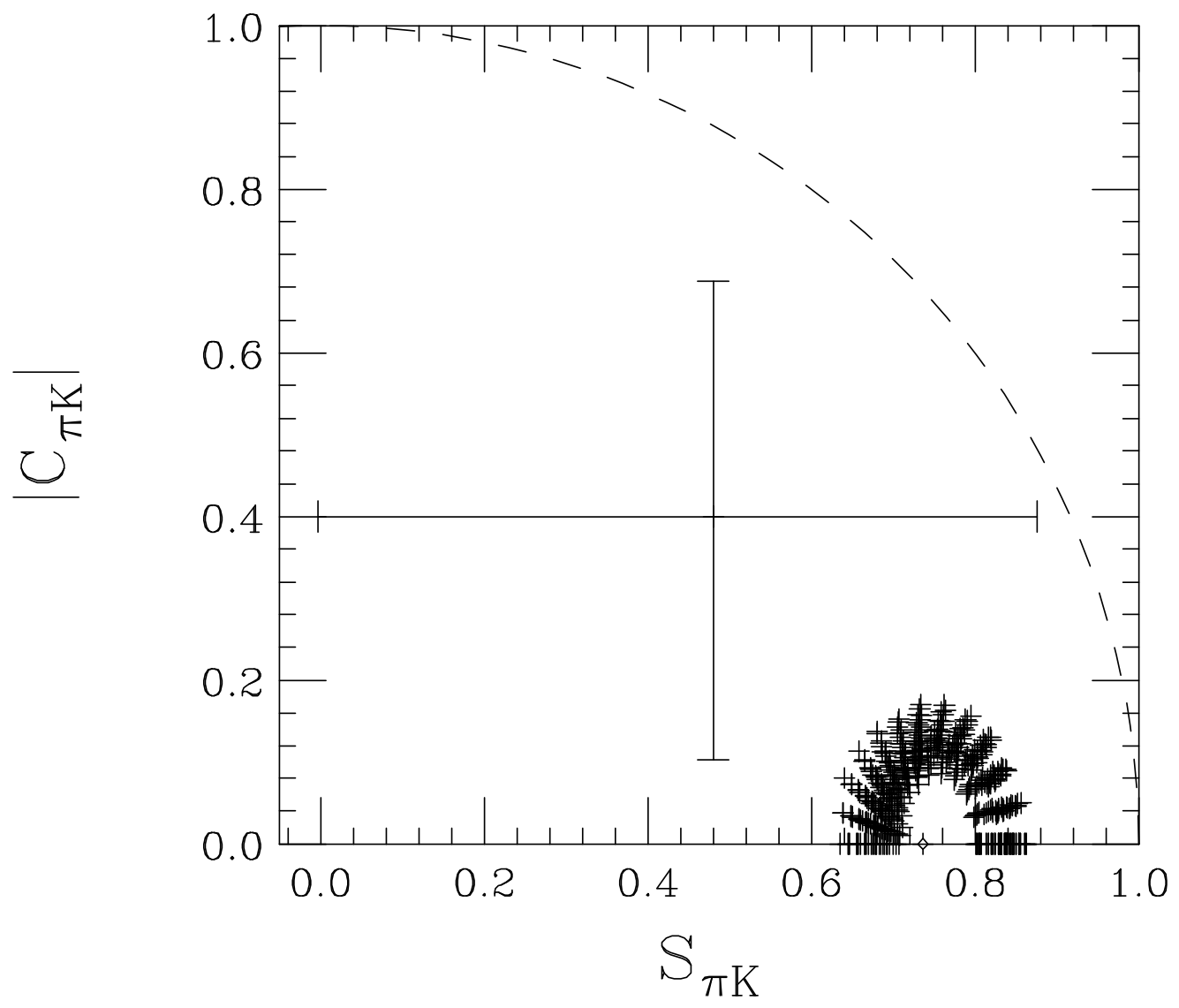

**Figure 3-20.** *Points in the $S_{\pi K}$–$|C_{\pi K}|$ plane allowed by the SU(3) relations. The small plotted point denotes the pure-penguin value $S_{\pi K} = \sin 2\beta$, $C_{\pi K} = 0$. The point with large error bars denotes the current experimental value. The dashed arc denotes the boundary of allowed values: $S^2_{\pi K} + C^2_{\pi K} \leq 1$.*

## Discussion

At present, only upper limits are available for many of the rates that enter into Eq. (3.77). Hence, this bound is probably a significant overestimate and will improve with further data. At the present state of the data, we do not consider it necessary to be concerned about SU(3) breaking corrections. Eventually, there may be sufficient data to fix all the amplitudes $a_f^{u,c}$. At that point, a much stronger bound can be expected, and allowance for SU(3) breaking corrections will have to be made.

We emphasize that using $U$ spin and SU(3) are equivalent; the size of the corrections due to breaking effects are expected to be the same. $U$ spin may be technically simpler, but, full SU(3) yields more relations: all $U$ spin relation are obtained with SU(3) but not the other way around. Note that while isospin is also a subgroup of SU(3), isospin is better, since the breaking effects are smaller.

In a Super $B$ Factory we can expect very precise measurement of the relevant asymmetries. At the same time we also expect measurement of all the needed rates, such that the SU(3) bounds will reach their limits. In some cases, where the SU(3) relations are dominated by one process, it will give an estimate of $\hat{\xi}_f$. In other cases, it may still be only a bound. If this bound is far larger then what is the true value of $\hat{\xi}_f$ we will have to relay on calculations to estimate it. Of course, in any event, we will be able to compare the different estimates of $|\xi_f|$.

Certainly, if deviations from $\sin 2\beta$ larger than these bounds are established, the case for New Physics would be convincing.

### 3.5.2 New physics: SUSY without R-parity

$\succ$– S. Oh –$\prec$

$CP$ violation in $B$ system has been confirmed in measurements of time-dependent $CP$ asymmetries in $B \to J/\psi K_S^0$ decay. The world average of the asymmetry in $B \to J/\psi K_S^0$ is given by [125]

$$\sin 2\beta_{J/\psi K_S^0} = 0.731 \pm 0.056 , \tag{3.79}$$

which is consistent with the Standard Model expectation. The measurements of time-dependent $CP$ asymmetries in $B^0 \to \phi K_S^0$ have been recently reported by Belle and *BABAR*, respectively [125]:

$$\sin 2\beta_{\phi K_S^0}^{\text{Belle}} = -0.96 \pm 0.50^{+0.09}_{-0.11}, \tag{3.80}$$

$$\sin 2\beta_{\phi K_S^0}^{BABAR} = 0.45 \pm 0.43 \pm 0.07 . \tag{3.81}$$

Because the Standard Model predicts that the value of $\sin 2\beta_{J/\psi K_S^0}$ should be the same as the value of $\sin 2\beta_{\phi K_S^0}$ to a good approximation, the Belle result or the average value of both Belle and *BABAR* results (possibly) indicates a deviation from the Standard Model prediction and may reveal New Physics effects on the internal quark-level penguin process $b \to s\bar{s}s$. However, it is interesting that the recent measurement of $\sin 2\beta_{\eta' K_S^0}$ in $B^0 \to \eta' K_S^0$ [125] agrees with $\sin 2\beta_{J/\psi K_S^0}$, even though the dominant quark-level process of the mode $B^0 \to \eta' K_S^0$ is also $b \to s\bar{s}s$. Therefore, any successful explanation invoking New Physics for understanding the Belle result (or the average of both Belle and *BABAR* results) on $\sin 2\beta_{\phi K_S^0}$ should simultaneously accommodate $\sin 2\beta_{\eta' K_S^0}$ as well as all the known data consistent with the Standard Model. In order to take $\sin 2\beta_{\eta' K_S^0}$ into account, one has to calculate the branching ratio for $B^0 \to \eta' K_S^0$. But it is well known that the branching ratios for $B^{+(0)} \to \eta' K^{+(0)}$ decays are found to be still larger than that expected within the Standard Model [126, 127, 128, 129]: *e.g.*, the experimental world average is $\mathcal{B}(B^+ \to \eta' K^+) = (77.6 \pm 4.6) \times 10^{-6}$.

In the framework of $R$ parity-violating (RPV) supersymmetry (SUSY), we focus on the recent measurement of $CP$ asymmetry in $B^0 \to \phi K_S^0$ and the large branching ratio for $B^\pm \to \eta' K^\pm$: both results appear to be (possibly) inconsistent with the Standard Model prediction. In RPV SUSY, the effects of RPV couplings on $B$ decays can appear at tree level and can be in some cases comparable to the Standard Model contribution.

The RPV part of the superpotential of the minimal supersymmetric Standard Model can contain terms of the form

$$\mathcal{W}_{\text{RPV}} = \kappa_i L_i H_2 + \lambda_{ijk} L_i L_j E_k^c + \lambda'_{ijk} L_i Q_j D_k^c + \lambda''_{ijk} U_i^c D_j^c D_k^c , \tag{3.82}$$

where $E_i$, $U_i$ and $D_i$ are, respectively, the $i$-th type of lepton, up quark and down quark singlet superfields, $L_i$ and $Q_i$ are the SU$(2)_L$ doublet lepton and quark superfields, and $H_2$ is the Higgs doublet with the appropriate hypercharge.

For our purpose, we will assume only $\lambda'-$type couplings to be present. Then, the effective Hamiltonian for charmless hadronic $B$ decay can be written as

$$H_{eff}^{\lambda'}(b \to \bar{d}_j d_k d_n) = d_{jkn}^R [\bar{d}_{n\alpha} \gamma_L^\mu d_{j\beta} \, \bar{d}_{k\beta} \gamma_{\mu R} b_\alpha] + d_{jkn}^L [\bar{d}_{n\alpha} \gamma_L^\mu b_\beta \, \bar{d}_{k\beta} \gamma_{\mu R} d_{j\alpha}] , \tag{3.83}$$

$$H_{eff}^{\lambda'}(b \to \bar{u}_j u_k d_n) = u_{jkn}^R [\bar{u}_{k\alpha} \gamma_L^\mu u_{j\beta} \, \bar{d}_{n\beta} \gamma_{\mu R} b_\alpha] ,$$

with $d_{jkn}^R = \sum_{i=1}^3 (\lambda'_{ijk} \lambda'^*_{in3})/(8 m_{\tilde{\nu}_{iL}}^2)$, $d_{jkn}^L = \sum_{i=1}^3 (\lambda'_{i3k} \lambda'^*_{inj})/(8 m_{\tilde{\nu}_{iL}}^2)$, $(j,k,n = 1,2)$ and $u_{jkn}^R = \sum_{i=1}^3 (\lambda'_{ijn} \lambda'^*_{ik3})/(8 m_{\tilde{e}_{iL}}^2)$, $(j,k = 1,\ n = 2)$, where $\alpha$ and $\beta$ are color indices and $\gamma_{R,L}^\mu \equiv \gamma^\mu (1 \pm \gamma_5)$. The leading order QCD correction to this operator is given by a scaling factor $f \simeq 2$ for $m_{\tilde{\nu}} = 200$ GeV.

The available data on low energy processes can be used to impose rather strict constraints on many of these couplings. The branching ratio of $B \to X_s \nu\nu$ can place bounds on $\lambda'_{322} \lambda'^*_{323}$ in certain limits. Using Ref. [130] and the experimental limit ($\mathcal{B} < 6.4 \times 10^{-4}$) on the branching ratio of $B \to X_s \nu\nu$ [131], we find that $|\lambda'| \leq 0.07$ (for $m_{\tilde{q}} = 200$ GeV). However, if we go to any realistic scenario, for example grand unified models (with $R$ parity violation), we find a natural hierarchy among the sneutrino and squark masses. The squark masses are much heavier than the sneutrino masses and the bound does not apply any more for $m_{\tilde{\nu}} = 200$ GeV.

**Table 3-8.** *$CP$ asymmetries in the decay modes $B^0 \to \phi K^0_S$ and $B^0 \to \eta' K^0_S$.*

| $\sin 2\tilde{\beta}$ | Case 1 | Case 2 |
|---|---|---|
| $\sin 2\tilde{\beta}_{\phi K^0_S}$ | 0 | $-0.82$ |
| $\sin 2\tilde{\beta}_{\eta' K^0_S}$ | 0.73 | 0.72 |

For the detailed discussion of our calculation, we refer to Ref. [129]. The Standard Model part and the RPV part of the decay amplitude for $B^- \to \phi K^-$ are, respectively, given by

$$\begin{aligned} \mathcal{A}^{\rm SM}_{\phi K} &= -\frac{G_F}{\sqrt{2}} V_{tb} V^*_{ts} (a_3 + a_4 + a_5 - \frac{1}{2} a_7 - \frac{1}{2} a_9 - \frac{1}{2} a_{10}) A_\phi \, , \\ \mathcal{A}^{\rm RPV}_{\phi K} &= \left( d^L_{222} + d^R_{222} \right) \, [\xi A_\phi] \, , \end{aligned} \tag{3.84}$$

where the effective coefficients $a_i$ are defined as $a_i = c^{eff}_i + \xi c^{eff}_{i+1}$ ($i =$ odd) and $a_i = c^{eff}_i + \xi c^{eff}_{i-1}$ ($i =$ even) with the effective WCs $c^{eff}_i$ at the scale $m_b$. The parameter $\xi \equiv 1/N_c$ ($N_c$ denotes the effective number of color) is treated as an adjustable parameter. The factorized form of the matrix element $A_\phi$ is defined as $A_\phi = \langle K | \bar{s} \gamma^\mu b | B \rangle \, \langle \phi | \bar{s} \gamma_\mu s | 0 \rangle$. This particular structure of $A_\phi$ is obtained from the operators $\overline{d}_{n\alpha} \gamma^\mu_{L(R)} d_{j\beta} \; \overline{d}_{k\beta} \gamma_{\mu R(L)} b_\alpha$, which are derived from the operators $\overline{s}_{L(R)} s_{R(L)} \overline{s}_{R(L)} b_{L(R)}$ by Fierz transformation.

The RPV SUSY part (relevant to the quark-level process $b \to s\bar{s}s$) of the decay amplitude of $B^- \to \eta' K^-$ is given by

$$\mathcal{A}^{\rm RPV}_{\eta' K} = \left( d^L_{222} - d^R_{222} \right) \, \left[ \frac{\overline{m}}{m_s} \left( A^s_{\eta'} - A^u_{\eta'} \right) \right] , \tag{3.85}$$

where $\overline{m} \equiv m^2_{\eta'} / (m_b - m_s)$ and $A^{u(s)}_{\eta'} = f^{u(s)}_{\eta'} F^{B \to K} (m^2_B - m^2_K)$. $F^{B \to \eta'}$ denotes the hadronic form factor for $B \to \eta'$ and $f^{u(s)}_{\eta'}$ is the decay constant of $\eta'$ meson.

Notice that $\mathcal{A}^{\rm RPV}_{\eta' K}$ is proportional to $(d^L_{222} - d^R_{222})$, while the RPV part of the decay amplitude of $B \to \phi K$ is proportional to $(d^L_{222} + d^R_{222})$. The opposite relative sign between $d^L_{222}$ and $d^R_{222}$ in the modes $B \to \eta' K$ and $B \to \phi K$ appears due to the different parity in the final state mesons $\eta'$ and $\phi$, and this different combination of $(d^L_{222} - d^R_{222})$ and $(d^L_{222} + d^R_{222})$ in these modes plays an important role to explain both the large negative value of $\sin(2\phi_1)_{\phi K^0_S}$ and the large branching ratios for $B \to \eta' K$ at the same time.

We consider the following two cases.

**Case 1:** The following input parameters are used: the $CP$ angle $\beta = 26^0$, $\gamma = 110^0$, and the $s$ quark mass $m_s$ (at $m_b$ scale) $= 85$ MeV. The other inputs are given in Ref. [129]. We set $d^L_{222} = k e^{-i\theta'}$ and $d^R_{222} = -k e^{i\theta'}$, where $k = |d^L_{222}| = |d^R_{222}|$ and $\theta'$ is a new weak phase defined by $\lambda'_{332(322)} \lambda'^*_{322(323)} = |\lambda'_{332(322)} \lambda'^*_{322(323)}| e^{i\theta'}$. In this choice of $d^L_{222}$ and $d^R_{222}$, $\mathcal{A}^{\rm RPV}_{\phi K}$ is purely imaginary and introduces a new weak phase to the decay amplitude for $B \to \phi K$, while $\mathcal{A}^{\rm RPV}_{\eta' K}$ introduces no new phase and gives a constructive contribution to the Standard Model part of the amplitude for $B \to \eta' K$.

For $|\lambda'_{322}| = |\lambda'_{332}| = |\lambda'_{323}| = 0.055$, $\tan\theta' = 0.52$, and $m_{\rm susy} = 200$ GeV, we find that $\sin 2\tilde{\beta}_{\phi K^0_S} = 0$ and $\sin 2\tilde{\beta}_{\eta' K^0_S} = 0.73$ for $\xi = 0.45$, as shown in Table 3-8 ($\tilde{\beta}$ denotes the effective $CP$ angle). This result on $\sin(2\tilde{\beta})_{\phi K^0_S}$ agrees with the average of the Belle and *BABAR* data, and the result on $\sin 2\tilde{\beta}_{\eta' K^0_S}$ also agrees with the data. In Table 3-9, we estimate the branchings ratios and $CP$ (rate) asymmetries $A_{CP}$ for $B \to \eta^{(\prime)} K^{(*)}$ and $B^+ \to \phi K^+$ modes. The estimated branching ratios are well within the experimental limits.

**Table 3-9.** *The branching ratios ($\mathcal{B}$) and CP rate asymmetries ($A_{CP}$) for $B \to \eta^{(\prime)} K^{(*)}$ and $B \to \phi K$.*

| mode | Case 1 $\mathcal{B} \times 10^6$ | $A_{CP}$ | Case 2 $\mathcal{B} \times 10^6$ | $A_{CP}$ |
|---|---|---|---|---|
| $B^+ \to \eta' K^+$ | 69.3 | $-0.01$ | 76.1 | $-0.01$ |
| $B^+ \to \eta K^{*+}$ | 27.9 | $-0.04$ | 35.2 | $-0.03$ |
| $B^0 \to \eta' K^0$ | 107.4 | 0.00 | 98.9 | 0.00 |
| $B^0 \to \eta K^{*0}$ | 20.5 | 0.71 | 11.7 | 0.15 |
| $B^+ \to \phi K^+$ | 8.99 | $-0.21$ | 8.52 | $-0.25$ |

Even though the particular form of $d^L_{222}$ and $d^R_{222}$ has been assumed above, it can be shown that one can still obtain good fits by using their general form [129]. Because $d^L_{222}$ and $d^R_{222}$ are relevant to the process $b \to s\bar{s}s$ only, the other observed $B \to PP$ and $B \to VP$ decay modes without $\eta^{(\prime)}$ or $\phi$ in the final state, such as $B \to \pi\pi$, $\pi K$, $\rho\pi$, $\omega\pi$, and so on, are not affected in this scenario. The estimated branching ratios for those modes by using the above input values are consistent with the experimental data for $\xi = 0.45$ [129].

**Case 2:** Now we try to generate a large negative value of $\sin 2\tilde{\beta}_{\phi K^0_S}$ which is consistent with the Belle result. In this case, the smaller values of $\gamma$ and $m_s$ are used: $\gamma = 80^0$ and $m_s$ (at $m_b$ scale) $= 75$ MeV.

For $|\lambda'_{322}| = |\lambda'_{332}| = |\lambda'_{323}| = 0.069$ and $\tan\theta' = 2.8$, a large value of $\sin 2\tilde{\beta}$ with the negative sign is possible for $B^0 \to \phi K^0_S$: $\sin 2\tilde{\beta}_{\phi K^0_S} = -0.82$ for $\xi = 0.25$, as shown in Table 3-8. The other predicted values presented in Table 3-8 and 3-9 agree with the experimental results. The BRs of other $B \to PP,\ VP$ modes also satisfy the experimental data for $\xi = 0.25$ [129].

To summarize, we have shown that in $R$ parity-violating SUSY, it is possible to consistently understand both the recently measured $CP$ asymmetry $\sin 2\beta$ in $B^0 \to \phi K^0_S$ decay and the large branching ratio of $B^{+(0)} \to \eta' K^{+(0)}$ decay, which appear to be (possibly) inconsistent with the Standard Model prediction. We have searched for possible parameter space and found that all the observed data for $B \to PP$ and $B \to VP$ decays can be accommodated for certain values of RPV couplings. For future experiment, more precise measurements of direct $CP$ asymmetries in $b \to s\bar{q}q$ ($q = s, u, d$) penguin processes, such as $B \to \phi K$ and $B \to \eta^{(\prime)} K$ decays, are expected to be very interesting to test the Standard Model and different New Physics predictions.

### 3.5.3 Measurement of the $CP$ asymmetry in $B \to K^+K^-K^0$ decays

≻–D. Dujmic–≺

**Introduction**

Studies of the $CP$ asymmetry in decays of neutral $B$ into three kaon final state have attracted considerable interest in recent months, as the current experiments *BABAR*, Belle [132] create enough $B\overline{B}$ pairs to allow time-dependent measurement of $CP$ asymmetry. In the Standard Model, neglecting CKM-suppressed contributions, such decays are $b \to s$ transitions that proceed through gluonic penguin decay diagrams, with possible contributions from electroweak penguins, and $b \to u(c)$ transitions followed by rescattering. A weak phase enters the final amplitudes through $B^0\overline{B}^0$ mixing, resulting in the same expectation for the $CP$-asymmetry parameters as in netural $B$ decays into the $J/\psi K^0_S$ final state. If the New Physics adds couplings with $CP$-violating phases, the measured $CP$ parameters in $s$ penguin decays can be different from those observed in charmonium decays, where New Physics amplitudes are shadowed by strong Standard Model tree amplitudes [119].

The $CP$ asymmetry is measured using the time difference, $\Delta t$ between decays of $B^0$ and $\overline{B}^0$ mesons

$$a_{CP}(\Delta t) \;=\; S\sin(\Delta m_d\Delta t) \;-\; C\cos(\Delta m_d\Delta t) \tag{3.86}$$

where $a_{CP}(\Delta t)$ is the ratio of the difference over sum of $B^0, \overline{B}^0$ decay rates. An overview of available measurements of sine and cosine terms is given in Table 3-10. Three-body $KKK^0_S$ decays exclude $\phi K^0_S$ events and assume that the rest of events are all $CP$-even.

**Table 3-10.** *Overview of current $CP$ asymmetry measurements [134].*

| Mode | | $-\eta_{CP}S$ | $C$ |
|---|---|---|---|
| $KKK^0_S$ | *BABAR* | $0.56 \pm 0.25$ | $-0.10 \pm 0.19$ |
| | Belle | $0.51 \pm 0.26$ | $0.17 \pm 0.16$ |
| $\phi K^0_S$ | *BABAR* | $0.45 \pm 0.43$ | $-0.38 \pm 0.37$ |
| | Belle | $-0.96 \pm 0.50$ | $0.15 \pm 0.29$ |
| $\phi K^0_L$ (*BABAR*) | *BABAR* | $1.16 \pm 0.67$ | $1.99 \pm 0.82$ |
| Charmonium average | | $0.736 \pm 0.049$ | $0.052 \pm 0.037$ |

**Results**

In this section, we give estimates for decays that have $K^+K^-K^0_S(K^0_L)$ in the final state. Other decays into pseudoscalars ($\eta' K^0_S$, $K^0_S\pi^0$...) as well as VV modes ($\phi K^*$, $\eta' K^*$...) can also contribute to the study of $CP$ asymmetry in $s$ penguin decays; these are covered in other reports.

In evaluating future errors, we have unknown performance of the time difference measurement between $B^0$ and $\overline{B}^0$ decays, and the tagging quality of the recoil $B$ meson, which are two main factors contributing to the errors on $CP$ asymmetry parameters. We study the influence of the vertex resolution by smearing the present resolution at *BABAR* (1.1 ps) and observing the change in the error of the sine parameter, $S$. We empirically find that $\Delta\sigma(S)/S \approx 0.34\Delta\sigma(\Delta t)$. In our studies we choose the time resolution for the future detector of 1.4 ps which increases relative error on $S$ by 10%.

The tagging quality, $Q$ is defined as product of the tagging efficiency, $\epsilon$ and dilution squared, $D^2$ summed over tagging categories, $Q = \sum \epsilon D^2$, and the error on $CP$ parameters is proportional to $1/\sqrt{Q}$. The current tagging quality at *BABAR* and Belle is around 28% and we pick 20% for our studies.

Predicted statistical errors are shown in Fig. 3-21. Expected number of signal events with 10 ab$^{-1}$ of data are taken as 22700 ($KKK_S^0$), 10000 ($KKK_L^0$), 6400 ($\phi K_S^0$), 3300 ($\phi K_L^0$) and 1500 ($\phi K_S^0(\pi^0\pi^0)$).

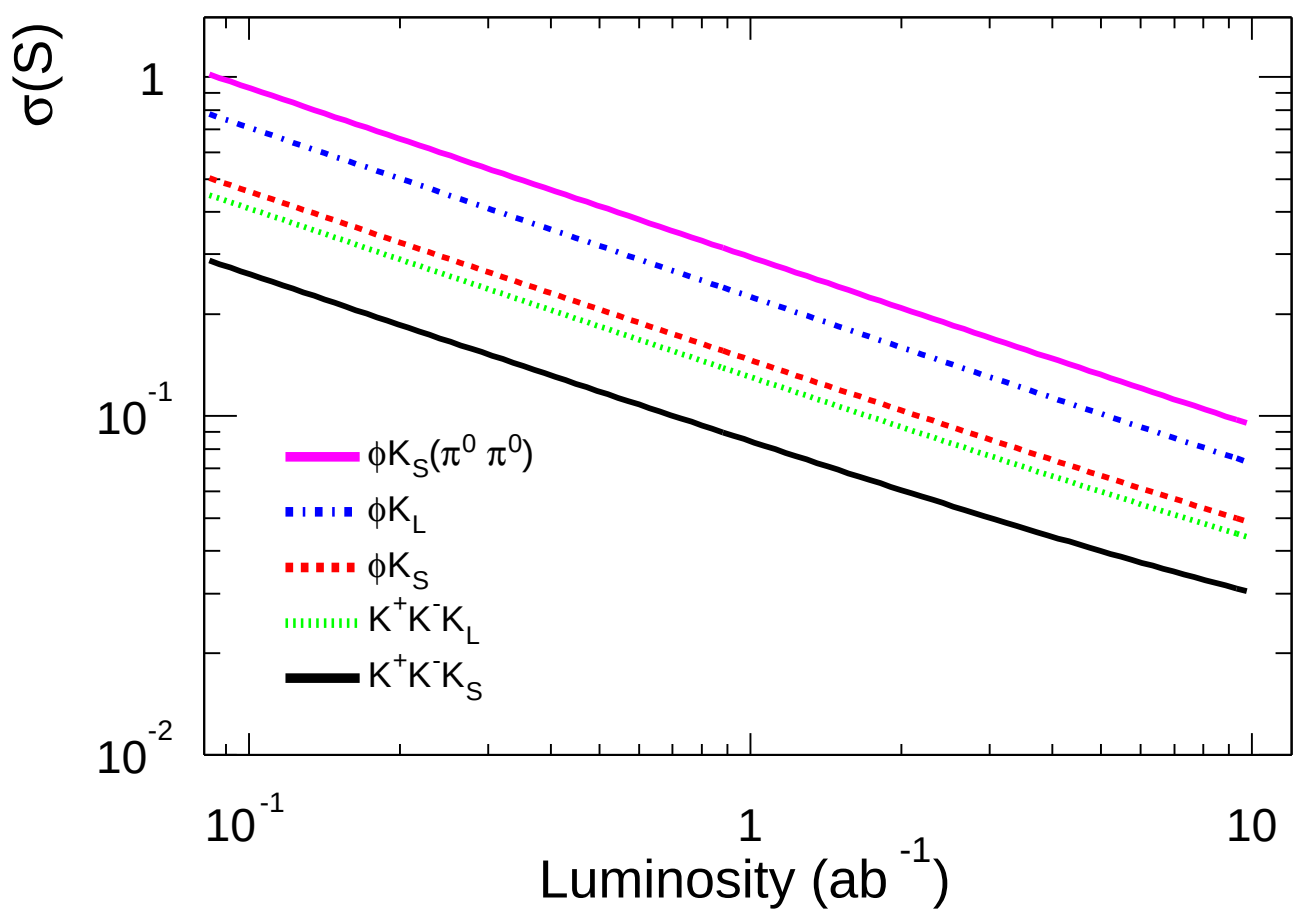

**Figure 3-21.** *Errors on $|S| \approx \sin 2\beta$ for different $KKK^0$ final states.*

Experimental systematic errors on $CP$ parameters coming from $\Delta t$ resolution and tagging are estimated to be less than 3%. Double CKM suppressed decays in the tagging side give an error of 0.013 in $S$ and 0.027 in $C$ [108].

We can estimate the fraction of $b \to u$ tree decays that enter with CKM phase $\gamma$ using SU(3) flavor symmetry and rates of modes that proceede predominantly through $b \to u$ transition [135]. Such an approach requires measurement of suppressed branching fractions that are possible only at a Super $B$ Factory. That is, an isospin decomposition of $\phi K_S^0$ decays gives an upper limit on $\sigma(S)$ due to $b \to u$ decays of 6% at 10 ab$^{-1}$ (Fig. 3-22). Currently estimates of the $b \to u$ contribution are based on less strict arguments [121].

The $CP$ content in $KKK_S^0$ decays is assumed to be purely $CP$-even, after excluding $\phi K_S^0$ events. We estimate the error on $S$ due to uncertainty in $CP$ content by using isospin symmetry and comparing the decay rate with $KK_S^0K_S^0$ decays. Assuming 10 ab$^{-1}$ of data we get an error on $S$ of 2%.

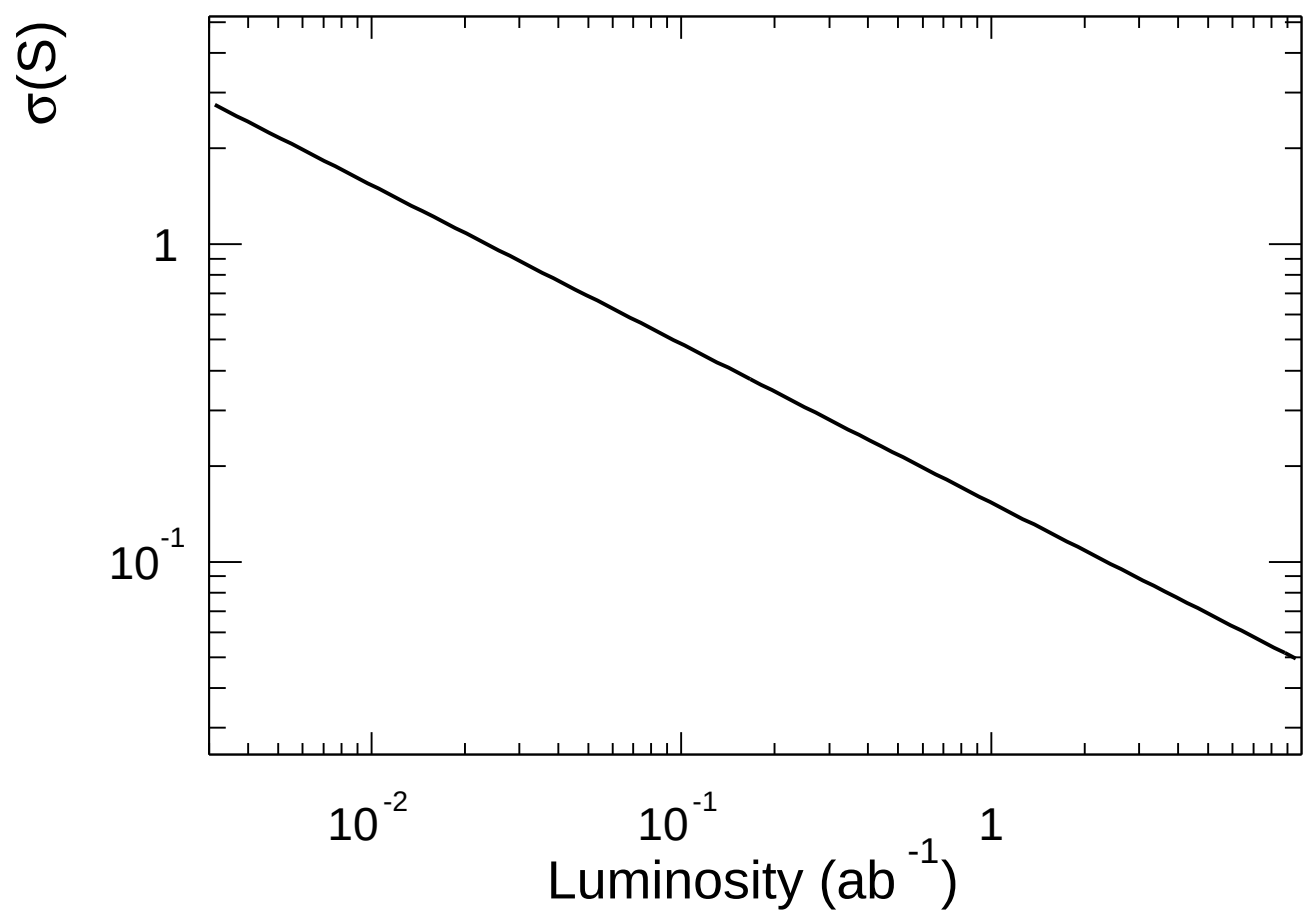

**Figure 3-22.** *Error on $S$ due to $b \to u$ tree amplitude in $\phi K_S^0$ decays.*

**Conclusion**

The time-dependent $CP$ asymmetry measurement in $B^0 \to KKK^0$ decays can reach a statistical precision of approximately 4% in three statistically dominant decay modes, $KKK^0_S$, $KKK^0_L$ and $\phi K^0_S$ within the first year of running of a Super $B$ Factory. At the same time, we can obtain a better understanding of these decay processes by measuring the contribution of CKM tree diagrams, and the $CP$ content in three-body decays.

### 3.5.4 $B \to D^{(*)}\overline{D}^{(*)}$

>–J. Albert–<

In the Standard Model, the time-dependent $CP$-violating asymmetries in $B \to D^{(*)}\overline{D}^{(*)}$ decays are related to the angle $\beta \equiv \arg[-V_{\rm cd}V^*_{\rm cb}/V_{\rm td}V^*_{\rm tb}]$. A Super $B$ Factory is needed in order to turn the current measurements of $CP$ asymmetries in $B \to D^{(*)}\overline{D}^{(*)}$ from *BABAR*[4] into precision measurements to make sensitive tests of the Standard Model, and to look for evidence of supersymmetry and other New Physics.

Decays involving $b \to c\bar{c}s$ transitions, such as $B^0 \to J/\psi K^0_S$, can be used to measure $\sin 2\beta$. The Standard Model also predicts that the time-dependent $CP$-violating asymmetries in $b \to c\bar{c}d$ decays, such as $B^0 \to D^{(*)}\overline{D^{(*)}}$ (see Fig. 3-23), can also measure $\sin 2\beta$. An independent measurement of $\sin 2\beta$ in these modes therefore provides a test of $CP$ violation in the Standard Model.

This is especially imperative, because very reasonable choices of SUSY parameters ($\tilde{b}$ and gluino masses in the range 100-300 GeV) can produce measurable differences in the values of $\sin 2\beta$ obtained from $b \to c\bar{c}s$ and $b \to c\bar{c}d$ [119]. In addition, a measurement of the $CP$ angle $\gamma$ can be obtained from combining information from $CP$ asymmetries in $B \to D^{(*)}\overline{D}^{(*)}$ with branching fraction information from $B \to D^{(*)}\overline{D_s}^{(*)}$ [136].

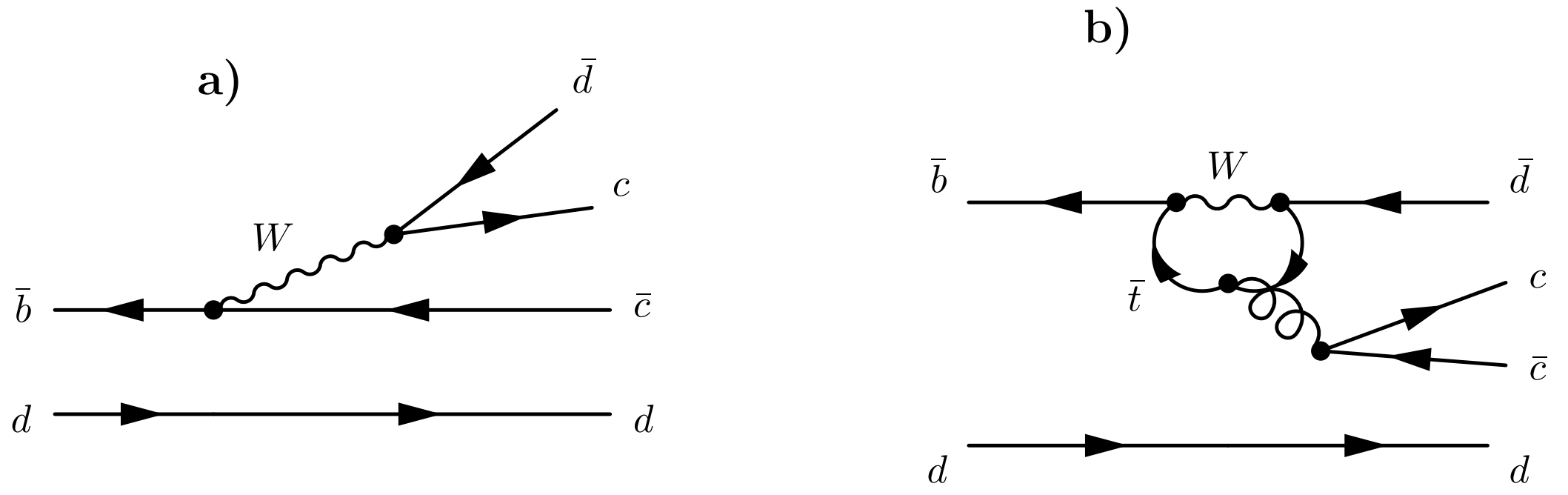

**Figure 3-23.** *The leading-order Feynman graphs for $B \to D^{(*)}\overline{D}^{(*)}$ decay: a) tree diagram and b) penguin diagram.*

This year, using a sample of $87.9 \pm 1.0$ million $B\overline{B}$ decays, the *BABAR* experiment reconstructed a signal yield of 156 $B \to D^{*+}D^{*-}$ events and 113 $B \to D^{*\pm}D^{\mp}$ events [137, 138] (see Figures 3-24 and 3-25). The time-dependent $CP$ asymmetries were measured, as well as the branching fractions, and also the time-integrated direct $CP$ asymmetry between the rates to $D^{*+}D^-$ and $D^{*-}D^+$ (which is physically independent of the time-dependent $CP$ asymmetries).

[4]presumably also soon from Belle

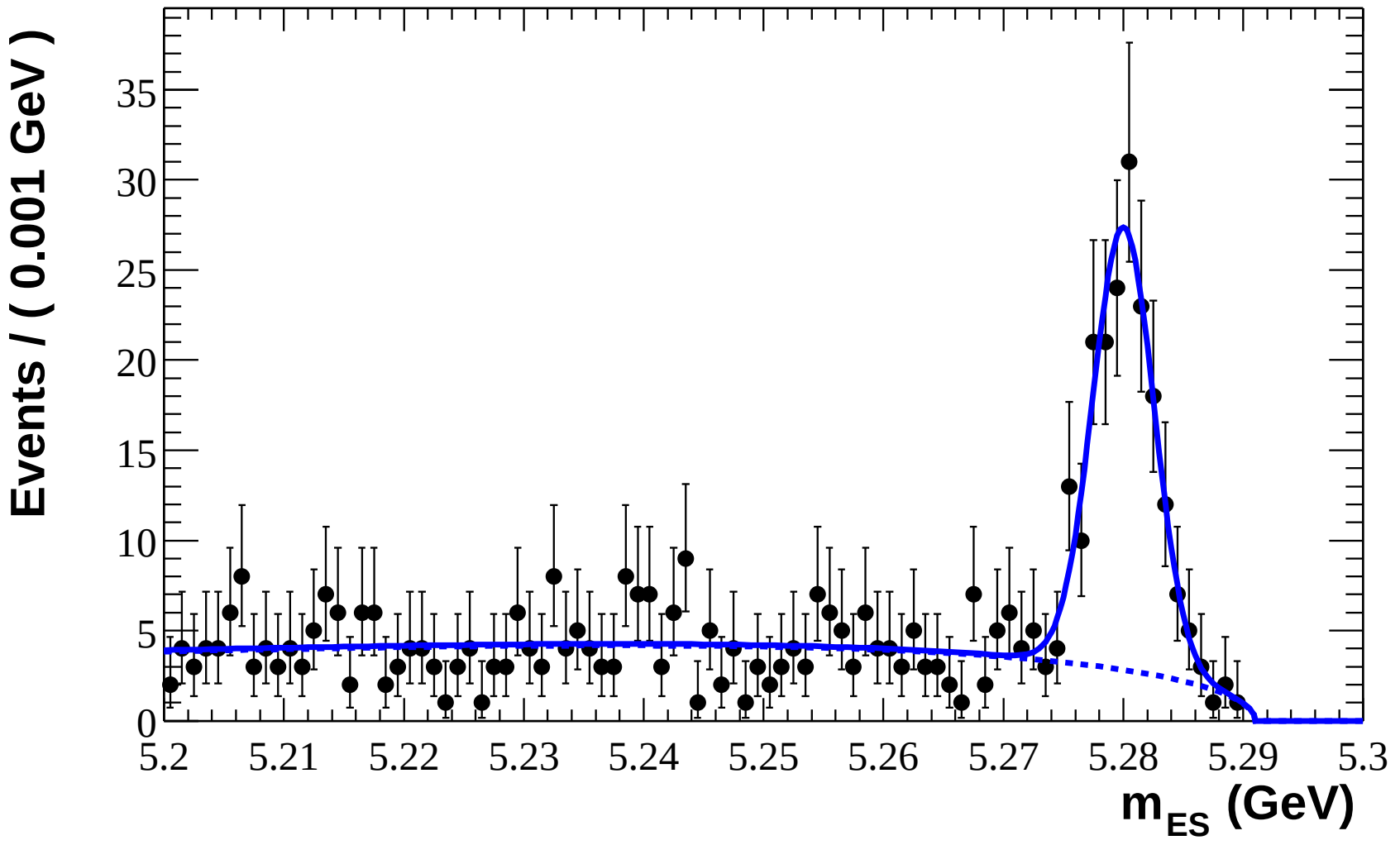

**Figure 3-24.** *The $m_{ES}$ distribution of $B^0 \to D^{*+}D^{*-}$ candidates with $-39 < \Delta E < 31$ MeV in 81 fb$^{-1}$. The fit includes a Gaussian distribution to model the signal and an ARGUS function to model the combinatoric background shape.*

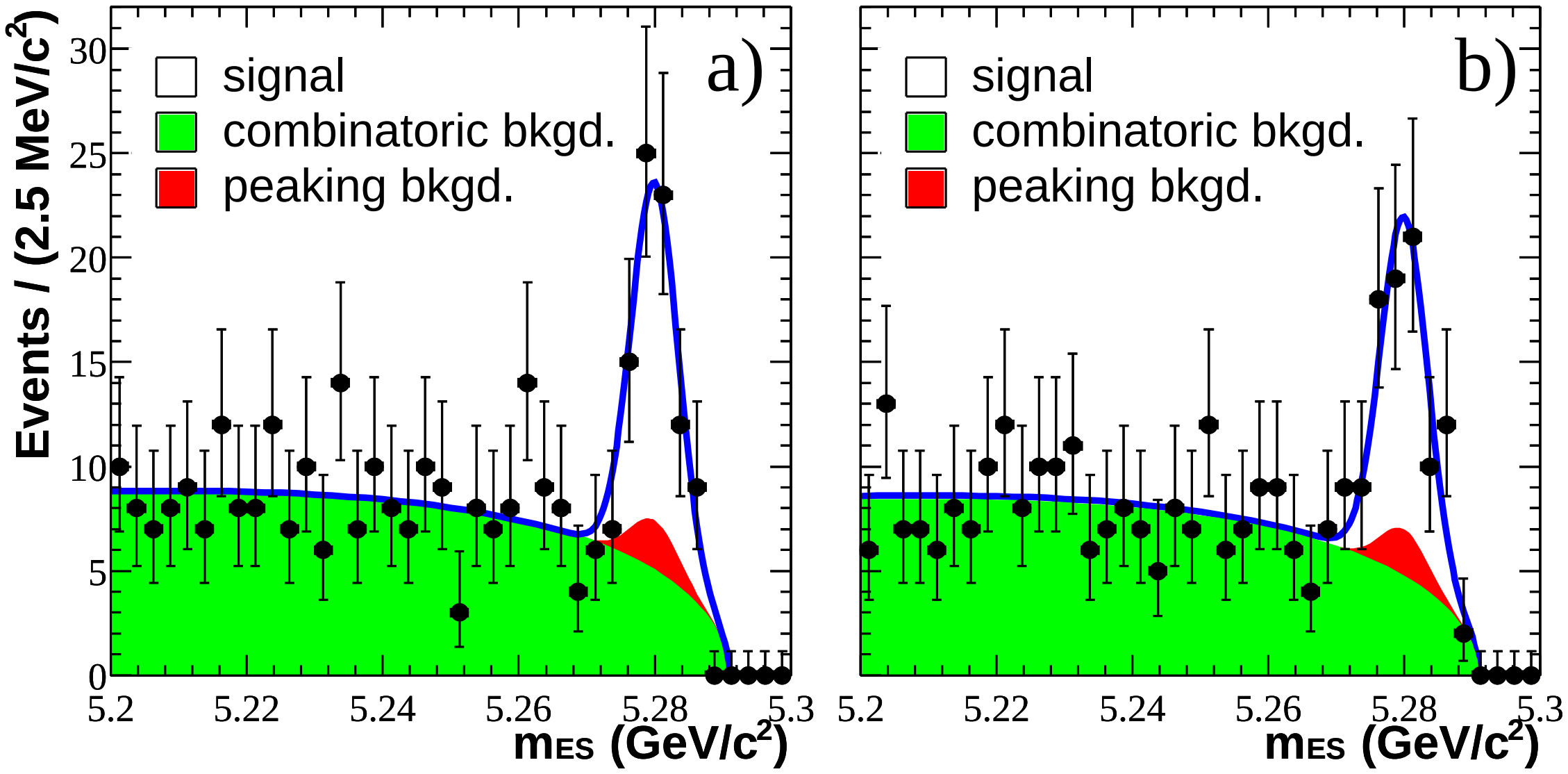

**Figure 3-25.** *The $m_{ES}$ distributions of a) $B \to D^{*-}D^{+}$ and b) $B \to D^{*+}D^{-}$ candidates with $|\Delta E| < 18$ MeV in 81 fb$^{-1}$. The fit includes Gaussian distributions to model the signal and a small peaking background component, and an ARGUS function to model the combinatoric background shape.*

Plots of the decay time difference ($\Delta t$) between the reconstructed and tag $B$, as well as the raw $CP$-violating asymmetry as a function of the decay time difference, are shown in Figures 3-26 and 3-27. These results are limited by small statistics, but they will soon begin to give us a window in the search for $CP$ violation beyond the Standard Model. Later this year, we intend to make the first measurements of and search for direct $CP$ violation in the $B^{\pm} \to D^{(*)\pm}D^{(*)0}$ modes, and also to find limits on (or branching fractions of) the yet-undiscovered color-suppressed $B^0 \to D^{(*)0}\overline{D}^{(*)0}$ modes.

The results for $D^{*+}D^{*-}$ are:

$$\begin{aligned} \Im(\lambda_+) &= 0.05 \pm 0.29(\text{stat.}) \pm 0.10(\text{syst.}), \\ |\lambda_+| &= 0.75 \pm 0.19(\text{stat.}) \pm 0.02(\text{syst.}), \end{aligned} \tag{3.87}$$

and for $D^{*\pm}D^{\mp}$ are:

$$\begin{aligned} S_{-+} &= -0.24 \pm 0.69(\text{stat.}) \pm 0.12(\text{syst.}), \\ C_{-+} &= -0.22 \pm 0.37(\text{stat.}) \pm 0.10(\text{syst.}), \\ S_{+-} &= -0.82 \pm 0.75(\text{stat.}) \pm 0.14(\text{syst.}), \\ C_{+-} &= -0.47 \pm 0.40(\text{stat.}) \pm 0.12(\text{syst.}). \end{aligned}$$

If the transitions proceed only through the $b \to c\bar{c}d$ tree amplitude, we expect that $\Im(\lambda_+) = S_{-+} = S_{+-} = -\sin 2\beta$, $|\lambda_+|$ = 1, and $C_{-+} = C_{+-} = 0$. In addition, we have measured the $D^{*\pm}D^{\mp}$ branching fraction to be

$$\mathcal{B}(B \to D^{*\pm}D^{\mp}) = (8.8 \pm 1.0(\text{stat.}) \pm 1.3(\text{syst.})) \times 10^{-4}$$

and the time integrated direct $CP$ asymmetry between rates to $D^{*+}D^{-}$ and $D^{*-}D^{+}$ is

$$A_{CP}^{\text{dir}} = -0.03 \pm 0.11(\text{stat.}) \pm 0.05(\text{syst.}).$$

Belle has made measurements of the branching fractions for $B \to D^{*+}D^{*-}$, $B \to D^{*\pm}D^{\mp}$, and $B \to D^{+}D^{-}$ [139].

Errors on the $CP$ asymmetries will scale approximately as the inverse square root of the integrated luminosity, as we will continue to be limited by statistics to data sets well into the tens of ab$^{-1}$. Thus a Super $B$ Factory will be necessary to take these results from initial discoveries to the realm of precision physics.

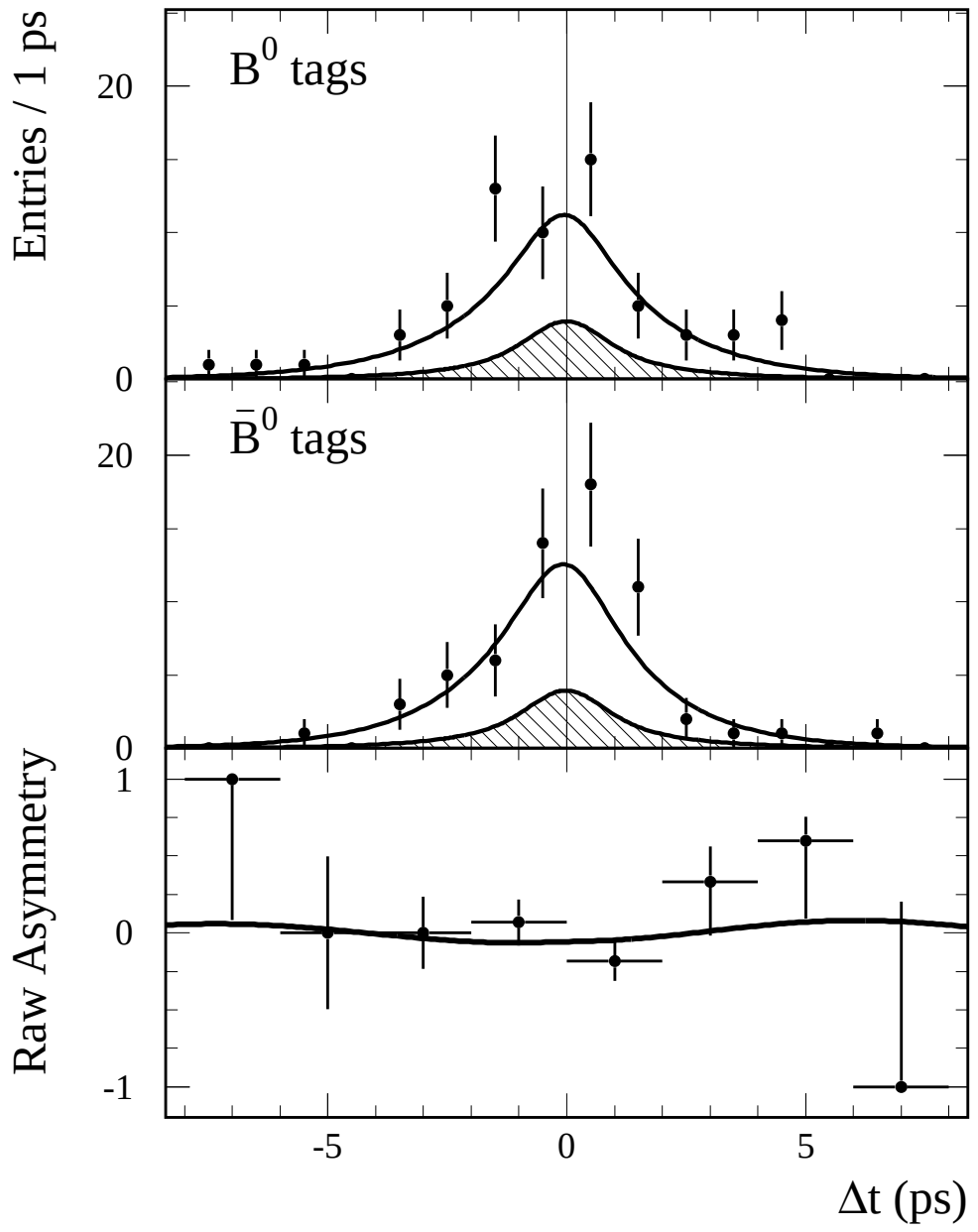

**Figure 3-26.** *Distributions of $\Delta t$ for $B \to D^{*+}D^{*-}$ candidates in the signal region in 81 fb$^{-1}$ with a) a $B^0$ tag ($N_{B^0}$) and b) with a $\overline{B}^0$ tag ($N_{\overline{B}^0}$), and c) the raw asymmetry $(N_{B^0} - N_{\overline{B}^0})/(N_{B^0} + N_{\overline{B}^0})$. The solid curves are the fit projections in $\Delta t$. The shaded regions represent the background contributions.*

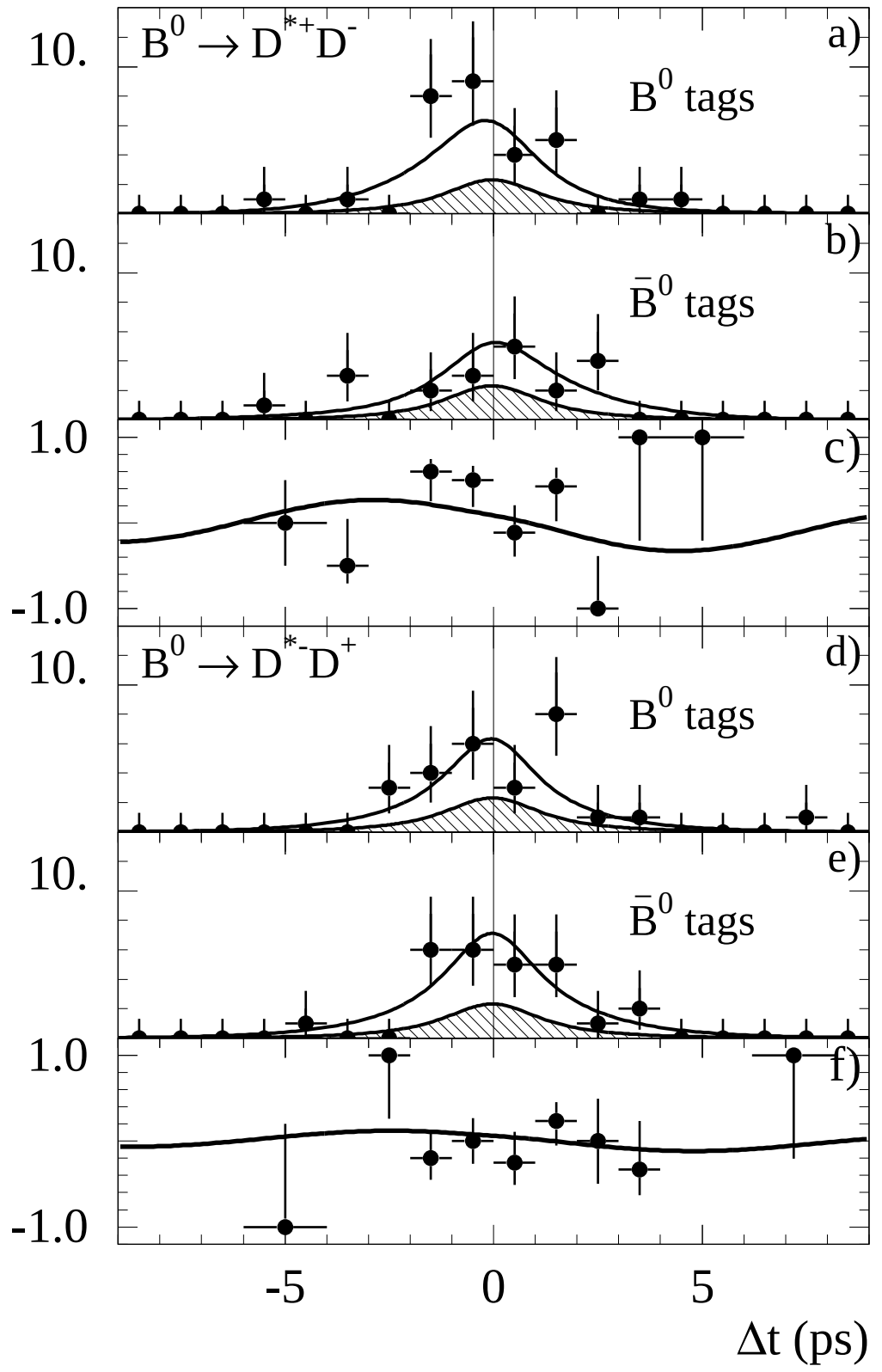

**Figure 3-27.** *Distributions of $\Delta t$ for $B \to D^{*+}D^-$ candidates in the signal region in $81\,fb^{-1}$ with a) a $B^0$ tag ($N_{B^0}$) and b) with a $\overline{B}^0$ tag ($N_{\overline{B}^0}$), and c) the raw asymmetry $(N_{B^0} - N_{\overline{B}^0})/(N_{B^0} + N_{\overline{B}^0})$. The solid curves are the fit projections in $\Delta t$. The shaded regions represent the background contributions. Figures d), e), and f) contain the corresponding information for $D^{*-}D^+$.*

# 3.6 $D^0\overline{D}^0$ Mixing

## 3.6.1 Theory

>– A. A. Petrov –<

Charm physics plays a unique dual role in the modern investigations of flavor physics, providing valuable supporting measurements for studies of $CP$ violation in $B$ decays, as well as outstanding opportunities for indirect searches for physics beyond the Standard Model. In many dynamical models of New Physics, the effects of new particles observed in $s$, $c$, and $b$ transitions are correlated, so such combined studies could yield the most stringent constraints on their parameters. In addition, charm physics studies could also be complimentary to the corresponding programs in bottom or strange systems. This is in part due to the fact that loop-dominated processes such as $D^0\overline{D}^0$ mixing or flavor-changing neutral current (FCNC) decays are influenced by the dynamical effects of *down-type particles*. From the practical point of view, charm physics experiments provide outstanding opportunities for studies of New Physics because of the availability of large statistical samples of data.

The low energy effects of New Physics particles can be naturally written in terms of a series of local operators of increasing dimension generating $\Delta C = 1$ (decays) or $\Delta C = 2$ (mixing) transitions. For $D^0\overline{D}^0$ mixing these operators, as well as the one loop Standard Model effects, generate contributions to the effective operators that change $D^0$ state into $\overline{D}^0$ state, leading to the mass eigenstates

$$|D_{\frac{1}{2}}\rangle = p|D^0\rangle \pm q|\overline{D}^0\rangle\,, \tag{3.88}$$

where the complex parameters $p$ and $q$ are obtained from diagonalizing the $D^0\overline{D}^0$ mass matrix with $|p|^2 + |q|^2 = 1$. If $CP$ violation in mixing is neglected, $p$ becomes equal to $q$, so $|D_{1,2}\rangle$ become $CP$ eigenstates, $CP|D_\pm\rangle = \pm|D_\pm\rangle$. The mass and width splittings between these eigenstates are given by

$$x \equiv \frac{m_2 - m_1}{\Gamma}, \;\; y \equiv \frac{\Gamma_2 - \Gamma_1}{2\Gamma}. \tag{3.89}$$

It is known experimentally that $D^0\overline{D}^0$ mixing proceeds extremely slowly, which in the Standard Model, is usually attributed to the absence of superheavy quarks that would destroy GIM cancellations.

It is instructive to see how New Physics can affect charm mixing. Since the lifetime difference $y$ is constructed from the decays of $D$ into physical states, it should be dominated by the Standard Model contributions, unless New Physics significantly modifies $\Delta C = 1$ interactions. On the contrary, the mass difference $x$ can receive contributions from all energy scales. Thus, it is usually conjectured that New Physics can significantly modify $x$ leading to the inequality [5] $x \gg y$.

Another possible manifestation of New Physics interactions in the charm system is associated with the observation of (large) $CP$ violation. This is due to the fact that all quarks that build up the hadronic states in weak decays of charm mesons belong to the first two generations. Since $2 \times 2$ Cabbibo quark mixing matrix is real, no $CP$ violation is possible in the dominant tree-level diagrams which describe the decay amplitudes. $CP$-violating amplitudes can be introduced in the Standard Model by including penguin or box operators induced by virtual $b$ quarks. However, their contributions are strongly suppressed by the small combination of CKM matrix elements $V_{cb}V^*_{ub}$. It is thus widely believed that the observation of (large) $CP$ violation in charm decays or mixing would be an unambiguous sign for New Physics. This fact makes charm decays a valuable tool in searching for New Physics, since the statistics available in charm physics experiment is usually quite large.

As in $B$ physics, $CP$-violating contributions in charm can be generally classified by three different categories: (I) $CP$ violation in the decay amplitudes. This type of $CP$ violation occurs when the absolute value of the decay amplitude for $D$ to decay to a final state $f$ ($A_f$) is different from the one of corresponding $CP$-conjugate amplitude ("direct $CP$-violation"); (II) $CP$ violation in the $D^0\overline{D}^0$ mixing matrix. This type of $CP$ violation is manifest when $R_m^2 = |p/q|^2 = (2M_{12} - i\Gamma_{12})/(2M_{12}^* - i\Gamma_{12}^*) \neq 1$; and (III) $CP$ violation in the interference of decays with and without mixing. This type of $CP$ violation is possible for a subset of final states to which both $D^0$ and $\overline{D}^0$ can decay.

[5] This signal for New Physics would be lost if a relatively large $y$, of the order of a percent, were to be observed [140, 141].

For a given final state $f$, $CP$-violating contributions can be summarized in the parameter

$$\lambda_f = \frac{q}{p}\frac{\overline{A}_f}{A_f} = R_m e^{i(\phi+\delta)} \left|\frac{\overline{A}_f}{A_f}\right|, \tag{3.90}$$

where $A_f$ and $\overline{A}_f$ are the amplitudes for $D^0 \to f$ and $\overline{D}^0 \to f$ transitions respectively and $\delta$ is the strong phase difference between $A_f$ and $\overline{A}_f$. Here $\phi$ represents the convention-independent weak phase difference between the ratio of decay amplitudes and the mixing matrix.

Current experimental information about the $D^0\overline{D}^0$ mixing parameters $x$ and $y$ comes from the time-dependent analyses that can roughly be divided into two categories. First, more traditional studies look at the time dependence of $D \to f$ decays, where $f$ is the final state that can be used to tag the flavor of the decayed meson. The most popular is the non-leptonic doubly Cabibbo suppressed decay $D^0 \to K^+\pi^-$. Time-dependent studies allow one to separate the DCSD from the mixing contribution $D^0 \to \overline{D}^0 \to K^+\pi^-$,

$$\Gamma[D^0 \to K^+\pi^-] = e^{-\Gamma t}|A_{K^-\pi^+}|^2 \left[R + \sqrt{R}R_m(y'\cos\phi - x'\sin\phi)\Gamma t + \frac{R_m^2}{4}(y^2+x^2)(\Gamma t)^2\right], \tag{3.91}$$

where $R$ is the ratio of DCS and Cabibbo-favored (CF) decay rates. Since $x$ and $y$ are small, the best constraint comes from the linear terms in $t$ that are also *linear* in $x$ and $y$. A direct extraction of $x$ and $y$ from Eq. (3.91) is not possible, due to unknown relative strong phase $\delta_D$ of DCS and CF amplitudes [142], as $x' = x\cos\delta_D + y\sin\delta_D$, $y' = y\cos\delta_D - x\sin\delta_D$. This phase can, however, be measured independently. The corresponding formula can also be written [140] for $\overline{D}^0$ decay with $x' \to -x'$ and $R_m \to R_m^{-1}$.

Second, $D^0$ mixing can be measured by comparing the lifetimes extracted from the analysis of $D$ decays into the $CP$-even and $CP$-odd final states. This study is also sensitive to a *linear* function of $y$ via

$$\frac{\tau(D \to K^-\pi^+)}{\tau(D \to K^+K^-)} - 1 = y\cos\phi - x\sin\phi\left[\frac{R_m^2 - 1}{2}\right]. \tag{3.92}$$

Time-integrated studies of the semileptonic transitions are sensitive to the *quadratic* form $x^2 + y^2$ and at the moment are not competitive with the analyses discussed above.

The construction of new $\tau$-charm factories CLEO-$c$ and BES-III will introduce new *time-independent* methods that are sensitive to a linear function of $y$. One can again use the fact that heavy meson pairs produced in the decays of heavy quarkonium resonances have the useful property that the two mesons are in the $CP$-correlated states [143]. For instance, by tagging one of the mesons as a $CP$ eigenstate, a lifetime difference may be determined by measuring the leptonic branching ratio of the other meson. Its semileptonic *width* should be independent of the $CP$ quantum number since it is flavor specific, yet its *branching ratio* will be inversely proportional to the total width of that meson. Since we know whether this $D(k_2)$ state is tagged as a ($CP$ eigenstate) $D_\pm$ from the decay of $D(k_1)$ to a final state $S_\sigma$ of definite $CP$ parity $\sigma = \pm$, we can easily determine $y$ in terms of the semileptonic branching ratios of $D_\pm$. This can be expressed simply by introducing the ratio

$$R_\sigma^L = \frac{\Gamma[\psi_L \to (H \to S_\sigma)(H \to Xl^\pm\nu)]}{\Gamma[\psi_L \to (H \to S_\sigma)(H \to X)]\; Br(H^0 \to Xl\nu)}, \tag{3.93}$$

where $X$ in $H \to X$ stands for an inclusive set of all final states. A deviation from $R_\sigma^L = 1$ implies a lifetime difference. Keeping only the leading (linear) contributions due to mixing, $y$ can be extracted from this experimentally obtained quantity,

$$y \cos\phi = (-1)^L \sigma \frac{R_\sigma^L - 1}{R_\sigma^L}. \tag{3.94}$$

The current experimental upper bounds on $x$ and $y$ are on the order of a few times $10^{-2}$, and are expected to improve significantly in the coming years. To regard a future discovery of nonzero $x$ or $y$ as a signal for New Physics, we would need high confidence that the Standard Model predictions lie well below the present limits. As was recently shown [141], in the Standard Model, $x$ and $y$ are generated only at second order in SU(3)$_F$ breaking,

$$x\,,\, y \sim \sin^2\theta_C \times [SU(3)\text{ breaking}]^2\,, \tag{3.95}$$

where $\theta_C$ is the Cabibbo angle. Therefore, predicting the Standard Model values of $x$ and $y$ depends crucially on estimating the size of SU(3)$_F$ breaking. Although $y$ is expected to be determined by the Standard Model processes, its value nevertheless affects significantly the sensitivity to New Physics of experimental analyses of $D$ mixing [140].

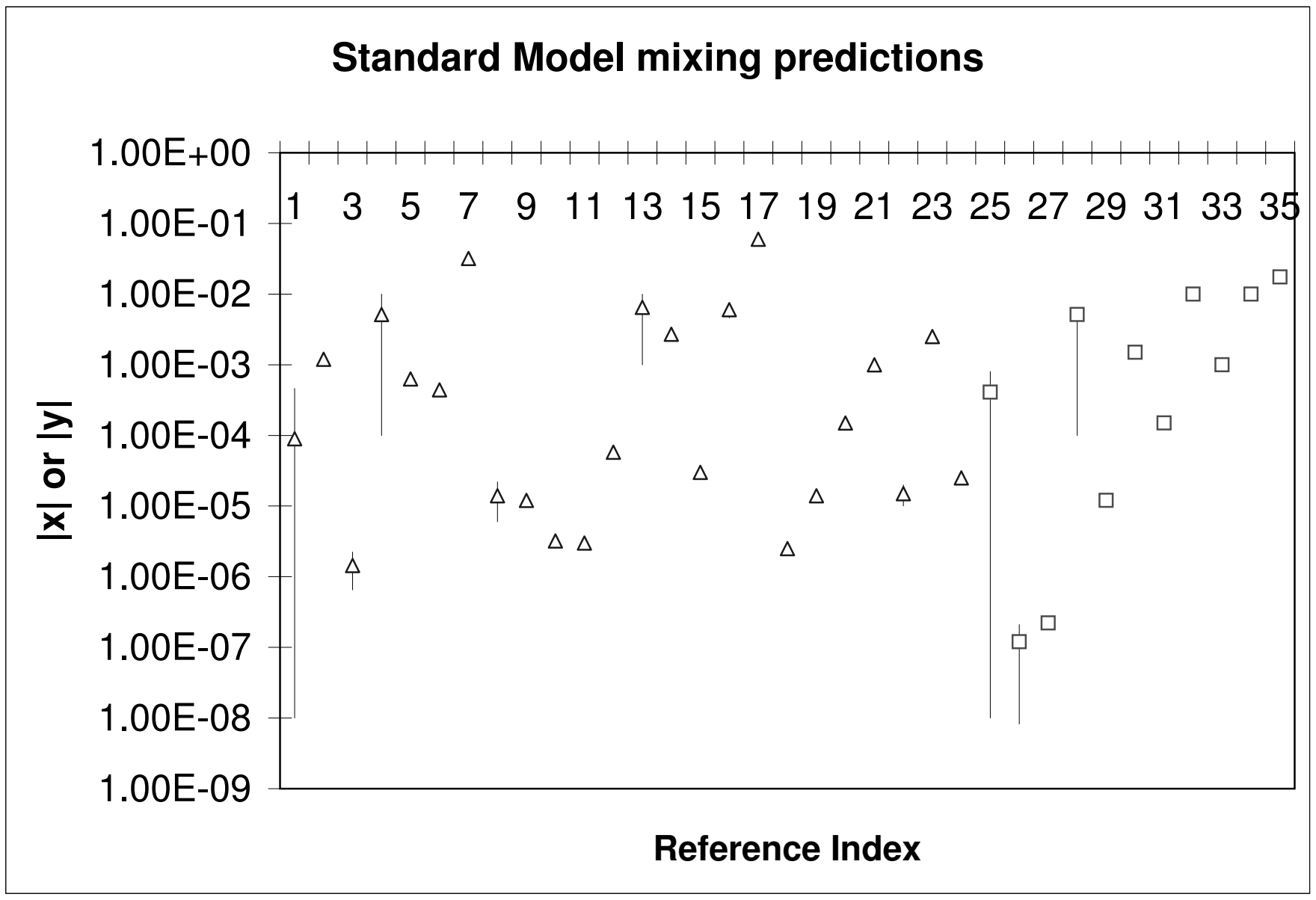

**Figure 3-28.** *Standard Model predictions for $|x|$ (open triangles) and $|y|$ (open squares). Horizontal line references are tabulated in Table 3-11.*

Theoretical predictions of $x$ and $y$ within and beyond the Standard Model span several orders of magnitude [144] (see Fig. 3-28). Roughly, there are two approaches, neither of which give very reliable results because $m_c$ is in some sense intermediate between heavy and light. The "inclusive" approach is based on the operator product expansion (OPE). In the $m_c \gg \Lambda$ limit, where $\Lambda$ is a scale characteristic of the strong interactions, $\Delta M$ and $\Delta\Gamma$ can be expanded in terms of matrix elements of local operators [145]. Such calculations yield $x, y < 10^{-3}$. The use of the OPE relies on local quark-hadron duality, and on $\Lambda/m_c$ being small enough to allow a truncation of the series after the first few terms. The charm mass may not be large enough for these to be good approximations, especially for nonleptonic $D$ decays. An observation of $y$ of order $10^{-2}$ could be ascribed to a breakdown of the OPE or of duality, but such a large value of $y$ is certainly not a generic prediction of OPE analyses. The "exclusive" approach sums over intermediate hadronic states, which may be modeled or fit to experimental data [146]. Since there are cancellations between states within a given SU(3) multiplet, one needs to know the contribution of each state with high precision. However, the $D$ is not light enough that its decays are dominated by a few final states. In the absence of sufficiently precise data on many decay rates and on strong phases, one is forced to use some assumptions. While most studies find $x, y < 10^{-3}$, Refs. [146]

obtain $x$ and $y$ at the $10^{-2}$ level by arguing that SU(3)$_F$ violation is of order unity. It was also shown that phase space effects alone provide enough SU(3)$_F$ violation to induce $y \sim 10^{-2}$ [141]. Large effects in $y$ appear for decays close to $D$ threshold, where an analytic expansion in SU(3)$_F$ violation is no longer possible. Thus, theoretical calculations of $x$ and $y$ are quite uncertain, and the values near the current experimental bounds cannot be ruled out. Therefore, it will be difficult to find a clear indication of physics beyond the Standard Model in $D^0\overline{D}^0$ mixing measurements alone. The only robust potential signal of New Physics in charm system at this stage is $CP$ violation.

$CP$ violation in $D$ decays and mixing can be searched for by a variety of methods. For instance, time-dependent decay widths for $D \to K\pi$ are sensitive to $CP$ violation in mixing (see Eq.(3.91)). Provided that the $x$ and $y$ are comparable to experimental sensitivities, a combined analysis of $D \to K\pi$ and $D \to KK$ can yield interesting constraints on $CP$-violating parameters [140].

Most of the techniques that are sensitive to $CP$ violation make use of the decay asymmetry,

$$A_{CP}(f) = \frac{\Gamma(D \to f) - \Gamma(\overline{D} \to \overline{f})}{\Gamma(D \to f) + \Gamma(\overline{D} \to \overline{f})} = \frac{1 - \left|\overline{A}_{\overline{f}}/A_f\right|^2}{1 + \left|\overline{A}_{\overline{f}}/A_f\right|^2}. \tag{3.96}$$

Most of the properties of Eq. (3.96), such as dependence on the strong final state phases, are similar to the ones in $B$ physics [147]. Current experimental bounds from various experiments, all consistent with zero within experimental uncertainties, can be found in [148].

Other interesting signals of $CP$ violation that are being discussed in connection with $\tau$-charm factory measurements exploit the quantum coherence of the initial state. An example of this type of signal is a decay $(D^0\overline{D}^0) \to f_1 f_2$ at $\psi(3770)$ with $f_1$ and $f_2$ being the different final $CP$ eigenstates of the same $CP$ parity. This type of signals are very easy to detect experimentally. The corresponding $CP$-violating decay rate for the final states $f_1$ and $f_2$ is

$$\begin{aligned} \Gamma_{f_1 f_2} &= \frac{1}{2R_m^2}\left[\left(2 + x^2 - y^2\right)|\lambda_{f_1} - \lambda_{f_2}|^2 \right. \\ &+ \left. \left(x^2 + y^2\right)|1 - \lambda_{f_1}\lambda_{f_2}|^2\right] \Gamma_{f_1}\Gamma_{f_2}. \end{aligned} \tag{3.97}$$

The result of Eq. (3.97) represents a generalization of the formula given in Ref. [149]. It is clear that both terms in the numerator of Eq. (3.97) receive contributions from $CP$ violation of Type I and III, while the second term is also sensitive to $CP$ violation of Type II. Moreover, for a large set of the final states the first term would be additionally suppressed by SU(3)$_F$ symmetry, as for instance, $\lambda_{\pi\pi} = \lambda_{KK}$ in the SU(3)$_F$ symmetry limit. This expression is of the *second* order in $CP$-violating parameters (it is easy to see that in the approximation where only $CP$ violation in the mixing matrix is retained, $\Gamma_{f_1 f_2} \propto \left|1 - R_m^2\right|^2 \propto A_m^2$). As it follows from the existing experimental constraints on rate asymmetries, $CP$-violating phases are quite small in charm system, regardless of whether they are produced by the Standard Model mechanisms or by some New Physics contributions. In that respect, it is unlikely that the Standard Model signals of $CP$ violation would be observed at CLEO-$c$ with this observable.

While searches for direct $CP$ violation via the asymmetry of Eq. (3.96) can be done with the charged $D$ mesons (which are self-tagging), investigations of the other two types of $CP$ violation require flavor tagging of the initial state, which severely reduces the available data set. It is therefore interesting to look for signals of $CP$ violation that do not require identification of the initial state. One possible $CP$-violating signal involves the observable obtained by summing over the initial states,

$$\Sigma_i = \Gamma_i(t) + \overline{\Gamma}_i(t) \tag{3.98}$$

for $i = f$ and $\overline{f}$. A $CP$-odd observable which can be formed out of $\Sigma_i$ is the asymmetry [150]

$$A_{CP}^U(f,t) = \frac{\Sigma_f - \Sigma_{\overline{f}}}{\Sigma_f + \Sigma_{\overline{f}}} \equiv \frac{N(t)}{D(t)}. \tag{3.99}$$

**Table 3-11.** *Theoretical predictions for mixing parameters (Standard Model). The notation "±" indicates the range of predictions.*

| Mass difference, $x$ | Reference Index | Citation |
|---|---|---|
| $(0.9 \pm 3.7) \times 10^{-4}$ | 1 | Phys. Rev. D 26, 143 (1982) |
| $1.2 \times 10^{-3}$ | 2 | Phys. Lett. B128, 240 (1983) |
| $(1.44 \pm 0.79) \times 10^{-6}$ | 3 | Z. Phys. C 27, 515 (1985) |
| $(0.01 - 10) \times 10^{-2}$ | 4 | Phys. Lett. B 164, 170 (1985) |
| $6.3 \times 10^{-4}$ | 5 | Phys. Rev. D 33, 179 (1986) |
| $4.4 \times 10^{-4}$ | 6 | Phys. Rev. D 35, 3484 (1987) |
| $3.2 \times 10^{-2}$ | 7 | Phys. Lett. B224, 71 (1990) |
| $(1.4 \pm 0.8) \times 10^{-5}$ | 8 | Nucl. Phys. B403, 71 (1993) |
| $1.2 \times 10^{-5}$ | 9 | hep-ph/9407378 |
| $3.2 \times 10^{-6}$ | 10 | Chin. J. Phys. 32, 1163 (1994) |
| $3.0 \times 10^{-6}$ | 11 | hep-ph/9409379 |
| $5.8 \times 10^{-5}$ | 12 | hep-ph/9508349 |
| $(1 - 10) \times 10^{-3}$ | 13 | hep-ph/9508349 |
| $2.7 \times 10^{-4}$ | 14 | hep-ph/9508349 |
| $3 \times 10^{-5}$ | 15 | Phys. Lett. B357, 151 (1995) |
| $(6.0 \pm 1.4) \times 10^{-3}$ | 16 | Phys. Lett. B357, 151 (1995) |
| $6 \times 10^{-2}$ | 17 | Phys. Lett. B357, 151 (1995) |
| $2.5 \times 10^{-6}$ | 18 | Phys. Rev. D 56, 1685 (1997) |
| $1.4 \times 10^{-5}$ | 19 | Phys. Lett. B 422, 265 (1998) |
| $1.5 \times 10^{-4}$ | 20 | Phys. Lett. B 427, 172 (1998) |
| $1.0 \times 10^{-3}$ | 21 | Nucl.Phys. B592, 92 (2001) |
| $(1.5 \pm 0.5) \times 10^{-5}$ | 22 | Phys. Lett. B297, 353 (1992) |
| $2.50 \times 10^{-3}$ | 23 | Phys. Rev. D 43, 1641 (1991) |
| $2.50 \times 10^{-5}$ | 24 | hep-ph/9706548 |
| Lifetime difference, $y$ | Reference Index | Citation |
| $-(0.06 - 8.0) \times 10^{-4}$ | 25 | Phys. Rev. D 26, 143 (1982) |
| $(0.082 - 2.1) \times 10^{-7}$ | 26 | Phys. Lett. B 128, 240 (1983) |
| $2.2 \times 10^{-7}$ | 27 | Z. Phys. C 27, 515 (1985) |
| $(0.01 - 10) \times 10^{-2}$ | 28 | Phys. Lett. B 164, 170 (1985) |
| $1.2 \times 10^{-5}$ | 29 | hep-ph/9407378 |
| $1.5 \times 10^{-3}$ | 30 | Phys. Lett. B 379, 249 (1996) |
| $1.0 \times 10^{-4}$ | 31 | Phys. Lett. B 427, 172 (1998) |
| $1.0 \times 10^{-2}$ | 32 | Phys. Rev. Lett. 83, 4005 (1999) |
| $1.0 \times 10^{-3}$ | 33 | Nucl.Phys. B592, 92 (2001) |
| $1.0 \times 10^{-2}$ | 34 | Phys. Rev. D65, 054034 (2002) |
| $(1.5 - 2.0) \times 10^{-2}$ | 35 | Phys. Rev. D 43, 1641 (1991) |

Note that this asymmetry does not require quantum coherence of the initial state and therefore is accessible in any $D$ physics experiment. The final states must be chosen such that $A^U_{CP}$ is not trivially zero. It is easy to see that decays of $D$ into the final states that are $CP$ eigenstates would result in zero asymmetry, while the final states like $K^+K^{*-}$ or $K^0_S\pi^+\pi^-$ would not. A non-zero value of $A^U_{CP}$ in Eq. (3.99) can be generated by both direct and indirect $CP$-violating contributions. These can be separated by appropriately choosing the final states. For instance, indirect $CP$ violating amplitudes are tightly constrained in the decays dominated by the Cabibbo-favored tree level amplitudes, while singly Cabibbo-suppressed amplitudes also receive contributions from direct $CP$ violating amplitudes. Choosing a transition $D \to K\pi$ as an example we find that

$$A^U_{CP}(K,\pi) = -y \sin\delta \sin\phi \sqrt{R} \tag{3.100}$$

for the time-integrated asymmetry. The asymmetry of Eq. (3.100) is clearly of the *first* order in $CP$-violating phase $\phi$. Time-dependent analysis could also be possible with huge statistics available at a Super $B$ Factory. For a generic final state it is expected that the numerator and denominator of Eq. (3.99) would have the form,

$$\begin{aligned} N(t) &= \Sigma_f - \Sigma_{\overline{f}} = e^{-\mathcal{T}} \left[A + B\mathcal{T} + C\mathcal{T}^2\right], \\ D(t) &= e^{-\mathcal{T}} \left[|A_f|^2 + \left|\overline{A}_{\overline{f}}\right|^2 + \left|A_{\overline{f}}\right|^2 + |\overline{A}_f|^2\right]. \end{aligned} \tag{3.101}$$

Integrating the numerator and denominator of Eq. (3.99) over time yields

$$A^U_{CP}(f) = \frac{1}{D}\left[A + B + 2C\right], \tag{3.102}$$

where $D = \Gamma \int_0^\infty dt\, D(t)$.

Both time-dependent and time-integrated asymmetries depend on the same parameters $A, B$, and $C$. Since $CP$ violation in the mixing matrix is expected to be small, we expand $R_m^{\pm 2} = 1 \pm A_m$. The result is

$$\begin{aligned} A &= \left(|A_f|^2 - \left|\overline{A}_{\overline{f}}\right|^2\right) - \left(\left|A_{\overline{f}}\right|^2 - |\overline{A}_f|^2\right) = |A_f|^2\left[\left(1 - \left|\overline{A}_{\overline{f}}\right|^2/|A_f|^2\right) + R\left(1 - \left|A_{\overline{f}}\right|^2/|\overline{A}_f|^2\right)\right], \\ B &= -2y\sqrt{R}\left[\sin\phi\sin\delta\left(|\overline{A}_f|^2 + \left|A_{\overline{f}}\right|^2\right) - \cos\phi\cos\delta\left(|\overline{A}_f|^2 - \left|A_{\overline{f}}\right|^2\right)\right] + \mathcal{O}(A_m x, r_f x, ...), \\ C &= \frac{x^2}{2}\left[\left(|A_f|^2 - \left|\overline{A}_{\overline{f}}\right|^2\right) - \left(\left|A_{\overline{f}}\right|^2 - |\overline{A}_f|^2\right)\right] = \frac{x^2}{2}A + \mathcal{O}(A_m x^2, A_m y^2). \end{aligned} \tag{3.103}$$

Here we neglect small corrections of the order of $\mathcal{O}(A_m x, r_f x, ...)$ and higher. It follows that Eq. (3.103) receives contributions from both direct and indirect $CP$-violating amplitudes. Those contributions have different time dependence and can be separated either by time-dependent analysis of Eq. (3.99) or by the "designer" choice of the final state.

In summary, charm physics, and in particular studies of $D^0\overline{D}^0$ mixing, could provide new and unique opportunities for indirect searches for New Physics at a Super $B$ Factory. Huge statistical samples of charm data will allow new sensitive measurements of charm mixing and $CP$-violating parameters.

# 3.7 SUSY $CP$ Violation

## 3.7.1 Measuring squark mixing angles and $CP$-violating phases at the LHC

$\succ$– K. Matchev –$\prec$

The Minimal Supersymmetric Standard Model (MSSM ) leads to a proliferation of theory parameters. In addition to the parameters already present in the Standard Model, the MSSM has 105 new parameters: 33 masses, 41 phases and 31 super-CKM angles [151]. Measuring all of them directly at high energy colliders represents a formidable experimental challenge. In this section we review some of the methods of measuring supersymmetric phases and mixing angles that have been discussed in the literature.

**L-R sfermion mixing angles**

Electroweak symmetry breaking induces mixing among the superpartners of the left-handed and the right-handed quarks and charged leptons of the Standard Model (*i.e.*, between the so-called "left-handed" and "right-handed" squarks and sleptons of a particular flavor). For example, the up-type squark mass matrix has the form

$$\begin{pmatrix} M_Q^2 + m_u^2 + g_{u_L} M_Z^2 c_{2\beta} & m_u\,(A_u + \mu\cot\beta) \\ m_u\,(A_u + \mu\cot\beta) & M_U^2 + m_u^2 + g_{u_R} M_Z^2 c_{2\beta} \end{pmatrix} \tag{3.104}$$

where we use the notation of [152]. Its diagonalization leads to a L-R mixing angle $\theta_u$ given by

$$\tan(2\theta_u) = \frac{2m_u\,(A_u + \mu\cot\beta)}{M_Q^2 - M_U^2 + (g_{u_L} - g_{u_R})M_Z^2 c_{2\beta}}. \tag{3.105}$$

The mixing angle $\theta_u$ is in general complex, since both $A_u$ and $\mu$ may have a complex phase. Similar expressions hold for the down-type squarks and charged sleptons as well.

We see from Eq. (3.105) that the L-R mixing is proportional to the *fermion* mass. Hence, L-R mixing is only significant for third generation sfermions: stops, sbottoms and staus. Conversely, the L-R mixing angles for the first two generations of squarks and sleptons are expected to be too small ever to be directly measured at a high energy collider.

In principle, there is also mixing among different generations (super-CKM angles). The amount of flavor violation in the squark and slepton sectors is indirectly constrained by various rare low-energy processes. For example, squark 1-2 mixing is severely constrained by $K^0\overline{K}^0$ mixing. Furthermore, we cannot identify light quark jets as such, so it is difficult to observe a direct signal of squark 1-2 mixing. Slepton flavor mixing is in turn constrained by processes such as $\mu \to e\gamma$, $\tau \to \mu\gamma$, $\tau \to e\gamma$ [153, 154, 155]. Nevertheless, it may yield interesting signals at both the LHC [156, 157] and NLC [158, 159]. In what follows we shall concentrate on L-R mixing angles only.

We will first discuss the possibility of measuring the L-R mixing angles of third generation squarks. At hadron colliders such as the Tevatron or the LHC, stops and sbottoms are predominantly strongly produced, and thus their production cross-sections are insensitive to the L-R mixing angles. We are therefore forced to concentrate on squark decays. To this end, Ref. [160] considers the process of gluino production and the subsequent decay chain through the lightest stop $\tilde{t}_1$:

$$\tilde{g} \to t\tilde{t}_1 \to tb\tilde{\chi}_j^{\pm} \to Wbb\tilde{\chi}_j^{\pm} \tag{3.106}$$

The left-right stop mixing affects the invariant mass distribution of the $b$ jet pair in each gluino decay. This can be seen from the following argument [160]. The top quark from the decay mode (3.106) will be polarized to be left-handed (right-handed) if $\tilde{t}_1$ is left-handed (right-handed). The top polarization is reflected in the angular distribution of the $b$-quark from top decay:

$$\frac{1}{\Gamma_t}\frac{d\Gamma_t}{d\cos\theta} \propto \left(\frac{m_t}{m_W}\right)^2 \sin^2\frac{\theta}{2} + 2\cos^2\frac{\theta}{2} \approx 4.78\sin^2\frac{\theta}{2} + 2\cos^2\frac{\theta}{2}, \tag{3.107}$$

where $\theta$ is the angle between the direction of the $b$ quark and the top quark spin in the rest frame of the top quark. Hence the $b$ quark tends to go in a direction opposite to the top quark *spin*. At the same time, the $b$ quark from the stop decay is preferentially in a direction opposite to the top quark *momentum*. Thus the $b\bar{b}$ system encodes information about the top quark polarization. In particular, the invariant mass distribution of the two $b$'s is harder (softer) for left-handed (right-handed) top quarks. This is illustrated in Fig. 3-29, which shows the result from a full numerical simulation with HERWIG. The statistical significance of the effect is about $3\sigma$ for $\mathcal{O}(100)$ events [160]. In order to convert this result into a measurement of the stop L-R mixing angle, one would have to match the measured distribution to a series of templates corresponding to different values for the mixing angle. In reality the measurement will be complicated by the presence of other supersymmetric decay chains besides (3.106), which were neglected here. They will certainly contribute to the $m_{bb}$ distribution and dilute the effect. The size of the degradation is however rather model dependent, as it depends on the gluino branching fractions.

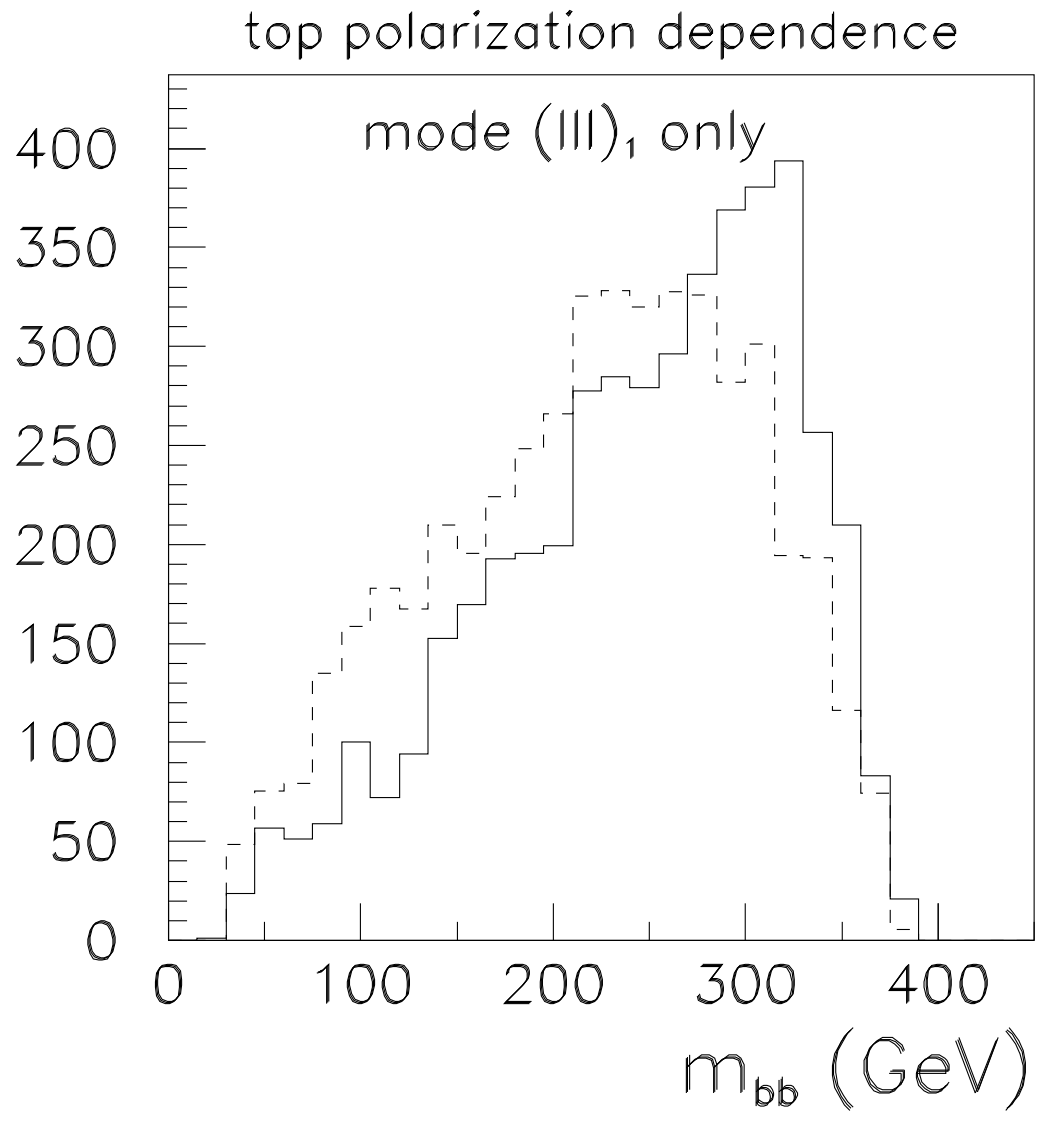

**Figure 3-29.** *The invariant mass distribution of the two $b$-jets from the gluino decay (3.106), for $\tilde{t}_1 = \tilde{t}_L$ (solid) and $\tilde{t}_1 = \tilde{t}_R$ (dashed). The superpartner masses are chosen to be $m_{\tilde{g}} = 707$ GeV, $m_{\tilde{t}_1} = 427$ GeV and $m_{\tilde{\chi}^\pm} = 220$ GeV. The normalization is arbitrary. (From Ref. [160].)*

The mixing angle for down-type sfermions (squarks and sleptons) is proportional to $m_d\mu\tan\beta$ and can be significant for the third generation (sbottoms and staus) at large $\tan\beta$. There have been no studies on the possibility to measure sbottom mixing at the LHC, so in the remainder of this subsection we will discuss stau mixing.

In unified models such as minimal supergravity, the stau lepton is often the next-to lightest supersymmetric particle and is abundantly produced in squark and gluino cascade decays:

$$\tilde{q}\bar{\tilde{q}} \to q\bar{q}\tilde{W}_1^+\tilde{W}_1^- \to q\bar{q}\nu\bar{\nu}\tilde{\tau}\tilde{\tau} \to q\bar{q}\nu\bar{\nu}\tilde{Z}_1\tilde{Z}_1\tau\tau, \tag{3.108}$$

$$\tilde{q}\bar{\tilde{q}} \to q\bar{q}\tilde{W}_1^\pm\tilde{Z}_2 \to q\bar{q}\nu\tau'\tilde{\tau}\tilde{\tau} \to q\bar{q}\nu\tilde{Z}_1\tilde{Z}_1\tau'\tau\tau. \tag{3.109}$$

The polarization of the tau lepton in stau decay is given by [161]

$$P_\tau = \frac{a_R^2 - a_L^2}{a_R^2 + a_L^2}, \tag{3.110}$$

where

$$a_R \sim g' N_{11} \cos\theta_\tau + \lambda_\tau N_{13} \sin\theta_\tau, \tag{3.111}$$
$$a_L \sim (g' N_{11} + g N_{12}) \sin\theta_\tau - \lambda_\tau N_{13} \cos\theta_\tau. \tag{3.112}$$

One can show that for any value of $\tan\beta$, $P_\tau$ is close to $+1$. At small $\tan\beta$, both the Yukawa couplings and the stau mixing are small and we have $a_R \sim g' >> a_L \sim 0$. At large $\tan\beta$, all terms in (3.111-3.112) are sizable, but because of a cancellation in (3.112), $a_R >> a_L$ still holds [161]. One can then employ the standard techniques in measuring tau polarization in order to test the $P_\tau = +1$ prediction of supersymmetry and perhaps even measure the stau mixing angle $\theta_\tau$.

By measuring separately the tau jet energy in the calorimeter and the momentum of the charged tracks, one can compute the $\tau$-jet momentum fraction

$$R = \frac{\text{momentum of charged tracks}}{\text{total jet energy in calorimeter}} \tag{3.113}$$

carried by the charged-prongs, which is sensitive to the tau polarization. Figure 3-30 shows the normalized $R$ distributions for hypothetical supersymmetric signals with $P_\tau = +1$, $P_\tau = 0$ and $P_\tau = -1$. An effect is seen, although it needs to be confirmed by a detailed study which would include all relevant backgrounds and a realistic detector simulation.

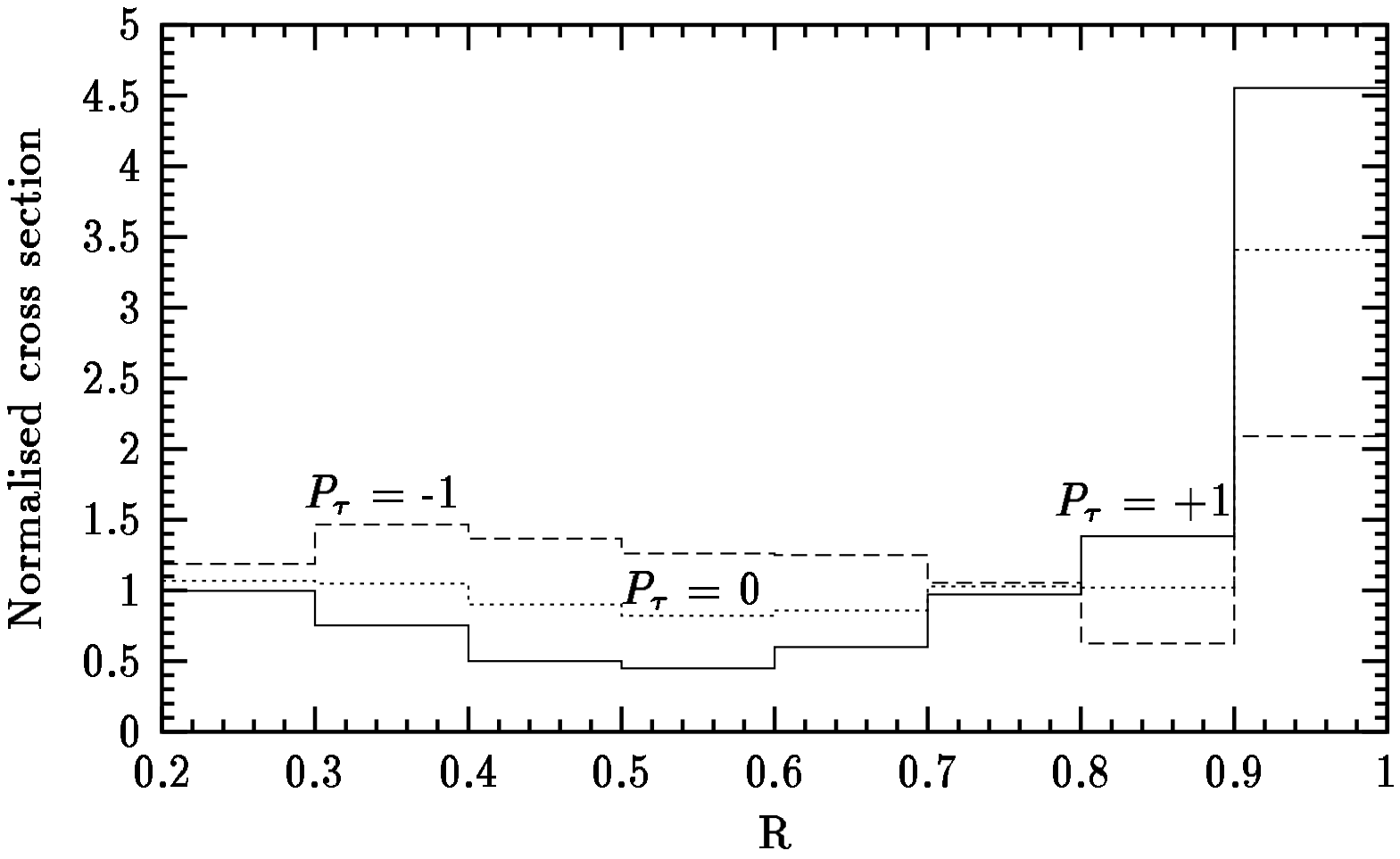

**Figure 3-30.** *Normalized supersymmetric signal cross-sections for $P_\tau = +1$ (solid line), 0 (dotted) and $-1$ (dashed), as a function of the $\tau$ jet momentum fraction R (3.113). (From Ref. [161].)*

### *CP*-violating phases

In spite of the large number of $CP$-violating phases in the MSSM Lagrangian, $CP$ violation might be difficult to observe directly at the LHC. In principle, the phases can manifest themselves in both $CP$-conserving and $CP$-violating observables. In the case of the former, the effect of the phases can be masked by a variation in the remaining SUSY parameters, while in case of the latter, the experimental precision may not be sufficient for a discovery.

The gluino provides an unique opprotunity for measuring a $CP$-violating SUSY phase, since the gluino does not mix with any other states, and therefore a mixing angle confusion is lacking. One can then consider gluino pair-production

$$\tilde{g}\tilde{g} \to q\bar{q}q\bar{q}\tilde{\chi}^0_1\tilde{\chi}^0_1 \tag{3.114}$$

and look for an observable effect of the phase $\phi$ of the gluino mass parameter $M_3$. Reference [162] observed that the $q\overline{q}$ invariant mass of the two quarks coming from each gluino decay is sensitive to $\phi$. The effect of varying $\phi$ from $0$ to $\pi$ is shown in the left panel of Fig. 3-31, where only the correct combination of jet pairs in (3.114) was used, and any backgrounds were ignored. We see that at this level there is an observable effect. However, once all jet pairings are used, the effect is washed out to a large extent [162]. It is therefore preferable to look for the impact of the phases on $CP$-violating observables.

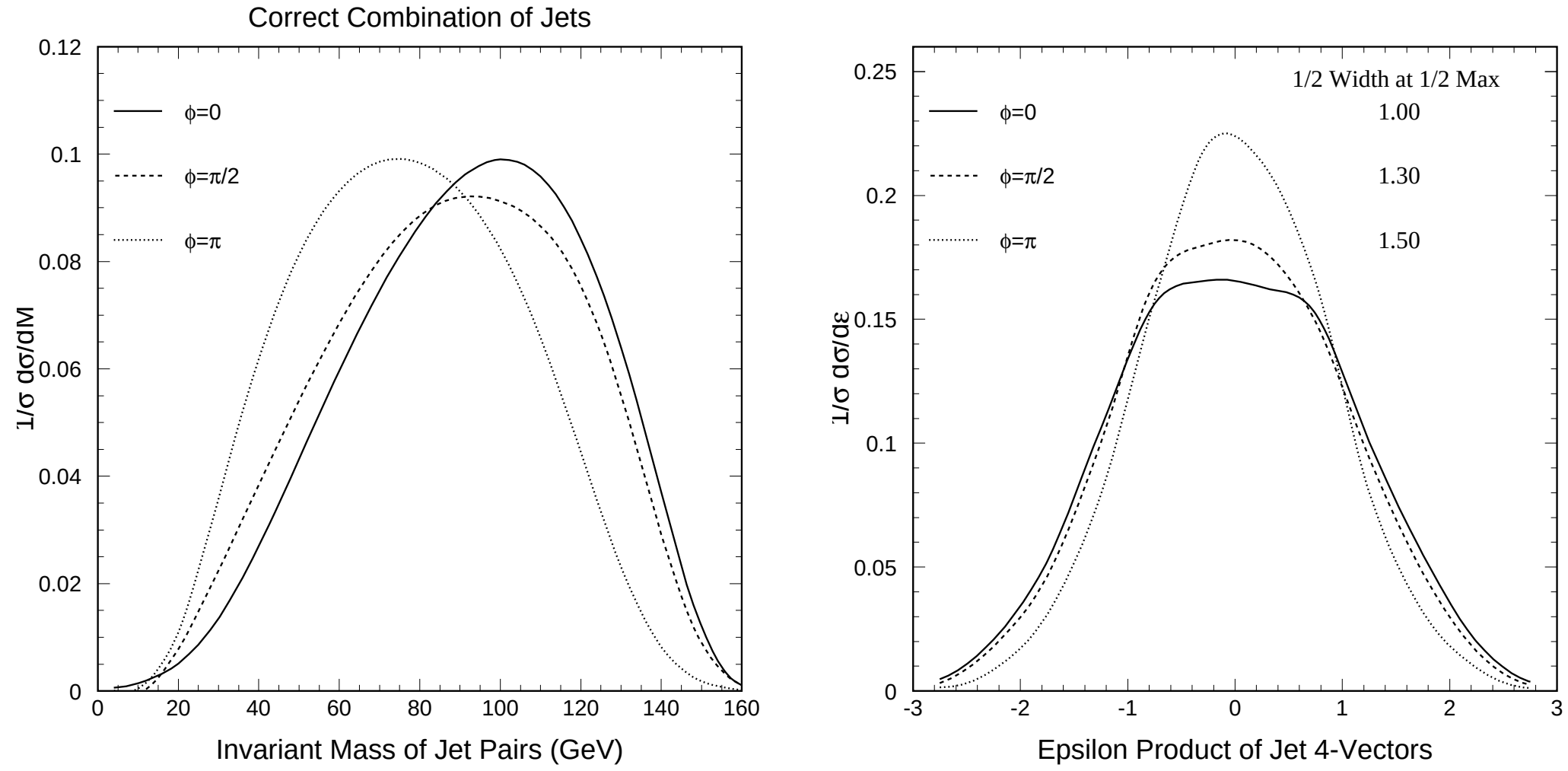

**Figure 3-31.** *Left: The invariant mass distribution of the two jets from each gluino decay in (3.114), assuming correct jet pairing, and for different values of the gluino phase $\phi$. The gluino (LSP) mass is $250$ ($105$) GeV. Right: The distribution of the $CP$-violating observable (3.115) for different values of the gluino phase $\phi$. (From Ref. [162].)*

Reference [162] proposed the following $CP$-violating observable:

$$\epsilon = \frac{\epsilon_{\mu\nu\rho\sigma} p_1^\mu p_2^\nu p_3^\rho p_4^\sigma}{E_1 E_2 E_3 E_4} \tag{3.115}$$

involving the energies and momenta of all four jets in (3.114) as measured in the lab frame. This distribution is shown in the right panel of Fig. 3-31. The distinguishing feature is the half width at half maximum, which varies from 1.0 to 1.5 as $\phi = 0 \to \pi$. It is worth repeating the analysis of [162] including all relevant backgrounds and a realistic detector simulation.

Reference [163] proposed a different $CP$-violating observable, which relies on the presence of a) large $CP$-violating phases in the slepton sector; b) large flavor-violation in the slepton sector; and c) sufficient degree of slepton degeneracy. Under those circumstances, slepton oscillations will result in a nonzero excess in the subtracted distribution

$$N(e^+\mu^-) - N(e^-\mu^+). \tag{3.116}$$

The main advantage of this observable is the very small Standard Model background.

# 3.8 Other Probes of $CP$ Violation

## 3.8.1 Triple Products

≻– A. Datta and D. London–≺

Most of the theoretical work on $CP$ violation in the $B$ system has focussed on indirect and direct $CP$-violating asymmetries in $B$ decays. However, there is another interesting class of $CP$-violating effects: triple-product correlations (TP's), which take the form $\vec{v}_1 \cdot (\vec{v}_2 \times \vec{v}_3)$, where each $v_i$ is a spin or momentum. These TP's are odd under time reversal ($T$). Assuming that $CPT$ is conserved, which is the case for all local and Lorentz-invariant field theories, $T$ and $CP$ violation are related. Thus, TP's correspond to potential signals of $CP$ violation.

One can establish a nonzero TP by measuring a nonzero value of the $T$-odd asymmetry

$$A_T \equiv \frac{\Gamma(\vec{v}_1 \cdot (\vec{v}_2 \times \vec{v}_3) > 0) - \Gamma(\vec{v}_1 \cdot (\vec{v}_2 \times \vec{v}_3) < 0)}{\Gamma(\vec{v}_1 \cdot (\vec{v}_2 \times \vec{v}_3) > 0) + \Gamma(\vec{v}_1 \cdot (\vec{v}_2 \times \vec{v}_3) < 0)} , \tag{3.117}$$

where $\Gamma$ is the decay rate for the process in question. By itself, $A_T$ does not measure $CP$ violation. Since it is only $T$-odd, and not $T$-violating, strong phases alone can generate $A_T \neq 0$, even if the weak $CP$-violating phases are absent. In order to find a true $CP$-violating effect, one has to compare $A_T$ with $\overline{A}_T$, where $\overline{A}_T$ is the $T$-odd asymmetry measured in the $CP$-conjugate decay process [164]. $A_T \neq \overline{A}_T$ is a true measure of $CP$ violation.

One class of processes in which triple products can appear are the decays $B \to V_1 V_2$, where $V_1$ and $V_2$ are vector mesons [164, 165, 166, 167]. In this case, the TP takes the form $\vec{q} \cdot (\vec{\varepsilon}_1 \times \vec{\varepsilon}_2)$, where $\vec{q}$ is the momentum of one of the final vector mesons, and $\vec{\varepsilon}_1$ and $\vec{\varepsilon}_2$ are the polarizations of $V_1$ and $V_2$. The amplitude for the decay $B(p) \to V_1(k_1, \varepsilon_1) V_2(k_2, \varepsilon_2)$ is given by [164]

$$M = a\, \varepsilon_1^* \cdot \varepsilon_2^* + \frac{b}{m_B^2} (p \cdot \varepsilon_1^*)(p \cdot \varepsilon_2^*) + i \frac{c}{m_B^2} \epsilon_{\mu\nu\rho\sigma} p^\mu q^\nu \varepsilon_1^{*\rho} \varepsilon_2^{*\sigma} , \tag{3.118}$$

where $q \equiv k_1 - k_2$. This amplitude contains three partial waves: $c$ is $P$-wave, while $a$ and $b$ are combinations of $S$- and $D$-wave. It is the last term above which interests us: in the rest frame of the $B$, $\epsilon_{\mu\nu\rho\sigma} p^\mu q^\nu \varepsilon_1^{*\rho} \varepsilon_2^{*\sigma} \to m_B\, \vec{q} \cdot (\vec{\varepsilon}_1^* \times \vec{\varepsilon}_2^*)$, which is the triple product. Thus, the TP is generated by the interference of the $c$ term with the $a$ and/or $b$ terms.

Of course, as discussed above, true $CP$ violation is found when one compares the triple product in $|M|^2$ with that in $|\overline{M}|^2$. For $B \to V_1 V_2$ decays, one has to *add* the two asymmetries [164]:

$$\mathcal{A}_T = \frac{1}{2} \left( A_T + \overline{A}_T \right) , \tag{3.119}$$

where $\mathcal{A}_T$ is the true $CP$-violating TP asymmetry. It is not difficult to show that $A_T \sim \sin(\phi + \delta)$, where $\phi$ and $\delta$ are, respectively, the weak and strong phase differences between the interfering amplitudes [167]. Similarly, $\overline{A}_T \sim -\sin(-\phi + \delta)$, so that

$$\mathcal{A}_T \sim \sin\phi \cos\delta \; . \tag{3.120}$$

Thus, the $CP$-violating TP asymmetry does not require a strong phase difference. This is unlike direct $CP$ asymmetries, which are proportional to $\sin\phi \sin\delta$.

Another difference between triple products and direct $CP$ asymmetries is that TP's are a kinematical effect [168]. That is, in order to generate a nonzero TP asymmetry, it is not enough to have two different amplitudes with a relative weak phase. Instead, one requires two *kinematically distinct* amplitudes with a relative weak phase. Two amplitudes which are kinematically identical may generate a direct $CP$ asymmetry (if the strong phase difference is nonzero), but will not generate a TP.

There are three helicity amplitudes in $B \to V_1 V_2$ decays. These are $A_0$ and $A_\parallel$, which are $CP$-even, and $A_\perp$ which is $CP$-odd. In terms of helicity amplitudes, the decay amplitude can be written

$$M = A_0 \varepsilon_1^{*L} \cdot \varepsilon_2^{*L} - \frac{1}{\sqrt{2}} A_\parallel \vec{\varepsilon}_1^{*T} \cdot \vec{\varepsilon}_2^{*T} - \frac{i}{\sqrt{2}} A_\perp \vec{\varepsilon}_1^{*T} \times \vec{\varepsilon}_2^{*T} \cdot \hat{q} , \tag{3.121}$$

where $\hat{q}$ is the unit vector along the direction of motion of $V_2$ in the rest frame of $V_1$. Here, $\varepsilon_i^{*L} = \vec{\varepsilon}_i^{\,*} \cdot \hat{q}$, and $\vec{\varepsilon}_i^{\,*T} = \vec{\varepsilon}_i^{\,*} - \varepsilon_i^{*L}\hat{q}$. The three helicity amplitudes can be expressed in terms of the parameters $a$, $b$ and $c$ of Eq. (3.118):

$$A_\parallel = \sqrt{2}a \;, \;\; A_0 = -ax - \frac{m_1 m_2}{m_B^2}(x^2-1)b \;, \;\; A_\perp = 2\sqrt{2}\,\frac{m_1 m_2}{m_B^2}\sqrt{(x^2-1)}c \;, \tag{3.122}$$

where $x = (k_1 \cdot k_2)/(m_1 m_2)$. From their $CP$ properties, it is obvious that there are two $CP$-violating TP terms, due to the interference of $A_\perp$ with $A_0$ or $A_\parallel$.

It is well known that one can perform an angular analysis of the decay $B \to V_1 V_2$ in order to separate the $CP$-even from $CP$-odd components. What is not as well known is that one can measure TP's in this way. When one squares the amplitude of Eq. (3.121), the time-integrated differential decay rate contains six terms. Assuming that both vector mesons decay into pseudoscalars, *i.e.*, $V_1 \to P_1 P_1'$, $V_2 \to P_2 P_2'$, two of these terms are [166]

$$-\frac{\mathrm{Im}(A_\perp A_0^*)}{2\sqrt{2}} \sin 2\theta_1 \sin 2\theta_2 \sin\phi - \frac{\mathrm{Im}(A_\perp A_\parallel^*)}{2} \sin^2\theta_1 \sin^2\theta_2 \sin 2\phi \;. \tag{3.123}$$

Here, $\theta_1$ ($\theta_2$) is the angle between the directions of motion of the $P_1$ ($P_2$) in the $V_1$ ($V_2$) rest frame and the $V_1$ ($V_2$) in the $B$ rest frame, and $\phi$ is the angle between the normals to the planes defined by $P_1 P_1'$ and $P_2 P_2'$ in the $B$ rest frame. Both of these terms involve the TP $\vec{\varepsilon}_1^{\,*T} \times \vec{\varepsilon}_2^{\,*T} \cdot \hat{q}$. Thus, as expected, TP's are generated due to the interference of $A_\perp$ with $A_0$ or $A_\parallel$.

We therefore see that triple products can be observed in the angular distribution of the decay $B \to V_1 V_2$. However, note that a full angular analysis is not necessary to measure TP's. There are two distinct TP's which can be measured:

$$A_T^1 = \frac{\mathrm{Im}(A_\perp A_0^*)}{|A_0|^2 + |A_\parallel|^2 + |A_\perp|^2} \quad , \quad A_T^2 = \frac{\mathrm{Im}(A_\perp A_\parallel^*)}{|A_0|^2 + |A_\parallel|^2 + |A_\perp|^2} \;. \tag{3.124}$$

$A_0$ is expected to be the largest helicity amplitude in $B \to V_1 V_2$ decays, *i.e.*, longitudinal polarization should dominate. We therefore expect $A_T^2$ to be suppressed relative to $A_T^1$.

We now turn to an examination of triple products in specific decays. There are many such decays which can be analyzed [167]. Here we discuss two processes: $B \to J/\psi K^*$ and $B \to \phi K^*$.

We begin with a study of the Standard Model prediction for triple products in $B \to J/\psi K^*$. In order to have a nonzero TP, one needs two (kinematically distinct) decay amplitudes with a relative weak phase. The decay $B \to J/\psi K^*$ (or $J/\psi K$) is dominated by a color-suppressed tree-level $\bar{b} \to \bar{c} c \bar{s}$ diagram $C$, described by the CKM matrix elements $V_{cb}^* V_{cs}$. There is also a penguin contribution to this decay with internal $t$, $c$ and $u$ quarks. (The penguins with internal $c$ and $u$ quarks come from the rescattering of tree operators.) We can use the unitarity of the CKM matrix to eliminate the $t$-quark penguin, $P_t$, in favor of the $c$ and $u$ pieces, $P_c$ and $P_u$. The amplitude for $B \to J/\psi K^*$ can then be written as

$$A(B \to J/\psi K^*) = V_{cb}^* V_{cs} (C + P_c - P_t) + V_{ub}^* V_{us} (P_u - P_t) \;. \tag{3.125}$$

A nonzero Standard Model TP in this decay then requires that both amplitudes be nonzero.

However, note that all of the penguin contributions require the gluon to transform into a $J/\psi$, and are hence OZI suppressed. Thus, the size of the second term relative to the first is approximately given by

$$\left| \frac{V_{ub}^* V_{us}}{V_{cb}^* V_{cs}} \right| \left| \frac{P_{OZI}}{C} \right| \sim 2\% \frac{P_{OZI}}{C} \;, \tag{3.126}$$

which is tiny. Thus, to a very good approximation, the decay $B \to J/\psi K^*$ is described by a single weak decay amplitude. Therefore, within the Standard Model, no TP's are predicted in this decay. The measurement of a nonzero TP would be a smoking-gun signal of New Physics. (If a time-dependent angular analysis of $B_d^0(t) \to J/\psi K^*$ can be done, there are many more tests of New Physics, see Ref. [169].)

We now turn to $B \to \phi K^*$. This is a pure $\bar{b} \to \bar{s}s\bar{s}$ penguin decay. Its amplitude is given by

$$A(B \to \phi K^*) = V_{cb}^* V_{cs}(P_c - P_t) + V_{ub}^* V_{us}(P_u - P_t) \ . \quad (3.127)$$

As before, both contributions must be nonzero in order to generate a TP in this decay. However, there is an important difference compared to $B \to J/\psi K^*$: here, the penguin amplitude contains pieces which are not OZI suppressed. The size of the second term relative to the first is therefore

$$\left| \frac{V_{ub}^* V_{us}}{V_{cb}^* V_{cs}} \right| \left| \frac{P_u - P_c}{P_c - P_t} \right| \sim 2\% \left| \frac{P_u - P_c}{P_c - P_t} \right| \ . \quad (3.128)$$

Within factorization, we expect that $P_{u,c} \le 0.5 P_t$. Thus, the second term is $\sim 2\%$. (The fact that this term is small is what leads to the conclusion that the indirect $CP$ asymmetry in $B_d^0(t) \to \phi K_S$ should be almost equal to that in $B_d^0(t) \to J/\psi K_S$.)

Based only on the relative sizes of the contributing amplitudes, we can expect TP's in $B \to \phi K^*$ to be small, $\mathcal{O}(5\%)$. However, as we have stressed above, it is not enough to have two amplitudes with a relative weak phase – the two amplitudes must be kinematically distinct. If both $(P_c - P_t)$ and $(P_u - P_t)$ are factorizable, then they each contribute to the *same* kinematical amplitude [167]. In this case, although direct $CP$ violation may be present (if the strong phase difference is nonzero), the triple products will vanish. In order to generate a TP in $B \to \phi K^*$ within the Standard Model, we require nonfactorizable corrections to the penguin amplitudes. Furthermore, these corrections must be different for the three helicity amplitudes [167]. Taking all these factors into account, it is likely that the TP's in $B \to \phi K^*$ are quite small in the Standard Model .

As has been mentioned several times in this workshop, the indirect $CP$ asymmetries in $B_d^0(t) \to J/\psi K_S$ and $B_d^0(t) \to \phi K_S$ have been found to be different:

$$A_{CP}(B_d^0(t) \to J/\psi K_S) = 0.73 \pm 0.054 \ \ , \quad A_{CP}(B_d^0(t) \to \phi K_S) = -0.15 \pm 0.33 \ . \quad (3.129)$$

Many models of New Physics have been proposed to explain this discrepancy [170]. In all cases, it is suggested that New Physics appears in $\bar{b} \to \bar{s}s\bar{s}$ transitions, thus altering the $CP$ asymmetry in $B_d^0(t) \to \phi K_S$.

If such New Physics is present, it may also generate TP's in the decay $B \to \phi K^*$. Within factorization, New Physics which involves only the left-handed $b$-quark produces the same kinematical amplitude as the Standard Model, so that no TP can be generated. However, some types of New Physics can couple to the right-handed $b$ quark. These New Physics operators will produce different kinematical amplitudes, giving rise to a TP asymmetry [167, 171]. Thus, the measurement of a nonzero TP in $B \to \phi K^*$ would not only indicate the presence of New Physics, but it would also yield partial information about the nature of the New Physics.

As an example of this, one model which explains the data in Eq. (3.129) is supersymmetry with $R$ parity violation [172]. In this model there are new operators involving both the left-handed and right-handed components of the $b$-quark:

$$L_{eff} = X_L \, \bar{s}\gamma_\mu\gamma_R s \, \bar{s}\gamma^\mu\gamma_L b + X_R \, \bar{s}\gamma_\mu\gamma_L s \, \bar{s}\gamma^\mu\gamma_R b \ . \quad (3.130)$$

These operators will contribute to both $B \to \phi K_S$ and $B \to \phi K^*$. For the case $X_L = X_R$, we find a sizable triple product asymmetry of about $-16\%$ in $B \to \phi K^*$. If the New Physics coupling is purely right-handed, the TP will be even larger. This example demonstrates the usefulness of TP's in searching for New Physics, and diagnosing its properties.

In summary, we have presented a review of $CP$-violating triple-product correlations (TP's). Since TP's do not require a strong-phase difference between the two interfering amplitudes, they are complementary to direct $CP$ asymmetries. A particularly useful class of decays in which to search for TP's is $B \to V_1 V_2$. Here we have studied two specific decays: $B \to J/\psi K^*$ and $B \to \phi K^*$. In the Standard Model, TP's in both of these processes are expected to be very small. They are therefore a good place to search for physics beyond the Standard Model. Several models of New Physics have been proposed to explain the discrepancy between the indirect $CP$ asymmetries found in $B_d^0(t) \to J/\psi K_S^0$ and $B_d^0(t) \to \phi K_S^0$. If there is New Physics in $B \to \phi K$, it will also affect $B \to \phi K^*$. If this New Physics contains significant couplings to the right-handed $b$ quark, TP's can be generated. The measurement of TP's in $B \to \phi K^*$ will therefore be a good way of confirming the presence of New Physics, and of ruling out certain models.

### 3.8.2 Direct $CP$ violation in $B \to \phi\phi X_s$

>–M. Hazumi–<

We discuss a novel method to search for a new $CP$-violating phase in the hadronic $b \to s$ transition using $B^\pm \to \phi\phi X_s^\pm$ decays [173], where $X_s^\pm$ represents a final state with a specific strange flavor such as $K^\pm$ or $K^{*\pm}$. These non-resonant direct decay amplitudes are dominated by the $b \to s\bar{s}s\bar{s}s$ transition. A contribution from the $b \to u\bar{u}s$ transition followed by rescattering into $s\bar{s}s$ is expected to be below 1% because of the CKM suppression and the OZI rule [173]. In these decays, when the invariant mass of the $\phi\phi$ system is within the $\eta_{\rm cut}$ resonance region, they interfere with the $B^\pm \to \eta_{\rm cut}(\to \phi\phi)X_s^\pm$ decay that is dominated by the $b \to c\bar{c}s$ transition. The decay width of $\eta_{\rm cut}$ is sufficiently large [107, 174] to provide a sizable interference. Within the Standard Model, this interference does not cause sizable direct $CP$ violation because there is no weak phase difference between the $b \to s\bar{s}s\bar{s}s$ and the $b \to c\bar{c}s$ transitions. On the other hand, a New Physics contribution with a new $CP$-violating phase can create a large weak phase difference. Thus large $CP$ asymmetries can appear only from New Physics amplitudes, and an observation of direct $CP$ violation in these decays is an unambiguous manifestation of physics beyond the Standard Model . Although the same argument so far is applicable to the $B^\pm \to \phi X_s^\pm$ decays, there is no guaranteed strong phase difference that is calculable reliably for these decays. In contrast, the Breit-Wigner resonance provides the maximal strong phase difference in the case of $B^\pm \to (\phi\phi)_{m\sim m_{\eta_{\rm cut}}} X_s^\pm$ decays.

The Belle Collaboration recently announced evidence for $B \to \phi\phi K$ decays [175]. The signal purity is close to 100% when the $\phi\phi$ invariant mass is within the $\eta_{\rm cut}$ mass region. Belle [174] has also reported the first observation of the $B^0 \to \eta_{\rm cut}K^{*0}$ decay. This implies that other modes such as $B^+ \to \eta_{\rm cut}K^{*+}$ will also be seen with a similar branching fraction, so that we will be able to study semi-inclusive $B^\pm \to \eta_{\rm cut}X_s^\pm$ transitions experimentally. The semi-inclusive branching fraction of $B^\pm \to \eta_{\rm cut}X_s^\pm$ is not yet measured, but is theoretically expected to be comparable to the branching fraction of the semi-inclusive decay $B^\pm \to J/\psi X_s^\pm$ [176].

We derive the rates and the asymmetry of the decays $B^\pm \to (\phi\phi)_{m\sim m_{\eta_{\rm cut}}} X_s^\pm$ based on the formalism described in the study of $B^\pm \to \eta_{\rm cut}(\chi_{c0})\pi^\pm$ decays [177]. The distribution of two $\phi$'s is determined with two kinematical variables; one is the invariant mass of the $\phi\phi$ system, $m$, and the other is the angle $\theta$ between the $B$-meson momentum and the momentum of one of two $\phi$'s in the center-of-mass frame of the $\phi\phi$ system. To have the interference between resonant and direct amplitudes, $m$ should be in the $\eta_{\rm cut}$ resonance region. To be specific, we require in this study that the difference between $m$ and $\eta_{\rm cut}$ mass ($M$) should satisfy $|m - M| < 3\Gamma$, where $\Gamma$ is the the width of the $\eta_{\rm cut}$ resonance. (In this study we take $M = 2980$ MeV/$c^2$ and $\Gamma = 29$ MeV, which are the values from the recent measurements by the Belle collaboration [174].) The differential decay rate normalized with the total $B^\pm$ decay rate is then given by the following equation:

$$\frac{1}{\Gamma_B}\frac{d\Gamma^\pm}{dz} = \int_{(M-3\Gamma)^2}^{(M+3\Gamma)^2} ds|R(s) + D^\pm(s,z)|^2 \,, \tag{3.131}$$

where $R(s)$ is the resonant amplitude, $D^\pm(s,z)$ is the direct amplitude of the $B^\pm \to \phi\phi X_s^\pm$ decay, $\Gamma_B$ is the total $B$ decay rate, $s \equiv m^2$ and $z \equiv \cos\theta$.

The resonant amplitude $R(s)$ is given by

$$R(s) \equiv A(B^\pm \to \eta_{\rm cut}X_s^\pm \to \phi\phi X_s^\pm) = \frac{a_{\rm R}\sqrt{M\Gamma}}{(s - M^2) + iM\Gamma} \,, \tag{3.132}$$

where $a_{\rm R}$ is a product of the weak decay amplitude of $B^\pm \to \eta_{\rm cut}X_s^\pm$ and the real part of the $\eta_{\rm cut}$ decay amplitude to $\phi\phi$.

The direct amplitude $D^\pm$ is separated into contributions from the Standard Model , $D_{\rm SM}$, and from New Physics, $D^\pm_{\rm NP}$,

$$D^\pm(s \approx M^2, z) \equiv D_{\rm SM}(s \approx M^2, z) + D^\pm_{\rm NP}(s \approx M^2, z), \tag{3.133}$$

$$D_{\rm SM}(s \approx M^2, z) \equiv \frac{a_{\rm D}(z)}{\sqrt{M\Gamma}}\, e^{i\delta}, \tag{3.134}$$

$$D^{\pm}_{\rm NP}(s \approx M^2, z) \equiv \frac{a_{\rm NP}(z)}{\sqrt{M\Gamma}}\, e^{i\delta'} e^{\pm i\theta_{\rm NP}}, \tag{3.135}$$

where $a_{\rm D}(z)$ is a real part of the Standard Model direct amplitude, $\delta$ ($\delta'$) is a strong phase difference between the resonant amplitude and the Standard Model (New Physics) direct amplitude, $a_{\rm NP}(z)$ is a real part of the New Physics amplitude and $\theta_{\rm NP}$ is a new $CP$-violating phase. If $\delta \neq \delta'$ holds, direct $CP$ violation can also occur from an interference between the Standard Model and New Physics direct amplitudes. We do not take this case in our study and assume $\delta = \delta'$ in the following discussion.

The difference between the decay rates of $B^+$ and $B^-$ is given by

$$\frac{1}{\Gamma_B}(\frac{d\Gamma^+}{dz} - \frac{d\Gamma^-}{dz}) \equiv \gamma^-(z) \cong -4\pi a_{\rm R} a_{\rm NP}(z)\cos\delta \cdot \sin\theta_{\rm NP}\ . \tag{3.136}$$

Similarly the sum of two decay rates is given by

$$\begin{aligned}\frac{1}{\Gamma_B}(\frac{d\Gamma^+}{dz} + \frac{d\Gamma^-}{dz}) \equiv \gamma^+(z) &\cong 2\pi a_{\rm R}^2 + 24 a_{\rm D}^2(z)(r^2 + 2r\cos\theta_{\rm NP} + 1)\\ &\quad -4\pi a_{\rm R} a_{\rm D}(z)(r\cos\theta_{\rm NP} + 1)\sin\delta,\end{aligned} \tag{3.137}$$

where $r \equiv a_{\rm NP}(z)/a_{\rm D}(z)$ is the amplitude ratio of New Physics to the Standard Model . The $z$ dependence of $r$ reflects the spin components of the $\phi\phi$ system, which can be determined at Super $B$ Factoriesin the future, from the differential decay rates in the mass-sideband region below the $\eta_{\rm cut}$ resonance. Although only a pseudoscalar component in the direct transition interferes with the $\eta_{\rm cut}$ resonance, the effect of other components can be estimated by such a measurement. Thus, for simplicity, we assume that the direct transition is dominated by a pseudo-scaler component and ignore the $z$ dependence of $r$ in the following discussion. The maximum asymmetry is realized when $\cos\delta \simeq 1$ is satisfied. Assuming that $\delta$ is small following the discussion by Eilam, Gronau and Mendel [177], the differential partial rate asymmetry is

$$A_{CP}(z) \equiv \frac{\gamma^-(z)}{\gamma^+(z)} \cong \frac{-4\pi a_{\rm R} a_{\rm NP}(z)\sin\theta_{\rm NP}}{2\pi a_{\rm R}^2 + 24 a_{\rm D}^2(z)(r^2 + 2r\cos\theta_{\rm NP} + 1)}\ . \tag{3.138}$$

As a measure of $CP$ violation, we define the following $CP$ asymmetry parameter:

$$A_{CP} \equiv \sqrt{\frac{\int_{-1}^{1} dz \gamma^-(z)^2}{\int_{-1}^{1} dz \gamma^+(z)^2}}. \tag{3.139}$$

The numerator of $A_{CP}$ can be expressed with the branching fraction of the resonance ($2\pi a_{\rm R}^2$) and that of New Physics in the resonance region ($\mathcal{B}_{\rm NP}$):

$$\int_{-1}^{1} dz \gamma^-(z)^2 = (2\pi a_{\rm R}^2)\cdot \mathcal{B}_{\rm NP} \cdot \frac{2\pi}{3}\sin^2\theta_{\rm NP}\ , \tag{3.140}$$

$$2\pi a_{\rm R}^2 \cong \mathcal{B}(B^\pm \to \eta_{\rm cut} X_s^\pm)\cdot \mathcal{B}(\eta_{\rm cut} \to \phi\phi) = (2 \sim 5)\times 10^{-5}\ , \tag{3.141}$$

and

$$\mathcal{B}_{\rm NP} \equiv \frac{1}{M\Gamma}\int_{(M-3\Gamma)^2}^{(M+3\Gamma)^2} ds \int_{-1}^{1} dz a_{\rm NP}^2(z) \leq 5\times 10^{-6}, \tag{3.142}$$

where the estimations are given in Ref. [173]. The bound on $\mathcal{B}_{\rm NP}$ corresponds to $r^2 \leq 5$. From (3.136)-(3.142), we obtain

$$\begin{aligned}A_{CP} \leq &\sqrt{\mathcal{B}_{\rm NP} \times \left(\mathcal{B}(B^\pm \to \eta_{\rm cut} X_s^\pm)\cdot \mathcal{B}(\eta_{\rm cut} \to \phi\phi) + 2(1 + 2r^{-1}\cos\theta_{\rm NP} + r^{-2})\mathcal{B}_{\rm NP}\right)^{-1}}\\ &\times\sqrt{\pi/3}\cdot|\sin\theta_{\rm NP}| \sim 0.40\cdot|\sin\theta_{\rm NP}|\ .\end{aligned} \tag{3.143}$$

A large *CP* asymmetry of 0.4 is allowed. The asymmetry is roughly proportional to $|r|$. Therefore it can be sizable even with $r^2 < 1$; for example, $A_{CP} \sim 0.1$ is allowed for $r^2 = 0.3$.

We perform a Monte Carlo simulation for the $B^{\pm} \to \phi\phi K^{\pm}$ decay and estimate statistical errors on the *CP* asymmetry parameter. For this decay mode, the background level is small enough to be neglected [175]. The reconstruction efficiency and the $\phi\phi$ mass resolution are estimated using a GEANT-based detector simulator for the Belle detector [178]. Assuming the branching fractions given in (3.141) and (3.142), we obtain ~300 events for $\mathrm{N}_B = 10^9$, where $\mathrm{N}_B$ is the number of charged $B$ mesons recorded by a detector. We perform an unbinned maximum-likelihood fit to the differential decay rate distribution, which is proportional to $|R(s) + D^{\pm}(s,z)|^2$, instead of integrating the distribution. We choose the following two free parameters in the fit: $\mathcal{A}^0_{CP} \equiv -2r(a_{\mathrm{D}}/a_{\mathrm{R}}) \sin\theta_{\mathrm{NP}}$ and $\mathcal{B} \equiv a^2_{\mathrm{D}}(r^2 + 2r\cos\theta_{\mathrm{NP}} + 1)$. $\mathcal{A}^0_{CP}$ is the *CP* asymmetry in the Breit-Wigner term. $\mathcal{B}$ is proportional to the branching ratio of the non-resonant $B^{\pm} \to \phi\phi K^{\pm}$ decay below the $\eta_{\mathrm{cut}}$ mass region. The statistical error for $\mathcal{A}^0_{CP}$ is estimated to be $\delta\mathcal{A}^0_{CP} \sim 0.06$. Figure 3-32 shows the $5\sigma$ search regions for $\mathrm{N}_B = 10^9$ (dotted line) and for $\mathrm{N}_B = 10^{10}$ (solid line), which will be accessible at next-generation high-luminosity $e^+e^-$ $B$ factories. Direct *CP* violation will be observed in a large parameter space above a $5\sigma$ significance. We also repeat the fit procedure described above with the branching fractions reported in Ref. [175]. Although the smaller value for $\mathcal{B}(\eta_{\mathrm{cut}} \to \phi\phi)$ results in the smaller number of signal events, the *CP* asymmetry from the interference between the resonant and the New Physics amplitudes becomes larger. We find that the change in $\mathcal{B}(\eta_{\mathrm{cut}} \to \phi\phi)$ does not largely affect the significance; the difference is less than 10% for $r^2 = 0.5$ and $\sin\theta_{\mathrm{NP}} = 1$.

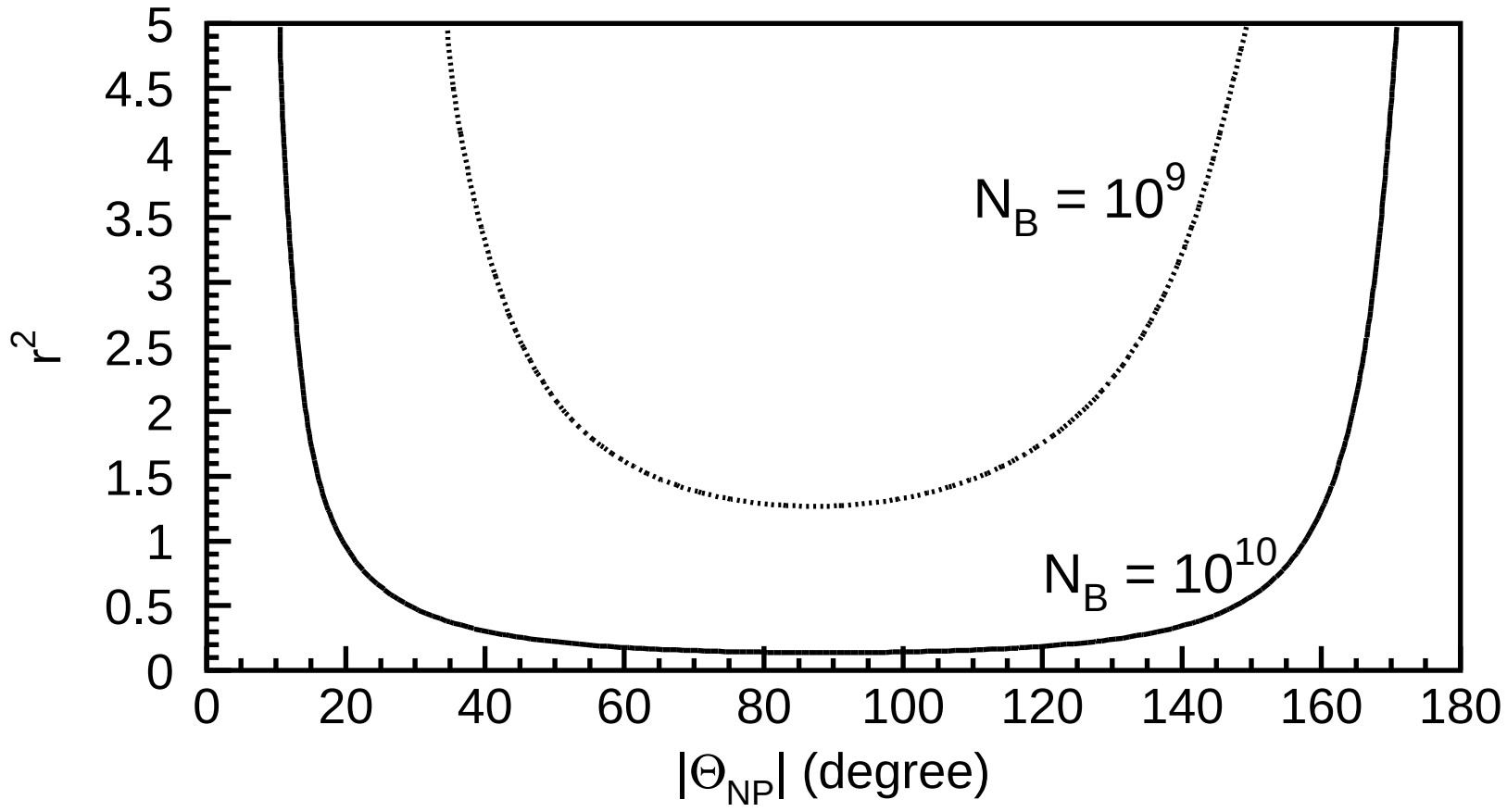

**Figure 3-32.** *Expected sensitivities on direct CP violation in the $B^{\pm} \to \phi\phi K^{\pm}$ decay for $10^9$ $B$ mesons (dotted line) and $10^{10}$ $B$ mesons (solid line). In the regions above the curves, direct CP violation can be measured with a $5\sigma$ significance or larger.*

The new *CP*-violating phase $\theta_{\mathrm{NP}}$ also affects time-dependent *CP*-violating asymmetries $A_{CP}(t) = \mathcal{S}\sin(\Delta m_d t) + \mathcal{A}\cos(\Delta m_d t)$ in $B^0 \to \phi K^0_S$ and related decays. Here $\Delta m_d$ is the mass difference between the two $B^0$ mass eigenstates, and $\mathcal{S}$ and $\mathcal{A}$ are parameters for mixing-induced *CP* violation and direct *CP* violation, respectively. Ignoring a strong phase difference between the amplitude of New Physics ($A_{\mathrm{NP}}$) and Standard Model ($A_{\mathrm{SM}}$), we obtain

$$\mathcal{S} = \frac{\sin 2\phi_1 + 2\rho\sin(2\phi_1 + \theta_{\mathrm{NP}}) + \rho^2\sin(2\phi_1 + 2\theta_{\mathrm{NP}})}{1 + \rho^2 + 2\rho\cos\theta_{\mathrm{NP}}}, \tag{3.144}$$

where $\rho \equiv A_{\mathrm{NP}}/A_{\mathrm{SM}}$ is an amplitude ratio of New Physics to the Standard Model and $\phi_1$ is one of the angles of the unitarity triangle. In particular, a difference in $\mathcal{S}$ between $B^0 \to \phi K^0_S$ and $B^0 \to J/\psi K^0_S$ decays, *i.e.*, $\Delta\mathcal{S} \equiv \mathcal{S}(\phi K^0_S) - \mathcal{S}(J/\psi K^0_S) \neq 0$, would be a clear signal of the new phase since $\mathcal{S}(J/\psi K^0_S) = \sin 2\phi_1$ is held to a good approximation. We define expected statistical significance of the deviation from the Standard Model by $\mathcal{A}^0_{CP}/\delta\mathcal{A}^0_{CP}$ for the $B^\pm \to \phi\phi K^\pm$ decay and by $\Delta\mathcal{S}/\delta\Delta\mathcal{S}$ for the $B^0 \to \phi K^0_S$ decay, where $\delta\Delta\mathcal{S}$ is an expected statistical error of $\Delta\mathcal{S}$ extrapolated from the latest result by the Belle experiment [179]. Although $r^2$ is not necessarily equal to $\rho^2$, both decays are governed by the same $b \to s\bar{s}s$ transition. Therefore it is reasonable to choose $r^2 = \rho^2$ for comparison. Figure 3-33 shows the resulting significance for $10^{10}$ $B$ mesons and with $r^2 = \rho^2 = 0.5$.

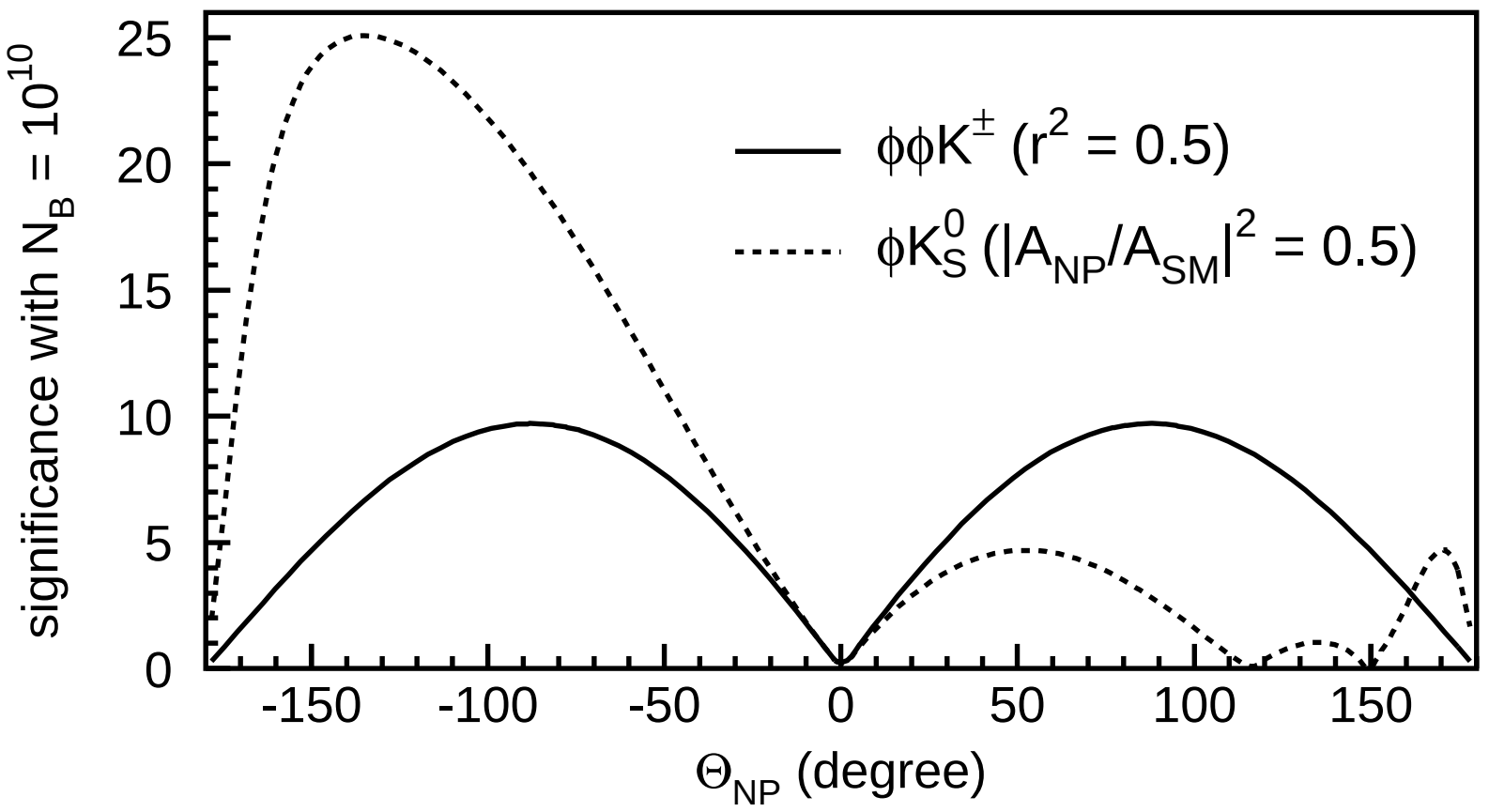

**Figure 3-33.** *Expected statistical significance of deviations from the Standard Model for direct CP violation in the $B^\pm \to \phi\phi K^\pm$ decay with $r^2 = 0.5$ (solid line) and for time-dependent CP violation in the $B^0 \to \phi K^0_S$ decay with $|A_{\rm NP}/A_{\rm SM}|^2 = 0.5$ (dashed line). For each case, significance is calculated with $10^{10}$ $B$ mesons.*

The significance for $\Delta\mathcal{S}$ largely depends on the sign of $\theta_{\rm NP}$, which is not the case for the $B^\pm \to \phi\phi K^\pm$ decay. The sign dependence arises from an asymmetric range for $\Delta\mathcal{S}$; to a good approximation, we have $-1 - \sin 2\phi_1 \leq \Delta\mathcal{S} \leq 1 - \sin 2\phi_1$ where $\sin 2\phi_1 = +0.736 \pm 0.049$ [95]. Therefore the $B^\pm \to \phi\phi K^\pm$ decay plays a unique role in searching for a new $CP$-violating phase.

In the above estimation, we use parameters that have uncertainties. However, they can in principle be measured precisely if a sufficient number of $B$ mesons are produced. In our estimation, we assume efficiencies and background levels that have been achieved with the Belle detector at the KEK $B$ Factory. They depend on the actual detector performance and beam conditions, which might be different at a Super $B$ Factory with higher luminosity. Detailed simulation studies as well as some extrapolation from data at current $B$ Factories will be needed for further quantitative evaluation.

Experimental sensitivities can be improved by adding more final states. The technique to reconstruct $X_s$, which has been successfully adopted for the measurements of semi-inclusive $B \to X_s\ell\ell$ transitions [180], can be used for this purpose. Flavor-specific neutral $B$ meson decays, such as $B^0 \to \phi\phi K^{*0}(\to K^+\pi^-)$, and other charmonia such as the $\chi_{c0} \to \phi\phi$ decay can also be included.

## 3.9 Hadron Machines

### 3.9.1 LHC*b*

>– U. Egede, on behalf of the LHC*b* Collaboration –<

With the hadronic $B$ factories currently under construction or in the design phase $B$ physics will enter a new era. The LHC$b$ experiment is planned to start taking data in April 2007. The aim of the experiment is to gain a comprehensive understanding of the CKM matrix for discovering physics beyond the Standard Model. The much larger statistics and the access to $B_s$ decays will allow to many cross checks of $CP$ violation that are not possible at the current $B$ factories.

By 2007 the current $e^+e^-$ $B$ factories will have collected samples of the order of $10^9$ $B$ meson decays. This, combined with the data from the Tevatron, will give a precision on the value of the CKM angle $\beta$ of $\sigma(\sin 2\beta) = \mathcal{O}(10^{-2})$ which is close to the systematic uncertainty from penguin pollution in the channel $B \to J/\psi K^0_S$. At the same time the anticipated measurement of $B_s$ mixing will improve the value of $|V_{td}|/|V_{ts}|$ from the partial cancellation of the $B^0$ and $B_s$ form factors, thus giving an improved measurement for the apex of the unitarity triangle. New Physics contributing to $B$ mixing will require independent measurements of the unitarity triangle to reveal itself; measurements of the $CP$ angle $\gamma$ are well suited for this.

**Table 3-12.** *A summary of the experimental conditions for the LHCb experiment.*

| | |
|---|---|
| Beam type | $p$-$p$ |
| $\sqrt{s}$ | 14 TeV |
| $\sigma_{b\bar{b}}$ | 500 $\mu$b |
| $\sigma_{c\bar{c}}$ | 3.5 mb |
| $\sigma_{\rm inelastic}$ | 80 mb |
| $B^+/B^0/B_s/\Lambda_b$ mixture | 40/40/10/10 |
| Bunch separation | 25 ns |
| Size of collision region | 5.3 cm |
| Pseudorapidity coverage | 2.1–5.3 |
| $\mathcal{L}$ | $2 \times 10^{32}$ cm$^{-2}$s$^{-1}$ |
| $<n>$ per bunch crossing | 0.5 |
| $n_{b\bar{b}}$ per $10^7$ s | $10^{12}$ |

In Table 3-12 the experimental conditions for LHC$b$ are summarized. Several comments are in order:

- At LHC the ratio between the $b\bar{b}$ cross section and the total inelastic cross section is very small but still equivalent to the $\sigma_{c\bar{c}}/\sigma_{\rm inelastic}$ ratio at earlier successful fixed-target charm experiments.
- The large production of $B_s$, $\Lambda_b$ and $B_c$ will open up entirely new areas of $B$ physics where the present data samples are very limited.
- The optimal luminosity for LHC$b$ is $2 \times 10^{32}$ cm$^{-2}$s$^{-1}$ where single interactions in the bunch crossings dominate. As this is much lower than even the initial LHC luminosity it will be reached quickly and after that kept constant through local detuning of the beams. We thus expect a fast exploitation of the full physics programme.
- The ATLAS and CMS experiments do not have $B$ physics as their primary goal; they have a much lower trigger bandwidth dedicated to $B$ physics and no dedicated system for kaon-pion separation.

### Detector design

For most studies of $CP$ violation in $B$ meson decays we must identify the flavor of the $B$ meson at production time. The dominant contribution to this flavor tagging is through identification of particles from the decay of the other $B$-hadron created in the event. Hence the detector needs to be designed such that a significant part of the produced pairs of $B$ hadrons both end up within the detector acceptance. The most cost-effective solution to this is to build a detector that sits as much in the forward region as technology allows. As both $B$ hadrons tend to be boosted in the same direction there is no synergetic effect from covering both forward regions. This leads to the design of the LHC$b$ detector as a single-arm forward spectrometer.

The overall design of the LHC$b$ detector is shown in Fig. 3-34. The most essential parts of the detector are: the trigger system which reduces the rate of events going to mass storage to an acceptable level; the vertex detector which provides the trigger with secondary vertex identification and the physics with the ability to resolve $B_s$ oscillations; and the particle identification system which provides the essential pion-kaon separation required for $CP$ violation studies.

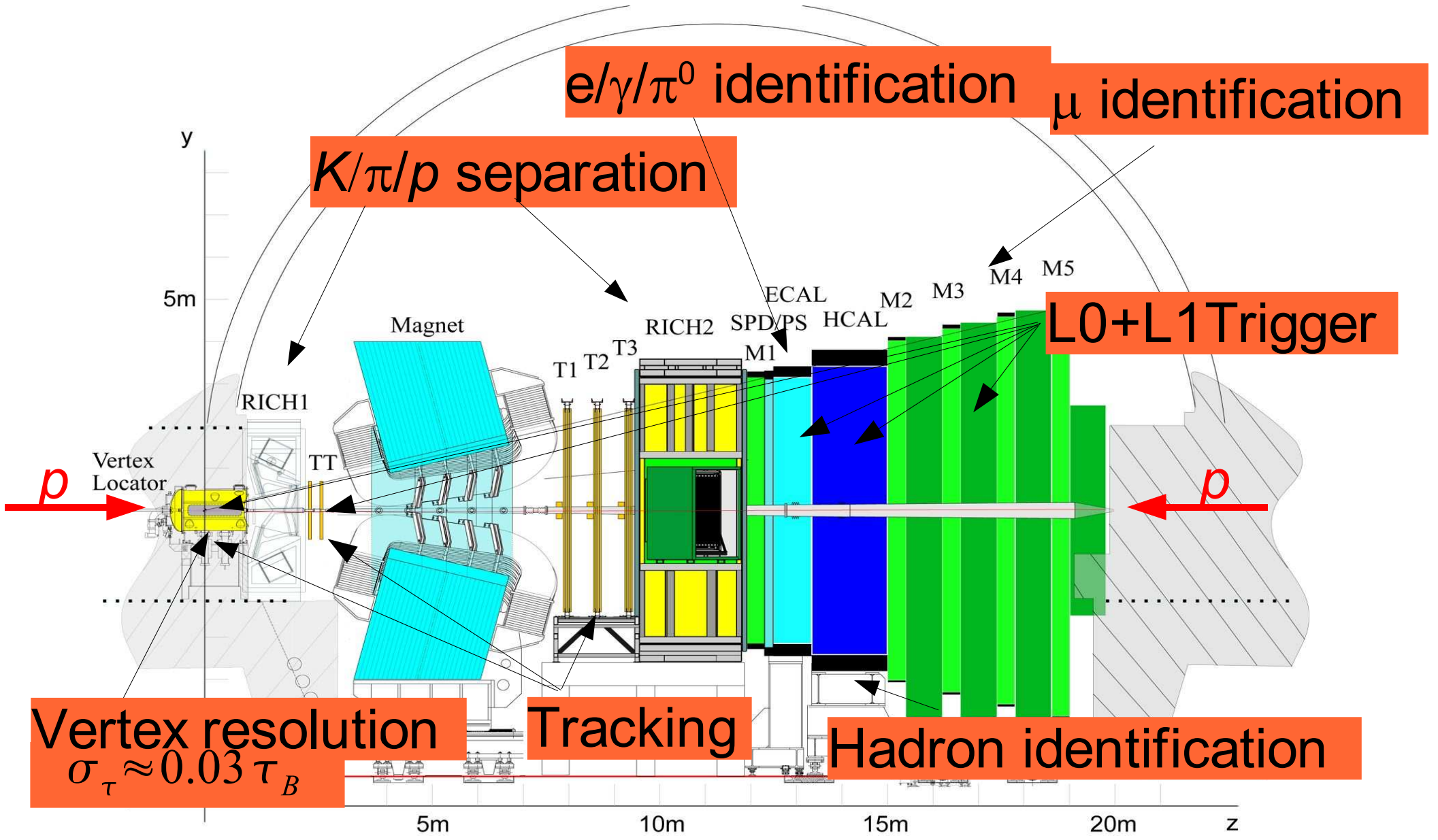

**Figure 3-34.** *The design of the LHC$b$ detector. The collision point for the protons is within the vertex detector to the left in this drawing.*

The single most demanding task for a hadronic $B$ physics experiment is the trigger. The combination of a cross section for minimum bias, which is orders of magnitude larger than the $b$ cross section, with the rare $B$ decays which are of interest, requires a sophisticated trigger that can suppress rates by many orders of magnitude. With a rate of around $10^{12}$ $B$ hadrons produced in a year the trigger also have to be selective. This is a very different situation to current $e^+e^-$ colliders where all $B$ decays are recorded.

There are three main elements that allow identification of events with a $B$ hadron:

- Large transverse energy or momenta with respect to the beam axis. This is simply an indicator of a high mass particle decaying.
- Vertices which are displaced from the primary vertex. This takes advantage of the long lifetime of $B$ hadrons compared to other hadrons produced ($K^0_S$ and $\Lambda$ live much longer, and do not interfere with the trigger).

- High energy leptons, produced either either singly from semileptonic $B$ decays or in pairs from $B$ hadrons with a $J/\psi$ in the decay chain. This will also be the trigger for rare $B \to \mu^+\mu^-$ decays.

In LHC$b$ the aim of the Level 0 trigger is to identify events with particles of high transverse energy or momentum with respect to the beam axis. The high $p_T$ and $E_T$ arises from the decay of high mass objects and thus favor $B$ hadrons to the background events with lower mass hadrons. In addition these are the type of $B$ mesons events that the final selection of events for physics analysis favor.

A combination of the impact parameters of the tracks and their $p_T$ is used at Level 1 to identify secondary vertices. The $p_T$ measurement is done with the use of the vertex detector and a tracking station placed inside the fringe field of the magnet. The impact parameter is calculated with respect to the primary vertex in the event.

The High Level Trigger identifies more specific classes of $B$ decays using the results of the online reconstruction. Table 3-13 contains an overview of the trigger rates at the different levels. Further information on the LHC$b$ trigger can be found in the recently published trigger TDR [181].

**Table 3-13.** *An overview of the expected rates at the different trigger levels of LHC$b$.*

| Trigger level | Main discriminator | Ingoing rate |
|---|---|---|
| Level 0 | High $p_T$, high $E_T$ | 40 MHz |
| Level 1 | Impact parameter, $m_{\mu\mu}$ | 1 MHz |
| High Level Trigger | Physics algorithms | 40 kHz |
| To mass storage | | 200 Hz |

To make hadronic final states useful for $CP$ violation studies, good separation power between pions and kaons is required. In LHC$b$ this is accomplished using a RICH detector system with three different radiators providing kaon-pion separation for tracks from 2–100 GeV/$c$.

An example illustrating the need for particle identification is the $B_s \to D_s^{\mp}K^{\pm}$ decay to be used for the extraction of the angle $\gamma$. The decay $B_s \to D_s^-\pi^+$ is expected to have a branching fraction 12 times larger than the same decay with a bachelor kaon, thus drowning the $B_s \to D_s^{\mp}K^{\pm}$ signal without any particle identification. In Fig. 3-35 we illustrate the particle identification capability of LHC$b$ to isolate the $B_s \to D_s^{\mp}K^{\pm}$ signal. For the two-body $B$ meson decays the kaon-pion separation is also essential for the extraction of the angle $\gamma$ from the individual measurements of $B^0 \to \pi^+\pi^-$ and $B_s \to K^+K^-$ decays.

In addition, kaon identification is one of the dominant sources for flavor tagging. This can either be through identifying the charge of a kaon from the decay of the other $B$ created in the event or for the tagging of $B_s$ decays from charged kaons created adjacent to the $B_s$ in the fragmentation. The current estimates for the effective flavor tagging efficiency is around 4% for $B^0$ decays and 6% for $B_s$ decays.

### Physics reach

The aim of giving numbers for the physics reach before the start-up of experiments is to assure that the detector design is able to give the promised results in a selection of channels that are thought to be representative of the physics that will be of interest in 2007 and beyond. No attempt has been made to be comprehensive. We show a summary of annual yields in Table 3-14. All numbers in this section are taken from [182].

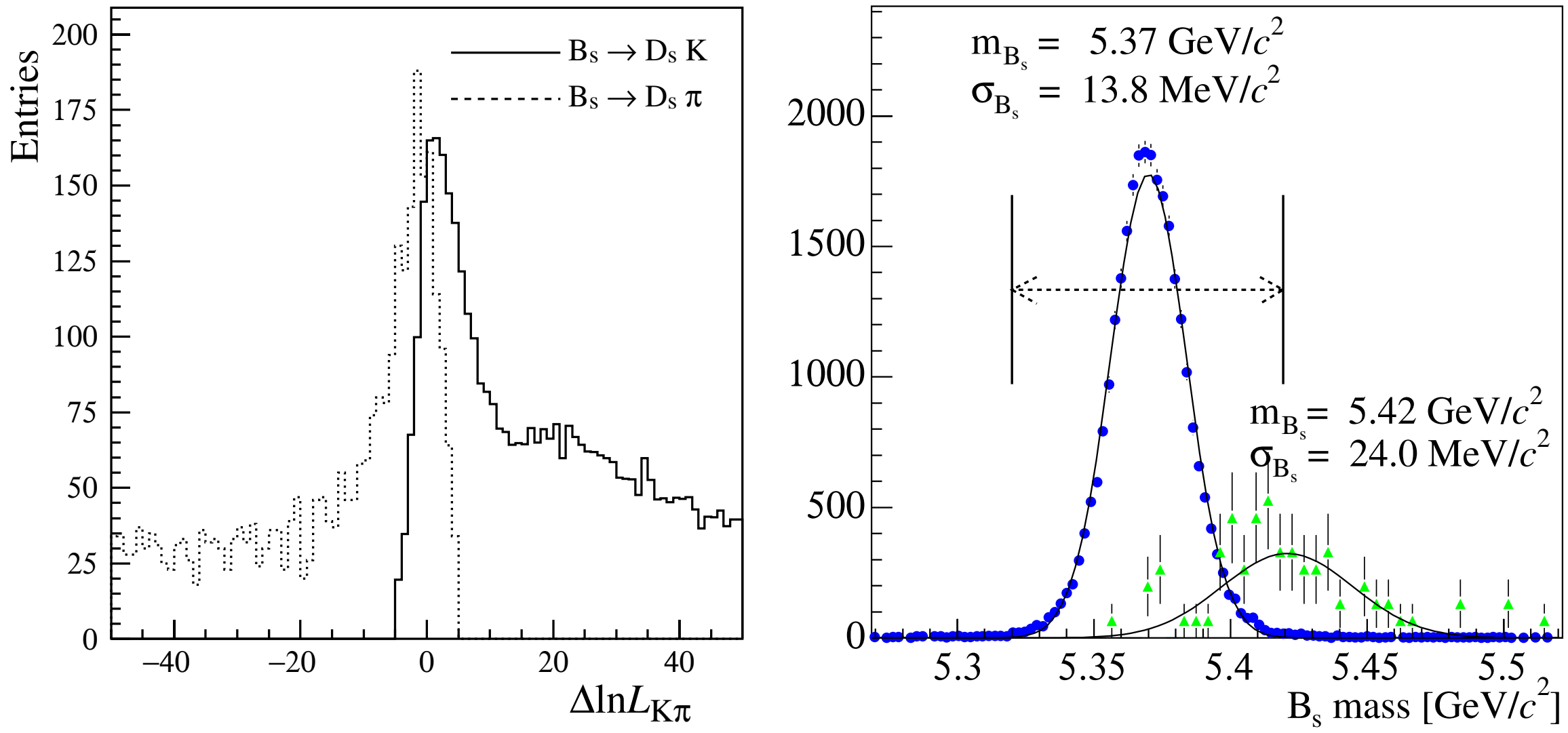

**Figure 3-35.** *To the left the log likelihood difference between a kaon and pion hypothesis of the bachelor kaon (full) and pion (dashed) from the $B_s \to D_s^{\mp} K^{\pm}$ decay with an arbitrary normalization between the two curves. To the right the resulting $B_s \to D_s^{\mp} K^{\pm}$ signal after a cut on the likelihood difference at 5 for the bachelor kaon. The correctly normalized $B_s \to D_s^- \pi^+$ background is seen as the small peak to the right.*

Within the Standard Model, the weak phase $\phi_s$ in $B_s$ mixing is given by the small value $-2\chi \equiv -2\lambda^2\eta$. This means that New Physics could easily show up as a larger value of *CP* violation in a decay like $B_s \to J/\psi\,\phi$, which is equivalent to the $B^0 \to J/\psi\,K_S^0$ decay for the measurement of the phase $2\beta$ in $B^0$ mixing. The precision in the angle $\phi_s$ will depend on how fast the oscillation frequency is for $B_s$ mixing; for $\Delta m_s = 20\ \mathrm{ps}^{-1}$ we estimate $\sigma(\phi_s) = 0.06$. LHC$b$ will be able to detect $B_s$ mixing at the $5\sigma$ level as long as $\Delta m_s < 68\ \mathrm{ps}^{-1}$ and have a resolution in $\Delta\Gamma_s/\Gamma_s$ of around 0.02.

Extraction of the angle $\gamma$ is possible through multiple decay modes at LHC$b$ each with their own advantages. An overview of the expected sensitivity for 3 different methods is given in Table 3-15.

The decay $B_s \to D_s^{\mp} K^{\pm}$ is sensitive to the angle $\phi_s + \gamma$, where the $\phi_s$ part comes from $B_s$ mixing and the $\gamma$ part from the phase of $V_{ub}$ in the tree level decay. If New Physics contributes to $\phi_s$ it will be the same contribution as for the direct measurement of $\phi_s$ and as such will not interfere with a clean measurement of $\gamma$ from the tree level decay. The decay $B^0 \to D^{*-}\pi^+$ is the equivalent decay for $B^0$, but suffers from the problem that one of the interfering decays is doubly-Cabibbo-suppressed with respect to the other; the increased statistics in this channel due to the large branching fraction will more or less cancel the deterioration in sensitivity from the doubly-Cabibbo-suppressed amplitude leading to a similar overall sensitivity to $\gamma$. A particular problem to deal with is the ambiguities from the extraction of $\gamma$ due to the fact that we measure $\sin(\phi + \gamma \pm \delta)$ where $\phi$ is the phase in the $B$ mixing and $\delta$ the strong phase difference. This gives an 8-fold ambiguity in the measurement potentially leading to a situation where New Physics cannot be distinguished from the Standard Model. In LHC$b$ where $B^0$ and $B_s$ decays can be combined, the ambiguity reduces to a two-fold ambiguity, since $\phi_s$ and $\phi_d$ are different. If $\Delta\Gamma_s/\Gamma_s$ is sufficiently large to be measured, the ambiguity will only be two-fold from measurements of the $B_s \to D_s^{\mp} K^{\pm}$ decay alone.

Another method for extracting $\gamma$ with high precision is from a comparison of $B_s \to K^+K^-$ and $B^0 \to \pi^+\pi^-$ [183] under the assumption of $U$ spin symmetry. This method is sensitive to new phases introduced in the penguin decays and as such might not measure the Standard Model value of $\gamma$.

The final method for $\gamma$ to be mentioned here is from the decay $B^0 \to D^0 K^{*0}$ where the decay rate into both the Cabibbo-favored $D^0 \to K^-\pi^+$ and the Cabibbo-suppressed *CP* eigenstate $D^0 \to K^+K^-$ are measured [1, 70]. The method benefits from that only six different decay rates need to be measured so no flavor tagging involved but on the other hand we will only see about 600 events per year reconstructed in the Cabibbo-suppressed channel.

**Table 3-14.** *Summary of the signal efficiencies, untagged annual signal yields and background-over-signal ($B/S$) ratios from inclusive $b\bar{b}$ events for LHCb. The detector efficiency $\varepsilon_{\rm det}$ includes the geometrical acceptance and material effects while $\varepsilon_{\rm tot}$ is the efficiency before flavor tagging. The annual signal yields include both the indicated decays and their charge conjugates. Quoted errors on $B/S$ are from the Monte Carlo statistics; estimates based on less than 10 Monte Carlo background events are quoted as 90% CL upper limits.*

| Decay channel | Efficiency (in %) $\varepsilon_{\rm det}$ | $\varepsilon_{\rm tot}$ | Assumed visible $\mathcal{B}$ (in $10^{-6}$) | Annual signal yield ($\times 10^3$) | $B/S$ ratio from inclusive $b\bar{b}$ background |
|---|---|---|---|---|---|
| $B^0 \to \pi^+\pi^-$ | 12.2 | 0.688 | 4.8 | 26. | $< 0.7$ |
| $B^0 \to K^+\pi^-$ | 12.2 | 0.94 | 18.5 | 135. | $0.16 \pm 0.04$ |
| $B_s \to \pi^+K^-$ | 12.0 | 0.548 | 4.8 | 5.3 | $< 1.3$ |
| $B_s \to K^+K^-$ | 12.0 | 0.988 | 18.5 | 37. | $0.31 \pm 0.10$ |
| $B^0 \to \rho\pi$ | 6.0 | 0.028 | 20. | 4.4 | $< 7.1$ |
| $B^0 \to D^{*-}\pi^+$ | 9.4 | 0.370 | 71. | 206. | $< 0.3$ |
| $B^0 \to \overline{D}^0(K\pi)K^{*0}$ | 5.3 | 0.354 | 1.2 | 3.4 | $< 0.5$ |
| $B^0 \to D^0_{CP}(KK)K^{*0}$ | 5.2 | 0.390 | 0.19 | 0.59 | $< 2.9$ |
| $B_s \to D_s^-\pi^+$ | 5.4 | 0.337 | 120. | 80. | $0.32 \pm 0.10$ |
| $B_s \to D_s^{\mp}K^{\pm}$ | 5.4 | 0.269 | 10. | 5.4 | $< 1.0$ |
| $B^0 \to J/\psi(\mu\mu)K^0_S$ | 6.5 | 1.39 | 19.8 | 216. | $0.80 \pm 0.10$ |
| $B^0 \to J/\psi(ee)K^0_S$ | 5.8 | 0.164 | 20.0 | 25.6 | $0.98 \pm 0.21$ |
| $B^0 \to J/\psi(\mu\mu)K^{*0}$ | 7.2 | 1.462 | 59. | 670. | $0.17 \pm 0.03$ |
| $B^+ \to J/\psi(\mu\mu)K^+$ | 11.9 | 3.28 | 68. | 1740. | $0.37 \pm 0.02$ |
| $B_s \to J/\psi(\mu\mu)\phi$ | 7.6 | 1.672 | 31. | 100. | $< 0.3$ |
| $B_s \to J/\psi(ee)\phi$ | 6.7 | 0.315 | 31. | 20. | $0.7 \pm 0.2$ |
| $B_s \to J/\psi(\mu\mu)\eta$ | 10.1 | 0.461 | 7.6 | 7.0 | $< 5.1$ |
| $B_s \to \eta_c\phi$ | 2.6 | 0.078 | 21. | 3.2 | $< 1.4$ |
| $B_s \to \phi\phi$ | 6.7 | 0.470 | 1.3 | 1.2 | $< 0.4$ |
| $B^0 \to \mu^+\mu^-K^{*0}$ | 7.2 | 0.704 | 0.8 | 4.4 | $< 2.0$ |
| $B^0 \to K^{*0}\gamma$ | 9.5 | 0.156 | 29. | 35. | $< 0.7$ |
| $B_s \to \phi\gamma$ | 9.7 | 0.220 | 21.2 | 9.3 | $< 2.4$ |
| $B_c^+ \to J/\psi(\mu\mu)\pi^+$ | 11.5 | 1.30 | 680. | 14.0 | $< 0.8$ |

**Table 3-15.** *The expected LHCb sensitivity to the angle $\gamma$ after one year. These numbers are for an expected angle $\gamma = 65^\circ$.*

| Channel | Sensitivity | Comment |
|---|---|---|
| $B_s \to D_s^{\mp}K^{\pm}$ | $14^\circ$ | $B_s$ equivalent of $B^0 \to D^{*\pm}\pi^{\mp}$ |
| $B_s \to K^+K^-$/ $B^0 \to \pi^+\pi^-$ | $5^\circ$ | Relies on $U$ spin symmetry |
| $B^0 \to D^0_{CP}(KK)K^{*0}$ | $8^\circ$ | Might be affected by New Physics in $D^0$ decays |

### Systematics

In LHC$b$ it is necessary to control all effects that can produce a flavor asymmetry and thereby fake $CP$ violation. There are several penitential sources for a flavor asymmetry:

- Since LHC is a proton-proton machine the angular distributions and relative ratio of $B$ and $\overline{B}$ hadrons for a given type will be different at the percent level which is larger than some of the effects we want to measure.
- The tracking efficiency for positive and negative particles will be different due to the magnetic dipole field (positive and negative particles go through different parts of the detector).
- Particle identification will be different for $K^+$ and $K^-$ due the the difference in nuclear cross sections.
- The flavor tagging will be different due to asymmetries in both the efficiency and mis-tag rates.

All these effects should be measured and corrected using the data. Separate control channels should be found for each of the different types of hadrons and care should be taken that there is no expected direct $CP$ violation in the control channels. As an example the $B_s \to D_s^- \pi^+$channel will act as a control channel for $B_s \to D_s^{\mp} K^{\pm}$.

### Conclusions

Starting from 2007 LHC$b$ will see $10^{12}$ $b\bar{b}$ pairs per year. A sophisticated trigger is required to reduce the background from the much larger production of minimum bias events and to select the specific $B$ decays of interest.

The LHC$b$ detector is optimized to cover a wide range of (semi)-leptonic and hadronic decays with high efficiency and the experiment will be able to make comprehensive measurements of the $CP$ violating effects in the quark sector. Hopefully we will from this see that the single $CP$-violating phase of the Standard Model is no longer sufficient to explain all the data and that New Physics is required.

Finally I would like to thank the organizers of the workshop for providing a good atmosphere for discussions, not only about a possible future $e^+e^-$ Super $B$ Factory, but also about new ideas for $B$ physics at LHC$b$.

### 3.9.2 *CP* violation at the Large Hadron Collider

>–S. Gopalakrishna and J. D. Wells–<

The Large Hadron Collider (LHC) is expected to begin taking data in 2007. It is a $pp$ collider with center-of-mass energy $14\,\mathrm{TeV}$. One of its main goals is to hunt for clues to the origin of electroweak symmetry breaking, and the Higgs boson is its primary quarry. Other ideas such as supersymmetry and extra dimensions also have an excellent chance of being discovered at the LHC, if either (or both) of these ingredients are what keeps the electroweak scale stable against higher scales (grand unification scale, Planck scale, *etc.*).

The LHC is likely to meet with spectacular success in the endeavors outlined above. However, shortly after the discovery of superpartners, for example, we will want to know the answer to new questions. The LHC is unlikely to be able to answer every question we can possibly formulate about the New Physics we will be witnessing. One area of challenge for the LHC is its ability to discover and confirm new sources of $CP$ violation.

To say that the LHC will have difficulty discovering new sources of $CP$ violation is not the same thing as saying that new $CP$-violating phases have no effect on LHC observables. New $CP$-violating phases, can, in fact, have an enormous impact on LHC observables. We might, however, find it difficult to know that $CP$-violating phases are at work. We can illustrate this point using supersymmetry, since it is a perturbative, well-defined calculational framework with known ways of incorporating new $CP$-violating phases.

One example of new $CP$-violating phases affecting observables is the chargino mass, whose value relative to the (generally) independent lightest neutralino mass can be determined in some circumstances to within a few percent at the LHC [184]. The chargino mass matrix is

$$M_{\chi^\pm} = \begin{pmatrix} M_2 & \sqrt{2} m_W \sin\beta \\ \sqrt{2} m_W \cos\beta & -\mu \end{pmatrix} . \tag{3.145}$$

Assume $\mu$ is complex, thus introducing a new source of $CP$ violation into the theory. The physical masses are the real eigenvalues of

$$MM^\dagger = \begin{pmatrix} M_2^2 + 2m_W^2 s_\beta^2 & \sqrt{2} m_W (M_2 c_\beta - \mu^* s_\beta) \\ \sqrt{2} m_W (M_2 c_\beta - \mu s_\beta) & |\mu|^2 + 2m_W^2 c_\beta^2 \end{pmatrix} . \tag{3.146}$$

The real characteristic equation to solve is,

$$\lambda^2 - \lambda T + D = 0, \tag{3.147}$$

where $T = \mathrm{Tr}\,(MM^\dagger)$ and $D = \det\,(MM^\dagger)$. It is clear that the the eigenvalues depend on the phase of $\mu$, and thus on a $CP$-violating phase.

We plot this effect by varying the $CP$-violating phase of $\mu$ assuming $|\mu| = 500\,\mathrm{GeV}$ and varying $\tan\beta$. We have fixed $M_2 = 250\,\mathrm{GeV}$ and assumed it to be real for illustration. The resulting lightest chargino mass is displayed in Fig. 3-36. We can see that there is a large effect on the chargino mass if $\phi_\mu$ is allowed to vary. For various values of $\tan\beta$ we compute the difference in the lightest chargino mass for $\phi_\mu = 0$ compared to $\phi_\mu = \pi$:

$$\tan\beta = 3 \implies \Delta m_{\chi_1^\pm} = 19.2\,\mathrm{GeV} \tag{3.148}$$

$$\tan\beta = 5 \implies \Delta m_{\chi_1^\pm} = 12.3\,\mathrm{GeV} \tag{3.149}$$

$$\tan\beta = 10 \implies \Delta m_{\chi_1^\pm} = 6.3\,\mathrm{GeV} \tag{3.150}$$

$$\tan\beta = 30 \implies \Delta m_{\chi_1^\pm} = 0.2\,\mathrm{GeV} \tag{3.151}$$

Only for very large values of $\tan\beta$ does $\phi_\mu$ not have a discernible effect on the chargino mass. By “discernible effect” of a parameter ($\phi_\mu$ in this case) on an observable (chargino mass in this case) we mean that the measurement of the observable will be sufficiently precise that a reasonable variation in the underlying parameter would predict a variation in the observable greater than the experimental uncertainty.

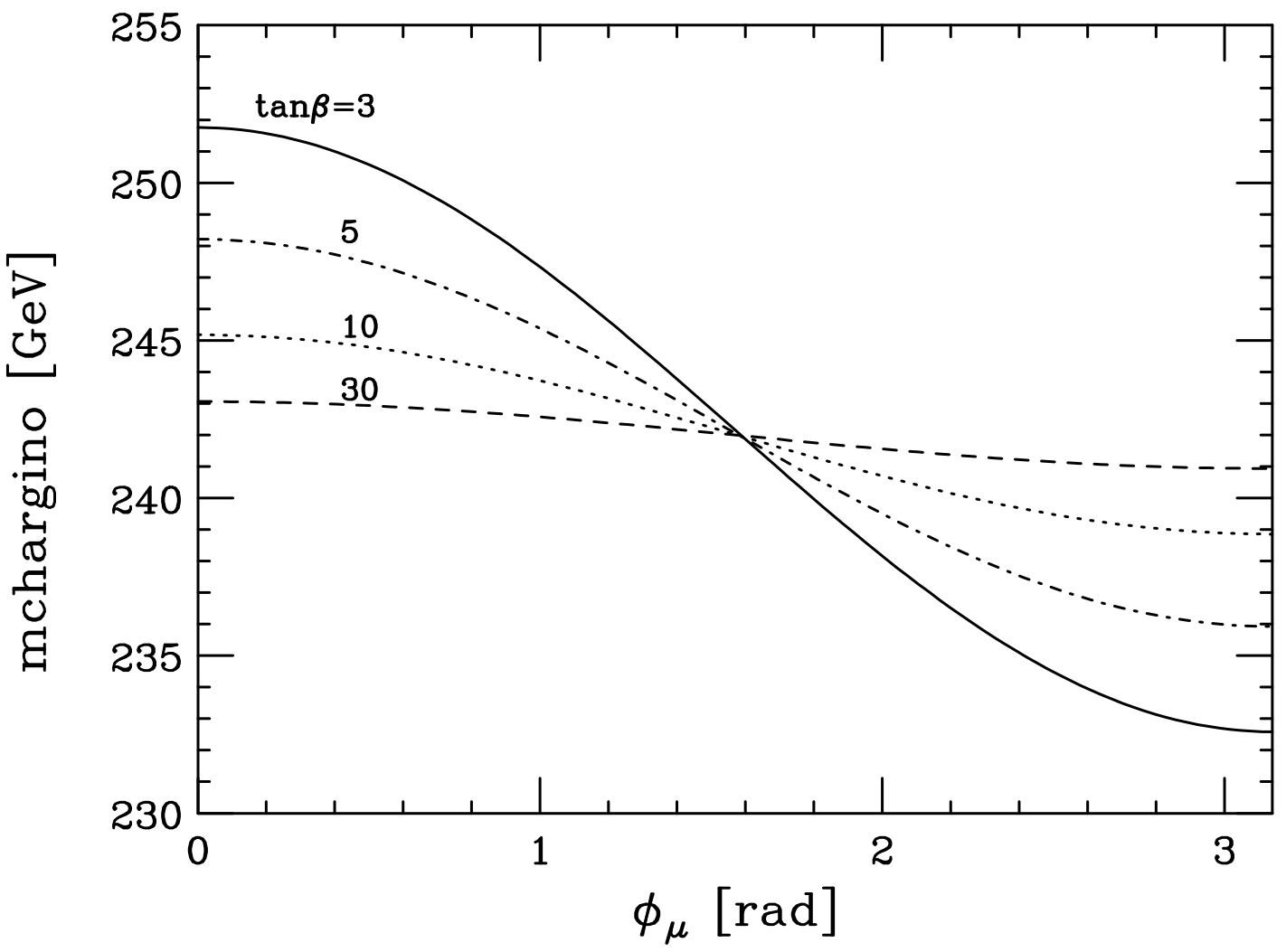

**Figure 3-36.** *Dependence of the $CP$-violating phase of $\mu$ on the mass of the lightest chargino for various values of* $\tan\beta$*.* $M_2 = 250$ *and* $|\mu| = 500\,\mathrm{GeV}$ *were assumed for this plot.*

However, it must be emphasized that when we say that a $CP$-violating parameter has a "discernible effect" on an observable, we do not mean that we will be able to determine experimentally what that parameter is, or even if it is nonzero. For the case of the chargino mass, there are several free parameters that ultimately determine the chargino mass: $M_2$, $\mu$, $\tan\beta$. We can reproduce any value of the lightest chargino mass we want by using purely real parameters (*i.e.,* no $CP$-violating phases). If we can determine the second eigenvalue of the chargino mass matrix, the parameter space of real values shrinks but can still accommodate any two values specified. When all the chargino and neutralino masses and mixings are known, if that ever becomes possible, the parameter space of real values might disappear. However, Brhlik and Kane claim [185] that the number of observables at the LHC that can be reasonably well measured for this purpose is too small and one will not ever be able to measure $\tan\beta$ well in the more general MSSM (*i.e.,* no dramatically simplifying assumptions) much less have confidence that a new source of $CP$ violation is at play. This claim deserves more scrutiny; however, we can find no publication in the literature that would dispute it either explicitly or implicitly.

The above discussion does not even take into account that non-minimal supersymmetric extensions can add many more parameters that would further increase the difficulty of establishing that $CP$-violating phases were affecting observables. For example, a new gauge group at the TeV scale would introduce new gauginos and higgsinos into the spectrum, increasing the complexity and number of parameters of the neutralino and possibly chargino mass matrix.

Since establishing that a new $CP$-violation source in particle physics is of utmost qualitative importance to our understanding of nature, establishing a non-zero $CP$-violating observable is crucial and probably cannot be replaced in value with any number of well-measured $CP$-conserving observables.

Perhaps the most promising $CP$-violating observables in high energy collisions at the LHC are those involving top quark production [186]-[191], where a non-zero $CP$-odd triple scalar product involving its decay products' momenta would unmask new $CP$ violation. One reason why top quarks are the focus of attention is that they are currently relatively unknown quarks—their interactions have not been measured extremely well yet and they could involve $CP$ violation. Second, and related to the first, the phases of New Physics that couple directly to top quarks (*e.g.,* top squarks) are generally less constrained than the phases of New Physics that couples directly to first and second generation quarks (*e.g.,* up, down, charm and strange squarks), which participate directly in the neutron and electron electric dipole moment (EDM) observables. Many studies have been performed by theorists, and it is difficult to judge how viable the $CP$-violating observables will be in the real collider detector environment. All studies apparently seem to agree that a small amount of new $CP$ violation will not be detectable in $t\bar{t}$ production—a large $\mathcal{O}(1)$ new phase directly inserted into the relevant interactions is needed.

Again, it is perhaps easiest to discuss the prospects of finding $CP$-violating signals of $t\bar{t}$ production and decay in the context of well-defined supersymmetry. The general hope is that a large $CP$-violating phase will enter in loop corrections to top quark production or decay but will not enter in loop calculations of well-constrained observables such as the neutron and electron EDMs. Supersymmetry has many sources of phases: the $\mu$ term, gaugino masses, $A$ terms, off-diagonal squark masses, *etc.* The $A$-term of the top squark is of particular importance. One can conceivably give it an $\mathcal{O}(1)$ $CP$-violating phase and declare all other $A$-terms involving other squarks to have zero $CP$-violating phase, thereby keeping within the EDM constraints. Assuming all supersymmetry masses as low as they can go and not be in conflict with direct experimental limits, and assuming the phase of $A_t$ is $\mathcal{O}(1)$, the resulting $CP$ asymmetries that can be constructed from $t\bar{t}$ production are at the few percent level at most [191]. This is at the edge of detectability at LHC.

The general two-Higgs doublet model has been a major focus of models that can contribute to $CP$-violating observables in $t\bar{t}$ production. The reason is that one can include extra sources of $CP$-violating phases via new Higgs boson couplings to the top quark. The relevant couplings can be safely large since the top quark mass is large. Recently, it has been determined that there may be a small region of parameter space in the two Higgs doublet model that the LHC has a chance of seeing $CP$-violating effects [188, 189], but these regions of parameter space are special, in that they must be chosen for no other reason than to maximize that signal. The parameter space does not appear to fit nicely into any wanted structure of a deeper theory with more explanatory power than the Standard Model.

As for the supersymmetric cases, there is a theoretical reason that casts some doubt on the most favorable set of parameters leading to $CP$-violating signals discussed above. It comes from renormalization group mixing of $CP$ phases [192]. Suppose that at some scale $\Lambda$ one sets the phase of $A_u$ to zero but $A_t$ phase is $\mathcal{O}(1)$. Thus, the $H_u\tilde{Q}_u\tilde{u}_R$ tri-scalar interaction strength $y_uA_u$ has no $CP$-violating phase in it. However, there is a one-loop correction to this interaction that scales as

$$(y_uA_u)_{1-\mathrm{loop}}(Q) \sim 6\frac{y_tA_t}{16\pi^2}y_uy_t^\dagger \log\frac{\Lambda}{Q}\ . \tag{3.152}$$

This is a large effect if $Q$ is more than a few orders of magnitude away from $\Lambda$, and the neutron EDM prediction is too high. The lesson is that it is not natural to isolate the third generation $CP$-violation phases from those of the first two generations. Of course, one could arrange cancellations of phases just so the $CP$-violating observables of $t\bar{t}$ would be borderline at LHC with no EDM problems, but it does not appear natural. One way to escape this situation is if the first two generation scalars are very heavy (a few TeV). In such a case, the EDM constraints are not violated by the contribution in Eq. (3.152), and it is possible that the phase of $A_t$, along with all other phases of the theory, could be $\mathcal{O}(1)$. Such an "effective supersymmetry" scenario seems to be the most promising one for the LHC to be able to see a $CP$-violating signal in the top sector.

Another possible $CP$-violating phase that could be large without violating EDM constraints is in the $23$ element of the squark mass matrix. The phase of $(\delta^d_{RR})_{23}$ could be probed by measuring $\Delta m$ of the $B_s$ meson. A complex $(\delta^d_{LR})_{23}$ mass insertion might be responsible for the anomalous $B\to\phi K^0_S$ $CP$ asymmetry [193]. This phase can also, interestingly, lead to a $CP$-asymmetry in the $b\to s\gamma$ decays at the level of a few percent, depending on the size of $(\delta^d_{LR})_{23}$. Probing much of this range would require a Super $B$ Factory.

Unfortunately, we know of no way that this complex squark flavor off-diagonal phase could be probed in high-energy collisions at the LHC. The closest hope would be that the phase would imply, in some emerging theory, similar $CP$-violating phases in the slepton sector. In that case, it might be possible with very favorable values of other parameters to see the effects of $CP$-violating slepton oscillations at the LHC [163]. Again, even under these most favorable circumstances it would be extremely challenging for the LHC.

We conclude by remarking that at the LHC, even in the most favorable situation, it might be quite challenging to observe direct $CP$-violating signals in high-energy collisions, even while the $B$ physics programs might be observing new $CP$-violating phenomena. The Super $B$ Factory has potentially unique prospects of probing $CP$-violating phases, for example through the modes $b \to s\gamma$ and $B \to \phi K_S^0$, as is emphasized in other chapters of this study.

# 4

# Semileptonic Decays and Sides of the Unitarity Triangle

**Conveners:** U. Langenegger, Z. Ligeti, I. Stewart

**Authors:** C. Bauer, C. Bernard, I. Bigi, M. Datta, D. del Re, B. Grinstein, S. Hashimoto, U. Langenegger, Z. Ligeti, M. Luke, E. Lunghi, P. Mackenzie, A. Manohar, T. Moore, D. Pirjol, S. Robertson, I. Rothstein, I. Stewart, M. Voloshin

## 4.1 Theory overview

### 4.1.1 Continuum methods (OPE, HQET, SCET)

>– A. Manohar –<

The elements of the CKM matrix enter the expressions for the decay rates and mixing amplitudes of hadrons. In some cases, the the theoretical expressions are free of strong interaction effects, for example the $CP$ asymmetry in $B \to J/\psi\, K_s^0$, so that measuring the $CP$ asymmetry directly gives the value of $\sin 2\beta$, with the error in the result given by the experimental error in the measurement. In most cases, however, the experimentally measured quantities depend on strong interactions physics, and it is absolutely essential to have accurate model-free theoretical calculations to compare with experiment. A number of theoretical tools have been developed over the years which now allow us to compute $B$ decays with great accuracy, sometimes at the level of a few percent or better. These calculations are done using effective theory methods applied to QCD, and do not rely on model assumptions.

Inclusive decays can be treated using the operator product expansion (OPE). The total decay rate is given by twice the imaginary part of the forward scattering amplitude, using the optical theorem. In heavy hadron decays, the intermediate states in the forward scattering amplitude can be integrated out, so that the decay rate can be written as an expansion in local operators. The expansion parameter is $1/m_B$, the mass of the decaying hadron. OPE techniques have been well-studied in the context of deep-inelastic scattering, where the expansion in powers of $1/Q^2$ is called the twist expansion. In inclusive $B$ decays, the leading term in the $1/m_B$ expansion gives the parton decay rate, and non-perturbative effects enter at higher orders in $1/m_B$.

The OPE can be combined with heavy quark effective theory (HQET) for greater predictive power in heavy hadron decays. HQET is an effective theory for heavy quarks at low energies, and the HQET Lagrangian has an expansion in powers of $1/m_b$, the inverse heavy quark mass. The HQET Lagrangian is written in terms of the field $b_v$, which annihilates a $b$ quark moving with velocity $v$. One usually works in the rest frame of the heavy quark $v = (1, 0, 0, 0)$. At leading order ($m_b \to \infty$), the heavy quark behaves like a static color source. As a result, the leading order HQET Lagrangian has heavy quark spin-flavor symmetry, since the color interactions of a static color source are spin and flavor independent. The $1/m_b$ terms in the Lagrangian break the spin and flavor symmetries, and are treated as perturbations. Since this is an effective theory, radiative corrections can be included in a systematic way. Most quantities of interest have been computed to $1/m_b^3$ in the $1/m_b$ expansion, and radiative corrections to the leading term are typically known to order $\alpha_s^2$ or $\alpha_s^2\beta_0$. In a few cases, the order $\alpha_s$ corrections are known for the $1/m_b$ terms. The calculations can be pushed to higher orders, if this is experimentally relevant.

The OPE can be combined in a natural way with HQET for inclusive heavy hadron decays, since both involve an expansion in $1/m_b$. This allows one to write the inclusive decay rates in terms of forward matrix elements of local operators. At leading order, the decay rate can be written in terms of the operator $\bar{b}\gamma^\mu b$, the $b$ quark number current in full QCD. The matrix element of this operator in $B$ hadrons is one to all orders in $\Lambda_{\rm QCD}/m_b$ and all orders in $\alpha_s$. At leading order, the inclusive decay rates of all $b$ hadrons is the same. At order $1/m_b$, the only operator allowed by dimensional analysis is the operator $\bar{b}_v(iv\cdot D)b_v$, whose matrix element vanishes by the equations of motion. This is an important result—non-perturbative corrections first enter at order $\Lambda_{\rm QCD}^2/m_B^2$, which is of order a few percent. At order $1/m_b^2$, the inclusive rate depends on two non-perturbative parameters $\lambda_1$ and $\lambda_2$ which are the heavy quark kinetic energy and hyperfine energy, respectively. The same parameters $\lambda_{1,2}$ enter other quantities such as the hadron masses. For example, the $B^*$–$B$ mass difference gives $\lambda_2 = 0.12$ GeV$^2$. As the data become more precise, various HQET parameters are pinned down with greater precision, increasing the accuracy with which the decay rates are known.

HQET can also be applied to study exclusive decays. Heavy quark spin-flavor symmetry puts constraints on the form factors; *e.g.*, heavy quark symmetry provides an absolute normalization of the form factor at zero-recoil for the semileptonic decay $B \to D^{(*)}$, up to corrections of order $1/m_b^2$. The reason is that at zero recoil the decay proceeds by a $b$ quark at rest turning into a $c$ quark at rest. Since the strong interactions at leading order in $1/m$ are flavor-blind, the form-factor at zero-recoil is unity. Corrections to this result follow from the $1/m$ symmetry breaking terms. It is known that there are no $1/m$ corrections, so the first corrections are order $1/m^2$. As for inclusive decays, the $1/m^2$ corrections are a few percent, so the exclusive decay can be used to obtain $V_{cb}$ to a few percent.

Heavy to light decays such as $B \to \pi\pi$, which is required for a determination of $\sin 2\alpha$, are more difficult to treat theoretically. Here the $B$ meson decays into two fast moving light hadrons, and it is difficult to treat strong interactions in this kinematic regime. A recently developed effective theory, soft-collinear effective theory (SCET) is being used to deal with this situation. SCET is an effective theory that describes fast moving quarks with momentum proportional to the light-like vector $n$ by collinear fields $\xi_n$. In the case of $B \to \pi\pi$, there are two back-to-back light-like vectors $n$ and $\overline{n}$ giving the directions of the two pions. The SCET fields needed to describe this process are collinear quarks and gluons in the $n$ and $\overline{n}$ directions, $\xi_n$, $A_n$, $\xi_{\overline{n}}$, $A_{\overline{n}}$, as well as soft quarks and gluons that describe the light degrees of freedom in the $B$ meson. The non-perturbative interactions of $\xi_n$ and $A_n$ produce the pion moving in the $n$ direction, and the interactions of $\xi_{\overline{n}}$ and $A_{\overline{n}}$ produce the pion moving in the $\overline{n}$ direction. If one neglects the soft fields, the $n$ and $\overline{n}$ fields do not interact, so there are no final state interactions between the pions in $B \to \pi\pi$, and the factorization approximation for the decay is valid. Soft gluons interact with all modes in the effective theory, and introduce final state interactions. The extent to which this affects the decay amplitude and final state interaction phase-shift is being investigated.

SCET is also applicable in inclusive decays where the final hadronic state has small invariant mass, and is jet-like. An example is the endpoint of the photon spectrum in $B \to X_s\gamma$ decay, or electron spectrum in $B \to X_u e\nu$ decay. SCET allows one to systematically resum the Sudakov double logarithmic radiative corrections which become very large in the endpoint region.

The effective theories discussed here will be used later in this chapter to obtain detailed predictions for decay rates and form factors.

### 4.1.2 Lattice QCD and systematic errors

>– C. Bernard, S. Hashimoto, P. Mackenzie –<

Systematic errors of lattice QCD computations come from a variety of sources. Many of these are associated with an extrapolation from a practical lattice calculation (at finite lattice spacing, unphysically heavy quark mass values, and finite spatial volume) to the real, continuum, infinite volume world, where the quark masses take their physical values. There are also lattice systematics that are not directly connected to an extrapolation. These include the perturbative error in connecting lattice currents to their continuum counterparts, and the "scale error" coming from the need to determine the lattice spacing in physical units.

One possible lattice systematic that will not be included below is quenching, the omission of virtual (sea) quark loops. Although the quenched approximation has been used in most lattice computations to date, one must remember that it is an *uncontrolled* approximation, not systematically improvable. Indications are that it produces errors of 10 to 20% on the phenomenologically interesting quantities we discuss in this report. However, these are uncontrolled and hence unreliable error estimates and completely unsuitable for use in connection with the precise experimental results that a Super $B$ Factory will make possible. We therefore consider only lattice computations in which the effects of three light flavors ($u$, $d$, and $s$) of virtual quarks are included.

It is important to distinguish here between quenching and "partial quenching." Partial quenching [1, 2, 3] is a somewhat misleading term in this context and simply means that the valence quark masses in the lattice simulation are not necessarily chosen equal to the sea quark masses. As emphasized by Sharpe and Shoresh [3], as long there are three light virtual flavors in a partially quenched simulation, real-world, full ("unquenched") QCD results can be extracted. This is not surprising, since the real-world situation is just a special case (valence masses = sea masses) of the partially quenched simulation. In fact, partial quenching is often *preferable* to simple unquenching because it separates the valence and sea mass contributions and allows one to use the information contained in the correlations, for fixed sea masses, of the results for different valence masses. Partial quenching will be assumed in lattice errors estimates given in Section 4.5.2 and 4.6.1.

We now discuss the relevant lattice systematic effects in more detail.

**Chiral extrapolation**

The computer time required for a lattice simulation rises as a large power of $1/m_{u,d}$ as these masses approach their physical values. One must therefore work with larger masses and extrapolate to the real world. Chiral perturbation theory ($\chi$PT) determines the functional form of the extrapolation and makes it possible to get good control of the associated systematic error. (In the partially quenched case one must use the corresponding "partially quenched chiral perturbation theory" (PQ$\chi$PT) [1].)

As the physical values of the $u, d$ quark masses are approached, there is significant curvature in essentially all interesting quantities that involve light quarks. The curvature comes from chiral logarithms that are proportional to $m_\pi^2 \ln(m_\pi^2/\Lambda_\chi^2)$. This implies that one must get to rather small quark mass (probably $m_{u,d} \sim m_s/4$ to $m_s/8$) to control the extrapolation. If only large masses are available ($m_{u,d} \gtrsim m_s/2$), we will be limited to 10% or even 20% errors, a point that has been emphasized recently by several groups [4, 5, 6].

Thus it does little good to include virtual quark loop effects unless the $u, d$ quark masses in the loops are significantly lighter than $m_s/2$. To achieve this goal in the near term appears to require use of staggered quarks, in particular an "improved staggered" [7] action. This fermion discretization is computationally very fast, and has a residual (non-singlet) chiral symmetry that prevents the appearance of "exceptional configurations"—thereby allowing simulation at much lighter quark masses than are currently accessible with other discretizations. There are, however, some theoretical and practical problems with staggered fermions, which we address in Section 4.1.2 below.

**Discretization**

The lattice takes continuous space-time and replaces it with discrete points separated by lattice spacing $a$. The leading $a$ dependence for small $a$ depends on the lattice action: improved staggered quarks have errors proportional to $\alpha_S a^2$. There are also formally subleading errors that can be quite important numerically. These are so-called "taste violations," discussed in Section 4.1.2, which are $\mathcal{O}(\alpha_S^2 a^2)$. Precise (few percent) lattice calculations with staggered light quarks will likely require detailed control of such taste violations.

Heavy quarks introduce additional discretization errors. We assume here that the heavy quarks are introduced with the standard Fermilab approach [8], which has $\mathcal{O}(\alpha_S a, a^2)$ errors. Improvement of the heavy quarks is also possible [9, 10], although it is not yet clear whether such actions will be practical in the near term. Introducing heavy quarks *via* nonrelativistic QCD (NRQCD) [11] is likely to produce comparable errors to the Fermilab approach, especially for $b$ quarks.

### Finite volume

Since we can simulate only a finite region in space-time, there will always be some finite volume errors. The size of such errors of course depends sensitively on the number of hadrons present. For this reason lattice computations with more than one hadron in the initial or final state are probably out of reach in the next five years for all but the most qualitative studies. Even on a longer time scale such calculations will continue to be very difficult.

For single hadrons, currently feasible volumes are enough to reduce finite volume errors to the few percent level (without major sacrifice on discretization errors). Typically a volume $V \gtrsim (2.5\,\mathrm{fm})^3$ is sufficient. We can do even better for single-particle quantities whose mass dependence is determined by $\chi$PT, which also predicts the volume dependence (for large volume). This allows us to correct for finite volume effects and reduce the errors to a negligible level. We will therefore ignore finite volume effects for single-particle states from here on.

### Setting the scale

In simulations, the lattice spacing $a$ is determined after the fact by comparing the result for a one dimensional quantity with experiment. (This is equivalent to fixing $\Lambda_{QCD}$ or $\alpha_S$.) Therefore, the lattice error in the quantity used to set the scale will infect all other dimensionful results. The best we can do today is probably from $\Upsilon(2S-1S)$ or $\Upsilon(1P-1S)$ splittings [12, 13], which lead to a roughly 2% scale error on other quantities, after extrapolation to the continuum [14]. The scale error is usually negligible on dimensionless quantities (like form factors or $f_{B_s}/f_B$), but is not strictly zero because the error can enter indirectly through the determination of quark masses or momenta.)

### Perturbation theory

Most interesting quantities require a weak-coupling perturbative calculation (or equivalent nonperturbative lattice computation) to match lattice currents (or, more generally, operators) to their continuum counterparts. The light-light leptonic decay constants (*e.g.*, $f_\pi$, $f_K$) are exceptions: staggered lattice PCAC implies that the lattice axial current is not renormalized, so lattice and continuum currents are the same. This is not true, however, for heavy-light quantities such as $f_B$ or semileptonic form factors. To date, all such matching calculations have been done only to one loop, leaving large errors (∼10%). Some reduction (perhaps by a factor of 2) in these errors may be possible using simple nonperturbative information [15]. However, it is not obvious that this technique will be successful in the current case of interest: light staggered quarks and heavy Fermilab quarks. So the range of possible errors from a one loop calculation is $\sim 5$–$10\%$. For simplicity, we use $7.5\%$ as the nominal one-loop error in Sections 4.5.2 and 4.6.1 below; one should keep in mind that the uncertainty on this error is significant.

To do better, two-loop perturbative calculations are required. But lattice perturbation theory is very messy, since the actions are complicated and there is no Lorentz invariance. "Automated perturbation theory" [16] is probably required. There do not appear to be any fundamental impediments to this approach; however, some practical problems still need to be overcome. In particular, the issue of infrared regulation is important. Currently, "twisted boundary conditions" on the lattice fields in finite volume are used to regulate the IR divergences. In order to match to the continuum, one should the use same twisted boundary conditions there. However continuum perturbation theory (*e.g.*, dimensional regularization) with twisted boundary conditions is difficult, especially beyond one loop. Since the time scale on which the two-loop calculations will become available is therefore not clear, we present future error estimates both with and without assuming the existence of two-loop matching. Luckily, many interesting quantities, *e.g.*, ratios like $f_{B_s}/f_B$, are independent or nearly independent of perturbation theory.

### Issues with staggered fermions

Staggered fermions carry an extra, unwanted quantum number, "taste," which is 4-fold remnant of the lattice doubling symmetry. Taste symmetry is believed to become an exact SU(4) in the continuum limit, but is broken at finite lattice spacing. The taste degree of freedom is not a problem for valence quarks, since one may choose specific tastes by hand. But for sea quark effects, the only known method for eliminating the taste degree of freedom in simulations is to take the fourth root of the staggered fermion determinant. Because of taste violations, this is not an exact reduction at finite lattice spacing and is a non-local operation. Therefore some authors worry that it could introduce non-universal behavior and lead to the wrong theory in the continuum limit. Although there is no proof that the fourth-root procedure is correct, there are several pieces of evidence in its favor [12]. In particular, if the taste symmetry does become exact in the continuum limit (which few doubt), then the fourth-root procedure is correct to all orders in perturbation theory.

There is also a practical issue with staggered fermions: It is difficult to control the chiral extrapolations unless one takes taste violations explicitly into account. Because taste violations are an artifact due to finite lattice spacing, this represents an entanglement of chiral and discretization errors. To help disentangle these errors, one can fit the lattice data to "staggered chiral perturbation theory" (S$\chi$PT) instead of ordinary continuum $\chi$PT. S$\chi$PT has been worked out for the $\pi$-$K$ system [17, 18, 19]; it is necessary to obtain precise results for $f_\pi$, $f_K$, and the $\mathcal{O}(p^4)$ chiral parameters [14]. S$\chi$PT for heavy-light mesons is being worked out [20]. It is not yet clear whether the number of new chiral parameters due to taste violations in the heavy-light case will be sufficiently small that it will be as useful as in the light-light case.

In estimating the expected precision of lattice computations (Sections 4.5.2 and 4.6.1), we give two versions: "S$\chi$PT" assumes that the heavy-light S$\chi$PT works as in the light-light case and is similarly useful; "No S$\chi$PT" assumes that S$\chi$PT is not useful because of a proliferation of parameters, and one must disentangle chiral and continuum extrapolations without its help (probably by extrapolating to the continuum first and then using ordinary $\chi$PT).

All estimates given below for the expected precision of lattice computations assume the that the staggered fermions with the fourth-root procedure produce standard QCD in the continuum limit. If this assumption turns out to be incorrect, there are safer but slower methods that could be used instead. The most likely choice appears to us to be domain wall fermions (DWF), which are of order 100 times slower. (The precise factor is not known, largely because DWF have not yet been used in extensive unquenched simulations.) From Moore's law alone, this could delay by as much as a decade the attainment of lattice computations with the desired level of precision. However, despite the fact that DWF have $\mathcal{O}(a^2)$ errors, formally larger than improved staggered fermion $\mathcal{O}(\alpha a^2)$ errors, the coefficient of $a^2$ seems quite small, giving discretization errors smaller than for improved staggered fermions. In addition, the DWF discretization errors are not entangled with chiral extrapolation errors. Therefore, a delay of order five years, not ten, seems to us a better estimate.

### Gold-plated quantities

Given the above issues and systematic errors, only a small number of hadronic quantities are likely to be computed with high (few percent) precision on the lattices in the next decade. Such quantities are called "gold plated" [12]. To be gold-plated, a quantity must involve:

- At most one hadron in initial and final state.
- Stable hadrons, not near thresholds. Unstable particles require very large volumes and untested techniques to treat decay products correctly; the same applies to the virtual decay products of stable particles near thresholds. Thus, for example, semileptonic form factors for $B \to \rho$ are excluded.
- Connected graphs only (valence quark lines connecting the initial and final state). Disconnected graphs are difficult and noisy. The $\eta$ is probably excluded, because one needs to include $\eta$-$\eta'$ mixing, which is governed by disconnected graphs.

- Low momenta only. Momenta $|\vec{p}|a \gtrsim 1$ lead to unacceptable discretization errors, so we are probably limited to $|\vec{p}| \lesssim 1\text{GeV}$. This implies $q^2 \gtrsim 17$ GeV$^2$ for $B \to \pi$ semileptonic form factors. (The minimum available lattice momentum for fixed lattice size may also require $|\vec{p}| \gtrsim 350$ MeV or more.)

- A controlled chiral extrapolation.

The gold-plated lattice quantities relevant to the Super $B$ Factory are heavy-light leptonic decay constants ($f_B$, $f_{B_s}$), bag parameters for $B - \overline{B}$ and $B_s$-$\overline{B}_s$ mixing ($B_B$ and $B_{B_s}$), and the semileptonic form factors for $B \to \pi$ and $B \to D$. In addition, the semileptonic form factors for $B \to D^*$ may also be possible because model dependence from the unstable $D^*$ multiplies $\mathcal{F}(1) - 1$ and may be negligible.

## 4.2 Experimental overview

For precision studies of (inclusive) semileptonic $B$ decays it is often necessary to apply an event selection procedure providing an event sample enriched in $B$ decays and suppressing events from continuum $q\overline{q}$ production (where $q = u, d, s, c$). Traditionally, this has been implemented with the requirement of a high-momentum lepton, *e.g.*, $p > 1.4$ GeV as measured in the center-of-mass system (CMS), indicating the semileptonic decay of a $B$-meson. With the arrival of $B$ factories, a new paradigm has become possible: event selection based on the fully reconstructed (hadronic or semileptonic) decay of one of the $B$ mesons [21]. In this approach, the fully reconstructed $B_{reco}$ meson constitutes a "tag", and—in the $\Upsilon(4S)$ CM frame—the signal decay is observed in the "recoil" of the $B_{reco}$ candidate. This approach yields lower backgrounds because of a cleaner environment and offers excellent possibilities to determine background control samples directly in data.

### 4.2.1 Recoil Physics

>– D. del Re –<

The study of semileptonic $B$ meson decays $B \to X\ell\overline{\nu}$ in the recoil of a fully reconstructed $B$ meson presents many advantages. First of all, it assures a very clean environment to study the properties of the recoil. One of the two $B$ mesons from the decay of the $\Upsilon(4S)$ is reconstructed either in a hadronic or semileptonic decay mode. The remaining particles of the event originate from the decay of the other (recoiling) $B$ meson. In the case of a semileptonic decay of the recoiling $B$, the only missing particle is a neutrino. This implies that a requirement on the net charge of the event (charge conservation) can be applied. In the case of hadronic tags, the missing mass (possibly scaled with the missing energy) of the entire event should be consistent with zero. Moreover, since the kinematics are over-constrained, the resolution on the reconstructed quantities, such as the mass of the hadronic system $m_X$, can be improved with kinematic fitting. The momentum of the recoiling $B$ is also known (up to a twofold ambiguity for the case of semileptonic tags) and therefore the lepton momentum can be boosted into the $B$ rest frame. The charge and the flavor of the $B$ is known. Decays of $B^0$ and the $B^+$ mesons can be studied separately. The correlation between the charge of the lepton and the flavor of the $B$ can be used to reduce backgrounds from $B \to \overline{D} \to \ell$ events.

The only drawback is that the overall efficiency of this method is very low and is dominated by the $B$ reconstruction efficiency, a problem that is not longer relevant at very high luminosities. For this reason, the recoil approach seems to be ideal in a Super $B$ Factory, since this is the method with the smallest experimental uncertainty.

#### Hadronic tags

The sum of a few, very pure fully reconstructed hadronic modes (as done, for instance, in the *BABAR* $B$ lifetime analysis [22]) assures very high purity with minimum event selection bias, albeit at a very low efficiency. On the other hand, a fully inclusive approach with high multiplicity reconstructed modes is not feasible since the level of combinatorics would be too high. A compromise implemented by the *BABAR* experiment (see Ref. [23]) considers only a restricted mode set with a limit on the number of particles used and employs an algorithm that is as inclusive as possible in combining the particles, neglecting the intermediate states, when possible.

$B$ mesons decay predominantly into hadronic final states involving $\overline{D}$ mesons. Because the dominant $B$ decay modes are $B^- \to D^{(*)0}Y^-$, $B^0 \to D^{(*)-}Y^+$, only these modes[1] are considered, where the $Y^\pm$ system consists of at most 5 charged tracks and two $\pi^0$ mesons. For each possible track and $\pi^0$ composition of the $Y^\pm$ system, several subsamples are identified depending on the possible resonant states in that sample. For instance, $B \to \overline{D}^{(*)}\pi^+\pi^0$ is subdivided into two kinematic region, one with $m(\pi^+\pi^0) < 1.5\,\mathrm{GeV}/c^2$, dominated by $B \to \overline{D}^{(*)}\rho^+$ decays and one containing the rest of the events. This allows us to isolate samples in which the signal is enhanced with respect to the combinatorial background (the $m(\pi^+\pi^0) < 1.5\mathrm{GeV}/c^2$ sample, in the example above). Enumerating the $D$ decay modes separately, we must consider 1153 different modes.

In order to discriminate fully-reconstructed $B$ candidates from the combinatorial background, two kinematic variables are used. The *energy difference* $\Delta E$ is defined as

$$\Delta E = E_B^* - \sqrt{s}/2\,, \tag{4.1}$$

where $E_B^*$ is the energy of the $B$ čandidate in the $\Upsilon(4S)$ CM frame and $\sqrt{s}$ is the CM energy. The $\Delta E$ distribution for signal decays peaks at zero, while the continuum and part of the $B\overline{B}$ background can be parameterized with a polynomial distribution. The resolution of this variable is affected by the detector momentum resolution and by the performance of particle identification (since a wrong mass assignment implies a shift in $\Delta E$). Therefore it depends strongly on the reconstructed $B$ mode and can vary from $20\ \mathrm{MeV}$ to $40\ \mathrm{MeV}$ depending on the charged track and $\pi^0$ multiplicity in the reconstructed mode. We therefore apply a mode-dependent $\Delta E$ selection, as tight as $-45 < \Delta E < 30\ \mathrm{MeV}$ for modes with charged tracks only and as loose as $-90 < \Delta E < 60\ \mathrm{MeV}$ for modes with two $\pi^0$ mesons.

The *beam energy-substituted mass* is defined as

$$m_{ES} = \sqrt{(\sqrt{s}/2)^2 - p_B^{*2}}\,, \tag{4.2}$$

where $\sqrt{s}$ is the total energy of the $e^+e^-$ system in the CMS and $p^*$ is the $B$ candidate momentum in the CMS. Since $|p_B^*| \ll \sqrt{s}/2$, the experimental resolution on $m_{ES}$ is dominated by beam energy fluctuations. To an excellent approximation, the shapes of the $m_{ES}$ distributions for $B$ meson reconstructed in a final state with charged tracks only are Gaussian. The presence of neutrals in the final state can introduce tails, due to preshowering in the material in front of the calorimeter or due to leakage outside the active detector volume.

Since the $m_{ES}$ resolution is dominated by beam energy uncertainty while momentum resolution dominates the $\Delta E$ resolution, the two variables are practically uncorrelated.

As an estimator of the quality of a reconstruction mode we define the purity as the ratio of the integral of the signal component in the $m_{ES}$ fit over the total number of events in the signal region ($\mathcal{P} = S/(S+B)$). We also define the integrated purity $\mathcal{P}_{\rm int}$ of a given mode as the purity of all the modes that have greater or equal $\mathcal{P}$. These quantities are computed before any other selection criteria and are to be considered as labels of the decay mode. In events with several $B_{reco}$ candidates differing only by their submode, we choose the one with the highest value of $\mathcal{P}$. If there are multiple candidates in the same submode, the minimum $\Delta E$ criterion is used and one candidate per submode is selected. The $\mathcal{P}$ variable is also utilized to choose which of the 1153 modes is actually used in the analysis; the final yields depend on this choice. For instance for the analysis presented in [23], a cut on $\mathcal{P}$ has been optimized and a large set of modes with low $\mathcal{P}$ have been removed. The resulting $m_{ES}$ distribution for an integrated luminosity of $80\,\mathrm{fb}^{-1}$ is shown in Fig. 4-1(a). In Table 4-1 the corresponding yields for four different levels of purity are summarized. As shown, this reconstruction method can provide close to $4000B/\,\mathrm{fb}^{-1}$ of fully reconstructed $B_{reco}$ mesons ($1500\,B^0/\,\mathrm{fb}^{-1}$ and $2500\,B^+/\,\mathrm{fb}^{-1}$). The corresponding purity is about 26%, which is not an important issue, as the combinatorial background depends strongly on the recoil itself. The situation improves a lot once requirements on the recoil are applied. For instance, the requirement of a lepton with a moderate momentum of $p > 1.0\ \mathrm{GeV}$ removes most of the non-$b\bar{b}$ events, while leaving the $m_{ES}$ signal shape essentially unchanged, as illustrated in Fig. 4-1(right).

[1]Charge conjugate states are implied throughout.

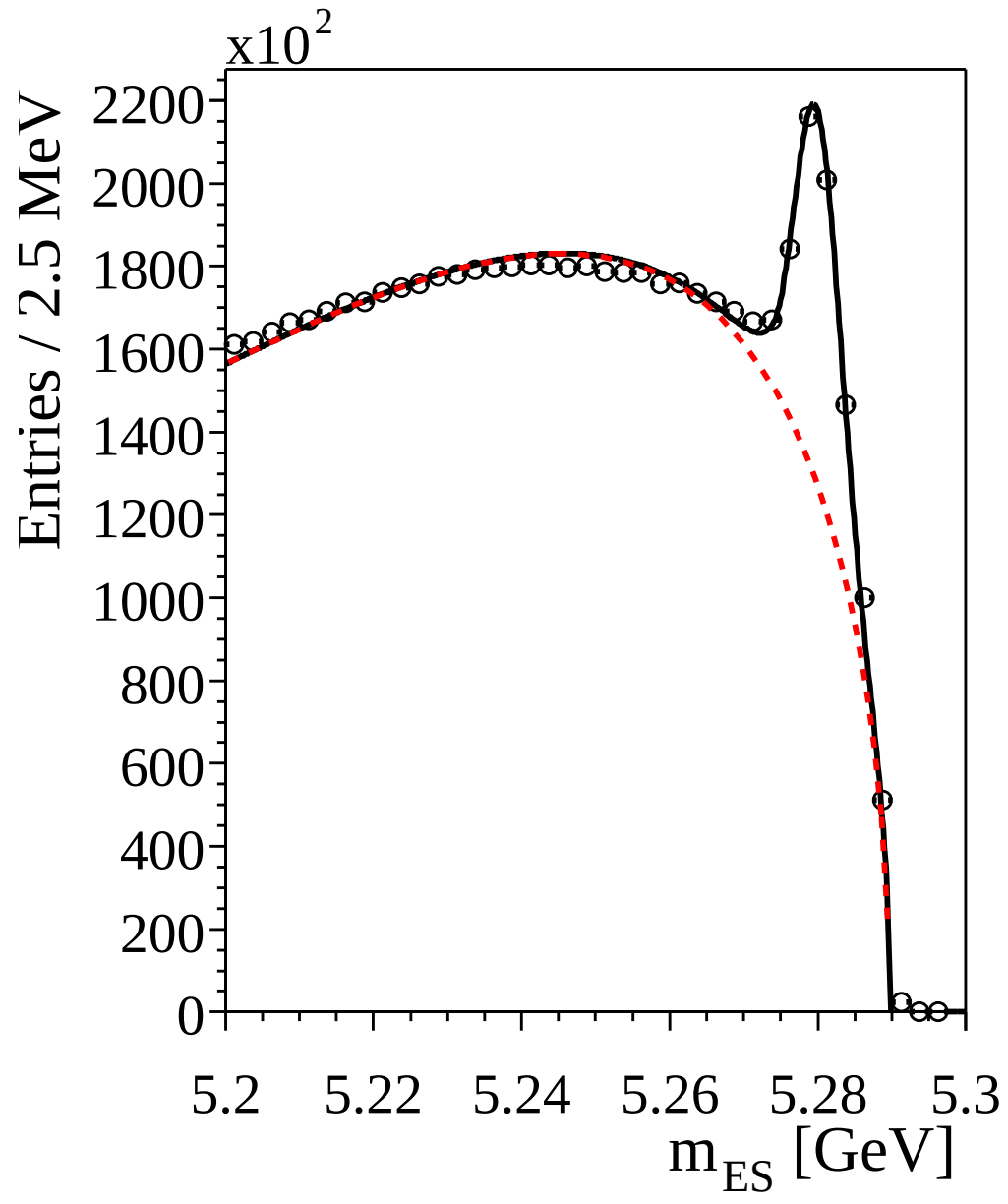

**Figure 4-1.** *Fit to the $m_{ES}$ distributions of fully reconstructed hadronic $B$ meson decays with (left) no requirement on the recoil and (right) the requirement of one lepton with $p^* > 1.0\ \mathrm{GeV}$ in the recoil. Both plots are for an integrated luminosity of $80\,\mathrm{fb}^{-1}$.*

In Fig. 4-2(left) we show the extrapolation of the number of fully reconstructed hadronic $B$ meson decays for large integrated luminosities. The corresponding plot with the measured signal yields for a few selected processes (assuming a rough estimate of the selection efficiency on the recoil) is displayed in Fig. 4-2 (right). With $10\ \mathrm{ab}^{-1}$, even rare decays such as $B \to K\nu\nu$ or $B \to \pi\tau\nu$ have sufficient statistics to be observed.

**Table 4-1.** *Yields for fully reconstructed hadronic $B$ decays for $80\,\mathrm{fb}^{-1}$ at different levels of the single mode purity $\mathcal{P}$ and integrated purity $\mathcal{P}_{\mathrm{int}}$.*

| Channel | $\mathcal{P}_{\mathrm{int}} > 80\%$ | $\mathcal{P}_{\mathrm{int}} > 50\%$ | $\mathcal{P} > 10\%$ | **Selection as in** [23] |
|---|---|---|---|---|
| $B^+ \to D^0 X$ | $19120 \pm 170$ | $54120 \pm 370$ | $95204 \pm 660$ | $100650 \pm 640$ |
| $B^0 \to D^+ X$ | $11070 \pm 130$ | $25720 \pm 260$ | $55830 \pm 480$ | $62960 \pm 550$ |
| $B^+ \to D^{*0} X$ | $18600 \pm 170$ | $44270 \pm 330$ | $75350 \pm 580$ | $82660 \pm 640$ |
| $B^0 \to D^{*+} X$ | $20670 \pm 170$ | $50300 \pm 340$ | $55560 \pm 390$ | $46380 \pm 310$ |
| Total $B^+$ | $37720 \pm 240$ | $98390 \pm 500$ | $170560 \pm 880$ | $183310 \pm 905$ |
| Total $B^0$ | $31740 \pm 210$ | $76020 \pm 430$ | $111390 \pm 620$ | $109340 \pm 630$ |
| Total | $69460 \pm 320$ | $174410 \pm 660$ | $281950 \pm 1080$ | $292650 \pm 1100$ |

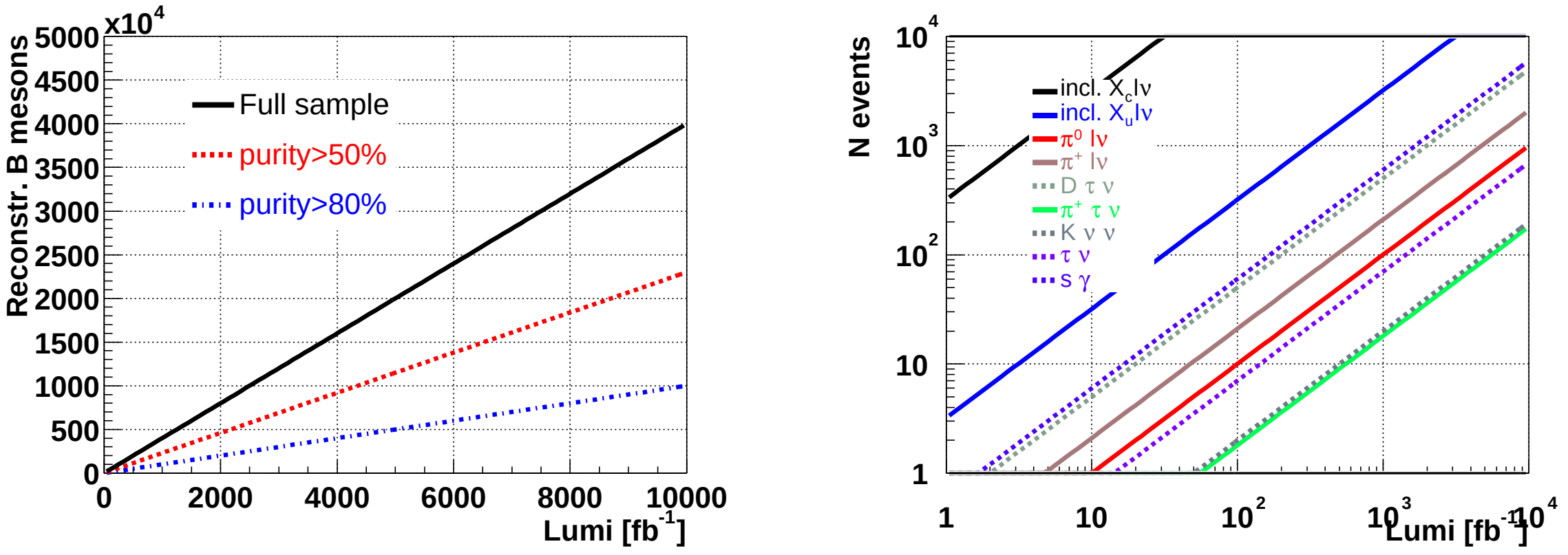

**Figure 4-2.** *Left: yields of fully reconstructed hadronic $B$ meson decays for different levels of purity as a function of the integrated luminosity. Right: number of selected signal events for different processes as a function of the integrated luminosity. We assume a rough estimate of the selection efficiency on the recoil. The purity of the selected sample can vary depending on the process.*

### Semileptonic tags

≻– D. del Re, M. Datta –≺

An alternative method of event tagging employs the reconstruction of semileptonic decays. The technique has a higher efficiency compared to the fully hadronic approach, but it has some disadvantages due to a smaller number of constraints. For instance, the presence of an extra neutrino does not allow the use of kinematic fits, and the momentum of the recoiling $B$ meson is thus known only with large uncertainty. Moreover, there is no equivalent of the $m_{ES}$ variable, and the fit of yields and the subtraction of the continuum is therefore not possible. Reconstruction efficiencies for both signal and combinatorial background must be estimated on Monte Carlo (MC) simulation and off-peak data, but can be calibrated with control samples. On the other hand, the method can still allow for a direct determination of the recoil (such as the invariant mass of the $X$ system in $B \to Xl\nu$ decays), since all visible particles are reconstructed.

In semi-exclusive semileptonic $B$ tags, excited neutral $D$ modes are not explicitly reconstructed, potentially leaving unassigned neutral energy in the event. $B^-$ candidates are reconstructed via the decay $B^- \to D^0\ell^-\overline{\nu}X$, where the $X$ system is either nothing, a $\pi^0$ meson or a $\gamma$ from the $D^{*0}$ meson or an unreconstructed higher $D$ meson resonance. After imposing kinematic requirements on the $D^0$-$\ell$ combination, the $X$ is usually either nothing or a soft transition pion or photon from a higher mass charm state. The subsequent $D$ meson decay is reconstructed as either $D^0 \to K^-\pi^+$, $D^0 \to K^-\pi^+\pi^+\pi^-$ or $D^0 \to K^-\pi^+\pi^0$. These $D^0$ decay modes are chosen, since they provide both the highest statistics hadronic decay modes and are the cleanest. The lepton $\ell$ denotes either an electron or a muon.

The exclusive semileptonic decays $B^- \to D^{*0}\ell^-\overline{\nu}_\ell$ are a cleaner subset of $D^0\ell^-\nu X$ tags. Due to the reconstruction of all the tag side visible particles, the recoil of this tagging mode is clean enough to search for signal decays with a less clean signature.

To study neutral modes $\overline{B}^0 \to D^+\ell^-\overline{\nu}$ we use the charged $D$ meson decay $D^+ \to K^-\pi^+\pi^+$. Also, although we do not require the reconstruction of a $D^{*+}\ \ell^-\overline{\nu}$, if an acceptable $D^{*+}$ candidate can be formed by combining a found $D^0$ with a soft pion, it is used in place of the $D^0$ candidate. If an acceptable $D^{*+}$ candidate can be reconstructed, it is considered a suitable $\overline{B}^0$ tag. As mentioned, missing particles in the tagging $B$ do not constitute a problem, as long as all measured particles are properly assigned. The efficiency on this method is $\sim 1\%$ of $\Upsilon(4S) \to B\overline{B}$ events ($\sim 0.35\%$ for $\overline{B}^0$ and $\sim 0.65\%$ for $B^-$). Even though the experimental systematic uncertainties are larger in this case, this method can provide larger statistics, and can be very useful for the study of many modes with small branching ratios. The overlap between this sample and the fully hadronic one is negligible; the two approaches can be considered uncorrelated.

### 4.2.2 Machine backgrounds

>– S. Robertson –<

Several analyses in which the signal decay mode contains one or more neutrinos rely heavily on the missing-energy signature of the unobserved neutrino(s) as part of the signal selection. In the case of $B \to X_u \ell \overline{\nu}$ decays, the neutrino four-vector may be explicitly reconstructed from the missing momentum and energy in the event, while for $B^+ \to \ell^+ \nu$, the neutrino is implicitly reconstructed by demanding that the four-vectors of all observed particle in the event other than the signal candidate lepton can be summed to form a four-vector consistent with a $B^-$ meson. For decay modes such as $B^+ \to \tau^+ \nu$ and $B^+ \to K^+ \nu \overline{\nu}$, in which there are more than one neutrino in the final state, the signal selection similarly requires that there is large missing energy, and that all observed particles in the event can be associated with either the signal decay, or a reconstructed $B^-$ against which it is recoiling. Two factors therefore strongly impact the performance of these analyses:

- Failure to reconstruct particles that pass outside of the geometric or kinematic acceptance of the detector.
- The presence of additional reconstructed energy in the event due to detector "noise" or reconstruction artifacts, due to physics effects such as bremsstrahlung and hadronic split-offs in the calorimeter, or due to cosmics or beam-related backgrounds.

In the next subsections, we discuss these two factors.

**Acceptance and Hermiticity**

Fiducial acceptance currently has the largest impact on missing energy reconstruction in *BABAR*, with an average of $\sim 1$ GeV of energy being missed per event. However, analyses suffering from backgrounds due to this mechanism can require that the missing momentum vector point into the detector acceptance (cf. Fig. 4-14). More problematic are backgrounds that have large missing energy due to a combination of sources, as is the case for $B^+ \to \tau^+ \nu$. In this analysis, backgrounds typically arise from events in which one or more particles pass outside of the geometrical acceptance, and additionally the event contains an unidentified $K^0_L$, in which case the missing momentum vector can point in any direction. For this analysis, improving the acceptance does not result in a dramatic reduction in the background rate. A study of the effect of instrumenting the *BABAR* forward B1 magnet with a "veto" detector to increase the effective geometric acceptance indicated only about a $15\%$ reduction of background, even assuming perfect reconstruction efficiency for this detector and no occupancy due to beam backgrounds or QED processes. Some gain would potentially be realized in the signal efficiency if the tracking and/or calorimeter acceptance were increased compared with the existing *BABAR* detector.

**Occupancy**

Issues related to reconstruction artifacts are likely to be similar at PEP-II/*BABAR* and at a Super $B$ Factory. Moreover, these are not expected to be the dominant source of extra energy in a high luminosity environment. Potentially the most serious issue is the presence of significant occupancy in the calorimeter, and to a lesser degree the tracking system, due to beam backgrounds and "non-physics" luminosity effects. *BABAR* data currently contains an average of $\sim 1$ spurious calorimeter cluster per event with a typical energy of $60 - 100$ MeV. Most of this energy is the result of single-beam lost-particle sources in the high-energy or low-energy rings, and linear scaling of these backgrounds to the anticipated Super $B$ Factory beam currents has the effect of increasing their contribution to the level of $\sim 300$ MeV/event, comparable to the total "extra energy" currently observed in *BABAR* data (*i.e.*, including beam backgrounds, bremsstrahlung, hadronic split-offs, *etc.*). Consequently, the missing energy resolution probably will not dramatically be degraded by this effect. However there is also currently a significant beam background component that scales with luminosity rather than with beam currents. This background is believed to be caused by extremely low angle radiative Bhabha scattering producing particles that scatter into machine elements in the vicinity of the Q2 septum chambers. A naive scaling of current background rates to Super $B$ Factory luminosity would then imply extremely high occupancy in portions of the calorimeter, potentially degrading the missing energy resolution to the point that some or all of these analyses would not be not possible. Additional studies of this effect are needed; missing energy reconstruction should be an important benchmark in designing machine elements in the vicinity of the interaction region.

### 4.2.3 Detector Simulations

>– M. Datta, T. Moore –<

A detailed simulation of the current *BABAR* detector has been employed for many of the high-luminosity studies presented here. The full *BABAR* simulation includes a detailed detector model using the GEANT4 toolkit [24]. GEANT 4 provides simulations for both electromagnetic and hadronic interactions. The full detector response is simulated in each sub-system so that the standard reconstruction algorithms may be applied to the simulated data. Machine backgrounds are included by overlaying random trigger events from the real data on top of the simulated events. The simulated samples of generic decays ($B\overline{B}$, $c\overline{c}$, $uds$, *etc.*) represent up to three times the existing data sample. Much larger equivalent samples have been produced for specific signal decay modes.

In order to study the large data samples possible at a Super $B$ Factory, a fast MC simulation called "Pravda" has also been developed. This simulation begins by running the same event generators that are used in the full simulation. Instead of employing the detector simulation and response code, however, the detector response to the final state particles (charged tracks and neutrals) is parametrized. The charged track parametrization includes track smearing and a full error matrix. The same *BABAR* analysis code that is run on real data may also be run with the Pravda simulation.

The Pravda simulation does have some shortcomings that may or may not be important depending on the analysis considered. Because the reconstructed objects are parametrized from the true generator-level particles, there is no simulation of fake tracks and calorimeter noise. Furthermore, beam-related backgrounds are not included. This could have a significant impact on results obtained with the Pravda simulation, since we expect beam backgrounds to increase with higher luminosity. We currently have no reliable estimate of this effect, but work is underway to improve the characterization of these backgrounds. Studies of the $B^+ \to \mu^+\nu_\mu$ analysis showed optimistic predictions for the signal efficiency due to better than expected resolution on the event total energy. Nevertheless, we believe the Pravda MC was adequate for these studies. The $B^+ \to \tau^+\nu_\tau$ analysis, however, is critically dependent on the neutral energy reconstruction which was found to be inadequate in the Pravda simulation (see Section 4.6.2 for more details).

## 4.3 $b \to c\ell\overline{\nu}$ Inclusive and Exclusive Decays

>– I. Bigi –<

$|V_{cb}|$ is known from inclusive and exclusive semileptonic $B$ decays with a few percent uncertainty. The error is likely to be reduced to the 1-2% level through more data and a refined analysis of energy and mass moments in semileptonic and radiative $B$ decays. Experimental cuts on energies and momenta introduce biases in the extracted values of the heavy quark parameters; keeping those biases under control such that one can correct for them requires low cuts. The recently proposed BPS expansion might open up a novel way to determine $|V_{cb}|$ from $B \to D\ell\nu$. If successful for $B \to De/\mu\nu$, one can use the ratio $\Gamma(B \to D\tau\nu)/\Gamma(B \to De/\mu\nu)$ as a sensitive probe for New Physics, where the BPS expansion is essential in treating the hadronic form factors. Extracting $|V_{cb}|$ from semileptonic $B_s$ decays in $e^+e^- \to \Upsilon(5S) \to B_s\overline{B}_s$ would constitute a powerful check on our theoretical control.

We are witnessing how the study of $B$ physics, which has been based on the paradigm of high sensitivity to subtle and potentially new features of fundamental dynamics, is now also acquiring the aspect of high numerical accuracy.

This development has been driven by two interrelated phenomena, namely the availability of magnificent experimental facilities that challenged theoretical technologies and, in doing so, inspired—actually pushed—them to become more powerful. There is every reason to expect that this fruitful interplay will continue: theoretical technologies will be further refined in response to even more detailed data.

One expression of this paradigm shift has the suggestion of a Super $B$ Factory , an asymmetric $e^+e^-$ collider operating near $B$ production threshold with a luminosity of close to $10^{36}\mathrm{s}^{-1}\mathrm{cm}^{-2}$. Its justification has to be different than that more than ten years ago for the current $B$ Factories: one has to learn to harness the much higher statistics to shape a Super $B$ Factory into a true precision tool for exploring dynamics. This means one has to strive for

- *more accuracy* in extracting the sides of the CKM unitarity triangle,
- analyzing *more decays* – like $B \to De/\mu\nu$, D $\tau\nu$ – and
- possibly cover *new territory*, namely $e^+e^- \to \Upsilon(5S) \to B_s\overline{B}_s$.

It also means that one should not apply if one is deterred by truly hard measurements.

The '1% challenge' is the following: can we learn to predict certain observables with an $\mathcal{O}(1\%)$ accuracy, measure them, interpret the results and diagnose what they tell us about specific features of the underlying dynamics with commensurate accuracy?

In taking up this challenge, we have to be aware that assumptions that are well justified on the $\mathcal{O}(10\%)$ accuracy level might no longer be adequate on the $\mathcal{O}(1\%)$ accuracy level. Furthermore, the most convincing way that we have established control over the systematics—be they experimental or theoretical—is to determine the same basic parameter in more than one independent way. Heavy quark theory [25, 26, 27, 28] is quite well positioned to satisfy this demand, as will be illustrated below.

We expect that $|V_{cb}|$ will be determined with 1-2% accuracy soon at the current $B$ Factories. We address it here in the Super $B$ Factory context mainly to describe what will be the status and to illustrate at the same time the new paradigm of heavy flavor physics, which is based on two pillars:

- building a rich database involving hard measurements;
- implementing overconstraints as much as possible.

In that spirit we briefly sketch important cross checks that could be performed in $\Upsilon(5S) \to B_s\overline{B}_s$.

### 4.3.1 On the Heavy Quark Expansion (HQE)

$B$ decays—mostly of the inclusive variety—can be described through an operator product expansion (OPE) in inverse powers of the heavy quark masses and of the $B$ meson expectation values of *local* quark and gluon field operators of increasing dimension. Those are referred to as heavy quark parameters (HQP): the heavy quark masses—$m_b$, $m_c$—on the leading level, the kinetic energy and chromomagnetic moments—$\mu^2_\pi$, $\mu^2_G$—to order $1/m^2_Q$ and the Darwin and $LS$ terms—$\rho^3_D$, $\rho^3_{LS}$—to order $1/m^3_Q$, *etc.*

The important point is that this set of HQP is 'universal' in the sense that it appears in the HQE of a host of transitions, namely $b \to c$ and $b \to u$ semileptonic, radiative and even nonleptonic ones. These HQP can be extracted from the *shape* of energy, mass, *etc.*distributions as conveniently encoded in various moments of different orders. In general there is *not* a one-to-one correspondence between these HQP and the moments; *i.e.*, the former are obtained from nontrivial linear combinations of the latter. Likewise the HQP can be determined from different types of moments, namely leptonic, hadronic or photonic moments. They can thus be greatly overconstrained, providing a high degree of quality control over systematics on the theoretical as well as experimental side. Once the HQP are obtained from moments of $B \to X_c\ell\nu$ transitions, they can be used perfectly well for $B \to X_u\ell\nu$ and $B \to X_s\gamma$. Claiming that one needs to measure moments of $b \to u$ decays to obtain the HQP for describing them would be incorrect.

More than one treatment of the HQE with different definitions of the HQP can be found in the literature. We use 'kinetic' masses and other HQP with a hard Wilsonian cut-off scale $\mu \sim 1$ GeV. Other authors [29] studied many schemes, such as the '1S' and 'PS' masses, using HQET quantities $\lambda_{1,2}$ and four non-local correlators $\mathcal{T}_{1-4}$ together with $\rho_D^3$, $\rho_{LS}^3$ in orders $1/m_Q^2$ and $1/m_Q^3$, respectively. In any schemes there are six hadronic matrix elements that need to be determined from the data, in addition to $|V_{cb}|$. For practical applications, where only a handful of HQP truly matter, there are simple expressions relating the two sets of HQP [30]. One should keep in mind the general caveat that the role and weight of perturbative corrections is quite different in the various schemes.

### 4.3.2 $|V_{cb}|$

Three methods for extracting $|V_{cb}|$ from semileptonic $B$ decays that can boast of a genuine connection to QCD have been suggested: namely the 'inclusive' one relying on $\Gamma_{SL}(B)$, the 'exclusive' one employing $B \to D^*\ell\nu$ at zero recoil, and a newcomer, namely treating $B \to D\ell\nu$ with the help of the "so-called BPS" expansion, may become competitive.

#### 'The Golden Way': $\Gamma(B \to X_c\ell\nu)$

In the first step, one sets out to express the total $b \to c$ semileptonic width in terms of *a priori* unknown HQP and perturbative corrections, in addition to the sought-after $|V_{cb}|$ in a way that the higher-order contributions not included cannot amount to more than 1 or 2%, which then denotes the theoretical uncertainty:

$$\Gamma(B \to X_c\ell\nu) = F(|V_{cb}|; \alpha_S, HQP : m_Q, \mu_\pi^2, ...) \pm (1-2)\%|_{th} \tag{4.3}$$

This step has been completed. As shown in Ref. [31], to achieve the set goal of no more than 1–2% theoretical uncertainty at this step the following features have been included:

- all order BLM together with an estimate of second-order *non*-BLM corrections to the leading term,
- contributions through order $1/m_Q^3$,
- without ignoring, as it is usually done, contributions from HQP of the type $\langle B|(\bar{b}...c)(\bar{c}...b)|B\rangle$—*i.e.,* with local operators containing a pair of charm fields explicitly—which could be labeled 'intrinsic charm'. For otherwise there would emerge a chain of higher-dimensional operators, whose contributions scale like $\overline{\Lambda}^n/m_b^3 m_c^{n-3}$ instead of $\overline{\Lambda}^n/m_b^n$.

The main stumbling block in decreasing the theoretical uncertainty is the fact that we do not know yet even the $\mathcal{O}(\alpha_S)$ perturbative corrections of the leading nonperturbative contributions to $\mu_G^2$ and $\rho_D^3$ (as well as $\mu_\pi^2$ for moments).

It had been customary for a number of years to impose a constraint on the $b$ and $c$ quark masses that relates their difference to that of the spin-averaged $B$ and $D$ meson masses:

$$m_b - m_c = \langle M_B\rangle - \langle M_D\rangle + \mu_\pi^2\left(\frac{1}{2m_c} - \frac{1}{2m_b}\right) + \frac{\rho_D^3 - \overline{\rho}^3}{4}\left(\frac{1}{m_c^2} - \frac{1}{m_b^2}\right) + \mathcal{O}(1/m_Q^3)\,, \tag{4.4}$$

where $\overline{\rho}^3$ denotes the sum of two positive nonlocal correlators [32].

This procedure was legitimate and appropriate when one had to allow for very sizable uncertainties in the $b$ quark mass and the aim was to extract $|V_{cb}|$ with no better than 10% accuracy. However now $m_b$ is known with at least 2% precision, and the aim for $|V_{cb}|$ is considerably higher. The relation of Eq. (4.4) then turns into a weak spot or even a liability. It should, therefore, no longer be imposed as an *a priori* constraint. One can, instead, check *a posteriori* to what degree it holds.

Using the measured value for $\Gamma_{SL}(B)$ one then obtains a value for $|V_{cb}|$ as a function of the HQP [31]:[2]

$$\begin{aligned}
\frac{|V_{cb}|}{0.0417} \cdot SF \;\simeq\; & (1+\delta_{\Gamma_{SL},th})[1+0.30(\alpha_S(m_b)-0.22)] \times \\
& \times[1-0.66(m_b(1\text{ GeV})-4.6\text{ GeV}) \\
& +0.39(m_c(1\text{ GeV})-1.15\text{ GeV}) \\
& +0.05(\mu_G^2-0.35\text{ GeV}^2)-0.013(\mu_\pi^2-0.40\text{ GeV}^2) \\
& -0.09(\rho_D^3-0.2\text{ GeV}^3)-0.01(\rho_{LS}^3+0.15\text{ GeV}^3) \qquad (4.5)\\
SF \;=\; & \sqrt{\frac{0.105}{\mathcal{B}_{SL}(B)}\frac{\tau_B}{1.55\text{ps}}}, \qquad (4.6)
\end{aligned}$$

where $\delta_{\Gamma_{SL},th}$ denotes the uncertainty in the theoretical expression for $\Gamma_{SL}(B)$. More specifically:

$$\delta_{\Gamma_{SL},th} = \pm 0.005|_{pert} \pm 0.012|_{hWc} \pm 0.004|_{hpc} \pm 0.007|_{IC} \;; \qquad (4.7)$$

the numbers on the right hand side refer to the remaining uncertainty in the Wilson coefficient of the leading $\bar{b}b$ operator, the as yet uncalculated perturbative corrections to the chromomagnetic and Darwin contributions—this is the leading source of the present theoretical error—higher order power corrections including limitations to quark-hadron duality [33] and possible nonperturbative effects in operators with charm fields, respectively.

As a matter of practicality, the value of the chromomagnetic moment $\mu_G^2$ is conveniently fixed by the $B^* - B$ mass splitting.

In the second step one determines the HQP from energy and/or hadronic mass moments of different orders measured in semileptonic $b \to c$ and radiative $B$ decays. They are of the types

$$\mathcal{M}_1(E_l) = \Gamma^{-1}\int dE_l E_l d\Gamma/dE_l \qquad (4.8)$$

$$\mathcal{M}_n(E_l) = \Gamma^{-1}\int dE_l [E_l - \mathcal{M}_1(E_l)]^n d\Gamma/dE_l \;,\; n>1 \qquad (4.9)$$

$$\mathcal{M}_1(M_X) = \Gamma^{-1}\int dM_X^2 [M_X^2 - \overline{M_D}^2] d\Gamma/dM_X^2 \qquad (4.10)$$

$$\mathcal{M}_n(M_X) = \Gamma^{-1}\int dM_X^2 [M_X^2 - \langle M_X^2\rangle]^n d\Gamma/dM_X^2 \;,\; n>1\;. \qquad (4.11)$$

The DELPHI and *BABAR* analyses [34, 36] demonstrate the value of relying on several lepton energy as well as hadronic mass moments, since they provide valuable overconstraints, and, in particular, $\mathcal{M}_2(M_X)$ as well as $\mathcal{M}_3(M_X)$ are sensitive to different combinations of the relevant HQP than the other moments. The results can be stated as follows:

$$|V_{cb}|_{incl} = \frac{0.0416}{SF} \times \left[1 \pm 0.017|_{exp} \pm 0.015|_{\Gamma(B)} \pm 0.015|_{HQP}\right] \qquad (4.12)$$

where the second and third errors reflect the theoretical uncertainties in Eq. (4.7) (when added in quadrature) and in the evaluations of the HQP from the moments.

One might think that the theoretical uncertainties given in Eq. (4.12) are grossly understated. For an uncertainty of $\sim 2\%$ in the value of $m_b$ that emerged from the DELPHI analysis should contribute an uncertainty of $\sim 5\%$ in $|V_{cb}|$; *i.e.*, this source alone should produce an error larger than allowed for in Eq. (4.12). The resolution of this apparent

[2] Analogous expressions in other schemes can be found in Ref. [29], yielding similar results. (Conveners)

paradox lies in the fact that the width and the low moments depend on practically the same combination of HQP. This can be made manifest by replacing $m_b$ in Eq. (4.6) with, say, the first lepton energy or hadronic mass moments $\langle E_l \rangle$ and $\langle M_X^2 \rangle$:

$$\frac{|V_{cb}|}{0.042} \cdot SF \simeq 1 - 1.70[\langle E_l \rangle - 1.383 \text{ GeV}] - 0.075[m_c(1 \text{ GeV}) - 1.15 \text{ GeV}]$$
$$+0.085[\mu_G^2 - 0.35 \text{ GeV}^2] - 0.07[\mu_\pi^2 - 0.40 \text{ GeV}^2]$$
$$-0.055[\rho_D^3 - 0.2 \text{ GeV}^3] - 0.005[\rho_{LS}^3 + 0.15 \text{ GeV}^3] \quad (4.13)$$
$$\simeq 1 - 0.14[\langle M_X^2 \rangle - 4.54 \text{ GeV}^2] - 0.03[m_c(1 \text{ GeV}) - 1.15 \text{ GeV}]$$
$$-0.01[\mu_G^2 - 0.35 \text{ GeV}^2] - 0.1[\mu_\pi^2 - 0.40 \text{ GeV}^2]$$
$$-0.1[\rho_D^3 - 0.2 \text{ GeV}^3] + 0.006[\rho_{LS}^3 + 0.15 \text{ GeV}^3] ; \quad (4.14)$$

*i.e.*, once this substitution has been made, the sensitivity to $m_c$ has been greatly reduced, while the one to the other HQP is still rather mild.

As a '*caveat emptor*' it should be noted that the relationship between the moments and the HQP has not been scrutinized to the same degree as the one between $\Gamma_{SL}(B)$ and the HQP. Yet there are some general lessons to be drawn from it:

- One has to allow $m_b$ and $m_c$ to float independently of each other rather than impose the constraint of Eq. (4.4).
- Harnessing different types and different order of moments is essential to obtain the overconstraints that provide a sensible measure for the theoretical as well as experimental control one has achieved.
- The values of the HQP inferred from this analysis can be used in describing other widths as well like for $B \to X_u \ell \nu$ and $B \to X\gamma$. The only difference is that one has to use a different linear combination of moments to obtain $m_b$ rather than $m_b - 0.65m_c$.

**The photon spectrum—cuts and biases**

When measuring spectra to evaluate moments, experimental cuts are imposed on energies or momenta for good practical reasons. Yet theoretically such cuts can have a significant nontrivial impact not reproduced by merely integrating the usual OPE expressions over the limited range in energy or momentum, since there are exponential contributions of the form $e^{-c\mathcal{Q}/\mu_{had}}$ that do not appear in the usual OPE expressions; $\mathcal{Q}$ denotes the 'hardness' of the transition, $\mu_{had}$ the scale of nonperturbative dynamic (and $c$ a dimensionless number). Such contributions are indeed quite irrelevant for $\mathcal{Q} \gg \mu_{had}$, in particular for $\mathcal{Q} \simeq m_b,\ m_b - m_c$. Yet the aforementioned cuts degrade the 'hardness' of the transition.

For $B \to X\gamma$ the first photon energy moment and the variance provide a measure of $m_b/2$ and $\mu_\pi^2/12$, respectively. Cutting off the lower end of the photon spectrum increases the former and reduces the latter in an obvious way. Yet the impact of such a cut is not fully described by the usual OPE expressions: for the degrading of the 'hardness' is not reflected there. One has $\mathcal{Q} \simeq m_b - 2E_{cut}^\gamma$; *i.e.*, for $E_{cut}^\gamma \simeq 2\,\text{GeV}$ one has $\mathcal{Q} < 1\,\text{GeV}$, making these exponential contributions significant; for higher cuts the OPE expressions quickly lose reliability and then even meaning.

A pilot study [35] of such effects has been performed, where it was found that they introduce a bias, *i.e.*, a systematic shift in the values of $m_b$ and $\mu_\pi^2$ extracted from the measured moments with a cut [37]. The good news is that this bias does not imply the need to increase the theoretical uncertainties, but can be corrected for; *e.g.*, for $E_{cut}^\gamma = 2\,\text{GeV}$ the bias corrected and thus 'true' $m_b$ is about 50 MeV lower than the bare value extracted from the moment using the usual OPE corresponding to a $\sim 1\%$ upward shift; likewise one finds a correction of about $0.1 - 0.15\ \text{GeV}^2$ for $\mu_\pi^2$, *i.e.*, a $\sim 25\,\%$ shift, which is much larger than for the leading HQP $m_b$.

A much more detailed study is now underway [38]. Since one is merely analyzing a correction, terms $\sim \mathcal{O}(1/m_Q^3)$ are irrelevant. Thus, there are only three relevant parameters with dimension—$M_B - m_b$, $\mu_\pi^2$ and $\mathcal{Q}$—and there must be simple scaling behavior for the correction.

Some general conclusions can already be drawn:

- One should strive hard in all moment analyses to keep the experimental cuts as lows as possible.
- Such biases in the experimentally truncated moments can be corrected for rather than be invoked to inflate the theoretical uncertainties.
- Measuring moments with cuts in a range where the biases can be handled provides important cross checks of our control over the systematics.

### 'The Gold-Plated Way': $B \longrightarrow D^*\ell\nu$ at zero recoil

The second method involves measuring the exclusive reaction $B \to D^*\ell\nu$, extrapolate it to the zero recoil point[3] for $D^*$ and extract $|V_{cb}F_{D^*}(0)|$. The zero-recoil form factor has the important property that it is normalized to unity for $m_Q \to \infty$ and has no correction linear in $1/m_Q$:

$$F_{D^*}(0) = 1 + \mathcal{O}(\alpha_S) + \mathcal{O}(1/m_Q^2)\ . \tag{4.15}$$

At finite quark masses there are corrections that lower the form factor. The drawbacks are that it contains an expansion in powers of $1/m_c$ rather than just $1/m_b$ or $1/(m_b - m_c)$ and that non-local operators appear in higher orders. Different estimates for $F_{D^*}(0)$ can be found in the literature:

$$F_{D^*}(0) = \begin{cases} 0.89 \pm 0.06 \text{ Sum Rules [39]} \\ 0.913 \pm 0.042 \text{ \textit{BABAR} Physics Book [40]} \\ 0.913^{+0.024+0.017}_{-0.017-0.030} \text{ Quenched Lattice QCD [41].} \end{cases} \tag{4.16}$$

The first value was obtained by applying the HQ sum rules and includes terms through $\mathcal{O}(1/m_Q^2)$; the uncertainty applies to adding errors linearly. The lattice result is obtained in the quenched approximation and includes terms through $\mathcal{O}(1/m_Q^3)$; keeping only terms through $\mathcal{O}(1/m_Q^2)$ reduces the central value to 0.89. One should also note that the lattice analysis assumes that one can rely on an expansion in powers of $1/m_c$ (an assumption that is partially checked for self-consistency).

With $|V_{cb}F_{D^*}(0)| = 0.0367 \pm 0.0013$ and using $F_{D^*}(0) = 0.90 \pm 0.05$ for convenience, one obtains

$$|V_{cb}|_{excl} = 0.0408 \cdot [1 \pm 0.035|_{exp} \pm 0.06|_{theor}]\ , \tag{4.17}$$

to be compared with

$$|V_{cb}|_{incl} = 0.0416 \cdot [1 \pm 0.017|_{exp} \pm 0.015|_{\Gamma_{SL}(B)} \pm 0.015|_{HQP}]\ . \tag{4.18}$$

The agreement between the two values represents a highly satisfying and quite non-trivial success of both the experimental and theoretical analysis. At the same time, it is our considered judgment that with $F_{D^*}(0)$ depending on an expansion in $1/m_c$, this exclusive method is running into a 'brick wall' for the theoretical uncertainty of about 5%.

[3]This extrapolation is actually quite nontrivial, and needs to be redone carefully with better data.

### ’The Cinderella Story’: $B \to D\ell\nu$

As is now well-known, QCD possesses heavy-flavor as well as spin symmetry for $m_Q \to \infty$. At finite values of $m_Q$ both are broken by terms $\sim \mathcal{O}(1/m_Q^2)$ in $\Gamma_{SL}(B)$ and in $F_{D^*}(0)$, as just discussed. The reaction $B \to D\ell\nu$ is usually seen as a ‘poor relative’ of the more glamorous $B \to D^*\ell\nu$, since its form factor has a contribution linear in $1/m_Q$ and thus $1/m_c$. It is also harder to measure, since the relevant rate is smaller, and one cannot benefit from the $D^* \to D\pi$ ‘trick’. Yet we might be seeing a ‘Cinderella story’ in the making, namely the emergence of a novel approach allowing us to calculate the nonperturbative contributions to the form factor $F_D$ quite reliably.

The role of the ‘good fairy’ could be played by the so-called ‘BPS’ approximation [42]. If $\mu_\pi^2 = \mu_G^2$ were to hold exactly,[4] one would have

$$\vec{\sigma}_Q \cdot \vec{\pi}_Q|B\rangle = 0 \ , \ \varrho^2 = \frac{3}{4} \tag{4.19}$$

where $\vec{\pi}_Q \equiv i\vec{\partial} + g_S\vec{A}$ denotes the covariant derivative and $\varrho^2$ the slope of the Isgur-Wise function.

The BPS limit cannot be exact in QCD. From the SV sum rules, we have inferred the general inequality $\mu_\pi^2 > \mu_G^2$; yet one expects the difference to be of quite moderate size. Experimentally we have, indeed,

$$\mu_G^2 = (0.35^{+0.03}_{-0.02}) \ \mathrm{GeV}^2 \quad vs. \quad \mu_\pi^2 = (0.45 \pm 0.1) \ \mathrm{GeV}^2 \ , \tag{4.20}$$

which provides a measure for the proximity of the BPS limit through the ratio $(\mu_\pi^2 - \mu_G^2)/\mu_\pi^2$. This can be parametrized through the dimensionless quantity

$$\gamma_{\mathrm{BPS}} \equiv \sqrt{\varrho^2 - 0.75} \ , \tag{4.21}$$

which is smaller than 1/2 for $\varrho^2 < 1$. There are further suppression factors, yet even so the BPS treatment might provides only a qualitative description for observables that receive contributions linear in $\gamma_{\mathrm{BPS}}$. Yet there is a whole class of quantities where the leading corrections are of order $\gamma_{\mathrm{BPS}}^2 \propto (\mu_\pi^2 - \mu_G^2)/\mu_\pi^2$. Among them is the form factor describing $B \to D\ell\nu$ at zero recoil, analogous to $F_{D^*}(0)$ described above:

$$\mathcal{F}_+ = \frac{2\sqrt{M_B M_D}}{M_B + M_D} f_+(0) \ , \tag{4.22}$$

with the usual definition:

$$\langle D(p_D)|(\bar{c}\gamma_\mu b)|B(p_B)\rangle = f_+(q^2)(p_B + p_D)_\mu + f_-(q^2)(p_B - p_D)_\mu \ , \ q = (p_B - p_D) \ . \tag{4.23}$$

In the BPS limit, $\mathcal{F}_+$ is normalized to unity: $\mathcal{F}_+ = 1 + \mathcal{O}(\gamma_{\mathrm{BPS}}^2(1/m_c - 1/m_b)^2)$. The power-suppressed contributions are then very small; the more significant effect is due to perturbative corrections which produce a slight excess over unity for $\mathcal{F}_+$ [42]:

$$\mathcal{F}_+ = 1.04 + 0.13 \cdot \frac{\mu_\pi^2(1 \ \mathrm{GeV}^2) - 0.43 \ \mathrm{GeV}^2}{1 \ \mathrm{GeV}^2} \pm \delta|_{expon} \tag{4.24}$$

The intrinsic limitation $\delta|_{expon}$ is due to ‘exponential’ terms

$$\delta|_{expon} \propto \left(e^{-m_c/\mu_{had}} - e^{-m_b/\mu_{had}}\right)^2 \tag{4.25}$$

that have to exist, yet do not appear in the usual HQE expressions. A reasonable estimate for it is in the 1–2% range; *i.e.*, at present it seems possible that one could extract $|V_{cb}|$ from $B \to D\ell\nu$ at zero recoil with a higher accuracy than from $B \to D^*\ell\nu$. This requires that $\mu_\pi^2(1 \ \mathrm{GeV}^2) \leq 0.45 \ \mathrm{GeV}^2$ holds, *i.e.*, its value falls into the lower part of the presently allowed range. In any case, this method has to be and can be validated by comparing the value of $|V_{cb}|$ thus obtained with the one from $\Gamma_{SL}(B)$.

[4]This is not a renormalization scale independent statement, yielding concerns that have not been fully addressed. (Conveners)

**"From Rags to Riches": $B \to D\tau\nu$**

A success of this method in extracting $|V_{cb}|$ opens up an intriguing avenue to search for the intervention of New Physics in $B \to D\tau\nu$. It has been noted [43] that the ratio $B \equiv \Gamma(B \to D\tau\nu)/\Gamma(B \to De/\mu\nu)$ could be changed significantly relative to its Standard Model value by a contribution from a charged Higgs exchange. Its impact can be parametrized in a two-Higgs-doublet model by the ratio $R = M_W \tan\beta/M_H$ with tan$\beta$ denoting the ratio of the two VEV's. The authors of Ref. [43] find sizable deviations from the Standard Model value of $B$ for $R \geq 10$, which could be realized even for $M_H$ as high as 200 – 300 GeV for sufficiently large tan$\beta$. There is a considerable 'fly in the ointment', though. The authors argued that in the infinite mass limit the hadronic form factors drops out from $B$. However that is not true at finite values of the heavy quark masses. In particular there are $1/m_c$ (and $1/m_b$) corrections that are likely to be sizable; furthermore the rate for $B \to De/\mu\nu$ depends on the single form factor $f_+$, whereas $B \to D\tau\nu$ is also sensitive to the second form factor $f_-$, since $m_\tau$ is *non*negligible on the scale of $M_B$.

Yet the BPS expansion—once it is validated by $|V_{cb}|$—allows us to relate these form factors, and thus predict the value of $\Gamma(B \to D\tau\nu)/\Gamma(B \to De/\mu\nu)$ in the Standard Model. A 'significant' deviation—'significant' probably means larger than 10 %—provides evidence for New Physics.

Measuring $B \to \tau\nu D$ appears feasible only at a Super $B$ Factory due to the small branching ratio of $B \to \tau\nu D$ relative to $B \to D^*\ell\nu$, the absence of the $D^*$ 'trick' and the complication of having to identify the $\tau$ lepton.

### 4.3.3 Quality control

The option to run at $\Upsilon(5S) \to B_s\overline{B}_s$ might turn out to be very valuable. The motivation would *not* be to perform measurements that can be done at LHC and the Tevatron, such as searching for $B_s - \overline{B}_s$ oscillations and *CP* asymmetries in $B_s(t) \to D_sK$, $J/\psi\,\phi$; instead one would perform measurements *uniquely* possible here. One is the extraction of $|V_{cb}|$ from $\Gamma_{SL}(B_s)$ and $B_s \to D_s^*\ell\nu$ at zero recoil in close analogy to nonstrange $B$ decays. This is another example of following Lenin's dictum "Trust is good—control is better!". For comparing $|V_{cb}|$ as inferred from $B_d$, $B_u$ and $B_s$ decays provides a powerful check of experimental systematics and even more of theoretical uncertainties like the often mentioned limitations to quark-hadron duality. Such limitations could be larger than predicted due to the accidental "nearby presence" of a hadronic resonance of appropriate quantum numbers. This would be a stroke of bad luck, but could happen. Due to the isospin invariance of the strong interactions it would affect $B_d \to X_c\ell\nu$ and $B_u \to X_c\ell\nu$ equally (unlike $B_d \to X_u\ell\nu$ *vs.* $B_u \to X_u\ell\nu$), but *not* $B_s \to X_{c\bar{s}}\ell\nu$. Such a scenario would reveal itself by yielding inconsistent values for $|V_{cb}|$ from $B_{u,d}$ and $B_s$ semileptonic decays.

### 4.3.4 Conclusions

The study of heavy flavor dynamics in the beauty sector has made tremendous progress in both the quantity and quality of data, and in the power of the theoretical tools available to treat them. This progress is well-illustrated by the determination of $|V_{cb}|$. The pieces are in place to extract it from $\Gamma(B \to X_c\ell\nu)$ with 1–2% accuracy. This is being achieved by fixing the HQP appearing in the HQE through the shape of distributions in semileptonic and radiative $B$ decays as encoded through their energy and mass moments. Analyzing $B \to D^*\ell\nu$ at zero recoil provides a valuable cross check; yet both the procedure for extrapolating to zero-recoil and the evaluation of the form factor $F_{D^*}(0)$ have to be scrutinized very carefully. Only dedicated lattice QCD studies hold out the promise to reduce the theoretical uncertainty below the 5% mark; however, that is truly a tall order, and requires a fully unquenched treatment, and a very careful evaluation of the scaling in powers of $1/m_c$.

These developments will happen irrespective of the existence of a Super $B$ Factory . However their description is highly relevant for discussions about a Super $B$ Factory :

- The HQP $m_b$, $\mu_\pi^2$ etc. extracted from moments of $B \to X_c\ell\nu$ and $B \to X_s\gamma$ are the basic parameters needed for describing $B \to X_u\ell\nu$, $B \to \gamma X_d$, $B \to X\ell^+\ell^-$ *etc.*, transitions.

- Reproducing $|V_{cb}|$ within the stated uncertainty of $1-2\%$ constitutes a valuable validation for Super $B$ Factory measurements.
- $B \to D\ell\nu$ has been put forward as a second theoretically clean exclusive mode for determining $|V_{cb}|$; to perform an accurate analysis close to the zero-recoil domain presumably requires data from a Super $B$ Factory .
- On a more general level, it demonstrates the 'high precision' paradigm that has to be at the core of such a program. For it illustrates how alleged high accuracy can be validated through overconstraints, namely determining the basic parameters in many systematically different ways in various decays. These lessons can be fully carried over to extractions of other CKM parameters like $|V_{ub}|$ and $|V_{td}|$.
- The huge statistics and hoped-for purity of Super $B$ Factory data are required to measure $B \to D\tau\nu$ as a sensitive probe for New Physics, most likely in the form of charged Higgs states.
- One should contemplate a run of $e^+e^- \to \Upsilon(4S) \to B_s\overline{B}_s$, not only for calibrating absolute $B_s$ branching ratios, but also to extract $|V_{cb}|$ from $B_s$ decays, as the final cross check of our theoretical control.

### 4.3.5 Experimental Prospects

$\succ$– U. Langenegger –$\prec$

Recent preliminary measurements of the lepton spectrum [21] and the mass moments of the hadronic system [44] presented by the *BABAR* and Belle collaborations using the recoil approach already show very competitive results compared to the the traditional $B$ tagging with high-momentum leptons. With statistics of 200–300 $\text{fb}^{-1}$, the analyses will probably become systematics-limited. At the moment, there are no prospects for substantial gains at higher luminosities in the study of these decays.

## 4.4 $b \to u$ Inclusive Decays

### 4.4.1 Theory

>– M. Luke –<

A precise and model independent determination of the magnitude of the Cabibbo-Kobayashi-Maskawa (CKM) matrix element $V_{ub}$ is important for testing the Standard Model at $B$ Factories via the comparison of the angles and the sides of the unitarity triangle.

$|V_{ub}|$ is notoriously difficult to measure in a model independent manner. The first extraction of $|V_{ub}|$ from experimental data relied on a study of the lepton energy spectrum in inclusive charmless semileptonic $B$ decay [45], a region in which (as will be discussed) the rate is highly model-dependent. $|V_{ub}|$ has also been measured from exclusive semileptonic $B \to \rho\ell\overline{\nu}$ and $B \to \pi\ell\overline{\nu}$ decay [46]. These exclusive determinations also suffer from model dependence, as they rely on form factor models (such as light-cone sum rules [47]) or quenched lattice calculations at the present time (for a review of recent lattice results, see [48]).

In contrast, inclusive decays are quite simple theoretically, and if it were not for the huge background from decays to charm, it would be straightforward to determine $|V_{ub}|$ from inclusive semileptonic decays. Inclusive $B$ decay rates can be computed model independently in a series in $\Lambda_{\rm QCD}/m_b$ and $\alpha_s(m_b)$ using an operator product expansion (OPE) [49]. At leading order, the $B$ meson decay rate is equal to the $b$ quark decay rate. The leading nonperturbative corrections of order $\Lambda^2_{\rm QCD}/m_b^2$ are characterized by two heavy quark effective theory (HQET) matrix elements, usually called $\lambda_1$ and $\lambda_2$,

$$\lambda_1 \equiv \frac{1}{2m_B}\langle B|\overline{h}_v(iD)^2 h_v|B\rangle, \qquad \lambda_2(\mu) \equiv \frac{1}{6m_B}\langle B|\overline{h}_v\sigma^{\mu\nu}G_{\mu\nu}h_v|B\rangle. \tag{4.26}$$

The $B-B^*$ mass splitting determines $\lambda_2(m_b) \simeq 0.12~\mathrm{GeV}^2$, while a recent fit to moments of the charged lepton spectrum in semileptonic $b \to c$ decay obtained [50]

$$m_b^{1S} = 4.82 \pm 0.07_E \pm 0.11_T~\mathrm{GeV}, \qquad \lambda_1 = -0.25 \pm 0.02_{ST} \pm 0.05_{SY} \pm 0.14_T~\mathrm{GeV}^2\,, \tag{4.27}$$

where $m_b^{1S}$ is the short-distance "$1S$ mass" of the $b$ quark [51, 52]. (Moments of other spectra give similar results [53].)

Since the parton level decay rate is proportional to $m_b^5$, the uncertainty in $m_b$ is a dominant source of uncertainty in the relation between $B \to X_u\ell\overline{\nu}_\ell$ and $|V_{ub}|$; an uncertainty in $m_b$ of 50 MeV corresponds to a $\sim 5\%$ determination of $|V_{ub}|$ [51, 54]. Unfortunately, the semileptonic $b \to u$ decay rate is difficult to measure experimentally, because of the large background from charmed final states. As a result, there has been much theoretical and experimental interest in the decay rate in restricted regions of phase space where the charm background is absent. Of particular interest have been the large lepton energy region, $E_\ell > (m_B^2 - m_D^2)/2m_B$, the low hadronic invariant mass region, $m_X \equiv \sqrt{s_H} < m_D$ [55], the large lepton invariant mass region $q^2 > (m_B - m_D)^2$ [56], and combinations of these [57]. Of these, the charged lepton cut is the easiest to implement experimentally, while the hadronic mass cut has the advantage that it contains roughly $80\%$ of the semileptonic rate [58]. However, in both of these cases, the kinematic cuts constrain the final hadronic state to consist of energetic, low invariant mass hadrons, and the local OPE breaks down. By contrast, in the large $q^2$ region the local OPE remains valid, although there are a number of other sources of theoretical uncertainty.

**The shape function region:** For the cuts $E_\ell > (m_B^2 - m_D^2)/2m_B$ and $m_X \equiv \sqrt{s_H} < m_D$, the local OPE breaks down and the relevant spectrum is instead determined at leading order in $\Lambda_{\rm QCD}/m_b$ by the light-cone distribution function of the $b$ quark in the meson [59],

$$f(\omega) \equiv \frac{\langle B|\overline{b}\;\delta(\omega + in\cdot D)\;b|B\rangle}{2m_B}\,, \tag{4.28}$$

where $n^\mu$ is a light-like vector. $f(\omega)$ is often referred to as the shape function, and corresponds to resumming an infinite series of local operators in the usual OPE. The physical spectra are determined by convoluting the shape function with the appropriate kinematic functions:

$$\frac{1}{\Gamma}\frac{\mathrm{d}\Gamma(B \to X_u \ell \overline{\nu}_\ell)}{\mathrm{d}E_\ell} = \frac{4}{m_b}\int \theta(m_b - 2E_\ell - \omega) f(\omega)\, d\omega + \dots \tag{4.29}$$

$$\frac{1}{\Gamma}\frac{\mathrm{d}\Gamma(B \to X_u \ell \overline{\nu}_\ell)}{\mathrm{d}s_H} = \frac{1}{m_b^3}\int \frac{2s_H^2(3\omega - 2s_H/m_b)}{\omega^4}\, \theta(\omega - s_H/m_b) f(\omega - \Delta)\, d\omega + \dots \tag{4.30}$$

where $m_b - 2E_\ell \lesssim \Lambda_{\rm QCD}$, $s_H \lesssim \Lambda_{\rm QCD} m_b$, $\Delta \equiv m_B - m_b$, and the ellipses denote terms suppressed by powers of $\alpha_s$ or $\Lambda_{\rm QCD}/m_b$. $f(\omega)$ is a nonperturbative function and cannot be calculated analytically, so the rate in this region is model-dependent even at leading order in $\Lambda_{\rm QCD}/m_b$.

However, $f(\omega)$ also determines the shape of the photon spectrum in $B \to X_s\gamma$ at leading order,

$$\frac{1}{\Gamma}\frac{\mathrm{d}\Gamma(B \to X_s\gamma)}{\mathrm{d}E_\gamma} = 2f(m_b - 2E_\gamma) + \dots \tag{4.31}$$

so $f(\omega)$ may be determined experimentally from the measured $B \to X_s\gamma$ spectrum and applied to semileptonic decay. The CLEO collaboration [60] recently used a variant of this approach to determine $|V_{ub}|$ from their measurements of the $B \to X_s\gamma$ photon spectrum and the charged lepton spectrum in $B \to X_u \ell \overline{\nu}_\ell$.

The relations (4.29–4.31) hold only at tree level and at leading order in $\Lambda_{\rm QCD}/m_b$, so a precision determination of $|V_{ub}|$ requires an understanding of the size of the corrections. The most important radiative corrections are the parametrically large Sudakov logarithms, which have been summed to subleading order [61]. In addition, contributions from additional operators which contribute to $B \to X_s\gamma$ have been calculated [62]. The perturbative corrections are typically included by convoluting the partonic rate with the shape function $f(\omega)$ [58]; however, the consistency of this approach has been questioned in [63].

The subleading twist corrections have been studied more recently [64, 65, 66, 67, 68]. In [65, 66], it was shown that there is a large $\mathcal{O}(\Lambda_{\rm QCD}/m_b)$ correction to the relation between the $B \to X_s\gamma$ spectrum and the charged lepton energy endpoint region, shifting the extracted value of $|V_{ub}|$ by $\sim 10 - 15\%$. Since this is a simple model estimate, the corresponding uncertainty is not clear. In Ref. [67] it was shown that the variation of this estimate in a number of models was quite small, suggesting a small uncertainty in $|V_{ub}|$. However, models that give larger effects do exist [68]. A second source of uncertainty arises because of the weak annihilation (WA) contribution, which will be discussed in more detail in the next section. These are formally sub-subleading twist effects, but are enhanced by a factor of $\sim 16\pi^2$ because there are only two particles in the final state. However, the relevant matrix elements vanish under the assumption of factorization; hence, as will be discussed in the next section, the size of the WA contribution is very difficult to determine reliably. The authors of [66] estimated the corresponding uncertainty in $|V_{ub}|$ to be at the $\sim 10\%$ level (with unknown sign) for a cut $E_\ell > 2.3$ GeV. For both subleading effects, the fractional uncertainty in $|V_{ub}|$ is reduced considerably as the cut on $E_\ell$ is lowered below 2.3 GeV.

Analogous corrections to the region between the $B \to X_s\gamma$ spectrum and the hadronic invariant mass spectrum were considered in [68], and found to be much smaller. In the range of models studied, the subleading effects were at the few percent level for a cut $m_X < 1.55\,\mathrm{GeV}$. The subleading effects are reduced as the cut is raised.

**Lepton $q^2$ cuts:** Another solution to the problem of the breakdown of the local OPE is to find a set of cuts which eliminate the charm background but do not destroy the convergence of the OPE, so that the distribution function $f(\omega)$ is not required. In Ref. [56] it was pointed out that this is the situation for a cut on the dilepton invariant mass. Decays with $q^2 > (m_B - m_D)^2$ must arise from $b \to u$ transition. Such a cut forbids the hadronic final state from moving fast in the $B$ rest frame, and simultaneously imposes $m_X < m_D$ and $E_X < m_D$. Thus, the region selected by a $q^2$ cut is entirely contained within the $m_X^2$ cut, but because the dangerous region of high energy, low invariant mass final states is not included, the OPE does not break down [69]. The price to be paid is that the relative size of the unknown $\Lambda^3_{\rm QCD}/m_b^3$ terms in the OPE grows as the $q^2$ cut is raised. Equivalently, as was stressed in [70], the effective expansion parameter for integrated rate inside the region $q^2 > (m_B - m_D)^2$ is $\Lambda_{\rm QCD}/m_c$, not $\Lambda_{\rm QCD}/m_b$. In addition, the integrated cut rate is very sensitive to $m_b$, with a $\pm 80$ MeV error in $m_b$ corresponding to a $\sim \pm 10\%$ uncertainty in $|V_{ub}|$ [70, 57].

An additional source of uncertainty arises from weak annihilation (WA) graphs [71]. WA arises at $\mathcal{O}(\Lambda^3_{\rm QCD}/m_b^3)$ in the local OPE, but, as previous mentioned, is enhanced by a factor of $\sim 16\pi^2$, but vanishes in factorization. Assuming factorization is violated at the $10\%$ level gives a corresponding uncertainty in $|V_{ub}|$ from a pure $q^2$ cut of $\sim 10\%$ [71]; however, this estimate is highly uncertain.[5] In addition, since the contribution is fixed at maximal $q^2$, the corresponding uncertainty grows as the cuts are tightened.

The theoretical uncertainties from a pure $q^2$ cut may be considerably reduced by considering more complicated kinematic cuts: in [57] it was proposed that by combining cuts on both the leptonic and hadronic invariant masses the theoretical uncertainty on $|V_{ub}|$ could be minimized. For a fixed cut on $m_X$, lowering the bound on $q^2$ increases the cut rate and decreases the relative size of the $1/m_b^3$ terms (including the WA terms), while introducing only a small dependence on $f(\omega)$. Since this dependence is so weak, a crude measurement of $f(\omega)$ suffices to keep the corresponding theoretical error negligible. The sensitivity to $m_b$ is also reduced. With the representative cuts $q^2 > 6\ {\rm GeV}^2$, $m_X < 1.86$ GeV, the overall theoretical uncertainty in $|V_{ub}|$ was estimated to be at the $\sim 8\%$ level, assuming a $\pm 80$ MeV uncertainty in $m_b$. Tightening these cuts further increases the overall theoretical uncertainty; estimates of the theoretical uncertainty for different cuts are given in Ref. [56].

### Nonfactorizable terms and the determination of $|V_{ub}|$

>– M. Voloshin –<

The well-known difficulty of determining the mixing parameter $|V_{ub}|$ from the inclusive semileptonic decay rate is the need to cope with the overwhelming background due to the transition $b \to c$. The suggested way to eliminate, or strongly suppress, this background is to measure the rate of the decays $B \to X_u\, \ell\, \nu$ in restricted regions of the phase space that are kinematically forbidden for $B \to X_c\, \ell\, \nu$. Such kinematical cuts however leave as 'usable' only a fraction of the total inclusive rate of the decays $B \to X_u\, \ell\, \nu$, and the nonperturbative effects discussed in this section become relatively enhanced in the restricted decay rate, while being quite small in the total probability of the decay. Namely, the discussed effects behave formally as a delta function located either at the lowest end of the spectrum of the hadronic recoil invariant mass $m_X$, or, equivalently, at the highest value of the $q^2$ for the lepton pair. In reality these effects are spread over interval determined by $\Lambda_{QCD}$, although resolving such smearing is beyond the current accuracy of the theoretical analysis.

The standard description [74, 75] of nonperturbative effects in the inclusive decay rates of a heavy hadron $H_Q$ containing a heavy quark $Q$ is based on the Operator Product Expansion (OPE) in inverse powers of the heavy quark mass $m_Q$ for the effective operator

$$\mathcal{L}_{eff} = 2\,{\rm Im}\,\left[ i \int d^4x\, e^{ipx}\, T\left\{ \mathcal{L}_W^\dagger(x), \mathcal{L}_W(0) \right\} \right], \tag{4.32}$$

constructed from the weak-interaction Lagrangian $\mathcal{L}_W$, in terms of which operator (at $p^2 = m_Q^2$) the total decay rate is given by[6]

$$\Gamma_H = \langle H_Q|\, \mathcal{L}_{eff}\, |H_Q\rangle\, . \tag{4.33}$$

Using in Eq. (4.32) the term

$$\mathcal{L}_{ub} = \frac{G_F\, V_{ub}}{\sqrt{2}}\, (\overline{u}\, \gamma_\mu\, (1-\gamma_5)\, b)\, \ell_\mu \tag{4.34}$$

[5]After completion of this report, it was observed that the $\mathcal{O}(\alpha_s)$ corrections to WA may actually dominate in the endpoint regions [72], as the $\alpha_s/(4\pi)$ suppression is compensated by a $m_b/\Lambda_{\rm QCD}$ enhancement. At present there is disagreement as to whether the $\mathcal{O}(0.1)$ suppression of the tree level term discussed after Eq. (4.38) is lifted at $\mathcal{O}(\alpha_s)$ [72] or not [73].

[6] The non-relativistic normalization for the *heavy* quark states is used here: $\langle Q|Q^\dagger Q|Q\rangle = 1$.

with $\ell_\mu = \bar{\ell}\,\gamma_\mu\,(1-\gamma_5)\,\nu$ in place of $\mathcal{L}_W$, one would find the total inclusive decay rate of $B \to X_u\,\ell\,\nu$. The effective operator (4.32) is evaluated using short-distance OPE. The leading term in the expansion describes the perturbative decay rate, while subsequent terms containing operators of higher dimension describe the nonperturbative contributions. The term of interest for the present discussion is the third one in this expansion, containing a four-quark operator [74, 75, 76, 77]

$$\mathcal{L}^{(3)}_{b\to u\ell\nu} = -\frac{2\,G_F^2\,|V_{ub}|^2\,m_b^2}{3\,\pi}\,\left(O^u_{V-A} - O^u_{S-P}\right), \tag{4.35}$$

where the following notation [78] is used for the relevant four-quark operators (normalized at $\mu = m_b$):

$$\begin{aligned} O^q_{V-A} &= (\bar{b}_L\gamma_\mu q_L)(\bar{q}_L\gamma_\mu b_L)\,, & O^q_{S-P} &= (\bar{b}_R q_L)(\bar{q}_L b_R)\,,\\ T^q_{V-A} &= (\bar{b}_L t^a\gamma_\mu q_L)(\bar{q}_L t^a\gamma_\mu b_L)\,, & T^q_{S-P} &= (\bar{b}_R t^a q_L)(\bar{q}_L t^a b_R)\,. \end{aligned} \tag{4.36}$$

(The operators $T$, containing the color generators $t^a$, will appear in further discussion.)

The matrix elements of the operators $O^u$ over the $B$ mesons can be parameterized in terms of the meson annihilation constant $f_B$ and of dimensionless coefficients $B$ (“bag constants”) as

$$\langle B^+|O^u_{V-A}|B^+\rangle = \frac{f_B^2\,m_B}{16}\,(B_1^s + B_1^{ns})\,, \qquad \langle B^+|O^u_{S-P}|B^+\rangle = \frac{f_B^2\,m_B}{16}\,(B_2^s + B_2^{ns})\,, \tag{4.37}$$

for the $B^+$ meson containing the same light quark ($u$) as in the operator, and

$$\langle B_d|O^u_{V-A}|B_d\rangle = \frac{f_B^2\,m_B}{16}\,(B_1^s - B_1^{ns})\,, \qquad \langle B_d|O^u_{S-P}|B_d\rangle = \frac{f_B^2\,m_B}{16}\,(B_2^s - B_2^{ns})\,, \tag{4.38}$$

for the $B_d$ meson where the light quark ($d$) is different from the one in the operator. In the limit of naive factorization the “bag constants”, both the flavor-singlet ($B^s$) and the flavor non-singlet ($B^{ns}$) ones are all equal to one: $B_1^s = B_1^{ns} = B_2^s = B_2^{ns} = 1$, and the matrix elements over the $B$ mesons of the difference of the operators entering Eq. (4.35) are vanishing. However the expected accuracy of the factorization is only about 10%, which sets the natural scale for the non-factorizable contributions, *i.e.*, for the deviations from the naive factorization. (Numerical estimates of non-factorizable terms can be found in [79, 80, 81].) After averaging the operator in Eq. (4.35) one finds the contribution of the non-factorizable terms to the rates of the $B \to X_u\,\ell\,\nu$ decays in the form

$$\delta\Gamma(B^\pm \to X_u\,\ell\,\nu) = \frac{G_F^2\,|V_{ub}|^2\,f_B^2\,m_b^3}{12\,\pi}\,\frac{\delta B^s + \delta B^{ns}}{2}\,, \qquad \delta\Gamma(B_d \to X_u\,\ell\,\nu) = \frac{G_F^2\,|V_{ub}|^2\,f_B^2\,m_b^3}{12\,\pi}\,\frac{\delta B^s - \delta B^{ns}}{2}\,, \tag{4.39}$$

where $\delta B^s = B_2^s - B_1^s$ and $\delta B^{ns} = B_2^{ns} - B_1^{ns}$. These contributions can be compared with the ‘bare’ total decay rate $\Gamma_0 = G_F^2\,|V_{ub}|^2 m_b^5/(192\pi^3)$:

$$\begin{aligned} \frac{\delta\Gamma(B^\pm)}{\Gamma_0} &\approx \frac{16\pi^2\,f_B^2}{m_b^2}\,\frac{\delta B^s + \delta B^{ns}}{2} \approx 0.03\left(\frac{f_B}{0.2\,GeV}\right)^2\,\frac{\delta B^s + \delta B^{ns}}{0.2}\,,\\ \frac{\delta\Gamma(B_d)}{\Gamma_0} &\approx \frac{16\pi^2\,f_B^2}{m_b^2}\,\frac{\delta B^s - \delta B^{ns}}{2} \approx 0.03\left(\frac{f_B}{0.2\,GeV}\right)^2\,\frac{\delta B^s - \delta B^{ns}}{0.2}\,. \end{aligned} \tag{4.40}$$

Thus non-factorizable terms may show up in the total decay rates only at the level of few percent. Nevertheless their relative contribution in a kinematically restricted decay rate can be substantial and generally limits the precision of determination of $|V_{ub}|$ at the level of uncertainty of about 10% [82], at least until a better quantitative understanding of such terms is available.

It should be emphasized that it would be incorrect to interpret the effects of the nonfactorizable terms as due to the ‘Weak Annihilation’ of the ‘constituent’ quarks: $b\,\overline{u} \to \ell\,\nu$, since the amplitude of such a process is essentially zero for obvious chiral reasons. Rather, one might think of the discussed effects as arising from the interference and annihilation processes involving the light ‘sea’ quarks in the $B$ mesons, for which the chiral suppression is not operative, and the expected smallness of order 10% arises due to the overall smallness of the ‘sea’ contribution.

In lieu of a good theory of the non-factorizable terms, these can be studied experimentally. One straightforward way of probing these terms is to measure the difference of the (similarly kinematically restricted) decay rates for the charged $B^{\pm}$ and the neutral $B_d$ mesons. According to equations (4.40), this would allow the extraction of the flavor non-singlet coefficient $\delta B^{ns}$. However, the most natural place to study these terms are the decays of $D$ mesons, where the relative contribution of the nonperturbative effects is greatly enhanced.

In particular, it is well-known that there is a noticeable difference between the lifetimes of the strange $D_s$ and the neutral $D^0$ mesons: $\tau(D_s)/\tau(D^0) = 1.20 \pm 0.025$, which cannot be described by spectator dependent effects in Cabibbo-suppressed decay channels, or by the flavor SU(3) breaking [77]. Although this discrepancy can be attributed merely to the overall inaccuracy of the OPE in the inverse of the charm quark mass[7], a more constructive approach would be to attempt to describe this difference in lifetimes as due to deviations from factorization (see also in [77, 83]). In the limit of flavor SU(3) symmetry, the difference of the dominant inclusive nonleptonic decay rates of $D^0$ and $D_s$ mesons can be written [75] in terms of matrix elements of four-quark operators (normalized at $\mu = m_c$) as

$$\Gamma(D^0) - \Gamma(D_s) = \frac{2\, G_F^2\, \cos^4\theta_c\, m_c^2\, f_D^2\, m_D}{9\pi}\, C_+\, C_- \left( -\delta B^{ns} - \frac{3}{4}\,\varepsilon_1^{ns} + \frac{3}{4}\,\varepsilon_2^{ns} \right), \tag{4.41}$$

where $\theta_c$ is the Cabibbo angle, $C_+$ and $C_-$ are the well known short-distance QCD renormalization coefficients for nonleptonic weak interaction: $C_- = C_+^{-2} = (\alpha_s(m_c)/\alpha_s(m_W))^{12/25}$, and the flavor non-singlet coefficients $B$ and $\varepsilon$ parameterize the following differences of the matrix elements:

$$\langle T^s_{V-A}\rangle_{D_s} - \langle T^s_{V-A}\rangle_{D^0} = \frac{f_D^2\, m_D}{8}\,\varepsilon_1^{ns}\,, \qquad \langle T^s_{S-P}\rangle_{D_s} - \langle T^s_{S-P}\rangle_{D^0} = \frac{f_D^2\, m_D}{8}\,\varepsilon_2^{ns}\,, \tag{4.42}$$

where the operators $T$ are the same as in Eq. (4.36) with the $b$ quark being replaced by $c$. (The parameters $\varepsilon_1$ and $\varepsilon_2$ both vanish in the limit of factorization.) It should be also mentioned that no attempt is being made here to allow for the breaking of the flavor SU(3) symmetry, thus no distinction is made between the annihilation constants or masses of the $D_s$ and $D_0$ mesons.

The expression (4.41) for the difference of the total decay rates corresponds numerically to

$$\Gamma(D^0) - \Gamma(D_s) \approx 3.3 \left(\frac{f_D}{0.22\, GeV}\right)^2 \left( -\delta B^{ns} - \frac{3}{4}\,\varepsilon_1^{ns} + \frac{3}{4}\,\varepsilon_2^{ns} \right) \mathrm{ps}^{-1}\,. \tag{4.43}$$

Comparing this estimate with the experimental value for the difference of the total decay rates: $0.41 \pm 0.05\,\mathrm{ps}^{-1}$, one arrives at an estimate of corresponding combination of the non-singlet factorization parameters:

$$-\,\delta B^{ns} - \frac{3}{4}\,\varepsilon_1^{ns} + \frac{3}{4}\,\varepsilon_2^{ns} \approx 0.12\,, \tag{4.44}$$

which agrees with the understanding that nonfactorizable contributions are at a level of about 10%.

The estimate (4.44) of the non-factorizable terms, however, can serve only as a semi-quantitative indicator of the magnitude of the spectator effects in the inclusive rate of the processes $B \to X_u\,\ell\,\nu$ described by a different combination of the factorization parameters in Eq. (4.39) than in Eq. (4.44). A somewhat more direct test of the relevant combination of the parameters would be possible from the difference of the total semileptonic decay rates of $D_s$ and $D^0$ mesons. Indeed, in the limit of the flavor SU(3) symmetry this difference arises only in the decays due to $c \to s\,\ell\,\nu$ and is given in terms of the operators normalized at $\mu = m_c$ as

$$\Gamma_{sl}(D^0) - \Gamma_{sl}(D_s) = \frac{G_F^2\, \cos^2\theta_c\, m_c^2\, f_D^2\, m_D}{12\pi}\,(-\delta B^{ns}) \approx 1.1 \left(\frac{f_D}{0.22\, GeV}\right)^2 (-\delta B^{ns})\ \mathrm{ps}^{-1}\,. \tag{4.45}$$

Given that the total semileptonic decay rate of the $D^0$ meson is approximately $0.16\,\mathrm{ps}^{-1}$, the discussed difference can easily amount to a quite sizable fraction of the semileptonic rate, provided that $|\delta B^{ns}| \sim 0.1$.

[7] In this respect the situation is no better for the expansion of a constrained inclusive rate of the decays $B \to X_u\,\ell\,\nu$ [82].

A measurement of the difference of the inclusive semileptonic decay rates of the $D^0$ and $D_s$ mesons would make it possible to more reliably predict the difference of the corresponding decay rates between $B^0$ and $B^\pm$ mesons: $\Gamma(B^0 \to X_u\,\ell\,\nu) - \Gamma(B^\pm \to X_u\,\ell\,\nu)$, which, according to the previous discussion, is dominantly concentrated in the upper part of the spectrum of the invariant mass of the lepton pair. At the level of accuracy of the present discussion the only difference between the theoretical expressions for $B$ and for $D$ mesons arises through a different normalization point of the four-quark operators in the equations (4.39) and (4.45). Taking into account the 'hybrid' evolution of the operators containing $b$ quark down to $\mu = m_c$ gives the relation between the non-singlet factorization constants:

$$\delta B^{ns}(m_b) = \frac{8\,\kappa^{1/2}+1}{9}\delta B^{ns}(m_c) + \frac{2\,(\kappa^{1/2}-1)}{3}\,\left[\varepsilon_1^{ns}(m_c) - \varepsilon_2^{ns}(m_c)\right], \tag{4.46}$$

where $\kappa = (\alpha_s(m_c)/\alpha_s(m_b))$. However, modulo the unlikely case that the difference of the constants $\varepsilon$ in this relation is much bigger than the difference between the constants $B$, the renormalization effect is quite small, and most likely is at the level of other uncertainties in the considered approach (such as the accuracy of the flavor SU(3) symmetry, higher QCD corrections, contribution of higher terms in $m_c^{-1}$, *etc.*). Thus with certain reservations, one can use the approximate relation $\delta B^{ns}(m_b) \approx \delta B^{ns}(m_c)$ to directly relate the differences in the inclusive semileptonic decay rates:

$$\Gamma(B^0 \to X_u\,\ell\,\nu) - \Gamma(B^\pm \to X_u\,\ell\,\nu) \approx \frac{|V_{ub}|^2}{|V_{cs}|^2}\,\frac{f_B^2}{f_d^2}\,\frac{m_b^3}{m_c^3}\,\left[\Gamma_{sl}(D^0) - \Gamma_{sl}(D_s)\right]. \tag{4.47}$$

A measurement of these differences of the semileptonic decay rates can provide information only on the flavor non-singlet part of the non-factorizable terms. In order to probe the singlet part of these terms one should gain insight into the absolute decay rate of individual particles rather than their differences. In doing this, it is also quite natural to discuss the semileptonic decay rates of the $D$ mesons, where the effect is larger than for the $B$ mesons. Neglecting the Cabibbo-suppressed transition $c \to d\,\ell\,\nu$, one can write the contribution of the non-factorizable terms to the semileptonic decay rate of either of the non-strange $D$ mesons as

$$\delta\Gamma_{sl}(D) = \frac{G_F^2\,f_D^2\,m_c^2\,m_D}{12\,\pi}\,\frac{\delta B^s - \delta B^{ns}}{2} \approx 0.08\,\mathrm{ps}^{-1}\left(\frac{m_c}{1.4\,GeV}\right)^2\left(\frac{f_D}{0.2\,GeV}\right)^2\,\frac{\delta B^s - \delta B^{ns}}{0.2}. \tag{4.48}$$

Thus with 'natural' values of the parameters the effect of the non-factorizable terms easily reaches about one half of the experimental semileptonic decay rate, *e.g.*, $\Gamma_{sl}(D^0) = 0.164 \pm 0.007\,\mathrm{ps}^{-1}$. Therefore an analysis of these rates necessarily should include the non-factorizable terms even at their expected suppressed level.

The 'full' formula for the semileptonic decay rate of a $D$ meson, that includes the QCD radiative corrections up to two loops [84], and the second term of the OPE of the effective operator (4.32) [85] reads as

$$\begin{aligned}\Gamma_{sl}(D) \;=\; & \frac{G_F^2\,m_c^5}{192\,\pi^3}\left[|V_{cs}|^2\left(1 - 8\frac{m_s^2}{m_c^2}\right) + |V_{cd}|^2\right] \\ & \times\left[1 - 2.413\,\frac{\alpha_s}{\pi} - 23.44\,\left(\frac{\alpha_s}{\pi}\right)^2\right]\left(1 + \frac{\lambda_1 + \mu_g^2}{2\,m_c^2}\right)\left(1 - \frac{\mu_g^2}{2\,m_c^2}\right) + \delta\Gamma_{sl}(D)\,, \end{aligned} \tag{4.49}$$

where $\alpha_s = \alpha_s(m_c)$, $\delta\Gamma_{sl}(D)$ is given by Eq. (4.48), and a certain inaccuracy has to be admitted in the treatment of the cross terms between, *e.g.*, the radiative corrections and the effect of the finite mass $m_s$ of the strange quark or between the radiative corrections and a part of the $\mathcal{O}(m_c^{-2})$ terms. This inaccuracy, however, is at the level of other uncertainties involved in Eq. (4.49), *e.g.*, due to higher perturbative terms, or the experimental uncertainties in the data, and can be safely neglected in the present discussion. Finally $\lambda_1$ and $\mu_g^2$ are the standard parameters of HQET. The 'chromo-magnetic' term $\mu_g^2$ is determined from the mass difference between the heavy vector and pseudoscalar mesons: $\mu_g^2 \approx 0.37\,GeV^2$, while the 'kinetic' term is less certain and should obey the inequality [86] $(-\lambda_1) \geq \mu_g^2$.

The contribution $\delta\Gamma_{sl}(D)$ of the non-factorizable terms could be estimated from comparison of Eq. (4.49) with the data, if not for the uncertainty of the first term, arising from the value of the charm quark mass $m_c$. A value of about $1.4$ GeV for the 'pole' mass of the charm quark originates from the charmonium sum rules [87]. If this value is used in Eq. (4.49), the first term accounts for only about one half of the experimental rate [79, 88]. In order to remedy this contradiction without involving a substantial nonfactorizable contribution it was suggested [79] that the 'pole' value of $m_c$ should be significantly larger, $m_c \approx 1.65$ GeV, which can hardly be reconciled with the rest of phenomenology of charmonium and charmed hadrons. In particular the mass parameter $m_c$, entering Eq. (4.49) can be deduced from the mass formula for a pseudoscalar meson:

$$M_P = m_Q + \overline{\Lambda} - \frac{\lambda_1 + \mu_g^2}{2\, m_Q} + O(m_Q^{-2})\,, \tag{4.50}$$

provided that the parameters $\overline{\Lambda}$ and $\lambda_1$ of the HQET can be determined. One way of experimentally determining these parameters is from a measurement of the moments of the lepton energy and of the hadronic recoil mass in the dominant semileptonic $B$ decays. This technique was recently pursued by the CLEO experiment [89]. An analysis [90] of their results in terms of Eq. (4.49) favors the 'pole' charm quark mass in the region around $1.4$ GeV, and thus suggests a large contribution of the non-factorizable term, reaching up to 0.5 – 0.6 (depending on the value of $\alpha_s(m_c)$) of the experimental semileptonic decay rate.

The discussion of the non-factorizable contribution to the semileptonic decays $B \to X_u\, \ell\, \nu$ presented in this subsection can be summarized by the following main points:

- The present poor knowledge of the non-factorizable terms can become a major source of uncertainty in determination of $|V_{ub}|$, limiting the accuracy of the knowledge of this mixing parameter at about 10%.
- The most favorable way of determining the flavor non-singlet part of these terms is from a measurement of the difference of the semileptonic decay rates of the strange $D_s$ meson and the non-strange $D$ mesons.
- The flavor singlet part of the non-factorizable terms can be estimated from the total semileptonic decay rate of the $D$ mesons with an improved knowledge of the parameters $\overline{\Lambda}$ and $\lambda_1$ of the HQET. The latter parameters can determined from moments of the spectra in semileptonic decays of the $B$ mesons.

**Constraining weak annihilation contributions with lattice QCD**

$\succ$ C. Bernard, S. Hashimoto, P. Mackenzie $\prec$

It may become feasible in the future to use lattice QCD calculaitons to constrain the size of non-factorizable amplitudes such as those due to weak annihilation. The necessary bag parameters $B_1$ and $B_2$ may be calculated using lattice QCD. There is an exploratory quenched lattice calculation by Di Pierro and Sachrajda [91]. They used the lattice HQET (static limit) and the matching between the continuum $\Delta B = 0$ four-quark operators and corresponding lattice HQET operators is done by one-loop calculation. Their results are

$$B_1(m_b) = 1.06(8), \qquad B_2(m_b) = 1.01(6), \tag{4.51}$$

which leads to $B_1 - B_2 = 0.05(10)$, assuming no error correlation. The result (4.51) is quite consistent with the vacuum saturation approximation (or the factorization).

The quantity $B_1 - B_2$ measures the violation of factorization. In the lattice calculation the sources of the violation are the perturbative matching and the non-perturbative lattice matrix elements. In the perturbative matching, the violation starts at one loop and thus the leading contribution to $B_1 - B_2$ is $\mathcal{O}(\alpha_s)$. To control the systematic error to better than 10% one needs a two loop matching calculation. The non-perturbative calculation seems completely consistent with the factorization assumption in the quenched approximation (for both $\Delta B$ =0 and 2 operators), as there is no hint of deviation in Eq. (4.51) from unity.

To improve the accuracy in the future one has to do (i) unquenching, (ii) two-loop matching, (iii) further improvement of lattice action and/or continuum extrapolation, just as in the lattice calculations of other quantities. (Note that the result (4.51) does not contain the quenching error.) We may expect that the error is similar to that for the $\Delta B = 2$ matrix element $B_B$, which is 8% for $\delta(f_B^2 B_B)$ (see Section 4.6.1). This means that the improvement over the current guess, $|B_1 - B_2| = \mathcal{O}(0.1)$, is unlikely to be significant enough in the near future to allow for either establishing $B_1 - B_2 \neq 0$ at a roboust level or to demostrate if $|B_1 - B_2|$ is smaller than expected.

### 4.4.2 The relevance of the decay $B \to X_s\gamma$ to the extraction of $V_{ub}$

$\succ$ I. Rothstein $\prec$

The extraction of $|V_{ub}|$ from inclusive $B \to X_u$ decays is complicated by the fact that in order to reject the overwhelming charm background one must cut the spectrum in a corner of phase space. This not only hurts statistically, but also makes the theory much more complicated. In particular, when one cuts the spectrum close to the endpoint, the rate becomes sensitive to the non-perturbative motion of the heavy quark inside the meson. This motion is described by a well-defined universal matrix element called the "shape function"[92], defined as

$$f(k_+) = \langle B(v) \mid \bar{b}_v \delta(k_+ - iD_+) b_v \mid B(v)\rangle. \tag{4.52}$$

This function is interpreted as the probability for the $b$ quark to carry light cone momentum fraction $k_+$ in the meson. The amount of sensitivity to this presently unknown function depends upon the choice of observable[93]. Cutting on the lepton energy is simplest from the experimental point of view, since in this case there is no need to reconstruct the neutrino momentum. This method has the disadvantage in that it only contains $\approx 10\%$ of the rate, whereas a cut on the hadronic mass [94] contains $70 - 80\%$ [8] of the rate. A cut on leptonic mass [95] is favored, since it is less sensitive to large energy, small mass hadronic states, and thus the error induced by ignoring the shape function is in the noise. The downsides of this cut are that the effective expansion parameter becomes $\Lambda/m_c$, and not $\Lambda/m_b$ [96], and that it captures only $10$ to $20\%$ of the rate. Hybrid cuts [97] have been proposed to minimize the uncertainties due to the ignorance of the shape function and formally sub-leading corrections.

We will only address the lepton energy and hadronic mass cuts, as these have order one sensitivity to the shape function. Since the shape function is universal, it can, in principle, be extracted from one decay for use in another. In

[8] These percentages are estimates based upon models.

particular, the cut rate for the decay $B \to X_S + \gamma$ may be written, at tree level, as

$$\Gamma_H\left[\frac{2E_{cut}}{M_B}\right] = \int_{2E_{cut}-m_b}^{\overline{\Lambda}} dk_+ f(k_+)\Gamma_p\left[\frac{2E_{cut}}{m_b+k_+}\right], \tag{4.53}$$

where $\Gamma_p\left[\frac{2E_{cut}}{m_b+k_+}\right]$ is the partonic rate with a cut on $x = 2E/m_b$ at $x_p = 2E_{cut}/(m_b + k_+)$. A similar expression can be derived for the semileptonic decay. Thus, one would hope to extract $f(k_+)$ by fitting the end-point spectrum in the radiative decay, and use it to predict $V_{ub}$. Indeed, most extractions to date follow the results in [98], where it was assumed that the radiative corrections can simply be incorporated in (4.53) by changing $\Gamma_p$ to include the one loop QCD corrections. Unfortunately, as pointed out by [99], this is incorrect, due to the fact the presumed relation between the moments of the shape function and matrix elements of local operators does not hold beyond tree level. When CLEO [100], *BABAR* [23] and Belle [101] performed their extraction, they assumed that the shape function was constrained to have certain properties; these constraints followed from the aforementioned erroneous relationship between moments and local operators. Thus, the true size of systematic errors for those measurements is not clear. We expect that extractions utilizing the hybrid cut will be less sensitive to this issue, and thus the errors made using this method of fitting the shape function will be diminished in amplitude, though it is not clear by how much.

Fortunately, there is no need to extract the shape function in the first place, since by taking the ratio of the moments of the radiative and semi-leptonic decay rates, we can eliminate the need for the shape function altogether [102]. It has been shown that we can write a closed form expression for $\mid V_{ub} \mid$ in terms of the cut lepton energy spectrum as [103, 104, 105]

$$\frac{|V_{ub}|^2}{|V_{ts}^* V_{tb}|^2} = \frac{3\,\alpha\, C_7^{(0)}(m_b)^2}{\pi}(1+\mathcal{H}_{mix}^{\gamma})\int_{x_B^c}^1 dx_B \frac{d\Gamma}{dx_B} \times \left\{\int_{x_B^c}^1 du_B W(u_B)\,\frac{d\Gamma^\gamma}{du_B}\right\}^{-1}, \tag{4.54}$$

where $\mathcal{H}_{mix}^{\gamma}$ represents the corrections due to interference coming from the operators $O_2$ and $O_8$ [106] .

$$\mathcal{H}_{mix}^{\gamma} = \frac{\alpha_s(m_b)}{2\pi C_7^{(0)}}\left[C_7^{(1)} + C_2^{(0)}\Re(r_2) + C_8^{(0)}\left(\frac{44}{9} - \frac{8\pi^2}{27}\right)\right], \tag{4.55}$$

and $x_B^c$ is the value of the cut. In Eq. (4.55), all the Wilson coefficients, evaluated at $m_b$, are "effective" as defined in [107], and $\Re(r_2) \approx -4.092 + 12.78(m_c/m_b - 0.29)$ [108]. The numerical values of the Wilson coefficients are: $C_2^{(0)}(m_b) \approx 1.11,\ C_7^{(0)}(m_b) \approx -0.31,\ C_7^{(1)}(m_b) \approx 0.48$, and $C_8^{(0)}(m_b) \approx -0.15$. The diagonal pieces from $O_2$ and $O_8$ are numerically insignificant. The function $W(u_B)$ is given by

$$W[u_B] = u_B^2 \int_{x_B^c}^{u_B} dx_B \left(1 - 3(1-x_B)^2 + \frac{\alpha_s}{\pi}(\frac{7}{2} - \frac{2\pi^2}{9} - \frac{10}{9} log(1 - \frac{x_B}{u_B}))\right). \tag{4.56}$$

This expression for $V_{ub}$ does is not afflicted by large end point logs which were resummed and shown to have a small effect on the rate [109, 103, 104, 110].

The expression for $V_{ub}$ for the case of the hadronic mass cut is given by [111]

$$\frac{|V_{ub}|^2}{|V_{ts}|^2} = \frac{6\,\alpha\, C_7(m_b)^2(1+\mathcal{H}_{mix}^{\gamma})\,\delta\Gamma(c)}{\pi\,[I_0(c) + I_+(c)]}. \tag{4.57}$$

The expressions for $\Gamma(c),\ I_0(c)$ and $I_+(c)$ can be found in [111]. The effect of resummation of the end-point logs in this case was again shown to be negligible [112],[113]. Note that the dominant source of errors in both of these extractions will come from sub-leading shape functions, which were studied in [114].

### 4.4.3 Experimental prospects

≻– D. del Re –≺

The *BABAR* experiment has already performed measurements of inclusive semileptonic $B$ decays with statistical errors comparable to the experimental systematic errors, while the theoretical error is already dominant. This is due to the fact that even Cabibbo-suppressed inclusive semileptonic $B$ decays are abundant at $B$ Factories, but also due to the large theoretical uncertainties affecting the study of inclusive decays in restricted regions of phase space. A substantial gain in the overall error will only be achieved if the theoretical error can be better controlled—more data and measurements in dedicated regions of phase can help in this regard.

The recoil approach should help in reaching this goal. It significantly reduces the experimental systematics, and, since the level of background is lower, permits looser cuts on the phase space and multiplicity, thereby reducing theoretical systematic uncertainties.

### Inclusive charmless semileptonic $B$ decays

In order to understand the level of sensitivity achievable in the study of inclusive charmless semileptonic $B$ decays, it is worth to briefly describe the measurement recently presented by the *BABAR* experiment [23]. It makes use of the recoil technique and it is the $|V_{ub}|$ measurement that, so far, obtained the smallest systematic uncertainty.

In this analysis, a semi-leptonic decay of one $B$ meson ($B_{recoil}$ ) is identified by the presence of a charged lepton in the recoil of a $B_{reco}$ candidate. In addition, the detection of missing energy and momentum in the event is taken as evidence for the presence of a neutrino. The $B \to X_u \ell\nu$ transitions are dominantly located in the low mass region $m_X < m_D$. Undetected particles and mis-measurement of detected particles distort the measured mass distribution and lead to a large background from the dominant $b \to c\ell\nu$ decays. To improve the resolution in the measurement of $m_X$, this analysis exploits the kinematic constraints and simplicity of the $B\overline{B}$ state and uses the measured momenta and energies of all particles in a 2C kinematic fit to the whole event. With the additional constraint that the missing particle should have zero mass the hadronic mass $m_X$ is determined, largely independent of the unfitted missing mass of the event. To extract the number of leptons from $b \to u\ell\nu$ transitions the data are divided into subsamples of events, one that is enriched in $b \to u$ transitions by a veto on the presence of kaons in the recoil system, and the rest of the sample, which is used to control the background. To derive the charmless semileptonic branching ratio, the observed number of events, corrected for background and efficiency, is normalized to the total number of semileptonic decays $b \to q\ell\nu$ (here $q$ stands for $c$ or $u$) in the $B_{reco}$ event sample. Additional selection criteria are imposed to select $b \to u\ell\nu$ decays. They include constraints on the sum of the charges of all observed particles in the events, correlations between the sign of the lepton and the flavor of the reconstructed $B$ meson, requirement on the missing momentum and mass, and most importantly a veto on strange particles. This *BABAR* analysis, based on 82 fb$^{-1}$, selects $\sim 170$ signal events signal events for $m_X < 1.55\,\mathrm{GeV}$ (see Fig. 4-3) , with a signal-to-background ratio that corresponds to $\sim 1.7$. The inclusive branching ratio comes out to be $\mathcal{B}(B \to X_u \ell\overline{\nu}\ ) = (2.24{\pm}0.27(\mathrm{stat}){\pm}0.26(\mathrm{syst}){\pm}0.39(\mathrm{theo})){\times}10^{-3}$, that can be translated into $|V_{ub}| = (4.62 \pm 0.28(\mathrm{stat}) \pm 0.27(\mathrm{syst}) \pm 0.48(\mathrm{theo})) \times 10^{-3}$. Even at these moderate luminosities, the systematic error is larger than the statistical error.

The experimental systematic error will be improved in the future. It is dominated by detector effects that will be better understood with more experience. A substantial component of this uncertainty is due to imperfect knowledge of semileptonic branching ratios ($B \to D^{(*,**)} l\nu$) and to the $D$ meson decay branching ratios ($D \to X$)—measurements which will improve with more data, leading to a reduction in the related systematics. A quite large error due to the MC statistics will decrease as soon as more simulated events become available. A reasonable estimate is that the total experimental systematic error can be below $5\%$ for the rate (and half of that for $|V_{ub}|$).

This measurement technique will be only limited by the theoretical uncertainty but even this error can be improved. The cleanliness of the technique allows a measurement of the $m_X$ spectrum with a good resolution. By adding statistics not only the $m_X$ integral but also the $m_X$ shape can be measured allowing the extraction of the theoretical parameters $m_b$ and $a$ (as suggested in [115]), reducing the uncertainty due the extrapolation to the full spectrum. Moreover new theoretical papers [57] suggest a different cut in the phase space. A combination of a cut on $q^2 > (m_B - m_D)^2$ (i.e. on the virtual $W$ invariant mass) and a cut on $m_X$ should decrease the theoretical error. Finally a combination of the $m_X$ spectrum and the photon spectrum in $b \to s\gamma$ decays [116, 104, 106] could be used to perform a $|V_{ub}|$ measurement with suppressed uncertainty related to the shape function.

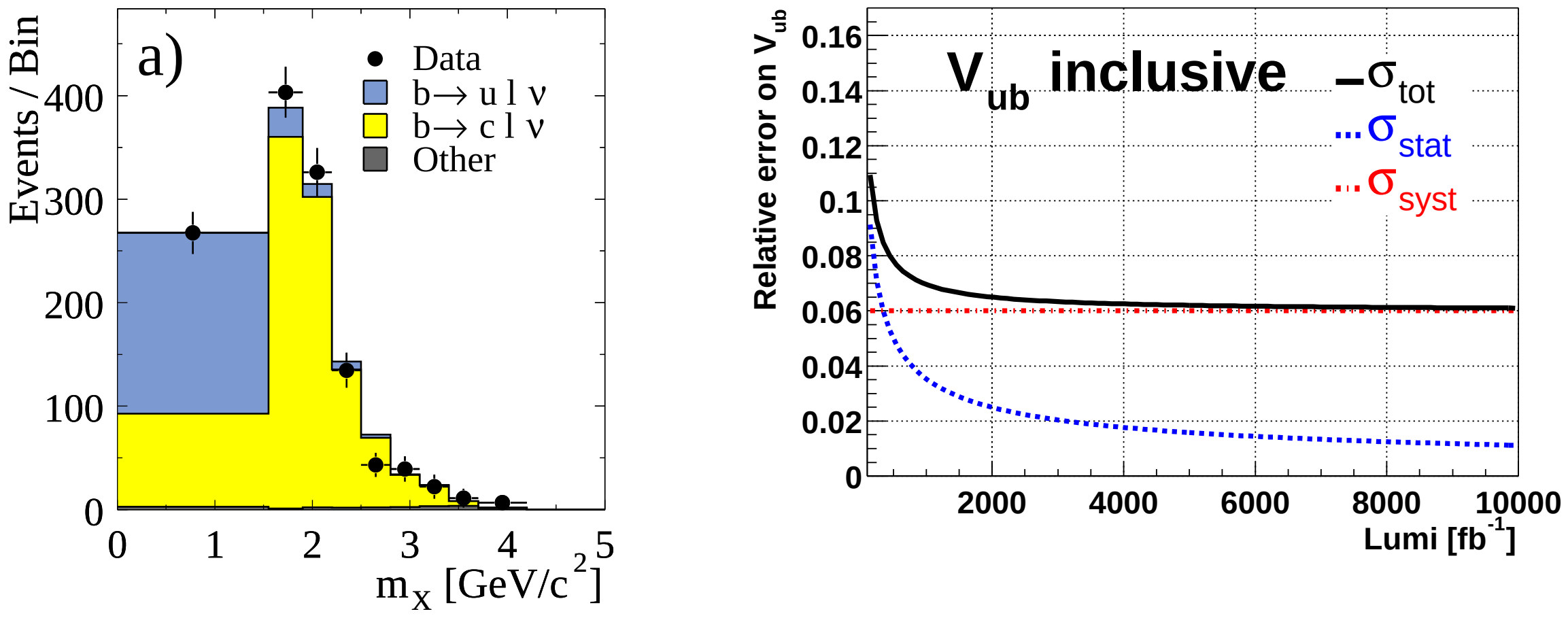

**Figure 4-3.** *Left: a $\chi^2$ fit to the $m_X$ distribution. Right: perspectives for the error on $|V_{ub}|$ as a function of the accumulated luminosity as described in text.*

In summary, we expect the total error on $|V_{ub}|$ to decrease down to 5–10% within several years. In Fig. 4-3 an extrapolation to higher luminosities is presented. The analysis method corresponds to that presented in [23], with the addition of a cut on $q^2 > 10$ GeV. We assume a systematic error of 6%. The plot clearly shows how this inclusive measurement cannot be improved by increasing the statistics above 1-2 ab$^{-1}$, unless systematic errors are further reduced.

**Inclusive rare radiative $B$ decays**

$\succ$– U. Langenegger –$\prec$

The measurement of the photon energy spectrum in inclusive radiative decays $B \to s\gamma$ provides a direct determination of the shape function of the $b$ quark. The first and second moment of this spectrum are related to the mass of the $b$ quark and HQET parameters describing its Fermi momentum within the hadron. From a theoretical point of view, it would be most desirable to measure the photon spectrum down to the lowest possible energies.

The experimental challenge here is on the one hand the small branching fraction of about $3 \times 10^{-4}$, and on the other hand, the very large background both from continuum $q\overline{q}$ production (where $q = u, d, s, c$) and from $B\overline{B}$ events. Both background spectra rise exponentially towards lower energies and therefore lead to an experimental spectrum truncated around $E_\gamma > 2$ GeV. There are two distinct types of analyses, semi-exclusive and inclusive.

In semi-exclusive analyses, the hadronic final state $X_s$ in $B \to X_s\gamma$ decays is reconstructed as the sum of several exclusive modes. This allows a measurement of the photon energy in the $B$ meson rest frame with an excellent $E_\gamma$ resolution, but is sensitive to only 50% of all $X_s$ states. The dependence on the modeling of the included hadronic final states constitutes the major difficulty in the analysis.

In inclusive analyses, the continuum $q\overline{q}$ background is rejected with high efficiency by selecting ("tagging") events based on $B$ decay signatures (see Section 4.2). This includes (1) high-momentum leptons and (2) a fully reconstructed hadronic $B$ decay.

In the first case, the photon energy is measured in the $\Upsilon(4S)$ restframe with a resolution of about $100 \text{ MeV}$. In the current *BABAR* analysis based on $82\,\text{fb}^{-1}$, it is expected to determine the mean photon energy with an error of about 1.2% (without background and efficiency contributions), dominated by the statistical error. Here, the spectrum is measured for energies $E_\gamma > 2.0 \text{ GeV}$.

In the second case, the photon can be boosted into the $B$ meson rest frame, and, due to the overconstrained kinematics, better resolution, compared to the lepton-tagged analysis, can be achieved. Because of the low efficiency for hadronic tags, the event yield is substantially lower: for $82\,\text{fb}^{-1}$ a total of about 60 events is expected. Comparable statistical errors to the lepton-tagged analysis are expected for an integrated luminosity of about $1\,\text{ab}^{-1}$. Nevertheless, this approach is very valuable as it offers the potential to lower the threshold for the photon energy and, more importantly, allows the best resolution in the measurement of the photon energy.

## 4.5 $b \to u$ Exclusive Decays ($\pi$, $\eta^{(\prime)}$, $\rho$, $\omega$, *etc.*)

### 4.5.1 Theory

⤚ C. Bauer, I. Stewart ⤙

Branching ratios of exclusive semileptonic $B$ decays proceed via the heavy-light current $\overline{u}\Gamma b$, and are proportional to the square of the magnitude of the CKM matrix element $V_{ub}$. However, the relevant matrix element of this $b \to u$ current for exclusive processes depends on non-perturbative hadronic physics parameterized by form factors, which are needed in order to extract CKM information from these decays. For decays to pseudoscalars $P$ or vectors $V$ these form factors are defined as

$$\begin{aligned}
\langle P(p)|\overline{q}\,\gamma^\mu b|\overline{B}(p_b)\rangle &= f_+(q^2)\left[p_b^\mu + p^\mu - \frac{m_B^2 - m_P^2}{q^2}\,q^\mu\right] + f_0(q^2)\,\frac{m_B^2-m_P^2}{q^2}\,q^\mu, \\
\langle V(p,\epsilon^*)|\overline{q}\gamma^\mu b|\overline{B}(p_b)\rangle &= \frac{2V(q^2)}{m_B+m_V}\,i\epsilon^{\mu\nu\rho\sigma}\epsilon^*_\nu\,(p_b)_\rho\,p_\sigma, \\
\langle V(p,\epsilon^*)|\overline{q}\gamma^\mu\gamma_5 b|\overline{B}(p_b)\rangle &= 2m_V A_0(q^2)\,\frac{\epsilon^*\cdot q}{q^2}\,q^\mu + (m_B+m_V)\,A_1(q^2)\left[\epsilon^{*\mu} - \frac{\epsilon^*\cdot q}{q^2}\,q^\mu\right] \\
&\quad - A_2(q^2)\,\frac{\epsilon^*\cdot q}{m_B+m_V}\left[p_b^\mu + p^\mu - \frac{m_B^2-m_V^2}{q^2}\,q^\mu\right]. \qquad (4.58)
\end{aligned}$$

where $q^\mu = p_B^\mu - p^\mu$ is the momentum transfer to the leptons. Decay rates to particular exclusive final states can be written in terms of these form factors. Decays to pseudoscalar mesons are given by

$$\frac{\mathrm{d}\Gamma(B\to P\ell\overline{\nu})}{\mathrm{d}q^2\,\mathrm{d}\cos\theta} = |V_{ub}|^2\,\frac{G_F^2|\vec{p}_P|^3}{32\pi^3}\,\sin^2\!\theta\,|f_+(q^2)|^2\,, \qquad (4.59)$$

where $\ell = \mu, e$ and an $f_0$ term would be proportional to $m_\ell^2$ and has been dropped. For the analogous decays to vector mesons one finds

$$\frac{\mathrm{d}\Gamma(B\to V\ell\overline{\nu})}{\mathrm{d}q^2\,\mathrm{d}\cos\theta} = |V_{ub}|^2\,\frac{G_F^2|\vec{p}_V|q^2}{768\pi^3 m_B^2}\left[(1+\cos\theta)^2|H_+|^2 + (1-\cos\theta)^2|H_-|^2 + 2\sin^2\theta|H_0|^2\right], \qquad (4.60)$$

where the three helicity amplitudes are given by

$$\begin{aligned}
H_\pm(q^2) &= (m_B+m_V)A_1(q^2) \mp \frac{2m_B|\vec{p}_V|}{(m_B+m_V)}\,V(q^2)\,, \\
H_0(q^2) &= \frac{(m_B+m_V)}{2m_V q^2}\left[\left(m_B^2 - m_V^2 - q^2\right)A_1(q^2) - \frac{4|\vec{p}_V|^2 m_B^2}{(m_B+m_V)^2}A_2(q^2)\right]. \qquad (4.61)
\end{aligned}$$

In Eqs. (4.59-4.61) the three momenta are related to $q^2$

$$4m_B^2|\vec{p}_{P,V}|^2 = (q^2 - m_B^2 - m_{P,V}^2)^2 - 4m_B^2 m_{P,V}^2\,. \qquad (4.62)$$

Given knowledge of the form factors, a measurement of the exclusive semileptonic branching ratios can be used to determine the CKM parameter $|V_{ub}|$.

Measurements of the heavy-to-light form factors themselves are also important ingredients in the description of many other exclusive $B$ meson decays. In addition to parameterizing the semileptonic decays they appear in rare radiative decays such as $B\to K^*\gamma$, $B\to\rho\gamma$, $B\to K^{(*)}\ell^+\ell^-$, and $B\to\pi\ell^+\ell^-$. They also play a crucial role in factorization theorems for nonleptonic $B\to MM'$ decays, with $M^{(\prime)}$ light pseudoscalar and vector mesons, which are important for measurements of $CP$ violation.

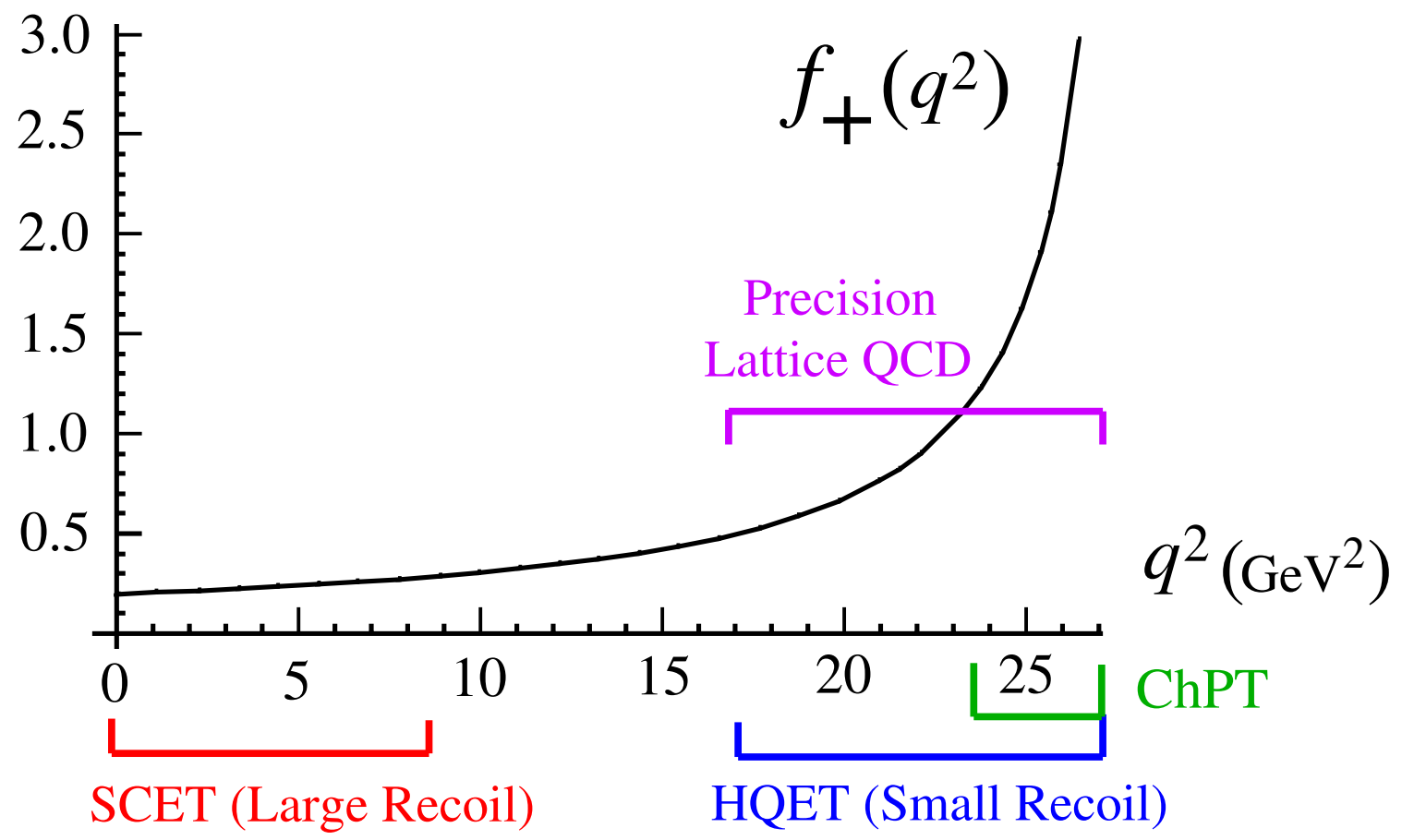

**Figure 4-4.** *Regions of validity in $q^2$ for different model independent methods for the $B \to \pi$ form factors. The abbreviations are Heavy Quark Effective Theory (HQET), Soft-Collinear Effective Theory (SCET), and Chiral Perturbation Theory (ChPT). The curve shown for $f_+(q^2)$ is for illustration only.*

For rare decays such as $B \to K^*\gamma$, $B \to K^*\ell^+\ell^-$, and $B \to K\ell^+\ell^-$ additional form factors appear via tensor currents. They can be defined by

$$\begin{aligned}
\langle P(p)|\bar{q}\, i\sigma^{\mu\nu} q_\nu b|\overline{B}(p_b)\rangle &= -\frac{f_T(q^2)}{m_B + m_P}\left[q^2(p_b^\mu + p^\mu) - (m_B^2 - m_P^2)\, q^\mu\right], \\
\langle V(p,\epsilon^*)|\bar{q}\, i\sigma^{\mu\nu} q_\nu b|\overline{B}(p_b)\rangle &= -2\, T_1(q^2)\, i\epsilon^{\mu\nu\rho\sigma}\epsilon^*_\nu\, (p_b)_\rho\, p_\sigma, \\
\langle V(p,\epsilon^*)|\bar{q}\, i\sigma^{\mu\nu}\gamma_5 q_\nu b|\overline{B}(p_b)\rangle &= T_2(q^2)\left[(m_B^2 - m_V^2)\,\epsilon^{*\mu} - (\epsilon^*\cdot q)\,(p_b^\mu + p^\mu)\right] \\
&\quad + T_3(q^2)\,(\epsilon^*\cdot q)\left[q^\mu - \frac{q^2}{m_B^2 - m_V^2}(p_b^\mu + p^\mu)\right].
\end{aligned} \tag{4.63}$$

Although the phenomenology and experimental methods for rare decays differ from the semileptonic decays, the theoretical description of the form factors in Eq. (4.63) is no more difficult than those in Eq. (4.58). Thus the theory techniques explored in this section apply equally well to both cases, and in certain kinematic cases actually provide useful relations between the two. For a detailed discussion of rare decays we refer the reader to Chapter 2.

Exclusive form factors depend in a complicated way on the details of the hadronic states, and their computation has been traditionally performed using QCD inspired phenomenological methods, such as quark models (for examples see [117]). Predictions for form factors can also be obtained with QCD sum rules [118, 119, 120, 121, 122, 123, 124], which we do not discuss here. For the level of precision obtainable from a high-luminosity asymmetric $B$-factory we expect that reliance on model independent methods with well understood theoretical uncertainty will be crucial. In this chapter we focus on results for form factors obtained with HQET, SCET, chiral perturbation theory, and lattice QCD over the regions of $q^2$ shown in Fig. 4-4. The best tool available to determine the heavy-light form factors directly from first principles QCD is the lattice. As illustrated in Fig. 4-4 precision control over the systematics of both the heavy $B$ and light meson is currently only projected for smaller recoils, where the light meson is not too energetic in the $B$'s rest frame. Lattice methods with a moving $B$ meson have recently been proposed [125, 126, 127] which have the potential to improve the precision of form factor determinations at lower values of $q^2$, however these methods are not included in the projections discussed here. The prospects for lattice determinations of the form factors are discussed in section 4.5.2.

Additional constraints on the form factors can be obtained with the help of expansion parameters derived from $\Lambda_{\rm QCD}$, $m_B$, and $E_M$. Here $q^2 = m_B^2 + m_M^2 - 2m_B E_M$, where $E_M$ is the energy of the light meson $M$ in the $B$-meson rest frame, and their is a one-to-one correspondence between values of $E_M$ and $q^2$. Different expansions are appropriate for different regions of $q^2$ and are made systematic with the help of several effective field theories as shown in Fig. 4-4. For the region where $E_M/\Lambda \ll 1$ is a good expansion parameter SU(2) heavy baryon chiral perturbation theory (ChPT) can be used to compute the form factors for $B \to \pi$ and SU(3) heavy baryon chiral perturbation theory can be used for $B \to K$ [128, 129, 130].[9] Here $\Lambda \sim 1\,\mathrm{GeV}$ is of order the chiral symmetry breaking scale. For example for $B \to \pi$ one obtains

$$f_+ - f_- = \frac{f_B}{f_\pi}\frac{g_\pi m_B}{E_\pi + \Delta}\,, \qquad f_+ + f_- = \frac{f_B}{f_\pi}\Big(1 - \frac{g_\pi E_\pi}{E_\pi + \Delta}\Big)\,, \tag{4.64}$$

where $f_- = (f_0 - f_+)(m_B^2 - m_\pi^2)/q^2$, $\Delta = m_{B^*} - m_B$, and $g_\pi$ is the $B^* B\pi$ coupling. Analysis beyond leading order can be found in Refs. [131, 132, 133].

The results in Eq. (4.64) are only valid in a very limited range at large $q^2$ or small $E_\pi$. For the larger region where $m_b \gg E_\pi \sim \Lambda_{\mathrm{QCD}}$ we can make use of heavy quark effective theory, HQET. Although HQET does not provide a normalization for any of the form factors it does give important relations between different form factors. The HQET form factor form factor relations are discusssed further in section 4.5.3.

For the other end of the spectrum, namely large recoil or small $q^2$, the power expansion in HQET breaks down since the light meson gets too energetic. In this region another effective theory is applicable, known as the soft-collinear effective theory (SCET) [134, 135, 136, 137]. The expansion parameters here are $\Lambda_{\mathrm{QCD}}/E_M$ and $\Lambda_{\mathrm{QCD}}/m_b$. In section 4.5.3 we discuss the LO predictions of SCET for heavy-to-light form factors, as well as reviewing the large recoil SCET form factor relations.

Finally, dispersion relations combined with analyticity provide important constraints on the shape of form factors over the entire region of $q^2$ [138, 139, 140, 141]. We do not review these methods here.

### 4.5.2 Lattice form factors ($|\vec{p}_M| \lesssim 1$ GeV)

$\succ$– C. Bernard, S. Hashimoto, P. Mackenzie –$\prec$

The estimates for future lattice precision presented in this section and Section 4.6.1 on leptonic decay constants are based largely on a DOE planning document prepared by S. Sharpe, C. Bernard, A. El-Khadra, P. Mackenzie, and R. Sugar.

We assume three levels of computation based on improved staggered simulations with $n_F = 3$ flavors of dynamical sea quarks:

- “MILC0.” These are existing configurations generated over the past four years by the MILC configurations. A complete analysis of heavy-light quantities on these lattices will probably take one to two years.
- “MILC1.” This level will take $\sim$6 Teraflop-years and require machines now being built under the DOE SciDAC project [142]: the Columbia QCDOC and large clusters at Fermilab and Jefferson Lab. We estimate that this level will be completed in three to five years from the present,including time for analysis of heavy-light quantities,
- “MILC2.” This level will take $\sim$ 50–100 Teraflop-years and require the next generation of machines. We estimate that this level will be completed in five to eight years from the present, including time for analysis of heavy-light quantities.

As mentioned in Section 4.1.2, dynamical domain wall fermions provide a safer, but slower, alternative to improved staggered. A level “DWF1” of dynamical domain wall fermions (or equivalent) at comparable mass and lattice spacings to MILC1 may have comparable precision to MILC2 because DWF have smaller discretization errors and are free from taste-violation issues. This may require $\sim$ 600-1000 Teraflop-years and the “next next” generation of machines, finishing perhaps ten or twelve years from the present. In other words, our guess is that use of DWF, as opposed to improved staggered fermions in lattice QCD computation, would delay the available lattice precision by roughly five years.

[9] For SU(3) it is obvious that precision results would require going beyond leading order in the chiral expansion.

Tables 4-2, 4-3, and 4-4 show estimates of precision attainable for lattice calculations of semileptonic form factors with data sets MILC0, MILC1, and MILC2, respectively. These are meant to be average errors for the form factors at fixed $q^2$ in the allowed range of momentum. We focus on the gold plated quantities $B \to \pi$ and $B \to D$; it is possible that the errors in $B \to D^*$ will not be much larger than for $B \to D$. As discussed in Section 4.1.2, we give two alternatives for perturbative errors (one-loop and two-loop) and two alternatives for chiral extrapolation errors: (no) S$\chi$PT assumes that staggered chiral perturbation theory is (is not) useful.

**Table 4-2.** *Estimated percent errors for form factors at MILC0 level: one to two years from the present. "Light $q$" includes light quark chiral and discretization errors. "Heavy $Q$" means heavy quark discretization errors. $B \to \pi$ form factors are for restricted range* $0.5\ \mathrm{GeV} \lesssim \vec{p}_\pi \lesssim 1\ \mathrm{GeV}$ *(in $B$ rest frame), but can have any bilinear current.*

| quantity | statist. | scale | light $q$ | | heavy $Q$ | pert. th. | |
|---|---|---|---|---|---|---|---|
| | | | no S$\chi$PT | S$\chi$PT | | 1-loop | 2-loop |
| $B \to \pi\ell\nu$ | 4.5 | 1 | 6 | 3 | 3 | 7.5 | 2 |
| $B \to D\ell\nu$ | 1 | 0.5 | 2 | 1 | 1 | 2.5 | 0.7 |

**Table 4-3.** *Same as Table 4-2, but for MILC1 level: three to five years from the present. $B \to \pi$ momentum range is slightly larger than for MILC0:* $0.35\ \mathrm{GeV} \lesssim \vec{p}_\pi \lesssim 1\ \mathrm{GeV}$ *(in $B$ rest frame).*

| quantity | statist. | scale | light $q$ | | heavy $Q$ | pert. th. | |
|---|---|---|---|---|---|---|---|
| | | | no S$\chi$PT | S$\chi$PT | | 1-loop | 2-loop |
| $B \to \pi\ell\nu$ | 3 | 0.7 | 4 | 2 | 2 | 7.5 | 2 |
| $B \to D\ell\nu$ | 0.6 | 0.5 | 2 | 1 | 0.6 | 2.5 | 0.7 |

**Table 4-4.** *Same as Table 4-3, but for MILC2 level: five to eight years from the present.*

| quantity | statist. | scale | light $q$ | | heavy $Q$ | pert. th. | |
|---|---|---|---|---|---|---|---|
| | | | no S$\chi$PT | S$\chi$PT | | 1-loop | 2-loop |
| $B \to \pi\ell\nu$ | 1.5 | 0.5 | 2.7 | 1.3 | 1.5 | 7.5 | 2 |
| $B \to D\ell\nu$ | 0.3 | 0.3 | 1.4 | 0.7 | 0.5 | 2.5 | 0.7 |

Table 4-5 shows total lattice form factor errors under various assumptions, together with our best guess of which alternatives are most likely to be realized in practice. It must be kept in mind that the errors themselves are uncertain, by a fractional amount which is at least $\sim 30\%$ and rises with time into the future.

**Table 4-5.** *Estimated total lattice errors in percent under various assumptions. Momentum ranges for $B \to \pi$ are same as in Tables 4-2, 4-3 and 4-4. Where there are four entries per column they correspond to: (1)* **no** *S$\chi$PT and 1-loop perturbation theory, (2) S$\chi$PT and 1-loop perturbation theory, (3)* **no** *S$\chi$PT and 2-loop perturbation theory, and (4) S$\chi$PT and 2-loop perturbation theory. Our best guesses of which alternative will in fact be realized are surrounded with boxes.*

| **quantity** | **now** | **1-2 yrs.** MILC0 | **3-5 yrs.** MILC1 | **5-8 yrs.** MILC2 |
|---|---|---|---|---|
| $B \to \pi\ell\nu$ | 15 | $\boxed{11}$, 10, 8, 7 | 9, 9, 6, $\boxed{5}$ | 8, 8, 4, $\boxed{3}$ |
| $B \to D\ell\nu$ | 6 | $\boxed{4}$, 3, 3, 2 | 3, 3, 2, $\boxed{1.6}$ | 3, 3, 2, $\boxed{1.2}$ |

### 4.5.3 Heavy-to-light form factors in SCET

$>\!\!-$ C. Bauer, D. Pirjol, I. Stewart $-\!\!<$

In the absence of perfect theoretical computations, it is of interest to exploit the existence of model-independent relations among form factors. Such relations can be established in two kinematical regions, corresponding to the limits of a) energetic and b) slow final light hadron. These two situations are described in terms of two effective theories: a) the Soft-Collinear Effective Theory (SCET) and b) the Heavy Quark Effective Theory (HQET). In this and the following section we consider these two types of predictions in turn.

In the large recoil region, the existence of symmetry relations for heavy-light form factors was first suggested by Charles *et al.* in Ref. [143], formalizing earlier results obtained in the quark model [144]. The derivation here was based on an effective theory, LEET [145], which unfortunately is flawed since LEET does not correctly capture the IR physics of QCD in the case of energetic mesons. An analysis of the leading order contributions in perturbation theory [146] showed the existence of calculable corrections to these "symmetry" relations. Rather than following the historical order of events, we review the results obtained from the all-order effective theory treatment based on SCET [135, 147, 148, 149, 150, 151, 152, 153, 154].

For small values of $q^2$ a weak current $\bar{q}P_R\gamma^\mu b$ can be matched onto the leading order SCET current

$$\bar{q}P_R\gamma^\mu b = \int \mathrm{d}\omega\, C_\Gamma^{(0)}\, [\bar{\xi}W]_\omega P_R \Gamma h_v \equiv \int \mathrm{d}\omega\, C_\Gamma^{(0)}\, J_{\Gamma,\omega}^{(0)}\,, \tag{4.65}$$

$h_v$ is the usual field in HQET and $[\xi W]_\omega$ is a gauge invariant collinear field with label momentum equal to $\omega$. There are only three independent Dirac structures $\Gamma$, since both the $\xi$ and the $h_v$ are two component spinors. The matrix element of this operator between a $B$ meson state and a collinear light meson state vanishes, since the interpolating field for a collinear light meson contains two collinear fermions. This fact on the one hand explains the suppression of the form factor in the large recoil region, but it also makes the SCET analysis difficult, since a good understanding of subleading effects are needed.

The analysis of the form factors is performed in a two step matching procedure, where one first matches QCD onto a theory called SCET$_\mathrm{I}$, containing collinear particles with off-shellness $p^2 \sim Q\lambda_{\mathrm{QCD}}$ and usoft particles with off-shellness $p^2 \sim \Lambda_{\mathrm{QCD}}^2$ [155]. In SCET$_\mathrm{I}$ the heavy to light current has to appear in a time-ordered product with an interactions which turn the soft spectator fermion in the $B$ meson into a collinear fermion. These interactions appear at subleading order in SCET [156]. It turns out that the first non-vanishing time-ordered product occurs two powers of $\lambda$ suppressed, and one therefore also requires the subleading heavy-light current in SCET, $J_{\omega_1\omega_2}^{(1)}$, which depends on two label momenta $\omega_1$ and $\omega_2$, as well as the subleading SCET Lagrangian. Combining these results, one conveniently divides the resulting time-ordered product into two terms

$$T_1^\Gamma(\omega) = i\int d^4y\, T[J_{\Gamma,\omega}^{(0)}(0), i\mathcal{L}_{\xi q}^2(y)] + i\int d^4y \int d^4z T[J_{\Gamma,\omega}^{(0)}(0), i\mathcal{L}^1(y), i\mathcal{L}_{\xi\xi}^1(z) + i\mathcal{L}_{cg}^1(z)]$$

$$T_2^\Gamma(\omega_1\omega_2) = i\int d^4 y T[J^{(1)}_{\Gamma,\omega_1\omega_2}(0), i\mathcal{L}^{(1)}_{\xi q}(y)] \tag{4.66}$$

To proceed, these time-ordered products are matched onto four quark operators in $\mathrm{SCET_{II}}$. The form factor is the matrix element of the resulting operator in $\mathrm{SCET_{II}}$.

$$F^{B\to M} = \int d\omega\, C_\Gamma^{(0)}(\omega)\langle M|T_1^\Gamma(\omega)|B\rangle + \int d\omega_1 \int d\omega_2\, C^{(1)}(\omega_1,\omega_2)\langle M|T_2^\Gamma(\omega_1,\omega_2)|B\rangle\,. \tag{4.67}$$

where in this equation it is understood that the $T_i^\Gamma$ are matched onto operators in $\mathrm{SCET_{II}}$ before taking the matrix element. There is still some discussion in the literature how to properly factorize $T_1$ and match it onto operators in $\mathrm{SCET_{II}}$. This can be avoided by simply defining the matrix element

$$\begin{aligned}\langle P|T_1^\Gamma(\omega)|B\rangle &= \overline{n}\cdot p\ \zeta(\overline{n}\cdot p)\delta(\omega - \overline{n}\cdot p)\,,\\ \langle V_{\perp,\|}|T_1^\Gamma|B\rangle &= \overline{n}\cdot p\ \zeta_{\perp,\|}(\overline{n}\cdot p)\delta(\omega-\overline{n}\cdot p)\end{aligned} \tag{4.68}$$

he functions $\zeta(\overline{n}\cdot p)$, $\zeta_\|(\overline{n}\cdot p)$, $\zeta_\perp(\overline{n}\cdot p)$ are called soft form factors, and the reason for there only being three soft form factors is due to the fact that each of the three independent Dirac structures in the SCET current gives rise to only one type of meson by parity and angular momentum.

For $T_2$, one integrates out the modes with $p^2 \sim Q\Lambda_{\rm QCD}$, which give rise to a jet function. The exact structure depends on what kind of meson and which Dirac structure appear in the matrix element. The general structure, however, of such a matrix element is

$$\langle M|T_2(\omega_1,\omega_2)|B\rangle = \frac{f_B f_M m_B}{\overline{n}\cdot p^2}\int_0^1 dx\int_0^\infty dk^+\ J_\Gamma(\omega_1,x,k^+)\phi_M(x)\phi_B^+(k^+)\delta(\omega_1+\omega_2-\overline{n}\cdot p) \tag{4.69}$$

In this expression, the jet function $J_\Gamma(\omega_1,x,k^+)$ depends on the Dirac structure of the suubleading current $J^{(1)}_{\Gamma,\omega_1,\omega_2}$ and can be expanded in a series in $\alpha_s(\sqrt{E\Lambda_{\rm QCD}})$. Inserting (4.68) and (4.69) into (4.67) we obtain the result for a general form factor

$$f_i(q^2) = C_{ij}^{(0)}(Q)\zeta_j^M(Q\Lambda,\Lambda^2) + \int dxdzdk_+ C_{ij}^{(1)}(z,Q^2)J_j(z,x,k_+)\phi_B^+(k_+)\phi_j^M(x) \tag{4.70}$$

As explained before, the coefficients $C_{ij}$ are calculable in an expansion in $\alpha_s(Q)$, the jet functions $J_j$ are calculable in an expansion in $\alpha_s(Q\Lambda)$ and the remaining elements in these expressions denote the non-perturbative parameters. They are the well known light cone wave functions of the $B$ meson and the pseudoscalar or vector meson, as well as the soft form factors explained earlier.

Below we summarize the factorization results for the $B\to P$ and $B\to V$ form factors (following the notation in Ref. [150] and Ref. [151]). We use below the notations of [150] for the Wilson coefficients of $\mathrm{SCET_I}$ operators $C_i(E), B_i(x,z)$. For decays to pseudoscalars

$$\begin{aligned} f_+(E) =& \left(C_1^{(v)} + \frac{E}{m_B}C_2^{(v)} + C_3^{(v)}\right)\zeta^P \\ &+N_0\int dxdzdl_+\left\{\frac{2E-m_B}{m_B}\left[B_1^{(v)} - \frac{E}{m_B-2E}B_2^{(v)} - \frac{m_B}{m_B-2E}B_3^{(v)}\right]\delta(x-z)\right.\\ &\left.+\frac{2E}{m_b}\left[B_{11}^{(v)} - \frac{E}{m_B}B_{12}^{(v)} - B_{13}^{(v)}\right]\right\}J_\|(x,z,l_+)\phi_\pi(x)\phi_B^+(l_+)\end{aligned} \tag{4.71}$$

$$\begin{aligned} \frac{m_B}{2E}f_0(q^2) =& \left(C_1^{(v)} + \frac{m_B-E}{m_B}C_2^{(v)} + C_3^{(v)}\right)\zeta^P \\ &+N_0\int dxdzdl_+\left\{\frac{m_B-2E}{m_B}\left[B_1 + \frac{m_B-E}{m_B-2E}B_2^{(v)} + \frac{m_B}{m_B-2E}B_3^{(v)}\right]\delta(x-z)\right.\end{aligned} \tag{4.72}$$

$$+\frac{2E}{m_b}\left[B_{11}^{(v)}-\frac{m_B-E}{m_B}B_{12}^{(v)}-B_{13}^{(v)}\right]\Bigg\} J_\|(x,z,l_+)\phi_\pi(x)\phi_B^+(l_+)$$

$$\frac{m_B}{m_B+m_P}f_T(q^2)=\left(C_1^{(t)}-C_2^{(t)}-C_4^{(t)}\right)\zeta^P \tag{4.73}$$

$$+N_0\int_0^1 dxdl_+\left\{-\left[B_1^{(t)}-B_2^{(t)}-2B_3^{(t)}+B_4^{(t)}\right]\delta(x-z)-\frac{2E}{m_b}[B_{15}^{(t)}+B_{16}^{(t)}-B_{18}^{(t)}]\right] J_\|(x,z,l_+)\phi_B^+(l_+)\phi(x)\,,$$

with $N_0=f_Bf_Pm_B/(4E^2)$. The corresponding results for the $B\to V$ form factors have a similar form

$$\frac{m_B}{m_B+m_V}V(q^2)=C_1^{(v)}\zeta_\perp^V$$

$$-N_\perp\int_0^1 dxdzdl_+\left[-\frac{1}{2}B_4^{(v)}\delta(x-z)+\frac{E}{m_b}(2B_{11}^{(v)}+B_{14}^{(v)})\right] J_\perp(x,z,l_+)\phi_B^+(l_+)\phi_\perp(x)$$

$$\frac{m_B+m_V}{2E}A_1(q^2)=C_1^{(a)}\zeta_\perp^V$$

$$-N_\perp\int_0^1 dxdl_+\left[-\frac{1}{2}B_4^{(a)}\delta(x-z)+\frac{E}{m_b}(2B_{11}^{(a)}+B_{14}^{(a)})\right] J_\perp(x,z,l_+)\phi_B^+(l_+)\phi_\perp(x)$$

$$A_0(q^2)=\left(C_1^{(a)}+\frac{m_B-E}{m_B}C_2^{(a)}+C_3^{(a)}\right)\zeta_\|^V \tag{4.74}$$

$$+N_\|\int_0^1 dxdzdl_+\left\{\left[\frac{m_B-2E}{m_B}B_1^{(a)}+\frac{m_B-E}{m_B}B_2^{(a)}+B_3^{(a)}\right]\delta(x-z)\right.$$

$$\left.-\frac{2E}{m_b}\left[-B_{11}^{(a)}+\frac{m_B-E}{m_B}B_{12}^{(a)}+B_{13}^{(a)}\right]\right\}\phi_B^+(l_+)\phi_\|(x)$$

$$\frac{m_BE}{m_B+m_V}A_2(q^2)-\frac{1}{2}(m_B+m_V)A_1(q^2)=-\left(C_1^{(a)}+\frac{E}{m_B}C_2^{(a)}+C_3^{(a)}\right)m_V\zeta_\|^V \tag{4.75}$$

$$+m_VN_\|\int_0^1 dxdzdl_+\left\{\left[\frac{m_B-2E}{m_B}B_1^{(a)}-\frac{E}{m_B}B_2^{(a)}-B_3^{(a)}\right]\delta(x-z)\right.$$

$$\left.-\frac{2E}{m_b}\left[B_{11}^{(a)}-\frac{E}{m_B}B_{12}^{(a)}-B_{13}^{(a)}\right]\right\} J_\|(x,z,l_+)\phi_B^+(l_+)\phi_\|(x)$$

$$T_1(q^2)=\frac{m_B}{2E}T_2(q^2)=\left\{C_1^{(t)}-\frac{m_B-E}{m_B}C_2^{(t)}-C_3^{(t)}\right\}\zeta_\perp^V \tag{4.76}$$

$$-\frac{1}{2}N_\perp\int_0^1 dxdzdl_+\left\{\left[B_5^{(t)}+\frac{m_B-E}{m_B}B_6^{(t)}\right]\delta(x-z)\right.$$

$$\left.-\frac{2E}{m_b}\left[2B_{15}^{(t)}+2B_{17}^{(t)}+B_{19}^{(t)}+B_{21}^{(t)}+\frac{m_B-E}{m_B}(2B_{16}^{(t)}+B_{20}^{(t)})\right]\right\} J_\perp(x,z,l_+)\phi_B^+(l_+)\phi_\perp(x)$$

$$ET_3(q^2)-\frac{m_B}{2}T_2(q^2)=-(C_1^{(t)}-C_2^{(t)}-C_4^{(t)})m_V\zeta_\|^V \tag{4.77}$$

$$+m_VN_\|\int_0^1 dxdzdl_+\left\{\left[B_1^{(t)}-B_2^{(t)}-2B_3^{(t)}+B_4^{(t)}\right]\delta(x-z)\right.$$

$$\left.+\frac{2E}{m_b}(B_{15}^{(t)}+B_{16}^{(t)}-B_{18}^{(t)})\right\} J_\|(x,z,l_+)\phi_B^+(l_+)\phi_\|(x)$$

where $N_\perp=m_B/(4E^2)f_Bf_V^T$ and $N_\|=m_B/(4E^2)f_Bf_V$. To all orders in $\alpha_s(\Lambda m_b)$ there are only 2 jet functions. One of them $J_\|$ contributes to $B\to P,V_\|$, and other one $J_\perp$ contributing only to $B\to V_\perp$. At tree level they are equal $J_{\|,\perp}(z,x,l_+)=\frac{\pi\alpha_sC_F}{N_c}\frac{1}{\bar{x}l_+}\delta(x-z)$, but in general they are different.

The Wilson coefficients satisfy $C_{1-3}^{(v)} = C_{1-3}^{(a)}$ and $B_{1-4}^{(v)} = B_{1-4}^{(a)}$ in the NDR scheme. Reparameterization invariance constrains them as $B_{1-3}^{(v,a,t)} = C_{1-3}^{(v,a,t)}$, $B_4^{(v,a)} = -2C_3^{(v,a)}$, $B_4^{(t)} = C_4^{(t)}$, $B_5^{(t)} = 2C_3^{(t)}$, $B_6^{(t)} = -2C_4^{(t)}$ [157, 150]. At tree level they are given by $C_1^{(v,a,t)} = 1$, $B_1^{(v,a,t)} = 1$, $B_{13}^{(v,a)} = -1$, $B_{17}^{(t)} = 1$.

From the above discussion it is clear that while SCET does not allow us to calculate the shape or normalization of the heavy-light form factors, it does give predictions amongst different form factors. In particular, relations between form factors arising in decays of $B$ mesons via tensor currents, such as $B \to K^*\gamma$ and form factors required for the extraction of $|V_{ub}|$ can be derived. This allows to get the necessary information about the form factors from decays which are independent of $|V_{ub}|$. First steps at understanding quark mass effects in SCET have been carried out in [158]. Model independent relations that survive including the leading SU(3) violation in the light-cone distribution functions were given in [159].

The generic structure of the SCET factorization theorem is

$$f_i(q^2) = C_{ij}^{(0)}(Q)\, \zeta_j^M(Q\Lambda, \Lambda^2) + \int dx dz dk_+ C_{ij}^{(1)}(z,Q) J_j(z,x,k_+) \phi_B^+(k_+) \phi_j^M(x)\,. \quad (4.78)$$

Both terms in the SCET factorization formula scale like $(\Lambda/Q)^{3/2}$, such that their relative numerical contributions could be comparable. In the absence of the factorizable term, all 10 $B \to P, V$ form factors are determined by only three unknown "soft" form factors $\zeta^P, \zeta_\|^V, \zeta_\perp^V$, and thus satisfy symmetry relations [143, 146, 135]. In general they are however broken by the factorizable terms, which are not spin-symmetric.

Two of these symmetry relations turn out to remain valid, even after including the factorizable terms. This can be seen by a simple helicity argument [160] or by examining the factorization theorems

$$V(q^2) = \frac{(m_B + m_V)^2}{2m_B E} A_1(q^2)\,, \qquad T_1(q^2) = \frac{m_B}{2E} T_2(q^2)\,. \quad (4.79)$$

These relations are broken only by power corrections of $O(\Lambda/Q)$, which can, however, be numerically sizable $\sim 30\%$.

An important point is related to the convergence of the convolutions appearing in the factorizable term in Eq. (4.78). This issue is connected to the asymptotic behaviour of the light-cone wave function $\phi_B(k_+)$ and of the jet functions $J(x,z,k_+)$, issues which were studied in Refs. [153] and [152], respectively.

We comment next on the important issue of the relative size of the two terms in Eq. (4.78). Due to the explicit factor of $\alpha_s(\mu_c)$ (with $\mu_c^2 \sim Q\Lambda$) appearing in the jet function $J$, one might be led to take the point of view that the factorizable term is a small correction to the nonfactorizable contribution [146], and therefore the symmetry relations would be satisfied to a good approximation. However, this point of view neglects the possibility of similar $O(\alpha_s(\Lambda Q))$ terms being present in the $\zeta$ functions, which in principle receive also contributions from the collinear scale $\mu_c^2 \sim Q\Lambda$. Recently, in Ref. [154] it was argued that no effects from the collinear scale are present in $\zeta$, which would indicate that the first term in Eq. (4.78) dominates. However, a more definitive conclusion requires the resummation of the Sudakov logs present in the coefficients $C^{(0,1)}$.

An extreme case of Sudakov suppression is assumed in the pQCD approach [161, 162]. Here one takes the point of view that the nonfactorizable term is suppressed as $m_b \to \infty$ by the Sudakov logs contained in the Wilson coefficients $C_{ij}^{(0)}$, which effectively renders the form factors calculable in perturbation theory. Such a conclusion could be invalidated by the fact that similar Sudakov logs (not yet computed) are present also in the factorizable term $C_{ij}^{(1)}$. See also Ref. [163] for a detailed discussion of Sudakov effects in this context.

In the following we will not make any assumptions about the relative size of the two terms in Eq. (4.78). Eventually the soft form factors $\zeta$ will be obtained from model computations or lattice QCD. However, even in the absence of such information, the factorization results have significant predictive power. For example, using as input the form factor $f_+(q^2)$ as measured in $B \to \pi e \nu$, the remaining $B \to \pi$ form factors can be computed using the explicit form of the factorization formulae in Ref. [150] and $\phi_B(k_+), \phi_\pi(x)$.

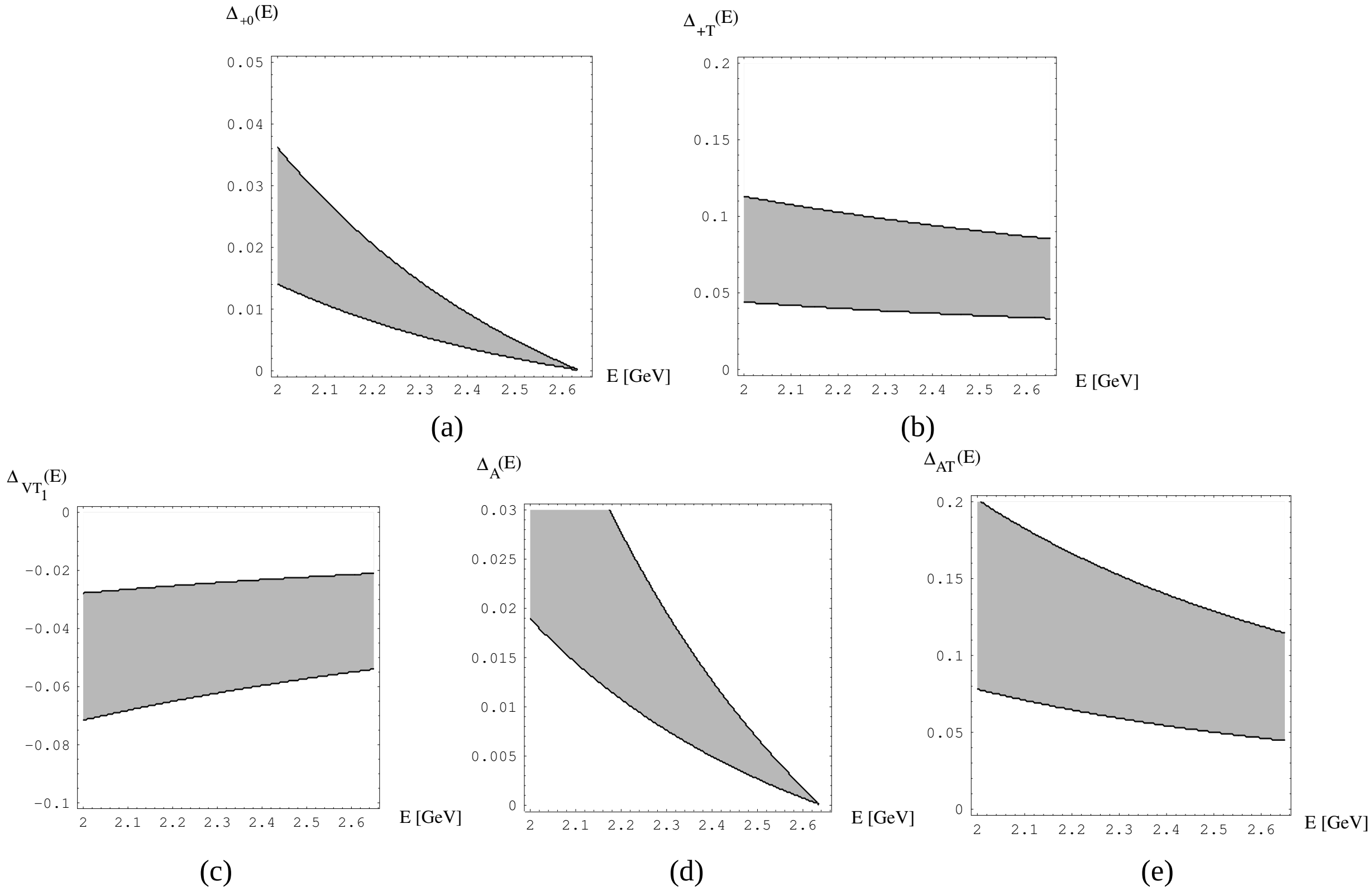

**Figure 4-5.** *Symmetry breaking corrections to the $B \to \pi$ form factor relations showing (a) $\Delta_{+0}(E)$ and (b) $\Delta_{+T}(E)$, and to the $B \to \rho$ form factor relations showing (c) $\Delta_{VT_1}(E)$, (d) $\Delta_A(E)$ and (e) $\Delta_{TA}(E)$. The shaded region corresponds to the variation in the collinear scale $\mu_c$ used to define the jet function between 0.54 and 2.18 GeV, with the choices of hadronic parameters defined in the text.*

To illustrate this approach, we present explicit results for form factor combinations from which the soft matrix elements $\zeta$ cancel out, and are therefore calculable. Working at tree level in matching at the scale $Q$, but to all orders in the jet function, there are 2 such combinations for the $B \to P$ form factors

$$\Delta_{+0}(E) = \frac{m_B}{2E} f_0(E) - f_+(E)\,, \qquad \Delta_{+T}(E) = f_+(E) - \frac{m_B}{m_B + m_P} f_T(E)\,. \tag{4.80}$$

and 3 combinations for the $B \to V$ form factors

$$\begin{aligned} \Delta_{VT_1}(E) &= \frac{m_B}{m_B + m_V} V(E) - T_1(E)\,, \\ \Delta_A(E) &= m_V A_0(E) + \frac{m_B E}{m_B + m_V} A_2(E) - \frac{1}{2}(m_B + m_V) A_1(E) \\ \Delta_{AT}(E) &= m_V A_0(E) + E T_3(E) - \frac{m_B}{2} T_2(E)\,. \end{aligned} \tag{4.81}$$

We show in Fig. 4-5 illustrative results for these form factor combinations, working at tree level in matching at the scale $Q$ and in the jet function. [10] In computing these results we used $f_B = 180$ MeV, $f_\pi = 131$ MeV, $f_\rho = 210$ MeV, $f_\rho^\perp(1.47\ \text{GeV}) = 152$ MeV, $\langle k_+^{-1} \rangle_B = (350\ \text{MeV})^{-1}$ and $a_2^\pi = a_2^\rho = a_2^{\rho\perp} = 0.2$.

[10] Editors note: Recently one-loop corrections to the jet functions became available [164], which substantially reduce the scale dependence shown in Fig. 4-5.

### 4.5.4 Form factor relations from HQET

$\succ\!\!-$ D. Pirjol $-\!\!\prec$

In the low recoil region for the final meson, corresponding to maximal $q^2 \sim (m_B - m_M)^2$, heavy quark symmetry can be applied to describe the transition process. For the heavy-to-heavy form factors, such as those parameterizing $B \to D^{(*)}\ell\nu$ decays, the normalization at zero recoil is fixed from the symmetry, with the leading power corrections of order $\Lambda/m_b$ vanishing for certain form factors [165]. No such information is available for heavy-to-light form factors, although some results can be established in a model-independent way.

The heavy mass scaling of the form factors can be straightforwardly derived from the mass dependence of the $|B\rangle$ states implicit in their relativistic normalization $|\overline{B}(p)\rangle \sim \sqrt{m_b}$. These relations are simpler when expressed in terms of the form factors defined in [166] (as opposed to the more commonly used form factors used in the preceding section). The scaling of the $B \to P$ form factors is

$$f_+(E) + f_-(E) \sim m_b^{-1/2}\,, \qquad f_+(E) - f_-(E) \sim m_b^{1/2}\,, \qquad s(E) \sim m_b^{1/2} \tag{4.82}$$

and for the $B \to V$ form factors

$$\begin{aligned} f(E) \sim m_b^{1/2}\,, \quad g(E) \sim m_b^{-1/2}\,, \quad a_+(E) - a_-(E) \sim m_b^{-1/2}\,, \quad a_+(E) + a_-(E) \sim m_b^{-3/2}\,, \\ g_+(E) - g_-(E) \sim m_b^{1/2}\,, \quad g_+(E) + g_-(E) \sim m_b^{-1/2}\,, \quad h(E) \sim m_b^{-3/2} \end{aligned} \tag{4.83}$$

We take the argument of the form factors as the light meson $M = P, V$ energy $E$ rather than $q^2 = m_B^2 + m_M^2 - 2m_B E$. In the low recoil region it scales as $E \sim \Lambda$.

Heavy quark spin symmetry implies also the existence of symmetry relations among form factors at fixed $E$ [166, 167]. There is one such relation for the $B \to P$ form factors

$$\text{(P-1)} : f_+(E) - f_-(E) - 2m_B s(E) \sim O(m_b^{-1/2}) \tag{4.84}$$

and three relations for the $B \to V$ form factors

$$\text{(V-1)} : \quad g_+(E) - g_-(E) + 2m_B g(E) \sim O(m_b^{-1/2}) \tag{4.85}$$

$$\text{(V-2)} : \quad g_+(E) + g_-(E) - 2Eg(E) - \frac{1}{m_B} f(E) \sim O(m_b^{-3/2}) \tag{4.86}$$

$$\text{(V-3)} : \quad a_+(E) - a_-(E) - 2g(E) + 2m_B h(E) \sim O(m_b^{-3/2})\,. \tag{4.87}$$

The leading power corrections to the heavy quark symmetry relations Eqs. (4.84)-(4.87) are also known from Ref. [168]. Contrary to naive expectations, they have a very simple form and depend only on the form factors of the dimension-4 currents $\overline{q}iD_\mu(\gamma_5)b$. We discuss in the following one possible application of these symmetry relations, and give a brief description of the $\Lambda/m_b$ improved form factor relations.

The HQET symmetry relations are relevant for a method discussed in Refs. [169, 170] for determining the CKM matrix element $|V_{ub}|$ from exclusive $B$ decays. This method combines data on semileptonic $B \to \rho\ell\nu$ and rare radiative decays $B \to K^*\ell^+\ell^-$ near the zero recoil point, and $|V_{ub}|$ is extracted from the ratio [169, 170]

$$\frac{\mathrm{d}\Gamma(B \to \rho e\nu)/\mathrm{d}q^2}{\mathrm{d}\Gamma(B \to K^*\ell^+\ell^-)/\mathrm{d}q^2} = \frac{8\pi^2}{\alpha^2} \frac{|V_{ub}|^2}{|V_{tb}V_{ts}^*|^2} \frac{1}{|C_9|^2 + |C_{10}|^2} \frac{|A_1^{B\to\rho}(q^2)|^2}{|A_1^{B\to K^*}(q^2)|^2} \frac{(m_B + m_\rho)^2}{(m_B + m_{K^*})^2} \frac{1}{1 + \Delta(q^2)} \tag{4.88}$$

The parameter $\Delta(q^2)$ contains the contribution of the radiative penguin $O_7$ to the $B \to K^*e^+e^-$ amplitude, and is computable at leading order in $1/m_b$ with the help of the symmetry relations Eqs. (4.85) and (4.86). The SU(3) breaking in the ratio of form factors on the right-hand side $A_1^{B\to\rho}(q^2)/A_1^{B\to K^*}(q^2)$ can be fixed using a Grinstein-type double ratio [171] and data on semileptonic $D \to K^*(\rho)e\overline{\nu}$ decays.

The leading power correction to the symmetry relations Eqs. (4.84)-(4.87) depends on the $B \to M$ matrix elements of dimension-4 currents. They are parameterized in terms of 2 form factors for $B \to P$

$$\langle P(p')|\bar{q}i\overleftarrow{D}_\mu h_v|\overline{B}(p)\rangle = \delta_+(E)(p+p')_\mu + \delta_-(E)(p-p')_\mu \tag{4.89}$$

and four form factors for $B \to V$ transitions

$$\langle V(p',\eta)|\bar{q}i\overleftarrow{D}_\mu b|\overline{B}(p)\rangle \; = \; d(E) i\epsilon^{\mu\nu\rho\sigma}\eta^*_\nu p_\rho p'_\sigma \tag{4.90}$$

$$\langle V(p',\eta)|\bar{q}i\overleftarrow{D}_\mu \gamma_5 b|\overline{B}(p)\rangle \; = \; d_1(E)\eta^*_\mu + d_+(E)(\eta^*\cdot p)(p_\mu + p'_\mu) + d_-(E)(\eta^*\cdot p)(p_\mu - p'_\mu)\,. \tag{4.91}$$

In the heavy quark limit, not all these form factors are independent; using the constraint $\not{v} h_v = h_v$ and the equation of motion for the heavy quark field $iv\cdot Dh_v = 0$, the number of independent subleading form factors is reduced to one for $B \to P$, and 3 for $B \to V$.

Furthermore, the $B \to \pi, K$ subleading form factors $\delta_\pm(E)$ can be computed in a model-independent way at leading order in the heavy mass and the chiral expansion [172, 168]. On the other hand, the corresponding $B \to V$ form factors have to estimated with the help of quark models or lattice QCD.

The improved HQET symmetry relations can be obtained from operator identities of the type

$$i\partial^\nu(\bar{q}i\sigma_{\mu\nu}b) = -(m_b+m_q)\bar{q}\gamma_\mu b - 2\bar{q}i\overleftarrow{D}_\mu b + i\partial_\mu(\bar{q}b)\,, \tag{4.92}$$

which follows from a simple application of the QCD equations of motion for the quark fields. Taking the $B \to V$ matrix element one finds the exact relation

$$g_+(q^2) = -(m_b+m_q)g(q^2) + d(q^2)\,. \tag{4.93}$$

Counting powers of $m_b$ and keeping only the leading order terms reproduces the symmetry relation (V-1) + (V-2) among vector and tensor form factors [166, 167]. Keeping also the subleading terms of $O(m_b^{-1/2})$ gives the improved version of the form factor relation Eq. (4.85)

$$(\text{V-1}') : g_+(E) - g_-(E) + 2m_B g(E) = -2(E-\overline{\Lambda})g(E) - \frac{1}{m_B}f(E) + 2d(E) + O(m_b^{-3/2}) \tag{4.94}$$

Similar improved versions of the other symmetry relations can be found in Ref. [168]. We quote here only the analog of (V-2) Eq. (4.86), which has implications for the method of determining $|V_{ub}|$ using exclusive decays (see Eq. (4.88))

$$\begin{aligned}(\text{V-2}') : \quad & g_+(E) + g_-(E) - 2Eg(E) - \frac{1}{m_B}f(E) = \frac{2}{m_B}\{(2E^2 - m_V^2) - E(\overline{\Lambda} - m_q)\}g(E) \\ & + \frac{1}{m_B^2}(2E - \overline{\Lambda} - m_q)f(E) - 2\frac{E}{m_B}d(E) + \frac{2}{m_B^2}d_1(E) + O(m_b^{-5/2})\end{aligned} \tag{4.95}$$

The improved symmetry relation Eq. (4.94) can be used to determine the tensor form factor $g_+(q^2)$ in terms of the vector and axial form factors $f(q^2), g(q^2)$ as measured in exclusive semileptonic $B \to V\ell\nu$ decays. Combining the symmetry relations Eqs. (4.85), (4.86) in order to extract $g_+$ at next-to-leading order in $\Lambda/m_b$ requires the knowledge of the leading correction of $O(m_b^{-1/2})$ to Eq. (4.85) (since the latter is of the same order as the terms shown on the RHS of Eq. (4.86)).

The relations Eqs. (4.94) and (4.95) were used in Ref. [173] to estimate the subleading corrections of $O(\Lambda/m_b)$ to the $|V_{ub}|$ determination using Eq. (4.88). These corrections can be as large as 5%, and are dominated by the unknown form factor $d_1(q^2)$ of $\bar{q}i\overleftarrow{D}_\mu\gamma_5 b$. Quark model estimates of this matrix element suggest that the correction is under a few percent, and more precise determinations (lattice QCD) could help to reduce or eliminate this source of uncertainty.

The rare $B$ decays $b \to s\gamma$ and $b \to se^+e^-$ receive significant long-distance effects arising from $c\bar{c}$ and $u\bar{u}$ quark loops. In Ref. [174] it was proposed to treat these effects in the small recoil region using an operator product expansion in $1/Q$, combined with HQET. This method is similar to the computation of $e^+e^- \to$ hadrons, and allows model-independent predictions of the $e^+e^-$ invariant spectrum in $B \to K^{(*)}\ell^+\ell^-$ decays in the small recoil region.

The results of [174] are applied to a method for determining $V_{ub}$ from combined exclusive $B$ decays, first proposed in [175, 176]. This method is improved here in two ways: a) combining the OPE method with recent results in the theory of $b \to se^+e^-$ decays, the complete next-to-leading perturbative corrections can be included; b) power corrections of order $\Lambda/Q$ and $m_c^2/m_b^2$ are included with the help of corrected heavy quark symmetry relations derived earlier in [177, 178]. The resulting uncertainty in $|V_{ub}|$ from this determination is dominated by scale dependence and is of the order of 15%.

### 4.5.5 SU(3) breaking in $B{\to}\rho/K^*\,\gamma, \rho/K^*\,\ell^+\ell^-$, double ratios, and $|V_{td}/V_{ts}|$

$\succ$– B. Grinstein –$\prec$

The radiative decays $b \to d\gamma$ and $b \to s\gamma$ are dominated by the short-distance top-quark penguin graph. Using SU(3) symmetry to relate the relevant form factors, it has been suggested to use a measurement of the ratio

$$\frac{\Gamma(B \to \rho\gamma)}{\Gamma(B \to K^*\gamma)} = \left|\frac{V_{td}}{V_{ts}}\right|^2 R_{\mathrm{SU(3)}}(1+\Delta) \tag{4.96}$$

to determine the CKM matrix element $V_{td}$. There are two theoretical sources of uncertainty in such a determination, coming from long distance effects (parameterized by $\Delta$) and SU(3) breaking in the form factor and kinematics (contained in $R_{\mathrm{SU(3)}}$). In Ref. [179] the different sources of long-distance contributions to the decays in Eq. (4.96) have been classified using a diagrammatic approach, essentially equivalent to a SU(3) flavor analysis.

The figure above defines the different long distance contributions as annihilation ($A$), $W$ exchange ($E$), penguin ($P_q^{(i)}$), penguin annihilation ($PA$) and gluonic t-penguin ($M^{(i)}$); the crosses indicate where the photon emission may take place at leading order in $1/m_b$, and the superscripts on $P_q$ and $M$ refer to whether the photon is emitted from the quark in the loop ("(1)") or not ("(2)"). Particular processes are affected by some, but not necessarily all, of these long distance "contamination." For example, the weak annihilation amplitude $A$ contribute only to the $B^\pm$ radiative decays,

$$\mathcal{A}(B^- \to \rho^-\gamma) = \lambda_u^{(d)}(P_u^{(1)} + Q_u P_u^{(2)} + A) + \lambda_c^{(d)}(P_c^{(1)} + Q_u P_c^{(2)}) + \lambda_t^{(d)}(\hat{P}_t + Q_u M^{(2)}), \tag{4.97}$$
$$\mathcal{A}(B^- \to K^{*-}\gamma) = \lambda_u^{(s)}(P_u^{(1)} + Q_u P_u^{(2)} + A) + \lambda_c^{(s)}(P_c^{(1)} + Q_u P_c^{(2)}) + \lambda_t^{(s)}(\hat{P}_t + Q_u M^{(2)}), \tag{4.98}$$

while W-exchange contributes only to $\overline{B}^0$ decays,

$$\sqrt{2}\mathcal{A}(\overline{B}^0 \to \rho^0\gamma) = \lambda_u^{(d)}(P_u^{(1)} + Q_d P_u^{(2)} - E - PA_u) + \lambda_c^{(d)}(P_c^{(1)} + Q_d P_c^{(2)} - PA_c) + \lambda_t^{(d)}(\hat{P}_t + Q_d M^{(2)}), \tag{4.99}$$

$$\sqrt{6}\mathcal{A}(\overline{B}^0 \to \phi^{(8)}\gamma) = -\lambda_u^{(d)}(P_u^{(1)} + Q_d P_u^{(2)} + E + PA_u) - \lambda_c^{(d)}(P_c^{(1)} + Q_d P_c^{(2)} + PA_c) - \lambda_t^{(d)}(\hat{P}_t + Q_d M^{(2)}). \tag{4.100}$$

Perhaps more interestingly, some amplitudes contain no annihilation or $W$ exchange contamination:

$$\mathcal{A}(\overline{B}^0 \to \overline{K}^{*0}\gamma) = \lambda_u^{(s)}(P_u^{(1)} + Q_d P_u^{(2)}) + \lambda_c^{(s)}(P_c^{(1)} + Q_d P_c^{(2)}) + \lambda_t^{(s)}(\hat{P}_t + Q_d M^{(2)}), \tag{4.101}$$
$$\mathcal{A}(\overline{B}_s \to K^{*0}\gamma) = -\lambda_u^{(d)}(P_u^{(1)} + Q_s P_u^{(2)}) - \lambda_c^{(d)}(P_c^{(1)} + Q_s P_c^{(2)}) - \lambda_t^{(d)}(\hat{P}_t + Q_s M^{(2)}). \tag{4.102}$$

We have used the shorthand $\lambda_{q'}^{(q)} = V_{q'b}V_{q'q}^*$ and, noting that $P_t$ and $M^{(1)}$ appear always in the same combination, we have defined $\hat{P}_t = P_t + M^{(1)}$.

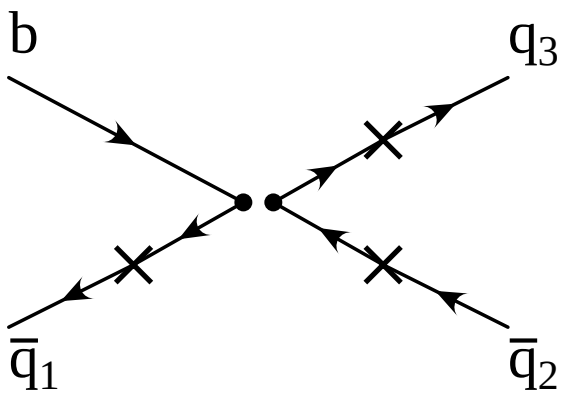

Weak Annihilation $A^{(i)}$

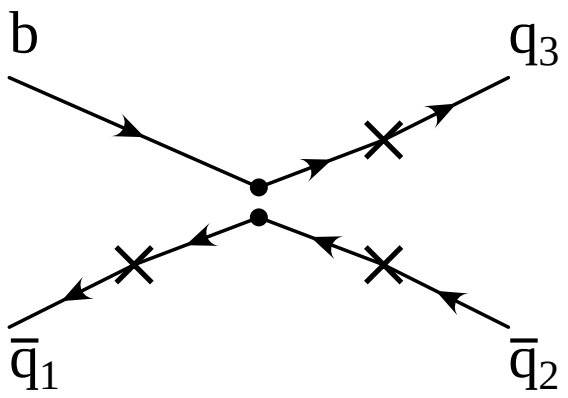

W-exchange $E^{(i)}$

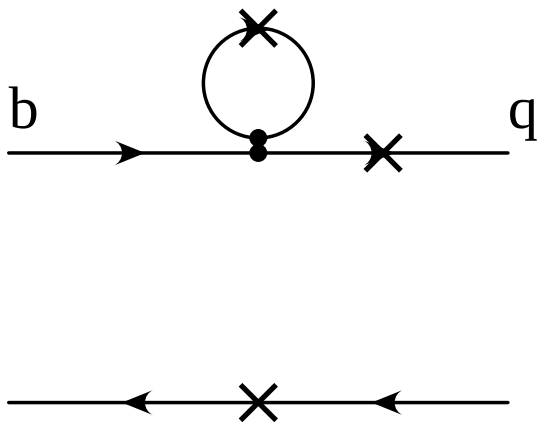

Penguin $P^{(1)}_{d,s}$ and $P^{(2)}_{u,c}$

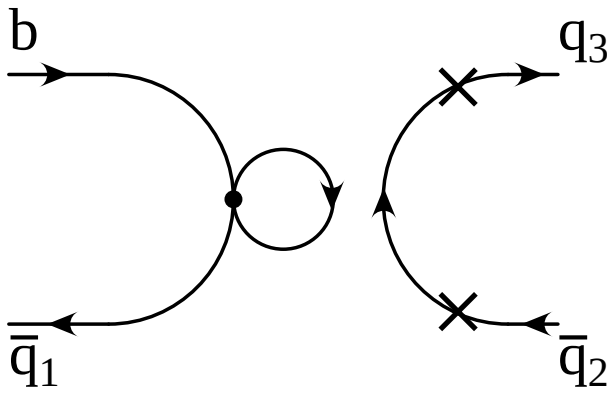

Penguin Annihilation PA

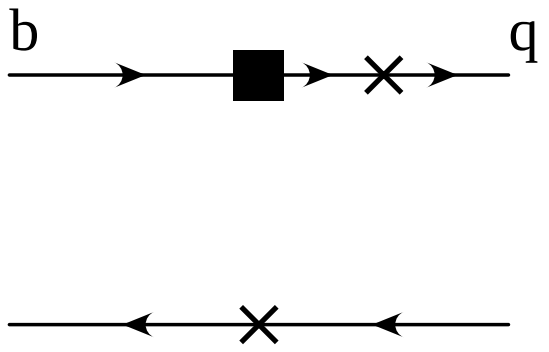

Gluonic t-Penguin $M^{(1)}$ and $M^{(2)}$

| Photon helicity | $\|P_{t\lambda}\|$ | $\|P_{c\lambda}\|$ | $\|P_{u\lambda}\|$ | $\|A_\lambda\|$ | $\|E_\lambda\|$ |
|---|---|---|---|---|---|
| $\lambda = L$ | 1.8 | 0.16 | 0.03 | 0.6 | 0.05 |
| $\lambda = R$ | 0 | 0.04 | 0.007 | 0.07 | 0.007 |

The table above shows an estimate of the individual amplitudes (in units of $10^{-6}$ MeV) contributing to $B \to \rho\gamma$ decays for different photon helicities. The $V - A$ structure of charged currents in the standard model gives a strong suppression to right handed helicities. This could be used as a probe of New Physics. The dominant amplitudes, with left handed photons, show an interesting pattern of magnitudes, $|P_{t\lambda}| > |A_\lambda| > |P_{c\lambda}| > |E_\lambda| \approx |P_{u\lambda}|$. As expected, the short distance contribution — the top-penguin — dominates.

Including the CKM factors, the weak annihilation amplitude contributes about 15% to the $B \to \rho\gamma$ decay amplitude. It is possible to show that the annihilation amplitude factorizes (to leading order in $1/m_b$) and the relevant hadronic matrix element can be related to the measurable decay rate of the radiative leptonic decay $B \to \gamma e\overline{\nu}$. Although this amplitude can be estimated theoretically [180], for a model-independent determination of $|V_{td}|$ it is preferable to use measurements of this process.

In order to determine the CKM ratio $|V_{td}/V_{ts}|$ the leading top-penguin amplitude, can be determined in terms of the form factors for $B \to \rho\ell\nu$ semileptonic decays using the form factor relations at large recoil (see the appropriate section in this report).

Keeping the dominant contributions in Eqs. (4.97)-(4.98) one can write for the amplitudes of the radiative decays

$$\mathcal{A}(B^- \to \rho^-\gamma_L) = V_{td}V^*_{tb}P(1 + \varepsilon_A e^{i(\alpha+\phi_A)}) \tag{4.103}$$
$$\mathcal{A}(B^- \to K^{*-}\gamma_L) = V_{td}V^*_{tb}P' \tag{4.104}$$

where the penguin amplitudes $P, P'$ include the effects of charm loops. The weak annihilation amplitude is negligible in $B^- \to K^{*-}\gamma$ because of its small CKM coefficient. Using these expressions, the factors appearing in the ratio Eq. (4.96) are given by

$$R_{\mathrm{SU}(3)} = \frac{|P|^2}{|P'|^2} \simeq \left(\frac{g_+^{(B\rho)}(0)}{g_+^{(BK^*)}(0)}\right)^2 = 0.76 \pm 0.22\,, \qquad \Delta = 2\varepsilon_A \cos\phi_A \cos\alpha + \varepsilon_A^2\,, \tag{4.105}$$

where the tensor form factor $g_+(q^2)$ is defined in Section 4.5.3. Model estimates give for the weak annihilation contribution $\varepsilon_A = 0.12$ which leads to an error of $12\%$ in $V_{td}$. The SU(3) breaking factor $R_{\mathrm{SU}(3)}$ has been computed using QCD sum rules and lattice QCD. The result quoted above is from the UKQCD collaboration [181].

The issue of SU(3) breaking in heavy-light form factors is also relevant for a method for determining $V_{ub}$ from rare radiative and semileptonic $B$ decays in the low recoil region. This has been discussed in some detail in Section 4.5.4; we comment here on the SU(3) breaking effects. This method requires the ratio of exclusive decay rates [182, 176, 183]

$$\frac{\mathrm{d}\Gamma(\overline{B} \to \rho e\nu)/\mathrm{d}q^2}{\mathrm{d}\Gamma(\overline{B} \to K^*\ell^+\ell^-)/\mathrm{d}q^2} = \frac{|V_{ub}|^2}{|V_{tb}V_{ts}^*|^2}\,\frac{8\pi^2}{\alpha^2}\,\frac{1}{|C_9|^2 + |C_{10}|^2}\,\frac{|f^{B\to\rho}(y)|^2}{|f^{B\to K^*}(y)|^2}\,\frac{1}{1+\Delta(y)} \tag{4.106}$$

where $y = E_V/M_V$ and $q^2$ is the invariant mass of the lepton pair. $C_i$ are coefficients of interactions in the effective Hamiltonian for $b \to see$ decays [184, 185, 186, 187]. In the SU(3) symmetry limit the ratio $f^{B\to\rho}(y)/f^{B\to K^*}(y)$ is unity. Since SU(3) is violated at the 30% level, a better approach is to measure the corresponding ratio in $D$ decays. The double ratio

$$R(y) \equiv \frac{|f^{B\to\rho}(y)/f^{B\to K^*}(y)|}{|f^{D\to\rho}(y)/f^{D\to K^*}(y)|} = 1 + \mathcal{O}\left(\frac{m_s}{\Lambda_\chi}\Big(\frac{\Lambda}{m_c} - \frac{\Lambda}{m_b}\Big)\right) \tag{4.107}$$

is protected *both* by heavy quark symmetry and by SU(3), so even if each of these is good only to about the 30% level, the ratio is unity to better than 10%. Calculations in heavy meson chiral perturbation theory [188, 189] show that double ratios are typically protected at the few percent level [132, 190, 191].

To summarize, the leading uncertainty in the extraction of CKM ratios from $\Gamma(B^- \to \rho^-\gamma_L)/\Gamma(B^- \to K^{*-}\gamma)$ is due to SU(3) symmetry breaking. The largest long distance correction, of order 15% in the amplitude, is from weak-annihilation, but can be computed reliably by measuring the photon energy spectrum in $B \to e\nu\gamma$. Form factor uncertainties are eliminated in $B \to K^* e^+ e^-$ using double ratios with the corresponding $D$ decays. A method for determining $V_{ub}$ using these decays contains SU(3) breaking effects which can be eliminated by combining $B$ and $D$ decays.

### 4.5.6 Experimental prospects

>– D. del Re –<

Exclusive charmless semileptonic $B$ decays have been previously studied by the CLEO [192], Belle [193] and *BABAR* [194] collaborations. All these measurements are performed by the reconstruction of one half of the event. One hard lepton in the event is identified and the charmless meson present in the semileptonic decay is reconstructed. Requirements on the missing mass of the event are also imposed. Since these requirements alone do not sufficiently reduce the background, significant restrictions on the lepton energy and other variables are applied. As a consequence, an extrapolation to the full phase space is needed thereby introducing large theoretical systematic uncertainties, that are already bigger than the statistical errors. If higher integrated luminosities are recorded, this approach will not allow us to improve the error on these branching ratios and on $|V_{ub}|$.

The recoil method can thus play an important role in the study of exclusive charmless semileptonic decays in the Super $B$ Factory era. This approach assures a sample with a much higher purity than in previous measurements. Since the level of background is low, no kinematic cuts are required, and nearly the full phase space is analyzed. Thus, the dependence on form factors and on the different decay models in the extraction of the branching ratios is largely eliminated. In terms of total error, the recoil method will surpass the traditional approach for an integrated luminosity of about $500\ \text{fb}^{-1}$, well before the projected advent of a Super $B$ Factory.

### $B \to X_u \ell\nu$ decays

In the following study, we propose a method very similar to the inclusive approach presented in section 4.4.3. A preliminary result based on this analysis has been already presented in [195]. As in the inclusive case, we select events with one or more leptons in the recoil. A very loose cut on the lepton momentum is applied ($p^* > 1$ GeV). We also apply cuts on the charge conservation and missing mass squared of the event. We inclusively reconstruct the invariant mass of the $X$ system and apply additional constraints on charged particle multiplicity, in order to select specific resonances. For instance, we require no tracks in the $B^- \to \pi^0\ell^+\nu$ case and two tracks for $B^- \to \rho^0\ell^+\nu$. Moreover, we apply cuts based on the neutral energy in the recoil to separate resonances with identical charged multiplicities (such as $\rho^0$ and $\omega$).

This technique selects a very clean sample of exclusive charmless decays. In Fig. 4-6 the result of a detailed generic Monte Carlo event sample of an equivalent integrated luminosity of $500\,\text{fb}^{-1}$ is shown for the modes $B^\pm \to \pi^0 l\nu$ , $B^\pm \to \omega l\nu$ and $B^\pm \to$ "$\rho^0$" $l\nu$ (here "$\rho^0$" indicates a combination of $\pi^+\ \pi^-$ with $m_{\pi^+\pi^-}$ in the window $0.65 < m_{\pi^+\pi^-} < 0.95\,\text{GeV}/c^2$ at generator level). The signal-to-background ratio is much better than in the standard exclusive analyses. The $B^\pm \to \pi^0 l\nu$ case, for instance, is basically background-free. A projection of the total error on the exclusive branching ratio as a function of the accumulated luminosity is shown in Fig. 4-7 for $B^\pm \to \pi^0 l\nu$ . A systematic uncertainty of $3\%$ for $B^\pm \to \pi^0 l\nu$ has been assumed. The extrapolation indicates how the error can be significantly reduced at a Super $B$ Factory .

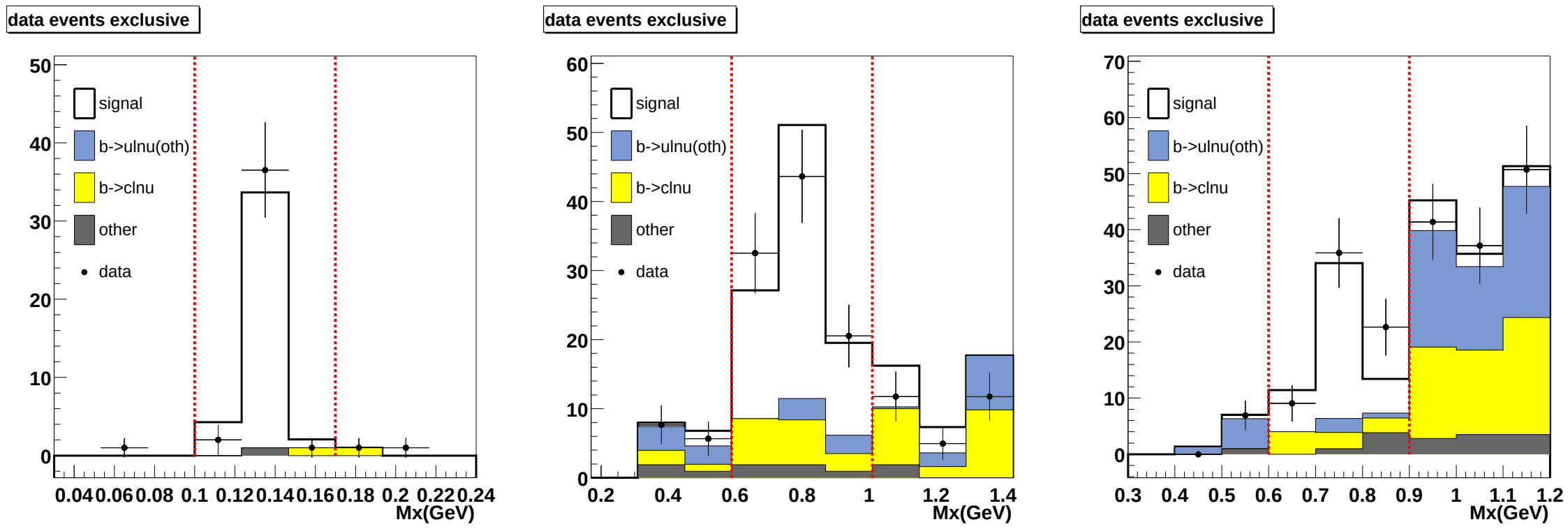

**Figure 4-6.** *Measurement of exclusive charmless semileptonic decays in the recoil of a fully reconstructed hadronic $B$ decay (detailed MC simulation for 500 $\text{fb}^{-1}$). Projections in the $m_X$ variable of the result. Vertical dotted lines represent the signal region. The plots show $B^\pm \to \pi^0 l\nu$ (left), $B^\pm \to$ "$\rho^0$" $l\nu$ (middle), and $B^\pm \to \omega l\nu$ (right).*

A study of the kinematic quantities can also be performed, as has been done by the CLEO collaboration [192], but the recoil approach offers the advantage of analyzing the full phase space. In Fig. 4-8 the measured $q^2$ spectrum for the $\overline{B}^0 \to \pi^+\ell^-\nu$ case on a MC sample equivalent to an equivalent integrated luminosity of $2\,\text{ab}^{-1}$ is compared with the distribution expected by using different models. With these statistics it is possible to have sufficient sensitivity to reject certain models. However, as mentioned in Section 4.5.2, lattice QCD should provide model-independent calculations for form factors on a timescale well-suited for this type of analysis.

This method can be further improved by performing a purely exclusive analysis on the recoil, and reconstructing the resonances, instead of inclusively reconstructing the $X$ system. A gain in efficiency is achievable using this technique, especially in $B^+ \to \pi^0\ell^+\nu$.

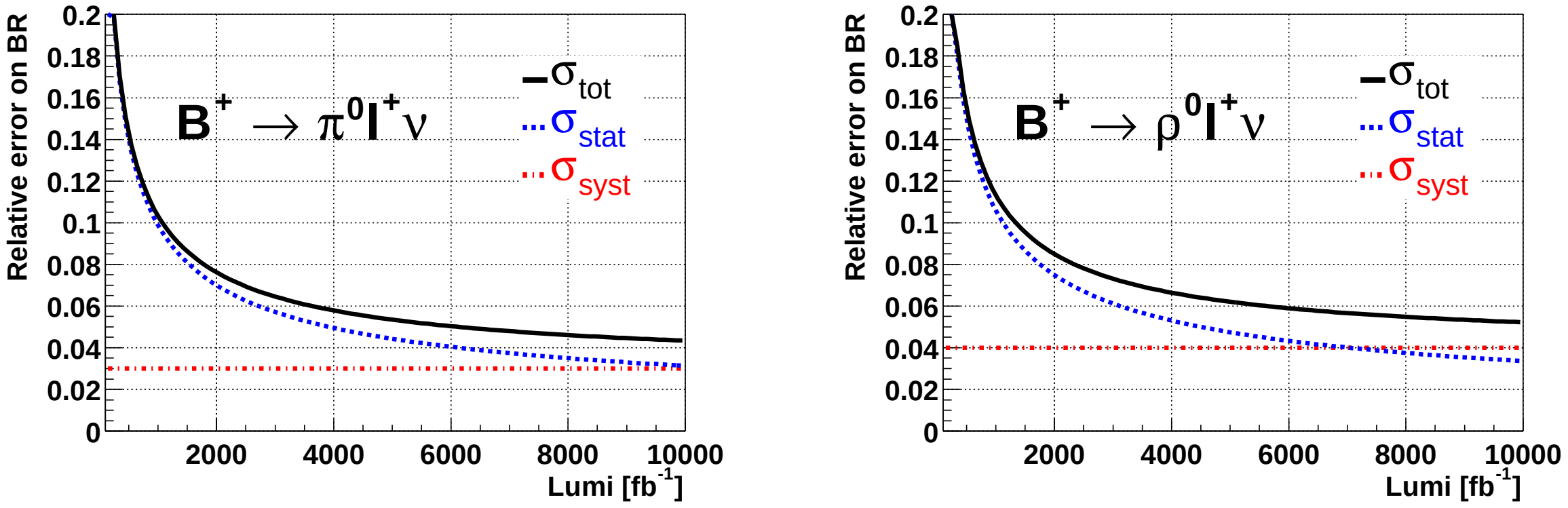

**Figure 4-7.** *Projections of the error on the exclusive branching ratio as a function of integrated luminosity.*

## $B \to X_u \tau \nu$ decays

The recoil technique, together with large data samples, also permits the study of more difficult exclusive decays, such as $B \to \pi\tau\nu$, which presents many additional challenges due the presence of a $\tau$. First the branching ratio for this decay should be 6 times smaller than the equivalent $e/\mu$ decays. In addition, instead of electrons and muons which can be simply identified, we have $\tau$ leptons whose decays involve additional neutrinos, thus destroying the powerful constraint from the missing mass squared. Preliminary studies show that, since the discrimination from $b \to c\ell\nu$ is much less effective in this case, additional efforts are needed to reduce the charm background, and make the analyses feasible. Furthermore, the background from Cabibbo-favored semileptonic decays should be studied with a full MC simulation (to account for the presence of, *e.g.*, $K_L^0$ in these decays).

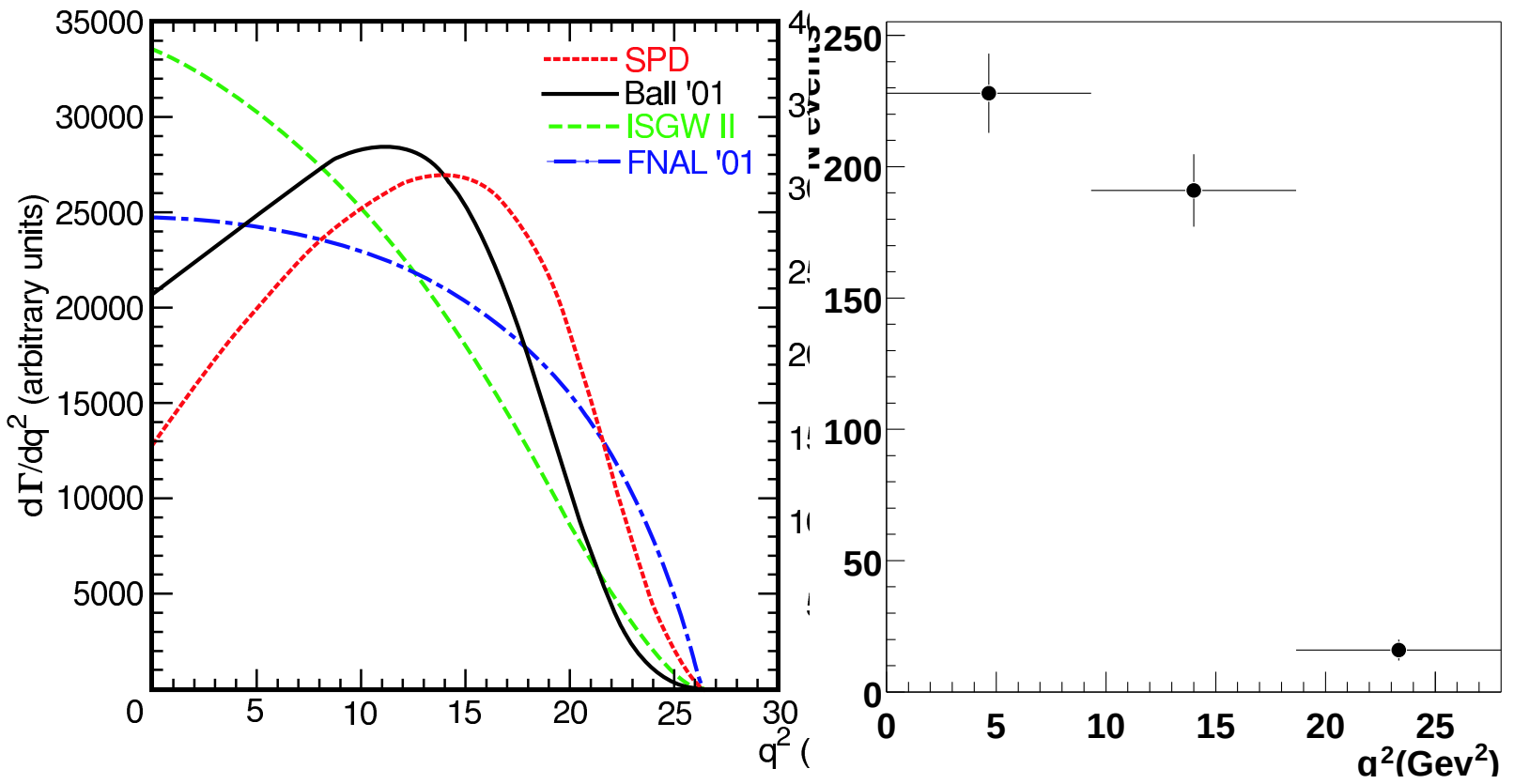

**Figure 4-8.** *Left: Theoretical expectations for the $q^2$-spectrum in $\overline{B}^0 \to \pi^+\ell^-\nu$ decays for different calculations [192]. Right: The $q^2$-spectrum in $\overline{B}^0 \to \pi^+\ell^-\nu$ decays (detailed MC simulation for an integrated luminosity of 2 ab$^{-1}$).*

# 4.6 Purely Leptonic Decays

## 4.6.1 $B \to \mu\overline{\nu}$ and $B \to \tau\overline{\nu}$ theory: $f_B$ from lattice QCD

>– C. Bernard, S. Hashimoto, P. Mackenzie –<

The estimates for future lattice precision presented in this section parallel those in Section 4.5.2 on semileptonic form factors. In addition to expected errors for the leptonic decay constants $f_B$, and $f_{B_s}$, we include estimates for errors on the combination relevant for $B$-$\overline{B}$ mixing, $f_B\sqrt{B_B}$, where $B_B$ is the bag parameter for $B$ mesons, as well as the ratios $f_{B_s}/f_B$ and

$$\xi \equiv \frac{f_{B_s}\sqrt{B_{B_s}}}{f_B\sqrt{B_B}} .$$

As in Section 4.5.2, we assume three levels of computation, MILC0, MILC1, and MILC2, based on improved staggered simulations with $n_F = 3$ flavors of dynamical sea quarks.

Tables 4-6, 4-7, and 4-8 show estimates of precision attainable for lattice calculations with data sets MILC0, MILC1, and MILC2, respectively. As in Section 4.5.2, we give two alternatives for perturbative errors (one-loop and two-loop) and two alternatives for chiral extrapolations errors: (no) S$\chi$PT assumes that staggered chiral perturbation theory is (is not) useful.

**Table 4-6.** *Estimated percentage errors for form factors at MILC0 level: one to two years from the present. "Light $q$" includes light quark chiral and discretization errors. "Heavy $Q$" means heavy quark discretization errors.*

| quantity | statist. | scale | light $q$ | | heavy $Q$ | pert. th. | |
|---|---|---|---|---|---|---|---|
| | | | no S$\chi$PT | S$\chi$PT | | 1-loop | 2-loop |
| $f_B$ | 3 | 2 | 5 | 2.5 | 3 | 7.5 | 2 |
| $f_B\sqrt{B_B}$ | 4 | 2 | 5.5 | 3 | 3 | 8.5 | 2.5 |
| $f_{B_s}/f_B$ | 1 | – | 5 | 2.5 | 1 | – | – |
| $\xi$ | 2 | – | 5.5 | 3 | 1 | – | – |

**Table 4-7.** *Same as Table 4-6, but for MILC1 level: three to five years from the present.*

| quantity | statist. | scale | light $q$ | | heavy $Q$ | pert. th. | |
|---|---|---|---|---|---|---|---|
| | | | no S$\chi$PT | S$\chi$PT | | 1-loop | 2-loop |
| $f_B$ | 2 | 1.5 | 3 | 1.5 | 2 | 7.5 | 2 |
| $f_B\sqrt{B_B}$ | 3 | 2 | 3.5 | 2 | 2 | 8.5 | 2.5 |
| $f_{B_s}/f_B$ | 0.8 | – | 3 | 1.5 | 0.8 | – | – |
| $\xi$ | 2 | – | 3.5 | 2 | 1 | – | – |

Table 4.6.1 shows the total lattice errors of the leptonic decay constants (and related quantities) under various assumptions, together with our best guess of which alternatives are most likely to be realized in practice. It must be kept in mind that the errors themselves are uncertain, by a fractional amount which is at least $\sim 30\%$ and rises with time into the future.

**Table 4-8.** *Same as Table 4-6, but for MILC2 level: five to eight years from present.*

| quantity | statist. | scale | light $q$ | | heavy $Q$ | pert. th. | |
|---|---|---|---|---|---|---|---|
| | | | no S$\chi$PT | S$\chi$PT | | 1-loop | 2-loop |
| $f_B$ | 1 | 1 | 2 | 1 | 1.5 | 7.5 | 2 |
| $f_B\sqrt{B_B}$ | 1.3 | 2 | 2.5 | 1 | 1.6 | 8.5 | 2.5 |
| $f_{B_s}/f_B$ | 0.5 | – | 2.5 | 1 | 0.5 | – | – |
| $\xi$ | 1 | – | 3 | 1 | 0.6 | – | – |

**Table 4-9.** *Estimated total lattice errors under various assumptions. Where there are four entries per column they correspond to: (1)* **no** *S$\chi$PT and 1-loop perturbation theory, (2) S$\chi$PT and 1-loop perturbation theory, (3)* **no** *S$\chi$PT and 2-loop perturbation theory, and (4) S$\chi$PT and 2-loop perturbation theory. Where there are two entries per column the quantity is free from perturbative errors, and the entries correspond to: (1)* **no** *S$\chi$PT and and (2) S$\chi$PT. Our best guesses of which alternative will in fact be realized are surrounded with boxes.*

| quantity | now | 1-2 yrs. MILC0 | 3-5 yrs. MILC1 | 5-8 yrs. MILC2 |
|---|---|---|---|---|
| $f_B$ | 15 | 10, $\boxed{9}$, 7, 6 | 9, 8, 5, $\boxed{4}$ | 8, 8, 4, $\boxed{3}$ |
| $f_B\sqrt{B_B}$ | 15-20 | $\boxed{12}$, 11, 8, 7 | 10, 10, 6, $\boxed{5}$ | 9, 9, 5, $\boxed{4}$ |
| $f_{B_s}/f_B$ | 6 | 5, $\boxed{3}$ | 3, $\boxed{2}$ | 3, $\boxed{1}$ |
| $\xi$ | 7 | $\boxed{6}$, 4 | 4, $\boxed{3}$ | 3, $\boxed{1.5}$ |

### 4.6.2 Experimental prospects

>– M. Datta, T. Moore –<

The purely leptonic decays $B^+ \to \ell^+\nu_\ell$ have not yet been observed experimentally. These decays are highly suppressed in the Standard Model due to their dependence on $|V_{ub}|^2$. Furthermore, helicity suppression introduces a dependence on $m_\ell^2$ where $m_\ell$ is the lepton mass. Assuming $|V_{ub}|$ = 0.0036 [196] and $f_B$ = 198 MeV [197], the Standard Model prediction for the $B^+ \to \tau^+\nu_\tau$ branching fraction is roughly $1 \times 10^{-4}$. Due to helicity suppression, $B^+ \to \mu^+\nu_\mu$ and $B^+ \to e^+\nu_e$ are further suppressed by factors of 225 and $10^7$, respectively. The Standard Model predictions have an uncertainty of about 50% from the uncertainties in $|V_{ub}|$ and $f_B$. The small Standard Model rate expected for $B^+ \to e^+\nu_e$ is even beyond the sensitivity of a Super $B$ Factory . Although searches for $B^+ \to e^+\nu_e$ are still interesting as tests of New Physics, only the $\tau$ and muon modes are discussed below.

$B^+ \to \ell^+\nu_\ell$ decays produce a mono-energetic lepton in the $B$ rest frame with a momentum given by

$$p_\ell = \frac{m_B^2 - m_\ell^2}{2m_B}. \tag{4.108}$$

In the case of $B^+ \to \mu^+\nu_\mu$, the muon momentum is approximately $m_B/2 = 2.645\ \text{GeV}/c$, which provides a strong experimental signature. By contrast, the decay of the $\tau^+$ lepton produced in $B^+ \to \tau^+\nu_\tau$ decays results in additional missing energy from the additional neutrino. The absence of strong kinematic constraints results in a more challenging experimental analysis. Thus, despite the larger branching fraction for $B^+ \to \tau^+\nu_\tau$, the two decay modes have comparable physics reach. Since very different analysis techniques have been developed for these searches, they will be discussed separately in the following sections.

### $B^+ \to \tau^+\nu_\tau$

In this section we briefly describe the analyses performed in the *BABAR* experiment for the search of the decay $B^+ \to \tau^+\nu_\tau$ and discuss the potential of similar analyses in the scenario of a Super $B$ Factory .

The $B^+ \to \tau^+\nu_\tau$ decay has very little experimental constraint, due to the presence of multiple neutrinos in the final state. Therefore, in the $\Upsilon(4S)$ CMS, the decay of one of the $B$ mesons (referred as the "tag" $B$ meson) is reconstructed and the signature of $B^+ \to \tau^+\nu_\tau$ decay is searched for in the recoil. In the *BABAR* experiment, both hadronic and semileptonic tags (cf. Section 4.2.1) have been used in analyses based on a data set of about $80\,\mathrm{fb}^{-1}$.

In the analysis with hadronic tags [198], the $\tau$ lepton is identified in both leptonic and hadronic decay modes: $\tau^+ \to e^+\nu_e\overline{\nu}_\tau$, $\tau^+ \to \mu^+\nu_\mu\overline{\nu}_\tau$, $\tau^+ \to \pi^+\overline{\nu}_\tau$, $\tau^+ \to \pi^+\pi^0\overline{\nu}_\tau$, $\tau^+ \to \pi^+\pi^-\pi^+\overline{\nu}_\tau$. This set is somewhat restricted in events with semi-exclusive semileptonic tags because of the higher background level (see below).

In the recoil all remaining particles are required to be consistent with coming from $B^+ \to \tau^+\nu_\tau$ decay. The selection criteria require that there be no extra charged particles besides one(three) track(s) from $\tau$ decay, and little neutral energy in the calorimeter, after excluding the energy of any neutrals coming from the decay of the tag $B$ and the $\tau$. Particle identification is used to identify leptonic and hadronic $\tau$ decays. Signal selection criteria vary among the analyses using different tag $B$ samples and $\tau$ decay modes. Continuum suppression cuts, $\gamma$ or(and) $\pi^0$ multiplicity requirements, *etc.* are also used in different analyses.

A GEANT4-based MC simulation is used to study the signal efficiency and to estimate backgrounds. The MC simulated events used for background estimation corresponds to roughly three times the luminosity of on-resonance data. The current analyses are optimized for $80\,\mathrm{fb}^{-1}$ on-resonance data luminosity. On larger data sets at a Super $B$ Factory , stricter selection criteria can be applied to improve the signal-to-background ratio. The main sources of background in all analyses are missing charged track(s) and undetected $K^0_L$'(s).

Signal selection efficiencies for range from 23% for $\tau \to e\overline{\nu}_e\nu_\tau$ to 7% for $\tau \to \pi^+\pi^-\pi^+\nu_\tau$ decay mode. The total signal selection efficiency is 11.3 %, which results in an overall selection efficiency of 0.028% when including the $B_{reco}$ tag efficiency. For a data set of $82\ \mathrm{fb}^{-1}$, we expect about 1.8 signal events and $38 \pm 5.0$ background events.

For semi-exclusive semileptonic $B$ tags [199], only the leptonic $\tau$ decay modes are identified. The signal selection efficiency is ~22.5% and the overall efficiency, including systematic corrections, is $(5.60 \pm 0.25(\mathrm{stat.}) \pm 0.44(\mathrm{syst.})) \times 10^{-4}$. With a data set of $82\,\mathrm{fb}^{-1}$, this leads to an expectation of 40 signal events with a background of 274 events. The analysis uses an unbinned maximum likelihood fit to extract signal and background yields. The probability density functions (PDFs) for signal and background are obtained from the distribution of the neutral energy remaining in the calorimeter, after excluding neutrals associated with the tag side ($E_{extra}$) in signal and background MC simulation, respectively. Figure 4-9 shows the $E_{extra}$ distributions in signal and background MC and in on-resonance data. The PDFs are shown in figure 4-10.

In the signal region $E_{extra} < 0.35$ GeV, the expected number of background from data sideband extrapolation is $39.9 \pm 2.8$ and the expected number of signal events is $\sim 5$, assuming a branching fraction of $\mathcal{B}(B^+ \to \tau^+\nu_\tau) = 10^{-4}$. With a luminosity of $82\,\mathrm{fb}^{-1}$, the observed number of events in the signal region is 47. The maximum likelihood fit to the data yields $10.9 \pm 7.5$ signal events and $258.1 \pm 17.4$ background events in the total fit region of $E_{extra} < 1.0$ GeV, consistent with signal and background expectations.

We next discuss expected signal and background for $B^+ \to \tau^+\nu_\tau$ decay at luminosities of $2\,\mathrm{ab}^{-1}$ in a Super $B$ Factory . The estimates are done under the assumption that the detector performance at a Super $B$ Factory will be same as the performance of the *BABAR* detector.

We take the expected numbers of background and signal events at $80\,\mathrm{fb}^{-1}$ of luminosity (see above) and extrapolate those numbers to a luminosity of $2\,\mathrm{ab}^{-1}$. For this estimate, $\tau^+ \to \pi^+\pi^0\overline{\nu}_\tau$ and $\tau^+ \to \pi^+\pi^-\pi^+\overline{\nu}_\tau$ decay modes are excluded, due to worse signal-to-background ratios in these two modes. The estimated number of signal and background events for different tag $B$ are listed in Table 4-10.

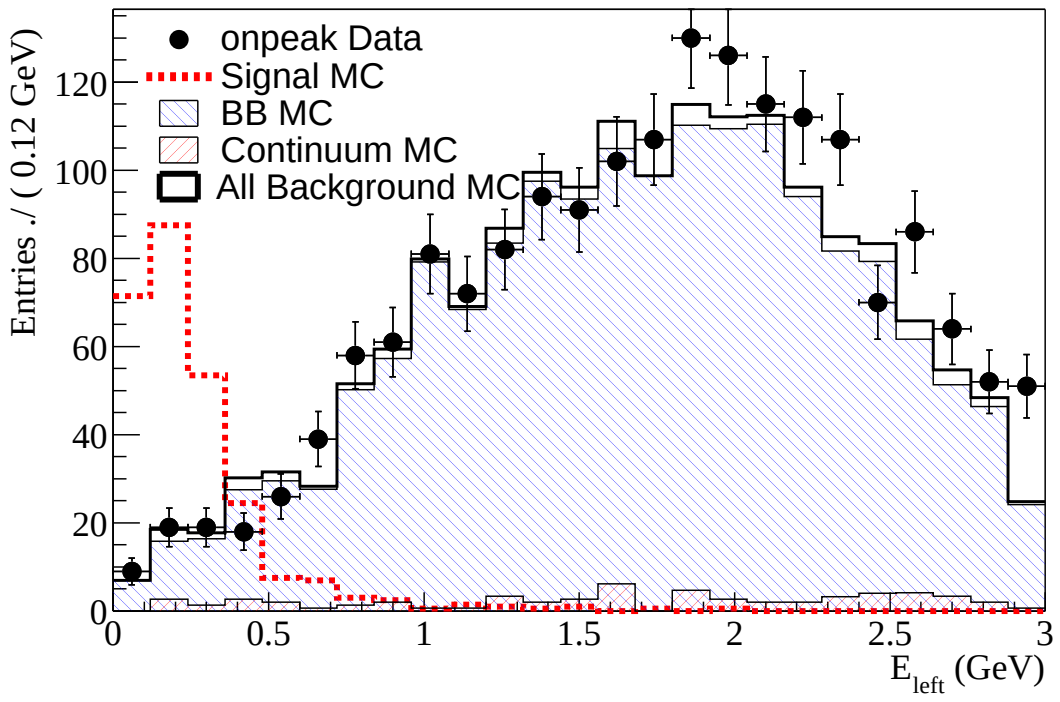

**Figure 4-9.** *$E_{extra}$, the neutral energy remaining in the calorimeter after excluding neutrals associated with the semi-leptonic side. In the above distribution all analysis selection criteria are applied. The normalization of the signal MC sample is arbitrary.*

**Table 4-10.** *Expected number of signal and background events at 2 ab$^{-1}$ of integrated luminosity, obtained by projecting estimations from current BABAR analyses.*

| Tag B decay mode | $\tau$ decay modes | Expected number of background events | Expected number of signal events for $\mathcal{B}(B^+ \to \tau^+\nu_\tau) = 10^{-4}$ |
|---|---|---|---|
| $B^- \to D^{(*)0}X^-$ | $\tau^+ \to e^+\nu_e\overline{\nu_\tau}$, $\mu^+\nu_\mu\overline{\nu_\tau}$, $\tau^+ \to \pi^+\overline{\nu_\tau}$ | 559 | 34 |
| $B^- \to D^0\ell^-\nu X^0$ ($E_{extra} < 0.35$ GeV) | $\tau^+ \to e^+\nu_e\overline{\nu_\tau}$, $\mu^+\nu_\mu\overline{\nu_\tau}$ | 974 | 122 |
| $B^- \to D^{*0}\ell^-\overline{\nu_\ell}$ | $\tau^+ \to e^+\nu_e\overline{\nu_\tau}$, $\mu^+\nu_\mu\overline{\nu_\tau}$, $\tau^+ \to \pi^+\overline{\nu_\tau}$ | 547 | 74 |

As discussed above, the analysis using semi-exclusive semileptonic tags ($B^- \to D^0\ell^-\nu X^0$) performs a maximum likelihood fit to extract signal and background yields. Toy MC experiments are used to study the signal sensitivity of the likelihood fit at a Super $B$ Factory . Toy MC samples are generated using the current PDFs (figure 4-10). By scaling the number of events in the fit region of $E_{extra} < 1.0$ GeV (see above), one expects about 6568 background events and about 151 signal events at 2 ab$^{-1}$. For each toy MC sample the number of generated background and signal events are obtained from Poisson fluctuation of those expected number of events. The same PDFs are used to fit the toy MC samples in order to obtain signal and background yields. The distributions of number of fitted signal and background events for 5000 such toy experiments are looked at. The mean and the rms of the distribution of number of fitted signal events from the toy experiments are 152 and 38 respectively, while for the distribution of the fitted number of background events, the mean and rms are 6568 and 38 respectively. Based on these studies, we expect about $4\sigma$ significance for the signal at 2 ab$^{-1}$.

A large sample of background and signal events also have been generated using the fast (Pravda) MC simulation. The Pravda MC does not presently have a realistic simulation of the detector response to neutral particles. Figure 4-11 shows a comparison of the distributions of quantities related to neutral simulation between detailed and fast MC simulation. Quantities related to neutral energy, such as number of $\pi^0$ mesons associated with the signal side, $E_{extra}$, *etc.*, are some of the major signal-defining quantities for identifying $B^+ \to \tau^+\nu_\tau$ signal. Since these distributions in fast MC simulation are quite different from those in the detailed MC simulation (which is in good agreement with data from the *BABAR* experiment) any estimation using fast MC simulation will not be realistic and reliable. Thus the fast MC sample has not been used.

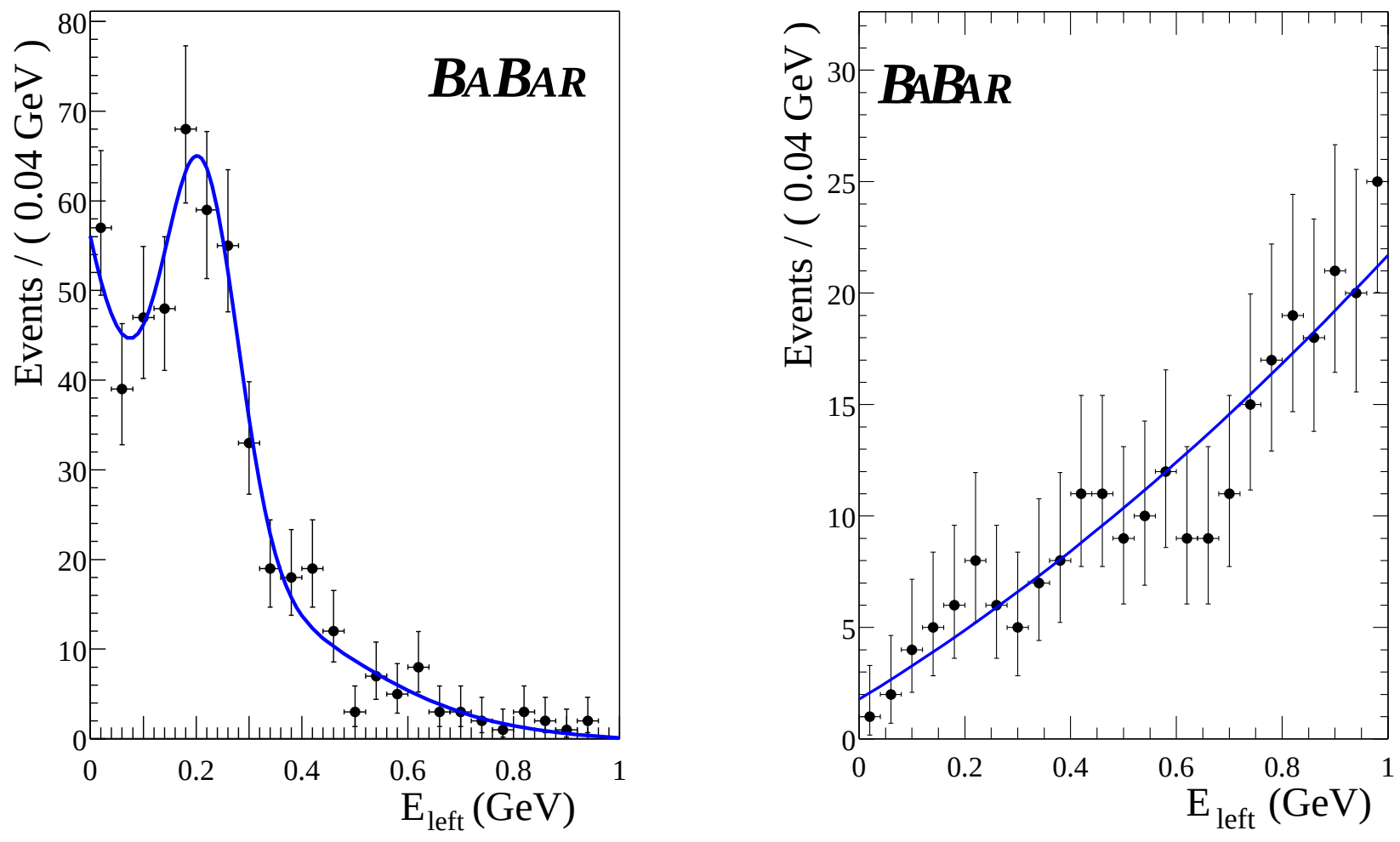

(a) (b)

**Figure 4-10.** *The signal PDF (left) fitted to $E_{extra}$ from the signal MC sample and the background PDF (right) fitted to $E_{extra}$ from the background MC sample. All selection criteria are applied to the events in signal and background MC samples. The normalization of the signal MC sample is arbitrary and the normalization of the background MC sample is fixed to the integrated luminosity of* $80\,\mathrm{fb}^{-1}$.

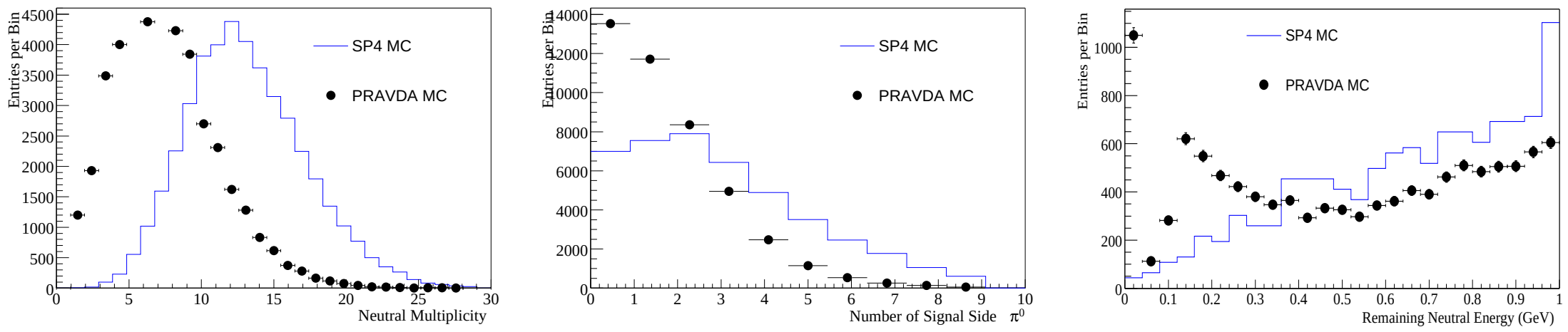

**Figure 4-11.** *(a) Distributions of number of reconstructed photons in the event, compared between detailed MC simulation (solid line) and fast MC simulation (dots) (b) Distributions of number of reconstructed $\pi^0$ associated with signal side, compared between detailed MC simulation (solid line) and fast MC simulation (dots). (c) Distributions of remaining neutral energy $E_{extra}$, compared between detailed MC simulation (solid line) and fast MC simulation (dots). The distributions related to simulation of neutrals are compared for detailed MC and fast (PRAVDA) MC simulations. The distributions for fast MC simulation are quite different than those for detailed MC simulation.*

From our studies, the potential of $B^+ \to \tau^+\nu_\tau$ decay in a Super $B$ Factory looks promising. The major issues concerning these analysis are the following.

- Search or observation of $B^+ \to \tau^+\nu_\tau$ signal are highly sensitive to quantities related to neutral particles. A detailed simulation of the calorimeter response, beam background at high luminosity environment etc. will be useful to get a more realistic estimation of the signal sensitivity at a Super $B$ Factory .

- Since the major source of background are from missing tracks and undetected $K_L^0$ mesons, detector coverage and neutral identification will affect the signal sensitivity.

With an integrated luminosity of $2\,\mathrm{ab}^{-1}$, we expect to observe $B^+ \to \tau^+\nu_\tau$ with $4\sigma$ significance.

### $B^+ \to \mu^+\nu_\mu$

The existing upper limits on the $B^+ \to \mu^+\nu_\mu$ branching fraction from CLEO [200], Belle [201], and *BABAR* [202] were all obtained using similar analysis techniques on data samples collected at the $\Upsilon(4S)$ resonance. In this section, we describe the existing *BABAR* measurement and estimate the sensitivity of a similar technique with a 5 ab$^{-1}$ sample collected at a Super $B$ Factory. The high luminosity study was carried out using the Pravda fast Monte Carlo described in section 4.2.3. We also briefly discuss the prospects for measuring $B^+ \to \mu^+\nu_\mu$, using a sample of events in which the other $B$ in the event has been fully reconstructed, similar to the $B^+ \to \tau^+\nu_\tau$ analysis.

As noted above, $B^+ \to \mu^+\nu_\mu$ is a two-body decay so the muon is monoenergetic in the $B$ rest frame. Since $B$ mesons from $\Upsilon(4S) \to B\overline{B}$ are produced with relatively low momenta ($\approx$ 320 MeV/$c$), the $\Upsilon(4S)$ rest frame is a good approximation to the $B$ rest frame. Therefore, the existing analysis begins by selecting well-identified muon candidates with momentum near $m_B/2$ in the $\Upsilon(4S)$ rest frame. The neutrino goes undetected so we can assume that all remaining particles are associated with the decay of the other $B$ in the event, which we denote the "companion" $B$. Signal decays can then be selected using the kinematic variables $\Delta E$ and energy-substituted mass $m_{ES}$ (see section 4.2.1).

After removing the muon candidate from the event, the companion $B$ can be reconstructed from the remaining visible energy. To aid the event energy resolution, only loose selection criteria are applied to the remaining charged tracks and neutral calorimeter clusters. In the *BABAR* analysis, the companion $B$ includes all charged tracks that are consistent with being produced at the interaction point and all neutral calorimeter clusters with energy greater than 30 MeV. Particle identification is applied to the charged tracks in order to select the appropriate mass hypothesis and thus improve the $\Delta E$ resolution. Events with additional identified leptons from the companion $B$ are discarded since they typically arise from semileptonic $B$ or charm decays and indicate the presence of additional neutrinos. Figure 4-12 shows distributions of $\Delta E$ and $m_{ES}$ for the *BABAR* on-resonance data, background MC and signal MC samples after muon candidate selection. For a properly reconstructed signal decay, we expect $m_{ES}$ to peak near the $B$ mass and the energy of the companion $B$ to be consistent with the beam energy so that $\Delta E$ peaks near 0. In practice, energy losses from detector acceptance, unreconstructed neutral hadrons and additional neutrinos result in the signal $\Delta E$ distribution being shifted toward negative $\Delta E$, while the $m_{ES}$ distribution develops a substantial tail below the $B$ mass.

Once the companion $B$ is reconstructed, we can calculate the muon momentum in the rest frame of the signal $B$. We assume the signal $B$ travels in the direction opposite that of the companion $B$ momentum in the $\Upsilon(4S)$ rest frame with a momentum determined by the two-body decay $\Upsilon(4S) \to B\overline{B}$. Figure 4-13 shows the muon candidate momentum distribution in the $B$ rest frame, $p_\mu$, for all muon candidates in the signal MC. The dashed curve is the momentum distribution of the same events in the $\Upsilon(4S)$ rest frame.

Backgrounds may arise from any process that produces charged tracks in the momentum range of the signal muon. The two most significant backgrounds are found to be $B$ semi-leptonic decays involving $b \to u\mu\overline{\nu}$ transitions where the endpoint of the muon spectrum approaches that of the signal, and non-resonant $q\overline{q}$ ("continuum") events where a charged pion is mistakenly identified as a muon. In order for continuum events to populate the signal region of $\Delta E$ and $m_{ES}$, there must be significant energy loss, mainly from particles outside the detector acceptance and unreconstructed neutral hadrons. We reduce these backgrounds by tightening the selection on the muon momentum. The momentum spectrum of the background decreases with increasing momentum, so we apply an asymmetric cut about the signal peak, $2.58 < p_\mu < 2.78$ GeV/$c$.

The continuum background is further suppressed using event-shape variables. These events tend to produce a jet-like event topology, as opposed to $B\overline{B}$ events, which tend to be spherical. We define a variable, $\theta_T^*$, which is the angle between the muon candidate momentum and the thrust axis of the companion $B$ in the $\Upsilon(4S)$ rest frame. For continuum background, $|\cos\theta_T^*|$ peaks sharply near one while the distribution is nearly flat for signal decays. The polar angle of the missing momentum vector in the laboratory frame, $\theta_\nu$, can also discriminate against continuum backgrounds. In continuum decays, the missing momentum is often due to undetected particles that were outside the detector acceptance. Therefore, we require that the missing momentum is directed into the detector's fiducial volume. Figure 4-14 shows the *BABAR* on-peak data and MC distributions of $|\cos\theta_T^*|$ and $|\cos\theta_\nu|$. For comparison, the signal MC is overlaid with an arbitrary normalization.

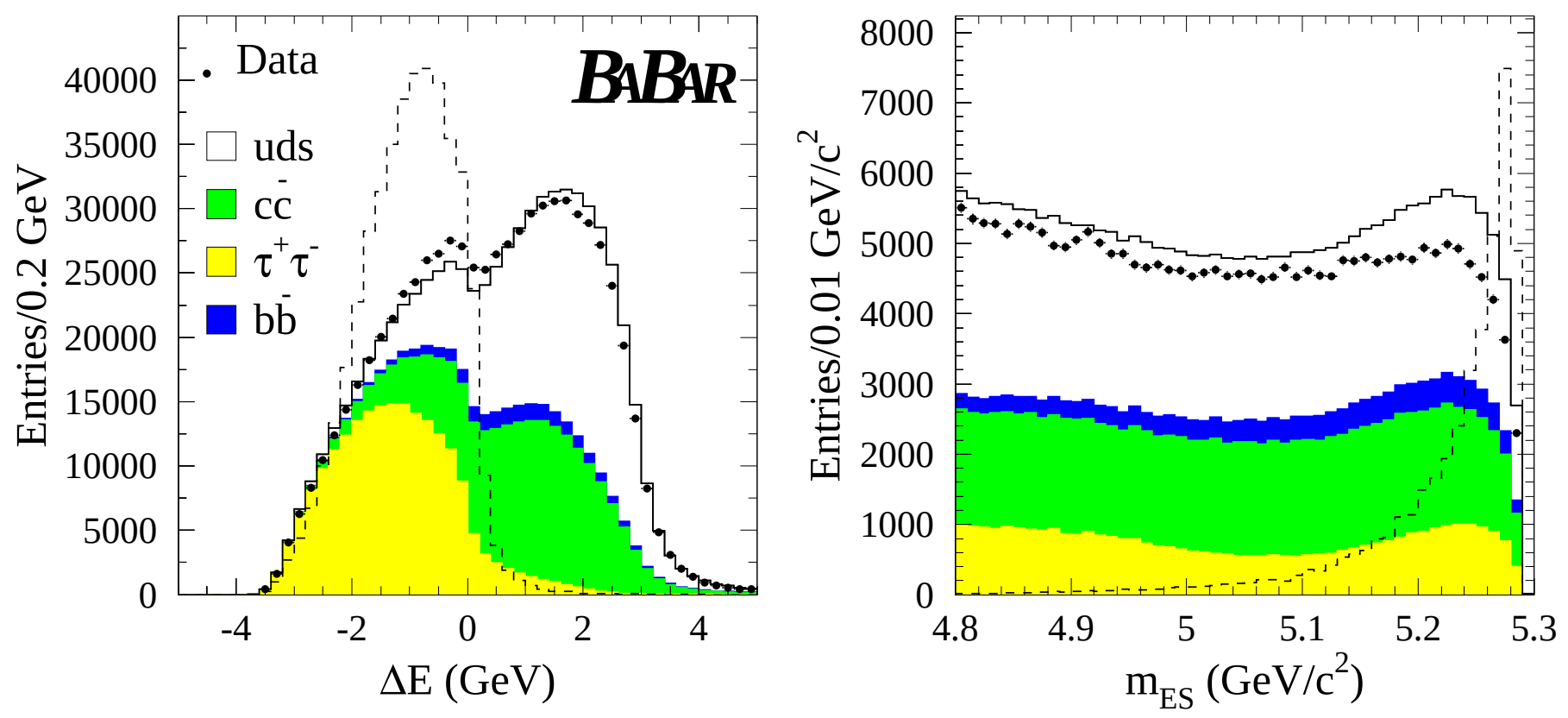

**Figure 4-12.** *The distributions of $\Delta E$ and $m_{ES}$ for on-peak data and MC samples after muon candidate selection. The signal distributions are overlaid (dashed histograms) with an arbitrary normalization.*

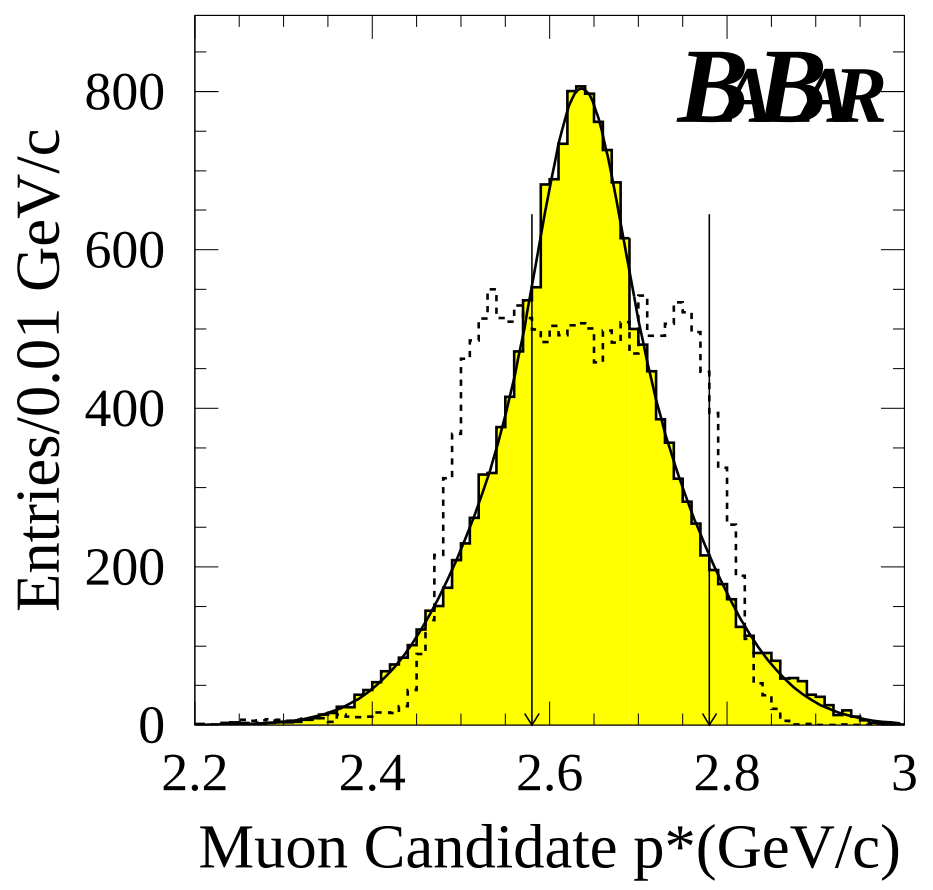

**Figure 4-13.** *The muon candidate momentum distribution in the reconstructed $B$ rest frame for all muon candidates in the signal MC. The dashed curve is the momentum distribution of the same events in the $\Upsilon(4S)$ rest frame. The arrows indicate the selected signal region.*

We select $B^+ \to \mu^+\nu_\mu$ signal events with simultaneous requirements on $\Delta E$ and $m_{ES}$, thus forming a "signal box" defined by $-0.75 < \Delta E < 0.5$ GeV and $m_{ES} > 5.27$ GeV$/c^2$. After applying all selection criteria, the $B^+ \to \mu^+\nu_\mu$ efficiency is determined from the simulation, after correcting for discrepancies between the data and MC, to be about 2.1%. The amount of background expected in the signal box is estimated to be $5.0^{+1.8}_{-1.4}$ events, by extrapolating from the signal box sidebands. From the MC simulation, we expect that this background is composed of approximately 57% light-quark ($u\overline{u}$, $d\overline{d}$, $s\overline{s}$), 23% $c\overline{c}$, and 20% $B\overline{B}$ events. In the on-resonance data we find 11 events in the signal box which results in an upper limit of $\mathcal{B}(B^+ \to \mu^+\nu_\mu) < 6.6 \times 10^{-6}$ at the 90% confidence level.

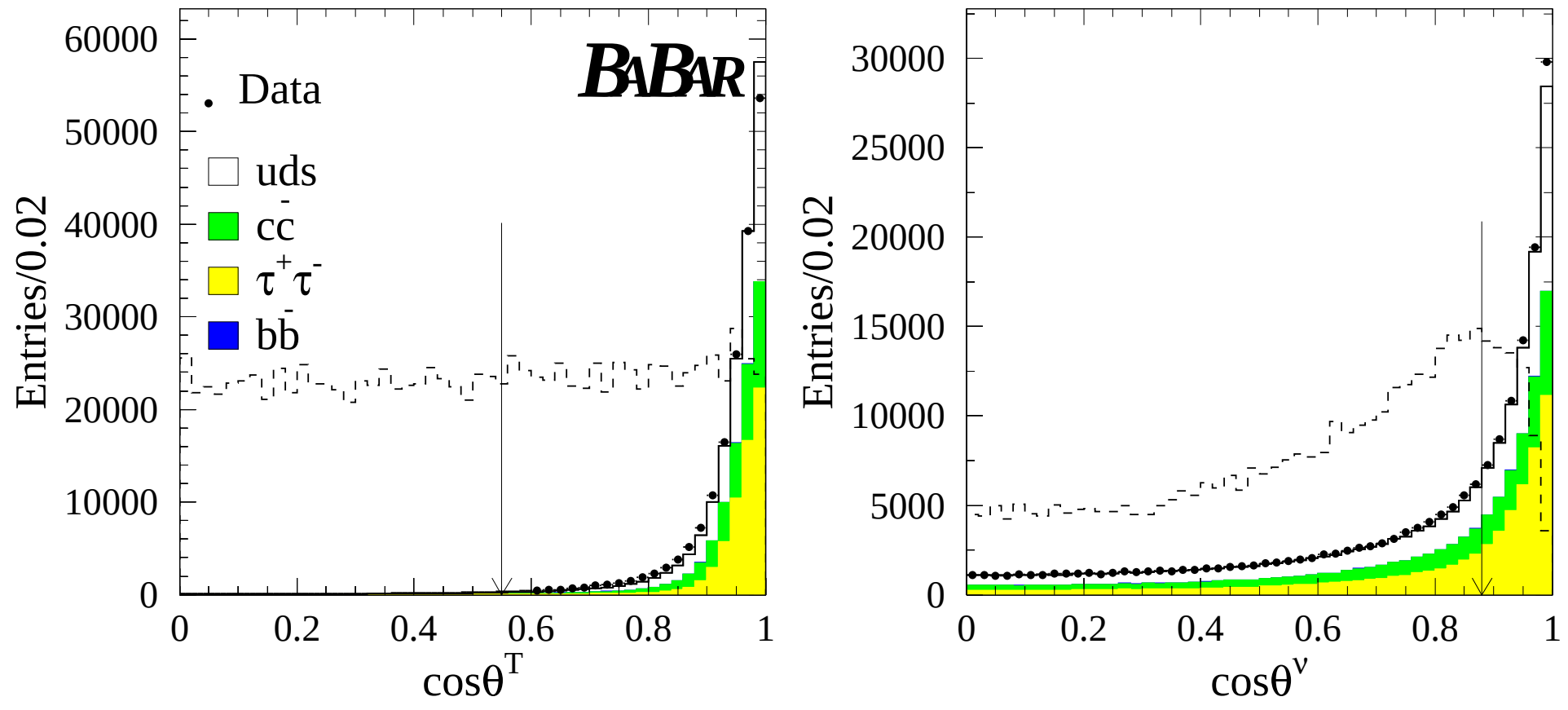

**Figure 4-14.** *The distributions of* $|\cos\theta_T^*|$ *and* $|\cos\theta_\nu|$ *for on-peak data and MC. The events in these plots have passed the requirement* $2.58 < p* < 2.78$ GeV/$c$. *The signal distributions are overlaid (dashed histograms) with an arbitrary normalization.*

To estimate the sensitivity to $B^+ \to \mu^+\nu_\mu$ at a Super $B$ Factory, this analysis has been repeated using a sample of approximately 5 ab$^{-1}$ simluated with the Pravda fast MC simulation. Here we have assumed 90% muon efficienciecy and 1% pion misidentification at the Super $B$ Factory .

The reliability of the Pravda simulation has been evaluated by comparing the event yields expected for the current analysis ($80\,\mathrm{fb}^{-1}$) with the full simulation. For these comparisons, we have applied the current BABAR muon identification performance to the Pravda simulation. In general, the results are in reasonable agreement. In the signal box, Pravda predicts 7.6 background events where we find 5.3 in the full simulation. In the "grand sideband" defined by $-3.0 < \Delta E < 1.5$ GeV and $m_{ES} > 5.23$ GeV/$c^2$, we see 257 Pravda background events as compared to 200 in the full simulation. Although, the background totals are in adequate agreement, we do observe some notable discrepancies in particular modes. For example, the $B^+ \to \mu^+\nu_\mu$ and $B^0 \to \pi^+\mu^-\overline{\nu}_\mu$ efficiencies are overestimated in Pravda by roughly a factor of 2. Furthermore, the Pravda simulation appears to neglect interactions of neutral hadrons in the calorimeter. Therefore, we see an enhanced background rate from processes involving neutral hadrons. The increase in the signal efficiency is likely due to the lack of detector related backgrounds such as fake charged tracks, calorimeter noise and beam backgrounds, which improves the event energy resolution. We actually expect these backgrounds to increase with luminosity but we currently have no estimate of this effect.

With higher luminosity, the optimum values of analysis cuts may change. Therefore, we have re-optimized the cut on $|\cos\theta_T^*|$ (the most effective variable for continuum rejection) using signal boxes of various sizes. The optimum combination was found to be $|\cos\theta_T^*| < 0.6$, $-0.5 < \Delta E < 0$ GeV and $m_{ES} > 5.27$ GeV/$c^2$. Therefore, the $\Delta E$ range of the signal box has decreased but all other cuts retain essentially the same value as in the current analysis. We also found a small benefit by requiring the total event charge to be 0. With this combination of cuts we find a signal efficiency of approximately 4% in the Pravda simulation. For a $5\,\mathrm{ab}^{-1}$ data sample, this simulation yields approximately 90 signal and 210 background events in the signal box. The background composition is significantly different than that found in the full simulation. Because we have assumed an improved muon identification probability, as well as a factor of two improvement in the pion misidentification rate, the background is now roughly half $B\overline{B}$ as opposed to being dominated by continuum. We also note that about 85% of the continuum background involves a neutral hadron.

Figure 4-15 shows the distributions of $m_{ES}$ and $p_\mu$ for signal and background MC. In each plot, all other cuts have been applied. Note that a large contribution from $b \to c\ell\nu$ decays would normally be evident in the $p_\mu$ distribution. However, those decays do not produce muons in the momentum range of the signal, so they have been neglected here. Also, the sharp peak in $m_{ES}$ due to $B\overline{B}$ events with fake muons is due mostly to decays such as $B^+ \to K_L^0\pi^+$. This decay mode is enhanced due to the lack of simulation of neutral hadrons in the Pravda MC.

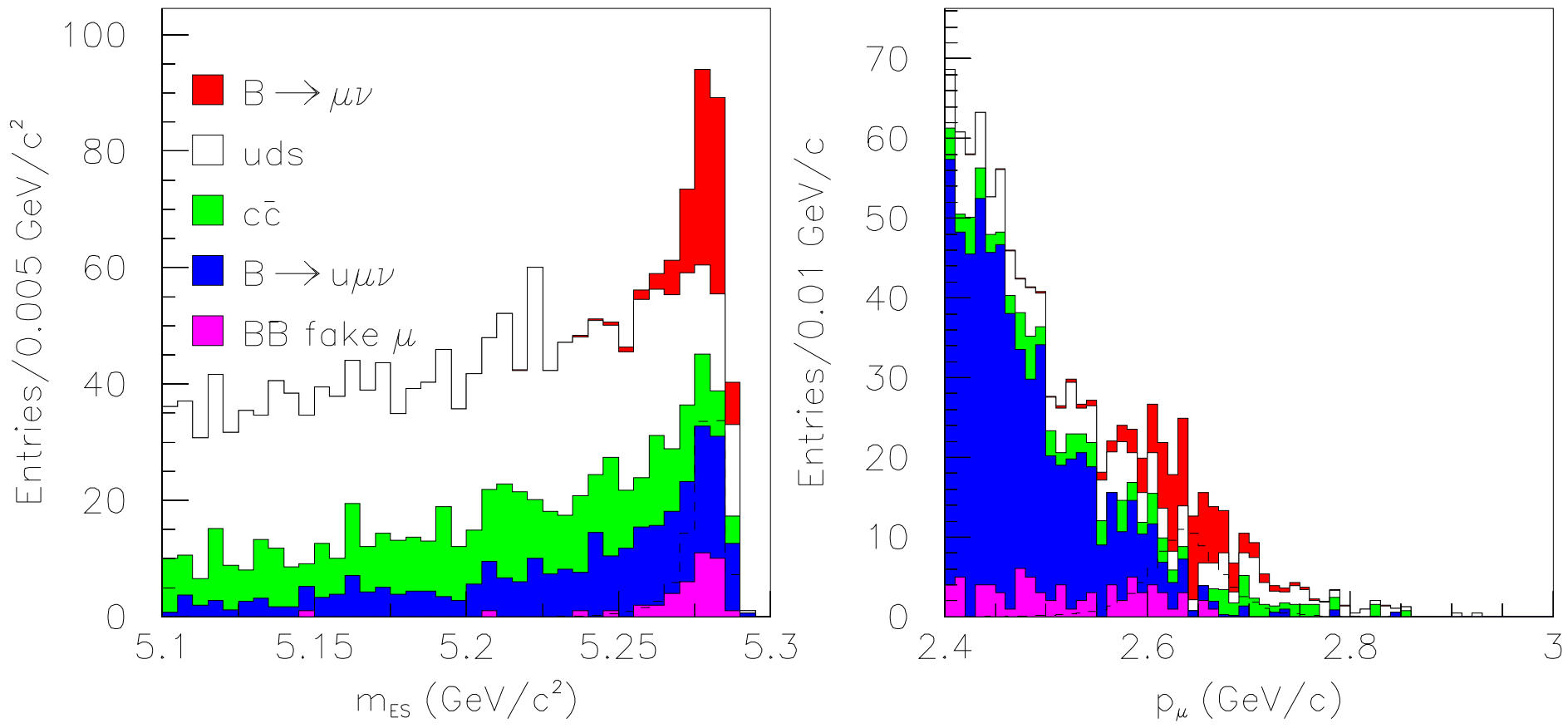

**Figure 4-15.** *The distributions of $m_{ES}$ and $p_\mu$ for signal and background Pravda MC simulation, normalized to $5\,\mathrm{ab}^{-1}$. In each plot, all other cuts have been applied.*

With a larger data sample we would likely extract the signal yield using a maximum likelihood analysis rather than the "cut and count" method employed so far. As a simple example, we have performed a binned likelihood fit to the $p_\mu$ distribution from the Pravda simulation. The background PDF was assumed to be a single Gaussian function while the signal distribution was fit to a double Gaussian. The parameters of the Gaussians were fixed and a fit was performed for the signal and background normalizations. For a sample of $5\,\mathrm{ab}^{-1}$, the signal yield is extracted with approximately 15% statistical uncertainty, assuming the Standard Model branching fraction. If the branching fraction (or, equivalently, the signal efficiency) were a factor of two larger(smaller), the statistical uncertainty is expected to be about 10(30)%. These results could likely be improved with a simultaneous fit to $p_\mu$, $\Delta E$, and $m_{ES}$. Based on these results, $|V_{ub}|$ could be extracted with a statistical uncertainty of 5-15% assuming $f_B$ has been calculated to the necessary precision.

We have also considered searching for $B^+ \to \mu^+\nu_\mu$ using a fully-reconstructed tag $B$ as described for the $B^+ \to \tau^+\nu_\tau$ analysis. The reconstruction efficiency for this type of analysis is too small to be useful with existing data samples but may become feasible for the larger data samples provided by a Super $B$ Factory. The primary benefit of this "recoil" method is that the backgrounds can be significantly reduced by requiring the existence of another fully reconstructed $B$. The $B^+ \to \tau^+\nu_\tau$ analyses have demonstrated $B$ tagging efficiencies of 0.25% for the hadronic modes and 0.31% for the semi-leptonic modes. Furthermore, due to the simplicity of the $B^+ \to \mu^+\nu_\mu$ signal side (1 charged track), we might expect some improvement in the tagging efficiency and reduction of the combinatorial background.

Given a good tag $B$, the signal-side selection for this analysis should be quite simple. We have considered, for example, requiring only one remaining charged track that passes muon identification and satisfies $2.6 < p_\mu < 2.7\ \mathrm{GeV}/c$. Note that the $p_\mu$ selection has been tightened, because having a fully reconstructed tag $B$ provides much better knowledge of the $B$ rest frame. Therefore, the $p_\mu$ resolution is significantly improved. We expect about 91% of the signal muons to be reconstructed due detector acceptance, about 90% to pass muon identification, and about 95% to pass the $p_\mu$ requirement yielding a total signal-side efficiency of about 78%.

Given the above tag-side and signal-side efficiencies, we expect a total signal efficiency of about 0.5% for a recoil anlysis. Therefore, in a sample of $5\,\mathrm{ab}^{-1}$, we expect about 10 signal events to pass all cuts, assuming a signal branching fraction of $4 \times 10^{-7}$. The expected background has been investigated by applying the above signal-side selection to the existing semi-leptonic $B^+ \to \tau^+\nu_\tau$ analysis. In roughly 200 fb$^{-1}$ of generic $B\overline{B}$ MC and 50 fb$^{-1}$ of continuum MC, both in the full simulation, we see no background events passing all cuts. If we optimistically assume that the backgrounds are negligible, the signal branching fraction could be measured with a statistical uncertainty of about 30%.

In conclusion, fast MC studies indicate that the branching fraction for $B^+ \to \mu^+\nu_\mu$ could be measured with a statistical uncertainty of 10-30% with a 5 ab$^{-1}$ data sample collected at a Super $B$ Factory . The measurement could be performed using either an inclusive reconstruction of the companion $B$, as in the current analysis, or a fully reconstructed companion $B$. At present, the inclusive analysis is better understood and appears to give a smaller statistical uncertainty. Assuming that the theoretical uncertainty in $f_B$ can be significantly improved in lattice QCD calculations, $|V_{ub}|$ could be determined to 5-15% in this mode. As the theoretical uncertainties here are very different from those in semileptonic $B$ decays, this provides a very powerful alternate route to $|V_{ub}|$. The critical considerations for the detector design are maximum hermiticity, neutral hadron identification, and, of course, muon identification. Finally, we do not expect these measurements to be possible at hadronic machines such as LHC$b$ and $B$TeV, due to the necessity of neutrino reconstruction. Therefore, a future Super $B$ Factory has the unique opportunity to observe leptonic $B$ decays, and thus constrain the Standard Model parameters $|V_{ub}|$ and $f_B$.

### 4.6.3 $B \to \gamma\ell\overline{\nu}$, $\gamma\ell^+\ell^-$, $\gamma\gamma$

>– E. Lunghi –<

The decays $B \to \gamma e\nu$, $B \to \gamma\gamma$ and $B \to \gamma ee$ are extremely rare modes that are nevertheless within the reach of a Super $B$ Factory. Rough estimates of their branching ratios give: $\mathcal{B}(B \to \gamma e\nu) \sim 10^{-6}$, $\mathcal{B}(B \to \gamma\gamma) \sim 3 \times 10^{-8}$, $\mathcal{B}(B \to \gamma ee) \sim 10^{-11} \div 10^{-10}$. The absence of hadrons in the final state facilitate the analysis of QCD effects; indeed, it can be shown that all these modes factorize up to power corrections.

#### $B \to \gamma e\nu$

The effective Hamiltonian responsible for this decay:

$$H_{\rm eff} = \frac{4G_F}{\sqrt{2}}\, V_{ub}\, (\overline{u}_L\gamma^\mu b_L)\, (\overline{e}_L\gamma_\mu\nu_L) \tag{4.109}$$

arises at tree level in the Standard Model. The amplitude for the $B \to \gamma e\nu$ transition can be exactly parameterized in terms of the following photonic form factors:

$$\frac{1}{e}\langle\gamma(q,\varepsilon)|\, \overline{u}\,\gamma_\mu\, b\, |\overline{B}(v)\rangle \;=\; i\epsilon_{\mu\alpha\beta\delta}\, \varepsilon^*_\alpha\, v_\beta\, q_\delta\, f_V(E_\gamma) \tag{4.110}$$

$$\frac{1}{e}\langle\gamma(q,\varepsilon)|\, \overline{u}\,\gamma_\mu\gamma_5\, b\, |\overline{B}(v)\rangle \;=\; \left[q_\mu(v\cdot\varepsilon^*) - \varepsilon^*_\mu(v\cdot q)\right] f_A(E_\gamma) + v_\mu\frac{v\cdot\varepsilon^*}{v\cdot q} f_B m_B\,, \tag{4.111}$$

where $\varepsilon$ is the photon polarization. The last term in (4.111) is a contact term that compensates the photon emission from the electron line. In Refs. [203, 204, 205] it was shown that, at leading order in $\Lambda_{\rm QCD}/E_\gamma$ and at all orders in $\alpha_s$, the form factors $f_{V,A}(E_\gamma)$ are equal at all orders in perturbation theory and factorize into the product of hard Wilson coefficients and a universal convolution of a jet function with the $B$ meson light cone distribution amplitude (LCDA):

$$f_V(E_\gamma) = f_A(E_\gamma) = C(E_\gamma)\int {\rm d}\xi\; J(E_\gamma,\xi)\; \phi_B(\xi) = C(E_\gamma)\; I(E_\gamma)\,, \tag{4.112}$$

where $C(E_\gamma)$ is the hard coefficient, $J$ is the jet-function containing terms of the type $(\log\xi)^n/\xi$ with $n \geq 0$ and $\phi_B(\xi)$ is the $B$ meson LCDA (see Ref. [204] for details).

Since we do not expect any sizable New Physics correction to a Standard Model tree level amplitude, this decay will provide us with valuable pieces of information on the first negative moment of the poorly known $B$ meson

LCDA[206]: $\lambda_b^{-1} = \int \phi_B(\xi)/\xi$. This quantity is important because it enters factorization formulas for several rare $B$ decays ($B \to (\rho, K^*)\gamma$, $B \to \rho e\nu$, $B \to K\pi$, $B \to \pi\pi$, ...). Unfortunately, the convolution $I(E_\gamma)$ evaluated at $\mathcal{O}(\alpha_s)$ depends on the first two logarithmic moments of $\phi_B$ as well ($\int \phi_B(\xi)\log\xi/\xi$ and $\int \phi_B(\xi)\log^2\xi/\xi$). This could limit the accuracy of the extraction of $\lambda_b$ from this measurement (See Ref. [207] for a detailed description of this problem).

Note that the above result are valid for large photon energy ($\Lambda_{\rm QCD}/E_\gamma << 1$); a cut in the photon spectrum ($E_{\gamma c} < E_\gamma < m_B/2$) is necessary to restrict to the theoretically clean region. Using the parametrization of the $B$ meson LCDA given in Ref. [206], we obtain the follow approximate expression, valid in the region $1{\rm GeV} < E_{\gamma c} < m_B/2$, for the integrated branching ratio:

$$\int_{E_{\gamma c}}^{m_B/2} {\rm d}E_\gamma \frac{{\rm d}\mathcal{B}(B\to\gamma e\nu)}{{\rm d}E_\gamma} = 10^{-2}\left|\frac{V_{ub}}{3.9\times10^{-3}}\right|^2 \left(\frac{\lambda_B^{-1}}{2.15\ {\rm GeV}}\right)^2 \left[5.97 - 4.08\,E_{\gamma c} + 0.65\,E_{\gamma c}^2\right]\,. \tag{4.113}$$

Using the QCD sum rules estimate $\lambda_b^{-1} = (2.15\pm0.5){\rm GeV}^{-1}$ [206] and a photon cut-off of $1\,{\rm GeV}$ we obtain $\mathcal{B}(B\to\gamma e\nu) \sim 1.8\times10^{-6}$ with O(100%) uncertainties coming mainly from $\lambda_b^{-1}$ and $V_{ub}$.

Note that a first principles computation of the $f_{V,A}$ form factors on the lattice would allow for a direct test of the relation $f_V(E_\gamma) = f_A(E_\gamma)$ and shed some light on the size of the incalculable power corrections.

### $B \to \gamma\gamma$

The decay $B\to\gamma\gamma$ arises, in the Standard Model, at loop level and is mediated by the same effective Hamiltonian that governs $b\to d\gamma$ transitions:

$$H_{\rm eff} = \frac{4G_F}{\sqrt{2}}\left(V_{tb}V_{td}^*\sum_{i=1}^{8} C_i O_i + V_{ub}V_{ud}^*\sum_{i=1}^{2} C_i O_i^u\right)\,, \tag{4.114}$$

where the most relevant operators are

$$O_2 = (\bar d_L\gamma^\mu c_L)(\bar c_L\gamma_\mu b_L)\,, \tag{4.115}$$
$$O_7 = \frac{e}{16\pi^2} m_b(\bar d_L\sigma^{\mu\nu}b_R)F_{\mu\nu}\,, \tag{4.116}$$
$$O_8 = \frac{g_s}{16\pi^2} m_b(\bar d_L T^a\sigma^{\mu\nu}b_R)G^a_{\mu\nu}\,. \tag{4.117}$$

The matrix element of $O_7$ can be parameterized in terms of the following tensor form factors:

$$\begin{aligned}\frac{1}{e}\langle\gamma(q,\varepsilon)|\bar u\sigma_{\mu\nu}b|\bar B(p)\rangle &= i\,\varepsilon_{\mu\nu\alpha\beta}\,\varepsilon^{*\alpha}\,(p+q)^\beta\,g_+(E_\gamma) + i\,\varepsilon_{\mu\nu\alpha\beta}\,\varepsilon^{*\alpha}\,(p-q)^\beta\,g_-(E_\gamma)\\ &\quad -2\,(\varepsilon^*\cdot p)\,h(E_\gamma) i\varepsilon_{\mu\nu\alpha\beta}p^\alpha q^\beta\,.\end{aligned} \tag{4.118}$$

From the results of Ref. [204] it follows that the three tensor form factors $g_\pm$ and $h$ factorize at all orders in $\alpha_s$ and are proportional to the convolution integral $I(E_\gamma)$. Therefore, the following ratios are clean of hadronic uncertainties up to power corrections:

$$\frac{g_+(E_\gamma)}{f_V(E_\gamma)} = \frac{1}{2}\frac{Q_d}{Q_u}\left(1 - \frac{\alpha_s C_F}{4\pi}\frac{E_\gamma}{E_\gamma - m_b/2}\log\frac{2E_\gamma}{m_b}\right) + O(\alpha_s^2) \tag{4.119}$$
$$g_-(E_\gamma) = -g_+(E_\gamma) + O(\alpha_s^2) \tag{4.120}$$
$$h(E_\gamma) = 0 + O(\alpha_s^2)\,. \tag{4.121}$$

The situation is more complicated for the matrix elements of other operators (the most relevant are $O_2$ and $O_8$), and the issue has not yet been addressed at all orders. In Ref. [207] the authors show explicitly that all diagrams that would lead to non-factorizable effects are indeed suppressed by at least one power of $\Lambda_{\rm QCD}/m_B$.

From a phenomenological point of view, it is more useful to normalize the $B \to \gamma\gamma$ branching ratio to $\mathcal{B}(B \to \gamma e\nu)$. This ratio allows for a determination of the Wilson coefficient $C_7^d$ with precision similar to the inclusive channel $B \to X_d\gamma$. In fact, the latter mode is plagued by non-perturbative contributions to the matrix elements of the four-quark operators induced by up quark loops [208].

Finally, note that some power suppressed contributions to the amplitude $B \to \gamma\gamma$ are nevertheless computable. They are responsible for the presence of a direct $CP$ asymmetry of order $-10\%$ (see Ref. [209] for further details).

### $B \to \gamma ee$

This mode is described by the effective Hamiltonian Eq. (4.114) with the inclusion of the semileptonic operators

$$O_9 = (\bar{d}_L\gamma^\mu b_L)\,(\bar{e}\gamma_\mu e)\,, \tag{4.122}$$
$$O_{10} = (\bar{d}_L\gamma^\mu b_L)\,(\bar{e}\gamma_\mu\gamma_5 e)\,. \tag{4.123}$$

The analysis of this decay follows closely that of $B \to \gamma\gamma$. In this case as well, a complete proof of factorization at all orders has not been completed yet. The shape of the dilepton invariant mass spectrum is very similar to the $B \to X_d ee$ case; in particular, the presence of non-perturbative $q\bar{q}$ rescattering results in the presence of resonant peaks corresponding to the tower of $c\bar{c}$ resonances ($J/\psi$, $\psi'$, ...). In analogy with $b \to (d,s)ee$ modes, it is, therefore, necessary to place cuts on the dilepton invariant mass distribution.

Moreover, factorization theorems are only valid in regions in which the photon energy is large or, equivalently, in which the dilepton invariant mass is small. This region is also free from effects induced by bremsstrahlung from the external leptons. The analysis of the high invariant mass region has to rely on other methods (see for instance Ref. [210]).

An important observable is the forward–backward asymmetry of the dilepton system. The measurement of a zero in the spectrum provides a determination of the sign of the Wilson coefficient $C_7^d$. In this case as well, considering the ratio to the leading $B \to \gamma e\nu$ mode allows to reach a precision comparable to the inclusive $B \to X_d ee$ channels.

### 4.6.4 Experiment

>– U. Langenegger –<

The leptonic decays modes $B \to \gamma\gamma$, and $B \to \gamma\ell^+\ell^-$, $B \to \ell\overline{\nu}\gamma$ are extremely rare and have not yet been observed experimentally; they come within reach at a Super $B$ Factory. The first two modes will not benefit from analyses on the recoil of a $B_{reco}$ candidate due to their very small expected branching fraction. Here progress will only come with a difficult improvement of the background rejection in the traditional reconstruction of the signal decay.

The best current experimental upper limit on $B \to \gamma\gamma$ has been determined at $\mathcal{B}(B \to \gamma\gamma) < 1.7\times10^{-6}$ by the *BABAR* collaboration [211]. Here, the dominant background processes are continuum $q\overline{q}$ production (where $q = u, d, s, c$). At some point, even the rare decay $B^0 \to \pi^0\pi^0$ will constitute a background for this decay mode.

No limits exist yet for the decay $\to \gamma\ell^+\ell^-$. Here, the backgrounds consist both of continuum processes and radiative $B$ meson decays (combined with a misidentified pion).

The study of $B \to \ell\overline{\nu}\gamma$ has a substantially larger expected branching fraction, but is complicated by the unmeasured neutrino. At a Super $B$ Factory, a large background in the electron channel is due to two-photon processes. This background is much reduced for the muon channel. Eventually, events tagged by the fully reconstructed hadronic decay of a $B$ meson will provide the best environment to measure this decay.

## 4.7 Summary

On the experimental side, the Super $B$ Factory will definitely establish the method of "recoil physics" as the primary approach for the precision study of semileptonic $B$ decays. Here, $B\overline{B}$ events are selected by the full reconstruction of a hadronic $B$ decays (serving as event tags), thus allowing the study of a semileptonic decay of the second $B$ meson in the event. While the overall efficiency for this approach is small, this is no longer a limiting factor at a Super $B$ Factory.

The *inclusive* determination of $|V_{ub}|$ will reach statistical and experimental systematic errors below the 3% level even before the arrival of a Super $B$ Factory and will be limited by the theoretical errors. With unquenched lattice QCD calculations for the form factors, the measurement of *exclusive* charmless semileptonic $B$ decays will provide a premier opportunity for the model-independent determination of $|V_{ub}|$. The statistical error of the recoil methods will approach the detector systematic error of about 2% only at the Super $B$ Factory, especially for those decay channels most amenable to lattice QCD calculations. The total error on $|V_{ub}|$ will be limited by theoretical uncertainties only after several years at a Super $B$ Factory.

The measurement of *leptonic* $B$ decays will provide complementary determinations of $|V_{ub}|$ at the Super $B$ Factory. The observation of $B \to \tau\overline{\nu}$ is expected to be achievable already at luminosities of around 2 ab$^{-1}$. It is difficult to predict the precision of the determination of $|V_{ub}|$ with this decay mode, as detailed background simulation studies are necessary for a reliable assessment of the experimental systematic errors. The decay $B \to \mu\overline{\nu}$ offers a much cleaner experimental environment, though at a much reduced rate due to helicity suppression. It allows for a statistics-limited determination of $|V_{ub}|$ at the level of about 10% at an integrated luminosity of about $5\,\text{ab}^{-1}$, if unquenched lattice QCD calculations provide $f_B$ with the necessary precision. Here, analyses based on the recoil method will surpass traditional analyses only after several years at a Super $B$ Factory.

# 5

# New Physics

**Conveners:** J. Hewett, Y. Okada

**Authors:** K. Agashe, A.J. Buras, G. Burdman, M. Ciuchini, A. Dedes, D.A. Demir, G. Eigen, T. Goto, J. Hewett, G. Hiller, D.G. Hitlin, J. Hisano, O. Igankina, G. Isidori, A.L. Kagan, P. Ko, C. Kolda, B. Lillie, D. London, Y. Okada, A. Poschenrieder, T.G. Rizzo, Y. Shimizu, T. Shindou, L. Silvistrini, N. Sinha, R. Sinha, M. Spranger, M. Tanaka, S.K. Vempati, O. Vives, A. Weiler

## 5.1 Overview

Particle physics is the study of the nature of matter, energy, space and time. Our goal is to reveal the innermost building blocks of matter and to understand the forces acting between them. Remarkable progress has been achieved towards this goal with the construction and verification of the Standard Model. However, we know that our current picture of nature's building blocks is incomplete. A missing ingredient is the mechanism responsible for the origin of mass and the breaking of the electroweak symmetry. This mechanism is related to a set of questions and puzzles which remain unanswered within the Standard Model, such as: $(i)$ the gauge hierarchy problem, $(ii)$ the flavor problem, $(iii)$ the strong $CP$ problem, $(iv)$ what is responsible for baryogenesis, $(v)$ how are neutrino masses generated, and $(vi)$ how is gravity incorporated? The electroweak symmetry breaking mechanism must manifest itself at the TeV scale and these questions indicate that it will be accompanied by New Physics, also present at the scale $\Lambda_{\rm NP} \sim 4\pi M_W \sim 1$ TeV. In addition, recent astro-physical observations of the presence of cold dark matter implies the existence of physics beyond the Standard Model. If the cold dark matter candidate is a weakly interacting massive particle, then it too must exist at the TeV scale in order to account for the measured dark matter density.

The Large Hadron Collider is currently under construction at CERN; it is expected to discover the mechanism of electroweak symmetry breaking and any New Physics which accompanies it. The International Linear Collider is being proposed as a microscopic tool for exploring the symmetry breaking sector, New Physics, and possibly dark matter particles. Together, these machines will unravel the underlying theory of the electroweak sector and will resolve the first question above, but will largely leave the remaining problems unanswered. String theory is the only possibility known at present for addressing the last question of incorporating gravity. The remaining puzzles $(ii) - (v)$ are questions regarding the flavor sector of particle physics and are best addressed by detailed studies of that sector. In particular, the flavor sector of the new TeV scale physics discovered at the LHC/ILC can be probed in heavy quark systems with ultra-precise data.

Heavy Flavor physics in the LHC/ILC era takes on a new context. The goal is not only to establish deviations from the Standard Model, but also to diagnose and interpret these signals in terms of the underlying theory. The discovery of New Physics at the LHC/ILC will lead to a determination of $\Lambda_{NP}$. Ultra-precise heavy flavor experiments are complementary in that they will probe the flavor violation associated with the New Physics and measure the new flavor parameters. Large data samples will be needed to explore the TeV scale and, in particular, the Super $B$ Factory is well-suited to determine the flavor structure of the new TeV physics. In the unlikely event that the LHC/ILC discovers nothing outside of a Standard Model Higgs, then the role of a Super $B$ Factory would be to confirm the Standard Model predictions, or find minute deviations in the flavor sector. Whatever transpires at the high energy colliders, Super $B$ Factories play an important role in elucidating the physics of the TeV scale.

A schematic drawing of the complementary nature of Super $B$ Factories and high energy colliders is given in Fig. 5-1. This displays a general parameter space of a New Physics model, in the plane of a typical phase (or flavor non-diagonal parameter) versus the mass scale associated with the new interactions. The LHC/ILC will be able to determine the mass scale to a fairly precise degree of accuracy, and explore this parameter space up to some vertical line located at a $\sim$ few TeV. The colliders will not, however, have the ability to perform measurements in the other direction of the plane, *i.e.*, on the phase or non-diagonal flavor parameters. All LHC/ILC measurements will be located on a vertical line in this plane. Super $B$ Factories will be able to probe a diagonal region of this plane, *i.e.*, they will be able to probe the phase or non-diagonal flavor parameters to a certain accuracy up to a particular mass scale. All such measurements will lie on a diagonal line in this plane. Only by working together can the high energy colliders and the high luminosity flavor machines pinpoint the spot in this plane occupied by New Physics.

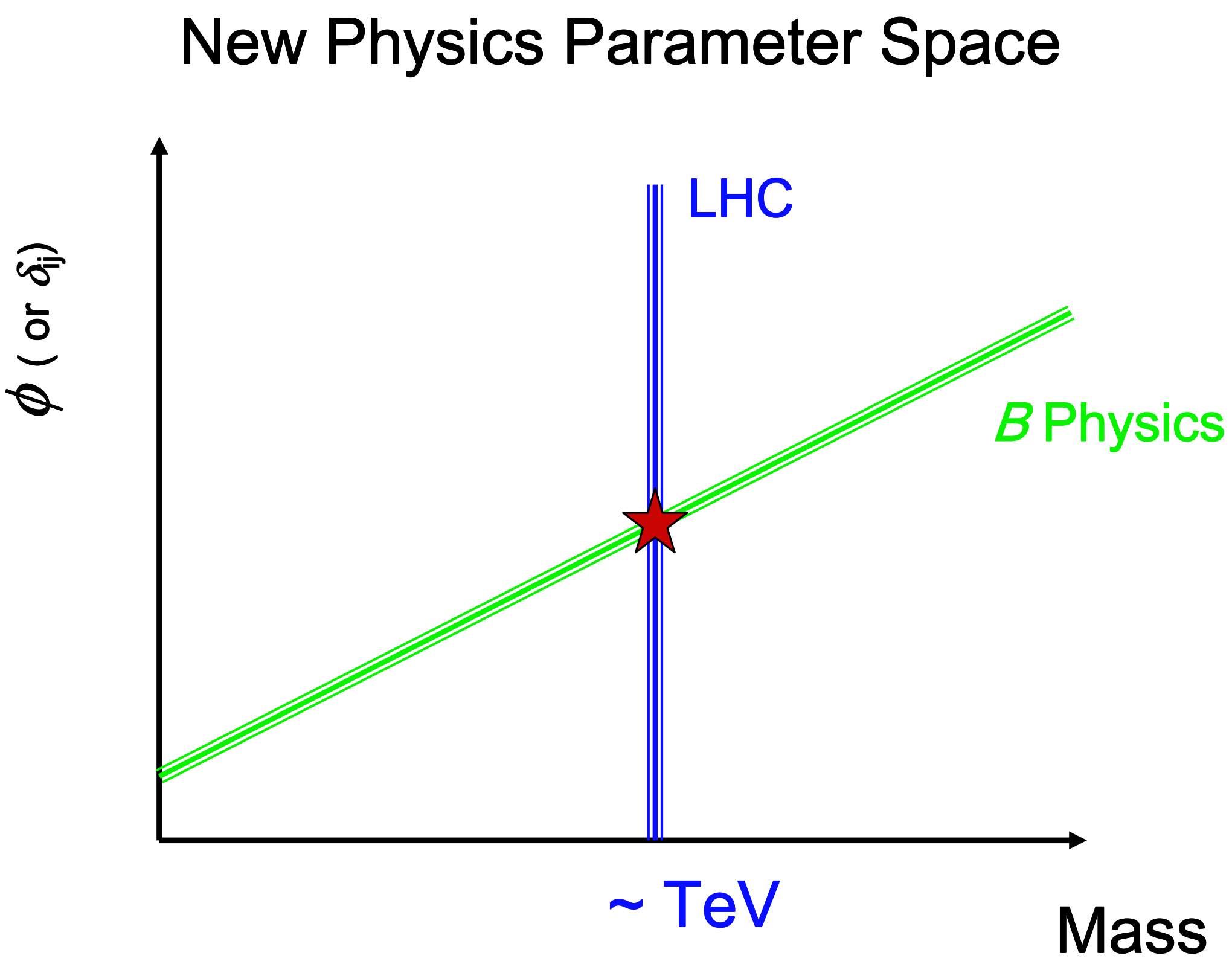

**Figure 5-1.** *Schematic representation of the general parameter space of a New Physics model in the plane of the phase (or non-diagonal flavor parameter) versus mass scale inherent to the new interactions. The LHC/ILC determines the mass scale as labeled and the Super $B$ Factory determines the diagonal line as indicated.*

At the Super $B$ Factory, there are a variety of methods to search for New Physics effects:

- Consistency tests of angle and side measurements of the $B_d$ unitarity triangle.

- Comparison among the time-dependent $CP$ asymmetries in different modes, such as $B \to J/\psi K_S$, $B \to \phi K_S$, and $B \to \eta' K_S$.
- Measurement of the branching ratios, $CP$ asymmetries, and various kinematic distributions of electroweak penguin processes such as $b \to s\gamma$, $b \to s\ell^+\ell^-$, and $b \to s\nu\overline{\nu}$ transitions.
- Measurement of $B$ decays to final states including a tau particle, for example, $B \to \tau\nu$ and $B \to D\tau\nu$ processes.
- Searches for leptonic decays such as $B \to \mu\mu$, $B \to \mu\nu$, and $B \to \mu e$.
- Searches for Lepton Flavor Violation in processes such as $\tau \to \mu\gamma$.
- Searches for meson mixing and $CP$ violation in the $D$ meson sector.

These measurements reveal different aspects of new interactions. In many cases, the first method is interpreted as a determination of new contributions in the $B_d - \overline{B_d}$ mixing amplitude. The second and third techniques are searches for new contributions in $b \to s$ transitions. The $B$ decay modes including a final tau particle is sensitive to the tree-level diagram with charged Higgs exchange. A combination of the above measurements, together with those in the $K$ and $B_s$ systems, offers a stringent test for new interactions as all of these processes are governed by the unique CKM matrix within the Standard Model. In addition, limits on tau lepton flavor violation, as well as $CP$ violation and rare decays in the charm sector can be significantly improved at the Super $B$ Factory, since this facility will be a Super-tau/charm Factory at the same time.

It is well-known that data in the $B$ sector already constrains models of New Physics[1]. For example, the rate for $b \to s\gamma$ places significant bounds in the common scalar - gaugino mass plane in Minimal SUGRA models of supersymmetry [2]. In another case, $B_d$ and $K$ meson mixing constrain the compactification radius of split fermion models of extra dimensions with gauge bosons in the bulk to be $R_c \geq 1 - 100$ TeV [3]. In particular, the flavor sector is important for distinguishing among models of supersymmetry since the effects of the supersymmetry breaking mechanism are manifested in flavor violating parameters. Once supersymmetry is discovered, we will want to determine the flavor structure of the squark mass matrices since they contain new sources of flavor mixing and $CP$ violation. This can only be accomplished by detailed explorations of the flavor sector, which can then reveal the underlying mechanism of supersymmetry breaking and probe the physics at the GUT scale.

The format of this chapter is as follows. We first discuss several techniques of exploring New Physics in $B$ decays in a model independent fashion. These range from an ultra-precise determination of the unitarity triangle, to global fits of rare electroweak penguin decay modes, to a thorough determination of various amplitudes which can contribute to $B \to V_1 V_2$ decays. We then consider the effects of supersymmetry and extra dimensions on the $B$ sector. Both of these theories address the hierarchy problem and contain a natural dark matter candidate. In both cases, data from Super $B$ Factories can distinguish between the possible classes of sub-models. We then discuss tests for lepton flavor violation with high data samples. In summary, we present a compilation of effects in numerous processes in the $B$ system within several models. This Table demonstrates that the pattern of effects within a particular model provides a powerful technique of identifying the source of new interactions.

## 5.2 Model-independent analyses

### 5.2.1 Sensitivity to New Physics from global fits of the Unitarity Triangle

>– G. Eigen –<

#### Introduction

The three-family Cabibbo-Kobayashi-Masakawa (CKM) quark-mixing matrix is a key ingredient of the Standard Model ( Standard Model ), as three real parameters and one phase are sufficient to completely specify the matrix. The Wolfenstein parameterization is a convenient representation of the CKM matrix, which to order $O(\lambda^5)$ is given by [4, 5]:

$$V = \begin{pmatrix} 1-\frac{\lambda^2}{2}-\frac{\lambda^4}{8} & \lambda & A\lambda^3(\rho-i\eta) \\ -\lambda+A^2\lambda^5(\frac{1}{2}-\rho-i\eta) & 1-\frac{\lambda^2}{2}-\frac{\lambda^4}{8}(1+4A^2) & A\lambda^2 \\ A\lambda^3(1-\overline{\rho}-i\overline{\eta}) & -A\lambda^2+A\lambda^4(\frac{1}{2}-\rho-i\eta) & 1-\frac{1}{2}A^2\lambda^4 \end{pmatrix} + O(\lambda^6). \quad (5.1)$$

The parameter $\lambda = 0.2235 \pm 0.0033$ [6], the sine of the Cabibbo angle, is the best measured; $A \simeq 0.82$, representing the deviation of $V_{cb}$ from $\lambda^2$, is known to $\sim \pm 5\%$, while $\overline{\rho} = \rho(1-\lambda^2/2)$ and the phase $\overline{\eta} = \eta(1-\lambda^2/2)$ are less well known. Unitarity of the CKM matrix yields six triangular relations, of which $V_{ud}V_{ub}^* + V_{cd}V_{cb}^* + V_{td}V_{tb}^* = 0$ is the most useful, since it specifies a triangle in the $\overline{\rho} - \overline{\eta}$ plane, often called the Unitarity Triangle, with apex $(\overline{\rho}, \overline{\eta})$ and nearly equal-length sides. To extract the CKM parameters and to explore New Physics domains, we perform global $\chi^2$ fits using measurements [7] that specify the sides and angles of the UT, as discussed in the next section.

One complication arises from non-probabilistic uncertainties appearing in the extraction of CKM parameters from measurements. Due to their non-probabilistic character, these uncertainties cannot be treated in the usual statistical way by adding them in quadrature with probabilistic errors, such as statistical errors and presumably experimental systematic errors. In order to treat non-probabilistic uncertainties in a coherent way, we have introduced the “scanning method”. Here, all significant non-probabilistic uncertainties are scanned within their allowed range providing a realistic treatment of non-probabilistic uncertainties and a robust method for reducing the sensitivity to fluctuations. This method was first used for Unitarity Triangle fits in the *BABAR* physics book [1]. Recently, we have refined our approach by separating coherently all theoretical parameters that are affected by non-probabilistic errors from measurements which presumably have probabilistic errors [8, 9]. A $\chi^2$ minimization is performed to determine the CKM parameters using a frequentist approach for specific values of the theoretical quantities, called a model. To accommodate the entire range of theoretical uncertainties we consider a representative set of models. Apart from New Physics parameters, we focus on the CKM parameters $\overline{\rho}$ and $\overline{\eta}$, since these are the least well known, and plot $95\%$ confidence level (CL) contours. The allowed range in the $\overline{\rho} - \overline{\eta}$ plane resulting from an overlay of all contours of the different models is typically wider than that obtained in a Bayesian approach [10]. In comparison to other frequentist approaches [11, 12], we have the ability to link specific CKM parameters to specific theoretical quantities. This is not trivial, as there is no one-to-one correspondence. In our approach, we can distinguish whether an inconsistency originates from the data or from non-probabilistic uncertainties of theoretical parameters. Our procedure is discussed in detail in reference [8].

In this study, we present results from the basic Standard Model analysis, a model-independent analysis that attributes New Physics to $B_d^0\overline{B}_d^0$ mixing and a model-independent analysis that looks for New Physics in $b \to s$ processes. For all studies we consider both present measurements as well as extrapolations to $10\ \text{ab}^{-1}$. In one case, we even consider an integrated luminosity of $50\ \text{ab}^{-1}$.

### Global fit method

Using the refined "scanning method" we perform global $\chi^2$ minimizations for different physics scenarios. In standard fits we include eight observables, of which two determine $|V_{cb}|$ and two determine $|V_{ub}|$. The $B \to D^*\ell\nu$ differential decay rate, extrapolated to zero recoil, yields the product $\langle |V_{cb}| \cdot F_{D^*}(1)\rangle$. The form factor $F_{D^*}(1)$, calculated using heavy quark effective theory (HQET) [13], involves non-probabilistic errors. The inclusive $B \to X_c\ell\nu$ branching fraction can be factorized in terms of $V_{cb}$, the reduced decay rate $\widetilde{\Gamma}^c_{inc}$, and the $b$-quark lifetime, $\mathcal{B}(B \to X_c\ell\nu) = |V_{cb}|^2\widetilde{\Gamma}^c_{inc}\cdot\tau_b$ [14, 15]. In a similar fashion, we can express the branching fractions $\mathcal{B}(B^0 \to \rho^-\ell^+\nu) = |V_{ub}|^2\widetilde{\Gamma}_\rho\cdot\tau_{B^0}$ and $\mathcal{B}(B \to X_u\ell\nu) = |V_{ub}|^2\widetilde{\Gamma}^u_{inc}\cdot\tau_b$. The inclusive reduced decay rates are calculated using the heavy quark expansion (HQE) [15, 16], while $\widetilde{\Gamma}_\rho$ is determined from the spread of different form-factor models [17]. The dominant non-probabilistic uncertainties here result from these reduced rates.

Mixing-induced $CP$ violation in the $K^0\overline{K}^0$ system is expressed by the parameter

$$|\epsilon_K| = C \cdot B_K\overline{\eta}|V_{cb}|^2\lambda^2\big\{[\eta_1 S_0(x_c) - \eta_3 S_0(x_c, x_t)] - |V_{cb}|^2(1-\overline{\rho})\eta_2 S_0(x_t)\big\}, \tag{5.2}$$

where $C$ is a collection of constants, $B_K$ denotes the "bag" factor [18] of the $K^0\overline{K}^0$ system, calculated using lattice QCD, $S_0(x)$ are the Inami-Lim functions [19] depending on the squared mass ratio of top (charm) quark to $W$-boson, $x_{t(c)} = m^2_{t(c)}/m^2_W$, and $\eta_1, \eta_2, \eta_3$ are QCD parameters [20, 18, 21]. The bag factor and the QCD parameters are affected by non-probabilistic uncertainties. The $B_{d(s)}\overline{B}_{d(s)}$ oscillation frequencies,

$$\Delta m_{B_{d(s)}} = \frac{G_F^2}{6\pi^2}\eta_B m_{B_{d(s)}} m_W^2 S_0(x_t) f^2_{B_{d(s)}} B_{B_{d(s)}} |V_{td(s)}V^*_{tb}|^2, \tag{5.3}$$

involve CKM parameters in the third row of the CKM matrix. Here, $G_F$ is the Fermi constant, $\eta_B$ is a QCD parameter, $m_{B_{d(s)}}$ is the $B_{d(s)}$ mass, $m_W$ is the $W$-mass, $f_{B_{d(s)}}$ is the $B_{d(s)}$ decay constant and $B_{B_{d(s)}}$ is the "bag" parameter in the $B_{d(s)}\overline{B}_{d(s)}$ system [18] , respectively. Lattice calculations yield the products $f_{B_{d(s)}}\sqrt{B_{B_{d(s)}}}$ with non-probabilistic errors. By considering the ratio

$$r_{\Delta m} = \frac{\Delta m_{B_s}}{\Delta m_{B_d}} = \frac{m_{B_s}}{m_{B_d}}\xi^2\frac{|V_{ts}V^*_{tb}|^2}{|V_{td}V^*_{tb}|^2} \tag{5.4}$$

instead of $\Delta m_{B_s}$, non-probabilistic errors now appearing in the ratio $\xi = f_{B_s}/f_{B_d}\cdot\sqrt{B_{B_s}/B_{B_d}}$ can be significantly reduced. Presently, only a lower limit of $\Delta m_{B_s} > 14.4\ \mathrm{ps}^{-1}$ exists [22]. We, however, expect a $\Delta m_{B_s}$ measurement at the Tevatron within the next couple of years.

Finally, we use the $CP$ asymmetry in $CP$ eigenstate decays of a $B^0$ ($\overline{B}^0$) to a charmonium state and a $K^0_S$ or a $K^0_L$, denoted by $a_{CP}(\psi K^0_S)$. For these tree-diagram-dominated processes $a_{CP}(\psi K^0_S)$, equaling $\sin 2\beta$, is presently the only observable that is not affected by non-probabilistic uncertainties. Other $CP$ asymmetries in the $B^0_d\overline{B}^0_d$ system, such as $a_{CP}(\phi K^0_S)$, $a_{CP}(\eta' K^0_S)$ and $a_{CP}(D^{(*)+}D^{(*)-})$, also measure $\sin 2\beta$ in the Standard Model. Apart from $a_{CP}(\phi K^0_S)$, however, the other $CP$ asymmetries may involve an additional weak phase that enters through a sizable second decay amplitude. Since the present precision of these $CP$ asymmetries is furthermore quite limited with respect to $\sin 2\beta$, we do not include them in our fits.

While $\sin 2\beta$ is becoming a precisely-measured observable [23, 24], the $B$ factories have started to measure also the other angles of the Unitarity Triangle. First measurements of the $CP$ asymmetry in $B \to \pi^+\pi^-$ have been reported by *BABAR* and Belle [25, 26]. Besides the $b \to u$ tree amplitude, we expect a sizable penguin amplitude here, since the penguin-dominated process $B^0 \to K^+\pi^-$ has a factor of four larger branching fraction than $\mathcal{B}(B^0 \to \pi^+\pi^-)$. Since the penguin amplitude carries a different weak phase, $a_{CP}(\pi^+\pi^-)$ measures $\sin 2\alpha_{eff}$. An isospin analysis is necessary to extract $\alpha$ from $\alpha_{eff}$ [27]. Despite the branching fraction measurement of $B \to \pi^0\pi^0$ by *BABAR* [28] and Belle [29], the present precision of $|\alpha - \alpha_{eff}| < 48^\circ$ @ $90\%$ CL, apart from a four-fold ambiguity, provides no useful constraint [30].

Rather promising, however, is the determination of $\sin 2\alpha$ from the recently observed $CP$ asymmetry in $B \to \rho^+\rho^-$ [28]. Using the Grossman-Quinn bound [27, 31] with recent *BABAR* measurements of $B \to \rho^\pm\rho^0, \rho^0\rho^0$ [32], yields a limit on $|\alpha_{eff} - \alpha| < 13^\circ$ (68% CL). From this we can extract $\alpha$ up to a four-fold ambiguity [33]. The solution closest to the Standard Model yields $\alpha = (102^{+16+5}_{-12-4} \pm 13)^\circ$. Extrapolations to $10\ \mathrm{ab}^{-1}$ yield an experimental uncertainty of $\sigma_\alpha = 1.5^\circ$ and a theoretical uncertainty of the phase of $\delta_\alpha = 5^\circ$.

The observed decay amplitudes of the modes $B^- \to D^0 K^{(*)-}$, $B^- \to \overline{D}^0 K^{(*)-}$ and $B^- \to D^0_{CP} K^{(*)-}$ and their charge-conjugate partners allow for a determination of the angle $\gamma$ [34]. The $CP$ asymmetry of $B^0 \to D^{(*)+}\pi^-$ modes measures $\sin(2\beta + \gamma)$ [35]. From the $DK$ modes we presently determine $\gamma = (95.1 \pm 28.1)^\circ$ [36, 37]. Extrapolations to $10\ \mathrm{ab}^{-1}$ yield an experimental uncertainty in $\gamma$ of $2^\circ - 3^\circ$, while the theoretical uncertainty is expected to be $\delta_\gamma \simeq 0.1^\circ$.

In some of our global fits, particular those using extrapolations to $10\ \mathrm{ab}^{-1}$, we include $\sin 2\alpha$ and $\gamma$ measurements. We assume that all ambiguities can be resolved by additional measurements and, therefore, include only the solution closest to the Standard Model. Table 5-1 summarizes the present average values of all considered observables and extrapolations to sensitivities expected at a Super $B$ Factory integrating $10\ \mathrm{ab}^{-1}$ annually.

In terms of Wolfenstein parameters, the three Unitarity Triangle angles satisfy the relations

$$\sin 2\beta = \frac{2\overline{\eta}(1-\overline{\rho})}{\{(1-\overline{\rho})^2 + \overline{\eta}^2\}}, \quad \sin 2\alpha = \frac{2\overline{\eta}(1-\overline{\rho})\{\overline{\eta}^2 + \overline{\rho}(\overline{\rho}-1)\}}{(\overline{\rho}^2+\overline{\eta}^2)\{(1-\overline{\rho})^2 + \overline{\eta}^2\}}, \quad \sin 2\gamma = \frac{2\overline{\eta}}{(\overline{\rho}^2+\overline{\eta}^2)}. \tag{5.5}$$

**Table 5-1.** *Input values of the observables used in the global fit. All other measured quantities are taken from [6].* * *The value in parentheses corresponds to* $50\ \mathrm{ab}^{-1}$. *The theoretical uncertainties in* $\mathcal{B}(B \to X_u(\rho)\ell\nu)$ *are accounted for in the reduced rates.*

| Observable | Present Value | Reference | Value for $10\ \mathrm{ab}^{-1}$ |
|---|---|---|---|
| $\lvert V_{cb}\rvert F_{D^*}(1)$ | $0.0367 \pm 0.0008$ | [22] | $0.0378 \pm 0.00037$ |
| $\Upsilon(4S)\ \mathcal{B}(B \to X_c\ell\nu)$ [%] | $10.9 \pm 0.23$ | [22] | $10.50 \pm 0.05$ |
| LEP $\mathcal{B}(B \to X_c\ell\nu)$ [%] | $10.42 \pm 0.26$ | [62] | - |
| $\Upsilon(4S)\ \mathcal{B}(B \to X_u\ell\nu)$ [$10^{-3}$] | $1.95 \pm 0.19_{exp} \pm 0.31_{th}$ | [65, 64, 66, 65] | $1.85 \pm 0.06_{exp} \pm 0.1_{th}$ |
| LEP $\mathcal{B}(B \to X_u\ell\nu)$ [$10^{-3}$] | $1.71 \pm 0.48_{exp} \pm 0.21_{th}$ | [67] | - |
| $\Upsilon(4S)\ \mathcal{B}(B \to \rho\ell\nu)$ [$10^{-4}$] | $2.68 \pm 0.43_{exp} \pm 0.5_{th}$ | [68, 69] | $3.29 \pm 0.14_{exp} \pm 0.16_{th}$ |
| $\Delta m_{B_d}$ [$\mathrm{ps}^{-1}$] | $0.502 \pm 0.007$ | [22] | $0.502 \pm 0.00104$ |
| $\Delta m_{B_s}$ [$\mathrm{ps}^{-1}$] | $> 14.4$@90% CL | [22] | $20 \pm 1$ |
| $\lvert\epsilon_K\rvert$ [$10^{-3}$] | $2.282 \pm 0.017$ | [6] | $2.282 \pm 0.017$ |
| $\sin 2\beta(\psi K^0_S)$ | $0.736 \pm 0.049$ | [22] | $0.736 \pm 0.007$ |
| $\sin 2\beta(\phi K^0_S)^*$ | $0.02 \pm 0.29$ | [22] | $0.6 \pm 0.03$ ($0.6 \pm 0.015$) |
| $\sin 2\alpha$ | $-0.42 \pm 0.44$ | [28] | $-0.42 \pm 0.047$ |
| $\gamma$ | $95.1^\circ \pm 28^\circ$ | [36, 37] | $53.0^\circ \pm 3.5^\circ$ |
| $\lambda$ | $0.2235 \pm 0.0033$ | [6] | $0.2235 \pm 0.0033$ |
| $m_t$ [GeV/c$^2$] | $169.3 \pm 5.1$ | [6] | $169.3 \pm 2.$ |
| $m_c$ [GeV/c$^2$] | $1.3 \pm 0.2$ | [6] | $1.3 \pm 0.1$ |

The theoretical parameters $F_{D^*}(1)$, $\widetilde{\Gamma}^c_{inc}$, $\widetilde{\Gamma}_\rho$, $\widetilde{\Gamma}^u_{inc}$, $B_K$, $f_{B_d}\sqrt{B_{B_d}}$, and $\xi$ involve both statistical and a non-probabilistic uncertainties. The statistical uncertainties contain errors that result from measurements such as the top-quark mass or the statistics in lattice gauge theory calculations. The non-probabilistic parts contain uncertainties such as high-order effects, scale dependence, QCD corrections, and quenching effects as well as other systematic uncertainties in lattice gauge calculations. While the statistical parts are represented by additional terms in the $\chi^2$ minimization, the full regions of the non-probabilistic parts are scanned. Eventually, fully unquenched lattice gauge calculations will be carried out and uncertainties from higher-order effects, QCD corrections and the scale dependence will be reduced significantly, because many precise measurements will become available from the $B$ factories and new calculations will be accomplished. Thus, we expect non-probabilistic theoretical errors of $F_{D^*}(1)$, $B_K$, $f_{B_d}\sqrt{B_{B_d}}$ and $\xi$ eventually to become small. In addition, we will focus on observables in the future that provide the most precise extraction of CKM parameters. For example, $B \to \pi\ell\nu$ may be better suited to determine $|V_{ub}|$ than $B \to \rho\ell\nu$, since lattice gauge calculation may achieve a more precise value for $\tilde{\Gamma}_\pi$ than for $\tilde{\Gamma}_\rho$. The determination of the UT angle $\alpha$ from $a_{CP}(\rho^+\rho^-)$ may remain more precise than that from $a_{CP}(\pi^+\pi^-)$. Table 5-2 summarizes the present values of the theoretical parameters and projections expected at the time scale of a Super $B$ Factory .

**Table 5-2.** *Range of the theory parameters scanned in the global fit for present values and extrapolations to* $10\ \mathrm{ab}^{-1}$ *[41]. For parameters calculated on the lattice the uncertainty is split into a statistical piece and a non-probabilistic piece.*

| Parameter | Present Range | $\sigma_{stat}$ | Extrapolation to $10\ \mathrm{ab}^{-1}$ | expected $\sigma_{stat}$ |
|---|---|---|---|---|
| $F_{D^*}(1)$ | $0.87 - 0.95$ | - | $0.90 - 0.92$ | - |
| $\Gamma(c\ell\nu)$ $[\mathrm{ps}^{-1}]$ | $34.1 - 41.2$ | - | $35.7 - 39.2$ | - |
| $\Gamma(\rho\ell\nu)$ $[\mathrm{ps}^{-1}]$ | $12.0 - 22.2$ | - | $11.0 - 13.4$ | - |
| $\Gamma(u\ell\nu)$ $[\mathrm{ps}^{-1}]$ | $54.6 - 80.2$ | - | $60.6 - 75.0$ | - |
| $B_K$ | $0.74 - 1.0$ | $\sigma_{B_K} = \pm 0.06$ | $0.805 - 0.935$ | $\sigma_{B_K} = \pm 0.03$ |
| $f_{B_d}\sqrt{B_{B_d}}$ [MeV] | $218 - 238$ | $\sigma_{f_B\sqrt{B_B}} = \pm 30$ | $223 - 233$ | $\sigma_{f_B\sqrt{B_B}} = \pm 10$ |
| $\xi$ | $1.16 - 1.26$ | $\sigma_\xi = \pm 0.05$ | $1.19 - 1.23$ | $\sigma_\xi = \pm 0.02$ |
| $\eta_1$ | $1.0 - 1.64$ | - | $1.0 - 1.64$ | - |
| $\eta_2$ | $0.564 - 0.584$ | - | $0.564 - 0.584$ | - |
| $\eta_3$ | $0.43 - 0.51$ | - | $0.43 - 0.51$ | - |
| $\eta_B$ | $0.54 - 0.56$ | - | $0.54 - 0.56$ | - |
| $\delta_\alpha$ | $13^\circ$ | - | $5^\circ$ | - |

### The fit function

For Standard Model global fits, the observables are expressed in terms of the CKM parameters $A, \lambda, \overline{\rho}, \overline{\eta}$. For global fits testing for New Physics effects, we include additional parameters as discussed in the next section. Furthermore, to account for various correlations among observables we include additional $\chi^2$ terms for $b$ lifetimes, $b$-hadron production fractions at the $Z^0$ and $\Upsilon(4S)$ as well as the masses of the $W$ boson, $c$ quark and $t$ quark. Since the non-probabilistic uncertainties in the QCD parameters are comparatively small, we do not scan them but include them as statistical errors. In global fits that involve measurements of $\sin 2\alpha$ we also scan the uncertainty of $\delta_\alpha$ expected in $B \to \rho^+\rho^-$ decays. Thus, in a Standard Model analysis we typically perform 17-parameter fits and scan up to eight theory parameters. In order to be independent of New Physics effects in the $K^0\overline{K}^0$ system and in $B_s^0\overline{B}_s^0$ mixing, we also perform global fits excluding the observables $|\epsilon_K|$ and $\Delta m_{B_s}$.

For the Standard Model global fits, the $\chi^2$ function is given by

$$\chi^2_{\mathcal{M}}(A,\overline{\rho},\overline{\eta}) = \left(\frac{\langle|V_{cb}\mathcal{F}_{D^*}(1)|\rangle - A^2\lambda^4\,|\mathcal{F}_{D^*}(1)|^2)}{\sigma_{V_{cb}\mathcal{F}_{D^*}(1)}}\right)^2 + \left(\frac{\langle B_{c\ell\nu}\rangle - \widetilde{\Gamma}_{c\ell\nu}A^2\lambda^4\tau_b}{\sigma_{B_{c\ell\nu}}}\right)^2$$
$$+ \left(\frac{\langle B_{\rho\ell\nu}\rangle - \widetilde{\Gamma}_{\rho\ell\nu}A^2\lambda^6\tau_{B^0}(\rho^2+\eta^2)}{\sigma_{B_{\rho\ell\nu}}}\right)^2 + \left(\frac{\langle B_{u\ell\nu}\rangle - \widetilde{\Gamma}_{u\ell\nu}A^3\lambda^6\tau_b(\rho^2+\eta^2)}{\sigma_{B_{u\ell\nu}}}\right)^2$$

$$+\left(\frac{\langle B_K\rangle - B_K}{\sigma_{B_K}}\right)^2 + \left(\frac{\langle|\varepsilon_K|\rangle - |\varepsilon_K|\,(A,\overline{\rho},\overline{\eta})}{\sigma_\varepsilon}\right)^2 + \left(\frac{\langle\Delta m_{B_d}\rangle - \Delta m_{B_d}(A,\overline{\rho},\overline{\eta})}{\sigma_{\Delta m}}\right)^2$$
$$+\chi^2_{\Delta m_{B_s}}(A,\overline{\rho},\overline{\eta}) + \left(\frac{\left\langle a_{\psi K^0_S}\right\rangle - \sin 2\beta(\overline{\rho},\overline{\eta})}{\sigma_{\sin 2\beta}}\right)^2 + \left(\frac{\langle f_B\sqrt{B_B}\rangle - f_B\sqrt{B_B}}{\sigma_{f_B\sqrt{B_B}}}\right)^2$$
$$+\left(\frac{\langle\xi\rangle - \xi}{\sigma_\xi}\right)^2 + \left(\frac{\langle\lambda\rangle - \lambda}{\sigma_\lambda}\right)^2 + \left(\frac{\langle m_t\rangle - m_t}{\sigma_{m_t}}\right)^2 + \left(\frac{\langle m_c\rangle - m_c}{\sigma_{m_c}}\right)^2 + \left(\frac{\langle m_W\rangle - m_W}{\sigma_{M_W}}\right)^2$$
$$+\left(\frac{\langle\tau_{B^0}\rangle - \tau_{B^0}}{\sigma_{\tau_{B^0}}}\right)^2 + \left(\frac{\langle\tau_{B^+}\rangle - \tau_{B^+}}{\sigma_{\tau_{B^+}}}\right)^2 + \left(\frac{\langle\tau_{B_s}\rangle - \tau_{B_s}}{\sigma_{\tau_{B_s}}}\right)^2 + \left(\frac{\langle\tau_{\Lambda_b}\rangle - \tau_{\Lambda_b}}{\sigma_{\tau_{\Lambda_b}}}\right)^2$$
$$+\left(\frac{\langle f_{B^+}\rangle - f_{B^+}}{\sigma_{f_{B^+}}}\right)^2 + \left(\frac{\langle f_{B_s}\rangle - f_{B_s}}{\sigma_{f_{B_s}}}\right)^2 + \left(\frac{\langle f_{B^{+,0}}\rangle - f_{B^{+,0}}}{\sigma_{f_{B^{+,0}}}}\right)^2. \tag{5.6}$$

The notation $\langle Y\rangle$ is used to denote the average value of observable $Y$, and $\mathcal{M}$ is used to denote a "model" that corresponds to a specific set of theoretical parameters within the range of non-probabilistic uncertainties. A $\chi^2$ minimization using a frequentist approach is performed for many different models, scanning over the entire space of allowed theoretical parameters. The minimization solution $(\lambda, A, \overline{\rho}, \overline{\eta})_{\mathcal{M}}$ for a particular model now depends only on measurement errors and other probabilistic uncertainties. A model is retained if the fit probability exceeds $5\%$. For accepted models, the central value and a $95\%$ CL contour in the $\overline{\rho} - \overline{\eta}$ plane are plotted. If no "model" were to survive we would have evidence of a consistency problem at the $2\sigma$ level between data and theory, independent of the calculations of the theoretical parameters or the choices of their uncertainties.

Since $\Delta m_{B_s}$ has not yet been measured, we use a $\chi^2$ term derived from the significance [8]

$$S = \sqrt{\frac{N}{2}} f_{B_s}(1-2w)e^{-\frac{1}{2}(\Delta m_s \sigma_t)^2}, \tag{5.7}$$

yielding

$$\chi^2_{\Delta m_{B_s}} = C_s^2\left(1 - \frac{\Delta}{\Delta m_{B_s}}\right)^2 e^{-(\Delta m_{B_s}\sigma_t)^2}, \tag{5.8}$$

where $\Delta$ is the best estimate according to experiment. The values of $(\Delta, C_s^2, \sigma_t)$ are chosen to give a minimum at $17\ \mathrm{ps}^{-1}$, and a $\chi^2$ probability of 5% at $\Delta m_{B_s} = 14.4\ \mathrm{ps}^{-1}$ [6]. For the extrapolations to $10\ \mathrm{ab}^{-1}$ we replace this $\chi^2$ term by the standard parabolic $\chi^2$ expression of a measurement, $\chi^2_{r_{\Delta m}} = \left(\frac{\langle r_{\Delta m}\rangle - r_{\Delta m}(A,\overline{\rho},\overline{\eta})}{\sigma_{r_{\Delta m}}}\right)^2$.

### Results in the Standard Model

Figure 5-2a shows the results of the Standard Model global fits for a representative number of models using present averages of the standard eight observables. For each accepted model, we plot in the $\overline{\rho}-\overline{\eta}$ plane the central value (black points) and the $95\%$ CL contour. The overlay of contours of different accepted models is shown by the dark-shaded (red) region. As an illustration, a contour of a typical model is highlighted by the light (yellow) error ellipse. The sizes of the contours vary and are typically correlated with the probability of that model. Global fits with high probabilities typically have larger contours than those with low probabilities. We can infer the size of large contours from the distance of the black central points from the dark-shaded (red) region on the left-hand side. For a specific model, we can give a frequentist interpretation. Thus, for a specific model all points inside the contour are distributed with known probability. However, there is no frequentist interpretation for comparing which models are to be "preferred", other than the statement that at most one model is correct. In particular, the density of contours seen in some of the fit results has no physics interpretation. In this analysis we cannot, and do not, give any relative probabilistic weighting among the contours, or their overlap regions; doing so would be equivalent to a Bayesian analysis. For a qualitative comparison, we also show the boundaries of the $|V_{ub}/V_{cb}|$, $|V_{td}/V_{cb}|$, $\sin 2\beta$, and $|\epsilon_K|$ bands that result from adding $1.96\sigma$ of the total experimental error for that observable linearly to the relevant non-probabilistic theoretical uncertainty.

Due to both the large non-probabilistic theoretical uncertainties and the limited precision of present measurements, many models are accepted. From Fig. 5-2a we extract ranges for the CKM parameters $\overline{\rho}$, $\overline{\eta}$, and $A$, as well as ranges for the angles $\alpha$, $\beta$ and $\gamma$. The results using the standard eight observables in the Standard Model global fits are summarized in Table 5-3. We distinguish between uncertainties that result from a spread of the models by quoting a range and those that originate from experimental uncertainties, listing in addition a statistical error. If we exclude $|\epsilon_K|$ and $\Delta m_{B_s}$ from the global fit, the overlay region of accepted models increases only slightly, as shown in Fig. 5-2b. The ranges of CKM parameters and UT angles are similar to those of the standard eight-observable global fits. The measurements of $|V_{ub}/V_{cb}|$, $\Delta m_{B_d}$ and $\sin 2\beta$ already provide useful constraints regarding the spread of the different models. Including measurements of $|\epsilon_K|$ and $\Delta m_{B_s}$ yields decreased sizes of contours and in turn a slight decrease in the overlay region of contours. It is important to note that present measurements in the $B_d^0$ system are sufficiently precise to establish $CP$ violation without using any $CP$ violation results in the $K^0\overline{K}^0$ system. Including the present results of the $\sin 2\alpha$ measurement from $B \to \rho^+\rho^-$ and a $\gamma$ measurement from $B \to DK$ [38, 37] have small effects on the region of overlays as shown in Figs. 5-2c,d.

For global fits using uncertainties extrapolated to $10\ \mathrm{ab}^{-1}$, the $B \to \rho\ell\nu$ and $b \to u\ell\nu$ branching fractions need to be tuned, since for present central values no model survives. The observed discrepancy between exclusive and inclusive $|V_{ub}|$ measurements may be a hint of limitations of the assumption of quark-hadron duality. Figure 5-3a shows the $\overline{\rho} - \overline{\eta}$ plane for global fits using the extrapolations to $10\ \mathrm{ab}^{-1}$ for all eight observables. The corresponding plots without $|\epsilon_K|$ and $\Delta m_{B_s}$ measurements are depicted in Fig. 5-3b. Again, the measurements of $|\epsilon_K|$ and $\Delta m_{B_s}$ reduce the allowed region of the overlay of different contours. Table 5-3 shows the ranges and experimental errors for CKM parameters and Unitarity Triangles angles using the standard eight-observable global fits for our extrapolations to $10\ \mathrm{ab}^{-1}$.

**Table 5-3.** *Precision of CKM parameters and Unitarity Triangle angles for present measurements and extrapolations to $10\ \mathrm{ab}^{-1}$ for Standard Model global fits using the standard eight observables plus $\sin 2\alpha$ and $\gamma$. The second and third columns summarize the range and the experimental uncertainty using present measurements, respectively. The fourth and fifth columns summarize the range and the experimental uncertainty using present extrapolations to $10\ \mathrm{ab}^{-1}$, respectively.*

| Parameter | Present Results | Error | Extrapolations to $10\ \mathrm{ab}^{-1}$ | Error |
|---|---|---|---|---|
| $\overline{\rho}$ | $0.120 - 0.332$, | $\sigma_{\overline{\rho}} =^{+0.029}_{-0.064}$ | $0.219 - 0.283$ | $\sigma_{\overline{\rho}} =^{+0.012}_{-0.016}$ |
| $\overline{\eta}$ | $0.272 - 0.407$ | $\sigma_{\overline{\eta}} =^{+0.028}_{-0.020}$ | $0.318 - 0.345$ | $\sigma_{\overline{\eta}} = \pm 0.02$ |
| $A$ | $0.80 - 0.89$ | $\sigma_A =^{+0.028}_{-0.024}$ | $0.76 - 0.83$ | $\sigma_A =^{+0.017}_{-0.015}$ |
| $\mathrm{m_c}$ | $1.05 - 1.29$ | $\sigma_{m_c} = \pm 0.18$ | $1.11 - 1.29$ | $\sigma_{m_c} =^{+0.017}_{-0.015}$ |
| $\beta$ | $(20.8 - 26.9)^\circ$ | $\sigma_\beta =^{+7.0^\circ}_{-2.1^\circ}$ | $(23.7 - 24.0)^\circ$ | $\sigma_\beta =^{+1.36^\circ}_{-1.22^\circ}$ |
| $\alpha$ | $(84.6 - 117.2)^\circ$ | $\sigma_\alpha =^{+5.4^\circ}_{-15.7^\circ}$ | $(98.8 - 107.7)^\circ$ | $\sigma_\alpha =^{+2.0^\circ}_{-2.8^\circ}$ |
| $\gamma$ | $(41.0 - 71.9)^\circ$ | $\sigma_\gamma =^{+8.2^\circ}_{-3.3^\circ}$ | $(48.3 - 57.5)^\circ$ | $\sigma_\gamma =^{+1.9^\circ}_{-1.4^\circ}$ |

In order to study the impact of the $\sin 2\beta$ measurement, we perform the same fits as above but leaving $\sin 2\beta$ out of the fit. The results are shown in Figs. 5-3c,d. The region of overlaid contours is significantly increased, demonstrating that $\sin 2\beta$ provides one of the most stringent constraints in the $\overline{\rho} - \overline{\eta}$ plane; this is already the case for present $\sin 2\beta$ results. In order to visualize the level of consistency, we have plotted the $\sin 2\beta$ bands for a value reduced by $1\sigma$ of the present precision (0.049) to $\sin 2\beta = 0.687 \pm 0.01$. Only a small region at high $\overline{\rho}$ and low $\overline{\eta}$ values is consistent at the $95\%$ CL value with the $\sin 2\beta$ measurement (Fig. 5-3d). For reduced values of $\sin 2\beta$ a conflict begins to emerge. If this value of $\sin 2\beta$ is actually included in the global fits for the standard, eight observables, we find no model consistent with the Standard Model . For global fits excluding the measurements of $|\epsilon_K|$ and $\Delta m_{B_s}$ the only models that survive lie inside the $\sin 2\beta$ band on the lower right-hand side. If we include, in addition, present $\sin 2\alpha$ and $\gamma$ measurements this small region of models is also excluded.

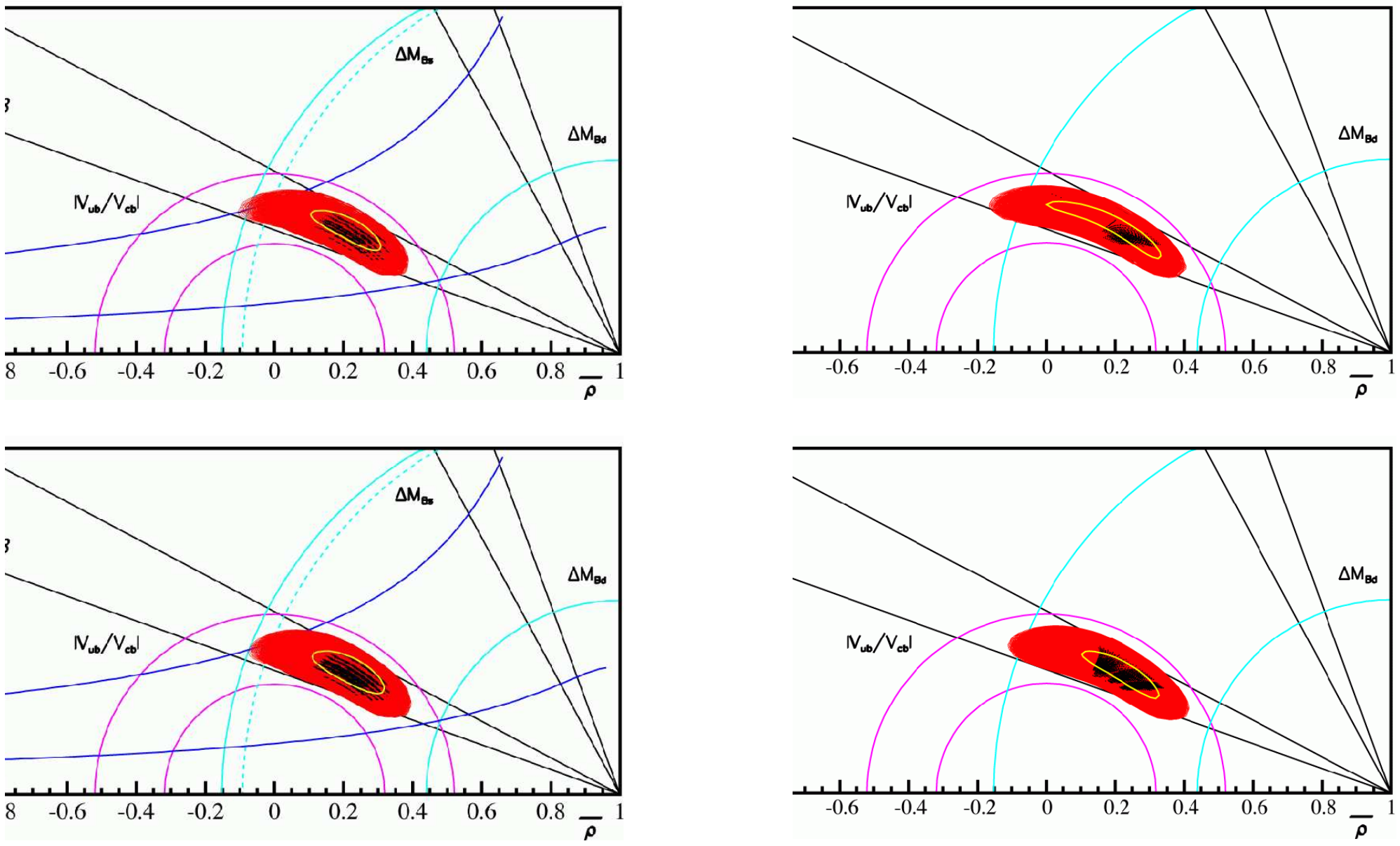

**Figure 5-2.** *Results of the Standard Model global fits in the $\overline{\rho}-\overline{\eta}$ plane, using present averages, a) for the eight standard observables, b) for the eight standard observables except for $|\epsilon_K|$ and $\Delta m_{B_s}$, c) for same observables used in a) plus $\sin 2\alpha$ and $\gamma$, and d) for the same observables used in b) plus $\sin 2\alpha$ and $\gamma$. The dark-shaded (red) region shows the overlay of individual contours of all accepted models. The black points represent the central values of accepted models and the light (yellow) ellipse shows a representative individual contour for a typical model.*

### Extensions of the Standard Model

In the Standard Model , the phase in the CKM matrix is the only source of $CP$ violation. In extension of the Standard Model, various new sources of $CP$ violation are expected, some of which may in fact contribute in $B$ decays. For example, in the minimal supersymmetric extensions of the Standard Model (called the MSSM), 124 new parameters enter, of which 44 are $CP$-violating. We consider two possible scenarios in the following.

### Model-independent analysis of $B^0_d\overline{B}^0_d$ mixing

In the first scenario we consider physics beyond the Standard Model that affects only $B^0_d\overline{B}^0_d$ mixing. For example, New Physics may introduce additional box diagrams that carry a different weak phase than the Standard Model box diagrams. This may change both the size of $B^0_d\overline{B}^0_d$ mixing and $CP$ asymmetries that result from an interference between direct decays and decays after mixing. Thus, this New Physics contribution would affect both $\Delta m_{B_d}$ and the $CP$ asymmetries $a_{CP}(\psi K^0_S)$ and $a_{CP}(\pi\pi)$ (or $a_{CP}(\rho\rho)$). In order to perform a model-independent analysis of the UT we make the following assumptions given in [1, 39, 40].

- In the presence of New Physics, $b \to c\bar{c}s$ and $b \to u\bar{u}d$ processes that respectively yield $CP$ asymmetries in $B \to J/\psi K^0_S$ and $B \to \rho^+\rho^-$ are mediated by Standard Model tree level diagrams.
- Though New Physics could yield significant contributions to $K^0\overline{K}^0$ mixing, the small value of $|\epsilon_K|$, however, forbids large deviations from the Standard Model.
- Unitarity of the three-family CKM matrix is maintained.

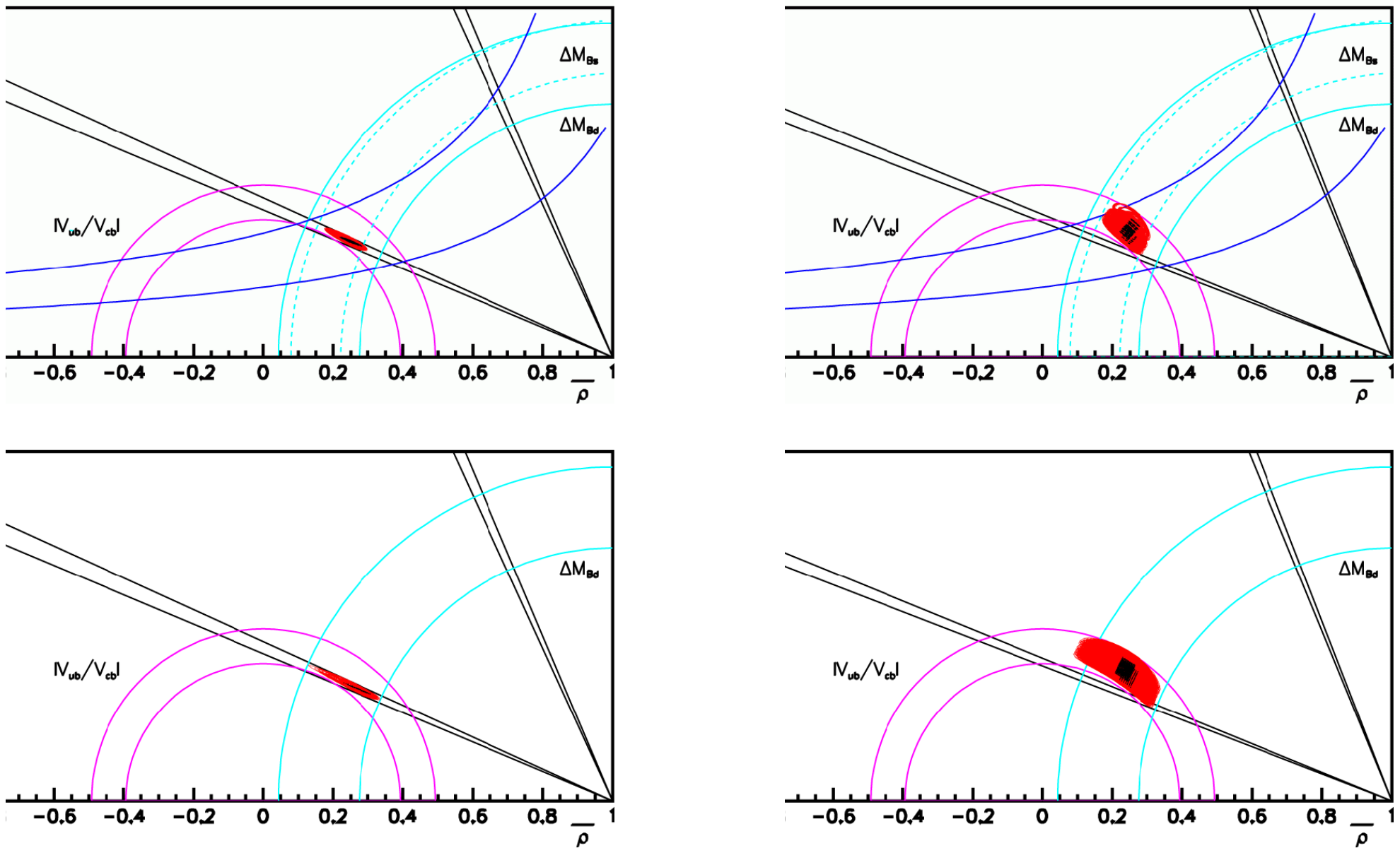

**Figure 5-3.** *Results of the Standard Model global fit in the* $\bar\rho - \bar\eta$ *plane, using extrapolations to* $10\,\mathrm{ab}^{-1}$*, a) for the eight standard observables, b) for the eight standard observables excluding* $|\epsilon_K|$ *and* $\Delta m_{B_s}$*, and c,d) for the same observables as in a,b) but excluding the* $\sin 2\beta$ *measurement from the fit. In the latter two plots the* $\sin 2\beta$ *bands have been shifted to* $\sin 2\beta = 0.687 \pm 0.049$.

Under these circumstances we can write an effective Hamiltonian that deviates from the Standard Model Hamiltonian by two new parameters, a scale parameter $r_d$ and a phase $\theta_d$ [42],

$$\frac{\langle B_d^0 \mid H_{eff}^{full} \mid \overline{B}_d^0\rangle}{\langle B_d^0 \mid H_{eff}^{StandardModel} \mid \overline{B}_d^0\rangle} = \left(r_d e^{i\theta_d}\right)^2 \tag{5.9}$$

These new parameters modify the $B_d^0\overline{B}_d^0$ oscillation frequency, $(\Delta m_{B_d})_{obs} = (\Delta m_{B_d})_{StandardModel} \cdot r_d^2$, and the *CP* asymmetries $a_{CP}(\psi K_S^0) = \sin(2\beta + 2\theta_d)$ and $a_{CP}(\rho\rho) = \sin(2\alpha - 2\theta_d)$, respectively. For $r_d = 1$ and $\theta_d = 0$ the Standard Model is retained. We have extended our method to perform 19-parameter global $\chi^2$ fits in the $\bar\rho - \bar\eta$ plane including the two new parameters $r_d$ and $\theta_d$. In order to ascertain the model independence we also perform our global fits by excluding $|\epsilon_K|$ and $\Delta m_{B_s}$ measurements.

Figure 5-4a shows the results of our 19-parameter global fits in the $\bar\rho - \bar\eta$ plane for a representative number of models using present averages of the standard eight observables. While the central values of the different models still lie inside the Standard Model region, the contours of some models are no longer constrained by the $\sin 2\beta$, $\Delta m_{B_d}$ and $\Delta m_{B_s}$ bounds. For those models the contours exhibit a banana shape rather than an elliptical shape, extending into a region with negative $\bar\rho$ that is not preferred by the Standard Model .

Figure 5-4b shows the corresponding $r_d - \theta_d$ contours. Since $\theta_d$ is specified up to a four-fold ambiguity, for the present $\sin 2\beta$ we expect phases near $0°, 42°, -137°$ and $-180°$. The figure, however, just displays contours with central values near zero; this is merely an artifact of the fit, depending on the starting value (here $0°$), the models investigated and the size of measurement errors with respect to the spacing of the ambiguous phases. By moving the starting value, we can also populate the other regions; for example for starting values of $45°$ and $-135°$, we dominantly populate contours around $42°$ and $-137°$, respectively. In order to find deviations from the Standard Model, we focus on the region near zero, the only one that includes the Standard Model point. Presently, due to the large non-probabilistic theoretical uncertainties and still quite sizable measurement errors, many models with large contours are accepted,

spanning a large overlay region in the $r_d - \theta_d$ plane including the Standard Model point ($r_d = 1, \theta_d = 0$). The part of the $\overline{\rho} - \overline{\eta}$ contours extending into the New Physics region correspond to small $r_d$ and large $\theta_d$ region in the $r_d - \theta_d$ plane. If we exclude the observables $|\epsilon_K|$ and $\Delta m_{B_s}$ from the global fits, we do not have enough sensitivity in the remaining six observables to determine all fit parameters. For example, fixing $r_d$ at specific values yields the expected result that all allowed contours lie within the $r_u = |V_{ub}|/|V_{cb}|/\lambda$ circle.

Adding the present measurements of $\sin 2\alpha$ and $\gamma$ to the global fits as listed in Table 5-1, yields moderate improvement. For the standard global fits with ten observables the size of overlay region in the in $\overline{\rho} - \overline{\eta}$ plane basically remains unchanged, as shown in Fig. 5-4c. The overlay region of contours in the $r_d - \theta_d$ plane depicted in Fig. 5-4c is only slightly reduced. However, the four-fold ambiguity becomes visible, although most contours are distributed among the two positive $\theta_d$ regions. The reason for populating more than one region here is the addition of a scan of the weak phase $\delta_\alpha$ in our global fits. For the present level of precision, the extra parameter provides sufficient flexibility for $\theta_d$ to jump from one region to the next. Though most acceptable fits retain the overlay region around zero, especially if $\delta_\alpha = 0$, the overlay region near $42^\circ$ looks rather similar to that near $0^\circ$.

Global fits that exclude $|\epsilon_K|$ and $\Delta m_{B_s}$ now have sufficient measurements to extract $r_d$ and $\theta_d$. The corresponding results in the $\overline{\rho} - \overline{\eta}$ plane and $r_d - \theta_d$ plane are depicted in Figs. 5-4e,f, respectively. Contours in the $\overline{\rho} - \overline{\eta}$ plane are still basically only constrained by the $r_u = |V_{ub}|/|V_{cb}|/\lambda$ circle, since present $\gamma$ measurements have experimental errors too large to have any significant impact. Though the majority of considered models still produces elliptical contours in the Standard Model region, some models exhibit banana-shaped contours that extend into New Physics regions, characterized either by large $\overline{\rho}$ and negative $\overline{\eta}$ or by negative $\overline{\rho}$ and small $\overline{\eta}$. Some contours lack a smooth shape, an artifact that is caused by joining a limited number of point across a large area. The $r_d - \theta_d$ contours again indicate the four-fold ambiguity. Since individual contours here are typically somewhat larger than equivalent ones of the ten-observable global fits, the two regions with positive $\theta_d$ central values are not disjoint any longer.

The $\overline{\rho} - \overline{\eta}$ contours of the global fits for extrapolations to $10\ \mathrm{ab}^{-1}$ using ten observables are shown in Fig. 5-5a; the corresponding $r_d - \theta_d$ contours are plotted in Fig. 5-5b. Both the $\overline{\rho} - \overline{\eta}$ and $r_d - \theta_d$ contours still cover a region consistent with the Standard Model. This is not too surprising, since, except for $|V_{ub}|_{exc}$, we have used present average values that are consistent with the Standard Model, but with reduced errors. The four-fold ambiguity for $\theta_d$ is still present. Similar results are obtained for global fits excluding $|\epsilon_K|$ and $\Delta m_{B_s}$ as depicted in Figs. 5-5c,d. If we shift the present central value of $\sin 2\beta$, for example, by $1\sigma$ to $0.687$ and use the extrapolated error of $0.007$, we obtain the results shown in Fig. 5-6a-d. The overlay region of contours in the $\overline{\rho} - \overline{\eta}$ is reduced and the overlay region contours in the $r_d - \theta_d$ plane cannot accommodate the Standard Model values of $r_d = 1$ and $\theta_d = 0$, indicating New Physics at the $> 2\sigma$ level.

**New phase in $b \to s\overline{s}s$ penguin processes**

The decay $B \to \phi K^0_S$ is a pure penguin process. As with the tree-dominated decay $B \to J/\psi K^0_S$, within the Standard Model the only phase comes from $B^0_d\overline{B}^0_d$ mixing. Thus, in the Standard Model, the *CP* asymmetry $a_{CP}(\phi K^0_S)$ should be dominated by the sine term, $S(\phi K^0_S)$, that should be equal to $a_{CP}(\psi K^0_S) = S(\psi K^0_S) = \sin 2\beta$ to within $\sim 4\%$ [43]. *BABAR* and Belle have measured the *CP* asymmetry of $B \to \phi K^0_S$, yielding a combined value of $S(\phi K^0_S) = 0.02 \pm 0.29$ [44, 45]. This represents a $2.7\sigma$ deviation from the observed $\sin 2\beta$ value. In physics beyond the Standard Model, however, new penguin contributions may be present that carry new particles in the loop. These contributions may introduce a new weak phase that may cause a deviation of $S(\phi K^0_S)$ from $\sin 2\beta$. Such contributions may be approximated by a mass insertion $(\delta_{23})$ in third-to-second family processes. In order to represent this New Physics effect in our global fit, we parameterize the *CP* asymmetry of $B \to \phi K^0_S$ with an additional phase $\theta_s$, $S(\phi K^0_S) = \sin 2(\beta + \theta_s)$. Since we have just one new parameter here, we determine our contours with the new phase in the $\overline{\rho} - \theta_s$ plane.

Using present average values of the standard eight observables plus $S(\phi K^0_S)$, Fig. 5-7a shows the results of 18-parameter global fits with extra phase $\theta_s$ in the $\overline{\rho} - \overline{\eta}$ plane. The overlay region of contours is very similar to that observed in the Standard Model analysis. The corresponding contours in the $\overline{\rho} - \theta_s$ plane are depicted in Fig. 5-7b. The four-fold ambiguity is clearly reproduced. For the present values of $S(\psi K^0_S)$ and $S(\phi K^0_S)$ the $\theta_s$ central values are expected to cluster near $23°, 113°, -67°$ and $-157°$. With present measurement precision, many models yield contours consistent with the Standard Model expectation. The difference between $S(\phi K^0_S)$ and $S(\psi K^0_S)$, which is consistent with zero, is absorbed by the phase $\theta_s$, leaving the $\overline{\rho}-\overline{\eta}$ plane unaffected. Including present $\sin 2\alpha$ and $\gamma$ measurements in the global fits produces contours shown in Figs. 5-7c,d. The results for global fits without $|\epsilon_K|$ and $\Delta m_{B_s}$ are plotted in Figs. 5-7e,f. For all three types of fits the overlay regions in the $\overline{\rho}-\overline{\eta}$ plane and in $\overline{\rho}-\theta_s$ plane are rather similar. In the ten-observables fits the $\overline{\rho}-\theta_s$ region seems to be slightly smaller than in the other fits. This difference is absorbed in slightly wider $\theta_s$ regions, respectively.

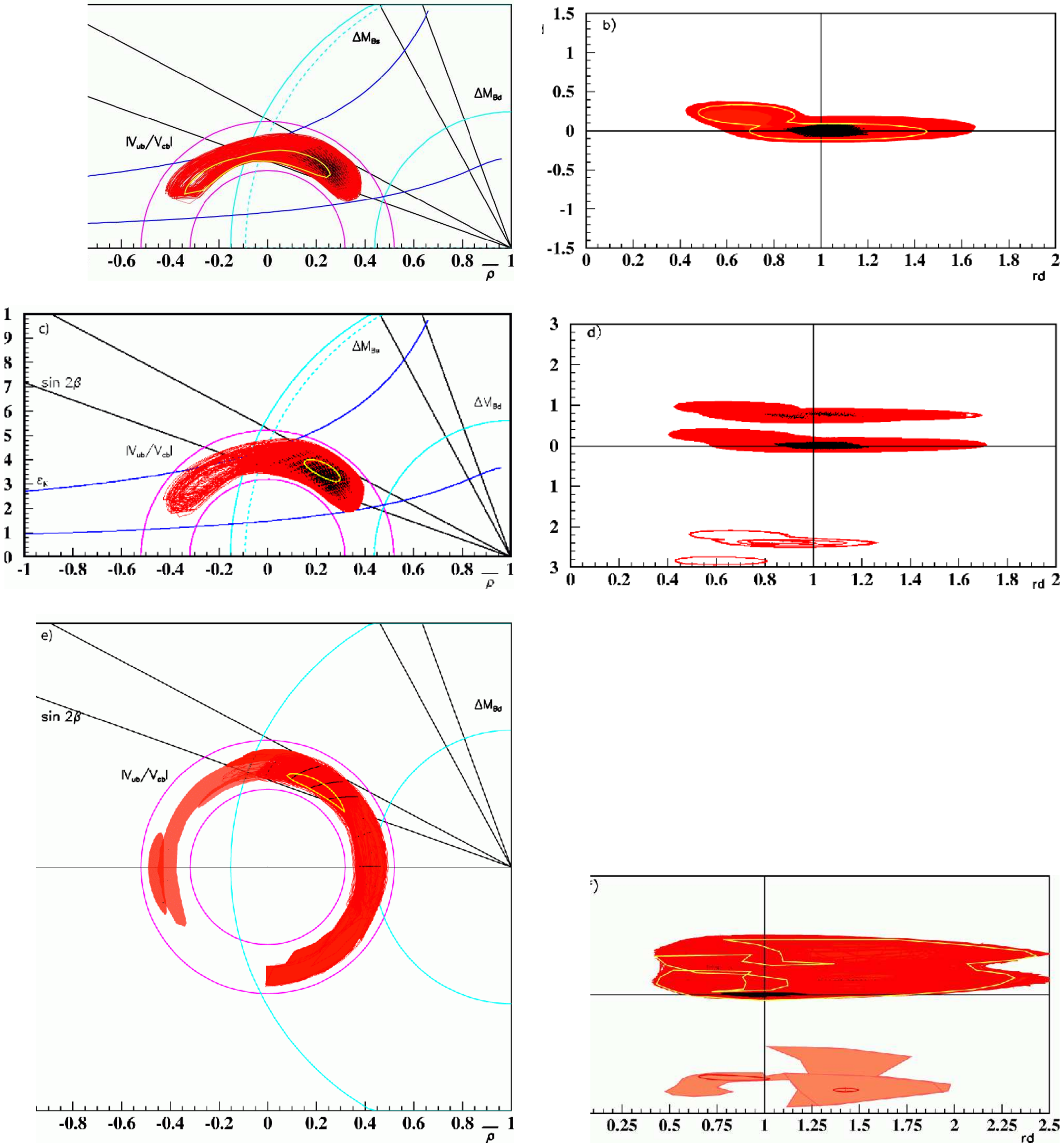

**Figure 5-4.** *Results of 19-parameter global fits with New Physics parameters $r_d$ and $\theta_d$ using present averages of all eight observables for a) the $\overline{\rho} - \overline{\eta}$ plane and b) for the $r_d - \theta_d$ plane. Plots c) and d) show corresponding results for including present $\sin 2\alpha$ and $\gamma$ measurements. Plots e) and f) show the corresponding results with excluding $|\epsilon_K|$ and $\Delta m_{B_s}$ but including present $\sin 2\alpha$ and $\gamma$ measurements. The Standard Model value corresponds to the cross point of the $r_d = 1$ line and the $\theta_d = 0$ line.*

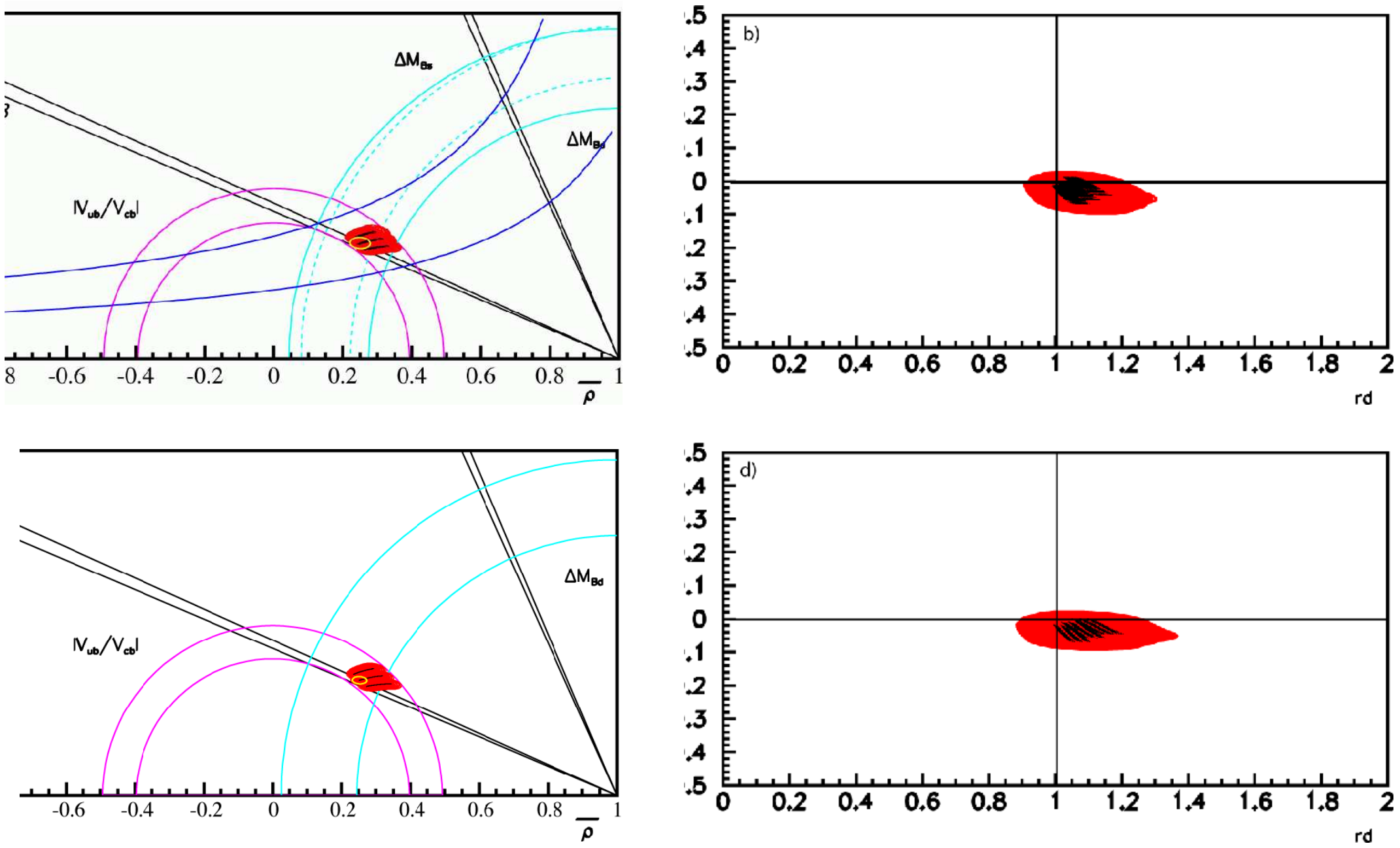

**Figure 5-5.** *Results of global fits with New Physics parameters* $r_d$ *and* $\theta_d$ *using extrapolations to* $10\ \mathrm{ab}^{-1}$ *for all eight standard observables plus* $\sin 2\alpha$ *and* $\gamma$ *for a) the* $\overline{\rho}-\overline{\eta}$ *plane and b) the* $r_d-\theta_d$ *plane. The corresponding results for global fits without* $|\epsilon_K|$ *and* $\Delta m_{B_s}$ *are shown in c) and d).*

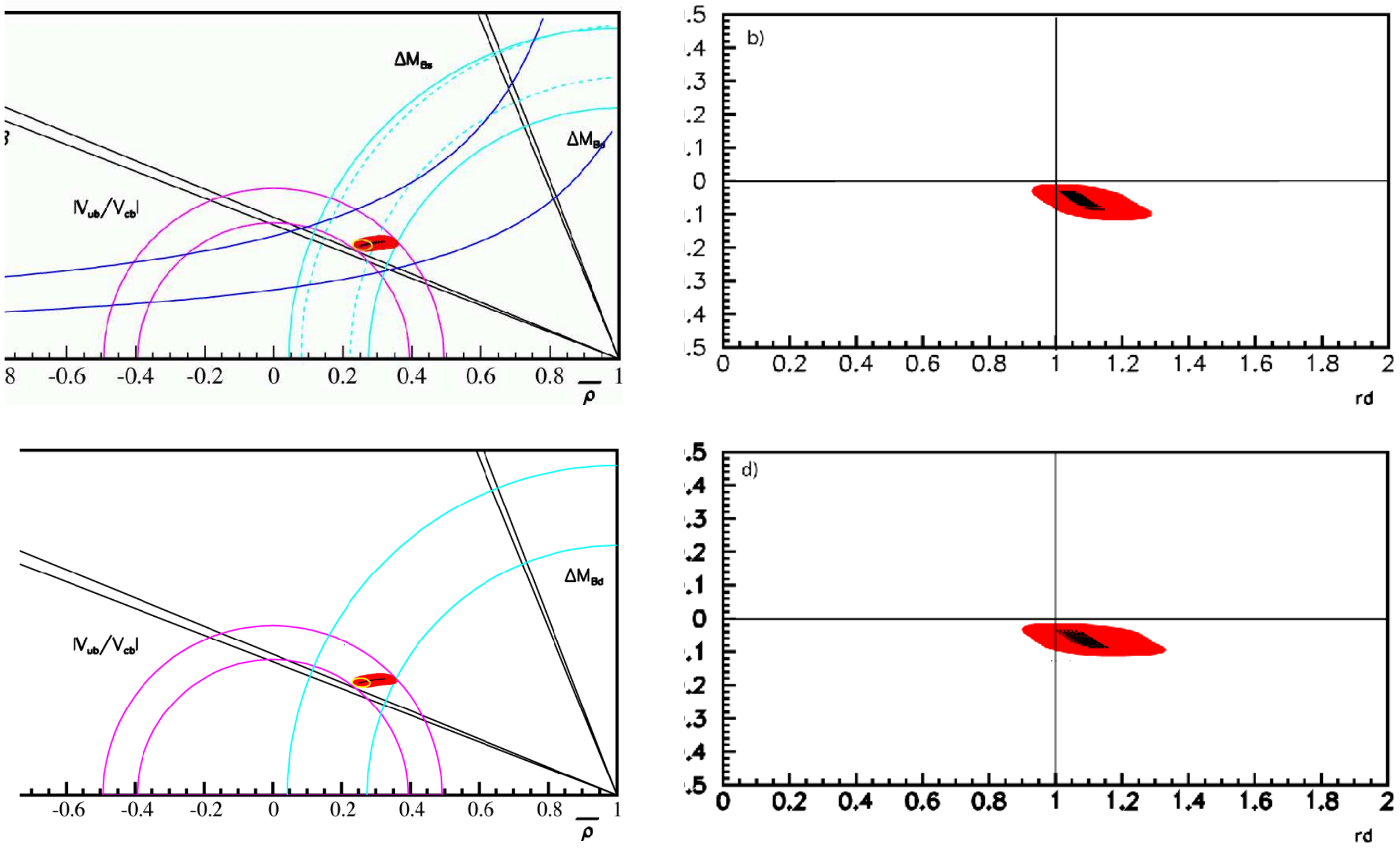

**Figure 5-6.** *Results of global fits with New Physics parameters* $r_d$ *and* $\theta_d$ *using the extrapolations to* $10\ \mathrm{ab}^{-1}$ *for all eight standard observables plus* $\sin 2\alpha$ *and* $\gamma$ *for a) the* $\overline{\rho}-\overline{\eta}$ *plane and b) the* $r_d-\theta_d$ *plane. The corresponding results for global fits without* $|\epsilon_K|$ *and* $\Delta m_{B_s}$ *are shown in c) and d). The central value of* $\sin 2\beta$ *has been reduced to* $0.687 \pm 0.007$.

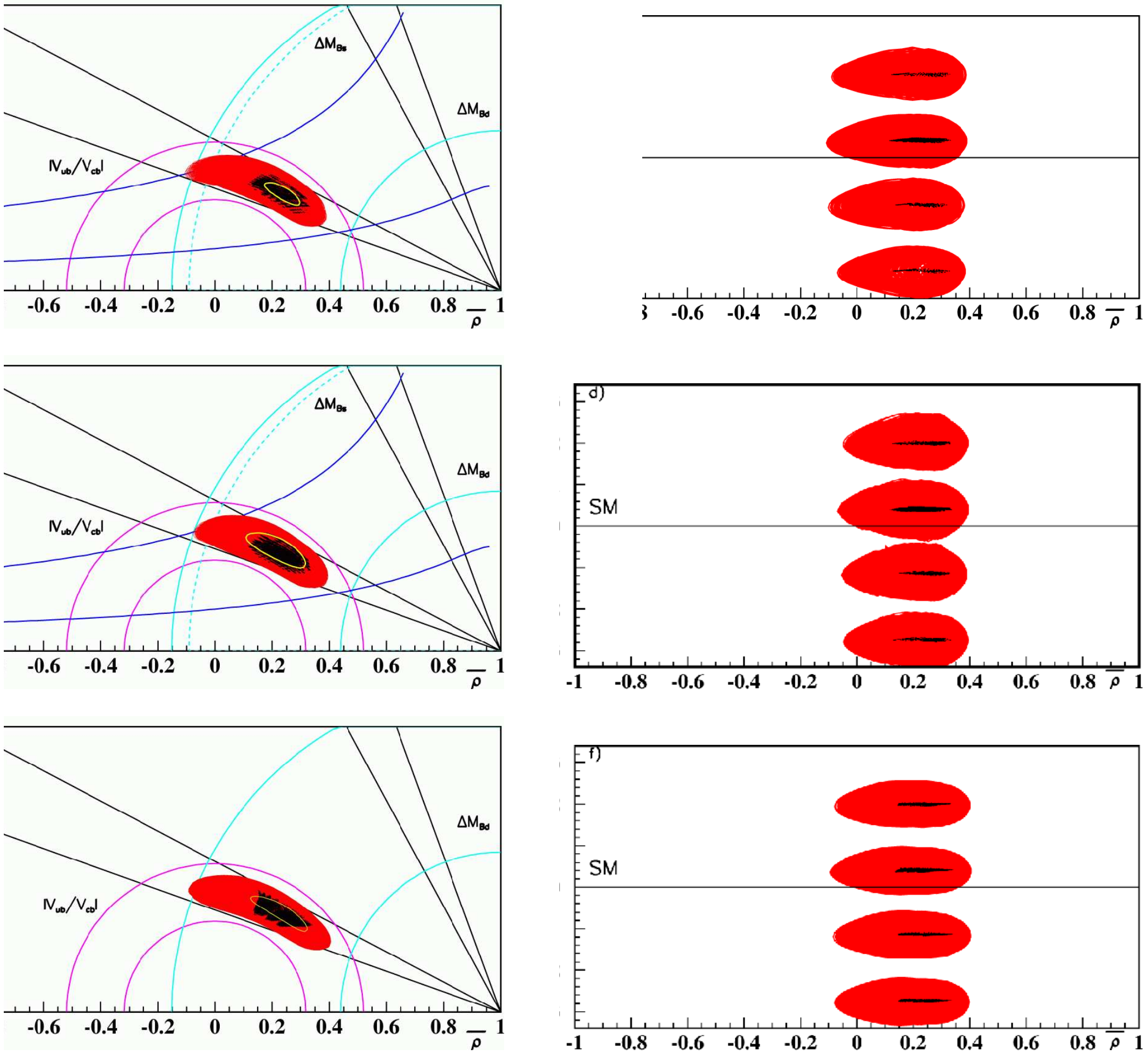

**Figure 5-7.** *Results of global fits with extra phase $\theta_s$ for present measurements of the eight standard observables, for a) the $\overline{\rho}-\overline{\eta}$ plane and b) the $\overline{\rho}-\theta_s$ plane. Corresponding results that include present $\sin 2\alpha$ and $\gamma$ measurements are depicted in plots c) and d), while those that exclude $|\epsilon_K|$ and $\Delta m_{B_s}$ but include $\sin 2\alpha$ and $\gamma$ are shown in e) and f), respectively.*

In order to explore the sensitivity of this approach at high luminosities we perform global fits with the extrapolations to $10\ \mathrm{ab}^{-1}$ specified in tables 5-1 and 5-2 as well as with $S(\phi K^0_S)$ extrapolated to $50\ \mathrm{ab}^{-1}$. In the latter case, we assume $S(\phi K^0_S) = 0.6 \pm 0.015$, representing a 7.5 $\sigma$ deviation from $S(\psi K^0_S) = 0.736 \pm 0.007$.

Figures 5-8a,b (c,d) show the results of our global fits for extrapolations to $10\ \mathrm{ab}^{-1}$ for the eight standard observables (excluding $|\epsilon_K|$ and $\Delta m_{B_s}$), while Figs. 5-9a,b (c,d) depict the equivalent results if extrapolated $\sin 2\alpha$ and $\gamma$ measurements are included. In the $\overline{\rho}-\overline{\eta}$ plane the overlay region of contours is reduced similarly to that of corresponding Standard Model global fits. In the $\overline{\rho}-\theta_s$ plane, where we plot only the region closest to zero, the phase $\theta_s$ of all

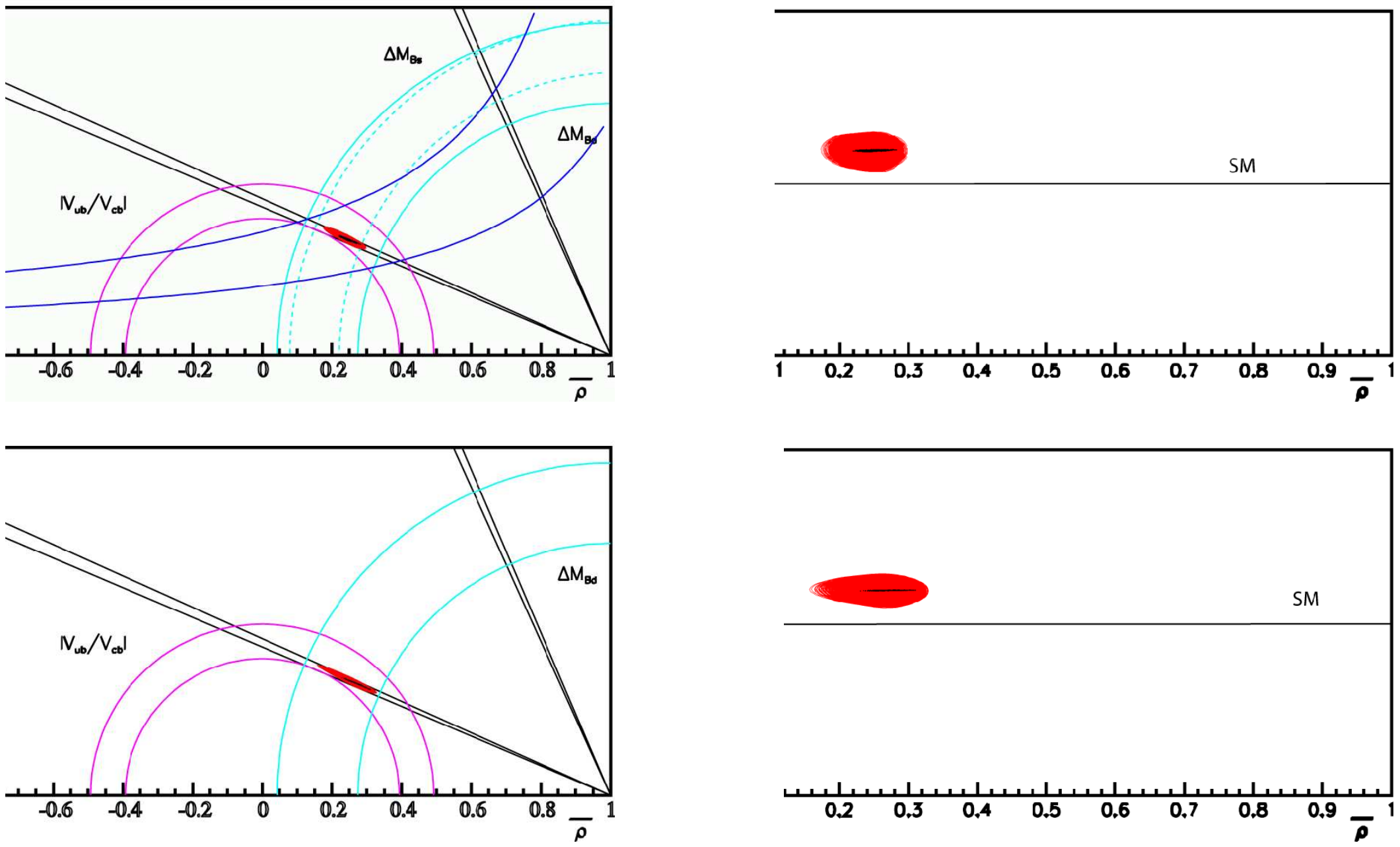

**Figure 5-8.** *Results of a global fit with an extra phase, $\theta_s$, using extrapolations of the eight standard observables to* $10\ \mathrm{ab}^{-1}$ *for a) the $\overline{\rho} - \overline{\eta}$ plane and for b) the $\overline{\rho} - \theta_s$ plane. The results without $|\epsilon_K|$ and $\Delta m_{B_s}$ are shown in c) and d), respectively.*

investigated models becomes inconsistent with the Standard Model value. For the global fits with ten observables, the segmentation of the overlay region of $\overline{\rho} - \theta_s$ contours into three parts is caused by scanning three values of $\delta_\alpha$. Since $\delta_\alpha$ enters only in the $\sin 2\alpha$ term it has an affect on $\overline{\rho}$ but not on $\theta_s$. In our parameterization, negative (positive) values of $\delta_\alpha$ yield large (small) values of $\overline{\rho}$, explaining the observed structure in Fig. 5-9d.

Figures 5-10a,b (c,d) respectively show the resulting $\overline{\rho} - \overline{\eta}$ contours and $\overline{\rho} - \theta_s$ contours for extrapolations to $50\ \mathrm{ab}^{-1}$. The number of accepted models is significantly reduced, as is the size of individual contours. While the contours in the $\overline{\rho} - \overline{\eta}$ plane are still in good agreement with the Standard Model-allowed range, the inconsistency between the observed phase $\theta_s$ and the Standard Model value increases. Such an observation would indicate the presence of New Physics.

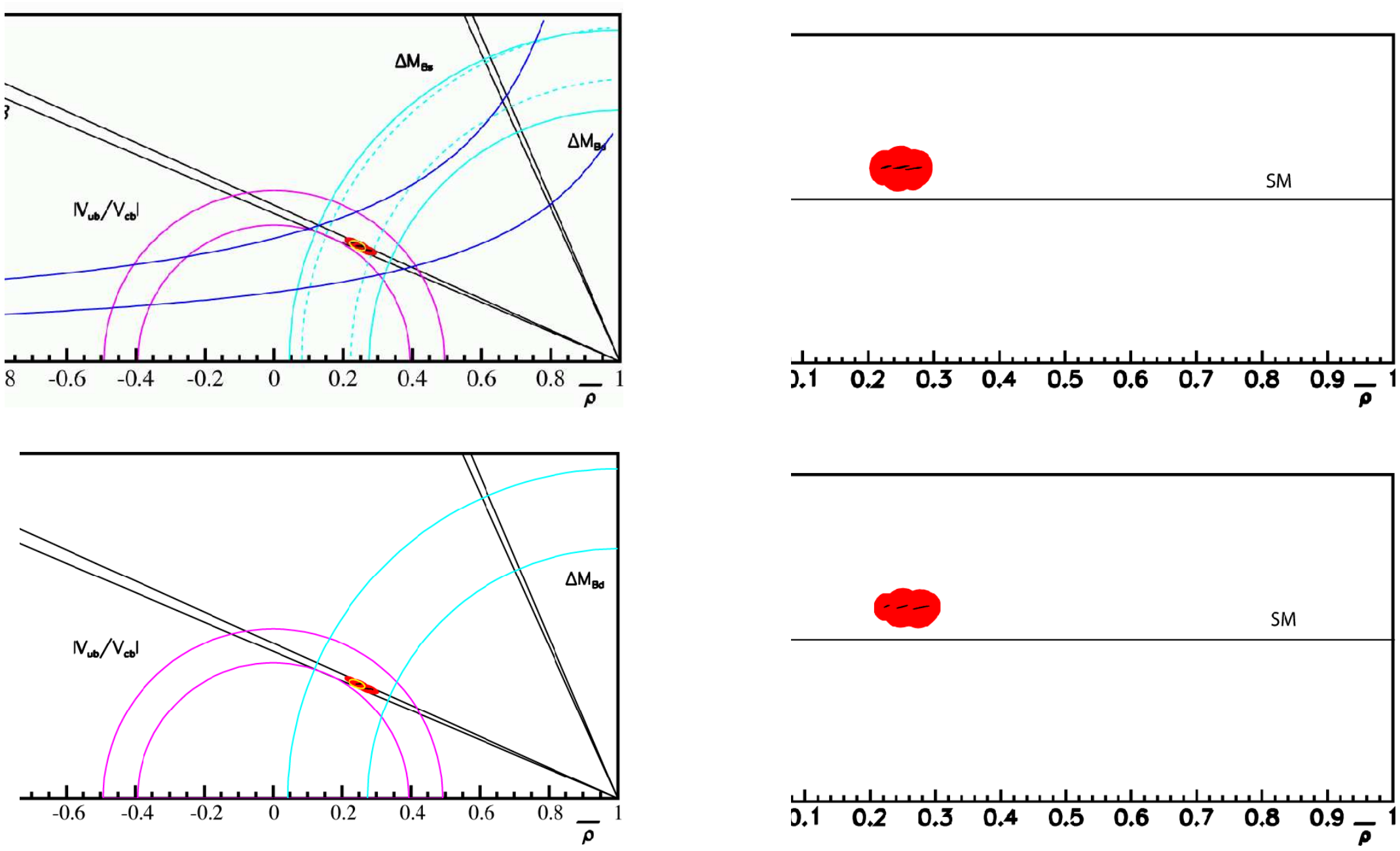

**Figure 5-9.** *Results of a global fit with an extra phase, $\theta_s$, using extrapolations of the eight standard observables plus* $\sin 2\alpha$ *and* $\gamma$ *to* $10\ \mathrm{ab}^{-1}$ *for a) the* $\overline{\rho} - \overline{\eta}$ *plane and for b) the* $\overline{\rho} - \theta_s$ *plane. The results without* $|\epsilon_K|$ *and* $\Delta m_{B_s}$ *are shown in c) and d), respectively.*

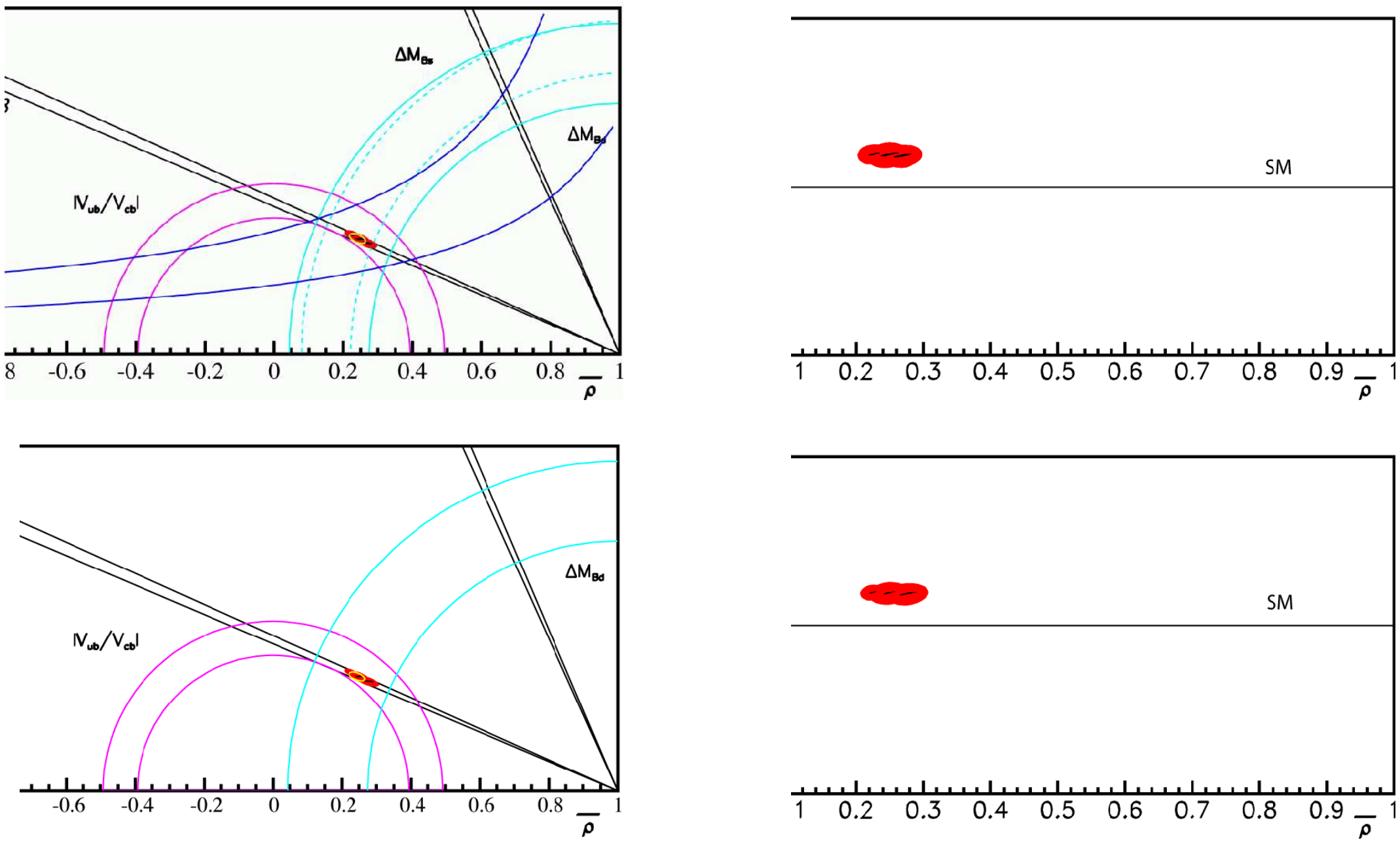

**Figure 5-10.** *Results of a global fit with an extra phase, $\theta_s$, using extrapolations of the eight standard observables plus* $\sin 2\alpha$ *and* $\gamma$ *to* $50\ \mathrm{ab}^{-1}$ *for a) the* $\overline{\rho} - \overline{\eta}$ *plane and for b) the* $\overline{\rho} - \theta_s$ *plane. The results without* $|\epsilon_K|$ *and* $\Delta m_{B_s}$ *are shown in c) and d), respectively.*

**Conclusions**

At the present level of precision, the CKM parameters $\overline{\rho}, \overline{\eta}$ are in good agreement with the Standard Model expectations. Due to large non-probabilistic theoretical uncertainties and still sizable experimental errors, the global fits still accept many models and yield wide contours in the $\overline{\rho} - \overline{\eta}$ plane leaving sufficient room for New Physics contributions. With the present level of precision, however, $CP$ violation in the $B$ system is clearly established without using any direct input of $CP$ violation in the $K^0\overline{K}^0$ system. For $10~\mathrm{ab}^{-1}$ we expect that both non-probabilistic theoretical uncertainties and experimental errors will be significantly reduced, such that Standard Model global fits may already achieve the necessary sensitivity to establish an inconsistency among the different observables.

In order to look for extensions of the Standard Model, model-independent analyses are an important tool. In this report we have explored New Physics scenarios in $B^0_d\overline{B}^0_d$ mixing and in $b \rightarrow s\overline{s}s$ decays. Presently, the inclusion of two new parameters, $r_d$ and $\theta_d$, in the global fits to represent New Physics in $B^0_d\overline{B}^0_d$ mixing does not have any significant effects on the $\overline{\rho} - \overline{\eta}$ plane. The overlay region of contours in the $r_d - \theta_d$ plane is rather large, and includes the Standard Model value. For our choice of central values and for the improved precisions expected at $10~\mathrm{ab}^{-1}$ several accepted models are still consistent with the Standard Model both in the $\overline{\rho} - \overline{\eta}$ plane and the $r_d$ and $\theta_d$ plane. In order to detect a deviation from the Standard Model, one observable must deviate from its present value. For example, if the central value of $\sin 2\beta$ at $10~\mathrm{ab}^{-1}$ is lowered from its present value by $1\sigma$, we expect to find an inconsistency with the Standard Model in $r_d$ and $\theta_d$ plane, whereas the $\overline{\rho} - \overline{\eta}$ plane still would not indicate a problem.

The observed deviation of $S(\psi K^0_S)$ from $S(\phi K^0_S)$ is interesting but not yet significant. If it is caused by New Physics, we can parameterize this effect in a model-independent way by including an extra phase $\theta_s$. Our analysis in the $\overline{\rho} - \theta_s$ plane would reveal a deviation from the Standard Model , if a significant difference between $S(\psi K^0_S)$ and $S(\phi K^0_S)$ remains at high luminosities. Depending upon the actual difference between $S(\psi K^0_S)$ and $S(\phi K^0_S)$, luminosities of $10 - 50~\mathrm{ab}^{-1}$ are necessary to establish an inconsistency with the Standard Model.

Our extrapolations to high luminosities show that significant deviations in measurements yield observable inconsistencies with the Standard Model . Additional measurements, such as $\sin 2\alpha$, $\sin\gamma$ and $\sin(2\beta + \gamma)$, provide further important constraints. Though we already have incorporated $\sin 2\alpha$ and $\sin\gamma$ measurements into our global fits, we expect to use additional observables in the future apart from a real measurement of $\Delta m_{B_s}$ expected for 2005. We further expect that both measurement errors and non-probabilistic uncertainties can be reduced according to our expectations to reach the anticipated precisions. Furthermore, our analysis is sufficiently flexible to incorporate other model-independent parameterizations of New Physics into our global fits.

In order to ascertain that measurements in the $B^0_d\overline{B}^0_d$ system are inconsistent with the CKM sector in the Standard Model, it is mandatory to perform precision measurements of the sides and angles of the Unitarity Triangle. For some observables, such as $S(\phi K^0_S)$ and $a_{CP}(\rho^+\rho^-)$, high precision measurements are only achievable at a Super $B$ Factory . Combining results in the $CP$ sector with those of rare $B$ decays allows us to establish a pattern that is characteristic for a particular extension of the Standard Model . For example in the case of SUSY, we actually should be able to ascertain which SUSY breaking scheme has been adopted by nature. Since we can track the influence of non-probabilistic uncertainties of theoretical parameters in our method (see discussion in [8]), we are capable to distinguish between effects caused by non-probabilistic uncertainties and discrepancies among measurements. This unique property distinguishes our approach from other techniques of determining CKM parameters and for searching for New Physics beyond the Standard Model in the CKM sector.

This work was supported by the Norwegian Research Council. I would like to thank D. Hitlin and Y. Okada for fruitful discussions.

### 5.2.2 Global fit to the Wilson Coefficients

$>\!\!-$ JoAnne Hewett $-\!\!<$

Rare decays of the $B$ meson which proceed through electroweak penguin diagrams provide a stringent test of the Standard Model by probing the indirect effects of new interactions in higher order processes. In particular, the exploration of loop-induced couplings examine the detailed structure of the Standard Model at the level of radiative corrections where the Glashow-Iliopoulos-Maiani cancellations are important. The flavor changing transitions $B \to X_s\gamma$, and $B \to X_s\ell^+\ell^-$, where $\ell = e, \mu, \tau$, are especially sensitive to new physics at the electroweak scale, and can yield information on the masses and couplings of new virtual particles participating in the loops. Observables which are associated with these decays (and their corresponding exclusive modes), such as rates, kinematic distributions, and $CP$ asymmetries, can be combined in a global fit to determine contributions due to new physics. The Standard Model expectations for these transitions are discussed in Chapter 2.

As discussed in Section 2.2, the effective field theory for these transitions is governed by the operator product expansion with the effective Hamiltonian

$$\mathcal{H}_{eff} = -\frac{4G_F}{\sqrt{2}} V_{tb} V_{ts}^* \sum_i^{10} C_i(\mu) O_i \,. \tag{5.10}$$

The operators are listed in [46], with $O_{1,2}$ being the current-current operators, $O_{3,4,5,6}$ being the 4-quark operators, and

$$\begin{aligned} O_7 = \frac{e}{g_s^2}\bar{s}_\alpha\sigma^{\mu\nu}(m_bP_R + m_sP_L)b_\alpha F_{\mu\nu}\,,\ O_8 \ &= \frac{1}{g_s}\bar{s}_\alpha\sigma^{\mu\nu}(m_bP_R + m_sP_L)T^a_{\alpha\beta}b_\beta G^a_{\mu\nu}\,, \\ O_9 = \frac{e^2}{g_s^2}\bar{s}_\alpha\gamma^\mu P_L b_\alpha\bar{\ell}\gamma_\mu\ell\,,\ O_{10} \ &= \frac{e^2}{g_s^2}\bar{s}_\alpha\gamma^\mu P_L b_\alpha\bar{\ell}\gamma_\mu\gamma_5\ell\,, \end{aligned} \tag{5.11}$$

being the electroweak penguin operators. Here, $\alpha, \beta$ are color indices, the chiral structure is specified by the projectors $P_{L,R} = (1 \mp \gamma_5)/2$, and $F_{\mu\nu}$ and $G^a_{\mu\nu}$ denote the QED and QCD field strength tensors, respectively. The $C_i$ represent the Wilson coefficients corresponding to each operator, and contain the relevant short-distance physics. They are evaluated perturbatively at the electroweak scale $\mu_{\rm EWK}$, where matching conditions are imposed, and then evolved down to the scale $\mu_b \sim m_b$ via the renormalization group equations (RGE). The status of the computation of the QCD corrections to the effective Wilson coefficients at the scale $\mu_b$ , $\tilde{C}_i^{eff}(\mu_b)$, is reviewed in Sections 2.2 – 2.4. The effective coefficient of the magnetic dipole operator, $\tilde{C}_7^{eff}(\mu_b)$, mediates the radiative transition $B \to X_s\gamma$, while $\tilde{C}_{7,9,10}^{eff}(\mu_b)$ all participate in the decay $B \to X_s\ell\ell$. Expressions for the effective Standard Model coefficients at the order of NLL can be found in [47].

A simultaneous experimental determination of the magnitude and sign of the Wilson coefficients provides a powerful and model-independent test of the Standard Model. The procedure for such a global fit is outlined in [48], [49], [50], [51], [52]. In general, the presence of new physics can affect a global fit to the Wilson coefficients in three ways:

- the numerical values for the coefficients are found to agree with Standard Model expectations; in this case the new physics is either decoupled or non-existent.
- A quality fit is obtained, but the fit values deviate from the Standard Model expectations.
- The $\chi^2$ value for the best parameter fit is found to be large and cannot be accounted for by an under estimation of systematic uncertainties. This case indicates the existence of additional non-Standard Model operators, such as right-handed operators [53], or new $CP$ phases.

The coefficients at the matching scale can be written in the form

$$C_i(\mu_{\rm EWK}) = C_i^{\rm SM}(\mu_{\rm EWK}) + \frac{\alpha_s}{4\pi} C_i^{\rm NP}(\mu_{\rm EWK})\,, \tag{5.12}$$

where $C_i^{\rm NP}$ represents the contributions from new interactions at the electroweak scale. Note that the factor of $\alpha_s/4\pi$ is present due to our choice of normalization for the electroweak penguin operators in Eq. (5.11). Higher order corrections to the new physics contributions to the matching conditions are negligible are usually neglected. Determination of the $C_i^{\rm NP}(\mu_{\rm EWK})$ is complicated since the RGE evolved coefficients $\tilde{C}_i^{eff}(\mu_b)$ are the quantities which mediate the decays and thus determined by experiment. The effects of operator mixing from the evolution to the scale $\mu_b$, charm penguin contributions, as well as the penguin contributions that restore the renormalization scheme independence of the matrix element must all be taken into account when placing constraints on the $C_i^{\rm NP}(\mu_{\rm EWK})$. Theoretical errors from missing higher order corrections, as well as the imprecisely known values of the charm-quark mass and the scale $\mu_b$ enter into the determination of the new contributions. Expressions for the modified NLL effective Wilson coefficients at the scale $\mu_b$, including new physics effects from the electroweak scale, are given in [48], [52].

Measurement of $B(B \to X_s\gamma)$ alone constrains the magnitude, but not the sign of $\tilde{C}_7^{eff}(\mu_b)$. Due to operator mixing, this effective coefficient contains the new physics contributions to $C_i^{\rm NP}(\mu_{\rm EWK})$ for both $i = 7, 8$. Inclusive radiative decays thus probe the possible values for new contributions to both the magnetic and chromomagnetic operators. The bounds from recent data ($B(B \to X_s\gamma) = (3.34 \pm 0.38) \times 10^{-4}$ with a photon energy cut of $E_\gamma.m_b/20$) are displayed in Fig. 5-11. In this figure, the Wilson coefficients have been normalized to the Standard Model expectations; $\xi_i(\mu_b)$ represents this ratio with $\xi_i^{\rm SM}(\mu_b) = 1$. The theoretical uncertainty from the prescription of the charm-quark mass has been taken into account. Note that from $B \to X_S\gamma$ alone, large values of $C_8^{\rm NP}(\mu_{\rm EWK})$ are allowed, even in the region where $C_7^{\rm NP}(\mu_{\rm EWK}) \simeq 0$. Bounds on $B(B \to X_s g)$ must then be employed to limit the size of the chromomagnetic contributions; these correspond to the horizontal lines in the figure. Future reductions in the experimental and theoretical errors associated with $B \to X_s\gamma$ will tighten these bounds somewhat, but other observables are needed to substantially improve the constraints.

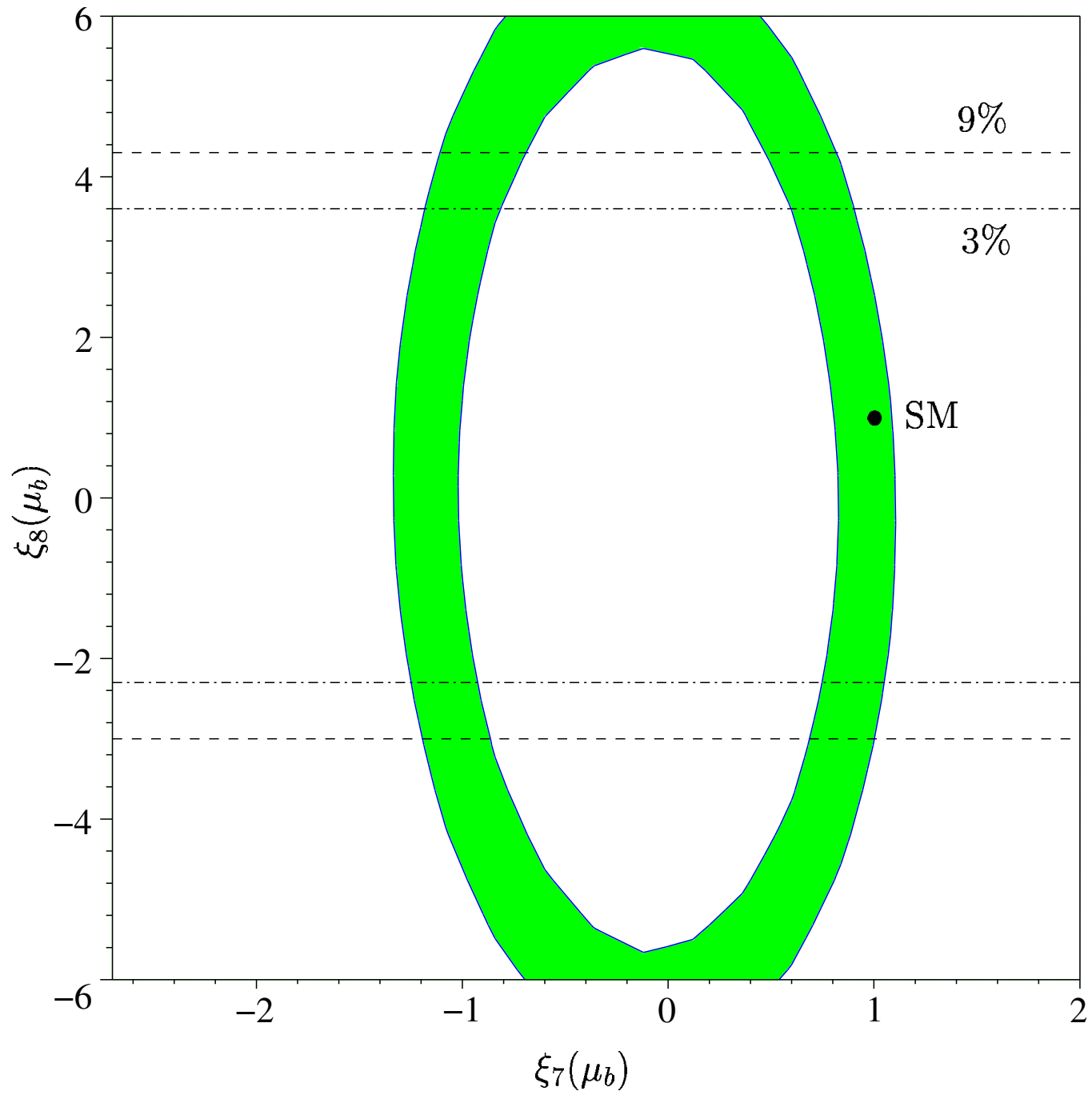

**Figure 5-11.** *Constraints on the scaled new physics contributions $\xi_{7,8}(\mu_b)$. The upper and lower bounds on $\xi_8(\mu_b)$ from the 90% CL experimental limit $B(B \to X_s g) < 9\%$ (dashed lines) and for a future limit of $3\%$ (dash-dotted lines) is also displayed. From [48].*

Measurement of the kinematic distributions associated with the final state lepton pair in $B \to X_S/K/K^* + \ell^+\ell^-$, as well as the rate for $B \to X_s\gamma$, allows for the determination of the sign and magnitude of all the Wilson coefficients for the contributing operators in a model-independent fashion. We note in passing that there is a trade-off in the utilization of the inclusive versus exclusive semi-leptonic modes: $B \to K/K^* + \ell\ell$ is plagued with form factor uncertainties, while $B \to X_s\ell\ell$ suffers from smaller statistics. A determination of $\tilde{C}_{9,10}^{eff}$ from present data on the rates for the exclusive decays alone is is given in Ref. [48], [52] for limiting values of $\tilde{C}_7^{eff}$.

Here, a Monte Carlo analysis is performed in order to ascertain how much quantitative information will be available at a Super $B$ Factory. In addition to $B(B \to X_s\gamma)$, the lepton pair invariant mass distribution and forward-backward asymmetry in $B \to X_s\ell\ell$ are considered for $\ell = e, \mu$. We note that ultra-large data samples are necessary in order to determine the invariant mass and forward-backward asymmetry distributions. The accuracy to which these can be determined is discussed in Chapter 2. Recall that the asymmetry has the form $A(q^2) \sim C_{10}(ReC_9 f_1(q^2) + C_7 f_2(q^2))$ with $f_{1,2}$ being kinematic form factors, and hence are extremely sensitive probes of the Wilson coefficients. Monte Carlo data is generated by dividing the lepton pair invariant mass spectrum into nine bins and assuming that the Standard Model is realized. Six of the bins are taken to lie below the $J/\psi$ resonance, and three bins are in the high $M_{\ell\ell}$ region above the $\psi'$ pole. The region near $q^2 = 0$ has been cut in order to avoid the photon pole. The generated data is statistically fluctuated using a normalized Gaussian distributed random number procedure. A 3-dimensional $\chi^2$ fit to the coefficients $\tilde{C}_{7,9,10}^{eff}(\mu_b)$ is performed over the observables. The 95% CL allowed regions as projected onto the $C_9^{eff}(\mu_b) - C_{10}^{eff}(\mu_b)$ and $C_7^{eff}(\mu_b) - C_{10}^{eff}(\mu_b)$ planes are shown in Fig. 5-12. The diamond represents the Standard Model expectations. The three curves correspond to 150, 500, and 3000 events in invariant mass and asymmetry distributions. It is clear that large luminosities are needed in order to focus on regions centered around the Standard Model. If deviations from the Standard Model expectations are found, then the coefficients must be evolved up to the matching scale to determine the value of $C_i^{NP}$ or the operator basis must be extended to incorporate new operators or new sources of $CP$ violation. An example of an extended operator basis is given in the next section.

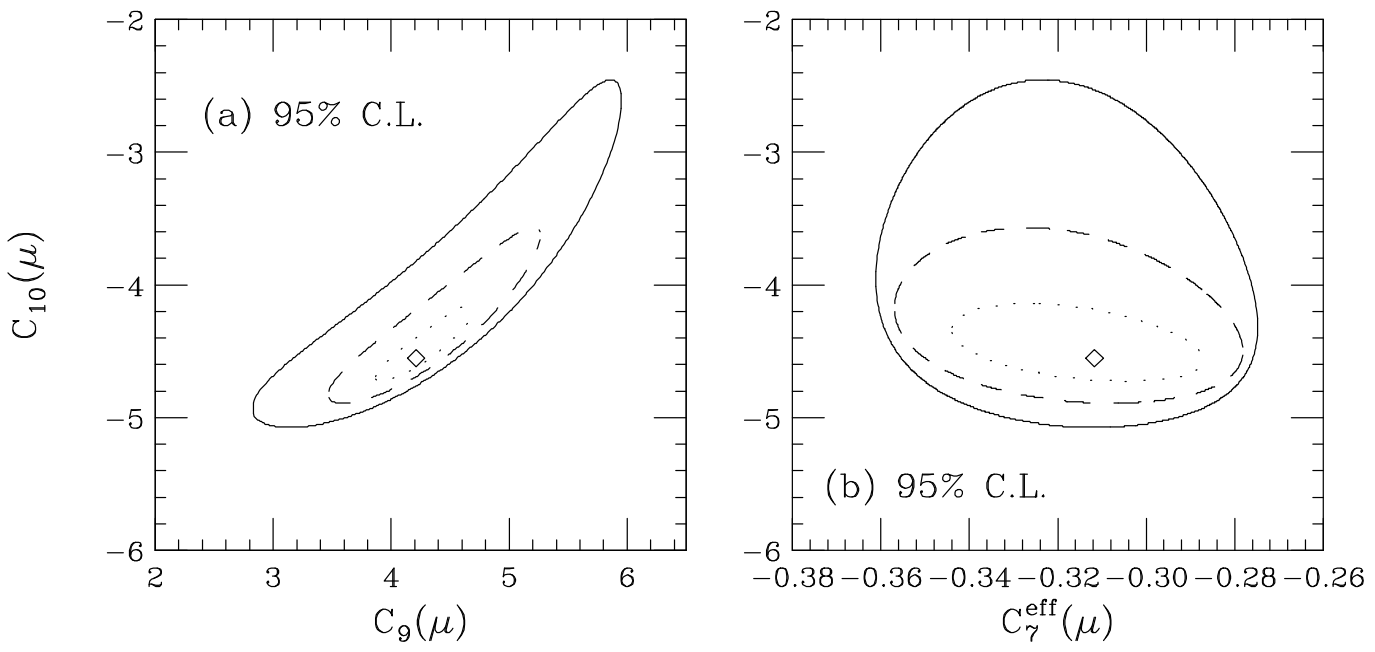

**Figure 5-12.** *Results of a global fit to the Wilson coefficients described in the text as projected into the $C_9^{eff}(\mu_b) - C_{10}^{eff}(\mu_b)$ and $C_7^{eff}(\mu_b) - C_{10}^{eff}(\mu_b)$ planes. The 95% CL allowed regions lie inside the curves. The solid, dashed, and dotted contours are for 150, 500, and 3000 events, respectively, in the kinematic distributions for $B \to X_s\ell\ell$.*

### 5.2.3 (More) Model-Independent Effects in Rare Decays

>– Gudrun Hiller –<

Section 5.2.2 discussed the “standard strategy” for extracting Wilson coefficients model-independently from rare radiative and semileptonic $b$-decays [49] [50]. Real $b \to s\gamma$, $b \to sg$ and $b \to s\ell^+\ell^-$ decay data based analyses have been performed by [51][52]. Here we briefly discuss the assumptions that go into such an analysis and propose ways to go beyond them. These assumptions are:

*i* no further operators than already present in the Standard Model,

*ii* no beyond-the-Standard Model $CP$ violation, and

*iii* New Physics contributions to four-quark operators are negligible.

Among these, dropping the requirement of no New Physics phases poses no difficulty in principle, but does invoke practical problems, since the number of independent parameters is thereby doubled. It requires sufficiently precise measurement of several observables, in particular $CP$ asymmetries in $b \to s\gamma$ decays, the forward-backward asymmetry in $B \to X_s \ell^+ \ell^-$ and $B \to K^* \ell^+ \ell^-$ decays or even better, doing a full angular analysis in $B \to (K^* \to K\pi)\ell\ell$ (see Section 2.17). In the following, we discuss operators from physics beyond the Standard Model in Section 5.2.3, the impact of New Physics on four-quark operators in Section 5.2.3, and an analysis in an extended operator basis in Section 5.2.3.

Hadronic two-body $b \to s$ decays also receive contributions from the very same kind of New Physics operators discussed here, although their matrix elements have a much larger theoretical uncertainty than those involved in radiative and semileptonic decays. Asymmetries such as $\sin 2\beta$ measurements (see, *e.g.*, Section 5.3.1, or polarisation measurements such as $\Gamma_L/\Gamma$ for $B \to \phi K^*$ (see Sections 5.2.4 and 5.2.5 ) are theoretically cleaner than branching ratios. Finally, model-independent analyses in $b \to d$ transitions allows us to test the CKM paradigm of flavor violation.

**Operators beyond the Standard Model basis**

Right-handed (RH) currents, *i.e.*, contributions to operators with flipped chirality $L \leftrightarrow R$, are suppressed in the Standard Model and in models with minimal flavor violation by $\sim m_s/m_b$. In some scenarios, however, *e.g.*, with Left-Right symmetry, they can be sizable, and affect the model independent analysis [53]. The working hypothesis of no RH contributions to FCNC‘s can be tested, *e.g.*, with polarization studies as in $B \to K^{**}\gamma$ decay or angular analysis in $B \to K^* \ell^+ \ell^-$ decays, see Sections 2.10 and 2.17, or with $\Lambda_b$ decays at high energy colliders, see [54] and references therein. (A method to search for $V+A$ handedness in scalar and pseudoscalar operators is discussed below.) In $R$-parity violating SUSY LFV contributions to operators such as $(\bar{s}b)(\bar{\ell}\ell')$ are induced. They do not interfere with the fit analysis presented here.

Neutral Higgs boson exchange generates scalar and pseudoscalar couplings

$$O_S = \frac{e^2}{16\pi^2}(\bar{s}_L b_R)(\bar{\ell}\ell), \quad O_P = \frac{e^2}{16\pi^2}(\bar{s}_L b_R)(\bar{\ell}\gamma_5\ell) \tag{5.13}$$

which are relevant for $b \to s\ell^+\ell^-$ processes. They are constrained by the upper limit on the $B_s \to \mu^+\mu^-$ branching ratio [55] [48] [56] [1]

$$\sqrt{|C_S(\mu)|^2 + |C_P(\mu) + \delta_{10}(\mu)|^2} \le 3.3 \left[\frac{\mathcal{B}(B_s \to \mu^+\mu^-)}{2.0 \times 10^{-6}}\right]^{1/2} \times \left[\frac{|V_{tb}V_{ts}^*|}{0.04}\right]^2 \left[\frac{m_b(\mu)}{4.4\,\mathrm{GeV}}\right]\left[\frac{238\,\mathrm{MeV}}{f_{B_s}}\right]\left[\frac{1/133}{\alpha}\right], \tag{5.14}$$

as well as by the experimental bound $R_K \le 1.2$ on the ratio of the $B \to K\mu^+\mu^-$ to $B \to Ke^+e^-$ branching ratios, see Section 2.16.3, which is of similar magnitude [48]. (The Higgs contribution in the dielectron mode is suppressed by the tiny electron Yukawa). Both bounds are complementary because contributions from RH scalar and pseudoscalar operators $O'_{S,P}$ can be included in the $B_s \to \ell^+\ell^-$ branching ratio by $C_{S,P} \to C_{S,P} - C'_{S,P}$ and into the $B \to K\ell^+\ell^-$ spectrum by $C_{S,P} \to C_{S,P} + C'_{S,P}$. The breakdown of correlations between $\mathcal{B}(B_s \to \mu^+\mu^-)$ and $R_K$ shown in Fig. 2-19 would indicate the presence of both chirality contributions $C_{S,P}$ and $C'_{S,P}$. Combining both bounds yields an upper bound on the individual coefficients of $|C_{S,P}^{(\prime)}| \le 4.3$ [48], excluding large cancellations.

The scalar and pseudoscalar operators are important for a model independent analysis using $b \to s\mu^+\mu^-$ decays. The enhancement of the dimuon w.r.t. the electron channel with the same cuts on the dilepton mass can be of order 10 % [48]. An analysis including $O_{S,P}$ is shown in Section 5.2.3.

[1] The latest CDF bound is $\mathcal{B}(B_s \to \mu^+\mu^-) < 5.8 \cdot 10^{-7}$ @ 90 % C.L. [57].

#### New Physics contributions to four-quark operators

The QCD penguins $O_{3-6}$ appear in the Standard Model and many extensions to lowest order only through operator mixing. They enter the matrix element of $b \to s\gamma$ and $b \to s\ell^+\ell^-$ decays at the loop level. Hence, their impact is subdominant, and New Physics effects in QCD penguins are negligible within the current theoretical precision of $A_7^{\rm SM}$ and $A_9^{\rm SM}$, see, *e.g.,* [52]. Note that isospin breaking in $B \to K^*\gamma$ [58] and $B \to K^*\ell^+\ell^-$ [59] decays is sensitive to New Physics in the QCD penguins.

It is conceivable that the dynamics which generates large couplings to dileptons, *i.e.*, to the operators $O_{S,P}$, induces contributions to 4-Fermi operators with diquarks as well ($f$ denotes a fermion)

$$O_L^f = (\overline{s}_L b_R)(\overline{f}_R f_L), \quad O_R^f = (\overline{s}_L b_R)(\overline{f}_L f_R) \tag{5.15}$$

For muons we identify $C_{L,R}^{\mu} = e^2/(16\pi^2)(C_S \mp C_P)$. We assume that that the coupling strength is proportional to the fermion mass, which naturally arises in models with Higgs-boson exchange. Hence, the corresponding Wilson coefficients for $b$ quarks can be large. The constraint given in Eq. (5.14) implies [48]

$$\sqrt{|C_L^b(m_W)|^2 + |C_R^b(m_W)|^2} \leq \frac{e^2}{16\pi^2}\frac{m_b(m_W)}{m_\mu}\sqrt{2(|C_S(m_W)|^2 + |C_P(m_W)|^2)} \lesssim 0.06 \tag{5.16}$$

The operators $O_{L,R}^b$ enter radiative and semileptonic rare $b \to s$ decays at one-loop. In particular, $O_L^b$ mixes onto $O_9$ and the QCD penguins [48], whereas $O_R^b$ mixes onto the dipole operators [60]. The complete lowest order anomalous dimensions are given in [48].

With the bound in Eq. (5.16) the New Physics effect from $O_L^b$ is small, of the order of one percent for $\widetilde{O}_9$ and up to several percent for the QCD penguins. The renormalization effect induced by $\mathcal{O}_R^b$ can be order one for the chromomagnetic and a few $\times\, \mathcal{O}(10\%)$ for the photon dipole operator, respectively. Hence, contributions to the dipole operators are in general non-negligible. This is further discussed in Section 5.2.3.

#### Analysis with (pseudo)scalar operators

We extend the “standard” analysis to model-independently analyse $b \to s\gamma$ and $b \to s\ell^+\ell^-$ decays given in Section 5.2.2 by allowing for Higgs type induced contributions (see *ii* below) to four-Fermi operators [48]. We make the following assumptions:

- *i* no RH currents,
- *ii* the coefficients of the scalar and pseudoscalar operators are proportional to the fermion mass, and
- *iii* no beyond-the-Standard Model *CP* violation. We give the Standard Model values for completeness: $A_7^{\rm SM}(2.5\ {\rm GeV}) = -0.330$, $A_9^{\rm SM}(2.5\ {\rm GeV}) = 4.069$ and $A_{10}^{\rm SM} = -4.213$.

We start with the dipole operators. We normalize the Wilson coefficients in the presence of New Physics to the ones in the Standard Model , and denote this ratio by $\xi$, such that $\xi^{\rm SM} = 1$. We obtain to next-to-leading order in the Standard Model operator basis and to leading logarithmic approximation in $C_R^b$, see Section 5.2.3

$$\xi_7(m_b) = 0.514 + 0.450\,\xi_7(m_W) + 0.035\,\xi_8(m_W) - 2.319\,C_R^b(m_W), \tag{5.17}$$

$$\xi_8(m_b) = 0.542 + 0.458\,\xi_8(m_W) + 19.790\,C_R^b(m_W). \tag{5.18}$$

With the upper bound in Eq. (5.16) corrections of up to $14\%$ and $119\%$ to $\xi_7$ and $\xi_8$ from four-quark operators can arise [48].

The correlations between $\xi_7$ and $\xi_8$ from $\mathcal{B}(B \to X_s\gamma) = (3.34 \pm 0.38) \times 10^{-4}$ [61] and $\mathcal{B}(B \to X_s g) < 9\%$ [70] at NLO are shown in Fig. 5-13. We give the allowed $90\%$ C.L. regions at the $\mu_b$ scale for $C_R^b(m_W) = 0$ (left plot) and $C_R^b(m_W) = 0.06$ (right plot). From Fig. 5-13 one sees that $A_7 = 0$ for $C_R^b(m_W) = 0.06$ is allowed by present data on the $b \to sg$ branching fraction. This particular scenario could be excluded by an improved experimental analysis of $b \to sg$. Also, if $C_R^b(m_W)$ is near its upper bound, it implies a contribution to the matching conditions for $\widetilde{C}_{7,8}(m_W)$ in order to be consistent with experimental data. The large renormalization effects from (pseudo)scalar in the dipole operators can be avoided assuming $C_R^b(m_W) \simeq 0$, *i.e.*,

$$C_S + C_P = 0. \tag{5.19}$$

The absence of logarithms in the matching conditions for $\widetilde{C}_{7,8}(m_W)$ from neutral Higgs-boson exchange in a two-Higgs-doublet model type II [76, 344] is consistent with the fact that in this model Eq. (5.19) is satisfied [345]. This is also the case for the MSSM at large $\tan\beta$ [55] [344].

The smallness of $C_S + C_P$ in the MSSM is a feature of its Higgs sector. It also holds with flavor violation beyond the CKM matrix. Radiative corrections to Eq. (5.19) are small $|C_S + C_P|_{\rm MSSM} < 0.08$ [71]. A model that does have contributions to $C_S + C_P$, and hence $C_R^b$ up to the experimental limit is the Next-to-minimal supersymmetric Standard Model (NMSSM ) [71], see [71] for further implications for flavor physics.

If $C_R^b$ is negligible the bounds on the dipole operators are the same as in the “standard” analysis. One obtains the ranges at $\mu_b = 2.5$ GeV [48]

$$-0.36 \le A_7 \le -0.17 \quad \text{or} \quad 0.21 \le A_7 \le 0.42. \tag{5.20}$$

The constraint on the semileptonic scalar and pseudoscalar operators is given in Eq. (5.14). Bounds on the couplings of the vector and axial vector operators $O_{9,10}$ in the presence of $O_{S,P}$ can be obtained from using $b \to s e^+ e^-$ modes alone where Higgs effects are tiny. In addition, a finite value of the $B \to X_s \mu^+\mu^-$ branching ratio together with the upper bound on $C_{S,P}$ in Eq. (5.14) gives a lower bound. The combined results of Belle [72] and *BABAR* [73] for the inclusive $b \to s\ell^+\ell^-$ decays yield at 90% C.L. [48]

$$2.8 \times 10^{-6} \le \mathcal{B}(B \to X_s e^+ e^-) \le 8.8 \times 10^{-6}, \tag{5.21}$$

$$3.5 \times 10^{-6} \le \mathcal{B}(B \to X_s \mu^+ \mu^-) \le 10.4 \times 10^{-6}. \tag{5.22}$$

The statistical significance of the Belle (*BABAR*) measurements of $\mathcal{B}(B \to X_s e^+ e^-)$ and $\mathcal{B}(B \to X_s \mu^+ \mu^-)$ is $3.4\sigma$ $(4.0\sigma)$ and $4.7\sigma$ $(2.2\sigma)$, respectively. To be conservative, we also use in our analysis the 90 % C.L. limit [74]

$$\mathcal{B}(B \to X_s e^+ e^-) < 10.1 \times 10^{-6} \tag{5.23}$$

Allowed 90 % C.L. contours in the $A_9$–$A_{10}$ plane are shown in Fig. 5-14 [48]. The shaded areas are obtained from the dielectron bound Eq. (5.23) (outer ring) and the dimuon lower limit Eq. (5.22) (inner ring). The two remaining contours are from the $\mathcal{B}(B \to X_s e^+ e^-)$ measurement Eq. (5.21).

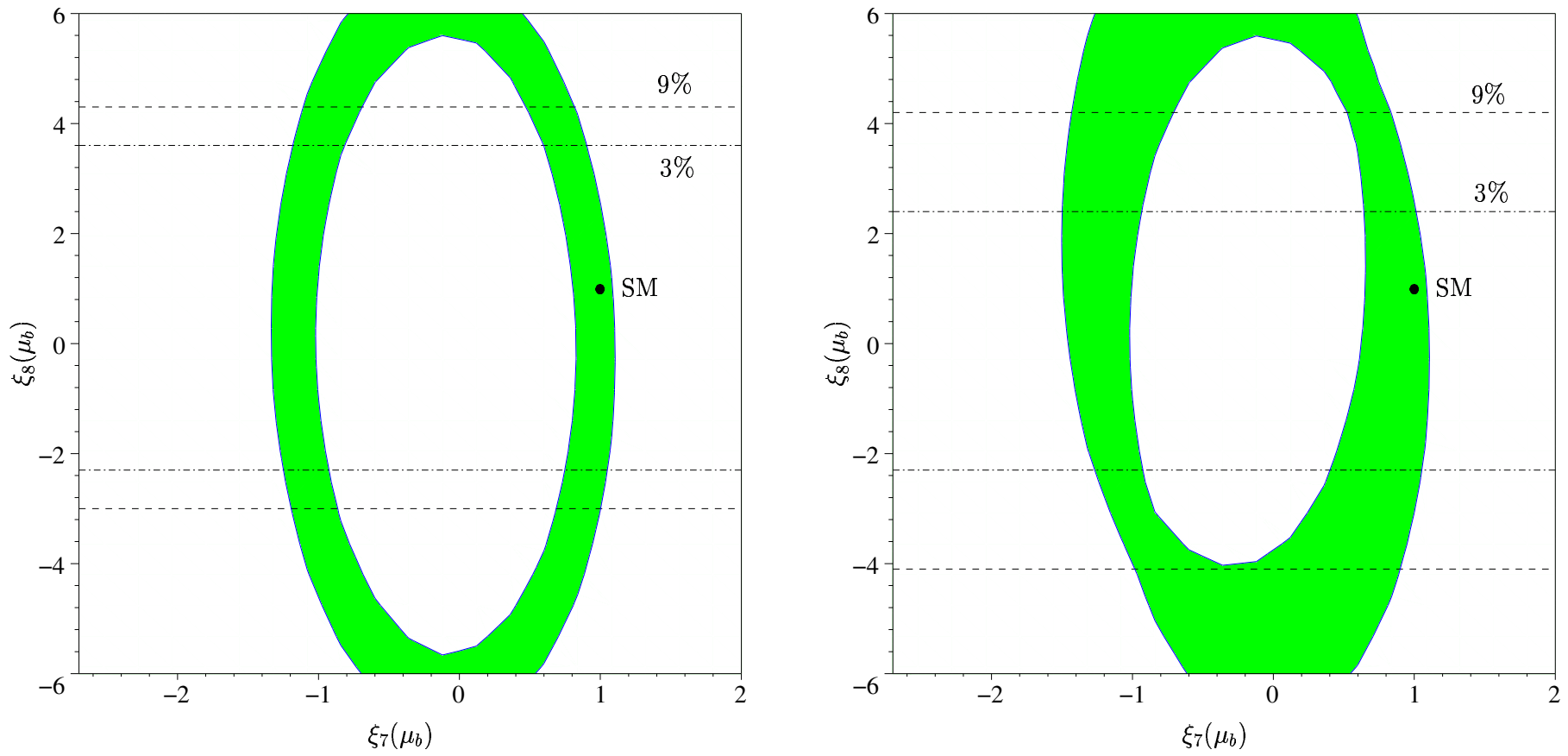

**Figure 5-13.** *Constraints on $\xi_{7,8}(\mu_b)$ from $\mathcal{B}(b \to s\gamma)$ for $C_R^b(m_W) = 0$ (left plot) and $C_R^b(m_W) = 0.06$ (right plot). Also shown are the bounds on $\xi_8(\mu_b)$ for the experimental limit $\mathcal{B}(B \to X_s g) < 9\%$ [70] (dashed) and for an assumed value of $\mathcal{B}(B \to X_s g) < 3\%$ (dash-dotted). Figure taken from [48].*

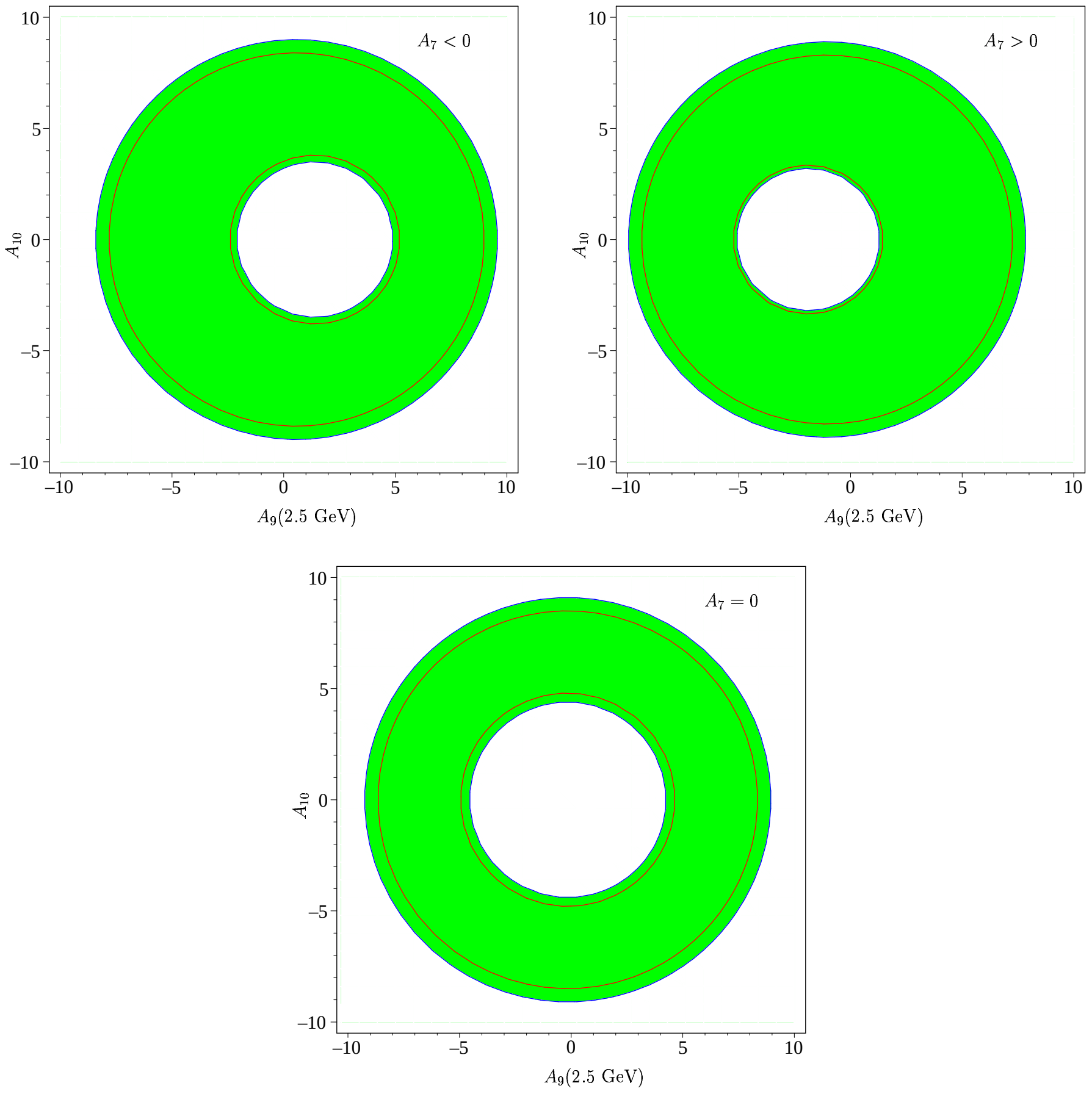

**Figure 5-14.** *Allowed 90 % CL regions in the $A_9$–$A_{10}$ plane in the presence of scalar and pseudoscalar operators from inclusive $b \to s\ell^+\ell^-$ and $b \to s\gamma$ decays for different values of $A_7$. Figure from [48].*

### 5.2.4 Probing New Physics in B $\to V_1V_2$ Decays

>— D. London, N. Sinha, R. Sinha —<

There are a great many tests for the presence of New Physics in $B$ decays [84]. Should a signal for New Physics be found, there are basically two ways to proceed. One can examine various models of physics beyond the Standard Model to see whether a particular model can account for the experimental results. Alternatively, one can perform a model-independent analysis to learn about general properties of the New Physics responsible for the signal. Most theoretical work has concentrated on the first approach.

For example, within the Standard Model, the $CP$-violating asymmetries in $B^0(t) \to J/\psi K_S^0$ and $B^0(t) \to \phi K_S^0$ both measure the $CP$ phase $\beta$, to a good approximation [85, 86]. However, although the *BABAR* measurement of the $CP$ asymmetry in $B^0(t) \to \phi K_S^0$ agrees with that found in $B^0(t) \to J/\psi K_S^0$ (within errors), the Belle measurement disagrees at the level of $3.5\sigma$ [87]. This suggests that physics beyond the Standard Model — specifically new decay amplitudes in $B \to \phi K$ — may be present. In light of this, many papers have been written to show how particular models of New Physics can account for this discrepancy [88, 89, 90, 91, 92, 93, 94, 95, 96, 97, 98, 99]. On the other hand, only two papers contain a model-independent analysis of $B^0(t) \to \phi K_S^0$ [100, 101] (and even here some theoretical numerical input is required).

Here we show how model-independent information about New Physics can be obtained from an angular analysis of $B \to V_1V_2$ decays, where $V_1$ and $V_2$ are vector mesons. This method is applicable to those $B \to V_1V_2$ decays in which

*i* $\overline{V}_1\overline{V}_2 = V_1V_2$, so that this final state is accessible to both $B^0$ and $\overline{B}^0$, and

*ii* a single decay amplitude dominates in the Standard Model.

The only theoretical assumption we make is that there is only a single New Physics amplitude, with a different weak phase from that of the Standard Model amplitude, contributing to these decays. In the event that a signal for New Physics is found, we demonstrate that one can place *lower* bounds on the New Physics parameters [102, 103].

If physics beyond the Standard Model contributes to $B^0(t) \to \phi K_S^0$, there should also be New Physics signals in the corresponding $B \to V_1V_2$ decay $B^0(t) \to \phi K^{*0}$. Our method can be used in this situation to get information about the New Physics. It can also be applied to $B^0(t) \to J/\psi K^{*0}$, $B^0(t) \to K^{*0}\overline{K}^{*0}$, $B_s(t) \to J/\psi\,\phi$, *etc.*, should New Physics signals be found in these decays. The analysis here treats only the situation where there are additional New Physics decay amplitudes; it does not apply to the case where the New Physics appears only in $B^0-\overline{B}^0$ mixing.

Any New Physics effects in $B$ decays are necessarily virtual. On the other hand, future experiments at the Large Hadron Collider (LHC) and at a linear $e^+e^-$ collider (ILC) will make direct searches for such New Physics. Should New Physics be found in both $B \to V_1V_2$ decays and at the LHC/ILC, the bounds from the angular analysis can tell us whether the New Physics seen at LHC/ILC can be responsible for the effects in $B \to V_1V_2$ decays.

We begin in Section 5.2.4 by describing the theoretical framework of our method. Signals of New Physics are examined in Section 5.2.4. The main results—how to place bounds on the theoretical New Physics parameters—are presented in Section 5.2.4. We discuss and summarize these results in Section 5.2.4.

#### Theoretical Framework

Consider a $B \to V_1V_2$ decay that is dominated by a single weak decay amplitude within the Standard Model. This holds for processes which are described by the quark-level decays $\bar{b} \to \bar{c}c\bar{s}$, $\bar{b} \to \bar{s}s\bar{s}$ or $\bar{b} \to \bar{s}d\bar{d}$. In all cases, the weak phase of the Standard Model amplitude is zero in the standard parametrization [104, 105, 106, 107]. Suppose now that there is a single New Physics amplitude, with a different weak phase, that contributes to the decay. The decay amplitude for each of the three possible helicity states may be written as

$$\begin{aligned} A_\lambda &\equiv Amp(B \to V_1V_2)_\lambda \,=\, a_\lambda e^{i\delta_\lambda^a} + b_\lambda e^{i\phi} e^{i\delta_\lambda^b} \,, \\ \overline{A}_\lambda &\equiv Amp\overline{B} \to (V_1V_2)_\lambda \,=\, a_\lambda e^{i\delta_\lambda^a} + b_\lambda e^{-i\phi} e^{i\delta_\lambda^b} \,, \end{aligned} \tag{5.24}$$

where $a_\lambda$ and $b_\lambda$ represent the Standard Model and New Physics amplitudes, respectively, $\phi$ is the New Physics weak phase, the $\delta_\lambda^{a,b}$ are the strong phases, and the helicity index $\lambda$ takes the values $\{0, \|, \perp\}$. Using $CPT$ invariance, the full decay amplitudes can be written as

$$\begin{aligned} \mathcal{A} &= Amp(B \to V_1V_2) = A_0 g_0 + A_\| g_\| + i\, A_\perp g_\perp \,, \\ \overline{\mathcal{A}} &= Amp(\overline{B} \to V_1V_2) = \overline{A}_0 g_0 + \overline{A}_\| g_\| - i\, \overline{A}_\perp g_\perp \,, \end{aligned} \tag{5.25}$$

where the $g_\lambda$ are the coefficients of the helicity amplitudes written in the linear polarization basis. The $g_\lambda$ depend only on the angles describing the kinematics [346].

Note that the assumption of a single New Physics amplitude is not terribly strong. First, the New Physics is expected to be heavy, so that all strong phases $\delta_\lambda$ should be small. In this case, since the $\delta_\lambda$ are all of similar size, our parametrization above is adequate. Second, if it happens that this is not the case, and there are several different contributing New Physics amplitudes, our analysis pertains to the dominant signal. Finally, if all the New Physics amplitudes are of the same size, our approach provides an order-of-magnitude estimate for the size of New Physics.

Equations (5.24) and (5.25) above enable us to write the time-dependent decay rates as

$$\Gamma(\overset{(-)}{B}(t) \to V_1V_2) = e^{-\Gamma t} \sum_{\lambda \le \sigma} \Big( \Lambda_{\lambda\sigma} \pm \Sigma_{\lambda\sigma} \cos(\Delta M t) \mp \rho_{\lambda\sigma} \sin(\Delta M t) \Big) g_\lambda g_\sigma \,. \tag{5.26}$$

Thus, by performing a time-dependent angular analysis of the decay $B(t) \to V_1V_2$, one can measure 18 observables. These are:

$$\begin{aligned} &\Lambda_{\lambda\lambda} = \frac{1}{2}(|A_\lambda|^2 + |\overline{A}_\lambda|^2), \quad \Sigma_{\lambda\lambda} = \frac{1}{2}(|A_\lambda|^2 - |\overline{A}_\lambda|^2), \\ &\Lambda_{\perp i} = -\mathrm{Im}(A_\perp A_i^* - \overline{A}_\perp \overline{A}_i{}^*), \quad \Lambda_{\|0} = \mathrm{Re}(A_\| A_0^* + \overline{A}_\| \overline{A}_0{}^*), \\ &\Sigma_{\perp i} = -\mathrm{Im}(A_\perp A_i^* + \overline{A}_\perp \overline{A}_i{}^*), \quad \Sigma_{\|0} = \mathrm{Re}(A_\| A_0^* - \overline{A}_\| \overline{A}_0{}^*), \\ &\rho_{\perp i} = \mathrm{Re}\Big(\frac{q}{p}[A_\perp^* \overline{A}_i + A_i^* \overline{A}_\perp]\Big), \quad \rho_{\perp\perp} = \mathrm{Im}\Big(\frac{q}{p}\, A_\perp^* \overline{A}_\perp\Big), \\ &\rho_{\|0} = -\mathrm{Im}\Big(\frac{q}{p}[A_\|^* \overline{A}_0 + A_0^* \overline{A}_\|]\Big), \quad \rho_{ii} = -\mathrm{Im}\Big(\frac{q}{p} A_i^* \overline{A}_i\Big), \end{aligned} \tag{5.27}$$

where $i = \{0, \|\}$. In the above, $q/p$ is the weak phase factor associated with $B^0 - \overline{B}^0$ mixing. For $B^0$ mesons, $q/p = \exp(-2\,i\beta)$, while $q/p = 1$ for $B_s$ mesons. Henceforth, we will concentrate on the decays of $B^0$ mesons, though our results can easily be adapted to $B_s$ decays. Note that $\beta$ may include New Physics effects in $B^0 - \overline{B}^0$ mixing. Note also that the signs of the various $\rho_{\lambda\lambda}$ terms depend on the $CP$-parity of the various helicity states. We have chosen the sign of $\rho_{00}$ and $\rho_{\|\|}$ to be $-1$, which corresponds to the final state $\phi K^*$.

Not all of the 18 observables are independent. There are a total of six amplitudes describing $B \to V_1V_2$ and $\overline{B} \to V_1V_2$ decays [Eq. (5.24)]. Thus, at best one can measure the magnitudes and relative phases of these six amplitudes, giving 11 independent measurements.

The 18 observables given above can be written in terms of 13 theoretical parameters: three $a_\lambda$'s, three $b_\lambda$'s, $\beta$, $\phi$, and five strong phase differences defined by $\delta_\lambda \equiv \delta_\lambda^b - \delta_\lambda^a$, $\Delta_i \equiv \delta_\perp^a - \delta_i^a$. The explicit expressions for the observables are

as follows:

$$
\begin{aligned}
\Lambda_{\lambda\lambda} &= a_\lambda^2 + b_\lambda^2 + 2a_\lambda b_\lambda \cos\delta_\lambda \cos\phi ~, \\
\Sigma_{\lambda\lambda} &= -2a_\lambda b_\lambda \sin\delta_\lambda \sin\phi ~, \\
\Lambda_{\perp i} &= 2\left[a_\perp b_i \cos(\Delta_i - \delta_i) - a_i b_\perp \cos(\Delta_i + \delta_\perp)\right] \sin\phi ~, \\
\Lambda_{\|0} &= 2\left[a_\| a_0 \cos(\Delta_0 - \Delta_\|) + a_\| b_0 \cos(\Delta_0 - \Delta_\| - \delta_0)\cos\phi \right. \\
&\quad \left. + a_0 b_\| \cos(\Delta_0 - \Delta_\| + \delta_\|)\cos\phi + b_\| b_0 \cos(\Delta_0 - \Delta_\| + \delta_\| - \delta_0)\right] ~, \\
\Sigma_{\perp i} &= -2\left[a_\perp a_i \sin\Delta_i + a_\perp b_i \sin(\Delta_i - \delta_i)\cos\phi \right. \\
&\quad \left. + a_i b_\perp \sin(\Delta_i + \delta_\perp)\cos\phi + b_\perp b_i \sin(\Delta_i + \delta_\perp - \delta_i)\right] ~, \\
\Sigma_{\|0} &= 2\left[a_\| b_0 \sin(\Delta_0 - \Delta_\| - \delta_0) - a_0 b_\| \sin(\Delta_0 - \Delta_\| + \delta_\|)\right] \sin\phi ~, \\
\rho_{ii} &= a_i^2 \sin 2\beta + 2a_i b_i \cos\delta_i \sin(2\beta + \phi) + b_i^2 \sin(2\beta + 2\phi) ~, \\
\rho_{\perp\perp} &= -a_\perp^2 \sin 2\beta - 2a_\perp b_\perp \cos\delta_\perp \sin(2\beta + \phi) - b_\perp^2 \sin(2\beta + 2\phi) ~, \\
\rho_{\perp i} &= 2\left[a_i a_\perp \cos\Delta_i \cos 2\beta + a_\perp b_i \cos(\Delta_i - \delta_i)\cos(2\beta + \phi) \right. \\
&\quad + ~ a_i b_\perp \cos(\Delta_i + \delta_\perp)\cos(2\beta + \phi) \\
&\quad \left. + b_i b_\perp \cos(\Delta_i + \delta_\perp - \delta_i)\cos(2\beta + 2\phi)\right] ~, \\
\rho_{\|0} &= 2\left[a_0 a_\| \cos(\Delta_0 - \Delta_\|)\sin 2\beta + a_\| b_0 \cos(\Delta_0 - \Delta_\| - \delta_0)\sin(2\beta + \phi) \right. \\
&\quad + ~ a_0 b_\| \cos(\Delta_0 - \Delta_\| + \delta_\|)\sin(2\beta + \phi) \\
&\quad \left. + b_0 b_\| \cos(\Delta_0 - \Delta_\| + \delta_\| - \delta_0)\sin(2\beta + 2\phi)\right] ~.
\end{aligned}
\tag{5.28}
$$

In subsequent sections, we will work extensively with these expressions.

It is straightforward to see that, in the presence of New Physics, one cannot extract the phase $\beta$. There are 11 independent observables, but 13 theoretical parameters. Since the number of measurements is fewer than the number of parameters, one cannot express any of the theoretical unknowns purely in terms of observables. In particular, it is impossible to extract $\beta$ cleanly. Nevertheless, we will show that the angular analysis does allow one to obtain significant *lower bounds* on the New Physics parameters, as well as on the deviation of $\beta$ from its measured value.

**Signals of New Physics**

As mentioned in the introduction, lower bounds on New Physics parameters are possible only if there is a signal of physics beyond the Standard Model. In this section, we discuss the possible New Physics signals in $B \to V_1 V_2$ decays.

In the absence of New Physics, the $b_\lambda$ are zero in Eq. (5.24). The number of parameters is then reduced from 13 to 6: three $a_\lambda$'s, two strong phase differences ($\Delta_i$), and $\beta$. It is straightforward to show that all six parameters can be determined cleanly in terms of observables [Eq. (5.28)]. However, there are a total of 18 observables. Thus, there must exist 12 relations among the observables in the absence of New Physics. These are:

$$
\begin{aligned}
& \Sigma_{\lambda\lambda} = \Lambda_{\perp i} = \Sigma_{\|0} = 0 ~, \\
& \frac{\rho_{ii}}{\Lambda_{ii}} = -\frac{\rho_{\perp\perp}}{\Lambda_{\perp\perp}} = \frac{\rho_{\|0}}{\Lambda_{\|0}} ~, \\
& \Lambda_{\|0} = \frac{1}{2\Lambda_{\perp\perp}} \left[ \frac{\Lambda_{\lambda\lambda}^2 \rho_{\perp 0}\rho_{\perp\|} + \Sigma_{\perp 0}\Sigma_{\perp\|}(\Lambda_{\lambda\lambda}^2 - \rho_{\lambda\lambda}^2)}{\Lambda_{\lambda\lambda}^2 - \rho_{\lambda\lambda}^2} \right] ~, \\
& \frac{\rho_{\perp i}^2}{4\Lambda_{\perp\perp}\Lambda_{ii} - \Sigma_{\perp i}^2} = \frac{\Lambda_{\perp\perp}^2 - \rho_{\perp\perp}^2}{\Lambda_{\perp\perp}^2} ~.
\end{aligned}
\tag{5.29}
$$

The key point is the following: *The violation of any of the above relations will be a smoking-gun signal of New Physics*. We therefore see that the angular analysis of $B \to V_1 V_2$ decays provides numerous tests for the presence of New Physics. One may wonder if, despite the many tests, New Physics can still remain hidden. If the three strong phase differences $\delta_\lambda$ vanish, and the ratio $r_\lambda \equiv b_\lambda/a_\lambda$ is the same for all helicities, *i.e.*, $r_0 = r_\parallel = r_\perp$, then all the relations in Eq. (5.29) are satisfied, even in the presence of New Physics. It is easy to show that the transformation $\tilde{a}_\lambda^2 = a_\lambda^2(1 + 2r\cos\phi + r^2)$ results in relations identical to the case with no New Physics but with $a_\lambda \to \tilde{a}_\lambda$. Thus, if these very special conditions happen to hold, the angular analysis of $B \to V_1 V_2$ would show no signal for New Physics even if it is present, and the measured value of $\beta$ would not correspond to its true ( Standard Model ) value. Still, we should stress that it is highly unlikely that the New Physics parameters should respect such a singular situation.

Since there are 11 independent observables and 6 parameters in the Standard Model , one might expect that only 5 tests are needed to verify the presence of New Physics. However, if the Standard Model parameters take certain special values, more tests are needed. For example, suppose that $b_\parallel = b_\perp = 0$ and $\delta_0 = 0$. Since $b_0 \neq 0$, New Physics is present. We have $\Sigma_{\lambda\lambda} = \Lambda_{\perp\parallel} = 0$. If $\Delta_0$ takes the (fine-tuned) value $\pi/2$, we will also find that $\Lambda_{\perp\parallel} = 0$. Thus, despite the presence of New Physics, 5 of the 12 tests above agree with the Standard Model . In this case, further tests are needed to confirm the fact that New Physics is present. In the most general case, *all* 12 tests above are needed to search for New Physics. (In any event, because it is not known *a priori* which observables will be measured, it is important to have a list of all New Physics tests.)

We should stress here that the list of New Physics signals is independent of the parametrization of New Physics. That is, even if there are several contributing amplitudes, the New Physics can still be discovered through the tests in Eq. (5.29). Furthermore, even in this general case, it is necessary to perform all 12 tests in order to show that New Physics is not present.

The observable $\Lambda_{\perp i}$ deserves special attention. It is the coefficient of the $T$-odd "triple product" in $B \to V_1 V_2$ decays, $\vec{q} \cdot (\vec{\varepsilon}_1 \times \vec{\varepsilon}_2)$, where $\vec{q}$ is the momentum of one of the final vector mesons in the rest frame of the $B$, and $\vec{\varepsilon}_{1,2}$ are the polarizations of $V_1$ and $V_2$ [108]. From Eq. (5.28), one sees that even if the strong phase differences vanish, $\Lambda_{\perp i}$ is nonzero in the presence of New Physics ($\phi \neq 0$), in contrast to the direct $CP$ asymmetries (proportional to $\Sigma_{\lambda\lambda}$). This is due to the fact that the $\perp$ helicity is $CP$-odd, while the 0 and $\parallel$ helicities are $CP$-even. Thus, $\perp$–0 and $\perp$–$\parallel$ interferences include an additional factor of '$i$' in the full decay amplitudes [Eq. (5.25)], which leads to the cosine dependence on the strong phases.

Although the reconstruction of the full $B^0(t)$ and $\overline{B}^0(t)$ decay rates in Eq. (5.26) requires both tagging and time-dependent measurements, the $\Lambda_{\lambda\sigma}$ terms remain even if the two rates for $B^0(t)$ and $\overline{B}^0(t)$ decays are added together. Note also that these terms are time-independent. Therefore, *no tagging or time-dependent measurements are needed to extract* $\Lambda_{\perp i}$. It is only necessary to perform an angular analysis of the final state $V_1V_2$. Thus, this measurement can even be made at a symmetric $B$ Factory.

The decays of charged $B$ mesons to vector-vector final states are even simpler to analyze, since no mixing is involved. One can in principle combine charged and neutral $B$ decays to increase the sensitivity to New Physics. For example, for $B \to J/\psi K^*$ decays, one simply performs an angular analysis on all decays in which a $J/\psi$ is produced accompanied by a charged or neutral $K^*$. A nonzero value of $\Lambda_{\perp i}$ would be a clear signal for New Physics [109].

**Bounds on the Theoretical Parameters**

In this section we explore the constraints on the size of New Physics, assuming that a New Physics signal is observed in $B \to V_1 V_2$. As we have shown, the amplitudes are written in terms of 13 theoretical parameters (including $\beta$), but there are only 11 independent observables. Since the number of unknowns is greater than the number of observables, naively one would think that it is not possible to obtain any information about the New Physics parameters. However, since the expressions for the observables in terms of the theoretical parameters are nonlinear [Eq. (5.28)], it is in fact possible to obtain *bounds* on the New Physics parameters. One can even put a lower bound on the difference between the measured value of $\beta$ (which is affected by the presence of New Physics) and its true (Standard Model) value.

The first step is to reduce the number of unknowns in the expressions for the observables. That is, even though one cannot solve for the theoretical parameters in terms of observables, one can obtain a partial solution, in which observables are written in terms of a smaller number of parameters plus other observables.

For $B \to V_1 V_2$ decays, the analogue of the usual direct $CP$ asymmetry $a_{dir}^{CP}$ is $a_\lambda^{dir} \equiv \Sigma_{\lambda\lambda}/\Lambda_{\lambda\lambda}$, which is helicity-dependent. We define the related quantity,

$$y_\lambda \equiv \sqrt{1 - \Sigma_{\lambda\lambda}^2/\Lambda_{\lambda\lambda}^2} \,. \tag{5.30}$$

The measured value of $\sin 2\beta$ can also depend on the helicity of the final state: $\rho_{\lambda\lambda}$ can be recast in terms of a measured weak phase $2\beta_\lambda^{\rm meas}$, defined as

$$\sin\, 2\,\beta_\lambda^{\rm meas} \equiv \frac{\pm\rho_{\lambda\lambda}}{\sqrt{\Lambda_{\lambda\lambda}^2 - \Sigma_{\lambda\lambda}^2}} \,, \tag{5.31}$$

where the $+\ (-)$ sign corresponds to $\lambda = 0, \|\ (\perp)$.

It is possible to express the 9 theoretical parameters $a_\lambda$, $b_\lambda$ and $\delta_\lambda$ in terms of the 9 observables $\Lambda_{\lambda\lambda}$, $\Sigma_{\lambda\lambda}$, and $\rho_{\lambda\lambda}$, and the parameters $\beta$ and $\phi$. The other observables can in turn be expressed in terms of $\Lambda_{\lambda\lambda}$, $\Sigma_{\lambda\lambda}$, and $\rho_{\lambda\lambda}$, along with the three theoretical parameters $\beta+\phi$ and $\Delta_i$. Using the expressions for $\Lambda_{\lambda\lambda}$, $\Sigma_{\lambda\lambda}$ and $\beta_\lambda^{\rm meas}$ above, one can express $a_\lambda$ and $b_\lambda$ as follows:

$$2\,a_\lambda^2\,\sin^2\phi \;=\; \Lambda_{\lambda\lambda}\Big(1 - y_\lambda \cos(2\beta_\lambda^{\rm meas} - 2\beta - 2\phi)\Big)\,, \tag{5.32}$$

$$2\,b_\lambda^2\,\sin^2\phi \;=\; \Lambda_{\lambda\lambda}\Big(1 - y_\lambda \cos(2\beta_\lambda^{\rm meas} - 2\beta)\Big)\,. \tag{5.33}$$

These expression will play a critical role in the derivation of bounds on the New Physics parameters.

The seemingly impossible task of eliminating 10 combinations of the theoretical parameters $a_\lambda$, $b_\lambda$, $\delta_\lambda$, $\beta$ and $\phi$ in terms of the observables $\Lambda_{\lambda\lambda}$, $\Sigma_{\lambda\lambda}$ and $\rho_{\lambda\lambda}$, and variable $\beta+\phi$ becomes possible by using the following relation:

$$\begin{aligned}\frac{b_\lambda}{a_\lambda}\cos\delta_\lambda\,\cos\phi &= \frac{-2\Lambda_{\lambda\lambda}\cos^2\phi + y_\lambda\,\Lambda_{\lambda\lambda}\left(\cos(2\beta_\lambda^{\rm meas} - 2\beta - 2\phi) + \cos(2\beta_\lambda^{\rm meas} - 2\beta)\right)}{2\Lambda_{\lambda\lambda}(1 - y_\lambda \cos(2\beta_\lambda^{\rm meas} - 2\beta - 2\phi))} \\ &= -\cos^2\phi\left(1 + \frac{y_\lambda \sin(2\beta_\lambda^{\rm meas} - 2\beta - 2\phi)\tan\phi}{1 - y_\lambda\cos(2\beta_\lambda^{\rm meas} - 2\beta - 2\phi)}\right)\,,\end{aligned} \tag{5.34}$$

where we have used the expression for $\Lambda_{\lambda\lambda}$ given in Eq. (5.28). We introduce a compact notation to express Eq. (5.34) by defining

$$P_\lambda^2 \;\equiv\; \Lambda_{\lambda\lambda}(1 - y_\lambda\, \cos(2\beta_\lambda^{\rm meas} - 2\beta - 2\phi))\,, \tag{5.35}$$

$$\xi_\lambda \;\equiv\; \frac{\Lambda_{\lambda\lambda}\, y_\lambda\, \sin(2\beta_\lambda^{\rm meas} - 2\beta - 2\phi)}{P_\lambda^2}\,. \tag{5.36}$$

This results in

$$\frac{b_\lambda}{a_\lambda}\cos\delta_\lambda\,\cos\phi = -\cos^2\phi - \cos\phi\sin\phi\,\xi_\lambda \tag{5.37}$$

Similarly, we define

$$\sigma_\lambda \equiv \frac{\Sigma_{\lambda\lambda}}{P_\lambda^2}\,, \tag{5.38}$$

which allows us to write

$$\frac{b_\lambda}{a_\lambda}\sin\delta_\lambda\,\sin\phi = -\sin^2\phi\,\sigma_\lambda\,. \tag{5.39}$$

We can now express the remaining 9 observables in terms of $\Delta_i$, $\beta+\phi$ and the newly-defined parameters $P_\lambda$, $\xi_\lambda$ and $\sigma_\lambda$ as follows:

$$\Sigma_{\perp i} \;=\; P_i P_\perp\left[\Big(\xi_\perp\,\sigma_i - \xi_i\,\sigma_\perp\Big)\,\cos\Delta_i - \Big(1 + \xi_i\,\xi_\perp + \sigma_i\,\sigma_\perp\Big)\,\sin\Delta_i\right]\,, \tag{5.40}$$

$$\Lambda_{\perp i} = P_i P_\perp \left[ \Big( \xi_\perp - \xi_i \Big) \cos\Delta_i - \Big( \sigma_i + \sigma_\perp \Big) \sin\Delta_i \right] , \tag{5.41}$$

$$\begin{aligned} \rho_{\perp i} = P_i P_\perp \Bigg[ & \Big( (-1 + \xi_i\,\xi_\perp + \sigma_i\,\sigma_\perp) \cos(2\beta + 2\phi) - (\xi_i + \xi_\perp) \sin(2\beta + 2\phi) \Big) \cos\Delta_i \\ & + \Big( (-\xi_i\,\sigma_\perp + \xi_\perp\,\sigma_i) \cos(2\beta + 2\phi) - (\sigma_i - \sigma_\perp) \sin(2\beta + 2\phi) \Big) \sin\Delta_i \Bigg] , \end{aligned} \tag{5.42}$$

$$\Sigma_{\parallel 0} = P_\parallel P_0 \left[ (\xi_\parallel - \xi_0) \sin(\Delta_0 - \Delta_\parallel) + (\sigma_\parallel + \sigma_0) \cos(\Delta_0 - \Delta_\parallel) \right] , \tag{5.43}$$

$$\Lambda_{\parallel 0} = P_\parallel P_0 \left[ (\xi_0 \sigma_\parallel - \sigma_0 \xi_\parallel) \sin(\Delta_0 - \Delta_\parallel) + (1 + \xi_0 \xi_\parallel + \sigma_\parallel \sigma_0) \cos(\Delta_0 - \Delta_\parallel) \right] , \tag{5.44}$$

$$\begin{aligned} \rho_{\parallel 0} = P_\parallel P_0 \Bigg[ & \Big( (-1 + \xi_\parallel\,\xi_0 + \sigma_\parallel\,\sigma_0) \sin(2\beta + 2\phi) \\ & \quad + (\xi_\parallel + \xi_0) \cos(2\beta + 2\phi) \Big) \cos(\Delta_0 - \Delta_\parallel) \\ & + \Big( (\xi_\parallel\,\sigma_0 - \xi_0\,\sigma_\parallel) \sin(2\beta + 2\phi) + (\sigma_0 - \sigma_\parallel) \cos(2\beta + 2\phi) \Big) \sin(\Delta_0 - \Delta_\parallel) \Bigg] . \end{aligned} \tag{5.45}$$

The notable achievement of the above relations is the expression of observables involving the interference of helicities in terms of only 3 theoretical parameters ($\Delta_i$, $\beta + \phi$), thereby reducing the complexity of the extremization problem. The above relations are extremely important in obtaining bounds on New Physics parameters.

We now turn to the issue of New Physics signals. The presence of New Physics is indicated by the violation of at least one of the relations given in Eq. (5.29). This in turn implies that $b_\lambda \neq 0$ and $|\beta_\lambda^{\rm meas} - \beta| \neq 0$ for at least one helicity $\lambda$. Clearly, any bounds on New Physics parameters will depend on the specific signal of New Physics. We therefore examine several different New Physics signals and explore the restrictions they place on New Physics parameter space.

Note that we do not present an exhaustive study of New Physics signals. The main point of the present paper is to show that it is possible to obtain bounds on the New Physics parameters, even though there are more unknowns than observables. Whenever possible, we present analytic bounds on the New Physics parameters. However, for certain New Physics signals, analytical bounds are either not easy to derive or not obtained as a simple analytical expression. In such a case we only obtain numerical bounds. Only in two cases does this become necessary. When only a single helicity-interference observable is measured and when considering bounds on $r_\lambda$. In all cases, the bounds are found without any approximations. This demonstrates the power of angular analysis and its usefulness in constraining New Physics parameters.

We will see that, while $b_\lambda$ and $b_\lambda/a_\lambda$ can be constrained with just one signal of New Physics, obtaining a bound on $|\beta_\lambda^{\rm meas} - \beta|$ requires at least two New Physics signals. Also, because the equations are nonlinear, there are often discrete ambiguities in the bounds. These can be reduced, leading to stronger bounds on New Physics, if a larger set of observables is used.

In the subsections below we present bounds for five different signals of New Physics.

### $\boldsymbol{\Sigma_{\lambda\lambda} \neq 0}$

Suppose first that one observes direct *CP* violation in at least one helicity, *i.e.*, $\Sigma_{\lambda\lambda} \neq 0$. The minimum value of $b_\lambda^2$ can be obtained by minimizing $b_\lambda^2$ [Eq. (5.33)] with respect to $\beta$ and $\phi$:

$$b_\lambda^2 \geq \frac{1}{2} \left[ \Lambda_{\lambda\lambda} - \sqrt{\Lambda_{\lambda\lambda}^2 - \Sigma_{\lambda\lambda}^2} \right] . \tag{5.46}$$

Thus, if direct $CP$ violation is observed, one can place a lower bound on the New Physics amplitude $b_\lambda$.

On the other hand, it follows from Eq. (5.33) that no upper bound can ever be placed on $b_\lambda^2$. One can always take $b_\lambda \to \infty$, as long as $\phi \to 0$ with $b_\lambda \sin\phi$ held constant. For the same reason, one can never determine the New Physics weak phase $\phi$, or place a lower bound on it.

It is possible, however, to place lower bounds on other New Physics quantities. Using Eqs. (5.32) and (5.33), it is straightforward to obtain the constraints

$$\begin{aligned} \tfrac{1}{2}\Lambda_{\lambda\lambda}\,(1-y_\lambda) \le b_\lambda^2\,\sin^2\phi \le \tfrac{1}{2}\Lambda_{\lambda\lambda}\,(1+y_\lambda)\ , \\ \frac{1-y_\lambda}{1+y_\lambda} \le r_\lambda^2 \le \frac{1+y_\lambda}{1-y_\lambda}\ , \end{aligned} \tag{5.47}$$

where

$$r_\lambda \equiv \frac{b_\lambda}{a_\lambda}\ . \tag{5.48}$$

If $\Sigma_{\lambda\lambda} \neq 0$, these give nontrivial lower bounds. The lower bound on $r_\lambda$ is very useful in estimating the magnitude of New Physics amplitudes or the scale of New Physics.

One interesting observation can be made regarding bounds on $b_\lambda^2$. Saying that New Physics is present implies that the New Physics amplitude $b_\lambda$ must be nonzero for at least one helicity; the other two helicities could have vanishing New Physics amplitudes. A nonzero direct asymmetry $a_{dir}^{CP} \neq 0$ (*i.e.*, $\Sigma_{\lambda\lambda} \neq 0$) implies a nonzero New Physics amplitude with a lower bound given by Eq. (5.46). Other New Physics signals [Eq. (5.29)] do not bound the New Physics amplitude $b_\lambda^2$ for a single helicity, but can bound combinations $(b_\lambda^2 \pm b_\sigma^2)$. This is perhaps surprising but may be understood as follows. Consider, for example, the New Physics signal $\Lambda_{\perp i} \neq 0$. Even in the presence of such a signal it is possible that one of either $b_i$ or $b_\perp$ is zero, but not both [see Eq. (5.28)]. Thus, one can only obtain a lower bound when simultaneously bounding $b_i^2$ and $b_\perp^2$. Hence, for $\Lambda_{\perp i} \neq 0$, we must consider bounds on sums and differences of the New Physics amplitudes, $b_i^2 \pm b_\perp^2$. A similar argument applies to all signals of New Physics in Eq. (5.29) involving two helicities. We will encounter such lower bounds in subsequent subsections.

### $\boldsymbol{\beta_\lambda^{\rm meas} \neq \beta_\sigma^{\rm meas}}$

Another signal of New Physics is differing measured values of $\beta$ in two different helicities, *i.e.*, $\beta_\lambda^{\rm meas} \neq \beta_\sigma^{\rm meas}$. We define

$$2\omega_{\sigma\lambda} \equiv 2\beta_\sigma^{meas} - 2\beta_\lambda^{meas}\quad ,\quad \eta_\lambda \equiv 2(\beta_\lambda^{\rm meas} - \beta)\ . \tag{5.49}$$

Using Eq. (5.33) we have

$$(b_\lambda^2 \pm b_\sigma^2)\sin^2\phi = \frac{\Lambda_{\lambda\lambda} \pm \Lambda_{\sigma\sigma}}{2} - \frac{y_\lambda \Lambda_{\lambda\lambda}\cos\eta_\lambda \pm y_\sigma \Lambda_{\sigma\sigma}\cos(2\omega_{\sigma\lambda}+\eta_\lambda)}{2}\ . \tag{5.50}$$

Extremizing this expression with respect to $\eta_\lambda$, we obtain a solution for $\eta_\lambda$:

$$\sin\eta_\lambda = \pm\frac{y_\sigma \Lambda_{\sigma\sigma}\sin 2\omega_{\sigma\lambda}}{\sqrt{y_\lambda^2\Lambda_{\lambda\lambda}^2 + y_\sigma^2\Lambda_{\sigma\sigma}^2 - 2\,y_\lambda y_\sigma \Lambda_{\lambda\lambda}\Lambda_{\sigma\sigma}\cos 2\omega_{\sigma\lambda}}}\ . \tag{5.51}$$

Taking into account the sign of the second derivative, we get the bounds

$$(b_\lambda^2 \pm b_\sigma^2)\sin^2\phi \ge \frac{\Lambda_{\lambda\lambda} \pm \Lambda_{\sigma\sigma}}{2} - \frac{\sqrt{y_\lambda^2\Lambda_{\lambda\lambda}^2 + y_\sigma^2\Lambda_{\sigma\sigma}^2 \pm 2\,y_\lambda y_\sigma \Lambda_{\lambda\lambda}\Lambda_{\sigma\sigma}\cos 2\omega_{\sigma\lambda}}}{2}\ , \tag{5.52}$$

$$(b_\lambda^2 \pm b_\sigma^2)\sin^2\phi \le \frac{\Lambda_{\lambda\lambda} \pm \Lambda_{\sigma\sigma}}{2} + \frac{\sqrt{y_\lambda^2\Lambda_{\lambda\lambda}^2 + y_\sigma^2\Lambda_{\sigma\sigma}^2 \pm 2\,y_\lambda y_\sigma \Lambda_{\lambda\lambda}\Lambda_{\sigma\sigma}\cos 2\omega_{\sigma\lambda}}}{2}\ . \tag{5.53}$$

Extremizing with respect to $\phi$ as well, one obtains the bounds

$$(b_\lambda^2 \pm b_\sigma^2) \ge \frac{\Lambda_{\lambda\lambda} \pm \Lambda_{\sigma\sigma}}{2} - \frac{\left|y_\lambda \Lambda_{\lambda\lambda} \pm y_\sigma \Lambda_{\sigma\sigma} e^{2i\omega_{\sigma\lambda}}\right|}{2}\ , \tag{5.54}$$

where it has been assumed that $\Lambda_{\lambda\lambda} > \Lambda_{\sigma\sigma}$, and that the right-hand side of the inequality is positive. (Note that an upper bound on $(b_\lambda^2 \pm b_\sigma^2)$ cannot be obtained.) We will see below that Eq. (5.54) plays a central role in deriving bounds for other signals of New Physics.

We emphasize that all of the above bounds are exact – no approximations or limits have been used. From the constraints on $(b_\lambda^2 \pm b_\sigma^2)$ one can obtain lower bounds on $b_\lambda^2$ and $b_\sigma^2$ individually.

Even without extremization, careful examination of Eq. (5.50) implies minimum and maximum possible values for $(b_\lambda^2 \pm b_\sigma^2)\sin^2\phi$. These can also be derived from Eq. (5.47) and are given by

$$\begin{aligned}(b_\lambda^2 \pm b_\sigma^2)\sin^2\phi &\geq \frac{\Lambda_{\lambda\lambda} \pm \Lambda_{\sigma\sigma}}{2} - \frac{y_\lambda \Lambda_{\lambda\lambda} + y_\sigma \Lambda_{\sigma\sigma}}{2} ,\\ (b_\lambda^2 \pm b_\sigma^2)\sin^2\phi &\leq \frac{\Lambda_{\lambda\lambda} \pm \Lambda_{\sigma\sigma}}{2} + \frac{y_\lambda \Lambda_{\lambda\lambda} + y_\sigma \Lambda_{\sigma\sigma}}{2} . \end{aligned} \tag{5.55}$$

Note that if $2\omega_{\sigma\lambda} = 0$, Eqs. (5.52) and (5.53) reproduce the bounds of Eq. (5.55) for $(b_\lambda^2 + b_\sigma^2)\sin^2\phi$; if $2\omega_{\sigma\lambda} = \pi$, one reproduces the bounds on $(b_\lambda^2 - b_\sigma^2)\sin^2\phi$. If one uses other New Physics signals to constrain the New Physics parameters, then unless these other signals result in constraining the value of $2\omega_{\sigma\lambda}$ to be other than 0 or $\pi$, one cannot obtain better bounds than those of Eq. (5.55). Note also that, while $2\omega_{\sigma\lambda}$ can be measured directly up to discrete ambiguities, additional measurements will result in the reduction of such ambiguities and lead to tighter bounds.

### $\Lambda_{\perp i} \neq 0$ with $\Sigma_{\lambda\lambda} = 0$

We now turn to the New Physics signal $\Lambda_{\perp i} \neq 0$. Here we assume that the phase of $B^0 - \overline{B}^0$ mixing has not been measured in any helicity, *i.e.,* the parameter $\omega_{\perp i}$ is unknown. This situation is plausible: as discussed above, $\Lambda_{\perp i}$ can be obtained without tagging or time-dependence, while the measurement of $\omega_{\perp i}$ requires both.

In order to obtain analytic bounds which depend on $\Lambda_{\perp i}$, it is simplest to consider the limit in which all direct $CP$-violating asymmetries vanish ($\Sigma_{\lambda\lambda} = 0$). In this limit, with a little algebra Eq. (5.41) reduces to

$$\frac{\Lambda_{\perp i}}{2\sqrt{\Lambda_{ii}\Lambda_{\perp\perp}}} = -\sin\omega_{\perp i}\cos\Delta_i , \tag{5.56}$$

where $\omega_{\perp i} \equiv \beta_\perp^{meas} - \beta_i^{meas}$. We solve the above for $\sin\omega_{\perp i}$ and substitute it into the expressions for $(b_i^2 \pm b_\perp^2)\sin^2\phi$ [Eq. (5.50)]. The resulting expressions are minimized straightforwardly with respect to $\cos\Delta_i$ and $\eta_i$ to obtain new bounds. The extrema with respect to $\Delta_i$ for both $(b_i^2 \pm b_\perp^2)$ occur at

$$\cos^2\Delta_i = \left\{1, \frac{\Lambda_{\perp i}^2}{4\Lambda_{ii}^2\Lambda_{\perp\perp}^2\cos^2(\eta_i/2)}, \frac{\Lambda_{\perp i}^2}{4\Lambda_{ii}^2\Lambda_{\perp\perp}^2\sin^2(\eta_i/2)}\right\} , \tag{5.57}$$

while that with respect to $\eta_i$ depends on $\Lambda_{\perp i}$, and occurs for both $(b_i^2 \pm b_\perp^2)$ at

$$\sin\eta_i = \pm\frac{2R\sqrt{1-R^2}\Lambda_{\perp\perp}}{\sqrt{\Lambda_{ii}^2 \pm 2(1-2R^2)\Lambda_{ii}\Lambda_{\perp\perp} + \Lambda_{\perp\perp}^2}} , \tag{5.58}$$

where

$$R = \frac{\Lambda_{\perp i}}{2\sqrt{\Lambda_{ii}\Lambda_{\perp\perp}}} . \tag{5.59}$$

These extrema yield new lower limits on $(b_i^2 \pm b_\perp^2)$:

$$2(b_i^2 \pm b_\perp^2) \geq \Lambda_{ii} \pm \Lambda_{\perp\perp} - \sqrt{(\Lambda_{ii} \pm \Lambda_{\perp\perp})^2 \mp \Lambda_{\perp i}^2} , \tag{5.60}$$

Interference terms such as $\Lambda_{\perp i}$ also allow us to obtain bounds for $\eta_\lambda$. Using Eqs. (5.50) and (5.60), one can easily derive the bound

$$(\Lambda_{ii} + \Lambda_{\perp\perp}\cos 2\omega_{\perp i})\cos\eta_i + \Lambda_{\perp\perp}\sin 2\omega_{\perp i}\sin\eta_i \leq \sqrt{(\Lambda_{ii} + \Lambda_{\perp\perp})^2 - \Lambda_{\perp i}^2} , \tag{5.61}$$

which can be rewritten as

$$\Lambda_{ii} \cos\eta_i + \Lambda_{\perp\perp} \cos\eta_\perp \le \sqrt{(\Lambda_{ii} + \Lambda_{\perp\perp})^2 - \Lambda_{\perp i}^2} \,. \tag{5.62}$$

Thus, if $\Lambda_{\perp i} \neq 0$, one cannot have $\eta_i = \eta_\perp = 0$. These constraints therefore place a lower bound on $|\beta_i^{\rm meas} - \beta|$ and/or $|\beta_\perp^{\rm meas} - \beta|$. This procedure can also be applied to $\Sigma_{\|0}$, and different lower bounds on $(b_\|^2 \pm b_0^2)$ and on $\eta_\|$, $\eta_0$ can be derived.

### $\Lambda_{\perp i} \neq 0$ with $\Sigma_{\lambda\lambda} \neq 0$

We now assume that both $\Lambda_{\perp i} \neq 0$ and $\Sigma_{\lambda\lambda} \neq 0$, but no measurement has been made of the parameter $\omega_{\perp i}$. In this case the procedure outlined in the previous subsection cannot be used to obtain analytic bounds on $(b_i^2 \pm b_\perp^2)$. The reason is that one does not find a simple solution for $\omega_{\perp i}$ such as that given in Eq. (5.56). In this case, we are forced to turn to numerical solutions. We use the same method as in the previous subsection—we solve Eq. (5.41) for $\omega_{\perp i}$ and substitute it into Eq. (5.50)—except that now the minimization is performed numerically with respect to the variables $\eta_i$, $\phi$ and $\Delta_i$ using the computer program MINUIT [110].

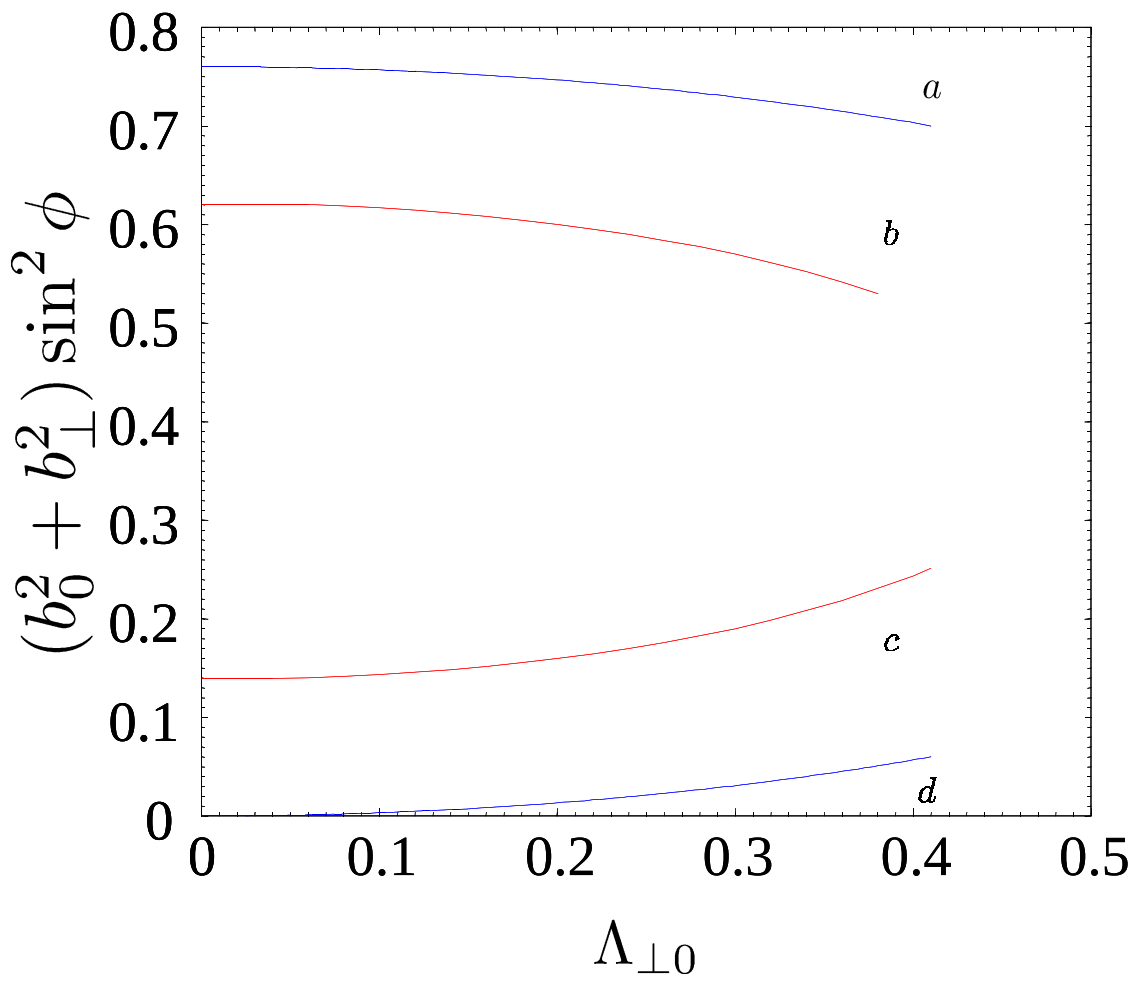

**Figure 5-15.** *The lower and upper bounds on* $(b_0^2 + b_\perp^2)\sin^2\phi$ *as a function of* $\Lambda_{\perp 0}$. *For curves* $b$ *and* $c$ *we have assumed the following values for the observables:* $\Lambda_{00} = 0.6$, $\Lambda_{\perp\perp} = 0.16$, $y_0 = 0.60$, $y_\perp = 0.74$. *Curves* $a$ *and* $d$ *represent the corresponding case with no direct CP asymmetry (i.e.,* $y_0 = y_\perp = 1.0$*).*

We assume the New Physics signal $\Lambda_{\perp 0} \neq 0$. In order to perform numerical minimization, we must choose values for the observables. Here and in the next subsection, we take $\Lambda_{00} = 0.6$, $\Lambda_{\perp\perp} = 0.16$, $y_0 = 0.60$ and $y_\perp = 0.74$.

In Fig. 5-15, we present the lower and upper bounds on $(b_0^2 + b_\perp^2)\sin^2\phi$ as a function of $\Lambda_{\perp 0}$. As in the previous subsection, these bounds are obtained by minimizing with respect to the variables $\Delta_i$ and $\eta_i$. Since the minimum value of $(b_0^2 + b_\perp^2)$ can be obtained from that of $(b_0^2 + b_\perp^2)\sin^2\phi$ by setting $\sin\phi = 1$ (its maximum value), the lower bound on $(b_0^2 + b_\perp^2)$ is identical to that of $(b_0^2 + b_\perp^2)\sin^2\phi$. However, upper bounds can only be derived for $(b_0^2 + b_\perp^2)\sin^2\phi$. For comparison, we include the bounds for the case of vanishing direct *CP* asymmetry, *i.e.*, $\Sigma_{00} = \Sigma_{\perp\perp} = 0$ [Eq. (5.60)]. It is clear that the bounds are stronger if there are more signals of New Physics.

As in the previous subsection, the constraints on $(b_0^2 + b_\perp^2)\sin^2\phi$ imply certain allowed regions for $\eta_0$ and $\eta_\perp$ (see Eq. (5.62) and the surrounding discussion). These are shown in Fig. 5-16. Recall that $\eta_\lambda \equiv 2(\beta_\lambda^{\rm meas} - \beta)$. Since it is not possible to simultaneously have $\eta_0 = \eta_\perp = 0$ (or $\pi$), this is a clear sign of New Physics (as is $\Lambda_{\perp 0} \neq 0$). However, since neither $\eta_0$ nor $\eta_\perp$ is constrained to lie within a certain range, no bounds on $\beta$ can be derived.

One can perform a similar numerical extremization for $(b_0^2 - b_\perp^2)\sin^2\phi$. However, for this particular data set, we simply reproduce the bounds of Eq. (5.55): $-0.02 \le (b_0^2 - b_\perp^2)\sin^2\phi \le 0.46$. Since this bound is independent of $\Lambda_{\perp 0}$, we have not plotted it.

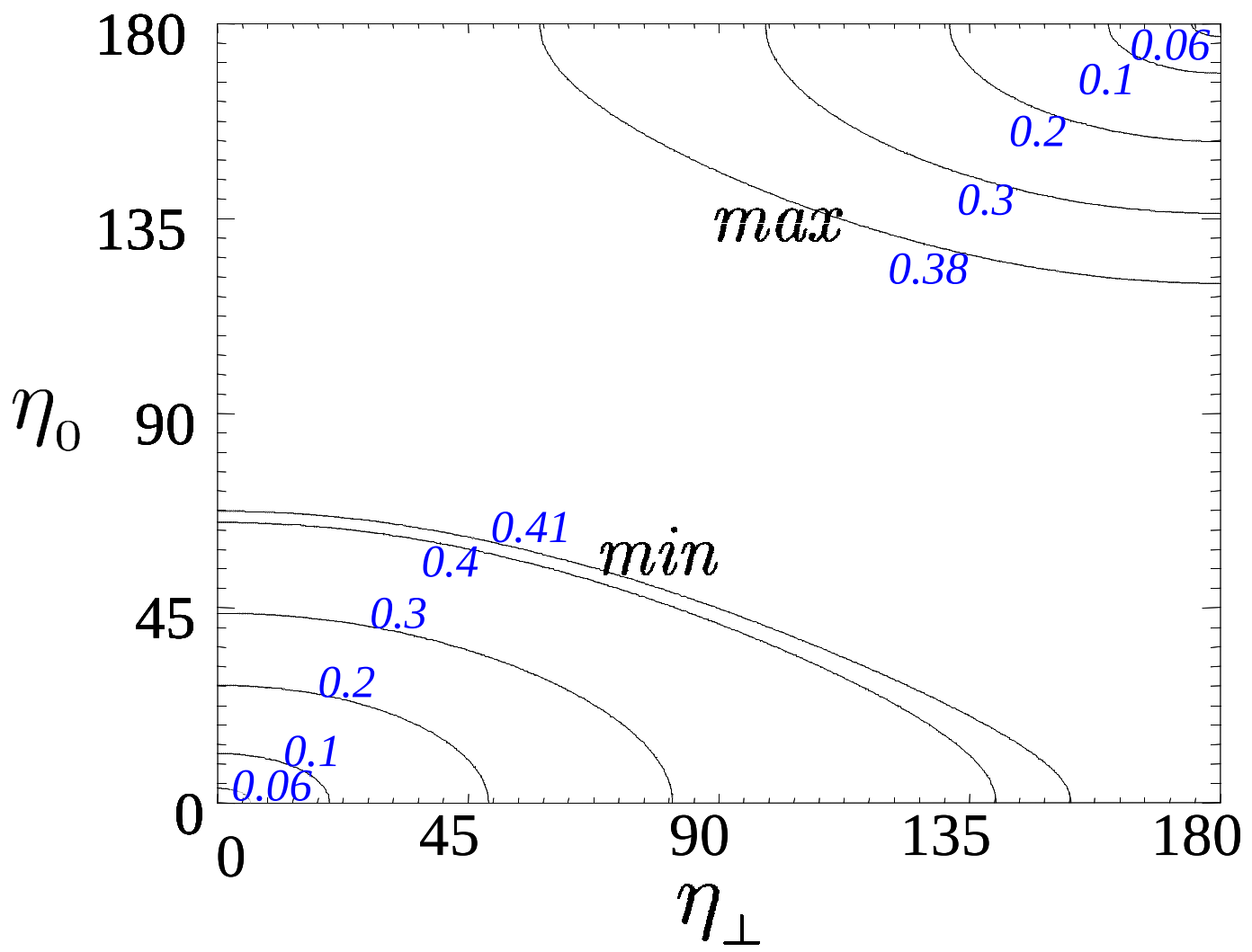

**Figure 5-16.** *Contours showing the (correlated) lower and upper bounds on $\eta_0$ and $\eta_\perp$, corresponding to the different values of $\Lambda_{\perp 0}$ shown on the figure. We have assumed the following values for the observables: $\Lambda_{00} = 0.6$, $\Lambda_{\perp\perp} = 0.16$, $y_0 = 0.60$, $y_\perp = 0.74$. Values of $\eta_0$ and $\eta_\perp$ above (below) and to the right (left) of the minimum (maximum) contours are allowed.*

The easiest way to see whether the numerical extremization of $(b_0^2 \pm b_\perp^2)\sin^2\phi$ depends on $\Lambda_{\perp 0}$ or not is as follows. We refer to Eq. (5.50), and note that $2\omega_{\perp 0} + \eta_0 = \eta_\perp$. The minimal [maximal] value of $(b_0^2 + b_\perp^2)\sin^2\phi$ occurs at the point $(\eta_0, \eta_\perp) = (0, 0)$ $[(\pi, \pi)]$. Thus, the minimal [maximal] value of $(b_0^2 + b_\perp^2)\sin^2\phi$ depends on $\Lambda_{\perp 0}$ only if the point $(0, 0)$ $[(\pi, \pi)]$ is excluded. Similarly, the minimal [maximal] value of $(b_0^2 - b_\perp^2)\sin^2\phi$ depends on $\Lambda_{\perp 0}$ only if the point $(0, \pi)$ $[(\pi, 0)]$ is excluded. Referring to Fig. 5-16, we note that the points $(\eta_0, \eta_\perp) = (0, 0)$, $(\pi, \pi)$ are excluded. Thus, the minimal and maximal values of $(b_0^2 + b_\perp^2)\sin^2\phi$ depend on $\Lambda_{\perp 0}$, as in Fig. 5-15. On the other hand, the points $(\eta_0, \eta_\perp) = (0, \pi)$ and $(\pi, 0)$ are allowed, so the minimal and maximal values of $(b_0^2 - b_\perp^2)\sin^2\phi$ are independent of $\Lambda_{\perp 0}$, as described above.

As noted previously, the minimal values for $(b_0^2 \pm b_\perp^2)$ are equal to those for $(b_0^2 \pm b_\perp^2)\sin^2\phi$. These values can then be combined to give individual minima on $b_0^2$ and $b_\perp^2$.

It is also possible to obtain numerical bounds on the combinations of ratios $r_0^2 \pm r_\perp^2$ [Eq. (5.48)]. The procedure is very similar to that used to obtain bounds on $(b_0^2 \pm b_\perp^2)\sin^2\phi$. The bounds on $r_0^2 \pm r_\perp^2$ are shown in Fig. 5-17. As was the case for $(b_0^2 - b_\perp^2)\sin^2\phi$, the bounds on $r_0^2 - r_\perp^2$ are independent of $\Lambda_{\perp i}$ and follow directly from Eq. (5.47): $-6.44 \le r_0^2 - r_\perp^2 \le 3.85$. However, unlike $b_0^2 \pm b_\perp^2$, upper bounds on $r_0^2 \pm r_\perp^2$ can also be obtained. This constrains the scale of New Physics, and so is quite significant.

**Observation of $\Lambda_{\perp 0}$ and $\Sigma_{\perp 0}$ with $\Sigma_{00} \neq 0$, $\Sigma_{\perp\perp} \neq 0$.**

In this subsection we assume that, in addition to $\Lambda_{\perp 0}$, $\Sigma_{\perp 0}$ is also known ($\omega_{\perp 0}$ is still assumed not to have been measured). We then see, from Eqs. (5.40) and (5.41), that both $\cos(\Delta_0)$ and $\sin(\Delta_0)$ can be determined in terms of these two observables. Thus, $\Delta_0$ can be obtained without ambiguity. Furthermore, using the relation $\cos^2(\Delta_0) + \sin^2(\Delta_0) = 1$, we can *solve* for $\omega_{\perp 0}$, up to an 8-fold discrete ambiguity (*i.e.*, a 4-fold ambiguity in $2\omega_{\perp 0}$). This is shown explicitly in [103]. Thus, $\omega_{\perp 0}$ does not take a range of values, as in the previous subsections, but instead takes specific values. (In fact, one can solve for $\omega_{\perp 0}$, up to discrete ambiguities, whenever two observables are known which involve the interference of two helicity amplitudes.)

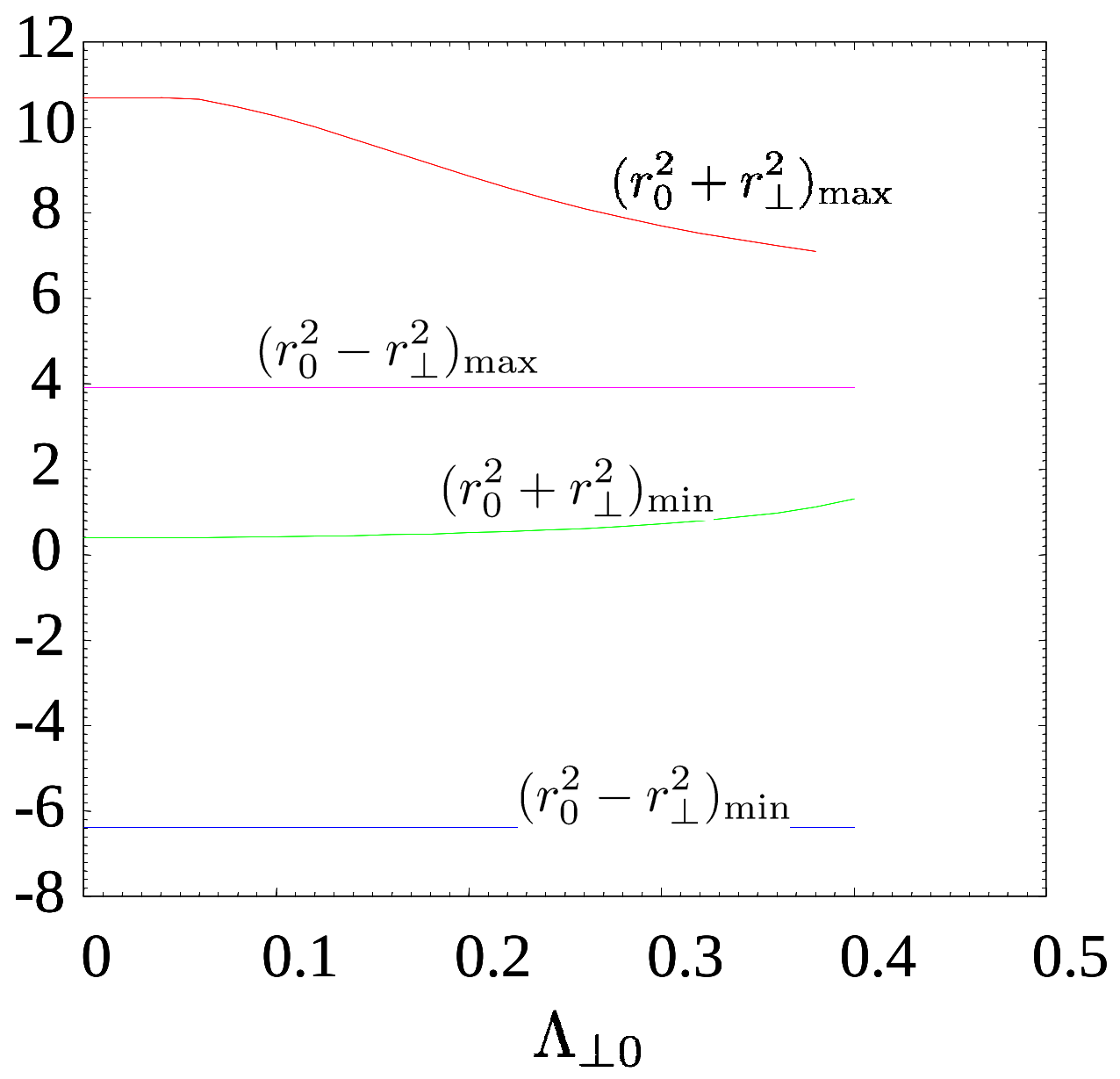

**Figure 5-17.** *Upper and lower bounds on $r_0^2 \pm r_\perp^2$ as a function of $\Lambda_{\perp 0}$. We have assumed the following values for the observables: $\Lambda_{00} = 0.6$, $\Lambda_{\perp\perp} = 0.16$, $y_0 = 0.60$, $y_\perp = 0.74$.*

The expressions and values for $\Delta_0$ and $\omega_{\perp 0}$ are then substituted into Eq. (5.50), and we use MINUIT to numerically minimize the resulting expression with respect to $\eta_i$ and $\phi$. As before, we take $\Lambda_{00} = 0.6$, $\Lambda_{\perp\perp} = 0.16$, $y_0 = 0.60$ and $y_\perp = 0.74$.

The numerical constraints on $(b_0^2 \pm b_\perp^2)\sin^2\phi$ and $(r_0^2 \pm r_\perp^2)$ are shown in Fig. 5-18. In these figures, we have only presented results for positive values of $\Lambda_{\perp 0}$. A point on a plot with a negative value of $\Lambda_{\perp 0}$ is equivalent to that with a positive $\Lambda_{\perp 0}$ and negative $\Sigma_{\perp 0}$. This interchange reverses the signs of $\cos(\Delta_0)$ and $\sin(\Delta_0)$, but does not change the value of $\omega_{\perp 0}$.

As noted above, the knowledge of both $\Lambda_{\perp 0}$ and $\Sigma_{\perp 0}$ allows us to fix the value of $\omega_{\perp 0}$, up to an 8-fold discrete ambiguity. In this case, we can use Eqs. (5.52), (5.53) and (5.54) to directly bound $(b_\lambda^2 \pm b_\sigma^2)\sin^2\phi$. This is illustrated in Fig. 5-19 for $\Lambda_{\perp 0} = 0.2$ and $\Sigma_{\perp 0} = 0.2$.

Of course, it is also possible to measure $2\omega_{\perp 0}$ directly [Eq. (5.31)], up to a 4-fold discrete ambiguity. As we show in Appendix 1, in general these four values only partially overlap with the four values obtained from the derivation of $2\omega_{\perp 0}$ from measurements of $\Lambda_{\perp 0}$ and $\Sigma_{\perp 0}$ – the discrete ambiguity in $2\omega_{\perp 0}$ is reduced to twofold. Thus, by combining the two ways of obtaining $2\omega_{\perp 0}$, the discrete ambiguity can be reduced. This will in turn improve the bounds on the New Physics parameters.

As in the previous subsection, one can also place (correlated) constraints on $\eta_0$ and $\eta_\perp$. In itself, this does lead to a bound on $\beta$. However, if in addition $2\,\beta_\lambda^{\rm meas}$ is measured directly [Eq. (5.31)], then $\beta$ can be constrained.

### Discussion and Summary

We have considered $B \to V_1 V_2$ decays in which $\overline{V}_1 \overline{V}_2 = V_1 V_2$, so that both $B^0$ and $\overline{B}^0$ can decay to the final state $V_1 V_2$. If a time-dependent angular analysis of $B^0(t) \to V_1 V_2$ can be performed, it is possible to extract 18 observables [Eq. (5.27)]. However, there are only six helicity amplitudes describing the decays $B \to V_1 V_2$ and $\overline{B} \to V_1 V_2$. There are therefore only 11 independent observables (equivalent to the magnitudes and relative phases of the six helicity amplitudes).

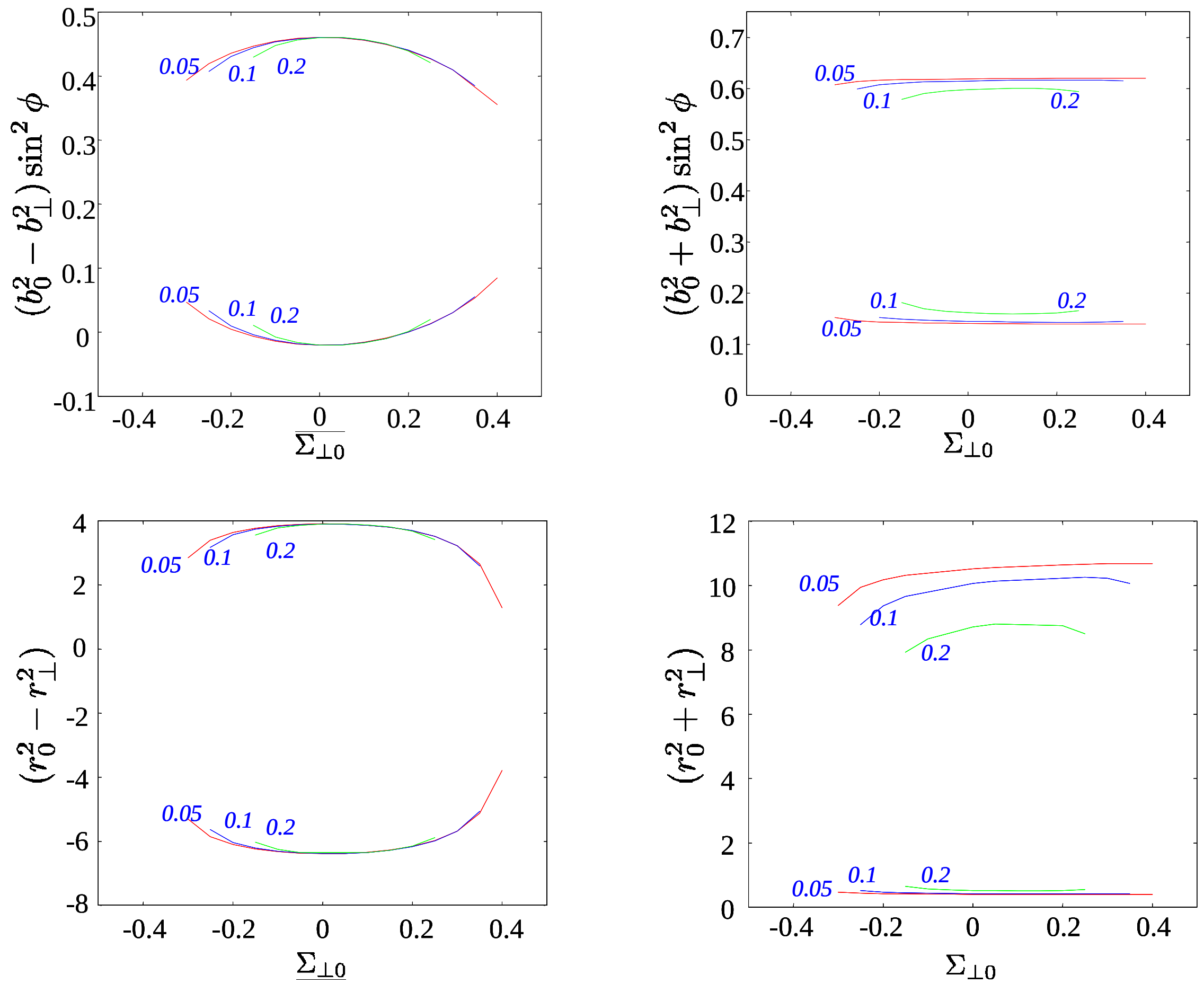

**Figure 5-18.** *The lower and upper bounds on* $(b_0^2 \pm b_\perp^2)\sin^2\phi$ *and* $(r_0^2 \pm r_\perp^2)$ *as a function of* $\Sigma_{\perp 0}$. *Each curve corresponds to a specific value of* $\Lambda_{\perp 0}$, *shown on the figure. We have assumed the following values for the observables:* $\Lambda_{00} = 0.6$, $\Lambda_{\perp\perp} = 0.16$, $y_0 = 0.60$, $y_\perp = 0.74$.

We assume that the $B \to V_1 V_2$ decays are dominated by a single decay amplitude in the Standard Model. The Standard Model parametrization of such decays contains six theoretical parameters: three helicity amplitudes $a_\lambda$, two relative strong phases, and the weak phase $\beta$ (the phase of $B^0 - \overline{B}^0$ mixing). Because there are 18 observables, one has a total of 12 relations to test for the presence of New Physics [Eq. (5.29)]. With 11 independent observables and six Standard Model parameters, one might expect that only five tests are necessary to search for New Physics. However, for certain (fine-tuned) values of the Standard Model parameters, some tests can agree with the Standard Model predictions, even in the presence of New Physics. To take this possibility into account, all 12 New Physics tests are needed to perform a complete search for New Physics.

We assume that a single New Physics amplitude contributes to $B \to V_1 V_2$ decays. In this case one finds a total of 13 theoretical parameters: in addition to the six Standard Model parameters, there are three New Physics helicity amplitudes $b_\lambda$, three additional relative strong phases, and one New Physics weak phase $\phi$. Suppose now that a New Physics signal is seen. With only 11 independent observables, it is clear that one cannot extract any of the New Physics parameters. However, because the equations relating the observables to the parameters are nonlinear, one can place *lower bounds* on the theoretical parameters. This is the main point of the paper.

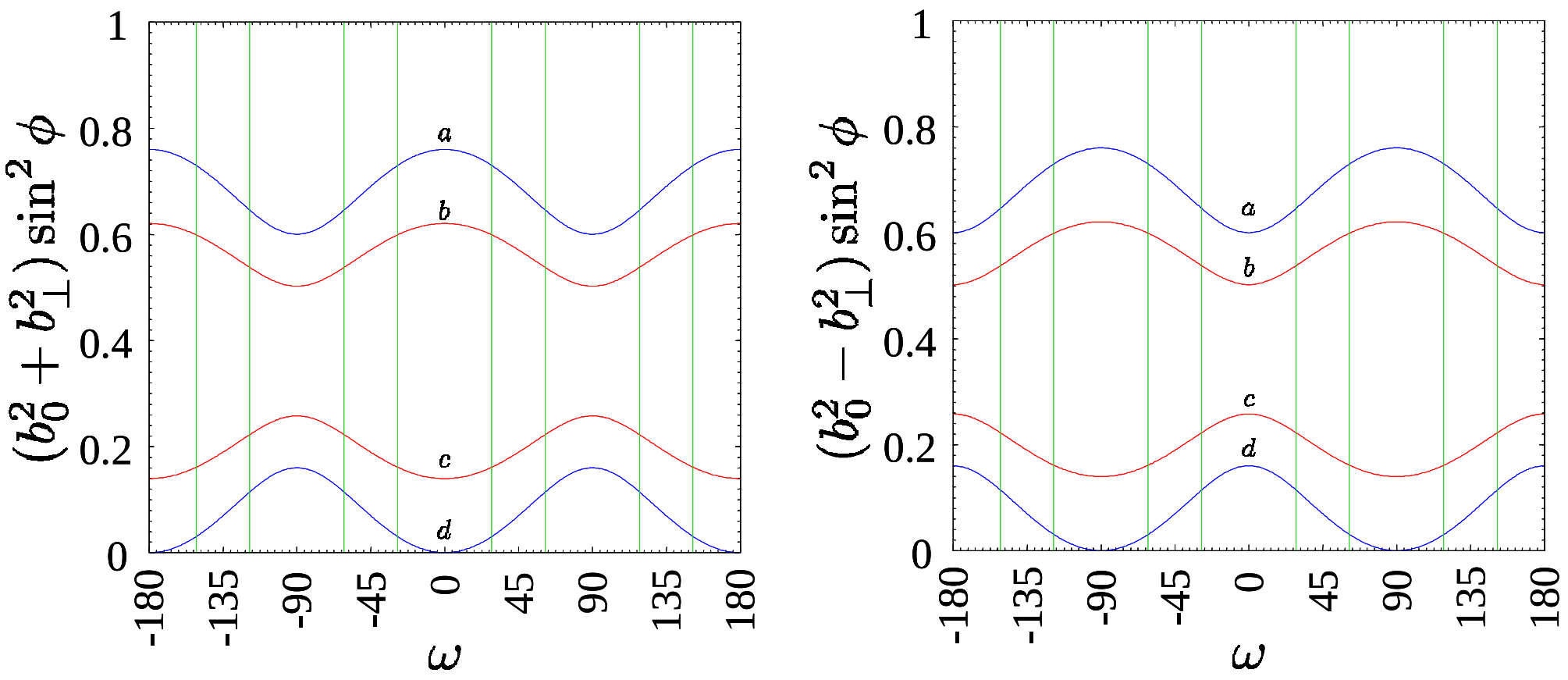

**Figure 5-19.** *The lower and upper bounds on* $(b_0^2 \pm b_\perp^2)\sin^2\phi$ *as a function of* $\omega_{\perp 0}$. *For curves* $b$ *and* $c$ *we have assumed the following values for the observables:* $\Lambda_{00} = 0.6$, $\Lambda_{\perp\perp} = 0.16$, $y_0 = 0.60$, $y_\perp = 0.74$. *Curves* $a$ *and* $d$ *represent the corresponding case with no direct CP asymmetry (i.e.,* $y_0 = y_\perp = 1.0$*). The solutions for* $\omega_{\perp 0}$ *for* $\Lambda_{\perp 0} = 0.2$ *and* $\Sigma_{\perp 0} = 0.2$ *are shown as vertical lines.*

In the previous section we presented several such constraints, which we summarize here. The form of the constraints depends on which observables have been measured. In some cases, it is possible to obtain analytic results; in other cases only numerical bounds are possible.

For example, three distinct New Physics signals are $\Sigma_{\lambda\lambda} \neq 0$, $\beta_\lambda^{\rm meas} \neq \beta_\sigma^{\rm meas}$, and $\Lambda_{\perp i} \neq 0$ (with $\Sigma_{\lambda\lambda} = 0$). In all three cases one can derive analytic lower bounds on the size of $b_\lambda$:

$$
\begin{aligned}
b_\lambda^2 &\geq \frac{1}{2}\left[\Lambda_{\lambda\lambda} - \sqrt{\Lambda_{\lambda\lambda}^2 - \Sigma_{\lambda\lambda}^2}\right], \\
(b_\lambda^2 \pm b_\sigma^2) &\geq \frac{\Lambda_{\lambda\lambda} \pm \Lambda_{\sigma\sigma}}{2} - \frac{\left|y_\lambda \Lambda_{\lambda\lambda} \pm y_\sigma \Lambda_{\sigma\sigma} e^{2i\omega_{\sigma\lambda}}\right|}{2}, \\
2(b_i^2 \pm b_\perp^2) &\geq \Lambda_{ii} \pm \Lambda_{\perp\perp} - \sqrt{(\Lambda_{ii} \pm \Lambda_{\perp\perp})^2 \mp \Lambda_{\perp i}^2}, \qquad (5.63)
\end{aligned}
$$

where $y_\lambda \equiv \sqrt{1 - \Sigma_{\lambda\lambda}^2/\Lambda_{\lambda\lambda}^2}$ and $2\omega_{\sigma\lambda} \equiv 2\beta_\sigma^{meas} - 2\beta_\lambda^{meas}$. One does not know, *a priori*, which of the above constraints will be the strongest – this will depend on the measured values of the observables and/or which New Physics signals are seen.

Constraints on other theoretical parameters are possible. For example, if one measures $\Lambda_{\perp i} \neq 0$ (with $\Sigma_{\lambda\lambda} = 0$), one finds

$$
\Lambda_{ii}\cos\eta_i + \Lambda_{\perp\perp}\cos\eta_\perp \leq \sqrt{(\Lambda_{ii} + \Lambda_{\perp\perp})^2 - \Lambda_{\perp i}^2}, \qquad (5.64)
$$

where $\eta_\lambda \equiv 2(\beta_\lambda^{\rm meas} - \beta)$. Thus, if $\Lambda_{\perp i} \neq 0$, one obtains correlated lower bounds on $|\beta_i^{\rm meas} - \beta|$ and $|\beta_\perp^{\rm meas} - \beta|$.

If more observables or New Physics signals are measured, then it is not possible to obtain analytic constraints – one must perform a numerical analysis. In Section 5.2.4 we presented numerical results for $\Lambda_{\perp 0} \neq 0$ with $\Sigma_{00} \neq 0$ and $\Sigma_{\perp\perp} \neq 0$. In Section 5.2.4 we assumed that in addition $\Sigma_{\perp 0}$ was measured. In both cases we were able to put lower bounds on $(b_0^2 \pm b_\perp^2)$. (Upper bounds are possible only for $(b_0^2 + b_\perp^2)\sin^2\phi$.) We also obtained bounds on $r_0^2 \pm r_\perp^2$ ($r_\lambda \equiv b_\lambda/a_\lambda$).

The bounds improve as more New Physics signals are included in the fits. This is logical. For a particular New Physics signal, the bounds are weakest if that signal is zero. (Indeed, the bounds vanish if all New Physics signals are zero.) If a nonzero value for that signal is found, the bound will improve. Similarly, the bounds generally improve if additional observables are measured, even if they are not signals of New Physics. This is simply because additional measurements imply additional constraints, which can only tighten bounds on the theoretical parameters.

In addition to the bounds on the $b_\lambda$ and $r_\lambda$, it is possible to find correlated numerical constraints on the $\eta_\lambda$, as in Fig. 5-16. If these are combined with a measurement of $2\,\beta_\lambda^{\rm meas}$, one can then obtain a bound on $\beta$, even though New Physics is present.

Finally, even if $2\omega_{\sigma\lambda}$ is not measured directly, one can obtain its value (up to a four-fold ambiguity) through measurements of two observables involving the interference of two helicity amplitudes (as well as the $\Lambda_{\lambda\lambda}$ and $\Sigma_{\lambda\lambda}$). These can be converted into bounds on the other New Physics parameters. If $2\omega_{\sigma\lambda}$ is measured directly, this reduces the discrete ambiguity to twofold, and improves the bounds.

We stress that we have not presented a complete list of constraints on the New Physics parameters – the main aim of this paper was simply to show that such bounds exist. Our results have assumed that only a subset of all observables has been measured, and the bounds vary depending on the New Physics signal found. In practice, the constraints will be obtained by performing a numerical fit using all measurements. If it is possible to measure all observables, one will obtain the strongest constraints possible.

As a specific application, we have noted the apparent discrepancy in the value of $\sin 2\beta$ as obtained from measurements of $B^0(t) \to J/\psi\, K_S^0$ and $B^0(t) \to \phi K_S^0$. In this case, the angular analyses of $B^0(t) \to J/\psi\, K^*$ and $B^0(t) \to \phi K^*$ would allow one to determine if New Physics is indeed present. If New Physics is confirmed, the method described in this paper would allow one to put constraints on the New Physics parameters. If New Physics is subsequently discovered in direct searches at the LHC or ILC, these bounds would indicate whether this New Physics could be responsible for that seen in $B$ decays.

N.S. and R.S. thank D.L. for the hospitality of the Université de Montréal, where part of this work was done. The work of D.L. was financially supported by NSERC of Canada. The work of Nita Sinha was supported by a project of the Department of Science and Technology, India, under the young scientist scheme.

## 5.2.5 Right-Handed Currents, *CP* Violation and $B \to VV$

>– A. L. Kagan –<

We discuss signals for right-handed currents in rare hadronic $B$ decays. Signals in radiative $B$ decays are discussed elsewhere in this Proceedings. Implications of right-handed currents for $CP$-violation phenomenology will be addressed in $\mathrm{SU}(2)_L \times \mathrm{SU}(2)_R \times \mathrm{U}(1)_{B-L} \times P$ symmetric models, and in the more general case of no left-right symmetry. We will see that it may be possible to distinguish between these scenarios at a Super $B$ Factory . Remarkably, the existence of SU(2)$_R$ symmetry could be determined even if it is broken at a scale many orders of magnitude larger than the weak scale, *e.g.*, $M_R \lesssim M_{\rm GUT}$ [111, 112].

A direct test for right-handed currents from polarization measurements in $B$ decays to light vector meson pairs will also be discussed [113]. Finally, in the event that non-Standard Model $CP$-violation is confirmed, *e.g.*, in the $B \to \phi K_S^0$ time-dependent $CP$ asymmetry, an important question will be whether it arises via New Physics contributions to the four-quark operators, the $b \to sg$ dipole operators, or both. We will see that this question can be addressed by comparing $CP$ asymmetries in the different transversity final states in pure penguin $B \to VV$ decays, *e.g.*, $B \to \phi K^*$. The underlying reason is large suppression of the *transverse* dipole operator matrix elements. It is well known that it is difficult to obtain new $\mathcal{O}(1)$ $CP$ violation effects at the *loop-level* from the *dimension-six* four-quark operators. Thus, this information could help discriminate between scenarios in which New Physics effects are induced via loops from those in which they occur at tree-level.

Extensions of the Standard Model often include new $b \to s_R$ right-handed currents. These are conventionally associated with opposite chirality effective operators $\tilde{Q}_i$ which are related to the Standard Model operators by parity transformations,

- QCD Penguin operators

$$
\begin{aligned}
Q_{3,5} &= (\bar{s}b)_{V-A}\,(\bar{q}q)_{V\mp A} &\to\ \tilde{Q}_{3,5} &= (\bar{s}b)_{V+A}\,(\bar{q}q)_{V\pm A}\\
Q_{4,6} &= (\bar{s}_i b_j)_{V-A}\,(\bar{q}_j q_i)_{V\mp A} &\to\ \tilde{Q}_{4,6} &= (\bar{s}_i b_j)_{V+A}\,(\bar{q}_j q_i)_{V\pm A}
\end{aligned}
$$

- Chromo/Electromagnetic Dipole Operators

$$
\begin{aligned}
Q_{7\gamma} &= \tfrac{e}{8\pi^2} m_b \bar{s}_i \sigma^{\mu\nu}(1+\gamma_5) b_i F_{\mu\nu} &\to\ \tilde{Q}_{7\gamma} &= \tfrac{e}{8\pi^2} m_b \bar{s}_i \sigma^{\mu\nu}(1-\gamma_5) b_i F_{\mu\nu}\\
Q_{8g} &= \tfrac{g_s}{8\pi^2} m_b \bar{s} \sigma^{\mu\nu}(1+\gamma_5) t^a b G^a_{\mu\nu} &\to\ \tilde{Q}_{8g} &= \tfrac{g_s}{8\pi^2} m_b \bar{s} \sigma^{\mu\nu}(1-\gamma_5) t^a b G^a_{\mu\nu}
\end{aligned}
$$

- Electroweak Penguin Operators

$$
\begin{aligned}
Q_{7,9} &= \tfrac{3}{2}(\bar{s}b)_{\rm V-A}\, e_q\, (\bar{q}q)_{\rm V\pm A} &\to\ \tilde{Q}_{7,9} &= \tfrac{3}{2}(\bar{s}b)_{\rm V+A}\, e_q\, (\bar{q}q)_{\rm V\mp A}\\
Q_{8,10} &= \tfrac{3}{2}(\bar{s}_i b_j)_{\rm V-A}\, e_q\, (\bar{q}_j q_i)_{\rm V\pm A} &\to\ \tilde{Q}_{8,10} &= \tfrac{3}{2}(\bar{s}_i b_j)_{\rm V+A}\, e_q\, (\bar{q}_j q_i)_{\rm V\mp A}
\end{aligned}
$$

Examples of New Physics which could give rise to right-handed currents include supersymmetric loops which contribute to the QCD penguin or chromomagnetic dipole operators. These are discussed at great length elsewhere in this report. Figure 5-20 illustrates the well known squark-gluino loops in the squark mass-insertion approximation. For example, the down-squark mass-insertion $\delta m^2_{\tilde{b}_R \tilde{s}_L}$ ($\delta m^{2\,*}_{\tilde{s}_R \tilde{b}_L}$) would contribute to $Q_{8g}$ ($\tilde{Q}_{8g}$), whereas $\delta m^2_{\tilde{b}_L \tilde{s}_L}$ ($\delta m^2_{\tilde{s}_R \tilde{b}_R}$) would contribute to $Q_{3,..6}$ ($\tilde{Q}_{3,..,6}$). Right-handed currents could also arise at tree-level via new contributions to the QCD or electroweak penguin operators, *e.g.*, due to flavor-changing $Z^{(\prime)}$ couplings, $R$-parity violating couplings, or color-octet exchange.

### Null Standard Model *CP* asymmetries

We will exploit the large collection of *pure-penguin* $B \to f$ decay modes, which in the Standard Model have

- null decay rate $CP$-asymmetries,

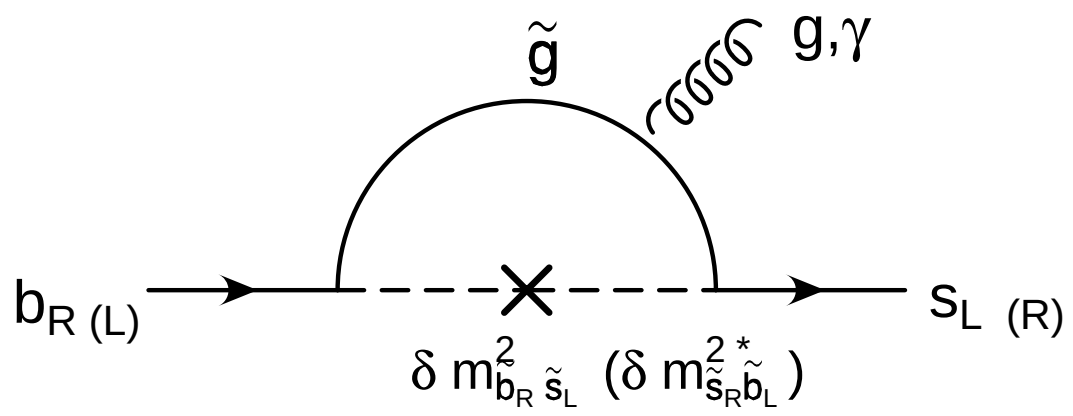

**Figure 5-20.** *Down squark-gluino loop contributions to the Standard Model and opposite chirality dipole operators in the squark mass insertion approximation.*

- null deviations of the time-dependent $CP$-asymmetry coefficient $S_{f_{CP}}$ from $(\sin 2\beta)_{J/\psi K^0_S}$ in decays to $CP$-eigenstates, $|(\sin 2\beta)_{J/\psi K^0_S} + (-)^{CP} S_{f_{\rm CP}}| \sim 1\%$, or
- null triple-product $CP$-asymmetries $A_T^{0,\|}(f) \sim 1\%$ in $B \to VV$ decays.

We recall that there are three helicity amplitudes $\overline{\mathcal{A}}^h$ ($h = 0, -, +$) in $\overline{B} \to VV$ decays: $\overline{\mathcal{A}}^0$, in which both vectors are longitudinaly polarized; $\overline{\mathcal{A}}^-$, in which both vectors have negative helicity; and $\overline{\mathcal{A}}^+$, in which both vectors have positive helicity. In the transversity basis [114], the amplitudes are given by,

$$\overline{\mathcal{A}}_{\perp,\|} = (\overline{\mathcal{A}}^- \mp \overline{\mathcal{A}}^+)/\sqrt{2}, \quad \overline{\mathcal{A}}_L = \overline{\mathcal{A}}^0 \tag{5.65}$$

(In $B$ decays, $\mathcal{A}_{\perp,\|} = (\mathcal{A}^+ \mp \mathcal{A}^-)/\sqrt{2}$). The $CP$-violating triple-products [115] ($\vec{q} \cdot \vec{\epsilon}_1 \times \vec{\epsilon}_2$) are then given by

$$A_T^{0\,(\|)} = \frac{1}{2}\left(\frac{\mathrm{Im}(\overline{\mathcal{A}}_{\perp\,(\|)}\overline{\mathcal{A}}_L^*)}{\sum|\overline{\mathcal{A}}_i|^2} - \frac{\mathrm{Im}(\mathcal{A}_{\perp\,(\|)}\mathcal{A}_L^*)}{\sum|\mathcal{A}_i|^2}\right). \tag{5.66}$$

The triple-products are discussed in detail in the contribution of A. Datta.

A partial list of null Standard Model $CP$ asymmetries in pure-penguin decays is given below [116],

- $A_{CP}(K^0\pi^\pm)$, $A_{CP}(\eta' K^\pm)$, $A_{CP}(\phi K^{*0,\pm})_{0,\|,\perp}$, $A_{CP}(K^{*0}\pi^\pm)$, $A_{CP}(K^{*0}\rho^\pm)_{0,\|,\perp}$, $A_{CP}(K_1\pi^\pm)$, $A_{CP}(K^0 a_1^\pm)$, $A_{CP}(\phi K^{0,\pm})$,...
- $S_{\phi K^0_S}$, $S_{\eta' K^0_S}$, $(S_{\phi K^{*0}})_{0,\|,\perp}$, $(S_{\phi K_1})_{0,\|,\perp}$, $S_{K^0_S K^0_S K^0_S}$,...
- $A_T^{0,\|}(\phi K^{*0,\pm})$, $A_T^{0,\|}(K^{*0}\rho^\pm)$,...

In addition, there are several modes that are penguin-dominated and are predicted to have approximately null or small Standard Model asymmetries, *e.g.*, $S_{K^+K^-K^0}$ ($\phi$ subtracted) [117, 118], $S_{K^0_S\pi^0}$ [119], and $S_{f^0 K^0_S}$.

### Right-handed currents and *CP* violation

Under parity, the effective operators transform as $Q_i \leftrightarrow \tilde{Q}_i$. The New Physics amplitudes, for final states $f$ with parity $P_f$, therefore satisfy

$$\langle f|Q_i|B\rangle = -(-)^{P_f}\langle f|\tilde{Q}_i|B\rangle \;\Rightarrow\; A_i^{NP}(B \to f) \propto C_i^{\rm NP}(\mu_b) - (-)^{P_f}\tilde{C}_i^{\rm NP}(\mu_b)\,, \tag{5.67}$$

where $C_i^{\rm NP}$ and $\tilde{C}_i^{\rm NP}$ are the new Wilson coefficient contributions to the *i*'th pair of Standard Model and opposite chirality operators [120]. It follows that for decays to $PP$, $VP$, and $SP$ final states, where $S$, $P$ and $V$ are scalar, pseudoscalar, and vector mesons, respectively, the New Physics amplitudes satisfy

$$A_i^{\rm NP}(B \to PP) \propto C_i^{\rm NP}(\mu_b) - \tilde{C}_i^{\rm NP}(\mu_b), \quad A_i^{\rm NP}(B \to VP) \propto C_i^{\rm NP}(\mu_b) + \tilde{C}_i^{\rm NP}(\mu_b)$$

$$A_i^{\rm NP}(B \to SP) \propto C_i^{\rm NP}(\mu_b) + \tilde{C}_i^{\rm NP}(\mu_b)\,. \tag{5.68}$$

In $B \to VV$ decays the $\perp$ transversity and $0,\ \|$ transversity final states are $P$-odd and $P$-even, respectively, yielding

$$A_i^{NP}(B \to VV)_{0,\|} \propto C_i^{\rm NP}(\mu_b) - \tilde{C}_i^{\rm NP}(\mu_b), \quad A_i^{NP}(B \to VV)_{\perp} \propto C_i^{\rm NP}(\mu_b) + \tilde{C}_i^{\rm NP}(\mu_b)\,. \tag{5.69}$$

Similarly, replacing one of the vector mesons with an axial-vector meson gives

$$A_i^{NP}(B \to VA)_{0,\|} \propto C_i^{\rm NP}(\mu_b) + \tilde{C}_i^{\rm NP}(\mu_b), \quad A_i^{NP}(B \to VA)_{\perp} \propto C_i^{\rm NP}(\mu_b) - \tilde{C}_i^{\rm NP}(\mu_b)\,. \tag{5.70}$$

It is useful to classify the null and approximately null Standard Model *CP* asymmetries listed above according to whether the final state is $P$-odd or $P$-even,

- $P$-even: $A_{CP}(K^0\pi^\pm)$, $A_{CP}(\eta' K^\pm)$, $A_{CP}(\phi K^{*\pm})_{0,\|}$, $S_{\eta' K^0_S}$, $(S_{\phi K^{*0}})_{0,\|}$, $A_{CP}(K^{*0}\rho^\pm)_{0,\|}$, $A_{CP}(K_1\pi^\pm)$, $A_{CP}(K^0 a_1^\pm)$, $(S_{\phi K_1})_\perp$,...
- $P$-odd: $A_{CP}(\phi K^\pm)$, $S_{\phi K^0_S}$, $A_{CP}(K^{*0}\pi^\pm)$, $A_{CP}(\phi K^{*\pm})_\perp$, $(S_{\phi K^{*0}})_\perp$, $(S_{\phi K_1})_{0,\|}$,...
- Modes with small Standard Model asymmetries: $S_{K^+K^-K^0}$ (approximately $P$-even), $S_{K^0_S\pi^0}$ ($P$-even), and $S_{f^0K^0_S}$ ($P$-odd).

We are now ready to discuss implications for *CP* violation phenomenology in the two classes of models mentioned earlier.

**Parity-symmetric New Physics**

In the limit in which New Physics is parity-symmetric at the weak scale the relation $C_i^{\rm NP}(\mu_W) = \tilde{C}_i^{\rm NP}(\mu_W)$ would hold. In light of (5.67), this would imply [120, 111]

- preservation of null *CP* asymmetry predictions in $P$-even final states. Similarly, the $\epsilon'/\epsilon$ constraint would be trivially satisfied.
- possibly large departures from null *CP* asymmetries in $P$-odd final states.

For example, no deviations in $S_{\eta' K^0_S}$, $(S_{\phi K^{*0}})_{0,\|}$, $A_{CP}(\phi K^{*\pm})$, $A_{CP}(K^0\pi^\pm)$ could be accompanied by significant deviations in $S_{\phi K^0_S}$, $A_{CP}(\phi K^\pm)$, $(S_{\phi K^{*0}})_\perp$, and $S_{f^0K^0_S}$. Both of the triple-products $A_T^0$ and $A_T^\|$ in (5.66) could be affected through a modification of $\mathcal{A}_\perp(VV)$. However, there would be no novel *CP* asymmetry in the interference of the parallel and longitudinal polarizations. Equivalently, the measurable quantities $\Delta_L$ and $\Delta_\|$ defined below

$$\Delta_{L\,(\|)} = ({\rm Arg}\,\overline{\mathcal{A}}_{L\,(\|)} - {\rm Arg}\,\overline{\mathcal{A}}_\perp) - ({\rm Arg}\,\mathcal{A}_{L\,(\|)} - {\rm Arg}\,\mathcal{A}_\perp) \tag{5.71}$$

would be equal.

Parity-symmetric New Physics requires ${\rm SU}(2)_L \times {\rm SU}(2)_R \times {\rm U}(1)_{B-L} \times P$ symmetry at high energies. Thus, exact weak scale parity can not be realized due to renormalization group effects below the ${\rm SU}(2)_R$ breaking scale, $M_R$. Potentially, the largest source of parity violation is the difference between the top and bottom quark Yukawa couplings. In particular, when $\lambda^t \neq \lambda^b$ the charged Higgs Yukawa couplings break parity. Two scenarios for the Yukawa couplings naturally present themselves:

- moderate $\tan\beta$, or $\lambda^t >> \lambda^b$
- maximal-parity: $\lambda^b = \lambda^t + \mathcal{O}(V_{cb})$ or $\tan\beta \cong m_t/m_b$ Small corrections to the limit of equal up and down Yukawa matrices are required in order to generate the observed CKM quark mixings and light quark masses. $V_{cb}$ therefore sets the scale for minimal parity violation in the Yukawa sector.

A large hierarchy between the SU(2)$_R$ breaking scale and the weak scale can be realized naturally in supersymmetric left-right symmetric models containing two Higgs bi-doublet superfields $\Phi_{1,2}(2_L, 2_R, 0_{B-L})$ (or four SU(2)$_L$ doublets). Via the 'doublet-doublet splitting mechanism' [121] two linear combinations of the Higgs doublets acquire masses of order $M_R$, leaving the two light Higgs doublets of the MSSM. Realization of approximately parity symmetric contributions to the *dipole* operators favors explicit *CP* violation. Spontaneous *CP* violation could lead to complex $P$-violating vacuum expectation values, which would feed into new loop contributions to the operators. For example, $P$ invariance above the weak scale would imply

$$C_{8g}^{\rm NP} = \kappa\langle\phi\rangle, \qquad \tilde{C}_{8g}^{\rm NP} = \kappa\langle\phi^\dagger\rangle, \tag{5.72}$$

where $\langle\phi\rangle$ breaks SU(2)$_L$ and $\kappa \sim 1/M_{\rm NP}^2$ is in general complex due to explicit *CP* violating phases. ($M_{\rm NP}$ is an order  TeVNew Physics scale, *e.g.*, the squark or gluino masses in Fig. 1). Thus, $\langle\phi\rangle$ would have to be real to good approximation in order to obtain $C_{8g}^{\rm NP} \approx \tilde{C}_{8g}^{\rm NP}$. Note that this requires real gaugino masses; otherwise RGE effects would induce a complex Higgs bilinear $B$ term in the scalar potential, thus leading to complex $\langle\phi\rangle$. Ordinary parity symmetry insures real $U(1)_{B-L}$ and SU(3)$_C$ gaugino masses. Real $\mathrm{SU}(2)_L \times \mathrm{SU}(2)_R$ gaugino masses would follow from the SO(10) generalization of parity [122]. All the VEVs entering new *four-quark* operator loops can, in principle, be parity neutral. Therefore, real VEVs are less crucial for obtaining approximately parity-symmetric four-quark operator contributions.

We have carried out a two-loop RGE analysis for down squark-gluino loop contributions to the dipole operators. Choosing parity symmetric boundary conditions at $M_R$, taking $M_R \leq M_{\rm GUT}$, and running to the weak scale we obtain

- Moderate $\tan\beta$, *e.g.*, $\tan\beta \sim 5$:

$$\frac{\mathrm{Re}[C_{8g}^{\rm NP}(m_W) - \tilde{C}_{8g}^{\rm NP}(m_W)]}{\mathrm{Re}[C_{8g}^{\rm NP}(m_W) + \tilde{C}_{8g}^{\rm NP}(m_W)]} \leq 10\%, \qquad \frac{\mathrm{Im}[C_{8g}^{\rm NP}(m_W) - \tilde{C}_{8g}^{\rm NP}(m_W)]}{\mathrm{Im}[C_{8g}^{\rm NP}(m_W) + \tilde{C}_{8g}^{\rm NP}(m_W)]} \leq 10\%$$

- Maximal parity, $\tan\beta \cong m_t/m_b$

$$\frac{\mathrm{Im}[C_{8g}^{\rm NP}(m_W) - \tilde{C}_{8g}^{\rm NP}(m_W)]}{\mathrm{Im}[C_{8g}^{\rm NP}(m_W) + \tilde{C}_{8g}^{\rm NP}(m_W)]} = O(1\%)$$

The above quantities give a measure of parity violation in the weak scale Wilson coefficients. Thus, we see that for $M_R \leq M_{\rm GUT}$, new *CP*-violating contributions to the low energy Lagrangian could *respect parity* to $\mathcal{O}(1\%)$. Precision *CP* violation measurements in $B$ decays which respect (violate) null Standard Model predictions in $P$-even ($P$-odd) final states could therefore provide evidence for $\mathrm{SU}(2)_L \times \mathrm{SU}(2)_R \times \mathrm{U}(1)_{B-L} \times P$ symmetry, even if SU(2)$_R$ is broken at the GUT scale. We expect similar results for survival of parity in the four-quark operators [112].

**The $^{199}$Hg mercury EDM constraint**

Any discussion of *dipole* operator phenomenology should consider the upper bound on the strange quark chromo-electric dipole moment $d_s^C$, obtained from the upper bound on the $^{199}$Hg mercury electric dipole moment (EDM) [177]. Correlations between $d_s^C$ and new *CP*-violating contributions to $C_{8g}$, $\tilde{C}_{8g}$ are most easily seen by writing the dipole operator effective Hamiltonian in the weak interaction basis,

$$\frac{G_F}{\sqrt{2}}\, V_{cb}V_{cs}\, C_{i_L j_R}\, \frac{g_s}{8\pi^2} m_b\, \bar{i}\, \sigma^{\mu\nu}(1+\gamma^5)\, j\, G_{\mu\nu} + h.c.\,. \tag{5.73}$$

$|i_L\rangle$ and $|i_R\rangle$ ($i = 1,2,3$) are the left-handed and right-handed down quark weak interaction eigenstates, respectively. The mass eigenstates can be written as $\left|d^i_{L(R)}\right\rangle = x_{ij}^{L(R)}\,|i_{L(R)}\rangle$, where $d^{1,2,3}$ stands for the $d, s, b$ quarks, respectively, and $x_{ii}^{L,R} \approx 1$. The bound on $d_s^C$ is $\mathrm{Im}\, C_{s_L s_R} \lesssim 4 \times 10^{-4}$, with large theoretical uncertainty, where $C_{s_L s_R}$ is the flavor-diagonal strange quark dipole operator coefficient (in the mass eigentate basis). It is given as

$$C_{s_L s_R} \approx C_{2_L 2_R} + x_{23}^{L*} C_{3_L 2_R} + x_{23}^R C_{2_L 3_R} + x_{23}^{L*} x_{23}^R C_{3_L 3_R} + \dots . \tag{5.74}$$

Similarly, the $b \to sg$ Wilson coefficients are given as

$$C_{8g} \approx C_{2_L 3_R} + x_{23}^{L*} C_{3_L 3_R} + \dots, \quad \tilde{C}_{8g} \approx C_{3_L 2_R}^* + x_{23}^{R*} C_{3_L 3_R}^* + \dots . \tag{5.75}$$

If significant contributions to the CKM matrix elements are generated in the down quark sector, then $x_{23}^L, x_{32}^L \sim V_{cb}$, $x_{13}^L, x_{31}^L \sim V_{ub}$, and $x_{12}^L, x_{21}^L \sim \theta_c$. In the absence of special flavor symmetries, similar magnitudes would be expected for the corresponding right-handed quark mixing coefficients, $x_{ij}^R$. We therefore expect $C_{s_L s_R} \sim V_{cb} C_{8g} + \dots$ to hold generically. $S_{\phi K_S^0} < 0$ would correspond to $\mathrm{Im}\,[C_{8g}(m_b) + \tilde{C}_{8g}(m_b)] \sim 1$. Thus, $\mathcal{O}(1)$ $CP$-violating effects generically correspond to a value for $d_s^C$ which is a factor of 100 too large. One way to evade this bound is by invoking some mechanism, *e.g.*, flavor symmetries, for generating the large hierarchies $x_{23}^R << x_{23}^L$ and $C_{3_L 2_R} << C_{2_L 3_R}$. An elegant alternative solution is provided by parity symmetry [123]. It is well known that EDM's must vanish in the parity symmetric limit, see *e.g.*, [122]. For example, in (5.74) exact parity would imply $x_{23}^L = x_{23}^R$, $C_{3_L 2_R} = C_{2_L 3_R}^*$ and real $C_{i_L i_R}$, thus yielding a real coefficient, $C_{s_L s_R}$. An RGE analysis along the lines discussed above is required in order to determine the extent to which this can be realized at low energies. We find that in both the maximal parity scenario ($\tan\beta \cong m_t/m_b$) and in moderate $\tan\beta$ scenarios it is possible to obtain $S_{\phi K_S^0} < 0$ and at the same time satisfy the bound on $d_s^C$ if $M_R \leq M_{\mathrm{GUT}}$ [112].

**Generic case: Right-handed currents without Parity**

In the parity-symmetric scenario, an unambiguous theoretical interpretation of the pattern of $CP$ violation is possible, because null predictions are maintained for the $P$-even final states. However, if new contributions to the $Q_i$ and $\tilde{Q}_i$ operators are unrelated, then $CP$ asymmetries in the the $P$-odd and $P$-even null Standard Model modes could differ significantly both *from each other, and from the null predictions*. This is due to the opposite relative sign between the left-handed and right-handed New Physics amplitudes for $P$-odd and $P$-even final states in Eqs. (5.67)–(5.70). For example, $S_{\phi K_S^0}$ and $S_{\eta' K_S^0}$ could be affected differently in the MSSM [124, 125]. An interesting illustration would be provided by models with $\mathcal{O}(1)$ contributions to the $\tilde{Q}_i$, and negligible new contributions to the $Q_i$. This could happen, for example, in supersymmetric models with large (negligible) $\tilde{s}_{R(L)} - \tilde{b}_{R(L)}$ squark mixing [125]–[128], or in models in which $R$-parity violation induces opposite chirality four-quark operators at the tree-level [129]. Unrelated right-handed currents could also arise in warped extra dimension models with bulk (custodial) left-right symmetry [130].

Unfortunately, $CP$ asymmetry predictions have large theoretical uncertainties due to $1/m$ power corrections, especially from the QCD penguin annihilation amplitudes. They are therefore difficult to interpret. An illustration is provided in Fig. 5-21, which compares predictions for $S_{\phi K_S^0}$ and $S_{\pi^0 K_S^0}$ arising from new contributions to $Q_{8g}$ and $\tilde{Q}_{8g}$ in QCD factorization [131, 132]. For $S_{\phi K_S^0}$ we take $C_{8g}^{\mathrm{NP}}(m_W) + \tilde{C}_{8g}^{\mathrm{NP}}(m_W) = e^{i\theta}$. For $S_{\pi^0 K_S^0}$ two corresponding cases are considered: (a) a purely left-handed current, $C_{8g}^{\mathrm{NP}}(m_W) = e^{i\theta}$, $\tilde{C}_{8g}^{\mathrm{NP}}(m_W) = 0$, (b) a purely right-handed current, $C_{8g}^{\mathrm{NP}}(m_W) = 0$, $\tilde{C}_{8g}^{\mathrm{NP}}(m_W) = e^{i\theta}$. The scatter plots scan over the input parameter ranges given in [132] (with the exception of the Gegenbauer moments of the light meson light-cone distribution amplitudes and $m_c/m_b$, which have been set to their default values). In addition, the branching ratios are required to lie within their 90% CL intervals.

Clearly, very different values for the two $CP$ asymmetries can be realized if the New Physics only appears in $Q_{8g}$. For example, for $\theta \sim 50°$, it is possible to obtain $S_{\phi K_S^0} \sim -0.5$ and $S_{\pi^0 K_S^0} \sim 0.4$. The theoretical uncertainty in $S_{\eta' K_S^0}$ is larger than for $S_{\pi^0 K_S^0}$. We therefore expect that even larger differences are possible between $S_{\eta' K_S^0}$ and $S_{\phi K_S^0}$, for purely left-handed currents. However, Fig. 5-21 suggests that $S_{\phi K_S^0} < 0$ and $S_{\pi^0 K_S^0} > (\sin 2\beta)_{J/\psi K_S^0}$ (corresponding to $\theta = 0$) could be a signal for right-handed currents [125]. More theoretical studies are needed in order to determine if this is indeed the case. In particular, a more thorough analysis of uncertainties due to $\mathcal{O}(1/m)$ effects needs to be undertaken. For example, power corrections to the dipole operator matrix elements remain to be included. Furthermore, the impact on $S_{\phi K_S^0}$, $S_{\pi^0 K_S^0}$ of New Physics in all of the 'left-handed' four-quark operators needs to be thoroughly studied.

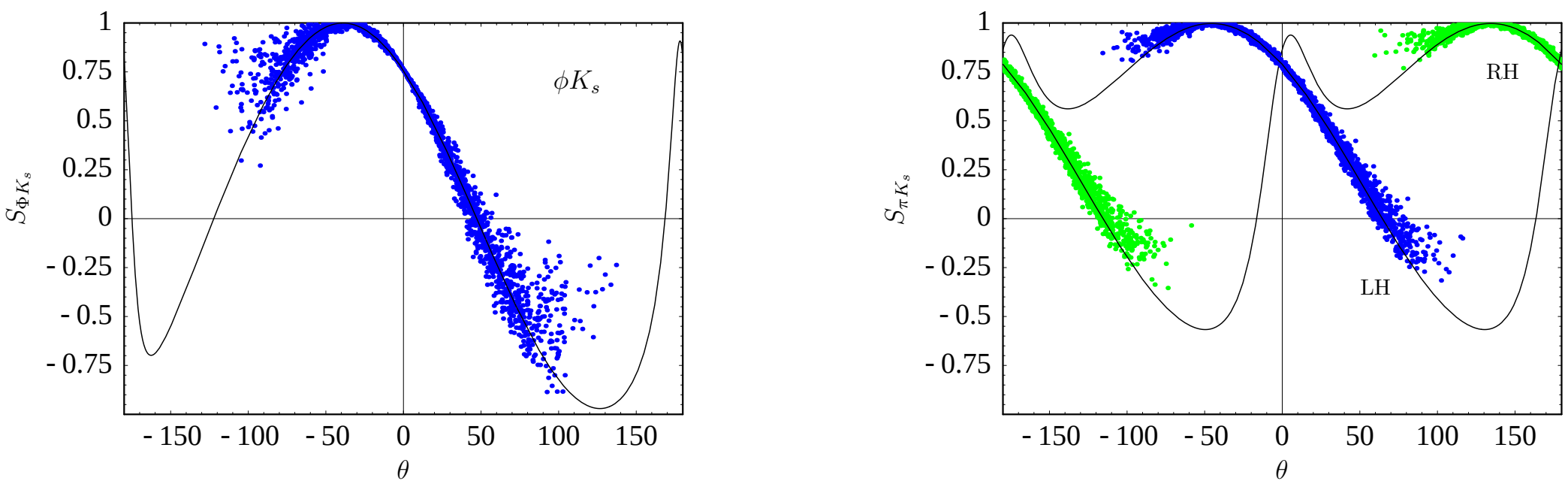

**Figure 5-21.** *Scatter plots in QCD factorization for $S_{\phi K^0_S}$ vs. $\theta$ for $C_{8g}^{\rm NP}(m_b)+\tilde{C}_{8g}^{\rm NP}(m_b) = e^{i\theta}$, and for $S_{\pi^0 K^0_S}$ versus $\theta$ for left-handed currents, $C_{8g}^{\rm NP}(m_b) = e^{i\theta}$, $\tilde{C}_{8g}^{\rm NP} = 0$ (blue), and for right-handed-currents, $C_{8g}^{\rm NP} = 0$, $\tilde{C}_{8g}^{\rm NP}(m_b) = e^{i\theta}$ (green).*

## Polarization and *CP* violation in $B \to VV$ decays

A discussion of polarization in $B \to VV$ decays has been presented in [113] in the framework of QCD factorization. Here we summarize some of the results. To begin with we note that the polarization should be sensitive to the $V-A$ structure of the Standard Model, due to the power suppression associated with the 'helicity-flip' of a collinear quark. For example, in the Standard Model the factorizable graphs for $\overline{B} \to \phi K^*$ are due to transition operators with chirality structures $(\overline{s}b)_{V-A}(\overline{s}s)_{V\mp A}$, see Fig. 5-22 . In the helicity amplitude $\overline{\mathcal{A}}^-$ a collinear $s$ or $\overline{s}$ quark with positive helicity ends up in the negatively polarized $\phi$, whereas in $\overline{\mathcal{A}}^+$ a second quark 'helicity-flip' is required in the form factor transition. Collinear quark helicity flips require transverse momentum, $k_\perp$, implying a suppression of $\mathcal{O}(\Lambda_{\rm QCD}/m_b)$ per flip. In the case of new right-handed currents, *e.g.*, $(\overline{s}b)_{V+A}(\overline{s}s)_{V\pm A}$, the helicity amplitude hierarchy would be inverted, with $\overline{\mathcal{A}}^+$ and $\overline{\mathcal{A}}^-$ requiring one and two helicity-flips, respectively.

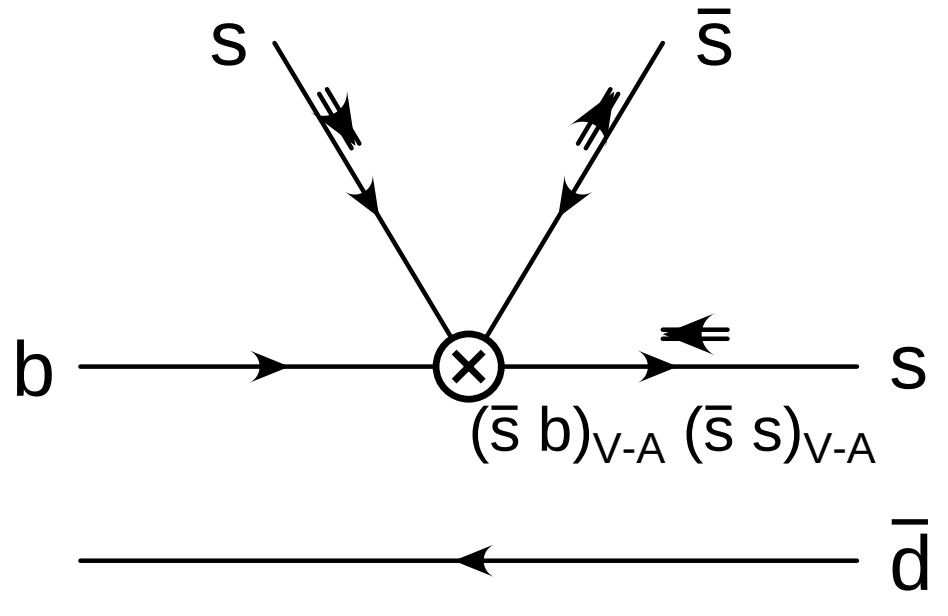

**Figure 5-22.** *Quark helicities (short arrows) for the $\overline{B} \to \phi K^*$ matrix element of the operator $(\overline{s}b)_{V-A}(\overline{s}s)_{V-A}$ in naive factorization. Upward lines form the $\phi$ meson.*

In naive factorization the $\overline{B} \to \phi K^*$ helicity amplitudes, supplemented by the large energy form factor relations [133], satisfy

$$\overline{\mathcal{A}}^0 \propto f_\phi m_B^2\, \zeta_\parallel^{K^*}, \quad \overline{\mathcal{A}}^- \propto -f_\phi m_\phi m_B\, 2\, \zeta_\perp^{K^*}, \quad \overline{\mathcal{A}}^+ \propto -f_\phi m_\phi m_B\, 2\, \zeta_\perp^{K^*} r_\perp^{K^*}. \tag{5.76}$$

$\zeta_\parallel^V$ and $\zeta_\perp^V$ are the $B \to V$ form factors in the large energy limit [133]. Both scale as $m_b^{-3/2}$ in the heavy quark limit, implying $\overline{\mathcal{A}}^- / \overline{\mathcal{A}}^0 = \mathcal{O}(m_\phi/m_B)$. $r_\perp$ parametrizes form factor helicity suppression. It is given by

$$r_\perp = \frac{(1+m_{V_1}/m_B)A_1^{V_1} - (1-m_{V_1}/m_B)V^{V_1}}{(1+m_{V_1}/m_B)A_1^{V_1} + (1-m_{V_1}/m_B)V^{V_1}}\,, \tag{5.77}$$

where $A_{1,2}$ and $V$ are the axial-vector and vector current form factors, respectively. The large energy relations imply that $r_\perp$ vanishes at leading power, reflecting the fact that helicity suppression is $\mathcal{O}(1/m_b)$. Thus, $\overline{\mathcal{A}}^{+}/\overline{\mathcal{A}}^{-} = O(\Lambda_{\rm QCD}/m_b)$. Light-cone QCD sum rules [134], and lattice form factor determinations scaled to low $q^2$ using the sum rule approach [135], give $r_\perp^{K^*} \approx 1-3\,\%$; QCD sum rules give $r_\perp^{K^*} \approx 5\,\%$ [136]; and the BSW model gives $r_\perp^{K^*} \approx 10\%$ [137].

The polarization fractions in the transversity basis (5.65) therefore satisfy

$$1 - f_L = \mathcal{O}\left(1/m_b^2\right), \quad f_\perp/f_\parallel = 1 + \mathcal{O}\left(1/m_b\right), \tag{5.78}$$

in naive factorization, where the subscript $L$ refers to longitudinal polarization, $f_i = \Gamma_i/\Gamma_{\rm total}$, and $f_L + f_\perp + f_\parallel = 1$. $\tilde{C}_i^{NP} \sim C_i^{\rm SM}$ The measured longitudinal fractions for $B \to \rho\rho$ are close to 1 [138, 139]. This is not the case for $B \to \phi K^{*0}$ for which full angular analyses yield

$$f_L = .43 \pm .09 \pm .04, \quad f_\perp = .41 \pm .10 \pm .04 \;\; [140] \tag{5.79}$$

$$f_L = .52 \pm .07 \pm .02, \quad f_\perp = .27 \pm .07 \pm .02 \;\; [141]. \tag{5.80}$$

Naively averaging the Belle and *BABAR* measurements (without taking correlations into account) also yields $f_\perp/f_\parallel = 1.39 \pm .69$. We must go beyond naive factorization in order to determine if the small value of $f_L(\phi K^*)$ could simply be due to the dominance of QCD penguin operators in $\Delta S = 1$ decays, rather than New Physics. In particular, it is necessary to determine if the power counting in (5.78) is preserved by non-factorizable graphs, *i.e.*, penguin contractions, vertex corrections, spectator interactions, annihilation graphs, and graphs involving higher Fock-state gluons. This question can be addressed in QCD factorization [113].

In QCD factorization exclusive two-body decay amplitudes are given in terms of convolutions of hard scattering kernels with meson light-cone distribution amplitudes [131, 132]. At leading power this leads to factorization of short and long-distance physics. This factorization breaks down at sub-leading powers with the appearance of logarithmic infrared divergences, *e.g.*, $\int_0^1 dx/x \sim \ln m_B/\Lambda_h$, where $x$ is the light-cone quark momentum fraction in a final state meson, and $\Lambda_h \sim \Lambda_{\rm QCD}$ is a physical infrared cutoff. Nevertheless, the power-counting for all amplitudes can be obtained. The extent to which it holds numerically can be determined by assigning large uncertainties to the logarithmic divergences. Fortunately, certain polarization observables are less sensitive to this uncertainty, particularly after experimental constraints, *e.g.*, total rate or total transverse rate, are imposed.

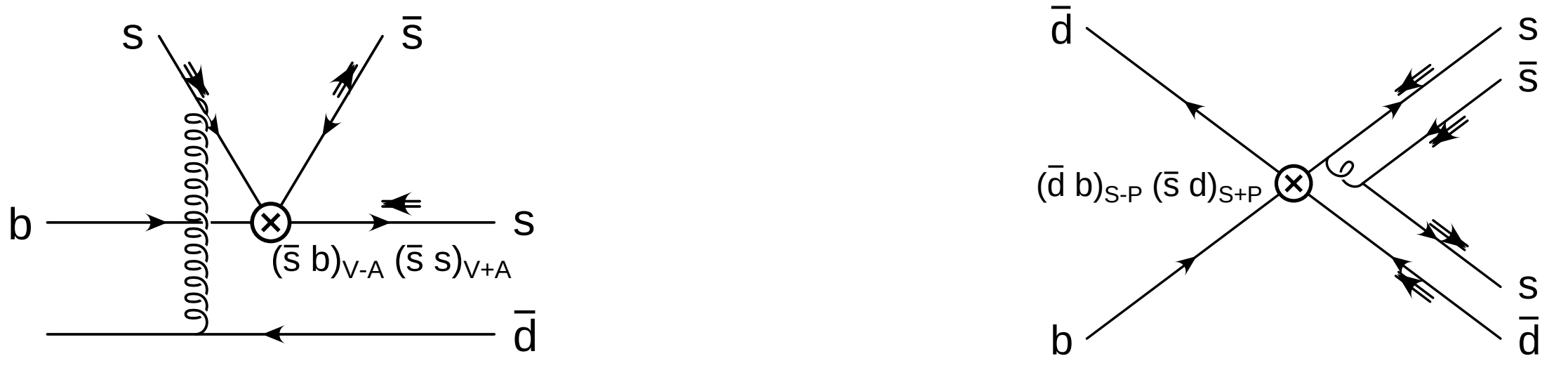

**Figure 5-23.** *Quark helicities in $\overline{B} \to \phi K^*$ matrix elements: the hard spectator interaction for the operator $(\bar{s}b)_{V-A}(\bar{s}s)_{V\mp A}$ (left), and annihilation graphs for the operator $(\bar{d}b)_{S-P}(\bar{s}d)_{S+P}$ with gluon emitted from the final state quarks (right).*

Examples of logarithmically divergent hard spectator interaction and QCD penguin annihilation graphs are shown in Fig. 5-23, with the quark helicities indicated. The power counting for the helicity amplitudes of the annihilation graph, including logarithmic divergences, is

$$\overline{\mathcal{A}}^0, \overline{\mathcal{A}}^- = \mathcal{O}\left(\frac{1}{m^2}\ln^2\frac{m}{\Lambda_h}\right), \qquad \overline{\mathcal{A}}^+ = \mathcal{O}\left(\frac{1}{m^4}\ln^2\frac{m}{\Lambda_h}\right). \tag{5.81}$$

The logarithmic divergences are associated with the limit in which both the $s$ and $\bar{s}$ quarks originating from the gluon are soft. Note that the annihilation topology implies one overall factor of $1/m_b$. Each remaining factor of $1/m_b$ is associated with a quark helicity flip. In fact, adding up all of the helicity amplitude contributions in QCD factorization *formally* preserves the naive factorization power counting in (5.78). However, as we will see below, this turns out not to be the case numerically, because of QCD penguin annihilation.

**Numerical results for polarization**

The numerical inputs are given in [113]. The logaritmic divergences are modeled as in [131, 132]. For example, in the annihilation amplitudes the quantities $X_A$ are introduced as

$$\int_0^1 \frac{dx}{x} \rightarrow X_A = (1 + \varrho_A e^{i\varphi_A}) \ln \frac{m_B}{\Lambda_h}\,; \quad \varrho_A \leq 1\,, \quad \Lambda_h \approx 0.5\,\mathrm{GeV}\,. \tag{5.82}$$

This parametrization reflects the physical $\mathcal{O}(\Lambda_{\mathrm{QCD}})$ cutoff, and allows for large strong phases $\varphi_A \in [0, 2\pi]$ from soft rescattering. The quantities $X_A$ (and the corresponding hard spectator interaction quantities $X_H$) are varied independently for unrelated convolution integrals.

The predicted longitudinal polarization fractions $f_L(\rho^-\rho^0)$ and $f_L(\rho^-\rho^+)$ are close to unity, in agreement with observation [139, 138] and with naive power counting (5.78). The theoretical uncertainties are small, particularly after imposing the branching ratio constraints, due to the absence of (for $\rho^-\rho^0$) or CKM suppression of (for $\rho^-\rho^+$) the QCD penguin amplitudes.

Averaging the Belle and *BABAR* $\overline{B} \rightarrow \phi K^{*0}$ measurements [140, 141, 139] yields $f_L^{\mathrm{exp}} = .49 \pm .06$ and $\mathcal{B}^{\mathrm{exp}} = 10.61 \pm 1.21$, or $\mathcal{B}_L^{\mathrm{exp}} = 5.18 \pm .86$ and $\mathcal{B}_T^{\mathrm{exp}} = 5.43 \pm .88$.In the absence of annihilation, the predicted rates are $10^6\,\mathcal{B}_L = 5.15^{+6.79+.88}_{-4.66-.81}$ and $10^6\,\mathcal{B}_T = .61^{+.60+.38}_{-.42-.29}$, where the second (first) set of error bars is due to variations of $X_H$ (all other inputs). However, the $(S+P)(S-P)$ QCD penguin annihilation graph in Fig. 5-23 can play an important role in both $\overline{\mathcal{A}}^0$ and $\overline{\mathcal{A}}^-$ due to the appearance of a logarithmic divergence squared ($X_A^2$), the large Wilson coefficient $C_6$, and a $1/N_c$ rather than $1/N_c^2$ dependence. Although formally $\mathcal{O}(1/m^2)$, see (5.81), these contributions can be $\mathcal{O}(1)$ numerically. This is illustrated in Fig. 5-24, where the ($CP$-averaged) longitudinal branching ratio, $\mathcal{B}_L$, and the total transverse branching ratio, $\mathrm{Br}_T = \mathcal{B}_\perp + \mathcal{B}_\parallel$, are plotted versus the quantities $\rho_A^0$ and $\rho_A^-$, respectively, for $\overline{B} \rightarrow \phi K^{*0}$. $\rho_A^0$ and $\rho_A^-$ enter the parametrizations (5.82) of the logarithmic divergences appearing in the longitudinal and negative helicity $(S+P)(S-P)$ annihilation amplitudes, respectively. As $\rho_A^{0,-}$ increase from 0 to 1, the corresponding annihilation amplitudes increase by more than an order of magnitude. The theoretical uncertainties on the rates are very large. Furthermore, the largest input parameter uncertainties in $\mathcal{B}_L$ and $\mathcal{B}_T$ are a priori unrelated. However, it is clear from Fig. 5-24 that the QCD penguin annihilation amplitudes can account for the $\phi K^{*0}$ measurements. Similarly, the *BABAR* measurement of $f_L(\phi K^{*-}) \approx 50\%$ [139] can be accounted for.

Do the QCD penguin annihilation amplitudes also imply large transverse polarizations in $B \rightarrow \rho K^*$ decays? The answer depends on the pattern of SU(3)$_F$ flavor symmetry violation in these amplitudes. For light mesons containing a single strange quark, *e.g.*, $K^*$, non-asymptotic effects shift the weighting of the meson distribution amplitudes towards larger strange quark momenta. As a result, the suppression of $s\bar{s}$ popping relative to light quark popping in annihilation amplitudes can be $\mathcal{O}(1)$, which is consistent with the the order of magnitude hierarchy between the $\overline{B} \rightarrow D^0\pi^0$ and $\overline{B} \rightarrow D_s^+ K^-$ rates [142]. In the present case, this implies that the longitudinal polarizations should satisfy $f_L(\rho^\pm K^{*0}) \lesssim f_L(\phi K^*)$ in the Standard Model [113]. Consequently, $f_L(\rho^\pm K^{*0}) \approx 1$ would indicate that $U$-spin violating New Physics entering mainly in the $b \rightarrow s\bar{s}s$ channel is responsible for the small values of $f_L(\phi K^*)$. One possibility would be right-handed vector currents; they could interfere constructively (destructively) in $\overline{\mathcal{A}}_\perp$ ($\overline{\mathcal{A}}_L$) transversity amplitudes, see (5.69). Alternatively, a parity-symmetric scenario would only affect $\overline{\mathcal{A}}_\perp$. A more exotic possibility would be tensor currents; they would contribute to the longitudinal and transverse amplitudes at sub-leading and leading power, respectively.

**A test for right-handed currents**

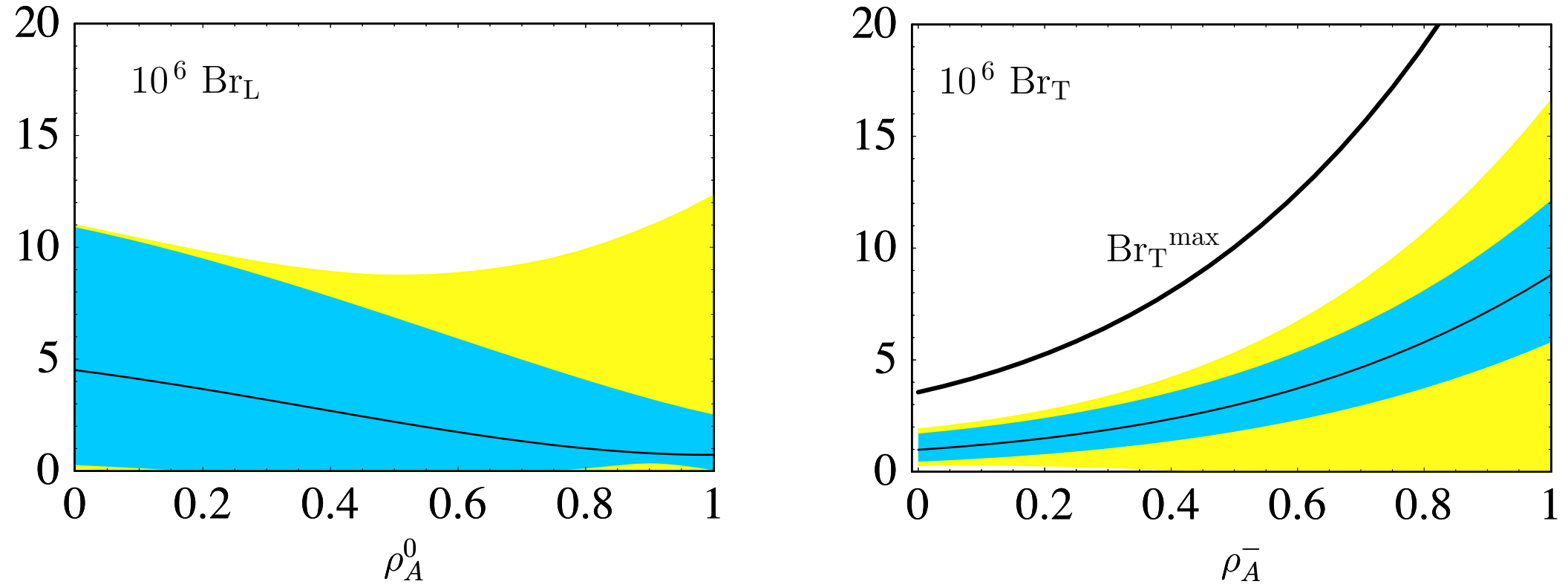

**Figure 5-24.** $\mathrm{Br}_L(\phi K^{*0})$ *vs.* $\rho_A^0$ *(left),* $\mathrm{Br}_T(\phi K^{*0})$ *vs.* $\rho_A^-$ *(right). Black lines: default inputs. Blue bands: input parameter variation uncertainties added in quadrature, keeping default annihilation and hard spectator interaction parameters. Yellow bands: additional uncertainties, added in quadrature, from variation of parameters entering logarithmically divergent annihilation and hard spectator interaction power corrections. Thick line:* $\mathrm{Br}_T^{\max}$ *under simultaneous variation of all inputs.*

Does the naive factorization relation $f_\perp/f_\| = 1 + \mathcal{O}(\Lambda_{\mathrm{QCD}}/m_b)$ (5.78) survive in QCD factorization? This ratio is very sensitive to the quantity $r_\perp$ defined in (5.77).As $r_\perp$ increases, $f_\perp/f_\|$ decreases. The range $r_\perp^{K^*} = .05 \pm .05$, spanning existing model determinations [134]–[137], is taken in [113]. In Fig. 5-25 (left) the resulting predictions for $f_\perp/f_\|$ and $\mathcal{B}_T$ are studied simultaneaously for $\overline{B} \to \phi K^{*0}$ in the Standard Model. Note that the theoretical uncertainty for $f_\perp/f_\|$ is much smaller than for $f_L$. Evidently, the above relation still holds, particularly at larger values of $\mathcal{B}_T$ where QCD penguin annihilation dominates both $\mathcal{B}_\perp$ and $\mathcal{B}_\|$.

A ratio for $f_\perp/f_\|$ in excess of the Standard Model range, *e.g.*, $f_\perp/f_\| > 1.5$ if $r_\perp > 0$, would signal the presence of new right-handed currents. This is due to the inverted hierarchy between $\overline{\mathcal{A}}^-$ and $\overline{\mathcal{A}}^+$ for right-handed currents, and is reflected in the relative sign with which the corresponding Wilson coefficients $\tilde{C}_i$ enter $\overline{\mathcal{A}}_\perp$ and $\overline{\mathcal{A}}_\perp$. For illustration, new contributions to the QCD penguin operators are considered in Fig. 5 (right). At the New Physics matching scale $M$, these can be parametrized as $\overset{(\sim)}{C}_4 = \overset{(\sim)}{C}_6 = -3\,\overset{(\sim)}{C}_5 = -3\,\overset{(\sim)}{C}_3 = \overset{(\sim)}{\kappa}$. For simplicity, we take $M \approx M_W$ and consider two cases: $\kappa = -.007$ or new left-handed currents (lower bands), and $\tilde{\kappa} = -.007$ or new right-handed currents (upper bands), corresponding to $C^{NP}_{4\,(5)}(m_b)$ or $\tilde{C}^{NP}_{4\,(5)}(m_b) \approx .18\, C^{\mathrm{SM}}_{4\,(5)}(m_b)$, and $C^{NP}_{6\,(3)}(m_b)$ or $\tilde{C}^{NP}_{6\,(3)}(m_b) \approx .25\, C^{\mathrm{SM}}_{6\,(3)}(m_b)$. Clearly, moderately sized right-handed currents could increase $f_\perp/f_\|$ well beyond the Standard Model range if $r_\perp \geq 0$. However, new left-handed currents would have little effect.

**Distinguishing four-quark and dipole operator effects**

The $\mathcal{O}(\alpha_s)$ penguin contractions of the chromomagnetic dipole operator $Q_{8g}$ are illustrated in Fig. 5-26. $a_4$ and $a_6$ are the QCD factorization coefficients of the transition operators $(\overline{q}b)_{V-A} \otimes (\overline{D}q)_{V-A}$ and $(\overline{q}b)_{S-P} \otimes (\overline{D}q)_{S+P}$, respectively, where $q$ is summed over $u, d, s$ [131, 132]. Only the contribution on the left ($a_4$) to the longitudinal helicity amplitude $\overline{\mathcal{A}}^0$ is non-vanishing [113]. In particular, the chromo- and electromagnetic dipole operators $Q_{8g}$ and $Q_{7\gamma}$ *do not contribute to the transverse penguin amplitudes* at $\mathcal{O}(\alpha_s)$ due to angular momentum conservation: the dipole tensor current couples to a transverse gluon, but a ‘helicity-flip’ for $q$ or $\overline{q}$ in Fig. 2 would require a longitudinal gluon coupling. Formally, this result follows from Wandura-Wilczek type relations among the vector meson distribution amplitudes, and the large energy relations between the tensor-current and vector-current form factors. Transverse amplitudes in which a vector meson contains a collinear higher Fock state gluon also vanish at $\mathcal{O}(\alpha_s)$, as can be seen from the vanishing of the corresponding partonic dipole operator graphs in the same momentum configurations. Furthermore, the transverse $\mathcal{O}(\alpha_s^2)$ contributions involving spectator interactions are highly suppressed.

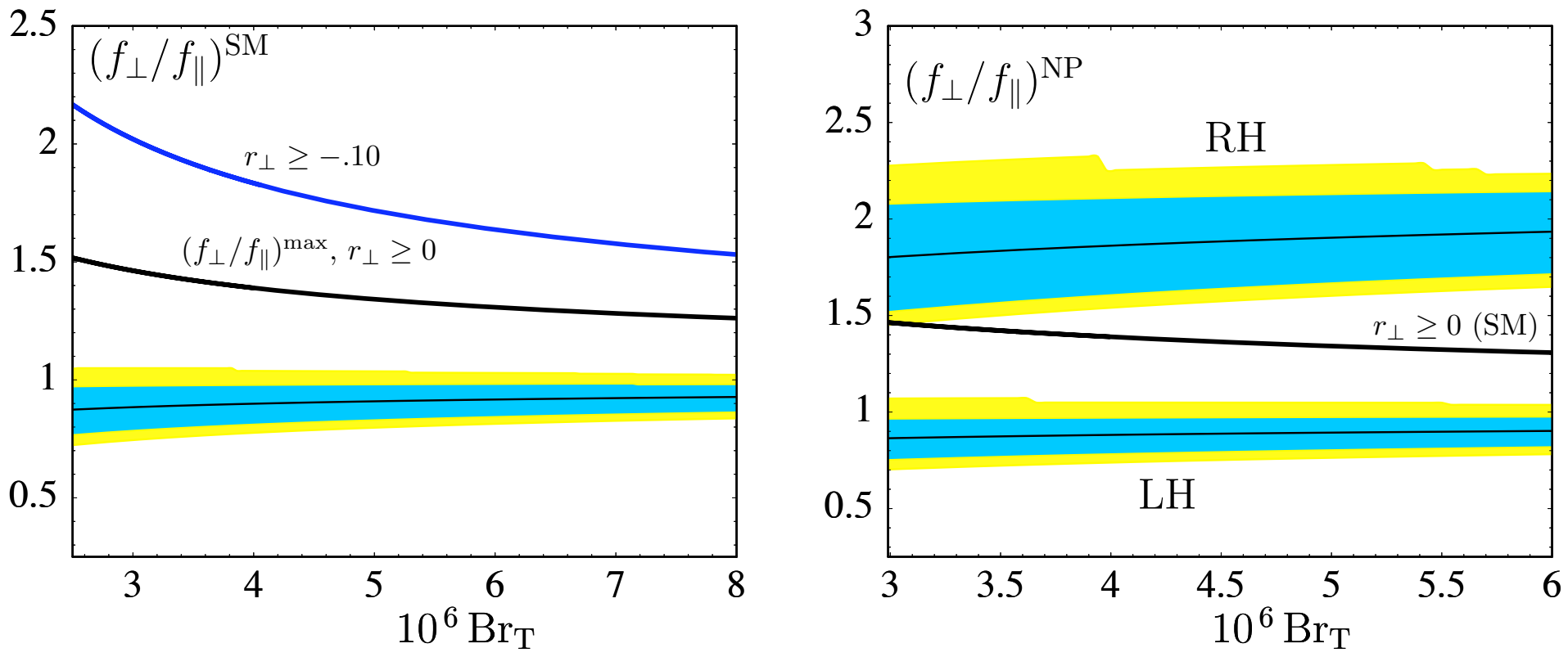

**Figure 5-25.** *$f_\perp/f_\parallel$ vs. ${\rm Br}_T$ in the Standard Model (left), and with new RH or LH currents (right). Black lines, blue bands, and yellow bands are as in Fig. 5-24. Thick lines: $(f_\perp/f_\parallel)^{\rm max}$ in the Standard Model for indicated ranges of $r_\perp^{K^*}$ under simultaneous variation of all inputs. Plot for $r_\perp^{K^*} > 0$ corresponds to ${\rm Br}_T^{\rm max}$ in Fig. 5-24.*

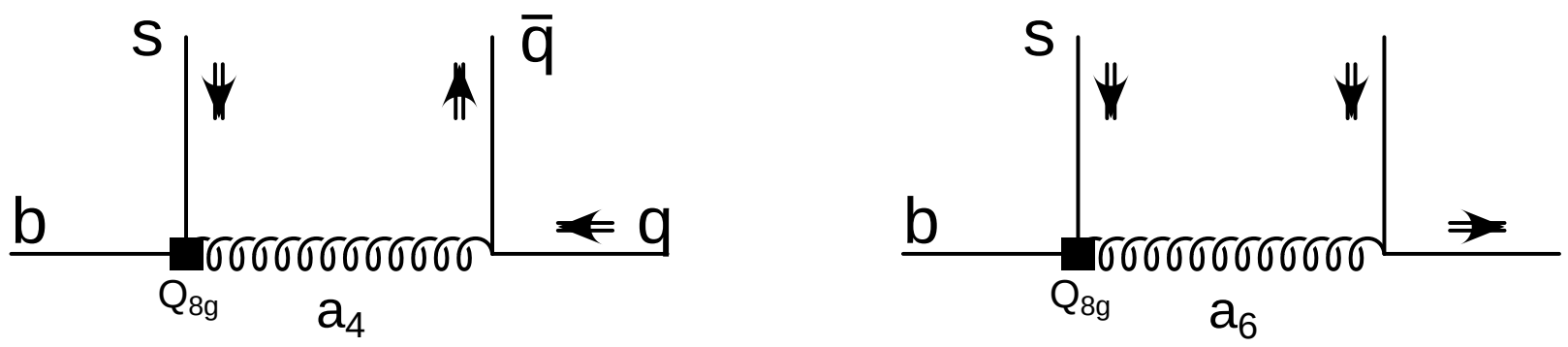

**Figure 5-26.** *Quark helicities for the $O(\alpha_s)$ penguin contractions of $Q_{8g}$. The upward lines form the $\phi$ meson in $\overline{B} \to \phi K^*$ decays.*

This has important implications for New Physics searches. For example, in pure penguin decays to $CP$-conjugate final states $f$, *e.g.*, $\overline{B} \to \phi\,(K^{*0} \to K_S^0\pi^0)$, if the transversity basis time-dependent $CP$ asymmetry parameters $(S_f)_\perp$ and $(S_f)_\parallel$ are consistent with $(\sin 2\beta)_{J/\psi K_S^0}$, and $(S_f)_0$ is not, then this would signal new $CP$ violating contributions to the chromomagnetic dipole operators. However, deviations in $(S_f)_\perp$ or $(S_f)_\parallel$ would signal new $CP$-violating four-quark operator contributions. If the triple-products $A_T^0$ and $A_T^\perp$ (5.66) do not vanish and vanish, respectively, in pure penguin decays, then this would also signal new $CP$ violating contributions to the chromomagnetic dipole operators. This assumes that a significant strong phase difference is measured between $\overline{\mathcal{A}}_\parallel$ and $\overline{\mathcal{A}}_\perp$, for which there is some experimental indication [141]. However, non-vanishing $A_T^\parallel$, or non-vanishing transverse direct $CP$ asymmetries would signal the intervention of four-quark operators. The above would help to discriminate between different explanations for an anomalous $S_{\phi K_S^0}$, which fall broadly into two categories: radiatively generated dipole operators, *e.g.*, supersymmetric loops; or tree-level four-quark operators, *e.g.*, flavor changing (leptophobic) $Z'$ exchange [146], $R$-parity violating couplings [129], or color-octet exchange [130]. Finally, a large $f_\perp/f_\parallel$ would be a signal for right-handed four-quark operators.

### Conclusion

There are a large number of penguin-dominated rare hadronic $B$ decay modes in the Standard Model in which departures from null $CP$ asymmetry predictions would be a signal for New Physics. We have seen that in order to detect the possible intervention of new $b \to s_R$ right-handed currents it is useful to organize these modes according to the parity of the final state. ${\rm SU}(2)_L \times {\rm SU}(2)_R \times {\rm U}(1)_{B-L} \times P$ symmetric models in which $CP$ violating contributions to the effective $\Delta B = 1$ Hamiltonian are, to good approximation, parity symmetric, would only give rise to deviations from null $CP$ asymmetries in parity-odd final states. For example, no deviations from the null Standard Model $CP$ asymmetry predictions in $S_{\eta' K_S^0}$, $(S_{\phi K^{*0}})_{0,\parallel}$, $A_{CP}(\phi K^{*\pm})$, $A_{CP}(K^0\pi^\pm)$ could be accompanied by significant deviations in $S_{\phi K_S^0}$, $A_{CP}(\phi K^\pm)$, $(S_{\phi K^{*0}})_\perp$, and $S_{f^0 K_S^0}$. This pattern would provide a clean signal for left-right symmetry. However, the precision of $CP$ asymmetry measurements necessary to discern its existence would require a Super $B$ Factory . Remarkably, approximate parity invariance in the $\Delta B = 1$ effective Hamiltonian can be realized even if the SU(2)$_R$ symmetry breaking scale $M_R$ is as large as $M_{\rm GUT}$. An explicit example in which large departures from the null predictions are possible, but in which deviations from parity invariance can be as small as $\mathcal{O}(1\%)$ for $M_R \leq M_{\rm GUT}$, is provided by squark-gluino loops in parity-symmetric SUSY models. It is noteworthy that, due to parity invariance, stringent bounds on new sources of $CP$ and flavor violation arising from the $^{199}$Hg mercury EDM are naturally evaded in such models.

More generally, in models in which new contributions to Standard Model (left-handed) and opposite chirality (right-handed) effective operators are unrelated, the $CP$ asymmetries in the the $P$-odd and $P$-even null Standard Model modes could differ substantially both from each other, and from the null predictions. This is because the right-handed operator Wilson coefficients enter with opposite sign in the amplitudes for decays to $P$-odd and $P$-even final states. Unfortunately, $CP$ asymmetry predictions have large theoretical uncertainties due to $1/m$ power corrections, especially from the QCD penguin annihilation amplitudes. We therefore can not rule out substantial differences between new $CP$-violating effects in parity-even and parity-odd modes arising solely from left-handed currents. However, very large differences, *e.g.*, $S_{\phi K^0_S} < 0$ and $S_{\pi^0 K^0_S} > (\sin 2\beta)_{J/\psi\, K^0_S}$, may provide a signal for $CP$-violating right-handed currents. More theoretical work will be required in order to make this statement more precise.

Polarization measurements in $B$ decays to light vector meson pairs offer a unique opportunity to probe the chirality structure of rare hadronic decays. A Standard Model analysis which includes all non-factorizable graphs in QCD factorization shows that the longitudinal polarization formally satisfies $1 - f_L = \mathcal{O}(1/m^2)$, as in naive factorization. However, the contributions of a particular QCD penguin annihilation graph which is formally $\mathcal{O}(1/m^2)$ can be $\mathcal{O}(1)$ numerically in longitudinal and negative helicity $\Delta S\!=\!1$ $\overline{B}$ decays. Consequently, the observation of $f_L(\phi K^{*0,-}) \approx 50\%$ can be accounted for, albeit with large theoretical errors. The expected pattern of SU(3)$_F$ violation in the QCD penguin annihilation graphs, *i.e.*, large suppression of $s\overline{s}$ relative to $u\overline{u}$ or $d\overline{d}$ popping, implies that the longitudinal polarizations should satisfy $f_L(\rho^{\pm} K^{*0}) \lesssim f_L(\phi K^*)$ in the Standard Model. Consequently, $f_L(\rho^{\pm} K^{*0}) \approx 1$ would indicate that $U$-spin violating New Physics entering mainly in the $b \rightarrow s\overline{s}s$ channel is responsible for the small values of $f_L(\phi K^*)$.

The ratio of transverse rates in the transversity basis satisfies $\Gamma_\perp/\Gamma_\parallel = 1 + \mathcal{O}(1/m)$, in agreement with naive power counting. A ratio in excess of the predicted Standard Model range would signal the presence of new right-handed currents in dimension-6 four-quark operators. The maximum ratio attainable in the Standard Model is sensitive to the $B \rightarrow V$ form factor combination $r_\perp$, see (5.77), which controls helicity suppression in form factor transitions. All existing model determinations give a positive sign for $r_\perp$, which would imply $\Gamma_\perp(\phi K^*)/\Gamma_\parallel(\phi K^*) < 1.5$ in the Standard Model. The magnitude, and especially the sign, of $r_\perp^{K^*}$ is clearly an important issue that should be clarified further with dedicated lattice QCD studies.

Contributions of the dimension-5 $b \rightarrow sg$ dipole operators to the transverse $B \rightarrow VV$ modes are highly suppressed, due to angular momentum conservation. Comparison of $CP$ violation involving the longitudinal modes with $CP$ violation only involving the transverse modes, in pure penguin $\Delta S = 1$ decays, could therefore distinguish between new contributions to the dipole and four-quark operators. More broadly, this could distinguish between scenarios in which New Physics effects are loop induced and scenarios in which they are tree-level induced, as it is difficult to obtain $\mathcal{O}(1)$ $CP$-violating effects from dimension-6 operators beyond tree-level. Again, this will require a Super $B$ Factory in order to obtain the necessary level of precision in $CP$ violation measurements.

### 5.2.6 Variation of *CP* Asymmetries Across Charmonium Resonances as a Signal of New Physics

$\succ$– J. Hewett, D.G. Hitlin, N. Sinha and R. Sinha –$\prec$

Several techniques have been proposed to isolate signals of New Physics. Comparing $\sin 2\beta$ by measuring the $CP$ asymmetry in a tree-dominated mode such as $B^0 \to J/\psi K_S^0$ with a penguin-dominated mode such as $B^0 \to \phi K_S^0$, is a classic example of a clean signal for New Physics [119, 86]. Another clean method, providing several signals of New Physics, involves angular analysis in modes like $B^0 \to \phi K^{*0}$ or $B^0 \to J/\psi K^{*0}$ [109]. These methods, however, cannot extract New Physics parameters. We propose herein a clean method that can not only provide a signal of New Physics if it exists, but also allows extraction of the New Physics parameters in a model-independent way. We explicitly show how the New Physics parameters can be determined up to a two-fold ambiguity. While the approach throughout is model-independent, the extraction of the New Physics parameters is demonstrated using the example of SUSY [143] motivated gluino-mediated $b \to sc\bar{c}$ transitions, induced by flavor mixing in the down-squark sector.

Consider the decay $B \to \psi K_S$, where $\psi$ is generic for any $c\bar{c}$ resonance, *i.e.* $J/\psi$, $\psi(2S)$, $\psi(3770)$, $\psi(4040)$, $\psi(4160)$ or $\psi(4415)$. The amplitudes for this mode and for the conjugate mode may be written as[2]

$$\begin{aligned} \mathcal{A}(s) &= a(s)e^{i\delta_a(s)} + be^{i\delta_b}e^{i\phi} \\ \overline{\mathcal{A}}(s) &= a(s)e^{i\delta_a(s)} + be^{i\delta_b}e^{-i\phi} \;. \end{aligned} \tag{5.83}$$

where $s$ is the invariant mass of the $\psi$ decay products, $a(s)$ and $\delta_a(s)$ are the Standard Model amplitude and the associated strong phase; $b$, $\delta_b$ and $\phi$ are the New Physics amplitude, strong phase and the weak phase respectively. We are interested in studying the variation of this amplitude as a function of $s$ over the $\psi$ line shape, since the relative strength's of $a$ and $b$ may vary across the charmonium resonance, yielding a unique handle on potential New Physics amplitudes. If there is New Physics (*i.e.*, if $b \neq 0$ and $\phi \neq 0$), the measured $B^0 - \overline{B^0}$ mixing phase $\sin 2\beta_{obs}$ will change as a function of $s$. Since the width of both the $J/\psi$ and the $\psi(2S)$ is mush less than the experimental resolution, it is not possible experimentally to measure the variation of the $CP$ asymmetry across the resonance. The proposed measurements may, however, be possible using the mode $B \to \psi(3770)K_S^0 \to (D^+D^-)_{\psi(3770)}K_S^0$.

The time dependent decay rate at each $s$ is given by

$$\Gamma(B^0(t) \to f) \propto B(s)\,(1 + C(s)\cos(\Delta m t) + S(s)\sin(\Delta m t)) \;. \tag{5.84}$$

For the amplitudes given in Eq. (5.83) it is easy to derive,

$$B(s) = a(s)^2 + b^2 + 2\,a(s)\,b\,\cos\phi\cos\delta(s) \;, \tag{5.85}$$

$$C(s) = \frac{-2\,a(s)\,b\,\sin\phi\sin\delta(s)}{B(s)} \;, \tag{5.86}$$

$$S(s) = -\sqrt{1 - C(s)^2}\,\sin 2\beta_{obs}(s) \;, \tag{5.87}$$

with

$$\sin 2\beta_{obs}(s) = \frac{a(s)^2 \sin 2\beta + b^2 \sin(2\beta + 2\phi) + 2a(s)\,b\,\sin(2\beta + \phi)\cos\delta(s)}{B(s)\sqrt{1 - C(s)^2}} \;, \tag{5.88}$$

where $\delta(s) = \delta_b - \delta_a(s)$. The lineshape $a(s)e^{i\delta_a(s)}$ is proportional to a Breit-Wigner function:

$$a(s)e^{i\delta_a(s)} = \frac{-a_N\, m_\psi \Gamma_\psi}{s - m_\psi^2 + i m_\psi \Gamma_\psi} \;, \tag{5.89}$$

where $\psi$ is the $c\bar{c}$ resonance being studied, $m_\psi$ and $\Gamma_\psi$ are the mass and total width of the resonance, and $a_N$ is the normalization factor. As a consequence of Eq. (5.88), we expect that $\sin 2\beta_{obs}$ will vary as a function of $s$ if $b \neq 0$ and $\phi \neq 0$. In Fig. 5-27 we show the variation of the number of events, the direct $CP$ asymmetry $C$ and $\sin 2\beta_{obs}$ as a function of $\sqrt{s}$. The conclusion that $\sin 2\beta_{obs}$ varies as a function of $\sqrt{s}$ is independent of the exact parameterization of the amplitudes and associated strong phases.

[2] A more general form of the amplitude is considered later, when considering explicit New Physics models

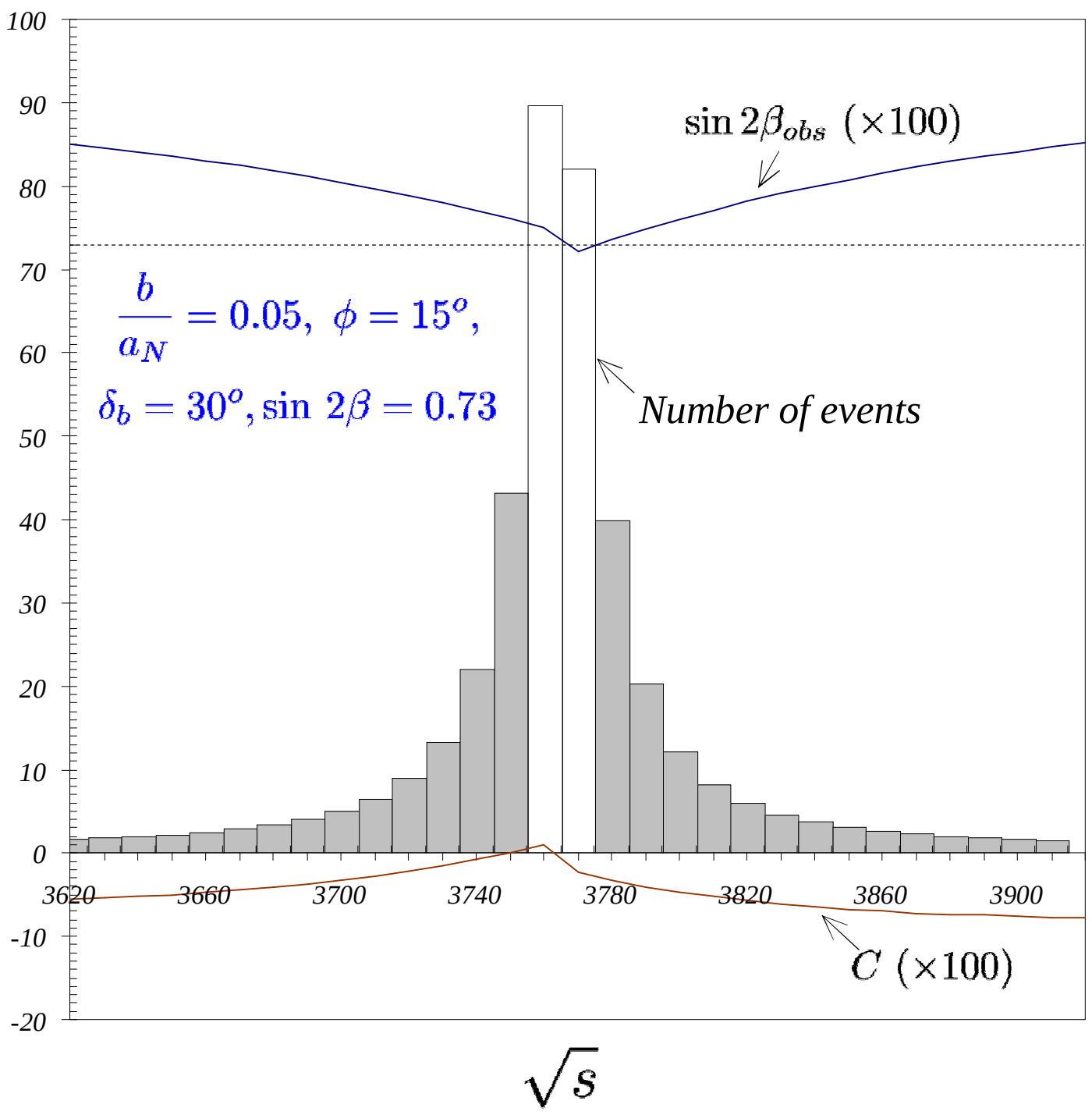

**Figure 5-27.** *The distribution of 400 events around* $\psi(3770)$ *in* $10\ MeV$ *bins, together with the variation of* $\sin(2\beta^{obs})$ *and* $C$ *as a function of the invariant mass. The white and shaded bins correspond to almost equal number of events. Clearly in the presence of New Physics, the average of* $\sin(2\beta^{obs})$ *will differ for the white and shaded bins.*

**Solution for the theoretical parameters.**

The amplitudes for the decay at two distinct points $s_1$ and $s_2$ in the resonance region may be written as

$$\begin{aligned}\mathcal{A}(s_1) &= a(s_1)e^{i\delta_a(s_1)} + be^{i\delta_b}e^{i\phi} \\ \mathcal{A}(s_2) &= a(s_2)e^{i\delta_a(s_2)} + be^{i\delta_b}e^{i\phi}\ .\end{aligned} \tag{5.90}$$

Given these two complex amplitudes and the corresponding two complex amplitudes for the conjugate process ($\overline{\mathcal{A}}(s_1)$ and $\overline{\mathcal{A}}(s_2)$), there are seven measurable quantities. These measurements are not independent, however, since they obey a complex relation:

$$\mathcal{A}(s_2) - \mathcal{A}(s_1) = \overline{\mathcal{A}}(s_2) - \overline{\mathcal{A}}(s_1)\ . \tag{5.91}$$

Thus there are five possible independent measurements. Since the number of theoretical parameters is also five: $a_N$, $b$, $\delta_b$, $\phi$ and $\beta$, we can solve for all of them. We now explicitly obtain the solutions and demonstrate that these *solutions are possible up to a two fold ambiguity*.

Using Eq.(5.83), an evaluation of $|\mathcal{A}(s) - \overline{\mathcal{A}}(s)|^2$ yields the following relation for the New Physics amplitude:

$$b^2 = \frac{B(s)}{2\sin^2\phi}[1 - \sqrt{1 - C(s)^2}\cos(2\beta^{obs}(s) - 2\beta)]\ . \tag{5.92}$$

Since the New Physics contribution is assumed constant over the $\psi$ pole, we have the relation,

$$B_1[1 - \sqrt{1 - C_1^2}\cos(2\beta_1^{obs} - 2\beta)] = B_2[1 - \sqrt{1 - C_2^2}\cos(2\beta_2^{obs} - 2\beta)]\ , \tag{5.93}$$

where $B_1(B_2)$, $C_1(C_2)$ and $\beta_1^{obs}(\beta_2^{obs})$ are the branching ratio, the direct asymmetry and the observed value of $\beta$ at $s_1(s_2)$. We can solve Eq. (5.93) for $\sin(2\beta)$ with a two-fold ambiguity,

$$\sin(2\beta) = \frac{(B_1 - B_2)X_1 \pm X_2\sqrt{X_1^2 + X_2^2 - (B_1 - B_2)^2}}{X_1^2 + X_2^2}\ , \tag{5.94}$$

where,

$$\begin{aligned} X_1 &= B_1\sqrt{1 - C_1^2}\sin(2\beta_1^{obs}) - B_2\sqrt{1 - C_2^2}\sin(2\beta_2^{obs})\ , \\ X_2 &= B_1\sqrt{1 - C_1^2}\cos(2\beta_1^{obs}) - B_2\sqrt{1 - C_2^2}\cos(2\beta_2^{obs}) \\ &= \pm B_1\sqrt{1 - C_1^2}\sqrt{1 - \sin^2(2\beta_1^{obs})} - B_2\sqrt{1 - C_2^2}\sqrt{1 - \sin^2(2\beta_2^{obs})}\ . \end{aligned} \tag{5.95}$$

Since, the measurement of the coefficients of the $\sin(\Delta mt)$ pieces at $s_1$ and $s_2$ only yield, $\sin(2\beta_1^{obs})$ and $\sin(2\beta_2^{obs})$, there is ambiguity in determining $\cos(2\beta_1^{obs})$ and $\cos(2\beta_2^{obs})$. Once the ambiguity in $\cos(2\beta_1^{obs})$ is included, the ambiguity in $\cos(2\beta_2^{obs})$ will only lead to an overall sign change of $X_2$. However, the ambiguity resulting from the overall sign change of $X_2$ has already been incorporated in the two allowed solutions of $\sin(2\beta)$ in Eq.(5.94). Hence, $\sin(2\beta)$ is determined up to an overall 4-fold ambiguity.

A calculation of $|\mathcal{A}(s)e^{-i\phi} - \overline{\mathcal{A}}(s)e^{i\phi}|^2$ shows that the Standard Model amplitude is,

$$a^2(s) = \frac{B(s)}{2\sin^2\phi}[1 - \sqrt{1 - C(s)^2}\cos(2\beta^{obs}(s) - 2\beta - 2\phi)]\ . \tag{5.96}$$

Since, $a(s)e^{i\delta_a(s)}$ has a Breit-Wigner shape, the ratio $r \equiv \dfrac{a^2(s_1)}{a^2(s_2)}$ is known. Hence $\sin(2\beta + 2\phi)$ can be determined from

$$B_1[1 - \sqrt{1 - C_1^2}\cos(2\beta_1^{obs} - 2\beta - 2\phi)] = B_2 r\,[1 - \sqrt{1 - C_2^2}\cos(2\beta_2^{obs} - 2\beta - 2\phi)]\ . \tag{5.97}$$

$\sin(2\beta + 2\phi)$ therefore has a form similar to $\sin(2\beta)$ in Eq.(5.94), with $B_2$ replaced by $r\,B_2$. Note that the solution of the quadratic equation deduced from Eq.(5.97) introduces an additional 2-fold ambiguity. Knowing $\beta$ and $\phi$, the size of New Physics amplitude is known from Eq.(5.92). Using the observables at $s_1$ for example:

$$\begin{aligned} b^2 &= \frac{B_1}{2\sin^2\phi}[1 - \sqrt{1 - C_1^2}\cos(2\beta_1^{obs} - 2\beta)] \\ &= B_1\frac{\left[1 - \sqrt{1 - C_1^2}\Big(\cos(2\beta_1^{obs})\cos(2\beta) + \sin(2\beta_1^{obs})\sin(2\beta)\Big)\right]}{\Big[1 - \cos(2\beta + 2\phi)\cos(2\beta) - \sin(2\beta + 2\phi)\sin(2\beta)\Big]}\ . \end{aligned} \tag{5.98}$$

The ambiguity in $b^2$ is expected to be 32-fold, due to the additional ambiguities in determination of $\cos(2\beta)$ and $\cos(2\beta + 2\phi)$, from the evaluated values of $\sin(2\beta)$ and $\sin(2\beta + 2\phi)$. $a^2$ can also be similarly calculated using Eq.(5.96) and will have the same ambiguity. However, a relation among the observables could help in reducing the ambiguities. Calculation of $a^2$ at $s = m_\psi^2$ gives the normalization coefficient $a_N^2$. The remaining parameter, $\delta_b$ can also be trivially determined. Using, Eqs.(5.85,5.86), we have

$$\tan\delta(s) = \frac{-C(s)B(s)}{(B(s) - a^2(s) - b^2)\tan\phi}\ . \tag{5.99}$$

Knowing $\tan\delta_a(s)$ from the Breit-Wigner form, we can thus evaluate $\tan\delta_b$. In particular, if $s_1 = m_\psi^2$, $\delta_a(s_1) = \pi/2$ and therefore

$$\begin{aligned} \tan\delta_b &= -\cot\delta(s_1) \\ &= \frac{(B_1 - a^2(s_1) - b^2)\tan\phi}{C_1 B_1}\ . \end{aligned} \tag{5.100}$$

**Table 5-4.** *The constraints of Eq.(5.101) from all possible 64 ambiguities is studied. It can be seen that only solutions 5 and 32 agree, reducing the* **64 fold ambiguity to 2 fold**. *The data set was generated assuming* $a_N = 0.9$, $b = 0.009$, $\delta_b = \pi/6$, $\phi = \pi/12$ *and* $\sin(2\beta) = 0.73$.

| | $\cos(2\beta_1^{obs})$ | $\cos(2\beta_2^{obs})$ | $\sin(2\beta{+}2\phi)$ | $\cos(2\beta)$ | $\cos(2\beta{+}2\phi)$ | $\tan\phi$ +ve LHS Eq.(19) | $\tan\phi$ −ve LHS Eq.(19) | $\tan\phi$ ±ve RHS Eq.(19) | valid Solutions |
|---|---|---|---|---|---|---|---|---|---|
| 1 | – | – | – | – | – | 0.659727539 | 1.51062478 | 0.0600762024 | |
| 2 | – | – | – | – | + | -0.275220063 | -3.65582671 | -0.0608450001 | |
| 3 | – | – | – | + | – | -0.0672433404 | -15.2276045 | 0.00211641553 | |
| 4 | – | – | – | + | + | 0.998428351 | 0.998441676 | $9.4290453510^{-7}$ | |
| 5 | – | – | + | – | – | -0.268788699 | -3.74377544 | -0.268788699 | + |
| 6 | – | – | + | – | + | -0.2732755 | -3.68197944 | -0.113566094 | |
| 7 | – | – | + | + | – | 0.986541493 | 1.01047169 | 0.00103105535 | |
| 8 | – | – | + | + | + | 0.994823458 | 1.00205966 | 0.000440104728 | |
| 9 | – | + | – | – | – | 0.995824756 | 1.0010521 | 0.000391657388 | |
| 10 | – | + | – | – | + | 0.995749206 | 1.00112805 | 0.0004036831 | |
| 11 | – | + | – | + | – | 0.996507713 | 1.00036604 | 0.000258479092 | |
| 12 | – | + | – | + | + | 0.99658332 | 1.00029014 | 0.000248863041 | |
| 13 | – | + | + | – | – | 1.01453657 | 0.982588612 | 0.0296120909 | |
| 14 | – | + | + | – | + | 1.01426946 | 0.982847389 | 0.00235802456 | |
| 15 | – | + | + | + | – | 0.987070047 | 1.00993063 | 0.000438141182 | |
| 16 | – | + | + | + | + | 0.987329986 | 1.00966475 | 0.000787830383 | |
| 17 | + | – | – | – | – | 0.996722888 | 1.00015007 | 0.000232158178 | |
| 18 | + | – | – | – | + | 0.996460496 | 1.00041344 | 0.000263957038 | |
| 19 | + | – | – | + | – | 0.995749206 | 1.00112805 | 0.0004036831 | |
| 20 | + | – | – | + | + | 0.996011412 | 1.0008645 | 0.000364029027 | |
| 21 | + | – | + | – | – | 0.987191722 | 1.00980615 | 0.000540840841 | |
| 22 | + | – | + | – | + | 0.987116822 | 1.00988277 | 0.000539738891 | |
| 23 | + | – | + | + | – | 1.01426946 | 0.982847389 | 0.00235802456 | |
| 24 | + | – | + | + | + | 1.01434642 | 0.982772813 | 0.00236815528 | |
| 25 | + | + | – | – | – | 0.99470151 | 1.00218251 | 0.000451748957 | |
| 26 | + | + | – | – | + | 0.986420551 | 1.01059558 | 0.00102917303 | |
| 27 | + | + | – | + | – | 1.49699286 | 0.665743119 | -0.0452017561 | |
| 28 | + | + | – | + | + | 1.51062478 | 0.659727539 | -0.0600762024 | |
| 29 | + | + | + | – | – | 0.998305962 | 0.998564081 | $1.81814410^{-5}$ | |
| 30 | + | + | + | – | + | -0.067304912 | -15.2133436 | 0.00206305917 | |
| 31 | + | + | + | + | – | 1.49114723 | 0.668356377 | -0.0332981457 | |
| 32 | + | + | + | + | + | -3.74377544 | -0.268788699 | 0.268788699 | – |

Having computed $\tan\delta_b$, the value of $\tan\delta$ at another point $s_2$, can also be determined. The product $\tan\delta(s_1)\tan\delta(s_2)$ can be evaluated and we have the relation

$$\frac{1-\tan\delta_a(s_2)/\tan\delta_b}{1+\tan\delta_a(s_2)\tan\delta_b} = \frac{C_1C_2B_1B_2}{(B_1-a^2(s_1)-b^2)(B_2-a^2(s_2)-b^2)\tan^2\phi} \tag{5.101}$$

The LHS in the above is known using the Breit-Wigner form for $\tan\delta_a(s_2)$ and Eq.(5.100). RHS is again known in terms of observables and evaluated parameters and is independent of an additional sign ambiguity in $\tan\phi$. A requirement that the solutions obtained for $b^2$ and $a^2$, and the corresponding observables obey this relation, helps in reducing the ambiguity in $b^2$ **from 32-fold to only 2-fold!** This is shown explicitly by a numerical calculation and the results are tabulated in Table 5.2.5. Several simulations using different data sets reveal the same reduction of ambiguity to 2 fold.

**Extraction of New Physics parameters using the mode $B \to (D^+D^-)_\psi K_S^0$.**

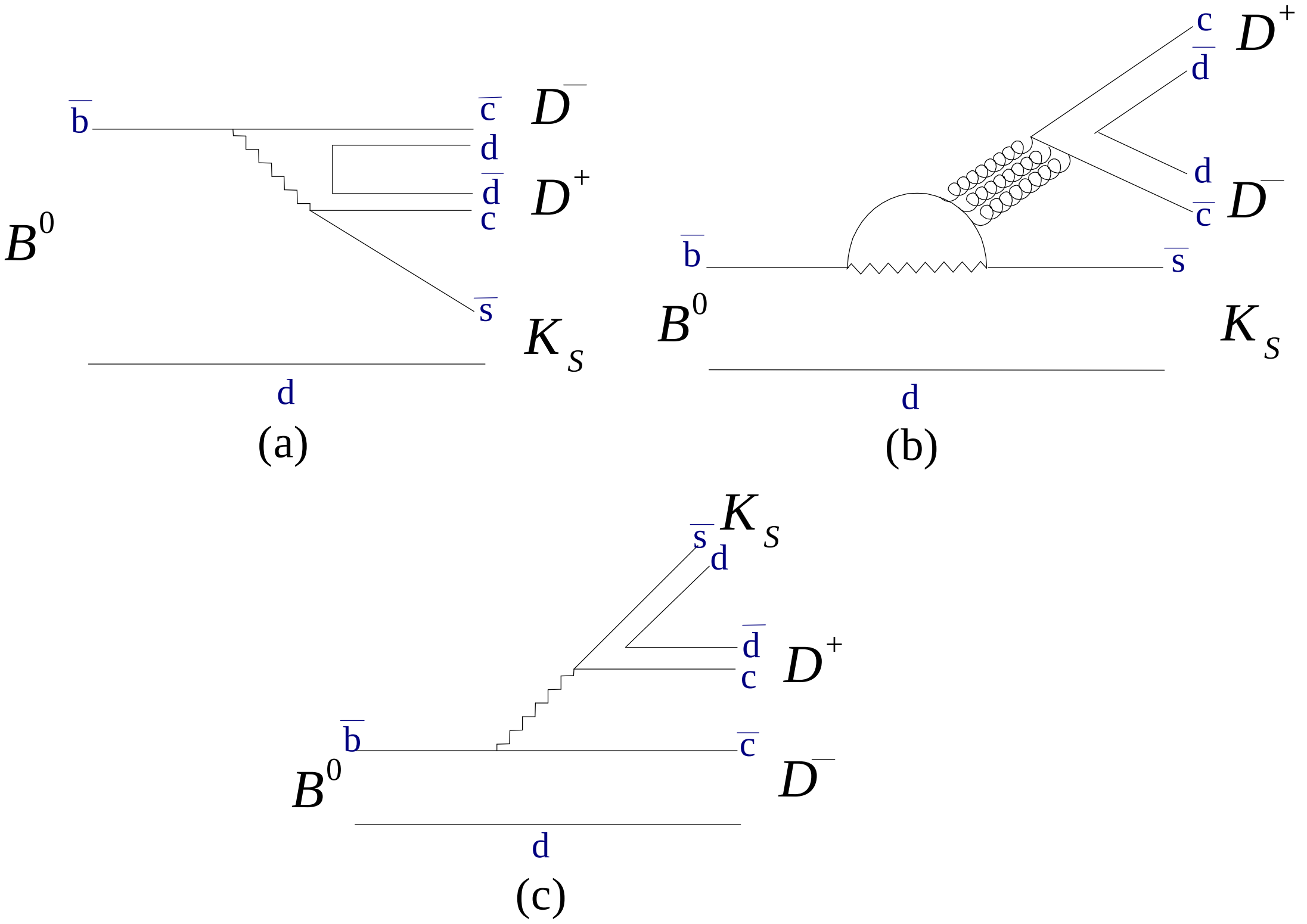

**Figure 5-28.** *Standard Model contributions to the decays $B \to (D^+D^-)K_S^0$. Fig (a) and (b) provide resonant contributions from $\psi(3770)$ to $D^+D^-$, whereas (c) gives non-resonant contributions.*

In a realistic scenario, there will be non-resonant contributions to the Standard Model, as well as resonant contributions to New Physics, invalidating the parameterization of Eq.(5.83). We extend the formulation developed previously to incorporate these additional contributions and show that we can still solve for the parameters of New Physics by considering more bins.

While the approach developed here is independent of the model of New Physics, we choose SUSY as an example to show how New Physics can contribute to the $B \to (D^+D^-)_{\psi(3770)}K_S^0$ mode. Although several classes of general SUSY models can contribute to this decay mode, following Kane et. al [144], we consider contributions from gluino-mediated $b \to sc\bar{c}$ transitions, induced by flavor mixing in the down-squark sector. This class of potentially important

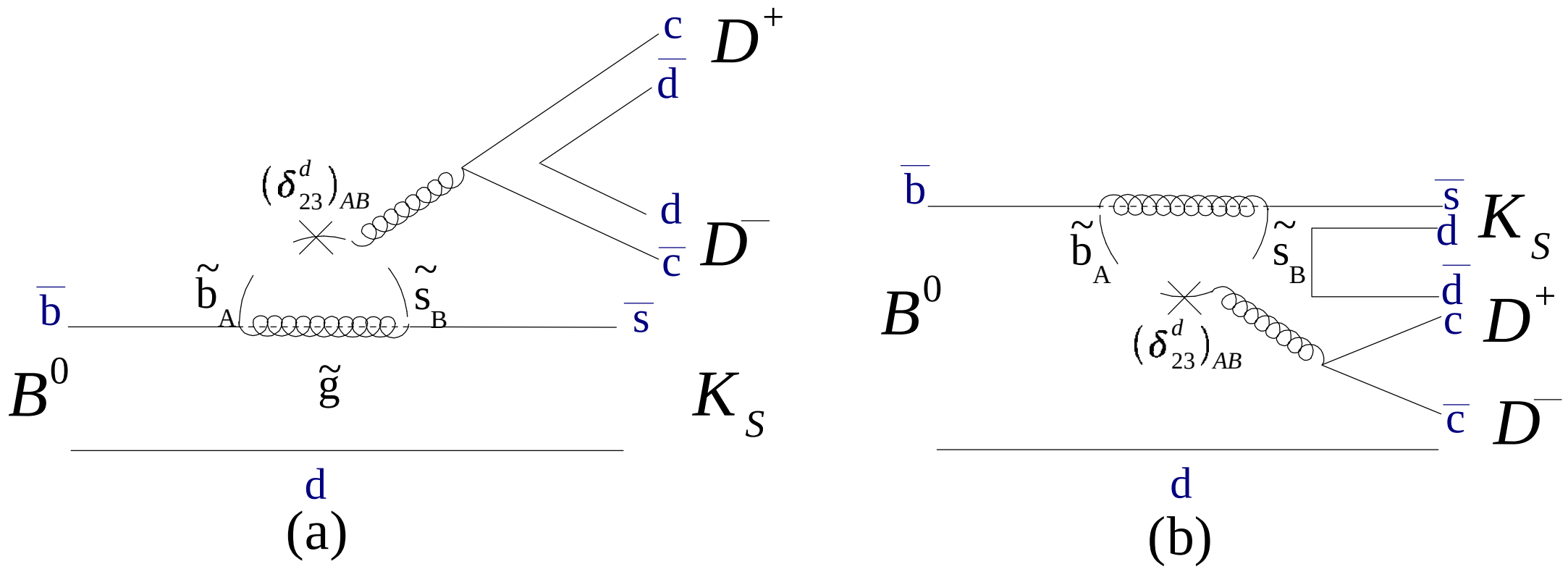

**Figure 5-29.** *New Physics contributions from SUSY [143] to the decays $B \to (D^+D^-)K^0_S$. Fig (a) provides resonant contribution from $\psi(3770)$ to $D^+D^-$, whereas (b) gives non-resonant contribution.*

SUSY contributions has been successful [144, 88, 89, 100, 90, 91, 92, 93, 95, 96, 97, 98, 99] in explaining the deviation of time-dependent asymmetry in $\phi K^0_S$ from that of $J/\psi K^0_S$. The decay mode being considered can get contributions from the SUSY mass insertion parameters $(\delta^d_{23})_{AB}$, where $A, B = \{L, R\}$. While the Standard Model contributions are depicted in Fig. 5-28, such contributions from SUSY are shown in Fig. 5-29.

The amplitude for the decay $B \to (D^+D^-)K^0_S$, in the vicinity of the $\psi(3770)$ resonance may be written as

$$\mathcal{A}(s) = \frac{a_1\, m_\psi \Gamma_\psi}{s - m_\psi^2 + i m_\psi \Gamma_\psi} + a_2\, e^{i\delta_2^a} + \frac{b_1\, m_\psi \Gamma_\psi}{s - m_\psi^2 + i m_\psi \Gamma_\psi}\, e^{i\phi} + b_2\, e^{i\delta_2^b}\, e^{i\phi}\ . \tag{5.102}$$

The first term with coefficient $a_1$ represents the resonant Standard Model contribution from Fig. 5-28(a) and (b); the second term with coefficient $a_2$ represents non-resonant Standard Model contribution from Fig. 5-28(c). The last two terms are the corresponding New Physics contributions coming from Fig. 5-29(a) and (b) respectively. We have assumed that the non-resonant contributions are constant across the $\psi$ resonance. This is not a critical assumption. A more complicated functional form for variation as a function of $s$ can be assumed. Solutions for New Physics parameters is still possible by increasing the number of bins further.

The parametrization of the amplitude in Eq. (5.102) involves 7 parameters, $a_1$, $a_2$, $b_1$, $b_2$, $\delta_2^a$, $\delta_2^b$ and $\phi$. In addition to these 7 variables, observables involve $\beta$. Hence we have a total of 8 variables that we need to determine. We will show that if the three observables $B$, $C$ and $S$ are measured at three different $s$, resulting in 9 observables $B_1$, $B_2$, $B_3$, $C_1$, $C_2$, $C_3$, $S_1$, $S_2$ and $S_3$, we can solve for all the 8 variables.

We once again evaluate $|\mathcal{A}(s) - \overline{\mathcal{A}}(s)|^2$, which now yields

$$\mathcal{O}^b(s) \equiv \frac{B(s)}{2\sin^2\phi}\left[1 - \sqrt{1 - C(s)^2}\cos(2\beta^{obs}(s) - 2\beta)\right] = \left|\frac{b_1\, m_\psi \Gamma_\psi}{s - m_\psi^2 + i m_\psi \Gamma_\psi} + b_2\, e^{i\delta_2^b}\right|^2 . \tag{5.103}$$

With the definitions $\epsilon = m_\psi \Gamma_\psi$, $w(s) = (s - m_\psi^2)$, $u = b_1^2 - 2b_1 b_2 \sin\delta_2^b$ and $v = 2b_1 b_2 \cos\delta_2^b$, we can express $\mathcal{O}^b$ as

$$\mathcal{O}^b(s) = \frac{u\,\epsilon^2 + v\, w(s)\,\epsilon}{w(s)^2 + \epsilon^2} + b_2^2\ . \tag{5.104}$$

We introduce $\mathcal{O}^b(s_j) \equiv \mathcal{O}^b_j$ and $w(s_j) \equiv w_j$, to simplify notation. Evaluating $\mathcal{O}^b$ at two points $s_1$ and $s_2$ in the resonance region, we can express $\mathcal{O}^b_1 - \mathcal{O}^b_2$ as

$$\mathcal{O}^b_1 - \mathcal{O}^b_2 = \frac{u\,\epsilon^2\,(w_2^2 - w_1^2) + v\,\epsilon(w_1\, w_2 - \epsilon^2)(w_2 - w_1)}{(w_1^2 + \epsilon^2)(w_2^2 + \epsilon^2)}\ , \tag{5.105}$$

which implies

$$\frac{(u\,\epsilon + v\,w_1)\,\epsilon}{(w_1^2+\epsilon^2)} = \frac{(\mathcal{O}_1^b - \mathcal{O}_2^b)(w_2^2+\epsilon^2)}{w_2^2-w_1^2} + \frac{v\epsilon}{w_1+w_2}\,, \tag{5.106}$$

leading to the solution for $b_2^2$:

$$\begin{aligned} b_2^2 &= \mathcal{O}_1^b - \frac{(\mathcal{O}_1^b - \mathcal{O}_2^b)(w_2^2+\epsilon^2)}{w_2^2-w_1^2} - \frac{v\epsilon}{w_1+w_2} \\ &= \frac{\mathcal{O}_2^b\,(w_2^2+\epsilon^2) - \mathcal{O}_1^b\,(w_1^2+\epsilon^2)}{w_2^2-w_1^2} - \frac{v\epsilon}{w_1+w_2} \\ &= \Lambda[2,1] - \frac{v\epsilon}{w_1+w_2}\,, \end{aligned} \tag{5.107}$$

where

$$\Lambda[i,j] \equiv \frac{\mathcal{O}_i^b\,(w_i^2+\epsilon^2) - \mathcal{O}_j^b\,(w_j^2+\epsilon^2)}{w_i^2-w_j^2}\,. \tag{5.108}$$

One could consider an additional point $s_3$ leading to a different solution for $b_2^2$:

$$b_2^2 = \frac{\mathcal{O}_3^b\,(w_3^2+\epsilon^2) - \mathcal{O}_1^b\,(w_1^2+\epsilon^2)}{w_3^2-w_1^2} - \frac{v\epsilon}{w_1+w_3} \equiv \Lambda[3,1] - \frac{v\epsilon}{w_1+w_3}\,. \tag{5.109}$$

Equating the two solutions in Eq. (5.107) and (5.109) for $b_2^2$, we get an equation for $v$ in terms of $\beta$, $\phi$ and observables to be

$$v = \frac{(\Lambda[2,1] - \Lambda[3,1])}{w_3-w_2}\,\frac{(w_1+w_2)(w_1+w_3)}{\epsilon}\,. \tag{5.110}$$

We thus have an expression for $b_2^2$ also in terms of $\beta$, $\phi$ and observables using Eqns. (5.109) and (5.110):

$$b_2^2 = \frac{\Lambda[3,1](w_1+w_3) - \Lambda[2,1](w_1+w_2)}{w_3-w_2}\,. \tag{5.111}$$

Using Eq. (5.105), $u$ can written as

$$u\epsilon^2 = \frac{(\mathcal{O}_1^b - \mathcal{O}_2^b)(w_1^2+\epsilon^2)(w_2^2+\epsilon^2)}{w_2^2-w_1^2} - \frac{v\epsilon(w_1\,w_2-\epsilon^2)}{w_1+w_2} \tag{5.112}$$

Note, that until now the only ambiguity in $b_2^2$, $v$ and $u$ solutions comes from the ambiguity in $2\beta^{obs}$.

We now solve for $b_1^2$ using the solutions for $b_2^2$, $v^2$ and $u^2$. Note that,

$$(u-b_1^2)^2 + v^2 = 4b_1^2 b_2^2\,, \tag{5.113}$$

which being a quadratic equation, yields $b_1^2$ with an additional two fold ambiguity. We also have the following relation for $\tan\delta_2^b$:

$$\tan\delta_2^b = \frac{b_1^2-u}{v}\,. \tag{5.114}$$

Using the amplitude in Eq. (5.102), a calculation of $|\mathcal{A}(s)e^{-i\phi} - \overline{\mathcal{A}}(s)e^{i\phi}|^2$, yields

$$\mathcal{O}^a(s) \equiv \frac{B(s)}{2\sin^2\phi}\Big[1 - \sqrt{1-C(s)^2}\cos(2\beta^{obs}(s) - 2\beta - 2\phi)\Big] = \Big|\frac{a_1\,m_\psi\Gamma_\psi}{s - m_\psi^2 + im_\psi\Gamma_\psi} + a_2\,e^{i\delta_2^a}\Big|^2\,. \tag{5.115}$$

Following a procedure similar to that laid out in Eqs. (5.104-5.114), we can also solve for $a_2^2$, $a_1^2$ and $\tan\delta_2^a$, in terms of $\beta$, $\phi$ and observables.

Hence, all observables can now be expressed in terms of only two parameters $\beta$ and $\phi$. The dependence of the observables on $\beta$ and $\phi$ is somewhat complicated to allow for a simple analytic solution. However, $\beta$ and $\phi$ can be solved numerically by minimizing the $\chi^2$ for the difference between experimentally observed values of the observables and the values of observables generated by simulating random values $\beta$ and $\phi$ using MINUIT.

We have thus demonstrated that by considering three points in the resonance region one can not only solve for $\beta$ but also for the New Physics amplitudes $b_1^2$, $b_2^2$ and weak phase $\phi$, even in the presence of non-resonant Standard Model and resonant New Physics contributions. Hence, New Physics parameters can be determined even in the general case with resonant as well as non-resonant contributions to both Standard Model and New Physics.

#### Measurement of the variation of the variation of the $CP$ asymmetry

Since the decay width of the two lowest charmonium resonances is narrower than the experimental resolution, it is not possible to make this measurement using the $J/\psi$ or the $\psi(2S)$. The width of the $\psi(3770)$ is, however, larger than the experimental resolution, so the measurement can, in principle, be done. Table 5-5 shows the PDG values of the $J/\psi$, $\psi(2S)$ and $\psi(3770)$ widths.

**Table 5-5.** *Measured widths of the charmonium resonances*

| Charmonium resonance | Width |
|---|---|
| $J/\psi$ | $87 \pm 5$ keV |
| $\psi(2S)$ | $300 \pm 25$ keV |
| $\psi(3770)$ | $23.6 \pm 2.7$ MeV |

Table 5-6 shows the pertinent measured branching fractions from *BABAR* and Belle, while Table 5-7 shows the measured ratio of decays of the $\psi(3770)$ to $D^0\overline{D}^0$ and $D^+D^-$.

**Table 5-6.** *Measured branching fractions*

| Experiment | Data Sample | Mode | Efficiency $\times 10^{-4}$ | Branching Fraction $\times 10^{-4}$ |
|---|---|---|---|---|
| *BABAR* | $82 \times 10^6 B\overline{B}$ pairs | $B^0 \to D^-D^+K^0$ | | $8\,^{+6}_{-5}$ |
| Phys. Rev. D **68**, | | $B^0 \to D^0\overline{D}^0 K^0$ | | $8 \pm 4$ |
| 092001, 2003 | | $B^+ \to D^0\overline{D}^0 K^+$ | | $19 \pm 3$ |
| | | $B^+ \to D^-D^+K^+$ | | $< 4$ @ $90\%\ CL$ |
| Belle | 88 fb$^{-1}$ | $B^+ \to D^0\overline{D}^0 K^+$ | 8.7 | $11.7 \pm 2.1$ |
| hep-ex/0307061 | | $B^+ \to D^-D^+K^+$ | 5 | $< 7.9$ @ $90\%\ CL$ |
| | | $B^+ \to \psi(3770)K^+$ | | $4.8 \pm 1.1$ |

**Table 5-7.** *Measured $\psi(3770)$ decay branching ratios.*

| Experiment | Ratio | Experimental Value $\times 10^{-4}$ |
|---|---|---|
| Belle | $\frac{\mathcal{B}(\psi(3770)\to D^0\overline{D}^0)}{\mathcal{B}(\psi(3770)\to D^+D^-)}$ | $2.43 \pm 1.50$ |
| Mark III | | $1.36 \pm 0.23$ |

These numbers are not entirely consistent, but they allow us to make some reasonable estimates. The branching ratio $\mathcal{B}(B^+ \to \psi(3770)K^+)$ is surprisingly large; the PDG values for decays involving the $J/\psi$ and $\psi(2S)$ are collected in Table 5-8:

**Table 5-8.** *PDG averages for measured $B$ decay branching ratios to charmonium resonances plus a kaon.*

| Decay mode | Branching Fraction $\times 10^{-4}$ |
|---|---|
| $J/\psi\, K^+$ | $10.1 \pm 0.5$ |
| $J/\psi\, K^0$ | $8.7 \pm 0.5$ |
| $\psi(2S)K^+$ | $6.6 \pm 0.6$ |
| $\psi(2S)K^0$ | $5.7 \pm 1.0$ |

Thus $B$ decays to $\psi(3770)K$ are nearly as large as decays to $J/\psi\, K$ or $\psi(2S)K$. This is a bit surprising, but can at least partly be explained by the fact that the mixing angle between the predominantly $2^3S_1$ $\psi(2S)$ and the predominantly $1^3D_1$ $\psi(3770)$ is about 35° [145].

Using these measurements, we can, using the following inputs, estimate the number of events and precision of the $CP$ asymmetry measurement:

**Table 5-9.** *Input parameters to the calculation.*

| Input | Value |
|---|---|
| $\mathcal{B}(B^+ \to \psi(3770)K^+)$ | $5 \times 10^{-4}$ |
| $\mathcal{B}(B^0 \to \psi(3770)K^0_S)$ | $2.5 \times 10^{-4}$ |
| $\frac{\mathcal{B}(\psi(3770)\to D^0\overline{D}^0)}{\mathcal{B}(\psi(3770)\to D^+D^-)}$ | 1.1 |
| $B^0 \to \psi(3770)K^0_S$ Reconstruction efficiency | $6 \times 10^{-4}$ |

With these inputs, we can estimate that in 10 ab$^{-1}$, with $10^{10}$ $B^0\overline{B}^0$ pairs produced, we will have $5 \times 10^6$ $B^0 \to \psi(3770)K^0_S$ decays, with 3,000 events reconstructed. With a tagging efficiency of 30%, we then have 900 tagged, reconstructed events in 10 ab$^{-1}$, yielding a statistical error on $A_{CP}$ of 10% integrated across the $\psi(3770)$ resonance.

The optimum strategy for demonstrating interference with a New Physics amplitude is likely to be to measure $A_{CP}$ above and below the peak of the $\psi(3770)$. The reconstructed mass resolution for the $\psi(3770)$ is small compared to the width of the resonance. In an actual experiment, a small correction would have to be made for the convolution of natural and experimental resolution, but for purposes of estimating the sensitivity, we can ignore this. Thus in 10 ab$^{-1}$ the statistical error on the upper and lower $A_{CP}$ measurements would be $\sim$14%for each sample. In 50 ab$^{-1}$, this would be reduced to 6%, which could yield interesting results if the squark mass scale is 500 GeV or below.

In conclusion, a study of the $CP$ asymmetries in the decay mode $B \to (D^+D^-)_{\psi(3770)}K^0_S$ across the charmonium resonance $\psi(3770)$ provides a clean signal of New Physics. This technique provides a method to determine the size and weak phase of New Physics as well as the $B^0 - \overline{B}^0$ mixing phase $\beta$, without any theoretical uncertainties.

### 5.2.7 Minimal Flavor Violation

>– G. Isidori –<

### The flavor problem

Despite the fact that the Standard Model provides a successful description of particle interactions, it is natural to consider it only as the low-energy limit of a more general theory, or as the renormalizable part of an effective field theory valid up to some still undetermined cut-off scale $\Lambda$. Since the Standard Model is renormalizable, we have no direct indications about the value of $\Lambda$. However, theoretical arguments based on a natural solution of the hierarchy problem suggest that $\Lambda$ should not exceed a few TeV.

One of the strategies to obtain additional clues about the value of $\Lambda$ is to constrain (or find evidence) of the effective non-renormalizable interactions, suppressed by inverse powers of $\Lambda$, that encode the presence of new degrees of freedom at high energies. These operators should naturally induce large effects in processes which are not mediated by tree-level Standard Model amplitudes, such as $\Delta F = 1$ and $\Delta F = 2$ flavor-changing neutral current (FCNC) transitions. Up to now there is no evidence of these effects and this implies severe bounds on the effective scale of dimension six FCNC operators. For instance, the good agreement between Standard Model expectations and experimental determinations of $K^0 - \overline{K}^0$ mixing leads to bounds above $10^3$ TeV for the effective scale of $\Delta S = 2$ operators, *i.e.*, well above the few TeV range suggested by the Higgs sector.

The apparent contradiction between these two determinations of $\Lambda$ is a manifestation of what in many specific frameworks (supersymmetry, techincolor, *etc.*) goes under the name of the *flavor problem*: if we insist with the theoretical prejudice that New Physics has to emerge in the TeV region, we have to conclude that the new theory possesses a highly non-generic flavor structure. Interestingly enough, this structure has not been clearly identified yet, mainly because the Standard Model, *i.e.*, the low-energy limit of the new theory, doesn't possess an exact flavor symmetry. Within a model-independent approach, we should try to deduce this structure from data, using the experimental information on FCNC transitions to constrain its form.

Recently the flavor problem has been considerably exacerbated by the new precision data from the $B$ Factories, which show no sizable deviations from Standard Model expectations in $B_d$–$\overline{B}_d$ mixing or in clean $\Delta B = 1$ FCNC processes such as $B \to X_s\gamma$. One could therefore doubt the need for new tests of the Standard Model in the quark sector of flavor physics. However, there are at least two arguments that show that the present status cannot be considered conclusive, and that a deeper study of FCNCs would be very useful:

- The information used at present to test the CKM mechanism and, in particular, to constrain the unitary triangle, is obtained only from charged currents (*i.e.*, from tree-level amplitudes) and $\Delta F = 2$ loop-induced processes. In principle, rare $B$ decays (and also rare $K$ decays) mediated by $\Delta F = 1$ FCNCs could be used to extract indirect information on the unitary triangle, or to constrain New Physics effects. However, with the exception of the $B \to X_s\gamma$ rate, the quality of this information is currently very poor, either because of experimental difficulties or because of theoretical problems. On the other hand, the number of higher-dimensional operators potentially affected by New Physics is much larger in the $\Delta F = 1$ sector. Therefore, New Physics could affect $\Delta F = 2$ and $\Delta F = 1$ loop-induced amplitudes in a very different manner, *e.g.*, with $\mathcal{O}(100\%)$ effects in the former and $\mathcal{O}(10\%)$ in the latter. It is thus mandatory to improve the quality of the information on the $\Delta F = 1$ decays, which have very small branching ratios.

- The most reasonable (but the most *pessimistic*) solution to the flavor problem is the so-called *Minimal Flavor Violation* (MFV) hypothesis. Within this framework, which will be discussed in detail in the next section, flavor- and $CP$-violating interactions are linked to the known structure of Yukawa couplings beyond the Standard Model . This implies that deviations from the Standard Model in FCNC amplitudes rarely exceed the $\mathcal{O}(10\%)$ level, or the level of irreducible theoretical errors in most of the presently available observables. Moreover, theoretically clean quantities such as $A_{CP}(B \to J/\psi K_S^0)$ and $\Delta M_{B_d}/\Delta M_{B_s}$, which measure only ratios of FCNC amplitudes, turn out to be insensitive to New Physics effects. Within this framework the need for additional clean and precise information on FCNC transitions is therefore even more important. The precise measurements of rare FCNC transitions in the $B$ sector would offer a unique opportunity in this respect.

### The Minimal Flavor Violation hypothesis

The pure gauge sector of the Standard Model is invariant under a large symmetry group of flavor transformations:

$$G_F = \mathrm{U}(3)^5 = \mathrm{SU}(3)_q^3 \otimes \mathrm{SU}(3)_\ell^2 \otimes \mathrm{U}(1)^5 \ , \tag{5.116}$$

where $\mathrm{SU}(3)_q^3 = \mathrm{SU}(3)_{Q_L} \otimes \mathrm{SU}(3)_{U_R} \otimes \mathrm{SU}(3)_{D_R}$ , $\mathrm{SU}(3)_\ell^2 = \mathrm{SU}(3)_{L_L} \otimes \mathrm{SU}(3)_{E_R}$ , and three of the five U(1) charges can be identified with baryon number, lepton number and hypercharge [75]. This large group, and in particular the $\mathrm{SU}(3)$ subgroups controlling flavor-changing transitions, is explicitly broken by the Yukawa interaction

$$\mathcal{L}_Y = \overline{Q}_L Y_D D_R H + \overline{Q}_L Y_U U_R H_c + \overline{L}_L Y_E E_R H \ + \ \mathrm{h.c.} \tag{5.117}$$

Since $G_F$ is already broken within the Standard Model, it would not be consistent to impose it as an exact symmetry beyond the Standard Model : even if absent at tree-level, the breaking of $G_F$ would reappear at the quantum level because of the Yukawa interaction. The most restrictive hypothesis we can make to *protect* the breaking of $G_F$ in a consistent way, is to assume that $Y_D$, $Y_U$ and $Y_E$ are the only source of $G_F$-breaking beyond the Standard Model. To implement and interpret this hypothesis in a consistent way, we can assume that $G_F$ is indeed a good symmetry, promoting the $Y$ to be dynamical fields with non-trivial transformation properties under $G_F$:

$$Y_U \sim (3, \overline{3}, 1)_{\mathrm{SU}(3)_q^3} \ , \qquad Y_D \sim (3, 1, \overline{3})_{\mathrm{SU}(3)_q^3} \ , \qquad Y_E \sim (3, \overline{3})_{\mathrm{SU}(3)_\ell^2} \ . \tag{5.118}$$

If the breaking of $G_F$ occurs at very high energy scales — well above the TeV region where the new degrees of freedom necessary to stabilize the Higgs sector appear — at low-energies we would only be sensitive to the background values of the $Y$, *i.e.*, to the ordinary Standard Model Yukawa couplings. Employing the effective-theory language, we then define that an effective theory satisfies the criterion of Minimal Flavor Violation if all higher-dimensional operators, constructed from Standard Model and $Y$ fields, are invariant under *CP* and (formally) under the flavor group $G_F$ [76].

According to this criterion, one should in principle consider operators with arbitrary powers of the (adimensional) Yukawa fields. However, a strong simplification arises by the observation that all the eigenvalues of the Yukawa matrices are small, but for the top one, and that the off-diagonal elements of the CKM matrix ($V_{ij}$) are very suppressed. It is then easy to realize that, similarly to the pure Standard Model case, the leading coupling ruling all FCNC transitions with external down-type quarks is [76]:

$$(\lambda_{\mathrm{FC}})_{ij} = \left\{ \begin{array}{ll} \left(Y_U Y_U^\dagger\right)_{ij} \approx \lambda_t^2 V_{3i}^* V_{3j} & \quad i \neq j \ , \\ 0 & \quad i = j \ . \end{array} \right\} \tag{5.119}$$

The number of relevant dimension six effective operators is then strongly reduced: the complete list can be found in Ref. [76]. In Table 5-10 we show a few representative examples. Note that the built-in CKM suppression leads to bounds on the effective scale of New Physics not far from the TeV scale.

Within this framework, the present bounds on FCNC operators are weaker, but not far from similar the bounds on flavor-conserving operators derived by precision electroweak tests. This observation reinforces the conclusion that a deeper study of rare decays is definitely needed in order to clarify the flavor problem. The experimental precision on the clean FCNC observables required to obtain competitive bounds, and, possibly, discover New Physics is typically in the $1\% - 10\%$ range [76].

### Comparison with other approaches

The idea that the CKM matrix also rules the strength of FCNC transitions beyond the Standard Model has become a very popular concept in the recent literature and has been implemented and discussed in several works (see *e.g.*, Refs. [77, 78, 79, 80, 81]). In particular, a detailed and updated review of the phenomenological consequences of this hypothesis can be found in Ref. [82].

**Table 5-10.** $99\%$ CL *bounds on the scale of representative dimension six operators in the MFV scenario [76]. The constraints are obtained on the single operator, with coefficient* $\pm 1/\Lambda^2$ *($+$ or $-$ denote constructive or destructive interference with the Standard Model amplitude). The* $*$ *signals the cases in which a significant increase in sensitivity can be achieved by future measurements of rare decays.*

| MFV dim-6 operators | Main observable | $\Lambda$ [TeV] | | |
|---|---|---|---|---|
| $\frac{1}{2}(\overline{Q}_L Y_U Y_U^\dagger \gamma_\mu Q_L)^2$ | $\epsilon_K, \quad \Delta m_{B_d}$ | 6.4 [+] | 5.0 [−] | |
| $eH^\dagger \left(\overline{D}_R Y_D^\dagger Y_U Y_U^\dagger \sigma_{\mu\nu} Q_L\right) F_{\mu\nu}$ | $B \to X_s \gamma$ | 5.2 [+] | 6.9 [−] | |
| $(\overline{Q}_L Y_U Y_U^\dagger \gamma_\mu Q_L)(\overline{L}_L \gamma_\mu L_L)$ | $B \to (X)\ell\bar{\ell}, \quad K \to \pi\nu\overline{\nu}, (\pi)\ell\bar{\ell}$ | 3.1 [+] | 2.7 [−] | $*$ |
| $(\overline{Q}_L Y_U Y_U^\dagger \gamma_\mu Q_L)(H^\dagger i D_\mu H)$ | $B \to (X)\ell\bar{\ell}, \quad K \to \pi\nu\overline{\nu}, (\pi)\ell\bar{\ell}$ | 1.6 [+] | 1.6 [−] | $*$ |

It is worth stressing that the CKM matrix represents only one part of the problem: a key role in determining the structure of FCNCs is also played by quark masses, or by the Yukawa eigenvalues. In this respect, the MFV criterion illustrated above provides the maximal protection of FCNCs (or the minimal violation of flavor symmetry), since the full structure of Yukawa matrices is preserved. At the same time, this criterion is based on a renormalization-group-invariant symmetry argument. Therefore, it can be implemented independently of any specific hypothesis about the dynamics of the New Physics framework. The only two assumptions are: i) the $G_F$ symmetry and its breaking sources; ii) the number of light degrees of freedom of the theory (identified with the Standard Model fields in the minimal case).

This model-independent structure does not hold in most of the alternative definitions of MFV models that can be found in the literature. For instance, the definition of Ref. [81, 82] contains the additional requirement that only the effective FCNC operators which play a significant role within the Standard Model are relevant. This condition is naturally realized within weakly coupled theories at the TeV scale with only one light Higgs doublet, such as the MSSM at small $\tan\beta$, or even in specific models with extra dimensions [83]. However, it does not hold in other frameworks, such as techincolor models or the MSSM at large $\tan\beta$, whose low-energy phenomenology could still be described using the general MFV criterion discussed above.

## 5.3 Supersymmetry

### 5.3.1 *CP* Asymmetries in Supersymmetry

>– C. Kolda –<

The Minimal Supersymmetric Standard Model (MSSM) provides an abundant opportunity for discovering New Physics in $CP$-violating and/or flavor-changing $b$ quark processes. In fact, the most general version of the MSSM provides an over-abundance, with 12 masses, 30 mixing angles and 27 phases in the (s)quark sector, beyond those of the Standard Model. Of these, the LHC has only limited ability to go beyond measurements of the masses, leaving 57 parameters unconstrained, even after finding and studying SUSY at hadron colliders.

The other 57 parameters are not, however, wholly unconstrained. If one were to take all phases and angles $\mathcal{O}(1)$ and all masses $\mathcal{O}(\mathrm{TeV})$, the MSSM would make predictions for $CP$ violation and FCNCs in the $K$ sector far beyond those observed. One therefore expects some organizing principle to be at work in the MSSM, constraining the masses and/or phases and/or mixing angles in order to avoid phenomenological trouble. This is the so-called "SUSY flavor problem".

The source of the problem is that quarks get their masses only by electroweak symmetry breaking, while squarks get masses also by SUSY breaking. The SUSY-breaking contributions have no reason to be diagonal in the same bases as the SU(2)-breaking pieces, and so quark and squark mass eigenstates are not simultaneously defined. Unitarity of the mixing matrices is enough to force the quark-quark-gauge and squark-squark-gauge interactions to be diagonal (flavor-conserving), but the quark-squark-gaugino interaction will not be diagonal and will generate quark flavor-changing (and $CP$ violation) through loops of squarks and gauginos.

There are three basic schemes which render the $CP$ violation and FCNCs in the MSSM small: decoupling, alignment and degeneracy. Decoupling [147] is nothing more than the statement that if the MSSM sparticles are very heavy, then processes generated by them will be small. However, the masses required in order to actually get decoupling can be in the 100 to 1000 TeV range, far above the range where the MSSM plays an important role in electroweak symmetry breaking. Thus such models create their own mini-hierarchy problems. In the alignment scheme [148], one forces the squark and quark mass matrices to be diagonal in the same basis. However realistic alignment models are difficult to construct and often lead to large flavor-changing in the charm ($D^0$) sector.

The last of the suggestions is the one most often considered: degeneracy. If all squarks with the same gauge charges are degenerate in mass, then their contributions to flavor-changing/$CP$-violating processes exactly cancel. The degeneracy constraint is far more severe between the first and second generation squarks than with the third generation, because the constraints from the kaon system are so stringent. However, degeneracy is also more natural between the first and second generation, where Yukawa-induced renormalizations of the squark masses are small. While the current constraints for the third generation are much less severe, there is also reason to believe that some non-degeneracy is inevitable: the large top (and possibly bottom) Yukawa couplings will split the third generation off from the other two and will generate 1-3 and 2-3 squark mixings proportional to CKM elements. This can be seen simply by examination of the soft mass renormalization group equations. For example [149]:

$$\frac{d}{d\log Q}\left(m^2_{\tilde{Q}}\right)_{ij} \propto \left[Y_u^\dagger m^2_{\tilde{Q}} Y_u + Y_d m^2_{\tilde{Q}} Y_d^\dagger + \cdots\right]_{ij} . \tag{5.120}$$

In the basis in which $Y_d$ is diagonal, the $Y_u$ terms are rotated away from the diagonal by the CKM matrix:

$$(\delta m^2_{\tilde{Q}})_{ij} \simeq \frac{1}{8\pi^2}\log(M_X/M_{\mathrm{SUSY}}) \times (V^\dagger_{\mathrm{KM}} V_{\mathrm{KM}})_{ij} . \tag{5.121}$$

This is simply the most trivial example of physics that could cause the third generation to behave differently than the first two, and in fact this example introduces no new phases by itself (it is an example of minimal flavor violation [150]). But more complicated models exist, particularly those which attempt to explain the quark mass hierachy. So even though kaon physics may strongly constrain $CP$ violation and FCNCs in the first two generations, there is still plenty of room for both in the third generation.

**Why SUSY is Special**

There is a hidden advantage to the scalar mass problems in SUSY. The lack of any strong flavor violation in kaons or $B$ mesons seems to demand that the ultimate scale at which the usual Standard Model flavor problem (namely, why is $m_u \ll m_t$?) is resolved lie above 100 TeV or more. In fact there is no reason at all to prefer a scale for flavor physics near our current experimental sensitivity rather than far into the ultraviolet. But even if the flavor scale is far above the weak scale, SUSY may provide a unique window into this world, for two reasons. First, because SUSY is associated with the weak-scale hierarchy problem, its spectrum must lie near the weak scale. Thus the precision measurements at a Super $B$ Factory will be sensitive to physics at the very scale at which SUSY is expected to be found.

But more importantly, it is the presence of so many scalar particles in SUSY that provides an extra sensitivity to high-scale flavor physics that would not normally be available. Scalar masses, through their renormalization, are sensitive to physics at all scales, from the weak scale to the far ultraviolet. In non-SUSY theories, quadratic divergences dominate this renormalization and it would not be clear how to interpret a scalar mass spectrum if one were observed. But in SUSY the scalar masses are only logarithmically renormalized, which allows the masses to be run up to high energies using the renormalization group. The presence of any non-trivial flavor physics anywhere below the SUSY-breaking scale tends to imprint itself on the spectrum of the scalar particles either through their renormalization group running or through threshold corrections at the flavor physics scale. In either case, flavor-violating operators which would normally be suppressed by powers of the flavor scale are instead suppressed only by powers of the SUSY mass scale (often with an additional large log enhancement); see Eq. (5.121). This idea has been particularly fruitful (at least theoretically) for probing the structure of the neutrino mass matrix and its correlations with $\tau \to \mu\gamma$ and $\mu \to e\gamma$. It is also the basic idea underlying several of the approaches [151] to $B \to \phi K_s^0$ that will be outlined in the next two sections.

Thus SUSY, which by itself provides no new insights into the question of flavor, may in fact be the mechanism by which we are finally able to gain experimental insights into flavor. It is for this reason that considerations of SUSY models and sensitivities will play an extremely important role in the future of high precision heavy flavor physics.

**“Flavors” of SUSY**

Because the MSSM requires some external organizing principle in order to keep the theory even remotely viable, the kinds of signals one expects at colliders depend sensitively on the organizing principle itself. In the simplest case in which degeneracy is enforced, all flavor violation is due to the CKM matrix. This is true even for the non-universal corrections generated by the renormalization group equations. Such models provide good examples of Minimal Flavor Violation [150] and one can refer to the section on MFV earlier in this chapter for a discussion of the relevant phenomenology.

However if the scale at which non-trivial flavor physics lies is below the scale at which SUSY is broken in the visible sector, evidence of the flavor physics should be imparted on the scalar spectrum in some way, even if suppressed. It would not be surprising to find that the strongest flavor violation among the scalars would occur where the Yukawa couplings are the greatest, namely in the third generation interactions. Thus a Super $B$ Factory is the natural place to search for these effects.

It is customary (for ease of calculation) to work in a basis in which the quark masses are diagonal as are the quark-squark-gluino interactions. This forces the $6 \times 6$ squark mass matrix to remain non-diagonal. In the limit of approximate degeneracy (or approximate alignment), we interpret the diagonal elements of the mass-squared matrix to be the left- and right-handed squark masses, and the off-diagonal elements as mass insertions denoted $(\Delta^d_{ij})_{AB}$ where $i \neq j$ are generation indices ($i, j = 1 \ldots 3$) and $A, B$ denote left(L) and right(R). We then define a mixing parameter:

$$(\delta^d_{ij})_{AB} = \frac{(\Delta^d_{ij})_{AB}}{\tilde{m}^2}\,, \tag{5.122}$$

where $\tilde{m}$ is a typical squark mass. Kaon physics constrains $(\delta^d_{12})_{AB}$ (for all $AB$) to be much, much smaller than one [152]. Experimental agreement of $B^0 - \overline{B}^0$ mixing with the Standard Model prediction likewise constrains $(\delta^d_{13})_{AB}$ [153]. Compared to these cases, constraints on $(\delta^d_{23})_{AB}$ are relatively weak. Specifically, the $LL$ and $RR$ insertions can be $\mathcal{O}(1)$ while the $LR$ and $RL$ can be $\mathcal{O}(10^{-2})$ due to constraints from $b \to s\gamma$.

Appearance of a sizable $(\delta^d_{23})_{AB}$ will generate non- Standard Model $b \to s$ transitions, affecting branching ratios and asymmetries in a number of processes including $B \to \phi K^0_S$, $B \to X_s \ell\ell$, $B \to X_s\gamma$, $B \to \eta^{(\prime)} K^0_S$, $B \to K^+K^-K^0_S$, $B_s \to \ell\ell$, $\Delta m_{B_s}$ and others. (We will assume that there is no large flavor violation in the 1-3 sector; such violation could enter $B^0-\overline{B}^0$ mixing, and from there affect $B \to \psi K^0_S$.)

Of these, the $CP$-violating phase in $B \to \phi K^0_S$, namely $\beta_{\phi K}$, is of particular importance. $\beta_{\phi K}$ has been measured by *BABAR* and Belle to an accuracy around 10%. As of this writing, the *BABAR* and Belle experiments are in disagreement about whether or not there is an anomaly in the experimental data on $\beta_{\phi K}$ (the data is reported in terms of the oscillation parameter $S_{\phi K}$). Because of the hint that there might be an anomaly, many groups have conducted analyses of the $B \to \phi K^0_S$ in the context of SUSY [151, 154, 155]. Regardless, decays like $B \to \phi K^0_S$ and other $b \to s$ processes are key testing grounds for SUSY flavor physics.

### $b \to s$ transitions in SUSY

The calculation of the short-distance SUSY contributions to $B \to \phi K^0_S$ is relatively straightforward. There are two classes of contributions which bear discussion, namely loops of charginos and loops of gluinos. Chargino loops contribute to the amplitude for $B \to \phi K^0_S$ with a structure that mimics the Standard Model. In particular, in models with minimal flavor violation, there is a SUSY contribution to the branching fraction for $B \to \phi K$ but not to the $CP$-violating asymmetries. If we extend minimal models to include arbitrary new phases (but not mixings) then the $CP$ asymmetries can receive new contributions, but these are generally small. It may be possible to push $S_{\phi K}$ down to zero, but it is appears to be difficult to go any lower [156].

In models with arbitrary phases *and* mixings, the chargino contributions can be even larger, but now they are typically dwarfed by gluino contributions. The gluino contributions are absent in minimally flavor-violating models, but dominate in the case of general mixings and phases. Two types of gluino-mediated diagrams typically dominate the amplitudes for $B \to \phi K$: the chromomagnetic moment and gluonic penguins. (For details of these calculations, see Ref. [154]).

There are two questions of particular importance in examining the SUSY contributions to the $CP$ asymmetries in $B \to \phi K^0_S$: can SUSY provide a large deviation from the Standard Model in $S_{\phi K}$, and what other observables would be correlated with a large deviation? In doing so, it is natural to consider four distinct cases or limits, with the understanding that a realistic model might contain elements of more than one case. Those cases are labelled by the chirality of the squark mixing: $LL$, $RR$, $LR$ and $RL$, where the first letter labels the $s$ squark and second the $b$ squark.

Of the four cases, the $LL$ insertion is particularly well motivated. In particular, one expects $(\delta^d_{23})_{LL} \sim V_{ts}$ even in models with minimal flavor violation. In models in which the SUSY breaking occurs at a high scale, the insertion can be enhanced by an additional large logarithm. The $RR$ insertion is less motivated in minimal SUSY models, but is naturally generated in grand unified (GUT) models with large neutrino mixing [151]. In this case, the large mixing in the neutrino sector (which is contained in the $\overline{5}$ of SU(5)) is transmitted by GUT and renormalization effects to the right-handed down quark sector, which is also part of the same GUT representation.

The physics consequences of the $LL$ and $RR$ insertions are very similar to one another. In both cases, measurable deviations in $S_{\phi K}$ can be obtained. A sizable deviation in $S_{\phi K}$, however, requires large $(\delta^d_{23})_{LL,RR} \sim \mathcal{O}(1)$ and a relatively light SUSY spectrum. In order to obtain a negative $S_{\phi K}$ using an $LL$ or $RR$ insertion one requires gluinos with mass below 300 GeV, for example. The strongest external constraints on such large insertions and light masses come from direct searches for gluinos and from $b \to s\gamma$; the latter only constrains the $Re(\delta^d_{23})_{LL}$ to be greater than about $-0.5$, while providing no constraint on the $RR$ insertion.

In order to determine that the New Physics in $S_{\phi K}$ would be coming from an $LL$ or $RR$ insertion, it must be correlated to other observables. Deviations in $S_{\phi K}$ are well correlated with deviations in $C_{\phi K}$: measurements of $S_{\phi K}$ below the

**Table 5-11.** *Correlated signatures for an observation of $S_{\phi K}$ much smaller than $S_{\psi K}$, assuming a single SUSY $d$-squark insertion of the type indicated. The $\pm$ signs represent the sign of the corresponding observable.*

| | $LL$ | $RR$ | $LR$ | $RL$ |
|---|---|---|---|---|
| $(\delta^d_{23})$ | $O(1)$ | $O(1)$ | $O(10^{-2})$ | $O(10^{-2})$ |
| SUSY masses | $\lesssim$ 300 GeV | $\lesssim$ 300 GeV | $\lesssim$ TeV | $\lesssim$ TeV |
| $C_{\phi K}$ | $-$, small | $-$, small | $-$, small | $-$, can be large |
| $\mathcal{B}(B \to \phi K)$ | SM-like | SM-like | varies | varies |
| $A_{\rm CP}^{b\to s\gamma}$ | $+$, few % | SM-like | $+$, $\mathcal{O}(10\%)$ | SM-like |
| $\Delta m_{B_s}$ | can be large | can be large | SM-like | SM-like |

Standard Model expectation correlate to negative values of $C_{\phi K}$. (Note that the calculation of $C_{\phi K}$ is very sensitive to the techniques used for calculating the long distance effects; these correlations are found using the BBNS [157] method.) However the deviations in $C_{\phi K}$ are at most $\mathcal{O}(10\%)$ and so will require a much larger data sample such as that available at a Super $B$ Factory. More striking is the correlation with $\Delta m_{B_s}$, the $B_s - -\overline{B}_s$ mass difference. Large deviations in $S_{\phi K}$ due to an $LL$ or $RR$ insertion correlate directly with very large mass differences, far outside the range that will be probed at Run II of the Tevatron. Mass differences of the order of 100 ps$^{-1}$ are not atypical in models with large $LL$ or $RR$ insertions, making their experimental measurement very difficult.

Specific to an $LL$ insertion (rather than an $RR$) will be deviations in the $CP$ asymmetry in $b \to s\gamma$. Large negative deviations in $S_{\phi K}$ correlate cleanly with positive $CP$ asymmetries of the order of a few percent. Measuring these asymmetries will require of order 10 ab$^{-1}$ of data and so call for a Super $B$ Factory.

The picture presented by the $LR$ and $RL$ insertions is quite different. First, the $LR$ and $RL$ insertions would generically be suppressed with respect to the $LL$ and $RR$ insertions, because they break SU(2) and must therefore scale as $M_W/M_{\rm SUSY}$. However they generate new contributions to the chromomagnetic operators which are enhanced by $M_{\rm SUSY}/m_q$ $(q = s, b)$ and are therefore very effective at generating large deviations in $S_{\phi K}$. The $LR$ insertion is the more well motivated, since one expects $\tilde{s}_L$–$\tilde{b}_R$ mixing to be proportional to the bottom Yukawa coupling, while $\tilde{s}_R$–$\tilde{b}_L$ mixing would come from the much smaller strange Yukawa. However it is possible to build reasonable flavor models in which this assumed hierarchy is not preserved and sizable $RL$ insertions are generated [154].

In either case, whether $LR$ or $RL$, strong constraints from the branching ratio of $b \to s\gamma$ force $(\delta^d_{23})_{LR,RL}$ to be $\mathcal{O}(10^{-2})$. Neither insertion generates an observable shift in $\Delta m_{B_s}$, but both can generate large shifts in the branching fraction for $B \to \phi K$. Of more interest are the correlations between $S_{\phi K}$, $C_{\phi K}$ and the $CP$ asymmetry in $b \to s\gamma$. For measurements of $S_{\phi K}$ below the Standard Model , the $LR$ insertion always predicts a negative $C_{\phi K}$, with values down to $-0.3$ when $S_{\phi K}$ goes as low as $-0.6$. On the other hand, negative contributions to $S_{\phi K}$ are associated with positive asymmetries in $b \to s\gamma$, often as large as 5% to 15%. These large asymmetries are a clear sign of the presence of an $LR$ insertion, as opposed to $LL$ insertions which give asymmetries of only a few percent.

For the $RL$ case, the phenomenology is much the same except: (1) the values of $C_{\phi K}$ implied by a down deviation in $S_{\phi K}$ are even more negative, all the way down to $C_{\phi K} = -1$; (2) though the $RL$ operators contribute to $b \to s\gamma$, they do not interfere with the Standard Model contribution and thus do not generate any new source of $CP$ violation. Thus $RL$ insertions predict no new observable $CP$ asymmetry in $b \to s\gamma$.

Table 5-11 summarizes the correlations for each type of insertion. Of course, more than one insertion may be present, so one could generate large deviations in $S_{\phi K}$ with an $LR$ insertion and large $\Delta m_{B_s}$ with an $LL$ insertion. Such combinations can be read off of the table.

In conclusion, observation of a significant (or any) deviation in the $CP$ asymmetries in $B \to \phi K_S^0$ could be an early and strong indication of SUSY flavor physics. But it is the correlations between the $\phi K_S^0$ signal and other observables that will lead us to a deeper understanding of flavor in the MSSM.

### 5.3.2 SUSY at the Super $B$ Factory

$\succ$– T. Goto, Y. Okada, Y. Shimizu, T. Shindou, and M. Tanaka –$\prec$

**The Unitarity triangle and rare decays in three SUSY models**

Among various candidates of physics beyond the Standard Model, SUSY is regraded as the most attractive possibility. The weak scale SUSY provides a solution of the hierarchy problem in the Standard Model. Although an extreme fine-tuning is necessary to keep the weak scale very small compared to the Planck scale within the Standard Model , SUSY theory does not have this problem, because of the cancellation of the quadratic divergence in the scalar mass renormalization. SUSY has attracted much attention since early 1990's, when three gauge coupling constants determined at LEP and SLC turned out to be consistent with the coupling unification predicted in SUSY GUT.

One of main motivations of the LHC experiment is a direct search for SUSY particles. The mass reach of colored SUSY particles will be about 2 TeV, an order-of-magnitude improvement from the present limit. It is quite likely that some signal of SUSY can be obtained in the early stage of the LHC experiment. It is therefore important to clarify the role of SUSY studies at a Super $B$ Factory in the LHC era.

In order to illustrate how $B$ physics can provide useful information to distinguish various SUSY models, we study SUSY effects in the length and angle measurements of the unitarity triangle and rare B decays for the following four cases in three SUSY models [158, 159, 99].

- Minimal supergravity model
- SU(5) SUSY GUT with right-handed neutrinos: Case 1 (degenerate case)
- SU(5) SUSY GUT with right-handed neutrinos: Case 2 (non-degenerate case)
- MSSM with a U(2) flavor symmetry

In the first model, all squarks and sleptons are assumed to be degenerate at a high energy scale such as the Planck scale, where SUSY breaking effects are transmitted to the observable sector from the hidden sector by the gravitational interaction. Flavor mixings and mass-splittings are induced by renormalization effects due to the ordinary quark Yukawa coupling constants, especially from a large top Yukawa coupling constant. The matrices which diagonalize the resulting squark mass matrices are approximately given by the CKM matrix, because this is the only source of the flavor mixing in the quark and squark sectors. In this sense, this model is a realization of so called "minimal flavor violation" scenario. We can consider new SUSY $CP$ phases for the trilinear scalar coupling (A parameter) and the higgsino mass term ($\mu$ parameter). There phases are, however constrained by various electric dipole moment (EDM) experiments in the context of the minimal supergravity model, so that effects on B physics are relatively small [160].

A SUSY GUT with right-handed neutrinos is a well-motivated candidate of the physics beyond the Standard Model. Here, we consider an SU(5) SUSY GUT model and incorporate the seesaw mechanism for neutrino mass generation by introducing an SU(5) singlet right-handed neutrinos. In this model, large flavor mixing in the neutrino sector can affect the flavor mixing in the squark sector through renormalization of sfermion mass matrices [161, 162, 163, 164]. Since the lepton weak doublet and the right-handed down-type quark are included in the same SU(5) multiplet, the renormalization induces the flavor mixing in the right-handed down-type squark sector. At the same time, lepton flavor violation (LFV) in the charged lepton sector is also induced.

We consider two specific cases for the right-handed neutrino mass matrix, since constraints from LFV processes, especially from the $\mu \to e\gamma$ process, depend on the matrix significantly. From the seesaw relation, the light neutrino mass matrix is given by $m_\nu = y_\nu^T (M_N)^{-1} y_\nu (v^2 \sin^2 \beta/2)$ in the basis where the charged Higgs Yukawa coupling is diagonal. Here, $y_\nu$, $M_N$, and $\beta$ are the neutrino Yukawa coupling constant, the right-handed neutrino mass matrix, and the vacuum angle, respectively. On the other hand, the LFV mass term for the left-handed slepton are given by $(\delta m_L^2)^{ij} \simeq -(y_\nu^\dagger y_\nu)^{ij}(3m_0^2 + |A_0|^2)\ln(M_P/M_R)/8\pi^2$, where $m_0$, $A_0$, $M_P$, and $M_R$ are the universal scalar mass, the trilinear scalar coupling, the Planck mass, and the right-handed mass scale. The first case is a degenerate case, where the right-handed neutrino mass matrix is proportional to a unit matrix. The off-diagonal element of the slepton mass matrix is related to the neutrino mixing matrix in this case, and therefore the large mixing angle suggested by the solar neutrino observation indicates a severe constraint on SUSY parameters from the $\mu \to e\gamma$ branching ratio. On the other hand, the constraint is relaxed, if we arrange the right-handed neutrino mass matrix such that the 1-2 and 1-3 mixings of $y^\dagger y$ vanish. We call this limiting case a non-degenerate case. Since the constraint to the SUSY parameter space is quite different in two cases, we calculate various observable quantities for both, and compare their phenomenological implications. Note that in the viewpoint of the LFV constraint the degenerate case represents a more generic situation than the non-degenerate case, because the $\mu \to e\gamma$ process generally puts a severe restriction to allowed ranges of SUSY parameters.

The minimal supersymmetric Standard Model (MSSM) with U(2) flavor symmetry was proposed sometime ago as a solution of the flavor problem in a general SUSY model [165, 166]. Unless the squark and slepton masses are in the multi-TeV range, there should be some suppression mechanism for flavor changing neutral current (FCNC) processes, especially for the squarks and sleptons of the first two generations. If we introduce a U(2) flavor symmetry, under which the first two generations are assigned to be doublets and the third generation is a singlet, we can explain the realistic pattern of the quark mass and suppress the unwanted FCNC at least for the kaon sector. FCNC of the bottom sector, on the other hand, can be interesting signals. We follow a specific model of this type according to Ref. [166]. In this model, there are many $\mathcal{O}(1)$ parameters in squark mass matrices, which we have scanned in a reasonable range.

We have calculated the following quantities for the above four cases in the three models.

- $CP$ violation parameter $\epsilon_K$ in the $K^0 - \overline{K^0}$ mixing.
- $B_d - \overline{B}_d$ mixing and $B_s - \overline{B}_s$ mixing.
- The mixing induced $CP$ violations in $B \to J/\psi K_S^0$ and $B \to \phi K_S^0$ modes.
- The mixing-induced $CP$ violation in $B \to M_s\gamma$, where $M_s$ is a $CP$ eigenstate with a strange quark such as $K^*(\to K_S^0\pi^0)$.
- Direct $CP$ violation in the inclusive $b \to s\gamma$ process.

These quantities provide several independent methods to look for New Physics. New Physics contributions in the mixing quantities may be identified from the consistency test of the unitarity triangle. The difference of the $CP$ asymmetries in $B \to J/\psi K_S^0$ and $B \to \phi K_S^0$ implies existence of a new $CP$ phase in the $b \to s$ transition amplitude. For the $b \to s\gamma$ process, a sizable direct $CP$ asymmetry means a new phase in the $b \to s\gamma$ amplitude, while the mixing-induced asymmetry arises from the interference between the amplitudes with $b \to s\gamma_L$ and $\overline{b} \to \overline{s}\gamma_L$. Although this is suppressed by $m_s/m_b$ in the Standard Model, New Physics effects can generate $O(1)$ asymmetry, if there is a $b \to s\gamma$ amplitude with the opposite chirality. Detailed description of our calculation is given in Ref. [158, 99].

The correlation among the $CP$ asymmetry of the $B \to J/\psi K_S^0$ mode, the phase of $V_{ub}^*$ element ($\phi_3$), and the ratio of the $B_s - \overline{B}_s$ mixing and $B_d - \overline{B}_d$ = mixing ($\Delta m(B_s)/\Delta m(B_d)$) is shown in Fig. 5-30. In this figure, we have taken into account theoretical uncertainties due to the kaon bag parameters ($\pm 15\%$) and $f_B\sqrt{B_d}$ ($\pm 20\%$) and take $|V_{ub}/V_{cb}| = 0.09 \pm 0.01$. In the calculation, we have imposed various phenomenological constraints to restrict SUSY parameter space. These includes constraints from the Higgs boson and SUSY particle searches in collider experiments, the branching ratio of the $b \to s\gamma$ process, and various EDM experiments. We updated the previous calculation given in Ref. [158, 99] by taking account of the constraint on parameter space from the strange quark EDM contribution to the Hg EDM, which was pointed out recently [167]. For the two cases of the SU(5) GUT with right-handed neutrinos, we included the $\mu \to e\gamma$ constraint, which is especially important in the degenerate case. The figure corresponds to $\tan\beta = 30$. For neutrino parameters in the GUT model, we take a hierarchal light neutrino mass spectrum with the large mixing angle MSW solution. The right-handed neutrino masses are taken to be $4 \times 10^{13}$ GeV for the degenerate case, and $5.7, 18, 45 \times 10^{13}$ GeV for the non-degenerate case.

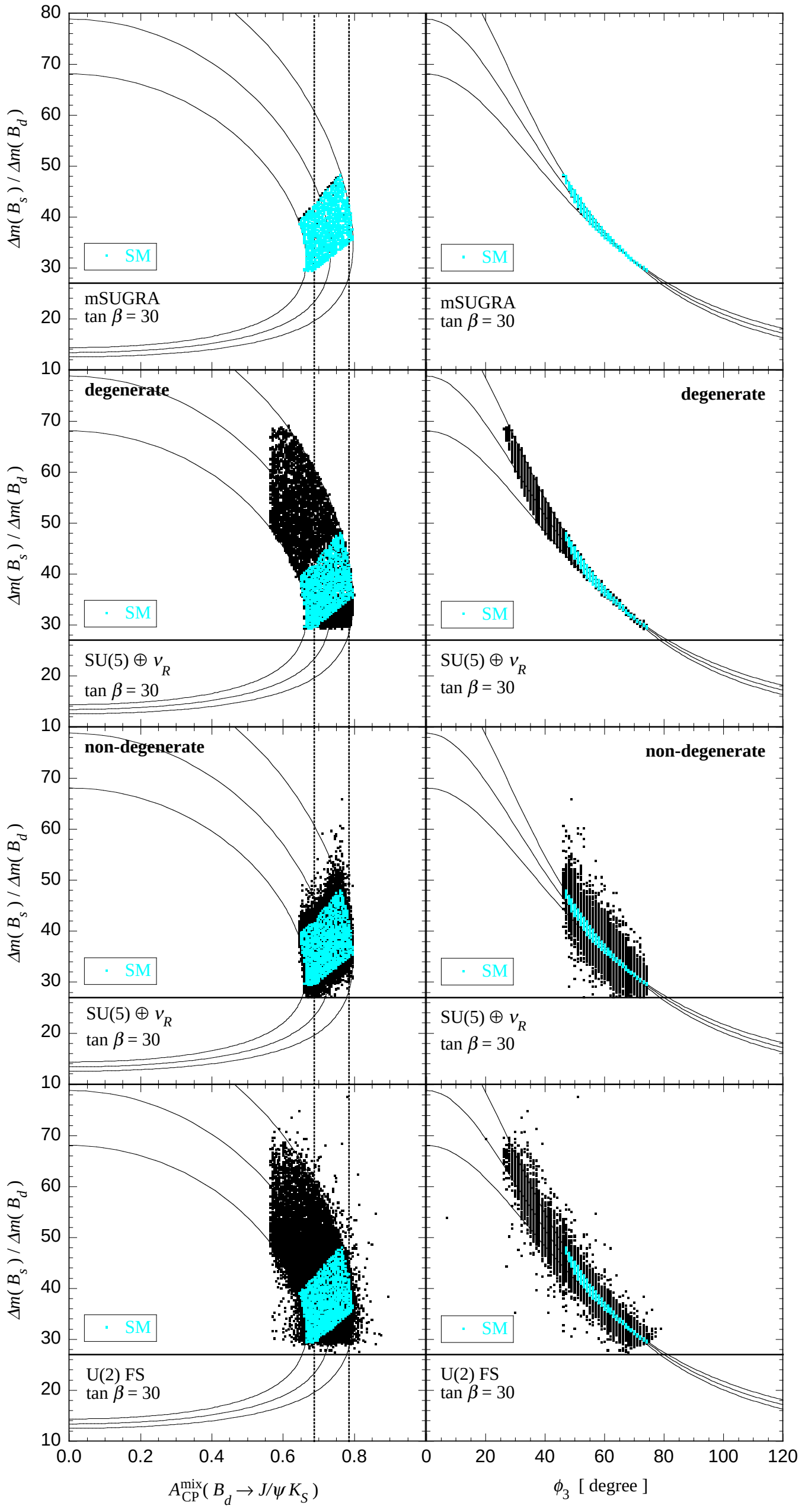

**Figure 5-30.** $\Delta m(B_s)/\Delta m(B_d)$ *versus the mixing-induced* $CP$ *asymmetry of* $B_d \to J/\psi K_S^0$ *and* $\phi_3$ *in the minimal supergravity model, SU(5) SUSY GUT with right-handed neutrinos for the degenerate and non-degenerate cases of the right-handed neutrino mass matrix, and the MSSM with a U(2) flavor symmetry. The light-colored regions show the allowed region in the Standard Model . The curves show the Standard Model values with* $|V_{ub}/V_{cb}| = 0.08$*, 0.09 and 0.10.*

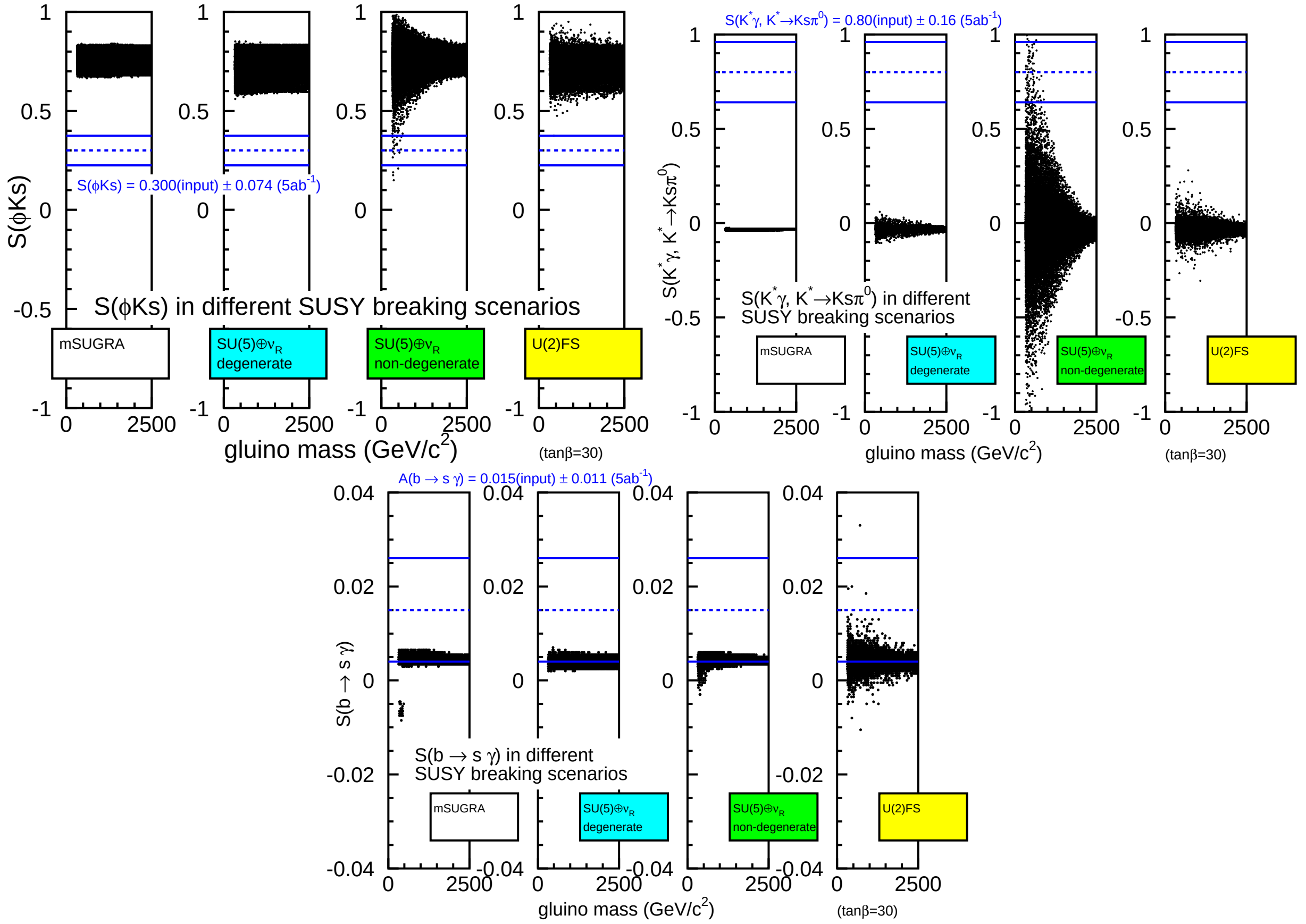

**Figure 5-31.** *Mixing-induced $CP$ asymmetry in $\phi K^0_S$ and $M_s\gamma$ modes and direct $CP$ asymmetry in $b \to s\gamma$ as a function of the gluino mass.*

We can see that possible deviations from the Standard Model prediction are small for the minimal supergravity model. In the SU(5) GUT with right-handed neutrinos, the pattern of the deviation is different for the two cases. For the degenerate case, $\Delta m(B_s)/\Delta m(B_d)$ can be enhanced from the Standard Model prediction, while the correct value of $\phi_3$ is smaller than that expected in the Standard Model; this deviation, in fact, arises from a large SUSY contribution to $\epsilon_K$. The deviation may become clear when the value of $\phi_3$ is precisely determined from $CP$ asymmetries of tree precesses such as $B \to DK$. For the non-degenerate case, the deviation can be seen only for $\Delta m(B_s)/\Delta m(B_d)$. We can conclude that SUSY contributions are large for the 1-2 generation mixing in the former case, and the 2-3 generation mixing in the latter case. More general type of deviations is possible for the MSSM with a U(2) flavor symmetry, because all three mixing diagrams can have large contributions.

The mixing-induced $CP$ asymmetries in the $B \to \phi K^0_S$ and $B \to M_s\gamma$ modes and the direct $CP$ asymmetry of $b \to s\gamma$ are shown in Fig. 5-31 for the four cases in the three models. Possible values of these observable quantities are plotted in terms of the gluino mass for $\tan\beta = 30$. These figures are updated from those in Ref. [99], taking into account the

**Table 5-12.** *Pattern of the deviation from the Standard Model predictions for unitarity triangle and rare decays. "$\surd$" means that the deviation can be large and "-" means a small deviation. "closed" in the first row of the $B_d$ unitarity means that the unitarity triangle is closed among observables related to $B_d$, and the second and the third rows show whether deviation is observed from consistency check between the $B_d$ unitarity and $\epsilon_K$ and $\Delta m(B_s)/\Delta m(B_d)$, respectively.*

| | $B_d$ unitarity | | | Rare Decays | | |
|---|---|---|---|---|---|---|
| | closure | $+\epsilon_K$ | $+\Delta m(B_s)$ | $A_{CP}^{\rm mix}(B \to \phi K_S^0)$ | $A_{CP}^{\rm mix}(B \to M_s \gamma)$ | $A_{CP}^{\rm dir}(B \to X_s \gamma)$ |
| mSUGRA | closed | - | - | - | - | - |
| SU(5) SUSY GUT (degenerate RHN) | closed | $\surd$ | - | - | - | - |
| SU(5) SUSY GUT (non-deg. RHN) | closed | - | $\surd$ | $\surd$ | $\surd$ | - |
| MSSM with U(2) | $\surd$ | $\surd$ | $\surd$ | $\surd$ | $\surd$ | $\surd$ |

Hg EDM constant. The expected experimental sensitivities at a Super $B$ Factory with integrated luminosity of 5 ab$^{-1}$ are also indicated based on the study for the Super KEKB LoI. The central values are chosen for illustrative purpose. We can see that SUSY effects are large for the mixing-induced $CP$ asymmetries for $B \to \phi K_S^0$ and $B \to M_s\gamma$ in the non-degenerate case of SU(5) SUSY GUT with right-handed neutrinos, whereas the corresponding deviations are small for the degenerate case. In the degenerate case the constraint from the $\mu \to e\gamma$ branching ratio is so strong that the effect in the 2-3 generation mixing is not sizable. In the non-degenerate case, the contribution from the sbottom-sdown mixing induces large effects in the $b \to s$ transition, because the $\mu \to e\gamma$ constraint is somewhat relaxed. For the U(2) case, we can see that all three deviations can be sizable.

Possible deviations from the Standard Model prediction in the consistency test of the unitarity triangle and rare decays are summarized in Table 5-12. The patterns of the deviations are different for these cases. For instance, observables related to the $B_d$ unitarity triangle, namely $\Delta m(B_d)$, $|V_{ub}|$, $\phi_1$ from the $B \to J/\psi_S$ mode, and $\phi_3$ from the $B \to DK$ mode are consistent with a single triangle for the first tree cases in the table, but deviation can be observed if we compare $\epsilon_K$ and $\Delta m(B_s)/\Delta m(B_d)$ with the Standard Model prediction for the second and third cases. The deviation patterns are also different for various rare decay observables. These features are useful to distinguish different SUSY models at a Super $B$ Factory.

### $B$ physics signals in the Snowmass Points & Slopes

It is expected that LHC experiments can significantly improve the search limit of SUSY particles. In a typical scenario like the minimal supergravity model, squarks and gluino will be found if their masses are below 2 TeV. Snowmass Points and Slopes (SPS) are proposed sets of benchmark parameters of SUSY parameter space [168]. Such model points and lines are selected as representative cases for phenomenological studies of SUSY theory, especially for SUSY particle searches in future collider experiments.

From the viewpoint of a Super $B$ Factory, it is interesting to study possible flavor physics signals in these benchmark scenarios, and compare them with collider signals. In order to illustrate how LHC and a Super $B$ Factory can be complementary to each other, we calculated FCNC processes and rare decays along several benchmark parameter lines for the two cases of SU(5) GUT with right-handed neutrinos. We should note that the benchmark points are mainly intended to select representative SUSY mass spectrum for physics analysis at collider experiments, whereas the flavor physics depends on how flavor off-diagonal terms in the squark/slepton mass matrices are generated. It is therefore conceivable that $B$ physics can distinguish different SUSY models, even if the SUSY spectrum looks very similar.

We consider the following model-parameter lines, corresponding to four cases in the SPS list. These lines are defined by input parameters of the minimal supergravity model, namely the universal scalar mass ($m_0$), the gaugino mass ($m_{1/2}$), the universal trilinear coupling ($A_0$), and the vacuum ratio ($\tan\beta$). The sign of the higgsino mass term ($\mu$) is taken to be positive.

- SPS 1a: $m_0 = -A_0 = 0.4 m_{1/2}, \tan\beta = 10$
- SPS 1b: $m_0 = 0.5 m_{1/2}, A_0 = 0, \tan\beta = 30$
- SPS 2: $m_0 = 2 m_{1/2} + 850\text{GeV}, \text{A}_0 = 0, \tan\beta = 10$
- SPS 3: $m_0 = 0.25 m_{1/2} - 10\text{GeV}, \text{A}_0 = 0, \tan\beta = 10$

The lines are defined by varying $m_{1/2}$ The first two cases represent typical parameter points in the minimal supergravity model. (SPS 2b was only defined for a point with $m_{1/2} = 400$ GeV in [168], but here we generalize it to a line by varying $m_{1/2}$.) SPS 2 corresponds to the focus point scenario, where squarks and sleptons are rather heavy [169]. SPS 3 is a line in the co-annihilation region, where a rapid co-annihilation between a lighter stau and a LSP neutralino allows acceptable relic abundance for LSP dark matter. We take these input SUSY parameters for the SUSY GUT model, although the precise mass spectrum is not exactly the same as the minimal supergravity case due to additional renormalization effects from the neutrino Yukawa coupling, *etc.*

The results of the calculation for $\epsilon_K/(\epsilon_K)_{SM}$, $\Delta m(B_s)/(\Delta m(B_s))_{SM}$, $A^{\text{mix}}_{CP}(B \to \phi K^0_S)$, and $A^{\text{mix}}_{CP}(B \to M_s\gamma)$ are summarized in Table 5-13. In this calculation, we take the right-handed neutrino mass scale around $10^{14}$ GeV as before, and new phases associated with GUT interactions are varied. The calculation procedure is the same as that in Ref. [158, 99]. The table lists magnitudes of maximal deviations from the Standard Model prediction for each quantity. We do not list $A^{\text{dir}}_{CP}(B \to X_s\gamma)$, because possible deviations are not large even in more general parameter space as described before. We see that the only sizable deviation appears for $\epsilon_K/(\epsilon_K)_{StandardModel}$ in the degenerate case of SPS 2. For other cases, it is difficult to distinguish these models from the prediction of the Standard Model or the minimal supergravity model with an integrated luminosity of 5 ab$^{-1}$. $\epsilon_K/(\epsilon_K)_{SM}$ is shown for the degenerate case as a function of the gluino mass in Fig. 5-32. For the case of SPS 2, the deviation of this size can be distinguished at a Super $B$ Factory by improved measurements of quantities related to the unitarity triangle, especially $\phi_3$. On the other hand, the $b \to s$ transition processes do not show large deviations even for the non-degenerate case for the selected model-lines. This is in contrast to the scatter plot in more general parameter space. We find that a large deviation occurs only for large values of the $A_0$ parameter, but the benchmark lines do not correspond to such cases. We should also note that a sizable deviation in the SPS 2 case can be seen even for a relatively heavy SUSY spectrum where squarks are 1 -2 TeV, which can be close to the discovery limit of SUSY at the LHC experiments.

**Table 5-13.** *Possible deviation from the Standard Model prediction for various observable quantities for benchmark parameter lines in the SU(5) SUSY GUT with right-handed neutrinos. The degenerate and non-degenerate cases for right-handed neutrinos are shown separately. SPS 1a, 1b, 2, and 3 are model-parameter lines defined in the text. The right-handed neutrino mass scale is taken to be $O(10^{14})$ GeV, and GUT phases are varied.*

| Degenerate case | $\epsilon_K/(\epsilon_K)_{SM}$ | $\Delta m(B_s)/(\Delta m(B_s))_{SM}$ | $A^{\text{mix}}_{CP}(B \to \phi K^0_S)$ | $A^{\text{mix}}_{CP}(B \to M_s\gamma)$ |
|---|---|---|---|---|
| SPS 1a | $\lesssim$ 10% | $\lesssim$ 2% | $\lesssim$ 0.1% | $\lesssim$ 0.1% |
| SPS 1b | $\lesssim$ 10% | $\lesssim$ 2% | $\lesssim$ 0.5% | $\lesssim$ 0.2% |
| SPS 2 | $\lesssim$ 100% | $\lesssim$ 2% | $\lesssim$ 0.3% | $\lesssim$ 0.5% |
| SPS 3 | $\lesssim$ 5% | $\lesssim$ 2% | $\lesssim$ 0.1% | $\lesssim$ 0.1% |
| Non-degenerate case | | | | |
| SPS 1a | $\lesssim$ 2% | $\lesssim$ 2% | $\lesssim$ 0.1% | $\lesssim$ 1% |
| SPS 1b | $\lesssim$ 1% | $\lesssim$ 2% | $\lesssim$ 0.5% | $\lesssim$ 2% |
| SPS 2 | $\lesssim$ 1% | $\lesssim$ 3% | $\lesssim$ 1% | $\lesssim$ 3% |
| SPS 3 | $\lesssim$ 2% | $\lesssim$ 2% | $\lesssim$ 0.1% | $\lesssim$ 1% |

In summary, we studied SUSY effects to various FCNC processes related to the unitarity triangle and rare $B$ decay processes with a $b \to s$ transition. We considered the minimal supergravity model, two cases for the SU(5) SUSY GUT with right-handed neutrinos, and the MSSM with a U(2) flavor symmetry. We found that large deviations are possible in observable quantities with either 1-2 or 2-3 generation transition depending on the choice of the right-handed neutrino mass matrices and the neutrino Yukawa coupling constants in the GUT model. Various New Physics signals are possible in the U(2) model, while the deviation is small for the minimal supergravity model. These features are useful to identify possible SUSY models at a Super $B$ Factory . We also consider SUSY parameter space based on benchmark scenarios of SPS. We observe that SUSY contribution can be large in $\epsilon_K$ for the case of SPS 2 (focus point scenario) with the degenerate right-handed neutrinos in SU(5) SUSY GUT. This example illustrates that a Super $B$ Factory can provide important insight to the flavor structure of SUSY theory, which is complementary to what will be obtained at energy frontier collider experiments.

### 5.3.3 Electric Dipole Moment for $^{199}$Hg atom and $B \to \phi K_S^0$ in Supersymmetric Models with Right-Handed Squark Mixing

>– J. Hisano and Y. Shimizu –<

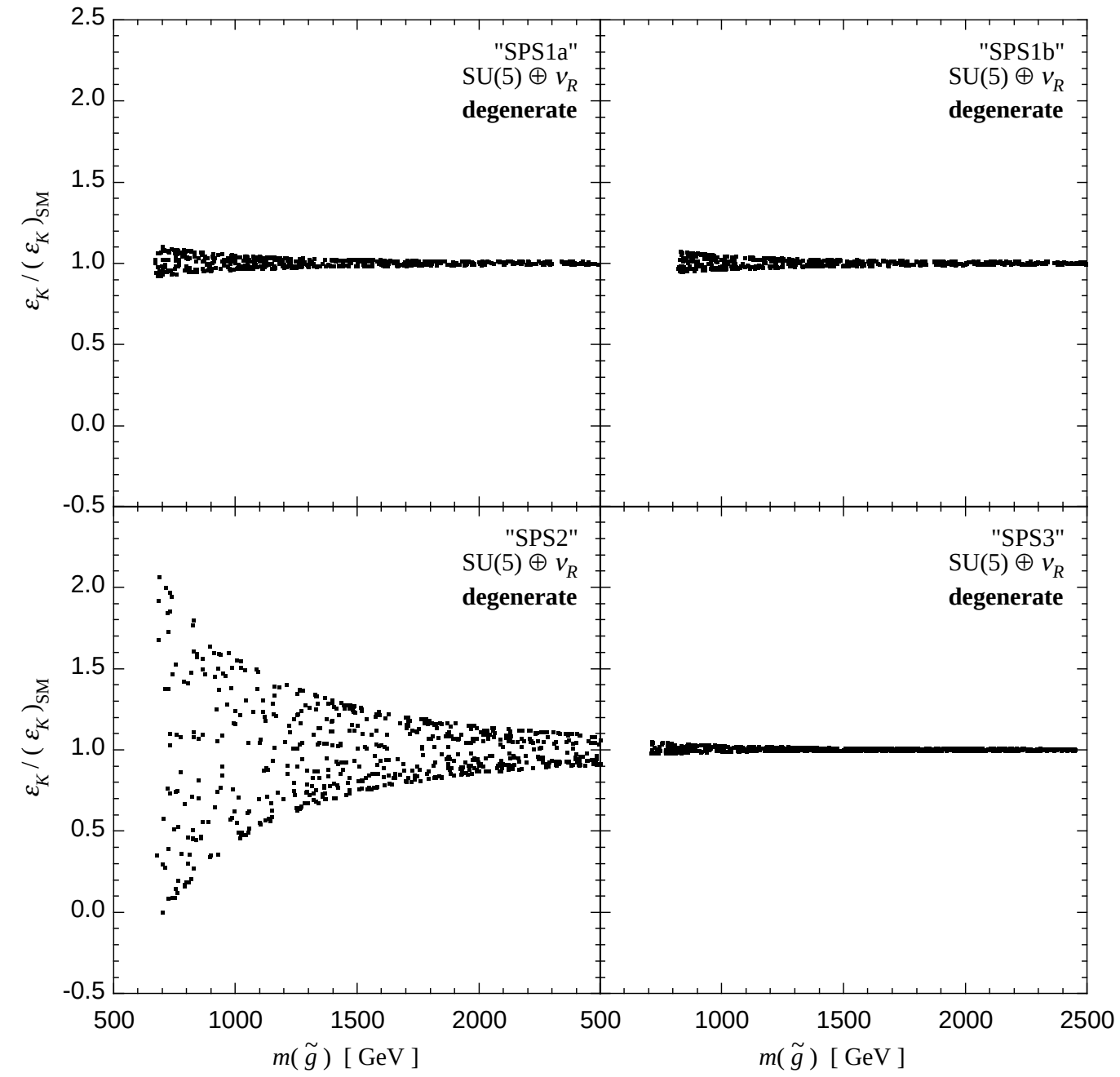

**Figure 5-32.** $\epsilon_K/(\epsilon_K)_{StandardModel}$ *for SPS 1a, 1b, 2 and 3 in SU(5) SUSY GUT with right-handed neutrinos for the degenerate right-handed neutrino mass case. Deviation is* $O(1)$ *only for SPS 2.*

### Introduction

The Belle experiment in the KEK $B$ Factory reported recently that the $CP$ asymmetry in $B \to \phi K_S^0$ ($S_{\phi K_S^0}$) is $-0.96 \pm 0.50^{+0.09}_{-0.11}$, and $3.5\sigma$ deviation from the Standard-Model prediction $0.731 \pm 0.056$ is found [170]. At present the *BABAR* experiment does not observe such a large deviation, finding $0.45 \pm 0.43 \pm 0.07$ [171]. The combined result is not yet significant, however, Belle's result might be a signature of New Physics.

The $CP$ violation in $B \to \phi K_S^0$ is sensitive to New Physics, since $b \to s\bar{s}s$ is a radiative process [172]. In fact, the SUSY models may predict a sizable deviation of the $CP$ violation in $B \to \phi K_S^0$ from the Standard Model prediction. If the right-handed bottom and strange squarks have a sizable mixing, the gluon-penguin diagram may give a non-negligible contribution to $b \to s\bar{s}s$ in a broad parameter space where the contribution to $b \to s\gamma$ is a sub-dominant. $B \to \phi K_S^0$ in SUSY models has been studied in many papers [173][174][175][167].

In this article the correlation between the $CP$ asymmetry in $B \to \phi K_S^0$ ($S_{\phi K_S^0}$) and the chromoelectric dipole moment (CEDM) of the strange quark ($d_s^C$) is discussed in SUSY models with right-handed squark mixing. In typical SUSY models, the left-handed squarks also have flavor mixing, due to the top-quark Yukawa coupling and the CKM mixing, and the left-handed bottom and strange squark mixing is as large as $\lambda^2 \sim 0.04$. When both the right-handed and left-handed squark mixings between the second and third generations are non-vanishing, a CEDM of the strange quark is generated. Since $S_{\phi K_S^0}$ and $d_s^C$ may have a strong correlation in the SUSY models with the right-handed squark mixing, the constraint on $d_s^C$ by the measurement of the EDM of $^{199}$Hg limits the gluon-penguin contribution from the right-handed squark mixing to $S_{\phi K_S^0}$ [167].

In next section we discuss the $^{199}$Hg EDM in SUSY models. In Section 5.3.3 the correlation between the $CP$ asymmetry in $B \to \phi K_S^0$ and the CEDM of strange quark in the SUSY models with the right-handed squark mixing is presented. Section 5.3.3 is devoted to discussion.

### The $^{199}$Hg EDM in SUSY models

The EDMs of electron, neutron and nuclei are extensively studied in the SUSY models, and it is found that the relative phases among the flavor-diagonal $A$ terms, $B$ term in the Higgs potential, and the gaugino mass terms should be suppressed. However, even in that case, the EDMs are generated if both the left- and right-handed sfermions are mixed. It is especially interesting that EDM's are enhanced by heavier fermion masses, while they are suppressed by the mixing angles. Thus, the EDMs provide a stringent constraint on the SUSY models with both left- and right-handed sfermion mixings.

Before deriving the constraint on the bottom and strange squark mixing, we discuss the EDM of the nuclei. The EDMs of the diamagnetic atoms, such as $^{199}$Hg, come from the $CP$-violating nuclear force by pion or eta meson exchange. The quark CEDMs,

$$H = \sum_{q=u,d,s} d_q^C \frac{i}{2} g_s \bar{q} \sigma^{\mu\nu} T^A \gamma_5 q G^A_{\mu\nu}, \tag{5.123}$$

generate the $CP$-violating meson-nucleon coupling, and the EDM of $^{199}$Hg is evaluated in Ref. [176] as

$$d_{\rm Hg} = -3.2 \times 10^{-2} e \times (d_d^C - d_u^C - 0.012 d_s^C). \tag{5.124}$$

Chiral perturbation theory implies that $\bar{s}s$ in the matrix element of the nucleon is not suppressed, leading to a non-vanishing contribution from the CEDM of the strange quark. The suppression factor in front of $d_s^C$ in Eq. (5.124) comes from the $\eta$ meson mass and the $CP$-conserving coupling of the $\eta$ meson and nucleon. From the current experimental bound on $d_{\rm Hg}$ ($d_{\rm Hg} < 2.1 \times 10^{-28} e\,{\rm cm}$) [177]:

$$e|d_d^C - d_u^C - 0.012 d_s^C| < 7 \times 10^{-27} e\,{\rm cm}\,. \tag{5.125}$$

If $d_d^C$ and $d_u^C$ are negligible in the equation,

$$e|d_s^C| < 7 \times 5.8 \times 10^{-25} e\,\text{cm}\,. \tag{5.126}$$

The neutron EDM should also be affected by the CEDM of the strange quark. However, it is argued in Ref. [178] that this is suppressed by Peccei-Quinn symmetry. It is not clear at present whether the contribution of the CEDM of the strange quark is completely decoupled from the neutron EDM under Peccei-Quinn symmetry. In the following, we adopt the constraint on CEDM of the strange quark from $^{199}$Hg.

In SUSY models, when the left-handed and right-handed squarks have mixings between the second and third generations, the CEDM of the strange quark is generated by a diagram in Fig. 5-33(a), and is enhanced by $m_b/m_s$. Using the mass insertion technique, $d_s^C$ is given as

$$d_s^C = \frac{\alpha_s}{4\pi}\frac{m_{\tilde g}}{m_{\tilde q}^2}\left(-\frac{1}{3}N_1(x) - 3N_2(x)\right)\text{Im}\left[(\delta_{LL}^{(d)})_{23}\,(\delta_{LR}^{(d)})_{33}\,(\delta_{RR}^{(d)})_{32}\right]\,, \tag{5.127}$$

up to the QCD correction, where $m_{\tilde g}$ and $m_{\tilde q}$ are the gluino and averaged squark masses. The functions $N_i$ are given as

$$N_1(x) = \frac{3 + 44x - 36x^2 - 12x^3 + x^4 + 12x(2+3x)\log x}{(x-1)^6}, \tag{5.128}$$

$$N_2(x) = -2\frac{10 + 9x - 18x^2 - x^3 + 3(1+6x+3x^2)\log x}{(x-1)^6}. \tag{5.129}$$

The mass insertion parameters $(\delta_{LL}^{(d)})_{23}$, $(\delta_{RR}^{(d)})_{32}$, and $(\delta_{LR}^{(d)})_{33}$ are given by

$$(\delta_{LL}^{(d)})_{23} = \frac{\left(m^2_{\tilde d_L}\right)_{23}}{m_{\tilde q}^2},\ (\delta_{RR}^{(d)})_{32} = \frac{\left(m^2_{\tilde d_R}\right)_{32}}{m_{\tilde q}^2},\ (\delta_{LR}^{(d)})_{33} = \frac{m_b\,(A_b - \mu\tan\beta)}{m_{\tilde q}^2}, \tag{5.130}$$

where $(m^2_{\tilde d_{L(R)}})$ is the left-handed (right-handed) down-type squark mass matrix. In typical SUSY models, $(\delta_{LL}^{(d)})_{23}$ is $O(\lambda^2) \simeq 0.04$. From this formula, $d_s^C$ is estimated in a limit of $x \to 1$ as

$$ed_s^C = e\frac{\alpha_s}{4\pi}\frac{m_{\tilde g}}{m_{\tilde q}^2}\left(-\frac{11}{30}\right)\text{Im}\left[(\delta_{LL}^{(d)})_{23}\,(\delta_{LR}^{(d)})_{33}\,(\delta_{RR}^{(d)})_{32}\right] \tag{5.131}$$

$$= -4.0 \times 10^{-23}\sin\theta\, e\,\text{cm}\left(\frac{m_{\tilde q}}{500\text{GeV}}\right)^{-3}\left(\frac{(\delta_{LL}^{(d)})_{23}}{0.04}\right)\left(\frac{(\delta_{RR}^{(d)})_{32}}{0.04}\right)\left(\frac{\mu\tan\beta}{5000\text{GeV}}\right)\,, \tag{5.132}$$

where $\theta = \arg[(\delta_{LL}^{(d)})_{23}\,(\delta_{LR}^{(d)})_{33}\,(\delta_{RR}^{(d)})_{32}]$. Here, we neglect the contribution proportional to $A_b$, since it is subdominant. From this formula, it is obvious that the right-handed squark mixing or the *CP*-violating phase should be suppressed. For example, for $m_{\tilde q} = 500$GeV, $\mu\tan\beta = 5000$GeV, and $(\delta_{LL}^{(d)})_{23} = 0.04$,

$$|\sin\theta(\delta_{RR}^{(d)})_{32}| < 5.8 \times 10^{-4}. \tag{5.133}$$

### Correlation between $d_s^C$ and $B \to \phi K_S^0$ in models with right-handed squark mixing

Let us discuss the correlation between $d_s^C$ and $S_{\phi K_S^0}$ in the SUSY models with right-handed squark mixing. As mentioned in Introduction, the right-handed bottom and strange squark mixing may lead to the sizable deviation of $S_{\phi K_S^0}$ from the Standard Model prediction by the gluon-penguin diagram, especially for large $\tan\beta$. The box diagrams with the right-handed squark mixing also contribute to $S_{\phi K^0_S}$, but they tend to be sub-dominant, and do not generate a large deviation of $S_{\phi K^0_S}$ from the Standard Model prediction. Thus, for simplicity, we will neglect the box contribution in this article.

The effective operator inducing the gluon-penguin diagram by the right-handed squark mixing is

$$H = -C_8^R \frac{g_s}{8\pi^2} m_b \overline{s_R} \sigma^{\mu\nu} T^A b_L G^A_{\mu\nu}. \quad (5.134)$$

When the right-handed squarks are mixed, the dominant contribution to $C_8^R$ is supplied by a diagram with the double mass insertion of $(\delta^{(d)}_{RR})_{32}$ and $(\delta^{(d)}_{RL})_{33}$ (Fig. 5-34(b)). This contribution is specially significant when $\mu \tan\beta$ is large. The contribution of Fig. 5-33(b) to $C_8^R$ is given as

$$C_8^R = \frac{\pi\alpha_s}{m^2_{\tilde{q}}} \frac{m_{\tilde{g}}}{m_b} (\delta^{(d)}_{LR})_{33} (\delta^{(d)}_{RR})_{32} (-\frac{1}{3} M_1(x) - 3M_2(x)) \quad (5.135)$$

up to QCD corrections. Here,

$$M_1(x) = \frac{1 + 9x - 9x^2 - x^3 + (6x + 6x^2)\log x}{(x-1)^5}, \quad (5.136)$$

$$M_2(x) = -2\frac{3 - 3x^2 + (1 + 4x + x^2)\log x}{(x-1)^5}. \quad (5.137)$$

In a limit of $x \to 1$, $C_8^R$ is reduced to

$$C_8^R = \frac{7\pi\alpha_s}{30 m_b m_{\tilde{q}}} (\delta^{(d)}_{LR})_{33} (\delta^{(d)}_{RR})_{32}. \quad (5.138)$$

Comparing Eq. (5.131) and Eq. (5.138), we see a strong correlation between $d^C_s$ and $C^R_8$:

$$d^C_s = -\frac{m_b}{4\pi^2} \frac{11}{7} \mathrm{Im}\left[ (\delta^{(d)}_{LL})_{23} C^R_8 \right], \quad (5.139)$$

up to QCD corrections. The coefficient $11/7$ in Eq. (5.139) changes from 3 to 1 for $0 < x < \infty$.

In Fig. 5-34 the correlation between $d^C_s$ and $S_{\phi K^0_S}$ is presented. Here, we assume $d^C_s = -m_b/(4\pi^2)\mathrm{Im}[(\delta^{(d)}_{LL})_{23} C^R_8]$, up to QCD corrections. Here, we take $(\delta^{(d)}_{LL})_{23} = -0.04$, $\arg[C^R_8] = \pi/2$ and $|C^R_8|$ corresponding to $10^{-5} < |(\delta^{(d)}_{RR})_{32}| < 0.5$. The matrix element of chromomagnetic moment in $B \to \phi K^0_S$ is

$$\langle \phi K^0_S | \frac{g_s}{8\pi^2} m_b (\overline{s}_i \sigma^{\mu\nu} T^a_{ij} P_R b_j) G^a_{\mu\nu} | \overline{B}_d \rangle = \kappa \frac{4\alpha_s}{9\pi} (\epsilon_\phi p_B) f_\phi m^2_\phi F_+(m^2_\phi), \quad (5.140)$$

and $\kappa = -1.1$ in the heavy-quark effective theory [175]. Since $\kappa$ may suffer from large hadronic uncertainties, we take $\kappa = -1$ and $-2$. From this figure, it is found that the deviation of $S_{\phi K^0_S}$ from the Standard Model prediction due to the gluon penguin contribution should be tiny when the constraint on $d^C_s$ in Eq. (5.126) is applied.

**Discussion**

In this article the correlation between the $CP$ asymmetry in $B \to \phi K^0_S$ and the chromoelectric dipole moment (CEDM) of strange quark has been discussed in SUSY models with right-handed squark mixing. While the gluon-penguin diagram might give a large deviation of $S_{\phi K^0_S}$ from the Standard Model prediction, the size is limited from the constraint on the CEDM of the strange quark. The constraint from the CEDM of the strange quark on the mixing between the right-handed strange and bottom squark is the most stringent at present, compared with other processes where the left-handed squarks are also mixed. For example, the $\mathcal{B} b \to s\gamma$ gives the constraint as

$$\left| (\delta^{(d)}_{RR})_{23} \right| \lesssim 0.27 \left( \frac{\mu \tan\beta}{5000 GeV} \right)^{-1} \left( \frac{m_{\tilde{q}}}{500 GeV} \right)^2, \quad (5.141)$$

which is looser. Also, the right-handed down-type squark mixing is related to the left-handed slepton mixing in the SUSY SU(5) GUT, and the experimental bound on $\mathcal{B}(\tau \to \mu\gamma)$ gives a constraint on the mixing between the right-handed strange and bottom squark [174]. While the current bound on $\mathcal{B}(\tau \to \mu\gamma)$ may exclude the possibility of a large deviation of $S_{\phi K^0_S}$, a sizable deviation is still allowed.

It has been argued recently in Ref. [179] that the measurement of the deuteron EDM may improve the bound on the $CP$-violating nuclear force by two orders of magnitude. If this is realized, it will be a stringent test of SUSY models with right-handed squark mixing, such as SUSY GUTs.

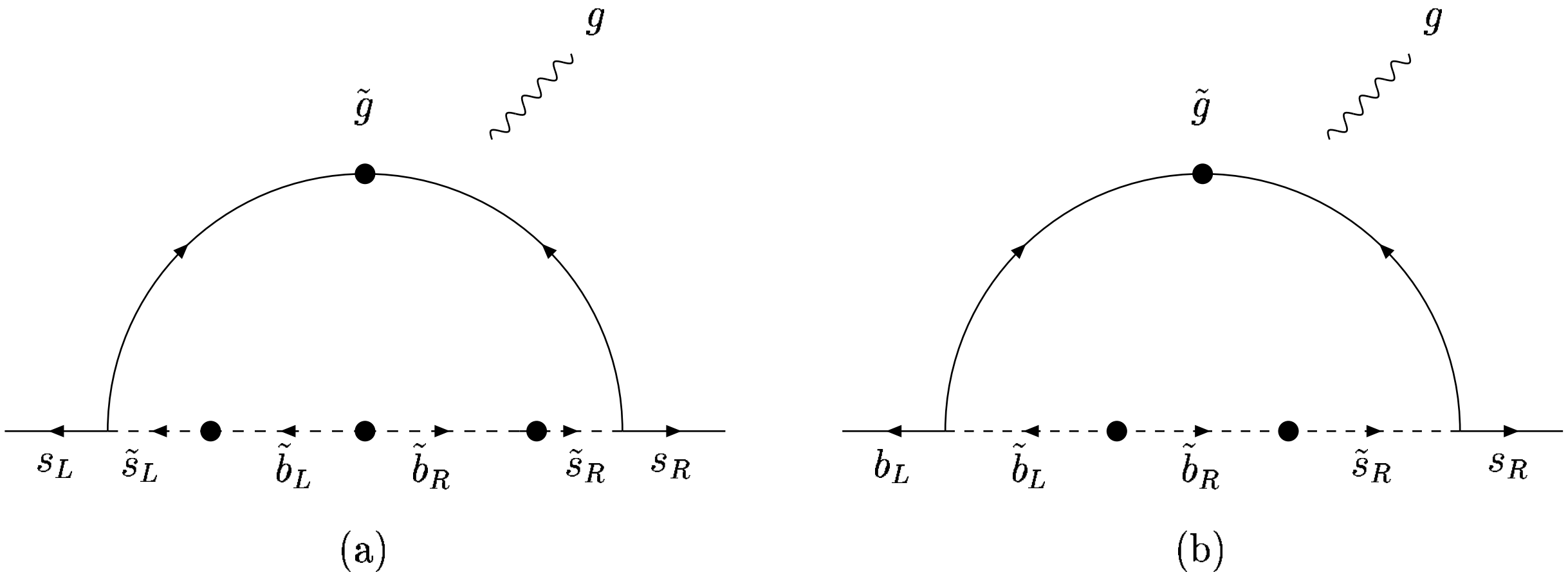

**Figure 5-33.** *a) The dominant diagram contributing to the CEDM of the strange quark when both the left-handed and right-handed squarks have mixings. b) The dominant SUSY diagram contributing to the $CP$ asymmetry in $B \to \phi K^0_S$ when the right-handed squarks have a mixing.*

### 5.3.4 SUSY Analysis in $B$ Decays: the Mass Insertion Approximation

>– M. Ciuchini, A. Masiero, L. Silvestrini, S. K. Vempati and O. Vives –<

#### Introduction

Our knowledge of the flavor sector in the Minimal Supersymmetric extension of the Standard Model (MSSM) is still very limited. Only after the discovery of SUSY particles and the measurement of the supersymmetric spectrum we will be able to explore in detail this fundamental piece of the MSSM. Nevertheless, we already have a lot of useful information on this sector from experiments looking for indirect effects of SUSY particles in low-energy experiments [180, 181].

To analyze flavor-violating constraints at the electroweak scale, the model-independent mass-insertion (MI) approximation is advantageous [182, 183, 184, 185]. In this method, the experimental limits lead to upper bounds on the parameters (or combinations of) $\delta^f_{ij} \equiv \Delta^f_{ij}/m^2_{\tilde{f}}$; where $\Delta^f_{ij}$ is the flavor-violating off-diagonal entry appearing in the $f = (u, d, l)$ sfermion mass matrices in the basis of diagonal Yukawa matrices and $m^2_{\tilde{f}}$ is the average sfermion mass. In addition, the mass-insertions are further sub-divided into $LL/LR/RL/RR$ types, labeled by the chirality of the corresponding Standard Model fermions. With the help of this MI formalism we can easily estimate the sensitivity of different processes to offdiagonal entries in the sfermion mass matrices. In this respect, it is instructive to compare the sensitivity of kaon and $B$ physics experiments.

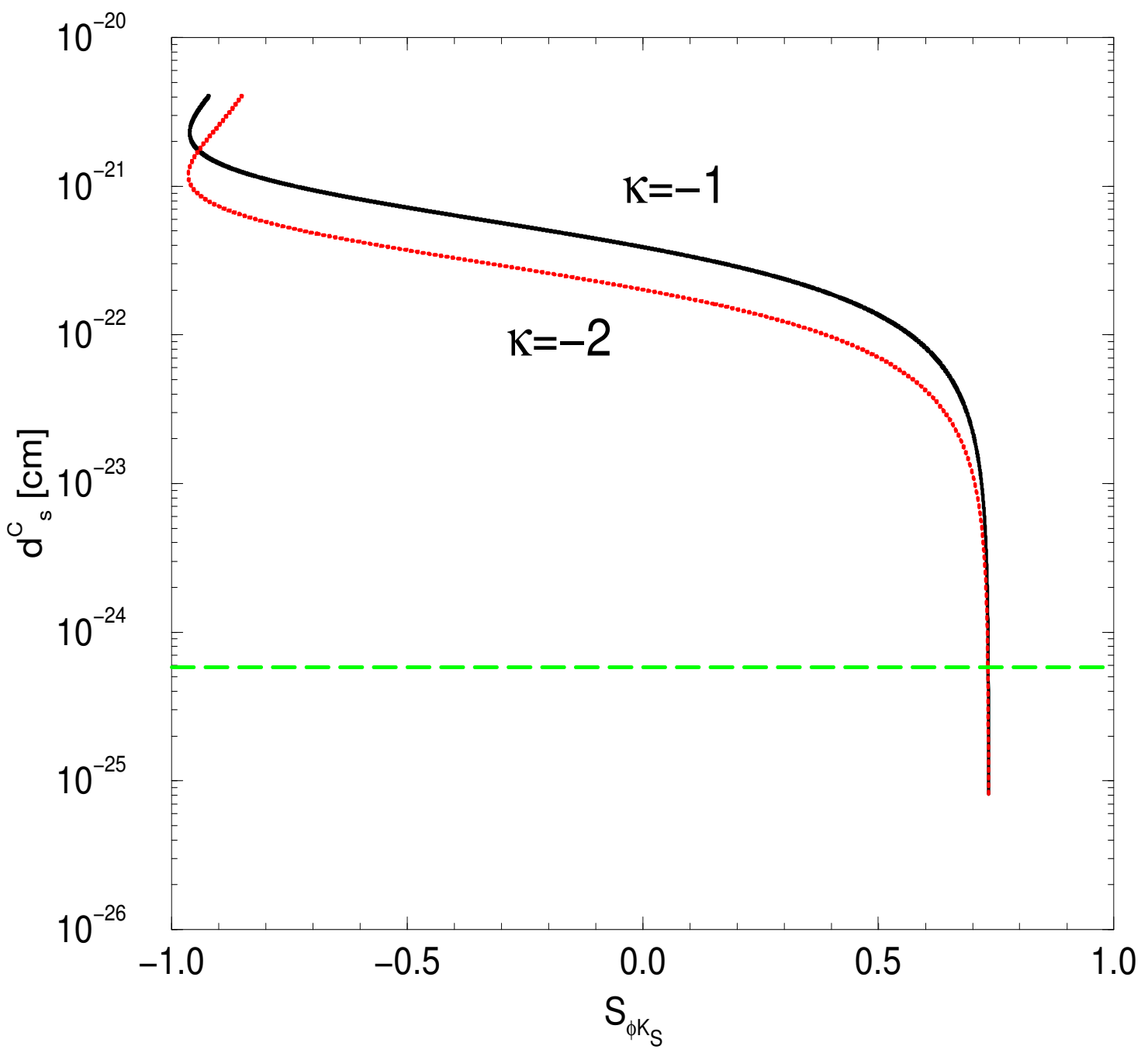

**Figure 5-34.** *The correlation between $d_s^C$ and $S_{\phi K_S^0}$ assuming $d_s^C = -m_b/(4\pi^2)\mathrm{Im}[(\delta_{LL}^{(d)})_{23} C_8^R]$. Here, $(\delta_{LL}^{(d)})_{23} = -0.04$ and $\arg[C_8^R] = \pi/2$. $\kappa$ comes from the matrix element of chromomagnetic moment in $B \to \phi K_S^0$. The dashed line is the upperbound on $d_s^C$ from the EDM of $^{199}$Hg atom.*

If we assume that indirect $CP$ violation in the kaon sector gets a sizable contribution from SUSY, while the kaon mass difference is mainly due to the Standard Model loops we have,

$$2.3 \times 10^{-3} \geq \varepsilon_K^{\text{SUSY}} = \frac{\text{Im}\, M_{12}|_{\text{SUSY}}}{\sqrt{2}\,\Delta M_K|_{\text{SM}}} \simeq \frac{\alpha_s^2}{\alpha_W^2}\,\frac{M_W^2}{M_{\text{SUSY}}^2}\,\frac{\text{Im}\left\{(\delta_{12}^d)^2_{LL}\right\}}{(V_{cd}V_{cs}^*)^2\,\frac{m_c^2}{M_W^2}}$$
$$\simeq 12.5 \times 0.026 \times \frac{\text{Im}\left\{(\delta_{12}^d)^2_{LL}\right\}}{1.5 \times 10^{-5}} \Rightarrow \sqrt{\text{Im}\left\{(\delta_{12}^d)^2_{LL}\right\}} \leq 3.3 \times 10^{-4}\,, \tag{5.142}$$

where we have assumed a SUSY mass scale of 500 GeV. In the Standard Model contribution, we have taken into account that quark masses must be present because of the GIM mechanism, and we have used the fact that the loop function in the W diagram, $S(x_c = m_c^2/M_W^2) \simeq x_c$ for $x_c \ll 1$ [186][3]. We have ignored factors of $\mathcal{O}(1)$, as well as the different loop functions in the gluino contributions. Hence, we can see that the SUSY contribution is suppressed by the heavy squark masses with respect to the $W$ boson mass. However, the SUSY gluino contribution is proportional to the strong coupling, while the Standard Model contribution is proportional to the weak coupling. Apart from these factors, we have to compare the mass insertion $(\delta_L^d)_{12}$ with $V_{cd}\frac{m_c}{M_W}V_{cs}^*$. Hence we can see that, due to the small fermion masses and/or mixing angles, $\varepsilon_K$ is in fact sensitive to MI at the level of a few $\times\, 10^{-4}$ [187].

[3]Here the charm–$W$ loop gives the main contribution for the kaon mass difference and this is sufficient for our estimate. Long distance (and top quark) effects are not included here, although they give a sizable ($\simeq 30\%$) contribution to $\Delta m_K$

**Table 5-14.** *Bounds on the mass insertions from $\varepsilon_K$, $\varepsilon'/\varepsilon$, BR($b \to s\gamma$) and $\Delta M_{B_{d,s}}$ for $m_{\tilde{q}} = 500$ GeV. For different squark masses, bounds on $(\delta_{LR})_{12}$ and $(\delta_{LR})_{13}$ scale as $(m_{\tilde{q}}(\text{GeV})/500)^2$, while other bounds scale as $(m_{\tilde{q}}(\text{GeV})/500)$. These bounds are equal under the exchange $L \leftrightarrow R$.*

| $x$ | $\sqrt{\left\|\text{Im}\left(\delta^d_{12}\right)^2_{LL}\right\|}$ | $\sqrt{\left\|\text{Im}\left(\delta^d_{12}\right)^2_{LR}\right\|}$ | $\sqrt{\left\|\text{Im}\left(\delta^d_{12}\right)_{LL}\left(\delta^d_{12}\right)_{RR}\right\|}$ | $\sqrt{\left\|\text{Re}\left(\delta^d_{13}\right)^2_{LL}\right\|}$ | $\sqrt{\left\|\text{Re}\left(\delta^d_{13}\right)^2_{LR}\right\|}$ |
|---|---|---|---|---|---|
| 0.3 | $2.9 \times 10^{-3}$ | $1.1 \times 10^{-5}$ | $1.1 \times 10^{-4}$ | $4.6 \times 10^{-2}$ | $5.6 \times 10^{-2}$ |
| 1.0 | $6.1 \times 10^{-3}$ | $2.0 \times 10^{-5}$ | $1.3 \times 10^{-4}$ | $9.8 \times 10^{-2}$ | $3.3 \times 10^{-2}$ |
| 4.0 | $1.4 \times 10^{-2}$ | $6.3 \times 10^{-5}$ | $1.8 \times 10^{-4}$ | $2.3 \times 10^{-1}$ | $3.6 \times 10^{-2}$ |
| | $\sqrt{\left\|\text{Re}\left(\delta^d_{23}\right)^2_{LL}\right\|}$ | $\left\|\left(\delta^d_{23}\right)_{LR}\right\|$ | $\sqrt{\left\|\text{Re}\left(\delta^d_{23}\right)_{LL}\left(\delta^d_{23}\right)_{RR}\right\|}$ | $\sqrt{\left\|\text{Re}\left(\delta^d_{13}\right)_{LL}\left(\delta^d_{13}\right)_{RR}\right\|}$ | |
| 0.3 | 0.21 | $1.3 \times 10^{-2}$ | $7.4 \times 10^{-2}$ | $1.6 \times 10^{-2}$ | |
| 1.0 | 0.45 | $1.6 \times 10^{-2}$ | $8.3 \times 10^{-2}$ | $1.8 \times 10^{-2}$ | |
| 4.0 | 1 | $3.0 \times 10^{-2}$ | $1.2 \times 10^{-1}$ | $2.5 \times 10^{-2}$ | |

Similarly, we analyse the MI in (1,3) transitions from the $B^0$ $CP$ asymmetries,

$$0.74 \geq a_{J/\psi}\big|_{\text{SUSY}} = \frac{\text{Im}\, M_{12}|_{\text{SUSY}}}{|M_{12}|_{\text{SM}}} \simeq \frac{\alpha_s^2}{\alpha_W^2}\,\frac{M_W^2}{M_{\text{SUSY}}^2}\,\frac{\text{Im}\left\{(\delta^d_{13})^2_{LL}\right\}}{(V_{tb}V_{td}^*)^2\,\frac{m_t^2}{M_W^2}}$$
$$\simeq 12.5 \times 0.026 \times \frac{\text{Im}\left\{(\delta^d_{13})^2_{LL}\right\}}{3 \times 10^{-4}} \Rightarrow \sqrt{\text{Im}\left\{(\delta^d_{13})^2_{LL}\right\}} \leq 0.0\,,3 \qquad (5.143)$$

where again we use $S(x_t) \simeq x_t$ [186]. In this case we have some differences with respect to the situation in the kaon sector. First, in this case the combination of fermion masses and mixing angles in the Standard Model contribution is larger by a factor of 20. So, if we could reach the same experimental sensitivity in $B$ $CP$ experiments as in kaon $CP$ violation experiments we would be able to explore MI a factor $\sqrt{20}$ larger. However, the main difference between both experiments is the different experimental sensitivity to the observables. In kaon physics we are sensitive to signals of $CP$ violation that are $10^3$ times smaller than the kaon mass difference. In $B$ physics we measure $CP$ violation effects of the same order as the $B$ mass difference, and we are sensitive to signals roughly one order of magnitude smaller. This is the main reason why $CP$ experiments in kaon physics are sensitive to much smaller entries in the sfermion mass matrices than experiments in $B$ physics [187]. We can compare these estimates with the actual bounds in Table 5-14 and we see that our simplified calculations are correct as order of magnitude estimates.

However, this does not mean that it is impossible to find signs of supersymmetry in $B$ physics experiments. We have several reasons to expect larger off diagonal entries in the elements associated with $b \to s$ or $b \to d$ transitions than in $s \to d$ transitions. In some grand unified models, large atmospheric neutrino mixing is associated with large right-handed down quark mixing [188, 189, 190]. Sizable mixing in transitions between the third and second generations is also generically expected in flavor models [191]. In fact, we have only weak experimental constraints on $b \to s$ transitions from the $\mathcal{B}(b \to s\gamma$ and $\Delta m_s$, as shown in Table 5-14. So, the question is now, are large SUSY effects possible in $B$ transitions?.

### FCNC in GUT supersymmetry

In a SUSY GUT, quarks and leptons are in the same multiplet. As long as the scale associated with the transmission of SUSY breaking to the visible sector is larger than the GUT breaking scale, the quark-lepton unification also seeps into the SUSY breaking soft sector, leading to squark-slepton mass-squared unification [192]. The exact relations between the mass matrices depend on the choice of the GUT gauge group. For instance, in SU(5) $(\Delta^d_{ij})_{RR}$ and $(\Delta^l_{ij})_{LL}$ are equal; in SO(10) all $\Delta_{ij}$ are equal at $M_{GUT}$ implying strong correlations within FCNCs at that scale that can have significant implications on flavor phenomenology.

To be specific, we concentrate on SUSY SU(5), with soft terms generated above $M_{GUT}$. We assume generic flavor-violating entries to be present in the sfermion matrices at the GUT scale [4]. The part of the superpotential relevant for quarks and charged lepton masses can be written as

$$W_{\mathrm{SU}(5)} = h^u_{ij}\, T_i\, T_j\, H + h^d_{ij} T_i\, \overline{F}_j\, \overline{H} + \mu\, H\, \overline{H}, \tag{5.144}$$

where we have used the standard notation, with $T$ transforming as a $10$ and $\overline{F}$ as a $\overline{5}$ under SU(5). The corresponding SU(5) invariant soft potential has now the form:

$$-\mathcal{L}_{soft} = m^2_{T_{ij}} \tilde{T}_i^\dagger \tilde{T}_j + m^2_{\overline{F}} \tilde{\overline{F}}_i^\dagger \tilde{\overline{F}}_j + m^2_H H^\dagger H + m^2_{\overline{H}} \overline{H}^\dagger \overline{H} + A^u_{ij} T_i T_j H + A^d_{ij} T_i \overline{F}_j \overline{H} + B\mu H \overline{H}\,. \tag{5.145}$$

Rewriting this in terms of the Standard Model representations, we have

$$-\mathcal{L}_{soft} = m^2_{Q_{ij}} \tilde{Q}_i^\dagger \tilde{Q}_j + m^2_{u^c_{ij}} \tilde{u^c}^\star_i \tilde{u^c}_j + m^2_{e^c_{ij}} \tilde{e^c}^\star_i \tilde{e^c}_j + m^2_{d^c_{ij}} \tilde{d^c}^\star_i \tilde{d^c}_j + m^2_{L_{ij}} \tilde{L}_i^\dagger \tilde{L}_j +$$
$$m^2_{H_1} H_1^\dagger H_1 + m^2_{H_2} H_2^\dagger H_2 + A^u_{ij}\, \tilde{Q}_i \tilde{u^c}_j H_2 + A^d_{ij}\, \tilde{Q}_i \tilde{d^c}_j H_1 + A^e_{ij}\, \tilde{L}_i \tilde{e^c}_j H_1 + \ldots \tag{5.146}$$

$$m^2_Q = m^2_{\tilde{e^c}} = m^2_{\tilde{u^c}} = m^2_T, \qquad m^2_{\tilde{d^c}} = m^2_L = m^2_{\overline{F}}, \qquad A^e_{ij} = A^d_{ji}\,. \tag{5.147}$$

Eqs. (5.147) are matrices in flavor space. These equations lead to relations within the slepton and squark flavor-violating off-diagonal entries $\Delta_{ij}$. These are:

$$(\Delta^u_{ij})_L = (\Delta^u_{ij})_R = (\Delta^d_{ij})_L = (\Delta^l_{ij})_R, \quad (\Delta^d_{ij})_R = (\Delta^l_{ij})_L, \quad (\Delta^d_{ij})_{LR} = (\Delta^l_{ji})_{LR} = (\Delta^l_{ij})^\star_{RL}\,. \tag{5.148}$$

These relations are exact at $M_{GUT}$; however, after SU(5) breaking, quarks and leptons suffer different renormalization effects and are thus altered at $M_W$. It is easy to see from the RG equations that off-diagonal elements in the squark mass matrices in the first two of Eqs. (5.148) are approximately not renormalized due to the smallness of CKM mixing angles and that the sleptonic entries in them are left unchanged (in the absence of right-handed neutrinos). On the other hand, the last equation receives corrections due to the different nature of the RG scaling of the $LR$ term ($A$-parameter). This correction can be roughly approximated as proportional to the corresponding fermion masses. Taking this into consideration, we can now rewrite the Eqs. (5.148) at the weak scale,

$$(\delta^d_{ij})_{RR} \;\approx\; \frac{m^2_L}{m^2_{d^c}} (\delta^l_{ij})_{LL}, \qquad (\delta^{u,d}_{ij})_{LL} \;\approx\; \frac{m^2_{e^c}}{m^2_Q} (\delta^l_{ij})_{RR},$$
$$(\delta^u_{ij})_{RR} \;\approx\; \frac{m^2_{e^c}}{m^2_{u^c}} (\delta^l_{ij})_{RR}, \qquad (\delta^d_{ij})_{LR} \;\approx\; \frac{m^2_{L_{avg}}}{m^2_{Q_{avg}}}\, \frac{m_b}{m_\tau}\, (\delta^l_{ij})^\star_{RL}, \tag{5.149}$$

where $m^2_{L_{avg}}$ ($m^2_{Q_{avg}}$ ) are given by the geometric average of left- and right-handed slepton (down-squark) masses $\sqrt{m^2_L\, m^2_{e^c}}$ $\left(\sqrt{m^2_Q\, m^2_{d^c}}\,\right)$, all defined at the weak scale.

To account for neutrino masses, we can use the seesaw mechanism by adding singlet right-handed neutrinos. In their presence, additional couplings occur in Eqs. (5.144 - 5.148) at the high scale which affect the RG evolution of slepton matrices. To understand the effect of these new couplings, one can envisage two scenarios [193]: (a) small couplings and/or small mixing in the neutrino Dirac Yukawa matrix, (b) large couplings and large mixing in the neutrino sector. In case (a), the effect on slepton mass matrices due to neutrino Dirac Yukawa couplings is very small and the above relations Eqs. (5.149) still hold. In case (b), however, large RG effects can significantly modify the slepton doublet flavor structure while keeping the squark sector and right handed charged slepton matrices essentially unmodified, thus breaking the GUT-symmetric relations. Even in this case, barring accidental cancellations among the mass insertions already present at $M_{GUT}$ and the radiatively generated mass insertions between $M_{GUT}$ and $M_{\nu_R}$, there exists an upper bound on the down quark $\delta$ parameters of the form:

$$|(\delta^d_{ij})_{RR}| \quad\leq\quad \frac{m^2_L}{m^2_{d^c}} |(\delta^l_{ij})_{LL}|\,, \tag{5.150}$$

[4]Note that even assuming complete universality of the soft breaking terms at $M_{Planck}$, as in mSUGRA, the RG effects on $M_{GUT}$ will induce flavor off-diagonal entries at the GUT scale [194].

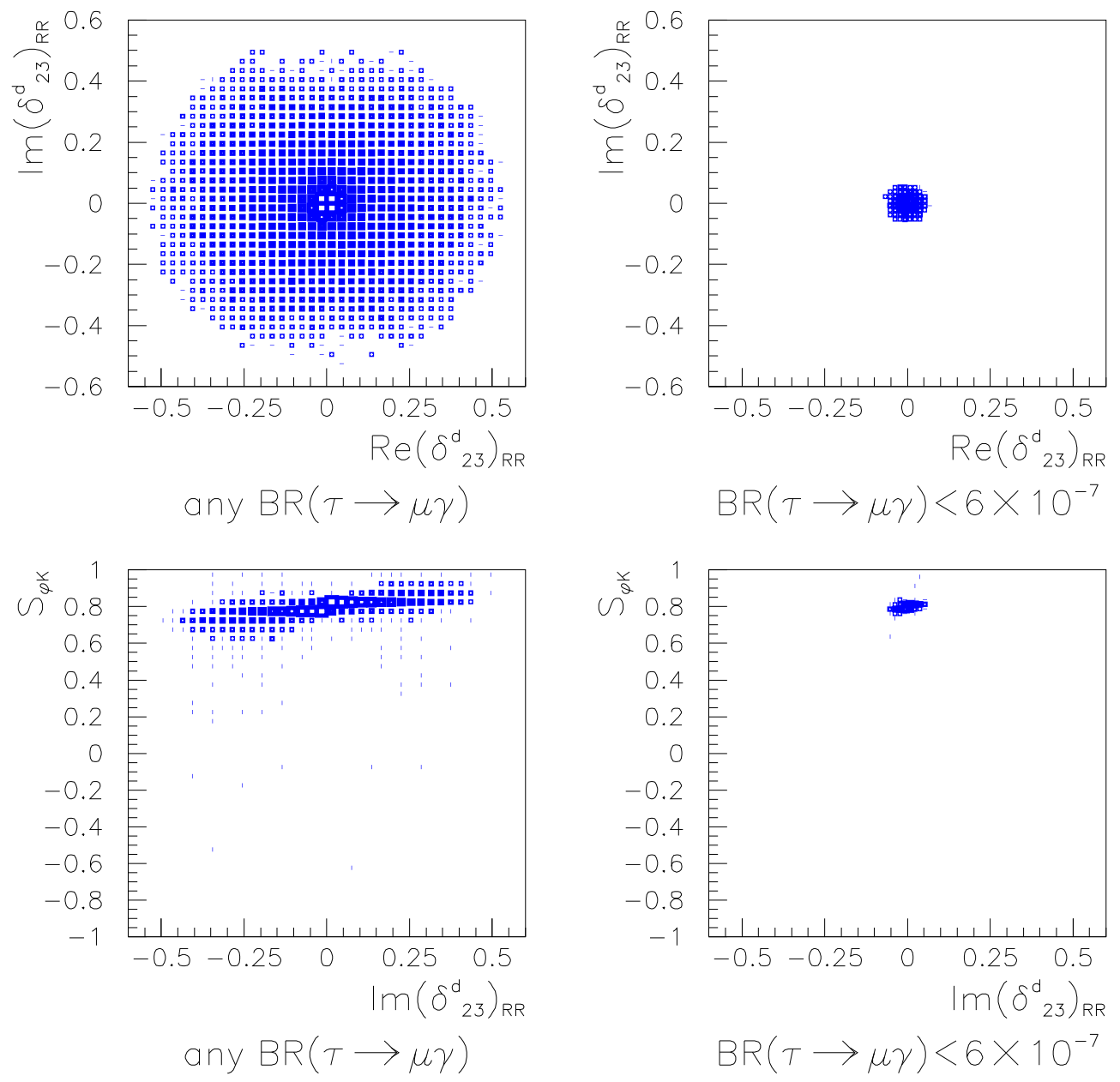

**Figure 5-35.** *Allowed regions in the Re$(\delta^d_{23})_{RR}$–Im$(\delta^d_{23})_{RR}$ plane and in the $S_{K\phi}$–Im$(\delta^d_{23})_{RR}$ plane. Constraints from $B \to X_s\gamma$, $BR(B \to X_s\ell^+\ell^-)$, and the lower bound on $\Delta m_s$ have been used.*

while the last three equations in Eq. (5.149) remain valid in this case.

The relations (5.149, 5.150) predict links between lepton and quark flavor-changing transitions at the weak scale. For example, we see that $\mu \to e\,\gamma$ can be related to $K^0 - \overline{K}^0$ mixing and to $D^0 - \overline{D}^0$ mixing. Similarly, one can expect correlations between $\tau \to e\,\gamma$ and $B_d - \overline{B}_d$ mixing, as well as between $\tau \to \mu\,\gamma$ and $b \to s$ transitions such as $B \to \phi K^0_S$.

To show the impact of these relations, let us assume that all the flavor diagonal sfermion masses are approximately universal at the GUT scale, with $m^2_{T} = m^2_{\tilde{F}} = m^2_H = m^2_{\tilde{H}} = m^2_0$, with flavor off-diagonal entries $m^2_{\tilde{f}} = m^2_0\mathbf{1} + \mathbf{\Delta^f_{ij}}$, with $|\Delta^f_{ij}| \leq m^2_0$. $\Delta^f_{ij}$ can be present either through the running from the Planck scale to the GUT scale [194] or through some flavor non-universality originally present [195, 196, 197]. All gaugino masses are unified to $M_{1/2}$ at $M_{GUT}$. For a given set of initial conditions ($M_{1/2}$, $m^2_0$, $A_0$, $\Delta_{ij}$, $\tan\beta$) we obtain the full spectrum at $M_W$ with the requirement of radiative symmetry breaking. We then apply limits from direct searches on SUSY particles. Finally, we calculate the contributions of different $\delta_{23}$ parameters to both leptonic and hadronic processes, considering the region in the $(m_0, M_{1/2})$ plane corresponding to a relatively light sparticle spectrum, with squark masses of roughly 350–550 GeV and slepton masses of about 150–400 GeV.

$b \to s$ transitions have recently received much attention, as it has been shown that the discrepancy with Standard Model expectations in the measurements of $A_{CP}(B \to \phi K^0_S)$ can be attributed to the presence of large neutrino mixing within SO(10) models [188, 190, 189]. Subsequently, a detailed analysis has been presented [198, 199] within the context of the MSSM. It has been shown that, for squark and gluino masses around $350$ GeV, the presence of a $\mathcal{O}(1)$ $(\delta^d_{23})_{LL,RR}$ could lead to significant discrepancies from the Standard Model expectations. Similar statements hold for a $\mathcal{O}(10^{-2})$ $LR$ or $RL$ MI. Here, we study the impact of LFV bounds on these $\delta$ parameters and subsequently the effect on $B$ physics observables. In Table 5-15, we present upper bounds on $\delta^d_{23}$ within the above mass ranges for three values of the limits on $\mathcal{B}(\tau \to \mu\,\gamma)$. There are no bounds on $(\delta^d_{23})_{LL}$ because, as is well known [200, 201, 202], large values of $(\delta^l_{ij})_{RR}$ are still allowed, due to the cancellations of bino and higgsino contributions in the decay amplitude.

**Table 5-15.** *Bounds on $(\delta^d_{23})$ from $\mathcal{B}(\tau \to \mu\,\gamma)$ for three values of the branching ratios for $\tan\beta$ = 10.*

| Type | $< 6 \cdot 10^{-7}$ | $< 1 \cdot 10^{-7}$ | $< 1 \cdot 10^{-8}$ |
|---|---|---|---|
| $LL$ | - | - | - |
| $RR$ | 0.070 | 0.030 | 0.010 |
| $RL$ | 0.080 | 0.035 | 0.010 |
| $LR$ | 0.080 | 0.035 | 0.010 |

At present, the constraints coming from $B$ physics are stronger than those obtained for the lepton sector in the cases of $(\delta^d_{23})_{LL,LR,RL}$. Therefore no impact on $B$ phenomenology is expected even if the present bound on $\mathcal{B}(\tau \to \mu\,\gamma)$ were pushed down to $1 \times 10^{-7}$. On the contrary, the bound on $(\delta^d_{23})_{RR}$ induced by $\mathcal{B}(\tau \to \mu\,\gamma)$ is already at present much stronger than the bounds from hadronic processes, reducing considerably the room left for SUSY effects in $B$ decays. To illustrate this point in detail, we repeat the analysis of Ref. [199] including the bounds coming from lepton processes. We therefore compute at the NLO branching ratios and $CP$ asymmetries for $B \to X_s\gamma$ and $B \to \phi K^0_S$, $\mathcal{B}(B \to X_s\ell^+\ell^-)$ and $\Delta m_s$ (see Ref. [199] for details). In the first row of Fig. 5-35, we plot the probability density in the Re$(\delta^d_{23})_{RR}$–Im$(\delta^d_{23})_{RR}$ plane for different upper bounds on $\mathcal{B}(\tau \to \mu\,\gamma)$. Note that making use of Eq. (5.149) with $|(\delta^l_{23})_{LL}| < 1$, implies $|(\delta^d_{23})_{RR}| \lesssim 0.5$ as the ratio $(m^2_L/m^2_{d^c})$ varies roughly between $(0.2 - 0.5)$ at the weak scale, for the chosen high scale boundary conditions. The effect on $(\delta^d_{23})_{RR}$ of the upper bound on $\mathcal{B}(\tau \to \mu\,\gamma)$ is dramatic already with the present experimental value. Correspondingly, as can be seen from the second row of Fig. 5-35, the possibility of large deviations from the Standard Model in the coefficient $S_{\phi K}$ of the sine term in the time-dependent $A_{CP}(B \to \phi K^0_S)$ is excluded in the $RR$ case. Hence, we conclude that in SUSY GUTs the most likely possibility to strongly depart from the Standard Model expectations for $S_{\phi K}$ relies on a sizable contribution from $(\delta^d_{23})_{LL}$ or $(\delta^d_{23})_{LR,RL}$, as long as they are small enough to be within the severe limits imposed by $\mathcal{B}(B \to X_s\gamma)$ [199]. These results would not change significantly if one started with a SO(10) theory instead of a SU(5) theory. The relation in Eq. (5.149) would be still valid however with the additional constraint: $(\delta^d_{ij})_{RR} = m^2_Q/m^2_d(\delta^d_{ij})_{LL} = m^2_L/m^2_d(\delta^l_{ij})_{LL}$. The results of our analysis are therefore valid also for SO(10), although stronger correlations are generally expected.

Exploiting the Grand Unified structure of the theory, we can obtain similar bounds on other $\delta^d_{ij}$ parameters. For example, considering the first two generations, the bound on $\delta^d_{12}$ from $\mathcal{B}(\mu \to e\,\gamma)$ can, in many cases, compete with the bound from $\Delta M_K$ [185]. Similar comparisons can be made for the $\delta^d_{13}$ from limits on $\mathcal{B}(\tau \to e\,\gamma)$ and $B^0_d - \overline{B}^0_d$ mixing.

**Example: SU(3) flavor theory**

Finally as an example, we discuss here the main features of a recent supersymmetric SU(3) flavor model [203, 204, 205] that successfully reproduces quark and lepton masses and mixing angles, and predicts the structure of the sfermion mass matrices. Under this SU(3) family symmetry, all left-handed fermions ($\psi_i$ and $\psi^c_i$) are triplets. To allow for the spontaneous symmetry breaking of SU(3), it is necessary to add several new scalar fields that are either triplets ($\overline{\theta}_3$, $\overline{\theta}_{23}$, $\overline{\theta}_2$) or anti-triplets ($\theta_3$, $\theta_{23}$). We assume that SU(3)$_F$ is broken in two steps. The first step occurs when $\theta_3$ gets a large vev breaking SU(3) to SU(2). Subsequently, a smaller vev of $\theta_{23}$ breaks the remaining symmetry. After this breaking, we obtain the effective Yukawa couplings at low energies through the Froggatt-Nielsen mechanism. In this theory, the Yukawa superpotential is

$$W_Y = H\psi_i\psi^c_j\left[\theta^i_3\theta^j_3 + \theta^i_{23}\theta^j_{23}\Sigma + \left(\epsilon^{ikl}\overline{\theta}_{23,k}\overline{\theta}_{3,l}\theta^j_{23}\left(\theta_{23}\overline{\theta_3}\right) + (i \leftrightarrow j)\right)\right], \quad (5.151)$$

and so the Yukawa textures are

$$Y^f \propto \begin{pmatrix} 0 & \varepsilon^3 e^{i\delta} X_1 & \varepsilon^3 e^{i(\delta+\beta_3)} X_2 \\ \ldots & \varepsilon^2 \frac{\Sigma}{|a_3|^2} & \varepsilon^2 e^{i\beta_3} \frac{\Sigma}{|a_3|^2} \\ \ldots & \ldots & e^{2i\chi} \end{pmatrix}, \tag{5.152}$$

where $\overline{\varepsilon} = \langle \theta_{23} \rangle / M \simeq 0.15$ with $M$ a mediator mass in terms of dimension greater than three, and the $X_a$ are $\mathcal{O}(1)$ coefficients. In the same way, the supergravity Kähler potential receives new contributions after SU(3)$_F$ breaking,

$$K = \psi_i^\dagger \psi_j \Big( \delta^{ij}(c_0 + d_0 X X^\dagger) + \frac{1}{M^2}[\theta_3^{i\dagger}\theta_3^j (c_1 + d_1 X X^\dagger) + \theta_{23}^{i\dagger}\theta_{23}^j (c_2 + d_2 X X^\dagger)] + \\ + (\epsilon^{ikl}\overline{\theta}_{3,k}\overline{\theta}_{23,l})^\dagger (\epsilon^{jmn}\overline{\theta}_{3,m}\overline{\theta}_{23,n})(c_3 + d_3 X X^\dagger)] \Big), \tag{5.153}$$

where $c_i, d_i$ are $\mathcal{O}(1)$ coefficients and we include a field $X$ with non-vanishing F-term. From here we obtain the structure of the sfermion mass matrices [203, 204, 205]. In the basis of diagonal quark Yukawa couplings (SCKM basis) we obtain for the down quarks, suppressing factors $\mathcal{O}(1)$,

$$(M^2_{\tilde{D}_R})^{\rm SCKM} \simeq \begin{pmatrix} 1+\overline{\varepsilon}^3 & -\overline{\varepsilon}^3 e^{-i\omega} & -\overline{\varepsilon}^3 e^{-i\omega} \\ -\overline{\varepsilon}^3 e^{i\omega} & 1+\overline{\varepsilon}^2 & \overline{\varepsilon}^2 \\ -\overline{\varepsilon}^3 e^{i\omega} & \overline{\varepsilon}^2 & 1+\overline{\varepsilon} \end{pmatrix} m_0^2 \,. \tag{5.154}$$

Thus, we can see that in $3 \to 2$ transitions we have off-diagonal entries of order $\overline{\varepsilon}^2$, although these must still be small to have large effects. In the case of lepton flavor violation, the slepton mass matrices are similar to Eq. (5.154) with different $\mathcal{O}(1)$ coefficients. However, the main advantage of leptonic processes is that the MI are not greatly reduced from $M_{GUT}$ to $M_W$. In this case, $(\delta^e_{LL})_{23} \simeq 2 \times 10^{-2}$ (except factors for order 1), contributing to $\tau \to \mu\gamma$ transitions, while the bound in Table 5-14 is only $3 \times 10^{-2}$. Therefore a $\tau \to \mu\gamma$ branching ratio close to the experimental bound is indeed possible.

#### Conclusions

We have introduced the mass insertion formalism and we have applied it to $CP$ violation in $B$ physics. We have seen that Super $B$ Factoriescan explore the flavor structure of the sfermion mass matrices, both in the squark and in the slepton sectors. Supersymmetric Grand Unification predicts links between various leptonic and hadronic FCNC observables. We have quantitatively studied a SU(5) model and the implications for transitions between the second and third generations. We have shown that the present limit on $\mathcal{B}(\tau \to \mu\,\gamma)$ significantly constrains the observability of SUSY in $CP$ violating $B$ decays. In these models, lepton flavor-violating decays may be closer to experimental bounds than quark FCNCs, although these decays measure slightly different flavor parameters. We have also seen that sizable contributions are possible in "realistic" flavor models. Thus, precision measurements in $B$ (and $\tau$) physics are necessary to understand flavor physics.

### 5.3.5 Effective Supersymmetry in $B$ Decays

>– P. Ko –<

#### Introduction

Generic SUSY models suffer from serious SUSY flavor and $CP$ problems, because the squark mass matrices and quark mass matrices need not be simultaneously diagonalizable in the flavor space. Therefore the $\tilde{g} - \tilde{q}_{iA} - q_{jB}$ vertices ($i, j = 1, 2, 3$ are flavor indices, and $A, B = L, R$ denote chiralities) are described by some unitary matrix $W^d_{ij,AB}$ in the down (s)quark sector, which is analogous to the CKM matrix in the Standard Model. Since this coupling has a root in strong interaction and can have $CP$ violating phases, it leads to too large flavor changing neutral current (FCNC) amplitudes through gluino-squark loop as well as too large $\epsilon_K$ and neutron electric dipole moment (EDM), which could easily dominate the Standard Model contributions and the data. Some examples are $K^0\overline{K}^0$ ($\Delta M_K$ and $\epsilon_K$) and $B^0\overline{B}^0$ mixing ($\Delta M_B$ and *CP* asymmetry in $B_d \to J/\psi K_S^0$) and $B \to X_s\gamma$, *etc...* The lepton sector has the same problem through the neutralino-slepton loop, and the most serious constraint comes from $\mathcal{B}(\mu \to e\gamma)$ and electron EDM.

One way out of these SUSY flavor and *CP* problems is to assume that the first and second generation squarks are very heavy ($\gtrsim \mathcal{O}(10)$ TeV) and almost degenerate [206]. The third generation squarks and gauginos should be relatively light ( $\lesssim 1$ TeV) in order that the quantum correction to Higgs mass parameter is still reasonably small. This scenario is called an effective SUSY model, or a decoupling scenario. In effective SUSY models, the $\tilde{b} - \tilde{g}$ loop can still induce a certain amount of flavor and *CP* violation in the quark sector through the mixing matrices $W^d_{ij,AB}$. In addition to the flavor mixing and *CP* violation from $W^d_{ij,AB}$'s, there could be flavor-conserving *CP* violation through the $\mu$ and $A_t$ parameters within the effective SUSY models. Note that this class of models, ignoring the gluino-mediated FCNC, are used in the context of electroweak baryogenesis within SUSY (see, *e.g.*, [207]). Although the phases in $\mu$ and $A_t$ are flavor-conserving, they can affect $K$ and $B$ physics through chargino/stop propagators and mixing angles.

Since there are no well-defined effective SUSY models as there are in gauge mediation or minimal supergravity scenarios, one has to assume that all the soft SUSY breaking terms have arbitrary *CP*-violating phases, as long as they satisfy the decoupling spectra and various experimental constraints. In order to make the analysis easy and transparent, we consider two extreme cases:

- Minimal flavor violation (MFV)
- Gluino-squark dominance in $b \to s(d)$ transition ($\tilde{g}$ dominance)

We will describe typical signatures of each scenario, keeping in mind that reality may involve a combination of these two extreme cases.

### *CP* violation from $\mu$ and $A_t$ phases

Let us first discuss minimal flavor violation models with effective SUSY spectra. In this model, flavor violation comes through the CKM matrix, whereas *CP* violation originates from the $\mu$ and $A_t$ phases, as well as the CKM phase. The one loop electric dipole moment (EDM) constraint is evaded in the effective SUSY model due to the decoupling of the first/second generation sfermions, but there are potentially large two loop contribution to electron/neutron EDM's through Barr-Zee type diagrams in the large $\tan\beta$ region [208]. Imposing this two-loop EDM constraint and direct search limits on Higgs and SUSY particles, we find that [209, 210]

- There are no new phase shifts in $B^0\overline{B}^0$ and $B_s^0\overline{B}_s^0$ mixing: Time-dependent *CP* asymmetries in $B_d \to J/\psi K_S^0$ still measure the CKM angle $\beta = \phi_1$ [Fig. 5-36 (a)]
- $\Delta M_{B_d}$ can be enhanced up to $\sim 80\%$ compared to the Standard Model prediction [Fig. 5-36 (b)]
- Direct *CP* asymmetry in $B \to X_s\gamma$ ($A_{CP}^{b\to s\gamma}$) can be as large as $\pm 15\%$ [see Fig. 5-37]
- $R_{\mu\mu}$ can be as large as 1.8
- $\epsilon_K$ can differ from the Standard Model value by $\sim 40\%$

One can therefore anticipate substantial deviations in certain observables in the $B$ system in SUSY models with minimal flavor violation and complex $\mu$ and $A_t$ parameters. This class of models include electroweak baryogenesis (EWBGEN) within the MSSM and some of its extensions (such as NMSSM), where the chargino and stop sectors are the same as in the MSSM. In the EWBGEN scenario within the MSSM, the current lower limit on the Higgs mass requires a large radiative correction from the stop loop. Since $\tilde{t}_R$ has to be light to have a sufficiently strong 1st order electroweak phase transition, one has to have heavy $\tilde{t}_L$ to induce a large $\Delta m_h^2$. After considering $B \to X_s\gamma$, one expects a very small deviations in $A_{CP}^{b\to s\gamma}$ and $\Delta M_{B_d}$ [211]. However, in some extensions of the MSSM, the tension between $m_h$ and $m_{\tilde{t}_L}$ becomes significantly diluted in EWBGEN scenarios, because there could be tree level contributions to $m_h^2$. Therefore, the predictions made in Refs. [209, 210] will be still valid in EWBGEN scenarios beyond the MSSM.

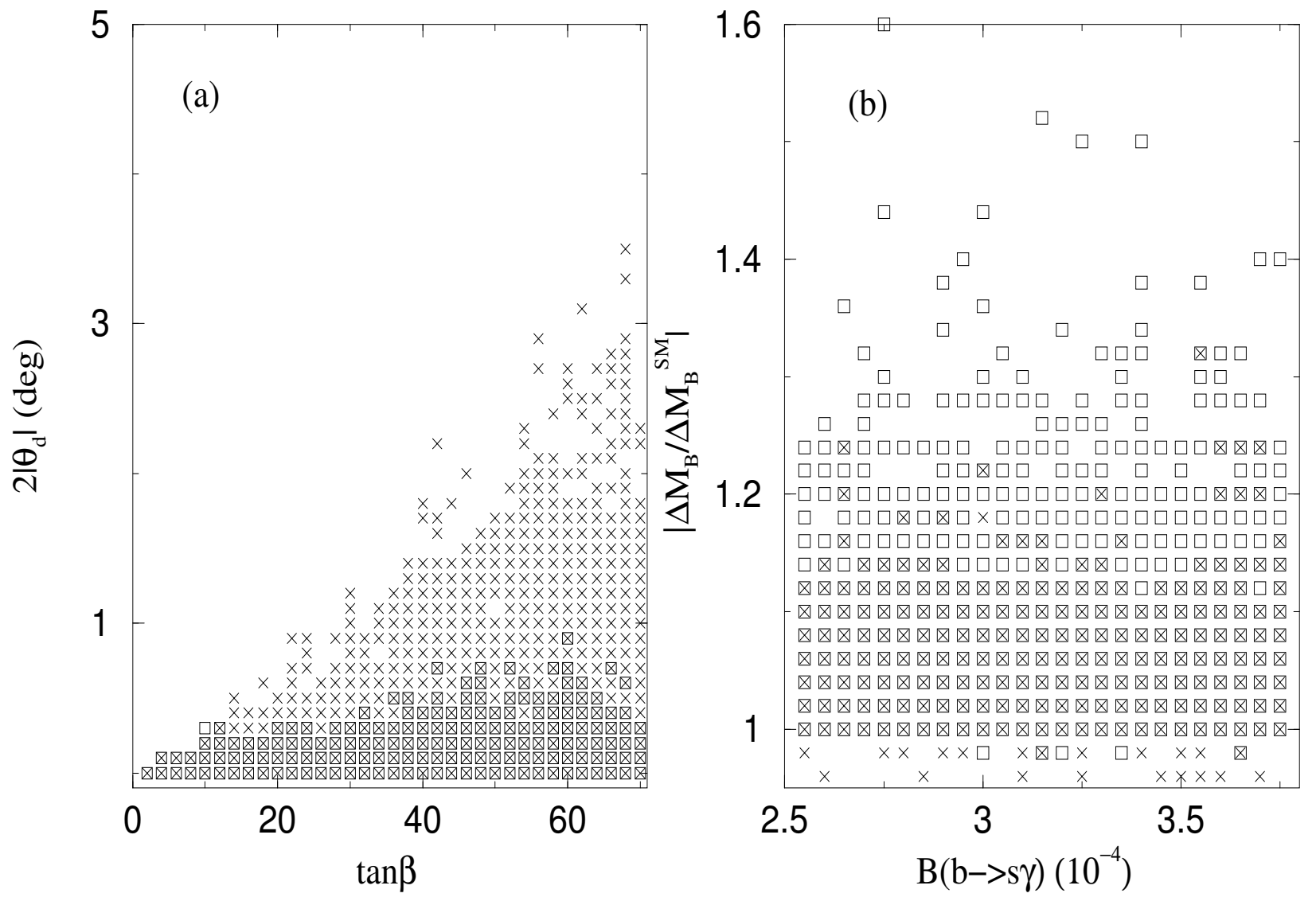

**Figure 5-36.** *Correlations between (a) $\tan\beta$ vs. the new phase shift in the $BzBzb$ mixing, and (b) $\mathcal{B}(B \to X_s\gamma)$ vs. $|\Delta M_{B_d}/\Delta M_{B_d}^{\rm SM}|$. The squares (the crosses) denote those which (do not) satisfy the two-loop EDM constraints.*

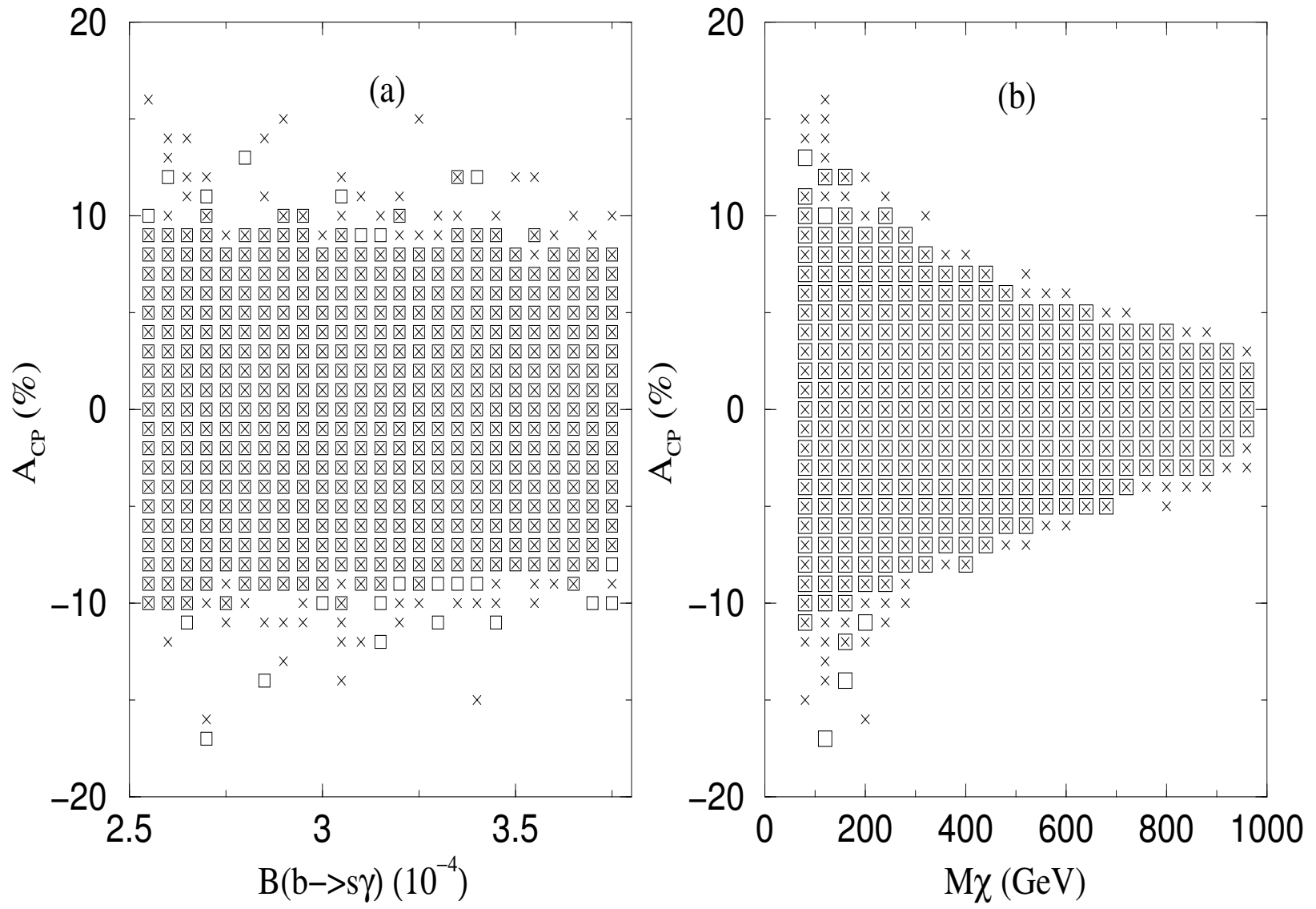

**Figure 5-37.** *Correlations of $A_{CP}^{b\to s\gamma}$ with (a) $\mathcal{B}(B \to X_s\gamma)$ and (b) the lighter chargino mass $M_{\chi^\pm}$. The squares (the crosses) denote those which (do not) satisfy the two-loop EDM constraints.*

Super $B$ Factoriesshould be able to measure $A_{CP}^{b\to s\gamma}$ to higher accuracy, and will impose a strong constraint on a new $CP$-violating phase that could appear in $B \to X_s\gamma$. Also the forward-backward asymmetries in $B \to X_s\ell^+\ell^-$ with $\ell = e$ or $\mu$ are equally important probes of new $CP$-violating phases, and important observables to be measured at Super $B$ Factories, for which LHC$b$ or $B$TeV cannot compete.

### $CP$ violation from gluino-squark loops

In effective SUSY scenarios, it is possible that the gluino-mediated $b \to s$ transition is dominant over other SUSY contributions. Cohen *et al.* have described qualitative features of such scenarios in $B$ physics [212]; a more quantitative analysis was presented by other groups. In Ref. [213], effects of possible new $CP$-violating phases on $B \to X_s\gamma$ and $B \to X_s\ell^+\ell^-$ were considered both in a model-independent manner, and in gluino mediation dominance scenario. In effective SUSY models, $A_{CP}^{b\to s\gamma}$ can be as large as $\pm 10\%$, if the third generation squarks are light enough $m_{\tilde{b}} \simeq (100-200)$ GeV (see Fig. 5-38), whereas $B \to X_s\ell^+\ell^-$ is almost the same as the Standard Model prediction [213].

### How to distinguish the $\mu$ or $A_t$ phase from the $\delta_{23}^d$ phase

Should we find deviations in $\sin 2\beta_{\phi K_S^0}$ or $A_{CP}^{b\to s\gamma}$, it will be very important to figure out the origin of new $CP$-violating phases. In an effective SUSY context, one has complex $A_t$, $\mu$ or $(\delta_{AB}^d)_{23}$ (with $A, B = L, R$). The effects of these new complex parameters on some oservables in the $B$ system are shown in Table 5-16. The only process which is not directly affected by gluino-mediated FCNC is $B \to X_s\nu\overline{\nu}$. All the other observables are basically affected by both the phases of $\mu, A_t$ and $(\delta_{AB}^d)_{23}$ parameters. In fact, this feature is not specific to the effective SUSY scenarios, but is rather generic within SUSY models. Therefore the measurement of $B \to X_s\nu\overline{\nu}$ branching ratio will play a crucial role to tell if the observed $CP$-violating phenomena comes from the $\mu$ or $A_t$ phase or $(\delta_{AB}^d)_{23}$. This can be done only at a $e^+e^-$ Super $B$ Factory, and not at hadron $B$ factories.

**Table 5-16.** *Possible effects of the phase of $\mu$ or $A_t$ for moderate $\tan\beta$ ($3 \le \tan\beta \le 6$) and the phase of $\delta_{i3}^d$ (with $i = 1, 2$) to various observables in the $B$ systems, and possibilities to probe these at various experiments*

| Observables | Arg ($\mu$) or Arg ($A_t$) | Arg ($\delta_{i3}^d$) | Super $B$ Factory | LHC$b$ |
|---|---|---|---|---|
| $\Delta m_d$ | Y | Y | O | O |
| $\sin 2\beta$ | N | Y | O | O |
| $\Delta m_s$ | Y | Y | X | O |
| $\sin 2\beta_s$ | N | Y | X | O |
| $A_{\rm CP}^{b\to s\gamma}$ | Y | Y | O | X |
| $A_{\rm CP}^{b\to d\gamma}$ | Y | Y | O | X |
| $B \to X_s\ell^+\ell^-$ | Y | Y | O | X |
| $B \to X_s\nu\overline{\nu}$ | Y | N | O | X |
| $B_d \to \phi K_S^0$ | Y | Y | O | O |

### Conclusion

We showed that there could be large deviations in certain observables in the $B$ system, which can be studied only in Super $B$ Factories. The most prominent deviations are the branching ratio of $B \to X_s\nu\overline{\nu}$, $A_{CP}^{b\to s\gamma}$, the forward-backward asymmetry in $B \to X_s\ell^+\ell^-$, and the branching ratio of $B \to X_d\gamma$ and $CP$ violation therein, without any conflict with our current understanding based on the CKM paradigm. These observables could reveal new sources of $CP$ and flavor violations that could originate from SUSY models, including effective SUSY models, and should be an important topic at Super $B$ Factories.

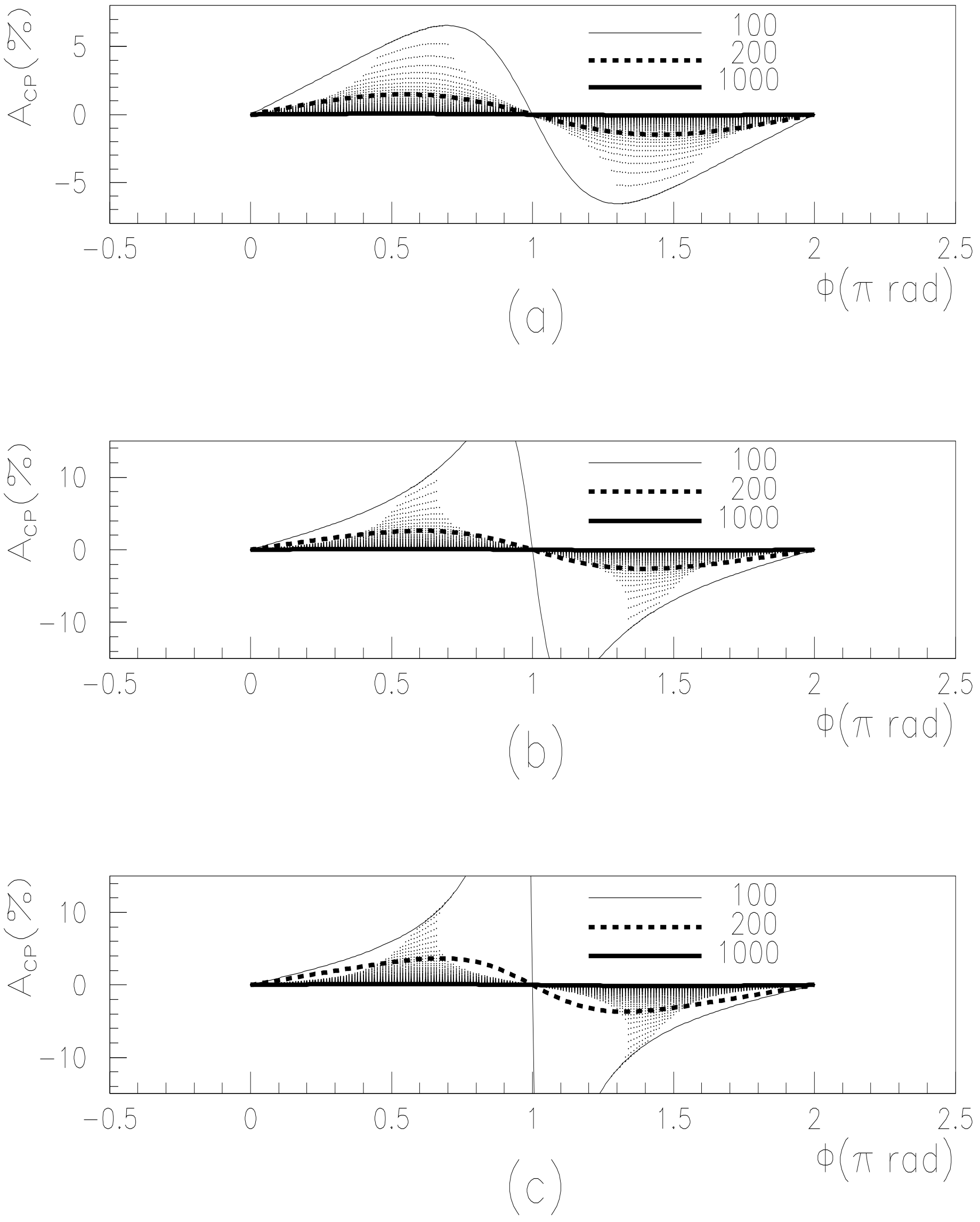

**Figure 5-38.** *$A_{\rm CP}^{b\to s\gamma}$ contours in the $(\tilde{m},\phi)$ plane for (a) $x=0.3$, (b) $x=1$ and (c) $x=3$ in the $(LL)$ insertion case using the vertex mixing method with $x=m_{\tilde{g}}^2/m_{\tilde{b}}^2$.*

I am grateful to S. Baek, Y. G. Kim and J. S. Lee for their collaboration on the work presented here.

### 5.3.6 Supersymmetric Flavor Violation: Higgs–Quark Interactions

>– D. A. Demir –<

The primary goal of the existing and planned hadron colliders and the $B$ meson factories is to test the Standard Model and determine possible New Physics effects on its least understood sectors: breakdown of $CP$, flavor and gauge symmetries. In the standard picture, both $CP$ and flavor violations are restricted to arise from the CKM matrix, and the gauge symmetry breaking is accomplished by introducing the Higgs field. However, the Higgs sector is badly behaved at quantum level; its stabilization against quadratic divergences requires supersymmetry (SUSY) or some other extension of the Standard Model. The soft breaking sector of the minimal SUSY model (MSSM) accommodates novel sources for $CP$ and flavor violations [214, 215] with testable signatures at present (PEP-II, KEK-$B$) or future (Super $B$ Factory or LHC) experiments. The Yukawa couplings, which are central to Higgs searches at the LHC, differ from all other couplings in one aspect: the radiative corrections from sparticle loops depend only on the ratio of the soft masses, and hence they do not decouple even if the SUSY-breaking scale lies far above the weak scale. In this sense, a non-standard hierarchy and texture of the Higgs-quark couplings, once confirmed experimentally, might provide direct access to sparticles, irrespective of how heavy they might be (though not too large to regenerate the gauge hierarchy problem). This section will summarize the results of recent work [216] that discusses the radiative corrections to Yukawa couplings from sparticle loops and their impact on flavor-changing neutral current (FCNC) observables and Higgs phenomenology.

The soft breaking sector mixes sfermions of different flavor via the off–diagonal entries of the sfermion mass–squared matrices. The $LR$ and $RL$ blocks are generated after the electroweak breaking with the maximal size $\mathcal{O}(m_t M_{SUSY})$, and their flavor-mixing potential is dictated by the Yukawa couplings $\mathbf{Y}_{u,d}$ and by the trilinear coupling matrices $\mathbf{Y}^A_{u,d}$ with $\left(\mathbf{Y}^A_{u,d}\right)_{ij} = (\mathbf{Y}_{u,d})_{ij}\,(A_{u,d})_{ij}$ where $A_{u,d}$ are not necessarily unitary, so that even their diagonal entries contribute to $CP$-violating observables. The flavor mixings in the $LL$ and $RR$ sectors, however, are insensitive to electroweak breaking; they are of pure SUSY origin. Clearly, $CP$ violation in the $LL$ and $RR$ sectors is restricted to the flavor-violating entries, due to hermiticity. In discussing the FCNC transitions, it is useful to work with the mass insertions [215]

$$\left(\delta^{d,u}_{ij}\right)_{RR,LL} = \frac{\left(M^2_{D,U}\right)^{ij}_{RR,LL}}{\overline{M^2_{D,U}}}\,, \tag{5.155}$$

where $\left(M^2_D\right)_{RR,LL}$ have the generic form

$$\left(M^2_D\right)_{LL} = \begin{pmatrix} M^2_{\tilde d_L} & M^2_{\tilde d_L \tilde s_L} & M^2_{\tilde d_L \tilde b_L} \\ M^2_{\tilde s_L \tilde d_L} & M^2_{\tilde s_L} & M^2_{\tilde s_L \tilde b_L} \\ M^2_{\tilde b_L \tilde d_L} & M^2_{\tilde b_L \tilde s_L} & M^2_{\tilde b_L} \end{pmatrix} \quad , \quad \left(M^2_D\right)_{RR} = \begin{pmatrix} M^2_{\tilde d_R} & M^2_{\tilde d_R \tilde s_R} & M^2_{\tilde d_R \tilde b_R} \\ M^2_{\tilde s_R \tilde d_R} & M^2_{\tilde s_R} & M^2_{\tilde s_R \tilde b_R} \\ M^2_{\tilde b_R \tilde d_R} & M^2_{\tilde b_R \tilde s_R} & M^2_{\tilde b_R} \end{pmatrix} \tag{5.156}$$

in the bases $\{\tilde d_L, \tilde s_L, \tilde b_L\}$ and $\{\tilde d_R, \tilde s_R, \tilde b_R\}$, respectively. The same structure repeats for the up sector. The mass insertions are defined in terms of $\overline{M^2_{D,U}}$ which stand for the mean of diagonal entries. The textures of the $LL$ and $RR$ blocks are dictated by the SUSY breaking pattern. In minimal SUGRA and its nonuniversal variants with $CP$ violation, for instance, the size and structure of flavor and $CP$ violation are dictated by the CKM matrix [214]. On the other hand, in SUSY GUTs with Yukawa unification *e.g.*, SO(10), implementation of the see–saw mechanism for neutrino masses implies sizable flavor violation in the $RR$ block, given the large mixing observed in atmospheric neutrino data [217].

The effective theory below the SUSY breaking scale $M_{SUSY}$ consists of a modified Higgs sector; in particular, the tree level Yukawa couplings receive sizable corrections from sparticle loops [216]. For instance, the $d$ quark Yukawa coupling relates to the physical Yukawas via

$$h_d = \frac{g_2 \overline{m}_d}{\sqrt{2} M_W \cos\beta} \frac{1 - a^2 \left(\delta^d_{23}\right)_{LR} \left(\delta^d_{32}\right)_{LR} - a A_{12} \frac{\overline{m}_s}{\overline{m}_d} - a A_{13} \frac{\overline{m}_b}{\overline{m}_d}}{1 - a^2 A_2 - a^3 A_3}, \qquad (5.157)$$

where $a = \epsilon \tan\beta / (1 + \epsilon \tan\beta)$, $A_{12} = \left[\left(\delta^d_{12}\right)_{LR} - a \left(\delta^d_{13}\right)_{LR} \left(\delta^d_{32}\right)_{LR}\right]$, $A_{13} = A_{12}(2 \leftrightarrow 3)$, $A_2 = \left|\left(\delta^d_{12}\right)_{LR}\right|^2 + \left|\left(\delta^d_{13}\right)_{LR}\right|^2 + \left|\left(\delta^d_{23}\right)_{LR}\right|^2$ and $A_3 = \left(\delta^d_{12}\right)_{LR} \left(\delta^d_{23}\right)_{LR} \left(\delta^d_{31}\right)_{LR} + \text{h.c.}$ Here $\epsilon = (\alpha_s/3\pi) e^{-i(\theta_\mu + \theta_g)}$, and

$$\left(\delta^d_{ij}\right)_{LR} = \frac{1}{6} \left(\delta^d_{ij}\right)_{RR} \left(\delta^d_{ji}\right)_{LL}, \qquad (5.158)$$

with the SUSY $CP$-odd phases defined as $\theta_g = \text{Arg}[M_g]$, $\theta_\mu = \text{Arg}[\mu]$, $\theta^d_{ij} = \text{Arg}[(A_d)_{ij}$, *etc.*As (5.157) suggests, in contrast to the minimal flavor violation (MFV) scheme, the Yukawa couplings acquire large corrections from those of the heavier ones. Indeed, the radiative corrections to $h_d/\overline{h}_d$, $h_s/\overline{h}_s$, $h_u/\overline{h}_u$ and $h_c/\overline{h}_c$ involve, respectively, the large factors $\overline{m}_b/\overline{m}_d \sim (\tan\beta)^2_{max}$, $\overline{m}_b/\overline{m}_s \sim (\tan\beta)_{max}$, $\overline{m}_t/\overline{m}_u \sim (\tan\beta)^3_{max}$, and $\overline{m}_t/\overline{m}_c \sim (\tan\beta)^2_{max}$ with $(\tan\beta)_{max} \lesssim \overline{m}_t/\overline{m}_b$. Unlike the light quarks, the top and bottom Yukawas remain close to their MFV values, to a good approximation. Therefore, the SUSY flavor-violation sources mainly influence the light quark sector, thereby modifying several processes in which they participate. These corrections are important even at low $\tan\beta$. As an example, consider $\left(\delta^d_{13}\right)_{LR} \sim 10^{-2}$ for which $h_d/h_d^{MFV} \simeq 0.02(2.11), -2.3(6.6), -4.6(17.7)$ for $\tan\beta = 5, 20, 40$ at $\theta_\mu + \theta_g \to 0(\pi)$. Note that the Yukawas are enhanced especially for $\theta_\mu + \theta_g \to \pi$, which is the point preferred by Yukawa–unified models such as SO(10). In general, as $\tan\beta \to (\tan\beta)_{max}$ the Yukawa couplings of down and strange quarks become amproximately degenerate with the bottom Yukawa for $\left(\delta^d_{13,23}\right)_{LR} \sim 0.1$ and $\theta_\mu + \theta_g \to \pi$. There is no $\tan\beta$ enhancement for up quark sector but still the large ratio $\overline{m}_t/\overline{m}_u$ sizably folds $h_u$ compared to its Standard Model value: $h_u \simeq 0.6\, e^{i(\theta^u_{11} - \theta_g)}\, \overline{h}_c$ for $(\delta^u_{13})_{LR} \sim 0.1$.

The SUSY flavor violation influences the Higgs-quark interactions by ($i$) modifying $H^a \overline{q} q$ couplings via sizable changes in Yukawa couplings as in (5.157), and by ($ii$) inducing large flavor changing couplings $H^a \overline{q} q'$:

$$\begin{aligned} &\frac{\overline{h}_{d^i}^{\,SM}}{\sqrt{2}} \left[\frac{h^i_d}{\overline{h}^i_d} \tan\beta\, C^d_a + \left(\frac{h^i_d}{\overline{h}^i_d} - 1\right) \left(e^{i(\theta^d_{ii} + \theta_\mu)}\, C^d_a - C^{u\star}_a\right)\right] \overline{d}^i_R\, d^i_L\, H_a \\ &+ \frac{\overline{h}_{d^i}^{SM}}{3\sqrt{2}} \epsilon \tan\beta \left[\frac{h^i_d}{\overline{h}_{d^i}} \left(\delta^d_{ij}\right)_{LL} + \frac{h^j_d}{\overline{h}^i_d} \left(\delta^d_{ij}\right)_{RR}\right] \left(\tan\beta\, C^d_a - C^{u\star}_a\right) \overline{d}^i_R\, d^j_L\, H_a \end{aligned} \qquad (5.159)$$

where $C^d_a \equiv \{-\sin\alpha, \cos\alpha, i\sin\beta, -i\cos\beta\}$ and $C^u_a \equiv \{\cos\alpha, \sin\alpha, i\cos\beta, i\sin\beta\}$ in the basis $H_a \equiv \{h, H, A, G\}$ if the $CP$ violation effects in the Higgs sector, which can be quite sizable [218] and add additional $CP$-odd phases [219] to Higgs-quark interactions, are neglected . Similar structures also hold for the up sector. The interactions contained in (5.159) have important implications for both FCNC transitions and Higgs decay modes. The FCNC processes are contributed by both the sparticle loops (*e.g.*, the gluino-squark box diagram for $K^0 \overline{K}^0$ mixing) and Higgs exchange amplitudes. The constraints on various mass insertions can be satisfied by a partial cancellation between these two contributions if $M_{SUSY}$ is close to the weak scale. On the other hand, if $M_{SUSY}$ is high, then the only surviving SUSY contribution is the Higgs exchange. In either of these extremes, or in-between, the main issue is to determine what size and phase the FCNC observables allow for the mass insertions. This certainly requires a global analysis of the existing FCNC data by incorporating the Higgs exchange effects to other SUSY contributions [220]. For example, in parameter regions where the latter are suppressed ($M_{SUSY} \gg m_t$), one can determine the allowed sizes of mass insertions by using (5.157) in (5.159). Doing so, one finds that the flavor–changing Higgs vertices $bsH^a$ and $bdH^a$ become vanishingly small for $\tan\beta \simeq 60$ when all MIs are $\mathcal{O}(1)$, for $\tan\beta \simeq 65$ when $\left(\delta^d_{12}\right)_{LL,RR} \simeq 0$, and, finally, for $\tan\beta \simeq 68$ ,when $\left(\delta^d_{12}\right)_{LL} \simeq -\left(\delta^d_{12}\right)_{RR}$, provided that $\phi_\mu + \phi_g \rightarrow \pi$ in all three cases. Therefore, in this parameter domain, though the flavor-changing Higgs decay channels are sealed up, the decays into similar quarks are highly enhanced. For instance, $\Gamma(h \rightarrow \overline{d}d)/\Gamma(h \rightarrow \overline{b}b) \simeq (\text{Re}\,[h_d/h_b])^2$ which is $\mathcal{O}(1)$ when $h_d \sim h_b$, as is the case with SUSY flavor violation. Such enhancements in light quark Yukawas induce significant reductions in $\overline{b}b$ branching fractions — which is a very important signal for hadron colliders to determine the non-standard nature of the Higgs boson ($h \rightarrow \overline{b}b$ has $\sim 90\%$ branching fraction in the Standard Model). If FCNC constraints are saturated without a strong suppression of the flavor-changing Higgs couplings (which requires $M_{SUSY}$ to be close to the weak scale) then Higgs decays into dissimilar quarks get significantly enhanced. For instance, $h \rightarrow \overline{b}s + \overline{s}b$ can be comparable to $h \rightarrow \overline{b}b$. (See [221] for a diagrammatic analysis of $\rightarrow \overline{b}s + \overline{s}b$ decay.) In conclusion, as fully detailed in [216], SUSY flavor and $CP$ violation sources significantly modify Higgs–quark interactions, thereby inducing potentially large effects that can be discovered at Super $B$ Factories, as well as at the hadron colliders.

The research was supported by DOE grant DE-FG02-94ER40823 at Minnesota.

# 5.4 Extra Dimensions

## 5.4.1 Large Extra Dimensions and Graviton Exchange in $b \to s\ell^+\ell^-$

>– T. G. Rizzo –<

### Introduction

The existence of extra space-like dimensions has been proposed as a possible solution to the gauge hierarchy problem. Although there are many models in the literature attempting to address this issue a common feature is the existence of a higher dimensional 'bulk' space in which gravity is free to propagate. In the Kaluza-Klein (KK) picture the reduction to four dimensions leads to the existence of a massive tower of gravitons that can be exchanged between Standard Model fields. The two most popular scenarios are those of Arkani-Hamed, Dimopolous and Dvali (ADD)[222] and of Randall and Sundrum (RS)[223]. The properies of the KK gravitons are significantly different in these two models. However, in either case the exchange of KK gravitons has been shown to lead to unique signatures that may be discovered at the LHC[224].

While high-$p_T$ measurements at hadron colliders may tell us some of the gross features of the extra-dimensional model, other sets of measurements will be necessary in order to disentangle its complete structure. For example, the LHC may observe the graviton resonances of the RS model in the Drell-Yan and/or dijet channels, but it will be very difficult, if not impossible, to examine the possible *flavor* structure of graviton couplings in such an enviornment[225]. While such determinations will certainly be possible at a future Linear Collider, provided it has suffient center-of-mass energy to sit on a graviton resonance, it may be a while between the LHC discovery and the data from the Linear Collider becoming available. It is possible, however, that at least some aspects of the flavor structure of the graviton KK couplings may be determined using precision data at lower energies, through rare decays such as $b \to s\ell^+\ell^-$. This is the subject of the discussion below.

Flavor dependence, as well as flavor violation in KK graviton couplings can be generated in models which attempt to explain the fermion mass hierarchy as well as the structure of the CKM matrix[226]. In such scenarios, fermions are localized in the extra dimensions either via scalar 'kink'-like solutions, or via their 5-d Dirac masses. A description of the details of such models is, however, beyond the scope of this discussion. In fact, wishing to be as model-independent as possible, we note that in all scenarios at low energies the exhange of gravitons between Standard Model fields can be described by the single dimension-8 operator

$$O_{grav} = \frac{1}{M^4}\, X\, T_{\mu\nu} T^{\mu\nu}\,, \tag{5.160}$$

where $M$ is a mass scale of order $\sim$ a few TeV, the $T_{\mu\nu}$ are the stress-energy tensors of the Standard Model fields, which can have complex flavor structures, and $X$ is a general coupling matrix. Operators such as these may be generated in either ADD-like or RS-like scenarios but we will not be interested here in the specific model details. Instead we focus on unique signatures for graviton exchange associated with the above operator.

### Analysis

How can $b \to s\ell^+\ell^-$probe such operators? To be specific, let us consider the case of ADD-like models; as we will see, our results are easily generalized to RS-type scenarios. In ADD, we identify $M \to M_H$, the cutoff scale in the theory, and $X \to \lambda X$, with $\lambda$ being an overall sign. Identifying the first(second) $T_{\mu\nu}$ with the $b\bar{s}(\ell^+\ell^-)$-effective graviton vertex, the new operator will lead to, *e.g.*, a modification of the $b \to s\ell^+\ell^-$differential decay distribution. Following the notation in [227], we find that this is now given by

$$\frac{d^2\Gamma}{dsdz} \sim [(C_9 + 2C_7/s)^2 + C_{10}^2][(1+s) - (1-s)z^2] - 2C_{10}(C_9 + 2C_7/s)sz\,,$$
$$+\ \frac{4}{s^2}C_7^2(1-s)^2(1-z^2) - \frac{4}{s}C_7(C_9 + 2C_7/s)(1-s)(1-z^2)\,,$$

$$+ DC_9(1-s)z[2s+(1-s)z^2] + DC_{10}s(1-s)(1-z^2)\,, \tag{5.161}$$

where the $C_i$ are the usual effective Standard Model Wilson coefficients, $s = q^2/m_b^2$ is the scaled momentum transfer, $z = \cos\theta$ is the dilepton pair decay angle, and

$$D = \frac{2m_b^2}{G_F\alpha}\sqrt{2}\pi\frac{1}{V_{tb}V_{ts}}\frac{\lambda X}{M_H^4} \simeq 0.062\frac{\lambda X}{M_H^4} \tag{5.162}$$

describes the strength of the graviton contribution with $M_H$ in TeV units. The terms proportional to $D$ in this expression result from the interference of the Standard Model and graviton KK tower exchange amplitudes; note that there is no term proportional to $DC_7$, as dipole and graviton exchanges do not interfere. Here we neglect the square of the pure graviton contribution in the rate, since it is expected to be small.

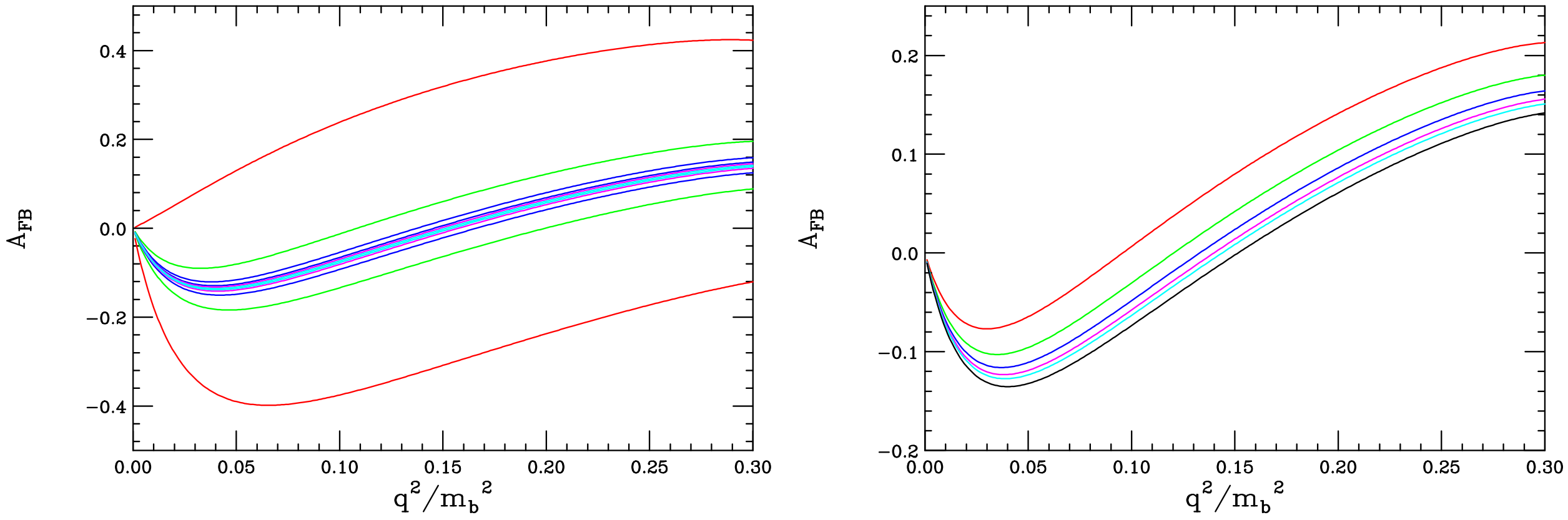

**Figure 5-39.** *$A_{FB}$ as a function of $s$ in the ADD(left) and RS(right) scenarios. In the ADD case, from outside to inside the curves are for $M_H = 1, 1.5, 2.....$ TeV with results for both signs of $\lambda$ shown. In the RS case, from left to right the curves correspond to masses of the lightest KK graviton being 600, 700,...GeV with $k/M_{pl} = 0.1$ being assumed. In either case we take $X = 1$ for purposes of demonstration; for other $X$ values the curves will scale as $M \to M/X^{1/4}$. The current collider bounds correspond to $M_H > 1$ TeV and $m_1 > 600$ GeV, respectively.*

How can graviton exchange be observed uniquely in $b \to s\ell^+\ell^-$? As is well known, many sources of New Physics can lead to modifications in the $b \to s\ell^+\ell^-$differential distribution[228]. In particular, one quantity of interest is the forward-backward asymmetry $A_{FB}$ and the location of its corresponding zero as a function of $s$[229]. That graviton KK tower exchange modifies the location of the zero is clear from the expression above. Fig. 5-39 shows the typical shifts in $A_{FB}$ and its zero in both ADD- and RS-like scenarios. Clearly any observable shifts due to graviton exchange are not by any means unique though they are signatures for New Physics.

Graviton exchange *does*, however, lead to a new effect which will be absent in all other cases of New Physics. The source of this new distinct signature is the $z^3$ term in the differential distribution above, which can be traced back to the spin-2 nature of graviton exchange. The existence of this type of term can be observed experimentally by using the moment method[230] previously employed to probe for KK graviton tower exchange in fermion pair production at the Linear Collider. To this end, we define the quantity

$$< P_3(s) >= \frac{\int \frac{d^2\Gamma}{dsdz}P_3(z)\,dz}{\frac{d\Gamma}{ds}} \tag{5.163}$$

where $P_3 = z(5z^2-3)/2$ is the third Legendre polynomial. Due to the orthogonality of the $P_n$, the presence of the $z^3$ term induces a non-zero value for this moment; the terms that go as $\sim z^{0,1,2}$ in the distribution yield zero for this observable. *Any* experimental observation of a non-zero value for this moment would signal the existence of flavor-changing gravitational interactions. Fig. 5-40 shows the typical $s-$dependence of this moment in both the ADD-like and RS-like scenarios. It now becomes an experimental issue as to whether or not such a non-zero moment is observable. Clearly a very large statistical sample will be required on the order of $\sim 50 - 100 \text{ ab}^{-1}$ or so. To reach this level, a Super $B$ Factory is required. An experimental simulation along these lines would be useful.

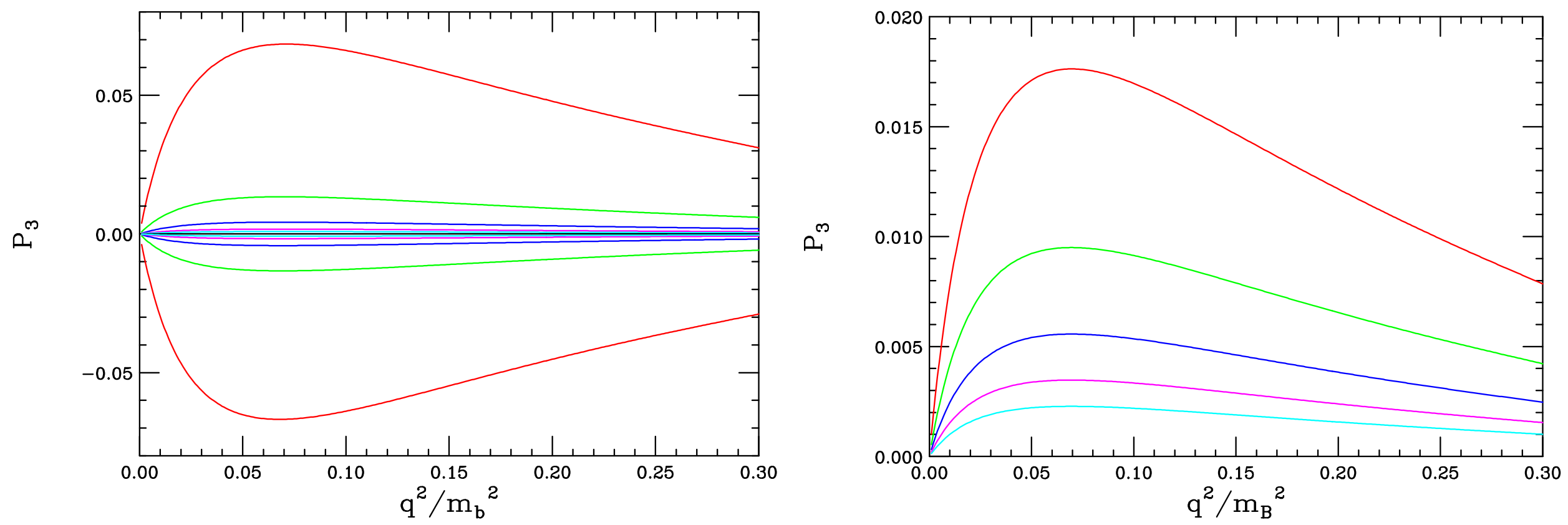

**Figure 5-40.** *Same as the previous figure but now showing the quantity* $< P_3 >$ *as a function of* $s$.

In conclusion, we have shown that flavor-changing KK graviton exchange can be probed via the $b \to s\ell^+\ell^-$decay. A unique signature for these contributions can be obtained through the use of the moment technique. A nonzero value of the third Legendre moment will prove the existence of spin-2 exchange in this process. A Super $B$ Factory is needed to reach the required level of statistics.

The author would like to thank J. L. Hewett for discussions related to this work.

## 5.4.2 TeV$^{-1}$-sized Extra Dimensions with Split Fermions

>– B. Lillie –<

Extra dimensions, in addition to the virtues already discussed, also present the possibility of understanding geometrically several dimensionless numbers that are observed to be very small. These include the small rate of proton decay, and the large ratios of fermion masses. This was first noted by Arkani-Hamed and Schmaltz [231]. They noticed that if the zero modes of the fermion fields were localized to Gaussians in the extra dimensions, then effective 4-d operators that contain fermions will be proportional to the overlaps of these Gaussians. If the localized fermions are separated from each other, these overlap integrals can be exponentially small. For example, separating quark and lepton fields by a distance $a$ in one extra dimension (Fig. 5-41) results in a suppression of the proton decay operator $qqq\ell$ by

$$\int_0^R dy e^{-3\frac{y^2}{\sigma^2}} e^{-\frac{(y-a)^2}{\sigma^2}} = e^{-\frac{3}{4}\frac{a^2}{\sigma^2}} \tag{5.164}$$

where $\sigma$ is the width of the fermions.

If the Higgs field lives in the bulk, then the fermion masses are generated by the flat zero mode of the Higgs, and are proportional to the overlap of the left- and right-handed fields. If the chiral components of different fermions are separated by different distances in the extra dimension, then exponentially different masses can be generated. The Yukawa coupling between the $i$-th left handed and $j$-th right-handed fermions is proportional to

$$\int_0^R dy e^{-\frac{(y-y_i)^2}{\sigma^2}} e^{-\frac{(y-y_j)^2}{\sigma^2}} = e^{-\frac{1}{2}\frac{(y_i-y_j)^2}{\sigma^2}} . \tag{5.165}$$

Thus, if this scenario were true, we could understand the large ratios of fermion masses as being due to order one differences in the parameters of the fundamental theory. It has been shown by explicit construction that the observed values of the fermion masses (the Yukawa hierarchies), as well as the CKM matrix elements can be obtained in this way [232]. In addition, several variant models have been proposed. The left and right-handed fermions could be exponentially localized to two different branes, and the Higgs field to the left-handed brane. The Yukawa hierarchies are then obtained from the exponentially small values of the right-handed fermion wavefunctions at the left-handed brane [233]. This scenario generalizes very nicely to the case of a warped extra dimension, where the Higgs is localized to the TeV brane and all the fermions, except the top, are localized near the Plank brane [234]. Finally, rather than fixed-width Gaussians separated by some distance, one could consider different width Gaussians localized to the same point. Instead of exponentially small Yukawa matrix elements, this scenario generates Yukawa matrix elements that are all approximately the same, realizing the democratic scenario of fermion masses [235].

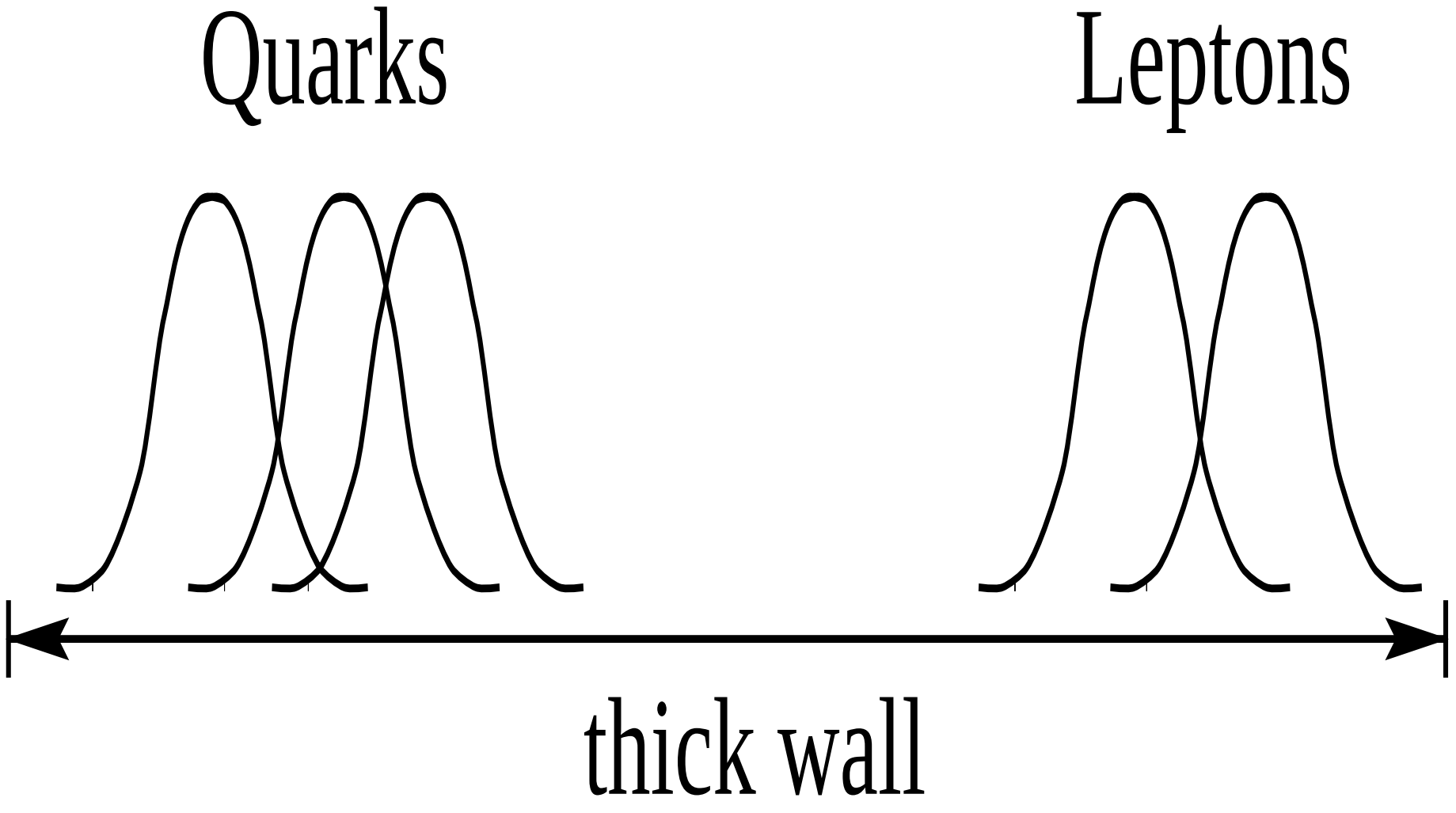

**Figure 5-41.** *Illustration of the concept of Split Fermions. Different species of fermions are localized to Gaussians at different locations in an extra dimension. This can be interpreted as a dimension compactified at the* TeV *scale, or a "brane" of* TeV *thickness embedded in a larger extra dimension. Figure taken from [231].*

Split fermion scenarios naturally suppress many dangerous operators, but they do not suppress flavor changing effects [236, 237]. It is for this reason that they are of interest to a Super $B$ Factory. To see this, note that, while fermions can be localized to different points in an extra dimension, the gauge fields must interact universally with the matter fields, and hence must posses a flat (or nearly flat) zero mode. This implies that they are delocalized at least on the scale of the separation of the fermions. They will then have excited KK modes with non-trivial wavefunctions. These will interact non-universally with the matter fields, and hence will produce flavor-changing effects. This is due to the fact that the non-universal couplings pick out a direction in flavor space, so that when the Yukawa matrices are diagonalized to find the mass eigenstates, non-trivial effects will be seen in the KK couplings. In the case of a single, flat, extra dimension, the coupling of the $n$-th excited gauge boson to a fermion localized to the point $\ell$ is proportional to

$$\int_0^R dy\ \cos(n\pi y) e^{-(y-\ell)^2 R^2/\sigma^2} \approx \cos(n\pi\ell) e^{-n^2\sigma^2/R^2}. \tag{5.166}$$

All excited gauge bosons, including the excited gluons, will have non-universal couplings and can generate flavor-changing neutral currents at tree-level. Hence, the model contains tree-level FCNC effects, suppressed only by the mass of the KK excitations. These can produce large effects, and thus already produce strong constraints from existing data.

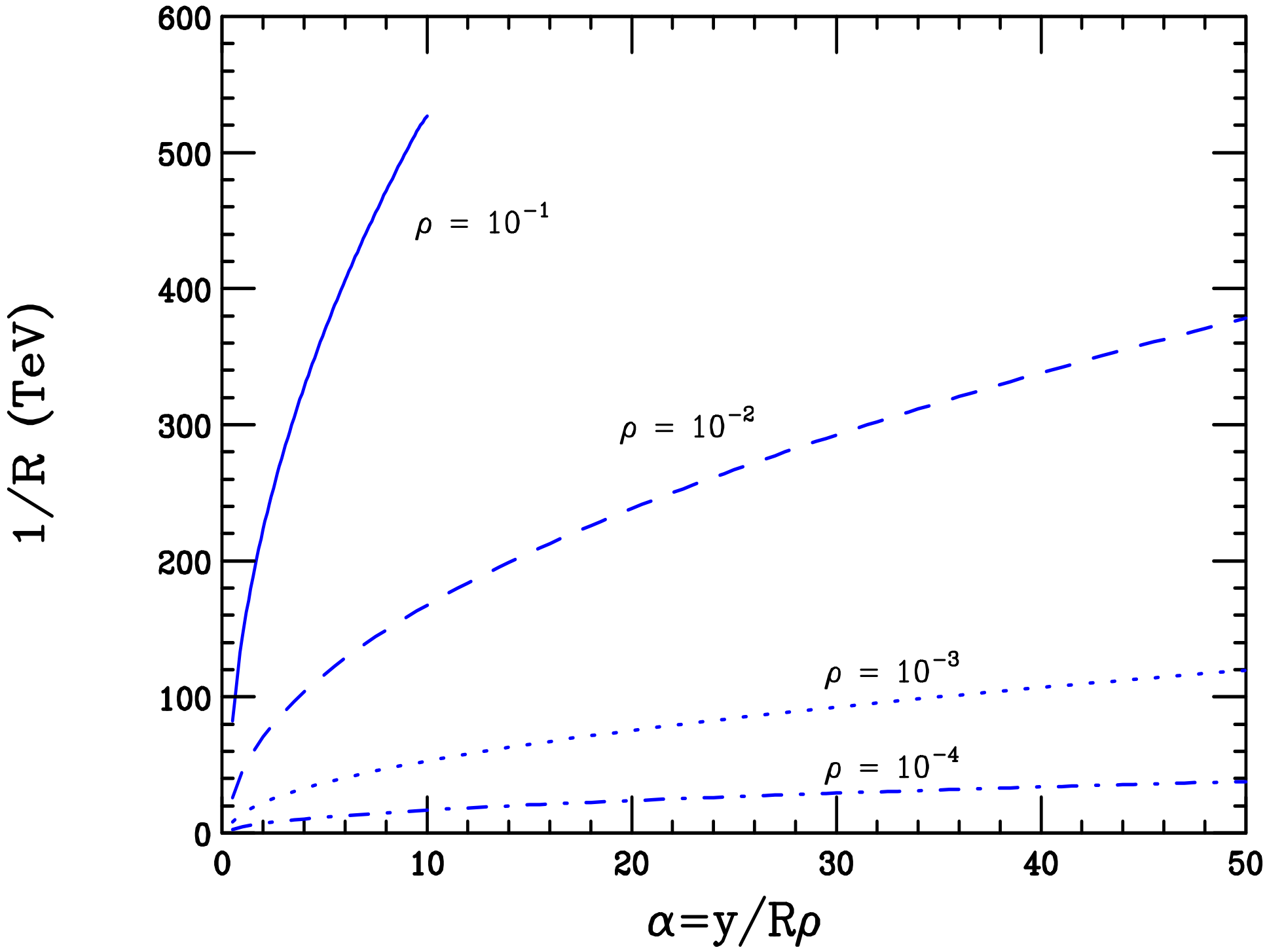

**Figure 5-42.** *Constraints on the compactification scale $1/R$ arising from the FCNC contribution to $\Delta m_K$, as a function of the separation of the fermions in units of the fermion width. The regions below the curves are excluded. Different curves are different values of $\rho = \sigma/R$, the ratio of the compactification scale to the fermion localization scale. Note that the size of the extra dimension in units of the fermion width is $1/\rho$, so for $\rho = 1/10$, $\alpha = 10$ corresponds to the fermions being localized at opposite ends of the dimension.*

Since all KK states contribute, any FCNC effect must be summed over all states. In one flat dimension, this sum is approximated by the distance between the two relevant flavors, in units of the size of the dimension. Note that the size of Yukawa matrix elements depends only on the separations in units of the fermion width, $\sigma$, and is independent of $R$. Hence, if we are interested in the flavor-changing effects in a scenario where the fermion masses are explained by localization, the relevant parameter is $\rho = \sigma/R$.

For example, the contribution to the mass splitting of an oscillating neutral meson system, $P$, from the separation of a pair of split fermions, takes the form

$$\Delta m_P = \frac{2}{9} g_s^2 f_P^2 m_P R^2 V_{qq',qq'}^4 F(\rho\alpha). \tag{5.167}$$

Here $F(\rho\alpha)$ is a function that depends on the extra dimensional splitting, $\alpha$ of the two quarks, $q$ and $q'$, that form the meson, and $V^4$ represents four mixing angles arising from the matrices that diagonalize the Yukawa matrices (each one is roughly the square root of a CKM matrix element). The overall scale is set by the factor of $R^2$, and hence this formula can be interpreted as a constraint on the compactification scale from the measured value of $\Delta m_P$. Fig. 5-42 shows this constraint from $\Delta m_K$ for several values of $\rho$. Note that extremely high scales can be probed.

Split fermion models are expected to live within a larger model that solves the hierarchy problem, and hence the cutoff of the theory is expected to be not much larger than $10$ TeV. The compactification scale must be even smaller. We see that one can achieve a small value of $1/R$ at the expense of going to a very small value of $\rho$. However, this implies that the fermion localization scale $1/\sigma$ is very large, of order $10^4$ TeV. Hence it looks like the models are essentially ruled out. This conclusion can be escaped in two ways. The bounds shown are for the kaon mass splitting, and hence are most sensitive to the separation of the $d$ and $s$ quarks. It could be that the down-type quarks are all aligned, and the CKM mixing is induced by splittings of the up-type quarks. To constrain this, one would need to look at $D^0 - \overline{D}^0$ mixing. Alternately, the mixing could be in both the up and the down sector, but such that the Yukawa matrices are diagonalized by mixing that is predominantly between the third and first or second generations, in such a way as to produce the CKM matrix when the product of up and down diagonalization matrices is taken. Constraining this requires measurement of both the $B_d$ and $B_s$ mixing parameters. Fig. 5-43 shows the constraints from all neutral meson mixings. A combination of these is needed to fully constrain the model.

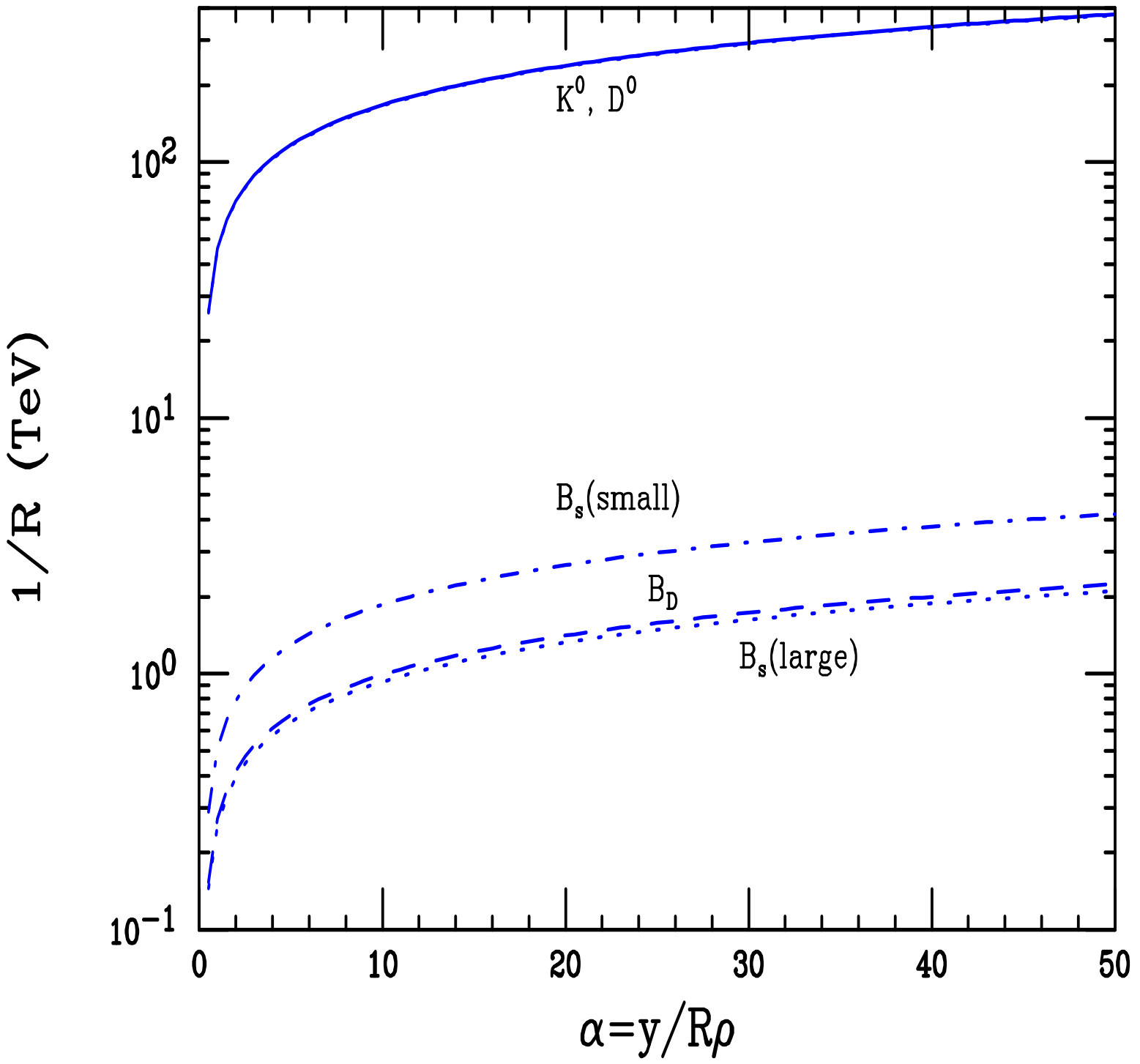

**Figure 5-43.** *Constraints on the compactification scale arising from the mass splitting of different neutral meson systems for $\rho = 1/100$. For $B_s$, since no upper bound is known, two different values were taken, the "small" value is about the size expected in the Standard Model, the "large" value about four times bigger. A combination of all of these measurement, and others, is needed to fully constrain the models.*

It is also possible that there may be interesting signatures in rare decays. In particular, lepton family number-violating decays can produce limits on the splittings of leptons, which are not available from the meson oscillation data. It is also possible in a small region of parameter space to have contributions to $B^0 \to J/\psi K^0_S$ or $B^0 \to \phi K^0_S$ that are near the same order as the Standard Model, leading to interesting effects in $CP$-violating observables.

Another, more attractive, possibility is to go from a flat to a warped geometry. In [234] it was shown that the same flavor constraints are much more mild in an RS model where the fermions live in the bulk, but are localized near the Plank brane. In that case, the fact that the gauge KK wavefunctions are nearly flat near the Plank brane helps to naturally suppress the non-universality of the couplings to fermions. As a result all scales in the model can be below $10\ \mathrm{TeV}$, but there are still effects predicted in rare decays that might be visible to future experiments. This case is the most promising for future study at a Super $B$ Factory.

### 5.4.3 Universal Extra Dimensions

≻— A. J. Buras, A. Poschenrieder, M. Spranger, and A. Weiler —≺

#### Introduction

Models with more than three spatial dimensions have been used to unify the forces of nature ever since the seminal papers of Kaluza and Klein [238]. More recently, extra dimensional models have been employed as an alternative explanation of the origin of the TeV scale [239].

A simple model of the so called universal type is the Appelquist, Cheng and Dobrescu (ACD) model [240] with one universal extra dimension. It is an extension of the Standard Model to a 5 dimensional orbifold $\mathcal{M}^4 \times S^1/Z_2$, where all the Standard Model fields live in all available 5 dimensions. In what follows we will briefly describe this model and subsequently report on the results of two papers [241, 83] relevant for these Proceedings, in which we investigated the impact of the KK modes on FCNC processes in this model. Further details can be found in Ref. [242].

#### The ACD Model

The full Lagrangian of this model includes both the boundary and the bulk Lagrangian. The coefficients of the boundary terms, although volume-suppressed, are free parameters and will get renormalized by bulk interactions. Flavor non-universal boundary terms would lead to large FCNCs. In analogy to a common practice in the MSSM, in which the soft supersymmetry breaking couplings are chosen to be flavor-universal, we assume negligible boundary terms at the cut-off scale. Now the bulk Lagrangian is determined by the Standard Model parameters, after an appropriate rescaling. With this choice, contributions from boundary terms are of higher order, and we only have to consider the bulk Lagrangian for the calculation of the impact of the ACD model.

Since all our calculations are cut-off-independent (see below) the only additional free parameter relative to the Standard Model is the compactification scale $1/R$. Thus, all the tree level masses of the KK particles and their interactions among themselves and with the Standard Model particles are described in terms of $1/R$ and the parameters of the Standard Model. This economy in new parameters should be contrasted with supersymmetric theories and models with an extended Higgs sector. All Feynman rules necessary for the evaluation of FCNC processes can be found in [241, 83].

A very important property of the ACD model is the conservation of KK parity that implies the absence of tree level KK contributions to low energy processes taking place at scales $\mu \ll 1/R$. In this context the flavor-changing neutral current(FCNC) processes like particle-antiparticle mixing, rare $K$ and $B$ decays and radiative decays are of particular interest. Since these processes first appear at one-loop in the Standard Model and are strongly suppressed, the one-loop contributions from the KK modes to them could in principle be important.

The effects of the KK modes on various processes of interest have been investigated in a number of papers. In [240, 243] their impact on the precision electroweak observables assuming a light Higgs ($m_H \leq 250$ GeV) and a heavy Higgs led to the lower bound $1/R \geq 300\,\mathrm{GeV}$ and $1/R \geq 250\,\mathrm{GeV}$, respectively. Subsequent analyses of the anomalous magnetic moment [244] and the $Z \to b\bar{b}$ vertex [245] have shown the consistency of the ACD model with the data for $1/R \geq 300\,\mathrm{GeV}$. The latter calculation has been confirmed in [241]. The scale of $1/R$ as low as $300\,\mathrm{GeV}$ would also lead to an exciting phenomenology in the next generation of colliders and could be of interest in connection with dark matter searches. The relevant references are given in [83].

The question then arises whether such low compactification scales are still consistent with the data on FCNC processes. This question has been addressed in detail in [241, 83]. Before presenting the relvant results of these papers let us recall the particle content of the ACD model that has been described in detail in [241].

In the effective four dimensional theory, in addition to the ordinary particles of the Standard Model, denoted as zero $(n = 0)$ modes, there are infinite towers of the KK modes $(n \geq 1)$. There is one such tower for each Standard Model boson and two for each Standard Model fermion, while there also exist physical neutral ($a^0_{(n)}$) and charged ($a^\pm_{(n)}$) scalars with $(n \geq 1)$ that do not have any zero mode partners. The masses of the KK particles are universally given by

$$(m^2_{(n)})_{\rm KK} = m_0^2 + \frac{n^2}{R^2}\ . \tag{5.168}$$

Here $m_0$ is the mass of the zero mode, as $M_W$, $M_Z$, $m_t$ respectively. For $a^0_{(n)}$ and $a^\pm_{(n)}$ this is $M_Z$ and $M_W$, respectively. In phenomenological applications it is more useful to work with the variables $x_t$ and $x_n$ defined through

$$x_t = \frac{m_t^2}{m_W^2}, \qquad x_n = \frac{m_n^2}{m_W^2}, \qquad m_n = \frac{n}{R} \tag{5.169}$$

than with the masses in (5.168).

### The ACD Model and FCNC Processes

As our analysis of [241, 83] shows, the ACD model with one extra dimension has a number of interesting properties from the point of view of FCNC processes discussed here. These are:

- The GIM mechanism [246] that significantly improves the convergence of the sum over the KK modes corresponding to the top quark, removing simultaneously to an excellent accuracy the contributions of the KK modes corresponding to lighter quarks and leptons. This feature removes the sensitivity of the calculated branching ratios to the scale $M_s \gg 1/R$ at which the higher dimensional theory becomes non-perturbative, and at which the towers of the KK particles must be cut off in an appropriate way. This should be contrasted with models with fermions localized on the brane, in which the KK parity is not conserved, and the sum over the KK modes diverges. In these models the results are sensitive to $m_s$ and, for instance, in $\Delta m_{s,d}$, the KK effects are significantly larger [247] than found by us. We expect similar behavior in other processes considered below.
- The low energy effective Hamiltonians are governed by local operators already present in the Standard Model. As flavor violation and $CP$ violation in this model is entirely governed by the CKM matrix, the ACD model belongs to the class of models with minimal flavor violation (MFV), as defined in [78]. This has automatically the following important consequence for the FCNC processes considered in [241, 83]: the impact of the KK modes on the processes in question amounts only to the modification of the Inami-Lim one-loop functions [248].
- Thus in the case of $\Delta m_{d,s}$ and of the parameter $\varepsilon_K$, that are relevant for the standard analysis of the unitarity triangle, these modifications have to be made in the function $S$ [249]. In the case of the rare $K$ and $B$ decays that are dominated by $Z^0$ penguins the functions $X$ and $Y$ [250] receive KK contributions. Finally, in the case of the decays $B \to X_s\gamma$, $B \to X_s$ gluon, $B \to X_s\mu\overline{\mu}$ and $K^0_S \to \pi^0 e^+e^-$ and the $CP$-violating ratio $\varepsilon'/\varepsilon$ the KK contributions to new short distance functions have to be computed. These are the functions $D$ (the $\gamma$ penguins), $E$ (gluon penguins), $D'$ ($\gamma$-magnetic penguins) and $E'$ (chromomagnetic penguins). Here we will only report on the decays relevant for Super $B$ Factories.

Thus, each function mentioned above, which in the Standard Model depends only on $m_t$, now also becomes a function of $1/R$:

$$F(x_t, 1/R) = F_0(x_t) + \sum_{n=1}^{\infty} F_n(x_t, x_n), \quad F = B, C, D, E, D', E', \tag{5.1}$$

with $x_n$ defined in (5.169). The functions $F_0(x_t)$ result from the penguin and box diagrams in the Standard Model and the sum represents the KK contributions to these diagrams.

In phenomenological applications, it is convenient to work with the gauge invariant functions [250]

$$X = C + B^{\nu\overline{\nu}}, \qquad Y = C + B^{\mu\overline{\mu}}, \qquad Z = C + \frac{1}{4}D. \tag{5.2}$$

The functions $F(x_t, 1/R)$ have been calculated in [241, 83] with the results given in Table 5-17. Our results for the function $S$ have been confirmed in [251]. For $1/R = 300$ GeV, the functions $S$, $X$, $Y$, $Z$ are enhanced by 8%, 10%, 15% and 23% relative to the Standard Model values, respectively. The impact of the KK modes on the function $D$ is negligible. The function $E$ is moderately enhanced but this enhancement plays only a marginal role in the phenomenological applications. The most interesting are very strong suppressions of $D'$ and $E'$, that for $1/R = 300\,\mathrm{GeV}$ amount to 36% and 66% relative to the Standard Model values, respectively. However, the effect of the latter suppressions is softened in the relevant branching ratios through sizable additive QCD corrections.

**Table 5-17.** *Values for the functions $S$, $X$, $Y$, $Z$, $E$, $D'$, $E'$, $C$ and $D$.*

| $1/R$ [GeV] | $S$ | $X$ | $Y$ | $Z$ | $E$ | $D'$ | $E'$ | $C$ | $D$ |
|---|---|---|---|---|---|---|---|---|---|
| 200 | 2.813 | 1.826 | 1.281 | 0.990 | 0.342 | 0.113 | −0.053 | 1.099 | −0.479 |
| 250 | 2.664 | 1.731 | 1.185 | 0.893 | 0.327 | 0.191 | 0.019 | 1.003 | −0.470 |
| 300 | 2.582 | 1.674 | 1.128 | 0.835 | 0.315 | 0.242 | 0.065 | 0.946 | −0.468 |
| 400 | 2.500 | 1.613 | 1.067 | 0.771 | 0.298 | 0.297 | 0.115 | 0.885 | −0.469 |
| Standard Model | 2.398 | 1.526 | 0.980 | 0.679 | 0.268 | 0.380 | 0.191 | 0.798 | −0.476 |

### The impact of the KK modes on specific decays

**The impact on the Unitarity Triangle.** The function $S$ plays the crucial role here. Consequently the impact of the KK modes on the Unitarity Triangle is rather small. For $1/R = 300\,\mathrm{GeV}$, $|V_{td}|$, $\overline{\eta}$ and $\gamma$ are suppressed by 4%, 5% and 5°, respectively. It will be difficult to see these effects in the $(\overline{\varrho}, \overline{\eta})$ plane. On the other hand, a 4% suppression of $|V_{td}|$ means an 8% suppression of the relevant branching ratio for rare decays sensitive to $|V_{td}|$ and this effect has to be taken into account. Similar comments apply to $\overline{\eta}$ and $\gamma$. Let us also mention that for $1/R = 300\,\mathrm{GeV}$, $\Delta m_s$ is enhanced by 8%; in view of the sizable uncertainty in $\hat{B}_{B_s}\sqrt{f_{B_s}}$, this will also be difficult to see.

**The impact on rare $B$ decays.** Here, the dominant KK effects enter through the function $C$, or equivalently, $X$ and $Y$, depending on the decay considered. In Table 5-18 we show seven branching ratios as functions of $1/R$ for central values of all remaining input parameters. The hierarchy of the enhancements of branching ratios can easily be explained by inspecting the enhancements of the functions $X$ and $Y$ that is partially compensated by the suppression of $|V_{td}|$ in decays sensitive to this CKM matrix element, but fully effective in decays governed by $|V_{ts}|$.

**Table 5-18.** *Branching ratios for rare $B$ decays in the ACD model and the Standard Model as discussed in the text.*

| $1/R$ | 200 GeV | 250 GeV | 300 GeV | 400 GeV | Standard Model |
|---|---|---|---|---|---|
| $Br(B \to X_s \nu\overline{\nu}) \times 10^5$ | 5.09 | 4.56 | 4.26 | 3.95 | 3.53 |
| $Br(B \to X_d \nu\overline{\nu}) \times 10^6$ | 1.80 | 1.70 | 1.64 | 1.58 | 1.47 |
| $Br(B_s \to \mu^+\mu^-) \times 10^9$ | 6.18 | 5.28 | 4.78 | 4.27 | 3.59 |
| $Br(B_d \to \mu^+\mu^-) \times 10^{10}$ | 1.56 | 1.41 | 1.32 | 1.22 | 1.07 |

For $1/R = 300\,\mathrm{GeV}$, the following enhancements relative to the Standard Model predictions are seen: $B \to X_d\nu\overline{\nu}$ (12%), $B \to X_s\nu\overline{\nu}$ (21%), $B_d \to \mu\overline{\mu}$ (23%) and $B_s \to \mu\overline{\mu}$ (33%). These results correspond to central values of the input parameters. The uncertainties in these parameters partly cover the differences between the ACD model and the Standard Model, and it is essential to considerably reduce these uncertainties if one wants to see the effects of the KK modes in the branching ratios in question.

**The Impact on $B \to X_s\gamma$ and $B \to X_s$ gluon.** Due to strong suppressions of the functions $D'$ and $E'$ by the KK modes, the $B \to X_s\gamma$ and $B \to X_s$ gluon decays are considerably suppressed compared to Standard Model estimates (consult [252, 253] for the most recent reviews). For $1/R = 300\,\mathrm{GeV}$, $\mathcal{B}(B \to X_s\gamma)$ is suppressed by $20\%$, while $\mathcal{B}(B \to X_s\ \mathrm{gluon})$ even by $40\%$. The phenomenological relevance of the latter suppression is unclear at present as $\mathcal{B}(B \to X_s\ \mathrm{gluon})$ suffers from large theoretical uncertainties, and its extraction from experiment is very difficult, if not impossible.

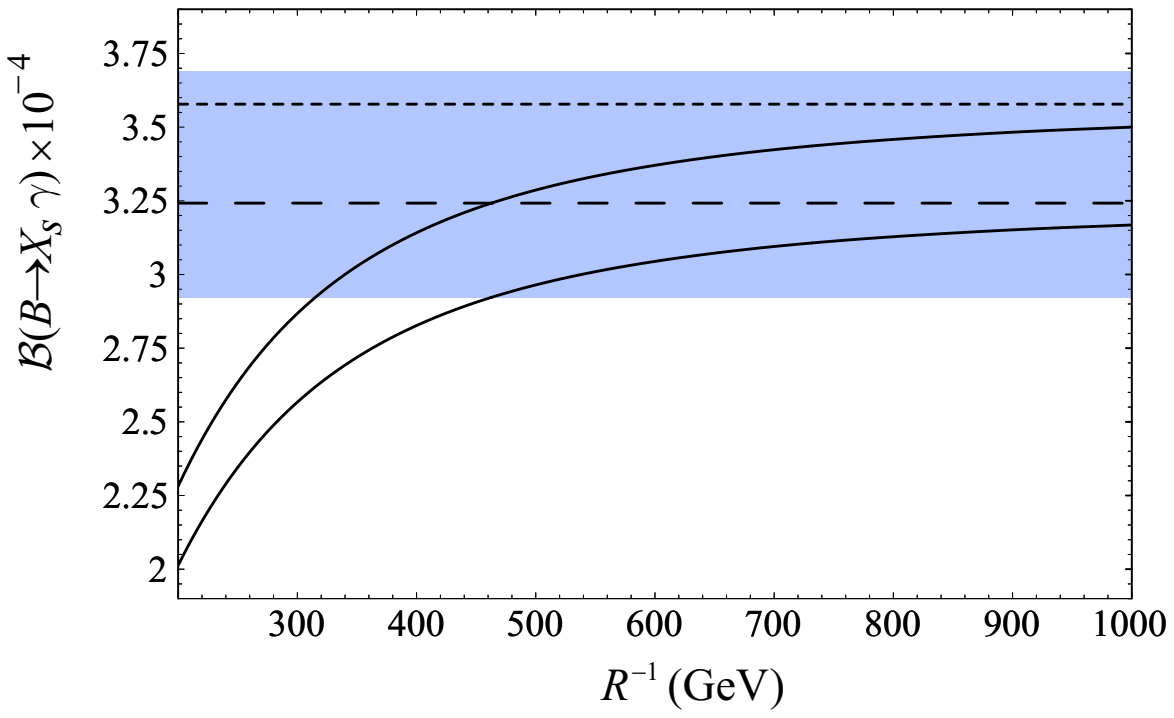

**Figure 5-44.** *The branching ratio for $B \to X_s\gamma$ and $E_\gamma > 1.6$ GeV as a function of $1/R$. See text for the meaning of various curves.*

In Fig. 5-44 we compare $\mathcal{B}(B \to X_s\gamma)$ in the ACD model with the experimental data and with the expectations of the Standard Model. The shaded region represents the data $\mathcal{B}(B \to X_s\gamma)_{E_\gamma>1.6\mathrm{GeV}} = (3.28^{+0.41}_{-0.36}) \cdot 10^{-4}$ [254] and the upper (lower) dashed horizontal line are the central values in the Standard Model for $m_c/m_b = 0.22$ $(m_c/m_b = 0.29)$ [255, 256]. The solid lines represent the corresponding central values in the ACD model. The theoretical errors, not shown in the plot, are roughly $\pm10\%$ for all curves.

We observe that in view of the sizable experimental error and considerable parametric uncertainties in the theoretical prediction, the strong suppression of $\mathcal{B}(B \to X_s\gamma)$ by the KK modes does not yet provide a powerful lower bound on $1/R$ and values of $1/R \geq 250\,\mathrm{GeV}$ are fully consistent with the experimental result. It should also be emphasized that $\mathcal{B}(B \to X_s\gamma)$ depends sensitively on the ratio $m_c/m_b$; the lower bound on $1/R$ is shifted above $400\,\mathrm{GeV}$ for $m_c/m_b = 0.29$, if other uncertainties are neglected. In order to reduce the dependence on $m_c/m_b$ a NNLO calculation is required [255, 256, 257]. Once it is completed, and the experimental uncertainties further reduced – a Super $B$ Factory could increase the experimental sensitivity up to a factor of three [258]–$\mathcal{B}(B \to X_s\gamma)$ may provide a very powerful bound on $1/R$ that is substantially stronger than the bounds obtained from the electroweak precision data. The suppression of $\mathcal{B}(B \to X_s\gamma)$ in the ACD model has already been found in [259]. The result presented above is consistent with the one obtained by these authors, but differs in details, as only the dominant diagrams have been taken into account in the latter paper, and the analysis was performed in the LO approximation.

**The Impact on $B \to X_s\mu^+\mu^-$ and $A_{FB}(\hat{s})$.** In Fig. 5-45 we show the branching ratio $\mathcal{B}(B \to X_s\mu^+\mu^-)$ as a function of $1/R$. The observed enhancement is mainly due to the function $Y$ that enters the Wilson coefficient of the operator $(\bar{s}b)_{V-A}(\bar{\mu}\mu)_A$. The Wilson coefficient of $(\bar{s}b)_{V-A}(\bar{\mu}\mu)_V$, traditionally denoted by $C_9$, is essentially unaffected by the KK contributions.

Of particular interest is the forward-backward asymmetry $A_{FB}(\hat{s})$ in $B \to X_s\mu^+\mu^-$ that, similar to the case of exclusive decays [260], vanishes at a particular value $\hat{s} = \hat{s}_0$. The fact that $A_{\rm FB}(\hat{s})$ and the value of $\hat{s}_0$ being sensitive to short distance physics are in addition subject to only very small non-perturbative uncertainties makes them particularly useful quantities to test physics beyond the Standard Model . A precise measurement however is a difficult task, but it could be performed at a Super $B$ Factory [261].

The calculations for $A_{\rm FB}(\hat{s})$ and of $\hat{s}_0$ have recently been done including NNLO corrections [262, 263] that turn out to be significant. In particular they shift the NLO value of $\hat{s}_0$ from $0.142$ to $0.162$ at NNLO. In Fig. 5-46 (a) we show the normalized forward-backward asymmetry that we obtained by means of the formulae and the computer program of [52, 262] modified by the KK contributions calculated in [83]. The dependence of $\hat{s}_0$ on $1/R$ is shown in Fig. 5-46 (b).

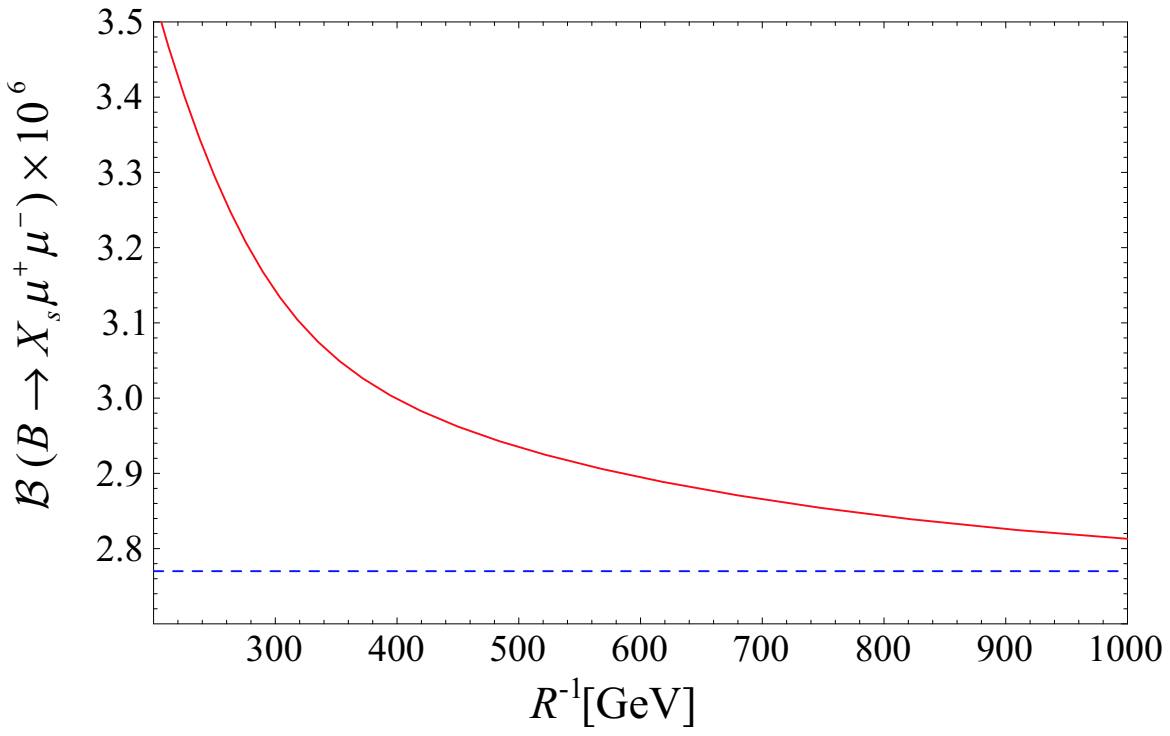

**Figure 5-45.** *$\mathcal{B}(B \to X_s\mu^+\mu^-)$ in the Standard Model (dashed line) [253, 52], and in the ACD model, where the dilepton mass spectrum has been integrated between the limits: $\left(\frac{2m_\mu}{m_b}\right)^2 \leq \hat{s} \leq \left(\frac{M_{J/\psi}-0.35\,\mathrm{GeV}}{m_b}\right)^2$ where $\hat{s} = (p_+ + p_-)^2/m_b^2$.*

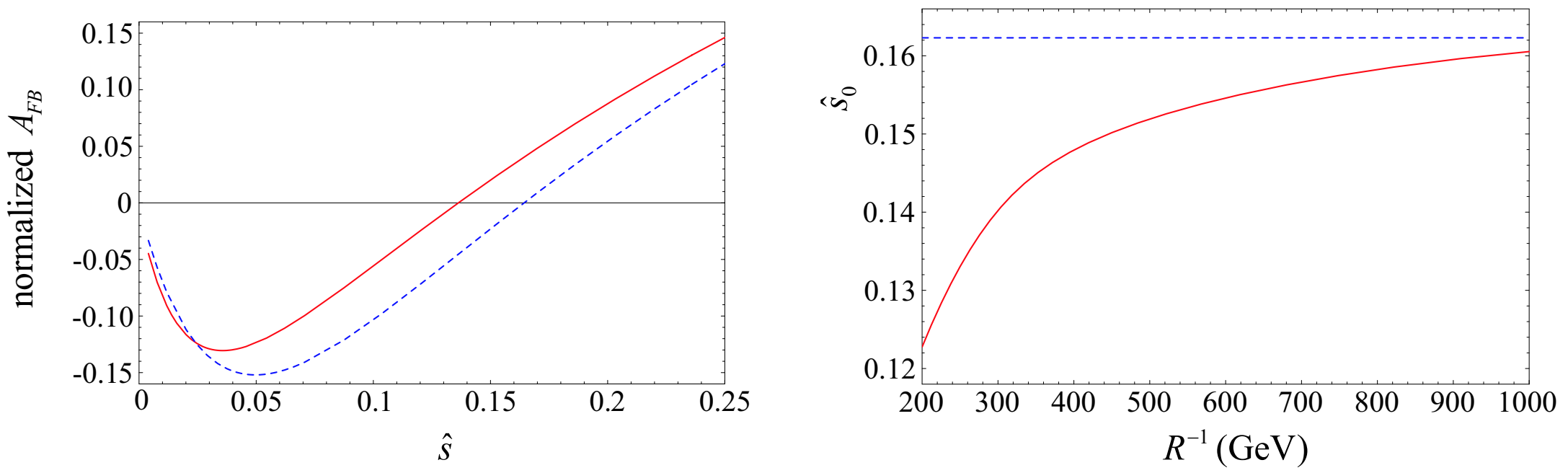

**Figure 5-46.** *(a) Normalized forward-backward asymmetry in the Standard Model (dashed line) and ACD for $R^{-1} = 250$ GeV. (b) Zero of the forward backward asymmetry $A_{FB}$ in the Standard Model (dashed line) and the ACD model.*

We observe that the value of $\hat{s}_0$ is considerably reduced relative to the Standard Model result obtained by including NNLO corrections [52, 262, 263]. This decrease is related to the decrease of $\mathcal{B}(B \to X_s\gamma)$ as discussed below. For $1/R = 300\,\mathrm{GeV}$ we find the value for $\hat{s}_0$ that is very close to the NLO prediction of the Standard Model. This result demonstrates very clearly the importance of the calculations of the higher order QCD corrections, in particular in quantities like $\hat{s}_0$ that are theoretically clean. We expect that the results in Figs. 5-46 (a) and (b) will play an important role in the tests of the ACD model in the future.

In MFV models there exist a number of correlations between different measurable quantities that do not depend on specific parameters of a given model [78, 264]. In [83] a correlation between $\hat{s}_0$ and $\mathcal{B}(B \to X_s\gamma)$ has been pointed out. It is present in the ACD model and in a large class of supersymmetric models discussed for instance in [52]. We show this correlation in Fig. 5-47. We refer to [83] for further details.

**Concluding Remarks**

Our analysis of the ACD model shows that all the present data on FCNC processes are consistent with $1/R$ as low as $250\,\mathrm{GeV}$, implying that the KK particles could, in principle, already be found at the Tevatron. Possibly, the most interesting results of our analysis is the enhancement of $\mathcal{B}(K^+ \to \pi^+\nu\overline{\nu})$ (see [241] for details), the sizable downward shift of the zero ($\hat{s}_0$) in the $A_{FB}$ asymmetry and the suppression of $\mathcal{B}(B \to X_s\gamma)$.

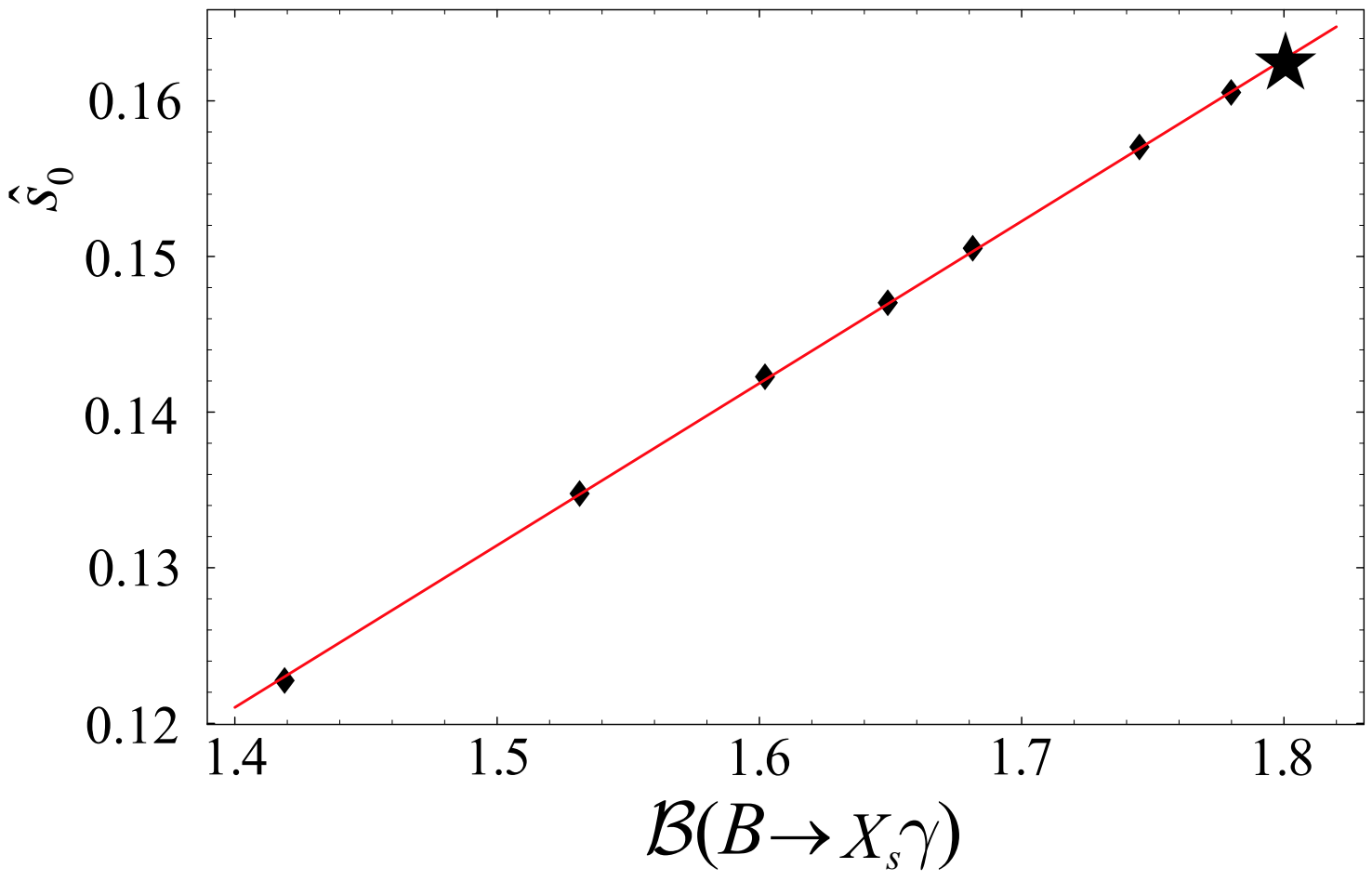

**Figure 5-47.** *Correlation between $\sqrt{\mathcal{B}(B \to X_s\gamma)}$ and $\hat{s}_0$. The straight line is a least square fit to a linear function. The dots are the results in the ACD model for $1/R = 200, 250, 300, 350, 400, 600$ and $1000$ GeV and the star denotes the Standard Model value.*

The nice feature of this extension of the Standard Model is the presence of only one additional parameter, the compactification scale. This feature allows a unique determination of various enhancements and suppressions relative to the Standard Model expectations. Together with a recent study [265] that found no significant difference of $S_{\phi K^0_S}$ in UED to the Standard Model prediction, we can summarize the relative deviations to the Standard Model in this model as follows

- Enhancements: $K^0_S \to \pi^0 e^+e^-$, $\Delta m_s$, $K^+ \to \pi^+\nu\overline{\nu}$, $K_{\rm L} \to \pi^0\nu\overline{\nu}$, $B \to X_d\nu\overline{\nu}$, $B \to X_s\nu\overline{\nu}$, $K^0_S \to \mu^+\mu^-$, $B_d \to \mu^+\mu^-$, $B \to X_s\mu^+\mu^-$ and $B_s \to \mu^+\mu^-$.
- Suppressions: $B \to X_s\gamma$, $B \to X_s$ gluon, the value of $\hat{s}_0$ in the forward-backward asymmetry and $\varepsilon'/\varepsilon$.

We would like to emphasize that violation of this pattern in future high statistics data will exclude the ACD model. For instance, a measurement of $\hat{s}_0$ higher than the Standard Model estimate would automatically exclude this model, as there is no compactification scale for which this could bhappen. Whether these enhancements and suppressions are required by the data, or whether they exclude the ACD model with a low compactification scale, will depend on the precision of the forthcoming experiments as well as on efforts to decrease theoretical uncertainties.

This research was partially supported by the German ‘Bundesministerium für Bildung und Forschung’ under contract 05HT1WOA3 and by the ‘Deutsche Forschungsgemeinschaft’ (DFG) under contract Bu.706/1-2.

### 5.4.4 Warped Extra Dimensions and Flavor Violation

≻– G. Burdman –≺

Randall and Sundrum have recently proposed the use of a non-factorizable geometry in five dimensions [266] as a solution of the hierarchy problem. The metric depends on the five dimensional coordinate $y$ and is given by

$$ds^2 = e^{-2\sigma(y)}\eta_{\mu\nu}dx^\mu dx^\nu - dy^2\,, \tag{5.3}$$

where $x^\mu$ are the four dimensional coordinates, $\sigma(y) = k|y|$, with $k \sim M_P$ characterizing the curvature scale. The extra dimension is compactified on an orbifold $S_1/Z_2$ of radius $r$ so that the bulk is a slice of $\mathrm{AdS}_5$ space between two four-dimensional boundaries. The metric on these boundaries generates two effective scales: $M_P$ and $M_P e^{-k\pi r}$. In this way, values of $r$ not much larger than the Planck length ($kr \simeq (11-12)$) can be used to generate a scale $\Lambda_r \simeq M_P e^{-k\pi r} \simeq \mathcal{O}(\mathrm{TeV})$ on one of the boundaries.

In the original RS scenario, only gravity was allowed to propagate in the bulk, with the Standard Model ( Standard Model ) fields confined to one of the boundaries. The inclusion of matter and gauge fields in the bulk has been extensively treated in the literature [267, 268, 269, 270, 271, 272, 273]. We are interested here in examining the situation when the Standard Model fields are allowed to propagate in the bulk. The exception is the Higgs field which must be localized on the TeV boundary in order for the $W$ and the $Z$ gauge bosons to get their observed masses [268]. The gauge content in the bulk may be that of the Standard Model, or it might be extended to address a variety of model building and phenomenological issues. For instance, the bulk gauge symmetries may correspond to Grand Unification scenarios, or they may be extensions of the Standard Model formulated to restore enough custodial symmetry and bring electroweak contributions in line with constraints. In addition, as was recognized in Ref. [270], it is possible to generate the fermion mass hierarchy from $\mathcal{O}(1)$ flavor breaking in the bulk masses of fermions. Since bulk fermion masses result in the localization of fermion zero-modes, lighter fermions should be localized toward the Planck brane, where their wave-function has exponentially suppressed overlap with the TeV-localized Higgs, whereas fermions with order one Yukawa couplings should be localized toward the TeV brane.

This creates an almost inevitable tension: since the lightest KK excitations of gauge bosons are localized toward the TeV brane, they tend to be strongly coupled to zero-mode fermions localized there. Thus, the flavor-breaking fermion localization leads to flavor-violating interactions of the KK gauge bosons. In particular, this is the case when one tries to obtain the correct top Yukawa coupling: the KK excitations of the various gauge bosons propagating in the bulk will have FCNC interactions with the third generation quarks. This results in interesting effects, most notably in the $CP$ asymmetries in hadronic $B$ decays [274].

In addition, the localization of the Higgs on the TeV brane expels the wave-function of the W and Z gauge bosons away from it resulting in a slightly non-flat profile in the bulk. This leads, for instance, to tree-level flavor changing interactions of the $Z^0$ [275], which in “Higgless” scenarios [276] can result in significant effects in $b \to s\ell^+\ell^-$ [277].

The KK decomposition for fermions can be written as [268, 269]

$$\Psi_{L,R}(x,y) = \frac{1}{\sqrt{2\pi r}} \sum_{n=0} \psi_n^{L,R}(x) e^{2\sigma} f_n^{L,R}(y) \, , \tag{5.4}$$

where $\psi_n^{L,R}(x)$ corresponds to the $nth$ KK fermion excitation and is a chiral four-dimensional field. The zero mode wave functions are

$$f_0^{R,L}(y) = \sqrt{\frac{2k\pi r\,(1 \pm 2c_{R,L})}{e^{k\pi r(1\pm 2c_{R,L})} - 1}} \; e^{\pm c_{R,L}\, k\, y} \, , \tag{5.5}$$

with $c_{R,L} \equiv M_f/k$ parametrizing the 5D bulk fermion mass in units of the inverse AdS radius $k$. The $Z_2$ orbifold projection is used, so that only one of these is actually allowed, either a left-handed or a right-handed zero mode. The Yukawa couplings of bulk fermions to the TeV brane Higgs can be written as

$$S_Y = \int d^4x\, dy\, \sqrt{-g}\, \frac{\lambda_{ij}^{5D}}{2\, M_5}\, \overline{\Psi}_i(x,y) \delta(y - \pi r) H(x) \Psi_j(x,y) \, , \tag{5.6}$$

where $\lambda_{ij}^{5D}$ is a dimensionless parameter and $M_5$ is the fundamental scale or cutoff of the theory. Naive dimensional analysis tells us that we should expect $\lambda_{ij}^{5D} \lesssim 4\pi$. Thus the 4D Yukawa couplings as a function of the bulk mass parameters are

$$Y_{ij} = \left( \frac{\lambda_{ij}^{5D}\, k}{M_5} \right) \sqrt{\frac{(1/2 - c_L)}{e^{k\pi r(1-2c_L)} - 1}} \sqrt{\frac{(1/2 - c_R)}{e^{k\pi r(1-2c_R)} - 1}} \; e^{k\pi r(1 - c_L - c_R)} \, . \tag{5.7}$$

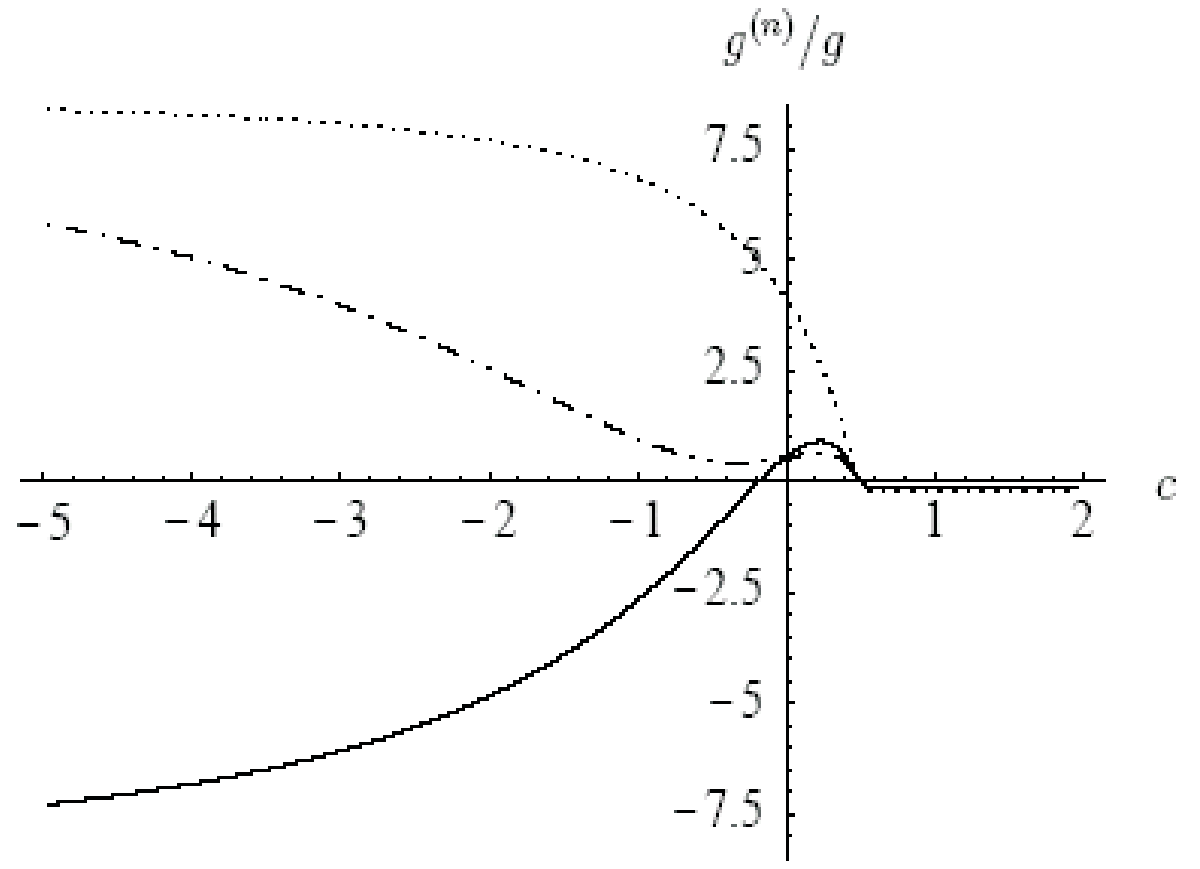

**Figure 5-48.** *Coupling of the first KK excitation of a gauge boson to a zero mode fermion vs. the bulk mass parameter c, normalized to the four-dimensional gauge coupling* $g$.

Given that we expect $k \lesssim M_5$ then the factor $\lambda_{ij}^{5D}\,k/M_5 \simeq \mathcal{O}(1)$. Thus, in order to obtain an $\mathcal{O}(1)$ Yukawa coupling, the bulk mass parameter $c_L$ should naturally be $c_L < 0.5$ and even negative. In other words, the left-handed zero-mode should also be localized toward the TeV brane. This however, posses a problem since it means that the left-handed doublet $q_L$, and therefore $b_L$ should have a rather strong coupling to the first KK excitations of gauge bosons. In Fig. 5-48 we plot the coupling of the first KK excitation of a gauge boson to a zero-mode fermion vs. the fermion's bulk mass parameter $c$ [270].

Thus, the localization of the third generation quark doublet $q_L$ leads to potentially large flavor violations, not only with the top quark, but also with $b_L$.

This induced flavor violation of KK gauge bosons with $b_L$ (we assume $b_R$ localized on the Planck brane) is, in principle, constrained by the precise measurement of the $Z^0 \to b\bar{b}$ interactions at the $Z^0$-pole. For instance, Ref. [307] considers a $\mathrm{SU}(3)_c \times \mathrm{SU}(2)_L \times \mathrm{SU}(2)_R \times \mathrm{U}(1)_{B-L}$ gauge theory in the bulk. After electroweak symmetry breaking the $Z^0$ mixes with its KK excitations, as well as with the KK modes of a $Z^0$'. This generates $\delta g_L^b \lesssim \mathcal{O}(1\%)g_L^b$, compatible with current bounds, as long as $c_L \gtrsim 0.3$. Even if this is considered, it still leaves a large flavor-violating coupling of the first KK excitations to the $b_L$, as we can see from Fig. 5-48. More generally, for instance, in the case of strong gauge coupling [277], the effects can be even larger.

### Signals in $b \to s$ and other hadronic processes

As discussed above, the flavor-changing exchange of KK gluons leads to four-fermion interactions contributing to the quark level processes $b \to d\bar{q}q$ and $b \to s\bar{q}q$, with $q = u, d, s$. We are interested in contributions that are typically of $\simeq \alpha_s$ strength due to the fact that in the product of a third generation current times a lighter quark current the enhancement in the former is (at least partially) canceled by the suppression of the latter. At low energies, the $b \to d_i\bar{q}q$ processes are described by the effective Hamiltonian [278]

$$\mathcal{H}\mathrm{eff.} = \frac{4G_F}{\sqrt{2}}\,V_{ub}V_{ui}^*\,[C_1(\mu)O_1 + C_2(\mu)O_2] - \frac{4G_F}{\sqrt{2}}\,V_{tb}V_{ti}^*\,\sum_{j=3}^{10} C_i(\mu)O_i + \mathrm{h.c.}, \tag{5.8}$$

where $i = d, s$ and the operator basis can be found in Ref. [278].

In the Standard Model, the operators $\{O_3 - O_6\}$ are generated from one-loop gluonic penguin diagrams, whereas operators $\{O_7 - O_{10}\}$ arise from one loop electroweak penguin diagrams. The Hamiltonian describing the $b \to s\bar{q}q$ decays is obtained by replacing $V_{ts}^*$ for $V_{td}^*$ in Eq. (5.8). Contributions from physics beyond the Standard Model affect the Wilson coefficients at some high energy scale. Additionally, New Physics could generate low energy interactions with the "wrong chirality" with respect to the Standard Model. This would expand the operator basis to include operators of the form $(\bar{s}_R \Gamma b_R)(\bar{q}_\lambda \Gamma q_\lambda)$, where $\Gamma$ reflects the Dirac and color structure and $\lambda = L, R$.

The exchange of color-octet gauge bosons such as KK gluons of the Randall-Sundrum scenario generate flavor-violating currents with the third generation quarks. Upon diagonalization of the Yukawa matrix, this results in FCNCs at tree level due to the absence of a complete GIM cancellation. The off-diagonal elements of the left and right, up and down quark rotation matrices $U_{L,R}$ and $D_{L,R}$ determine the strength of the flavor violation. In the Standard Model, only the left-handed rotations are observable through $V_{\rm CKM} = U_L^\dagger D_L$. Here, $D^{bs}_{L,R}$, $D^{bd}_{L,R}$, $U^{tc}_{L,R}$, *etc.*, become actual observables.

The tree level flavor-changing interactions induced by the color-octet exchange are described by a new addition to the effective Hamiltonian that can, in general, be written, for $b \to s$ transitions, as

$$\delta\mathcal{H}_{\rm eff.} = \frac{4\pi\alpha_s}{M_G^2} D_L^{bb*} D_L^{bs} |D_L^{qq}|^2 e^{-i\omega} \chi \left(\bar{s}_L \gamma_\mu T^a b_L\right) \left(\bar{q}_L \gamma^\mu T^a q_L\right) + \text{h.c.}\,. \tag{5.9}$$

where $\omega$ is the phase relative to the Standard Model contribution; and $\chi \simeq \mathcal{O}(1)$ is a model-dependent parameter. For instance, $\chi = 1$ corresponds to the choice of $c_L \simeq 0$ that gives a coupling of the KK gauge boson about five times larger than the corresponding Standard Model value for that gauge coupling. An expression analogous to (5.9) is obtained by replacing $d$ for $s$ in it. This would induce effects in $b \to d$ processes.

From Eq. (5.9) we can see that the color-octet exchange generates contributions to all gluonic penguin operators. Assuming that the diagonal factors obey $|D_L^{qq}| \simeq 1$, these will have the form

$$\delta C_i = -\pi\alpha_s(M_G) \left(\frac{v}{M_G}\right)^2 \left|\frac{D_L^{bs*}}{V_{tb}V_{ts}^*}\right| e^{-i\omega} f_i\, \chi\,, \tag{5.10}$$

where $f_3 = f_5 = -1/3$ and $f_4 = f_6 = 1$, and $v = 246$ GeV. This represents a shift in the Wilson coefficients at the high scale. We then must evolve the new coefficients down to $\mu = m_b$ by making use of renormalization group evolution [278]. The effects described by Eq. (5.10) are somewhat diluted in the final answer due to a large contribution from the mixing with $O_2$. Still, potentially large effects remain.

The phase $\omega$ in Eq. (5.9) is, in principle, a free parameter in most models and could be large. This is even true in the left-handed sector, since, in general, $V_{\rm CKM}$ comes from both the up and down quark rotation. Only if we were to argue that all of the CKM matrix comes from the down sector we could guarantee that $\omega = 0$. Furthermore, there is no such constraint in the right-handed quark sector.

We now examine what kind of effects these flavor-violating terms could produce. Their typical strength is given by $\alpha_s$ times a CKM-like factor coming from the $D_{L,R}$ off-diagonal elements connecting to $b$. The fact that the coupling is tree-level is somewhat compensated by the suppression factor $(v/M_G)^2$, for $M_G \simeq \mathcal{O}(1)$ TeV. Still, the contributions of Eq. (5.9) are typically larger than the Standard Model Wilson coefficients at the scale $M_W$, and, in fact, are comparable to the Wilson coefficients at the scale $m_b$. They could therefore significantly affect both the rates and the *CP* asymmetries.

The $b \to s\bar{s}s$ and $b \to s\bar{d}d$ pure penguin processes, such as $B_d \to \phi K^0_S$, $B_d \to \eta' K^0_S$ and $B_d \to \pi^0 K^0_S$, contain only small tree-level contamination of Standard Model amplitudes. These decays thus constitute a potentially clean test of the Standard Model, since their *CP* asymmetries are predicted to be a measurement of $\sin 2\beta_{J/\psi\, K^0_S}$, the same angle of the unitarity triangle as in the $b \to c\bar{c}s$ tree level processes such as $B_d \to J/\psi\, K^0_S$, up to small corrections. In order to estimate these effects and compare them to the current experimental information on these decay modes, we will compute the matrix elements of $\mathcal{H}_{\rm eff.}$ in the factorization approximation [279] as described in Ref. [280]. Although the predictions for the branching ratios suffer from significant uncertainties, we expect that these largely cancel when considering the effects in the *CP* asymmetries. Thus the *CP* asymmetries in non-leptonic $b \to s$ penguin-dominated processes constitute a suitable set of observables to test the effects of these color-octet states.

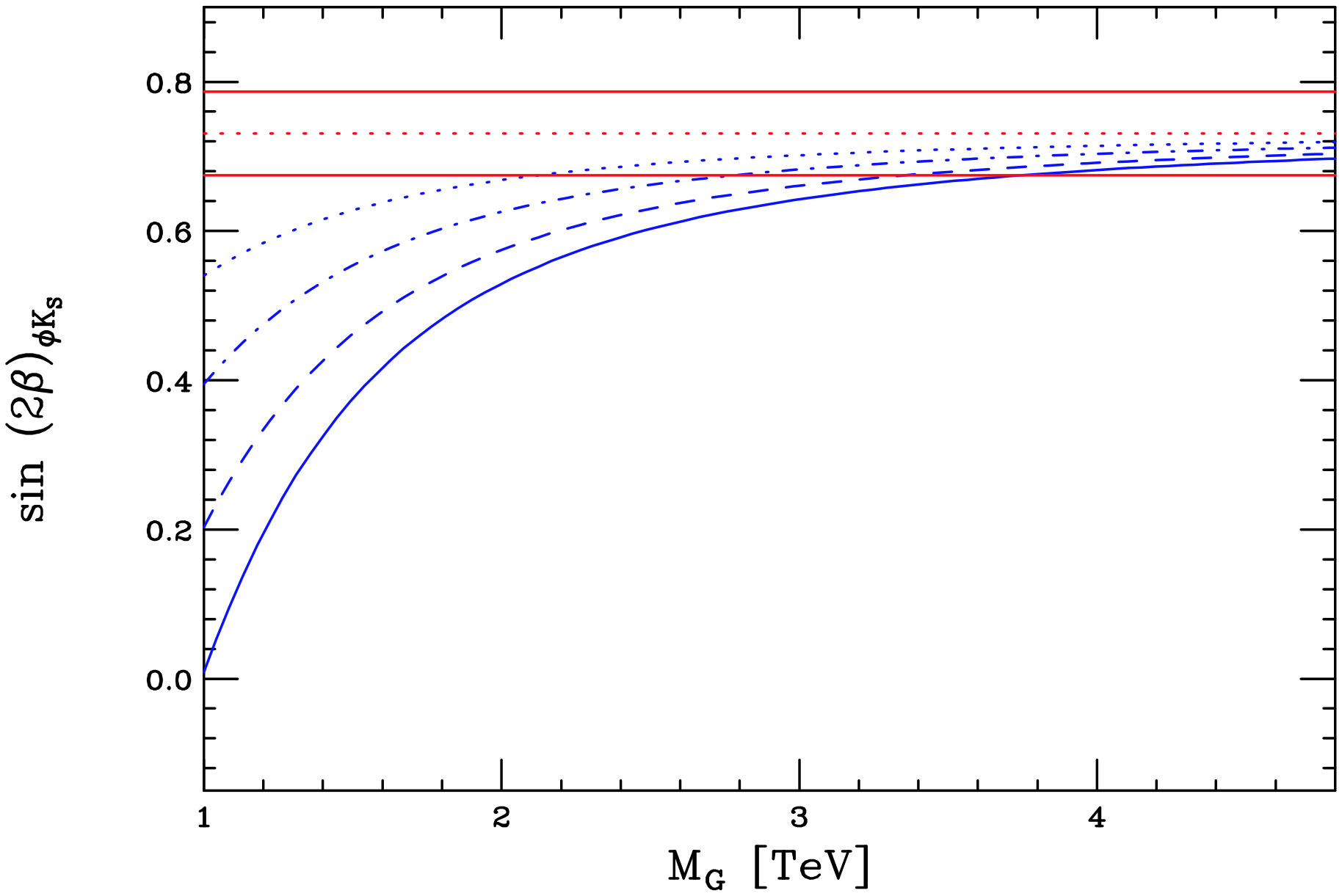

**Figure 5-49.** *The quantity to be extracted from the CP violation asymmetry in $B_d^0 \to \phi K_S^0$ vs. the heavy gluon mass and for various values of the decay amplitude phase $\omega$. The curves correspond to $\pi/3$ (solid), $\pi/4$ (dashed) and $\pi/6$ (dot-dash), and $\pi/10$ (dotted). The horizontal band corresponds to the world average value [281] as extracted from $B_d \to J/\psi K_S^0$, $sin(2\beta)_{\psi K_S^0} = 0.731 \pm 0.056$. From Ref. [274].*

In Fig. 5-49 we plot $\sin 2\beta_{\phi K_S^0}$ *vs.* the KK gluon mass for various values of the phase $\omega$. Here, for concreteness, we have taken $|D_L^{bs}| = |V_{tb}^* V_{ts}|$, assumed $b_R$ is localized on the Planck brane, and $\chi = 1$ in order to illustrate the size of the effect. The horizontal band corresponds to the $B_d \to J/\psi\, K_S^0$ measurement, $\sin 2\beta_{J/\psi\, K_S^0} = 0.731 \pm 0.056$ [281]. Only positive values of $\omega$ are shown, as negative values increase $\sin 2\beta$, contrary to the trend in the data. We see that there are sizable deviations from the Standard Model expectation for values in the region of interest $M_G \gtrsim 1$ TeV. This will be the case as long as $|D_L^{bs}| \simeq |V_{ts}|$, and $\chi \simeq \mathcal{O}(1)$, both natural assumptions.

For $D_L^{bs}$, this is valid as long as a significant fraction of the corresponding CKM elements comes from the down quark rotation. On the other hand, $\chi \simeq \mathcal{O}(1)$ in all the models considered here. In addition, we have not considered the effects of of $D_R^{bs}$, which could make the effects even larger.

Similar effects are present in $B_d \to \eta' K_S^0$, and $B_d \to K^+ K^- K_S^0$ also dominated by the $b \to s\bar{s}s$ penguin contribution; as well as in the $b \to s\bar{d}d$ mode $B_d \to \pi^0 K_S^0$ [274].

The flavor-violating exchange of the KK gluon also induces an extremely large contribution to $B_s$–$\overline{B}_s$ mixing, roughly given by

$$\Delta m_{B_s} \simeq 200\text{ps}^{-1} \left(\frac{|D_L^{bs}|}{\lambda^2}\right)^2 \left(\frac{2\text{ TeV}}{M_G}\right)^2 \left(\frac{g_{10}}{5}\right)^2 , \tag{5.11}$$

where $\lambda \simeq 0.22$ is the Cabibbo angle, and $g_{10} \equiv g_1/g$ represents the enhancement of the zero-mode fermion coupling to the first KK gluon with respect to the four-dimensional gauge coupling, as plotted in Fig. 5-48. The contribution of Eq. (5.11) by itself is about 10 times larger than the Standard Model one for this natural choice of parameters, and would produce $B_s$ oscillations too rapid for observation at the Tevatron or in similar experiments.

There are also similar contributions to $\Delta m_{B_d}$, when $D_L^{bs}$ is replaced by $D_L^{bd}$. These were examined in Ref. [282] in the context of topcolor assisted technicolor, a much more constrained brand of topcolor than the one we consider here. The bounds found in Ref. [282] can be accommodated, as long as $|D_L^{bd}| \lesssim |V_{td}|$, which is not a very strong constraint.

Thus, we see that the flavor violation effects of the first KK gluon excitation in Randall-Sundrum scenarios where the SU(3)$_c$ fields propagate in the bulk can be significant in non-leptonic $B$ decays, specifically in their $CP$ asymmetries. The dominance of these effects over those induced by "weak" KK excitations, such as KK $Z^0$ and $Z^0$'s, due to the larger coupling, would explain the absence of any effects in $b \to s\ell^+\ell^-$ processes, where up to now, the data is consistent with Standard Model expectations [283]. Deviations in the $CP$ asymmetries of $b \to s$ nonleptonic processes would naturally be the first signal of New Physics in these scenarios. These very same effects can be obtained by the exchange of the heavy gluons present in generic topcolor models.

These effects can also be obtained in generic topcolor models. It is not possible to distinguish these two sources using $B$ physics alone. This is true of any color-octet flavor-violating gauge interaction that couples strongly to the third generation. Other model building avenues addressing fermion masses might result in similar effects. In addition, the large contributions to $B_s$ mixing, perhaps rendering $\Delta m_{B_s}$ too large to be observed, is an inescapable prediction in this scenario, as it can be seen in Eq. (5.11), but it is also present in many other New Physics scenarios that produce large effects in $b \to s$ non-leptonic decays [306].

There will also be contributions from the heavy gluons to other non-leptonic $B$ decays, such as $B \to \pi\pi$, *etc.*These modes have less clean Standard Model predictions. However, if the deviations hinted in the current data are confirmed by data samples of $500\,\text{fb}^{-1}$, to be accumulated in the next few years, it might prove of great importance to confirm the existence of these effects in less clean modes, perhaps requiring even larger data samples. Even if the heavy gluons are directly observed at the LHC, their flavor-violating interactions will be less obvious there than from large enough $B$ physics samples. Thus, if flavor-violating interactions are observed at the LHC, high precision $B$ physics experiments could prove crucial to elucidate their role in fermion mass generation.

#### Signals in $b \to s\ell^+\ell^-$

Since the wave-function of the $Z^0$ is pushed away from the IR brane by the boundary conditions (or a large vev), there will be non-negligible tree-level FCNC couplings of $q^T = (t_L\; b_L)^T$ and $t_R$ with the $Z^0$, since these must be localized not too far from this brane. We define the effective $Zbs$ coupling by

$$\mathcal{L}_{Zbs} = \frac{g^2}{4\pi^2}\,\frac{g}{2\cos\theta_W}\left(Z_{bs}\,\bar{b}_L\gamma^\mu s_L + Z'_{bs}\,\bar{b}_R\gamma^\mu s_R\right)Z_\mu, \tag{5.12}$$

where $Z_{bs}$ and $Z'_{bs}$ encode both the one loop Standard Model as well as New Physics contributions. Up to a factor of order one, the tree-level FCNC vertex induced by the flavor-violating coupling results in [277]

$$\delta Z_{bs} \simeq -\Big(-\frac{1}{2}+\frac{1}{3}\sin^2\theta_W\Big)D_L^{bs}\,\frac{8\pi^2}{g^2}\left(\frac{v^2}{m_1^2}\right)\left(\frac{g_L^2}{\pi R\,g^2}\right)f \simeq \frac{D_L^{bs}}{2}\,\frac{g_L^2}{\pi R\,g^2}N, \tag{5.13}$$

where $f$ is defined in Ref. [277], and in order to respect the bounds from $Z \to b\bar{b}$, $|f| \lesssim \mathcal{O}(1)$. With the natural assumption $D_L^{bs} \simeq V_{tb}^* V_{ts}$, and reasonably small brane couplings $g_L^2/\pi R g^2 = O(1)$, the correction is of the same order as the Standard Model contribution to this vertex, which is [284] $Z_{bs}^{\rm SM} \simeq -0.04$ ($Z_{bs}^{\prime\rm SM} \simeq 0$). This leads to potentially observable effects in $b \to s\ell^+\ell^-$ decays, although the current experimental data, $|Z_{bs}| \lesssim 0.08$ [284], is not greatly constraining. The effects, however, could be larger in the case of strong bulk gauge couplings [277], which may require somewhat smaller mixing angles. The effect of Eq. (5.13) also contributes to hadronic modes, such as $B \to \phi K_S^0$, although there it must compete with the parametrically larger contributions from gluonic penguins.

#### $D^0\overline{D}^0$ mixing

Finally, the large flavor-violating coupling of the top quark, particularly $t_R$, may lead to a large contribution to $D^0\overline{D}^0$ mixing. This has contributions both from KK gluon and $Z^0$ exchanges and has the form [277]

$$\Delta m_D \simeq 4\pi\alpha_s\,\frac{\chi(c_R)}{2m_1^2}\,\frac{(U_R^{tu*}\,U_R^{tc})^2}{2m_D}\,\langle D^0|(\bar{c}_R\gamma_\mu u_R)(\bar{c}_R\gamma^\mu u_R)|\overline{D}^0\rangle, \tag{5.14}$$

for the KK gluon exchange. Here, $U_R$ is the rotation matrix for right-handed up quarks, and $\chi(c_R)$ is a function of $c_R$ which gives the enhancement due to the strong coupling of the KK gluons to $t_R$. For instance, for $c_R \simeq 0$ and small brane couplings, $\chi \simeq 16$. To estimate the contribution to $\Delta m_D$, we need the quark rotation matrix elements. If we take $U_R^{tu*} U_R^{tc} \simeq \sin^5\theta_C$, with $\sin\theta_C \simeq 0.2$ the Cabibbo angle, then the current experimental limit on $\Delta m_D$ translates[5] into $m_1 \gtrsim 2\ \mathrm{TeV}$.In the strong bulk coupling case, $\chi(c_R)$ can be enhanced and somewhat larger $c_R$ or smaller mixing angles may be required. The contribution from the $Z^0$ is generically the same order, but somewhat smaller. We thus find that the effect can be consistent with, but naturally close to, the current experimental limit. Similar contributions come from the interactions of $t_L$, but they are typically smaller than those from $t_R$, because of larger values of $c$.

[5]Unlike for $U_L$ and $D_L$, there is, in principle, no reason why $U_R$ must have such scaling with the Cabibbo angle.

### 5.4.5 Warped Extra Dimensions Signatures in $B$ Decays

$\succ$ K. Agashe $\prec$

#### Introduction

This section is based on [285], where the reader is referred for further details and for references.

Consider the Randall-Sundrum (RS1) model which is a compact slice of AdS$_5$,

$$ds^2 = e^{-2k|\theta|r_c}\eta^{\mu\nu}dx_\mu dx_\nu + r_c^2 d\theta^2, \ -\pi \le \theta \le \pi, \tag{5.15}$$

where the extra-dimensional interval is realized as an orbifolded circle of radius $r_c$. The two orbifold fixed points, $\theta = 0, \pi$, correspond to the "UV" (or "Planck) and "IR" (or "TeV") branes respectively. In warped spacetimes the relationship between 5D mass scales and 4D mass scales (in an effective 4D description) depends on location in the extra dimension through the warp factor, $e^{-k|\theta|r_c}$. This allows large 4D mass hierarchies to naturally arise without large hierarchies in the defining 5D theory, whose mass parameters are taken to be of order the observed Planck scale, $M_{Pl} \sim 10^{18}$ GeV. For example, the 4D massless graviton mode is localized near the UV brane, while Higgs physics is taken to be localized on the IR brane. In the 4D effective theory one then finds

$$\text{Weak Scale} \sim M_{Planck} e^{-k\pi r_c}. \tag{5.16}$$

A modestly large radius, *i.e.*, $k\pi r_c \sim \log\left(M_{Planck}/\text{TeV}\right) \sim 30$, can then accommodate a TeV-size weak scale. Kaluza-Klein (KK) graviton resonances have $\sim ke^{-k\pi r_c}$, *i.e.*, TeV-scale masses since their wave functions are also localized near the IR brane.

In the original RS1 model, it was assumed that the entire Standard Model (*i.e.*, including gauge and fermion fields) is localized on TeV brane. Thus, the effective UV cut-off for gauge and fermion fields and hence the scale suppressing higher-dimensional operators, is $\sim$ TeV, *e.g.*, the same as for Higgs. However, bounds from electroweak (EW) precision data on this cut-off are $\sim 5 - 10$ TeV, whereas those from flavor-changing neutral currents (FCNC's) (for example, $K^0\overline{K}^0$ mixing) are $\sim 1000$ TeV. Thus, to stabilize the electroweak scale requires fine-tuning, *e.g.*, even though RS1 explains the big hierarchy between Planck and electroweak scale, it has a "little" hierarchy problem.

#### Bulk fermions

A solution to this problem is to move Standard Model gauge and fermion fields into the bulk. Let us begin with how bulk fermions enable us to evade flavor constraints. The localization of the wavefunction of the massless chiral fermion mode is controlled by the $c$-parameter. In the warped scenario, for $c > 1/2$ ($c < 1/2$) the zero mode is localized near the Planck (TeV) brane, whereas for $c = 1/2$, the wave function is *flat*.

Therefore, we choose $c > 1/2$ for light fermions, so that the effective UV cut-off is $\gg$ TeV, and thus FCNC's are suppressed. Also this naturally results in a small $4D$ Yukawa coupling to the Higgs on TeV brane without any hierarchies in the fundamental $5D$ Yukawa. Similarly, we choose $c \ll 1/2$ for the top quark to obtain an $\mathcal{O}(1)$ Yukawa. If *left*-handed top is near TeV brane, then there are FCNC's involving $b_L$, as follows.

Since fermions are in the bulk, we also have $5D$ gauge fields and we can show that in this set-up *high*-scale unification can be accommodated.

#### Couplings of fermion to gauge KK mode

The flavor violation involving $b_L$ is due to KK modes of gauge fields, so that we need to consider couplings of these modes to fermions. We can show that wave functions of gauge KK modes are peaked near TeV brane (just like for graviton KK modes) so that their coupling to TeV brane fields (for example, the Higgs) is enhanced compared to that of zero mode (which has a flat profile) by $\approx \sqrt{2k\pi r_c}$. Thus, the coupling of gauge KK modes to zero mode fermions, denoted by $g^{(n)}$, in terms of $g^{(0)}$ (the coupling of zero mode of gauge field) has the form:

$$\begin{aligned} g^{(n)} & \sim g^{(0)} \times \sqrt{k\pi r_c},\ c \ll 1/2 \text{ (as for the Higgs)} \\ & = 0,\ c = 1/2 \text{ (fermion profile is flat)} \\ & \sim \frac{g^{(0)}}{\sqrt{k\pi r_c}},\ c \gtrsim 1/2 \text{ (independent of } c). \end{aligned} \tag{5.17}$$

Due to this coupling, there is a shift in coupling of fermions to the physical $Z^0$ from integrating out gauge KK modes (see Fig. 2 in Ref. [286]):

$$\delta g^{phys} \approx \sum_n g^{(n)} (-1)^n \sqrt{2k\pi r_c} M_Z^2 / m_{KK}^{(n)\,2}. \tag{5.18}$$

This shift is universal for light fermions, since light fermions have $c > 1/2$ and thus can absorbed into the $S$ parameter.

### Choice of $b_L$ localization

It is clear that we prefer $c$ for $(t,b)_L \ll 1/2$ inn order to obtain a top Yukawa of $\sim 1$ without a too large $5D$ Yukawa, but this implies a large shift in the coupling of $b_L$ to $Z^0$ (relative to that for light fermions). Thus, there is a tension between obtaining top Yukawa and not shifting the coupling of $b_L$ to $Z^0$. As a compromise, we choose $c$ for $(t,b)_L$ $\sim 0.4 - 0.3$ (corresponding to a coupling of $b_L$ to gauge KK modes $g^{(n)}/g^{(0)} \sim \mathcal{O}(1)$: see Eq. (5.17)) so that with KK masses $\sim 3 - 4$ TeV, the shift in coupling of $b_L$ to the $Z^0$ is $\sim 1\%$ (see Eq. (5.18)) which is allowed by precision electroweak data.

In order to obtain a top Yukawa $\sim 1$, we choose $c$ for $t_R \ll 1/2$ and $c$ for $b_R > 1/2$ to obtain $m_b \ll m_t$. We can further show that $3 - 4$ TeV KK masses are consistent with electroweak data ($S$ and $T$ parameters) provided we gauge SU(2)$_R$ in the bulk [286].

### Flavor violation from gauge KK modes

The flavor-violating couplings of zero-mode fermions to gauge KK modes are a result of going from a weak/gauge to a mass eigenstate basis:

$$D_L^\dagger \text{diag} \left[ g^{(n)} \left( c_{L\ d} \right), g^{(n)} \left( c_{L\ s} \right), g^{(n)} \left( c_{L\ b} \right) \right] D_L, \tag{5.19}$$

where $D_L$ is the unitary transformation from weak to mass eigenstate basis for left-handed down quarks.

Tree level KK gluon exchange contributes to $B^0\overline{B}^0$ mixing: the coefficient of $\left(\bar{b}_L \gamma^\mu d_L\right)^2$ is

$$\left[ (D_L)_{13} \right]^2 \sum_n g^{(n)\,2} / m_{KK}^{(n)\,2},$$

whereas the Standard Model box diagram contribution has coefficient

$$\sim \left(V_{tb}^* V_{td}\right)^2 g^4 / \left(16\pi^2\right) \times 1/m_W^2 \sim \left(V_{tb}^* V_{td}\right)^2 / (4 \text{ TeV})^2 .$$

Since $c$ for $(t,b)_L \sim 0.3 - 0.4$, *e.g.*, $g^{(n)}/g^{(0)} \sim \mathcal{O}(1)$, the KK gluon exchange contribution to $B^0\overline{B}^0$ mixing is comparable to the Standard Model box diagram for $m_{KK} \sim 3 - 4$ TeV. Such a large contribution is allowed, since tree level measurements of CKM matrix elements, combined with unitarity, do not really constrain $V_{td}$ in the Standard Model, so that the Standard Model contribution to $B^0\overline{B}^0$ mixing has a large uncertainty. Explicitly, the Standard Model contribution occurs at loop level with $g/m_W \sim v$ suppression, whereas the KK gluon contribution to $B^0\overline{B}^0$ mixing is at the tree level (with $\mathcal{O}(1)$ coupling of $b_L$ to the gauge KK mode), but suppressed by $3 - 4$ TeV $\sim 4\pi v$ KK masses, so that the two contributions are of the same size.

In other words, due to the large top mass, $c$ for $b_L$ is smaller (coupling to the gauge KK modes, $g^{(n)}$, is larger) than expected from $m_b$. This induces a large deviation from universality of the KK gluon coupling to left-handed down-type quarks. This is similar to the Standard Model, where there is no GIM suppression (due to the large $m_t$) in $b \to s, d$ or in the imaginary part of $s \to d$ (as opposed to the real part of $s \to d$)

This also shows that for $c$ for $(t, b)_L \stackrel{<}{\sim} 0.3$ (*e.g.*, coupling of KK gluon to $b_L \stackrel{>}{\sim} O(1)$), $B^0\overline{B}^0$ mixing requires $m_{KK} \stackrel{>}{\sim} 4$ TeV—this is an independent (of $Z \to b\bar{b}$) lower limit on $c$ for $(t, b)_L$, given that $m_{KK} \stackrel{<}{\sim} 4$ TeV by naturalness.

We next consider $b \to s\bar{s}s$. The KK gluon coupling to $s$ is suppressed by $\sim 1/\sqrt{k\pi r_c}$. Hence we see that the effect of tree level exchange of a KK gluon in this decay is smaller than the Standard Model QCD penguin, for a choice of parameters for which the effect in $B^0\overline{B}^0$ mixing is comparable to the Standard Model value.

**Flavor-violating coupling to the $Z^0$: $b \to s\ell^+\ell^-$**

The direct effect of KK $Z$ exchange is suppressed (compared to KK gluon exchange) by $\sim g_Z^2/g_s^2$. However, there is an indirect effect of KK $Z^0$ exchange: a shift in the coupling of $b_L$ to *physical* $Z^0$ by $\sim 1\%$. In turn, this results in a flavor-violating coupling to the $Z^0$ after going to a mass eigenstate basis:

$$D_L^\dagger \text{diag} \left[ \delta \left( g_Z^{d_L} \right), \delta \left( g_Z^{s_L} \right), \delta \left( g_Z^{b_L} \right) \right] D_L. \tag{5.20}$$

Since we have to allow $\delta \left( g_Z^{b_L} \right) \sim 1\%$, we get (relative to the the standard coupling of $d_L$ to the $Z^0$)

$$b_L s_L Z \sim 1\ \%\ V_{ts}\,. \tag{5.21}$$

Using these couplings, we see that there are contributions to $b \to s f\bar{f}$ that are comparable to the Standard Model $Z^0$ penguin, with a coefficient $\sim V_{ts}\ g^2/\left(16\pi^2\right) g_Z^2/m_Z^2$, (roughly $1\%$ in Eq. (5.21) comparable to the loop factor in the Standard Model $Z^0$ penguin)

This leads to a smoking gun signal in $b \to s\ell^+\ell^-$: the error in the theory prediction is $\sim 15\%$ (since $V_{ts}$ in the Standard Model is constrained by tree level measurements of CKM matrix elements and unitarity, unlike $V_{td}$), so that the $\mathcal{O}(1)$ effect (relative to the Standard Model) is observable (current experiment error on measurement of this branching ratio is $\sim 30\%$). What is interesting is that the coupling of charged leptons to the $Z^0$ is almost axial, whereas the coupling to photons is vector; the coefficient of the operator with only axial coupling of leptons gets a new contribution, so that the angular distributions (forward-backward asymmetry) and spectrum of $\ell^+\ell^-$ are affected.

In $b \to s\bar{s}s$, the contribution of the Standard Model QCD penguin is larger than the Standard Model $Z^0$ penguin (roughly by $\sim g_Z^2/g_s^2$) and so the effect of $b_L s_L Z$ coupling (which is comparable to the Standard Model $Z^0$ penguin) is less than $\mathcal{O}(1)$ (roughly $20\%$). This might be observable.

**Conclusions**

To summarize, we have shown that bulk fermion profiles in RS1 can explain the hierarchies of fermion masses. With only first and second generations, FCNC's are small. However, including the third generation produces interesting effects. There is tension between obtaining a large top mass and not affecting the coupling of $b_L$ to the $Z^0$: as a result, we have to compromise, and allow a shift in the coupling of $b_L$ to $Z^0$ by $\sim 1\%$. This, in turn, leads to a flavor-violating coupling to the $Z^0$, and a smoking gun signal in $b \to s\ell^+\ell^-$. Finally, using the AdS/CFT correspondence, this RS1 model is dual to a $4D$ composite Higgs model; thus a strongly interacting Higgs sector can address flavor issues.

## 5.5 Lepton Flavor Violation

### 5.5.1 Lepton flavor-violating decays of the $\tau$ at a Super $B$ Factory

>– O. Igonkina –<

#### Motivation

The lepton flavor-violating decays of the $\tau$ (LFV) are an excellent base for testing modern theoretical models such as supersymmetry, technicolor or models with extra dimensions.

Recent results from the neutrino oscillations experiments [287],[288],[289],[290] suggest that LFV decays do occur. However, the branching ratios expected in the charged lepton decays in the Standard Model with neutrino mixing alone is not more than $10^{-14}$[291], while many other theories predict values of the order of $10^{-10} - 10^{-7}$, which should be within a reach of the Super $B$ Factory. The predictions are summarized in Table 5-19. Among them are SUSY with different types of symmetry breaking, models with additional heavy neutrinos and models with extra gauge boson $Z^0$'.

Different LFV decays such as $\tau \to \ell\gamma$, $\tau \to \ell\ell\ell$, $\tau \to \ell hh$ (where $\ell$ is $e$ or $\mu$, and $h$ is a hadron) have different importance for these models. Therefore, by studying each of these channels, one can discriminate between the models ,and extract or restrict their parameters.

**Table 5-19.** *Predictions for the branching ratios of $\tau \to \ell\gamma$ and $\tau \to \ell\ell\ell$ in different models.*

| Model | $\tau \to \ell\gamma$ | $\tau \to \ell\ell\ell$ | Ref. |
|---|---|---|---|
| SM with lepton CKM | $10^{-40}$ | $10^{-14}$ | [291] |
| SM with left-handed heavy Dirac neutrino | $< 10^{-18}$ | $< 10^{-18}$ | [308] |
| SM with right-handed heavy Majorana neutrino | $< 10^{-9}$ | $< 10^{-10}$ | [309] |
| SM with left- and right-handed neutral singlets | $10^{-8}$ | $10^{-9}$ | [309] |
| MSSM with right-handed heavy Majorana neutrino | $10^{-10}$ | $10^{-9}$ | [310] |
| MSSM with seesaw | $10^{-7}$ | | [311] |
| left-right SUSY | $10^{-10}$ | $10^{-10}$ | [310] |
| SUSY SO(10) | $10^{-8}$ | | [193] |
| SUSY-GUT | $10^{-8}$ | | [312] |
| SUSY with neutral Higgs | $10^{-10}$ | $10^{-10} - 10^{-7}$ | [313],[314],[293] |
| SUSY with Higgs triplet | | $10^{-7}$ | [315] |
| gauge mediated SUSY breaking | $10^{-8}$ | | [316] |
| MSSM with universal soft SUSY breaking | $10^{-7}$ | $10^{-9}$ | [317] |
| MSSM with non-universal soft SUSY breaking | $10^{-10}$ | $10^{-6}$ | [318] |
| Non universal $Z'$ (technicolor) | $10^{-9}$ | $10^{-8}$ | [319] |
| two Higgs doublet III | $10^{-15}$ | $10^{-17}$ | [320] |
| seesaw with extra dimensions | $10^{-11}$ | | [321] |

#### The experimental situation

No signature for $\tau$ LFV decays has been yet found. The strictest upper limits, of order $10^{-7} - 10^{-6}$ (see Table 5-20), are limited by the size of the accumulated data samples; a data sample of 0.5 or 10 ab$^{-1}$ will significantly improve our understanding of the mechanism of lepton flavor violation. We present the prospects for measuring $\tau$ LFV decays at the future Super $B$ Factory, assuming that the center-of-mass energy and the detector performance similar to *BABAR*. The reconstruction of $\tau \to \ell\ell\ell$ and $\tau \to \ell\gamma$ is based on the unique topology of the $\tau\tau$ events at $\sqrt{s} \sim 10$ GeV, where $\tau$'s have a significant boost, and the decay products are easily separated. The former decays are selected using a 1-3 topology, while the latter satisfy a 1-1 topology. Additional requirements are: a positive identification of the leptons from the signal decay and several kinematic cuts on the tracks in the event. The determination of the number of signal events (or the setting of an upper limit) is based on the reconstructed invariant mass and the energy of the candidates on the signal side.

**Table 5-20.** *The current strictest upper limits on the branching ratios of $\tau \to \ell\gamma$ and $\tau \to \ell\ell\ell$. The $\bar{b}$ $\tau \to \ell\ell\ell$results were published while these Proceedings were in preparation. The Belle $\tau \to \mu\gamma$, result following [292], is shown.*

| | | | | |
|---|---|---|---|---|
| $\mathcal{B}(\tau \to \ell\gamma)$ | $< 3 \cdot 10^{-6}$ | CLEO | (4.8 $\text{fb}^{-1}$) | [322], [323] |
| $\mathcal{B}(\tau \to \mu\gamma)$ | $< 2 \cdot 10^{-6}$ | *BABAR*(preliminary) | (63 $\text{fb}^{-1}$) | [324] |
| $\mathcal{B}(\tau \to \mu\gamma)$ | $< 5 \cdot 10^{-7}$ | Belle | (86.3 $\text{fb}^{-1}$) | [325] |
| $\mathcal{B}(\tau \to \ell\ell\ell)$ | $< 2 \cdot 10^{-6}$ | CLEO | (4.8 $\text{fb}^{-1}$) | [326] |
| $\mathcal{B}(\tau \to \ell\ell\ell)$ | $< 3 \cdot 10^{-7}$ | Belle (preliminary) | (48.6 $\text{fb}^{-1}$) | [327] |
| $\mathcal{B}(\tau \to \ell\ell\ell)$ | $< 1 - 3 \cdot 10^{-7}$ | *BABAR* | (82 $\text{fb}^{-1}$) | [328] |
| $\mathcal{B}(\tau \to \ell hh)$ | $< 2 - 15 \cdot 10^{-6}$ | CLEO | (4.8 $\text{fb}^{-1}$) | [326] |

**Table 5-21.** *Expected signal efficiency, expected background level and the sensitivity to upper limit on LFV $\tau$ decays at 90% CL for different sizes of data samples.*

| | $\tau \to \ell\ell\ell$ | | |
|---|---|---|---|
| | 90 $\text{fb}^{-1}$ | 0.5 $\text{ab}^{-1}$ | 10 $\text{ab}^{-1}$ |
| Efficiency | 8.5% | 8% | 7% |
| Background | 0.4 | 1 | 1 |
| UL Sensitivity | $2 \cdot 10^{-7}$ | $4 \cdot 10^{-8}$ | $3 \cdot 10^{-9}$ |
| | $\tau \to \ell\gamma$ | | |
| | 63 $\text{fb}^{-1}$ | 0.5 $\text{ab}^{-1}$ | 10 $\text{ab}^{-1}$ |
| Efficiency | 5% | 4% | 4% |
| Background | 8 | 8 | 180 |
| UL Sensitivity | $1 \cdot 10^{-6}$ | $2 \cdot 10^{-7}$ | $3 \cdot 10^{-8}$ |

The main backgrounds for $\tau \to \ell\ell\ell$ are hadronic events resulting from hadron misidentification. The study shows that the $\tau \to \ell\ell\ell$ decay is well-controlled by cuts. The required suppression of the background for $0.5\,\text{ab}^{-1}$ is achieved by additional kinematic cuts on the 1-prong side; strengthening lepton identification is essential for the analysis of the $10\,\text{ab}^{-1}$ sample. The decays $\tau \to \ell\gamma$ are contaminated with non-LFV process $\tau \to \ell\nu\nu\gamma$, which is more difficult to suppress. However, the gain due to the large statistics sample is still significant, in spite of high level of background. Table 5-21 shows the upper limit sensitivity if no signal is observed. The calculation of upper limits is done following [292]. Further improvement can be made if a new detector has better lepton identification (in particular for soft leptons, with momenta below 0.5 GeV), more accurate momentum reconstruction and larger acceptance. Precise reconstruction of the photon energy is a key issue for the analysis of the $\tau \to \ell\gamma$ decay.

By comparing Tables 5-19 and 5-21 one can see that a sample of 10 $\text{ab}^{-1}$ will provide extremely interesting measurements. Such a measurement of $\tau \to \mu\gamma$ will be sensitive to a GUT scale $m_0$ up to 200 GeV, while observation of $\tau \to \mu\mu\mu$ will be sensitive to the slepton mass. It is interesting to notice that according to [293] the $\mathcal{B}(\tau \to \ell\ell\ell)$ is of the order of $10^{-7}$ if supersymmetric particles are heavier than 1 TeV. In such a scenario, the Super $B$ Factory would play an essential complementary role to the LHC and ILC in exploring supersymmetry.

### 5.5.2 Lepton flavor-violating $\tau$ Decays in the Supersymmetric Seesaw Model

>– J. Hisano and Y. Shimizu –<

#### Introduction

The discovery of atmospheric neutrino oscillation by the SuperKamiokande experiment [294] showed that the lepton sector has a much different flavor structure than the quark sector [295][296]. The mixing angles between the first and second and between second and third generations of neutrinos are almost maximal, and these are different from the naive expectation in the grand unified theories. Many attempts to understand those mismatches between quark and lepton mixing angles have been made.

Lepton-flavor violation (LFV) in the charged lepton sector is an important tool to probe the origin of neutrino masses, if the Standard Model is supersymmetric. The finite but small neutrino masses do not predict accessible event rates for the charged LFV processes in experiments in the near future, since these processes are suppressed by the small neutrino masses, in other words, by the scale for the origin of the neutrino masses. However, if the SUSY breaking terms are generated at the higher energy scale than that for the origin of the neutrino masses, imprints may be generated in the SUSY breaking slepton mass terms, and may induce sizable event rates for charged LFV processes, which are not necessarily suppressed by the scale for the origin of the neutrino masses. Charged LFV is a thus window into the origin of neutrino masses and can lead to an understanding within supersymmetry of the observed large neutrino mixing angles.

In this article, we review LFV $\tau$ decays in the minimal SUSY seesaw model. The seesaw model is the most fascinating model to explain the small neutrino masses in a natural way [297]. This model should be supersymmetric, since it introduces a hierarchical structure between the Standard Model and the right-handed neutrino mass scale. Thus, charged LFV processes, including LFV $\tau$ decays, may be experimentally accessible in the near future.

In the next section, we review the relations between neutrino oscillation and charged LFV processes in the minimal SUSY seesaw model. In Section 5.5.2 we discuss the LFV $\tau$ decays in this model. Section 5.5.2 is devoted to discussion.

#### The Minimal SUSY Seesaw Model

We consider the minimal SUSY Standard Model with three additional heavy singlet-neutrino superfields ${N^c}_i$, constituting the minimal SUSY seesaw model. The relevant leptonic part of its superpotential is

$$W = N_i^c (Y_\nu)_{ij} L_j H_2 - E_i^c (Y_e)_{ij} L_j H_1 + \frac{1}{2} {N^c}_i \mathcal{M}_{ij} N_j^c + \mu H_2 H_1 \,, \tag{5.22}$$

where the indexes $i, j$ run over three generations and $(M_N)_{ij}$ is the heavy singlet-neutrino mass matrix. In addition to the three charged-lepton masses, this superpotential has eighteen physical parameters, including six real mixing angles and six $CP$-violating phases. Nine parameters associated with the heavy-neutrino sector cannot be directly measured. The exception is the baryon number in the universe, if the leptogenesis hypothesis is correct [298].

At low energies the effective theory, after integrating out the right-handed neutrinos, is given by the effective superpotential

$$W_{\text{eff}} = E_i^c (Y_e)_i L_j H_1 + \frac{1}{2v^2 \sin^2\beta} (\mathcal{M}_\nu)_{ij} (L_i H_2)(L_j H_2) \,, \tag{5.23}$$

where we work in a basis in which the charged-lepton Yukawa coupling constants are diagonal. The second term in (5.23) leads to light neutrino masses and to mixing. The explicit form of the small neutrino mass matrix $\mathcal{M}_\nu$ is given by

$$(\mathcal{M}_\nu)_{ij} = \sum_k \frac{(Y_\nu)_{ki}(Y_\nu)_{kj}}{M_{N_k}} v^2 \sin^2\beta \,. \tag{5.24}$$

The light neutrino mass matrix $\mathcal{M}_\nu$ (5.24) is symmetric, with nine parameters, including three real mixing angles and three $CP$-violating phases. It can be diagonalized by a unitary matrix $U$ as

$$U^T \mathcal{M}_\nu U = \mathcal{M}_\nu \,. \tag{5.25}$$

By redefinition of fields, one can rewrite $U \equiv VP$, where $P \equiv \mathrm{diag}(e^{i\phi_1}, e^{i\phi_2}, 1)$ and $V$ is the MNS matrix, with the three real mixing angles and the remaining $CP$-violating phase.

If the SUSY breaking parameters are generated above the right-handed neutrino mass scale, the renormalization effects may induce sizable LFV slepton mass terms, which lead to charged LFV processes [299]. If the SUSY breaking parameters at the GUT scale are universal, off-diagonal components in the left-handed slepton mass matrix $m^2_{\tilde{L}}$ and the trilinear slepton coupling $A_e$ take the approximate forms

$$\begin{aligned}(\delta m^2_{\tilde{L}})_{ij} &\simeq -\frac{1}{8\pi^2}(3m_0^2 + A_0^2)H_{ij} \,, \\ (\delta A_e)_{ij} &\simeq -\frac{1}{8\pi^2} A_0 Y_{e_i} H_{ij} \,,\end{aligned} \tag{5.26}$$

where $i \neq j$, and the off-diagonal components of the right-handed slepton mass matrix are suppressed. Here, the Hermitian matrix $H$, whose diagonal terms are real and positive, is defined in terms of $Y_\nu$ and the heavy neutrino masses $M_{N_k}$ by

$$H_{ij} = \sum_k (Y_\nu^\dagger)_{ki}(Y_\nu)_{kj} \log\frac{M_G}{M_{N_k}} , \tag{5.27}$$

with $M_G$ the GUT scale. In Eq. (5.26) the parameters $m_0$ and $A_0$ are the universal scalar mass and trilinear coupling at the GUT scale. We ignore terms of higher order in $Y_e$, assuming that $\tan\beta$ is not extremely large. Thus, the parameters in $H$ may, in principle, be determined by the LFV processes of charged leptons [300].

The Hermitian matrix $H$ has nine parameters, including three phases, which are clearly independent of the parameters in $\mathcal{M}_\nu$. Thus $\mathcal{M}_\nu$ and $H$ together provide the required eighteen parameters, including six $CP$-violating phases, by which we can parameterize the minimal SUSY seesaw model.

Our ability to measure three phases in the Hermitian matrix $H$, in addition to the Majorana phases $e^{i\phi_1}$ and $e^{i\phi_2}$, arelimited at present. Only a phase in $H$ might be determined by $T$-odd asymmetries in $\tau \to \ell\ell\ell$ or $\mu \to eee$ [301], since they are proportional to a Jarlskog invariant obtainable from $H$,

$$J = \mathrm{Im}[H_{12}H_{23}H_{31}] \,. \tag{5.28}$$

However, the asymmetries, arising from interference between phases in $H$ and $\mathcal{M}_\nu$, must be measured in order to determine other two phases in $H$. A possibility to determine them might be the EDMs of the charged leptons [302]. The threshold correction due to non-degeneracy of the right-handed neutrino masses might enhance the imaginary parts of the diagonal component in $A_e$, which contribute to the EDMs of the charged leptons. These depend on all the phases in $\mathcal{M}_\nu$ and $H$. However, detailed studies show that the electron and muon EDMs are smaller than $10^{-27}\,e$ cm and $10^{-29}\,e$ cm, respectively, in the parameter space where the charged LFV processes are suppressed below the experimental bounds [300].

**LFV $\tau$ decays in the SUSY Seesaw Model**

As explained in the previous section, we have shown that charged LFV processes give information about the minimal SUSY seesaw model which is independent of neutrino oscillation experiments. In this section we demonstrate this by considering LFV $\tau$ decays.

In the SUSY models, the LFV processes of the charged leptons are radiative, due to $R$ parity. Thus, the largest LFV decay processes are $\tau \to \mu\gamma$ or $\tau \to e\gamma$, which come from diagrams such as those shown in Fig. 5-50. Other processes are suppressed by $\mathcal{O}(\alpha)$. These are discussed later.

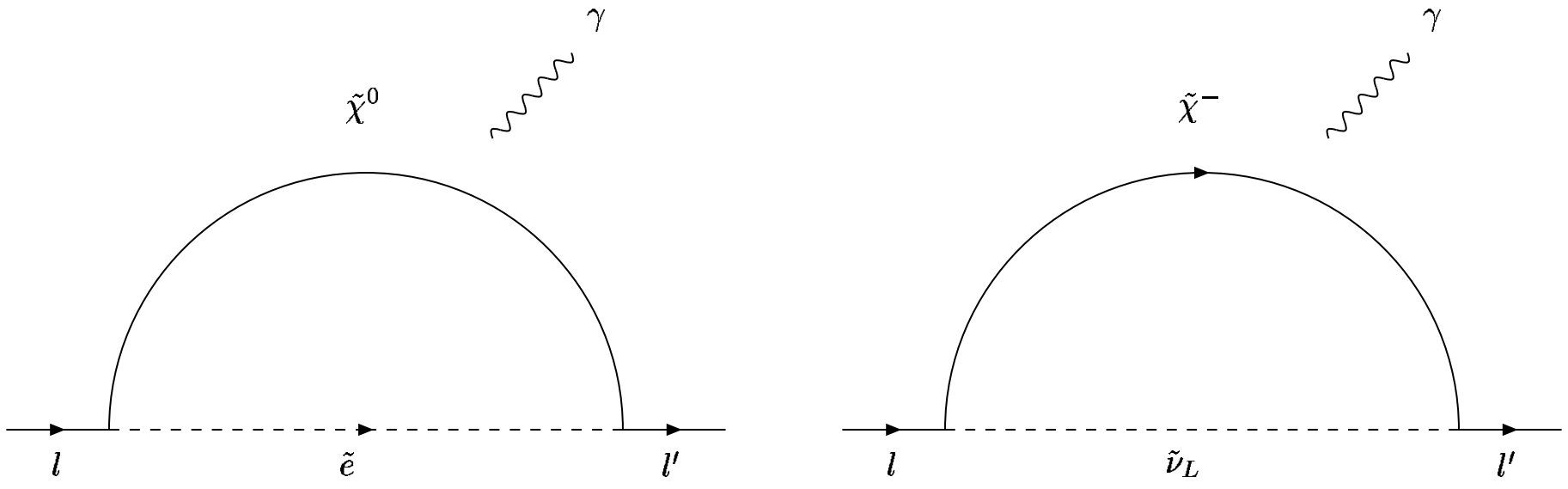

**Figure 5-50.** $l \to l'\gamma$ *processes in SUSY models.*

We will study LFV $\tau$ decays in two different limits of the parameter matrix $H$, of the form

$$H_1 = \begin{pmatrix} a & 0 & 0 \\ 0 & b & d \\ 0 & d^\dagger & c \end{pmatrix}, \tag{5.29}$$

and

$$H_2 = \begin{pmatrix} a & 0 & d \\ 0 & b & 0 \\ d^\dagger & 0 & c \end{pmatrix}, \tag{5.30}$$

where $a, b, c$ are real and positive, and $d$ is a complex number. The non-vanishing $(2,3)$ component in $H_1$ leads to $\tau \to \mu\gamma$ while the $(1,3)$ component in $H_2$ leads to $\tau \to e\gamma$.

In the above *ansatz*, we take $H_{12} = 0$ and $H_{13}H_{32} = 0$ because these conditions suppress $\mathcal{B}(\mu \to e\gamma)$. It is found from the numerical calculation that $\mathcal{B}(\mu \to e\gamma)$ is suppressed in a broad range of parameters with the chosen forms $H_1$ and $H_2$ [300]. From the viewpoint of model-building, the matrix $H_1$ is favored, since it is easier to explain the large mixing angles observed in the neutrino oscillation experiments by the structure of the Yukawa coupling $Y_\nu$. If we adopt $H_2$, we might have to require some conspiracy between $Y_\nu$ and $\mathcal{M}$. However, from the viewpoint of a bottom-up approach, we can always find parameters consistent with the observed neutrino mixing angles for both $H_1$ and $H_2$, as explained in the previous section.

In Fig. 5-51 we show $\mathcal{B}(\tau \to \mu\gamma)$ for the *ansatz* $H_1$ and $\mathcal{B}(\tau \to e\gamma)$ for $H_2$ as functions of the lightest stau mass. We take the SU(2) gaugino mass to be 200 GeV, $A_0 = 0$, $\mu > 0$, and $\tan\beta = 30$ for the SUSY breaking parameters in the SUSY Standard Model. We sample the parameters in $H_1$ or $H_2$ randomly in the range $10^{-2} < a, b, c, |d| < 10$, with distributions that are flat on a logarithmic scale. Also, we require the Yukawa coupling-squared to be smaller than $4\pi$, so that $Y_\nu$ remains perturbative up to $M_G$.

In order to fix $\mathcal{M}_\nu$, we fix the light neutrino parameters: $\Delta m^2_{32} = 3 \times 10^{-3}$ eV$^2$, $\Delta m^2_{21} = 4.5 \times 10^{-5}$ eV$^2$, $\tan^2\theta_{23} = 1$, $\tan^2\theta_{12} = 0.4$, $\sin\theta_{13} = 0.1$ and $\delta = \pi/2$. The Majorana phases $e^{i\phi_1}$ and $e^{i\phi_2}$ are chosen randomly. In Fig. 5-51 we assume the normal hierarchy for the light neutrino mass spectrum; as expected, the branching ratios are insensitive to the structure of the light neutrino mass matrix [300].

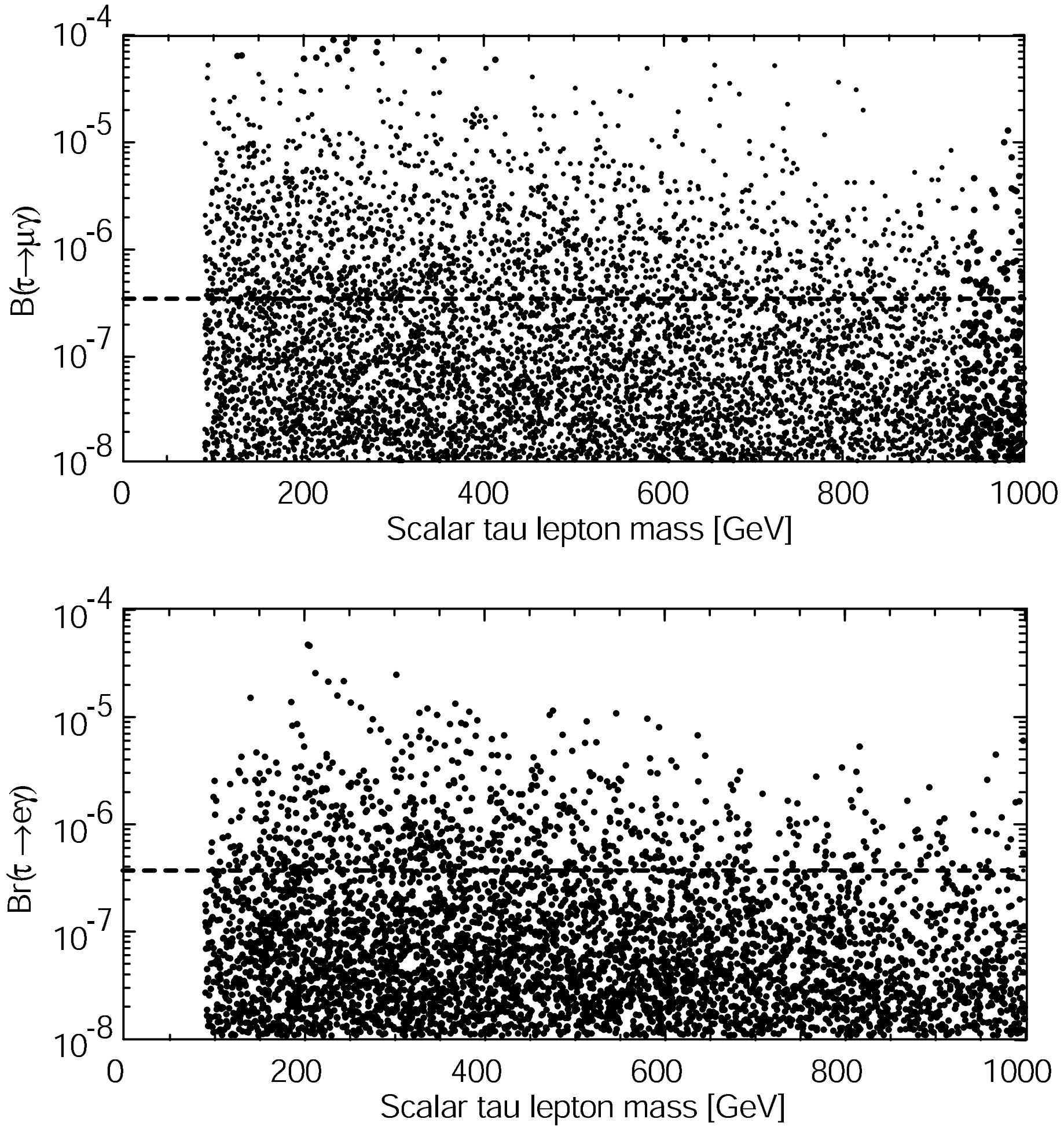

**Figure 5-51.** *$\mathcal{B}(\tau \to \mu\gamma)$ for $H_1$ and $\mathcal{B}(\tau \to e\gamma)$ for $H_2$. The input parameters are given in the text.*

Current experimental bounds for branching ratios of the LFV tau decays are derived by the Belle experiment, and $\mathcal{B}(\tau \to \mu(e)\gamma) < 3.2(3.6) \times 10^{-7}$ [303]. These results already exclude a fraction of the parameter space of the minimal SUSY seesaw model. These bounds can be improved to $10^{-8}$ at Super $B$ Factories[303].[6]

## Discussion

In this article we discussed LFV $\tau$ decays in the minimal SUSY seesaw model. We showed that the these processes provide information about the structure of this model which is independent of the neutrino oscillation experiments. Current experimental searches for $\tau \to \mu\gamma$ and $\tau \to e\gamma$ in the existing $B$ factories already exclude a portion of the parameter space.

Let us now discuss the other LFV $\tau$ decay processes. If the SUSY breaking scale is not extremely large, $\mathcal{B}(\tau \to \mu(e)ll)$ is strongly correlated with $\mathcal{B}(\tau \to \mu(e)\gamma)$, since the on-shell photon penguin diagrams dominate

[6]If sleptons are found in future collider experiments, $\tau$ flavor violation might be found in the signal. The cross sections for $e^+e^-(\mu^+\mu^-) \to \tilde{l}^+\tilde{l}^- \to \tau^\pm\mu^\mp(\tau^\pm e^\mp) + X$ are suppressed by at most the mass difference over the widths for the sleptons. The searches for these processes in the collider experiments have more sensitivity for a small $\tan\beta$ region compared with the search for the LFV tau decays [304].

over these processes. As the result,

$$\mathcal{B}(\tau \to \mu ee(3e))/\mathcal{B}(\tau \to \mu(e)\gamma) \simeq 1/94, \tag{5.31}$$

$$\mathcal{B}(\tau \to 3\mu(e2\mu))/\mathcal{B}(\tau \to \mu(e)\gamma) \simeq 1/440, \tag{5.32}$$

where the difference between the two relations above comes from phase space. If $\tau \to \mu\gamma$ or $\tau \to e\gamma$ are found, we can perform a non-trivial test.

When the SUSY particles are very heavy, $\mathcal{B}(\tau \to \mu\gamma)$ and $\mathcal{B}(\tau \to e\gamma)$ are suppressed by the masses. However, anomalous LFV Higgs boson couplings are then generated, and these can lead to LFV processes in charged lepton decay, such as $\tau \to 3\mu$ and $\tau \to \mu\eta$ [305]. The branching ratios are limited by the muon or strange quark Yukawa coupling constant, They might, however, be observable in future experiments when $\tan\beta$ is very large and the heavier Higgs bosons are relatively light.

### 5.5.3 Higgs-mediated lepton flavor-violating $B$ and $\tau$ decays in the Seesaw MSSM

>– A. Dedes –<

#### Introduction

Possible lepton number violation by two units ($\Delta L = 2$), can arise in the Standard Model or the Minimal Supersymmetric Standard Model (MSSM) from a dimension 5 operator [329, 330]

$$\Delta\mathcal{L} = -\frac{1}{4\Lambda} C^{AB} (\epsilon_{ab} H_a l_b^A) (\epsilon_{cd} H_c l_d^B) + \text{h.c.} , \tag{5.33}$$

where $l^A = (\nu^A, e^A)^T$, $H = (H^+, H^0)^T$ and $A = e, \mu, \tau$ and $a, b, c, d = 1, 2$ denote lepton, Higgs doublets and SU(2) doublet indices respectively. The operator (5.33) is generated at a scale $\Lambda$, and after electroweak symmetry breaking, results in neutrino masses

$$\mathcal{L}^{\nu}_{\text{mass}} = -\frac{1}{2} (\frac{v^2}{4\Lambda}) C^{AB} \nu^A \nu^B + \text{h.c.} , \tag{5.34}$$

of the order $m_\nu \simeq m_t^2/\Lambda$ and for $m_\nu = 10^{-3} - 1$ eV the scale $\Lambda$ should lie in the region $10^{13} - 10^{15}$ GeV. It should be emphasized here that there is no reason for the matrix $C^{AB}$ to be diagonal. In a basis where the charged lepton mass matrix is diagonal, $C^{AB}$ is the infamous MNS matrix with, as atmospheric neutrino oscillation data suggest, a maximal '$23'$' or '$32'$' element. There is a mechanism which explains the existence of the operator (5.33), namely the seesaw mechanism. One can add one or more SU(2)$_L\times$ U(1)$_Y$ singlet leptonic fields $N$ to the Lagrangian [297]

$$\Delta\mathcal{L} = -\epsilon_{ab} H_a N^A {Y_\nu}^{AB} l_b^B - \frac{1}{2} M_{\text{Maj}}^{AB} N^A N^B + \text{h.c.} , \tag{5.35}$$

which via the diagram

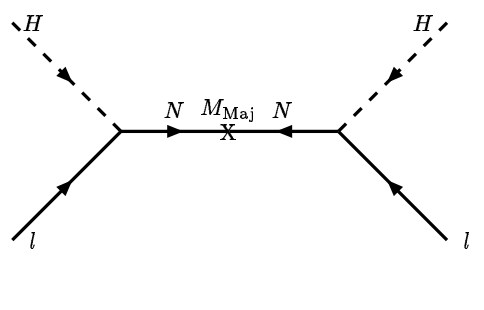

reproduces the operator (5.33) with

$$\Lambda^{-1} C = {Y_\nu}^T (M_{\text{Maj}})^{-1} Y_\nu . \tag{5.36}$$

Thus, the scale $\Lambda$ is identified with the heavy Majorana mass scale $M_{\text{Maj}}$, which decouples at low energies, leaving only the operator (5.33), leading to neutrino masses and possibly a non-trivial mixing matrix.

In the MSSM, the influence of the renormalizable Yukawa coupling $Y_\nu$ in the slepton mass $[m_{\tilde{L}}]$ renormalisation group equations (RGEs) running from the Planck (or GUT) scale down to the scale $M_{\rm Maj}$ induces a mixing among sleptons at low energies

$$(\Delta m^2_{\tilde{L}})_{ij} \simeq -\frac{1}{8\pi^2}(3m_0^2 + A_0^2)(Y_\nu{}^\dagger \ \ln \frac{M_{\rm GUT}}{M_{\rm Maj}} Y_\nu) \ , \tag{5.37}$$

and thus flavor-changing insertions such as,

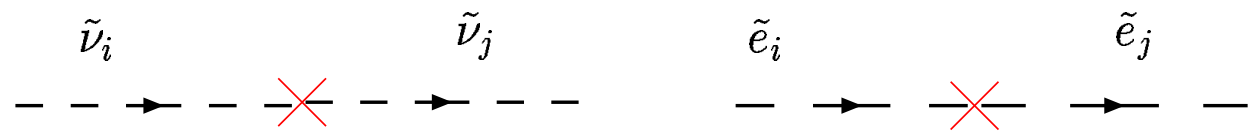

for sneutrinos ($\tilde{\nu}$) and charged sleptons ($\tilde{e}$) are produced. In (5.37), $m_0$ and $A_0$ are common supersymmetry breaking masses for sleptons and trilinear couplings at the Grand Unification scale (GUT), $M_{\rm GUT}$. These insertions enter in loops and result in lepton flavor-violating (LFV) processes such as $\tau \to \mu\gamma$ [see Fig.(5-52)].

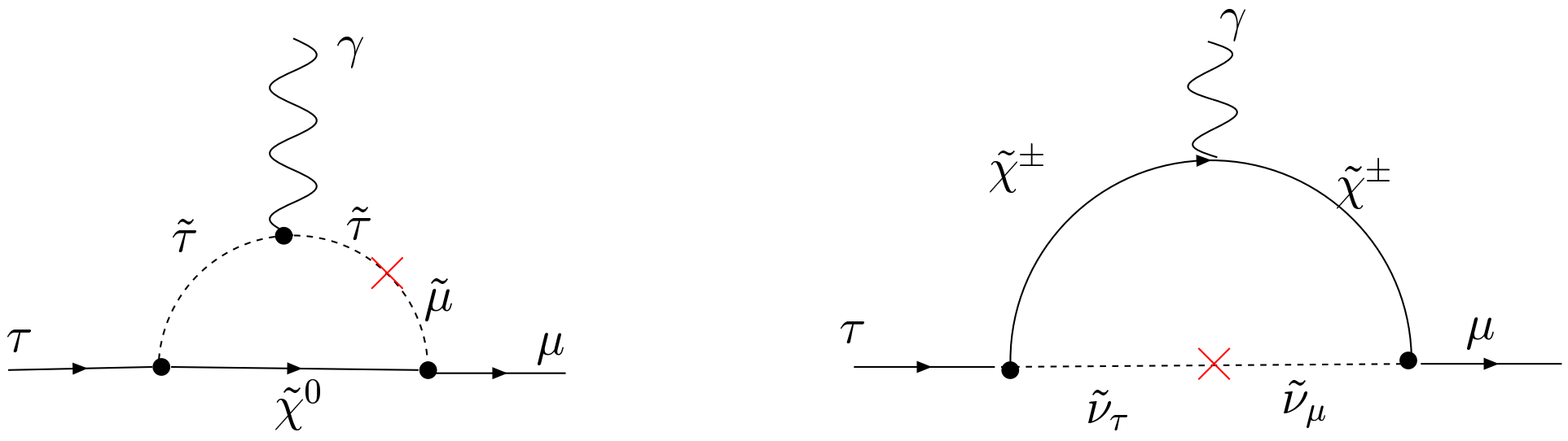

**Figure 5-52.** *Supersymmetric contributions to the LFV process $\tau \to \mu\gamma$.*

Equation (5.37) is the only source of lepton-flavor violation in supersymmetric seesaw models with flavor-universal soft mass terms. Since it is induced by heavy-neutrino Yukawa interactions, it relates the LFV processes to low-energy neutrino data. At the one-loop level, also gives rise to flavor violation in non-holomorphic interactions of the form $\overline{E}LH_u^*$ and leads to Higgs-mediated LFV processes in the charged-lepton sector. For technical details on deriving an effective Lagrangian for a lepton flavor-violating Higgs penguin $H - l' - l$, the reader is referred to Refs. [331, 332, 333]. Schematically, the LFV Higgs penguin is depicted in Fig. (5-53). The lepton flavor-violating Higgs penguin exists even in the Standard Model with the exchange of a $W$-gauge boson and the neutrino [diagram in Fig. (5-53)]. The result is proportional to the neutrino mass squared, and thus negligible. The situation changes when supersymmetrizing the $W - \nu$ loop in Fig. (5-53). Then the $W$-boson becomes a chargino and the neutrino becomes a sneutrino. If we calculate the chargino-sneutrino loop, we find that it is proportional to $\tan\beta$, denoted with the green dot in Fig. (5-53), and is proportional to the $\tau$-lepton mass. In addition, the $\tau - \tau - H$ vertex is also enhanced by $\tan\beta$. Furthermore, the single neutral Higgs boson in the Standard Model becomes three neutral Higgs-bosons in the MSSM, two $CP$-even $(h, H)$ and a $CP$-odd $(A)$. There are, of course, other Standard Model and MSSM contributions to the penguin $\tau - \mu - H$; for simplicity these are not shown in Fig. (5-53). The last step in the derivation of the $\tau - \mu - H$ effective Lagrangian is to integrate out all the heavy particles (charginos and sneutrinos, in our case), and include the possibility of Higgs mixing (see Ref. [334] for more details), where the Higgs bosons become three $CP$-indistinguishable particles, $H_1$, $H_2$ and $H_3$, with the heaviest being the $H_3$. Finally, the LFV Higgs penguin and the total amplitude are enhanced by two powers of $\tan\beta$. From now on we can use the "yellow" blob (Higgs penguin) of Fig.(5-53) in the amplitude calculations for physical processes. For a complete list of applications of the Higgs penguin the reader is referred to Ref. [335]. We shall present few of them below.

### Applications : $\tau \to \mu\mu\mu$, $B^0_{s,d} \to \mu\tau$, $\tau \to \mu\eta$

We shall focus on the lepton flavor-violating processes of Fig. (5-54). To quantify the above statements, we adopt, for the moment, the approximation in which we take all the supersymmetry-breaking mass parameters of the model to be equal at low scales, use heavy-neutrino masses that are degenerate with $M_N = 10^{14}$ GeV, and assume that $(Y_\nu^\dagger Y_\nu)_{32,33} = 1$. These are inputs for Eq. (5.37). This approximation is not realistic, but it is useful for comparing the sensitivities of different processes to New Physics: we shall present in the next section results from a more complete treatment $\tau \to 3\mu$ or related processes. In this simplified case, we obtain [332] for (5-54a)

$$Br(\tau \to 3\mu) \simeq 1.6 \times 10^{-8} \left[\frac{\tan\beta}{60}\right]^6 \left[\frac{100\,\mathrm{GeV}}{M_A}\right]^4 . \tag{5.38}$$

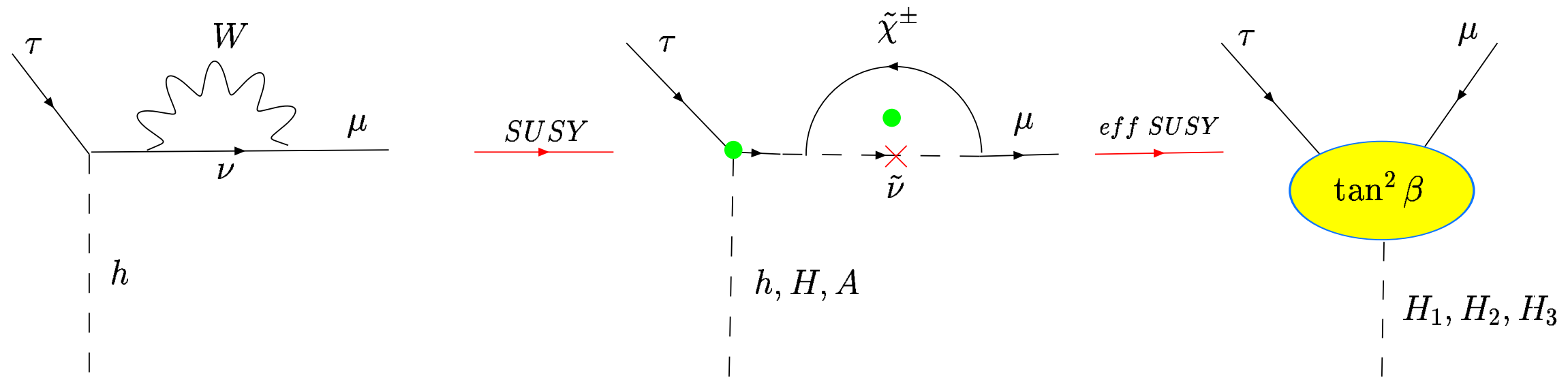

**Figure 5-53.** *The LFV Higgs penguin.*

This should be compared with the corresponding estimate

$$Br(\tau \to \mu\gamma) \simeq 1.3 \times 10^{-3} \left[\frac{\tan\beta}{60}\right]^2 \left[\frac{100\,\mathrm{GeV}}{M_S}\right]^4 . \tag{5.39}$$

Both equations (5.38) and (5.39) are valid in the large $\tan\beta$ limit. Whereas (5.38) is two orders of magnitude *below* the present experimental bound on $\tau \to 3\mu$, (5.39) is three orders of magnitude *above* the present bound on $\tau \to \mu\gamma$. There is also the photonic penguin contribution to the decay $\tau \to 3\mu$, which is related to $\mathcal{B}(\tau \to \mu\gamma)$ by

$$\mathcal{B}(\tau \to 3\mu)_\gamma = \frac{\alpha}{3\pi}\left(\ln\frac{m_\tau^2}{m_\mu^2} - \frac{11}{4}\right)\mathcal{B}(\tau \to \mu\gamma). \tag{5.40}$$

Numerically, (5.39) leads to

$$\mathcal{B}(\tau \to 3\mu)_\gamma \simeq 3.0 \times 10^{-6} \left[\frac{\tan\beta}{60}\right]^2 \left[\frac{100\,\mathrm{GeV}}{M_S}\right]^4 , \tag{5.41}$$

which is a factor of 100 larger than (5.38). Notice also that suppressing (5.39) by postulating large slepton masses would suppress (5.38) at the same time, since sleptons enter into both loops. However, suppressing (5.38) by a large Higgs mass would not affect $\tau \to \mu\gamma$. It has recently been shown [336] that with the same assumptions, the branching ratio of the process $\tau \to \mu\eta$ [see Fig. (5-54b)], is related to that for $\tau \to 3\mu$ of Eq. (5.38)

$$\mathcal{B}(\tau \to \mu\eta)^{\mathrm{Higgs}} = 8.4 \times \mathcal{B}(\tau \to 3\mu)^{\mathrm{Higgs}} , \tag{5.42}$$

and for the double Higgs penguin diagram of Fig. (5-54c) [332]

$$\mathcal{B}(B_s^0 \to \tau\mu) \simeq 3.6 \times 10^{-7} \left[\frac{\tan\beta}{60}\right]^8 \left[\frac{100\,\mathrm{GeV}}{M_A}\right]^4 , \tag{5.43}$$

where only the leading $\tan\beta$ dependence is presented. These have to be compared with the upper limits in Table 5-22 below. In the case of $B_d$ mesons, one should just multiply (5.43) with $|V_{td}/V_{ts}|^2 \simeq 0.05$. As expected, the Higgs-mediated branching ratio for $B_s^0 \to \tau\mu$ and $\tau \to \mu\eta$ can be larger than the one for $\tau \to 3\mu$.

Recently [337] the effect of the Higgs-exchange diagram for the lepton flavor-violating muon-electron conversion process, $\mu N \to eN$, in the supersymmetric seesaw model has been studied. The ratio of $\mathcal{B}(\mu N \to eN)/\mathcal{B}(\mu \to e\gamma)$ is enhanced at large $\tan\beta$ and for a relatively light Higgs sector. For reviews on $\tau$-lepton and $B$-meson lepton flavor-violating decays, the reader should consult Refs. [338, 339].

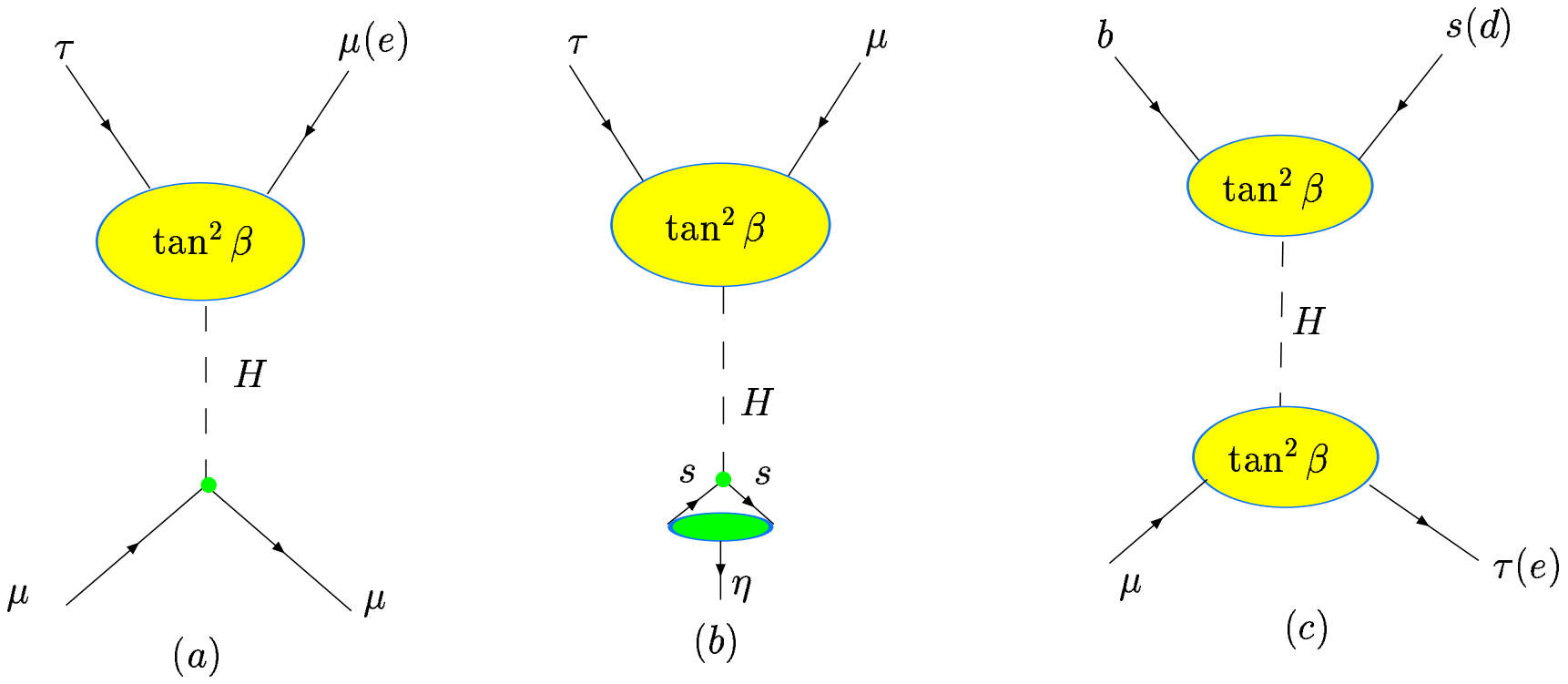

**Figure 5-54.** *(a) $\tau \to \mu(e)\mu\mu$ (b) $\tau \to \mu\eta$ (c) $B_{s(d)} \to \mu\tau$ Higgs-mediated processes in the MSSM. The green dot indicates an additional enhancement of $\tan\beta$.*

### Results

The purpose of this work is to study in a complete way the allowed rates for the Higgs-mediated LFV processes in supersymmetric seesaw models in which the only source of LFV is the renormalization of the soft supersymmetry-breaking mass parameters [see Eq. (5.37)] above $M_{N_i}$, due to the singlet-neutrino Yukawa couplings of Eq. (5.35). We follow the analysis of Ref. [332] where the more general flavor-universal MSSM case was considered in which the universal masses for squarks, sleptons and the Higgs doublets $H_d$ and $H_u$ are different from each other. This permits different mass scales for squarks and sleptons which, in turn, are independent of the Higgs boson masses. However, we always require that squark and slepton mass matrices at the GUT scale are each proportional to unit matrices. If one goes beyond this assumption, arbitrary sources of flavor violation appear in the soft supersymmetry-breaking sector, and the model loses all predictive power, in particular the connection between the LFV and the neutrino masses and mixing. We parametrise the singlet-neutrino Yukawa couplings $Y_\nu$ and masses $M_{N_i}$ in terms of low energy neutrino data according to Ref. [340]. We generate all the free parameters of the model randomly and calculate the low-energy sparticle masses and mixing by numerically solving the one-loop renormalization-group equations and imposing the requirement of electroweak symmetry breaking. For the decays $l_i \to l_j\gamma$, we use the exact diagrammatic formulae in [341]. The experimental CLEO and CDF bounds on $\tau \to \mu\gamma$ $B_s \to \mu\mu$ respectively, are reached first, and place upper allowed limits on $B_{s,d} \to \ell\ell'$, $\tau \to \ell\ell\ell'$ and $\tau \to \mu\eta$, that are summarized in Table 5-22.[7]

In summary, Higgs boson mediated FCNC (LFV) processes are interesting in the seesaw-MSSM. They are enhanced by powers of $\tan\beta$ and maybe orders of magnitude larger than the Standard Model predictions. By looking at Table 5-22 and comparing with the sensitivity of a Super $B$ Factory [342] we conclude that

- $\tau \to \mu\gamma$ is (at the moment) the only LFV decay which can saturate the experimental bound. Searching for this mode at a Super $B$ Factory is compulsory.
- LFV modes like $B \to \ell'\ell, \tau \to \ell'\ell\ell$, $\tau \to \mu\eta$, (relevant to $B$ Factories) turn out to be small, with $\mathcal{B}$'s at $10^{-9} - 10^{-10}$, if current experimental constraints from $B_s \to \mu\mu$ and $\tau \to \mu\gamma$ are imposed.
- Searching for LFV modes like $B \to \ell'\ell, \tau \to \ell'\ell\ell$, $\tau \to \mu\eta$, could distinguish among various models for neutrino masses, for example: the$R$ parity-violating MSSM [343] and the seesaw MSSM.

[7] After this talk was presented, newly improved bounds on $\tau \to \mu\gamma$ and $\tau \to 3\mu$ appeared [342]. These bounds set further constraints on the other processes of Table 5-22.

**Table 5-22.** *Experimental bounds and maximal value predictions for B- and $\tau$ decays discussed in this section. Results on $\mathcal{B} \to \ell^{+'}\ell^-$, $\tau^- \to \ell^-\ell^+\ell^-$ are based on the analysis of Ref. [332] described in the text. For $\tau \to \mu\eta$, we use the relation (5.42) from Ref. [339].*

| $\mathcal{B}$(Channel) | Expt. | Bound (90% CL) | Higgs med. MSSM |
|---|---|---|---|
| $B_s \to e^+\mu^-$ | CDF | $< 6.1 \times 10^{-6}$ | $\lesssim 10^{-11}$ |
| $B_s \to e^+\tau^-$ | – | $--$ | $\lesssim 4 \times 10^{-9}$ |
| $B_s \to \mu^+\tau^-$ | – | $--$ | $\lesssim 4 \times 10^{-9}$ |
| $B_d \to e^+\mu^-$ | BABAR | $< 2.0 \times 10^{-7}$ | $\lesssim 6 \times 10^{-13}$ |
| $B_d \to e^+\tau^-$ | CLEO | $< 5.3 \times 10^{-4}$ | $\lesssim 2 \times 10^{-10}$ |
| $B_d \to \mu^+\tau^-$ | CLEO | $< 8.3 \times 10^{-4}$ | $\lesssim 2 \times 10^{-10}$ |
| $\tau^- \to \mu^-\mu^+\mu^-$ | Belle | $< 3.8 \times 10^{-7}$ | $\lesssim 4 \times 10^{-10}$ |
| $\tau^- \to e^-\mu^+\mu^-$ | Belle | $< 3.1 \times 10^{-7}$ | $\lesssim 4 \times 10^{-10}$ |
| $\tau^- \to \mu^-\eta$ | CLEO | $< 9.6 \times 10^{-6}$ | $\lesssim 3 \times 10^{-9}$ |
| $\tau^- \to \mu^-\gamma$ | CLEO | $< 1.1 \times 10^{-6}$ | $< 1.1 \times 10^{-6}$ |

## 5.6 Conclusions and Patterns of New Physics Contributions

As discussed in the overview of this chapter, there are important questions that remain unanswered within the Standard Model. Various theoretical ideas have been proposed to answer these questions, and experimental efforts have been undertaken and will continue in the future to search for New Physics. The goal of the Super $B$ Factory is therefore to look for New Physics effects in the Heavy Flavor sector and furthermore, to provide critical information from the measurement of new sources of $CP$ violation and flavor mixing which will distinguish between various theoretical models.

In this chapter, we have first discussed model-independent approaches for analyzing New Physics contributions in $B$ decays. In this case, a set of new parameters is introduced at the level of the effective Lagrangian or decay/mixing amplitude, and various observable quantities are used to constrain these parameters. Model-independent approaches have been very successful in the past, such as the determination of the Michel parameters in muon decays and the parametrization of oblique corrections in electroweak precision measurements. In $B$ decays, depending on the processes considered, parameters representing New Physics contributions to the $B_d - \overline{B_d}$ amplitude and hadronic and electroweak $b \to s$ transitions can be introduced. In the decay $B \to V_1V_2$, where $V_{1,2}$ represent two vector particles, various angular distributions can be used to separate and measure different amplitudes, including those with $CP$ phases. These model-independent techniques are particularly useful for a global analysis with several observables, or in comparing experimental data with the predictions of several theoretical models.

In this chapter, $B$ physics signals in specific theoretical models are also examined. The focus is mainly on models with supersymmetry and large extra dimensions. These are leading candidates of physics beyond the Standard Model from the viewpoint of the gauge hierarchy problem. At the same time, these models have new sources of $CP$ violation and flavor mixing structure.

- **Supersymmetry**
  Squark mass matrices contain new sources of flavor mixing and $CP$ violation. These matrices depend on the supersymmetry breaking mechanism and flavor-dependent interactions at the high energy (GUT) scale and through renormalization the supersymmetric contributions to various $B$ observables vary for different supersymmetry breaking scenarios. The deviations from the Standard model predictions are small in the minimal supergravity model, except for the case of $B \to D\tau\nu$, $B \to s\ell^+\ell^-$ as well as $B_s \to \mu\mu$ with possible Higgs exchange with a large ratio of the two Higgs-doublet vacuum expectation values ($\tan\beta$). The flavor mixing in the neutrino sector can be an additional source of the squark flavor mixing in the context of SUSY GUT models. In this case, large deviations from the Standard Model prediction are possible in the time-dependent $CP$ asymmetries in the $B \to \phi K_S^0$ and $B \to K^*\gamma$ modes. This model also predicts possible observable effects in other processes such as $B_s$ mixing, lepton flavor violation in $\tau \to \mu\gamma$, and hadronic electric dipole moments. In the scenario of effective supersymmetry where the squarks of the first two generations are very heavy, various signals with a particular pattern are expected in rare decay processes.

- **Large extra dimensions**
  Although the introduction of large extra dimensions is not directly related to flavor physics, new flavor signals are expected, particularly in the case where the observed pattern of the fermion mass spectrum is explained from some geometrical setting. The impact on $B$ physics therefore depends on the details of the fermion mass generation mechanism. A generic signal of these theories is, however, the existence of Kaluza-Klein gravitons, and the exchange of these modes generates a set of higher dimensional operators. Such operators can produce characteristic signals in the angular distribution of the decay $b \to s\ell^+\ell^-$. The geometric construction of the fermion mass hierarchy could generate tree level flavor-changing couplings for Kaluza-Klein gauge bosons both in flat TeV$^{-1}$ scale extra dimensions with split fermions, and in warped extra dimensions. These effects can induce large deviations from the Standard Model prediction for meson mixing as well as flavor transition diagrams. On the other hand, such deviations are not very large in the case of universal extra dimensions, where the source of flavor mixing resides solely in the CKM matrix.

Patterns of deviations from the Standard Model predictions in the various models studied in this chapter are summarized in Table 5-23. We can see that New Physics effects can appear in different processes depending on different assumptions for the origin of flavor structure in these models. It is therefore important to clarify these patterns to distinguish various models. Although comparison of various processes in $B_d$, $B_s$, $K$, $D$ and lepton flavor violation are important, it is remarkable that the Super-$B$ factory itself can provide several ways to look for New Physics contributions.

This example shows the complementary nature of flavor physics and energy frontier physics. At the LHC and ILC, direct searches for SUSY particles or Kaluza-Klein modes is essential to establish the existence of New Physics. On the other hand, there are a variety of possibilities for the origin of flavor structure within supersymmetry or models with extra dimensions. Flavor physics provides an important tool with which fundamental questions, such as how supersymmetry is broken, or how fermions propagate in extra dimensions, can be addressed, and the Super $B$ Factory will play a central role in answering these questions.

**Table 5-23.** *Pattern of deviations from the Standard Model predictions in various models of supersymmetry and extra dimensions. Processes with possible large deviations are indicated. "-" means that the deviation is not expected to be large enough for observation, or not yet studied completely.*

| Model | $B_d$ Unitarity | Time-dep. $CPV$ | Rare $B$ decay | Other signals |
|---|---|---|---|---|
| mSUGRA(moderate $\tan\beta$) | - | - | - | - |
| mSUGRA(large $\tan\beta$) | $B_d$ mixing | - | $B \to (D)\tau\nu$<br>$b \to s\ell^+\ell^-$ | $B_s \to \mu\mu$<br>$B_s$ mixing |
| SUSY GUT with $\nu_R$ | - | $B \to \phi K_S$<br>$B \to K^*\gamma$ | - | $B_s$ mixing<br>$\tau$ LFV, $n$ EDM |
| Effective SUSY | $B_d$ mixing | $B \to \phi K_S$ | $A_{CP}^{b\to s\gamma}$, $b \to s\ell^+\ell^-$ | $B_s$ mixing |
| KK graviton exchange | - | - | $b \to s\ell^+\ell^-$ | - |
| Split fermions<br>in large extra dimensions | $B_d$ mixing | - | $b \to s\ell^+\ell^-$ | $K^0\overline{K}^0$ mixing<br>$D^0\overline{D}^0$ mixing |
| Bulk fermions<br>in warped extra dimensions | $B_d$ mixing | $B \to \phi K_S$ | $b \to s\ell^+\ell^-$ | $B_s$ mixing<br>$D^0\overline{D}^0$ mixing |
| Universal extra dimensioins | - | - | $b \to s\ell^+\ell^-$<br>$b \to s\gamma$ | $K \to \pi\nu\overline{\nu}$ |

# 6

# Workshop Participants

Kaustubh Agashe
Johns Hopkins University

Justin Albert
Caltech

Ahmed Ali
DESY

David Aston
SLAC

David Atwood
University of Iowa

Henry Band
University of Wisconsin, Madison

Debbie Bard
University of Edinburgh

Christian Bauer
University of California, San Diego

Johannes Bauer
University of Mississippi

Prafulla Behera
University of Pennsylvania

Claude Bernard
Washington University, St. Louis

Jeffrey Berryhill
University of California, Santa Barbara

Adrian Bevan
University of Liverpool

Maria Biagini
INFN Laboratori Nazionali Frascati

Ikaros Bigi
University of Notre Dame du Lac

Alexander Bondar
Budker Institute of Nuclear Physics

Gerard Bonneaud
CNRS

Anders Borgland
Lawrence Berkeley National Laboratory

Sylvie Brunet
Université de Montréal

Alexander Bukin
Budker Institute of Nuclear Physics

Gustavo Burdman
Lawrence Berkeley National Laboratory

Robert Cahn
Lawrence Berkeley National Laboratory

Alessandro Calcaterra
INFN Laboratori Nazionali Frascati

Giovanni Calderini
INFN Sezione di Pisa

Massimo Carpinelli
Universita’ di Pisa and INFN

Shenjian Chen
University of Colorado

Baosen Cheng
University of Wisconsin, Madison

Farrukh Chishtie
Cornell University

Stefan Christ
University of Rostock

Jonathon Coleman
University of Liverpool

Mark Convery
SLAC

David Côté
Université de Montréal

Markus Cristinziani
SLAC

Sridhara Dasu
University of Wisconsin, Madison

Mousumi Datta
University of Wisconsin, Madison

Alakabha Datta
University of Toronto

Riccardo De Sangro
INFN Laboratori Nazionali Frascati

Athanasios Dedes
Technical University, Munich

Daniele del Re
University of California, San Diego

Durmush Demir
University of Minnesota

Francesca Di Lodovico
University of Wisconsin

Lance Dixon
SLAC

Danning Dong
SLAC

Gregory Dubois-Felsmann
Caltech

Denis Dujmic
SLAC

Eric Eckhart
Colorado State University

Ulrik Egede
Imperial College, London

Andrew Eichenbaum
University of Wisconsin, Madison

Gerald Eigen
University of Bergen

Eckhard Elsen
SLAC

Ivo Eschrich
Imperial College, London

Adam Falk
Johns Hopkins University

Thorsten Feldmann
CERN, Theory Division

Fabio Ferrarotto
INFN Sezione di Roma

Fernando Ferroni
Universita' di Roma La Sapienza

Clive Field
SLAC

Robert Flack
Royal Holloway, Bedford College

Francesco Forti
INFN Sezione di Pisa

Mandeep Gill
Lawrence Berkeley National Laboratory

Marcello Giorgi
Universita' di Pisa

Romulus Godang
University of Mississippi

Maarten Golterman
San Francisco State University

Stephen Gowdy
SLAC

Giacomo Graziani
CEA Saclay, DAPNIA

Benjamin Grinstein
University of California, San Diego

Andrei Gritsan
Lawrence Berkeley National Laboratory

Junji Haba
KEK

Thomas Hadig
SLAC

Paul Harrison
Queen Mary, University of London

Shoji Hashimoto
KEK

Masashi Hazumi
KEK

JoAnne Hewett
SLAC

Richard Hill
SLAC

Gudrun Hiller
University of Munich

Junji Hisano
ICRR, University of Tokyo

David Hitlin
Caltech

Mark Hodgkinson
University of Manchester

George Hou
National Taiwan University

Tobias Hurth
CERN, Theory Division

Olya Igonkina
University of Oregon

Gino Isidori
INFN Laboratori Nazionali Frascati

Yoshihito Iwasaki
KEK

Joseph Izen
University of Texas at Dallas

Alex Kagan
University of Cincinnati

Nobu Katayama
KEK

Peter Kim
SLAC

Pyungwon Ko
KAIST

Christopher Kolda
University of Notre Dame du Lac

Patrick Koppenburg
KEK

Witold Kozanecki
CEA Saclay and SLAC

Frank Krüger
Technical University, Munich

Chung Khim Lae
University of Maryland

Livio Lanceri
Universita’ and INFN, Trieste

Bjorn Lange
Cornell University

Urs Langenegger
SLAC

Ben Lau
Princeton University

David Leith
SLAC

Adam Lewandowski
SLAC

Jim Libby
SLAC

Zoltan Ligeti
Lawrence Berkeley National Laboratory

Ben Lillie
SLAC

David London
Université de Montréal

John LoSecco
University of Notre Dame du Lac

Xinchou Lou
University of Texas at Dallas

Steffen Luitz
SLAC

Enrico Lunghi
University of Zurich

Vera Lüth
SLAC

Harvey Lynch
SLAC

David MacFarlane
University of California, San Diego

Giampiero Mancinelli
University of Cincinnati

Dominique Mangeol
McGill University

Aneesh Manohar
University of California, San Diego

Helmut Marsiske
SLAC

Michael Mazur
University of California, Santa Barbara

Robert Messner
SLAC

Gagan Mohanty
Queen Mary, University of London

Ajit Mohapatra
University of Wisconsin, Madison

Remi Mommsen
University of California, Irvine

Thomas Moore
University of Massachusetts

Ilya Narsky
Caltech

Nicola Neri
Universita' di Pisa and INFN

Sechul Oh
KEK

Yukiyoshi Ohnishi
KEK

Yasuhiro Okada
KEK

Alexei Onuchin
Budker Institute of Nuclear Physics

Stefano Passaggio
INFN Sezione di Genova

Popat M. Patel
McGill University

Nello Paver
Dept. of Theoretical Physics, Trieste and INFN

David Payne
University of Liverpool

Gilad Perez
Lawrence Berkeley National Laboratory

Alexey Petrov
Wayne State University

Aaron Pierce
SLAC

Maurizio Pierini
Universita' di Roma, La Sapienza

Giancarlo Piredda
INFN Roma

Steve Playfer
University of Edinburgh

Frank Porter
Caltech

Chris Potter
University of Oregon

Teela Pulliam
Ohio State University

Milind Purohit
University of South Carolina

Helen Quinn
SLAC

Matteo Rama
Universita' di Pisa

Blair Ratcliff
SLAC

Thomas Rizzo
SLAC

Douglas Roberts
University of Maryland

Steven Robertson
SLAC

Steven Robertson
McGill University

Enrico Robutti
INFN Sezione di Genova

Aaron Roodman
SLAC

Ira Rothstein
Carnegie Mellon University

Anders Ryd
Caltech

Alam Saeed
SUNY, Albany

Muhammad Saleem
SUNY, Albany

Alex Samuel
Caltech

Asish Satpathy
University of Texas at Austin

Rafe Schindler
SLAC

Alan Schwartz
University of Cincinnati

John Seeman
SLAC

Vasia Shelkov
Lawrence Berkeley National Laboratory

Yasuhiro Shimizu
Nagoya University

Maria Chiara Simani
Lawrence Livermore National Laboratory

Nita Sinha
Institute of Mathematical Sciences, Chennai

Rahul Sinha
Institute of Mathematical Sciences, Chennai

Jim Smith
University of Colorado

Arthur Snyder
SLAC

Abner Soffer
Colorado State University

Michael Sokoloff
University of Cincinnati

Evgeni Solodov
Budker Institute of Nuclear Physics

Amarjit Soni
Brookhaven National Laboratory

Iain Stewart
MIT

Dong Su
SLAC

Hiroyasu Tajima
SLAC

Makoto Tobiyama
SLAC

Walter Toki
Colorado State University

Silvano Tosi
INFN Sezione di Genova

Christos Touramanis
University of Liverpool

Georges Vasseur
CEA Saclay, DAPNIA

Jaroslav Va'vra
SLAC

Oscar Vives
University of Valencia

Steven Wagner
SLAC

Steven Weber
SLAC

Achim Weidemann
University of South Carolina

Andreas Weiler
Technical University, Munich

James Wells
University of Michigan

Guy Wilkinson
University of Oxford

David C. Williams
University of California, Santa Cruz

Stephane Willocq
University of Massachusetts

Jinwei Wu
University of Wisconsin, Madison

Daniel Wyler
University of Zurich

Masa Yamauchi
KEK

Aaron Yarritu
SLAC

Jong Yi
Iowa State University

Charles Young
SLAC

Francisco Yumiceva
University of South Carolina

Qinglin Zeng
Colorado State University

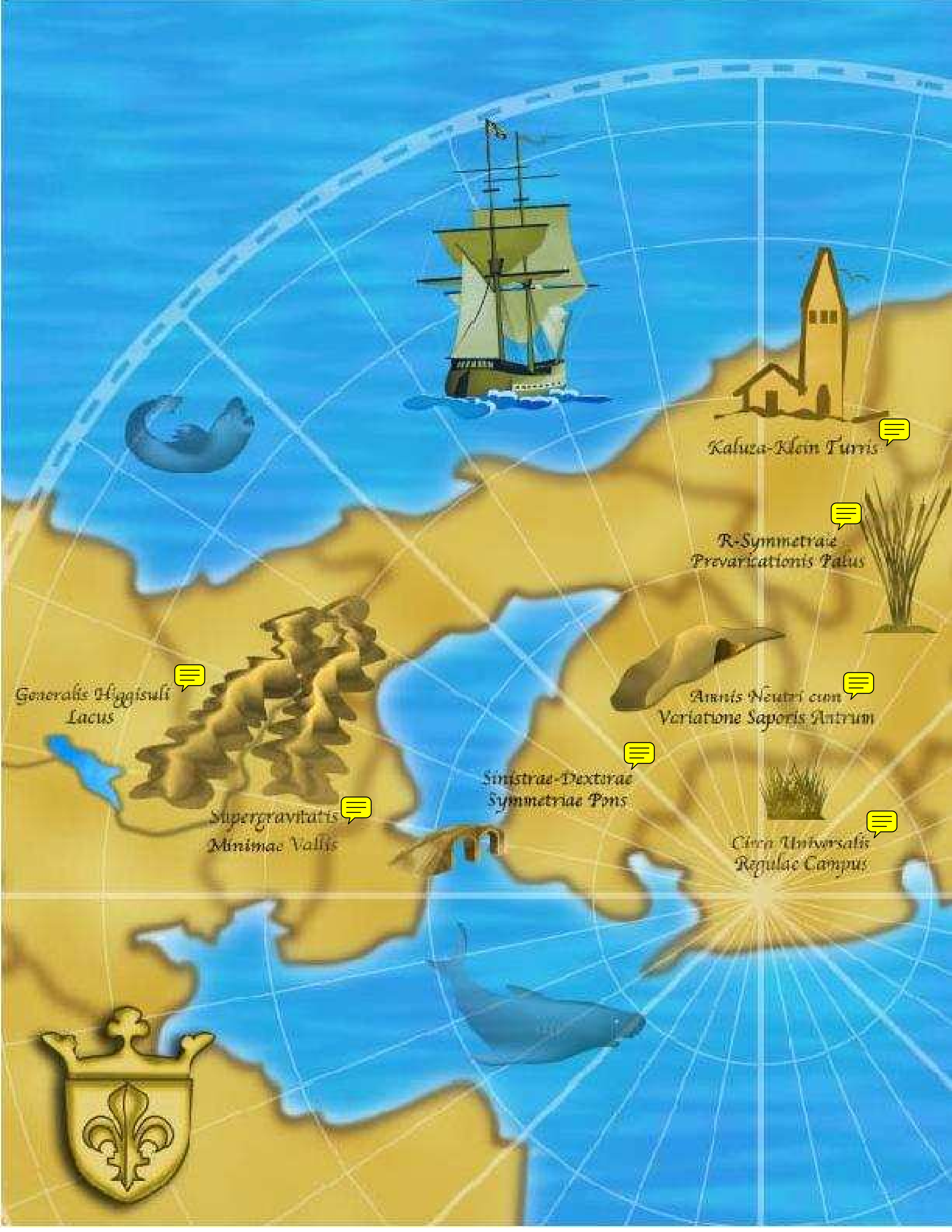
Kaluza-Klein Turris
R-Symmetriae Prevaricationis Palus
Amnis Neutri cum Variatione Saporis Antrum
Generalis Higgisuli Lacus
Sinistrae-Dextrae Symmetriae Pons
Supergravitatis Minimae Vallis
Circa Universalis Regulae Campus